\pdfoutput=1
\documentclass[12pt,twoside]{article}
\usepackage[utf8]{inputenc}
\usepackage[T1]{fontenc}
\usepackage{amsmath}
\usepackage[nottoc,notlot,notlof]{tocbibind}
\usepackage{calc}
\usepackage[pdftex]{graphicx} 
\usepackage[figuresright]{rotating}
\usepackage{authblk}
\usepackage{caption}
\usepackage{array}
\usepackage{longtable}
\usepackage{float}

\usepackage{adjustbox}
\usepackage{tabularx}
\usepackage{ltablex}
\usepackage{multirow}
\usepackage{booktabs}
\usepackage{color, colortbl}
\usepackage{footmisc}
\usepackage{afterpage}
\usepackage{relsize}                                      

\definecolor{Gray}{gray}{0.9}
\definecolor{LightGray}{gray}{0.97}

\usepackage{hyperref} 
\usepackage[all]{hypcap} 
\usepackage{cite}    
\usepackage{mciteplus}

\newcommand{\mysection}[1]{\section{\boldmath #1}}
\newcommand{\mysubsection}[1]{\subsection[#1]{\boldmath #1}}
\newcommand{\mysubsubsection}[1]{\subsubsection[#1]{\boldmath #1}}
\newcommand{\mysubsubsubsection}[1]{\subsubsubsection{\boldmath #1}}

\newcommand{\lesssim}{\ensuremath{\raise-.5ex\hbox{$\buildrel<\over\sim$}\,}} 

\def\dof{{\rm dof}}

\newcommand\VCKM{{V}}
\newcommand\etacpf{{\eta_f}}
\newcommand\etacp{{\eta}}

\renewcommand\Im{{\rm Im}} 
\renewcommand\Re{{\rm Re}}

\newcommand\Abar{\kern 0.18em\overline{\kern -0.18em A}{}}
\newcommand\Af{A_f}
\newcommand\Abarf{\Abar_f}
\newcommand\Afbar{A_{\bar f}}
\newcommand\Abarfbar{\Abar_{\bar f}}
\newcommand\Acp{{\cal A}}
\newcommand\Adirnoncp{\ensuremath{\langle{\cal A}_{f\bar f}\rangle}\xspace}
\newcommand\mc{\multicolumn}

\newcommand {\cbf}{\ensuremath{{\cal B}}}

\newcommand {\vcb}{\ensuremath{|V_{cb}|}}
\newcommand {\vub}{\ensuremath{|V_{ub}|}}

\def\Bp      {\ensuremath{B^{+}}}
\def\Bm      {\ensuremath{B^{-}}}
\def\Bz      {\ensuremath{B^{0}}}
\def\Bs      {\ensuremath{B_{s}}}

\newcommand{\BzbDplnu}    {\ensuremath{\Bzb \to D^{+}\ell^{-}\nub}}
\newcommand{\BzbDstarlnu} {\ensuremath{\Bzb \to D^{*+}\ell^{-}\nub}}

\newcommand {\rhoz} {\ensuremath{\rho^0}\hbox{ }}

\usepackage{relsize}

\def\beq{\begin{equation}}
\def\eeq#1{\label{#1}\end{equation}}
\def\eeqn{\end{equation}}

\def\beqa{\begin{eqnarray}}
\def\eeqa#1{\label{#1}\end{eqnarray}}
\def\eeqan{\end{eqnarray}}

\let\bar=\overbar

\def\etal{{\it et al.}}
\def\ie{{\it i.e.}}
\def\eg{{\it e.g.}}
\def\etc{{\it etc.}}
\def\cf{{\it cf.}}

\def\Dslash{\ensuremath{\not{\hbox{\kern-4pt $D$}}}\xspace}
\def\dslash{\not{\hbox{\kern-2pt $\del$}}}

\def\BR{\mbox{\rm BR}}
\def\ee{e^+e^-}

\def\alphas{\alpha_s}
\def\msb{{\bar{\ssstyle M \kern -1pt S}}}

\RequirePackage{xspace}

\usepackage{relsize}
\def\babar{\mbox{\slshape B\kern-0.1em{\smaller A}\kern-0.1em
    B\kern-0.1em{\smaller A\kern-0.2em R}}\xspace}
\def\belle{\mbox{\normalfont Belle}\xspace}

\def\dzero{\mbox{\normalfont D0}\xspace} 
\def\lhcb{\mbox{\normalfont LHCb}\xspace}

\def\ee         {\ensuremath{e^-e^-}\xspace}

\def\mup        {\ensuremath{\mu^+}\xspace}
\def\mun        {\ensuremath{\mu^-}\xspace}    
\def\mumu       {\ensuremath{\mu^+\mu^-}\xspace}
\def\mtau       {\ensuremath{\tau}\xspace}

\def\nub        {\ensuremath{\overline{\nu}}\xspace}

\def\nub        {\ensuremath{\overline{\nu}}\xspace}

\def\nut        {\ensuremath{\nu_\tau}\xspace}

\def\nul        {\ensuremath{\nu_\ell}\xspace}

\def\Z      {\ensuremath{Z^0}\xspace}

\def\ubar  {\ensuremath{\overline u}\xspace}

\def\dbar  {\ensuremath{\overline d}\xspace}
\def\ddbar {\ensuremath{d\overline d}\xspace}

\def\sbar  {\ensuremath{\overline s}\xspace}

\def\b  {\ensuremath{b}\xspace}
\def\bbar  {\ensuremath{\overline b}\xspace}

\def\piz   {\ensuremath{\pi^0}\xspace}

\def\pip   {\ensuremath{\pi^+}\xspace}
\def\pim   {\ensuremath{\pi^-}\xspace}
\def\pipi  {\ensuremath{\pi^+\pi^-}\xspace}

\def\pimp  {\ensuremath{\pi^\mp}\xspace}

\def\etapr {\ensuremath{\eta^{\prime}}\xspace}

\def\Kbar  {\kern 0.2em\overline{\kern -0.2em K}{}\xspace}

\def\Kpm   {\ensuremath{K^\pm}\xspace}
\def\Kmp   {\ensuremath{K^\mp}\xspace}
\def\Kp    {\ensuremath{K^+}\xspace}
\def\Km    {\ensuremath{K^-}\xspace}
\def\KS    {\ensuremath{K^0_{\scriptscriptstyle S}}\xspace} 
\def\KL    {\ensuremath{K^0_{\scriptscriptstyle L}}\xspace} 
\def\Kstarz  {\ensuremath{K^{*0}}\xspace}
\def\Kstarzb  {\ensuremath{\Kbar^{*0}}\xspace}
\def\Kstar   {\ensuremath{K^*}\xspace}

\def\Kstarpm   {\ensuremath{K^{*\pm}}\xspace}
\def\Kstarmp   {\ensuremath{K^{*\mp}}\xspace}
\def\Kz   {\ensuremath{K^0}\xspace}
\def\Kzb   {\ensuremath{\Kbar^0}\xspace}
\def\KzKzb {\ensuremath{K^0 \kern -0.16em \Kzb}\xspace}

\def\KorKstarpm {\ensuremath{K^{(*)\pm}}\xspace}

\def\Dz    {\ensuremath{D^0}\xspace}
\def\Dbar  {\kern 0.2em\overline{\kern -0.2em D}{}\xspace}

\def\Dzb   {\ensuremath{\Dbar^0}\xspace}
\def\DzDzb {\ensuremath{D^0 {\kern -0.16em \Dzb}}\xspace}
\def\Dp    {\ensuremath{D^+}\xspace}

\def\Dstar   {\ensuremath{D^*}\xspace}

\def\Dstarp  {\ensuremath{D^{*+}}}
\def\Dstarm  {\ensuremath{D^{*-}}}
\def\DorDstar   {\ensuremath{D^{(*)}}\xspace}
\def\DorDstarz  {\ensuremath{D^{(*)0}}\xspace}
\def\DorDstarzb {\ensuremath{\Dbar^{(*)0}}\xspace}

\def\Ds    {\ensuremath{D^+_s}\xspace}
\def\Dsp   {\ensuremath{D^+_s}\xspace}
\def\Dsm   {\ensuremath{D^-_s}\xspace}

\def\Bz    {\ensuremath{B^0}\xspace}
\def\B     {\ensuremath{B}\xspace}
\def\Bbar  {\kern 0.18em\overline{\kern -0.18em B}{}\xspace}
\def\Bb    {\ensuremath{\Bbar}\xspace}
\def\Bzb   {\ensuremath{\Bbar^0}\xspace}
\def\Bu    {\ensuremath{B^+}\xspace}

\def\Bpm   {\ensuremath{B^\pm}\xspace}
\def\Bmp   {\ensuremath{B^\mp}\xspace}
\def\Bs    {\ensuremath{B_s}\xspace}
\def\Bsb   {\ensuremath{\Bbar_s^0}\xspace}
\def\BB    {\ensuremath{B\Bbar}\xspace} 
\def\BzBzb {\ensuremath{B^0 {\kern -0.16em \Bzb}}\xspace}

\def\jpsi  {\ensuremath{{J\mskip -3mu/\mskip -2mu\psi\mskip 2mu}}\xspace}

\mathchardef\Upsilon="7107
\def\Y#1S{\ensuremath{\Upsilon{(#1S)}}\xspace}

\def\proton      {\ensuremath{p}\xspace}

\mathchardef\Deltares="7101
\mathchardef\Xi="7104
\mathchardef\Lambda="7103
\mathchardef\Sigma="7106
\mathchardef\Omega="710A
\def\Deltabar   {\kern 0.25em\overline{\kern -0.25em \Deltares}{}\xspace}
\def\Lbar {\kern 0.2em\overline{\kern -0.2em\Lambda\kern 0.05em}\kern-0.05em{}\xspace}
\def\Sigbar{\kern 0.2em\overline{\kern -0.2em \Sigma}{}\xspace}
\def\Xibar{\kern 0.2em\overline{\kern -0.2em \Xi}{}\xspace}
\def\Obar{\kern 0.2em\overline{\kern -0.2em \Omega}{}\xspace}
\def\Nbar{\kern 0.2em\overline{\kern -0.2em N}{}\xspace}
\def\Xb{\kern 0.2em\overline{\kern -0.2em X}{}}

\def\BR{{\ensuremath{\cal B}}}

\newcommand{\tev}{\ensuremath{\mathrm{Te\kern -0.1em V}}\xspace}
\newcommand{\gev}{\ensuremath{\mathrm{Ge\kern -0.1em V}}\xspace}
\newcommand{\mev}{\ensuremath{\mathrm{Me\kern -0.1em V}}\xspace}
\newcommand{\kev}{\ensuremath{\mathrm{ke\kern -0.1em V}}\xspace}
\newcommand{\ev}{\ensuremath{\mathrm{e\kern -0.1em V}}\xspace}
\newcommand{\gevc}{\ensuremath{{\mathrm{Ge\kern -0.1em V\!/}c}}\xspace}
\newcommand{\mevc}{\ensuremath{{\mathrm{Me\kern -0.1em V\!/}c}}\xspace}
\newcommand{\gevcc}{\ensuremath{{\mathrm{Ge\kern -0.1em V\!/}c^2}}\xspace}
\newcommand{\mevcc}{\ensuremath{{\mathrm{Me\kern -0.1em V\!/}c^2}}\xspace}

\def\pb {\ensuremath{\rm \,pb}\xspace}

\def\fb   {\ensuremath{\mbox{\,fb}}\xspace}
\def\invfb   {\ensuremath{\mbox{\,fb}^{-1}}\xspace}
\def\mus  {\ensuremath{\rm \,\mus}\xspace}

\def\ps   {\ensuremath{\rm \,ps}\xspace}

\def\mus        {\ensuremath{\,\mu{\rm s}}\xspace}    
\def\ps         {\ensuremath{{\rm \,ps}}\xspace}  
\def\gsim{{~\raise.15em\hbox{$>$}\kern-.85em
          \lower.35em\hbox{$\sim$}~}\xspace}
\def\lsim{{~\raise.15em\hbox{$<$}\kern-.85em
          \lower.35em\hbox{$\sim$}~}\xspace}

\def\CP                 {\ensuremath{C\!P}\xspace}
\def\CPT                {\ensuremath{C\!PT}\xspace}
\def\ra                 {\ensuremath{\to}\xspace}

\def\pep2{PEP-II}

\newcommand{\chisq}{\ensuremath{\chi^2}\xspace}

\def\rhobar {\ensuremath{\overline{\rho}}\xspace}
\def\etabar {\ensuremath{\overline{\eta}}\xspace}

\def\Vud  {\ensuremath{|V_{ud}|}\xspace}

\def\Vus  {\ensuremath{|V_{us}|}\xspace}

\def\Vub  {\ensuremath{|V_{ub}|}\xspace}

\def\stwob{\ensuremath{\sin\! 2 \beta   }\xspace}

\def\deltamd{\ensuremath{{\rm \Delta}m_d}\xspace}

\xspace
\newcommand{\fds}{\ensuremath{f_{D_s}}\xspace}

\def\jetset74   {\mbox{\tt Jetset \hspace{-0.5em}7.\hspace{-0.2em}4}}

\newcommand{\aerr}[4]   {\mbox{${{#1}^{+ #2}_{- #3}\pm #4}$}}
\newcommand{\berr}[4]   {\mbox{${{#1}\pm #2^{+ #3}_{- #4}}$}}
\newcommand{\cerr}[3]   {\mbox{${{#1}^{+ #2}_{- #3}}$}}
\newcommand{\aerrsy}[5] {\mbox{${{#1}^{+ #2 + #4}_{- #3 - #5}}$}}

\newcommand{\err}[3]   {\mbox{${{#1}\pm{#2}\pm{#3}}$}}

\newcommand{\nodata}{$$}
\newcommand{\vs}{\mbox{$vs.$}}
\def\etapr{{\eta^{\prime}}}

\def\sgline{\noalign{\vskip 0.10truecm\hrule\vskip 0.10truecm}}
\def\sglinespt{\noalign{\vskip 0.05truecm\hrule}}
\def\sglinespb{\noalign{\hrule\vskip 0.05truecm}}

\newcommand{\ks}    {\mbox{$K^0_S$}}
\newcommand{\kz}    {\mbox{$K^0$}}
\newcommand{\kzb}   {\mbox{$\overline{K^0}$}}

\renewcommand{\mysection}[1]{\section[#1]{#1}} 

\newcommand\red[1]{{\color{red}#1}}
\newcommand\blue[1]{{\color{blue}#1}}

\begin{document}

\setcounter{page}{1}
\thispagestyle{empty}
\renewcommand\Affilfont{\itshape\small}

\title{
  Averages of $b$-hadron, $c$-hadron, and $\tau$-lepton properties
  as of summer 2014
\vskip0.20in
\large{\it Heavy Flavor Averaging Group (HFAG):}
\vspace*{-0.20in}}
\author[1]{Y.~Amhis}\affil[1]{LAL, Universit\'{e} Paris-Sud, France}
\author[2]{Sw.~Banerjee}\affil[2]{University of Victoria, Canada}
\author[3]{E.~Ben-Haim}\affil[3]{LPNHE, Universit\'e Pierre et Marie Curie-Paris 6, Universit\'e Denis Diderot-Paris7, CNRS/IN2P3, France}
\author[4]{S.~Blyth}\affil[4]{National United University, Taiwan}
\author[5]{A.~Bozek}\affil[5]{H. Niewodniczanski Institute of Nuclear Physics, Poland}
\author[6]{C.~Bozzi}\affil[6]{INFN Ferrara, Italy}
\author[7,8]{A.~Carbone}\affil[7]{INFN Bologna, Italy}\affil[8]{Universit\'{a} di Bologna, Italy}
\author[9]{R.~Chistov}\affil[9]{Institute for Theoretical and Experimental Physics, Russia}
\author[5,10]{M.~Chrz\k{a}szcz}\affil[10]{Universit\"at Z\"urich, Switzerland}
\author[6]{G.~Cibinetto}
\author[11]{J.~Dingfelder}\affil[11]{Bonn University, Germany}
\author[12]{M.~Gelb}\affil[12]{Karlsruher Institut f\"{u}r Technologie, Germany}
\author[13]{M.~Gersabeck}\affil[13]{The University of Manchester, UK}
\author[14]{T.~Gershon}\affil[14]{University of Warwick, UK}
\author[15]{L.~Gibbons}\affil[15]{Cornell University, USA}
\author[16,17]{B.~Golob}\affil[16]{University of Ljubljana, Slovenia}\affil[17]{J. Stefan Institute, Slovenia}
\author[18]{R.~Harr}\affil[18]{Wayne State University, USA}
\author[19]{K.~Hayasaka}\affil[19]{Nagoya University, Japan}
\author[20]{H.~Hayashii}\affil[20]{Nara Women's University, Japan}
\author[12]{T.~Kuhr}
\author[21]{O.~Leroy}\affil[21]{CPPM, Aix-Marseille Universit\'{e}, CNRS/IN2P3, Marseille, France}
\author[22]{A.~Lusiani}\affil[22]{Scuola Normale Superiore and INFN, Pisa, Italy}
\author[20]{K.~Miyabayashi}
\author[23]{P.~Naik}\affil[23]{University of Bristol, UK}
\author[24]{S.~Nishida}\affil[24]{KEK, Tsukuba, Japan}
\author[25]{A.~Oyanguren Campos}\affil[25]{IFIC, University of Valencia, Spain}
\author[26]{M.~Patel}\affil[26]{Imperial College London, UK}
\author[27]{D.~Pedrini} \affil[27]{INFN Milano-Bicocca, Italy}
\author[17]{M.~Petri\v{c}}
\author[28]{M.~Rama}\affil[28]{INFN Frascati, Italy}
\author[2]{M.~Roney}
\author[29]{M.~Rotondo}\affil[29]{INFN Padova, Italy}
\author[30]{O.~Schneider}\affil[30]{Ecole Polytechnique F\'{e}d\'{e}rale de Lausanne (EPFL), Switzerland}
\author[31]{C.~Schwanda}\affil[31]{Austrian Academy of Sciences, Austria}
\author[32]{A.~J.~Schwartz}\affil[32]{University of Cincinnati, USA}
\author[33]{B.~Shwartz}\affil[33]{Budker Institute of Nuclear Physics, Russia}
\author[34]{J.~G.~Smith}\affil[34]{University of Colorado, USA}
\author[35]{R.~Tesarek}\affil[35]{Fermilab, USA}
\author[24,30]{K.~Trabelsi}
\author[36]{P.~Urquijo}\affil[36]{University of Melbourne, Australia}
\author[37]{R.~Van Kooten}\affil[37]{Indiana University, USA}
\author[17]{A.~Zupanc}

\date{\today} 
\maketitle

\begin{abstract}
This article reports world averages of measurements of $b$-hadron, $c$-hadron,
and $\tau$-lepton properties obtained by the Heavy Flavor Averaging Group (HFAG)
using results available through summer 2014.
For the averaging, common input parameters used in the various analyses
are adjusted (rescaled) to common values, and known correlations are taken
into account.
The averages include branching fractions, lifetimes,
neutral meson mixing parameters, \CP~violation parameters,
parameters of semileptonic decays and CKM matrix elements.
\end{abstract}

\newpage
\tableofcontents
\newpage



\mysection{Introduction}
\label{sec:intro}

Flavor dynamics is an important element in understanding the nature of
particle physics.  The accurate knowledge of properties of heavy flavor
hadrons, especially $b$ hadrons, plays an essential role for
determining the elements of the Cabibbo-Kobayashi-Maskawa (CKM)
weak-mixing matrix~\cite{Cabibbo:1963yz,Kobayashi:1973fv}. 
The operation of the \belle\ and \babar\ $e^+e^-$ $B$ factory 
experiments led to a large increase in the size of available 
$B$-meson, $D$-hadron and $\tau$-lepton samples, 
enabling dramatic improvement in the accuracies of related measurements.
The CDF and \dzero\ experiments at the Fermilab Tevatron 
have also provided important results in heavy flavour physics,
most notably in the $B^0_s$ sector.
Run~I of the CERN Large Hadron Collider delivered high luminosity, 
enabling the collection of even higher statistics samples of $b$ 
and $c$ hadrons, and thus a further leap in precision in many areas, at the
ATLAS, CMS, and (especially) LHCb experiments. 
 
The Heavy Flavor Averaging Group (HFAG) was formed in 2002 to 
continue the activities of the LEP Heavy Flavor Steering 
group~\cite{Abbaneo:2000ej_mod,*Abbaneo:2001bv_mod_cont}. 
This group was responsible for calculating averages of 
measurements of $b$-flavor related quantities. HFAG has evolved 
since its inception and currently consists of seven subgroups:
\begin{itemize}
\item the ``$B$ Lifetime and Oscillations'' subgroup provides 
averages for $b$-hadron lifetimes, $b$-hadron fractions in 
$\Upsilon(4S)$ decay and $pp$ or $p\bar{p}$ collisions, and various 
parameters governing $\Bz$-$\Bzb$ and $\Bs$--$\Bsb$ mixing;

\item the ``Unitarity Triangle Parameters'' subgroup provides
averages for time-dependent $\CP$ asymmetry parameters and studies of $B \to DK$ decays, and 
resulting determinations of the angles of the CKM unitarity triangle;

\item the ``Semileptonic $B$ Decays'' subgroup provides averages
for inclusive and exclusive $B$-decay branching fractions, and
subsequent determinations of the CKM matrix elements 
$|V_{cb}|$ and $|V_{ub}|$;

\item the ``$B$ to Charm Decays'' subgroup provides averages of 
branching fractions for $B$ decays to final states involving open 
charm or charmonium mesons;

\item the ``Rare Decays'' subgroup provides averages of branching 
fractions and $\CP$ asymmetries for charmless, radiative, 
leptonic, and baryonic $B$-meson and \b-baryon decays;

\item the ``Charm Physics'' subgroup provides averages of numerous quantities in the charm sector, including branching fractions 
properties of charm baryons and of excited $D^{**}$ and $D^{}_{sJ}$ mesons, 
averages of $D^0$-$\Dzb$ mixing and $\CP$ and $T$ violation parameters, 
and an average value for the $D^{}_s$ decay constant~$f^{}_{D_s}$.

\item the ``Tau Physics'' subgroup provides documentation and
averages for a selection of \mtau lepton quantities that most profit
from the adoption of the HFAG prescriptions. In particular, the \mtau
lepton branching fractions, uncertainties and correlations are
obtained from a global fit of the experimental results, and this
information is further elaborated to compute several lepton
universality tests and the CKM matrix element $|V_{us}|$. The \mtau
lepton-flavor-violating decays are documented and, starting with this
edition, combinations of such upper limits are also computed.
\end{itemize}

The ``Lifetime and Oscillations'' and ``Semileptonic'' subgroups were formed from the merger of four LEP working groups.
The ``Unitary Triangle,'' ``$B$ to Charm Decays,'' and ``Rare Decays''
subgroups were formed to provide averages for new results obtained
from the $B$ factory experiments (and now also from the Fermilab 
Tevatron and CERN LHC experiments).
The ``Charm'' and ``Tau''  subgroups were formed more recently in 
response to the wealth of new data concerning $D$ and $\tau$ physics. 
Subgroups typically include representatives from \belle, \babar\ and LHCb, 
plus, when relevant, CLEO, CDF and \dzero. 

This article is an update of the last HFAG preprint,
which used results available at least through early 2012~\cite{Amhis:2012bh}. 
Here we report world averages using results available at least through summer 2014.
In some cases results made available in the latter part of 2014 have been
included.\footnote{*
Particularly important new results have been included whenever possible.
The precise cut-off date for including results in the averages varies 
between subgroups.}
In general, we use all publicly available results that have written documentation. 
These include preliminary results presented at conferences or workshops.
However, we do not use preliminary results that remain unpublished 
for an extended period of time, or for which no publication is planned. 
Close contacts have been established between representatives from
the experiments and members of subgroups that perform averaging 
to ensure that the data are prepared in a form suitable for 
combinations.  

In the case of obtaining a world average for which $\chi^2/\dof > 1$,
where $\dof$ is the number of degrees of freedom in the average
calculation, we do not usually scale the resulting error, as is presently 
done by the Particle Data Group~\cite{PDG_2014}.
Rather, 
we examine the systematics of each measurement to better understand them. 
Unless we find possible systematic discrepancies between the measurements, 
we do not apply any additional correction to the calculated error. 
We provide the confidence level of the fit as an indicator for the 
consistency of the measurements included in the average. In case some
special treatment was necessary to calculate an average, or if an
approximation used in an average calculation might not be 
sufficiently accurate 
(\eg, assuming Gaussian errors when the likelihood function indicates 
non-Gaussian behavior), we include a warning message.

Chapter~\ref{sec:method} describes the methodology used for calculating
averages. In the averaging procedure, common input parameters used in 
the various analyses are adjusted (rescaled) to common values, and, 
where possible, known correlations are taken into account. 
Chapters~\ref{sec:life_mix}--\ref{sec:tau} present world 
average values from each of the subgroups listed above. 
A brief 
summary of the averages presented is given in Chapter~\ref{sec:summary}.   
A complete listing of the averages and plots,
including updates since this document was prepared,
are also available on the HFAG web site:
\vskip0.15in\hskip0.75in
\vbox{
  \href{http://www.slac.stanford.edu/xorg/hfag}{\tt http://www.slac.stanford.edu/xorg/hfag} 
}

\clearpage
\section{Methodology } 
\label{sec:method} 

The general averaging problem that HFAG faces is to combine 
information provided by different measurements of the same parameter
to obtain our best estimate of the parameter's value and
uncertainty. The methodology described here focuses on the problems of
combining measurements performed with different systematic assumptions
and with potentially-correlated systematic uncertainties. Our methodology
relies on the close involvement of the people performing the
measurements in the averaging process.

Consider two hypothetical measurements of a parameter $x$, which might
be summarized as
\begin{align*}
x &= x_1 \pm \delta x_1 \pm \Delta x_{1,1} \pm \Delta x_{2,1} \ldots \\
x &= x_2 \pm \delta x_2 \pm \Delta x_{1,2} \pm \Delta x_{2,2} \ldots
\; ,
\end{align*}
where the $\delta x_k$ are statistical uncertainties, and
the $\Delta x_{i,k}$ are contributions to the systematic
uncertainty. One popular approach is to combine statistical and
systematic uncertainties in quadrature
\begin{align*}
x &= x_1 \pm \left(\delta x_1 \oplus \Delta x_{1,1} \oplus \Delta
x_{2,1} \oplus \ldots\right) \\
x &= x_2 \pm \left(\delta x_2 \oplus \Delta x_{1,2} \oplus \Delta
x_{2,2} \oplus \ldots\right)
\end{align*}
and then perform a weighted average of $x_1$ and $x_2$, using their
combined uncertainties, as if they were independent. This approach
suffers from two potential problems that we attempt to address. First,
the values of the $x_k$ may have been obtained using different
systematic assumptions. For example, different values of the \Bz
lifetime may have been assumed in separate measurements of the
oscillation frequency $\deltamd$. The second potential problem is that
some contributions of the systematic uncertainty may be correlated
between experiments. For example, separate measurements of $\deltamd$
may both depend on an assumed Monte-Carlo branching fraction used to
model a common background.

The problems mentioned above are related since, ideally, any quantity $y_i$
that $x_k$ depends on has a corresponding contribution $\Delta x_{i,k}$ to the
systematic error which reflects the uncertainty $\Delta y_i$ on $y_i$
itself. We assume that this is the case and use the values of $y_i$ and
$\Delta y_i$ assumed by each measurement explicitly in our
averaging (we refer to these values as $y_{i,k}$ and $\Delta y_{i,k}$
below). Furthermore, since we do not lump all the systematics
together,
we require that each measurement used in an average have a consistent
definition of the various contributions to the systematic uncertainty.
Different analyses often use different decompositions of their systematic
uncertainties, so achieving consistent definitions for any potentially
correlated contributions requires close coordination between HFAG and
the experiments. In some cases, a group of
systematic uncertainties must be combined to obtain a coarser
description that is consistent between measurements. Systematic uncertainties
that are uncorrelated with any other sources of uncertainty appearing
in an average are lumped together with the statistical error, so that the only
systematic uncertainties treated explicitly are those that are
correlated with at least one other measurement via a consistently-defined
external parameter $y_i$. When asymmetric statistical or systematic
uncertainties are quoted, we symmetrize them since our combination
method implicitly assumes parabolic likelihoods for each measurement.

\begin{figure}
\begin{center}
\includegraphics[width=6.0in]{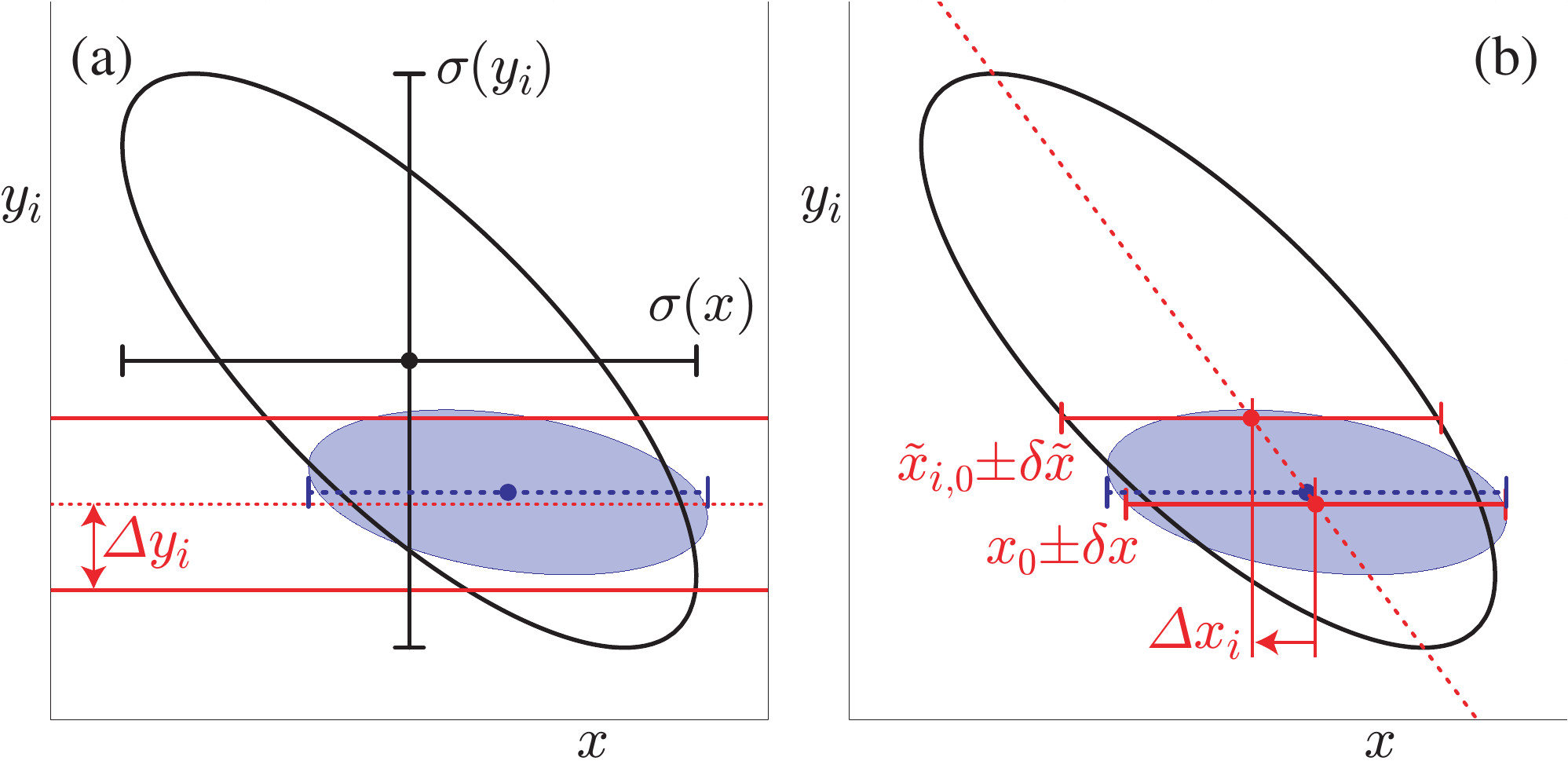}
\end{center}
\caption{The left-hand plot (a) compares the 68\% confidence-level
  contours of a
  hypothetical measurement's unconstrained (large ellipse) and
  constrained (filled ellipse) likelihoods, using the Gaussian
  constraint on $y_i$ represented by the horizontal band. The solid
  error bars represent the statistical uncertainties $\sigma(x)$ and
  $\sigma(y_i)$ of the unconstrained likelihood. The dashed
  error bar shows the statistical error on $x$ from a
  constrained simultaneous fit to $x$ and $y_i$. The right-hand plot
  (b) illustrates the method described in the text of performing fits
  to $x$ with $y_i$ fixed at different values. The dashed
  diagonal line between these fit results has the slope
  $\rho(x,y_i)\sigma(y_i)/\sigma(x)$ in the limit of a parabolic
  unconstrained likelihood. The result of the constrained simultaneous
  fit from (a) is shown as a dashed error bar on $x$.}
\label{fig:singlefit}
\end{figure}

The fact that a measurement of $x$ is sensitive to the value of $y_i$
indicates that, in principle, the data used to measure $x$ could
equally-well be used for a simultaneous measurement of $x$ and $y_i$, as
illustrated by the large contour in Fig.~\ref{fig:singlefit}(a) for a hypothetical
measurement. However, we often have an external constraint $\Delta
y_i$ on the value of $y_i$ (represented by the horizontal band in
Fig.~\ref{fig:singlefit}(a)) that is more precise than the constraint
$\sigma(y_i)$ from
our data alone. Ideally, in such cases we would perform a simultaneous
fit to $x$ and $y_i$, including the external constraint, obtaining the
filled $(x,y)$ contour and corresponding dashed one-dimensional estimate of
$x$ shown in Fig.~\ref{fig:singlefit}(a). Throughout, we assume that
the external constraint $\Delta y_i$ on $y_i$ is Gaussian.

In practice, the added technical complexity of a constrained fit with
extra free parameters is not justified by the small increase in
sensitivity, as long as the external constraints $\Delta y_i$ are
sufficiently precise when compared with the sensitivities $\sigma(y_i)$
to each $y_i$ of the data alone. Instead, the usual procedure adopted
by the experiments is to perform a baseline fit with all $y_i$ fixed
to nominal values $y_{i,0}$, obtaining $x = x_0 \pm \delta
x$. This baseline fit neglects the uncertainty due to $\Delta y_i$, but
this error can be mostly recovered by repeating the fit separately for
each external parameter $y_i$ with its value fixed at $y_i = y_{i,0} +
\Delta y_i$ to obtain $x = \tilde{x}_{i,0} \pm \delta\tilde{x}$, as
illustrated in Fig.~\ref{fig:singlefit}(b). The absolute shift,
$|\tilde{x}_{i,0} - x_0|$, in the central value of $x$ is what the
experiments usually quote as their systematic uncertainty $\Delta x_i$
on $x$ due to the unknown value of $y_i$. Our procedure requires that
we know not only the magnitude of this shift but also its sign. In the
limit that the unconstrained data is represented by a parabolic
likelihood, the signed shift is given by
\begin{equation}
\Delta x_i = \rho(x,y_i)\frac{\sigma(x)}{\sigma(y_i)}\,\Delta y_i \;,
\end{equation}
where $\sigma(x)$ and $\rho(x,y_i)$ are the statistical uncertainty on
$x$ and the correlation between $x$ and
$y_i$ in the unconstrained data.
While our procedure is not
equivalent to the constrained fit with extra parameters, it yields (in
the limit of a parabolic unconstrained likelihood) a central value
$x_0$ that agrees 
to ${\cal O}(\Delta y_i/\sigma(y_i))^2$ and an uncertainty $\delta x
\oplus \Delta x_i$ that agrees to ${\cal O}(\Delta y_i/\sigma(y_i))^4$.

In order to combine two or more measurements that share systematics
due to the same external parameters $y_i$, we would ideally perform a
constrained simultaneous fit of all data samples to obtain values of
$x$ and each $y_i$, being careful to only apply the constraint on each
$y_i$ once. This is not practical since we generally do not have
sufficient information to reconstruct the unconstrained likelihoods
corresponding to each measurement. Instead, we perform the two-step
approximate procedure described below.

\begin{figure}
\begin{center}
\includegraphics[width=6.0in]{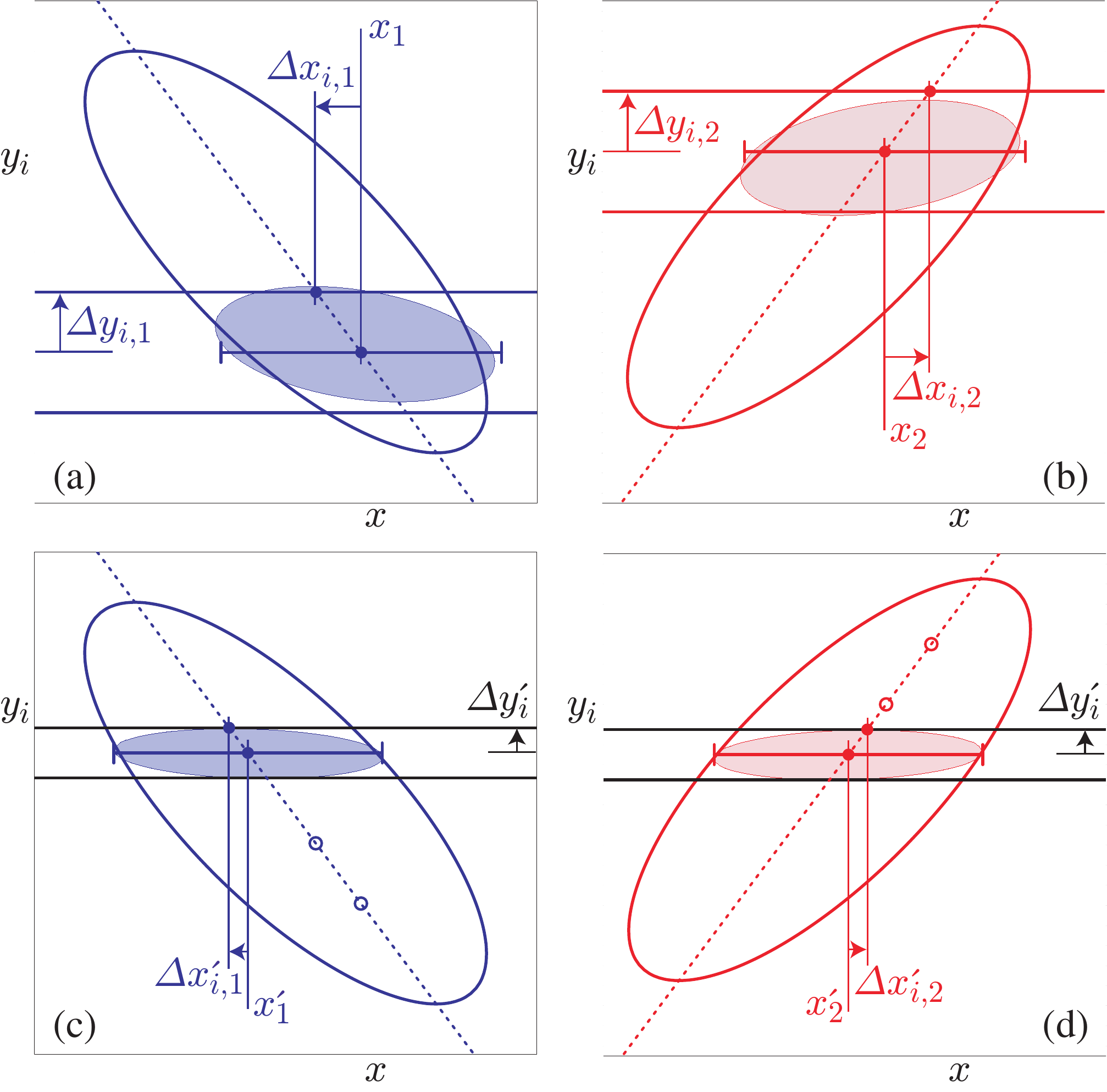}
\end{center}
\caption{The upper plots (a) and (b) show examples of two individual
  measurements to be combined. The large ellipses represent their
  unconstrained likelihoods, and the filled ellipses represent their
  constrained likelihoods. Horizontal bands indicate the different
  assumptions about the value and uncertainty of $y_i$ used by each
  measurement. The error bars show the results of the approximate
  method described in the text for obtaining $x$ by performing fits
  with $y_i$ fixed to different values. The lower plots (c) and (d)
  illustrate the adjustments to accommodate updated and consistent
  knowledge of $y_i$ as described in the text. Open circles mark the
  central values of the unadjusted fits to $x$ with $y$ fixed; these
  determine the dashed line used to obtain the adjusted values. }
\label{fig:multifit}
\end{figure}

Figs.~\ref{fig:multifit}(a,b) illustrate two
statistically-independent measurements, $x_1 \pm (\delta x_1 \oplus
\Delta x_{i,1})$ and $x_2\pm(\delta x_i\oplus \Delta x_{i,2})$, of the same
hypothetical quantity $x$ (for simplicity, we only show the
contribution of a single correlated systematic due to an external
parameter $y_i$). As our knowledge of the external parameters $y_i$
evolves, it is natural that the different measurements of $x$ will
assume different nominal values and ranges for each $y_i$. The first
step of our procedure is to adjust the values of each measurement to
reflect the current best knowledge of the values $y_i'$ and ranges
$\Delta y_i'$ of the external parameters $y_i$, as illustrated in
Figs.~\ref{fig:multifit}(c,b). We adjust the
central values $x_k$ and correlated systematic uncertainties $\Delta
x_{i,k}$ linearly for each measurement (indexed by $k$) and each
external parameter (indexed by $i$):
\begin{align}
x_k' &= x_k + \sum_i\,\frac{\Delta x_{i,k}}{\Delta y_{i,k}}
\left(y_i'-y_{i,k}\right)\\
\Delta x_{i,k}'&= \Delta x_{i,k}\cdot \frac{\Delta y_i'}{\Delta
  y_{i,k}} \; .
\end{align}
This procedure is exact in the limit that the unconstrained
likelihoods of each measurement is parabolic.

\begin{figure}
\begin{center}
\includegraphics[width=3.5in]{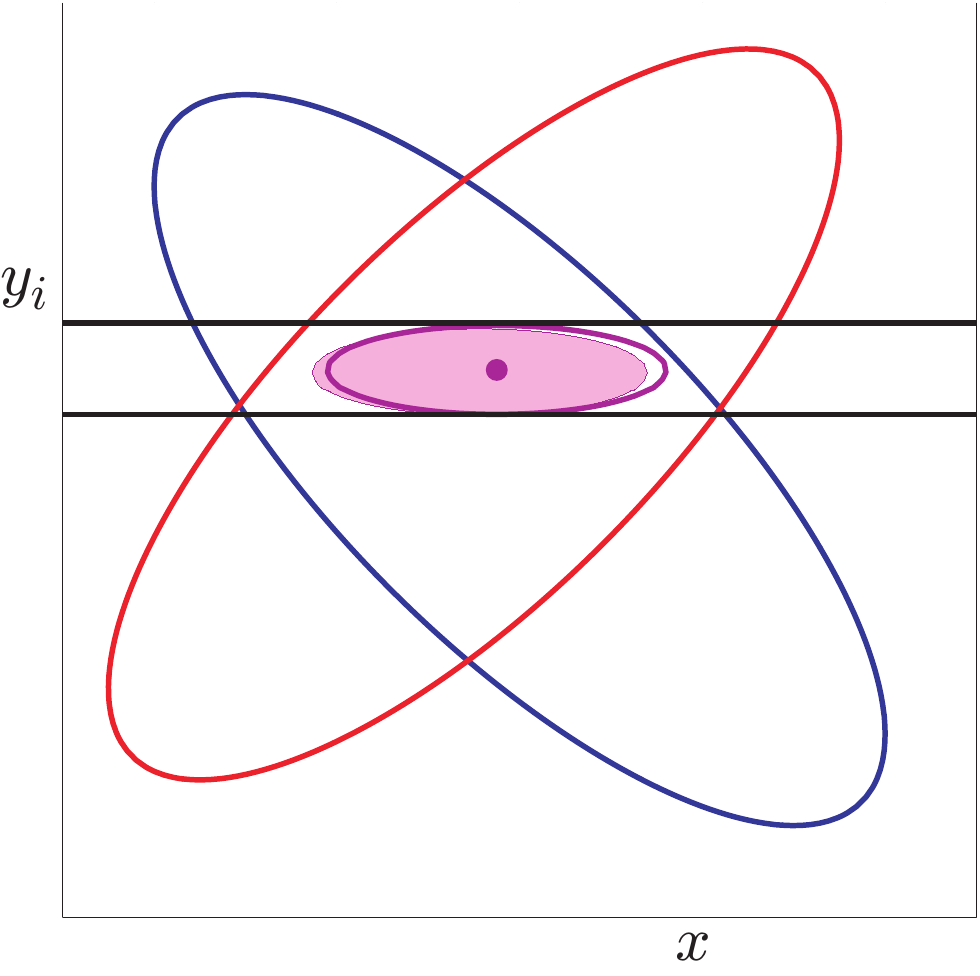}
\end{center}
\caption{An illustration of the combination of two hypothetical
  measurements of $x$ using the method described in the text. The
  ellipses represent the unconstrained likelihoods of each measurement,
  and the horizontal band represents the latest knowledge about $y_i$ 
  that is used to adjust the individual measurements. The filled small
  ellipse shows the result of the exact method using 
  ${\cal L}_{\text{comb}}$, and the hollow small ellipse and dot show 
  the result of the approximate method using $\chi^2_{\text{comb}}$.}
\label{fig:fit12}
\end{figure}

The second step of our procedure is to combine the adjusted
measurements, $x_k'\pm (\delta x_k\oplus \Delta x_{k,1}'\oplus \Delta
x_{k,2}'\oplus\ldots)$ using the chi-square 
\begin{equation}
\chi^2_{\text{comb}}(x,y_1,y_2,\ldots) \equiv \sum_k\,
\frac{1}{\delta x_k^2}\left[
x_k' - \left(x + \sum_i\,(y_i-y_i')\frac{\Delta x_{i,k}'}{\Delta y_i'}\right)
\right]^2 + \sum_i\,
\left(\frac{y_i - y_i'}{\Delta y_i'}\right)^2 \; ,
\end{equation}
and then minimize this $\chi^2$ to obtain the best values of $x$ and
$y_i$ and their uncertainties, as illustrated in
Fig.~\ref{fig:fit12}. Although this method determines new values for
the $y_i$, we do not report them since the $\Delta x_{i,k}$ reported
by each experiment are generally not intended for this purpose (for
example, they may represent a conservative upper limit rather than a
true reflection of a 68\% confidence level).

For comparison, the exact method we would
perform if we had the unconstrained likelihoods ${\cal L}_k(x,y_1,y_2,\ldots)$
available for each
measurement is to minimize the simultaneous constrained likelihood
\begin{equation}
{\cal L}_{\text{comb}}(x,y_1,y_2,\ldots) \equiv \prod_k\,{\cal
  L}_k(x,y_1,y_2,\ldots)\,\prod_{i}\,{\cal 
  L}_i(y_i) \; ,
\end{equation}
with an independent Gaussian external constraint on each $y_i$
\begin{equation}
{\cal L}_i(y_i) \equiv \exp\left[-\frac{1}{2}\,\left(\frac{y_i-y_i'}{\Delta
 y_i'}\right)^2\right] \; .
\end{equation}
The results of this exact method are illustrated by the filled ellipses
in Figs.~\ref{fig:fit12}(a,b) and agree with our method in the limit that
each ${\cal L}_k$ is parabolic and that each $\Delta
y_i' \ll \sigma(y_i)$. In the case of a non-parabolic unconstrained
likelihood, experiments would have to provide a description of ${\cal
  L}_k$ itself to allow an improved combination. In the case of
$\sigma(y_i)\simeq \Delta y_i'$, experiments are advised to perform a
simultaneous measurement of both $x$ and $y$ so that their data will
improve the world knowledge about $y$. 

 The algorithm described above is used as a default in the averages
reported in the following sections.  For some cases, somewhat simplified
or more complex algorithms are used and noted in the corresponding 
sections. Some examples for extensions of the standard method for extracting
averages are given here. These include the case where measurement errors
depend on the measured value, \ie\ are relative errors, unknown
correlation coefficients and the breakdown of error sources.

For measurements with Gaussian errors, the usual estimator for the
average of a set of measurements is obtained by minimizing the following
$\chi^2$:
\begin{equation}
\chi^2(t) = \sum_i^N \frac{\left(y_i-t\right)^2}{\sigma^2_i} ,
\label{eq:chi2t}
\end{equation}
where $y_i$ is the measured value for input $i$ and $\sigma_i^2$ is the
variance of the distribution from which $y_i$ was drawn.  The value $\hat{t}$
of $t$ at minimum $\chi^2$ is our estimator for the average.  (This
discussion is given for independent measurements for the sake of
simplicity; the generalization to correlated measurements is
straightforward, and has been used when averaging results.) 
The true $\sigma_i$ are unknown but typically the error as assigned by the
experiment $\sigma_i^{\mathrm{raw}}$ is used as an estimator for it.
Caution is advised,
however, in the case where $\sigma_i^{\mathrm{raw}}$
depends on the value measured for $y_i$. Examples of this include
an uncertainty in any multiplicative factor (like
an acceptance) that enters the determination of $y_i$, \ie\ the $\sqrt{N}$
dependence of Poisson statistics, where $y_i \propto N$
and $\sigma_i \propto \sqrt{N}$.
Failing to account for this type of
dependence when averaging leads to a biased average.
Biases in the average can be avoided (or at least reduced)
by minimizing the following
$\chi^2$:
\begin{equation}
\chi^2(t) = \sum_i^N \frac{\left(y_i-t\right)^2}{\sigma^2_i(\hat{t})} .
\label{eq:chi2that}
\end{equation}
In the above $\sigma_i(\hat{t})$ is the uncertainty
assigned to input $i$ that includes the assumed dependence of the
stated error on the value measured.  As an example, consider 
a pure acceptance error, for which
$\sigma_i(\hat{t}) = (\hat{t} / y_i)\times\sigma_i^{\mathrm{raw}}$ .
It is easily verified that solving Eq.~\ref{eq:chi2that} 
leads to the correct behavior, namely
$$ 
\hat{t} = \frac{\sum_i^N y_i^3/(\sigma_i^{\mathrm{raw}})^2}{\sum_i^N y_i^2/(\sigma_i^{\mathrm{raw}})^2},
$$
\ie\ weighting by the inverse square of the 
fractional uncertainty, $\sigma_i^{\mathrm{raw}}/y_i$.
It is sometimes difficult to assess the dependence of $\sigma_i^{\mathrm{raw}}$ on
$\hat{t}$ from the errors quoted by experiments.  


Another issue that needs careful treatment is the question of correlation
among different measurements, \eg\ due to using the same theory for
calculating acceptances.  A common practice is to set the correlation
coefficient to unity to indicate full correlation.  However, this is
not a ``conservative'' thing to do, and can in fact lead to a significantly
underestimated uncertainty on the average.  In the absence of
better information, the most conservative choice of correlation coefficient
between two measurements $i$ and $j$
is the one that maximizes the uncertainty on $\hat{t}$
due to that pair of measurements:
\begin{equation}
\sigma_{\hat{t}(i,j)}^2 = \frac{\sigma_i^2\,\sigma_j^2\,(1-\rho_{ij}^2)}
   {\sigma_i^2 + \sigma_j^2 - 2\,\rho_{ij}\,\sigma_i\,\sigma_j} ,
\label{eq:correlij}
\end{equation}
namely
\begin{equation}
\rho_{ij} = \mathrm{min}\left(\frac{\sigma_i}{\sigma_j},\frac{\sigma_j}{\sigma_i}\right) ,
\label{eq:correlrho}
\end{equation}
which corresponds to setting $\sigma_{\hat{t}(i,j)}^2=\mathrm{min}(\sigma_i^2,\sigma_j^2)$.
Setting $\rho_{ij}=1$ when $\sigma_i\ne\sigma_j$ can lead to a significant
underestimate of the uncertainty on $\hat{t}$, as can be seen
from Eq.~\ref{eq:correlij}.

Finally, we carefully consider the various sources of error
contributing to the overall uncertainty of an average.
The overall covariance matrix is constructed from a number of
individual sources, \eg\
$\mathbf{V} = \mathbf{V_{stat}+V_{sys}+V_{th}}$.
The variance on the average $\hat{t}$ can be written
\begin{eqnarray}
\sigma^2_{\hat{t}} 
 &=& 
\frac{ \sum_{i,j}\left(\mathbf{V^{-1}}\, 
\mathbf{[V_{stat}+V_{sys}+V_{th}]}\, \mathbf{V^{-1}}\right)_{ij}}
{\left(\sum_{i,j} V^{-1}_{ij}\right)^2}
= \sigma^2_{stat} + \sigma^2_{sys} + \sigma^2_{th} .
\end{eqnarray}
Written in this form, one can readily determine the 
contribution of each source of uncertainty to the overall uncertainty
on the average.  This breakdown of the uncertainties is used 
in the following sections.

Following the prescription described above, the central values and
errors are rescaled to a common set of input parameters in the averaging
procedures according to the dependency on any of these input parameters.
We try to use the most up-to-date values for these common inputs and 
the same values among the HFAG subgroups. For the parameters whose
averages are produced by HFAG, we use the values in the current 
update cycle.  For other external parameters, we use the most
recent PDG values available (usually Ref.~\cite{PDG_2014}). 
The parameters and values used are listed in each subgroup section.

\clearpage
%
%
%
%

%

%
%
%
%

%

\renewcommand{\floatpagefraction}{0.8}
\renewcommand{\topfraction}{0.9}

\newcommand{\comment}[1]{}

\newcommand{\auth}[1]{#1,}
\newcommand{\coll}[1]{#1 Collaboration,}
\newcommand{\authcoll}[2]{#1 \etal\ (#2 Collaboration),}
\newcommand{\authgrp}[2]{#1 \etal\ (#2),}
\newcommand{\titl}[1]{``#1'',} 
\newcommand{\J}[4]{{#1} {\bf #2}, #3 (#4)}
\newcommand{\subJ}[1]{submitted to #1}
\newcommand{\PRL}[3]{\J{Phys.\ Rev.\ Lett.}{#1}{#2}{#3}}
\newcommand{\subPRL}{\subJ{Phys.\ Rev.\ Lett.}}
\newcommand{\PRD}[3]{\J{Phys.\ Rev.\ D}{#1}{#2}{#3}}
\newcommand{\subPRD}{\subJ{Phys.\ Rev.\ D}}
\newcommand{\PREP}[3]{\J{Phys.\ Reports}{#1}{#2}{#3}}
\newcommand{\ZPC}[3]{\J{Z.\ Phys.\ C}{#1}{#2}{#3}}
\newcommand{\PLB}[3]{\J{Phys.\ Lett.\ B}{#1}{#2}{#3}}
\newcommand{\subPLB}{\subJ{Phys.\ Lett.\ B}}
\newcommand{\EPJC}[3]{\J{Eur.\ Phys.\ J.\ C}{#1}{#2}{#3}}
\newcommand{\NPB}[3]{\J{Nucl.\ Phys.\ B}{#1}{#2}{#3}}
\newcommand{\subNPB}{\subJ{Nucl.\ Phys.\ B}}
\newcommand{\NIMA}[3]{\J{Nucl.\ Instrum.\ Methods A}{#1}{#2}{#3}}
\newcommand{\subNIMA}{\subJ{Nucl.\ Instrum.\ Methods A}}
\newcommand{\JHEP}[3]{\J{J.\ of High Energy Physics }{#1}{#2}{#3}}
\newcommand{\JPG}[3]{\J{J.\ of Physics G}{#1}{#2}{#3}}
\newcommand{\ARNS}[3]{\J{Ann.\ Rev.\ Nucl.\ Sci.}{#1}{#2}{#3}}
\newcommand{\newref}{\\}

\newcommand{\particle}[1]{\ensuremath{#1}\xspace}
\renewcommand{\ee}{\particle{e^+e^-}}
\newcommand{\Ups}{\particle{\Upsilon(4S)}}
\newcommand{\Upsfive}{\particle{\Upsilon(5S)}}
\renewcommand{\b}{\particle{b}}
\renewcommand{\B}{\particle{B}}
\newcommand{\Bd}{\particle{B^0}}
\renewcommand{\Bs}{\particle{B^0_s}}
\renewcommand{\Bu}{\particle{B^+}}
\newcommand{\Bc}{\particle{B^+_c}}
\newcommand{\Bdbar}{\particle{\bar{B}^0}}
\newcommand{\Bsbar}{\particle{\bar{B}^0_s}}
\newcommand{\Lb}{\particle{\Lambda_b^0}}
\newcommand{\Xib}{\particle{\Xi_b}}
\newcommand{\Xibd}{\particle{\Xi_b^-}}
\newcommand{\Xibu}{\particle{\Xi_b^0}}
\newcommand{\Omegab}{\particle{\Omega_b^-}}
\newcommand{\Lc}{\particle{\Lambda_c^+}}

\newcommand{\fBs}{\ensuremath{f_{\particle{s}}}\xspace}
\newcommand{\fBd}{\ensuremath{f_{\particle{d}}}\xspace}
\newcommand{\fBu}{\ensuremath{f_{\particle{u}}}\xspace}
\newcommand{\fbb}{\ensuremath{f_{\rm baryon}}\xspace}
\newcommand{\fLb}{\ensuremath{f_{\Lb}}\xspace}
\newcommand{\fXib}{\ensuremath{f_{\Xi_{b}}}\xspace}
\newcommand{\fOb}{\ensuremath{f_{\Omega_{b}}}\xspace}

\newcommand{\dmd}{\ensuremath{\Delta m_{\particle{d}}}\xspace}
\newcommand{\dms}{\ensuremath{\Delta m_{\particle{s}}}\xspace}
\newcommand{\xd}{\ensuremath{x_{\particle{d}}}\xspace}
\newcommand{\xs}{\ensuremath{x_{\particle{s}}}\xspace}
\newcommand{\yd}{\ensuremath{y_{\particle{d}}}\xspace}
\newcommand{\ys}{\ensuremath{y_{\particle{s}}}\xspace}
\newcommand{\chibar}{\ensuremath{\overline{\chi}}\xspace}
\newcommand{\chid}{\ensuremath{\chi_{\particle{d}}}\xspace}
\newcommand{\chis}{\ensuremath{\chi_{\particle{s}}}\xspace}
\newcommand{\Gd}{\ensuremath{\Gamma_{\particle{d}}}\xspace}
\newcommand{\DGd}{\ensuremath{\Delta\Gd}\xspace}
\newcommand{\DGGd}{\ensuremath{\DGd/\Gd}\xspace}
\newcommand{\Gs}{\ensuremath{\Gamma_{\particle{s}}}\xspace}
\newcommand{\DGs}{\ensuremath{\Delta\Gs}\xspace}
\newcommand{\DGGs}{\ensuremath{\Delta\Gs/\Gs}\xspace}
\newcommand{\ASLd}{\ensuremath{{\cal A}_{\rm SL}^\particle{d}}\xspace}
\newcommand{\ASLs}{\ensuremath{{\cal A}_{\rm SL}^\particle{s}}\xspace}
\newcommand{\ASLb}{\ensuremath{{\cal A}_{\rm SL}^\particle{b}}\xspace}

\newcommand{\DG}{\ensuremath{\Delta\Gamma}\xspace}
\newcommand{\phiccbars}{\ensuremath{\phi_s^{c\bar{c}s}}\xspace}

\renewcommand{\BR}[1]{\particle{{\cal B}(#1)}}
\newcommand{\CL}[1]{#1\%~\mbox{CL}}
\newcommand{\Qjet}{\ensuremath{Q_{\rm jet}}\xspace}

\newcommand{\labe}[1]{\label{equ:#1}}
\newcommand{\labs}[1]{\label{sec:#1}}
\newcommand{\labf}[1]{\label{fig:#1}}
\newcommand{\labt}[1]{\label{tab:#1}}
\newcommand{\refe}[1]{\ref{equ:#1}}
\newcommand{\refs}[1]{\ref{sec:#1}}
\newcommand{\reff}[1]{\ref{fig:#1}}
\newcommand{\reft}[1]{\ref{tab:#1}}
\newcommand{\Ref}[1]{Ref.~\cite{#1}}
\newcommand{\Refs}[1]{Refs.~\cite{#1}}
\newcommand{\Refss}[2]{Refs.~\cite{#1} and \cite{#2}}
\newcommand{\Refsss}[3]{Refs.~\cite{#1}, \cite{#2} and \cite{#3}}
\newcommand{\eq}[1]{(\refe{#1})}
\newcommand{\Eq}[1]{Eq.~(\refe{#1})}
\newcommand{\Eqs}[1]{Eqs.~(\refe{#1})}
\newcommand{\Eqss}[2]{Eqs.~(\refe{#1}) and (\refe{#2})}
\newcommand{\Eqssor}[2]{Eqs.~(\refe{#1}) or (\refe{#2})}
\newcommand{\Eqsss}[3]{Eqs.~(\refe{#1}), (\refe{#2}), and (\refe{#3})}
\newcommand{\Figure}[1]{Figure~\reff{#1}}
\newcommand{\Figuress}[2]{Figures~\reff{#1} and \reff{#2}}
\newcommand{\Fig}[1]{Fig.~\reff{#1}}
\newcommand{\Figs}[1]{Figs.~\reff{#1}}
\newcommand{\Figss}[2]{Figs.~\reff{#1} and \reff{#2}}
\newcommand{\Figsss}[3]{Figs.~\reff{#1}, \reff{#2}, and \reff{#3}}
\newcommand{\Section}[1]{Section~\refs{#1}}
\newcommand{\Sec}[1]{Sec.~\refs{#1}}
\newcommand{\Secs}[1]{Secs.~\refs{#1}}
\newcommand{\Secss}[2]{Secs.~\refs{#1} and \refs{#2}}
\newcommand{\Secsss}[3]{Secs.~\refs{#1}, \refs{#2}, and \refs{#3}}
\newcommand{\Table}[1]{Table~\reft{#1}}
\newcommand{\Tables}[1]{Tables~\reft{#1}}
\newcommand{\Tabless}[2]{Tables~\reft{#1} and \reft{#2}}
\newcommand{\Tablesss}[3]{Tables~\reft{#1}, \reft{#2}, and \reft{#3}}

\newcommand{\subsubsubsection}[1]{\vspace{2ex}\par\noindent {\bf\boldmath\em #1} \vspace{2ex}\par}


\newcommand{\definemath}[2]{\newcommand{#1}{\ensuremath{#2}\xspace}}

\definemath{\hfagCHIBARLEPval}{0.1259}
\definemath{\hfagCHIBARLEPerr}{\pm0.0042}
\definemath{\hfagTAUBDval}{1.520}
\definemath{\hfagTAUBDerr}{\pm0.004}
\definemath{\hfagTAUBUval}{1.638}
\definemath{\hfagTAUBUerr}{\pm0.004}
\definemath{\hfagRTAUBUval}{1.076}
\definemath{\hfagRTAUBUerr}{\pm0.004}
\definemath{\hfagTAUBSval}{1.509}
\definemath{\hfagTAUBSerr}{\pm0.004}
\definemath{\hfagRTAUBSval}{0.993}
\definemath{\hfagRTAUBSerr}{\pm0.004}
\definemath{\hfagTAULBval}{1.467}
\definemath{\hfagTAULBerr}{\pm0.010}
\definemath{\hfagTAULBSval}{1.245}
\definemath{\hfagTAULBSerp}{^{+0.071}}
\definemath{\hfagTAULBSern}{_{-0.069}}
\definemath{\hfagTAULBEval}{1.471}
\definemath{\hfagTAULBEerr}{\pm0.010}
\definemath{\hfagTAUXBDval}{1.559}
\definemath{\hfagTAUXBDerr}{\pm0.037}
\definemath{\hfagTAUXBUval}{1.465}
\definemath{\hfagTAUXBUerr}{\pm0.031}
\definemath{\hfagTAUOBval}{1.57}
\definemath{\hfagTAUOBerp}{^{+0.23}}
\definemath{\hfagTAUOBern}{_{-0.20}}
\definemath{\hfagTAUBCval}{0.507}
\definemath{\hfagTAUBCerr}{\pm0.009}
\definemath{\hfagTAUBSSLval}{1.511}
\definemath{\hfagTAUBSSLerr}{\pm0.014}
\definemath{\hfagTAUBSMEANCval}{1.509}
\definemath{\hfagTAUBSMEANCerr}{\pm0.004}
\definemath{\hfagTAUBSJFval}{1.478}
\definemath{\hfagTAUBSJFerr}{\pm0.012}
\definemath{\hfagRTAUBSSLval}{0.994}
\definemath{\hfagRTAUBSSLerr}{\pm0.010}
\definemath{\hfagRTAUBSMEANCval}{0.993}
\definemath{\hfagRTAUBSMEANCerr}{\pm0.004}
\definemath{\hfagRTAUBSMEANCsig}{1.9}
\definemath{\hfagONEMINUSRTAUBSMEANCpercent}{(0.7\pm0.4)\%}
\definemath{\hfagRTAULBval}{0.965}
\definemath{\hfagRTAULBerr}{\pm0.007}
\definemath{\hfagTAUBVTXval}{1.572}
\definemath{\hfagTAUBVTXerr}{\pm0.009}
\definemath{\hfagTAUBLEPval}{1.537}
\definemath{\hfagTAUBLEPerr}{\pm0.020}
\definemath{\hfagTAUBJPval}{1.516}
\definemath{\hfagTAUBJPerr}{\pm0.028}
\definemath{\hfagNSIGMATAULBCDFTWO}{2.5}
\definemath{\hfagSDGDGDval}{0.001}
\definemath{\hfagSDGDGDerr}{\pm0.010}
\definemath{\hfagTAUBSJPSIPIPIval}{1.656}
\definemath{\hfagTAUBSJPSIPIPIerr}{\pm0.033}
\definemath{\hfagTAUBSJPSIKSHORTval}{1.75}
\definemath{\hfagTAUBSJPSIKSHORTerr}{\pm0.14}
\definemath{\hfagTAUBSLONGval}{1.656}
\definemath{\hfagTAUBSLONGerr}{\pm0.033}
\definemath{\hfagTAUBSKKval}{1.408}
\definemath{\hfagTAUBSKKerr}{\pm0.017}
\definemath{\hfagTAUBSDSDSval}{1.379}
\definemath{\hfagTAUBSDSDSerr}{\pm0.031}
\definemath{\hfagTAUBSSHORTval}{1.379}
\definemath{\hfagTAUBSSHORTerr}{\pm0.031}
\definemath{\hfagBRDSDSval}{0.033}
\definemath{\hfagBRDSDSerr}{\pm0.006}
\definemath{\hfagDGSGSBRDSDSval}{+0.069}
\definemath{\hfagDGSGSBRDSDSerr}{\pm0.012}
\definemath{\hfagGSval}{0.6629}
\definemath{\hfagGSerr}{\pm0.0020}
\definemath{\hfagTAUBSMEANval}{1.509}
\definemath{\hfagTAUBSMEANerr}{\pm0.005}
\definemath{\hfagDGSGSval}{+0.116}
\definemath{\hfagDGSGSerr}{\pm0.010}
\definemath{\hfagDGSval}{+0.077}
\definemath{\hfagDGSerr}{\pm0.007}
\definemath{\hfagRHOGSDGS}{-0.353}
\definemath{\hfagTAUBSLval}{1.426}
\definemath{\hfagTAUBSLerr}{\pm0.006}
\definemath{\hfagTAUBSHval}{1.602}
\definemath{\hfagTAUBSHerr}{\pm0.011}
\definemath{\hfagGSCOval}{0.6623}
\definemath{\hfagGSCOerr}{\pm0.0020}
\definemath{\hfagTAUBSMEANCOval}{1.510}
\definemath{\hfagTAUBSMEANCOerr}{\pm0.004}
\definemath{\hfagDGSGSCOval}{+0.123}
\definemath{\hfagDGSGSCOerr}{\pm0.009}
\definemath{\hfagDGSCOval}{+0.082}
\definemath{\hfagDGSCOerr}{\pm0.006}
\definemath{\hfagRHOGSDGSCO}{-0.323}
\definemath{\hfagTAUBSLCOval}{1.422}
\definemath{\hfagTAUBSLCOerr}{\pm0.006}
\definemath{\hfagTAUBSHCOval}{1.609}
\definemath{\hfagTAUBSHCOerr}{\pm0.011}
\definemath{\hfagGSCONval}{0.6628}
\definemath{\hfagGSCONerr}{\pm0.0019}
\definemath{\hfagTAUBSMEANCONval}{1.509}
\definemath{\hfagTAUBSMEANCONerr}{\pm0.004}
\definemath{\hfagDGSGSCONval}{+0.122}
\definemath{\hfagDGSGSCONerr}{\pm0.009}
\definemath{\hfagDGSCONval}{+0.081}
\definemath{\hfagDGSCONerr}{\pm0.006}
\definemath{\hfagRHOGSDGSCON}{-0.271}
\definemath{\hfagTAUBSLCONval}{1.422}
\definemath{\hfagTAUBSLCONerr}{\pm0.006}
\definemath{\hfagTAUBSHCONval}{1.607}
\definemath{\hfagTAUBSHCONerr}{\pm0.010}
\definemath{\hfagFCWval}{0.514}
\definemath{\hfagFCWerr}{\pm0.006}
\definemath{\hfagFNWval}{0.486}
\definemath{\hfagFNWerr}{\pm0.006}
\definemath{\hfagFFWval}{1.058}
\definemath{\hfagFFWerr}{\pm0.024}
\definemath{\hfagNSIGMAFFW}{2.4}
\definemath{\hfagFCNval}{0.513}
\definemath{\hfagFCNerr}{\pm0.013}
\definemath{\hfagFNNval}{0.487}
\definemath{\hfagFNNerr}{\pm0.013}
\definemath{\hfagFFNval}{1.053}
\definemath{\hfagFFNerr}{\pm0.054}
\definemath{\hfagFCval}{0.514}
\definemath{\hfagFCerr}{\pm0.006}
\definemath{\hfagFNval}{0.486}
\definemath{\hfagFNerr}{\pm0.006}
\definemath{\hfagFFval}{1.059}
\definemath{\hfagFFerr}{\pm0.027}
\definemath{\hfagNSIGMAFF}{2.2}
\definemath{\hfagFPRODval}{0.516}
\definemath{\hfagFPRODerr}{\pm0.019}
\definemath{\hfagFSUMval}{1.003}
\definemath{\hfagFSUMerr}{\pm0.029}
\definemath{\hfagFSFIVEOSval}{0.206}
\definemath{\hfagFSFIVEOSsta}{\pm0.010}
\definemath{\hfagFSFIVEOSsys}{\pm0.024}
\definemath{\hfagFSFIVEOSerr}{\pm0.027}
\definemath{\hfagFSFIVERLval}{0.215}
\definemath{\hfagFSFIVERLerr}{\pm0.031}
\definemath{\hfagFUDFIVEval}{0.761}
\definemath{\hfagFUDFIVEerp}{^{+0.027}}
\definemath{\hfagFUDFIVEern}{_{-0.042}}
\definemath{\hfagFSFIVEval}{0.200}
\definemath{\hfagFSFIVEerp}{^{+0.030}}
\definemath{\hfagFSFIVEern}{_{-0.031}}
\definemath{\hfagFSFUDFIVEval}{0.263}
\definemath{\hfagFSFUDFIVEerp}{^{+0.052}}
\definemath{\hfagFSFUDFIVEern}{_{-0.044}}
\definemath{\hfagFNBFIVEval}{0.039}
\definemath{\hfagFNBFIVEerp}{^{+0.050}}
\definemath{\hfagFNBFIVEern}{_{-0.004}}
\definemath{\hfagRBSTEVNOCONval}{0.211}
\definemath{\hfagRBSTEVNOCONerr}{\pm0.054}
\definemath{\hfagRLBTEVNOCONval}{0.212}
\definemath{\hfagRLBTEVNOCONerr}{\pm0.058}
\definemath{\hfagRBSLHCBNOCONval}{0.131}
\definemath{\hfagRBSLHCBNOCONerr}{\pm0.009}
\definemath{\hfagRLBLHCBNOCONval}{0.223}
\definemath{\hfagRLBLHCBNOCONerr}{\pm0.022}
\definemath{\hfagZFSFACTOR}{}
\definemath{\hfagZFBSNOMIXval}{0.088}
\definemath{\hfagZFBSNOMIXerr}{\pm0.013}
\definemath{\hfagZFBBNOMIXval}{0.083}
\definemath{\hfagZFBBNOMIXerr}{\pm0.011}
\definemath{\hfagZFBDNOMIXval}{0.414}
\definemath{\hfagZFBDNOMIXerr}{\pm0.008}
\definemath{\hfagWFSFACTOR}{1.0}
\definemath{\hfagWFBSNOMIXval}{0.104}
\definemath{\hfagWFBSNOMIXerr}{\pm0.006}
\definemath{\hfagWFBBNOMIXval}{0.084}
\definemath{\hfagWFBBNOMIXerr}{\pm0.011}
\definemath{\hfagWFBDNOMIXval}{0.406}
\definemath{\hfagWFBDNOMIXerr}{\pm0.005}
\definemath{\hfagTFSFACTOR}{}
\definemath{\hfagTFBSNOMIXval}{0.097}
\definemath{\hfagTFBSNOMIXerr}{\pm0.012}
\definemath{\hfagTFBBNOMIXval}{0.207}
\definemath{\hfagTFBBNOMIXerr}{\pm0.046}
\definemath{\hfagTFBDNOMIXval}{0.348}
\definemath{\hfagTFBDNOMIXerr}{\pm0.020}
\definemath{\hfagLFSFACTOR}{}
\definemath{\hfagLFBSNOMIXval}{0.090}
\definemath{\hfagLFBSNOMIXerr}{\pm0.006}
\definemath{\hfagLFBBNOMIXval}{0.216}
\definemath{\hfagLFBBNOMIXerr}{\pm0.020}
\definemath{\hfagLFBDNOMIXval}{0.347}
\definemath{\hfagLFBDNOMIXerr}{\pm0.009}
\definemath{\hfagCHIBARTEVval}{0.127}
\definemath{\hfagCHIBARTEVerr}{\pm0.008}
\definemath{\hfagCHIBARval}{0.1260}
\definemath{\hfagCHIBARerr}{\pm0.0037}
\definemath{\hfagWFBSMIXval}{0.109}
\definemath{\hfagWFBSMIXerr}{\pm0.010}
\definemath{\hfagTFBSMIXval}{0.110}
\definemath{\hfagTFBSMIXerr}{\pm0.019}
\definemath{\hfagZFBSMIXval}{0.108}
\definemath{\hfagZFBSMIXerr}{\pm0.011}
\definemath{\hfagCHIDUval}{0.182}
\definemath{\hfagCHIDUerr}{\pm0.015}
\definemath{\hfagCHIDWUval}{0.1875}
\definemath{\hfagCHIDWUerr}{\pm0.0017}
\definemath{\hfagXDWval}{0.775}
\definemath{\hfagXDWerr}{\pm0.006}
\definemath{\hfagXDWUval}{0.775}
\definemath{\hfagXDWUerr}{\pm0.006}
\definemath{\hfagDMDWval}{0.510}
\definemath{\hfagDMDWsta}{\pm0.003}
\definemath{\hfagDMDWsys}{\pm0.002}
\definemath{\hfagDMDWerr}{\pm0.003}
\definemath{\hfagDMDWUval}{0.510}
\definemath{\hfagDMDWUerr}{\pm0.003}
\definemath{\hfagDMDLval}{0.514}
\definemath{\hfagDMDLsta}{\pm0.005}
\definemath{\hfagDMDLsys}{\pm0.003}
\definemath{\hfagDMDLerr}{\pm0.006}
\definemath{\hfagZFBSval}{0.100}
\definemath{\hfagZFBSerr}{\pm0.008}
\definemath{\hfagZFBBval}{0.080}
\definemath{\hfagZFBBerr}{\pm0.010}
\definemath{\hfagZFBDval}{0.410}
\definemath{\hfagZFBDerr}{\pm0.007}
\definemath{\hfagZRHOFBBFBS}{+0.053}
\definemath{\hfagZRHOFBDFBS}{-0.646}
\definemath{\hfagZRHOFBDFBB}{-0.797}
\definemath{\hfagWFBSval}{0.105}
\definemath{\hfagWFBSerr}{\pm0.005}
\definemath{\hfagWFBBval}{0.083}
\definemath{\hfagWFBBerr}{\pm0.010}
\definemath{\hfagWFBDval}{0.406}
\definemath{\hfagWFBDerr}{\pm0.005}
\definemath{\hfagWRHOFBBFBS}{-0.096}
\definemath{\hfagWRHOFBDFBS}{-0.350}
\definemath{\hfagWRHOFBDFBB}{-0.899}
\definemath{\hfagTFBSval}{0.100}
\definemath{\hfagTFBSerr}{\pm0.010}
\definemath{\hfagTFBBval}{0.199}
\definemath{\hfagTFBBerr}{\pm0.044}
\definemath{\hfagTFBDval}{0.350}
\definemath{\hfagTFBDerr}{\pm0.020}
\definemath{\hfagTRHOFBBFBS}{-0.459}
\definemath{\hfagTRHOFBDFBS}{+0.255}
\definemath{\hfagTRHOFBDFBB}{-0.976}
\definemath{\hfagLFBSval}{0.093}
\definemath{\hfagLFBSerr}{\pm0.006}
\definemath{\hfagLFBBval}{0.213}
\definemath{\hfagLFBBerr}{\pm0.020}
\definemath{\hfagLFBDval}{0.347}
\definemath{\hfagLFBDerr}{\pm0.009}
\definemath{\hfagLRHOFBBFBS}{-0.396}
\definemath{\hfagLRHOFBDFBS}{+0.105}
\definemath{\hfagLRHOFBDFBB}{-0.955}
\definemath{\hfagZFBSBDval}{0.243}
\definemath{\hfagZFBSBDerr}{\pm0.023}
\definemath{\hfagWFBSBDval}{0.259}
\definemath{\hfagWFBSBDerr}{\pm0.013}
\definemath{\hfagTFBSBDval}{0.286}
\definemath{\hfagTFBSBDerr}{\pm0.029}
\definemath{\hfagLFBSBDval}{0.266}
\definemath{\hfagLFBSBDerr}{\pm0.018}
\definemath{\hfagDMDHval}{0.496}
\definemath{\hfagDMDHsta}{\pm0.010}
\definemath{\hfagDMDHsys}{\pm0.009}
\definemath{\hfagDMDHerr}{\pm0.013}
\definemath{\hfagDMDBval}{0.509}
\definemath{\hfagDMDBsta}{\pm0.003}
\definemath{\hfagDMDBsys}{\pm0.003}
\definemath{\hfagDMDBerr}{\pm0.005}
\definemath{\hfagDMDTWODval}{0.509}
\definemath{\hfagDMDTWODsta}{\pm0.004}
\definemath{\hfagDMDTWODsys}{\pm0.004}
\definemath{\hfagDMDTWODerr}{\pm0.006}
\definemath{\hfagTAUBDTWODval}{1.527}
\definemath{\hfagTAUBDTWODsta}{\pm0.006}
\definemath{\hfagTAUBDTWODsys}{\pm0.008}
\definemath{\hfagTAUBDTWODerr}{\pm0.010}
\definemath{\hfagRHOstaDMDTAUBD}{-0.19}
\definemath{\hfagRHOsysDMDTAUBD}{-0.25}
\definemath{\hfagRHODMDTAUBD}{-0.23}
\definemath{\hfagZRHOTAUHTAUL}{-0.464}
\definemath{\hfagTAUBZCALCval}{1.567}
\definemath{\hfagTAUBZCALCerr}{\pm0.003}
\definemath{\hfagQPDBval}{1.0009}
\definemath{\hfagQPDBerr}{\pm0.0013}
\definemath{\hfagQPDDval}{1.0000}
\definemath{\hfagQPDDerr}{\pm0.0010}
\definemath{\hfagQPDWval}{1.0005}
\definemath{\hfagQPDWerr}{\pm0.0009}
\definemath{\hfagQPDAval}{1.0005}
\definemath{\hfagQPDAerr}{\pm0.0009}
\definemath{\hfagASLDBval}{-0.0019}
\definemath{\hfagASLDBerr}{\pm0.0027}
\definemath{\hfagASLDDval}{+0.0001}
\definemath{\hfagASLDDerr}{\pm0.0020}
\definemath{\hfagASLDWval}{-0.0010}
\definemath{\hfagASLDWerr}{\pm0.0018}
\definemath{\hfagASLDAval}{-0.0010}
\definemath{\hfagASLDAerr}{\pm0.0018}
\definemath{\hfagREBDBval}{-0.0005}
\definemath{\hfagREBDBerr}{\pm0.0007}
\definemath{\hfagREBDDval}{+0.0000}
\definemath{\hfagREBDDerr}{\pm0.0005}
\definemath{\hfagREBDWval}{-0.0002}
\definemath{\hfagREBDWerr}{\pm0.0004}
\definemath{\hfagREBDAval}{-0.0002}
\definemath{\hfagREBDAerr}{\pm0.0004}
\definemath{\hfagASLSDval}{-0.0048}
\definemath{\hfagASLSDerr}{\pm0.0048}
\definemath{\hfagASLSWval}{-0.0083}
\definemath{\hfagASLSWsta}{\pm0.0027}
\definemath{\hfagASLSWsys}{\pm0.0021}
\definemath{\hfagASLSWerr}{\pm0.0034}
\definemath{\hfagQPSWval}{1.0042}
\definemath{\hfagQPSWsta}{\pm0.0014}
\definemath{\hfagQPSWsys}{\pm0.0010}
\definemath{\hfagQPSWerr}{\pm0.0017}
\definemath{\hfagASLSval}{-0.0075}
\definemath{\hfagASLSerr}{\pm0.0041}
\definemath{\hfagQPSval}{1.0038}
\definemath{\hfagQPSerr}{\pm0.0021}
\definemath{\hfagASLDval}{-0.0015}
\definemath{\hfagASLDerr}{\pm0.0017}
\definemath{\hfagQPDval}{1.0007}
\definemath{\hfagQPDerr}{\pm0.0009}
\definemath{\hfagRHOASLSASLD}{-0.158}
\definemath{\hfagREBDval}{-0.0004}
\definemath{\hfagREBDerr}{\pm0.0004}
\definemath{\hfagASLDASLSNSIGMA}{1.5}
\definemath{\hfagASLDASLSPVALPERCENT}{12.4}
\definemath{\hfagTANPHIval}{-1.7}
\definemath{\hfagTANPHIerr}{\pm0.9}
\definemath{\hfagDMSval}{17.757}
\definemath{\hfagDMSsta}{\pm0.020}
\definemath{\hfagDMSsys}{\pm0.007}
\definemath{\hfagDMSerr}{\pm0.021}
\definemath{\hfagXSval}{26.79}
\definemath{\hfagXSerr}{\pm0.08}
\definemath{\hfagCHISval}{0.499307}
\definemath{\hfagCHISerr}{\pm0.000004}
\definemath{\hfagRATIODMDDMSval}{0.02870}
\definemath{\hfagRATIODMDDMSerr}{\pm0.00020}
\definemath{\hfagVTDVTSval}{0.2166}
\definemath{\hfagVTDVTSexx}{\pm0.0007}
\definemath{\hfagVTDVTSthe}{\pm0.0108}
\definemath{\hfagVTDVTSerr}{\pm0.0108}
\definemath{\hfagXIval}{1.268}
\definemath{\hfagXIerr}{\pm0.063}
\definemath{\hfagBETASCOMBval}{+0.007}
\definemath{\hfagBETASCOMBerp}{^{+0.018}}
\definemath{\hfagBETASCOMBern}{_{-0.018}}
\definemath{\hfagPHISCOMBval}{-0.015}
\definemath{\hfagPHISCOMBerr}{\pm0.035}
\definemath{\hfagDGSCOMBval}{+0.081}
\definemath{\hfagDGSCOMBerr}{\pm0.007}
\definemath{\hfagNSIGMASM}{XXX}
\definemath{\hfagBETASCOMBCONval}{+4999.500}
\definemath{\hfagBETASCOMBCONerp}{^{+-4999.500}}
\definemath{\hfagBETASCOMBCONern}{_{--4999.500}}
\definemath{\hfagPHISCOMBCONval}{-9999.000}
\definemath{\hfagPHISCOMBCONerp}{^{+-9999.000}}
\definemath{\hfagPHISCOMBCONern}{_{--9999.000}}
\definemath{\hfagDGSCOMBCONval}{-9999.000}
\definemath{\hfagDGSCOMBCONerp}{^{+-9999.000}}
\definemath{\hfagDGSCOMBCONern}{_{--9999.000}}
\definemath{\hfagNSIGMASMCON}{-9999}
\definemath{\hfagPHISSMval}{-0.0363}
\definemath{\hfagPHISSMerp}{^{+0.0012}}
\definemath{\hfagPHISSMern}{_{-0.0014}}
\definemath{\hfagPHISTWELVESMval}{0.0038}
\definemath{\hfagPHISTWELVESMerr}{\pm0.0010}
\definemath{\hfagPHISTWELVEval}{0.025}
\definemath{\hfagPHISTWELVEerr}{\pm0.035}

\newcommand{\unit}[1]{~\ensuremath{\rm #1}\xspace}
\renewcommand{\ps}{\unit{ps}}
\newcommand{\invps}{\unit{ps^{-1}}}
\newcommand{\TeV}{\unit{TeV}}
\newcommand{\MeVcc}{\unit{MeV/\mbox{$c$}^2}}
\newcommand{\MeV}{\unit{MeV}}

\definemath{\hfagCHIBARLEP}{\hfagCHIBARLEPval\hfagCHIBARLEPerr}
\definemath{\hfagTAUBD}{\hfagTAUBDval\hfagTAUBDerr\ps}
\definemath{\hfagTAUBDnounit}{\hfagTAUBDval\hfagTAUBDerr}
\definemath{\hfagTAUBU}{\hfagTAUBUval\hfagTAUBUerr\ps}
\definemath{\hfagTAUBUnounit}{\hfagTAUBUval\hfagTAUBUerr}
\definemath{\hfagRTAUBU}{\hfagRTAUBUval\hfagRTAUBUerr}
\definemath{\hfagTAUBS}{\hfagTAUBSval\hfagTAUBSerr\ps}
\definemath{\hfagTAUBSnounit}{\hfagTAUBSval\hfagTAUBSerr}
\definemath{\hfagRTAUBS}{\hfagRTAUBSval\hfagRTAUBSerr}
\definemath{\hfagTAULB}{\hfagTAULBval\hfagTAULBerr\ps}
\definemath{\hfagTAULBnounit}{\hfagTAULBval\hfagTAULBerr}
\definemath{\hfagTAULBSerr}{\hfagTAULBSerp\hfagTAULBSern}
\definemath{\hfagTAULBS}{\hfagTAULBSval\hfagTAULBSerr\ps}
\definemath{\hfagTAULBSnounit}{\hfagTAULBSval\hfagTAULBSerr}
\definemath{\hfagTAULBE}{\hfagTAULBEval\hfagTAULBEerr\ps}
\definemath{\hfagTAULBEnounit}{\hfagTAULBEval\hfagTAULBEerr}
\definemath{\hfagTAUXBD}{\hfagTAUXBDval\hfagTAUXBDerr\ps}
\definemath{\hfagTAUXBDnounit}{\hfagTAUXBDval\hfagTAUXBDerr}
\definemath{\hfagTAUXBU}{\hfagTAUXBUval\hfagTAUXBUerr\ps}
\definemath{\hfagTAUXBUnounit}{\hfagTAUXBUval\hfagTAUXBUerr}
\definemath{\hfagTAUOBerr}{\hfagTAUOBerp\hfagTAUOBern}
\definemath{\hfagTAUOB}{\hfagTAUOBval\hfagTAUOBerr\ps}
\definemath{\hfagTAUOBnounit}{\hfagTAUOBval\hfagTAUOBerr}
\definemath{\hfagTAUBC}{\hfagTAUBCval\hfagTAUBCerr\ps}
\definemath{\hfagTAUBCnounit}{\hfagTAUBCval\hfagTAUBCerr}
\definemath{\hfagTAUBSSL}{\hfagTAUBSSLval\hfagTAUBSSLerr\ps}
\definemath{\hfagTAUBSSLnounit}{\hfagTAUBSSLval\hfagTAUBSSLerr}
\definemath{\hfagTAUBSMEANC}{\hfagTAUBSMEANCval\hfagTAUBSMEANCerr\ps}
\definemath{\hfagTAUBSMEANCnounit}{\hfagTAUBSMEANCval\hfagTAUBSMEANCerr}
\definemath{\hfagTAUBSJF}{\hfagTAUBSJFval\hfagTAUBSJFerr\ps}
\definemath{\hfagTAUBSJFnounit}{\hfagTAUBSJFval\hfagTAUBSJFerr}
\definemath{\hfagRTAUBSSL}{\hfagRTAUBSSLval\hfagRTAUBSSLerr}
\definemath{\hfagRTAUBSMEANC}{\hfagRTAUBSMEANCval\hfagRTAUBSMEANCerr}
\definemath{\hfagRTAULB}{\hfagRTAULBval\hfagRTAULBerr}
\definemath{\hfagTAUBVTX}{\hfagTAUBVTXval\hfagTAUBVTXerr\ps}
\definemath{\hfagTAUBVTXnounit}{\hfagTAUBVTXval\hfagTAUBVTXerr}
\definemath{\hfagTAUBLEP}{\hfagTAUBLEPval\hfagTAUBLEPerr\ps}
\definemath{\hfagTAUBLEPnounit}{\hfagTAUBLEPval\hfagTAUBLEPerr}
\definemath{\hfagTAUBJP}{\hfagTAUBJPval\hfagTAUBJPerr\ps}
\definemath{\hfagTAUBJPnounit}{\hfagTAUBJPval\hfagTAUBJPerr}
\definemath{\hfagSDGDGD}{\hfagSDGDGDval\hfagSDGDGDerr}
\definemath{\hfagTAUBSJPSIPIPI}{\hfagTAUBSJPSIPIPIval\hfagTAUBSJPSIPIPIerr\ps}
\definemath{\hfagTAUBSJPSIPIPInounit}{\hfagTAUBSJPSIPIPIval\hfagTAUBSJPSIPIPIerr}
\definemath{\hfagTAUBSJPSIKSHORT}{\hfagTAUBSJPSIKSHORTval\hfagTAUBSJPSIKSHORTerr\ps}
\definemath{\hfagTAUBSJPSIKSHORTnounit}{\hfagTAUBSJPSIKSHORTval\hfagTAUBSJPSIKSHORTerr}
\definemath{\hfagTAUBSLONG}{\hfagTAUBSLONGval\hfagTAUBSLONGerr\ps}
\definemath{\hfagTAUBSLONGnounit}{\hfagTAUBSLONGval\hfagTAUBSLONGerr}
\definemath{\hfagTAUBSKK}{\hfagTAUBSKKval\hfagTAUBSKKerr\ps}
\definemath{\hfagTAUBSKKnounit}{\hfagTAUBSKKval\hfagTAUBSKKerr}
\definemath{\hfagTAUBSDSDS}{\hfagTAUBSDSDSval\hfagTAUBSDSDSerr\ps}
\definemath{\hfagTAUBSDSDSnounit}{\hfagTAUBSDSDSval\hfagTAUBSDSDSerr}
\definemath{\hfagTAUBSSHORT}{\hfagTAUBSSHORTval\hfagTAUBSSHORTerr\ps}
\definemath{\hfagTAUBSSHORTnounit}{\hfagTAUBSSHORTval\hfagTAUBSSHORTerr}
\definemath{\hfagBRDSDS}{\hfagBRDSDSval\hfagBRDSDSerr}
\definemath{\hfagDGSGSBRDSDS}{\hfagDGSGSBRDSDSval\hfagDGSGSBRDSDSerr}
\definemath{\hfagGS}{\hfagGSval\hfagGSerr\invps}
\definemath{\hfagGSnounit}{\hfagGSval\hfagGSerr}
\definemath{\hfagTAUBSMEAN}{\hfagTAUBSMEANval\hfagTAUBSMEANerr\ps}
\definemath{\hfagTAUBSMEANnounit}{\hfagTAUBSMEANval\hfagTAUBSMEANerr}
\definemath{\hfagDGSGS}{\hfagDGSGSval\hfagDGSGSerr}
\definemath{\hfagDGS}{\hfagDGSval\hfagDGSerr\invps}
\definemath{\hfagDGSnounit}{\hfagDGSval\hfagDGSerr}
\definemath{\hfagTAUBSL}{\hfagTAUBSLval\hfagTAUBSLerr\ps}
\definemath{\hfagTAUBSLnounit}{\hfagTAUBSLval\hfagTAUBSLerr}
\definemath{\hfagTAUBSH}{\hfagTAUBSHval\hfagTAUBSHerr\ps}
\definemath{\hfagTAUBSHnounit}{\hfagTAUBSHval\hfagTAUBSHerr}
\definemath{\hfagGSCO}{\hfagGSCOval\hfagGSCOerr\invps}
\definemath{\hfagGSCOnounit}{\hfagGSCOval\hfagGSCOerr}
\definemath{\hfagTAUBSMEANCO}{\hfagTAUBSMEANCOval\hfagTAUBSMEANCOerr\ps}
\definemath{\hfagTAUBSMEANCOnounit}{\hfagTAUBSMEANCOval\hfagTAUBSMEANCOerr}
\definemath{\hfagDGSGSCO}{\hfagDGSGSCOval\hfagDGSGSCOerr}
\definemath{\hfagDGSCO}{\hfagDGSCOval\hfagDGSCOerr\invps}
\definemath{\hfagDGSCOnounit}{\hfagDGSCOval\hfagDGSCOerr}
\definemath{\hfagTAUBSLCO}{\hfagTAUBSLCOval\hfagTAUBSLCOerr\ps}
\definemath{\hfagTAUBSLCOnounit}{\hfagTAUBSLCOval\hfagTAUBSLCOerr}
\definemath{\hfagTAUBSHCO}{\hfagTAUBSHCOval\hfagTAUBSHCOerr\ps}
\definemath{\hfagTAUBSHCOnounit}{\hfagTAUBSHCOval\hfagTAUBSHCOerr}
\definemath{\hfagGSCON}{\hfagGSCONval\hfagGSCONerr\invps}
\definemath{\hfagGSCONnounit}{\hfagGSCONval\hfagGSCONerr}
\definemath{\hfagTAUBSMEANCON}{\hfagTAUBSMEANCONval\hfagTAUBSMEANCONerr\ps}
\definemath{\hfagTAUBSMEANCONnounit}{\hfagTAUBSMEANCONval\hfagTAUBSMEANCONerr}
\definemath{\hfagDGSGSCON}{\hfagDGSGSCONval\hfagDGSGSCONerr}
\definemath{\hfagDGSCON}{\hfagDGSCONval\hfagDGSCONerr\invps}
\definemath{\hfagDGSCONnounit}{\hfagDGSCONval\hfagDGSCONerr}
\definemath{\hfagTAUBSLCON}{\hfagTAUBSLCONval\hfagTAUBSLCONerr\ps}
\definemath{\hfagTAUBSLCONnounit}{\hfagTAUBSLCONval\hfagTAUBSLCONerr}
\definemath{\hfagTAUBSHCON}{\hfagTAUBSHCONval\hfagTAUBSHCONerr\ps}
\definemath{\hfagTAUBSHCONnounit}{\hfagTAUBSHCONval\hfagTAUBSHCONerr}
\definemath{\hfagFCW}{\hfagFCWval\hfagFCWerr}
\definemath{\hfagFNW}{\hfagFNWval\hfagFNWerr}
\definemath{\hfagFFW}{\hfagFFWval\hfagFFWerr}
\definemath{\hfagFCN}{\hfagFCNval\hfagFCNerr}
\definemath{\hfagFNN}{\hfagFNNval\hfagFNNerr}
\definemath{\hfagFFN}{\hfagFFNval\hfagFFNerr}
\definemath{\hfagFC}{\hfagFCval\hfagFCerr}
\definemath{\hfagFN}{\hfagFNval\hfagFNerr}
\definemath{\hfagFF}{\hfagFFval\hfagFFerr}
\definemath{\hfagFPROD}{\hfagFPRODval\hfagFPRODerr}
\definemath{\hfagFSUM}{\hfagFSUMval\hfagFSUMerr}
\definemath{\hfagFSFIVEOS}{\hfagFSFIVEOSval\hfagFSFIVEOSerr}
\definemath{\hfagFSFIVEOSfull}{\hfagFSFIVEOSval\hfagFSFIVEOSsta\hfagFSFIVEOSsys}
\definemath{\hfagFSFIVERL}{\hfagFSFIVERLval\hfagFSFIVERLerr}
\definemath{\hfagFUDFIVEerr}{\hfagFUDFIVEerp\hfagFUDFIVEern}
\definemath{\hfagFUDFIVE}{\hfagFUDFIVEval\hfagFUDFIVEerr}
\definemath{\hfagFSFIVEerr}{\hfagFSFIVEerp\hfagFSFIVEern}
\definemath{\hfagFSFIVE}{\hfagFSFIVEval\hfagFSFIVEerr}
\definemath{\hfagFSFUDFIVEerr}{\hfagFSFUDFIVEerp\hfagFSFUDFIVEern}
\definemath{\hfagFSFUDFIVE}{\hfagFSFUDFIVEval\hfagFSFUDFIVEerr}
\definemath{\hfagFNBFIVEerr}{\hfagFNBFIVEerp\hfagFNBFIVEern}
\definemath{\hfagFNBFIVE}{\hfagFNBFIVEval\hfagFNBFIVEerr}
\definemath{\hfagRBSTEVNOCON}{\hfagRBSTEVNOCONval\hfagRBSTEVNOCONerr}
\definemath{\hfagRLBTEVNOCON}{\hfagRLBTEVNOCONval\hfagRLBTEVNOCONerr}
\definemath{\hfagRBSLHCBNOCON}{\hfagRBSLHCBNOCONval\hfagRBSLHCBNOCONerr}
\definemath{\hfagRLBLHCBNOCON}{\hfagRLBLHCBNOCONval\hfagRLBLHCBNOCONerr}
\definemath{\hfagZFBSNOMIX}{\hfagZFBSNOMIXval\hfagZFBSNOMIXerr}
\definemath{\hfagZFBBNOMIX}{\hfagZFBBNOMIXval\hfagZFBBNOMIXerr}
\definemath{\hfagZFBDNOMIX}{\hfagZFBDNOMIXval\hfagZFBDNOMIXerr}
\definemath{\hfagWFBSNOMIX}{\hfagWFBSNOMIXval\hfagWFBSNOMIXerr}
\definemath{\hfagWFBBNOMIX}{\hfagWFBBNOMIXval\hfagWFBBNOMIXerr}
\definemath{\hfagWFBDNOMIX}{\hfagWFBDNOMIXval\hfagWFBDNOMIXerr}
\definemath{\hfagTFBSNOMIX}{\hfagTFBSNOMIXval\hfagTFBSNOMIXerr}
\definemath{\hfagTFBBNOMIX}{\hfagTFBBNOMIXval\hfagTFBBNOMIXerr}
\definemath{\hfagTFBDNOMIX}{\hfagTFBDNOMIXval\hfagTFBDNOMIXerr}
\definemath{\hfagLFBSNOMIX}{\hfagLFBSNOMIXval\hfagLFBSNOMIXerr}
\definemath{\hfagLFBBNOMIX}{\hfagLFBBNOMIXval\hfagLFBBNOMIXerr}
\definemath{\hfagLFBDNOMIX}{\hfagLFBDNOMIXval\hfagLFBDNOMIXerr}
\definemath{\hfagCHIBARTEV}{\hfagCHIBARTEVval\hfagCHIBARTEVerr}
\definemath{\hfagCHIBAR}{\hfagCHIBARval\hfagCHIBARerr}
\definemath{\hfagWFBSMIX}{\hfagWFBSMIXval\hfagWFBSMIXerr}
\definemath{\hfagTFBSMIX}{\hfagTFBSMIXval\hfagTFBSMIXerr}
\definemath{\hfagZFBSMIX}{\hfagZFBSMIXval\hfagZFBSMIXerr}
\definemath{\hfagCHIDU}{\hfagCHIDUval\hfagCHIDUerr}
\definemath{\hfagCHIDWU}{\hfagCHIDWUval\hfagCHIDWUerr}
\definemath{\hfagXDW}{\hfagXDWval\hfagXDWerr}
\definemath{\hfagXDWU}{\hfagXDWUval\hfagXDWUerr}
\definemath{\hfagDMDW}{\hfagDMDWval\hfagDMDWerr\invps}
\definemath{\hfagDMDWnounit}{\hfagDMDWval\hfagDMDWerr}
\definemath{\hfagDMDWfull}{\hfagDMDWval\hfagDMDWsta\hfagDMDWsys\invps}
\definemath{\hfagDMDWnounitfull}{\hfagDMDWval\hfagDMDWsta\hfagDMDWsys}
\definemath{\hfagDMDWU}{\hfagDMDWUval\hfagDMDWUerr\invps}
\definemath{\hfagDMDWUnounit}{\hfagDMDWUval\hfagDMDWUerr}
\definemath{\hfagDMDL}{\hfagDMDLval\hfagDMDLerr\invps}
\definemath{\hfagDMDLnounit}{\hfagDMDLval\hfagDMDLerr}
\definemath{\hfagDMDLfull}{\hfagDMDLval\hfagDMDLsta\hfagDMDLsys\invps}
\definemath{\hfagDMDLnounitfull}{\hfagDMDLval\hfagDMDLsta\hfagDMDLsys}
\definemath{\hfagZFBS}{\hfagZFBSval\hfagZFBSerr}
\definemath{\hfagZFBB}{\hfagZFBBval\hfagZFBBerr}
\definemath{\hfagZFBD}{\hfagZFBDval\hfagZFBDerr}
\definemath{\hfagWFBS}{\hfagWFBSval\hfagWFBSerr}
\definemath{\hfagWFBB}{\hfagWFBBval\hfagWFBBerr}
\definemath{\hfagWFBD}{\hfagWFBDval\hfagWFBDerr}
\definemath{\hfagTFBS}{\hfagTFBSval\hfagTFBSerr}
\definemath{\hfagTFBB}{\hfagTFBBval\hfagTFBBerr}
\definemath{\hfagTFBD}{\hfagTFBDval\hfagTFBDerr}
\definemath{\hfagLFBS}{\hfagLFBSval\hfagLFBSerr}
\definemath{\hfagLFBB}{\hfagLFBBval\hfagLFBBerr}
\definemath{\hfagLFBD}{\hfagLFBDval\hfagLFBDerr}
\definemath{\hfagZFBSBD}{\hfagZFBSBDval\hfagZFBSBDerr}
\definemath{\hfagWFBSBD}{\hfagWFBSBDval\hfagWFBSBDerr}
\definemath{\hfagTFBSBD}{\hfagTFBSBDval\hfagTFBSBDerr}
\definemath{\hfagLFBSBD}{\hfagLFBSBDval\hfagLFBSBDerr}
\definemath{\hfagDMDH}{\hfagDMDHval\hfagDMDHerr\invps}
\definemath{\hfagDMDHnounit}{\hfagDMDHval\hfagDMDHerr}
\definemath{\hfagDMDHfull}{\hfagDMDHval\hfagDMDHsta\hfagDMDHsys\invps}
\definemath{\hfagDMDHnounitfull}{\hfagDMDHval\hfagDMDHsta\hfagDMDHsys}
\definemath{\hfagDMDB}{\hfagDMDBval\hfagDMDBerr\invps}
\definemath{\hfagDMDBnounit}{\hfagDMDBval\hfagDMDBerr}
\definemath{\hfagDMDBfull}{\hfagDMDBval\hfagDMDBsta\hfagDMDBsys\invps}
\definemath{\hfagDMDBnounitfull}{\hfagDMDBval\hfagDMDBsta\hfagDMDBsys}
\definemath{\hfagDMDTWOD}{\hfagDMDTWODval\hfagDMDTWODerr\invps}
\definemath{\hfagDMDTWODnounit}{\hfagDMDTWODval\hfagDMDTWODerr}
\definemath{\hfagDMDTWODfull}{\hfagDMDTWODval\hfagDMDTWODsta\hfagDMDTWODsys\invps}
\definemath{\hfagDMDTWODnounitfull}{\hfagDMDTWODval\hfagDMDTWODsta\hfagDMDTWODsys}
\definemath{\hfagTAUBDTWOD}{\hfagTAUBDTWODval\hfagTAUBDTWODerr\ps}
\definemath{\hfagTAUBDTWODnounit}{\hfagTAUBDTWODval\hfagTAUBDTWODerr}
\definemath{\hfagTAUBDTWODfull}{\hfagTAUBDTWODval\hfagTAUBDTWODsta\hfagTAUBDTWODsys\ps}
\definemath{\hfagTAUBDTWODnounitfull}{\hfagTAUBDTWODval\hfagTAUBDTWODsta\hfagTAUBDTWODsys}
\definemath{\hfagTAUBZCALC}{\hfagTAUBZCALCval\hfagTAUBZCALCerr\ps}
\definemath{\hfagTAUBZCALCnounit}{\hfagTAUBZCALCval\hfagTAUBZCALCerr}
\definemath{\hfagQPDB}{\hfagQPDBval\hfagQPDBerr}
\definemath{\hfagQPDD}{\hfagQPDDval\hfagQPDDerr}
\definemath{\hfagQPDW}{\hfagQPDWval\hfagQPDWerr}
\definemath{\hfagQPDA}{\hfagQPDAval\hfagQPDAerr}
\definemath{\hfagASLDB}{\hfagASLDBval\hfagASLDBerr}
\definemath{\hfagASLDD}{\hfagASLDDval\hfagASLDDerr}
\definemath{\hfagASLDW}{\hfagASLDWval\hfagASLDWerr}
\definemath{\hfagASLDA}{\hfagASLDAval\hfagASLDAerr}
\definemath{\hfagREBDB}{\hfagREBDBval\hfagREBDBerr}
\definemath{\hfagREBDD}{\hfagREBDDval\hfagREBDDerr}
\definemath{\hfagREBDW}{\hfagREBDWval\hfagREBDWerr}
\definemath{\hfagREBDA}{\hfagREBDAval\hfagREBDAerr}
\definemath{\hfagASLSD}{\hfagASLSDval\hfagASLSDerr}
\definemath{\hfagASLSW}{\hfagASLSWval\hfagASLSWerr}
\definemath{\hfagASLSWfull}{\hfagASLSWval\hfagASLSWsta\hfagASLSWsys}
\definemath{\hfagQPSW}{\hfagQPSWval\hfagQPSWerr}
\definemath{\hfagQPSWfull}{\hfagQPSWval\hfagQPSWsta\hfagQPSWsys}
\definemath{\hfagASLS}{\hfagASLSval\hfagASLSerr}
\definemath{\hfagQPS}{\hfagQPSval\hfagQPSerr}
\definemath{\hfagASLD}{\hfagASLDval\hfagASLDerr}
\definemath{\hfagQPD}{\hfagQPDval\hfagQPDerr}
\definemath{\hfagREBD}{\hfagREBDval\hfagREBDerr}
\definemath{\hfagTANPHI}{\hfagTANPHIval\hfagTANPHIerr}
\definemath{\hfagDMS}{\hfagDMSval\hfagDMSerr\invps}
\definemath{\hfagDMSnounit}{\hfagDMSval\hfagDMSerr}
\definemath{\hfagDMSfull}{\hfagDMSval\hfagDMSsta\hfagDMSsys\invps}
\definemath{\hfagDMSnounitfull}{\hfagDMSval\hfagDMSsta\hfagDMSsys}
\definemath{\hfagXS}{\hfagXSval\hfagXSerr}
\definemath{\hfagCHIS}{\hfagCHISval\hfagCHISerr}
\definemath{\hfagRATIODMDDMS}{\hfagRATIODMDDMSval\hfagRATIODMDDMSerr}
\definemath{\hfagVTDVTS}{\hfagVTDVTSval\hfagVTDVTSerr}
\definemath{\hfagVTDVTSfull}{\hfagVTDVTSval\hfagVTDVTSexx\hfagVTDVTSthe}
\definemath{\hfagXI}{\hfagXIval\hfagXIerr}
\definemath{\hfagBETASCOMBerr}{\hfagBETASCOMBerp\hfagBETASCOMBern}
\definemath{\hfagBETASCOMB}{\hfagBETASCOMBval\hfagBETASCOMBerr}
\definemath{\hfagPHISCOMB}{\hfagPHISCOMBval\hfagPHISCOMBerr}
\definemath{\hfagDGSCOMB}{\hfagDGSCOMBval\hfagDGSCOMBerr\invps}
\definemath{\hfagDGSCOMBnounit}{\hfagDGSCOMBval\hfagDGSCOMBerr}
\definemath{\hfagBETASCOMBCONerr}{\hfagBETASCOMBCONerp\hfagBETASCOMBCONern}
\definemath{\hfagBETASCOMBCON}{\hfagBETASCOMBCONval\hfagBETASCOMBCONerr}
\definemath{\hfagPHISCOMBCONerr}{\hfagPHISCOMBCONerp\hfagPHISCOMBCONern}
\definemath{\hfagPHISCOMBCON}{\hfagPHISCOMBCONval\hfagPHISCOMBCONerr}
\definemath{\hfagDGSCOMBCONerr}{\hfagDGSCOMBCONerp\hfagDGSCOMBCONern}
\definemath{\hfagDGSCOMBCON}{\hfagDGSCOMBCONval\hfagDGSCOMBCONerr\invps}
\definemath{\hfagDGSCOMBCONnounit}{\hfagDGSCOMBCONval\hfagDGSCOMBCONerr}
\definemath{\hfagPHISSMerr}{\hfagPHISSMerp\hfagPHISSMern}
\definemath{\hfagPHISSM}{\hfagPHISSMval\hfagPHISSMerr}
\definemath{\hfagPHISTWELVESM}{\hfagPHISTWELVESMval\hfagPHISTWELVESMerr}
\definemath{\hfagPHISTWELVE}{\hfagPHISTWELVEval\hfagPHISTWELVEerr}


\mysection{\b-hadron production fractions, lifetimes and mixing parameters}
\labs{life_mix}


Quantities such as \b-hadron production fractions, \b-hadron lifetimes, 
and neutral \B-meson oscillation frequencies have been studied
in the nineties at LEP and SLC 
(\ee colliders at $\sqrt{s}=m_{\particle{Z}}$) 
as well as at the 
first version of the Tevatron
(\particle{p\bar{p}} collider at $\sqrt{s}=1.8\TeV$). 
Since then 
precise measurements of the \Bd and \Bu mesons
have also been performed at the 
asymmetric \B factories, KEKB and PEPII
(\ee colliders at $\sqrt{s}=m_{\Ups}$) while measurements related 
to the other \b hadrons, in particular \Bs, \Bc and \Lb, 
have been performed at the upgraded Tevatron ($\sqrt{s}=1.96\TeV$)
and are continuing at the LHC ($pp$ collider at $\sqrt{s}=7$ and $8\TeV$).
In most cases, these basic quantities, although interesting by themselves,
became necessary ingredients for the more complicated and 
refined analyses at the asymmetric \B factories, 
the Tevatron and the LHC,
in particular the time-dependent \CP asymmetry measurements.
It is therefore important that the best experimental
values of these quantities continue to be kept up-to-date and improved. 

In several cases, the averages presented in this chapter are 
needed and used as input for the results given in the subsequent chapters. 
Within this chapter, some averages need the knowledge of other 
averages in a circular way. This coupling, which appears through the 
\b-hadron fractions whenever inclusive or semi-exclusive measurements 
have to be considered, has been reduced drastically in the past several years 
with increasingly precise exclusive measurements becoming available
and dominating practically all averages. 

In addition to \b-hadron fractions, lifetimes and 
mixing parameters, this chapter also deals with the 
\CP-violating phase $\phiccbars\simeq -2\beta_s$, which is the phase 
difference between the \Bs mixing amplitude and the 
$b\to c\bar{c}s$ decay amplitude, as well as the parameters of \CP violation
in the $B$ mixing amplitudes. 
The angle $\beta$, which is the equivalent of $\beta_s$ for the \Bd 
system, is discussed in Chapter~\ref{sec:cp_uta}. 

\mysubsection{\b-hadron production fractions}
\labs{fractions}
 
We consider here the relative fractions of the different \b-hadron 
species found in an unbiased sample of weakly decaying \b hadrons 
produced under some specific conditions. The knowledge of these fractions
is useful to characterize the signal composition in inclusive \b-hadron 
analyses, to predict the background composition in exclusive analyses, 
or to convert (relative) observed rates into (relative) branching fraction 
measurements. 
Many \B-physics analyses need these fractions as input. We distinguish 
here the following three conditions: \Ups decays, \Upsfive decays, and 
high-energy collisions (including \Z decays). 

\mysubsubsection{\b-hadron production fractions in \Ups decays}
\labs{fraction_Ups4S}

Only pairs of the two lightest (charged and neutral) \B mesons 
can be produced in \Ups decays, 
and it is enough to determine the following branching 
fractions:
\begin{eqnarray}
f^{+-} & = & \Gamma(\Ups \to \particle{B^+B^-})/
             \Gamma_{\rm tot}(\Ups)  \,, \\
f^{00} & = & \Gamma(\Ups \to \particle{B^0\bar{B}^0})/
             \Gamma_{\rm tot}(\Ups) \,.
\end{eqnarray}
In practice, most analyses measure their ratio
\begin{equation}
R^{+-/00} = f^{+-}/f^{00} = \Gamma(\Ups \to \particle{B^+B^-})/
             \Gamma(\Ups \to \particle{B^0\bar{B}^0}) \,,
\end{equation}
which is easier to access experimentally.
Since an inclusive (but separate) reconstruction of 
\Bu and \Bd is difficult, specific exclusive decay modes, 
${\Bu} \to x^+$ and ${\Bd} \to x^0$, are usually considered to perform 
a measurement of $R^{+-/00}$, whenever they can be related by 
isospin symmetry (for example \particle{\Bu \to \jpsi K^+} and 
\particle{\Bd \to \jpsi K^0}).
Under the assumption that $\Gamma(\Bu \to x^+) = \Gamma(\Bd \to x^0)$, 
\ie\ that isospin invariance holds in these \B decays,
the ratio of the number of reconstructed
$\Bu \to x^+$ and $\Bd \to x^0$ mesons, after correcting for efficiency, is
proportional to
\begin{equation}
\frac{f^{+-}\, \BR{\Bu\to x^+}}{f^{00}\, \BR{\Bd\to x^0}}
= \frac{f^{+-}\, \Gamma({\Bu}\to x^+)\, \tau(\Bu)}%
{f^{00}\, \Gamma({\Bd}\to x^0)\,\tau(\Bd)}
= \frac{f^{+-}}{f^{00}} \, \frac{\tau(\Bu)}{\tau(\Bd)}  \,, 
\end{equation} 
where $\tau(\Bu)$ and $\tau(\Bd)$ are the \Bu and \Bd 
lifetimes respectively.
Hence the primary quantity measured in these analyses 
is $R^{+-/00} \, \tau(\Bu)/\tau(\Bd)$, 
and the extraction of $R^{+-/00}$ with this method therefore 
requires the knowledge of the $\tau(\Bu)/\tau(\Bd)$ lifetime ratio. 

\begin{table}
\caption{Published measurements of the $\Bu/\Bd$ production ratio
in \Ups decays, together with their average (see text).
Systematic uncertainties due to the imperfect knowledge of 
$\tau(\Bu)/\tau(\Bd)$ are included. The latest \babar result~\cite{Aubert:2004rz}
supersedes the earlier \babar measurements~\cite{Aubert:2001xs,Aubert:2004ur}.}
\labt{R_data}
\begin{center}
\begin{tabular}{lccll}
\hline
Experiment & Ref. & Decay modes & Published value of & Assumed value \\
and year & & or method & $R^{+-/00}=f^{+-}/f^{00}$ & of $\tau(\Bu)/\tau(\Bd)$ \\
\hline
CLEO,   2001 & \cite{Alexander:2000tb}  & \particle{\jpsi K^{(*)}} 
             & $1.04 \pm0.07 \pm0.04$ & $1.066 \pm0.024$ \\
\babar, 2002 & \cite{Aubert:2001xs} & \particle{(c\bar{c})K^{(*)}}
             & $1.10 \pm0.06 \pm0.05$ & $1.062 \pm0.029$\\ 
CLEO,   2002 & \cite{Athar:2002mr}  & \particle{D^*\ell\nu}
             & $1.058 \pm0.084 \pm0.136$ & $1.074 \pm0.028$\\
\belle, 2003 & \cite{Hastings:2002ff} & dilepton events 
             & $1.01 \pm0.03 \pm0.09$ & $1.083 \pm0.017$\\
\babar, 2004 & \cite{Aubert:2004ur} & \particle{\jpsi K}
             & $1.006 \pm0.036 \pm0.031$ & $1.083 \pm0.017$ \\
\babar, 2005 & \cite{Aubert:2004rz} & \particle{(c\bar{c})K^{(*)}}
             & $1.06 \pm0.02 \pm0.03$ & $1.086 \pm0.017$\\ 
\hline
Average      & & & \hfagFF~(tot) & \hfagRTAUBU \\
\hline
\end{tabular}
\end{center}
\end{table}

The published measurements of $R^{+-/00}$ are listed 
in \Table{R_data} together with the corresponding assumed values of 
$\tau(\Bu)/\tau(\Bd)$.
All measurements are based on the above-mentioned method, 
except the one from \belle, which is a by-product of the 
\Bd mixing frequency analysis using dilepton events
(but note that it also assumes isospin invariance, 
namely $\Gamma(\Bu \to \ell^+{\rm X}) = \Gamma(\Bd \to \ell^+{\rm X})$).
The latter is therefore treated in a slightly different 
manner in the following procedure used to combine 
these measurements:
\begin{itemize} 
\item each published value of $R^{+-/00}$ from CLEO and \babar
      is first converted back to the original measurement of 
      $R^{+-/00} \, \tau(\Bu)/\tau(\Bd)$, using the value of the 
      lifetime ratio assumed in the corresponding analysis;
\item a simple weighted average of these original
      measurements of $R^{+-/00} \, \tau(\Bu)/\tau(\Bd)$ from 
      CLEO and \babar (which do not depend on the assumed value 
      of the lifetime ratio) is then computed, assuming no 
      statistical or systematic correlations between them;


\item the weighted average of $R^{+-/00} \, \tau(\Bu)/\tau(\Bd)$ 
      is converted into a value of $R^{+-/00}$, using the latest 
      average of the lifetime ratios, $\tau(\Bu)/\tau(\Bd)=\hfagRTAUBU$ 
      (see \Sec{lifetime_ratio});
\item the \belle measurement of $R^{+-/00}$ is adjusted to the 
      current values of $\tau(\Bd)=\hfagTAUBD$ and 
      $\tau(\Bu)/\tau(\Bd)=\hfagRTAUBU$ (see \Sec{lifetime_ratio}),
      using the quoted systematic uncertainties due to these parameters;
\item the combined value of $R^{+-/00}$ from CLEO and \babar is averaged 
      with the adjusted value of $R^{+-/00}$ from \belle, assuming a 100\% 
      correlation of the systematic uncertainty due to the limited 
      knowledge on $\tau(\Bu)/\tau(\Bd)$; no other correlation is considered. 
\end{itemize} 
The resulting global average, 
\begin{equation}
R^{+-/00} = \frac{f^{+-}}{f^{00}} =  \hfagFF \,,
\labe{Rplusminus}
\end{equation}
is consistent with equal production of charged and neutral \B mesons, 
although only at the $\hfagNSIGMAFF\,\sigma$ level.

On the other hand, the \babar collaboration has 
performed a direct measurement of the $f^{00}$ fraction 
using an original method, which neither relies on isospin symmetry nor requires 
the knowledge of $\tau(\Bu)/\tau(\Bd)$. Its analysis, 
based on a comparison between the number of events where a single 
$B^0 \to D^{*-} \ell^+ \nu$ decay could be reconstructed and the number 
of events where two such decays could be reconstructed, yields~\cite{Aubert:2005bq}
\begin{equation}
f^{00}= 0.487 \pm 0.010\,\mbox{(stat)} \pm 0.008\,\mbox{(syst)} \,.
\labe{fzerozero}
\end{equation}

The two results of \Eqss{Rplusminus}{fzerozero} are of very different natures 
and completely independent of each other. 
Their product is equal to $f^{+-} = \hfagFPROD$, 
while another combination of them gives $f^{+-} + f^{00}= \hfagFSUM$, 
compatible with unity.
Assuming\footnote{A few non-$\B\bar{B}$
decay modes of the $\Upsilon(4S)$ 
($\Upsilon(1S)\pi^+\pi^-$,
$\Upsilon(2S)\pi^+\pi^-$, $\Upsilon(1S)\eta$) 
have been observed with branching fractions
of the order of $10^{-4}$~\cite{Aubert:2006bm,*Sokolov:2006sd,*Aubert_mod:2008bv},
corresponding to a partial
width several times larger than that in the \ee channel.
However, this can still be
neglected and the assumption $f^{+-}+f^{00}=1$ remains valid
in the present context of the determination of $f^{+-}$ and $f^{00}$.}
 $f^{+-}+f^{00}= 1$, also consistent with 
CLEO's observation that the fraction of \Ups decays 
to \BB pairs is larger than 0.96 at \CL{95}~\cite{Barish:1995cx},
the results of \Eqss{Rplusminus}{fzerozero}
can be averaged (first converting \Eq{Rplusminus} 
into a value of $f^{00}=1/(R^{+-/00}+1)$) 
to yield the following more precise estimates:
\begin{equation}
f^{00} = \hfagFNW  \,,~~~ f^{+-} = 1 -f^{00} =  \hfagFCW \,,~~~
\frac{f^{+-}}{f^{00}} =  \hfagFFW \,.
\end{equation}
The latter ratio differs from one by $\hfagNSIGMAFFW\,\sigma$.

\mysubsubsection{\b-hadron production fractions in \Upsfive decays}
\labs{fraction_Ups5S}

\newcommand{\fsfive}{\ensuremath{f^{\Upsfive}_{s}}}
\newcommand{\fudfive}{\ensuremath{f^{\Upsfive}_{u,d}}}
\newcommand{\fnBfive}{\ensuremath{f^{\Upsfive}_{B\!\!\!\!/}}}

Hadronic events produced in $e^+e^-$ collisions at the \Upsfive (also known as
$\Upsilon(10860)$) energy can be classified into three categories: 
light-quark ($u$, $d$, $s$, $c$) continuum events, $b\bar{b}$ continuum events,
and \Upsfive events. The latter two cannot be distinguished and will be called
$b\bar{b}$ events in the following. These $b\bar{b}$ events, which also include 
$b\bar{b}\gamma$ events because of possible initial-state radiation, 
can hadronize in different final states.
We define \fudfive\ as
the fraction of $b\bar{b}$ events with a pair of non-strange 
bottom mesons 
($B\bar{B}$, $B\bar{B}^*$, $B^*\bar{B}$, $B^*\bar{B}^*$,
$B\bar{B}\pi$, $B\bar{B}^*\pi$, $B^*\bar{B}\pi$,
$B^*\bar{B}^*\pi$, and $B\bar{B}\pi\pi$ final states, 
where
$B$ denotes a $B^0$ or $B^+$ meson and 
$\bar{B}$ denotes a $\bar{B}^0$ or $B^-$ meson), \fsfive\ as
the fraction of $b\bar{b}$ events with a pair of strange bottom mesons
($B_s^0\bar{B}_s^0$, $B_s^0\bar{B}_s^{*0}$, $B_s^{*0}\bar{B}_s^0$, and
$B_s^{*0}\bar{B}_s^{*0}$ final states), and 
\fnBfive\ as the fraction of $b\bar{b}$ events without 
any bottom meson in the final state.
Note that the excited bottom-meson states decay via $B^* \to B \gamma$ and
$B_s^{*0} \to B_s^0 \gamma$.
These fractions satisfy
\begin{equation}
\fudfive + \fsfive + \fnBfive = 1 \,.
\labe{sum_frac_five}
\end{equation} 

\begin{table}
\caption{Published measurements of \fsfive.
All values have been obtained assuming $\fnBfive=0$. 
They are quoted 
as in the original publications, except for the most
recent measurement which is quoted as 
$1-\fudfive$, with \fudfive\ from \Ref{Drutskoy:2010an}.
The last line gives our average of \fsfive\ assuming $\fnBfive=0$.}
\labt{fsFiveS}
\begin{center}
\begin{tabular}{lll}
\hline
Experiment, year, dataset                 & Decay mode or method    & Value of \fsfive\  \\
\hline
CLEO, 2006, 0.42\invfb~\cite{Huang:2006em_mod}     & $\Upsfive\to D_{s}X$     & $0.168 \pm 0.026^{+0.067}_{-0.034}$  \\
             & $\Upsfive \to \phi X$    & $0.246 \pm 0.029^{+0.110}_{-0.053}$ \\
             & $\Upsfive \to B\bar{B}X$ & $0.411 \pm 0.100 \pm 0.092$ \\  
             & CLEO average of above 3  & $0.21^{+0.06}_{-0.03}$      \\  \hline
Belle, 2006, 1.86\invfb~\cite{Drutskoy:2006fg} & $\Upsfive \to D_s X$     & $0.179 \pm 0.014 \pm 0.041$ \\
             & $\Upsfive \to D^0 X$     & $0.181 \pm 0.036 \pm 0.075$ \\  
             & Belle average of above 2 & $0.180 \pm 0.013 \pm 0.032$ \\  \hline 
Belle, 2010, 23.6\invfb~\cite{Drutskoy:2010an} & $\Upsfive \to B\bar{B}X$ & $0.263 \pm 0.032 \pm 0.051$ 
\\ \hline
\multicolumn{2}{l}{Average of all above 
after adjustments to inputs of \Table{fsFiveS_external}} & 
\hfagFSFIVERL             \\  \hline 
\end{tabular}
\end{center}
\end{table}

\begin{table}
\caption{External inputs on which the \fsfive\ averages are based.}
\labt{fsFiveS_external}
\begin{center}
\begin{tabular}{lcl}
\hline
Branching fraction   & Value     & Explanation and reference \\
\hline
${\cal B}(B\to D_s X)\times {\cal B}(D_s \to \phi\pi)$ & 
$0.00374\pm 0.00014$ & derived from~\cite{PDG_2012} \\
${\cal B}(B^0_s \to D_s X)$ & 
$0.92\pm0.11$ & model-dependent estimate~\cite{Artuso:2005xw} \\
${\cal B}(D_s \to \phi\pi)$ & 
$0.045\pm0.004$ & \cite{PDG_2012} \\
${\cal B}(B\to D^0 X)\times {\cal B}(D^0 \to K\pi)$ & 
$0.0243\pm0.0011$ & derived from~\cite{PDG_2012} \\
${\cal B}(B^0_s \to D^0 X)$ & 
$0.08\pm0.07$ & model-dependent estimate~\cite{Drutskoy:2006fg,Artuso:2005xw} \\
${\cal B}(D^0 \to K\pi)$ & 
$0.0387\pm0.0005$ & \cite{PDG_2012} \\
${\cal B}(B \to \phi X)$ & 
$0.0343\pm0.0012$ & world average~\cite{PDG_2012,Huang:2006em_mod} \\
${\cal B}(B^0_s \to \phi X)$ &
$0.161\pm0.024$ & model-dependent estimate~\cite{Huang:2006em_mod} \\
\hline
\end{tabular}
\end{center}
\end{table}

The CLEO and Belle collaborations have published 
measurements of several inclusive \Upsfive branching fractions, 
${\cal B}(\Upsfive\to D_s X)$, 
${\cal B}(\Upsfive\to \phi X)$ and 
${\cal B}(\Upsfive\to D^0 X)$, 
from which they extracted the
model-dependent estimates of \fsfive\
reported in \Table{fsFiveS}.\footnote{%
  \label{foot:life_mix:Esen:2012yz_mod}
  It was realized just before finalizing this document that
  more recent results from Belle~\cite{Esen:2012yz}, 
  $\fsfive = 0.172 \pm 0.030$, have been overlooked. These results are 
  not included in \Table{fsFiveS} nor in the averages presented here. 
} 
This extraction was performed 
under the implicit assumption  
$\fnBfive=0$, using the relation 
\begin{equation}
\frac12{\cal B}(\Upsfive\to D_s X)=\fsfive\times{\cal B}(B_s^0\to D_s X) + 
\left(1-\fsfive-\fnBfive\right)\times{\cal B}(B\to D_s X) \,,
\labe{Ds_correct}
\end{equation}
and similar relations for
${\cal B}(\Upsfive\to D^0 X)$ and ${\cal B}(\Upsfive\to \phi X)$.
In \Table{fsFiveS} we list also
the values of \fsfive\ derived from measurements of
$\fudfive={\cal B}(\Upsfive\to B\bar BX)$~\cite{Huang:2006em_mod,Drutskoy:2010an}, as well as our average value of  \fsfive,
all obtained under the assumption $\fnBfive=0$.

However, the assumption $\fnBfive=0$ is known to be invalid since the observation of
the following final states in $e^+e^-$ collisions at the \Upsfive\ energy:
$\Upsilon(1S)\pi^+\pi^-$,
$\Upsilon(2S)\pi^+\pi^-$,
$\Upsilon(3S)\pi^+\pi^-$
and
$\Upsilon(1S)K^+K^-$~\cite{Abe:2007tk,Garmash:2014dhx_mod},
$h_b(1P)\pi^+\pi^-$ and 
$h_b(2P)\pi^+\pi^-$~\cite{Adachi:2011ji},
and more recently 
$\Upsilon(1S)\pi^0\pi^0$,
$\Upsilon(2S)\pi^0\pi^0$ 
and
$\Upsilon(3S)\pi^0\pi^0$~\cite{Krokovny:2013mgx}.
The sum of the measurements of the corresponding visible cross-sections,
adding also the contributions of the unmeasured
$\Upsilon(1S)K^0\bar{K}^0$, $h_b(1P)\pi^0\pi^0$ and $h_b(2P)\pi^0\pi^0$ final states
assuming isospin conservation, amounts to
$$
\sigma^{\rm vis}(e^+e^-\to (\b\bar{\b})hh) = 13.2\pm1.4~{\rm pb} \,,
~~\mbox{for $(\b\bar{\b})=\Upsilon(1S,2S,3S),h_b(1P,2P)$ and $hh=\pi\pi,KK$}\,.
$$
We divide this by the $\b\bar{\b}$ production cross section, 
$\sigma(e^+e^- \to \b\bar{\b} X) = 337 \pm 15$~pb, obtained as the average of the 
CLEO~\cite{Artuso:2005xw} and Belle~\cite{Esen:2012yz}\footnote{%
\label{foot:life_mix:Esen:2012yz}
The results of Ref.~\cite{Esen:2012yz} supersede the $\sigma(e^+e^- \to \b\bar{\b} X)$ and \fsfive\ results of Ref.~\cite{Drutskoy:2006fg}.
} measurements, to obtain
$$
{\cal B}(\Upsfive\to (\b\bar{\b})hh) = 0.039\pm0.004 \,,
~~\mbox{for $(\b\bar{\b})=\Upsilon(1S,2S,3S),h_b(1P,2P)$ and $hh=\pi\pi,KK$}\,,
$$
which is to be considered as a lower bound for \fnBfive. 


Following the method described in \Ref{thesis_Louvot}, 
we perform a $\chi^2$ fit of the original 
measurements of the \Upsfive\ branching fractions of
\Refs{Huang:2006em_mod,Drutskoy:2006fg,Drutskoy:2010an},
using the inputs of \Table{fsFiveS_external},
the relations of \Eqss{sum_frac_five}{Ds_correct} and the
one-sided Gaussian constraint $\fnBfive \ge {\cal B}(\Upsfive \to (\b\bar{\b}) hh)$,
to simultaneously extract \fudfive, \fsfive\ and \fnBfive. Taking all known 
correlations into account, the best fit values are
\begin{eqnarray}
\fudfive &=& \hfagFUDFIVE \,, \labe{fudfive} \\
\fsfive  &=& \hfagFSFIVE  \,, \labe{fsfive}  \\
\fnBfive &=& \hfagFNBFIVE \,, \labe{fnBfive}
\end{eqnarray}
where the strongly asymmetric uncertainty on \fnBfive\ is due to the one-sided constraint
from the observed $(\b\bar{\b}) hh$ decays. These results, together with their correlation, 
imply
\begin{eqnarray}
\fsfive/\fudfive  &=& \hfagFSFUDFIVE  \,, \labe{fsfudfive} 
\end{eqnarray}
in fair agreement with the results of a \babar
analysis~\cite{Lees:2011ji}, performed as a function 
of centre-of-mass energy.\footnote{%
\label{foot:life_mix:Lees:2011ji}
This has not been included in the average, since 
no numerical value is given for $\fsfive/\fudfive$ in 
\Ref{Lees:2011ji}.
}

The production of $B^0_s$ mesons at the \Upsfive
is observed to be dominated by the $B_s^{*0}\bar{B}_s^{*0}$
channel, 
with $\sigma(e^+e^- \to B_s^{*0}\bar{B}_s^{*0})/%
\sigma(e^+e^- \to B_s^{(*)0}\bar{B}_s^{(*)0})
= (87.0\pm 1.7)\%$~\cite{Li:2011pg,Louvot:2008sc}.
The proportions of the various production channels 
for non-strange $B$ mesons have also been measured~\cite{Drutskoy:2010an}.

\mysubsubsection{\b-hadron production fractions at high energy}
\labs{fractions_high_energy}
\labs{chibar}

At high energy, all species of weakly decaying \b hadrons 
may be produced, either directly or in strong and electromagnetic 
decays of excited \b hadrons. It is often assumed that the fractions 
of these different species are the same in unbiased samples of 
high-$p_{\rm T}$ \b jets originating from \particle{Z^0} decays, 
from \particle{p\bar{p}} collisions at the Tevatron, or from 
\particle{p p} collisions at the LHC.
This hypothesis is plausible under the condition that the square of
the momentum transfer to the produced \b quarks, $Q^2$, is large compared 
with the square of the hadronization energy scale, 
$Q^2 \gg \Lambda_{\rm QCD}^2$.
On the other hand, there is no strong argument that the
fractions at different machines should be strictly equal, so 
this assumption should be checked experimentally. Although the 
available data is not sufficient at this time to perform a definitive
check, it is expected that more refined analyses of the Tevatron Run~II data 
and new analyses from LHC 
experiments may improve this situation and allow one to confirm or 
disprove this assumption with reasonable confidence. Meanwhile, the 
attitude adopted here is that these fractions are assumed to be equal 
at all high-energy colliders until demonstrated otherwise by 
experiment.
Both CDF and LHCb report a $p_{\rm T}$ dependence for \Lb
production relative to \Bu and \Bd; the number of \Lb baryons
observed at low $p_{\rm T}$ is enhanced with respect to that 
seen at LEP's higher $p_{\rm T}$.
Therefore we present 
three sets of complete averages: one set including only measurements 
performed at LEP, a second set including only measurements performed 
at the Tevatron, a third  set including measurements performed at LEP, 
Tevatron and LHCb.  The LHCb production fractions results by themselves 
are still incomplete, lacking measurements of the production of other
weakly decaying heavy-flavour baryons, $\Xi_b$ and $\Omega_b$, and a measurement of 
$\overline{\chi}$ giving an extra constraint between \fBd and \fBs.

Contrary to what happens in the charm sector where the fractions of 
\particle{D^+} and \particle{D^0} are different, the relative amount 
of \Bu and \Bd is not affected by the electromagnetic decays of 
excited $B^{*+}$ and $B^{*0}$ states and strong decays of excited 
$B^{**+}$ and $B^{**0}$ states. Decays of the type 
\particle{B_s^{**0} \to B^{(*)}K} also contribute to the \Bu and \Bd rates, 
but with the same magnitude if mass effects can be neglected.  
We therefore assume equal production of \Bu and \Bd mesons. We also  
neglect the production of weakly decaying states
made of several heavy quarks (like \Bc and doubly heavy baryons) 
which is known to be very small. Hence, for the purpose of determining 
the \b-hadron fractions, we use the constraints
\begin{equation}
\fBu = \fBd ~~~~\mbox{and}~~~ \fBu + \fBd + \fBs + \fbb = 1 \,,
\labe{constraints}
\end{equation}
where \fBu, \fBd, \fBs and \fbb
are the unbiased fractions of \Bu, \Bd, \Bs and \b baryons, respectively.

We note that there are many measurements of the production cross-sections of
different species of \b hadrons.
In principle these could be included in a global fit to determine the
production fractions.
We do not perform such a fit at the current time, and instead average only the
measured production fractions.

The LEP experiments have measured
$\fBs \times \BR{\Bs\to\particle{D_s^-} \ell^+ \nu_\ell \mbox{$X$}}$~\cite{Abreu:1992rv,*Acton:1992zq,*Buskulic:1995bd}, 
$\BR{\b\to\Lb} \times \BR{\Lb\to\Lc\ell^-\bar{\nu}_\ell \mbox{$X$}}$~\cite{Abreu:1995me,Barate:1997if}
and $\BR{\b\to\Xib^-} \times \BR{\Xi_b^- \to \Xi^-\ell^-\overline\nu_\ell 
\mbox{$X$}}$~\cite{Buskulic:1996sm,Abdallah:2005cw}\footnote{
  \label{foot:life_mix:Abdallah:2005cw}
  The DELPHI result of \Ref{Abdallah:2005cw} is considered to supersede an older one~\cite{Abreu:1995kt}.
} 
from partially reconstructed final states including a lepton, \fbb
from protons identified in \b events~\cite{Barate:1997ty}, and the 
production rate of charged \b hadrons~\cite{Abdallah:2003xp}. 
Ratios of \b-hadron fractions have been measured at CDF using 
lepton+charm final 
states~\cite{Affolder:1999iq,Aaltonen:2008zd,Aaltonen:2008eu}\footnote{
  \label{foot:life_mix:Affolder:1999iq}
  CDF updated their measurement of \fLb/\fBd~\cite{Affolder:1999iq} to account 
  for a measured $p_{\rm T}$ dependence between exclusively reconstructed 
  \Lb and $B^0$~\cite{Aaltonen:2008eu}.
}, double semileptonic decays 
with \particle{K^*\mu\mu} and \particle{\phi\mu\mu}
final states~\cite{Abe:1999ta},
and fully reconstructed $\Bs\to\jpsi\phi$ decays~\cite{CDFnote10795:2012}.
Measurements of the production of other heavy 
flavour baryons at the Tevatron are included in the determination of 
\fbb~\cite{Abazov_mod:2007ub,Abazov:2008qm,Aaltonen:2009ny}\footnote{
  \label{foot:life_mix:Abazov:2008qm}
  \dzero reports $f_{\Omega_b^-}/f_{\Xi_b^-}$.  We use the CDF+\dzero average of 
  $f_{\Xi_b^-}/f_{\Lb}$ to obtain $f_{\Omega_b^-}/f_{\Lb}$ and then 
  combine it with the CDF result.
} using the constraint
\begin{eqnarray}
\fbb & = & f_{\Lb} + f_{\Xi_b^0} + f_{\Xi_b^-} + f_{\Omega_b^-} 
     \nonumber \\
     & = & f_{\Lb}\left(1 + 2\frac{f_{\Xi_b^-}}{f_{\Lb}} 
           + \frac{f_{\Omega_b^-}}{f_{\Lb}}\right),
\end{eqnarray}
where isospin invariance is assumed in the production of $\Xi_b^0$ and 
$\Xi_b^-$. Other \b baryons are expected to decay strongly or 
electromagnetically to those baryons listed. For the production 
measurements, both CDF and \dzero\ reconstruct their \b baryons exclusively 
to final states which include a $\jpsi$ and a hyperon 
($\Lb\to \jpsi \Lambda$, 
$\Xi_b^- \rightarrow \jpsi \Xi^-$ and 
$\Omega_b^- \rightarrow \jpsi \Omega^-$).  
We assume that the partial decay width of a \b baryon to a $\jpsi$ and the 
corresponding hyperon is equal to the partial width of any other \b baryon to 
a $\jpsi$ and the corresponding hyperon.  LHCb has also measured
ratios of \b-hadron fractions in charm+lepton final states~\cite{Aaij:2011jp} 
and in fully reconstructed hadronic two-body decays $\Bd \to D^-\pi^+$, $\Bs \to D_s^- \pi^+$ and 
$\Lb \to \Lambda_c^+ \pi^-$~\cite{Aaij:2013qqa,Aaij:2014jyk}.%
\footnote{
  \label{foot:life_mix:Aaij:2013qqa}
  The results of Ref.~\cite{Aaij:2013qqa} supersede those of Ref.~\cite{Aaij:2011hi}.
}

Both CDF and LHCb observe that the ratio $\fLb/\fBd$ depends on the $p_{\rm T}$
of the charm+lepton system~\cite{Aaltonen:2008eu,Aaij:2011jp}.
\footnote{
  \label{foot:life_mix:Aaltonen:2008eu}
  CDF compares the $p_{\rm T}$ distribution of fully reconstructed 
  $\Lb\to\Lambda_c^+ \pi^-$ 
  with $\Bzb\rightarrow D^+\pi^-$, which 
  gives $\fLb/\fBd$ up to a scale factor. LHCb compares the $p_{\rm T}$ 
  in the charm+lepton system between \Lb and \Bd and \Bu, giving
  $R_{\Lb}/2 = \fLb/(\fBu+\fBd) = \fLb/2\fBd$.}
CDF chose to correct an older result to account for the $p_{\rm T}$ dependence.
In a second result, CDF binned their data in $p_{\rm T}$ of the charm+electron 
system~\cite{Aaltonen:2008zd}.
The more recent LHCb measurement using hadronic decays~\cite{Aaij:2014jyk} 
obtains the scale for $R_{\Lb} = \fLb/\fBd$ from their previous 
charm + lepton data~\cite{Aaij:2011jp}.  The LHCb measurement using hadronic
data also bins the same data in pseudorapidity ($\eta$) and sees a 
linear dependence of $R_{\Lb}$.  Since $\eta$ is not entirely
independent of $p_{\rm T}$ it is impossible to tell at this time whether 
this dependence is just an artifact of the $p_{\rm T}$ dependence.
\Figure{rlb_comb} shows the ratio $R_{\Lb}$ as a function of 
$p_{\rm T}$ for the \b hadron, as measured by LHCb.  LHCb fits their
scaled hadronic data to obtain
\begin{equation}
R_{\Lb} = (0.151\pm 0.030) + 
  \exp{\left\{-(0.57\pm 0.11) - 
  (0.095\pm 0.016)[{\rm GeV}/c]^{-1} \times p_{\rm T}\right\}}.
\end{equation}
A value of
$R_{\Lb}$ is also calculated for LEP and placed at the approximate $p_{\rm T}$ for the charm+lepton
system, but this value does not participate in any fit.\footnote{
  \label{foot:life_mix:Aaltonen:2008zd}
  The CDF semileptonic data would require significant corrections to obtain the $p_{\rm T}$ of the \b hadron and be included on the same plot with the LHCb data.
  We do not have these corrections at this time.}
Because the two LHCb results for $R_{\Lb}$ are not 
independent, we use only their semileptonic data for the averages.
Note that the $p_{\rm T}$ dependence
of $R_{\Lb}$ combined with the constraint from \Eq{constraints} implies
a compensating $p_{\rm T}$ dependence in one or more of the production fractions, \fBu, \fBd,
or \fBs.

\begin{figure}
 \begin{center}
  \includegraphics[width=\textwidth]{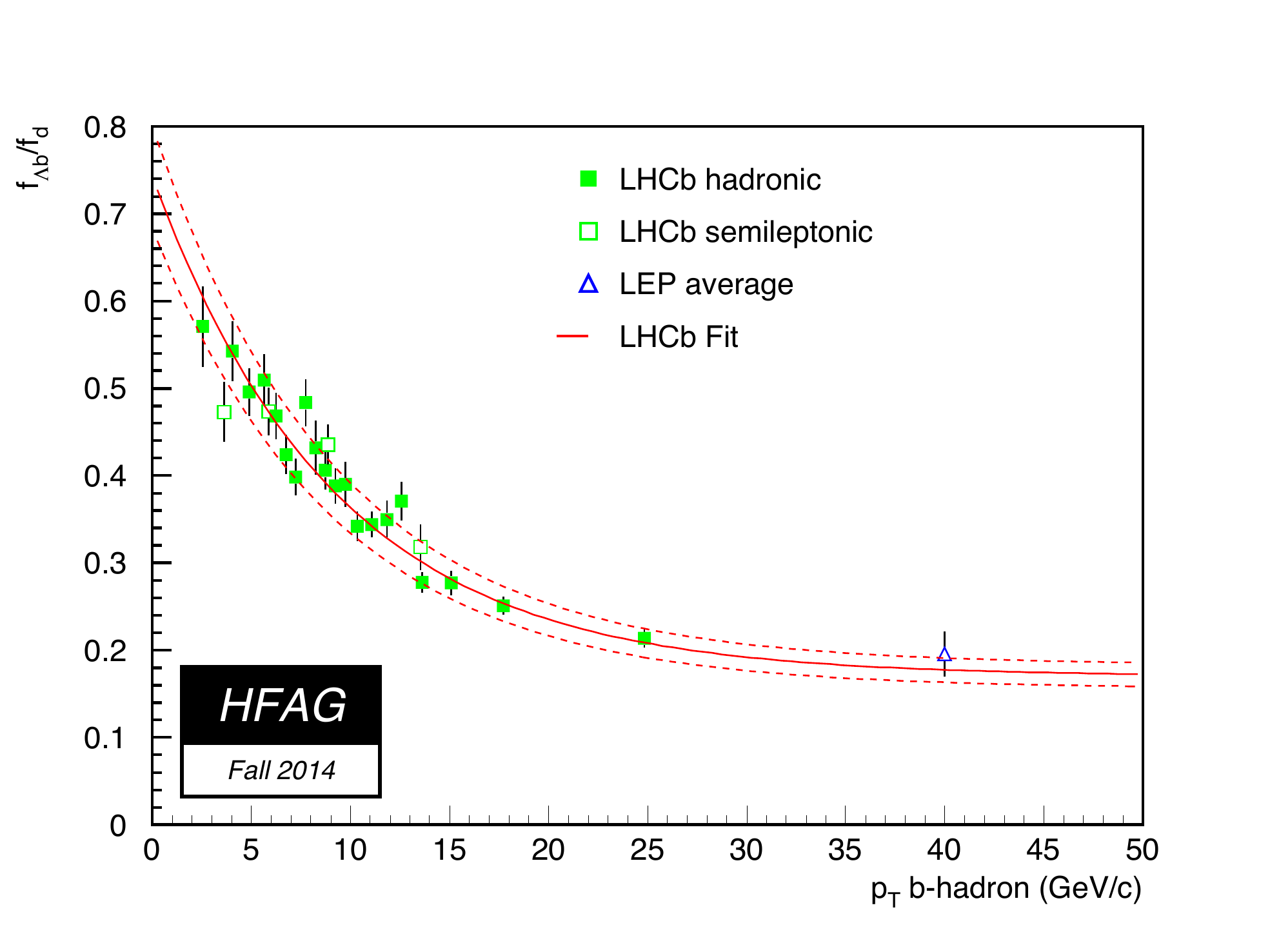}
  \caption{Ratio of production fractions $\fLb/\fBd$ as 
   a function of $p_{\rm T}$ of the \b hadron from
   LHCb data for \b hadrons decaying semileptonically~\cite{Aaij:2011jp}
   and fully reconstructed in hadronic decays~\cite{Aaij:2014jyk}. 
   The curve represents a fit to the LHCb hadronic data~\cite{Aaij:2014jyk}.
   The computed LEP ratio is included at an approximate $p_{\rm T}$ 
   in $Z$ decays, but does not participate in any fit.}
  \labf{rlb_comb}
 \end{center}
\end{figure}

Both CDF and LHCb have investigated the $p_{\rm T}$ dependence of $\fBs/\fBd$ using fully 
reconstructed $\Bs$ and $\Bd$ decays.  The CDF analysis reconstructed decays that include 
a $J/\psi$ in the final state~\cite{CDFnote10795:2012} and reports no significant $p_{\rm T}$ dependence 
on the ratio.  However, their result is dominated by an $18$\% scale uncertainty
from preliminary measurements of the branching ratios of the $\BR{\Bs\to J/\psi \phi}$ and 
$\BR{\Bd\to J/\psi K^{*}(892)}$.
LHCb reported $3\sigma$ evidence that the ratio $\fBs/\fBd$ decreases with 
$p_{\rm T}$ using fully reconstructed $\Bs$ and $\Bd$ decays and theoretical predictions for branching ratios~\cite{Aaij:2013qqa}\footref{foot:life_mix:Aaij:2013qqa}. \Figure{rs_comb} shows
the ratio $R_s = \fBs/\fBu$ as a function of $p_{\rm T}$ measured by CDF and LHCb.
Two fits are performed. 
The first fit, using a linear parameterization, yields
$R_s = (0.2760\pm 0.0068) - (0.00191\pm 0.00059)[{\rm GeV}/c]^{-1} \times p_{\rm T}$.  
A second fit, using a simple exponential, yields
$R_s = \exp\left\{(-1.293\pm 0.028) - (0.0077\pm 0.0025)[{\rm GeV}/c]^{-1} \times p_{\rm T}\right\}$.  
The two fits are nearly indistinguishable over the $p_{\rm T}$ range of the results,
but the second gives a physical value for all $p_{\rm T}$.  $R_s$ is also calculated
for LEP and placed at the approximate $p_{\rm T}$ for the \b hadron, though the LEP result
doesn't participate in the fit.  Our world average for $R_s$ is also included in the
figure for reference.

\begin{figure}
 \begin{center}
  \includegraphics[width=\textwidth]{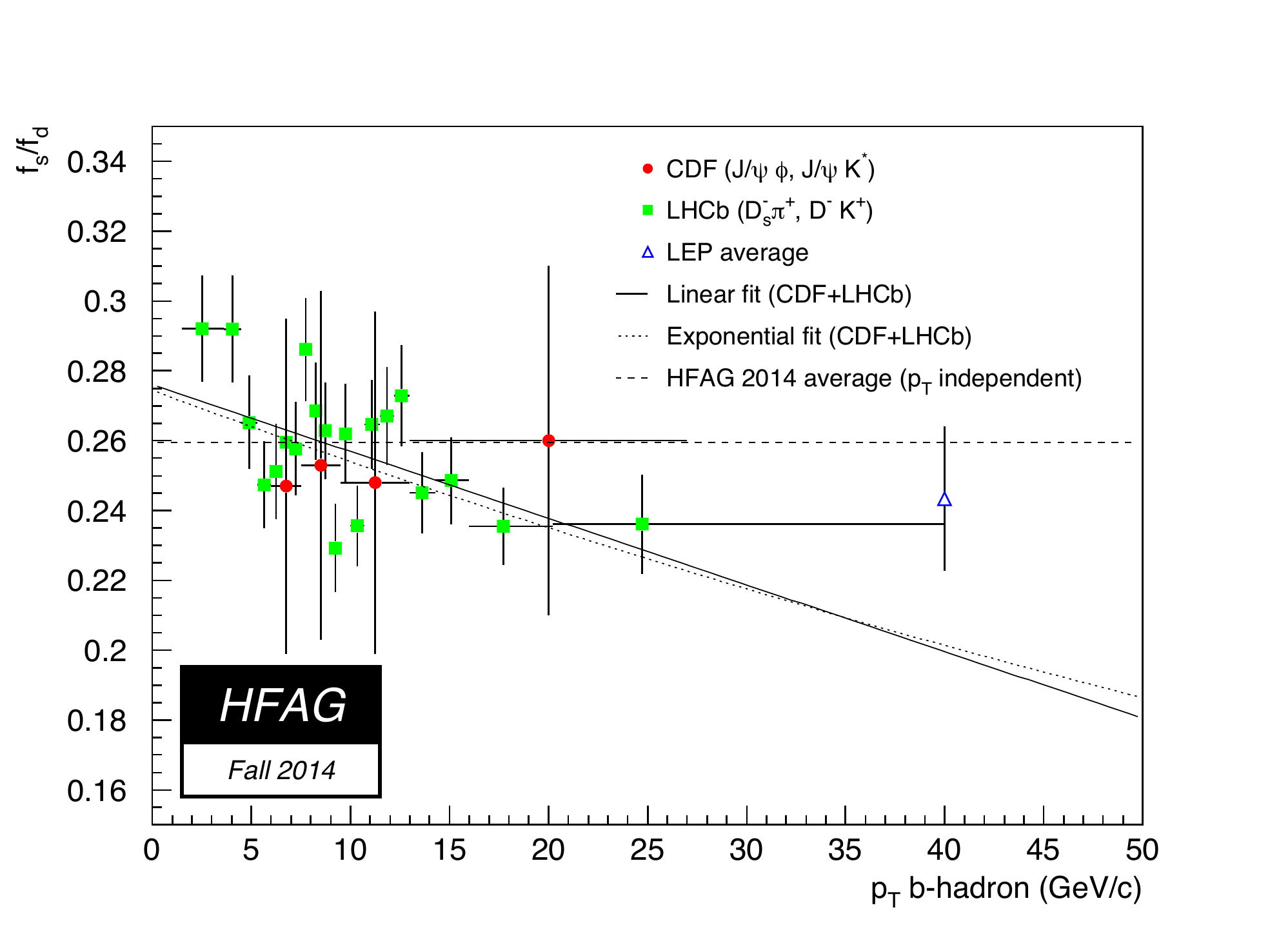}
  \caption{Ratio of production fractions $\fBs/\fBd$ as 
   a function of $p_{\rm T}$ of the reconstructed \b hadrons for the 
   CDF~\cite{CDFnote10795:2012} and LHCb~\cite{Aaij:2013qqa}\footref{foot:life_mix:Aaij:2013qqa}
   data. Note the suppressed zero for the vertical axis.
   The curves represent fits to the data:
   a linear fit (solid), and an exponential fit described in the text (dotted).
   The $p_{\rm T}$ independent value average of $R_s$ (dashed) is shown for 
   comparison.
   The computed LEP ratio is included at an 
   approximate $p_{\rm T}$ in $Z$ decays, but does not participate in any fit.}
  \labf{rs_comb}
 \end{center}
\end{figure}

In order to combine or compare LHCb results with other experiments,
the $p_{\rm T}$-dependent $\fLb/(\fBu + \fBd)$ is weighted by the $p_{\rm T}$ spectrum.\footnote{
  \label{foot:life_mix:Aaij:2011jp}
  In practice the LHCb data are given in 14 bins in $p_{\rm T}$ and $\eta$ with a full covariance matrix~\cite{Aaij:2011jp}. 
  The weighted average is calculated as
  $D^T C^{-1} M/\sigma$, where $\sigma = D^T C^{-1} D$, $M$ is a vector 
  of measurements, $C^{-1}$ is the inverse covariance matrix and $D^T$ is the 
  transpose of the design matrix (vector of 1's).}
\Table{LHCBcomp} compares 
the $p_{\rm T}$-weighted LHCb data with comparable averages from CDF. 
The average CDF 
and LHCb data are in good agreement despite the \b hadrons being produced 
in different kinematic regimes.

\begin{table}
 \caption{Comparison of average production fraction ratios from CDF and LHCb.
 The kinematic regime of the charm+lepton system reconstructed in each
 experiment is also shown.}
 \labt{LHCBcomp}
 \begin{center}
  \begin{tabular}{lccc}
   \hline
   Quantity                         & CDF               & LHCb \\
   \hline
   $\fBs/(\fBu + \fBd)$             & \hfagRBSTEVNOCON  & \hfagRBSLHCBNOCON   \\
   $\fLb/(\fBu + \fBd)$             & \hfagRLBTEVNOCON  & \hfagRLBLHCBNOCON   \\
   Average charm+lepton $p_{\rm T}$ & $\sim 13$~GeV/$c$ & $\sim 7$~GeV/$c$ \\
   Pseudorapidity range             & $-1 < \eta < 1$   & $2 < \eta < 5$      \\
   \hline
  \end{tabular}
 \end{center}
\end{table}

All these published results have been combined
following the procedure and 
assumptions described in \Ref{Abbaneo:2000ej_mod,*Abbaneo:2001bv_mod_cont},
to yield $\fBu=\fBd=\hfagWFBDNOMIX$, 
$\fBs=\hfagWFBSNOMIX$ and $\fbb=\hfagWFBBNOMIX$
under the constraints of \Eq{constraints}.  
Repeating the combinations for LEP and the Tevatron, we obtain 
$\fBu=\fBd=\hfagZFBDNOMIX$,
$\fBs=\hfagZFBSNOMIX$ and $\fbb=\hfagZFBBNOMIX$ when using the LEP data only, and
$\fBu=\fBd=\hfagTFBDNOMIX$, $\fBs=\hfagTFBSNOMIX$ and
$\fbb = \hfagTFBBNOMIX$ when using the Tevatron data only.  
As noted previously,
the LHCb data are insufficient to determine a complete set of \b-hadron production
fractions. The world averages (LEP, Tevatron and LHCb) for the various fractions 
are presented here for comparison with previous averages.  Significant differences
exist between the LEP and Tevatron fractions, therefore use of the world averages
should be taken with some care.
For these combinations other external inputs are used, 
\eg\ the branching ratios of \B mesons to final states with a \particle{D}, 
\particle{D^*} or \particle{D^{**}} in semileptonic decays, which are needed 
to evaluate the fraction of semileptonic \Bs decays with a \particle{D_s^-} 
in the final state.

Time-integrated mixing analyses performed with lepton pairs 
from \particle{b\bar{b}} 
events produced at high-energy colliders measure the quantity 
\begin{equation}
\chibar = f'_{\particle{d}} \,\chid + f'_{\particle{s}} \,\chis \,,
\labe{chibar}
\end{equation}
where $f'_{\particle{d}}$ and $f'_{\particle{s}}$ are 
the fractions of \Bd and \Bs hadrons 
in a sample of semileptonic \b-hadron decays, and where \chid and \chis 
are the \Bd and \Bs time-integrated mixing probabilities.
Assuming that all \b hadrons have the same semileptonic decay width implies 
$f'_i = f_i R_i$, where $R_i = \tau_i/\tau_{\b}$ is the ratio of the lifetime 
$\tau_i$ of species $i$ to the average \b-hadron lifetime 
$\tau_{\b} = \sum_i f_i \tau_i$.
Hence measurements of the mixing probabilities
\chibar, \chid and \chis can be used to improve our 
knowledge of \fBu, \fBd, \fBs and \fbb.
In practice, the above relations yield another determination of 
\fBs obtained from \fbb and mixing information, 
\begin{equation}
\fBs = \frac{1}{R_{\particle{s}}}
\frac{(1+r)\overline{\chi}-(1-\fbb R_{\rm baryon}) \chid}{(1+r)\chis - \chid} \,,
\labe{fBs-mixing}
\end{equation}
where $r=R_{\particle{u}}/R_{\particle{d}} = \tau(\Bu)/\tau(\Bd)$.

The published measurements of \chibar performed by the LEP
experiments have been combined by the LEP Electroweak Working Group to yield 
$\chibar = \hfagCHIBARLEP$~\cite{LEPEWWG:2005ema_mod}.
This can be compared with the Tevatron average, $\chibar = \hfagCHIBARTEV$,
obtained from \dzero~\cite{Abazov:2006qw} and CDF~\cite{CDFnote10335:2011}
measurements with Run~II data.\footnote{
  \label{foot:life_mix:Acosta:2003ie_mod}
  As explained in \Ref{CDFnote10335:2011}, a previous CDF analysis~\cite{Acosta:2003ie_mod}
  performed with Run~I data overlooked a background component, so the corresponding result is not 
  included in the average.}
The two averages agree, showing no evidence that the production fractions
of \Bd and \Bs mesons at the \particle{Z} peak or at the Tevatron are different.
We combine these two results in a simple weighted average,
assuming no correlations, and obtain 
$\chibar = \hfagCHIBAR$.

\begin{table}
\caption{Time-integrated mixing probability \chibar (defined in \Eq{chibar}), 
and fractions of the different \b-hadron species in an unbiased sample of 
weakly decaying \b hadrons, obtained from both direct
and mixing measurements. The correlation coefficients between the fractions are 
also given.
The last column includes measurements performed at LEP, Tevatron and LHCb.}
\labt{fractions}
\begin{center}
\begin{tabular}{@{}l@{~}l@{~}cccc@{}}
\hline
Quantity            &                      & $Z$ decays      & Tevatron       & LHCb~\cite{Aaij:2013qqa}\footref{foot:life_mix:Aaij:2013qqa} & all    \\
\hline
Mixing probability  & $\overline{\chi}$    & \hfagCHIBARLEP  & \hfagCHIBARTEV &         & \hfagCHIBAR \\
\Bu or \Bd fraction & $\fBu = \fBd$        & \hfagZFBD       & \hfagTFBD      &         & \hfagWFBD   \\
\Bs fraction        & $\fBs$               & \hfagZFBS       & \hfagTFBS      &         & \hfagWFBS   \\
\b-baryon fraction  & $\fbb$               & \hfagZFBB       & \hfagTFBB      &         & \hfagWFBB   \\
$\Bs/\Bd$ ratio     & $\fBs/\fBd$          & \hfagZFBSBD     & \hfagTFBSBD    & $0.256 \pm 0.020$ & \hfagWFBSBD \\
\multicolumn{2}{@{}l}{$\rho(\fBs,\fBu) = \rho(\fBs,\fBd)$} & \hfagZRHOFBDFBS & \hfagTRHOFBDFBS &         & \hfagWRHOFBDFBS \\
\multicolumn{2}{@{}l}{$\rho(\fbb,\fBu) = \rho(\fbb,\fBd)$} & \hfagZRHOFBDFBB & \hfagTRHOFBDFBB &         & \hfagWRHOFBDFBB \\
\multicolumn{2}{@{}l}{$\rho(\fbb,\fBs)$}                   & \hfagZRHOFBBFBS & \hfagTRHOFBBFBS &         & \hfagWRHOFBBFBS \\
\hline
\end{tabular}
\end{center}
\end{table}

Introducing the \chibar average in \Eq{fBs-mixing}, together with our world average 
$\chid = \hfagCHIDWU$ (see \Eq{chid} of \Sec{dmd}), the assumption $\chis= 1/2$ 
(justified by \Eq{chis} in \Sec{dms}), the 
best knowledge of the lifetimes (see \Sec{lifetimes}) and the estimate of \fbb given above, 
yields $\fBs = \hfagWFBSMIX$ 
(or $\fBs = \hfagZFBSMIX$ using only LEP data, 
or $\fBs = \hfagTFBSMIX$ using only Tevatron data),
an estimate dominated by the mixing information. 
Taking into account all known correlations (including that introduced by \fbb), 
this result is then combined with the set of fractions obtained from direct measurements 
(given above), to yield the 
improved estimates of \Table{fractions}, 
still under the constraints of \Eq{constraints}.
As can be seen, our knowledge on the mixing parameters 
substantially reduces the uncertainty on \fBs.
It should be noted that the results 
are correlated, as indicated in \Table{fractions}.

Although no recent measurements of the fractions have become available, 
the averages of \Table{fractions} (and most notably the \b-baryon fraction) 
have significantly improved in precision as compared to those given in our 
previous report~\cite{Amhis:2012bh}. This is mostly due to a new and precise 
model-independent measurement of the $\Lambda_c^+ \to p K^-\pi^+$
branching fraction from Belle~\cite{Zupanc:2013iki}, 
which has been used to adjust the fractions obtained from direct measurements. 


%
%

\mysubsection{\b-hadron lifetimes}
\labs{lifetimes}

In the spectator model the decay of \b-flavoured hadrons $H_b$ is
governed entirely by the flavour changing \particle{b\to Wq} transition
($\particle{q}=\particle{c,u}$).  For this very reason, lifetimes of all
\b-flavoured hadrons are the same in the spectator approximation
regardless of the (spectator) quark content of the $H_b$.  In the early
1990's experiments became sophisticated enough to start seeing the
differences of the lifetimes among various $H_b$ species.  The first
theoretical calculations of the spectator quark effects on $H_b$
lifetime emerged only few years earlier.

Currently, most such calculations are performed in the framework of
the Heavy Quark Expansion, HQE.  In the HQE, under certain assumptions
(the most important of which is that of
quark-hadron duality~\cite{Shifman:2000jv}), the decay
rate of an $H_b$ to an inclusive final state $f$ is expressed as the sum
of a series of expectation values of operators of increasing dimension,
multiplied by the correspondingly higher powers of $\Lambda_{\rm
QCD}/m_b$:
\begin{equation}
\Gamma_{H_b\to f} = |{\rm CKM}|^2\sum_n c_n^{(f)}
\Bigl(\frac{\Lambda_{\rm QCD}}{m_b}\Bigr)^n\langle H_b|O_n|H_b\rangle,
\labe{hqe}
\end{equation}
where $|{\rm CKM}|^2$ is the relevant combination of the CKM matrix elements.
The coefficients $c_n^{(f)}$ of this expansion, known as the Operator Product
Expansion~\cite{Shifman:1986mx,*Chay:1990da,*Bigi:1992su,*Bigi:1992su_erratum},
can be calculated perturbatively.  Hence, the HQE
predicts $\Gamma_{H_b\to f}$ in the form of an expansion in both
$\Lambda_{\rm QCD}/m_{\b}$ and $\alpha_s(m_{\b})$.  The precision of
current experiments makes it mandatory to go to the next-to-leading
order in QCD, \ie\ to include corrections of the order of
$\alpha_s(m_{\b})$ to the $c_n^{(f)}$ terms.  All non-perturbative physics
is shifted into the expectation values $\langle H_b|O_n|H_b\rangle$ of
operators $O_n$.  These can be calculated using lattice QCD or QCD sum
rules, or can be related to other observables via the
HQE~\cite{Bigi:1995jr,*Bellini:1996ra}.  One may reasonably expect that powers of
$\Lambda_{\rm QCD}/m_{\b}$ provide enough suppression that only the
first few terms of the sum in \Eq{hqe} matter.

Theoretical predictions are usually made for the ratios of the lifetimes
(with $\tau(\Bd)$ chosen as the common denominator) rather than for the
individual lifetimes, for this allows several uncertainties to cancel.
The precision of the current HQE calculations (see
\Refs{Ciuchini:2001vx,*Beneke:2002rj,*Franco:2002fc,Tarantino:2003qw,*Gabbiani:2003pq,Gabbiani:2004tp} for the latest updates)
is in some instances already surpassed by the measurements,
\eg\ in the case of $\tau(\Bu)/\tau(\Bd)$.  Also, HQE calculations are
not assumption-free.  More accurate predictions are a matter of progress
in the evaluation of the non-perturbative hadronic matrix elements and
verifying the assumptions that the calculations are based upon.
However, the HQE, even in its present shape, draws a number of important
conclusions, which are in agreement with experimental observations:
\begin{itemize}
\item The heavier the mass of the heavy quark, the smaller is the
  variation in the lifetimes among different hadrons containing this
  quark, which is to say that as $m_{\b}\to\infty$ we retrieve the
  spectator picture in which the lifetimes of all $H_b$ states are the same.
   This is well illustrated by the fact that lifetimes are rather
   similar in the \b sector, while they differ by large factors
   in the \particle{c} sector ($m_{\particle{c}}<m_{\b}$).
\item The non-perturbative corrections arise only at the order of
  $\Lambda_{\rm QCD}^2/m_{\b}^2$, which translates into 
  differences among $H_b$ lifetimes of only a few percent.
\item It is only the difference between meson and baryon lifetimes that
  appears at the $\Lambda_{\rm QCD}^2/m_{\b}^2$ level.  The splitting of the
  meson lifetimes occurs at the $\Lambda_{\rm QCD}^3/m_{\b}^3$ level, yet it is
  enhanced by a phase space factor $16\pi^2$ with respect to the leading
  free \b decay.
\end{itemize}

To ensure that certain sources of systematic uncertainty cancel, 
lifetime analyses are sometimes designed to measure
ratios of lifetimes.  However, because of the differences in decay
topologies, abundance (or lack thereof) of decays of a certain kind,
{\em etc.}, measurements of the individual lifetimes are also 
common.  In the following section we review the most common
types of lifetime measurements.  This discussion is followed by the
presentation of the averaging of the various lifetime measurements, each
with a brief description of its particularities.



\mysubsubsection{Lifetime measurements, uncertainties and correlations}

In most cases, the lifetime of an $H_b$ state is estimated from a flight
distance measurement
and a $\beta\gamma$ factor which is used to convert the geometrical
distance into the proper decay time.  Methods of accessing lifetime
information can roughly be divided in the following five categories:
\begin{enumerate}
\item {\bf\em Inclusive (flavour-blind) measurements}.  These
  measurements are aimed at extracting the lifetime from a mixture of
  \b-hadron decays, without distinguishing the decaying species.  Often
  the knowledge of the mixture composition is limited, which makes these
  measurements experiment-specific.  Also, these
  measurements have to rely on Monte Carlo simulation for estimating the
  $\beta\gamma$ factor, because the decaying hadrons are not fully
  reconstructed.  On the bright side, these are usually the largest
  statistics \b-hadron lifetime measurements that are accessible to a
  given experiment, and can, therefore, serve as an important
  performance benchmark.
\item {\bf\em Measurements in semileptonic decays of a specific
  {\boldmath $H_b$\unboldmath}}.  The \particle{W} boson from \particle{\b\to Wc}
  produces a $\ell\nu_l$ pair (\particle{\ell=e,\mu}) in about 21\% of the
  cases.  The electron or muon from such decays provides a clean and efficient
  trigger signature.
  The \particle{c} quark from the \particle{b\to Wc} transition and the other
  quark(s) making up the decaying $H_b$ combine into a charm hadron,
  which is reconstructed in one or more exclusive decay channels.
  Knowing what this charmed hadron is allows one to separate, at least
  statistically, different $H_b$ species.  The advantage of these
  measurements is in statistics, which is usually superior to the case of
  exclusively reconstructed $H_b$ decays.  Some of the main
  disadvantages are related to the difficulty of estimating the lepton+charm
  sample composition and to the Monte Carlo reliance for
  the momentum (and hence $\beta\gamma$ factor) estimate.
\item {\bf\em Measurements in exclusively reconstructed hadronic decays}.
  These
  have the advantage of complete reconstruction of the decaying $H_b$ state, 
  which allows one to infer the decaying species as well as to perform precise
  measurement of the $\beta\gamma$ factor.  Both lead to generally
  smaller systematic uncertainties than in the above two categories.
  The downsides are smaller branching ratios and larger combinatorial
  backgrounds, especially in $H_b\rightarrow H_c\pi(\pi\pi)$ and
  multi-body $H_c$ decays, or in a hadron collider environment with
  non-trivial underlying event.  Decays of the type $H_b\to \jpsi H_s$ are
  relatively clean and easy to trigger, due to the $\jpsi\to \ell^+\ell^-$
  signature, but their branching fraction is only about 1\%.
\item {\bf\em Measurements at asymmetric B factories}. 
  In the $\Ups\rightarrow B \bar{B}$ decay, the \B mesons (\Bu or \Bd) are
essentially at rest in the \Ups frame.  This makes direct lifetime
measurements impossible in experiments at symmetric colliders producing 
\Ups at rest. 
At asymmetric \B factories the \Ups meson is boosted
resulting in \B and \particle{\bar{B}} moving nearly parallel to each 
other with the same boost. The lifetime is inferred from the distance $\Delta z$        
separating the \B and \particle{\bar{B}} decay vertices along the beam axis 
and from the \Ups boost known from the beam energies. This boost is equal to 
$\beta \gamma \approx 0.55$ (0.43) in the \babar (\belle) experiment,
resulting in an average \B decay length of approximately 250~(190)~$\mu$m. 
In order to determine the charge of the \B mesons in each event, one of them is
fully reconstructed in a semileptonic or hadronic decay mode.
The other \B is typically not fully reconstructed, only the position
of its decay vertex is determined from the remaining tracks in the event.
These measurements benefit from large statistics, but suffer from poor proper time 
resolution, comparable to the \B lifetime itself. This resolution is dominated by the 
uncertainty on the decay vertices, which is typically 50~(100)~$\mu$m for a
fully (partially) reconstructed \B meson. 
With very large future statistics,
the resolution and purity could be improved (and hence the systematics reduced)
by fully reconstructing both \B mesons in the event. 
 
\item {\bf\em Direct measurement of lifetime ratios}.  This method, 
  initially applied 
  in the measurement of $\tau(\Bu)/\tau(\Bd)$, is now also used for other 
  \b-hadron species at the LHC. 
  The ratio of the lifetimes is extracted from the proper time 
  dependence of the ratio of the observed yields of 
  of two different \b-hadron species, 
  both reconstructed in decay modes with similar topologies. 
  The advantage of this method is that subtle efficiency effects
  (partially) cancel in the ratio. 
\end{enumerate}

In some of the latest analyses, measurements of two (\eg\ $\tau(\Bu)$ and
$\tau(\Bu)/\tau(\Bd)$) or three (\eg\ $\tau(\Bu)$,
$\tau(\Bu)/\tau(\Bd)$, and \dmd) quantities are combined.  This
introduces correlations among measurements.  Another source of
correlations among the measurements are the systematic effects, which
could be common to an experiment or to an analysis technique across the
experiments.  When calculating the averages, such correlations are taken
into account following the general procedure described in
\Ref{LEPBOSC:1996}.

\mysubsubsection{Inclusive \b-hadron lifetimes}

The inclusive \b hadron lifetime is defined as $\tau_{\b} = \sum_i f_i
\tau_i$ where $\tau_i$ are the individual species lifetimes and $f_i$ are
the fractions of the various species present in an unbiased sample of
weakly decaying \b hadrons produced at a high-energy
collider.\footnote{In principle such a quantity could be slightly
different in \particle{Z} decays, at the Tevatron or at the LHC, 
in case the
fractions of \b-hadron species are not exactly the same; see the
discussion in \Sec{fractions_high_energy}.}  This quantity is certainly
less fundamental than the lifetimes of the individual species, the
latter being much more useful in comparisons of the measurements with
the theoretical predictions.  Nonetheless, we perform the averaging of
the inclusive lifetime measurements for completeness as well as for the
reason that they might be of interest as ``technical numbers.''

\begin{table}[!htb]
\caption{Measurements of average \b-hadron lifetimes.}
\labt{lifeincl}
\begin{center}
\begin{tabular}{lcccl} \hline
Experiment &Method           &Data set & $\tau_{\b}$ (ps)       &Ref.\\
\hline
ALEPH  &Dipole               &91     &$1.511\pm 0.022\pm 0.078$ &\cite{Buskulic:1993gj}\\
DELPHI &All track i.p.\ (2D) &91--92 &$1.542\pm 0.021\pm 0.045$ &\cite{Abreu:1994dr}$^a$\\
DELPHI &Sec.\ vtx            &91--93 &$1.582\pm 0.011\pm 0.027$ &\cite{Abreu:1996hv}$^a$\\
DELPHI &Sec.\ vtx            &94--95 &$1.570\pm 0.005\pm 0.008$ &\cite{Abdallah:2003sb}\\
L3     &Sec.\ vtx + i.p.     &91--94 &$1.556\pm 0.010\pm 0.017$ &\cite{Acciarri:1997tt}$^b$\\
OPAL   &Sec.\ vtx            &91--94 &$1.611\pm 0.010\pm 0.027$ &\cite{Ackerstaff:1996as}\\
SLD    &Sec.\ vtx            &93     &$1.564\pm 0.030\pm 0.036$ &\cite{Abe:1995rm}\\ 
\hline
\multicolumn{2}{l}{Average set 1 (\b vertex)} && \hfagTAUBVTXnounit &\\
\hline\hline
ALEPH  &Lepton i.p.\ (3D)    &91--93 &$1.533\pm 0.013\pm 0.022$ &\cite{Buskulic:1995rw}\\
L3     &Lepton i.p.\ (2D)    &91--94 &$1.544\pm 0.016\pm 0.021$ &\cite{Acciarri:1997tt}$^b$\\
OPAL   &Lepton i.p.\ (2D)    &90--91 &$1.523\pm 0.034\pm 0.038$ &\cite{Acton:1993xk}\\ 
\hline
\multicolumn{2}{l}{Average set 2 ($\b\to\ell$)} && \hfagTAUBLEPnounit &\\
\hline\hline
CDF1   &\particle{\jpsi} vtx&92--95 &$1.533\pm 0.015^{+0.035}_{-0.031}$ &\cite{Abe:1997bd} \\ 
ATLAS &\particle{\jpsi} vtx& 2010 & $1.489\pm 0.016 \pm 0.043$ & \cite{ATLAS-CONF-2011-145}$^p$ \\
\hline
\multicolumn{2}{l}{Average set 3 (\particle{\b\to \jpsi})} && \hfagTAUBJPnounit & \\ 
\hline
\multicolumn{5}{l}{$^a$ \footnotesize The combined DELPHI result quoted in
\cite{Abreu:1996hv} is 1.575 $\pm$ 0.010 $\pm$ 0.026 ps.} \\[-0.5ex]
\multicolumn{5}{l}{$^b$ \footnotesize The combined L3 result quoted in \cite{Acciarri:1997tt} 
is 1.549 $\pm$ 0.009 $\pm$ 0.015 ps.} \\[-0.5ex]
\multicolumn{5}{l}{$^p$ \footnotesize Preliminary.}
\end{tabular}
\end{center}
\end{table}

In practice, an unbiased measurement of the inclusive lifetime is
difficult to achieve, because it would imply an efficiency which is
guaranteed to be the same across species.  So most of the measurements
are biased.  In an attempt to group analyses which are expected to
select the same mixture of \b hadrons, the available results (given in
\Table{lifeincl}) are divided into the following three sets:
\begin{enumerate}
\item measurements at LEP and SLD that accept any \b-hadron decay, based 
      on topological reconstruction (secondary vertex or track impact
      parameters);
\item measurements at LEP based on the identification
      of a lepton from a \b decay; and
\item measurements at the Tevatron based on inclusive 
      \particle{H_b\to \jpsi X} reconstruction, where the
      \particle{\jpsi} is fully reconstructed.
\end{enumerate}

The measurements of the first set are generally considered as estimates
of $\tau_{\b}$, although the efficiency to reconstruct a secondary
vertex most probably depends, in an analysis-specific way, on the number
of tracks coming from the vertex, thereby depending on the type of the
$H_b$.  Even though these efficiency variations can in principle be
accounted for using Monte Carlo simulations (which inevitably contain
assumptions on branching fractions), the $H_b$ mixture in that case can
remain somewhat ill-defined and could be slightly different among
analyses in this set.

On the contrary, the mixtures corresponding to the other two sets of
measurements are better defined in the limit where the reconstruction
and selection efficiency of a lepton or a \particle{\jpsi} from an
$H_b$ does not depend on the decaying hadron type.  These mixtures are
given by the production fractions and the inclusive branching fractions
for each $H_b$ species to give a lepton or a \particle{\jpsi}.  In
particular, under the assumption that all \b hadrons have the same
semileptonic decay width, the analyses of the second set should measure
$\tau(\b\to\ell) = (\sum_i f_i \tau_i^3) /(\sum_i f_i \tau_i^2)$ which is
necessarily larger than $\tau_{\b}$ if lifetime differences exist.
Given the present knowledge on $\tau_i$ and $f_i$,
$\tau(\b\to\ell)-\tau_{\b}$ is expected to be of the order of 0.003\ps.
On the other hand, the third set measuring $\tau(\b\to\particle{\jpsi})$
is expected to give an average smaller than $\tau_{\b}$ because 
of the \Bc meson which has a significantly
larger probability to decay to a \particle{\jpsi}
than other \b-hadron species. 

Measurements by SLC and LEP experiments are subject to a number of
common systematic uncertainties, such as those due to (lack of knowledge
of) \b and \particle{c} fragmentation, \b and \particle{c} decay models,
\BR{B\to\ell}, \BR{B\to c\to\ell}, \BR{c\to\ell}, $\tau_{\particle{c}}$,
and $H_b$ decay multiplicity.  In the averaging, these systematic
uncertainties are assumed to be 100\% correlated.  The averages for the
sets defined above (also given in \Table{lifeincl}) are
\begin{eqnarray}
\tau(\b~\mbox{vertex}) &=& \hfagTAUBVTX \,, \labe{TAUBVTX} \\
\tau(\b\to\ell) &=& \hfagTAUBLEP  \,, \\
\tau(\b\to\particle{\jpsi}) &=& \hfagTAUBJP\,.
\end{eqnarray}

\mysubsubsection{\Bd and \Bu lifetimes and their ratio}
\labs{taubd}
\labs{taubu}
\labs{lifetime_ratio}

After a number of years of dominating these averages the LEP experiments
yielded the scene to the asymmetric \B~factories and
the Tevatron experiments. The \B~factories have been very successful in
utilizing their potential -- in only a few years of running, \babar and,
to a greater extent, \belle, have struck a balance between the
statistical and the systematic uncertainties, with both being close to
(or even better than) the impressive 1\%.  In the meanwhile, CDF and
\dzero have emerged as significant contributors to the field as the
Tevatron Run~II data flowed in. Recently, the LHCb experiment reached 
a further step in precision, improving by a factor $\sim 2 $ 
over the previous best measurement. 


At present time we are in an interesting position of having three sets
of measurements (from LEP/SLC, \B factories and the Tevatron) that
originate from different environments, obtained using substantially
different techniques and are precise enough for incisive comparison.


\begin{table}[!htb]
\caption{Measurements of the \Bd lifetime.}
\labt{lifebd}
\begin{center}
\begin{tabular}{lcccl} \hline
Experiment &Method                    &Data set &$\tau(\Bd)$ (ps)                  &Ref.\\
\hline
ALEPH  &\particle{D^{(*)} \ell}       &91--95 &$1.518\pm 0.053\pm 0.034$          &\cite{Barate:2000bs}\\
ALEPH  &Exclusive                     &91--94 &$1.25^{+0.15}_{-0.13}\pm 0.05$     &\cite{Buskulic:1996hy}\\
ALEPH  &Partial rec.\ $\pi^+\pi^-$    &91--94 &$1.49^{+0.17+0.08}_{-0.15-0.06}$   &\cite{Buskulic:1996hy}\\
DELPHI &\particle{D^{(*)} \ell}       &91--93 &$1.61^{+0.14}_{-0.13}\pm 0.08$     &\cite{Abreu:1995mc}\\
DELPHI &Charge sec.\ vtx              &91--93 &$1.63 \pm 0.14 \pm 0.13$           &\cite{Adam:1995mb}\\
DELPHI &Inclusive \particle{D^* \ell} &91--93 &$1.532\pm 0.041\pm 0.040$          &\cite{Abreu:1996gb}\\
DELPHI &Charge sec.\ vtx              &94--95 &$1.531 \pm 0.021\pm0.031$          &\cite{Abdallah:2003sb}\\
L3     &Charge sec.\ vtx              &94--95 &$1.52 \pm 0.06 \pm 0.04$           &\cite{Acciarri:1998uv}\\
OPAL   &\particle{D^{(*)} \ell}       &91--93 &$1.53 \pm 0.12 \pm 0.08$           &\cite{Akers:1995pa}\\
OPAL   &Charge sec.\ vtx              &93--95 &$1.523\pm 0.057\pm 0.053$          &\cite{Abbiendi:1998av}\\
OPAL   &Inclusive \particle{D^* \ell} &91--00 &$1.541\pm 0.028\pm 0.023$          &\cite{Abbiendi:2000ec}\\
SLD    &Charge sec.\ vtx $\ell$       &93--95 &$1.56^{+0.14}_{-0.13} \pm 0.10$    &\cite{Abe:1997ys}$^a$\\
SLD    &Charge sec.\ vtx              &93--95 &$1.66 \pm 0.08 \pm 0.08$           &\cite{Abe:1997ys}$^a$\\
CDF1   &\particle{D^{(*)} \ell}       &92--95 &$1.474\pm 0.039^{+0.052}_{-0.051}$ &\cite{Abe:1998wt}\\
CDF1  &Excl.\ \particle{\jpsi K^{*0}}&92--95 &$1.497\pm 0.073\pm 0.032$          &\cite{Acosta:2002nd}\\
CDF2   &Excl.\ \particle {\jpsi K_S^0}, \particle{\jpsi K^{*0}} &02--09 &$1.507\pm 0.010\pm0.008$           &\cite{Aaltonen:2010pj,*Abulencia:2006dr_mod_cont} \\
\dzero &Excl.\ \particle{\jpsi K^{*0}}&03--07 &$1.414\pm0.018\pm0.034$ &\cite{Abazov:2008jz,*Abazov:2005sa_mod_cont}\\ 
\dzero &Excl.\ \particle {\jpsi K_S^0} &02--11 &$1.508 \pm0.025 \pm0.043$  &\cite{Abazov:2012iy,*Abazov:2007sf_mod_cont,*Abazov:2004bn_mod_cont} \\
\dzero &Inclusive \particle {D^-\mu^+} &02--11 &$1.534 \pm0.019 \pm0.021$  & \cite{Abazov:2014rua,*Abazov:2006cb_cont} \\ 
\babar &Exclusive                     &99--00 &$1.546\pm 0.032\pm 0.022$          &\cite{Aubert:2001uw}\\
\babar &Inclusive \particle{D^* \ell} &99--01 &$1.529\pm 0.012\pm 0.029$          &\cite{Aubert:2002gi,*Aubert:2002gi_erratum}\\
\babar &Exclusive \particle{D^* \ell} &99--02 &$1.523^{+0.024}_{-0.023}\pm 0.022$ &\cite{Aubert:2002sh}\\
\babar &Incl.\ \particle{D^*\pi}, \particle{D^*\rho} 
                                      &99--01 &$1.533\pm 0.034 \pm 0.038$         &\cite{Aubert:2002ms}\\
\babar &Inclusive \particle{D^* \ell}
&99--04 &$1.504\pm0.013^{+0.018}_{-0.013}$  &\cite{Aubert:2005kf} \\ 
\belle & Exclusive                     & 00--03 & $1.534\pm 0.008\pm0.010$        & \cite{Abe:2004mz}\\
ATLAS & Excl.\ \particle{\jpsi K^{*0}} & 2010 & $1.51 \pm0.04 \pm0.04$ & \cite{ATLAS-CONF-2011-092}$^p$ \\
LHCb  & Excl.\ \particle{\jpsi K^{*0}} & 2011 & $1.524 \pm0.006 \pm 0.004$ & \cite{Aaij:2014owa} \\
LHCb  & Excl.\ \particle {\jpsi K_S^0}   & 2011 & $1.499 \pm0.013 \pm 0.005$ & \cite{Aaij:2014owa} \\
LHCb    & \particle{K^+\pi^-}   & 2011 & $1.524 \pm 0.011 \pm 0.004$ & \cite{Aaij:2014fia,*Aaij:2012ns_cont} \\
\hline
Average&                               &        & \hfagTAUBDnounit & \\
\hline\hline           
\multicolumn{5}{l}{$^a$ \footnotesize The combined SLD result 
quoted in \cite{Abe:1997ys} is 1.64 $\pm$ 0.08 $\pm$ 0.08 ps.}\\[-0.5ex]
\multicolumn{5}{l}{$^p$ {\footnotesize Preliminary.}}
\end{tabular}
\end{center}
\end{table}

\afterpage{\clearpage}

\begin{table}[tbp]
\caption{Measurements of the \Bu lifetime.}
\labt{lifebu}
\begin{center}
\begin{tabular}{lcccl} \hline
Experiment &Method                 &Data set &$\tau(\Bu)$ (ps)                 &Ref.\\
\hline
ALEPH  &\particle{D^{(*)} \ell}    &91--95 &$1.648\pm 0.049\pm 0.035$          &\cite{Barate:2000bs}\\
ALEPH  &Exclusive                  &91--94 &$1.58^{+0.21+0.04}_{-0.18-0.03}$   &\cite{Buskulic:1996hy}\\
DELPHI &\particle{D^{(*)} \ell}    &91--93 &$1.61\pm 0.16\pm 0.12$             &\cite{Abreu:1995mc}$^a$\\
DELPHI &Charge sec.\ vtx           &91--93 &$1.72\pm 0.08\pm 0.06$             &\cite{Adam:1995mb}$^a$\\
DELPHI &Charge sec.\ vtx           &94--95 &$1.624\pm 0.014\pm 0.018$          &\cite{Abdallah:2003sb}\\
L3     &Charge sec.\ vtx           &94--95 &$1.66\pm  0.06\pm 0.03$            &\cite{Acciarri:1998uv}\\
OPAL   &\particle{D^{(*)} \ell}    &91--93 &$1.52 \pm 0.14\pm 0.09$            &\cite{Akers:1995pa}\\
OPAL   &Charge sec.\ vtx           &93--95 &$1.643\pm 0.037\pm 0.025$          &\cite{Abbiendi:1998av}\\
SLD    &Charge sec.\ vtx $\ell$    &93--95 &$1.61^{+0.13}_{-0.12}\pm 0.07$     &\cite{Abe:1997ys}$^b$\\
SLD    &Charge sec.\ vtx           &93--95 &$1.67\pm 0.07\pm 0.06$             &\cite{Abe:1997ys}$^b$\\
CDF1   &\particle{D^{(*)} \ell}    &92--95 &$1.637\pm 0.058^{+0.045}_{-0.043}$ &\cite{Abe:1998wt}\\
CDF1   &Excl.\ \particle{\jpsi K} &92--95 &$1.636\pm 0.058\pm 0.025$          &\cite{Acosta:2002nd}\\
CDF2   &Excl.\ \particle{\jpsi K} &02--09 &$1.639\pm 0.009\pm 0.009$          &\cite{Aaltonen:2010pj,*Abulencia:2006dr_mod_cont}\\ 
CDF2   &Excl.\ \particle{D^0 \pi}  &02--06 &$1.663\pm 0.023\pm0.015$           &\cite{Aaltonen:2010ta}\\
\babar &Exclusive                  &99--00 &$1.673\pm 0.032\pm 0.023$          &\cite{Aubert:2001uw}\\
\belle &Exclusive                  &00--03 &$1.635\pm 0.011\pm 0.011$          &\cite{Abe:2004mz}\\
LHCb  & Excl.\ \particle{\jpsi K} & 2011 & $1.637 \pm0.004 \pm 0.003$ & \cite{Aaij:2014owa} \\
\hline
Average&                           &       &\hfagTAUBUnounit &\\
\hline\hline
\multicolumn{5}{l}{$^a$ \footnotesize The combined DELPHI result quoted 
in~\cite{Adam:1995mb} is $1.70 \pm 0.09$ ps.} \\[-0.5ex]
\multicolumn{5}{l}{$^b$ \footnotesize The combined SLD result 
quoted in~\cite{Abe:1997ys} is $1.66 \pm 0.06 \pm 0.05$ ps.}\\[-0.5ex]
\end{tabular}
\end{center}
\end{table}

\begin{table}[tb]
\caption{Measurements of the ratio $\tau(\Bu)/\tau(\Bd)$.}
\labt{liferatioBuBd}
\begin{center}
\begin{tabular}{lcccl} 
\hline
Experiment &Method                 &Data set &Ratio $\tau(\Bu)/\tau(\Bd)$      &Ref.\\
\hline
ALEPH  &\particle{D^{(*)} \ell}    &91--95 &$1.085\pm 0.059\pm 0.018$          &\cite{Barate:2000bs}\\
ALEPH  &Exclusive                  &91--94 &$1.27^{+0.23+0.03}_{-0.19-0.02}$   &\cite{Buskulic:1996hy}\\
DELPHI &\particle{D^{(*)} \ell}    &91--93 &$1.00^{+0.17}_{-0.15}\pm 0.10$     &\cite{Abreu:1995mc}\\
DELPHI &Charge sec.\ vtx           &91--93 &$1.06^{+0.13}_{-0.11}\pm 0.10$     &\cite{Adam:1995mb}\\
DELPHI &Charge sec.\ vtx           &94--95 &$1.060\pm 0.021 \pm 0.024$         &\cite{Abdallah:2003sb}\\
L3     &Charge sec.\ vtx           &94--95 &$1.09\pm 0.07  \pm 0.03$           &\cite{Acciarri:1998uv}\\
OPAL   &\particle{D^{(*)} \ell}    &91--93 &$0.99\pm 0.14^{+0.05}_{-0.04}$     &\cite{Akers:1995pa}\\
OPAL   &Charge sec.\ vtx           &93--95 &$1.079\pm 0.064 \pm 0.041$         &\cite{Abbiendi:1998av}\\
SLD    &Charge sec.\ vtx $\ell$    &93--95 &$1.03^{+0.16}_{-0.14} \pm 0.09$    &\cite{Abe:1997ys}$^a$\\
SLD    &Charge sec.\ vtx           &93--95 &$1.01^{+0.09}_{-0.08} \pm0.05$     &\cite{Abe:1997ys}$^a$\\
CDF1   &\particle{D^{(*)} \ell}    &92--95 &$1.110\pm 0.056^{+0.033}_{-0.030}$ &\cite{Abe:1998wt}\\
CDF1   &Excl.\ \particle{\jpsi K} &92--95 &$1.093\pm 0.066 \pm 0.028$         &\cite{Acosta:2002nd}\\
CDF2   &Excl.\ \particle{\jpsi K^{(*)}} &02--09 &$1.088\pm 0.009 \pm 0.004$   &\cite{Aaltonen:2010pj,*Abulencia:2006dr_mod_cont}\\ 
\dzero &\particle{D^{*+} \mu} \particle{D^0 \mu} ratio
	                           &02--04 &$1.080\pm 0.016\pm 0.014$          &\cite{Abazov:2004sa}\\
\babar &Exclusive                  &99--00 &$1.082\pm 0.026\pm 0.012$          &\cite{Aubert:2001uw}\\
\belle &Exclusive                  &00--03 &$1.066\pm 0.008\pm 0.008$          &\cite{Abe:2004mz}\\
LHCb  & Excl.\ \particle{\jpsi K^{(*)}} & 2011 & $1.074 \pm0.005 \pm 0.003$ & \cite{Aaij:2014owa} \\
\hline
Average&                           &       & \hfagRTAUBU & \\   
\hline\hline
\multicolumn{5}{l}{$^a$ \footnotesize The combined SLD result quoted
	   in~\cite{Abe:1997ys} is $1.01 \pm 0.07 \pm 0.06$.}
\end{tabular}
\end{center}
\end{table}

The averaging of $\tau(\Bu)$, $\tau(\Bd)$ and $\tau(\Bu)/\tau(\Bd)$
measurements is summarized\footnote{%
We do not include the old unpublished measurements of Refs.~\cite{CDFnote7514:2005,CDFnote7386:2005}.}
in \Tablesss{lifebd}{lifebu}{liferatioBuBd}.
For $\tau(\Bu)/\tau(\Bd)$ we average only the measurements of this
quantity provided by experiments rather than using all available
knowledge, which would have included, for example, $\tau(\Bu)$ and
$\tau(\Bd)$ measurements which did not contribute to any of the ratio
measurements.

The following sources of correlated (within experiment/machine)
systematic uncertainties have been considered:
\begin{itemize}
\item for SLC/LEP measurements -- \particle{D^{**}} branching ratio uncertainties~\cite{Abbaneo:2000ej_mod,*Abbaneo:2001bv_mod_cont},
momentum estimation of \b mesons from \particle{Z^0} decays
(\b-quark fragmentation parameter $\langle X_E \rangle = 0.702 \pm 0.008$~\cite{Abbaneo:2000ej_mod,*Abbaneo:2001bv_mod_cont}),
\Bs and \b-baryon lifetimes (see \Secss{taubs}{taulb}),
and \b-hadron fractions at high energy (see \Table{fractions}); 
\item for \babar measurements -- alignment, $z$ scale, PEP-II boost,
sample composition (where applicable);
\item for \dzero and CDF Run~II measurements -- alignment (separately
within each experiment).
\end{itemize}
The resultant averages are:
\begin{eqnarray}
\tau(\Bd) & = & \hfagTAUBD \,, \\
\tau(\Bu) & = & \hfagTAUBU \,, \\
\tau(\Bu)/\tau(\Bd) & = & \hfagRTAUBU \,.
\end{eqnarray}
%
%
%

\mysubsubsection{\Bs lifetimes}
\labs{taubs}

Like neutral kaons, neutral \B mesons contain
short- and long-lived components, since the
light (L) and heavy (H)
eigenstates, $\B_{\rm L}$ and $\B_{\rm H}$, differ not only
in their masses, but also in their total decay widths,  
with a decay width difference defined as 
$\DG = \Gamma_{\rm L} - \Gamma_{\rm H}$. 
Neglecting \CP violation in $\B-\Bbar$ mixing, 
which is expected to be very
small~\cite{Lenz:2011ti,*Lenz:2006hd,Beneke:1998sy} (see also \Sec{qpd}),
the mass eigenstates are also \CP eigenstates,
with the light $\B_{\rm L}$ state being \CP-even 
and the heavy $\B_{\rm H}$ state being \CP-odd. 
While the decay width difference \DGd can be neglected in the \Bd system, 
the \Bs system exhibits a significant
value of \DGs: the sign of \DGs is known 
to be positive~\cite{Aaij:2012eq}, \ie\
the heavy eigenstate lives longer than the light eigenstate. 
Specific measurements of \DGs and 
$\Gs = (\Gamma_{\rm L} + \Gamma_{\rm H})/2$ are explained
and averaged in \Sec{DGs}, but the results for
$1/\Gamma_{\rm L}$, $1/\Gamma_{\rm H}$ and
the mean \Bs lifetime, defined as $\tau(\Bs) = 1/\Gs$,
are also quoted at the end of this section. 

Many \Bs lifetime analyses, in particular the early 
ones performed before the non-zero value of \DGs was 
firmly established, ignore \DGs and fit the proper time 
distribution of a sample of \Bs candidates 
reconstructed in a certain final state $f$
with a model assuming a single exponential function 
for the signal. We denote such {\rm effective lifetime}
measurements~\cite{Fleischer:2011cw} as $\tau_{\rm single}(\Bs\to f)$; 
their true values may lie {\em a priori} anywhere
between $1/\Gamma_{\rm L} = 1/(\Gs+\DGs/2)$ and
$1/\Gamma_{\rm H}= 1/(\Gs-\DGs/2)$, 
depending on the proportion of $\B_{\rm L}$ and $\B_{\rm H}$
in the final state $f$. 
More recent determinations of effective lifetimes may be interpreted as
measurements of the relative composition of $\B_{\rm L}$ and $\B_{\rm H}$
decaying to the final state $f$. 
\Table{lifebs} summarizes the effective 
lifetime measurements.

Averaging measurements of $\tau_{\rm single}(\Bs\to f)$
over several final states $f$ will yield a result 
corresponding to an ill-defined observable
when the proportions of $\B_{\rm L}$ and $\B_{\rm H}$ differ. 
Therefore,
the effective \Bs lifetime measurements are broken down into
several categories and averaged separately.

\begin{table}[t]
\caption{Measurements of the effective \Bs lifetimes obtained from single exponential fits.}
\labt{lifebs}
\begin{center}
\resizebox{\textwidth}{!}{
\begin{tabular}{l@{}c@{}c@{}c@{}rc@{}l} \hline
Experiment & \multicolumn{2}{c}{Final state $f$}           & \multicolumn{2}{c}{Data set} & $\tau_{\rm single}(\Bs\to f)$ (ps) & Ref. \\
\hline \hline
ALEPH  & \particle{D_s^- \ell^+}  & flavour-specific & 91--95 & & $1.54^{+0.14}_{-0.13}\pm 0.04$   & \cite{Buskulic:1996ei}          \\
CDF1   & \particle{D_s^- \ell^+}  & flavour-specific & 92--96 & & $1.36\pm 0.09 ^{+0.06}_{-0.05}$  & \cite{Abe:1998cj}           \\
DELPHI & \particle{D_s^- \ell^+}  & flavour-specific & 91--95 & & $1.42^{+0.14}_{-0.13}\pm 0.03$   & \cite{Abreu:2000sh}          \\
OPAL   & \particle{D_s^- \ell^+}  & flavour-specific & 90--95 & & $1.50^{+0.16}_{-0.15}\pm 0.04$   & \cite{Ackerstaff:1997qi}  \\
\dzero & \particle{D_s^-\mu^+X}   & flavour-specific & Run II & 10.4 fb$^{-1}$ & $1.479 \pm 0.010 \pm 0.021$   & \cite{Abazov:2014rua,*Abazov:2006cb_cont} \\
CDF2   & \particle{D_s^- \pi^+ (X)} 
                              & flavour-specific & 02--06 & 1.3 fb$^{-1}$ & $1.518 \pm 0.041 \pm 0.027     $   & \cite{Aaltonen:2011qsa,*Aaltonen:2011qsa_cont} \\ 
LHCb   &  \particle{D_s^- D^+} & flavour-specific & 11--12 & 3 fb$^{-1}$ & $1.52 \pm 0.15 \pm 0.01$ & \cite{Aaij:2013bvd} \\
LHCb   &  \particle{D_s^- \pi^+} & flavour-specific & 11 & 1 fb$^{-1}$ & $1.535 \pm 0.015 \pm 0.014$ & \cite{Aaij:2014sua} \\
\multicolumn{5}{l}{Average of above 8 flavour-specific lifetime measurements} &  \hfagTAUBSSLnounit & \\  
\hline\hline
LHCb    & \particle{\pi^+K^-}   &  flavour-specific & 11 & 1.0 fb$^{-1}$ & $1.60 \pm 0.06 \pm 0.01$ & \cite{Aaij:2014fia,*Aaij:2012ns_cont} \\
\hline
ALEPH  & \particle{D_s h}     & ill-defined & 91--95 & & $1.47\pm 0.14\pm 0.08$           & \cite{Barate:1997ua}          \\
DELPHI & \particle{D_s h}     & ill-defined & 91--95 & & $1.53^{+0.16}_{-0.15}\pm 0.07$   & \cite{Abreu:2000ev} \\
OPAL   & \particle{D_s} incl. & ill-defined & 90--95 & & $1.72^{+0.20+0.18}_{-0.19-0.17}$ & \cite{Ackerstaff:1997ne}          \\ 
\hline
CDF1     & \particle{\jpsi\phi} & \CP even+odd & 92--95 &  & $1.34^{+0.23}_{-0.19}    \pm 0.05$ & \cite{Abe:1997bd} \\
\dzero   & \particle{\jpsi\phi} & \CP even+odd & 02--04 &  & $1.444^{+0.098}_{-0.090} \pm 0.02$ & \cite{Abazov:2004ce}  \\
ATLAS & \particle{\jpsi\phi} & \CP even+odd & 10 & 40 pb$^{-1}$ & $1.41 \pm0.08 \pm0.05$ & \cite{ATLAS-CONF-2011-092}$^p$ \\
LHCb  & \particle{\jpsi\phi} & \CP even+odd & 11 & 1 fb$^{-1}$ & $1.480 \pm0.011 \pm 0.005$ & \cite{Aaij:2014owa} \\
\multicolumn{5}{l}{Average of above 4 \particle{\jpsi \phi} lifetime measurements} &  \hfagTAUBSJFnounit & \\ 
\hline\hline
ALEPH    & \particle{D_s^{(*)+}D_s^{(*)-}} & mostly \CP even & 91--95 & & $1.27 \pm 0.33 \pm 0.08$ & \cite{Barate:2000kd} \\
LHCb    & \particle{K^+K^-}   &  \CP-even & 10 & 0.037 fb$^{-1}$ & $1.440 \pm 0.096 \pm 0.009$ & \cite{Aaij:2011kn} \\
LHCb    & \particle{K^+K^-}   &  \CP-even & 11 & 1.0 fb$^{-1}$ & $1.407 \pm 0.016 \pm 0.007$ & \cite{Aaij:2014fia,*Aaij:2012ns_cont} \\
\multicolumn{5}{l}{Average of above 2 \particle{K^+K^-} lifetime measurements} &  \hfagTAUBSKKnounit & \\ 
LHCb   &  \particle{D_s^+ D_s^-} & \CP-even & 11--12 & 3 fb$^{-1}$ & $1.379 \pm 0.026 \pm 0.017$ & \cite{Aaij:2013bvd} \\
\multicolumn{5}{l}{Average of above 1 measurement of $1/\Gamma_{\rm L}$} &  \hfagTAUBSSHORTnounit & \\ \hline \hline
LHCb     & \particle{\jpsi K^0_{\rm S}} & \CP-odd & 11   & 1.0 fb$^{-1}$ & $1.75 \pm 0.12 \pm 0.07$ & \cite{Aaij:2013eia} \\
CDF2     & \particle{\jpsi f_0(980)} & \CP-odd & 02--08 & 3.8 fb$^{-1}$ & $1.70^{+0.12}_{-0.11} \pm 0.03$ & \cite{Aaltonen:2011nk} \\
LHCb     & \particle{\jpsi \pi^+\pi^-} & \CP-odd & 11   & 1.0 fb$^{-1}$ & $1.652 \pm 0.024 \pm 0.024$ & \cite{Aaij:2013oba,*LHCb:2011aa_mod,*LHCb:2012ad_mod,*LHCb:2011ab_mod,*Aaij:2012nta_mod} \\
\multicolumn{5}{l}{Average of above 2 measurements of $1/\Gamma_{\rm H}$} &  \hfagTAUBSLONGnounit & \\ \hline \hline
\multicolumn{5}{l}{$^p$ \footnotesize Preliminary.}
\end{tabular}
}
\end{center}
\end{table}

\afterpage{\clearpage}

\begin{itemize}
\item 
{\bf\em Decays to a flavour-specific final state}
without \CP violation in the decay amplitude,
such as $\Bs \to \particle{D_s^- \ell^+ \nu}$
or $\Bs\to \particle {D_s^- \pi^+}$, have equal 
fractions of $\B_{\rm L}$ and $\B_{\rm H}$ at time zero.\footnote{%
The assumption that such decays are flavour-specific is valid to an excellent approximation in the SM.
However, there are few experimental tests of it.}
If the resulting superposition of two exponential distributions
is fitted with a single exponential function, 
one obtains a measure of the so-called {\em flavour-specific lifetime}~\cite{Hartkorn:1999ga}:
\begin{equation}
\tau_{\rm single}(\Bs\to \mbox{flavour specific}) = \frac{1}{\Gs}
\frac{{1+\left(\frac{\DGs}{2\Gs}\right)^2}}{{1-\left(\frac{\DGs}{2\Gs}\right)^2}
}.
\labe{fslife}
\end{equation}

The average of all flavour-specific 
\Bs lifetime measurements\footnote{%
An old unpublished measurement~\cite{CDFnote7757:2005} is not included.}
is
\begin{equation}
\tau_{\rm single}(\Bs\to \mbox{flavour specific}) = \hfagTAUBSSL \,.
\labe{tau_fs}
\end{equation}
This average does not include an effective lifetime measurement of 
$\Bs \to \pi^+K^-$ decays~\cite{Aaij:2014fia,*Aaij:2012ns_cont}.

\item
{\bf\em \boldmath $\Bs\to D_s^{\mp} X$ decays}
include flavour-specific decays but also decays 
with an unknown mixture of light and heavy components. 
Measurements performed with such inclusive states are
no longer used in averages. 

\item
{\bf\em 
{\boldmath $\Bs \to \jpsi\phi$ \unboldmath}decays}
contain a well-measured mixture of \CP-even and \CP-odd states.
There are no known correlations
between the existing 
\particle{\Bs\to \jpsi\phi}
effective lifetime measurements; these are combined  
into the average\footnote{%
An old unpublished measurement~\cite{CDFnote8524:2007,*CDFnote8524:2007_cont} is not included.}
$\tau_{\rm single}(\Bs\to \jpsi \phi) = \hfagTAUBSJF$. 
A caveat is that different experimental acceptances
may lead to different admixtures of the 
\CP-even and \CP-odd states, and simple fits to a single
exponential may result in inherently different 
values of $\tau_{\rm single}(\Bs\to \jpsi \phi)$.
Analyses that separate the \CP-even and \CP-odd components in
this decay through a full angular study, outlined in \Sec{DGs},
provide directly precise measurements of $1/\Gs$ and $\DGs$ (see \Table{phisDGsGs}).

\item
{\bf\em Decays to \boldmath\CP eigenstates} have also 
been measured, in the \CP-even modes 
$\Bs \to D_s^{(*)+}D_s^{(*)-}$ by ALEPH~\cite{Barate:2000kd},
$\Bs \to K^+ K^-$ by LHCb~\cite{Aaij:2011kn,Aaij:2014fia,*Aaij:2012ns_cont}%
\footnote{An old unpublished measurement of the $\Bs \to K^+ K^-$
effective lifetime by CDF~\cite{Tonelli:2006np} is no longer considered.}
and $\Bs \to D_s^+D_s^-$ by LHCb~\cite{Aaij:2013bvd}, as well as in the \CP-odd modes 
$\Bs \to \jpsi f_0(980)$ by CDF~\cite{Aaltonen:2011nk}, 
$\Bs \to \jpsi \pi^+\pi^-$ by LHCb~\cite{Aaij:2013oba,*LHCb:2011aa_mod,*LHCb:2012ad_mod,*LHCb:2011ab_mod,*Aaij:2012nta_mod}
and $\Bs \to \jpsi K^0_{\rm S}$ by LHCb~\cite{Aaij:2013eia}.
If these 
decays are dominated by a single weak phase and if \CP violation 
can be neglected, then $\tau_{\rm single}(\Bs \to \mbox{\CP-even}) = 1/\Gamma_{\rm L}$ 
and  $\tau_{\rm single}(\Bs \to \mbox{\CP-odd}) = 1/\Gamma_{\rm H}$ 
(see \Eqss{tau_KK_approx}{tau_Jpsif0_approx} for approximate relations in presence of
\CP violation in the mixing). 
However, not all these modes can be considered as pure \CP eigenstates:
a small \CP-odd component is most probably present
in $\Bs \to D_s^{(*)+}D_s^{(*)-}$ decays. Furthermore the decays
$\Bs \to K^+ K^-$ and $\Bs \to \jpsi K^0_{\rm S}$ 
may suffer from \CP violation due to interfering tree and loop amplitudes. 
The averages for the effective lifetimes obtained for decays to
pure \CP-even ($D_s^+D_s^-$) and \CP-odd ($\jpsi f_0(980)$, $\jpsi \pi^+\pi^-$)
final states, where \CP conservation can be assumed, are
\begin{eqnarray}
\tau_{\rm single}(\Bs \to \mbox{\CP-even}) & = & \hfagTAUBSSHORT \,,
\labe{tau_KK}
\\
\tau_{\rm single}(\Bs \to \mbox{\CP-odd}) & = & \hfagTAUBSLONG \,.
\labe{tau_Jpsif0}
\end{eqnarray}

\end{itemize}

As described in \Sec{DGs}, 
the effective lifetime averages of \Eqsss{tau_fs}{tau_KK}{tau_Jpsif0}
are used as ingredients to improve the 
determination of $1/\Gs$ and \DGs obtained from the full angular analyses
of $\Bs\to \jpsi\phi$ and $\Bs\to \jpsi K^+K^-$ decays. 
The resulting world averages for the \Bs lifetimes are
\begin{eqnarray}
\tau(\B_{s\rm L}) =
\frac{1}{\Gamma_{\rm L}} = \frac{1}{\Gs+\DGs/2} & = & \hfagTAUBSLCON \,, \\
\tau(\B_{s\rm H}) =
\frac{1}{\Gamma_{\rm H}} = \frac{1}{\Gs-\DGs/2} & = & \hfagTAUBSHCON \,, \\
\tau(\Bs) = \frac{1}{\Gs} = \frac{2}{\Gamma_{\rm L}+\Gamma_{\rm H}} & = & \hfagTAUBSMEANCON \,.
\labe{oneoverGs}
\end{eqnarray}

\mysubsubsection{\Bc lifetime}
\labs{taubc}

Early measurements of the \Bc meson lifetime,
from CDF~\cite{Abe:1998wi,CDFnote9294:2008,Abulencia:2006zu} and \dzero~\cite{Abazov:2008rba},
use the semileptonic decay mode \particle{\Bc \to \jpsi \ell^+ \nu} 
and are based on a 
simultaneous fit to the mass and lifetime using the vertex formed
with the leptons from the decay of the \particle{\jpsi} and
the third lepton. Correction factors
to estimate the boost due to the missing neutrino are used.
Correlated systematic errors include the impact
of the uncertainty of the \Bc $p_T$ spectrum on the correction
factors, the level of feed-down from $\psi(2S)$ decays, 
Monte Carlo modeling of the decay model varying from phase space
to the ISGW model, and mass variations.
With more statistics, CDF2 was able to perform the first \Bc lifetime 
based on fully reconstructed
$\Bc \to J/\psi \pi^+$ decays~\cite{Aaltonen:2012yb},
which does not suffer from a missing neutrino. Recent measurements from 
LHCb, both with  
\particle{\Bc \to \jpsi \mu^+ \nu}~\cite{Aaij:2014bva} and 
\particle{\Bc \to \jpsi \pi^+}~\cite{Aaij:2014gka} decays, achieve the 
highest level of precision. 

All the measurements\footnote{We do not list (nor include in the average) an unpublished result from CDF2~\cite{CDFnote9294:2008}.}
are summarized in 
\Table{lifebc} and the world average, dominated by the LHCb measurements, is
determined to be
\begin{equation}
\tau(\Bc) = \hfagTAUBC \,.
\end{equation}

\begin{table}[tb]
\caption{Measurements of the \Bc lifetime.}
\labt{lifebc}
\begin{center}
\begin{tabular}{lccrcl} \hline
Experiment & Method                    & \multicolumn{2}{c}{Data set}  & $\tau(\Bc)$ (ps)
      & Ref.\\   \hline
CDF1       & \particle{\jpsi \ell} & 92--95 & 0.11 fb$^{-1}$ & $0.46^{+0.18}_{-0.16} \pm
 0.03$   & \cite{Abe:1998wi}  \\ 
CDF2       & \particle{\jpsi e} & 02--04 & 0.36 fb$^{-1}$ & $0.463^{+0.073}_{-0.065} \pm 0.036$   & \cite{Abulencia:2006zu} \\
 \dzero & \particle{\jpsi \mu} & 02--06 & 1.3 fb$^{-1}$  & $0.448^{+0.038}_{-0.036} \pm 0.032$
   & \cite{Abazov:2008rba}  \\
CDF2       & \particle{\jpsi \pi} & & 6.7 fb$^{-1}$ & $0.452 \pm 0.048 \pm 0.027$  & \cite{Aaltonen:2012yb} \\
LHCb & \particle{\jpsi \mu} & 12 & 2 fb$^{-1}$  & $0.509 \pm 0.008 \pm 0.012$ & \cite{Aaij:2014bva}  \\
LHCb & \particle{\jpsi \pi} & 11--12 & 3 fb$^{-1}$  & $0.5134 \pm 0.0110 \pm 0.0057$ & \cite{Aaij:2014gka} \\
\hline
  \multicolumn{2}{l}{Average} & &  &  \hfagTAUBCnounit
                 &    \\   \hline
\end{tabular}
\end{center}
\end{table}

\mysubsubsection{\Lb and \b-baryon lifetimes}
\labs{taulb}

The first measurements of \b-baryon lifetimes, performed at LEP,
originate from two classes of partially reconstructed decays.
In the first class, decays with an exclusively 
reconstructed \Lc baryon
and a lepton of opposite charge are used. These products are
more likely to occur in the decay of \Lb baryons.
In the second class, more inclusive final states with a baryon
(\particle{p}, \particle{\bar{p}}, $\Lambda$, or $\bar{\Lambda}$) 
and a lepton have been used, and these final states can generally
arise from any \b baryon.  With the large \b-hadron samples available
at the Tevatron and the LHC, the most precise measurements of \b baryons now
come from fully reconstructed exclusive decays.

The following sources of correlated systematic uncertainties have 
been considered:
experimental time resolution within a given experiment, \b-quark
fragmentation distribution into weakly decaying \b baryons,
\Lb polarization, decay model,
and evaluation of the \b-baryon purity in the selected event samples.
In computing the averages
the central values of the masses are scaled to 
$M(\Lb) = 5620 \pm 2\MeVcc$~\cite{Acosta:2005mq} and
$M(\mbox{\b-baryon}) = 5670 \pm 100\MeVcc$.

For the semi-inclusive lifetime measurements, 
the meaning of the decay model
systematic uncertainties
and the correlation of these uncertainties between measurements
are not always clear.
Uncertainties related to the decay model are dominated by
assumptions on the fraction of $n$-body semileptonic decays.
To be conservative, it is assumed
that these are 100\%  correlated whenever given as an error.
DELPHI varies the fraction of 4-body decays from 0.0 to 0.3. 
In computing the average, the DELPHI
result is corrected to a value of  $0.2 \pm 0.2$ for this fraction.

Furthermore, in computing the average,
the semileptonic decay results from LEP are corrected for a polarization of 
$-0.45^{+0.19}_{-0.17}$~\cite{Abbaneo:2000ej_mod,*Abbaneo:2001bv_mod_cont} and  a 
\Lb fragmentation parameter
$\langle X_E \rangle =0.70\pm 0.03$~\cite{Buskulic:1995mf}.




\begin{table}[!t]
\caption{Measurements of the \b-baryon lifetimes.
}
\labt{lifelb}
\begin{center}
\begin{tabular}{lcccl} 
\hline
Experiment&Method                &Data set& Lifetime (ps) & Ref. \\\hline\hline
ALEPH  &$\Lambda\ell$         & 91--95 &$1.20 \pm 0.08 \pm 0.06$ & \cite{Barate:1997if}\\
DELPHI &$\Lambda\ell\pi$ vtx  & 91--94 &$1.16 \pm 0.20 \pm 0.08$        & \cite{Abreu:1999hu}$^b$\\
DELPHI &$\Lambda\mu$ i.p.     & 91--94 &$1.10^{+0.19}_{-0.17} \pm 0.09$ & \cite{Abreu:1996nt}$^b$ \\
DELPHI &\particle{p\ell}      & 91--94 &$1.19 \pm 0.14 \pm 0.07$        & \cite{Abreu:1999hu}$^b$\\
OPAL   &$\Lambda\ell$ i.p.    & 90--94 &$1.21^{+0.15}_{-0.13} \pm 0.10$ & \cite{Akers:1995ui}$^c$  \\
OPAL   &$\Lambda\ell$ vtx     & 90--94 &$1.15 \pm 0.12 \pm 0.06$        & \cite{Akers:1995ui}$^c$ \\ 
\hline
ALEPH  &$\Lc\ell$             & 91--95 &$1.18^{+0.13}_{-0.12} \pm 0.03$ & \cite{Barate:1997if}$^a$\\
ALEPH  &$\Lambda\ell^-\ell^+$ & 91--95 &$1.30^{+0.26}_{-0.21} \pm 0.04$ & \cite{Barate:1997if}$^a$\\
DELPHI &$\Lc\ell$             & 91--94 &$1.11^{+0.19}_{-0.18} \pm 0.05$ & \cite{Abreu:1999hu}$^b$\\
OPAL   &$\Lc\ell$, $\Lambda\ell^-\ell^+$ 
                                 & 90--95 & $1.29^{+0.24}_{-0.22} \pm 0.06$ & \cite{Ackerstaff:1997qi}\\ 
CDF1   &$\Lc\ell$             & 91--95 &$1.32 \pm 0.15        \pm 0.07$ & \cite{Abe:1996df}\\
\dzero &$\Lc\mu$              & 02--06 &$1.290^{+0.119+0.087}_{-0.110-0.091}$ & \cite{Abazov_mod:2007tha} \\
\multicolumn{3}{l}{Average of above 6 (semileptonic \Lb decays)} & \hfagTAULBSnounit & \\
CDF2   &$\Lc\pi$              & 02--06 &$1.401 \pm 0.046 \pm 0.035$ & \cite{Aaltonen:2009zn} \\
CDF2   &$\jpsi \Lambda$      & 02--11 &$1.565 \pm 0.035 \pm 0.020$ & \cite{Aaltonen:2014wfa,*Aaltonen:2014wfa_cont} \\
\dzero &$\jpsi \Lambda$      & 02--11 &$1.303 \pm 0.075 \pm 0.035$ & \cite{Abazov:2012iy,*Abazov:2007sf_mod_cont,*Abazov:2004bn_mod_cont} \\
ATLAS  &$\jpsi \Lambda$      & 2011   &$1.449 \pm 0.036 \pm 0.017$ & \cite{Aad:2012sh} \\
CMS    &$\jpsi \Lambda$      & 2011   &$1.503 \pm 0.052 \pm 0.031$ & \cite{Chatrchyan:2013sxa} \\ 
LHCb   &$\jpsi \Lambda$      & 2011   &$1.415 \pm 0.027 \pm 0.006$ & \cite{Aaij:2014owa} \\
LHCb   &$\jpsi pK$           & 11--12 &$1.479 \pm 0.009 \pm 0.010$ & \cite{Aaij:2014zyy,*Aaij:2013oha_cont} \\ 
\multicolumn{3}{l}{Average of above 7 (fully reconstructed \Lb decays)} & \hfagTAULBEnounit & \\
\multicolumn{3}{l}{Average of above 13: \hfill \Lb lifetime $=$} & \hfagTAULBnounit & \\
\hline\hline
ALEPH  &$\Xi\ell$             & 90--95 &$1.35^{+0.37+0.15}_{-0.28-0.17}$ & \cite{Buskulic:1996sm}\\
DELPHI &$\Xi\ell$             & 91--93 &$1.5 ^{+0.7}_{-0.4} \pm 0.3$     & \cite{Abreu:1995kt}$^d$ \\
DELPHI &$\Xi\ell$             & 92--95 &$1.45 ^{+0.55}_{-0.43} \pm 0.13$     & \cite{Abdallah:2005cw}$^d$ \\
\hline
CDF2   &$\jpsi \Xi^-$        & 02--11 &$1.32 \pm 0.14 \pm 0.02$ & \cite{Aaltonen:2014wfa,*Aaltonen:2014wfa_cont} \\ 
LHCb   &$\jpsi \Xi^-$         & 11--12 &$1.55 ^{+0.10}_{-0.09} \pm 0.03$ & \cite{Aaij:2014sia} \\ 
LHCb   &$\Xi_c^0\pi^-$        & 11--12 &$1.599 \pm 0.041 \pm 0.022$ & \cite{Aaij:2014lxa} \\ 
\multicolumn{3}{l}{Average of above 3: \hfill \Xibd lifetime $=$} & \hfagTAUXBDnounit & \\
\hline\hline
LHCb   &$\Xi_c^+\pi^-$        & 11--12 &$1.477 \pm 0.026 \pm 0.019$ & \cite{Aaij:2014esa} \\ 
\multicolumn{3}{l}{Average of above 1: \hfill \Xibu lifetime $=$} & \hfagTAUXBUnounit & \\
\hline\hline
CDF2   &$\jpsi \Omega^-$     & 02--11 & $1.66 ^{+0.53}_{-0.40} \pm 0.02$ & \cite{Aaltonen:2014wfa,*Aaltonen:2014wfa_cont} \\ 
LHCb   &$\jpsi \Omega^-$     & 11--12 &$1.54 ^{+0.26}_{-0.21} \pm 0.05$ & \cite{Aaij:2014sia} \\ 
\multicolumn{3}{l}{Average of above 2: \hfill \Omegab lifetime $=$} & \hfagTAUOBnounit & \\
\hline\hline
\multicolumn{5}{l}{$^a$ \footnotesize The combined ALEPH result quoted 
in \cite{Barate:1997if} is $1.21 \pm 0.11$ ps.} \\[-0.5ex]
\multicolumn{5}{l}{$^b$ \footnotesize The combined DELPHI result quoted 
in \cite{Abreu:1999hu} is $1.14 \pm 0.08 \pm 0.04$ ps.} \\[-0.5ex]
\multicolumn{5}{l}{$^c$ \footnotesize The combined OPAL result quoted 
in \cite{Akers:1995ui} is $1.16 \pm 0.11 \pm 0.06$ ps.} \\[-0.5ex]
\multicolumn{5}{l}{$^d$ \footnotesize The combined DELPHI result quoted 
in \cite{Abdallah:2005cw} is $1.48 ^{+0.40}_{-0.31} \pm 0.12$ ps.}
\end{tabular}
\end{center}
\end{table}

Inputs to the averages are given in \Table{lifelb}.
For the \Lb lifetime average, we only include measurements obtained
with inclusive \particle{\Lambda^{\pm}_c \ell^{\mp}}, inclusive
$\Lambda \ell^- \ell^+$, and fully exclusive
final states.
The CDF $\Lb \to \jpsi \Lambda$
lifetime result~\cite{Aaltonen:2014wfa,*Aaltonen:2014wfa_cont} 
is larger than the world average computed excluding this result
by $\hfagNSIGMATAULBCDFTWO\,\sigma$. 
It is nonetheless combined with the rest 
without adjustment of input errors.
The world average \Lb lifetime is then
\begin{equation}
\tau(\Lb) = \hfagTAULB \,. 
\end{equation}
It turns out that the average obtained using only measurements performed 
with semileptonic \Lb decays (\hfagTAULBS) is significantly different 
from the one using only measurements performed with exclusively reconstructed 
\Lb decays (\hfagTAULBE). The latter is much more precise
(and less prone to systematic uncertainties) than the former. 
This discrepancy can only be attributed to a systematic experimental effect
or to a statistical fluctuation. 

For the strange \b baryons, we no longer include measurements based on
$\Xi^{\mp} \ell^{\mp}$~\cite{Buskulic:1996sm,Abdallah:2005cw,Abreu:1995kt} 
final states which consist of a mixture of 
$\Xib^0$ and $\Xib^-$ baryons. Instead we only average results obtained with 
fully exclusive modes, and obtain
\begin{eqnarray}
\tau(\Xibd) &=& \hfagTAUXBD \,, \\
\tau(\Xibu) &=& \hfagTAUXBU \,, \\
\tau(\Omegab) &=& \hfagTAUOB \,. 
\end{eqnarray}

\mysubsubsection{Summary and comparison with theoretical predictions}
\labs{lifesummary}

Averages of lifetimes of specific \b-hadron species are collected
in \Table{sumlife}.
\begin{table}[t]
\caption{Summary of the lifetime averages for the different \b-hadron species.}
\labt{sumlife}
\begin{center}
\begin{tabular}{lrc} \hline
\multicolumn{2}{l}{\b-hadron species} & Measured lifetime \\ \hline
\Bu &                       & \hfagTAUBU   \\
\Bd &                       & \hfagTAUBD   \\
\Bs & $1/\Gs =$               & \hfagTAUBSMEANC \\
~~ $\B_{s\rm L}$ & $1/\Gamma_{\rm L}=$  & \hfagTAUBSLCON \\
~~ $\B_{s\rm H}$ & $1/\Gamma_{\rm H}=$  & \hfagTAUBSHCON \\
\Bc     &                   & \hfagTAUBC   \\ 
\Lb     &                   & \hfagTAULB   \\
\Xibd   &                   & \hfagTAUXBD  \\
\Xibu   &                   & \hfagTAUXBU  \\
\Omegab &                   & \hfagTAUOB   \\
\hline
\end{tabular}
\end{center}
\caption{Measured ratios of \b-hadron lifetimes relative to
the \Bd lifetime and ranges predicted
by theory~\cite{Tarantino:2003qw,*Gabbiani:2003pq,Gabbiani:2004tp}.}
\labt{liferatio}
%
%
%
\begin{center}
\begin{tabular}{lcc} \hline
Lifetime ratio & Measured value & Predicted range \\ \hline
$\tau(\Bu)/\tau(\Bd)$ & \hfagRTAUBU & 1.04 -- 1.08 \\
$\tau(\Bs)/\tau(\Bd)$ & \hfagRTAUBSMEANC & 0.99 -- 1.01 \\
$\tau(\Lb)/\tau(\Bd)$ & \hfagRTAULB & 0.86 -- 0.95    \\
\hline
\end{tabular}
\end{center}
\end{table}
As described in the introduction to \Sec{lifetimes},
the HQE can be employed to explain the hierarchy of
$\tau(\Bc) \ll \tau(\Lb) < \tau(\Bs) \approx \tau(\Bd) < \tau(\Bu)$,
and used to predict the ratios between lifetimes.
Typical predictions are compared to the measured 
lifetime ratios in \Table{liferatio}.
The prediction of the ratio between the \Bu and \Bd lifetimes,
$1.06 \pm 0.02$~\cite{Tarantino:2003qw,*Gabbiani:2003pq}, 
is in good agreement with experiment. 

The total widths of the \Bs and \Bd mesons
are expected to be very close and differ by at most 
1\%~\cite{Beneke:1996gn,*Keum:1998fd,Gabbiani:2004tp}.
This prediction is consistent with the
experimental ratio $\tau(\Bs)/\tau(\Bd)=\Gd/\Gs$,
which is smaller than 1 by 
\hfagONEMINUSRTAUBSMEANCpercent. 

The ratio $\tau(\Lb)/\tau(\Bd)$ has particularly
been the source of theoretical
scrutiny since earlier calculations using the HQE
~\cite{Shifman:1986mx,*Chay:1990da,*Bigi:1992su,*Bigi:1992su_erratum,Voloshin:1999pz,*Guberina:1999bw,*Neubert:1996we,*Bigi:1997fj}
predicted a value larger than 0.90, almost $2\,\sigma$ 
above the world average at the time. 
Many predictions cluster around a most likely central value
of 0.94~\cite{Uraltsev:1996ta,*Pirjol:1998ur,*Colangelo:1996ta,*DiPierro:1999tb}.
More recent calculations
of this ratio that include higher-order effects predict a
lower ratio between the
\Lb and \Bd lifetimes~\cite{Tarantino:2003qw,*Gabbiani:2003pq,Gabbiani:2004tp}
and reduce this difference.
References~\cite{Tarantino:2003qw,*Gabbiani:2003pq,Gabbiani:2004tp} present probability density functions
of their predictions with a variation of theoretical inputs, and the
indicated ranges in \Table{liferatio}
are the RMS of the distributions from the most probable values, and for 
$\tau(\Lb)/\tau(\Bd)$, also encompass the earlier theoretical predictions%
~\cite{Shifman:1986mx,*Chay:1990da,*Bigi:1992su,*Bigi:1992su_erratum,Voloshin:1999pz,*Guberina:1999bw,*Neubert:1996we,*Bigi:1997fj,Uraltsev:1996ta,*Pirjol:1998ur,*Colangelo:1996ta,*DiPierro:1999tb}.
Note that in contrast to the $B$ mesons, complete NLO QCD
corrections and
fully reliable lattice
determinations of the matrix elements for $\Lb$ are not
yet available.
As already mentioned, the CDF measurement of the \Lb lifetime
in the exclusive decay mode $\jpsi \Lambda$~\cite{Aaltonen:2014wfa,*Aaltonen:2014wfa_cont} 
is significantly 
higher than the world average before inclusion, with a ratio
to the $\tau(\Bd)$ world average of 
$\tau(\Lb)/\tau(\Bd) = 1.012 \pm 0.031$, 
%
resulting in continued interest in lifetimes of \b baryons.

The lifetimes of the most abundant \b-hadron species are now all known to sub-percent precision. Neglecting the 
contributions of the rarer species (\Bc meson and \b baryons other than the \Lb), one can compute the average 
\b-hadron lifetime from the individual lifetimes and production fractions as 
\begin{equation}
\tau_b = \frac%
{\fBd \tau(\Bd)^2+ \fBu \tau(\Bu)^2+0.5 \fBs \tau(B_{s\rm H})^2+0.5 \fBs \tau(B_{s\rm L})^2+ \fbb \tau(\Lb)^2}%
{\fBd \tau(\Bd)  + \fBu \tau(\Bu)  +0.5 \fBs \tau(B_{s\rm H})  +0.5 \fBs \tau(B_{s\rm L})  + \fbb \tau(\Lb)  } \,.
\end{equation}
Using the lifetimes of \Table{sumlife} and the fractions in $Z$ decays of \Table{fractions},
taking into account the correlations between the fractions (\Table{fractions}) as well as the correlation 
between $\tau(B_{s\rm H})$ and $\tau(B_{s\rm L})$ (\hfagZRHOTAUHTAUL), one obtains
\begin{equation}
\tau_b(Z) = \hfagTAUBZCALC \,.
\end{equation}
This is in very good agreement with (and three times more precise than)
the average of \Eq{TAUBVTX} for the inclusive measurements performed at LEP. 

\mysubsection{Neutral \B-meson mixing}
\labs{mixing}

The $\Bd-\Bdbar$ and $\Bs-\Bsbar$ systems
both exhibit the phenomenon of particle-antiparticle mixing. For each of them, 
there are two mass eigenstates which are linear combinations of the two flavour states,
\B and $\bar{\B}$. 
The heaviest (lightest) of these mass states is denoted
$\B_{\rm H}$ ($\B_{\rm L}$),
with mass $m_{\rm H}$ ($m_{\rm L}$)
and total decay width $\Gamma_{\rm H}$ ($\Gamma_{\rm L}$). We define
\begin{eqnarray}
\Delta m = m_{\rm H} - m_{\rm L} \,, &~~~~&  x = \Delta m/\Gamma \,, \labe{dm} \\
\Delta \Gamma \, = \Gamma_{\rm L} - \Gamma_{\rm H} \,, ~ &~~~~&  y= \Delta\Gamma/(2\Gamma) \,, \labe{dg}
\end{eqnarray}
where 
$\Gamma = (\Gamma_{\rm H} + \Gamma_{\rm L})/2 =1/\bar{\tau}(\B)$ 
is the average decay width.
$\Delta m$ is positive by definition, and 
$\Delta \Gamma$ is expected to be positive within
the Standard Model.\footnote{
  \label{foot:life_mix:Eqdg}
  For reasons of symmetry in \Eqss{dm}{dg}, 
  $\Delta \Gamma$ is sometimes defined with the opposite sign. 
  The definition adopted here, \ie\
  \Eq{dg}, is the one used by most experimentalists and many
  phenomenologists in \B physics.}

There are four different time-dependent probabilities describing the 
case of a neutral \B meson produced as a flavour state and decaying without
\CP violation to a flavour-specific final state. 
If \CPT is conserved (which  
will be assumed throughout), they can be written as 
\begin{equation}
\left\{
\begin{array}{rcl}
{\cal P}(\B\to\B) & = &  \frac{e^{-\Gamma t}}{2} 
\left[ \cosh\!\left(\frac{\Delta\Gamma}{2}t\right) + \cos\!\left(\Delta m t\right)\right]  \\
{\cal P}(\B\to\bar{\B}) & = &  \frac{e^{-\Gamma t}}{2} 
\left[ \cosh\!\left(\frac{\Delta\Gamma}{2}t\right) - \cos\!\left(\Delta m t\right)\right] 
\left|\frac{q}{p}\right|^2 \\
{\cal P}(\bar{\B}\to\B) & = &  \frac{e^{-\Gamma t}}{2} 
\left[ \cosh\!\left(\frac{\Delta\Gamma}{2}t\right) - \cos\!\left(\Delta m t\right)\right] 
\left|\frac{p}{q}\right|^2 \\
{\cal P}(\bar{\B}\to\bar{\B}) & = &  \frac{e^{-\Gamma t}}{2} 
\left[ \cosh\!\left(\frac{\Delta\Gamma}{2}t\right) + \cos\!\left(\Delta m t\right)\right] 
\end{array} \right. \,,
\labe{oscillations}
\end{equation}
where $t$ is the proper time of the system (\ie\ the time interval between the production 
and the decay in the rest frame of the \B meson). 
At the \B factories, only the proper-time difference $\Delta t$ between the decays
of the two neutral \B mesons from the \Ups can be determined, but, 
because the two \B mesons evolve coherently (keeping opposite flavours as long
as neither of them has decayed), the 
above formulae remain valid 
if $t$ is replaced with $\Delta t$ and the production flavour is replaced by the flavour 
at the time of the decay of the accompanying \B meson in a flavour-specific state.
As can be seen in the above expressions,
the mixing probabilities 
depend on three mixing observables:
$\Delta m$, $\Delta\Gamma$,
and $|q/p|^2$, which signals \CP violation in the mixing if $|q/p|^2 \ne 1$.

In the next sections we review in turn the experimental knowledge
on the \Bd decay-width and mass differences, 
the \Bs decay-width and mass differences,  
\CP violation in \Bd and \Bs mixing, and mixing-induced \CP violation in \Bs decays. 

\mysubsubsection{\Bd mixing parameters \DGd and \dmd}
\labs{DGd} \labs{dmd}

\begin{table}
\caption{Time-dependent measurements included in the \dmd average.
The results obtained from multi-dimensional fits involving also 
the \Bd (and \Bu) lifetimes
as free parameter(s)~\cite{Aubert:2002sh,Aubert:2005kf,Abe:2004mz} 
have been converted into one-dimensional measurements of \dmd.
All the measurements have then been adjusted to a common set of physics
parameters before being combined.}
\labt{dmd}
\begin{center}
\begin{tabular}{@{}rc@{}cc@{}c@{}cc@{}c@{}c@{}}
\hline
Experiment & \multicolumn{2}{c}{Method} & \multicolumn{3}{l}{\dmd in\invps}   
                                        & \multicolumn{3}{l}{\dmd in\invps}     \\
and Ref.   &  rec. & tag                & \multicolumn{3}{l}{before adjustment} 
                                        & \multicolumn{3}{l}{after adjustment} \\
\hline
 ALEPH~\cite{Buskulic:1996qt}  & \particle{ \ell  } & \particle{ \Qjet  } & $  0.404 $ & $ \pm  0.045 $ & $ \pm  0.027 $ & & & \\
 ALEPH~\cite{Buskulic:1996qt}  & \particle{ \ell  } & \particle{ \ell  } & $  0.452 $ & $ \pm  0.039 $ & $ \pm  0.044 $ & & & \\
 ALEPH~\cite{Buskulic:1996qt}  & \multicolumn{2}{c}{above two combined} & $  0.422 $ & $ \pm  0.032 $ & $ \pm  0.026 $ & $  0.440 $ & $ \pm  0.032 $ & $ ^{+  0.020 }_{-  0.019 } $ \\
 ALEPH~\cite{Buskulic:1996qt}  & \particle{ D^*  } & \particle{ \ell,\Qjet  } & $  0.482 $ & $ \pm  0.044 $ & $ \pm  0.024 $ & $  0.482 $ & $ \pm  0.044 $ & $ \pm  0.024 $ \\
 DELPHI~\cite{Abreu:1997xq}  & \particle{ \ell  } & \particle{ \Qjet  } & $  0.493 $ & $ \pm  0.042 $ & $ \pm  0.027 $ & $  0.497 $ & $ \pm  0.042 $ & $ \pm  0.024 $ \\
 DELPHI~\cite{Abreu:1997xq}  & \particle{ \pi^*\ell  } & \particle{ \Qjet  } & $  0.499 $ & $ \pm  0.053 $ & $ \pm  0.015 $ & $  0.500 $ & $ \pm  0.053 $ & $ \pm  0.015 $ \\
 DELPHI~\cite{Abreu:1997xq}  & \particle{ \ell  } & \particle{ \ell  } & $  0.480 $ & $ \pm  0.040 $ & $ \pm  0.051 $ & $  0.496 $ & $ \pm  0.040 $ & $ ^{+  0.042 }_{-  0.040 } $ \\
 DELPHI~\cite{Abreu:1997xq}  & \particle{ D^*  } & \particle{ \Qjet  } & $  0.523 $ & $ \pm  0.072 $ & $ \pm  0.043 $ & $  0.518 $ & $ \pm  0.072 $ & $ \pm  0.043 $ \\
 DELPHI~\cite{Abdallah:2002mr}  & \particle{ \mbox{vtx}  } & \particle{ \mbox{comb}  } & $  0.531 $ & $ \pm  0.025 $ & $ \pm  0.007 $ & $  0.525 $ & $ \pm  0.025 $ & $ \pm  0.006 $ \\
 L3~\cite{Acciarri:1998pq}  & \particle{ \ell  } & \particle{ \ell  } & $  0.458 $ & $ \pm  0.046 $ & $ \pm  0.032 $ & $  0.467 $ & $ \pm  0.046 $ & $ \pm  0.028 $ \\
 L3~\cite{Acciarri:1998pq}  & \particle{ \ell  } & \particle{ \Qjet  } & $  0.427 $ & $ \pm  0.044 $ & $ \pm  0.044 $ & $  0.438 $ & $ \pm  0.044 $ & $ \pm  0.042 $ \\
 L3~\cite{Acciarri:1998pq}  & \particle{ \ell  } & \particle{ \ell\mbox{(IP)}  } & $  0.462 $ & $ \pm  0.063 $ & $ \pm  0.053 $ & $  0.470 $ & $ \pm  0.063 $ & $ \pm  0.044 $ \\
 OPAL~\cite{Ackerstaff:1997iw}  & \particle{ \ell  } & \particle{ \ell  } & $  0.430 $ & $ \pm  0.043 $ & $ ^{+  0.028 }_{-  0.030 } $ & $  0.466 $ & $ \pm  0.043 $ & $ ^{+  0.017 }_{-  0.016 } $ \\
 OPAL~\cite{Ackerstaff:1997vd}  & \particle{ \ell  } & \particle{ \Qjet  } & $  0.444 $ & $ \pm  0.029 $ & $ ^{+  0.020 }_{-  0.017 } $ & $  0.481 $ & $ \pm  0.029 $ & $ \pm  0.013 $ \\
 OPAL~\cite{Alexander:1996id}  & \particle{ D^*\ell  } & \particle{ \Qjet  } & $  0.539 $ & $ \pm  0.060 $ & $ \pm  0.024 $ & $  0.544 $ & $ \pm  0.060 $ & $ \pm  0.023 $ \\
 OPAL~\cite{Alexander:1996id}  & \particle{ D^*  } & \particle{ \ell  } & $  0.567 $ & $ \pm  0.089 $ & $ ^{+  0.029 }_{-  0.023 } $ & $  0.572 $ & $ \pm  0.089 $ & $ ^{+  0.028 }_{-  0.022 } $ \\
 OPAL~\cite{Abbiendi:2000ec}  & \particle{ \pi^*\ell  } & \particle{ \Qjet  } & $  0.497 $ & $ \pm  0.024 $ & $ \pm  0.025 $ & $  0.496 $ & $ \pm  0.024 $ & $ \pm  0.025 $ \\
 CDF1~\cite{Abe:1997qf,*Abe:1998sq_mod_cont}  & \particle{ D\ell  } & \particle{ \mbox{SST}  } & $  0.471 $ & $ ^{+  0.078 }_{-  0.068 } $ & $ ^{+  0.033 }_{-  0.034 } $ & $  0.470 $ & $ ^{+  0.078 }_{-  0.068 } $ & $ ^{+  0.033 }_{-  0.034 } $ \\
 CDF1~\cite{Abe:1999pv}  & \particle{ \mu  } & \particle{ \mu  } & $  0.503 $ & $ \pm  0.064 $ & $ \pm  0.071 $ & $  0.513 $ & $ \pm  0.064 $ & $ ^{+  0.070 }_{-  0.069 } $ \\
 CDF1~\cite{Abe:1999ds}  & \particle{ \ell  } & \particle{ \ell,\Qjet  } & $  0.500 $ & $ \pm  0.052 $ & $ \pm  0.043 $ & $  0.544 $ & $ \pm  0.052 $ & $ \pm  0.036 $ \\
 CDF1~\cite{Affolder:1999cn}  & \particle{ D^*\ell  } & \particle{ \ell  } & $  0.516 $ & $ \pm  0.099 $ & $ ^{+  0.029 }_{-  0.035 } $ & $  0.523 $ & $ \pm  0.099 $ & $ ^{+  0.028 }_{-  0.035 } $ \\
 \dzero~\cite{Abazov:2006qp}  & \particle{ D^{(*)}\mu  } & \particle{ \mbox{OST}  } & $  0.506 $ & $ \pm  0.020 $ & $ \pm  0.016 $ & $  0.506 $ & $ \pm  0.020 $ & $ \pm  0.016 $ \\
 \babar~\cite{Aubert:2001te,*Aubert:2002rg_cont}  & \particle{ \Bd  } & \particle{ \ell,K,\mbox{NN}  } & $  0.516 $ & $ \pm  0.016 $ & $ \pm  0.010 $ & $  0.521 $ & $ \pm  0.016 $ & $ \pm  0.008 $ \\
 \babar~\cite{Aubert:2001tf}  & \particle{ \ell  } & \particle{ \ell  } & $  0.493 $ & $ \pm  0.012 $ & $ \pm  0.009 $ & $  0.487 $ & $ \pm  0.012 $ & $ \pm  0.006 $ \\
 \babar~\cite{Aubert:2005kf}  & \particle{ D^*\ell\nu\mbox{(part)}  } & \particle{ \ell  } & $  0.511 $ & $ \pm  0.007 $ & $ \pm  0.007 $ & $  0.513 $ & $ \pm  0.007 $ & $ \pm  0.007 $ \\
 \babar~\cite{Aubert:2002sh}  & \particle{ D^*\ell\nu  } & \particle{ \ell,K,\mbox{NN}  } & $  0.492 $ & $ \pm  0.018 $ & $ \pm  0.014 $ & $  0.493 $ & $ \pm  0.018 $ & $ \pm  0.013 $ \\
 \belle~\cite{Zheng:2002jv}  & \particle{ D^*\pi\mbox{(part)}  } & \particle{ \ell  } & $  0.509 $ & $ \pm  0.017 $ & $ \pm  0.020 $ & $  0.513 $ & $ \pm  0.017 $ & $ \pm  0.019 $ \\
 \belle~\cite{Hastings:2002ff}  & \particle{ \ell  } & \particle{ \ell  } & $  0.503 $ & $ \pm  0.008 $ & $ \pm  0.010 $ & $  0.506 $ & $ \pm  0.008 $ & $ \pm  0.008 $ \\
 \belle~\cite{Abe:2004mz}  & \particle{ \Bd,D^*\ell\nu  } & \particle{ \mbox{comb}  } & $  0.511 $ & $ \pm  0.005 $ & $ \pm  0.006 $ & $  0.513 $ & $ \pm  0.005 $ & $ \pm  0.006 $ \\
 LHCb~\cite{LHCb-CONF-2011-010,*LHCb-CONF-2011-010_published}  & \particle{ \Bd  } & \particle{ \mbox{OST}  } & $  0.499 $ & $ \pm  0.032 $ & $ \pm  0.003 $ & $  0.499 $ & $ \pm  0.032 $ & $ \pm  0.003 $ \\
 LHCb~\cite{Aaij:2012nt}  & \particle{ \Bd  } & \particle{ \mbox{OST,SST}  } & $  0.516 $ & $ \pm  0.005 $ & $ \pm  0.003 $ & $  0.516 $ & $ \pm  0.005 $ & $ \pm  0.003 $ \\
 LHCb~\cite{Aaij:2013gja}  & \particle{ D\mu  } & \particle{ \mbox{OST,SST}  } & $  0.503 $ & $ \pm  0.011 $ & $ \pm  0.013 $ & $  0.503 $ & $ \pm  0.011 $ & $ \pm  0.013 $ \\
 \hline \\[-2.0ex]
 \multicolumn{6}{l}{World average (all above measurements included):} & $  0.510 $ & $ \pm  0.003 $ & $ \pm  0.002 $ \\

\\[-2.0ex]
\multicolumn{6}{l}{~~~ -- ALEPH, DELPHI, L3, OPAL and CDF1 only:}
     & \hfagDMDHval & \hfagDMDHsta & \hfagDMDHsys \\
\multicolumn{6}{l}{~~~ -- \babar and \belle only:}
     & \hfagDMDBval & \hfagDMDBsta & \hfagDMDBsys \\
\multicolumn{6}{l}{~~~ -- LHCb only:} & \hfagDMDLval & \hfagDMDLsta & \hfagDMDLsys \\
\hline
\end{tabular}
\end{center}
\end{table}

A large number of time-dependent \Bd--\Bdbar oscillation analyses
have been performed since almost 20 years by the 
ALEPH, DELPHI, L3, OPAL, CDF, \dzero, \babar, \belle and  LHCb collaborations. 
The corresponding measurements of \dmd are summarized in 
\Table{dmd},
where only the most recent results
are listed (\ie\ measurements superseded by more recent ones are omitted\footnote{
  \label{foot:life_mix:CDFnote8235:2006}
  Two old unpublished CDF2 measurements~\cite{CDFnote8235:2006,CDFnote7920:2005}
  are also omitted from our averages, \Table{dmd} and \Fig{dmd}.}).
Although a variety of different techniques have been used, the 
individual \dmd
results obtained at different colliders have remarkably similar precision.
The systematic uncertainties are comparable to the statistical uncertainties;
they are often dominated by sample composition, mistag probability,
or \b-hadron lifetime contributions.
Before being combined, the measurements are adjusted on the basis of a 
common set of input values, including the averages of the 
\b-hadron fractions and lifetimes given in this report 
(see \Secss{fractions}{lifetimes}).
Some measurements are statistically correlated. 
Systematic correlations arise both from common physics sources 
(fractions, lifetimes, branching ratios of \b hadrons), and from purely 
experimental or algorithmic effects (efficiency, resolution, flavour tagging, 
background description). Combining all published measurements
listed in \Table{dmd}
and accounting for all identified correlations
as described in \Ref{Abbaneo:2000ej_mod,*Abbaneo:2001bv_mod_cont} yields $\dmd = \hfagDMDWfull$.

On the other hand, ARGUS and CLEO have published 
measurements of the time-integrated mixing probability 
\chid~\cite{Albrecht:1992yd,*Albrecht:1993gr_cont,Bartelt:1993cf,Behrens:2000qu}, 
which average to $\chid =\hfagCHIDU$.
Following \Ref{Behrens:2000qu}, 
the decay width difference \DGd could 
in principle be extracted from the
measured value of $\Gd=1/\tau(\Bd)$ and the above averages for 
\dmd and \chid 
(provided that \DGd has a negligible impact on 
the \dmd and $\tau(\Bd)$ analyses that have assumed $\DGd=0$), 
using the relation
\begin{equation}
\chid = \frac{\xd^2+\yd^2}{2(\xd^2+1)} ~~~ \mbox{with} ~~ \xd=\frac{\dmd}{\Gd} 
~~ \mbox{and} ~~ \yd=\frac{\DGd}{2\Gd} \,.
\labe{chid_definition}
\end{equation}
However, direct time-dependent studies provide much stronger constraints: 
$|\DGd|/\Gd < 18\%$ at \CL{95} from DELPHI~\cite{Abdallah:2002mr},
$-6.8\% < {\rm sign}({\rm Re} \lambda_{\CP}) \DGGd < 8.4\%$
at \CL{90} from \babar~\cite{Aubert:2003hd,*Aubert:2004xga_mod_cont},
and ${\rm sign}({\rm Re} \lambda_{\CP})\DGGd = (1.7 \pm 1.8 \pm 1.1)\%$~\cite{Higuchi:2012kx}
from Belle, 
where $\lambda_{\CP} = (q/p)_{\particle{d}} (\bar{A}_{\CP}/A_{\CP})$
is defined for a \CP-even final state 
(the sensitivity to the overall sign of 
${\rm sign}({\rm Re} \lambda_{\CP}) \DGGd$ comes
from the use of \Bd decays to \CP final states).
In addition, the \dzero collaboration has recently extracted a value of 
$\DGGd=(0.50 \pm 1.38)\%$\cite{Abazov:2013uma,*Abazov:2011yk_mod,*Abazov:2010hv_mod_cont,*Abazov:2010hj_mod_cont,*Abazov:2011yk_cont}
from their measurements of the same-sign dimuon charge asymmetry, 
under the interpretation that 
the observed asymmetries are due to \CP violation in neutral $B$-meson mixing and interference.
More recently LHCb has obtained $\DGGd=(-0.044 \pm 0.025 \pm 0.011)\%$~\cite{Aaij:2014owa}
by comparing measurements of the $\Bd \to \jpsi K^{*0}$ and $\Bd \to \jpsi K^0_{\rm S}$
decays, following the method of Ref.~\cite{Gershon:2010wx}.
Assuming ${\rm Re} \lambda_{\CP} > 0$, as expected from the global fits
of the Unitarity Triangle within the Standard Model~\cite{Charles:2011va_mod,*Bona:2006ah},
a combination of these five results (after adjusting the DELPHI and \babar results to  
$1/\Gd=\tau(\Bd)=\hfagTAUBD$) yields
\begin{equation}
\DGGd  = \hfagSDGDGD \,.
\end{equation}

Assuming $\DGd=0$ 
and using $1/\Gd=\tau(\Bd)=\hfagTAUBD$,
the \dmd and \chid results are combined through \Eq{chid_definition} 
to yield the 
world average
\begin{equation} 
\dmd = \hfagDMDWU \,,
\labe{dmd}
\end{equation} 
or, equivalently,
\begin{equation} 
\xd= \hfagXDWU ~~~ \mbox{and} ~~~ \chid=\hfagCHIDWU \,.  
\labe{chid}
\end{equation}
\Figure{dmd} compares the \dmd values obtained by the different experiments.

\begin{figure}
\begin{center}
\includegraphics[width=\textwidth]{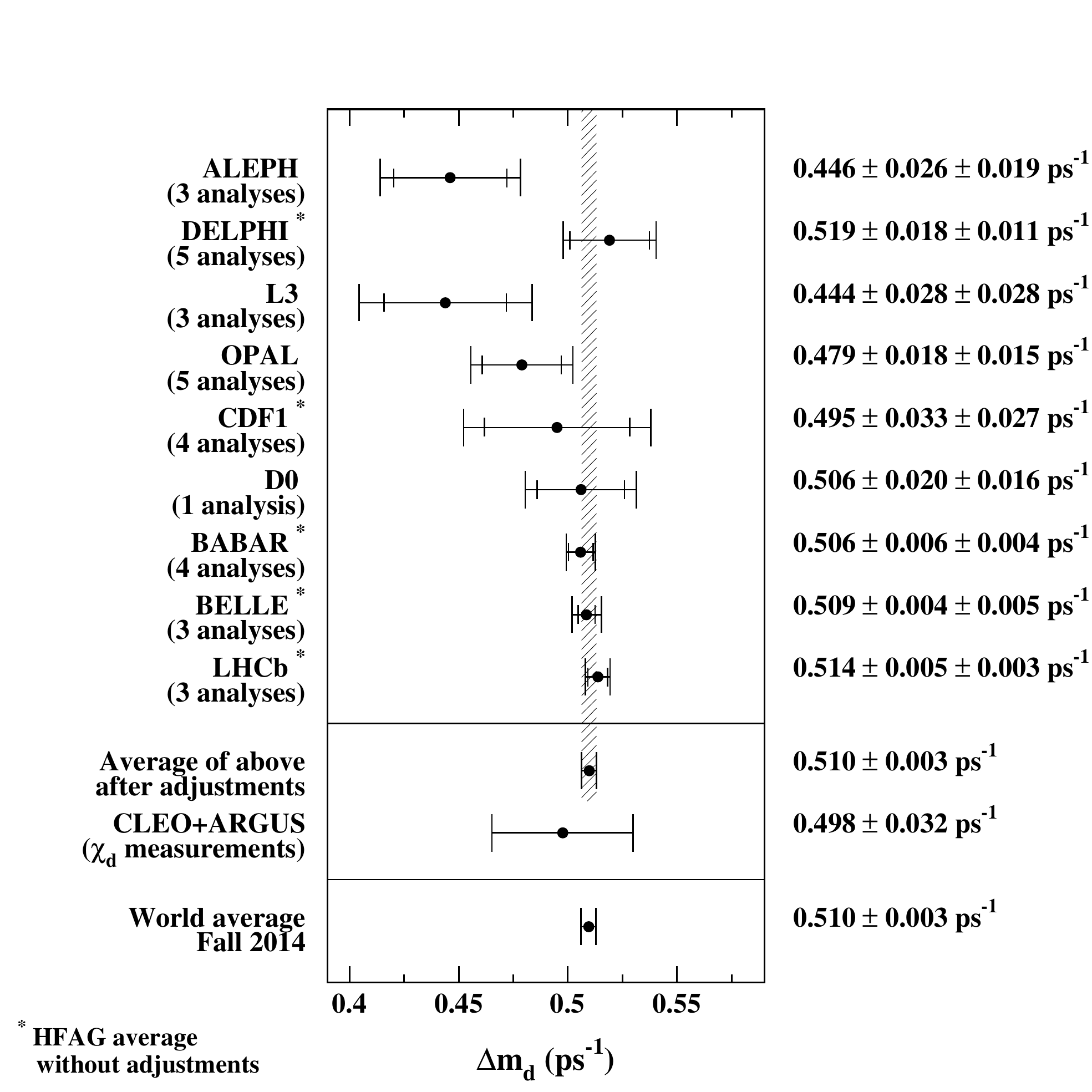}
\caption{The \Bd--\Bdbar oscillation frequency \dmd as measured by the different experiments. 
The averages quoted for ALEPH, L3 and OPAL are taken from the original publications, while the 
ones for DELPHI, CDF, \babar, \belle and LHCb have been computed from the individual results 
listed in \Table{dmd} without performing any adjustments. The time-integrated measurements 
of \chid from the symmetric \B factory experiments ARGUS and CLEO have been converted 
to a \dmd value using $\tau(\Bd)=\hfagTAUBD$. The two global averages have been obtained 
after adjustments of all the individual \dmd results of \Table{dmd} (see text).}
\labf{dmd}
\end{center}
\end{figure}

The \Bd mixing averages given in \Eqss{dmd}{chid}
and the \b-hadron fractions of \Table{fractions} have been obtained in a fully 
consistent way, taking into account the fact that the fractions are computed using 
the \chid value of \Eq{chid} and that many individual measurements of \dmd
at high energy depend on the assumed values for the \b-hadron fractions.
Furthermore, this set of averages is consistent with the lifetime averages 
of \Sec{lifetimes}.

\begin{table}
\caption{Simultaneous measurements of \dmd and $\tau(\Bd)$, and their average.
The \belle analysis also 
measures $\tau(\Bu)$ at the same time, but it is converted here into a two-dimensional measurement 
of \dmd and $\tau(\Bd)$, for an assumed value of $\tau(\Bu)$. 
The first quoted error on each measurement is statistical
and the second is systematic; in the case of adjusted measurements, the 
latter includes a contribution obtained from the variation of $\tau(\Bu)$ or 
$\tau(\Bu)/\tau(\Bd)$ in the indicated range. Units are\invps\ for \dmd
and\unit{ps} for lifetimes. 
The three different values of $\rho(\dmd,\tau(\Bd))$ correspond 
to the statistical, systematic and total correlation coefficients
between the adjusted measurements of \dmd and $\tau(\Bd)$.}
\labt{dmd2D}
\begin{center}
\begin{tabular}{@{}r@{~}c@{}c@{}c@{~}c@{}c@{}c@{~}c@{}c@{}c@{\hspace{0ex}}c@{}}
\hline
Exp.\ \& Ref.
& \multicolumn{3}{c}{Measured \dmd}   
& \multicolumn{3}{c}{Measured $\tau(\Bd)$}   
& \multicolumn{3}{c}{Measured $\tau(\Bu)$}   
&  Assumed $\tau(\Bu)$ \\
\hline
\babar \cite{Aubert:2002sh}  
      & $0.492$ & $\pm 0.018$ & $\pm 0.013$ 
      & $1.523$ & $\pm 0.024$ & $\pm 0.022$ 
      & \multicolumn{3}{c}{---}
      & $(1.083\pm 0.017)\tau(\Bd)$ \\  
\babar \cite{Aubert:2005kf}  
      & $0.511$ & $\pm 0.007$ & $^{+0.007}_{-0.006}$ 
      & $1.504$ & $\pm 0.013$ & $^{+0.018}_{-0.013}$
      & \multicolumn{3}{c}{---}
      & $1.671\pm 0.018$ \\  
\belle \cite{Abe:2004mz}  
      & $0.511$ & $\pm 0.005$ & $\pm 0.006$
      & $1.534$ & $\pm 0.008$ & $\pm 0.010$
      & $1.635$ & $\pm 0.011$ & $\pm 0.011$
      & --- \\  
\cline{2-10}
& \multicolumn{3}{c}{Adjusted \dmd}   
& \multicolumn{3}{c}{Adjusted $\tau(\Bd)$}   
& \multicolumn{3}{c}{$\rho(\dmd,\Bd)$} 
&  Assumed $\tau(\Bu)$ \\
\cline{2-10}
\babar \cite{Aubert:2002sh}  
      & $0.492$ & $\pm 0.018$ & $\pm 0.013$  
      & $1.523$ & $\pm 0.024$ & $\pm 0.022$  
      & $-0.22$ & $+0.71$ & $+0.16$ 
      & $(\hfagRTAUBUval$$\hfagRTAUBUerr)\tau(\Bd)$ \\  
\babar \cite{Aubert:2005kf} 
      & $0.512$ & $\pm 0.007$ & $\pm 0.007$  
      & $1.506$ & $\pm 0.013$ & $\pm 0.018$ 
      & $+0.01$ & $-0.85$ & $-0.48$ 
      & $\hfagTAUBUval$$\hfagTAUBUerr$ \\  
\belle \cite{Abe:2004mz}  
      & $0.511$ & $\pm 0.005$ & $\pm 0.006$ 
      & $1.535$ & $\pm 0.008$ & $\pm 0.011$ 
      & $-0.27$ & $-0.14$ & $-0.19$ 
      & $\hfagTAUBUval$$\hfagTAUBUerr$ \\  
\hline
\multicolumn{1}{l}{Average} 
      & \hfagDMDTWODval   & \hfagDMDTWODsta   & \hfagDMDTWODsys
      & \hfagTAUBDTWODval & \hfagTAUBDTWODsta & \hfagTAUBDTWODsys
      & \hfagRHOstaDMDTAUBD & \hfagRHOsysDMDTAUBD & \hfagRHODMDTAUBD 
      & $\hfagTAUBUval$$\hfagTAUBUerr$ \\  
\hline 
\end{tabular}
\end{center}
\end{table}
\begin{figure}
\begin{center}
\vspace{-0.5cm}
\includegraphics[width=0.6\textwidth]{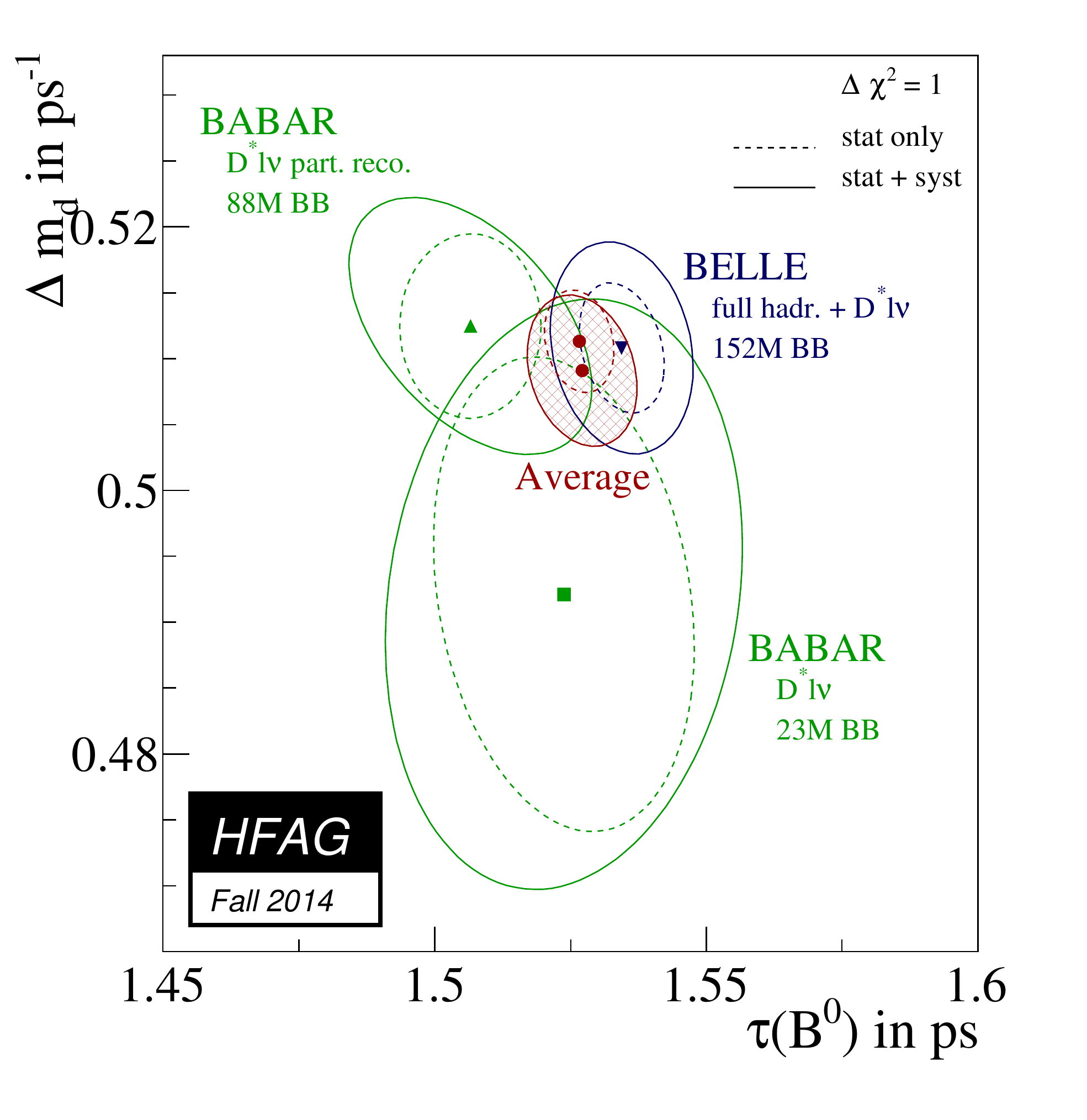}
\vspace{-0.5cm}
\caption{Simultaneous measurements of
\dmd and $\tau(\Bd)$~\cite{Aubert:2002sh,Aubert:2005kf,Abe:2004mz}, 
after adjustment to a common set of parameters (see text). 
Statistical and total uncertainties are represented as dashed and
solid contours respectively.
The average of the three measurements
is indicated by a hatched ellipse.}
\labf{dmd2D}
\end{center}
\end{figure}

It should be noted that the most recent (and precise) analyses at the 
asymmetric \B factories measure \dmd
as a result of a multi-dimensional fit. 
Two \babar analyses~\cite{Aubert:2002sh,Aubert:2005kf},  
based on fully and partially reconstructed $\Bd \to D^*\ell\nu$ decays
respectively, 
extract simultaneously \dmd and $\tau(\Bd)$
while the latest \belle analysis~\cite{Abe:2004mz},  
based on fully reconstructed hadronic \Bd decays and $\Bd \to D^*\ell\nu$ decays, 
extracts simultaneously \dmd, $\tau(\Bd)$ and $\tau(\Bu)$.
The measurements of \dmd and $\tau(\Bd)$ of these three analyses 
are displayed in \Table{dmd2D} and in \Fig{dmd2D}. Their two-dimensional average, 
taking into account all statistical and systematic correlations, and expressed
at $\tau(\Bu)=\hfagTAUBU$, is
\begin{equation}
\left.
\begin{array}{r@{}l}
\dmd = \hfagDMDTWODnounit & \invps \\
\tau(\Bd) = \hfagTAUBDTWODnounit & \ps
\end{array}
\right\}
~\mbox{with a total correlation of \hfagRHODMDTAUBD.}
\end{equation}

\mysubsubsection{\Bs mixing parameters \DGs and \dms}
\labs{DGs} \labs{dms}

%
%
%



Definitions and an introduction to \DGs have been given in \Sec{taubs}.
Neglecting \CP violation, the mass eigenstates are
also \CP eigenstates, with the short-lived state being
\CP-even and the long-lived state being \CP-odd.

The best sensitivity to \DGs is currently achieved 
by the recent time-dependent measurements
of the $\Bs\to\jpsi\phi$ (or more generally $\Bs\to\jpsi K^+K^-$) decay rates performed at
CDF~\cite{Aaltonen:2012ie,*CDF:2011af,*Aaltonen:2007he_mod,*Aaltonen:2007gf_mod},
\dzero~\cite{Abazov:2011ry,*Abazov_mod:2008fj,*Abazov:2007tx_mod_cont}, 
ATLAS~\cite{Aad:2014cqa,*Aad:2012kba_cont}, CMS~\cite{CMS-PAS-BPH-11-006,CMS-PAS-BPH-13-012}
and LHCb~\cite{Aaij:2014zsa,*Aaij:2013oba_supersede2},
where the \CP-even and \CP-odd
amplitudes are statistically separated through a full angular analysis
(see last two columns of \Table{phisDGsGs}). 
With the exception of the first CMS analysis~\cite{CMS-PAS-BPH-11-006},
these studies use both untagged and tagged \Bs\ candidates and 
are optimized for the measurement of the \CP-violating 
phase \phiccbars, defined later in \Sec{phasebs}.
The LHCb collaboration analyzed the $\Bs \to \jpsi K^+K^-$
decay, considering that the $K^+K^-$ system can be in a $P$-wave or $S$-wave state, 
and measured the dependence of the strong phase difference between the 
$P$-wave and $S$-wave amplitudes as a function of the $K^+K^-$ invariant
mass~\cite{Aaij:2012eq}. 
This allowed, for the first time, the unambiguous determination of the sign of 
$\DGs$, which was found to be positive at the $4.7\,\sigma$ level. 
The following averages present only the $\DGs > 0$ solutions.


The available results~\cite{Aaltonen:2012ie,*CDF:2011af,*Aaltonen:2007he_mod,*Aaltonen:2007gf_mod,Abazov:2011ry,*Abazov_mod:2008fj,*Abazov:2007tx_mod_cont,Aad:2014cqa,*Aad:2012kba_cont,CMS-PAS-BPH-11-006,CMS-PAS-BPH-13-012,Aaij:2014zsa,*Aaij:2013oba_supersede2}
are shown in \Table{GsDGs}. They are combined, taking into account, in each analysis, the correlation between \DGs and \Gs.
The results, displayed as the red contours labelled ``$\Bs \to \jpsi KK$ measurements'' in the
plots of \Fig{DGs}, are given in the first column of numbers of \Table{tabtauLH}.

\begin{table}
\caption{Measurements of \DGs and \Gs using
$\Bs\to\jpsi\phi$ and $\Bs\to\jpsi K^+K^-$ decays.
Only the solution with $\DGs > 0$ is shown, since the two-fold ambiguity has been
resolved in \Ref{Aaij:2012eq}. The first error is due to 
statistics, the second one to systematics. The last line gives our average.}
\labt{GsDGs}
\begin{center}
\begin{tabular}{llrlll} 
\hline
Exp.\ & Mode & Dataset
      & \multicolumn{1}{c}{\DGs (\!\!\invps)}
      & \multicolumn{1}{c}{\Gs  (\!\!\invps)}
      & Ref.\ \\
\hline
CDF    & $\jpsi\phi$ & $9.6\invfb$
       & $0.068\pm0.026\pm0.009$
       & $0.654\pm0.008\pm0.004$ 
       & \cite{Aaltonen:2012ie,*CDF:2011af,*Aaltonen:2007he_mod,*Aaltonen:2007gf_mod} \\
\dzero & $\jpsi\phi$ & $8.0\invfb$
       & $0.163^{+0.065}_{-0.064}$ 
       & $0.693^{+0.018}_{-0.017}$
       & \cite{Abazov:2011ry,*Abazov_mod:2008fj,*Abazov:2007tx_mod_cont} \\
ATLAS  & $\jpsi\phi$ & $4.9\invfb$
       & $0.053 \pm0.021 \pm0.010$
       & $0.677 \pm0.007 \pm0.004$
       & \cite{Aad:2014cqa,*Aad:2012kba_cont} \\
CMS    & $\jpsi\phi$ & $5.0\invfb$ 
       & $0.048\pm0.024\pm0.003$
       & $0.655\pm0.008\pm0.003$
       & \cite{CMS-PAS-BPH-11-006}$^p$ \\
CMS    & $\jpsi\phi$ & $20\invfb$ 
       & $0.096\pm0.014\pm0.007$
       & $0.670 \pm0.004 \pm0.005$
       & \cite{CMS-PAS-BPH-13-012}$^p$ \\
LHCb   & $\jpsi K^+K^-$ & $3.0\invfb$
       & $0.0805\pm0.0091\pm0.0033$
       & $0.6603\pm0.0027\pm0.0015$
       & \cite{Aaij:2014zsa,*Aaij:2013oba_supersede2} \\
\hline
\multicolumn{3}{l}{All combined} & \hfagDGSnounit & \hfagGSnounit & \\ 
\hline
\multicolumn{6}{l}{$^p$ {\footnotesize Preliminary.}}
\end{tabular}
\end{center}
\end{table}

\begin{figure}
\begin{center}
\includegraphics[width=0.99\textwidth]{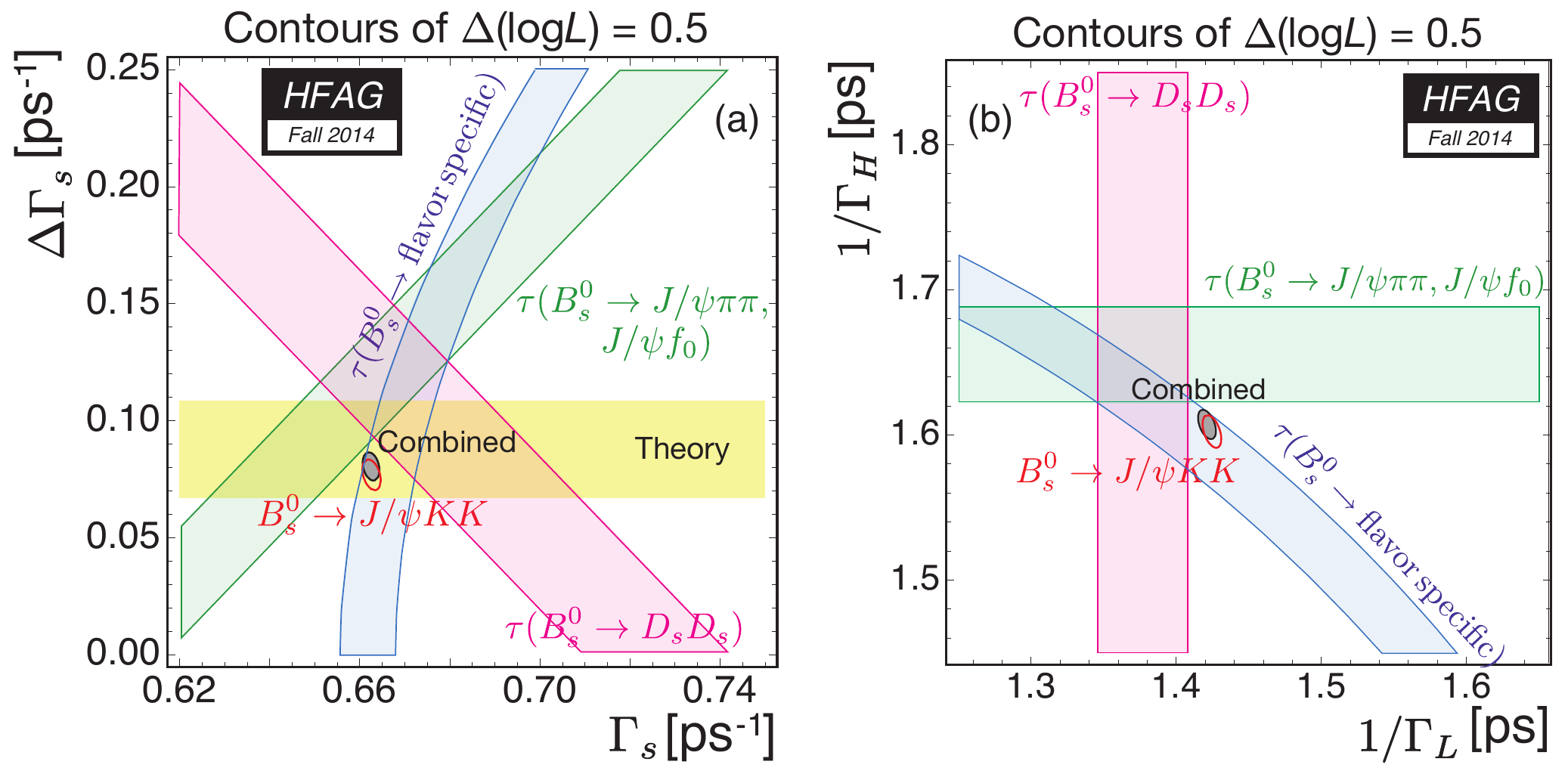}
\caption{Contours of $\Delta \ln L = 0.5$ (39\% CL for the enclosed 2D regions, 68\% CL for the bands)
shown in the $(\Gs,\,\DGs)$ plane on the left
and in the $(1/\Gamma_{\rm L},\,1/\Gamma_{\rm H})$ plane on the right. 
The average of all the $\Bs \to \jpsi\phi$ and $\Bs\to \jpsi K^+K^-$ 
results is shown as the red contour,
and the constraints given by the effective lifetime measurements of
\Bs\ to flavour-specific, pure \CP-odd and pure \CP-even final states
are shown as the blue, green and purple bands, 
respectively. The average taking all constraints into account is shown as the gray-filled contour.
The yellow band is a theory prediction
$\DGs = 0.087 \pm 0.021~\hbox{ps}^{-1}$~\cite{Lenz:2011ti,*Lenz:2006hd}
that assumes no new physics in \Bs\ mixing.}
\labf{DGs}
\end{center}
\end{figure}

\begin{table}
\caption{Averages of \DGs, $\Gs$ and related quantities, obtained from
$\Bs\to\jpsi\phi$ and $\Bs\to\jpsi K^+K^-$ alone (first column),
adding the constraints from the effective lifetimes measured in pure \CP modes
$\Bs\to D_s^+D_s^-$ and $\Bs \to \jpsi f_0(980), \jpsi \pi^+\pi^-$ (second column),
and adding the constraint from the effective lifetime measured in flavour-specific modes
$\Bs\to D_s^-\ell^+\nu X, \, D_s^-\pi^+, \, D_s^-D^+$ (third column, recommended world averages).}
\labt{tabtauLH}
\begin{center}
\begin{tabular}{c|c|c|c}
\hline
& $\Bs\to\jpsi K^+K^-$ modes & $\Bs\to\jpsi K^+K^-$ modes & $\Bs\to\jpsi K^+K^-$ modes \\
& only (see \Table{GsDGs}) & + pure \CP modes & + pure \CP modes \\
&                          &                  & + flavour-specific modes \\
\hline
\Gs                & \hfagGS        &  \hfagGSCO        &  \hfagGSCON        \\
$1/\Gs$            & \hfagTAUBSMEAN &  \hfagTAUBSMEANCO &  \hfagTAUBSMEANCON \\
$1/\Gamma_{\rm L}$ & \hfagTAUBSL    &  \hfagTAUBSLCO    &  \hfagTAUBSLCON    \\
$1/\Gamma_{\rm H}$ & \hfagTAUBSH    &  \hfagTAUBSHCO    &  \hfagTAUBSHCON    \\
\DGs               & \hfagDGS       &  \hfagDGSCO       &  \hfagDGSCON       \\
\DGs/\Gs           & \hfagDGSGS     &  \hfagDGSGSCO     &  \hfagDGSGSCON     \\
$\rho(\Gs,\DGs)$   & \hfagRHOGSDGS  &  \hfagRHOGSDGSCO  &  \hfagRHOGSDGSCON  \\
\hline
\end{tabular}
\end{center}
\end{table}


An alternative approach, which is directly sensitive to first order in 
$\DGs/\Gs$, 
is to determine the effective lifetime of untagged \Bs\ candidates
decaying to 
pure \CP eigenstates; we use here measurements with
$\Bs \to D_s^+D_s^-$~\cite{Aaij:2013bvd}, 
$\Bs \to \jpsi f_0(980)$~\cite{Aaltonen:2011nk}
and $\Bs\to \jpsi \pi^+\pi^-$~\cite{Aaij:2013oba,*LHCb:2011aa_mod,*LHCb:2012ad_mod,*LHCb:2011ab_mod,*Aaij:2012nta_mod} decays.
The precise extraction of $1/\Gs$ and $\DGs$
from such measurements, discussed in detail in \Ref{Fleischer:2011cw}, 
requires additional information 
in the form of theoretical assumptions or
external inputs on weak phases and hadronic parameters. 
If $f$ designates a final state in which both \Bs and \Bsbar can decay,
the ratio of the effective \Bs lifetime decaying to $f$ relative to the mean
\Bs lifetime is~\cite{Fleischer:2011cw}%
\footnote{%
\label{foot:life_mix:ADG-def}
The definition of $A_f^{\DG}$ given in \Eq{ADG} has the sign opposite to that given in \Ref{Fleischer:2011cw}.}
\begin{equation}
  \frac{\tau_{\rm single}(\Bs \to f)}{\tau(\Bs)} = \frac{1}{1-y_s^2} \left[ \frac{1 - 2A_f^{\DG} y_s + y_s^2}{1 - A_f^{\DG} y_s}\right ] \,,
\labe{tauf_fleisch}
\end{equation}
where
\begin{equation}
A_f^{\DG} = -\frac{2 \Re(\lambda_f)} {1+|\lambda_f|^2} \,.
\labe{ADG}
\end{equation}
To include the measurements of the effective
$\Bs \to D_s^+D_s^-$ (\CP-even), $\Bs \to \jpsi f_0(980)$ (\CP-odd) and
$\Bs \to \jpsi\pi^+\pi^-$ (\CP-odd) 
lifetimes as constraints in the \DGs fit,\footnote{%
The effective lifetimes measured in $\Bs\to K^+ K^-$ (mostly \CP-even) and  $\Bs \to \jpsi K_{\rm S}^0$ (mostly \CP-odd) are not used because we can not quantify the penguin contributions in those modes.}
we neglect sub-leading penguin contributions and possible direct \CP violation. 
Explicitly, in \Eq{ADG}, we set
$A_{\mbox{\scriptsize \CP-even}}^{\DG} = \cos \phiccbars$
and $A_{\mbox{\scriptsize \CP-odd}}^{\DG} = -\cos \phiccbars$.
Given the small value of $\phiccbars$, we have, to first order in $y_s$:
\begin{eqnarray}
\tau_{\rm single}(\Bs \to \mbox{\CP-even})
& \approx & \frac{1}{\Gamma_{\rm L}} \left(1 + \frac{(\phiccbars)^2 y_s}{2} \right) \,,
\labe{tau_KK_approx}
\\
\tau_{\rm single}(\Bs \to \mbox{\CP-odd})
& \approx & \frac{1}{\Gamma_{\rm H}} \left(1 - \frac{(\phiccbars)^2 y_s}{2} \right) \,.
\labe{tau_Jpsif0_approx}
\end{eqnarray}
The numerical inputs are taken from \Eqss{tau_KK}{tau_Jpsif0}
and the resulting averages, combined with the $\Bs\to\jpsi K^+K^-$ information,
are indicated in the second column of numbers of \Table{tabtauLH}. 
These averages assume $\phiccbars = 0$, which is compatible with
the \phiccbars average presented in \Sec{phasebs}.

Information on \DGs can also be obtained from the study of the
proper time distribution of untagged samples
of flavour-specific \Bs decays~\cite{Hartkorn:1999ga}, where
the flavour (\ie\ \Bs or \Bsbar) at the time of decay can be determined by
the decay products. In such decays,
\eg\ semileptonic \Bs decays, there is
an equal mix of the heavy and light mass eigenstates at time zero.
The proper time distribution is then a superposition 
of two exponential functions with decay constants
$\Gamma_{\rm L,H} = \Gs \pm \DGs/2$.
This provides sensitivity to both $1/\Gs$ and 
$(\DGs/\Gs)^2$. Ignoring \DGs and fitting for 
a single exponential leads to an estimate of \Gs with a 
relative bias proportional to $(\DGs/\Gs)^2$, as shown in \Eq{fslife}. 
Including the constraint from the world-average flavour-specific \Bs 
lifetime, given in \Eq{tau_fs}, leads to the results shown in the last column 
of \Table{tabtauLH}.
These world averages are displayed as the gray contours labelled ``Combined'' in the
plots of \Fig{DGs}. 
They correspond to the lifetime averages
$1/\Gs=\hfagTAUBSMEANCON$,
$1/\Gamma_{\rm L}=\hfagTAUBSLCON$,
$1/\Gamma_{\rm H}=\hfagTAUBSHCON$,
and to the decay-width difference
\begin{equation}
\DGs = \hfagDGSCON ~~~~\mbox{and} ~~~~~ \DGs/\Gs = \hfagDGSGSCON \,, 
\labe{DGs_DGsGs}
\end{equation}
which is in good agreement with the Standard Model prediction 
$\DGs = 0.087 \pm 0.021~\hbox{ps}^{-1}$~\cite{Lenz:2011ti,*Lenz:2006hd}.

Independent estimates of $\DGs/\Gs$ obtained from measurements of the 
$\Bs \to D_s^{(*)+} D_s^{(*)-}$ branching fraction~\cite{Barate:2000kd,Esen:2010jq_mod,Abazov:2008ig,Abulencia:2007zz}\footnote{
  \label{foot:life_mix:Abazov:2008ig}
  The result of Ref.~\cite{Abazov:2008ig} supersedes that of Ref.~\cite{Abazov:2007rb}.
}
have not been used,
since they are based on the questionable~\cite{Lenz:2011ti,*Lenz:2006hd}
assumption that these decays account for all \CP-even final states.
The results of early lifetime analyses attempting
to measure $\DGs/\Gs$~\cite{Acciarri:1998uv,Abreu:2000sh,Abreu:2000ev,Abe:1997bd}
have not been used either. 

\comment{



Numerical results of the combination of the CDF2 and \dzero inputs
of \Table{dgammat} are:
\begin{eqnarray}
\DGGs &=& \hfagDGSGS \,, \\
\DGs &=& \hfagDGS \,, \\
\bar{\tau}(\Bs) = 1/\Gs &=& \hfagTAUBSMEAN \,, \\
1/\Gamma_{\rm L} = \tau_{\rm short} &=& \hfagTAUBSL \,, \\
1/\Gamma_{\rm H} = \tau_{\rm long}  &=& \hfagTAUBSH \,. 
\end{eqnarray}

Flavour-specific lifetime measurements are of an equal mix
of \CP-even and \CP-odd states at time zero, and  
if a single exponential function is used in the likelihood
lifetime fit of such a sample~\cite{Hartkorn:1999ga}, 
\begin{equation}
\tau(\Bs)_{\rm fs} = \frac{1}{\Gs}
\frac{{1+\left(\frac{\DGs}{2\Gs}\right)^2}}{{1-\left(\frac{\DGs}{2\Gs}\right)^2}
} \,.
\labe{fslife_const}
\end{equation}
Using the world average flavour-specific 
lifetime of \Eq{tau_fs} in \Sec{taubs}
the one-sigma blue bands shown in \Fig{DGs} are obtained. 
Higher-order corrections were checked to be negligible in the
combination.

When the flavour-specific lifetime measurements 
are combined with the 
CDF2 and \dzero measurements of \Table{dgammat}, the solid-outline
shaded
regions of \Fig{DGs} are obtained, with numerical results:
\begin{eqnarray}
\DGGs &=& \hfagDGSGSCON \,, \labe{DGGs_ave} \\
\DGs &=& \hfagDGSCON \,, \\
\bar{\tau}(\Bs) = 1/\Gs &=& \hfagTAUBSMEANC \,, \labe{oneoverGs} \\
1/\Gamma_{\rm L} = \tau_{\rm short} &=& \hfagTAUBSLCON \,, \\
1/\Gamma_{\rm H} = \tau_{\rm long}  &=& \hfagTAUBSHCON \,. 
\end{eqnarray}
These results can
be compared with the theoretical prediction of 
$\DGs = 0.096 \pm 0.039\invps$
(or $\DGs = 0.088 \pm 0.017\invps$ if there is no new physics in
\dms)~\cite{Lenz:2011ti,*Lenz:2006hd,Beneke:1998sy}.

Measurements of $\BR{B^0_s \to D_s^{(*)+} D_s^{(*)-}}$ can 
also be sensitive to \DGs.
The decay $\Bs \to D_s^{+} D_s^{-}$ is into
a final state that is purely \CP even. 
Under various theoretical assumptions~\cite{Aleksan:1993qp,Dunietz:2000cr}, the
inclusive decay into this plus the excited states
$\Bs \to D_s^{(*)+} D_s^{(*)-}$ is also \CP even
to within 5\%, and 
$\Bs \to D_s^{(*)+} D_s^{(*)-}$ saturates
$\Gs^{\CP \thinspace {\rm even}}$.
Under these assumptions, for no \CP violation, we have: 
\begin{equation}
\DGGs \approx
\frac{2 \BR{\Bs \to D_s^{(*)+} D_s^{(*)-}}}
{1 - \BR{\Bs \to D_s^{(*)+} D_s^{(*)-}}} \,.
\labe{dGsBr}
\end{equation}
However, there are concerns~\cite{Nierste_private:2006} 
that the assumptions needed
for the above are overly restrictive and that the inclusive branching
ratio may be \CP even to only 30\%.
In the application of the constraint as a Gaussian penalty
function, the theoretical uncertainty is dealt with in two ways:
the fraction of the \CP-odd component of the decay~\cite{Dunietz:2000cr} 
is taken
to be a uniform distribution ranging from 0 to 0.05 and
convoluted in the Gaussian, and the fractional uncertainty on the
average measured value is increased in quadrature by 
30\%.

\begin{table}
\caption{Measurements of $\BR{\Bs \to D_s^{(*)+} D_s^{(*)-}}$.}
\labt{dGsBr}
\begin{center}
\begin{tabular}{l|c|c|c}
\hline
Experiment & Method & Value & Ref.  \\
\hline
ALEPH         & $\phi$-$\phi$ correlations              
           & $0.115 \pm 0.050^{+0.095}_{-0.045}$  & \cite{Barate:2000kd}$^a$     \\
\dzero        & $D_s \to \phi \pi$, $D_s \to \phi \mu \nu$            
           & $0.035 \pm 0.010 \pm 0.011$  & \cite{Abazov:2008ig}\footref{foot:life_mix:Abazov:2008ig} \\
\belle      & full reco.\ in 6 excl.\ $D_s$ modes 
            & $0.0685 ^{+0.0153}_{-0.0130} {}^{+0.0179}_{-0.0180}$ & \cite{Esen:2010jq_mod}$^{~}$ \\
	 \hline
\multicolumn{2}{l}{Average of above 3} &   \hfagBRDSDS  &   \\
      \hline
\multicolumn{4}{l}{
$^a$ \footnotesize The value quoted in this table is half of 
$\BR{\Bs{\rm(short)} \to D_s^{(*)+} D_s^{(*)-}}$
given in \Ref{Barate:2000kd}.} \\[-0.5ex]
\multicolumn{4}{l}{$^{~}$ \footnotesize Before averaging, it has been adjusted the latest values
of \fBs at LEP and \BR{\Ds \to \phi X}.} 
\end{tabular}
\end{center}
\end{table}

Measurements for the branching fraction for this
decay channel are shown in \Table{dGsBr}.
Using their average value of \hfagBRDSDS with \Eq{dGsBr} yields
\begin{equation}
\DGGs = \hfagDGSGSBRDSDS \,,
\end{equation}
consistent with the value given in \Eq{DGGs_ave}. 

As described in \Sec{taubs}
and \Eq{tau_CPeven}, the average of the lifetime
measurements with \Bs $\to K^+ K^-$ and
$\Bs \to D_s^{(*)} D_s^{(*)}$ decays
can be used to measure the lifetime
of the \CP-even (or ``light" mass) eigenstate
$\tau(\Bs \to \CP\mbox{-even}) = \tau_L = 1/\Gamma_{\rm L} =
\hfagTAUBSSHORT$. These decays are assumed to be 100\% \CP even, with
a 5\% theoretical uncertainty on this assumption added in quadrature
for the combination.


CDF has also measured the exclusive branching fraction 
$\BR{B^0_s \to D^+_s D^-_s} = 
(9.4^{+4.4}_{-4.2}) \times 10^{-3}$~\cite{Abulencia:2007zz}, and
they use this to set a lower bound of
$\DGs^{\CP}/\Gs \geq 0.012$ at \CL{95} (since
on its own it does not saturate the \CP-even states).

} 


The strength of \Bs mixing is known to be large since more than 20 years. 
Indeed the time-integrated measurements of \chibar (see \Sec{chibar}),
when compared to our knowledge
of \chid and the \b-hadron fractions, indicated that 
\chis should be close to its maximal possible value of $1/2$.
Many searches of the time dependence of this mixing 
were performed by ALEPH~\cite{Heister:2002gk},
CDF (Run~I)~\cite{Abe:1998qj},
DELPHI~\cite{Abreu:2000sh,Abreu:2000ev,Abdallah:2002mr,Abdallah:2003we},
OPAL~\cite{Abbiendi:1999gm,Abbiendi:2000bh} and
SLD~\cite{Abe:2002ua,Abe:2002wfa,Abe:2000gp},
but did not have enough statistical power
and proper time resolution to resolve 
the small period of the \Bs\ oscillations.

\Bs oscillations have been observed for the first time in 2006
by the CDF collaboration~\cite{Abulencia:2006ze,*Abulencia:2006mq_mod_cont},
based on samples of flavour-tagged hadronic and semileptonic \Bs decays
(in flavour-specific final states), partially or fully reconstructed in 
$1\invfb$ of data collected during Tevatron's Run~II. 
This was shortly followed by independent evidence obtained by the \dzero collaboration
with $2.4\invfb$ of
data~\cite{D0note5618:2008,*D0note5474:2007,*D0note5254:2006,*Abazov:2006dm_mod_cont}.
More recently the LHCb collaboration obtained the most precise results using fully reconstructed 
$\Bs \to D_s^-\pi^+$ and $\Bs \to D_s^-\pi^+\pi^-\pi^+$ decays at the 
LHC~\cite{Aaij:2011qx,Aaij:2013mpa}.
LHCb has also observed \Bs oscillations with 
$\Bs\to\jpsi K^+K^-$ decays~\cite{Aaij:2014zsa,*Aaij:2013oba_supersede2}
and with semileptonic $\Bs \to D_s^-\mu^+ X$ decays~\cite{Aaij:2013gja}.
The measurements of \dms are summarized in \Table{dms}. 

\begin{table}[t]
\caption{Measurements of \dms.}
\labt{dms}
\begin{center}
\resizebox{\textwidth}{!}{
\begin{tabular}{l@{}c@{}crl@{\,}l@{\,}ll} \hline
Experiment & Method           & \multicolumn{2}{c}{Data set} & \multicolumn{3}{c}{\dms (\!\!\invps)} & Ref. \\
\hline
CDF2   & \particle{D_s^{(*)-} \ell^+ \nu}, \particle{D_s^{(*)-} \pi^+}, \particle{D_s^{-} \rho^+}
       & & 1 \invfb & $17.77$ & $\pm 0.10$ & $\pm 0.07~$
       & \cite{Abulencia:2006ze,*Abulencia:2006mq_mod_cont} \\
\dzero & \particle{D_s^- \ell^+ X}, \particle{D_s^- \pi^+ X}
       &  & 2.4 \invfb & $18.53$ & $\pm 0.93$ & $\pm 0.30~$ 
       & \cite{D0note5618:2008,*D0note5474:2007,*D0note5254:2006,*Abazov:2006dm_mod_cont}$^u$ \\
LHCb   & \particle{D_s^- \pi^+}, \particle{D_s^- \pi^+\pi^-\pi^+}
       & 2010 & 0.034 \invfb & $17.63$ & $\pm 0.11$ & $\pm 0.02~$   
       & \cite{Aaij:2011qx} \\
LHCb   & \particle{D_s^- \mu^+ X}
       & 2011 & 1.0 \invfb & $17.93$ & $\pm 0.22$ & $\pm 0.15$ 
       & \cite{Aaij:2013gja}  \\
LHCb   & \particle{D_s^- \pi^+}
       & 2011 & 1.0 \invfb & $17.768$ & $\pm 0.023$ & $\pm 0.006$ 
       & \cite{Aaij:2013mpa}  \\
LHCb   & \particle{\jpsi K^+K^-}
       & 2011--2012 & 3.0 \invfb & $17.711$ & $^{+0.055}_{-0.057}$ & $\pm 0.011$ 
       & \cite{Aaij:2014zsa,*Aaij:2013oba_supersede2}  \\
\hline
\multicolumn{4}{l}{Average of CDF and LHCb measurements} & $\hfagDMSval$ & $\hfagDMSsta$ & $\hfagDMSsys$ & \\  
\hline
\multicolumn{5}{l}{$^u$ \footnotesize Unpublished.} 
\end{tabular}
}
\end{center}
\end{table}

\begin{figure}
\begin{center}
\includegraphics[width=0.8\textwidth]{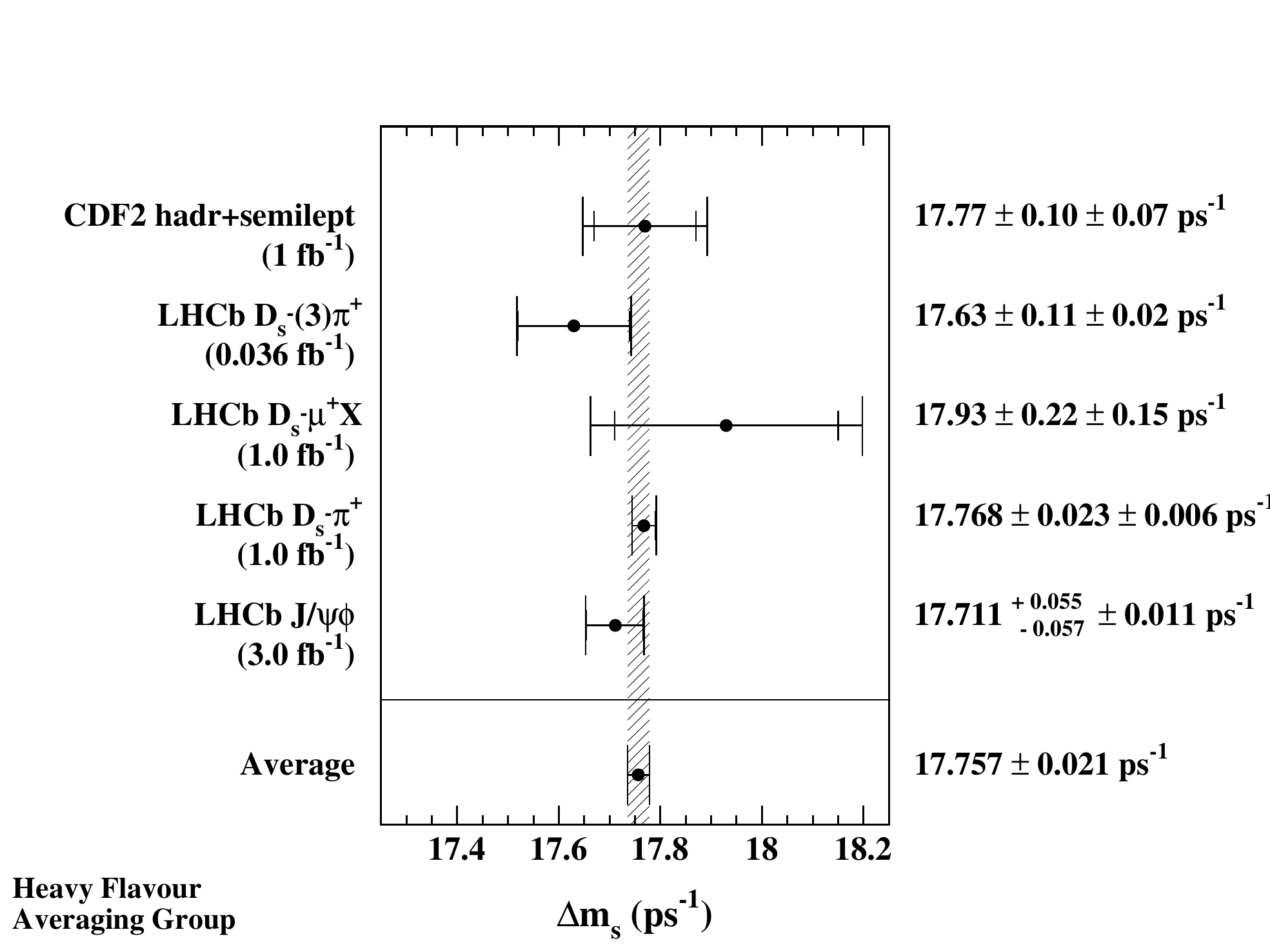}
\caption{Published 
measurements of \dms, together with their average.} 
\labf{dms}
\end{center}
\end{figure}

A simple average of the CDF and LHCb results\footnote{
  \label{foot:life_mix:D0note5618:2008}
  We do not include the old unpublished
  \dzero~\cite{D0note5618:2008,*D0note5474:2007,*D0note5254:2006,*Abazov:2006dm_mod_cont}
  result in the average.},
taking into account the correlated systematic uncertainties between the three 
LHCb measurements, yields 
\begin{equation}
\dms = \hfagDMSfull = \hfagDMS \labe{dms}
\end{equation}
and is illustrated in \Figure{dms}.
Multiplying this result with the 
mean \Bs lifetime of \Eq{oneoverGs}, $1/\Gs=\hfagTAUBSMEANCON$,
yields
\begin{equation}
\xs = \frac{\dms}{\Gs} = \hfagXS \,. \labe{xs}
\end{equation}
With $2\ys = \DGGs=\hfagDGSGSCON$ 
(see \Eq{DGs_DGsGs})
and under the assumption of no \CP violation in \Bs mixing,
this corresponds to
\begin{equation}
\chis = \frac{\xs^2+\ys^2}{2(\xs^2+1)} = \hfagCHIS \,. \labe{chis}
\end{equation}
The ratio of the \Bd and \Bs oscillation frequencies, 
obtained from \Eqss{dmd}{dms}, 
\begin{equation}
\frac{\dmd}{\dms} = \hfagRATIODMDDMS \,, \labe{dmd_over_dms}
\end{equation}
can be used to extract the following ratio of CKM matrix elements, 
\begin{equation}
\left|\frac{V_{td}}{V_{ts}}\right| =
\xi \sqrt{\frac{\dmd}{\dms}\frac{m(\Bs)}{m(\Bd)}} = 
\hfagVTDVTSfull \,, \labe{Vtd_over_Vts}
\end{equation}
where the first quoted error is from experimental uncertainties 
(with the masses $m(\Bs)$ and $m(\Bd)$ taken from \Ref{PDG_2012}),
and where the second quoted error is from theoretical uncertainties 
in the estimation of the SU(3) flavour-symmetry breaking factor
$\xi 
= \hfagXI$,
obtained from unquenched lattice QCD calculations~\cite{Aoki:2013ldr_mod}.
Note that \Eq{Vtd_over_Vts} assumes that \dms and \dmd only receive 
Standard Model contributions. An alternative approach would be to 
take $V_{td}/V_{ts}$ from global fits to predict $\dmd/\dms$, and then 
compare the prediction with the measurement
of \Eq{dmd_over_dms} to set limits on 
new physics effects. 

\mysubsubsection{\CP violation in \Bd and \Bs mixing}
\labs{qpd} \labs{qps}


Evidence for \CP violation in \Bd mixing
has been searched for,
both with flavour-specific and inclusive \Bd decays, 
in samples where the initial 
flavour state is tagged. In the case of semileptonic 
(or other flavour-specific) decays, 
where the final state tag is 
also available, the following asymmetry
\begin{equation} 
\ASLd = \frac{
N(\hbox{\Bdbar}(t) \to \ell^+      \nu_{\ell} X) -
N(\hbox{\Bd}(t)    \to \ell^- \bar{\nu}_{\ell} X) }{
N(\hbox{\Bdbar}(t) \to \ell^+      \nu_{\ell} X) +
N(\hbox{\Bd}(t)    \to \ell^- \bar{\nu}_{\ell} X) } 
= \frac{|p/q|_{\particle{d}}^2 - |q/p|_{\particle{d}}^2}%
{|p/q|_{\particle{d}}^2 + |q/p|_{\particle{d}}^2}
\labe{ASL}
\end{equation} 
has been measured, either in time-integrated analyses at 
CLEO~\cite{Behrens:2000qu,Jaffe:2001hz,*Jaffe:2001hz_cont},
\babar~\cite{Lees:2014qma,*Lees:2014qma_cont},
CDF~\cite{Abe:1996zt}%
\footnote{
  \label{foot:life_mix:CDFnote9015:2007}
  We do not include the unpublished measurement of Ref.~\cite{CDFnote9015:2007} in our average.}
and \dzero~\cite{Abazov:2013uma,*Abazov:2011yk_mod,*Abazov:2010hv_mod_cont,*Abazov:2010hj_mod_cont,*Abazov:2011yk_cont,Abazov:2012uia},
or in time-dependent analyses at 
OPAL~\cite{Ackerstaff:1997vd}, ALEPH~\cite{Barate:2000uk}, 
\babar~\cite{Aubert:2003hd,*Aubert:2004xga_mod_cont,Lees:2013sua,*Margoni:2013qx,*Aubert:2006sa_mod,Aubert:2006nf,*Aubert:2002mn_mod_cont}
and \belle~\cite{Nakano:2005jb}.
Note that the asymmetry of time-dependent decay rates in \Eq{ASL} is time-independent.
In the inclusive case, also investigated and published
at ALEPH~\cite{Barate:2000uk} and OPAL~\cite{Abbiendi:1998av},
no final state tag is used, and the asymmetry~\cite{Beneke:1996hv,*Dunietz:1998av}
\begin{equation} 
\frac{
N(\hbox{\Bd}(t) \to {\rm all}) -
N(\hbox{\Bdbar}(t) \to {\rm all}) }{
N(\hbox{\Bd}(t) \to {\rm all}) +
N(\hbox{\Bdbar}(t) \to {\rm all}) } 
\simeq
\ASLd \left[ \frac{\dmd}{2\Gd} \sin(\dmd \,t) - 
\sin^2\left(\frac{\dmd \,t}{2}\right)\right] 
\labe{ASLincl}
\end{equation} 
must be measured as a function of the proper time to extract information 
on \CP violation.

On the other hand, LHCb has studied the time-dependence of the 
charge asymmetry of $B^0 \to D^{(*)-}\mu^+\nu_{\mu}X$ decays
without tagging the initial state~\cite{Aaij:2014nxa}, 
which would be equal to 
\begin{equation} 
\frac{N(D^{(*)-}\mu^+\nu_{\mu}X)-N(D^{(*)+}\mu^-\bar{\nu}_{\mu}X)}%
{N(D^{(*)-}\mu^+\nu_{\mu}X)+N(D^{(*)+}\mu^-\bar{\nu}_{\mu}X)} =
\ASLd \left[ 1- \cos(\dmd \,t)\right]
\label{eq:untagged_ASL}
\end{equation}
in absence of detection and production asymmetries.

\Table{qoverp} summarizes the different measurements: 
in all cases asymmetries compatible with zero have been found,  
with a precision limited by the available statistics. 

\begin{table}
\caption{Measurements\footref{foot:life_mix:Abe:1996zt}
of \CP violation in \Bd mixing and their average
in terms of both \ASLd and $|q/p|_{\particle{d}}$.
The individual results are listed as quoted in the original publications, 
or converted\footref{foot:life_mix:epsilon_B}
to an \ASLd value.
When two errors are quoted, the first one is statistical and the 
second one systematic. The ALEPH and OPAL 
results assume no \CP violation in \Bs mixing.}
\labt{qoverp}
\begin{center}
\resizebox{\textwidth}{!}{
\begin{tabular}{@{}rcl@{$\,\pm$}l@{$\pm$}ll@{$\,\pm$}l@{$\pm$}l@{}}
\hline
Exp.\ \& Ref. & Method & \multicolumn{3}{c}{Measured \ASLd} 
                       & \multicolumn{3}{c}{Measured $|q/p|_{\particle{d}}$} \\
\hline
CLEO   \cite{Behrens:2000qu} & partial hadronic rec. 
                             & $+0.017$ & 0.070 & 0.014 
                             & \multicolumn{3}{c}{} \\
CLEO   \cite{Jaffe:2001hz,*Jaffe:2001hz_cont}   & dileptons 
                             & $+0.013$ & 0.050 & 0.005 
                             & \multicolumn{3}{c}{} \\
CLEO   \cite{Jaffe:2001hz,*Jaffe:2001hz_cont}   & average of above two 
                             & $+0.014$ & 0.041 & 0.006 
                             & \multicolumn{3}{c}{} \\
\babar \cite{Aubert:2003hd,*Aubert:2004xga_mod_cont}   & full hadronic rec. 
                             & \multicolumn{3}{c}{}  
                             & 1.029 & 0.013 & 0.011  \\
\babar \cite{Lees:2013sua,*Margoni:2013qx,*Aubert:2006sa_mod} & part.\ rec.\ $D^{*}X\ell\nu$ 
                             & $+0.0006$ & \multicolumn{2}{@{}l}{$0.0017 ^{+0.0038}_{-0.0032}$} 
                             & $0.99971$ & $0.00084$ & $0.00175$ \\ 
\babar \cite{Lees:2014qma,*Lees:2014qma_cont}  & dileptons
                             & $-0.0039$ & 0.0035 & 0.0019 
                             & \multicolumn{3}{c}{} \\
\belle \cite{Nakano:2005jb}  & dileptons 
                             & $-0.0011$ & 0.0079 & 0.0085 
                             & 1.0005 & 0.0040 & 0.0043 \\
\multicolumn{2}{l}{Average of above 6 \B factory results} & \multicolumn{3}{l}{\hfagASLDB\ (tot)} 
                             & \multicolumn{3}{l}{\hfagQPDB\  (tot)} \\ 
\hline
\dzero \cite{Abazov:2012uia} & $B^0 \to D^{(*)-}\mu+X$
                            & $+0.0068$ & 0.0045 & 0.0014 & \multicolumn{3}{c}{} \\
LHCb \cite{Aaij:2014nxa} & $B^0 \to D^{(*)-}\mu+X$
                            & $-0.0002$ & 0.0019 & 0.0030 & \multicolumn{3}{c}{} \\
\multicolumn{2}{l}{Average of above 8 pure $B^0$ results} & \multicolumn{3}{l}{\hfagASLDD\ (tot)}
                             & \multicolumn{3}{l}{\hfagQPDD\  (tot)} \\
\hline
\dzero  \cite{Abazov:2013uma,*Abazov:2011yk_mod,*Abazov:2010hv_mod_cont,*Abazov:2010hj_mod_cont,*Abazov:2011yk_cont}  & dimuons  
                             & $-0.0062$ & \multicolumn{2}{@{\hspace{0.26em}}l}{0.0043 (tot)}
                             & \multicolumn{3}{c}{} \\
\multicolumn{2}{l}{Average of above 9 direct measurements} & \multicolumn{3}{l}{\hfagASLDW\ (tot)} 
                             & \multicolumn{3}{l}{\hfagQPDW\  (tot)} \\ 
\hline
OPAL   \cite{Ackerstaff:1997vd}   & leptons     
                             & $+0.008$ & 0.028 & 0.012 
                             & \multicolumn{3}{c}{} \\
OPAL   \cite{Abbiendi:1998av}   & inclusive (\Eq{ASLincl}) 
                             & $+0.005$ & 0.055 & 0.013 
                             & \multicolumn{3}{c}{} \\
ALEPH  \cite{Barate:2000uk}       & leptons 
                             & $-0.037$ & 0.032 & 0.007 
                             & \multicolumn{3}{c}{} \\
ALEPH  \cite{Barate:2000uk}       & inclusive (\Eq{ASLincl}) 
                             & $+0.016$ & 0.034 & 0.009 
                             & \multicolumn{3}{c}{} \\
ALEPH  \cite{Barate:2000uk}       & average of above two 
                             & $-0.013$ & \multicolumn{2}{@{\hspace{0.26em}}l}{0.026 (tot)} 
                             & \multicolumn{3}{c}{} \\
\multicolumn{2}{l}{Average of above 14 results} & \multicolumn{3}{l}{\hfagASLDA\ (tot)} 
                             & \multicolumn{3}{l}{\hfagQPDA\  (tot)} \\ 
\hline
\multicolumn{5}{l}{Best fit value from 2D combination of} \\
\multicolumn{2}{l}{\ASLd and \ASLs results (see \Eq{ASLD})} & \multicolumn{3}{l}{\hfagASLD\ (tot)} 
                             & \multicolumn{3}{l}{\hfagQPD\  (tot)} \\ 
\hline
\end{tabular}
}
\end{center}
\end{table}

A simple average of all measurements performed at 
\B factories~\cite{Behrens:2000qu,Jaffe:2001hz,*Jaffe:2001hz_cont,Aubert:2003hd,*Aubert:2004xga_mod_cont,Lees:2013sua,*Margoni:2013qx,*Aubert:2006sa_mod,Lees:2014qma,*Lees:2014qma_cont,Nakano:2005jb}
yields $\ASLd = \hfagASLDB$; adding also the \dzero~\cite{Abazov:2012uia}
and LHCb~\cite{Aaij:2014nxa} measurements obtained with reconstructed 
semileptonic \Bd decays yields
\begin{equation}
\ASLd = \hfagASLDD  ~~~ \Longleftrightarrow ~~~ |q/p|_{\particle{d}} = \hfagQPDD \,,
\labe{ASLDD}
\end{equation}
where the relation between \ASLd and $|q/p|_{\particle{d}}$ is given in \Eq{ASL}.
As discussed in more detail later in this section, 
the latest dimuon \dzero analysis~\cite{Abazov:2013uma,*Abazov:2011yk_mod,*Abazov:2010hv_mod_cont,*Abazov:2010hj_mod_cont,*Abazov:2011yk_cont}
separates the \Bd and \Bs contributions by exploiting the dependence on the muon impact parameter cut; combining the 
\ASLd result quoted by \dzero with the above \Bd average of \Eq{ASLDD} yields
$\ASLd = \hfagASLDW$. 

All the other \Bd analyses performed at high energy, either at LEP or at the Tevatron,
did not separate the contributions from the \Bd and \Bs mesons.
Under the assumption of no \CP violation in \Bs mixing, a number of 
these analyses~\cite{Abazov:2006qw,Ackerstaff:1997vd,Barate:2000uk,Abbiendi:1998av}
quote a measurement of $\ASLd$ or $|q/p|_{\particle{d}}$ for the \Bd meson. Including also 
these results%
\footnote{
  \label{foot:life_mix:Abe:1996zt}
  A low-statistics result published by CDF using the Run I data~\cite{Abe:1996zt} and 
an unpublished result by CDF using Run II data~\cite{CDFnote9015:2007} 
are not included in our averages, nor in \Table{qoverp}.}
in the previous average 
under the assumption $\ASLs =0$
leads to 
$\ASLd = \hfagASLDA$ 
(\ie\ no change).
The latter assumption makes sense within the Standard Model, 
since \ASLs is predicted to be much smaller than \ASLd~\cite{Lenz:2011ti,*Lenz:2006hd},
but may not be suitable in the presence of new physics. 


The following constraints on a combination of \ASLd and \ASLs
(or equivalently $|q/p|_{\particle{d}}$ and $|q/p|_{\particle{s}}$)
have been obtained by the Tevatron 
experiments, using inclusive semileptonic decays of \b hadrons:
\begin{eqnarray}
\frac{1}{4}\left(f'_{\particle{d}} \,\chid \ASLd +
                 f'_{\particle{s}} \,\chis \ASLs \right) &=& 
+0.0015 \pm 0.0038 \mbox{(stat)} \pm 0.0020 \mbox{(syst)}
~~~ \mbox{CDF1~\cite{Abe:1996zt}} \,, ~
\labe{CDF_ASLDS} \\
\ASLb = \frac{f'_{\particle{d}}\chid \ASLd + f'_{\particle{s}}\chis \ASLs}%
{f'_{\particle{d}}\chid + f'_{\particle{s}}\chis} &=&
 -0.00496 \pm 0.00153 \mbox{(stat)} \pm 0.00072 \mbox{(syst)}
~~ \mbox{\dzero~\cite{Abazov:2013uma,*Abazov:2011yk_mod,*Abazov:2010hv_mod_cont,*Abazov:2010hj_mod_cont,*Abazov:2011yk_cont}} \,, ~
\labe{Dzero_ASLDS}
\end{eqnarray}
where the fractions $f'$ have been defined in \Eq{chibar}.
While the imprecise CDF1 result is compatible with no \CP violation%
\footnote{
  \label{foot:life_mix:CDFnote9015:2007-2}
  A measurement from CDF2, 
$ \ASLb = +0.0080 \pm 0.0090 \mbox{(stat)} \pm 0.0068 \mbox{(syst)}$~\cite{CDFnote9015:2007},
more precise than the \dzero measurement,
is also compatible with no \CP violation,
but since it is unpublished since 2007 
we no longer include it in our averages, nor in \Fig{ASLs}.},
the \dzero result of \Eq{Dzero_ASLDS}, obtained by measuring
the charge asymmetry of like-sign dimuons, differs by 2.8 standard
deviations from the Standard Model expectation of
$\ASLb({\mathrm{SM}}) = (-2.3\pm 0.4) \times 10^{-4}$%
~\cite{Abazov:2013uma,*Abazov:2011yk_mod,*Abazov:2010hv_mod_cont,*Abazov:2010hj_mod_cont,*Abazov:2011yk_cont,Lenz:2011ti,*Lenz:2006hd}.
With a more sophisticated analysis in bins of the
muon impact parameters, \dzero conclude that the overall deviation of the 
their measurements from the SM is at the level of $3.6\,\sigma$.

Using the average $\ASLd = \hfagASLDD$ of \Eq{ASLDD},
obtained from pure \Bd measurements,
the two results of \Eqss{CDF_ASLDS}{Dzero_ASLDS}
are turned\footnote{
  \label{foot:life_mix:f'f}
  For simplicity, we set $f'_{\particle{q}}=f_{\particle{q}}$.} 
into the measurements of \ASLs displayed in the top part of \Fig{ASLs}.
\begin{figure}[t]
\begin{center}
\includegraphics[width=0.8\textwidth]{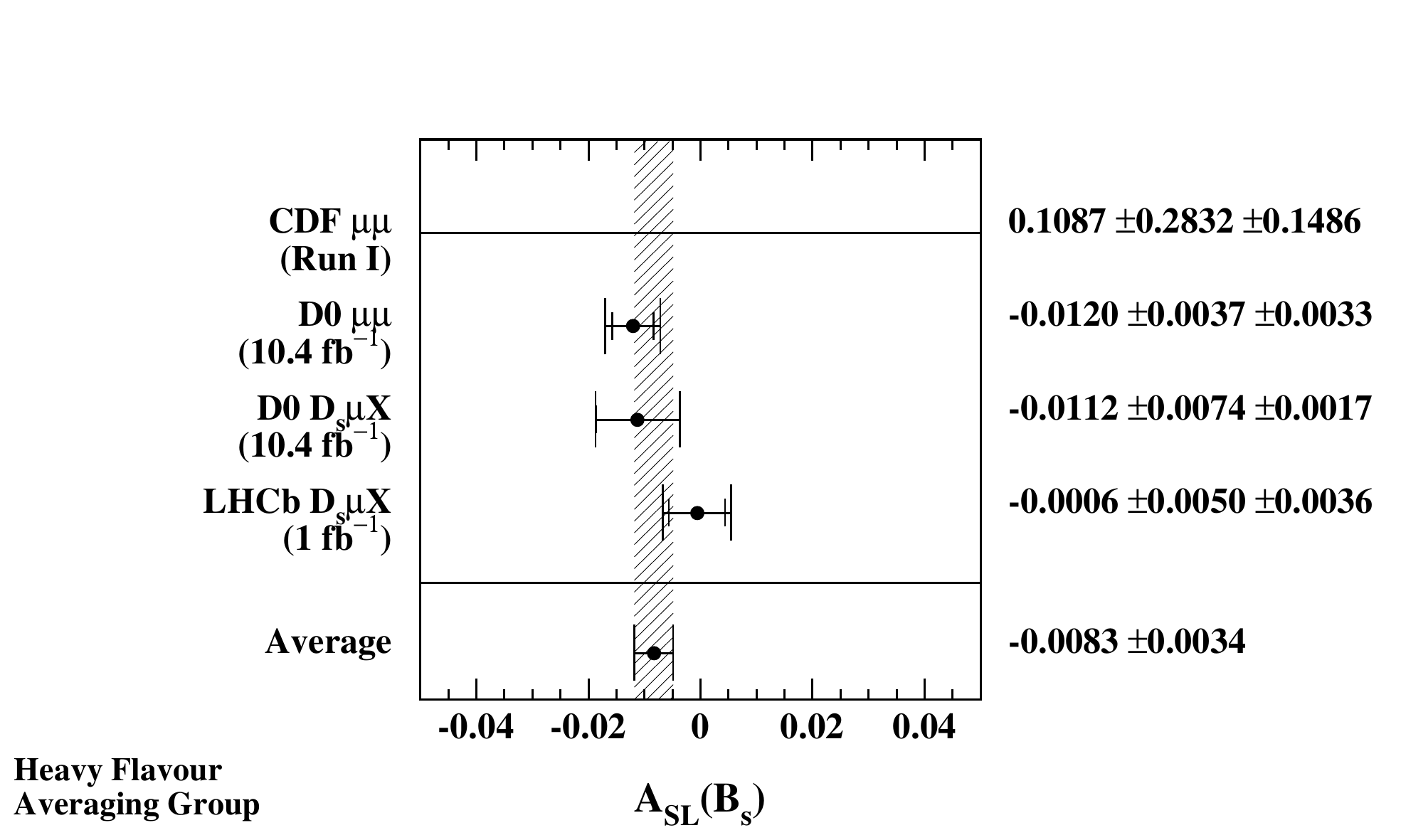}
\caption{Measurements of \ASLs,
derived from CDF~\cite{Abe:1996zt},\footref{foot:life_mix:CDFnote9015:2007-2}
\dzero~\cite{Abazov:2013uma,*Abazov:2011yk_mod,*Abazov:2010hv_mod_cont,*Abazov:2010hj_mod_cont,*Abazov:2011yk_cont,Abazov:2012zz,*Abazov:2009wg_mod_cont,*Abazov:2007nw_mod_cont}
and LHCb~\cite{Aaij:2013gta}
analyses, adjusted to the pure \Bd 
average of \ASLd. 
The combined value of \ASLs is also shown.}
\labf{ASLs}
\end{center}
\end{figure}
Taking into account the uncertainties on
the \b-hadron fractions and mixing parameters, 
the value derived from the \dzero analysis does not show evidence
of \CP violation in the \Bs system.
In addition, the third and fourth lines of \Fig{ASLs} show direct determination of \ASLs
obtained by \dzero~\cite{Abazov:2012zz,*Abazov:2009wg_mod_cont,*Abazov:2007nw_mod_cont}
and LHCb~\cite{Aaij:2013gta}
by measuring the time-integrated charge asymmetry of
untagged $\Bs \to D_s \mu X$ decays.
The four results of \Fig{ASLs} are
combined to yield
$\ASLs = \hfagASLSWval\hfagASLSWsta\mbox{(stat)}\hfagASLSWsys\mbox{(syst)} = \hfagASLSW$
or, equivalently through \Eq{ASL},
$|q/p|_{\particle{s}} = \hfagQPSWval\hfagQPSWsta\mbox{(stat)}\hfagQPSWsys\mbox{(syst)} = \hfagQPSW$.
The quoted systematic errors include experimental systematics as well as the correlated dependence on external 
parameters. 

As mentioned above, the \dzero like-sign dimuon analysis investigates 
the dependence of the charge asymmetry 
as a function of the muon impact parameters. 
Interpreting the observed asymmetries in terms of \CP violation in $B$-meson mixing and interference, 
and using 
the mixing parameters and the world \b-hadron fractions 
of \Ref{Amhis:2012bh}, the \dzero collaboration
extracts~\cite{Abazov:2013uma,*Abazov:2011yk_mod,*Abazov:2010hv_mod_cont,*Abazov:2010hj_mod_cont,*Abazov:2011yk_cont}
values for \ASLd and \ASLs and their correlation
coefficient\footnote{
  \label{foot:life_mix:Abazov:2013uma}
  They also extract at the same time a value for \DGGd (see \Sec{DGd}).}, 
as shown in \Table{ASLs_ASLd}.
However, the individual 
contributions to the total quoted errors from this analysis and from the
external inputs are not given, so the adjustment of these results to different
or more recent values of the external inputs cannot (easily) be done. 
Using a two-dimensional fit, these values are combined with the pure
\Bd average of \Eq{ASLDD} and with the results from the 
$\Bs \to D_s \mu X$ analyses~\cite{Abazov:2012zz,*Abazov:2009wg_mod_cont,*Abazov:2007nw_mod_cont,Aaij:2013gta},
assumed to be independent and also shown in \Table{ASLs_ASLd}.
The result, shown graphically in \Fig{ASLs_ASLd}, is 
\begin{table}
\caption{Direct measurements of \CP violation in \Bs and \Bd mixing, together 
with their two-dimensional average. Only total errors are quoted.}
\labt{ASLs_ASLd}
\begin{center}
\begin{tabular}{ccccc}
\hline
Exp.\ \& Ref.\ & Method & Measured \ASLs & Measured \ASLd & $\rho(\ASLs,\ASLd)$ \\
\hline
\dzero  \cite{Abazov:2012zz,*Abazov:2009wg_mod_cont,*Abazov:2007nw_mod_cont}  & $\Bs \to D_s \mu X$ 
       & $-0.0112 \pm 0.0076$ 
       & & \\
LHCb \cite{Aaij:2013gta} & $\Bs \to D_s \mu X$ & $-0.0006 \pm 0.0062$ & & \\
\hline
\multicolumn{2}{l}{Average of above \Bs results}
       & \hfagASLSD & & \\ 
\multicolumn{2}{l}{Average of \Bd results (\Eq{ASLDD})} 
       & & \hfagASLDD & \\ 
\dzero  \cite{Abazov:2013uma,*Abazov:2011yk_mod,*Abazov:2010hv_mod_cont,*Abazov:2010hj_mod_cont,*Abazov:2011yk_cont}  & dimuons  
       & $-0.0082 \pm 0.0099$ 
       & $-0.0062 \pm 0.0043$ 
       & $-0.61$ \\          
\hline
\multicolumn{2}{l}{Average of all above}
       & \hfagASLS & \hfagASLD & $\hfagRHOASLSASLD$ \\ 
\hline
\end{tabular}
\end{center}
\end{table}
\begin{figure}
\begin{center}
\includegraphics[width=0.6\textwidth]{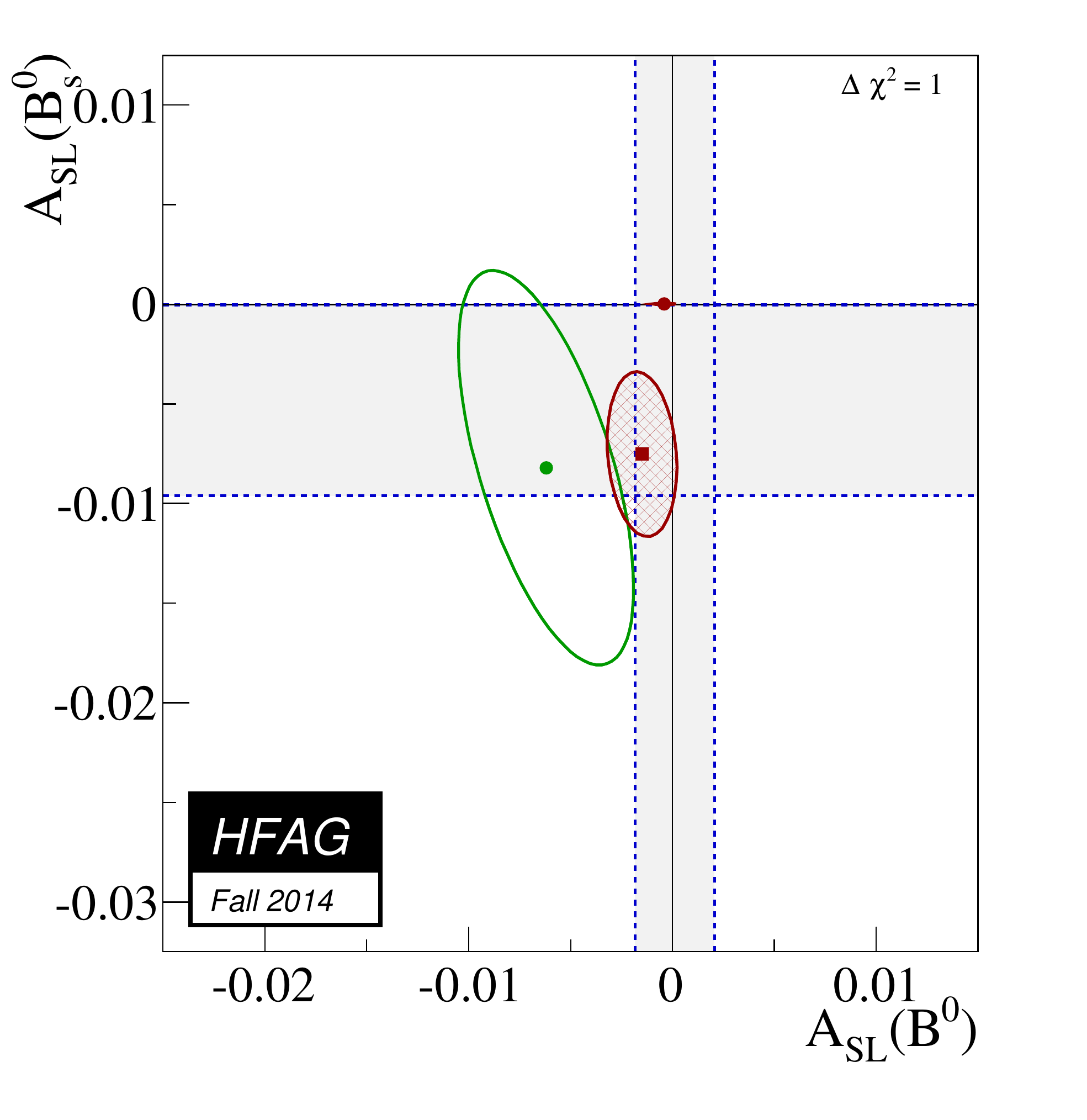}
\end{center}
\vspace{-5mm}
\caption{
Direct measurements of \ASLs and \ASLd listed in \Table{ASLs_ASLd}
(\Bd average as the vertical band, \Bs average as the horizontal band,
\dzero dimuon result as the green ellipse),
together with their two-dimensional average (red hatched ellipse).
The red point close to $(0,0)$ is the Standard Model prediction
of \Ref{Lenz:2011ti,*Lenz:2006hd} with error bars multiplied by 10.
The prediction and the experimental average deviate from each other by $\hfagASLDASLSNSIGMA\,\sigma$.}
\labf{ASLs_ASLd}
\end{figure}
\begin{eqnarray}
\ASLd & = & \hfagASLD ~~~ \Longleftrightarrow ~~~ |q/p|_{\particle{d}} = \hfagQPD \,,
\labe{ASLD}
\\
\ASLs & = & \hfagASLS ~~~ \Longleftrightarrow ~~~ |q/p|_{\particle{s}} = \hfagQPS \,,
\labe{ASLS}
\\
\rho(\ASLd , \ASLs) & = & \hfagRHOASLSASLD \,.
\labe{rhoASLDASLS}
\end{eqnarray}

The average of \Fig{ASLs} 
ignores the impact parameter study of \dzero.
The average of \Eq{ASLS} ignores the CDF1 
result, which has a very large uncertainty anyway.
We choose the results of \Eqsss{ASLD}{ASLS}{rhoASLDASLS}
as our final averages,\footnote{
  \label{foot:life_mix:epsilon_B}
  Early analyses and (perhaps hence) the PDG use the complex
  parameter $\epsilon_{\B} = (p-q)/(p+q)$; if \CP violation in the mixing is small,
  $\ASLd \cong 4 {\rm Re}(\epsilon_{\B})/(1+|\epsilon_{\B}|^2)$ and the averages of
  \Eqss{ASLDD}{ASLD} 
  correspond to ${\rm Re}(\epsilon_{\B})/(1+|\epsilon_{\B}|^2)=\hfagREBDD$ 
  and $\hfagREBD$, respectively.}
since they incorporate better the available published data. 

The above averages show no evidence of \CP violation in \Bd and \Bs mixing.
They deviate by $\hfagASLDASLSNSIGMA\,\sigma$ from the very small predictions of the Standard Model, 
${\ASLd}^{\rm SM} = -(4.1\pm 0.6)\times 10^{-4}$ and 
${\ASLs}^{\rm SM} = +(1.9\pm 0.3)\times 10^{-5}$~\cite{Lenz:2011ti,*Lenz:2006hd}.
Given the current size of the experimental uncertainties,
there is still significant room for a possible new physics contribution, especially in the \Bs system. 
In this respect, the deviation of the \dzero dimuon
asymmetry~\cite{Abazov:2013uma,*Abazov:2011yk_mod,*Abazov:2010hv_mod_cont,*Abazov:2010hj_mod_cont,*Abazov:2011yk_cont}
from expectation has generated a lot of excitement, however recent results from \dzero and LHCb 
have not yet settled the issue, and more experimental data (especially from LHCb) is awaited eagerly. 

At the more fundamental level, \CP violation in \Bs
mixing\footnote{
  \label{foot:life_mix:CPVinBd}
  Of course, a similar formalism exists for the \Bd system; for 
  simplicity we omit here the subscript $s$ for $\phi_{12}$, $M_{12}$ and $\Gamma_{12}$.}
is caused by the weak phase difference 
\begin{equation}
\phi_{12} = \arg \left[ -{M_{12}}/{\Gamma_{12}} \right], 
\end{equation}
where $M_{12}$ and $\Gamma_{12}$ are the off-diagonal
elements of the mass and decay matrices of the $\Bs-\Bsbar$ system.
This is related to the observed decay-width difference through the relation
\begin{equation}
\DGs = 2|\Gamma_{12}|\cos\phi_{12}+
{\cal O} \left( \left|\frac{\Gamma_{12}}{M_{12}}\right|^2 \right) \,,
\end{equation}
where quadratic (or higher-order) terms in the small quantity
$|\Gamma_{12}/M_{12}| \sim {\cal O}(m_b^2/m_t^2)$ can be neglected. 
The SM prediction for this phase is tiny~\cite{Lenz:2011ti,*Lenz:2006hd},
\begin{equation}
\phi_{12}^{\rm SM} = \hfagPHISTWELVESM \,;
\labe{phis12SM}
\end{equation}
however, new physics in \Bs mixing could change this observed phase to
\begin{equation}
\phi_{12} = \phi_{12}^{\rm SM} + \phi_{12}^{\rm NP} \,.
\labe{phi12NP}
\end{equation}
The \Bs semileptonic asymmetry can be expressed as~\cite{Beneke:2003az}
\begin{equation}
\ASLs = 
\Im \left(\frac{\Gamma_{12}}{M_{12}} \right) +
{\cal O} \left( \left|\frac{\Gamma_{12}}{M_{12}}\right|^2 \right) =
\frac{\DGs}{\dms}\tan\phi_{12} +
{\cal O} \left( \left|\frac{\Gamma_{12}}{M_{12}}\right|^2 \right) \,.
\labe{ASLS_tanphi12}
\end{equation}
Using this relation, the current knowledge of \ASLs, \DGs and \dms, 
given in \Eqsss{ASLS}{DGs_DGsGs}{dms} respectively, yield an
experimental determination of $ \phi_{12}$,
\begin{equation}
\tan\phi_{12} = \ASLs \frac{\dms}{\DGs} = \hfagTANPHI \,,
\labe{tanphi12}
\comment{ 
from math import *
asls = -0.010460 ; easls = 0.006400
dms = 17.719032  ; edms = 0.042701
dgs = 0.0951919  ; edgs = 0.01362480381803716 
tanphi12 = asls*dms/dgs
etanphi12 = tanphi12*sqrt((easls/asls)**2+(edms/dms)**2+(edgs/dgs)**2)
print tanphi12, etanphi12
} 
\end{equation}
which represents only a very weak constraint at present.


\mysubsubsection{Mixing-induced \CP violation in \Bs decays}

%
%
%

\labs{phasebs}


\CP violation induced by $\Bs-\Bsbar$ mixing
has been a field of 
very active study and fast experimental progress 
in the past couple of years.
The main observable is the 
\CP-violating phase \phiccbars, defined as 
the weak phase difference between
the $\Bs-\Bsbar$ mixing amplitude
and the $b \to c\bar{c}s$ decay amplitude.

The golden mode for such studies is 
$\Bs \to \jpsi\phi$, followed by $\jpsi \to \mu^+\mu^-$ and 
$\phi\to K^+K^-$, for which a full angular 
analysis of the decay products is performed to 
separate statistically the \CP-even and \CP-odd
contributions in the final state. As already mentioned in 
\Sec{DGs},
CDF~\cite{Aaltonen:2012ie,*CDF:2011af,*Aaltonen:2007he_mod,*Aaltonen:2007gf_mod},
\dzero~\cite{Abazov:2011ry,*Abazov_mod:2008fj,*Abazov:2007tx_mod_cont},
ATLAS~\cite{Aad:2014cqa,*Aad:2012kba_cont}, CMS~\cite{CMS-PAS-BPH-13-012}
and LHCb~\cite{Aaij:2014zsa,*Aaij:2013oba_supersede2}
have used both untagged and tagged $\Bs \to \jpsi\phi$ (and $\Bs \to \jpsi K^+K^-$) events 
for the measurement of \phiccbars.
LHCb~\cite{Aaij:2014dka,*Aaij:2013oba_supersede}
has used $\Bs \to \jpsi \pi^+\pi^-$ events, 
analyzed with a full amplitude model
including several $\pi^+\pi^-$ resonances (\eg $f_0(980)$),
although the
$\jpsi \pi^+\pi^-$ final state had already been shown
to be almost \CP pure with a \CP-odd fraction
larger than 0.977 at 95\% CL~\cite{LHCb:2012ae}. 
In addition, LHCb has used the $\Bs \to \Dsp\Dsm$ channel~\cite{Aaij:2014ywt} to measure \phiccbars.

All CDF, \dzero, ATLAS and CMS analyses provide 
two mirror solutions related by the transformation 
$(\DGs, \phiccbars) \to (-\DGs, \pi-\phiccbars)$. However, the
LHCb analysis of $\Bs \to \jpsi K^+K^-$ resolves this ambiguity and 
rules out the solution with negative \DGs~\cite{Aaij:2012eq},
a result in agreement with the Standard Model expectation.
Therefore, in what follows, we only consider the solution with $\DGs > 0$.

\begin{table}
\caption{Direct experimental measurements of \phiccbars, \DGs and \Gs using
$\Bs\to\jpsi\phi$, $\jpsi K^+K^-$, $\jpsi\pi^+\pi^-$ and $D_s^+D_s^-$ decays.
Only the solution with $\DGs > 0$ is shown, since the two-fold ambiguity has been
resolved in \Ref{Aaij:2012eq}. The first error is due to 
statistics, the second one to systematics. The last line gives our average.}
\labt{phisDGsGs}
\begin{center}
\begin{tabular}{llrlll} 
\hline
Exp.\ & Mode & Dataset & \multicolumn{1}{c}{\phiccbars}
                     & \multicolumn{1}{c}{\DGs (\!\!\invps)} & Ref.\ \\
\hline
CDF    & $\jpsi\phi$ & $9.6\invfb$
       & $[-0.60,\, 0.12]$, 68\% CL & $0.068\pm0.026\pm0.009$
       & \cite{Aaltonen:2012ie,*CDF:2011af,*Aaltonen:2007he_mod,*Aaltonen:2007gf_mod} \\
\dzero & $\jpsi\phi$ & $8.0\invfb$
       & $-0.55^{+0.38}_{-0.36}$ & $0.163^{+0.065}_{-0.064}$ 
       & \cite{Abazov:2011ry,*Abazov_mod:2008fj,*Abazov:2007tx_mod_cont} \\
ATLAS  & $\jpsi\phi$ & $4.9\invfb$
       & $+0.12 \pm 0.25 \pm 0.05$ & $0.053 \pm0.021 \pm0.010$ 
       & \cite{Aad:2014cqa,*Aad:2012kba_cont} \\
CMS    & $\jpsi\phi$ & $20\invfb$ 
       & $-0.03 \pm 0.11 \pm 0.03$ & $0.096\pm0.014\pm0.007$ 
       & \cite{CMS-PAS-BPH-13-012}$^p$ \\
LHCb   & $\jpsi K^+K^-$ & $3.0\invfb$
       & $-0.058\pm0.049\pm0.006$ & $0.0805\pm0.0091\pm0.0033$ 
       & \cite{Aaij:2014zsa,*Aaij:2013oba_supersede2} \\
LHCb   & $\jpsi\pi^+\pi^-$ & $3.0\invfb$
       & $+0.070 \pm0.068 \pm 0.008$ & --- 
       & \cite{Aaij:2014dka,*Aaij:2013oba_supersede} \\
LHCb   & $\jpsi h^+h^-$ & $3.0\invfb$
       & $-0.010\pm0.039(\rm tot)$ & --- 
       & \cite{Aaij:2014zsa,*Aaij:2013oba_supersede2}$^a$ \\
LHCb   & $D_s^+D_s^-$ & $3.0\invfb$
       & $+0.02 \pm0.17 \pm 0.02$ & --- 
       & \cite{Aaij:2014ywt} \\
\hline
\multicolumn{3}{l}{All combined} & \hfagPHISCOMB & \hfagDGSCOMBnounit & \\ 
\hline
\multicolumn{6}{l}{$^a$ {\footnotesize LHCb combination of $\jpsi K^+K^-$~\cite{Aaij:2014zsa,*Aaij:2013oba_supersede2} and $\jpsi\pi^+\pi^-$~\cite{Aaij:2014dka,*Aaij:2013oba_supersede}.}}\\[-0.8ex]
\multicolumn{6}{l}{$^p$ {\footnotesize Preliminary.}}
\end{tabular}
\end{center}
\end{table}

\begin{figure}
\begin{center}
\includegraphics[width=0.49\textwidth]{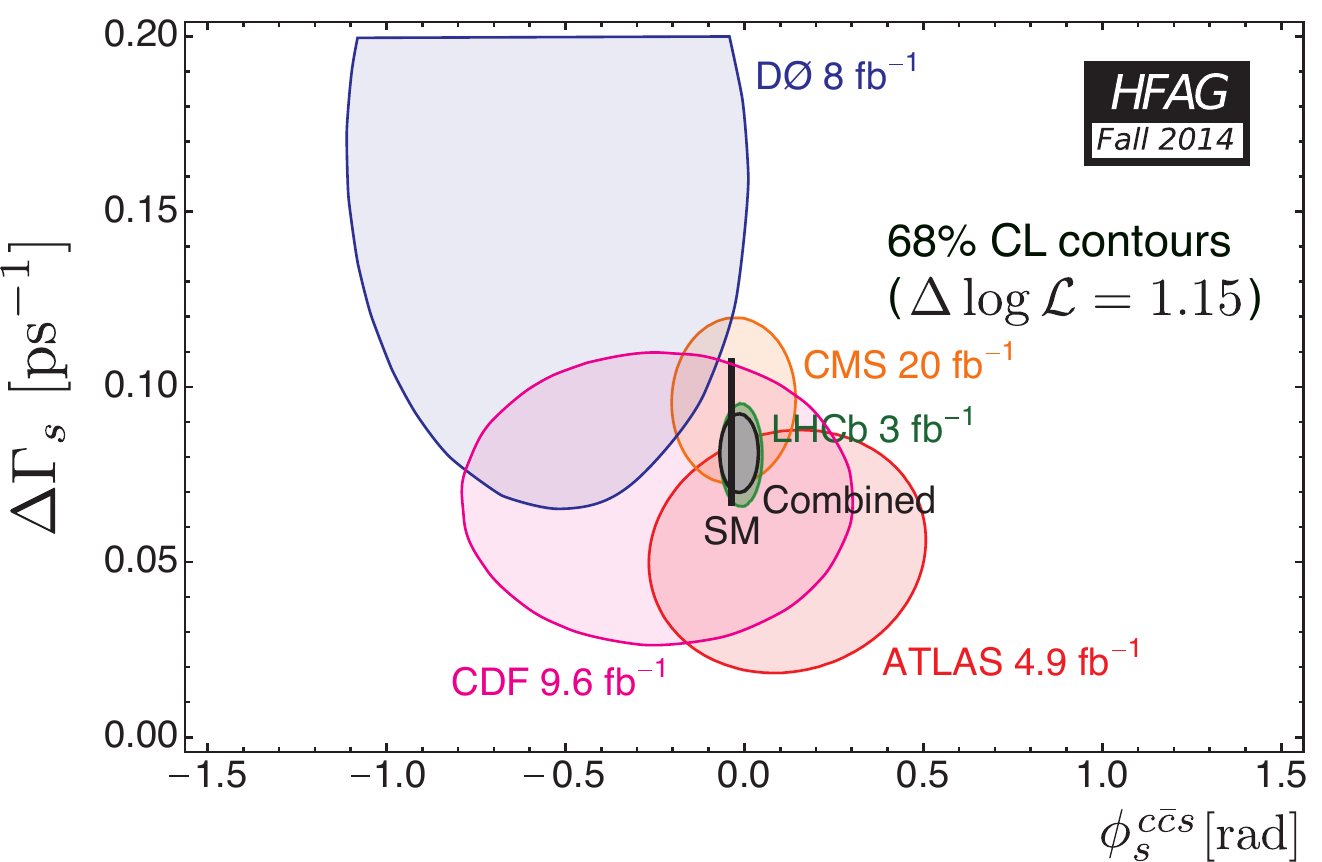}
\includegraphics[width=0.49\textwidth]{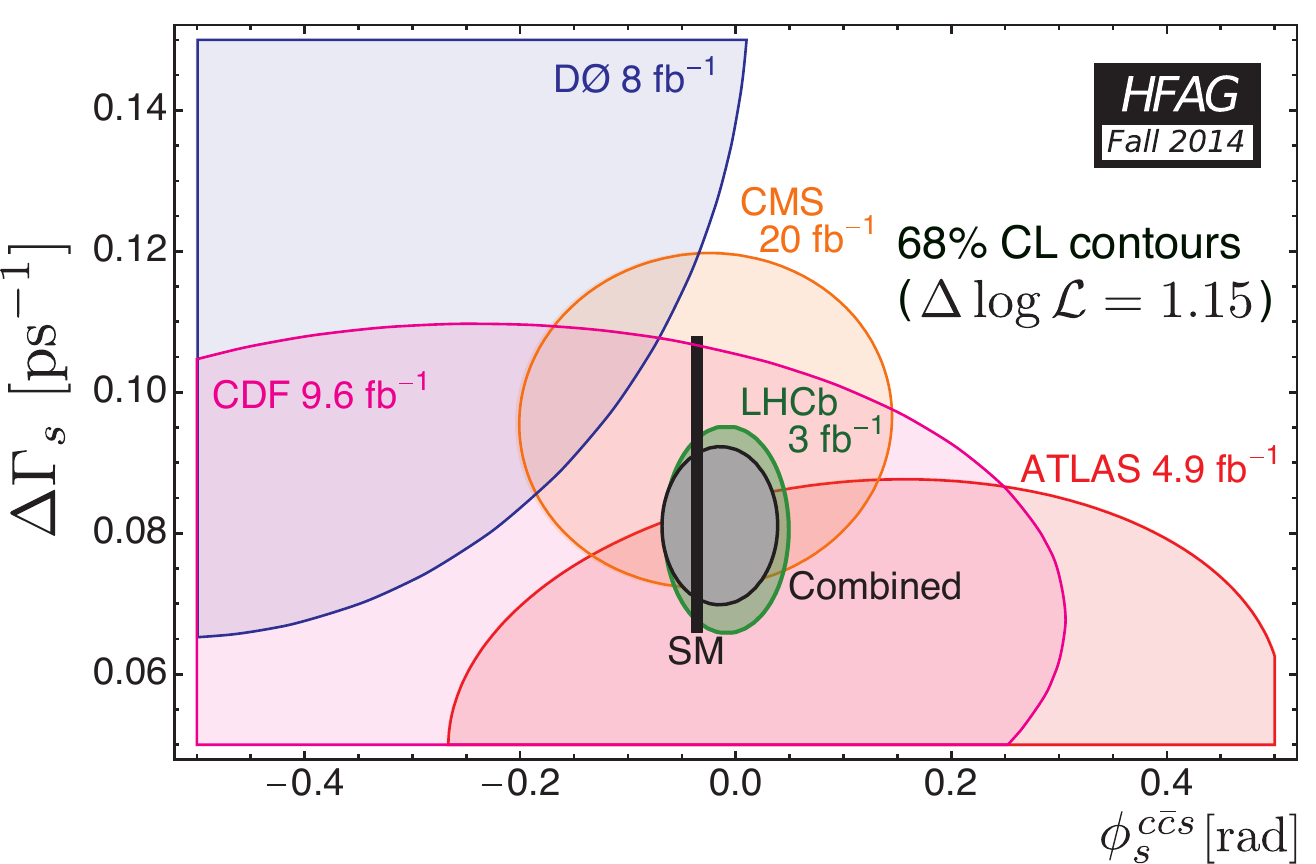}
\caption{
Left: 68\% CL regions in \Bs width difference \DGs and weak phase \phiccbars
obtained from individual and combined CDF~\cite{Aaltonen:2012ie,*CDF:2011af,*Aaltonen:2007he_mod,*Aaltonen:2007gf_mod},
\dzero~\cite{Abazov:2011ry,*Abazov_mod:2008fj,*Abazov:2007tx_mod_cont}, ATLAS~\cite{Aad:2014cqa,*Aad:2012kba_cont}, 
CMS~\cite{CMS-PAS-BPH-13-012}
and LHCb~\cite{Aaij:2014zsa,*Aaij:2013oba_supersede2,Aaij:2014dka,*Aaij:2013oba_supersede}
likelihoods of 
$\Bs\to \jpsi\phi$, $\Bs\to \jpsi K^+K^-$ and $\Bs\to\jpsi\pi^+\pi^-$. 
The expectation within the Standard Model~\cite{Charles:2011va_mod,Lenz:2011ti,*Lenz:2006hd}
is shown as the black rectangle.
Right: same, but zoomed on the region of interest.
}
\labf{DGs_phase}
\end{center}
\end{figure}

We perform a combination of the CDF~\cite{Aaltonen:2012ie,*CDF:2011af,*Aaltonen:2007he_mod,*Aaltonen:2007gf_mod},
\dzero~\cite{Abazov:2011ry,*Abazov_mod:2008fj,*Abazov:2007tx_mod_cont},
ATLAS~\cite{Aad:2014cqa,*Aad:2012kba_cont}, CMS~\cite{CMS-PAS-BPH-13-012}
and LHCb~\cite{Aaij:2014zsa,*Aaij:2013oba_supersede2,Aaij:2014dka,*Aaij:2013oba_supersede}
results summarized in \Table{phisDGsGs}.
This is done by adding the two-dimensional log profile-likelihood scans of
\DGs and \phiccbars from the four $\Bs\to\jpsi\phi$ ($\Bs\to\jpsi K^+K^-$) analyses and 
a one-dimensional log profile-likelihood of \phiccbars
from the $\Bs\to\jpsi\pi^+\pi^-$ and $\Bs \to D_s^+ D_s^-$ analyses; 
the combined likelihood is then maximized with respect to \DGs and \phiccbars.

In the $\Bs\to\jpsi\phi$ and $\Bs\to\jpsi K^+K^-$ analyses, \phiccbars and \DGs 
come from a simultaneous fit that determines also the \Bs lifetime,
the polarisation amplitudes and strong phases.
While the correlation between \phiccbars and all other parameters is small,
the correlations between \DGs and the polarisation amplitudes are sizeable.
However, since the various experiments use different conventions
for the amplitudes and phases, a full combination including all
correlations is not performed. Instead, our average only takes
into account the correlation between \phiccbars and \DGs.


In the recent LHCb $\Bs \to \jpsi K^+K^-$ analysis~\cite{Aaij:2014zsa,*Aaij:2013oba_supersede2}, the \phiccbars values are measured for the first time for each polarization of the final state. Since those values are compatible within each other, we still use the unique value of \phiccbars for our world average, corresponding to the one measured by the other-than-LHCb analyses. 
In the same analysis, the statistical correlation coefficient between \phiccbars and $|\lambda|$
(which signals \CP violation in the decay if different from unity) 
is measured to be very small ($-0.02$). We neglect this correlation in our average. 
Furthermore, the statistical correlation coefficient between \phiccbars and \DGs\ is measured to be small $(-0.08)$. When averaging LHCb results of 
$\Bs \to \jpsi K^+K^-$,  $\Bs \to \jpsi \pi^+\pi^-$ and $\Bs \to D_s^+ D_s^-$, we neglect this correlation coefficient (putting it to zero). 
Given the increasing experimental precision, we have also stopped using the two-dimensional $\DGs-\phiccbars$ histograms provided by the CDF and \dzero collaborations: we are now approximating those with two-dimensional Gaussian likelihoods. 

We obtain the individual and combined contours shown in \Fig{DGs_phase}. 
Maximizing the likelihood, we find, as summarized in \Table{phisDGsGs}:  
\begin{eqnarray}
\DGs &=& \hfagDGSCOMB \,, \\    
\phiccbars &=& \hfagPHISCOMB \,.
\labe{phis}
\end{eqnarray}
The above \DGs average is consistent, but highly correlated with the average
of \Eq{DGs_DGsGs}. Our
final recommended average for \DGs is the one of \Eq{DGs_DGsGs}, which 
includes all available information on \DGs. 

In the Standard Model and ignoring sub-leading penguin contributions, 
\phiccbars is expected to be equal to $-2\beta_s$, 
where
$\beta_s = \arg\left[-\left(V_{ts}V^*_{tb}\right)/\left(V_{cs}V^*_{cb}\right)\right]$ 
is a phase analogous to the angle $\beta$ of the usual CKM
unitarity triangle (aside from a sign change). 
An indirect determination via global fits to experimental data
gives~\cite{Charles:2011va_mod}
\begin{equation}
(\phiccbars)^{\rm SM} = -2\beta_s = \hfagPHISSM \,.
\labe{phisSM}
\end{equation}
The average value of \phiccbars from \Eq{phis} is consistent with this
Standard Model expectation.

New physics could contribute to \phiccbars. Assuming that new physics only 
enters in $M_{12}$ (rather than in $\Gamma_{12}$),
one can write~\cite{Lenz:2011ti,*Lenz:2006hd}
\begin{equation}
\phiccbars = -  2\beta_s + \phi_{12}^{\rm NP} \,,
\end{equation}
where the new physics phase $\phi_{12}^{\rm NP}$ is the same as that appearing in \Eq{phi12NP}.
In this case
\begin{equation}
\phi_{12} = 
\phi_{12}^{\rm SM} +2\beta_s + \phiccbars = \hfagPHISTWELVE \,,
\end{equation}
where the numerical estimation was performed with the values of \Eqsss{phis12SM}{phisSM}{phis}.
This can serve as a reference value to which the measurement of \Eq{tanphi12} can be compared.


%


\comment{

For non-zero $|\Gamma_{12}|$, analysis of the time-dependent
decay \particle{\Bs \to \jpsi\phi} can measure
the weak phase.  Including information on the \Bs flavour at production
time via flavour tagging improves precision and also resolves the 
sign ambiguity on the weak phase angle for a given \DGs.
Both CDF~\cite{CDF:2011af,*Aaltonen:2007he_mod,*Aaltonen:2007gf_mod} 
and \dzero~\cite{Abazov:2011ry,*Abazov_mod:2008fj,*Abazov:2007tx_mod_cont} have performed 
such analyses and measure the same observed phase that we denote
$\phi_s^{\jpsi \phi} = -2\beta_s^{\jpsi \phi}$ to reflect
the different conventions of the experiments.

Under the assumption of non-zero $\phi_s^{\jpsi\phi}$, 
in addition to the result listed
in \Table{dgammat}, 
the \dzero collaboration~\cite{Abazov:2011ry,*Abazov_mod:2008fj,*Abazov:2007tx_mod_cont}  has also made simultaneous
fits allowing $\phi_s^{\jpsi\phi}$ to float while weakly 
constraining the strong phases, $\delta_i$ to find: 
\begin{eqnarray}
\DGs &=& +0.19 \pm 0.07 ^{+0.02}_{-0.01}~{\mathrm{ps}}^{-1}\,,  \\ 
\bar{\tau}(\Bs) &= &1/\Gs = 1.52 \pm 0.06~{\mathrm{ps}}\,,  \\
\phi_s^{\jpsi\phi} &=& -0.57 ^{+0.24+0.07}_{-0.30-0.02} \,. 
\end{eqnarray}
If the SM value of $\phi_s^{\jpsi\phi} = -0.04$ is assumed, a probability of 
6.6\% to obtain a value of $\phi_s^{\jpsi\phi}$ lower than $-0.57$ is
found.

The CDF 
analysis~\cite{CDF:2011af,*Aaltonen:2007he_mod,*Aaltonen:2007gf_mod} 
reports confidence regions
in the two-dimensional space of $2\beta_s^{\jpsi\phi}$ and \DGs.
They present a Feldman-Cousins confidence interval of $2\beta_s^{\jpsi\phi}$
where \DGs is treated as a nuisance parameter:
\begin{equation}
2\beta_s^{\jpsi\phi} = -\phi_s^{\jpsi\phi} \in [0.56,2.58]~{\mathrm{at~68\%~CL}}.
\end{equation}
Only a confidence range is quoted and a  point 
estimate is not given since biases were observed in the analysis.
Assuming the SM predictions for $2\beta_s$ and \DGs, they find
that the probability of a deviation as large as the level of the 
observed data is 7\%.
Note that CDF has very recently made a preliminary update~\cite{CDFnote10778:2012,*CDFnote10778:2012_cont}
to their
\particle{\Bs \to \jpsi\phi} analysis to an
integrated luminosity of 5.2~fb$^{-1}$ indicating a best-fit
confidence interval of:
\begin{equation}
2\beta_s^{\jpsi\phi} = -\phi_s^{\jpsi\phi} 
\in [0.04,1.04] \cup [2.16,3.10]~{\mathrm{at~68\%~CL}},
\end{equation}
where the probability
of a larger deviation from the SM prediction is 44\% or $0.8\,\sigma$.
However, this new result has not yet been used in the combinations
below.

Given the consistency of these two measurements of the weak phase,
as well as their
deviations from the SM, there is interest in combining the results and
using in global fits, \eg\ see \Ref{Bona:2008jn}.
To allow a combination on equal footing, the \dzero collaboration
has redone their fits~\cite{D0web:2009} 
allowing  strong phase values, $\delta_i$, to float
as in the CDF analysis.
Ensemble studies to test confidence level coverage were performed 
by both collaborations and used to adjust likelihood
values to correspond to the usual Gaussian confidence levels. 
Two-dimensional likelihoods were 
combined~\cite{CDFnote9787:2009,D0Note5928:2009}
with the result shown in 
\Fig{DGs_phase}(a).  
After the combination, consistency  
of the best fit values for $\phi_s^{\jpsi\phi} = -2\beta_s^{\jpsi\phi}$ with
SM predictions is at the level of $\hfagNSIGMASM\,\sigma$, with numerical results
for the two solutions given below.
Despite possible biases in the CDF input, point estimates are still
presented and the confidence level regions are straight projections
onto the \DGs or phase angle axes.
\begin{eqnarray}
\DGs &=& \hfagDGSCOMB \,, \\
\phi_s^{\jpsi\phi} = -2\beta_s^{\jpsi\phi} &=& \hfagPHISCOMB \,.
\end{eqnarray}

A comparison between
the above sum of the
CDF and \dzero likelihoods 
and the world average \Bs semileptonic asymmetry of
\Eq{ASLS} through~\cite{Beneke:2003az}:
\begin{equation}
\ASLs = 
\frac{|\Gamma^{12}_s|}{|M^{12}_s|}\sin\phi_s = \frac{\DGs}{\dms}\tan\phi_s
\end{equation}
is also made and shown in 
\Fig{DGs_phase}(a).
Consistency between the two is observed, and the value
of \ASLs is applied as a constraint
resulting in the
confidence level regions 
shown in \Fig{DGs_phase}(b)
including the region delineated by new physics traced by 
the relation of \Eq{new_phys_phase}. Numerical results for the 
two solutions are:
\begin{eqnarray}
\DGs &=& \hfagDGSCOMBCON \,, \\
\phi_s^{\jpsi\phi} = -2\beta_s^{\jpsi\phi} &=& \hfagPHISCOMBCON \,.
\end{eqnarray}
with a consistency
of the best fit values with
SM predictions of $2\beta_s$ at the level of $\hfagNSIGMASMCON\,\sigma$.

}

\clearpage
\mysection{Measurements related to Unitarity Triangle angles
}
\label{sec:cp_uta}

The charge of the ``$\CP(t)$ and Unitarity Triangle angles'' group
is to provide averages of measurements 
from time-dependent asymmetry analyses,
and other quantities that are related 
to the angles of the Unitarity Triangle (UT).
In cases where considerable theoretical input is required to 
extract the fundamental quantities, no attempt is made to do so at 
this stage. However, straightforward interpretations of the averages 
are given, where possible.

In Sec.~\ref{sec:cp_uta:introduction} 
a brief introduction to the relevant phenomenology is given.
In Sec.~\ref{sec:cp_uta:notations}
an attempt is made to clarify the various different notations in use.
In Sec.~\ref{sec:cp_uta:common_inputs}
the common inputs to which experimental results are rescaled in the
averaging procedure are listed. 
We also briefly introduce the treatment of experimental errors. 
In the remainder of this section,
the experimental results and their averages are given,
divided into subsections based on the underlying quark-level decays.

\mysubsection{Introduction
}
\label{sec:cp_uta:introduction}

The Standard Model Cabibbo-Kobayashi-Maskawa (CKM) quark mixing matrix $\VCKM$ 
must be unitary. A $3 \times 3$ unitary matrix has four free parameters,\footnote{
  In the general case there are nine free parameters,
  but five of these are absorbed into unobservable quark phases.}
and these are conventionally written by the product
of three (complex) rotation matrices~\cite{Chau:1984fp}, 
where the rotations are characterised by the Euler angles 
$\theta_{12}$, $\theta_{13}$ and $\theta_{23}$, which are the mixing angles
between the generations, and one overall phase $\delta$,
\begin{equation}
\label{eq:ckmPdg}
\VCKM =
        \left(
          \begin{array}{ccc}
            V_{ud} & V_{us} & V_{ub} \\
            V_{cd} & V_{cs} & V_{cb} \\
            V_{td} & V_{ts} & V_{tb} \\
          \end{array}
        \right)
        =
        \left(
        \begin{array}{ccc}
        c_{12}c_{13}    
                &    s_{12}c_{13}   
                        &   s_{13}e^{-i\delta}  \\
        -s_{12}c_{23}-c_{12}s_{23}s_{13}e^{i\delta} 
                &  c_{12}c_{23}-s_{12}s_{23}s_{13}e^{i\delta} 
                        & s_{23}c_{13} \\
        s_{12}s_{23}-c_{12}c_{23}s_{13}e^{i\delta}  
                &  -c_{12}s_{23}-s_{12}c_{23}s_{13}e^{i\delta} 
                        & c_{23}c_{13} 
        \end{array}
        \right)
\end{equation}
where $c_{ij}=\cos\theta_{ij}$, $s_{ij}=\sin\theta_{ij}$ for 
$i<j=1,2,3$. 

Following the observation of a hierarchy between the different
matrix elements, the Wolfenstein parametrisation~\cite{Wolfenstein:1983yz}
is an expansion of $\VCKM$ in terms of the four real parameters $\lambda$
(the expansion parameter), $A$, $\rho$ and $\eta$. Defining to 
all orders in $\lambda$~\cite{Buras:1994ec}
\begin{eqnarray}
  \label{eq:burasdef}
  s_{12}             &\equiv& \lambda\,,\nonumber \\ 
  s_{23}             &\equiv& A\lambda^2\,, \\
  s_{13}e^{-i\delta} &\equiv& A\lambda^3(\rho -i\eta)\,,\nonumber
\end{eqnarray}
and inserting these into the representation of Eq.~(\ref{eq:ckmPdg}), 
unitarity of the CKM matrix is achieved to all orders.
A Taylor expansion of $\VCKM$ leads to the familiar approximation
\begin{equation}
  \label{eq:cp_uta:ckm}
  \VCKM
  = 
  \left(
    \begin{array}{ccc}
      1 - \lambda^2/2 & \lambda & A \lambda^3 ( \rho - i \eta ) \\
      - \lambda & 1 - \lambda^2/2 & A \lambda^2 \\
      A \lambda^3 ( 1 - \rho - i \eta ) & - A \lambda^2 & 1 \\
    \end{array}
  \right) + {\cal O}\left( \lambda^4 \right) \, .
\end{equation}
At order $\lambda^{5}$, the obtained CKM matrix in this extended
Wolfenstein parametrisation is:
{\small
  \begin{equation}
    \label{eq:cp_uta:ckm_lambda5}
    \VCKM
    =
    \left(
      \begin{array}{ccc}
        1 - \frac{1}{2}\lambda^{2} - \frac{1}{8}\lambda^4 &
        \lambda &
        A \lambda^{3} (\rho - i \eta) \\
        - \lambda + \frac{1}{2} A^2 \lambda^5 \left[ 1 - 2 (\rho + i \eta) \right] &
        1 - \frac{1}{2}\lambda^{2} - \frac{1}{8}\lambda^4 (1+4A^2) &
        A \lambda^{2} \\
        A \lambda^{3} \left[ 1 - (1-\frac{1}{2}\lambda^2)(\rho + i \eta) \right] &
        -A \lambda^{2} + \frac{1}{2}A\lambda^4 \left[ 1 - 2(\rho + i \eta) \right] &
        1 - \frac{1}{2}A^2 \lambda^4
      \end{array} 
    \right) + {\cal O}\left( \lambda^{6} \right)\,.
  \end{equation}
}

\vspace{-5mm}
\noindent
The non-zero imaginary part of the CKM matrix,
which is the origin of $\CP$ violation in the Standard Model,
is encapsulated in a non-zero value of $\eta$.



The unitarity relation $\VCKM^\dagger\VCKM = {\mathit 1}$
results in a total of nine expressions,
that can be written as
$\sum_{i=u,c,t} V_{ij}^*V_{ik} = \delta_{jk}$,
where $\delta_{jk}$ is the Kronecker symbol.
Of the off-diagonal expressions ($j \neq k$),
three can be transformed into the other three 
leaving six relations, in which three complex numbers sum to zero,
which therefore can be expressed as triangles in the complex plane.
More details about unitarity triangles can be found in Refs.~\cite{Jarlskog:1985ht,Jarlskog:2005uq,Bjorken:2005rm,Harrison:2009bz,Frampton:2010ii,Frampton:2010uq}.

\begin{figure}[t]
  \begin{center}
    \resizebox{0.55\textwidth}{!}{\includegraphics{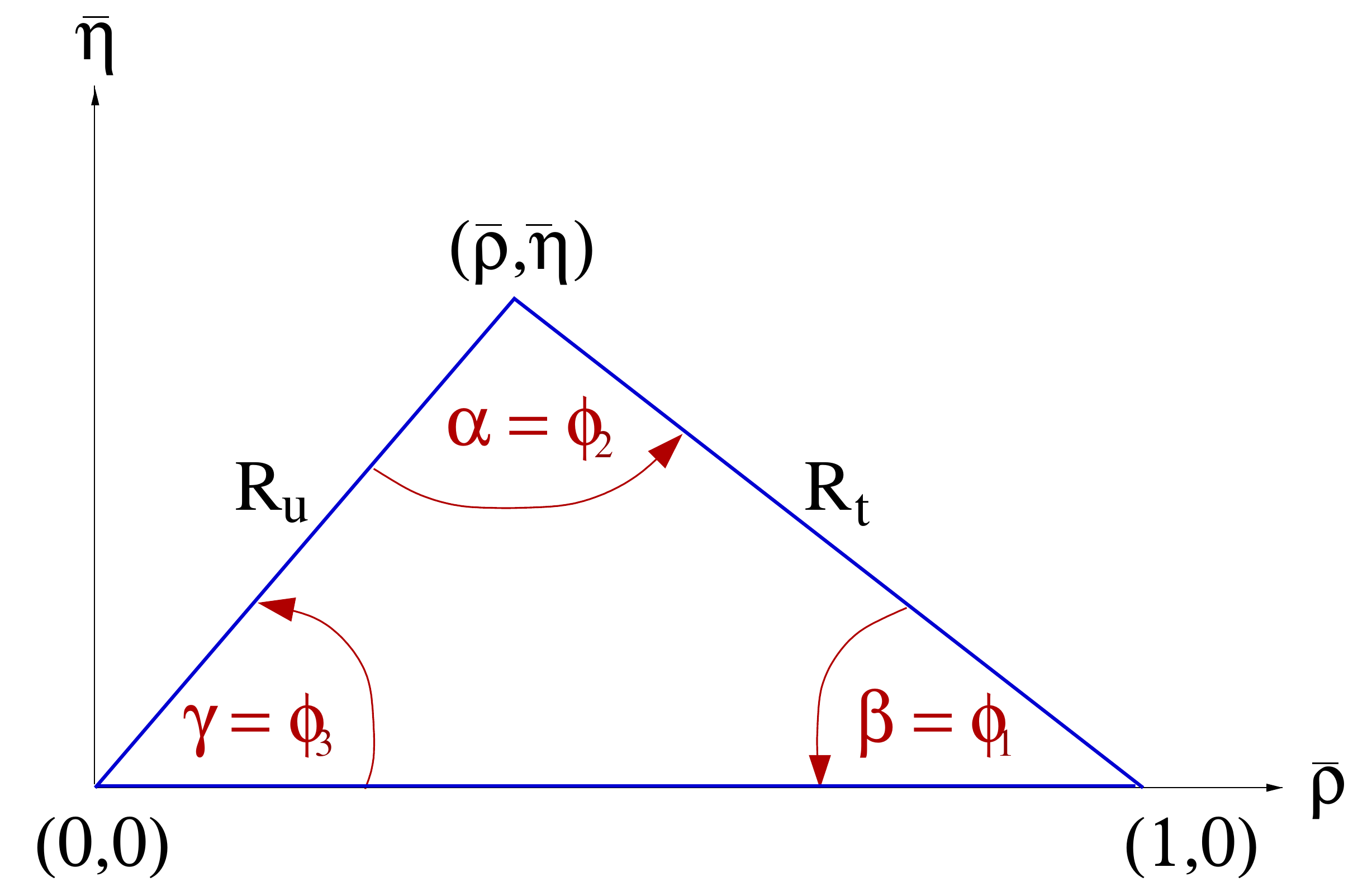}}
    \caption{The Unitarity Triangle.}
    \label{fig:cp_uta:ut}
  \end{center}
\end{figure}

One of these relations,
\begin{equation}
  \label{eq:cp_uta:ut}
  V_{ud}V_{ub}^* + V_{cd}V_{cb}^* + V_{td}V_{tb}^* = 0\,,
\end{equation}
is of particular importance to the $\B$ system, 
being specifically related to flavour changing 
neutral current $b \to d$ transitions.
The three terms in Eq.~(\ref{eq:cp_uta:ut}) are of the same order 
(${\cal O}\left( \lambda^3 \right)$),
and this relation is commonly known as the Unitarity Triangle.
For presentational purposes,
it is convenient to rescale the triangle by $(V_{cd}V_{cb}^*)^{-1}$,
as shown in Fig.~\ref{fig:cp_uta:ut}.

Two popular naming conventions for the UT angles exist in the literature:
\begin{equation}
  \label{eq:cp_uta:abc}
  \alpha  \equiv  \phi_2  = 
  \arg\left[ - \frac{V_{td}V_{tb}^*}{V_{ud}V_{ub}^*} \right]\,,
  \hspace{0.5cm}
  \beta   \equiv   \phi_1 =  
  \arg\left[ - \frac{V_{cd}V_{cb}^*}{V_{td}V_{tb}^*} \right]\,,
  \hspace{0.5cm}
  \gamma  \equiv   \phi_3  =  
  \arg\left[ - \frac{V_{ud}V_{ub}^*}{V_{cd}V_{cb}^*} \right]\,.
\end{equation}
In this document the $\left( \alpha, \beta, \gamma \right)$ set is used.\footnote{
  The relevant unitarity triangle for the $\Bs$ system is obtained 
  by replacing $d \leftrightarrow s$ in Eq.~(\ref{eq:cp_uta:ut}).
  Definitions of the set of angles $( \alpha_s, \beta_s, \gamma_s )$ 
  can be obtained using equivalent relations to those of Eq.~(\ref{eq:cp_uta:abc}),
  for example $\beta_s = \arg\left[ - (V_{cs}V_{cb}^*) / (V_{ts}V_{tb}^*) \right]$.
  This definition gives a value of $\beta_s$ that is negative in the Standard Model,
  so that the sign is often flipped in the literature.
}
The sides $R_u$ and $R_t$ of the Unitarity Triangle 
(the third side being normalised to unity) 
are given by
\begin{equation}
  \label{eq:ru_rt}
  R_u =
  \left|\frac{V_{ud}V_{ub}^*}{V_{cd}V_{cb}^*} \right|
  = \sqrt{\rhobar^2+\etabar^2} \,,
  \hspace{0.5cm}
  R_t = 
  \left|\frac{V_{td}V_{tb}^*}{V_{cd}V_{cb}^*}\right| 
  = \sqrt{(1-\rhobar)^2+\etabar^2} \,.
\end{equation} 
where $\rhobar$ and $\etabar$ define 
the apex of the Unitarity Triangle~\cite{Buras:1994ec} 
\begin{equation}
  \label{eq:rhoetabar}
  \rhobar + i\etabar
  \equiv -\frac{V_{ud}V_{ub}^*}{V_{cd}V_{cb}^*}
  \equiv 1 + \frac{V_{td}V_{tb}^*}{V_{cd}V_{cb}^*}
  = \frac{\sqrt{1-\lambda^2}\,(\rho + i \eta)}{\sqrt{1-A^2\lambda^4}+\sqrt{1-\lambda^2}A^2\lambda^4(\rho+i\eta)} \, .
\end{equation}
The exact relation between $\left( \rho, \eta \right)$ and 
$\left( \rhobar, \etabar \right)$ is
\begin{equation}
  \label{eq:rhoetabarinv}
  \rho + i\eta \;=\; 
  \frac{ 
    \sqrt{ 1-A^2\lambda^4 }(\rhobar+i\etabar) 
  }{
    \sqrt{ 1-\lambda^2 } \left[ 1-A^2\lambda^4(\rhobar+i\etabar) \right]
  } \, .
\end{equation}

By expanding in powers of $\lambda$, several useful approximate expressions
can be obtained, including
\begin{equation}
  \label{eq:rhoeta_approx}
  \rhobar = \rho (1 - \frac{1}{2}\lambda^{2}) + {\cal O}(\lambda^4) \, ,
  \hspace{0.5cm}
  \etabar = \eta (1 - \frac{1}{2}\lambda^{2}) + {\cal O}(\lambda^4) \, ,
  \hspace{0.5cm}
  V_{td} = A \lambda^{3} (1-\rhobar -i\etabar) + {\cal O}(\lambda^6) \, .
\end{equation}

\mysubsection{Notations
}
\label{sec:cp_uta:notations}

Several different notations for $\CP$ violation parameters
are commonly used.
This section reviews those found in the experimental literature,
in the hope of reducing the potential for confusion, 
and to define the frame that is used for the averages.

In some cases, when $\B$ mesons decay into 
multibody final states via broad resonances ($\rho$, $\Kstar$, \etc),
the experimental analyses ignore the effects of interference 
between the overlapping structures.
This is referred to as the quasi-two-body (Q2B) approximation
in the following.

\mysubsubsection{$\CP$ asymmetries
}
\label{sec:cp_uta:notations:pra}

The $\CP$ asymmetry is defined as the difference between the rate 
involving a $b$ quark and that involving a $\bar b$ quark, divided 
by the sum. For example, the partial rate (or charge) asymmetry for 
a charged $\B$ decay would be given as 
\begin{equation}
  \label{eq:cp_uta:pra}
  \Acp_{f} \;\equiv\; 
  \frac{\Gamma(\Bm \to f)-\Gamma(\Bp \to \bar{f})}{\Gamma(\Bm \to f)+\Gamma(\Bp \to \bar{f})}.
\end{equation}

\mysubsubsection{Time-dependent \CP asymmetries in decays to $\CP$ eigenstates
}
\label{sec:cp_uta:notations:cp_eigenstate}

If the amplitudes for $\Bz$ and $\Bzb$ to decay to a final state $f$, 
which is a $\CP$ eigenstate with eigenvalue $\etacpf$,
are given by $\Af$ and $\Abarf$, respectively, 
then the decay distributions for neutral $\B$ mesons, 
with known flavour at time $\Delta t =0$,
are given by
\begin{eqnarray}
  \label{eq:cp_uta:td_cp_asp1}
  \Gamma_{\Bzb \to f} (\Delta t) & = &
  \frac{e^{-| \Delta t | / \tau(\Bz)}}{4\tau(\Bz)}
  \left[ 
    1 +
    \frac{2\, \Im(\lambda_f)}{1 + |\lambda_f|^2} \sin(\Delta m \Delta t) -
    \frac{1 - |\lambda_f|^2}{1 + |\lambda_f|^2} \cos(\Delta m \Delta t)
  \right], \\
  \label{eq:cp_uta:td_cp_asp2}
  \Gamma_{\Bz \to f} (\Delta t) & = &
  \frac{e^{-| \Delta t | / \tau(\Bz)}}{4\tau(\Bz)}
  \left[ 
    1 -
    \frac{2\, \Im(\lambda_f)}{1 + |\lambda_f|^2} \sin(\Delta m \Delta t) +
    \frac{1 - |\lambda_f|^2}{1 + |\lambda_f|^2} \cos(\Delta m \Delta t)
  \right].
\end{eqnarray}
Here $\lambda_f = \frac{q}{p} \frac{\Abarf}{\Af}$ 
contains terms related to $\Bz$\textendash$\Bzb$ mixing and to the decay amplitude
(the eigenstates of the effective Hamiltonian in the $\BzBzb$ system 
are $\left| B_\pm \right> = p \left| \Bz \right> \pm q \left| \Bzb \right>$).
This formulation assumes $\CPT$ invariance, 
and neglects possible lifetime differences 
(between the eigenstates of the effective Hamiltonian;
see Sec.~\ref{sec:mixing} where the mass difference $\Delta m$ is also defined)
in the neutral $\B$ meson system.
The case where non-zero lifetime differences are taken into account is 
discussed in Sec.~\ref{sec:cp_uta:notations:Bs}.
Note that the notation and normalisation used here is that which is relevant for the $e^+e^-$ $B$ factory experiments.
At hadron collider experiments, the flavour tagging is done at production ($\Delta t = t = 0$), and therefore $t$ is usually used in place of $\Delta t$.
Moreover, since negative values of $t$ are not allowed, the normalisation is such that 
$\int_0^{+\infty} \left( 
\Gamma_{\Bzb \to f} (t) + \Gamma_{\Bz \to f} (t) \right) dt = 1$,
rather than 
$\int_{-\infty}^{+\infty} \left( 
\Gamma_{\Bzb \to f} (\Delta t) + \Gamma_{\Bz \to f} (\Delta t) \right) d(\Delta t) = 1$,
as in Eqs.~(\ref{eq:cp_uta:td_cp_asp1}) and~(\ref{eq:cp_uta:td_cp_asp2}).

The time-dependent $\CP$ asymmetry,
again defined as the difference between the rate 
involving a $b$ quark and that involving a $\bar b$ quark,
is then given by
\begin{equation}
  \label{eq:cp_uta:td_cp_asp}
  \Acp_{f} \left(\Delta t\right) \; \equiv \;
  \frac{
    \Gamma_{\Bzb \to f} (\Delta t) - \Gamma_{\Bz \to f} (\Delta t)
  }{
    \Gamma_{\Bzb \to f} (\Delta t) + \Gamma_{\Bz \to f} (\Delta t)
  } \; = \;
  \frac{2\, \Im(\lambda_f)}{1 + |\lambda_f|^2} \sin(\Delta m \Delta t) -
  \frac{1 - |\lambda_f|^2}{1 + |\lambda_f|^2} \cos(\Delta m \Delta t).
\end{equation}

While the coefficient of the $\sin(\Delta m \Delta t)$ term in 
Eq.~(\ref{eq:cp_uta:td_cp_asp}) is everywhere\footnote
{
  Occasionally one also finds Eq.~(\ref{eq:cp_uta:td_cp_asp}) written as
  $\Acp_{f} \left(\Delta t\right) = 
  {\cal A}^{\rm mix}_f \sin(\Delta m \Delta t) + {\cal A}^{\rm dir}_f \cos(\Delta m \Delta t)$,
  or similar.
} denoted $S_f$:
\begin{equation}
  \label{eq:cp_uta:s_def}
  S_f \;\equiv\; \frac{2\, \Im(\lambda_f)}{1 + \left|\lambda_f\right|^2},
\end{equation}
different notations are in use for the
coefficient of the $\cos(\Delta m \Delta t)$ term:
\begin{equation}
  \label{eq:cp_uta:c_def}
  C_f \;\equiv\; - A_f \;\equiv\; \frac{1 - \left|\lambda_f\right|^2}{1 + \left|\lambda_f\right|^2}.
\end{equation}
The $C$ notation is used by the \babar\  collaboration 
(see \eg\ Ref.~\cite{Aubert:2001sp}), 
and also in this document.
The $A$ notation is used by the \belle\ collaboration
(see \eg\ Ref.~\cite{Abe:2001xe}).

Neglecting effects due to $\CP$ violation in mixing 
(by taking $|q/p| = 1$),
if the decay amplitude contains terms with 
a single weak (\ie\ $\CP$ violating) phase
then $\left|\lambda_f\right| = 1$ and one finds
$S_f = -\etacpf \sin(\phi_{\rm mix} + \phi_{\rm dec})$, $C_f = 0$,
where $\phi_{\rm mix}=\arg(q/p)$ and $\phi_{\rm dec}=\arg(\Abarf/\Af)$.
Note that the $\Bz$--$\Bzb$ mixing phase $\phi_{\rm mix}\approx2\beta$
in the Standard Model (in the usual phase convention)~\cite{Carter:1980tk,Bigi:1981qs}. 

If amplitudes with different weak phases contribute to the decay, 
no clean interpretation of $S_f$ is possible without further input. 
If the decay amplitudes have in addition different $\CP$ conserving strong phases, then $\left| \lambda_f \right| \neq 1$ and additional input is required for interpretation.
The coefficient of the cosine term becomes non-zero,
indicating $\CP$ violation in decay.

Due to the fact that $\sin(\Delta m \Delta t)$ and $\cos(\Delta m \Delta t)$
are respectively odd and even functions of $\Delta t$, only small correlations
(that can be induced by backgrounds, for example) between $S_f$ and $C_f$ are
expected at an $e^+e^-$ $B$ factory experiment, where the range of $\Delta t$
is $-\infty < \Delta t < +\infty$. 
The situation is different for measurements at hadron collider experiments, where the range of the time variable is $0 < \Delta t < +\infty$, so that more sizable correlations can be expected.  
We include the correlations in the averages where available.

Frequently, we are interested in combining measurements 
governed by similar or identical short-distance physics,
but with different final states
(\eg, $\Bz \to \jpsi \KS$ and $\Bz \to \jpsi \KL$).
In this case, we remove the dependence on the $\CP$ eigenvalue 
of the final state by quoting $-\etacp S_f$.
In cases where the final state is not a $\CP$ eigenstate but has
an effective $\CP$ content (see below),
the reported $-\etacp S$ is corrected by the effective $\CP$.

\mysubsubsection{Time-dependent distributions with non-zero decay width difference}
\label{sec:cp_uta:notations:Bs}

A complete analysis of the time-dependent decay rates of 
neutral $B$ mesons must also take into account the lifetime difference
between the eigenstates of the effective Hamiltonian, 
denoted by $\Delta \Gamma$.
This is particularly important in the $B_s$ system,
since a non-negligible value of $\Delta \Gamma_s$ has been established
(see Sec.~\ref{sec:mixing} for the latest experimental constraints).
Neglecting $\CP$ violation in mixing,
the relevant replacements for 
Eqs.~(\ref{eq:cp_uta:td_cp_asp1})~and~(\ref{eq:cp_uta:td_cp_asp2}) 
are~\cite{Dunietz:2000cr}
\begin{equation}
  \label{eq:cp_uta:td_cp_bs_asp1}
  \begin{array}{lcr}
    \mc{2}{l}{
      \Gamma_{\Bsbar \to f} (\Delta t) = 
      {\cal N} 
      \frac{e^{-| \Delta t | / \tau(\Bs)}}{4\tau(\Bs)}
      \Big[ 
      \cosh(\frac{\Delta \Gamma \Delta t}{2}) +
    } & \hspace{40mm} \\
    \hspace{40mm} &
    \mc{2}{r}{
      \frac{2\, \Im(\lambda_f)}{1 + |\lambda_f|^2} \sin(\Delta m \Delta t) -
      \frac{1 - |\lambda_f|^2}{1 + |\lambda_f|^2} \cos(\Delta m \Delta t) -
      \frac{2\, \Re(\lambda_f)}{1 + |\lambda_f|^2} \sinh(\frac{\Delta \Gamma \Delta t}{2})
      \Big],
    } \\
  \end{array}
\end{equation}
and
\begin{equation}
  \label{eq:cp_uta:td_cp_bs_asp2}
  \begin{array}{lcr}
    \mc{2}{l}{
      \Gamma_{\Bs \to f} (\Delta t) =
      {\cal N} 
      \frac{e^{-| \Delta t | / \tau(\Bs)}}{4\tau(\Bs)}
      \Big[ 
      \cosh(\frac{\Delta \Gamma \Delta t}{2}) -
    } & \hspace{40mm} \\
    \hspace{40mm} & 
    \mc{2}{r}{
      \frac{2\, \Im(\lambda_f)}{1 + |\lambda_f|^2} \sin(\Delta m \Delta t) +
      \frac{1 - |\lambda_f|^2}{1 + |\lambda_f|^2} \cos(\Delta m \Delta t) -
      \frac{2\, \Re(\lambda_f)}{1 + |\lambda_f|^2} \sinh(\frac{\Delta \Gamma \Delta t}{2})
      \Big]. 
    } \\
  \end{array}
\end{equation}

To be consistent with our earlier notation,\footnote{
  As ever, alternative and conflicting notations appear in the literature.
  One popular alternative notation for this parameter is 
  ${\cal A}_{\Delta \Gamma}$.
  Particular care must be taken over the signs.
}
we write here the coefficient of the $\sinh$ term as
\begin{equation}
  A^{\Delta \Gamma}_f = - \frac{2\, \Re(\lambda_f)}{1 + |\lambda_f|^2} \, .
\end{equation}
A time-dependent analysis of \CP asymmetries in flavour-tagged
$B_s$ decays to a \CP eigenstate $f$ can thus obtain the parameters 
$S_f$, $C_f$ and $A^{\Delta \Gamma}_f$.
Note that, by definition, 
\begin{equation}
  \left( S_f \right)^2 + \left( C_f \right)^2 + \left( A^{\Delta \Gamma}_f \right)^2 = 1 \, ,
\end{equation}
and this constraint can be imposed or not in the fits.
Since these parameters have sensitivity to both
$\Im(\lambda_f)$ and $\Re(\lambda_f)$,
alternative choices of parametrisation, 
including those directly involving \CP violating phases (such as $\beta_s$), 
are possible.
These can also be adopted for vector-vector final states.

The {\it untagged} time-dependent decay rate is given by
\begin{equation}
  \Gamma_{\Bsbar \to f} (\Delta t) + \Gamma_{\Bs \to f} (\Delta t)
  = 
  {\cal N} 
  \frac{e^{-| \Delta t | / \tau(\Bs)}}{2\tau(\Bs)}
  \Big[ 
  \cosh\left(\frac{\Delta \Gamma \Delta t}{2}\right) -
  \frac{2\, \Re(\lambda_f)}{1 + |\lambda_f|^2} \sinh\left(\frac{\Delta \Gamma \Delta t}{2}\right)
  \Big] \, .
\end{equation}
With the requirement
$\int_{-\infty}^{+\infty} \left\{ 
  \Gamma_{\Bsbar \to f} (\Delta t) + \Gamma_{\Bs \to f} (\Delta t)
\right\} d(\Delta t) = 1$,
the normalisation factor ${\cal N}$ 
is fixed to $1 - (\frac{\Delta \Gamma}{2\Gamma})^2$.
Note that an untagged time-dependent analysis can probe
$\lambda_f$, through $\Re(\lambda_f)$, when $\Delta \Gamma \neq 0$.
This is equivalent to determining the ``{\it effective lifetime}''~\cite{Fleischer:2011cw}, as discussed in Sec.~\ref{sec:taubs}.
The tagged analysis is, of course, more sensitive.

Other expressions can be similarly modified to take into account 
non-zero lifetime differences.
Note that when the final state contains 
a mixture of $\CP$-even and $\CP$-odd states
(as, for example, for vector-vector or multibody self-conjugate states),
that $\Re(\lambda_f)$ contains terms proportional to 
both the sine and cosine of the weak phase difference, 
albeit with rather different sensitivities.

\mysubsubsection{Time-dependent \CP asymmetries in decays to vector-vector final states
}
\label{sec:cp_uta:notations:vv}

Consider \B decays to states consisting of two spin-1 particles,
such as $\jpsi K^{*0}(\to\KS\piz)$, $\jpsi\phi$, $D^{*+}D^{*-}$ and $\rho^+\rho^-$,
which are eigenstates of charge conjugation but not of parity.\footnote{
  \noindent
  This is not true of all vector-vector final states,
  \eg, $D^{*\pm}\rho^{\mp}$ is clearly not an eigenstate of 
  charge conjugation.
}
In fact, for such a system, there are three possible final states;
in the helicity basis these can be written $h_{-1}, h_0, h_{+1}$.
The $h_0$ state is an eigenstate of parity, and hence of $\CP$;
however, $\CP$ transforms $h_{+1} \leftrightarrow h_{-1}$ (up to 
an unobservable phase). In the transversity basis, these states 
are transformed into  $h_\parallel =  (h_{+1} + h_{-1})/2$ and 
$h_\perp = (h_{+1} - h_{-1})/2$.
In this basis all three states are $\CP$ eigenstates, 
and $h_\perp$ has the opposite $\CP$ to the others.

The amplitudes to these states are usually given by $A_{0,\perp,\parallel}$
(here we use normalisation such that 
$| A_0 |^2 + | A_\perp |^2 + | A_\parallel |^2 = 1$).
Then the effective $\CP$ of the vector-vector state is known if 
$| A_\perp |^2$ is measured.
An alternative strategy is to measure just the longitudinally polarised 
component,  $| A_0 |^2$
(sometimes denoted by $f_{\rm long}$), 
which allows a limit to be set on the effective $\CP$ since
$| A_\perp |^2 \leq | A_\perp |^2 + | A_\parallel |^2 = 1 - | A_0 |^2$.
The most complete treatment for 
neutral $\B$ decays to vector-vector final states
is time-dependent angular analysis 
(also known as time-dependent transversity analysis).
In such an analysis, 
the interference between the $\CP$-even and $\CP$-odd states 
provides additional sensitivity to the weak and strong phases involved.

In most analyses of time-dependent \CP asymmetries in decays to 
vector-vector final states carried out to date,
an assumption has been made that each helicity (or transversity) amplitude
has the same weak phase.
This is a good approximation for decays that are dominated by 
amplitudes with a single weak phase, such $\Bz \to \jpsi K^{*0}$,
and is a reasonable approximation in any mode for which only 
very limited statistics are available.
However, for modes that have contributions from amplitudes with different 
weak phases, the relative size of these contributions can be different 
for each helicity (or transversity) amplitude,
and therefore the time-dependent \CP asymmetry parameters can also differ.
The most generic analysis, suitable for modes with sufficient statistics,
would allow for this effect;
an intermediate analysis can allow different parameters for the 
$\CP$-even and $\CP$-odd components.
Such an analysis has been carried out by \babar\ for the decay
$\Bz \to D^{*+}D^{*-}$~\cite{Aubert:2008ah}.
The independent treatment of each helicity (or transversity) amplitude, as in the latest result on $\Bs \to \jpsi\phi$ (discussed in Sec.~\ref{sec:life_mix}), becomes increasingly important for high precision measurements.

\mysubsubsection{Time-dependent asymmetries: self-conjugate multiparticle final states
}
\label{sec:cp_uta:notations:dalitz}

Amplitudes for neutral \B decays into 
self-conjugate multiparticle final states
such as $\pi^+\pi^-\pi^0$, $K^+K^-\KS$, $\pi^+\pi^-\KS$,
$\jpsi \pi^+\pi^-$ or $D\pi^0$ with $D \to \KS\pi^+\pi^-$
may be written in terms of \CP-even and \CP-odd amplitudes.
As above, the interference between these terms 
provides additional sensitivity to the weak and strong phases
involved in the decay,
and the time-dependence depends on both the sine and cosine
of the weak phase difference.
In order to perform unbinned maximum likelihood fits,
and thereby extract as much information as possible from the distributions,
it is necessary to select a model for the multiparticle decay,
and therefore the results acquire some model dependence
(binned, model independent methods are also possible,
though are not as statistically powerful).
The number of observables depends on the final state (and on the model used);
the key feature is that as long as there are regions where both
\CP-even and \CP-odd amplitudes contribute,
the interference terms will be sensitive to the cosine 
of the weak phase difference.
Therefore, these measurements allow distinction between multiple solutions
for, \eg, the four values of $\beta$ from the measurement of $\sin(2\beta)$.

We now consider the various notations which have been used 
in experimental studies of
time-dependent asymmetries in decays to self-conjugate multiparticle final states.

\newpage 
\mysubsubsubsection{$\Bz \to D^{(*)}h^0$ with $D \to \KS\pi^+\pi^-$
}
\label{sec:cp_uta:notations:dalitz:dh0}

The states $D\pi^0$, $D^*\pi^0$, $D\eta$, $D^*\eta$, $D\omega$
are collectively denoted $D^{(*)}h^0$.
When the $D$ decay model is fixed,
fits to the time-dependent decay distributions can be performed
to extract the weak phase difference.
However, it is experimentally advantageous to use the sine and cosine of 
this phase as fit parameters, since these behave as essentially 
independent parameters, with low correlations and (potentially)
rather different uncertainties.
A parameter representing $\CP$ violation in the $B$ decay 
can also be floated.  
For consistency with other analyses, this could be chosen to be $C_f$,
but could equally well be $\left| \lambda_f \right|$, or other possibilities.

\belle\ performed an analysis of these channels
with $\sin(2\phi_1)$ and $\cos(2\phi_1)$ as free parameters~\cite{Krokovny:2006sv}.
\babar\ has performed an analysis floating also $\left| \lambda_f \right|$~\cite{Aubert:2007rp}
(and, of course, replacing $\phi_1$ with $\beta$).

\mysubsubsubsection{$\Bz \to D^{*+}D^{*-}\KS$
}
\label{sec:cp_uta:notations:dalitz:dstardstarks}

The hadronic structure of the $\Bz \to D^{*+}D^{*-}\KS$ decay
is not sufficiently well understood to perform a full 
time-dependent Dalitz plot analysis.
Instead, following Ref.~\cite{Browder:1999ng},
\babar~\cite{Aubert:2006fh} and \belle~\cite{Dalseno:2007hx} divide the Dalitz plane in two:
$m(D^{*+}\KS)^2 > m(D^{*-}\KS)^2$ $(\eta_y = +1)$ and 
$m(D^{*+}\KS)^2 < m(D^{*-}\KS)^2$ $(\eta_y = -1)$;
and then fit to a decay time distribution with asymmetry given by
\begin{equation}
  \Acp_{f} \left(\Delta t\right) =
  \eta_y \frac{J_c}{J_0} \cos(\Delta m \Delta t) -  
  \left[ 
    \frac{2J_{s1}}{J_0} \sin(2\beta) + \eta_y \frac{2J_{s2}}{J_0} \cos(2\beta) 
  \right] \sin(\Delta m \Delta t) \, .
\end{equation}
The measured values are $\frac{J_c}{J_0}$, $\frac{2J_{s1}}{J_0} \sin(2\beta)$
and $\frac{2J_{s2}}{J_0} \cos(2\beta)$, 
where the parameters $J_0$, $J_c$, $J_{s1}$ and $J_{s2}$ are the integrals 
over the half Dalitz plane $m(D^{*+}\KS)^2 < m(D^{*-}\KS)^2$ 
of the functions $|a|^2 + |\bar{a}|^2$, $|a|^2 - |\bar{a}|^2$, 
$\Re(\bar{a}a^*)$ and $\Im(\bar{a}a^*)$ respectively, 
where $a$ and $\bar{a}$ are the decay amplitudes of 
$\Bz \to D^{*+}D^{*-}\KS$ and $\Bzb \to D^{*+}D^{*-}\KS$ respectively. 
The parameter $J_{s2}$ (and hence $J_{s2}/J_0$) is predicted to be positive;
with this assumption it is possible to determine the sign of $\cos(2\beta)$.

\mysubsubsubsection{$\Bz \to K^+K^-\Kz$
}
\label{sec:cp_uta:notations:dalitz:kkk0}

Studies of $\Bz \to K^+K^-\Kz$~\cite{Aubert:2007sd,Nakahama:2010nj,Lees:2012kx} 
and of the related decay 
$\Bp \to K^+K^-K^+$~\cite{Garmash:2004wa,Aubert:2006nu,Lees:2012kx},
show that the decay is dominated by a large nonresonant contribution
with significant components from the 
intermediate $K^+K^-$ resonances $\phi(1020)$, $f_0(980)$,
and other higher resonances,\footnote{
  The broad structure that peaks near 
  $m(K^+K^-) \sim 1550 \ {\rm MeV}/c^2$ and was denoted $X_0(1550)$ 
  is now believed to originate from interference effects.
}
as well a contribution from $\chi_{c0}$.

The full time-dependent Dalitz plot analysis allows 
the complex amplitudes of each contributing term to be determined from data,
including $\CP$ violation effects
(\ie\ allowing the complex amplitude for the $\Bz$ decay to be independent
from that for $\Bzb$ decay), although one amplitude must be fixed 
to give a reference point.
There are several choices for parametrisation of the complex amplitudes 
(\eg\ real and imaginary part, or magnitude and phase).
Similarly, there are various approaches to include $\CP$ violation effects.
Note that positive definite parameters such as magnitudes are
disfavoured in certain circumstances 
(they inevitably lead to biases for small values).
In order to compare results between analyses,
it is useful for each experiment to present results in terms of the 
parameters that can be measured in a Q2B analysis
(such as $\Acp_{f}$, $S_f$, $C_f$, 
$\sin(2\beta^{\rm eff})$, $\cos(2\beta^{\rm eff})$, \etc)

In the \babar\ analysis of $\Bz \to K^+K^-\Kz$~\cite{Lees:2012kx},
the complex amplitude for each resonant contribution is written as
\begin{equation}
  A_f = c_f ( 1 + b_f ) e^{i ( \phi_f + \delta_f )} 
  \ , \ \ \ \ 
  \bar{A}_f = c_f ( 1 - b_f ) e^{i ( \phi_f - \delta_f )} \, ,
\end{equation}
where $b_f$ and $\delta_f$ introduce $\CP$ violation in the magnitude 
and phase respectively.
Belle~\cite{Nakahama:2010nj} use the same parametrisation but with a different notation for the parameters.\footnote{
  $(c, b, \phi, \delta) \leftrightarrow (a, c, b, d)$.
}
[The weak phase in $B^0$--$\bar{B}^0$ mixing ($2\beta$) also appears 
in the full formula for the time-dependent decay distribution.]
The Q2B parameter of $\CP$ violation in decay is directly related to $b_f$
\begin{equation}
  \Acp_{f} = \frac{-2b_f}{1+b_f^2} \approx C_f \, ,
\end{equation}
and the mixing-induced $\CP$ violation parameter can be used to obtain
$\sin(2\beta^{\rm eff})$
\begin{equation}
  -\eta_f S_f \approx \frac{1-b_f^2}{1+b_f^2}\sin(2\beta^{\rm eff}_f) \, ,
\end{equation}
where the approximations are exact in the case that $\left| q/p \right| = 1$.

Both \babar~\cite{Lees:2012kx} and \belle~\cite{Nakahama:2010nj} present results for $c_f$ and $\phi_f$,
for each resonant contribution,
and in addition present results for $\Acp_{f}$ and $\beta^{\rm eff}_{f}$ for $\phi(1020) \Kz$, $f_0(980) \Kz$ and for the remainder of the contributions to the $K^+K^-\Kz$ Dalitz plot combined.\footnote{
  \babar also present results for the Q2B parameter $S_{f}$ for these channels.
}
The models used to describe the resonant structure of the Dalitz plot differ, however.  Both analyses suffer from multiple solutions, from which we select only one for averaging.

\mysubsubsubsection{$\Bz \to \pi^+\pi^-\KS$
}
\label{sec:cp_uta:notations:dalitz:pipik0}

Studies of $\Bz \to \pi^+\pi^-\KS$~\cite{Aubert:2009me,:2008wwa}
and of the related decay
$\Bp \to \pi^+\pi^-K^+$~\cite{Garmash:2004wa,Garmash:2005rv,Aubert:2005ce,Aubert:2008bj}
show that the decay is dominated by components from intermediate resonances 
in the $K\pi$ ($K^*(892)$, $K^*_0(1430)$) 
and $\pi\pi$ ($\rho(770)$, $f_0(980)$, $f_2(1270)$) spectra,
together with a poorly understood scalar structure that peaks near 
$m(\pi\pi) \sim 1300 \ {\rm MeV}/c^2$ and is denoted $f_X(1300)$
(that could be identified as either the $f_0(1370)$ or $f_0(1500)$),
and a large nonresonant component.
There is also a contribution from the $\chi_{c0}$ state.

The full time-dependent Dalitz plot analysis allows 
the complex amplitudes of each contributing term to be determined from data,
including $\CP$ violation effects.
In the \babar\ analysis~\cite{Aubert:2009me}, 
the magnitude and phase of each component (for both $\Bz$ and $\Bzb$ decays) 
are measured relative to $\Bz \to f_0(980)\KS$, using the following
parametrisation
\begin{equation}
  A_f = \left| A_f \right| e^{i\,{\rm arg}(A_f)}
  \ , \ \ \ \ 
  \bar{A}_f = \left| \bar{A}_f \right| e^{i\,{\rm arg}(\bar{A}_f)} \, .
\end{equation}
In the \belle\ analysis~\cite{:2008wwa}, the $\Bz \to K^{*+}\pi^-$ amplitude
is chosen as the reference, and the amplitudes are parametrised as 
\begin{equation}
  A_f = a_f ( 1 + c_f ) e^{i ( b_f + d_f )} 
  \ , \ \ \ \ 
  \bar{A}_f = a_f ( 1 - c_f ) e^{i ( b_f - d_f )} \, .
\end{equation}
In both cases, the results are translated into Q2B parameters 
such as $2\beta^{\rm eff}_f$, $S_f$, $C_f$ for each \CP\ eigenstate $f$,
and parameters of \CP\ violation in decay for each flavour-specific state.
Relative phase differences between resonant terms are also extracted.

\mysubsubsubsection{$\Bz \to \pi^+\pi^-\pi^0$
}
\label{sec:cp_uta:notations:dalitz:pipipi0}

The $\Bz \to \pi^+\pi^-\pi^0$ decay is dominated by 
intermediate $\rho$ resonances.
Though it is possible, as above, 
to determine directly the complex amplitudes for each component,
an alternative approach~\cite{Snyder:1993mx,Quinn:2000by}
has been used by both \babar~\cite{Aubert:2007jn,Lees:2013nwa}
and \belle~\cite{Kusaka:2007dv,:2007mj}.
The amplitudes for $\Bz$ and $\Bzb$ to $\pi^+\pi^-\pi^0$ are written
\begin{equation}
  A_{3\pi} = f_+ A_+ + f_- A_- + f_0 A_0
  \ , \ \ \ 
  \bar{A}_{3\pi} = f_+ \bar{A}_+ + f_- \bar{A}_- + f_0 \bar{A}_0
\end{equation}
respectively.
The symbols $A_+$, $A_-$ and $A_0$
represent the complex decay amplitudes for 
$\Bz \to \rho^+\pi^-$, $\Bz \to \rho^-\pi^+$ and $\Bz \to \rho^0\pi^0$
while 
$\bar{A}_+$, $\bar{A}_-$ and $\bar{A}_0$
represent those for 
$\Bzb \to \rho^+\pi^-$, $\Bzb \to \rho^-\pi^+$ and $\Bzb \to \rho^0\pi^0$
respectively.
The terms $f_+$, $f_-$ and $f_0$ incorporate kinematic and dynamical factors
and depend on the Dalitz plot coordinates.
The full time-dependent decay distribution can then be written 
in terms of 27 free parameters,
one for each coefficient of the form factor bilinears,
as listed in Table~\ref{tab:cp_uta:pipipi0:uandi}.
These parameters are sometimes referred to as ``the $U$s and $I$s'',
and can be expressed in terms of 
$A_+$, $A_-$, $A_0$, $\bar{A}_+$, $\bar{A}_-$ and $\bar{A}_0$.
If the full set of parameters is determined,
together with their correlations,
other parameters, such as weak and strong phases,
parameters of $\CP$ violation in decay, \etc, 
can be subsequently extracted.
Note that one of the parameters (typically $U_+^+$)
is often fixed to unity to provide a reference point;
this does not affect the analysis.


\begin{table}[htbp]
  \begin{center}
    \caption{
      Definitions of the $U$ and $I$ coefficients.
      Modified from Ref.~\cite{Aubert:2007jn}.
    }
    \label{tab:cp_uta:pipipi0:uandi}
    \setlength{\tabcolsep}{0.3pc}
    \begin{tabular}{l@{\extracolsep{5mm}}l}
      \hline
      Parameter   & Description \\
      \hline
      $U_+^+$          & Coefficient of $|f_+|^2$ \\
      $U_0^+$          & Coefficient of $|f_0|^2$ \\
      $U_-^+$          & Coefficient of $|f_-|^2$ \\
      [0.15cm]
      $U_0^-$          & Coefficient of $|f_0|^2\cos(\Delta m\Delta t)$ \\
      $U_-^-$          & Coefficient of $|f_-|^2\cos(\Delta m\Delta t)$ \\
      $U_+^-$          & Coefficient of $|f_+|^2\cos(\Delta m\Delta t)$ \\
      [0.15cm]
      $I_0$            & Coefficient of $|f_0|^2\sin(\Delta m\Delta t)$ \\
      $I_-$            & Coefficient of $|f_-|^2\sin(\Delta m\Delta t)$ \\
      $I_+$            & Coefficient of $|f_+|^2\sin(\Delta m\Delta t)$ \\
      [0.15cm]
      $U_{+-}^{+,\Im}$ & Coefficient of $\Im[f_+f_-^*]$ \\
      $U_{+-}^{+,\Re}$ & Coefficient of $\Re[f_+f_-^*]$ \\
      $U_{+-}^{-,\Im}$ & Coefficient of $\Im[f_+f_-^*]\cos(\Delta m\Delta t)$ \\
      $U_{+-}^{-,\Re}$ & Coefficient of $\Re[f_+f_-^*]\cos(\Delta m\Delta t)$ \\
      $I_{+-}^{\Im}$   & Coefficient of $\Im[f_+f_-^*]\sin(\Delta m\Delta t)$ \\
      $I_{+-}^{\Re}$   & Coefficient of $\Re[f_+f_-^*]\sin(\Delta m\Delta t)$ \\
      [0.15cm]
      $U_{+0}^{+,\Im}$ & Coefficient of $\Im[f_+f_0^*]$ \\
      $U_{+0}^{+,\Re}$ & Coefficient of $\Re[f_+f_0^*]$ \\
      $U_{+0}^{-,\Im}$ & Coefficient of $\Im[f_+f_0^*]\cos(\Delta m\Delta t)$ \\
      $U_{+0}^{-,\Re}$ & Coefficient of $\Re[f_+f_0^*]\cos(\Delta m\Delta t)$ \\
      $I_{+0}^{\Im}$   & Coefficient of $\Im[f_+f_0^*]\sin(\Delta m\Delta t)$ \\
      $I_{+0}^{\Re}$   & Coefficient of $\Re[f_+f_0^*]\sin(\Delta m\Delta t)$ \\
      [0.15cm]
      $U_{-0}^{+,\Im}$ & Coefficient of $\Im[f_-f_0^*]$ \\
      $U_{-0}^{+,\Re}$ & Coefficient of $\Re[f_-f_0^*]$ \\
      $U_{-0}^{-,\Im}$ & Coefficient of $\Im[f_-f_0^*]\cos(\Delta m\Delta t)$ \\
      $U_{-0}^{-,\Re}$ & Coefficient of $\Re[f_-f_0^*]\cos(\Delta m\Delta t)$ \\
      $I_{-0}^{\Im}$   & Coefficient of $\Im[f_-f_0^*]\sin(\Delta m\Delta t)$ \\
      $I_{-0}^{\Re}$   & Coefficient of $\Re[f_-f_0^*]\sin(\Delta m\Delta t)$ \\     
      \hline
    \end{tabular}
  \end{center}
\end{table}

\mysubsubsection{Time-dependent \CP asymmetries in decays to non-$\CP$ eigenstates
}
\label{sec:cp_uta:notations:non_cp}

Consider a non-$\CP$ eigenstate $f$, and its conjugate $\bar{f}$. 
For neutral $\B$ decays to these final states,
there are four amplitudes to consider:
those for $\Bz$ to decay to $f$ and $\bar{f}$
($\Af$ and $\Afbar$, respectively),
and the equivalents for $\Bzb$
($\Abarf$ and $\Abarfbar$).
If $\CP$ is conserved in the decay, then
$\Af = \Abarfbar$ and $\Afbar = \Abarf$.


The time-dependent decay distributions can be written in many different ways.
Here, we follow Sec.~\ref{sec:cp_uta:notations:cp_eigenstate}
and define $\lambda_f = \frac{q}{p}\frac{\Abarf}{\Af}$ and
$\lambda_{\bar f} = \frac{q}{p}\frac{\Abarfbar}{\Afbar}$.
The time-dependent \CP asymmetries that are sensitive to mixing-induced
$\CP$ violation effects then follow Eq.~(\ref{eq:cp_uta:td_cp_asp}):
\begin{eqnarray}
\label{eq:cp_uta:non-cp-obs}
  {\cal A}_f (\Delta t) \; \equiv \;
  \frac{
    \Gamma_{\Bzb \to f} (\Delta t) - \Gamma_{\Bz \to f} (\Delta t)
  }{
    \Gamma_{\Bzb \to f} (\Delta t) + \Gamma_{\Bz \to f} (\Delta t)
  } & = & S_f \sin(\Delta m \Delta t) - C_f \cos(\Delta m \Delta t), \\
  {\cal A}_{\bar{f}} (\Delta t) \; \equiv \;
  \frac{
    \Gamma_{\Bzb \to \bar{f}} (\Delta t) - \Gamma_{\Bz \to \bar{f}} (\Delta t)
  }{
    \Gamma_{\Bzb \to \bar{f}} (\Delta t) + \Gamma_{\Bz \to \bar{f}} (\Delta t)
  } & = & S_{\bar{f}} \sin(\Delta m \Delta t) - C_{\bar{f}} \cos(\Delta m \Delta t),
\end{eqnarray}
with the definitions of the parameters 
$C_f$, $S_f$, $C_{\bar{f}}$ and $S_{\bar{f}}$,
following Eqs.~(\ref{eq:cp_uta:s_def}) and~(\ref{eq:cp_uta:c_def}).

The time-dependent decay rates are given by
\begin{eqnarray}
  \label{eq:cp_uta:non-CP-TD1}
  \Gamma_{\Bzb \to f} (\Delta t) & = &
  \frac{e^{-\left| \Delta t \right| / \tau(\Bz)}}{8\tau(\Bz)} 
  ( 1 + \Adirnoncp ) 
  \left\{ 
    1 + S_f \sin(\Delta m \Delta t) - C_f \cos(\Delta m \Delta t) 
  \right\},
  \\
  \label{eq:cp_uta:non-CP-TD2}
  \Gamma_{\Bz \to f} (\Delta t) & = &
  \frac{e^{-\left| \Delta t \right| / \tau(\Bz)}}{8\tau(\Bz)} 
  ( 1 + \Adirnoncp ) 
  \left\{ 
    1 - S_f \sin(\Delta m \Delta t) + C_f \cos(\Delta m \Delta t) 
  \right\},
  \\
  \label{eq:cp_uta:non-CP-TD3}
  \Gamma_{\Bzb \to \bar{f}} (\Delta t) & = &
  \frac{e^{-\left| \Delta t \right| / \tau(\Bz)}}{8\tau(\Bz)} 
  ( 1 - \Adirnoncp ) 
  \left\{ 
    1 + S_{\bar{f}} \sin(\Delta m \Delta t) - C_{\bar{f}} \cos(\Delta m \Delta t) 
  \right\},
  \\
  \label{eq:cp_uta:non-CP-TD4}
  \Gamma_{\Bz \to \bar{f}} (\Delta t) & = &
    \frac{e^{-\left| \Delta t \right| / \tau(\Bz)}}{8\tau(\Bz)} 
  ( 1 - \Adirnoncp ) 
  \left\{ 
    1 - S_{\bar{f}} \sin(\Delta m \Delta t) + C_{\bar{f}} \cos(\Delta m \Delta t) 
  \right\},
\end{eqnarray}
where the time-independent parameter \Adirnoncp
represents an overall asymmetry in the production of the 
$f$ and $\bar{f}$ final states,\footnote{
  This parameter is often denoted ${\cal A}_f$ (or ${\cal A}_{\CP}$),
  but here we avoid this notation to prevent confusion with the
  time-dependent $\CP$ asymmetry.
}
\begin{equation}
  \Adirnoncp = 
  \frac{
    \left( 
      \left| \Af \right|^2 + \left| \Abarf \right|^2
    \right) - 
    \left( 
      \left| \Afbar \right|^2 + \left| \Abarfbar \right|^2
    \right)
  }{
    \left( 
      \left| \Af \right|^2 + \left| \Abarf \right|^2
    \right) +
    \left( 
      \left| \Afbar \right|^2 + \left| \Abarfbar \right|^2
    \right)
  }.
\end{equation}
Assuming $|q/p| = 1$,
the parameters $C_f$ and $C_{\bar{f}}$
can also be written in terms of the decay amplitudes as follows:
\begin{equation}
  C_f = 
  \frac{
    \left| \Af \right|^2 - \left| \Abarf \right|^2 
  }{
    \left| \Af \right|^2 + \left| \Abarf \right|^2
  }
  \hspace{5mm}
  {\rm and}
  \hspace{5mm}
  C_{\bar{f}} = 
  \frac{
    \left| \Afbar \right|^2 - \left| \Abarfbar \right|^2
  }{
    \left| \Afbar \right|^2 + \left| \Abarfbar \right|^2
  },
\end{equation}
giving asymmetries in the decay amplitudes of $\Bz$ and $\Bzb$
to the final states $f$ and $\bar{f}$ respectively.
In this notation, the conditions for absence of $\CP$ violation in decay are
$\Adirnoncp = 0$ and $C_f = - C_{\bar{f}}$.
Note that $C_f$ and $C_{\bar{f}}$ are typically non-zero;
\eg, for a flavour-specific final state, 
$\Abarf = \Afbar = 0$ ($\Af = \Abarfbar = 0$), they take the values
$C_f = - C_{\bar{f}} = 1$ ($C_f = - C_{\bar{f}} = -1$).

The coefficients of the sine terms
contain information about the weak phase. 
In the case that each decay amplitude contains only a single weak phase
(\ie, no $\CP$ violation in decay),
these terms can be written
\begin{equation}
  S_f = 
  \frac{ 
    - 2 \left| \Af \right| \left| \Abarf \right| 
    \sin( \phi_{\rm mix} + \phi_{\rm dec} - \delta_f )
  }{
    \left| \Af \right|^2 + \left| \Abarf \right|^2
  } 
  \hspace{5mm}
  {\rm and}
  \hspace{5mm}
  S_{\bar{f}} = 
  \frac{
    - 2 \left| \Afbar \right| \left| \Abarfbar \right| 
    \sin( \phi_{\rm mix} + \phi_{\rm dec} + \delta_f )
  }{
    \left| \Afbar \right|^2 + \left| \Abarfbar \right|^2
  },
\end{equation}
where $\delta_f$ is the strong phase difference between the decay amplitudes.
If there is no $\CP$ violation, the condition $S_f = - S_{\bar{f}}$ holds.
If decay amplitudes with different weak and strong phases contribute,
no clean interpretation of $S_f$ and $S_{\bar{f}}$ is possible.

Since two of the $\CP$ invariance conditions are 
$C_f = - C_{\bar{f}}$ and $S_f = - S_{\bar{f}}$,
there is motivation for a rotation of the parameters:
\begin{equation}
\label{eq:cp_uta:non-cp-s_and_deltas}
  S_{f\bar{f}} = \frac{S_{f} + S_{\bar{f}}}{2},
  \hspace{4mm}
  \Delta S_{f\bar{f}} = \frac{S_{f} - S_{\bar{f}}}{2},
  \hspace{4mm}
  C_{f\bar{f}} = \frac{C_{f} + C_{\bar{f}}}{2},
  \hspace{4mm}
  \Delta C_{f\bar{f}} = \frac{C_{f} - C_{\bar{f}}}{2}.
\end{equation}
With these parameters, the $\CP$ invariance conditions become
$S_{f\bar{f}} = 0$ and $C_{f\bar{f}} = 0$. 
The parameter $\Delta C_{f\bar{f}}$ gives a measure of the ``flavour-specificity''
of the decay:
$\Delta C_{f\bar{f}}=\pm1$ corresponds to a completely flavour-specific decay,
in which no interference between decays with and without mixing can occur,
while $\Delta C_{f\bar{f}} = 0$ results in 
maximum sensitivity to mixing-induced $\CP$ violation.
The parameter $\Delta S_{f\bar{f}}$ is related to the strong phase difference 
between the decay amplitudes of $\Bz$ to $f$ and to $\bar f$. 
We note that the observables of Eq.~(\ref{eq:cp_uta:non-cp-s_and_deltas})
exhibit experimental correlations 
(typically of $\sim 20\%$, depending on the tagging purity, and other effects)
between $S_{f\bar{f}}$ and  $\Delta S_{f\bar{f}}$, 
and between $C_{f\bar{f}}$ and $\Delta C_{f\bar{f}}$. 
On the other hand, 
the final state specific observables of Eqs.~(\ref{eq:cp_uta:non-CP-TD1})--(\ref{eq:cp_uta:non-CP-TD4}) tend to have low correlations.

Alternatively, if we recall that the $\CP$ invariance
conditions at the decay amplitude level are
$\Af = \Abarfbar$ and $\Afbar = \Abarf$,
we are led to consider the parameters~\cite{Charles:2004jd}
\begin{equation}
  \label{eq:cp_uta:non-cp-directcp}
  {\cal A}_{f\bar{f}} = 
  \frac{
    \left| \Abarfbar \right|^2 - \left| \Af \right|^2 
  }{
    \left| \Abarfbar \right|^2 + \left| \Af \right|^2
  }
  \hspace{5mm}
  {\rm and}
  \hspace{5mm}
  {\cal A}_{\bar{f}f} = 
  \frac{
    \left| \Abarf \right|^2 - \left| \Afbar \right|^2
  }{
    \left| \Abarf \right|^2 + \left| \Afbar \right|^2
  }.
\end{equation}
These are sometimes considered more physically intuitive parameters
since they characterise $\CP$ violation in decay
in decays with particular topologies.
For example, in the case of $\Bz \to \rho^\pm\pi^\mp$
(choosing $f =  \rho^+\pi^-$ and $\bar{f} = \rho^-\pi^+$),
${\cal A}_{f\bar{f}}$ (also denoted ${\cal A}^{+-}_{\rho\pi}$)
parametrises $\CP$ violation
in decays in which the produced $\rho$ meson does not contain the 
spectator quark,
while ${\cal A}_{\bar{f}f}$ (also denoted ${\cal A}^{-+}_{\rho\pi}$)
parametrises $\CP$ violation in decays in which it does.
Note that we have again followed the sign convention that the asymmetry 
is the difference between the rate involving a $b$ quark and that
involving a $\bar{b}$ quark, \cf\ Eq.~(\ref{eq:cp_uta:pra}). 
Of course, these parameters are not independent of the 
other sets of parameters given above, and can be written
\begin{equation}
  {\cal A}_{f\bar{f}} =
  - \frac{
    \Adirnoncp + C_{f\bar{f}} + \Adirnoncp \Delta C_{f\bar{f}} 
  }{
    1 + \Delta C_{f\bar{f}} + \Adirnoncp C_{f\bar{f}} 
  }
  \hspace{5mm}
  {\rm and}
  \hspace{5mm}
  {\cal A}_{\bar{f}f} =
  \frac{
    - \Adirnoncp + C_{f\bar{f}} + \Adirnoncp \Delta C_{f\bar{f}} 
  }{
    - 1 + \Delta C_{f\bar{f}} + \Adirnoncp C_{f\bar{f}}  
  }.
\end{equation}
They usually exhibit strong correlations.

We now consider the various notations which have been used 
in experimental studies of
time-dependent $\CP$ asymmetries in decays to non-$\CP$ eigenstates.

\mysubsubsubsection{$\Bz \to D^{*\pm}D^\mp$
}
\label{sec:cp_uta:notations:non_cp:dstard}

The ($\Adirnoncp$, $C_f$, $S_f$, $C_{\bar{f}}$, $S_{\bar{f}}$)
set of parameters was used in early publications by both \babar~\cite{Aubert:2007pa} and \belle~\cite{Aushev:2004uc} (albeit with slightly different notations) in the $D^{*\pm}D^{\mp}$ system ($f = D^{*+}D^-$, $\bar{f} = D^{*-}D^+$).
In their most recent paper on this topic \belle~\cite{Rohrken:2012ta} instead used the parametrisation ($A_{D^*D}$, $S_{D^*D}$, $\Delta S_{D^*D}$, $C_{D^*D}$, $\Delta C_{D^*D}$), while \babar~\cite{Aubert:2008ah} give results in both sets of parameters.
We therefore use the ($A_{D^*D}$, $S_{D^*D}$, $\Delta S_{D^*D}$, $C_{D^*D}$, $\Delta C_{D^*D}$) set.

\mysubsubsubsection{$\Bz \to \rho^{\pm}\pi^\mp$
}
\label{sec:cp_uta:notations:non_cp:rhopi}

In the $\rho^\pm\pi^\mp$ system, the 
($\Adirnoncp$, $C_{f\bar{f}}$, $S_{f\bar{f}}$, $\Delta C_{f\bar{f}}$, 
$\Delta S_{f\bar{f}}$)
set of parameters has been used 
originally by \babar~\cite{Aubert:2003wr} and \belle~\cite{Wang:2004va}, 
in the Q2B approximation; 
the exact names\footnote{
  \babar\ has used the notations
  $A_{\CP}^{\rho\pi}$~\cite{Aubert:2003wr} and 
  ${\cal A}_{\rho\pi}$~\cite{Aubert:2007jn}
  in place of ${\cal A}_{\CP}^{\rho\pi}$.
}
used in this case are
$\left( 
  {\cal A}_{\CP}^{\rho\pi}, C_{\rho\pi}, S_{\rho\pi}, \Delta C_{\rho\pi}, \Delta S_{\rho\pi}
\right)$,
and these names are also used in this document.

Since $\rho^\pm\pi^\mp$ is reconstructed in the final state $\pi^+\pi^-\pi^0$,
the interference between the $\rho$ resonances
can provide additional information about the phases 
(see Sec.~\ref{sec:cp_uta:notations:dalitz}).
Both \babar~\cite{Aubert:2007jn} 
and \belle~\cite{Kusaka:2007dv,:2007mj}
have performed time-dependent Dalitz plot analyses, 
from which the weak phase $\alpha$ is directly extracted.
In such an analysis, the measured Q2B parameters are 
also naturally corrected for interference effects.

\mysubsubsubsection{$\Bz \to D^{\mp}\pi^{\pm}, D^{*\mp}\pi^{\pm}, D^{\mp}\rho^{\pm}$
}
\label{sec:cp_uta:notations:non_cp:dstarpi}

Time-dependent $\CP$ analyses have also been performed for the
final states $D^{\mp}\pi^{\pm}$, $D^{*\mp}\pi^{\pm}$ and $D^{\mp}\rho^{\pm}$.
In these theoretically clean cases, no penguin contributions are possible,
so there is no $\CP$ violation in decay.
Furthermore, due to the smallness of the ratio of the magnitudes of the 
suppressed ($b \to u$) and favoured ($b \to c$) amplitudes (denoted $R_f$),
to a very good approximation, $C_f = - C_{\bar{f}} = 1$
(using $f = D^{(*)-}h^+$, $\bar{f} = D^{(*)+}h^-$ $h = \pi,\rho$),
and the coefficients of the sine terms are given by
\begin{equation}
  S_f = - 2 R_f \sin( \phi_{\rm mix} + \phi_{\rm dec} - \delta_f )
  \hspace{5mm}
  {\rm and}
  \hspace{5mm}
  S_{\bar{f}} = - 2 R_f \sin( \phi_{\rm mix} + \phi_{\rm dec} + \delta_f ).
\end{equation}
Thus weak phase information can be cleanly obtained from measurements
of $S_f$ and $S_{\bar{f}}$, 
although external information on at least one of $R_f$ or $\delta_f$ is necessary.
(Note that $\phi_{\rm mix} + \phi_{\rm dec} = 2\beta + \gamma \equiv 2\phi_1 + \phi_3$ for all the decay modes 
in question, while $R_f$ and $\delta_f$ depend on the decay mode.)

Again, different notations have been used in the literature.
\babar~\cite{Aubert:2006tw,Aubert:2005yf}
defines the time-dependent probability function by
\begin{equation}
  f^\pm (\eta, \Delta t) = \frac{e^{-|\Delta t|/\tau}}{4\tau} 
  \left[  
    1 \mp S_\zeta \sin (\Delta m \Delta t) \mp \eta C_\zeta \cos(\Delta m \Delta t) 
  \right],
\end{equation} 
where the upper (lower) sign corresponds to 
the tagging meson being a $\Bz$ ($\Bzb$). 
Note here that a tagging $\Bz$ ($\Bzb$) corresponds to $-S_\zeta$ ($+S_\zeta$).
The parameters $\eta$ and $\zeta$ take the values $+1$ and $+$ ($-1$ and $-$) 
when the final state is, \eg, $D^-\pi^+$ ($D^+\pi^-$). 
However, in the fit, the substitutions $C_\zeta = 1$ and 
$S_\zeta = a \mp \eta b_i - \eta c_i$ are made.\footnote{
  The subscript $i$ denotes tagging category.
}
(Note that, neglecting $b$ terms, $S_+ = a - c$ and $S_- = a + c$, 
so that $a = (S_+ + S_-)/2$, $c = (S_- - S_+)/2$, in analogy to 
the parameters of Eq.~(\ref{eq:cp_uta:non-cp-s_and_deltas}).)
These are motivated by the possibility of 
$\CP$ violation on the tag side~\cite{Long:2003wq}, 
which is absent for semileptonic $\B$ decays (mostly lepton tags). 
The parameter $a$ is not affected by tag side $\CP$ violation. 
The parameter $b$ only depends on tag side $\CP$ violation parameters 
and is not directly useful for determining UT angles.
A clean interpretation of the $c$ parameter is only possible for 
lepton-tagged events,
so the \babar\ measurements report $c$ measured with those events only.

The parameters used by \belle\ in the analysis using 
partially reconstructed $\B$ decays~\cite{Bahinipati:2011yq}, 
are similar to the $S_\zeta$ parameters defined above. 
However, in the \belle\ convention, 
a tagging $\Bz$ corresponds to a $+$ sign in front of the sine coefficient; 
furthermore the correspondence between the super/subscript 
and the final state is opposite, so that $S_\pm$ (\babar) = $- S^\mp$ (\belle). 
In this analysis, only lepton tags are used, 
so there is no effect from tag side $\CP$ violation. 
In the \belle\ analysis using 
fully reconstructed $\B$ decays~\cite{Ronga:2006hv}, 
this effect is measured and taken into account using $\Dstar \ell \nu$ decays; 
in neither \belle\ analysis are the $a$, $b$ and $c$ parameters used. 
In the latter case, the measured parameters are 
$2 R_{D^{(*)}\pi} \sin( 2\phi_1 + \phi_3 \pm \delta_{D^{(*)}\pi} )$; 
the definition is such that 
$S^\pm$ (\belle) = $- 2 R_{\Dstar \pi} \sin( 2\phi_1 + \phi_3 \pm \delta_{\Dstar \pi} )$. 
However, the definition includes an 
angular momentum factor $(-1)^L$~\cite{Fleischer:2003yb}, 
and so for the results in the $D\pi$ system, 
there is an additional factor of $-1$ in the conversion.

Explicitly, the conversion then reads as given in 
Table~\ref{tab:cp_uta:notations:non_cp:dstarpi}, 
where we have neglected the $b_i$ terms used by \babar
(which are zero in the absence of tag side $\CP$ violation).
For the averages in this document,
we use the $a$ and $c$ parameters,
and give the explicit translations used in 
Table~\ref{tab:cp_uta:notations:non_cp:dstarpi2}.
It is to be fervently hoped that the experiments will
converge on a common notation in future.

\begin{table}
  \begin{center} 
    \caption{
      Conversion between the various notations used for 
      $\CP$ violation parameters in the 
      $D^{\pm}\pi^{\mp}$, $D^{*\pm}\pi^{\mp}$ and $D^{\pm}\rho^{\mp}$ systems.
      The $b_i$ terms used by \babar\ have been neglected.
      Recall that $\left( \alpha, \beta, \gamma \right) = \left( \phi_2, \phi_1, \phi_3 \right)$.
    }
    \vspace{0.2cm}
    \setlength{\tabcolsep}{0.0pc}
    \begin{tabular*}{\textwidth}{@{\extracolsep{\fill}}cccc} \hline 
      & \babar\ & \belle\ partial rec. & \belle\ full rec. \\
      \hline
      $S_{D^+\pi^-}$    & $- S_- = - (a + c_i)$ &  N/A  &
      $2 R_{D\pi} \sin( 2\phi_1 + \phi_3 + \delta_{D\pi} )$ \\
      $S_{D^-\pi^+}$    & $- S_+ = - (a - c_i)$ &  N/A  &
      $2 R_{D\pi} \sin( 2\phi_1 + \phi_3 - \delta_{D\pi} )$ \\
      $S_{D^{*+}\pi^-}$ & $- S_- = - (a + c_i)$ & $S^+$ &   
      $- 2 R_{\Dstar \pi} \sin( 2\phi_1 + \phi_3 + \delta_{\Dstar \pi} )$ \\
      $S_{D^{*-}\pi^+}$ & $- S_+ = - (a - c_i)$ & $S^-$ &
      $- 2 R_{\Dstar \pi} \sin( 2\phi_1 + \phi_3 - \delta_{\Dstar \pi} )$ \\
      $S_{D^+\rho^-}$    & $- S_- = - (a + c_i)$ &  N/A  &  N/A  \\
      $S_{D^-\rho^+}$    & $- S_+ = - (a - c_i)$ &  N/A  &  N/A  \\
      \hline 
    \end{tabular*}
    \label{tab:cp_uta:notations:non_cp:dstarpi}
  \end{center}
\end{table}
   
\begin{table}
  \begin{center} 
    \caption{
      Translations used to convert the parameters measured by \belle
      to the parameters used for averaging in this document.
      The angular momentum factor $L$ is $-1$ for $\Dstar\pi$ and $+1$ for $D\pi$.
      Recall that $\left( \alpha, \beta, \gamma \right) = \left( \phi_2, \phi_1, \phi_3 \right)$.
    }
    \vspace{0.2cm}
    \setlength{\tabcolsep}{0.0pc}
    \begin{tabular*}{\textwidth}{@{\extracolsep{\fill}}ccc} \hline 
        & $\Dstar\pi$ partial rec. & $D^{(*)}\pi$ full rec. \\
        \hline
        $a$ & $- (S^+ + S^-)$ &
        $\frac{1}{2} (-1)^{L+1}
        \left(
          2 R_{D^{(*)}\pi} \sin( 2\phi_1 + \phi_3 + \delta_{D^{(*)}\pi} ) + 
          2 R_{D^{(*)}\pi} \sin( 2\phi_1 + \phi_3 - \delta_{D^{(*)}\pi} )
        \right)$ \\
        $c$ & $- (S^+ - S^-)$ & 
        $\frac{1}{2} (-1)^{L+1}
        \left(
          2 R_{D^{(*)}\pi} \sin( 2\phi_1 + \phi_3 + \delta_{D^{(*)}\pi} ) -
          2 R_{D^{(*)}\pi} \sin( 2\phi_1 + \phi_3 - \delta_{D^{(*)}\pi} )
        \right)$ \\
        \hline 
      \end{tabular*}
    \label{tab:cp_uta:notations:non_cp:dstarpi2}
  \end{center}
\end{table}

\mysubsubsubsection{$\Bs \to D_s^{\mp}K^\pm$}
\label{sec:cp_uta:notations:non_cp:dsk}

The phenomenology of $\Bs \to D_s^{\mp}K^\pm$ decays is similar to that for $\Bz \to D^{\mp}\pi^{\pm}$, with some important caveats.
The larger size of the ratio $R$ of the magnitudes of the suppressed and favoured amplitudes allows it to be determined from the data, as the deviation of $C_f$ and $C_{\bar{f}}$ from unity (in magnitude) can be observed.
Moreover, the non-zero value of $\Delta \Gamma_s$ allows the determination of additional terms, $A^{\Delta\Gamma}_f$ and $A^{\Delta\Gamma}_{\bar{f}}$ (see Sec.~\ref{sec:cp_uta:notations:Bs}), that break ambiguities in the solutions for $\phi_{\rm mix} + \phi_{\rm dec}$, which for $\Bs \to D_s^{\mp}K^\pm$ decays is equal to $\gamma-2\beta_s$.

LHCb~\cite{Aaij:2014fba} have performed such an analysis with $\Bs \to D_s^{\mp}K^\pm$ decays.
The absence of \CP violation in decay is assumed, and the parameters that are determined from the fit are labelled $C$, $A^{\Delta\Gamma}$, $\bar{A}{}^{\Delta\Gamma}$, $S$, $\bar{S}$.
These are trivially related to the definitions used in this Section.

\mysubsubsubsection{Time-dependent asymmetries in radiative $\B$ decays
}
\label{sec:cp_uta:notations:non_cp:radiative}

As a special case of decays to non-$\CP$ eigenstates,
let us consider radiative $\B$ decays.
Here, the emitted photon has a distinct helicity,
which is in principle observable, but in practise is not usually measured.
Thus the measured time-dependent decay rates 
are given by~\cite{Atwood:1997zr,Atwood:2004jj}
\begin{eqnarray}
  \Gamma_{\Bzb \to X \gamma} (\Delta t) & = &
  \Gamma_{\Bzb \to X \gamma_L} (\Delta t) + \Gamma_{\Bzb \to X \gamma_R} (\Delta t) \\ \nonumber
  & = &
  \frac{e^{-\left| \Delta t \right| / \tau(\Bz)}}{4\tau(\Bz)} 
  \left\{ 
    1 + 
    \left( S_L + S_R \right) \sin(\Delta m \Delta t) - 
    \left( C_L + C_R \right) \cos(\Delta m \Delta t) 
  \right\},
  \\
  \Gamma_{\Bz \to X \gamma} (\Delta t) & = & 
  \Gamma_{\Bz \to X \gamma_L} (\Delta t) + \Gamma_{\Bz \to X \gamma_R} (\Delta t) \\ \nonumber 
  & = &
  \frac{e^{-\left| \Delta t \right| / \tau(\Bz)}}{4\tau(\Bz)} 
  \left\{ 
    1 - 
    \left( S_L + S_R \right) \sin(\Delta m \Delta t) + 
    \left( C_L + C_R \right) \cos(\Delta m \Delta t) 
  \right\},
\end{eqnarray}
where in place of the subscripts $f$ and $\bar{f}$ we have used $L$ and $R$
to indicate the photon helicity.
In order for interference between decays with and without $\Bz$-$\Bzb$ mixing
to occur, the $X$ system must not be flavour-specific,
\eg, in case of $\Bz \to K^{*0}\gamma$, the final state must be $\KS \pi^0 \gamma$.
The sign of the sine term depends on the $C$ eigenvalue of the $X$ system.
At leading order, the photons from 
$b \to q \gamma$ ($\bar{b} \to \bar{q} \gamma$) are predominantly
left (right) polarised, with corrections of order of $m_q/m_b$,
thus interference effects are suppressed.
Higher order effects can lead to corrections of order 
$\Lambda_{\rm QCD}/m_b$~\cite{Grinstein:2004uu,Grinstein:2005nu},
though explicit calculations indicate such corrections are small
for exclusive final states~\cite{Matsumori:2005ax,Ball:2006cva}.
The predicted smallness of the $S$ terms in the Standard Model
results in sensitivity to new physics contributions.

The formalism discussed above is valid from any radiative decay to a final state where the hadronic system is an eigenstate of $C$.
In addition to $\KS\piz\gamma$, experiments have presented results using $\Bz$ decays to $\KS\eta\gamma$, $\KS\rho\gamma$ and $\KS\phi\gamma$.
For the case of the $\KS\rho\gamma$ final state, particular care is needed, as due to the non-negligible width of the $\rho^0$ meson, decays selected as $\Bz \to \KS\rho^0\gamma$ can include a significant contribution from $K^{*\pm}\pimp\gamma$ decays, which are flavour-specific and do not have the same oscillation phenomenology. 
It is therefore necessary to correct the fitted asymmetry parameter for a ``dilution factor''.

\mysubsubsection{Asymmetries in $\B \to \DorDstar K^{(*)}$ decays
}
\label{sec:cp_uta:notations:cus}

$\CP$ asymmetries in $\B \to \DorDstar K^{(*)}$ decays are sensitive to $\gamma$.
The neutral $D^{(*)}$ meson produced 
is an admixture of $\DorDstarz$ (produced by a $b \to c$ transition) and 
$\DorDstarzb$ (produced by a colour-suppressed $b \to u$ transition) states.
If the final state is chosen so that both $\DorDstarz$ and $\DorDstarzb$ 
can contribute, the two amplitudes interfere,
and the resulting observables are sensitive to $\gamma$, 
the relative weak phase between 
the two $\B$ decay amplitudes~\cite{Bigi:1988ym}.
Various methods have been proposed to exploit this interference,
including those where the neutral $D$ meson is reconstructed 
as a $\CP$ eigenstate (GLW)~\cite{Gronau:1990ra,Gronau:1991dp},
in a suppressed final state (ADS)~\cite{Atwood:1996ci,Atwood:2000ck},
or in a self-conjugate three-body final state, 
such as $\KS \pi^+\pi^-$ (Dalitz)~\cite{Giri:2003ty,Poluektov:2004mf}.
It should be emphasised that while each method 
differs in the choice of $D$ decay,
they are all sensitive to the same parameters of the $B$ decay,
and can be considered as variations of the same technique.

Consider the case of $\Bmp \to D \Kmp$,
with $D$ decaying to a final state $f$,
which is accessible to both $\Dz$ and $\Dzb$.
We can write the decay rates for $\Bm$ and $\Bp$ ($\Gamma_\mp$), 
the charge averaged rate ($\Gamma = (\Gamma_- + \Gamma_+)/2$)
and the charge asymmetry 
(${\cal A} = (\Gamma_- - \Gamma_+)/(\Gamma_- + \Gamma_+)$, see Eq.~(\ref{eq:cp_uta:pra})) as 
\begin{eqnarray}
  \label{eq:cp_uta:dk:rate_def}
  \Gamma_\mp  & \propto & 
  r_B^2 + r_D^2 + 2 r_B r_D \cos \left( \delta_B + \delta_D \mp \gamma \right), \\
  \label{eq:cp_uta:dk:av_rate_def}
  \Gamma & \propto &  
  r_B^2 + r_D^2 + 2 r_B r_D \cos \left( \delta_B + \delta_D \right) \cos \left( \gamma \right), \\
  \label{eq:cp_uta:dk:acp_def}
  {\cal A} & = & 
  \frac{
    2 r_B r_D \sin \left( \delta_B + \delta_D \right) \sin \left( \gamma \right)
  }{
    r_B^2 + r_D^2 + 2 r_B r_D \cos \left( \delta_B + \delta_D \right) \cos \left( \gamma \right),  
  }
\end{eqnarray}
where the ratio of $\B$ decay amplitudes\footnote{
  Note that here we use the notation $r_B$ to denote the ratio
  of $\B$ decay amplitudes, 
  whereas in Sec.~\ref{sec:cp_uta:notations:non_cp:dstarpi} 
  we used, \eg, $R_{D\pi}$, for a rather similar quantity.
  The reason is that here we need to be concerned also with 
  $D$ decay amplitudes,
  and so it is convenient to use the subscript to denote the decaying particle.
  Hopefully, using $r$ in place of $R$ will reduce the potential for confusion.
} 
is usually defined to be less than one,
\begin{equation}
  \label{eq:cp_uta:dk:rb_def}
  r_B = 
  \frac{
    \left| A\left( \Bm \to \Dzb K^- \right) \right|
  }{
    \left| A\left( \Bm \to \Dz  K^- \right) \right|
  },
\end{equation}
and the ratio of $D$ decay amplitudes is correspondingly defined by
\begin{equation}
  \label{eq:cp_uta:dk:rd_def}
  r_D = 
  \frac{
    \left| A\left( \Dz  \to f \right) \right|
  }{
    \left| A\left( \Dzb \to f \right) \right|
  }.
\end{equation}
The strong phase differences between the $\B$ and $D$ decay amplitudes 
are given by $\delta_B$ and $\delta_D$, respectively.
The values of $r_D$ and $\delta_D$ depend on the final state $f$:
for the GLW analysis, $r_D = 1$ and $\delta_D$ is trivial (either zero or $\pi$),
in the Dalitz plot analysis $r_D$ and $\delta_D$ vary across the Dalitz plot,
and depend on the $D$ decay model used,
for the ADS analysis, the values of $r_D$ and $\delta_D$ are not trivial.

Note that, for given values of $r_B$ and $r_D$, 
the maximum size of ${\cal A}$ (at $\sin \left( \delta_B + \delta_D \right) = 1$)
is $2 r_B r_D \sin \left( \gamma \right) / \left( r_B^2 + r_D^2 \right)$.
Thus even for $D$ decay modes with small $r_D$, 
large asymmetries, and hence sensitivity to $\gamma$, 
may occur for $B$ decay modes with similar values of $r_B$.
For this reason, the ADS analysis of the decay $B^\mp \to D \pi^\mp$ 
is also of interest.

In the GLW analysis, the measured quantities are the 
partial rate asymmetry and the charge averaged rate,
which are measured both for $\CP$-even and $\CP$-odd $D$ decays.
The latter is defined as 
\begin{equation}
  \label{eq:cp_uta:dk:glw-rdef}
  R_{\CP} = 
  \frac{2 \, \Gamma \left( \Bm \to D_{\CP} \Km  \right)}
  {\Gamma\left( \Bm \to \Dz \Km \right)} \, .
\end{equation}
It is experimentally convenient to measure $R_{\CP}$ using a double ratio,
\begin{equation}
  \label{eq:cp_uta:dk:double_ratio}
  R_{\CP} = 
  \frac{
    \Gamma\left( \Bm \to D_{\CP} \Km  \right) \, / \, \Gamma\left( \Bm \to \Dz \Km \right)
  }{
    \Gamma\left( \Bm \to D_{\CP} \pim \right) \, / \, \Gamma\left( \Bm \to \Dz \pim \right)
  }
\end{equation}
that is normalised both to the rate for the favoured $\Dz \to \Km\pip$ decay, 
and to the equivalent quantities for $\Bm \to D\pim$ decays
(charge conjugate processes are implicitly included in 
Eq.~(\ref{eq:cp_uta:dk:glw-rdef}) and~(\ref{eq:cp_uta:dk:double_ratio})).
In this way the constant of proportionality drops out of 
Eq.~(\ref{eq:cp_uta:dk:av_rate_def}).
Eq.~(\ref{eq:cp_uta:dk:double_ratio}) is exact in the limit that the
contribution of the $b \to u$ decay amplitude to $\Bm \to D \pim$ vanishes and
when the flavour-specific rates $\Gamma\left( \Bm \to \Dz h^- \right)$ ($h =
\pi,K$) are determined using appropriately flavour-specific $D$ decays.
In reality, the decay $D \to K\pi$ is invariable used, leading to a small source of systematic uncertainty.
The \CP\ asymmetry is defined as
\begin{equation}
  \label{eq:cp_uta:dk:glw-adef}
  A_{\CP} = \frac{
    \Gamma\left(\Bm\to D_{\CP}\Km\right) - \Gamma\left(\Bp\to D_{\CP}\Kp\right)
  }{
    \Gamma\left(\Bm\to D_{\CP}\Km\right) + \Gamma\left(\Bp\to D_{\CP}\Kp\right)
  } \, .
\end{equation}

For the ADS analysis, using a suppressed $D \to f$ decay,
the measured quantities are again the partial rate asymmetry, 
and the charge averaged rate.
In this case it is sufficient to measure the rate in a single ratio
(normalised to the favoured $D \to \bar{f}$ decay)
since detection systematics cancel naturally;
the observed quantity is then
\begin{equation}
  \label{eq:cp_uta:dk:r_ads}
  R_{\rm ADS} = 
  \frac{
    \Gamma \left( \Bm \to \left[ f \right]_D \Km \right) + 
    \Gamma \left( \Bp \to \left[ \bar{f} \right]_D \Kp \right)
  }{
    \Gamma \left( \Bm \to \left[ \bar{f} \right]_D \Km \right) +
    \Gamma \left( \Bp \to \left[ f \right]_D \Kp \right)
  } \, ,
\end{equation}
where the inclusion of charge conjugate modes has been made explicit.
The \CP\ asymmetry is defined as
\begin{equation}
  \label{eq:cp_uta:dk:a_ads}
  A_{\rm ADS} = 
  \frac{
    \Gamma\left(\Bm\to\left[f\right]_D\Km\right)-
    \Gamma\left(\Bp\to\left[f\right]_D\Kp\right)
  }{
    \Gamma\left(\Bm\to\left[f\right]_D\Km\right)+
    \Gamma\left(\Bp\to\left[f\right]_D\Kp\right)
  } \, .
\end{equation}
Since the uncertainty of $A_{\rm ADS}$ depends on the central value of $R_{\rm ADS}$, for some statistical treatments it is preferable to use an alternative pair of parameters~\cite{Bondar:2004bi})
\begin{equation}
  R_- = \frac{
    \Gamma \left( \Bm \to \left[ f \right]_D \Km \right)
  }{
    \Gamma \left( \Bm \to \left[ \bar{f} \right]_D \Km \right)
  } \, 
  \hspace{5mm}
  R_+ = \frac{
    \Gamma \left( \Bp \to \left[ \bar{f} \right]_D \Kp \right)
  }{
    \Gamma \left( \Bp \to \left[ f \right]_D \Kp \right)
  } \, ,
\end{equation}
where there is no inclusion of charge conjugated processes.
We use the $(R_{\rm ADS}, A_{\rm ADS})$ set in our compilation.

In the ADS analysis, there are an additional two unknowns ($r_D$ and $\delta_D$)
compared to the GLW case.  
However, the value of $r_D$ can be measured using 
decays of $D$ mesons of known flavour, and $\delta_D$ can be measured from interference effects in decays of quantum-correlated $D\bar{D}$ pairs produced at the $\psi(3770)$ resonance.
More generally, one needs access to two different linear admixtures of $D^0$ and $\bar{D}{}^0$ states in order to determine the relative phase: one such sample can be flavour tagged $D$ mesons which are available in abundant quantities in many experiments; the other can be \CP-tagged $D$ mesons from $\psi(3770)$ decays or could be mixed $D$ mesons, or could for that matter be the combination of $D^0$ and $\bar{D}{}^0$ that is found in $B \to DK$ decays.
In fact, the most precise information on both $r_D$ and $\delta_D$ currently comes from global fits on charm mixing parameters, as discussed in Sec.~\ref{sec:charm:mixcpv}.

The relation of ${\cal A}_{\rm ADS}$ to the underlying parameters given in Eq.~(\ref{eq:cp_uta:dk:acp_def}) and Table~\ref{tab:cp_uta:notations:dk} is exact for a two-body $D$ decay.  
For multibody decays, a similar formalism can be used with the introduction of a coherence factor~\cite{Atwood:2003mj}.
This is most appropriate for doubly-Cabibbo-suppressed decays to non-self-conjugate final states, but can also be modified for use with singly-Cabibbo-suppressed decays~\cite{Grossman:2002aq}.
For multibody self-conjugate final states, such as $\KS\pi^+\pi^-$, a Dalitz plot analysis (discussed below) is often more appropriate.

Additional coherence factors enter the expressions when the $B$ decay is to a multibody final state.
In particular, experiments have studied $B^+ \to DK^*(892)^+$, $B^0 \to DK^*(892)^0$ and $B^+ \to DK^+\pi^+\pi^-$ decays.
The non-negligible width of the $K^*(892)$ resonance implies that contributions from other $B \to DK\pi$ decays can pass the selection requirements.
Their effect on the Q2B analysis can be accounted for with a coherence factor~\cite{Gronau:2002mu}.
An alternative approach, not yet pursued by experiments, but in certain cases potentially more advantageous~\cite{Gershon:2008pe,Gershon:2009qc}, is Dalitz plot analysis of the full $B \to DK\pi$ phase space.

In the Dalitz plot analysis of $D$ decays to multibody self-conjugate final states,
once a model is assumed for the $D$ decay, 
which gives the values of $r_D$ and $\delta_D$ across the Dalitz plot,
it is possible to perform a simultaneous fit to the $B^+$ and $B^-$ samples 
and directly extract $\gamma$, $r_B$ and $\delta_B$.
However, the uncertainties on the phases depend approximately inversely on $r_B$.
Furthermore, $r_B$ is positive definite (and small), 
and therefore tends to be overestimated,
which leads to an underestimation of the uncertainty on $\gamma$ that must be
corrected statistically. 
An alternative approach is to extract from the data the ``Cartesian''
variables
\begin{equation}
  \left( x_\pm, y_\pm \right) = 
  \left( \Re(r_B e^{i(\delta_B\pm\gamma)}), \Im(r_B e^{i(\delta_B\pm\gamma)}) \right) = 
  \left( r_B \cos(\delta_B\pm\gamma), r_B \sin(\delta_B\pm\gamma) \right).
\end{equation}
These variables are approximately statistically uncorrelated and almost Gaussian.
The pairs of variables $\left( x_\pm, y_\pm \right)$ can be extracted
from independent fits of the $B^\pm$ data samples.
Use of these variables makes the combination of results much simpler.

The assumption of a model for the $D$ decay can, however, lead to a non-negligible, and hard to quantify, source of uncertainty.
To obviate this, it is possible to use instead a model-independent approach, in which the Dalitz plot (or, more generally, the phase-space) is binned~\cite{Giri:2003ty,Bondar:2005ki,Bondar:2008hh}.
In this case, hadronic parameters describing the average strong phase difference in each bin between the suppressed and favoured decay amplitudes enter the equations.
These parameters can be determined from interference effects in decays of quantum-correlated $D\bar{D}$ pairs produced at the $\psi(3770)$ resonance.

However, if the Dalitz plot is effectively dominated by one $\CP$ state,
there will be additional sensitivity to $\gamma$ in the numbers of events
in the $B^\pm$ data samples.
This can be taken into account in various ways.
One possibility is to extract GLW-like variables 
in addition to the $\left( x_\pm, y_\pm \right)$ parameters.
An alternative approach proceeds by defining $z_\pm = x_\pm + i y_\pm$
and $x_0 = - \int \Re \left[ f(s_1,s_2)f^*(s_2,s_1) \right] ds_1ds_2$,
where $s_1, s_2$ are the coordinates of invariant mass squared that
define the Dalitz plot and $f$ is the complex amplitude for $D$ decay
as a function of the Dalitz plot coordinates.\footnote{
  The $x_0$ parameter is closely related to the $c_i$ parameters of 
  the model dependent Dalitz plot analysis~\cite{Giri:2003ty,Bondar:2005ki,Bondar:2008hh},
  and the coherence factor of inclusive ADS-type analyses~\cite{Atwood:2003mj},
  integrated over the entire Dalitz plot.
}
The fitted parameters ($\rho^\pm, \theta^\pm$) are then defined by
\begin{equation}
  \rho^\pm e^{i \theta^\pm} = z_\pm - x_0 \, .
\end{equation}
Note that the yields of $B^\pm$ decays are proportional 
to $1 + (\rho^\pm)^2 - (x_0)^2$. 
This choice of variables has been used by \babar\ in the analysis of
$\Bmp \to D\Kmp$ with $D \to \pi^+\pi^-\pi^0$~\cite{Aubert:2007ii};
for this $D$ decay, $x_0 = 0.850$.
More recently, it has been noted that $D \to \pi^+\pi^-\pi^0$ can be used in a
GLW-like analysis~\cite{Nayak:2014tea}.

The relations between the measured quantities and the
underlying parameters are summarised in Table~\ref{tab:cp_uta:notations:dk}.
Note carefully that the hadronic factors $r_B$ and $\delta_B$ 
are different, in general, for each $\B$ decay mode.

\begin{table}[htbp]
  \begin{center} 
    \caption{
      Summary of relations between measured and physical parameters 
      in GLW, ADS and Dalitz analyses of $\B \to \DorDstar K^{(*)}$.
    }
    \vspace{0.2cm}
    \setlength{\tabcolsep}{1.0pc}
    \begin{tabular}{cc} \hline 
      \mc{2}{c}{GLW analysis} \\
      $R_{\CP\pm}$ & $1 + r_B^2 \pm 2 r_B \cos \left( \delta_B \right) \cos \left( \gamma \right)$ \\
      $A_{\CP\pm}$ & $\pm 2 r_B \sin \left( \delta_B \right) \sin \left( \gamma \right) / R_{\CP\pm}$ \\
      \hline
      \mc{2}{c}{ADS analysis} \\
      $R_{\rm ADS}$ & $r_B^2 + r_D^2 + 2 r_B r_D \cos \left( \delta_B + \delta_D \right) \cos \left( \gamma \right)$ \\
      $A_{\rm ADS}$ & $2 r_B r_D \sin \left( \delta_B + \delta_D \right) \sin \left( \gamma \right) / R_{\rm ADS}$ \\
      \hline
      \mc{2}{c}{Dalitz analysis ($D \to \KS \pi^+\pi^-$)} \\
      $x_\pm$ & $r_B \cos(\delta_B\pm\gamma)$ \\
      $y_\pm$ & $r_B \sin(\delta_B\pm\gamma)$ \\
      \hline
      \mc{2}{c}{Dalitz analysis ($D \to \pi^+\pi^-\pi^0$)} \\
      $\rho^\pm$ & $|z_\pm - x_0|$ \\
      $\theta^\pm$ & $\tan^{-1}(\Im(z_\pm)/(\Re(z_\pm) - x_0))$ \\
      \hline
    \end{tabular}
    \label{tab:cp_uta:notations:dk}
  \end{center}
\end{table}


\mysubsection{Common inputs and error treatment
}
\label{sec:cp_uta:common_inputs}

The common inputs used for rescaling are listed in 
Table~\ref{tab:cp_uta:common_inputs}.
The $\Bz$ lifetime ($\tau(\Bz)$), mixing parameter ($\Delta m_d$) and relative width difference ($\Delta\Gamma_d / \Gamma_d$)
averages are provided by the HFAG Lifetimes and Oscillations 
subgroup (Sec.~\ref{sec:life_mix}).
The fraction of the perpendicularly polarised component 
($\left| A_{\perp} \right|^2$) in $\B \to \jpsi \Kstar(892)$ decays,
which determines the $\CP$ composition in these decays, 
is averaged from results by 
\babar~\cite{Aubert:2007hz}, \belle~\cite{Itoh:2005ks}, CDF~\cite{Acosta:2004gt}, D0~\cite{Abazov:2008jz} and LHCb~\cite{Aaij:2013cma}.
See also HFAG $B$ to Charm Decay Parameters subgroup (Sec.~\ref{sec:b2c}).

At present, we only rescale to a common set of input parameters
for modes with reasonably small statistical errors
($b \to c\bar{c}s$ transitions).
Correlated systematic errors are taken into account
in these modes as well.
For all other modes, the effect of such a procedure is 
currently negligible.

\begin{table}[htbp]
  \begin{center}
    \caption{
      Common inputs used in calculating the averages.
    }
    \vspace{0.2cm}
    \setlength{\tabcolsep}{1.0pc}
    \begin{tabular}{cc} \hline 
      $\tau(\Bz)$ $({\rm ps})$  & $1.519 \pm 0.005$  \\
      $\Delta m_d$ $({\rm ps}^{-1})$ & $0.510 \pm 0.003$ \\
      $\Delta\Gamma_d / \Gamma_d$ & $0.001 \pm 0.010$ \\
      $\left| A_{\perp} \right|^2 (\jpsi \Kstar)$ & $0.209 \pm 0.006$ \\
      \hline
    \end{tabular}
    \label{tab:cp_uta:common_inputs}
  \end{center}
\end{table}

As explained in Sec.~\ref{sec:intro},
we do not apply a rescaling factor on the error of an average
that has $\chi^2/\dof > 1$ 
(unlike the procedure currently used by the PDG~\cite{PDG_2014}).
We provide a confidence level of the fit so that
one can know the consistency of the measurements included in the average,
and attach comments in case some care needs to be taken in the interpretation.
Note that, in general, results obtained from data samples with low statistics
will exhibit some non-Gaussian behaviour.
We average measurements with asymmetric errors 
using the PDG~\cite{PDG_2014} prescription.
In cases where several measurements are correlated
(\eg\ $S_f$ and $C_f$ in measurements of time-dependent $\CP$ violation
in $B$ decays to a particular $\CP$ eigenstate)
we take these into account in the averaging procedure
if the uncertainties are sufficiently Gaussian.
For measurements where one error is given, 
it represents the total error, 
where statistical and systematic uncertainties have been added in quadrature.
If two errors are given, the first is statistical and the second systematic.
If more than two errors are given,
the origin of the additional uncertainty will be explained in the text.

\mysubsection{Time-dependent asymmetries in $b \to c\bar{c}s$ transitions
}
\label{sec:cp_uta:ccs}

\mysubsubsection{Time-dependent $\CP$ asymmetries in $b \to c\bar{c}s$ decays to $\CP$ eigenstates
}
\label{sec:cp_uta:ccs:cp_eigen}

\begin{figure}[htbp]
  \begin{center}
    \resizebox{0.51\textwidth}{!}{
      \includegraphics{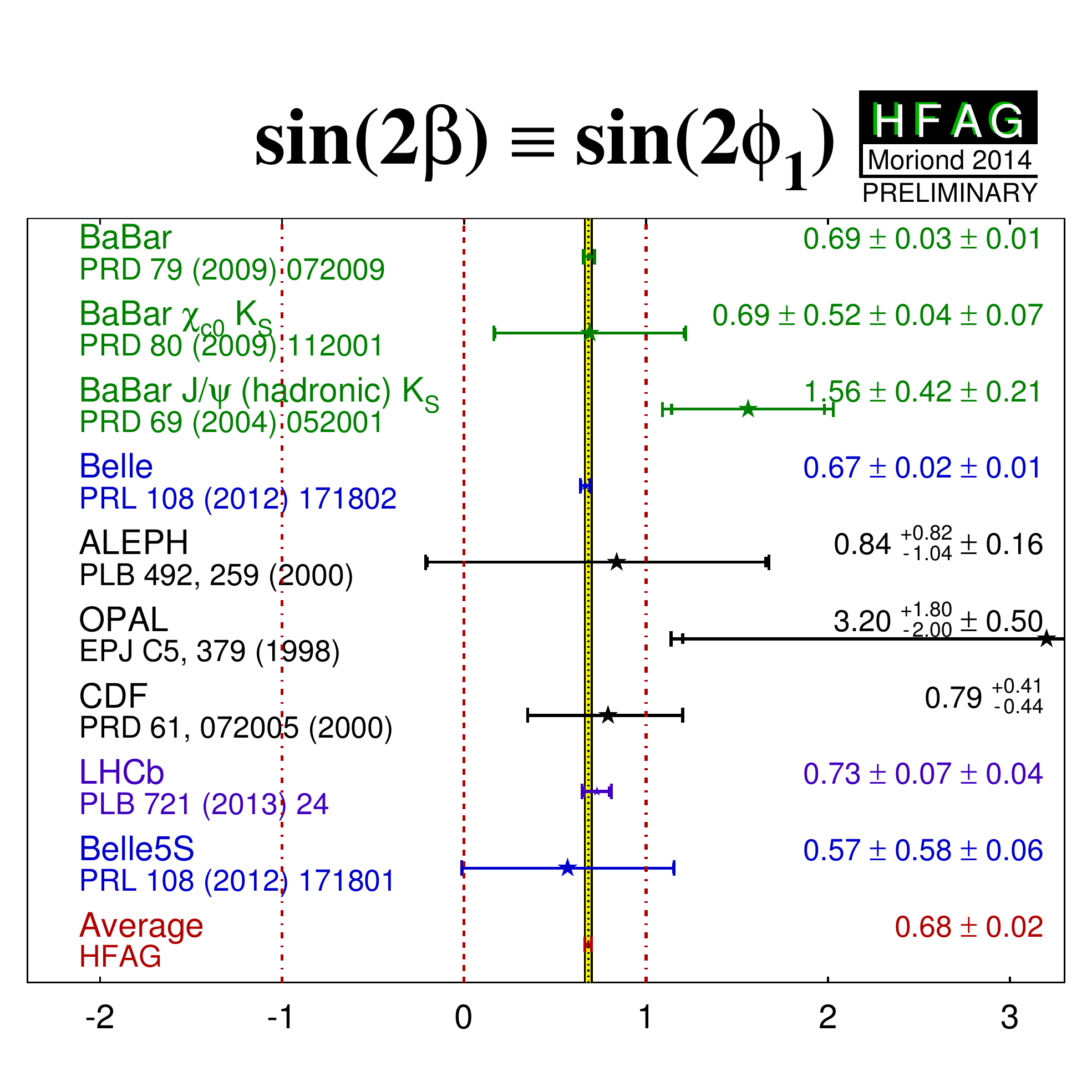}
    }
    \hfill
    \resizebox{0.48\textwidth}{!}{
      \includegraphics{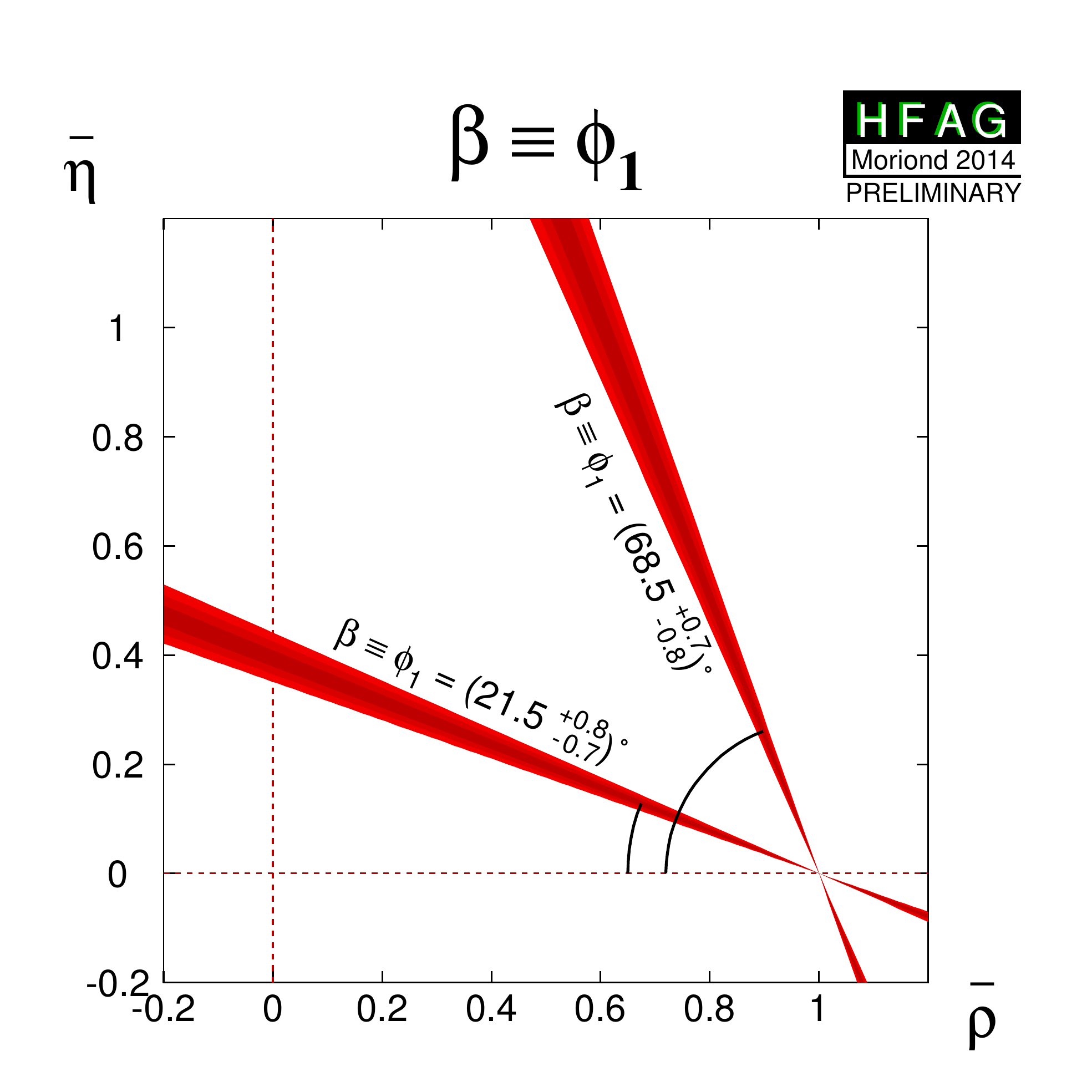}
    }
  \end{center}
  \vspace{-0.5cm}
  \caption{
    (Left) Average of measurements of $S_{b \to c\bar c s}$.
    (Right) Constraints on the $(\rhobar,\etabar)$ plane,
    obtained from the average of $-\etacp S_{b \to c\bar c s}$ 
    and Eq.~(\ref{eq:cp_uta:sin2beta}).
  }
  \label{fig:cp_uta:ccs}
\end{figure}

In the Standard Model, the time-dependent parameters for
$b \to c\bar c s$ transitions are predicted to be: 
$S_{b \to c\bar c s} = - \etacp \sin(2\beta)$,
$C_{b \to c\bar c s} = 0$ to very good accuracy.
The averages for $-\etacp S_{b \to c\bar c s}$ and $C_{b \to c\bar c s}$
are provided in Table~\ref{tab:cp_uta:ccs}.
The averages for $-\etacp S_{b \to c\bar c s}$ 
are shown in Fig.~\ref{fig:cp_uta:ccs}.

Both \babar\  and \belle\ have used the $\etacp = -1$ modes
$\jpsi \KS$, $\psi(2S) \KS$, $\chi_{c1} \KS$ and $\eta_c \KS$, 
as well as $\jpsi \KL$, which has $\etacp = +1$
and $\jpsi K^{*0}(892)$, which is found to have $\etacp$ close to $+1$
based on the measurement of $\left| A_\perp \right|$ 
(see Sec.~\ref{sec:cp_uta:common_inputs}).
The most recent \belle\ result does not use $\eta_c \KS$ or $\jpsi K^{*0}(892)$ decays.\footnote{
  Previous analyses from \belle\ did include these channels~\cite{Abe:2004mz},
  but it is not possible to obtain separate results for those modes from the
  published information.
}
ALEPH, OPAL, CDF and LHCb have used only the $\jpsi \KS$ final state.
\babar\ has also determined the \CP violation parameters of the
$\Bz\to\chi_{c0} \KS$ decay from the time-dependent Dalitz plot analysis of
$\Bz \to \pi^+\pi^-\KS$ (see Sec.~\ref{sec:cp_uta:qqs:dp}).
In addition, \belle\ has performed a measurement with data accumulated at the $\Upsilon(5S)$ resonance, using the $\jpsi\KS$ final state -- this involves a different flavour tagging method compared to the measurements performed with data accumulated at the $\Upsilon(4S)$ resonance.
A breakdown of results in each charmonium-kaon final state is given in 
Table~\ref{tab:cp_uta:ccs-BF}.

\begin{table}[htb]
	\begin{center}
		\caption{
                        $S_{b \to c\bar c s}$ and $C_{b \to c\bar c s}$.
                }
		\vspace{0.2cm}
		\setlength{\tabcolsep}{0.0pc}
		\begin{tabular*}{\textwidth}{@{\extracolsep{\fill}}lrccc} \hline
      \mc{2}{l}{Experiment} & Sample size & $- \etacp S_{b \to c\bar c s}$ & $C_{b \to c\bar c s}$ \\
      \hline
	\babar & \cite{:2009yr} & $N(B\bar{B})$ = 465M & $0.687 \pm 0.028 \pm 0.012$ & $0.024 \pm 0.020 \pm 0.016$ \\
	\babar\ $\chi_{c0} \KS$ & \cite{Aubert:2009me} & $N(B\bar{B})$ = 383M & $0.69 \pm 0.52 \pm 0.04 \pm 0.07$ & $-0.29 \,^{+0.53}_{-0.44} \pm 0.03 \pm 0.05$ \\
	\babar\ $J/\psi \KS$ ($^{*}$) & \cite{Aubert:2003xn} & $N(B\bar{B})$ = 88M & $1.56 \pm 0.42 \pm 0.21$ &  \textendash{} \\
	\belle & \cite{Adachi:2012et} & $N(B\bar{B})$ = 772M & $0.667 \pm 0.023 \pm 0.012$ & $-0.006 \pm 0.016 \pm 0.012$ \\
	\mc{3}{l}{\bf \boldmath $\B$ factory average} & $0.679 \pm 0.020$ & $0.005 \pm 0.017$ \\
	\mc{3}{l}{\small Confidence level} & {\small $0.28$} & {\small $0.47$} \\
        \hline
        ALEPH & \cite{Barate:2000tf} & \textendash{} & $0.84 \, ^{+0.82}_{-1.04} \pm 0.16$ &  \textendash{} \\
        OPAL  & \cite{Ackerstaff:1998xz} & \textendash{} & $3.2 \, ^{+1.8}_{-2.0} \pm 0.5$ &  \textendash{} \\
        CDF   & \cite{Affolder:1999gg} & \textendash{} & $0.79 \, ^{+0.41}_{-0.44}$ &  \textendash{} \\
	LHCb & \cite{Aaij:2012ke} & $1.0\ {\rm fb}^{-1}$ & $0.73 \pm 0.07 \pm 0.04$ &  $0.03 \pm 0.09 \pm 0.01$ \\
	Belle $\Upsilon(5S)$ & \cite{Sato:2012hu} & $121\ {\rm fb}^{-1}$ & $0.57 \pm 0.58 \pm 0.06$ &  \textendash{} \\
        \mc{3}{l}{\bf Average} & $0.682 \pm 0.019$ & $0.005 \pm 0.017$ \\
		\hline
		\end{tabular*}
                \label{tab:cp_uta:ccs}
        \end{center}
$^{*}$ {\small This result uses ``{\it hadronic and previously unused muonic decays of the $J/\psi$}''. We neglect a small possible correlation of this result with the main \babar\ result~\cite{:2009yr} that could be caused by reprocessing of the data.}
\end{table}

\begin{table}[htb]
	\begin{center}
		\caption{
                        Breakdown of $B$ factory results on $S_{b \to c\bar c s}$ and $C_{b \to c\bar c s}$.
                }
		\vspace{0.2cm}
		\setlength{\tabcolsep}{0.0pc}
		\begin{tabular*}{\textwidth}{@{\extracolsep{\fill}}lrccc} \hline
        \mc{2}{l}{Mode} & $N(B\bar{B})$ & $- \etacp S_{b \to c\bar c s}$ & $C_{b \to c\bar c s}$ \\
        \hline
        \mc{5}{c}{\babar} \\
        $J/\psi \KS$ & \cite{:2009yr} & 465M & $0.657 \pm 0.036 \pm 0.012$ & $\phantom{-}0.026 \pm 0.025 \pm 0.016$ \\
        $J/\psi \KL$ & \cite{:2009yr} & 465M & $0.694 \pm 0.061 \pm 0.031$ & $-0.033 \pm 0.050 \pm 0.027$ \\
        {\bf \boldmath $J/\psi K^0$} & \cite{:2009yr} & 465M & $0.666 \pm 0.031 \pm 0.013$ & $\phantom{-}0.016 \pm 0.023 \pm 0.018$ \\
        $\psi(2S) \KS$ & \cite{:2009yr} & 465M & $0.897 \pm 0.100 \pm 0.036$ & $\phantom{-}0.089 \pm 0.076 \pm 0.020$ \\
        $\chi_{c1} \KS$ & \cite{:2009yr} & 465M & $0.614 \pm 0.160 \pm 0.040$ & $\phantom{-}0.129 \pm 0.109 \pm 0.025$ \\
        $\eta_c \KS$ & \cite{:2009yr} & 465M & $0.925 \pm 0.160 \pm 0.057$ & $\phantom{-}0.080 \pm 0.124 \pm 0.029$ \\
        $\jpsi K^{*0}(892)$ & \cite{:2009yr} & 465M & $0.601 \pm 0.239 \pm 0.087$ & $\phantom{-}0.025 \pm 0.083 \pm 0.054$ \\
        {\bf All} & \cite{:2009yr} & 465M & $0.687 \pm 0.028 \pm 0.012$ & $\phantom{-}0.024 \pm 0.020 \pm 0.016$ \\
	\hline
	\mc{5}{c}{\bf \belle} \\
        $J/\psi \KS$ & \cite{Adachi:2012et} & 772M & $0.670 \pm 0.029 \pm 0.013$ & $\phantom{-}0.015 \pm 0.021 \,^{+0.023}_{-0.045}$ \\
        $J/\psi \KL$ & \cite{Adachi:2012et} & 772M & $0.642 \pm 0.047 \pm 0.021$ & $-0.019 \pm 0.026 \,^{+0.041}_{-0.017}$ \\
	$\psi(2{\rm S}) \KS$ & \cite{Adachi:2012et} & 772M & $0.738 \pm 0.079 \pm 0.036$ & $-0.104 \pm 0.055 \,^{+0.027}_{-0.047}$ \\
	$\chi_{c1} \KS$ & \cite{Adachi:2012et} & 772M & $0.640 \pm 0.117 \pm 0.040$ & $\phantom{-}0.017 \pm 0.083 \,^{+0.026}_{-0.046}$ \\
        {\bf All} & \cite{Adachi:2012et} & 772M & $0.667 \pm 0.023 \pm 0.012$ & $-0.006 \pm 0.016 \pm 0.012$ \\
	\hline
	\mc{5}{c}{\bf Averages} \\
        \mc{3}{l}{$J/\psi \KS$} & $0.665 \pm 0.024$ & $\phantom{-}0.024 \pm 0.026$ \\
        \mc{3}{l}{$J/\psi \KL$} & $0.663 \pm 0.041$ & $-0.023 \pm 0.030$ \\
        \mc{3}{l}{$\psi(2{\rm S}) \KS$} & $0.807 \pm 0.067$ & $-0.009 \pm 0.055$ \\
        \mc{3}{l}{$\chi_{c1} \KS$} & $0.632 \pm 0.099$ & $\phantom{-}0.066 \pm 0.074$ \\
		\hline
		\end{tabular*}
                \label{tab:cp_uta:ccs-BF}
        \end{center}
\end{table}

It should be noted that, while the uncertainty in the average for 
$-\etacp S_{b \to c\bar c s}$ is still limited by statistics,
the uncertainty for $C_{b \to c\bar c s}$ is close to being dominated by systematics.
This occurs due to the possible effect of tag side interference on the
$C_{b \to c\bar c s}$ measurement, an effect which is correlated between
different $e^+e^- \to \Upsilon(4S) \to B\bar{B}$ experiments.
Understanding of this effect may continue to improve in future,
allowing the uncertainty to reduce.

From the average for $-\etacp S_{b \to c\bar c s}$ above, 
we obtain the following solutions for $\beta$
(in $\left[ 0, \pi \right]$):
\begin{equation}
  \beta = \left( 21.5 \,^{+0.8}_{-0.7} \right)^\circ
  \hspace{5mm}
  {\rm or}
  \hspace{5mm}
  \beta = \left( 68.5 \,^{+0.7}_{-0.8} \right)^\circ
  \label{eq:cp_uta:sin2beta}
\end{equation}
In radians, these values are 
$\beta = \left( 0.375 \pm 0.013 \right)$, 
$\beta = \left( 1.196 \pm 0.013 \right)$.

This result gives a precise constraint on the $(\rhobar,\etabar)$ plane,
as shown in Fig.~\ref{fig:cp_uta:ccs}.
The measurement is in remarkable agreement with other constraints from 
$\CP$ conserving quantities, 
and with $\CP$ violation in the kaon system, in the form of the parameter $\epsilon_K$.
Such comparisons have been performed by various phenomenological groups,
such as CKMfitter~\cite{Charles:2004jd} 
and UTFit~\cite{Bona:2005vz}.

\mysubsubsection{Time-dependent transversity analysis of $\Bz \to J/\psi K^{*0}$
}
\label{sec:cp_uta:ccs:vv}

\begin{figure}[htbp]
  \begin{center}
    \resizebox{0.46\textwidth}{!}{
      \includegraphics{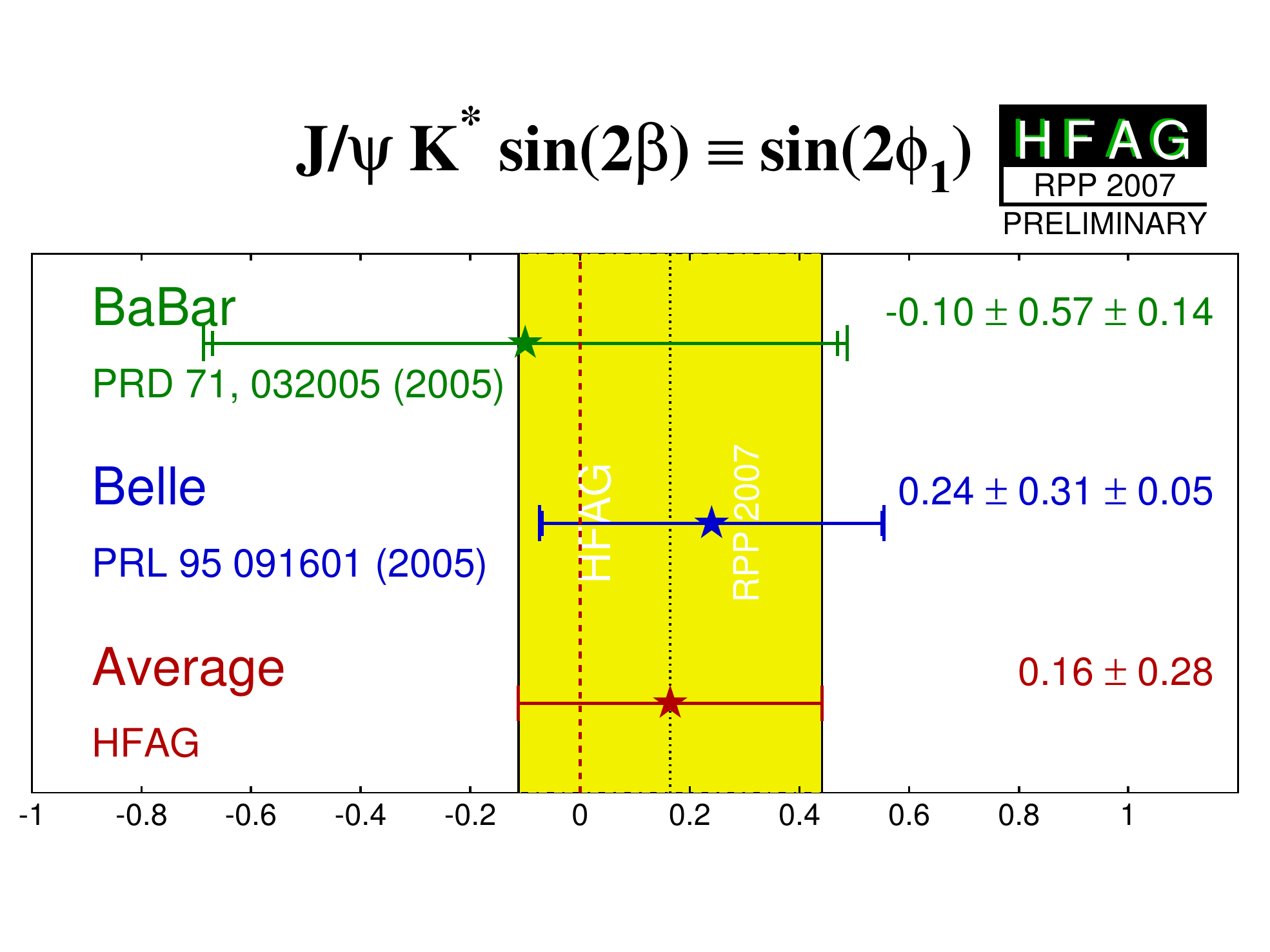}
    }
    \hfill
    \resizebox{0.46\textwidth}{!}{
      \includegraphics{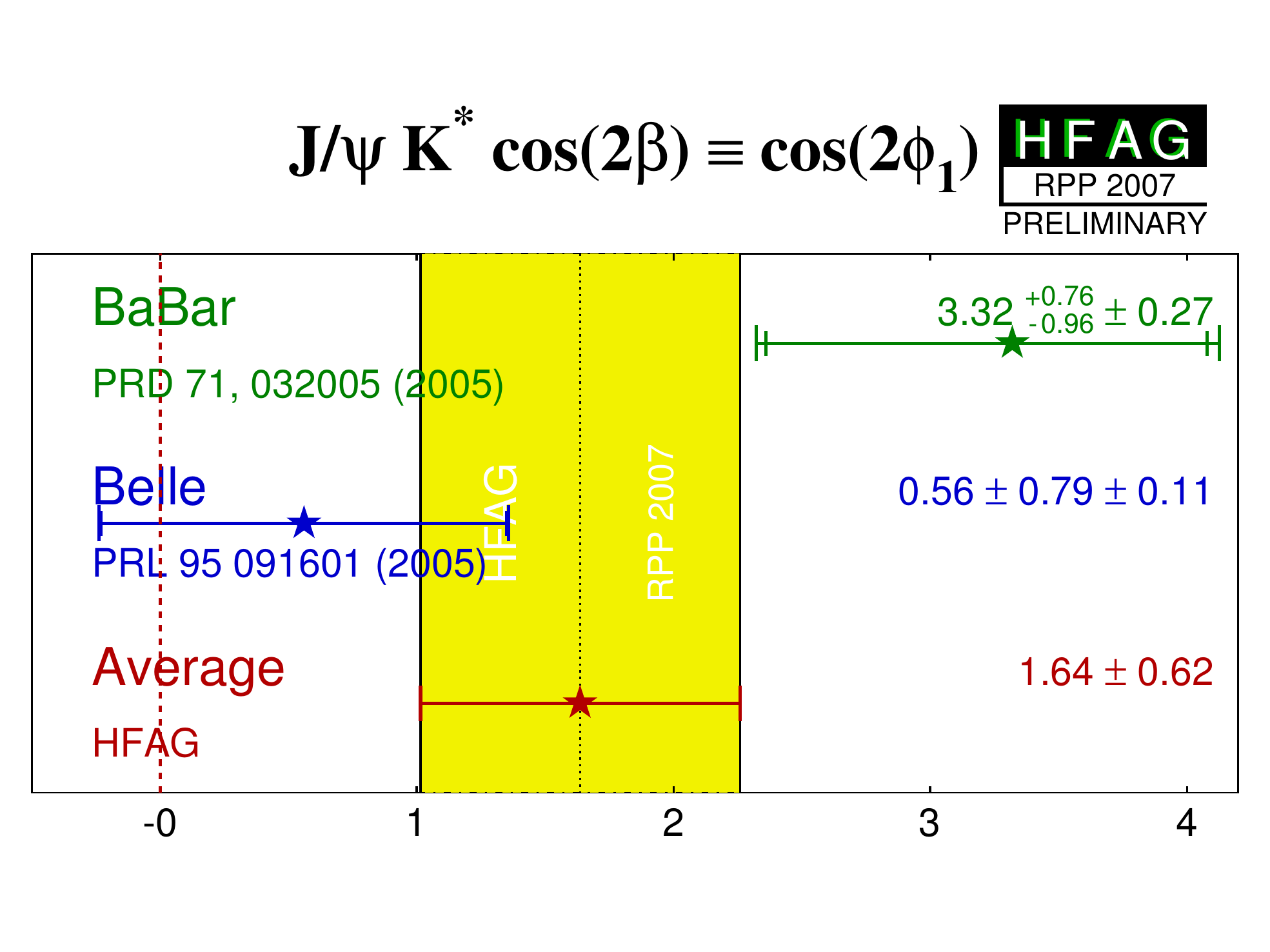}
    }
  \end{center}
  \vspace{-0.5cm}
  \caption{
    Averages of 
    (left) $\sin(2\beta) \equiv \sin(2\phi_1)$ and
    (right) $\cos(2\beta) \equiv \cos(2\phi_1)$
    from time-dependent analyses of $\Bz \to \jpsi K^{*0}$ decays.
  }
  \label{fig:cp_uta:JpsiKstar}
\end{figure}

$\B$ meson decays to the vector-vector final state $J/\psi K^{*0}$
are also mediated by the $b \to c \bar c s$ transition.
When a final state that is not flavour-specific ($K^{*0} \to \KS \pi^0$) is used,
a time-dependent transversity analysis can be performed 
allowing sensitivity to both 
$\sin(2\beta)$ and $\cos(2\beta)$~\cite{Dunietz:1990cj}.
Such analyses have been performed by both $\B$ factory experiments.
In principle, the strong phases between the transversity amplitudes
are not uniquely determined by such an analysis, 
leading to a discrete ambiguity in the sign of $\cos(2\beta)$.
The \babar\ collaboration resolves 
this ambiguity using the known variation~\cite{Aston:1987ir}
of the P-wave phase (fast) relative to the S-wave phase (slow) 
with the invariant mass of the $K\pi$ system 
in the vicinity of the $K^*(892)$ resonance. 
The result is in agreement with the prediction from 
$s$ quark helicity conservation,
and corresponds to Solution II defined by Suzuki~\cite{Suzuki:2001za}.
We use this phase convention for the averages given in 
Table~\ref{tab:cp_uta:ccs:psi_kstar} and Fig.~\ref{fig:cp_uta:JpsiKstar}.

\begin{table}[htb]
	\begin{center}
		\caption{
			Averages from $\Bz \to J/\psi K^{*0}$ transversity analyses.
		}
		\vspace{0.2cm}
		\setlength{\tabcolsep}{0.0pc}
		\begin{tabular*}{\textwidth}{@{\extracolsep{\fill}}lrcccc} \hline
		\mc{2}{l}{Experiment} & $N(B\bar{B})$ & $\sin 2\beta$ & $\cos 2\beta$ & Correlation \\
		\hline
	\babar & \cite{Aubert:2004cp} & 88M & $-0.10 \pm 0.57 \pm 0.14$ & $3.32 ^{+0.76}_{-0.96} \pm 0.27$ & $-0.37$ \\
	\belle & \cite{Itoh:2005ks} & 275M & $0.24 \pm 0.31 \pm 0.05$ & $0.56 \pm 0.79 \pm 0.11$ & $0.22$ \\
	\mc{3}{l}{\bf Average} & $0.16 \pm 0.28$ & $1.64 \pm 0.62$ &  \hspace{-8mm} {\small uncorrelated averages}  \\
        \mc{3}{l}{\small Confidence level} & {\small $0.61~(0.5\sigma)$} & {\small $0.03~(2.2\sigma)$} & \\
		\hline
		\end{tabular*}
		\label{tab:cp_uta:ccs:psi_kstar}
	\end{center}
\end{table}

At present the results are dominated by 
large and non-Gaussian statistical errors,
and exhibit significant correlations.
We perform uncorrelated averages, 
the interpretation of which has to be done with the greatest care. 
Nonetheless, it is clear that $\cos(2\beta)>0$ is preferred 
by the experimental data in $J/\psi \Kstarz$ 
(for example, \babar~\cite{Aubert:2004cp} 
find a confidence level for $\cos(2\beta)>0$ of $89\%$).

\mysubsubsection{Time-dependent $\CP$ asymmetries in $\Bz \to \Dstarp \Dstarm \KS$ decays
}
\label{sec:cp_uta:ccs:DstarDstarKs}

\begin{figure}[htbp]
  \begin{center}
    \begin{tabular}{c@{\hspace{-1mm}}c@{\hspace{-1mm}}c}
      \resizebox{0.32\textwidth}{!}{
        \includegraphics{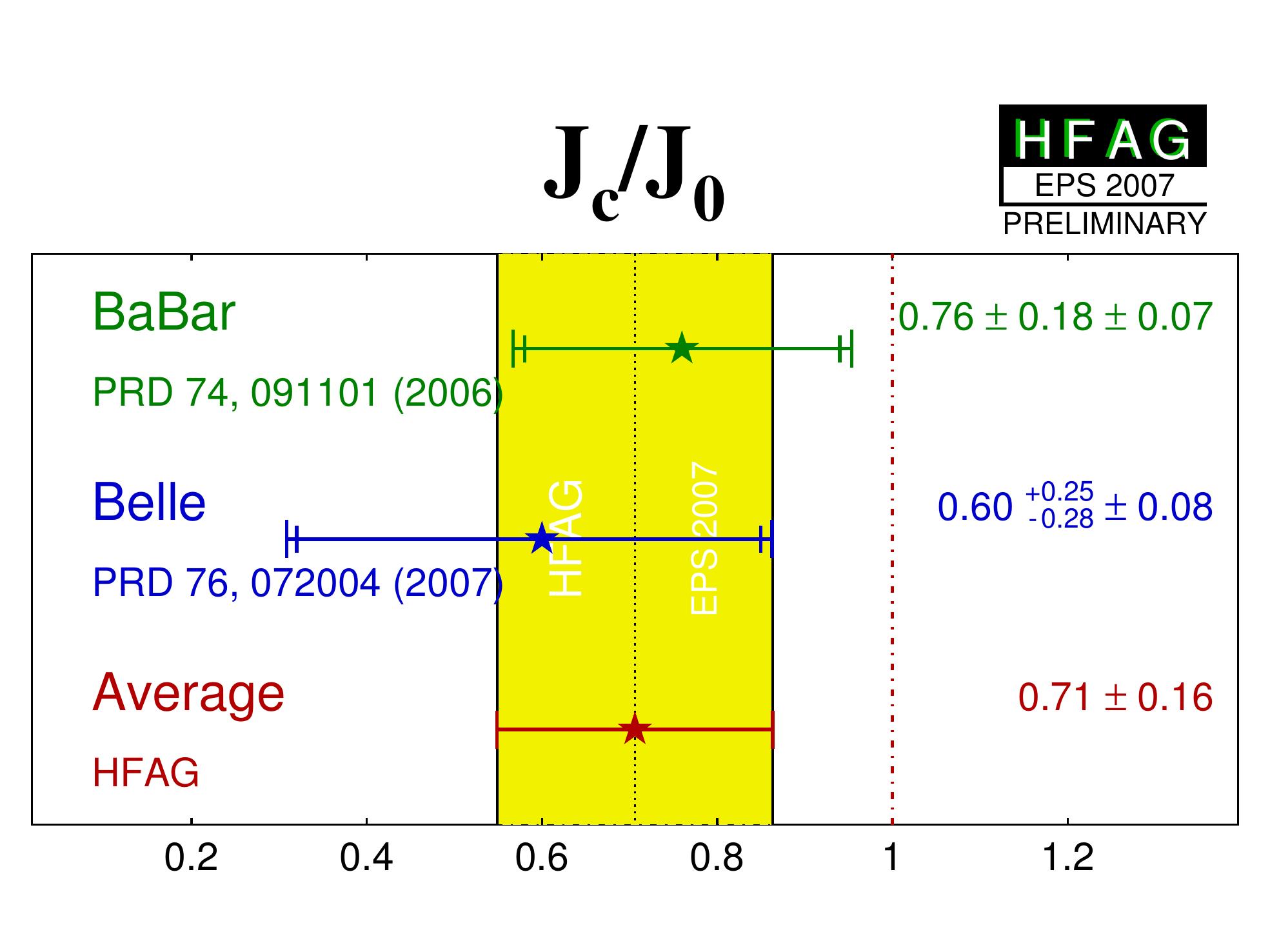}
      }
      &
      \resizebox{0.32\textwidth}{!}{
        \includegraphics{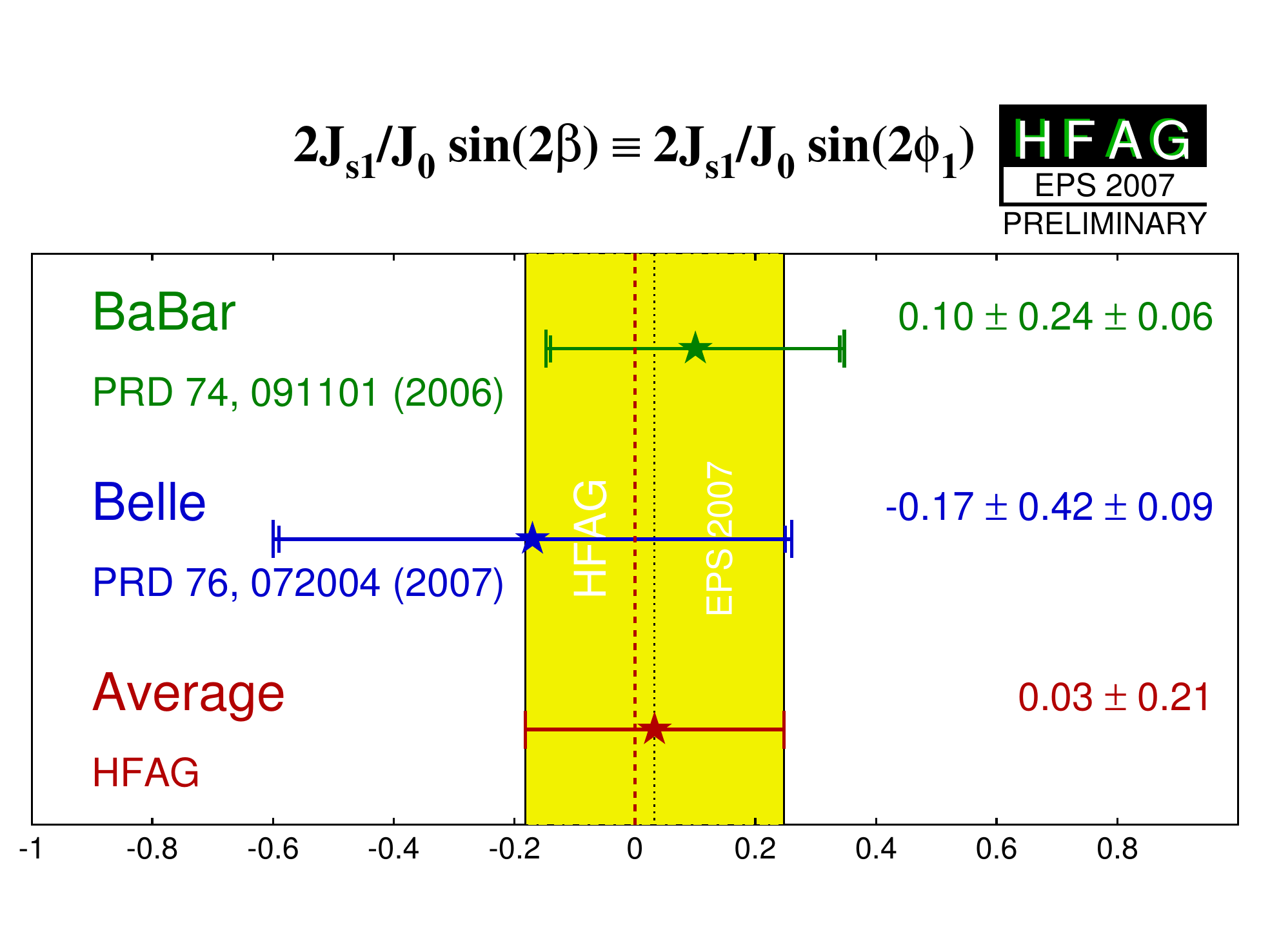}
      }
      &
      \resizebox{0.32\textwidth}{!}{
        \includegraphics{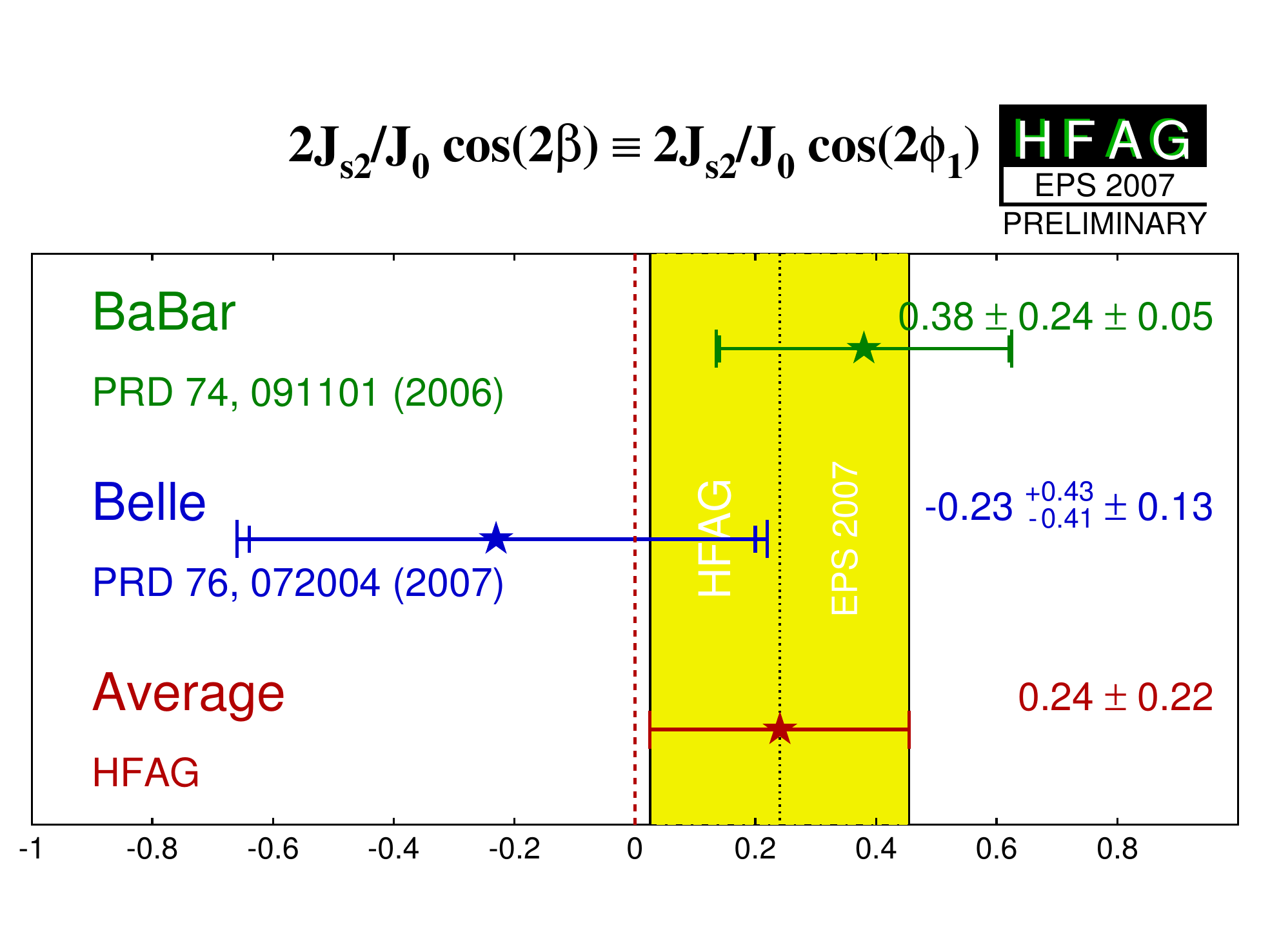}
      }
    \end{tabular}
  \end{center}
  \vspace{-0.8cm}
  \caption{
    Averages of 
    (left) $(J_c/J_0)$, (middle) $(2J_{s1}/J_0) \sin(2\beta)$ and 
    (right) $(2J_{s2}/J_0) \cos(2\beta)$
    from time-dependent analyses of $\Bz \to \Dstarp \Dstarm \KS$ decays.
  }
  \label{fig:cp_uta:ccs:dstardstarks}
\end{figure}

Both \babar~\cite{Aubert:2006fh} and \belle~\cite{Dalseno:2007hx} have performed
time-dependent analyses of the $\Bz \to \Dstarp \Dstarm \KS$ decay,
to obtain information on the sign of $\cos(2\beta)$.
More information can be found in 
Sec.~\ref{sec:cp_uta:notations:dalitz:dstardstarks}.
The results are shown in Table~\ref{tab:cp_uta:ccs:dstardstarks}, 
and Fig.~\ref{fig:cp_uta:ccs:dstardstarks}.

\begin{table}[htb]
	\begin{center}
		\caption{
                        Results from time-dependent analysis of $\Bz \to \Dstarp \Dstarm \KS$.
		}
		\vspace{0.2cm}
		\setlength{\tabcolsep}{0.0pc}
		\begin{tabular*}{\textwidth}{@{\extracolsep{\fill}}lrcccc} \hline
                \mc{2}{l}{Experiment} & $N(B\bar{B})$ & $\frac{J_c}{J_0}$ & $\frac{2J_{s1}}{J_0} \sin(2\beta)$ &  $\frac{2J_{s2}}{J_0} \cos(2\beta)$ \\
		\hline
	\babar & \cite{Aubert:2006fh} & 230M & $0.76 \pm 0.18 \pm 0.07$ & $0.10 \pm 0.24 \pm 0.06$ & $0.38 \pm 0.24 \pm 0.05$ \\
	\belle & \cite{Dalseno:2007hx} & 449M & $0.60 \,^{+0.25}_{-0.28} \pm 0.08$ & $-0.17 \pm 0.42 \pm 0.09$ & $-0.23 \,^{+0.43}_{-0.41} \pm 0.13$ \\
	\mc{3}{l}{\bf Average} & $0.71 \pm 0.16$ & $0.03 \pm 0.21$ & $0.24 \pm 0.22$ \\
	\mc{3}{l}{\small Confidence level} & {\small $0.63~(0.5\sigma)$} & {\small $0.59~(0.5\sigma)$} & {\small $0.23~(1.2\sigma)$} \\
		\hline
		\end{tabular*}
		\label{tab:cp_uta:ccs:dstardstarks}
	\end{center}
\end{table}

From the above result and the assumption that $J_{s2}>0$, 
\babar\ infer that $\cos(2\beta)>0$ at the $94\%$ confidence level~\cite{Aubert:2006fh}.

\mysubsubsection{Time-dependent analysis of $\Bs$ decays through the $b \to c\bar{c}s$ transition
}
\label{sec:cp_uta:ccs:jpsiphi}

As described in Sec.~\ref{sec:cp_uta:notations:Bs},
time-dependent analysis of decays such as $\Bs \to J/\psi \phi$ probes the 
$\CP$ violating phase of $\Bs$--$\Bsbar$ oscillations, $\phi_s$.
Within the Standard Model, this parameter is predicted to be small.\footnote{
   We make the approximation $\phi_s = -2 \beta_s$, 
   where $\phi_s \equiv \arg\left[ -M_{12}/\Gamma_{12} \right]$ 
   and $2\beta_s \equiv 2 \arg\left[ -(V_{ts}V_{tb}^*)/(V_{cs}V_{cb}^*) \right]$
   (see Sec.~\ref{sec:cp_uta:introduction}). 
   This is a reasonable approximation since, 
   although the equality does not hold in the Standard Model~\cite{Lenz:2011ti,*Lenz:2006hd}, 
   both are much smaller than the current experimental resolution, 
   whereas new physics contributions add a phase $\phi_{\rm NP}$ to $\phi_s$
   and subtract the same phase from $2\beta_s$, 
   such that the approximation remains valid.
}
The combination of results on $\Bs \to \jpsi \phi$ decays, including also results from $\Bs \to \jpsi \pi^+\pi^-$ and $\Bs \to D_s^+D_s^-$ decays, is performed by the HFAG Lifetimes and Oscillations group, see Sec.~\ref{sec:life_mix}.

\mysubsection{Time-dependent $\CP$ asymmetries in colour-suppressed $b \to c\bar{u}d$ transitions
}
\label{sec:cp_uta:cud_beta}

\mysubsubsection{Time-dependent $\CP$ asymmetries: $b \to c\bar{u}d$ decays to
  \CP eigenstates
}
\label{sec:cp_uta:cud_beta:cp}

Decays of $\B$ mesons to final states such as $D\pi^0$ are 
governed by $b \to c\bar{u}d$ transitions. 
If the final state is a $\CP$ eigenstate, \eg\ $D_{\CP}\pi^0$, 
the usual time-dependence formulae are recovered, 
with the sine coefficient sensitive to $\sin(2\beta)$. 
Since there is no penguin contribution to these decays, 
there is even less associated theoretical uncertainty 
than for $b \to c\bar{c}s$ decays such as $\B \to \jpsi \KS$.
Such measurements therefore allow to test the Standard Model prediction
that the $\CP$ violation parameters in $b \to c\bar{u}d$ transitions
are the same as those in $b \to c\bar{c}s$~\cite{Grossman:1996ke}.

Note that there is an additional contribution from CKM suppressed
$b \to u \bar{c} d$ decays.
The effect of this contribution is small, and can be taken into 
account in the analysis~\cite{Fleischer:2003ai,Fleischer:2003aj}.

Results of such an analysis are available from \babar~\cite{Aubert:2007mn}.
The decays $\Bz \to D\pi^0$, $\Bz \to D\eta$, $\Bz \to D\omega$,
$\Bz \to D^*\pi^0$ and $\Bz \to D^*\eta$ are used.
In the latter two modes, the daughter decay $D^* \to D\pi^0$ is used.
The $\CP$-even $D$ decay to $K^+K^-$ is used for all decay modes,
with the $\CP$-odd $D$ decay to $\KS\omega$ also used in $\Bz \to D^{(*)}\pi^0$
and the additional $\CP$-odd $D$ decay to $\KS\pi^0$ 
also used in $\Bz \to D\omega$.
Results are presented separately for $\CP$-even and $\CP$-odd 
$D^{(*)}$ decays (denoted $D^{(*)}_+ h^0$ and $D^{(*)}_- h^0$ respectively),
and for both combined, taking into account the different $\CP$ factors
(denoted $D^{(*)}_{\CP} h^0$).
The results are summarised in Table~\ref{tab:cp_uta:cud_cp_beta}.

\begin{table}[htb]
	\begin{center}
		\caption{
			Results from analyses of $\Bz \to D^{(*)}h^0$, $D \to CP$ eigenstates decays.
		}
		\vspace{0.2cm}
		\setlength{\tabcolsep}{0.0pc}
		\begin{tabular*}{\textwidth}{@{\extracolsep{\fill}}lrcccc} \hline
	\mc{2}{l}{Experiment} & $N(B\bar{B})$ & $S_{CP}$ & $C_{CP}$ & Correlation \\
	\hline
        \mc{6}{c}{$D^{(*)}_+ h^0$}  \\
	\babar & \cite{Aubert:2007mn} & 383M & $-0.65 \pm 0.26 \pm 0.06$ & $-0.33 \pm 0.19 \pm 0.04$ & $0.04$ \\
	\hline

        \mc{6}{c}{$D^{(*)}_- h^0$} \\
	\babar & \cite{Aubert:2007mn} & 383M & $-0.46 \pm 0.46 \pm 0.13$ & $-0.03 \pm 0.28 \pm 0.07$ & $-0.14$ \\
	\hline

        \mc{6}{c}{$D^{(*)}_{CP} h^0$} \\
	\babar & \cite{Aubert:2007mn} & 383M & $-0.56 \pm 0.23 \pm 0.05$ & $-0.23 \pm 0.16 \pm 0.04$ & $-0.02$ \\
	\hline
		\end{tabular*}
		\label{tab:cp_uta:cud_cp_beta}
	\end{center}
\end{table}

\mysubsubsection{Time-dependent Dalitz plot analyses of $b \to c\bar{u}d$ decays
}
\label{sec:cp_uta:cud_beta:dalitz}

When multibody $D$ decays, such as $D \to \KS\pi^+\pi^-$ are used, 
a time-dependent analysis of the Dalitz plot of the neutral $D$ decay 
allows for a direct determination of the weak phase: $2\beta$. 
(Equivalently, both $\sin(2\beta)$ and $\cos(2\beta)$ can be measured.)
This information can be used to resolve the ambiguity in the 
measurement of $2\beta$ from $\sin(2\beta)$~\cite{Bondar:2005gk}.

\begin{figure}[htbp]
  \begin{center}
    \begin{tabular}{cc}
      \resizebox{0.46\textwidth}{!}{
        \includegraphics{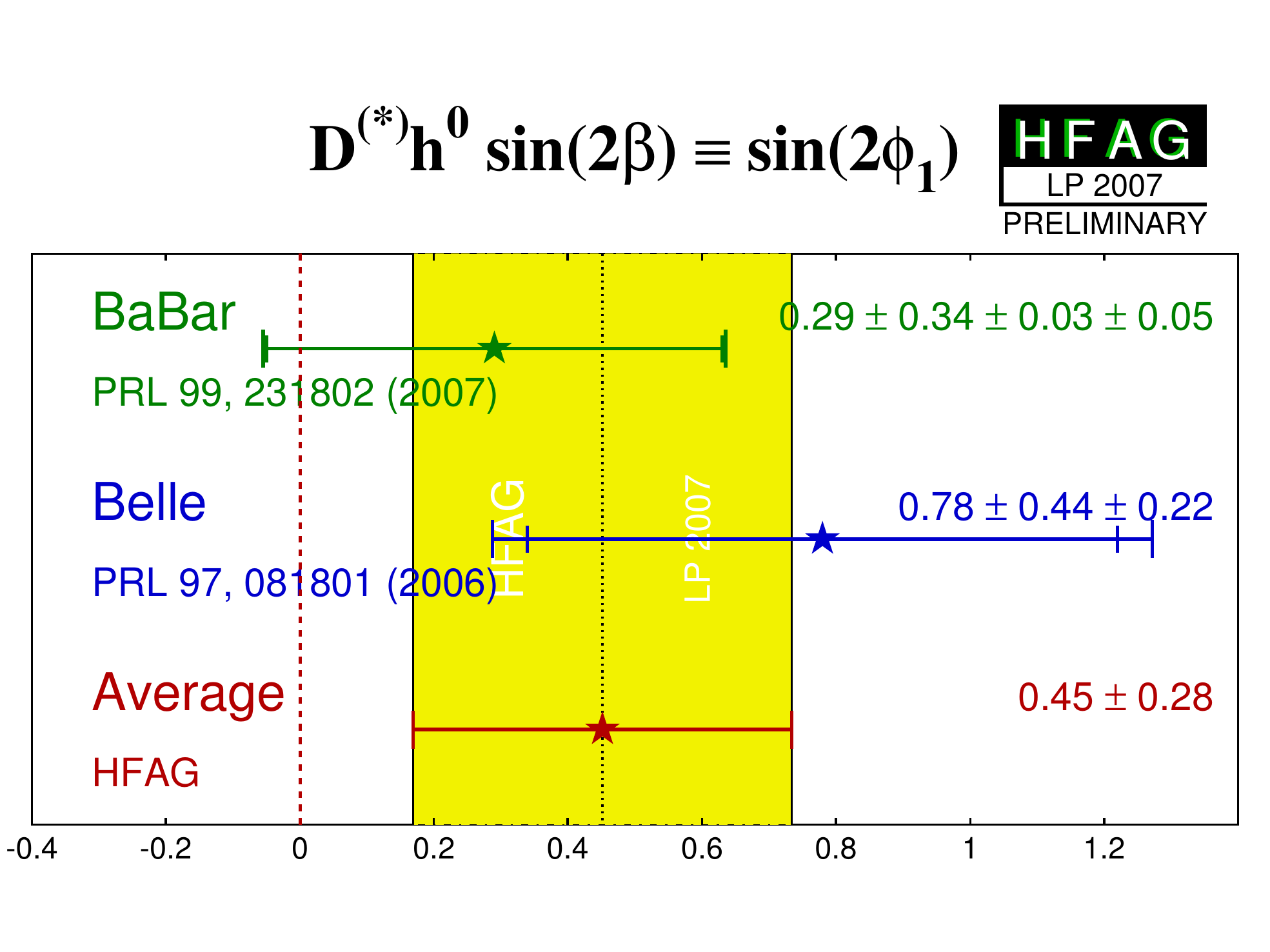}
      }
      &
      \resizebox{0.46\textwidth}{!}{
        \includegraphics{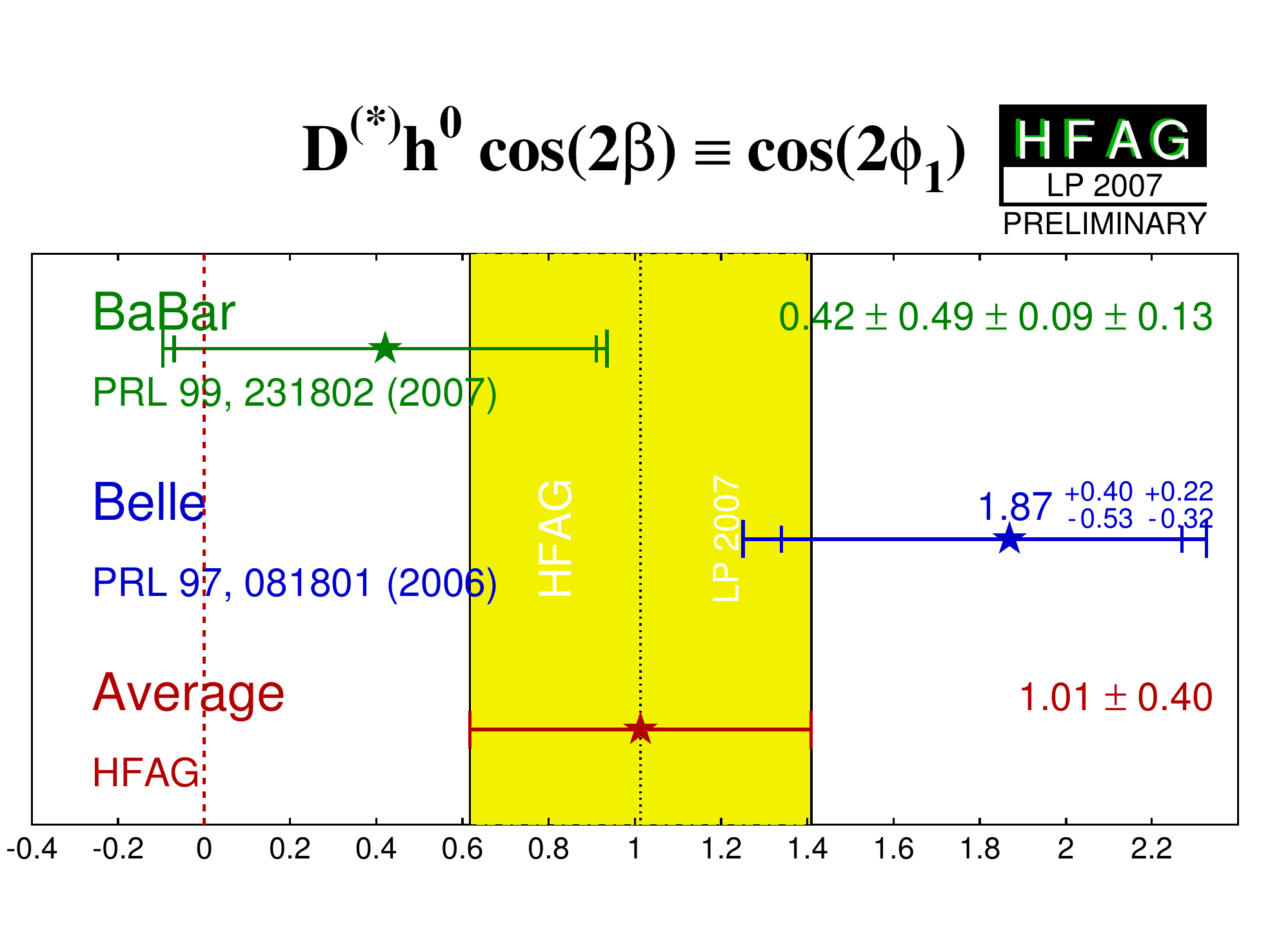}
      }
    \end{tabular}
  \end{center}
  \vspace{-0.8cm}
  \caption{
    Averages of 
    (left) $\sin(2\beta)$ and (right) $\cos(2\beta)$
    measured in colour-suppressed $b \to c\bar{u}d$ transitions.
  }
  \label{fig:cp_uta:cud_beta}
\end{figure}

Results of such analyses are available from both 
\belle~\cite{Krokovny:2006sv} and \babar~\cite{Aubert:2007rp}.
The decays $\B \to D\pi^0$, $\B \to D\eta$, $\B \to D\omega$, 
$\B \to D^*\pi^0$ and $\B \to D^*\eta$ are used. 
(This collection of states is denoted by $D^{(*)}h^0$.)
The daughter decays are $D^* \to D\pi^0$ and $D \to \KS\pi^+\pi^-$.
The results are shown in Table~\ref{tab:cp_uta:cud_beta},
and Fig.~\ref{fig:cp_uta:cud_beta}.
Note that \babar\ quote uncertainties due to the $D$ decay model 
separately from other systematic errors as a third source of uncertainty, while \belle\ do not.

\begin{table}[htb]
	\begin{center}
		\caption{
			Averages from $\Bz \to D^{(*)}h^0$, $D \to K_S\pi^+\pi^-$ analyses.
		}
		\vspace{0.2cm}
		\setlength{\tabcolsep}{0.0pc}
    \resizebox{\textwidth}{!}{
      		\begin{tabular*}{\textwidth}{@{\extracolsep{\fill}}lrcccc} \hline
	\mc{2}{l}{Experiment} & $N(B\bar{B})$ & $\sin 2\beta$ & $\cos 2\beta$ & $|\lambda|$ \\
		\hline
	\babar & \cite{Aubert:2007rp} & 383M & $0.29 \pm 0.34 \pm 0.03 \pm 0.05$ & $0.42 \pm 0.49 \pm 0.09 \pm 0.13$ & $1.01 \pm 0.08 \pm 0.02$ \\
	\belle & \cite{Krokovny:2006sv} & 386M & $0.78 \pm 0.44 \pm 0.22$ & $1.87 \,^{+0.40}_{-0.53} \,^{+0.22}_{-0.32}$ & \textendash{} \\
	\mc{3}{l}{\bf Average} & $0.45 \pm 0.28$ & $1.01 \pm 0.40$ & $1.01 \pm 0.08$ \\
	\mc{3}{l}{\small Confidence level} & {\small $0.59~(0.5\sigma)$} & {\small $0.12~(1.6\sigma)$} & \textendash{} \\
		\hline
		\end{tabular*}
    }
		\label{tab:cp_uta:cud_beta}
	\end{center}
\end{table}

Again, it is clear that the data prefer $\cos(2\beta)>0$.
Indeed, \belle~\cite{Krokovny:2006sv} 
determine the sign of $\cos(2\phi_1)$ to be positive at $98.3\%$ confidence level,
while \babar~\cite{Aubert:2007rp} 
favour the solution of $\beta$ with $\cos(2\beta)>0$ at $87\%$ confidence level.
Note, however, that the Belle measurement has strongly non-Gaussian behaviour. 
Therefore, we perform uncorrelated averages, 
from which any interpretation has to be done with the greatest care. 

\mysubsection{Time-dependent $\CP$ asymmetries in charmless $b \to q\bar{q}s$ transitions
}
\label{sec:cp_uta:qqs}

The flavour changing neutral current $b \to s$ penguin
can be mediated by any up-type quark in the loop, 
and hence the amplitude can be written as
\begin{equation}
  \label{eq:cp_uta:b_to_s}
  \begin{array}{ccccc}
    A_{b \to s} & = & 
    \mc{3}{l}{F_u V_{ub}V^*_{us} + F_c V_{cb}V^*_{cs} + F_t V_{tb}V^*_{ts}} \\
    & = & (F_u - F_c) V_{ub}V^*_{us} & + & (F_t - F_c) V_{tb}V^*_{ts} \\
    & = & {\cal O}(\lambda^4) & + & {\cal O}(\lambda^2) \\
  \end{array}
\end{equation}
using the unitarity of the CKM matrix.
Therefore, in the Standard Model, 
this amplitude is dominated by $V_{tb}V^*_{ts}$, 
and to within a few degrees 
($\delta\beta \lesssim 2^\circ$ for $\beta \approx 20^\circ$) 
the time-dependent parameters can be written\footnote
{
  The presence of a small (${\cal O}(\lambda^2)$) weak phase in 
  the dominant amplitude of the $s$ penguin decays introduces 
  a phase shift given by
  $S_{b \to q\bar q s} = -\eta\sin(2\beta)\cdot(1 + \Delta)$. 
  Using the CKMfitter results for the Wolfenstein 
  parameters~\cite{Charles:2004jd}, one finds: 
  $\Delta \simeq 0.033$, which corresponds to a shift of 
  $2\beta$ of $+2.1$ degrees. 
  Nonperturbative contributions can alter this result.
}
$S_{b \to q\bar q s} \approx - \etacp \sin(2\beta)$,
$C_{b \to q\bar q s} \approx 0$,
assuming $b \to s$ penguin contributions only ($q = u,d,s$).

Due to the suppression of the Standard Model amplitude, contributions of additional diagrams from physics beyond the Standard Model,
with heavy virtual particles in the penguin loops, may have observable effects.
In general, these contributions will affect the values of 
$S_{b \to q\bar q s}$ and $C_{b \to q\bar q s}$.
A discrepancy between the values of 
$S_{b \to c\bar c s}$ and $S_{b \to q\bar q s}$
can therefore provide a clean indication of new physics~\cite{Grossman:1996ke,Fleischer:1996bv,London:1997zk,Ciuchini:1997zp}.

However, there is an additional consideration to take into account.
The above argument assumes that only the $b \to s$ penguin contributes
to the $b \to q\bar q s$ transition.
For $q = s$ this is a good assumption, which neglects only rescattering effects.
However, for $q = u$ there is a colour-suppressed $b \to u$ tree diagram
(of order ${\cal O}(\lambda^4)$), 
which has a different weak (and possibly strong) phase.
In the case $q = d$, any light neutral meson that is formed from $d \bar{d}$ 
also has a $u \bar{u}$ component, and so again there is ``tree pollution''. 
The \Bz decays to $\piz\KS$, $\rho^0\KS$ and $\omega\KS$ belong to this category.
The mesons $\phi$, $f_0$ and $\etapr$ are expected to have predominant
$s\bar{s}$ parts, which reduces the relative size of the possible tree
pollution. 
If the inclusive decay $\Bz\to\Kp\Km\Kz$ (excluding $\phi\Kz$) is dominated by
a nonresonant three-body transition, 
an OZI-rule suppressed tree-level diagram can occur 
through insertion of an $s\sbar$ pair. 
The corresponding penguin-type transition 
proceeds via insertion of a $u\ubar$ pair, which is expected
to be favoured over the $s\sbar$ insertion by fragmentation models.
Neglecting rescattering, the final state $\Kz\Kzb\Kz$ 
(reconstructed as $\KS\KS\KS$) has no tree pollution~\cite{Gershon:2004tk}.
Various estimates, using different theoretical approaches,
of the values of $\Delta S = S_{b \to q\bar q s} - S_{b \to c\bar c s}$
exist in the literature~\cite{Grossman:2003qp,Gronau:2003ep,Gronau:2003kx,Gronau:2004hp,Cheng:2005bg,Gronau:2005gz,Buchalla:2005us,Beneke:2005pu,Engelhard:2005hu,Cheng:2005ug,Engelhard:2005ky,Gronau:2006qh,Silvestrini:2007yf,Dutta:2008xw}.
In general, there is agreement that the modes
$\phi\Kz$, $\etapr\Kz$ and $\Kz\Kzb\Kz$ are the cleanest,
with values of $\left| \Delta S \right|$ at or below the few percent level 
($\Delta S$ is usually positive).

\mysubsubsection{Time-dependent $\CP$ asymmetries: $b \to q\bar{q}s$ decays to $\CP$ eigenstates
}
\label{sec:cp_uta:qqs:cp_eigen}

\begin{figure}[htbp]
  \begin{center}
    \resizebox{0.45\textwidth}{!}{
      \includegraphics{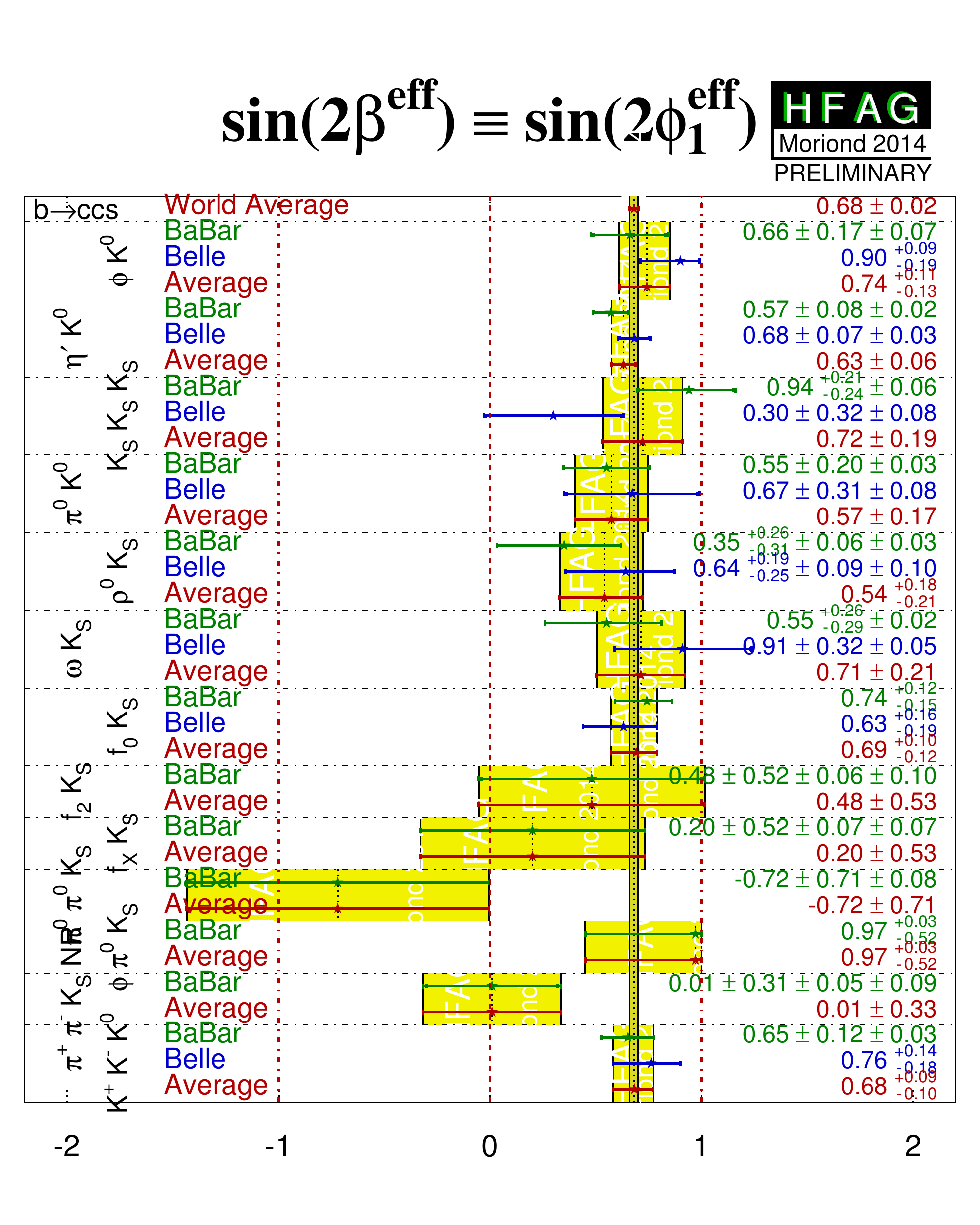}
    }
    \hfill
    \resizebox{0.45\textwidth}{!}{
      \includegraphics{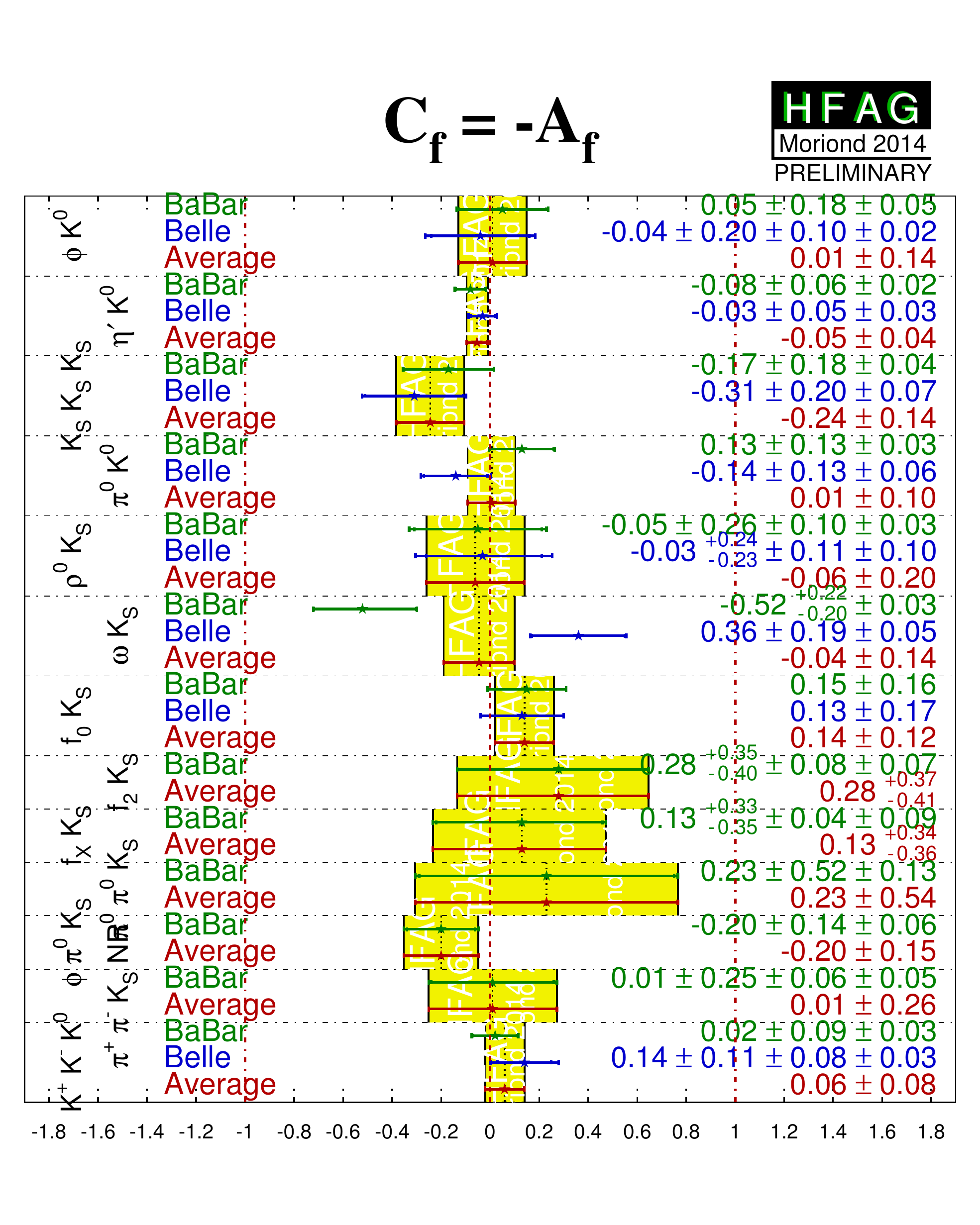}
    }
    \\
    \resizebox{0.45\textwidth}{!}{
      \includegraphics{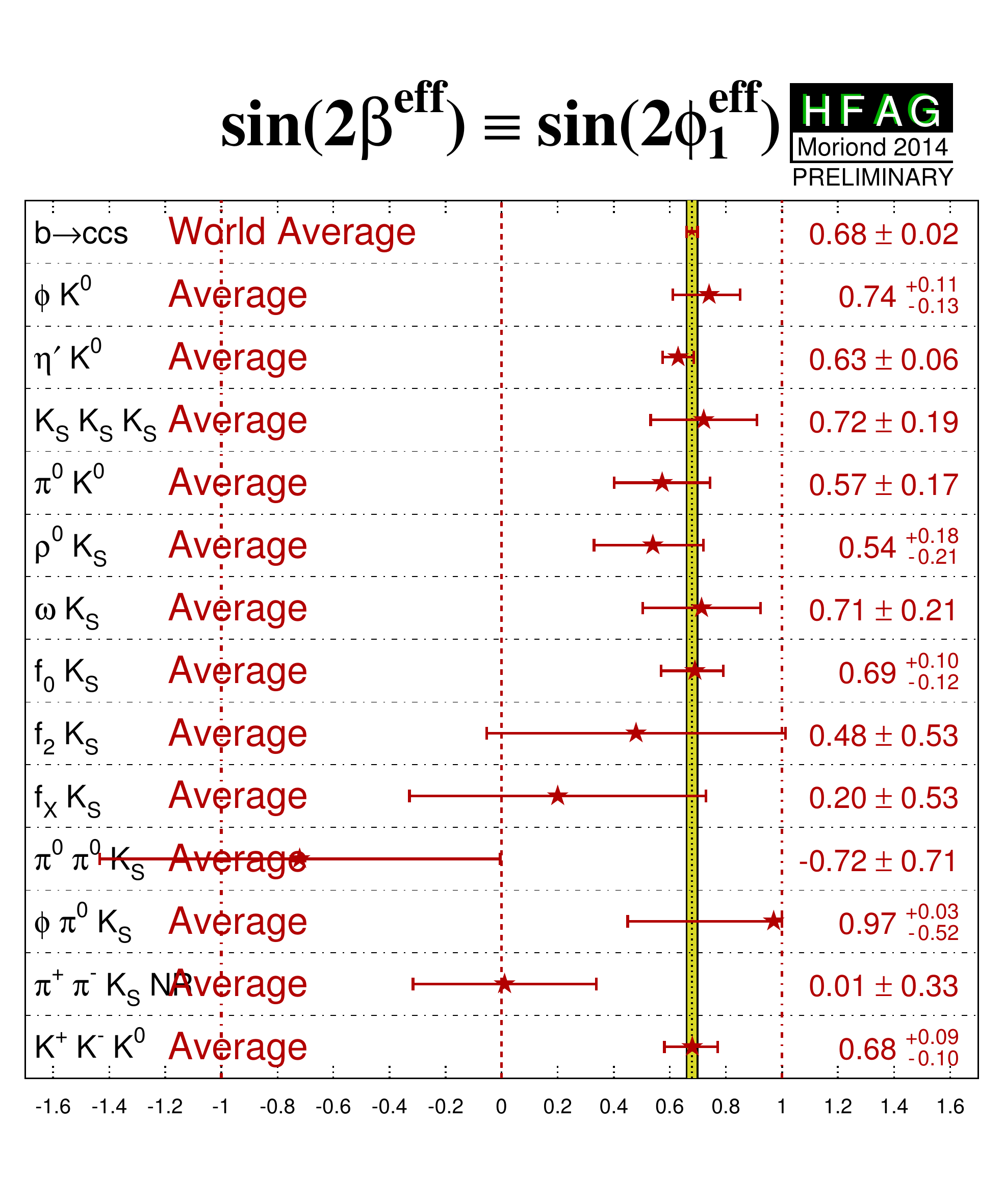}
    }
    \hfill
    \resizebox{0.45\textwidth}{!}{
      \includegraphics{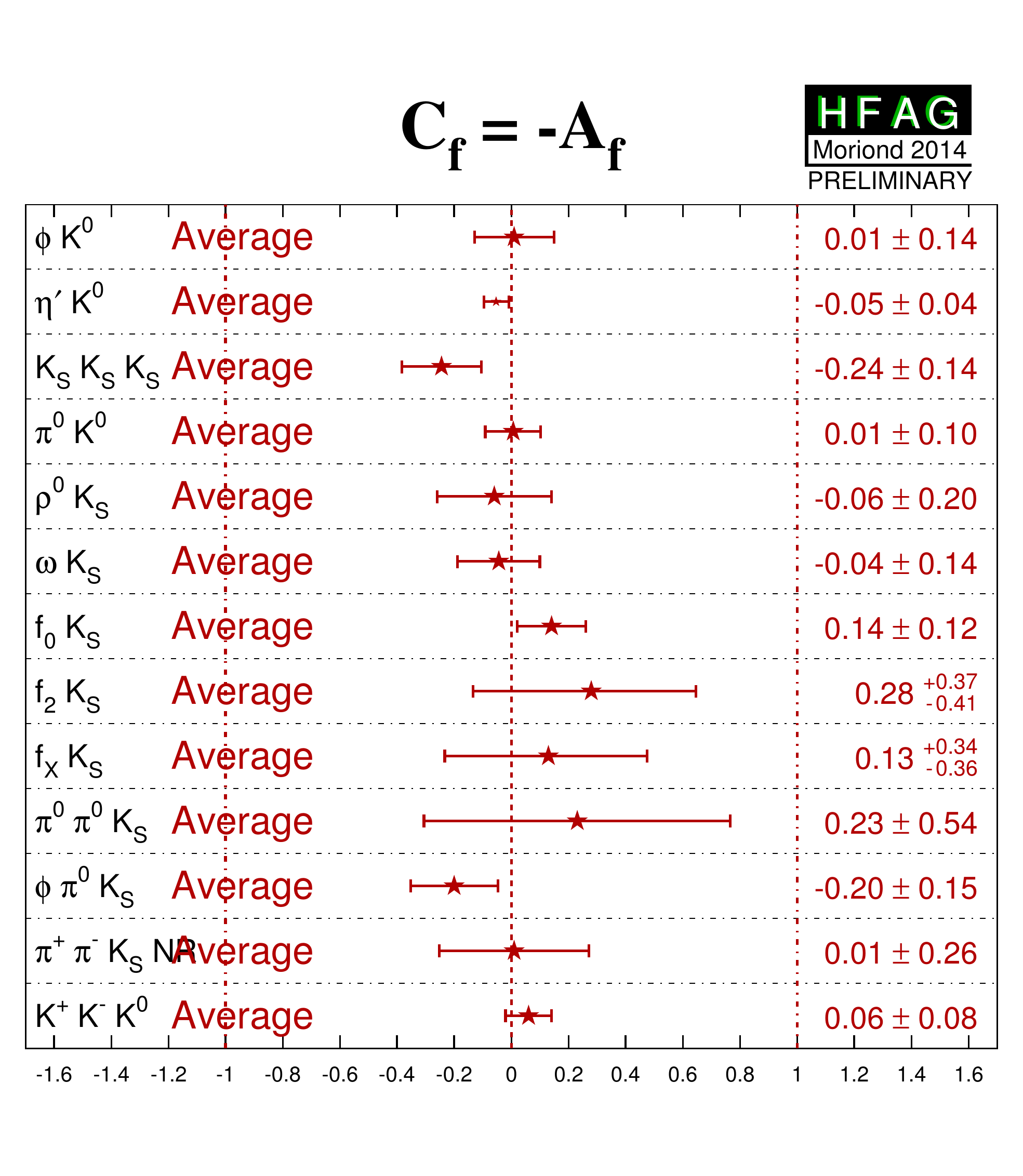}
    }
  \end{center}
  \vspace{-0.8cm}
  \caption{
    (Top)
    Averages of 
    (left) $-\etacp S_{b \to q\bar q s}$ and (right) $C_{b \to q\bar q s}$.
    The $-\etacp S_{b \to q\bar q s}$ figure compares the results to 
    the world average 
    for $-\etacp S_{b \to c\bar c s}$ (see Sec.~\ref{sec:cp_uta:ccs:cp_eigen}).
    (Bottom) Same, but only averages for each mode are shown.
    More figures are available from the HFAG web pages.
  }
  \label{fig:cp_uta:qqs}
\end{figure}

\begin{figure}[htbp]
  \begin{center}
    \resizebox{0.33\textwidth}{!}{
      \includegraphics{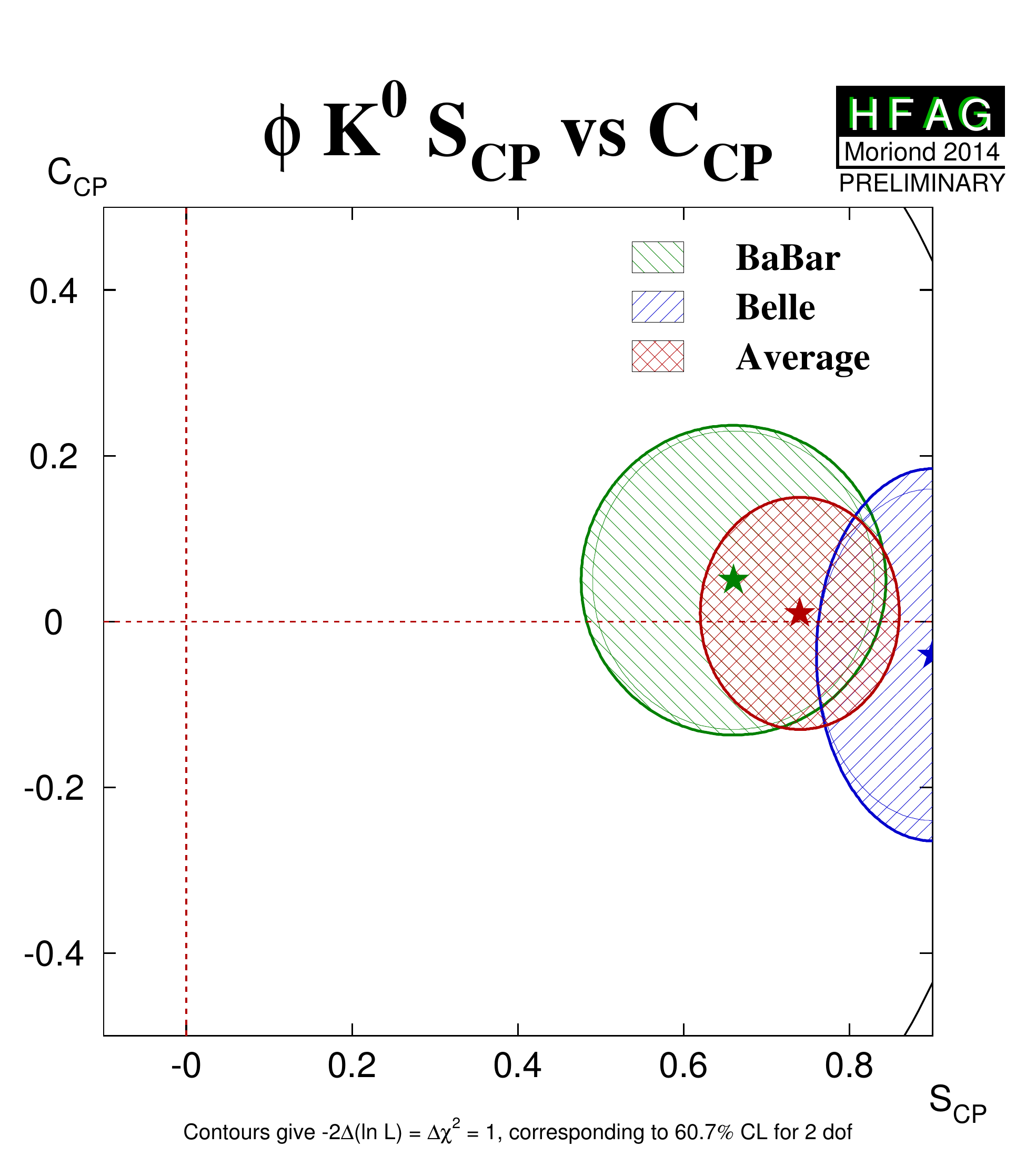}
    }
    \hspace{0.08\textwidth}
    \resizebox{0.33\textwidth}{!}{
      \includegraphics{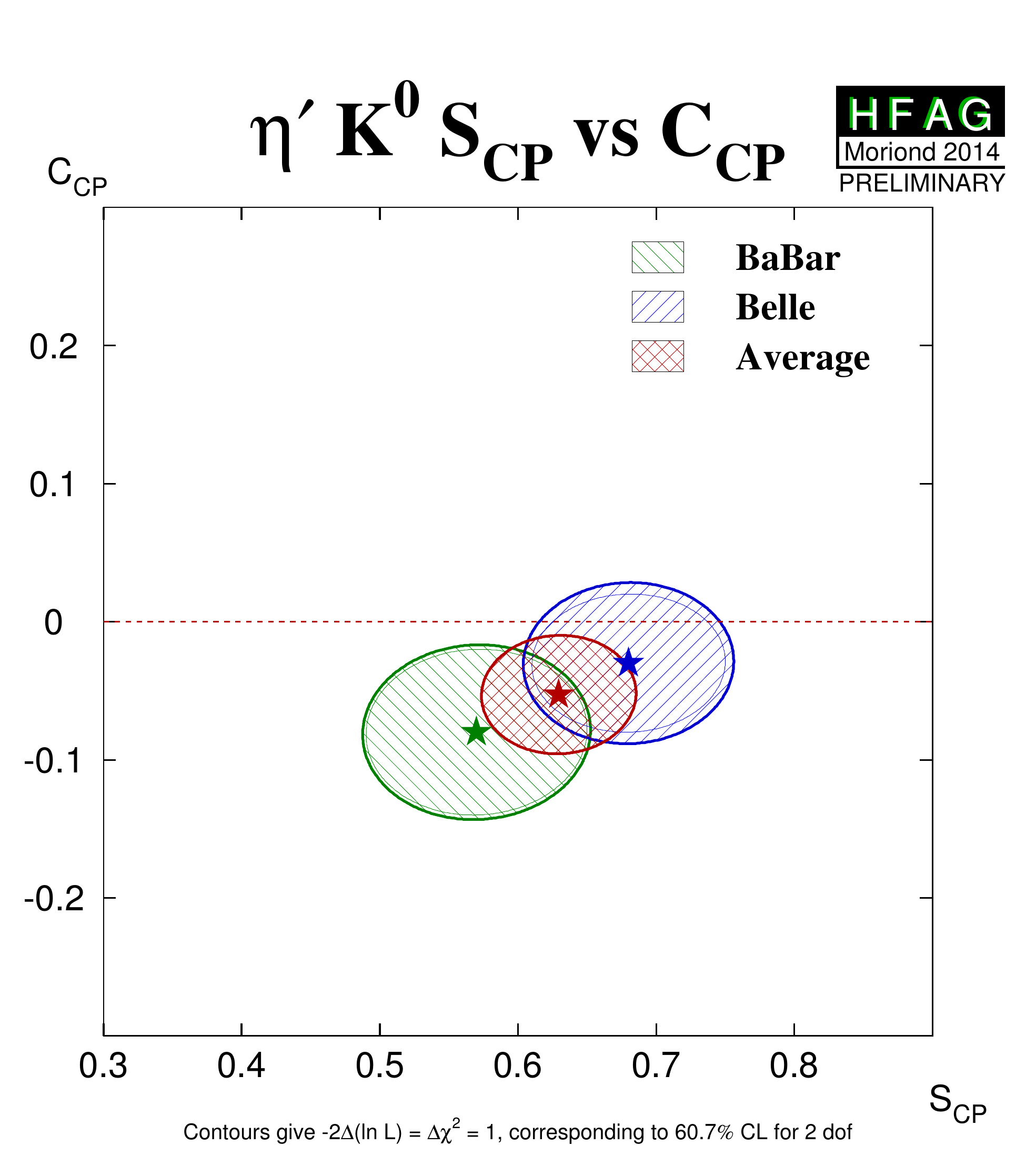}
    }
    \\
    \resizebox{0.33\textwidth}{!}{
      \includegraphics{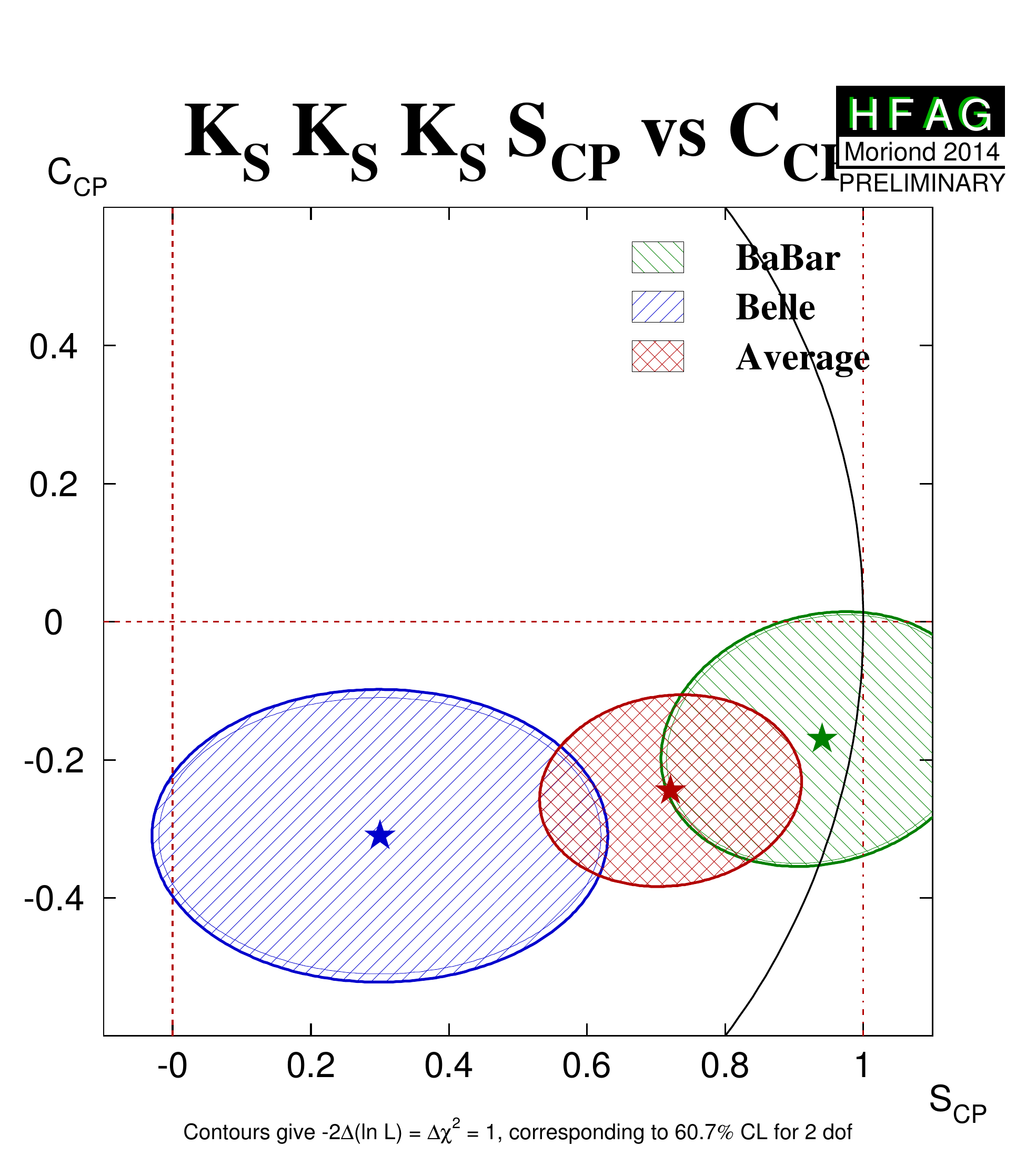}
    }
    \hspace{0.08\textwidth}
    \resizebox{0.33\textwidth}{!}{
      \includegraphics{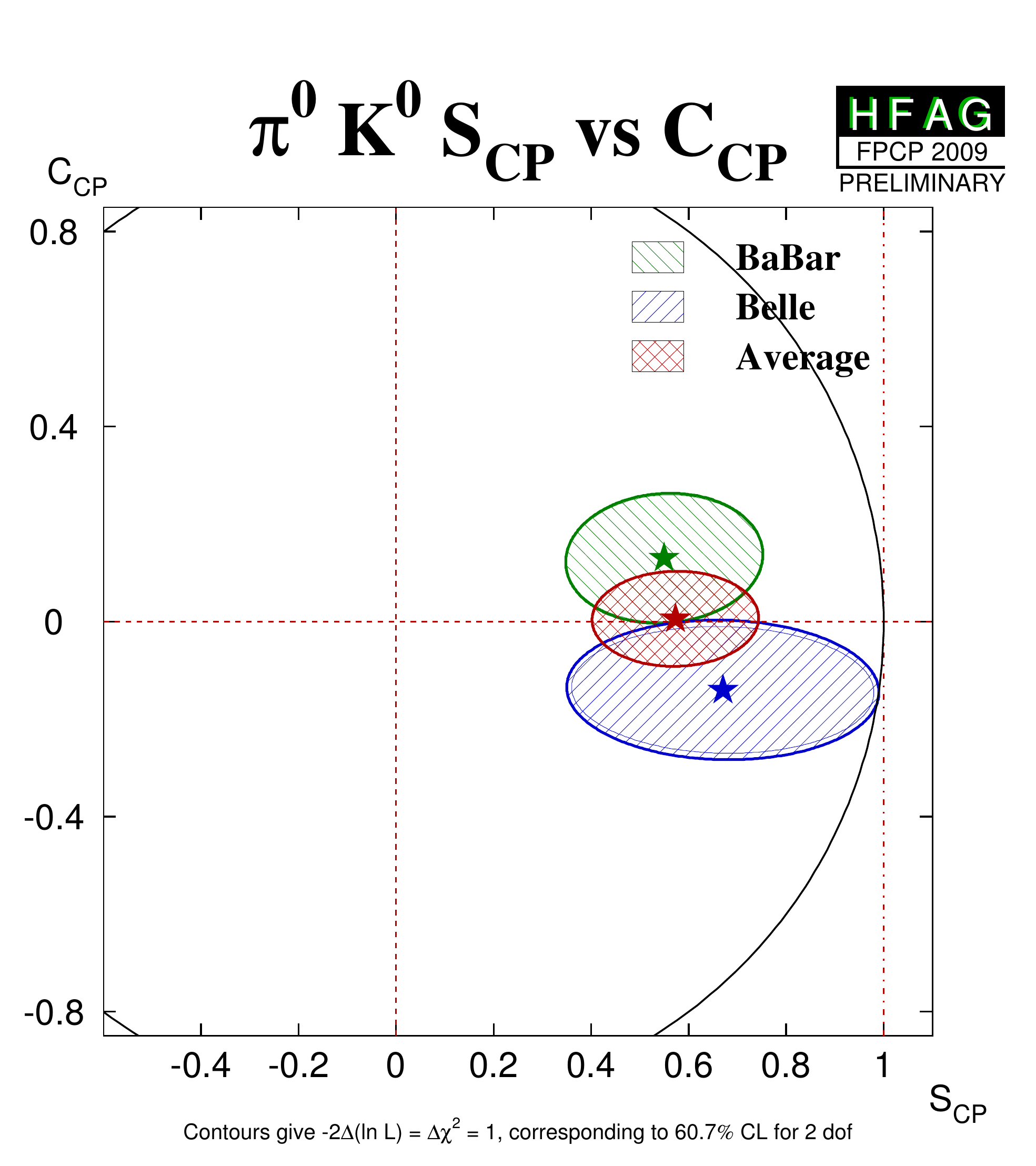}
    }
  \end{center}
  \vspace{-0.5cm}
  \caption{
    Averages of four $b \to q\bar q s$ dominated channels,
    for which correlated averages are performed,
    in the $S_{\CP}$ \vs\ $C_{\CP}$ plane,
    where $S_{\CP}$ has been corrected by the $\CP$ eigenvalue to give
    $\sin(2\beta^{\rm eff})$.
    (Top left) $\Bz \to \phi\Kz$,
    (top right) $\Bz \to \eta^\prime\Kz$,
    (bottom left) $\Bz \to \KS\KS\KS$,
    (bottom right) $\Bz \to \pi^0\KS$.
    More figures are available from the HFAG web pages.
  }
  \label{fig:cp_uta:qqs_SvsC}
\end{figure}

\begin{figure}[htbp]
  \begin{center}
    \resizebox{0.66\textwidth}{!}{
      \includegraphics{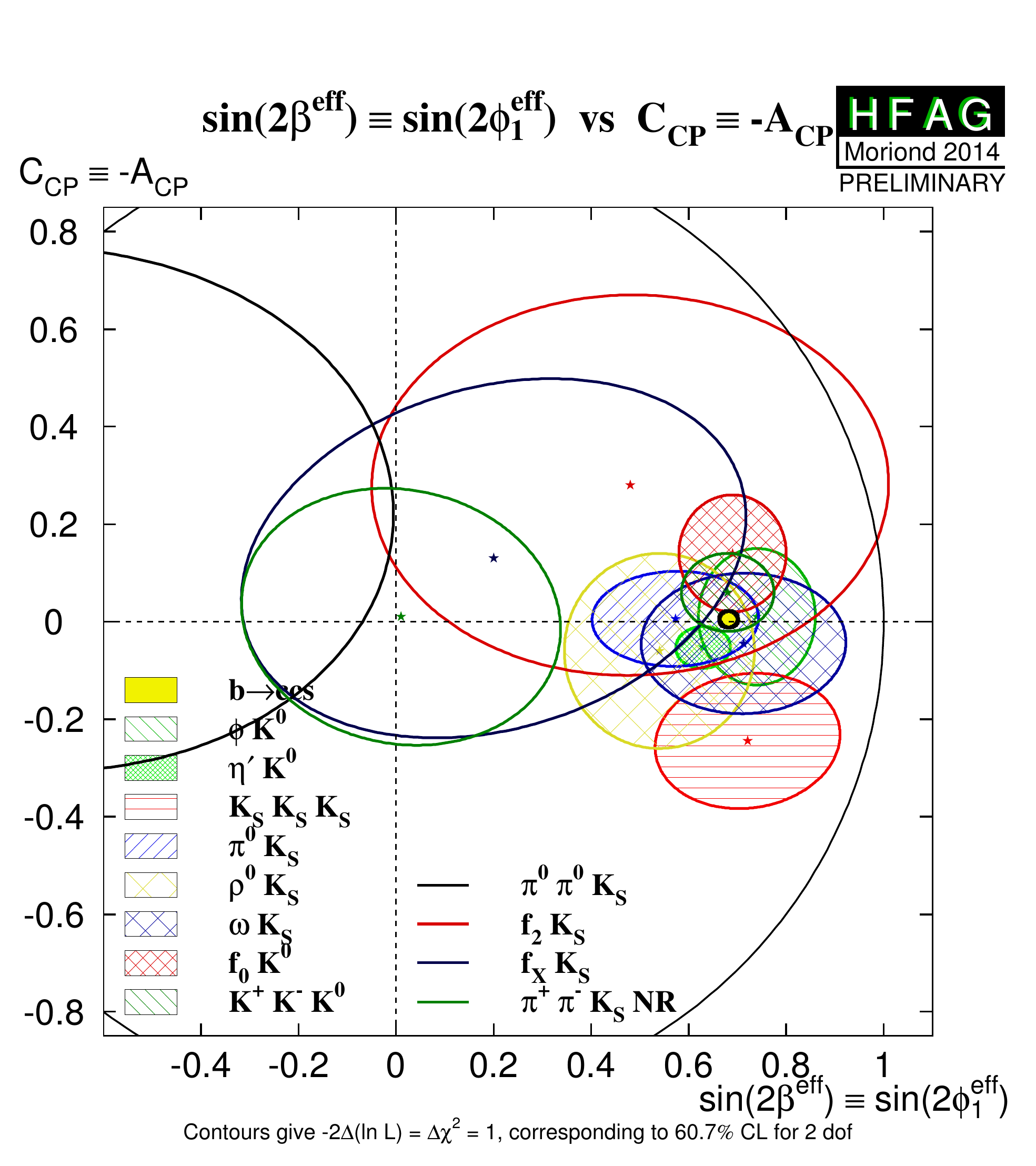}
    }
  \end{center}
  \vspace{-0.8cm}
  \caption{
    Compilation of constraints in the 
    $-\etacp S_{b \to q\bar q s}$ \vs\ $C_{b \to q\bar q s}$ plane.
  }
  \label{fig:cp_uta:qqs_SvsC-all}
\end{figure}

The averages for $-\etacp S_{b \to q\bar q s}$ and $C_{b \to q\bar q s}$
can be found in Table~\ref{tab:cp_uta:qqs},
and are shown in Figs.~\ref{fig:cp_uta:qqs},~\ref{fig:cp_uta:qqs_SvsC} 
and~\ref{fig:cp_uta:qqs_SvsC-all}.
Results from both \babar\  and \belle\ are averaged for the modes
$\etapr\Kz$ ($\Kz$ indicates that both $\KS$ and $\KL$ are used)
$\KS\KS\KS$, $\pi^0 \KS$ and $\omega\KS$.\footnote{
  \belle~\cite{Fujikawa:2008pk} include the $\pi^0\KL$ final state together with $\pi^0 \KS$ in order to improve the constraint on the parameter of \CP\ violation in decay; these events cannot be used for time-dependent analysis.
}
Results on $\phi\KS$ and $\Kp\Km\KS$ (implicitly excluding $\phi\KS$ and $f_0\KS$) are taken from time-dependent Dalitz plot analyses of $\Kp\Km\KS$;
results on $\rho^0\KS$, $f_2\KS$, $f_{\rm X}\KS$ and $\pip\pim\KS$ nonresonant are taken from time-dependent Dalitz plot analyses of $\pip\pim\KS$ (see
subsection~\ref{sec:cp_uta:qqs:dp}).
The results on $f_0\KS$ are from combinations of both Dalitz plot analyses.
\babar\ has also presented results with the final states
$\pi^0\pi^0\KS$,\footnote{
  We do not include a preliminary result from \belle~\cite{:2007xd}, which
  remains unpublished after more than two years.
}
and $\phi \KS \pi^0$. 

Of these final states,
$\phi\KS$, $\etapr\KS$, $\pi^0 \KS$, $\rho^0\KS$, $\omega\KS$ and $f_0\KL$
have $\CP$ eigenvalue $\etacp = -1$, 
while $\phi\KL$, $\etapr\KL$, $\KS\KS\KS$, $f_0 \KS$, $f_2 \KS$, 
$f_{\rm X} \KS$,\footnote{ 
  The $f_{\rm X}$ is assumed to be spin even.
} $\pi^0\pi^0\KS$ and $\pi^+ \pi^- \KS$ nonresonant have $\etacp = +1$.
The final state $K^+K^-\KS$ (with $\phi\KS$ and $f_0\KS$ implicitly excluded)
is not a $\CP$ eigenstate, but the \CP-content can be absorbed in the amplitude analysis to allow the determination of a single effective $S$ parameter.
(In earlier analyses of the $K^+K^-\Kz$ final state,
its $\CP$ composition was determined using an isospin argument~\cite{Abe:2006gy}
and a moments analysis~\cite{Aubert:2005ja}.)

\begin{table}[!htb]
	\begin{center}
		\caption{
      Averages of $-\etacp S_{b \to q\bar q s}$ and $C_{b \to q\bar q s}$.
      Where a third source of uncertainty is given, it is due to model
      uncertainties arising in Dalitz plot analyses.
		}
		\vspace{0.2cm}
    \resizebox{\textwidth}{!}{
		\begin{tabular}{@{\extracolsep{2mm}}lrccc@{\hspace{-3pt}}c} \hline
        \mc{2}{l}{Experiment} & $N(B\bar{B})$ & $- \etacp S_{b \to q\bar q s}$ & $C_{b \to q\bar q s}$ & Correlation \\
	\hline
      \mc{6}{c}{$\phi \Kz$} \\
	\babar & \cite{Lees:2012kx} & 470M & $0.66 \pm 0.17 \pm 0.07$ & $0.05 \pm 0.18 \pm 0.05$ & \textendash{} \\
	\belle & \cite{Nakahama:2010nj} & 657M & $0.90 \,^{+0.09}_{-0.19}$ & $-0.04 \pm 0.20 \pm 0.10 \pm 0.02$ & \textendash{} \\
	\mc{3}{l}{\bf Average} & $0.74 \,^{+0.11}_{-0.13}$ & $0.01 \pm 0.14$ & {\small uncorrelated averages} \\
		\hline

      \mc{6}{c}{$\etapr \Kz$} \\
	\babar & \cite{:2008se} & 467M & $0.57 \pm 0.08 \pm 0.02$ & $-0.08 \pm 0.06 \pm 0.02$ & $0.03$ \\
	\belle & \cite{Santelj:2014sja} & 772M & $0.68 \pm 0.07 \pm 0.03$ & $-0.03 \pm 0.05 \pm 0.03$ & $0.03$ \\
	\mc{3}{l}{\bf Average} & $0.63 \pm 0.06$ & $-0.05 \pm 0.04$ & $0.02$ \\
	\mc{3}{l}{\small Confidence level} & \mc{2}{c}{\small $0.53~(0.6\sigma)$} & \\
		\hline

      \mc{6}{c}{$\KS\KS\KS$} \\
	\babar & \cite{Lees:2011nf} & 468M & $0.94 \,^{+0.21}_{-0.24} \pm 0.06$ & $-0.17 \pm 0.18 \pm 0.04$ & $0.16$ \\
	\belle & \cite{Chen:2006nk} & 535M & $0.30 \pm 0.32 \pm 0.08$ & $-0.31 \pm 0.20 \pm 0.07$ & \textendash{} \\
	\mc{3}{l}{\bf Average} & $0.72 \pm 0.19$ & $-0.24 \pm 0.14$ & $0.09$ \\
	\mc{3}{l}{\small Confidence level} & \mc{2}{c}{\small $0.26~(1.1\sigma)$} & \\
		\hline

      \mc{6}{c}{$\pi^0 K^0$} \\
	\babar & \cite{:2008se} & 467M & $0.55 \pm 0.20 \pm 0.03$ & $0.13 \pm 0.13 \pm 0.03$ & $0.06$ \\
	\belle & \cite{Fujikawa:2008pk} & 657M & $0.67 \pm 0.31 \pm 0.08$ & $-0.14 \pm 0.13 \pm 0.06$ & $-0.04$ \\
	\mc{3}{l}{\bf Average} & $0.57 \pm 0.17$ & $0.01 \pm 0.10$ & $0.02$ \\
	\mc{3}{l}{\small Confidence level} & \mc{2}{c}{\small $0.37~(0.9\sigma)$} & \\
		\hline

		\hline
      \mc{6}{c}{$\rho^0 \KS$} \\
	\babar & \cite{Aubert:2009me} & 383M & $0.35 \,^{+0.26}_{-0.31} \pm 0.06 \pm 0.03$ & $-0.05 \pm 0.26 \pm 0.10 \pm 0.03$ & \textendash{} \\
	\belle & \cite{:2008wwa} & 657M & $0.64 \,^{+0.19}_{-0.25} \pm 0.09 \pm 0.10$ & $-0.03 \,^{+0.24}_{-0.23} \pm 0.11 \pm 0.10$ & \textendash{} \\
	\mc{3}{l}{\bf Average} & $0.54 \,^{+0.18}_{-0.21}$ & $-0.06 \pm 0.20$ & {\small uncorrelated averages} \\
		\hline

      \mc{6}{c}{$\omega \KS$} \\
	\babar & \cite{:2008se} & 467M & $0.55 \,^{+0.26}_{-0.29} \pm 0.02$ & $-0.52 \,^{+0.22}_{-0.20} \pm 0.03$ & $0.03$ \\
	\belle & \cite{Chobanova:2013ddr} & 772M & $0.91 \pm 0.32 \pm 0.05$ & $0.36 \pm 0.19 \pm 0.05$ & $-0.00$ \\
	\mc{3}{l}{\bf Average} & $0.71 \pm 0.21$ & $-0.04 \pm 0.14$ & $0.01$ \\
	\mc{3}{l}{\small Confidence level} & \mc{2}{c}{\small $0.007~(2.7\sigma)$} & \\
		\hline

      \mc{6}{c}{$f_0 \Kz$} \\
	\babar & \cite{Lees:2012kx,Aubert:2009me} & \textendash{} & $0.74 \,^{+0.12}_{-0.15}$ & $0.15 \pm 0.16$ & \textendash{} \\
	\belle & \cite{Nakahama:2010nj,:2008wwa} & \textendash{} & $0.63 \,^{+0.16}_{-0.19}$ & $0.13 \pm 0.17$ & \textendash{} \\
	\mc{3}{l}{\bf Average} & $0.69 \,^{+0.10}_{-0.12}$ & $0.14 \pm 0.12$ & {\small uncorrelated averages} \\
		\hline

      \mc{6}{c}{$f_2 \KS$} \\
	\babar & \cite{Aubert:2009me} & 383M & $0.48 \pm 0.52 \pm 0.06 \pm 0.10$ & $0.28 \,^{+0.35}_{-0.40} \pm 0.08 \pm 0.07$ & \textendash{} \\
		\hline

      \mc{6}{c}{$f_{\rm X} \KS$} \\
	\babar & \cite{Aubert:2009me} & 383M & $0.20 \pm 0.52 \pm 0.07 \pm 0.07$ & $0.13 \,^{+0.33}_{-0.35} \pm 0.04 \pm 0.09$ & \textendash{} \\
		\hline

 		\end{tabular}
}
		\label{tab:cp_uta:qqs}
	\end{center}
\end{table}

\begin{table}[!htb]
	\begin{center}
		\caption{
      Averages of $-\etacp S_{b \to q\bar q s}$ and $C_{b \to q\bar q s}$ (continued).
      Where a third source of uncertainty is given, it is due to model
      uncertainties arising in Dalitz plot analyses.
		}
		\vspace{0.2cm}
		\setlength{\tabcolsep}{0.0pc}
		\begin{tabular*}{\textwidth}{@{\extracolsep{\fill}}lrccc@{\hspace{-3pt}}c} \hline
        \mc{2}{l}{Experiment} & $N(B\bar{B})$ & $- \etacp S_{b \to q\bar q s}$ & $C_{b \to q\bar q s}$ & Correlation \\
	\hline
      \mc{6}{c}{$\pi^0 \pi^0 \KS$} \\
	\babar & \cite{Aubert:2007ub} & 227M & $-0.72 \pm 0.71 \pm 0.08$ & $0.23 \pm 0.52 \pm 0.13$ & $-0.02$ \\
		\hline

      \mc{6}{c}{$\phi \KS \pi^0$} \\
	\babar & \cite{Aubert:2008zza} & 465M & $0.97 \,^{+0.03}_{-0.52}$ & $-0.20 \pm 0.14 \pm 0.06$ & \textendash{} \\
 		\hline

      \mc{6}{c}{$\pi^+ \pi^- \KS$ nonresonant} \\
	\babar & \cite{Aubert:2009me} & 383M & $0.01 \pm 0.31 \pm 0.05 \pm 0.09$ & $0.01 \pm 0.25 \pm 0.06 \pm 0.05$ & \textendash{} \\
 		\hline

      \mc{6}{c}{$K^+K^- \Kz$} \\
	\babar & \cite{Lees:2012kx} & 470M & $0.65 \pm 0.12 \pm 0.03$ & $0.02 \pm 0.09 \pm 0.03$ & \textendash{} \\
	\belle & \cite{Nakahama:2010nj} & 657M & $0.76 \,^{+0.14}_{-0.18}$ & $0.14 \pm 0.11 \pm 0.08 \pm 0.03$ & \textendash{} \\
	\mc{3}{l}{\bf Average} & $0.68 \,^{+0.09}_{-0.10}$ & $0.06 \pm 0.08$ & {\small uncorrelated averages} \\
		\hline



		\hline
		\end{tabular*}
		\label{tab:cp_uta:qqs2}
	\end{center}
\end{table}

The final state $\phi \KS \pi^0$ is also not a \CP eigenstate but its
\CP-composition can be determined from an angular analysis.
Since the parameters are common to the $\Bz\to\phi \KS \pi^0$ and
$\Bz\to \phi \Kp\pim$ decays (because only $K\pi$ resonances contribute),
\babar\ perform a simultaneous analysis of the two final
states~\cite{Aubert:2008zza} (see Sec.~\ref{sec:cp_uta:qqs:vv}).

It must be noted that Q2B parameters extracted from Dalitz plot analyses 
are constrained to lie within the physical boundary ($S_{\CP}^2 + C_{\CP}^2 < 1$)
and consequently the obtained errors are highly non-Gaussian when
the central value is close to the boundary.  
This is particularly evident in the \babar\ results for 
$\Bz \to f_0\Kz$ with $f_0 \to \pi^+\pi^-$~\cite{Aubert:2009me}.
These results must be treated with extreme caution.

As explained above,
each of the modes listed in Table~\ref{tab:cp_uta:qqs} has
potentially different subleading contributions within the Standard Model,
and thus each may have a different value of $-\etacp S_{b \to q\bar q s}$.
Therefore, there is no strong motivation to make a combined average
over the different modes.
We refer to such an average as a ``na\"\i ve $s$-penguin average.''
It is na\"\i ve not only because of the theoretical uncertainties are neglected,
but also since possible correlations of systematic effects 
between different modes are not included.
In spite of these caveats there remains interest in the value of this quantity
and therefore it is given here:
$\langle -\etacp S_{b \to q\bar q s} \rangle = 0.655 \pm 0.032$,
with confidence level $0.77~(0.3\sigma)$.
This value is in agreement with the average 
$-\etacp S_{b \to c\bar c s}$ given in Sec.~\ref{sec:cp_uta:ccs:cp_eigen}.
(The average for $C_{b \to q\bar q s}$ is 
$\langle C_{b \to q\bar q s} \rangle = -0.006 \pm 0.026$
with confidence level $0.53~(0.6\sigma)$.)
We emphasise again that we do not advocate the use of these averages.

From Table~\ref{tab:cp_uta:qqs} it may be noted 
that the averages for $-\etacp S_{b \to q\bar q s}$ in 
$\phi\KS$, $\etapr \Kz$, $f_0\KS$ and $\Kp\Km\KS$
are all now more than $5\sigma$ away from zero, 
so that $\CP$ violation in these modes can be considered well established.
There is no evidence (above $2\sigma$) for $\CP$ violation in any $b \to q \bar q s$ decay.

\mysubsubsection{Time-dependent Dalitz plot analyses: $\Bz \to K^+K^-\Kz$ and $\Bz \to \pi^+\pi^-\KS$}
\label{sec:cp_uta:qqs:dp}

\begin{figure}[htbp]
  \begin{center}
    \resizebox{0.45\textwidth}{!}{
      \includegraphics{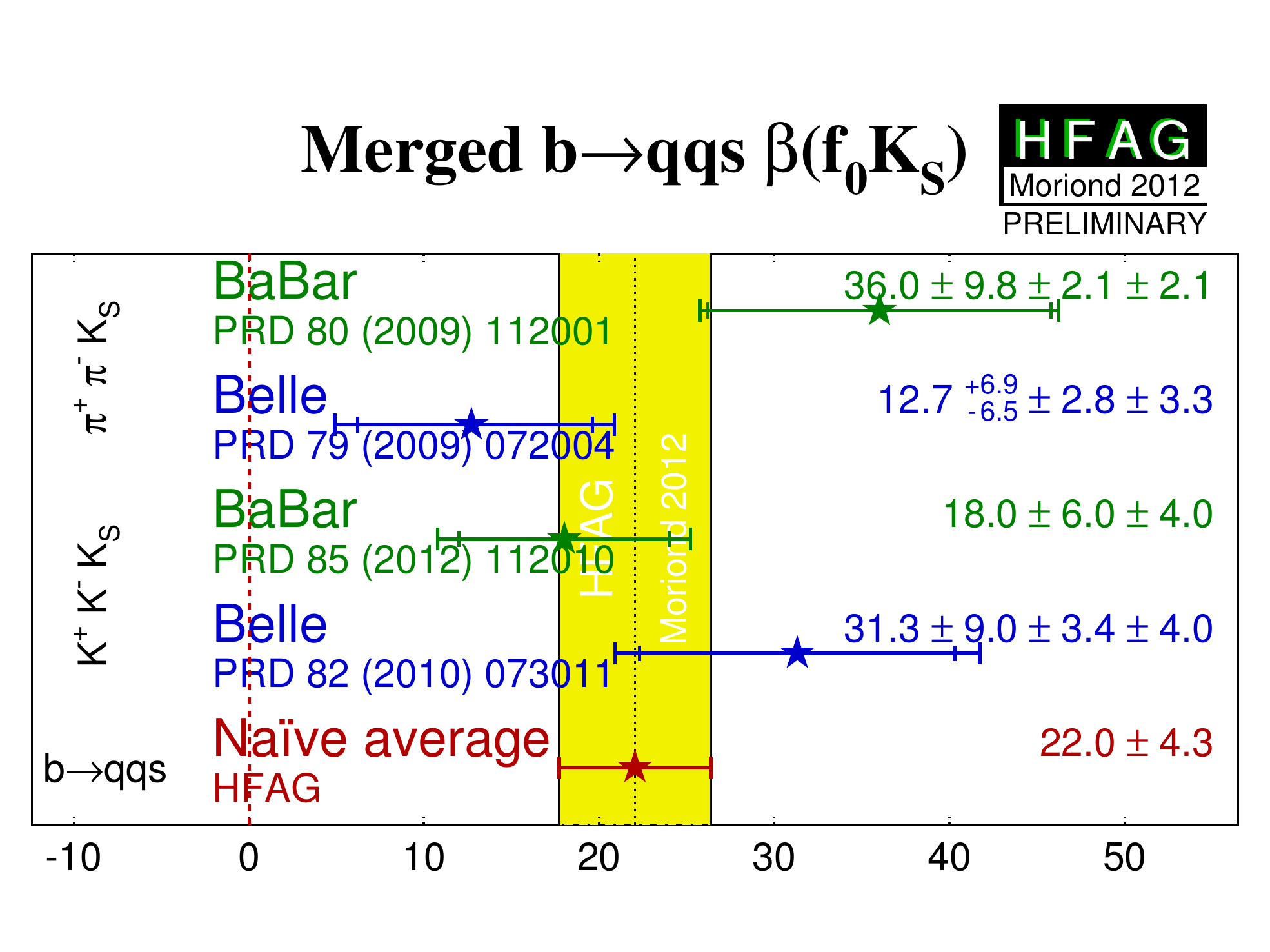}
    }
    \hfill
    \resizebox{0.45\textwidth}{!}{
      \includegraphics{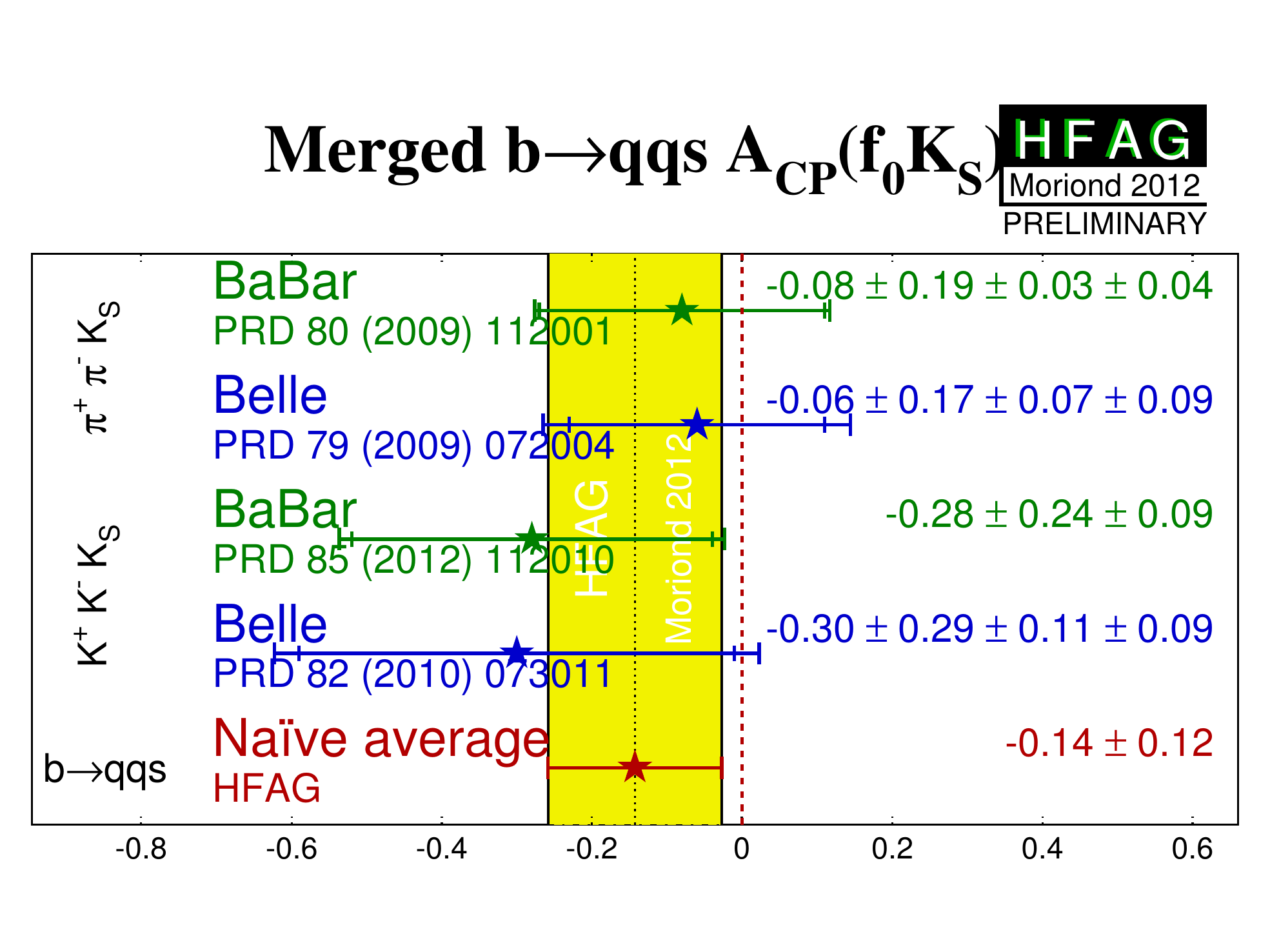}
    }
  \end{center}
  \vspace{-0.8cm}
  \caption{
    Averages of 
    (left) $\beta^{\rm eff} \equiv \phi_1^{\rm eff}$ and (right) $A_{\CP}$
    for the $\Bz\to f_0\KS$ decay including measurements from Dalitz plot analyses of both $\Bz\to K^+K^-\KS$ and $\Bz\to \pi^+\pi^-\KS$.
  }
  \label{fig:cp_uta:qqs:f0KS}
\end{figure}

As mentioned in Sec.~\ref{sec:cp_uta:notations:dalitz:kkk0} and above,
both \babar\ and \belle\ have performed time-dependent Dalitz plot analysis of
$\Bz \to K^+K^-\Kz$ and $\Bz \to \pi^+\pi^-\KS$ decays.
The results are summarised in Tables~\ref{tab:cp_uta:kkk0_tddp} 
and~\ref{tab:cp_uta:pipik0_tddp}.
Averages for the $\Bz\to f_0 \KS$ decay, which contributes to both Dalitz
plots, are shown in Fig.~\ref{fig:cp_uta:qqs:f0KS}.
Results are presented in terms of the effective weak phase (from mixing and
decay) difference $\beta^{\rm eff}$ and the parameter of $\CP$ violation in decay
$\Acp$ ($\Acp = -C$) for each of the resonant contributions.
Note that Dalitz plot analyses, including all those included in these
averages, often suffer from ambiguous solutions -- we quote the results
corresponding to those presented as solution 1 in all cases.
Results on flavour specific amplitudes that may contribute to these Dalitz
plots (such as $K^{*+}\pi^-$) are averaged by the HFAG Rare Decays subgroup 
(Sec.~\ref{sec:rare}).

For the $\Bz \to K^+K^-\Kz$ decay, both \babar\ and \belle\ measure the \CP
violation parameters for the $\phi\Kz$, $f_0\Kz$ and ``other $\Kp\Km\Kz$''
amplitudes, where the latter includes all remaining resonant and nonresonant
contributions to the charmless three-body decay.
For the $\Bz \to \pi^+\pi^-\KS$ decay, \babar\ report \CP violation parameters
for all of the \CP eigenstate components in the Dalitz plot model ($\rhoz\KS$,
$f_0\KS$, $f_2\KS$, $f_X\KS$ and nonresonant decays\footnote{
  The $f_X$ resonance included in the model is a poorly understood excess in
  the $m(\pi^+\pi^-)$ distribution, that may originate from interference
  between other states in this region and nonresonant amplitudes.
}), while \belle\ report the
\CP violation parameters for only the $\rhoz\KS$ and $f_0\KS$ amplitudes,
although the used Dalitz plot model is rather similar.

\begin{sidewaystable}
  \begin{center}
    \caption{
      Results from time-dependent Dalitz plot analyses of 
      the $\Bz \to K^+K^-\Kz$ decay.
      Correlations (not shown) are taken into account in the average.
    }
    \vspace{0.2cm}
    \setlength{\tabcolsep}{0.0pc}
    \resizebox{\textwidth}{!}{
      \begin{tabular}{l@{\hspace{2mm}}r@{\hspace{2mm}}c@{\hspace{2mm}}|@{\hspace{2mm}}c@{\hspace{2mm}}c@{\hspace{2mm}}|@{\hspace{2mm}}c@{\hspace{2mm}}c|@{\hspace{2mm}}c@{\hspace{2mm}}c} 
        \hline 
        \mc{2}{l}{Experiment} & $N(B\bar{B})$ &
        \mc{2}{c}{$\phi\KS$} & \mc{2}{c}{$f_0\KS$} & \mc{2}{c}{$K^+K^-\KS$} \\
        & & & $\beta^{\rm eff}\,(^\circ)$ & $\Acp$ & $\beta^{\rm eff}\,(^\circ)$ & $\Acp$ & $\beta^{\rm eff}\,(^\circ)$ & $\Acp$ \\
	\babar & \cite{Lees:2012kx} & 470M & $21 \pm 6 \pm 2$ & $-0.05 \pm 0.18 \pm 0.05$ & $18 \pm 6 \pm 4$ & $-0.28 \pm 0.24 \pm 0.09$ & $20.3 \pm 4.3 \pm 1.2$ & $-0.02 \pm 0.09 \pm 0.03$ \\
	\belle & \cite{Nakahama:2010nj} & 657M & $32.2 \pm 9.0 \pm 2.6 \pm 1.4$ & $0.04 \pm 0.20 \pm 0.10 \pm 0.02$ & $31.3 \pm 9.0 \pm 3.4 \pm 4.0$ & $-0.30 \pm 0.29 \pm 0.11 \pm 0.09$ & $24.9 \pm 6.4 \pm 2.1 \pm 2.5$ & $-0.14 \pm 0.11 \pm 0.08 \pm 0.03$ \\
	\mc{2}{l}{\bf Average} & & $24 \pm 5$ & $-0.01 \pm 0.14$ & $22 \pm 6$ & $-0.29 \pm 0.20$ & $21.6 \pm 3.7$ & $-0.06 \pm 0.08$ \\
	\mc{3}{l}{\small Confidence level} & \mc{6}{c}{\small $0.93~(0.1\sigma)$} \\
        \hline
      \end{tabular}
    }
    
    \label{tab:cp_uta:kkk0_tddp}
  \end{center}
\end{sidewaystable}

\begin{sidewaystable}
  \begin{center}
    \caption{
      Results from time-dependent Dalitz plot analysis of 
      the $\Bz \to \pi^+\pi^-\KS$ decay.
      Correlations (not shown) are taken into account in the average.
    }
    \vspace{0.2cm}
    \setlength{\tabcolsep}{0.0pc}
    \resizebox{\textwidth}{!}{
      \begin{tabular}{l@{\hspace{2mm}}r@{\hspace{2mm}}c@{\hspace{2mm}}|@{\hspace{2mm}}c@{\hspace{2mm}}c|@{\hspace{2mm}}c@{\hspace{2mm}}c} 
        \hline 
        \mc{2}{l}{Experiment} & $N(B\bar{B})$ & 
        \mc{2}{c}{$\rho^0\KS$} & \mc{2}{c}{$f_0\KS$} \\
        & & & $\beta^{\rm eff}$ & $\Acp$ & $\beta^{\rm eff}$ & $\Acp$ \\
        \hline
        \babar & \cite{Aubert:2009me} & 383M & $(10.2 \pm 8.9 \pm 3.0 \pm 1.9)^\circ$ & $0.05 \pm 0.26 \pm 0.10 \pm 0.03$ & $(36.0 \pm 9.8 \pm 2.1 \pm 2.1)^\circ$ & $-0.08 \pm 0.19 \pm 0.03 \pm 0.04$ \\
        \belle & \cite{:2008wwa} & 657M & $(20.0 \,^{+8.6}_{-8.5} \pm 3.2 \pm 3.5)^\circ$ & $0.03 \,^{+0.23}_{-0.24} \pm 0.11 \pm 0.10$ & $(12.7 \,^{+6.9}_{-6.5} \pm 2.8 \pm 3.3)^\circ$ & $-0.06 \pm 0.17 \pm 0.07 \pm 0.09$ \\
        \hline
        \mc{2}{l}{\bf Average} & & $16.4 \pm 6.8$ & $0.06 \pm 0.20$ & $20.6 \pm 6.2$ & $-0.07 \pm 0.14$  \\
        \mc{3}{l}{\small Confidence level} & \mc{4}{c}{\small $0.39~(0.9\sigma)$} \\
        \hline
      \end{tabular}
    }

    \vspace{2ex}

    \setlength{\tabcolsep}{0.0pc}
    \resizebox{\textwidth}{!}{
      \begin{tabular}{l@{\hspace{2mm}}r@{\hspace{2mm}}c@{\hspace{2mm}}|@{\hspace{2mm}}c@{\hspace{2mm}}c|@{\hspace{2mm}}c@{\hspace{2mm}}c} 
        \hline 
        \mc{2}{l}{Experiment} & $N(B\bar{B})$ & 
        \mc{2}{c}{$f_2\KS$} & \mc{2}{c}{$f_{\rm X}\KS$} \\
        & & & $\beta^{\rm eff}$ & $\Acp$ & $\beta^{\rm eff}$ & $\Acp$ \\
        \babar & \cite{Aubert:2009me} & 383M & $(14.9 \pm 17.9 \pm 3.1 \pm 5.2)^\circ$ & $-0.28 \,^{+0.40}_{-0.35} \pm 0.08 \pm 0.07$ & $(5.8 \pm 15.2 \pm 2.2 \pm 2.3)^\circ$ & $-0.13 \,^{+0.35}_{-0.33} \pm 0.04 \pm 0.09$ \\
        \hline
      \end{tabular}
    }

    \vspace{2ex}

    \setlength{\tabcolsep}{0.0pc}
    \resizebox{\textwidth}{!}{
      \begin{tabular}{l@{\hspace{2mm}}r@{\hspace{2mm}}c@{\hspace{2mm}}|@{\hspace{2mm}}c@{\hspace{2mm}}c|@{\hspace{2mm}}c@{\hspace{2mm}}c} 
        \hline 
        \mc{2}{l}{Experiment} & $N(B\bar{B})$ & 
        \mc{2}{c}{$\Bz \to \pi^+\pi^-\KS$ nonresonant} & \mc{2}{c}{$\chi_{c0}\KS$} \\
        & & & $\beta^{\rm eff}$ & $\Acp$ & $\beta^{\rm eff}$ & $\Acp$ \\
        \babar & \cite{Aubert:2009me} & 383M & $(0.4 \pm 8.8 \pm 1.9 \pm 3.8)^\circ$ & $-0.01 \pm 0.25 \pm 0.06 \pm 0.05$ & $(23.2 \pm 22.4 \pm 2.3 \pm 4.2)^\circ$ & $0.29 \,^{+0.44}_{-0.53} \pm 0.03 \pm 0.05$ \\
        \hline
      \end{tabular}
    }

    \label{tab:cp_uta:pipik0_tddp}
  \end{center}
\end{sidewaystable}


\mysubsubsection{Time-dependent analyses of $\Bz \to \phi \KS \pi^0$}
\label{sec:cp_uta:qqs:vv}

The final state in the decay $\Bz \to \phi \KS \pi^0$ is a mixture of \CP-even
and \CP-odd amplitudes. However, since only $\phi K^{*0}$ resonant states
contribute (in particular, $\phi K^{*0}(892)$, $\phi K^{*0}_0(1430)$ and $\phi
K^{*0}_2(1430)$ are seen), the composition can be determined from the analysis
of $B \to \phi K^+ \pi^-$, assuming only that the ratio of branching fractions
${\cal B}(K^{*0} \to \KS \pi^0)/{\cal B}(K^{*0} \to K^+ \pi^-)$ is the same
for each exited kaon state. 

\babar~\cite{Aubert:2008zza} have performed a simultaneous analysis of 
$\Bz \to \phi \KS \pi^0$ and $\Bz \to \phi K^+ \pi^-$ that is time-dependent
for the former mode and time-integrated for the latter. Such an analysis
allows, in principle, all parameters of the $\Bz \to \phi K^{*0}$ system to be
determined, including mixing-induced \CP violation effects. The latter is
determined to be $\Delta\phi_{00} = 0.28 \pm 0.42 \pm 0.04$, where
$\Delta\phi_{00}$ is half the weak phase difference between $\Bz$ and $\Bzb$
decays to $\phi K^{*0}_0(1430)$. As discussed above, this can also be
presented in terms of the Q2B parameter $\sin(2\beta^{\rm eff}_{00}) =
\sin(2\beta+2\Delta\phi_{00}) = 0.97 \,^{+0.03}_{-0.52}$. The highly asymmetric
uncertainty arises due to the conversion from the phase to the sine of the
phase, and the proximity of the physical boundary. 

Similar $\sin(2\beta^{\rm eff})$ parameters can be defined for each of the
helicity amplitudes for both $\phi K^{*0}(892)$ and $\phi
K^{*0}_2(1430)$. However, the relative phases between these decays are
constrained due to the nature of the simultaneous analysis of $\Bz \to \phi
\KS \pi^0$ and $\Bz \to \phi K^+ \pi^-$, and therefore these measurements are
highly correlated. Instead of quoting all these results, \babar provide an
illustration of their measurements with the following differences: 
\begin{eqnarray}
  \sin(2\beta - 2\Delta\delta_{01}) - \sin(2\beta) & = & -0.42\,^{+0.26}_{-0.34} \, \\
  \sin(2\beta - 2\Delta\phi_{\parallel1}) - \sin(2\beta) & = & -0.32\,^{+0.22}_{-0.30} \, \\
  \sin(2\beta - 2\Delta\phi_{\perp1}) - \sin(2\beta) & = & -0.30\,^{+0.23}_{-0.32} \, \\
  \sin(2\beta - 2\Delta\phi_{\perp1}) - \sin(2\beta - 2\Delta\phi_{\parallel1})
  & = & 0.02 \pm 0.23 \, \\
  \sin(2\beta - 2\Delta\delta_{02}) - \sin(2\beta) & = & -0.10\,^{+0.18}_{-0.29} \,
\end{eqnarray}
where the first subscript indicates the helicity amplitude and the second
indicates the spin of the kaon resonance. For the complete definitions of the
$\Delta\delta$ and $\Delta\phi$ parameters, please refer to the \babar\ paper~\cite{Aubert:2008zza}.

Parameters of \CP violation in decay for each of the contributing helicity
amplitudes can also be measured. Again, these are determined from a
simultaneous fit of $\Bz \to \phi \KS \pi^0$ and $\Bz \to \phi K^+ \pi^-$,
with the precision being dominated by the statistics of the latter mode. 
Direct \CP violation measurements are tabulated by HFAG - Rare Decays (Sec.~\ref{sec:rare}). 

\mysubsubsection{Time-dependent \CP asymmetries in $\Bs \to \Kp\Km$}
\label{sec:cp_uta:qqs:BstoKK}

The decay $\Bs \to \Kp\Km$ involves a $b \to u\bar{u}s$ transition, and hence has both penguin and tree contributions. Both mixing-induced and \CP violation in decay effects may arise, and additional input is needed to disentangle the contributions and determine $\gamma$ and $\beta_s^{\rm eff}$. For example, the observables in $\Bd \to \pip\pim$ can be related using U-spin, as proposed by Fleischer~\cite{Fleischer:1999pa}.

The observables are $A_{\rm mix} = S_{\CP}$, $A_{\rm dir} = -C_{\CP}$, and $A_{\Delta\Gamma}$. They can all be treated as free parameters, but are physically constrained to satisfy $A_{\rm mix}^2 + A_{\rm dir}^2 + A_{\Delta\Gamma}^2 = 1$. Note that the untagged decay distribution, from which an ``effective lifetime'' can be measured, retains sensitivity to $A_{\Delta\Gamma}$. Averages of effective lifetimes are performed by the HFAG Lifetimes and Oscillations group, see Sec.~\ref{sec:life_mix}.

The observables in $\Bs \to \Kp\Km$ have been measured by LHCb, who impose the constraint mentioned above to eliminate $A_{\rm \Delta\Gamma}$. 

\begin{table}[!htb]
	\begin{center}
		\caption{
      Results from time-dependent analysis of the $\Bs \to K^{+} K^{-}$ decay.
		}
		\vspace{0.2cm}
		\setlength{\tabcolsep}{0.0pc}
		\begin{tabular*}{\textwidth}{@{\extracolsep{\fill}}lrcccc} \hline
	\mc{2}{l}{Experiment} & Sample size & $A_{\rm mix}$ & $A_{\rm dir}$ & Correlation \\
	\hline
	LHCb & \cite{Aaij:2013tna} & $1.0 \ {\rm fb}^{-1}$ & $0.30 \pm 0.12 \pm 0.04$ & $0.14 \pm 0.11 \pm 0.03$ & 0.02 \\
	\hline
		\end{tabular*}
		\label{tab:cp_uta:BstoKK}
	\end{center}
\end{table}

\mysubsubsection{Time-dependent \CP asymmetries in $\Bs \to \phi\phi$}
\label{sec:cp_uta:qqs:Bstophiphi}

 The decay $\Bs \to \phi\phi$ involves a $b \to s\bar{s}s$ transition, and hence is a ``pure penguin'' mode (in the limit that the $\phi$ meson is treated as a pure $s\bar{s}$ state). Since the mixing phase and the decay phase are expected to cancel in the Standard Model, the prediction for the phase from the interference of mixing and decay is predicted to be $\phi_s(\phi\phi) = 0$ with low uncertainty~\cite{Raidal:2002ph}. Due to the vector-vector nature of the final state, angular analysis is needed to separate the \CP-even and \CP-odd contributions. Such an analysis also makes it possible to fit directly for $\phi_s(\phi\phi)$.

A constraint on $\phi_s(\phi\phi)$ has been obtained by LHCb using $3.0 \,{\rm fb}^{-1}$ of data~\cite{Aaij:2014kxa}.
The result is $\phi_s(\phi\phi) = -0.17 \pm 0.15 \pm 0.03 \, {\rm rad}$ where the first uncertainty is statistical and the second is systematic.

\mysubsection{Time-dependent $\CP$ asymmetries in $b \to c\bar{c}d$ transitions
}
\label{sec:cp_uta:ccd}

The transition $b \to c\bar c d$ can occur via either a $b \to c$ tree
or a $b \to d$ penguin amplitude.  
Similarly to Eq.~(\ref{eq:cp_uta:b_to_s}), the amplitude for 
the $b \to d$ penguin can be written
\begin{equation}
  \label{eq:cp_uta:b_to_d}
  \begin{array}{ccccc}
    A_{b \to d} & = & 
    \mc{3}{l}{F_u V_{ub}V^*_{ud} + F_c V_{cb}V^*_{cd} + F_t V_{tb}V^*_{td}} \\
    & = & (F_u - F_c) V_{ub}V^*_{ud} & + & (F_t - F_c) V_{tb}V^*_{td} \\
    & = & {\cal O}(\lambda^3) & + & {\cal O}(\lambda^3). \\
  \end{array}
\end{equation}
From this it can be seen that the $b \to d$ penguin amplitude 
contains terms with different weak phases at the same order of
CKM suppression.

In the above, we have followed Eq.~(\ref{eq:cp_uta:b_to_s}) 
by eliminating the $F_c$ term using unitarity.
However, we could equally well write
\begin{equation}
  \label{eq:cp_uta:b_to_d_alt}
  \begin{array}{ccccc}
    A_{b \to d} 
    & = & (F_u - F_t) V_{ub}V^*_{ud} & + & (F_c - F_t) V_{cb}V^*_{cd}, \\
    & = & (F_c - F_u) V_{cb}V^*_{cd} & + & (F_t - F_u) V_{tb}V^*_{td}. \\
  \end{array}
\end{equation}
Since the $b \to c\bar{c}d$ tree amplitude 
has the weak phase of $V_{cb}V^*_{cd}$,
either of the above expressions allow the penguin to be decomposed into 
parts with weak phases the same and different to the tree amplitude
(the relative weak phase can be chosen to be either $\beta$ or $\gamma$).
However, if the tree amplitude dominates,
there is little sensitivity to any phase 
other than that from $\Bz$\textendash$\Bzb$ mixing.

\begin{figure}[htbp]
  \begin{center}
    \begin{tabular}{cc}
      \resizebox{0.46\textwidth}{!}{
        \includegraphics{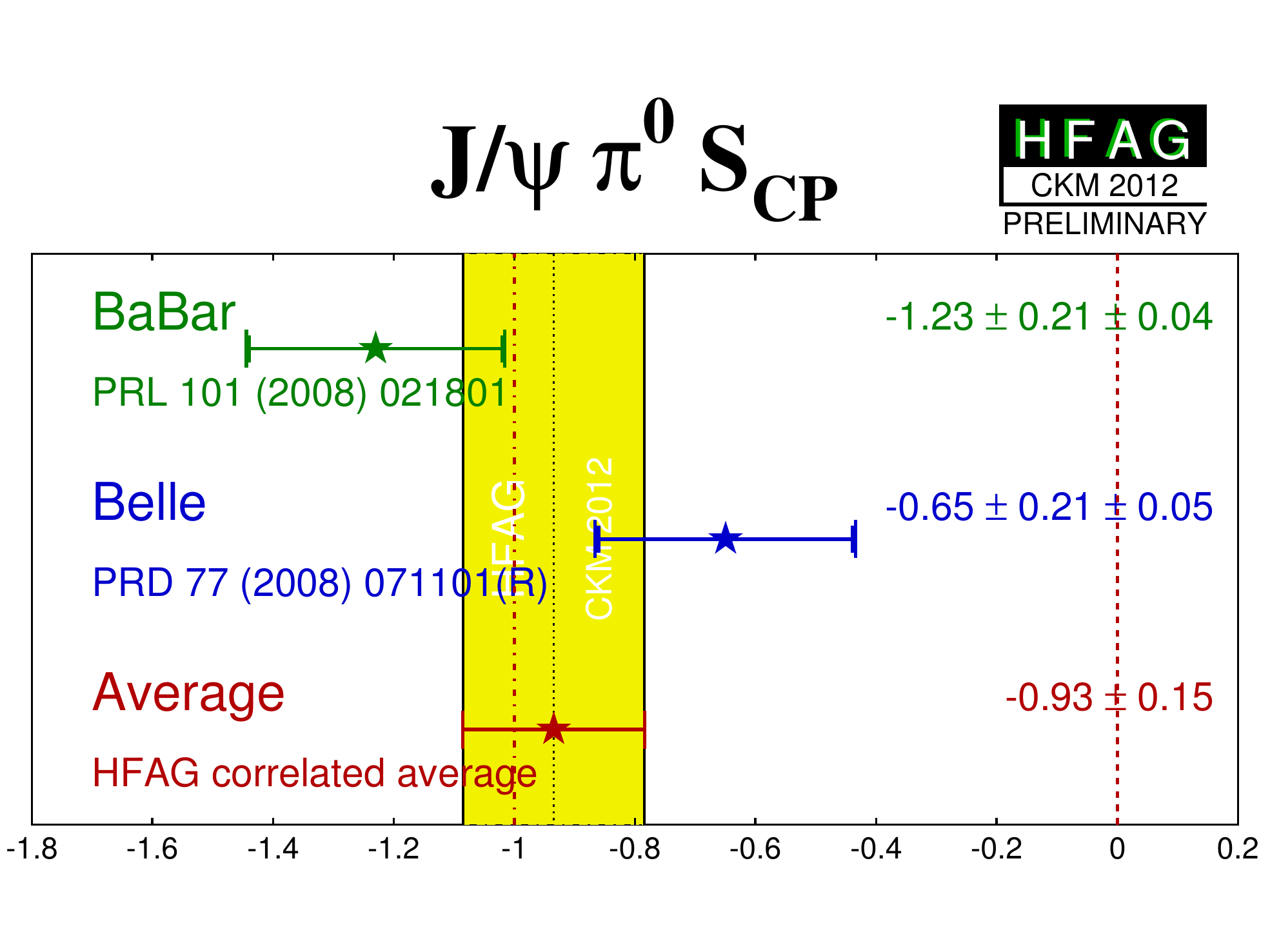}
      }
      &
      \resizebox{0.46\textwidth}{!}{
        \includegraphics{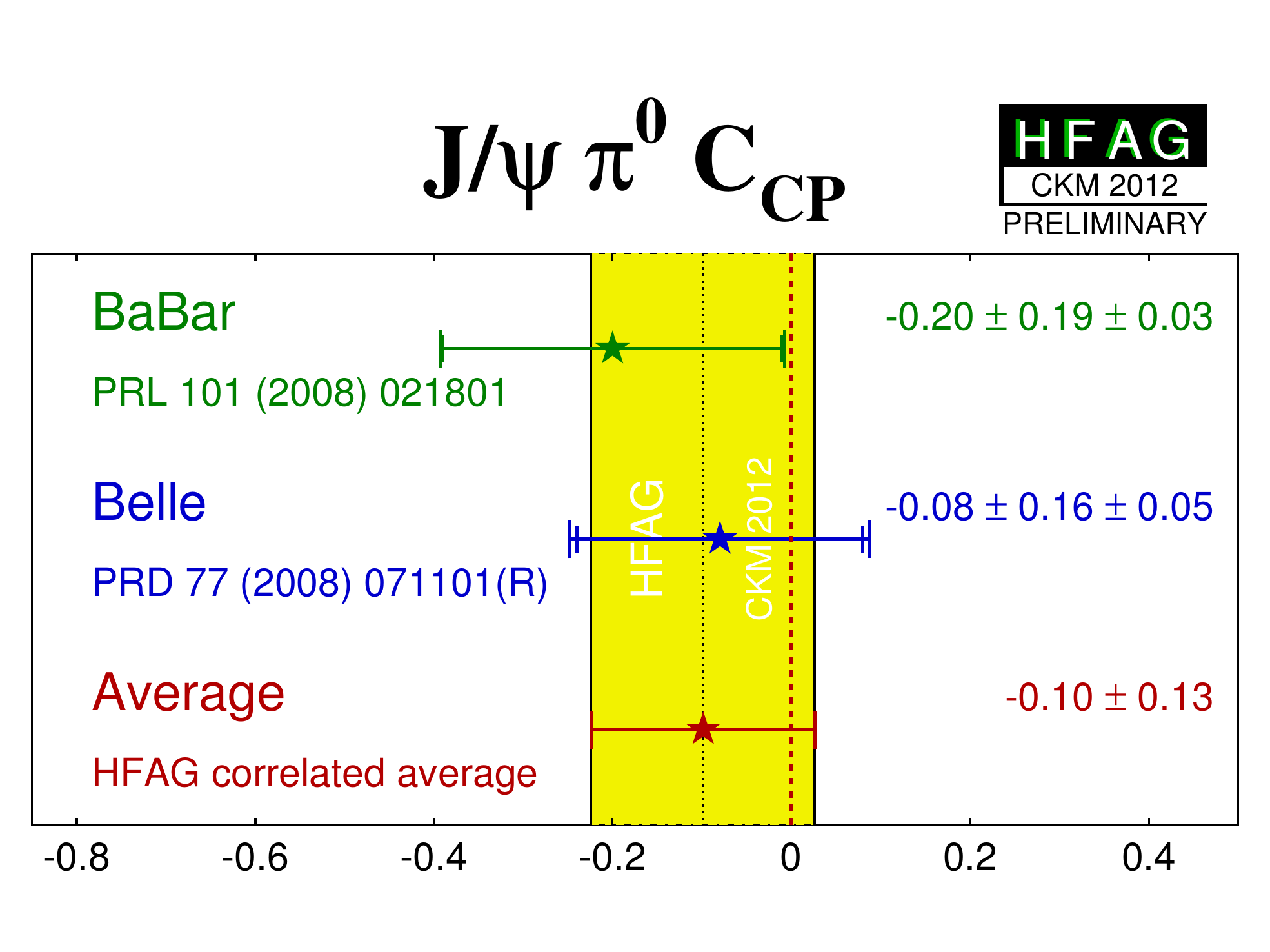}
      }
    \end{tabular}
  \end{center}
  \vspace{-0.8cm}
  \caption{
    Averages of 
    (left) $S_{b \to c\bar c d}$ and (right) $C_{b \to c\bar c d}$ 
    for the mode $\Bz \to J/ \psi \pi^0$.
  }
  \label{fig:cp_uta:ccd:psipi0}
\end{figure}

\begin{figure}[htbp]
  \begin{center}
    \begin{tabular}{cc}
      \resizebox{0.46\textwidth}{!}{
        \includegraphics{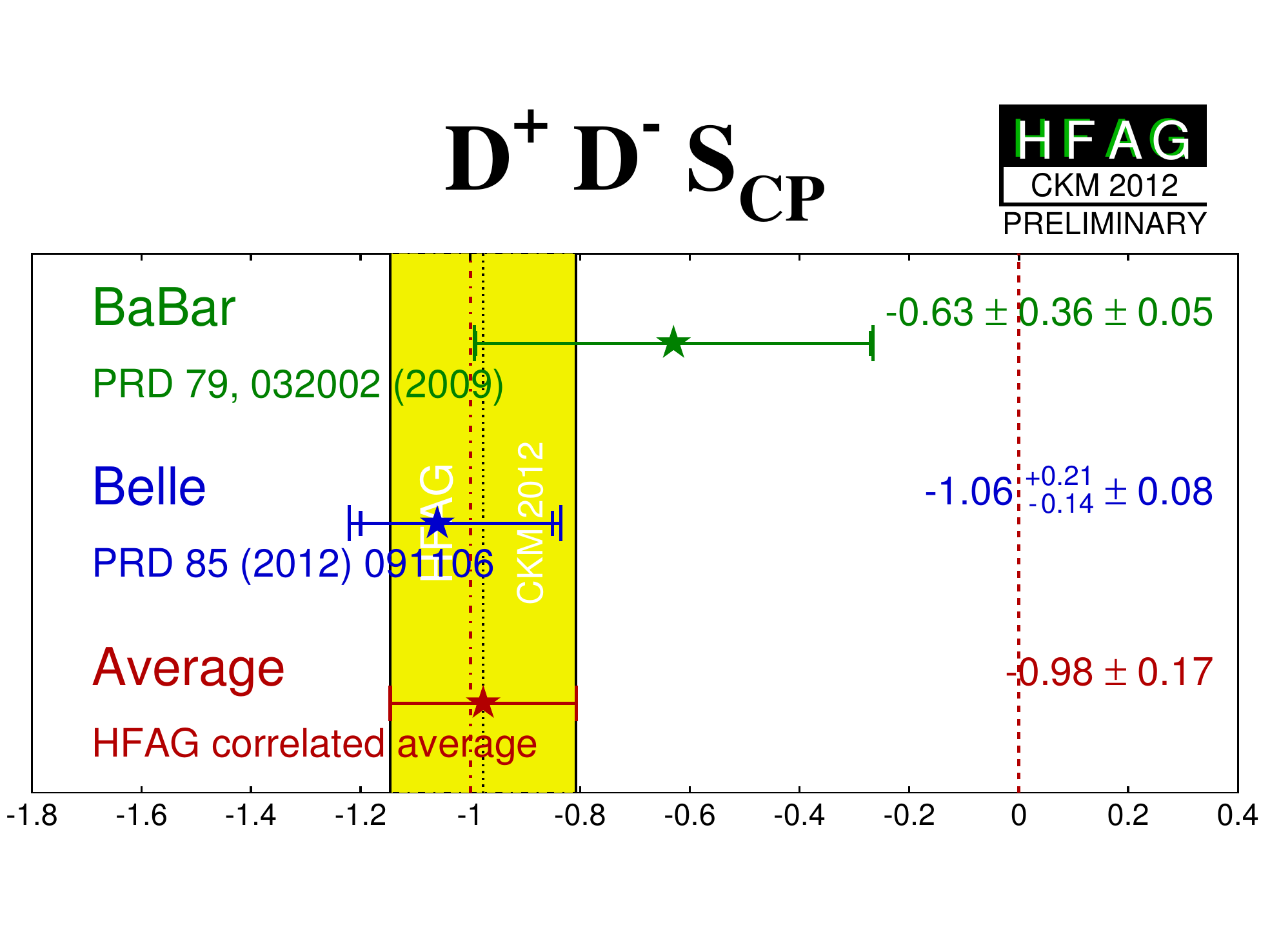}
      }
      &
      \resizebox{0.46\textwidth}{!}{
        \includegraphics{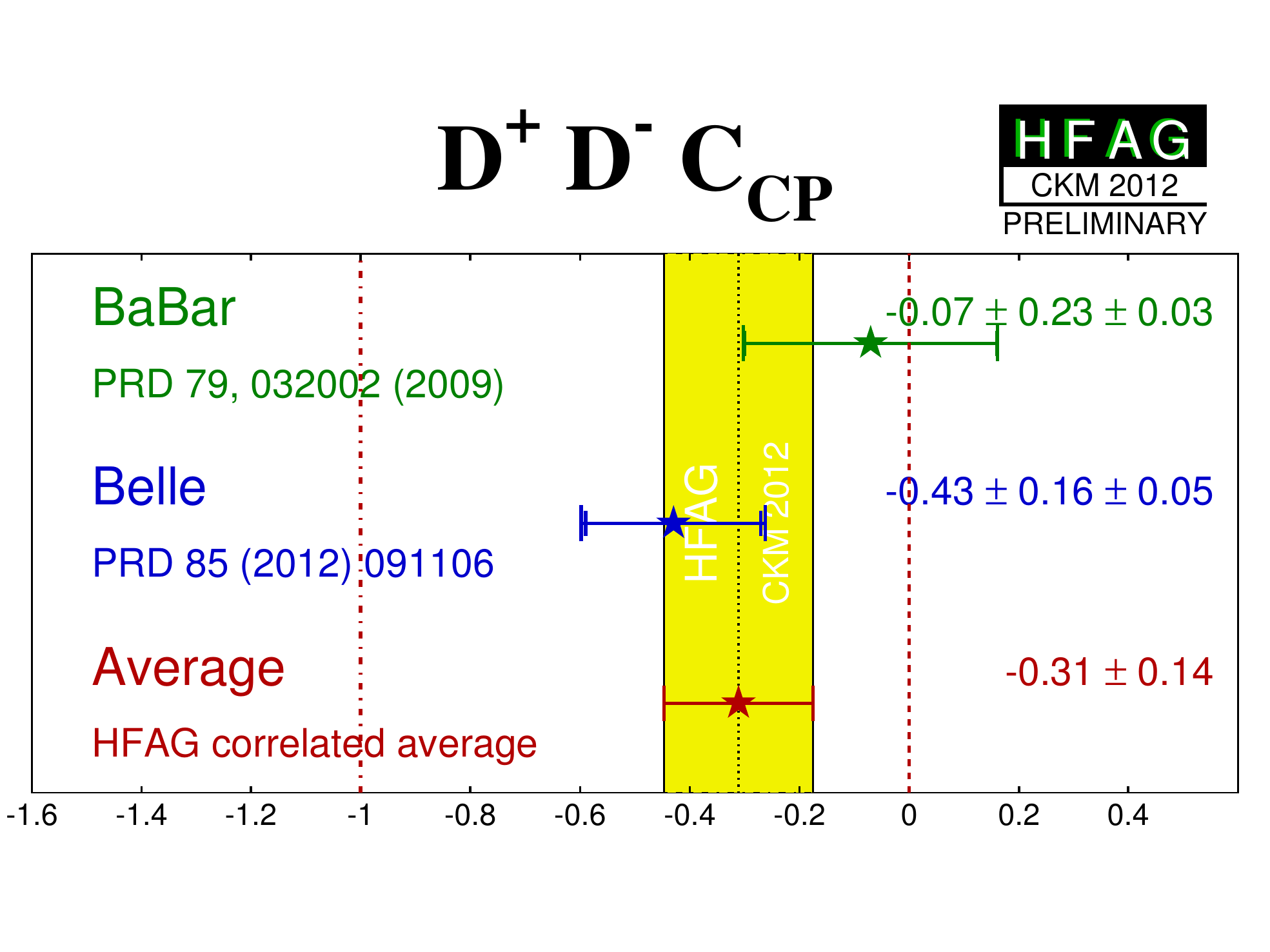}
      }
    \end{tabular}
  \end{center}
  \vspace{-0.8cm}
  \caption{
    Averages of 
    (left) $S_{b \to c\bar c d}$ and (right) $C_{b \to c\bar c d}$ 
    for the mode $\Bz \to D^+D^-$.
  }
  \label{fig:cp_uta:ccd:dd}
\end{figure}

\begin{figure}[htbp]
  \begin{center}
    \begin{tabular}{cc}
      \resizebox{0.46\textwidth}{!}{
        \includegraphics{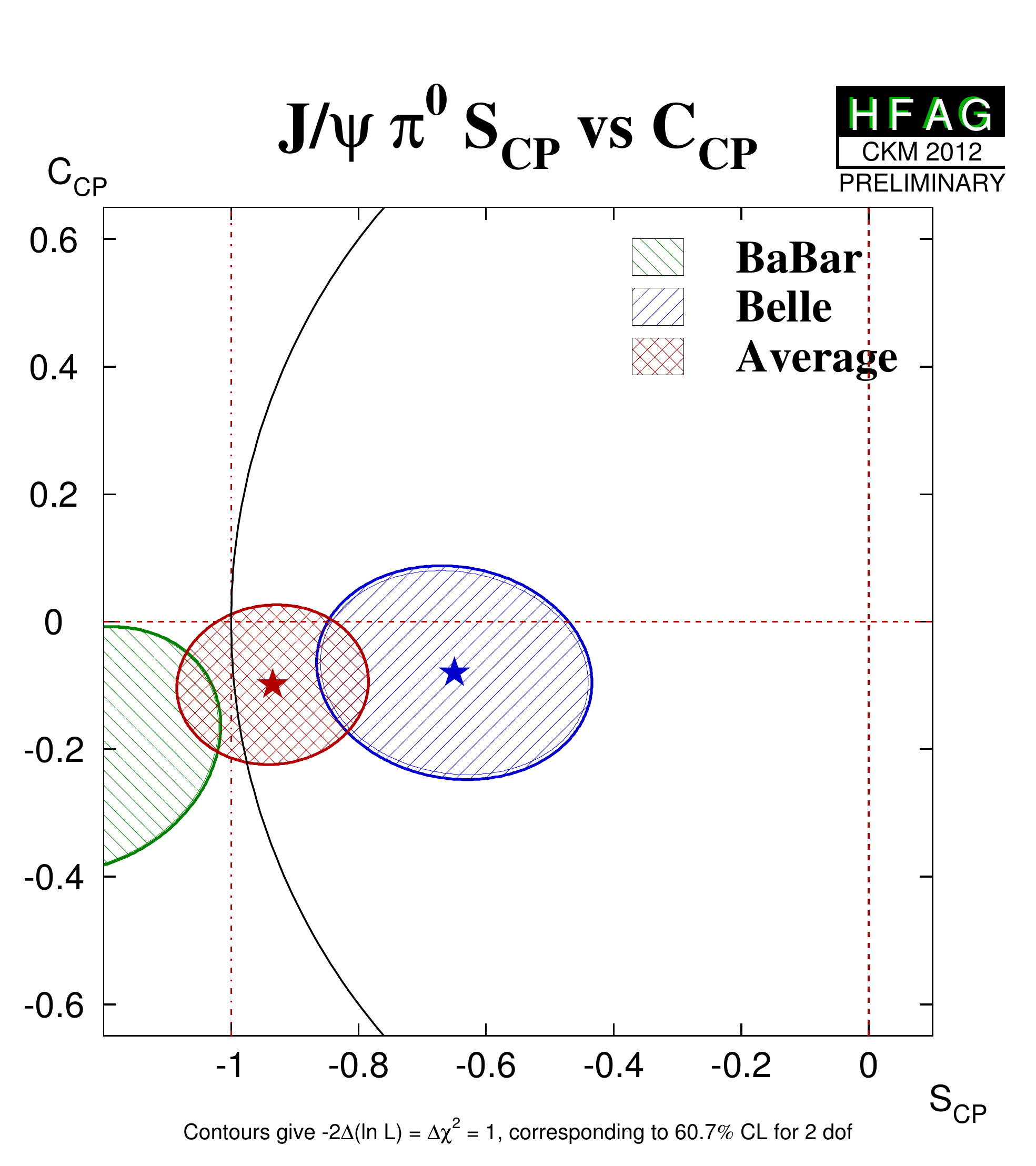}
      }
      &
      \resizebox{0.46\textwidth}{!}{
        \includegraphics{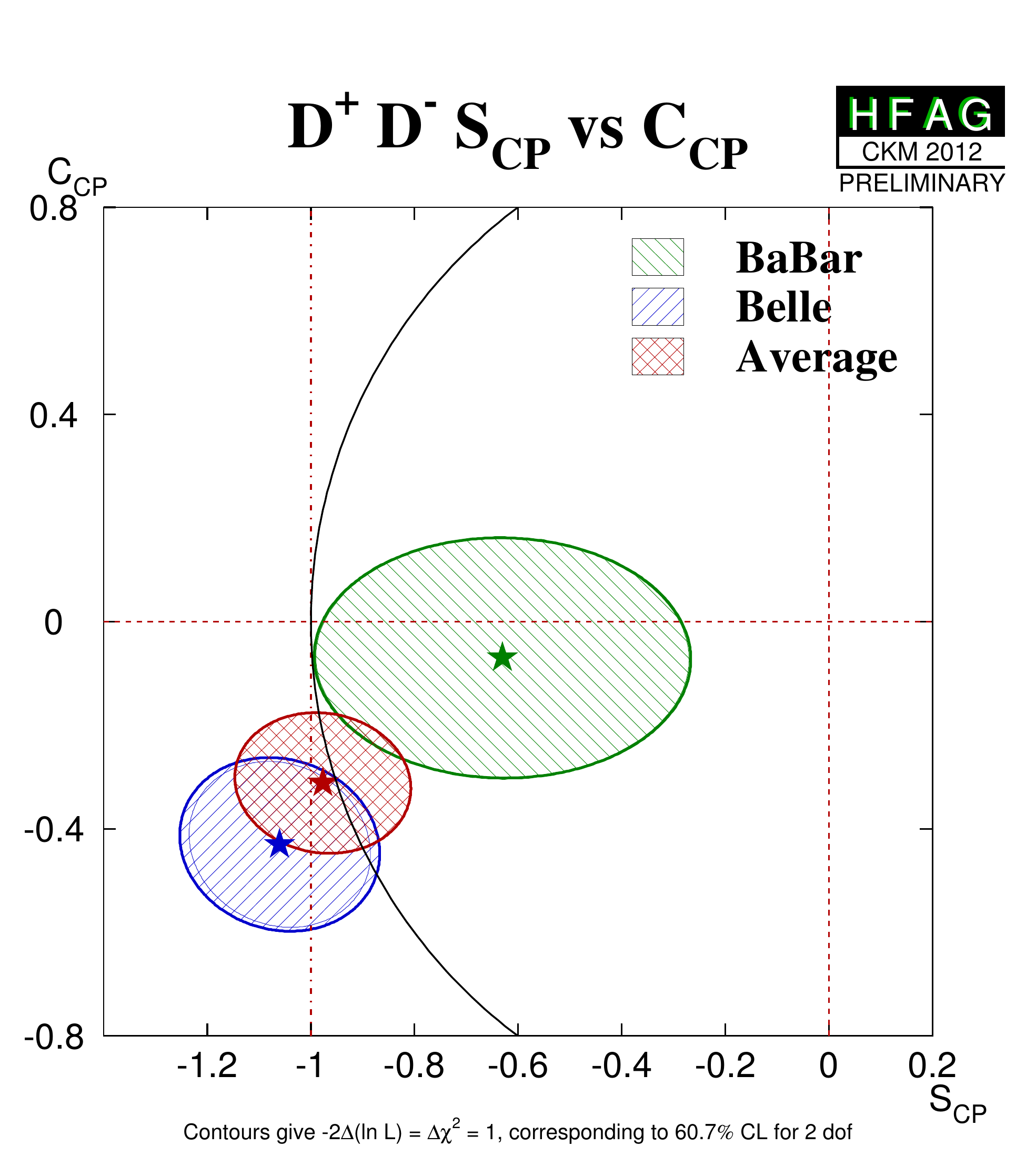}
      }
    \end{tabular}
  \end{center}
  \vspace{-0.8cm}
  \caption{
    Averages of two $b \to c\bar c d$ dominated channels,
    for which correlated averages are performed,
    in the $S_{\CP}$ \vs\ $C_{\CP}$ plane.
    (Left) $\Bz \to J/ \psi \pi^0$ and (right) $\Bz \to D^+D^-$.
  }
  \label{fig:cp_uta:ccd_SvsC}
\end{figure}

The $b \to c\bar{c}d$ transitions can be investigated with studies 
of various different final states. 
Results are available from both \babar\  and \belle\ 
using the final states $\jpsi \pi^0$, $D^+D^-$, 
$D^{*+}D^{*-}$ and $D^{*\pm}D^{\mp}$,
the averages of these results are given in Tables~\ref{tab:cp_uta:ccd1} and~\ref{tab:cp_uta:ccd2}.
The results using the $\CP$ eigenstate ($\etacp = +1$) modes
$\jpsi \pi^0$ and $D^+D^-$
are shown in Fig.~\ref{fig:cp_uta:ccd:psipi0} and 
Fig.~\ref{fig:cp_uta:ccd:dd} respectively,
with two-dimensional constraints shown in Fig.~\ref{fig:cp_uta:ccd_SvsC}.

\begin{figure}[htbp]
  \begin{center}
    \begin{tabular}{cc}
      \resizebox{0.46\textwidth}{!}{
        \includegraphics{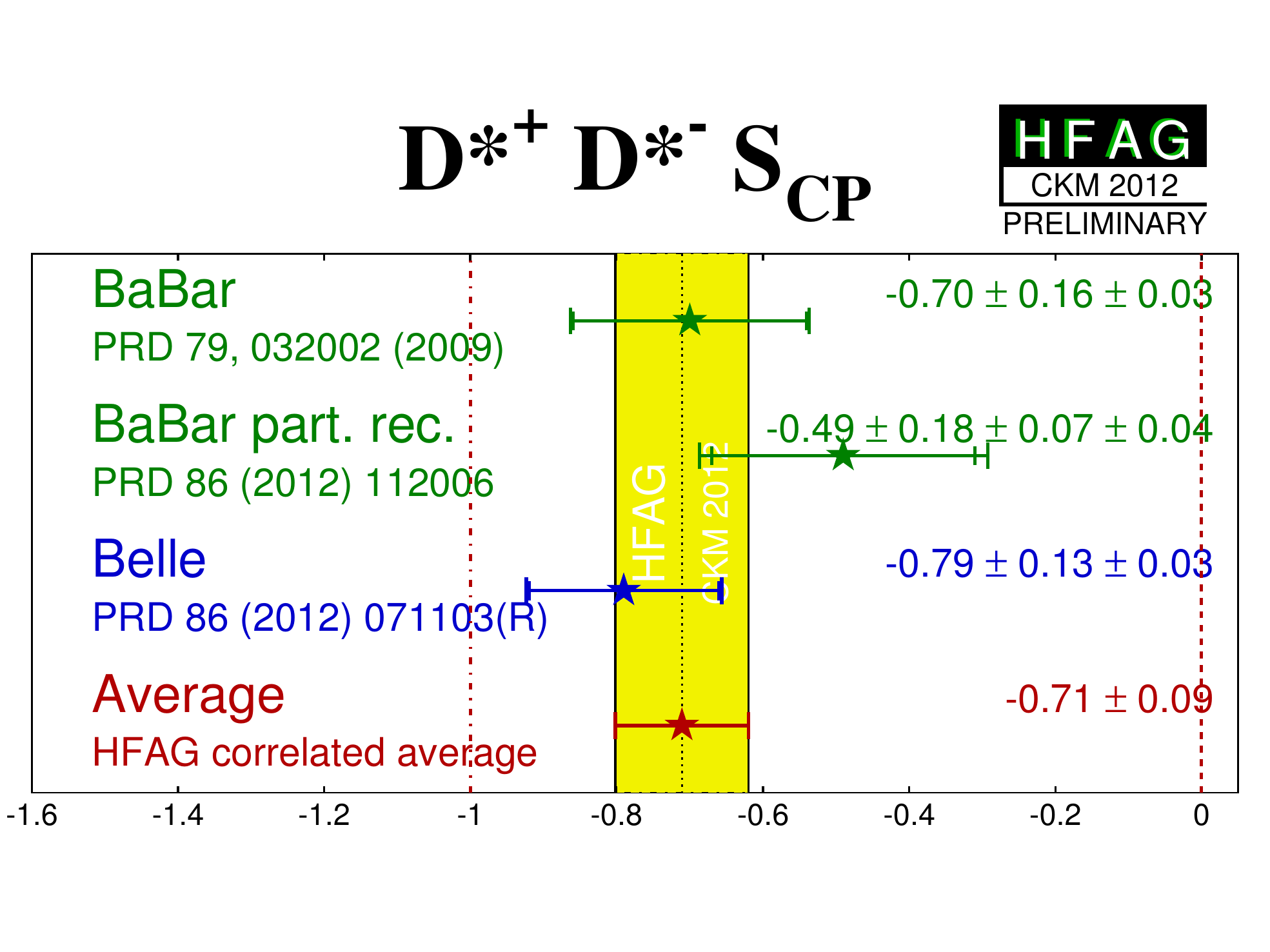}
      }
      &
      \resizebox{0.46\textwidth}{!}{
        \includegraphics{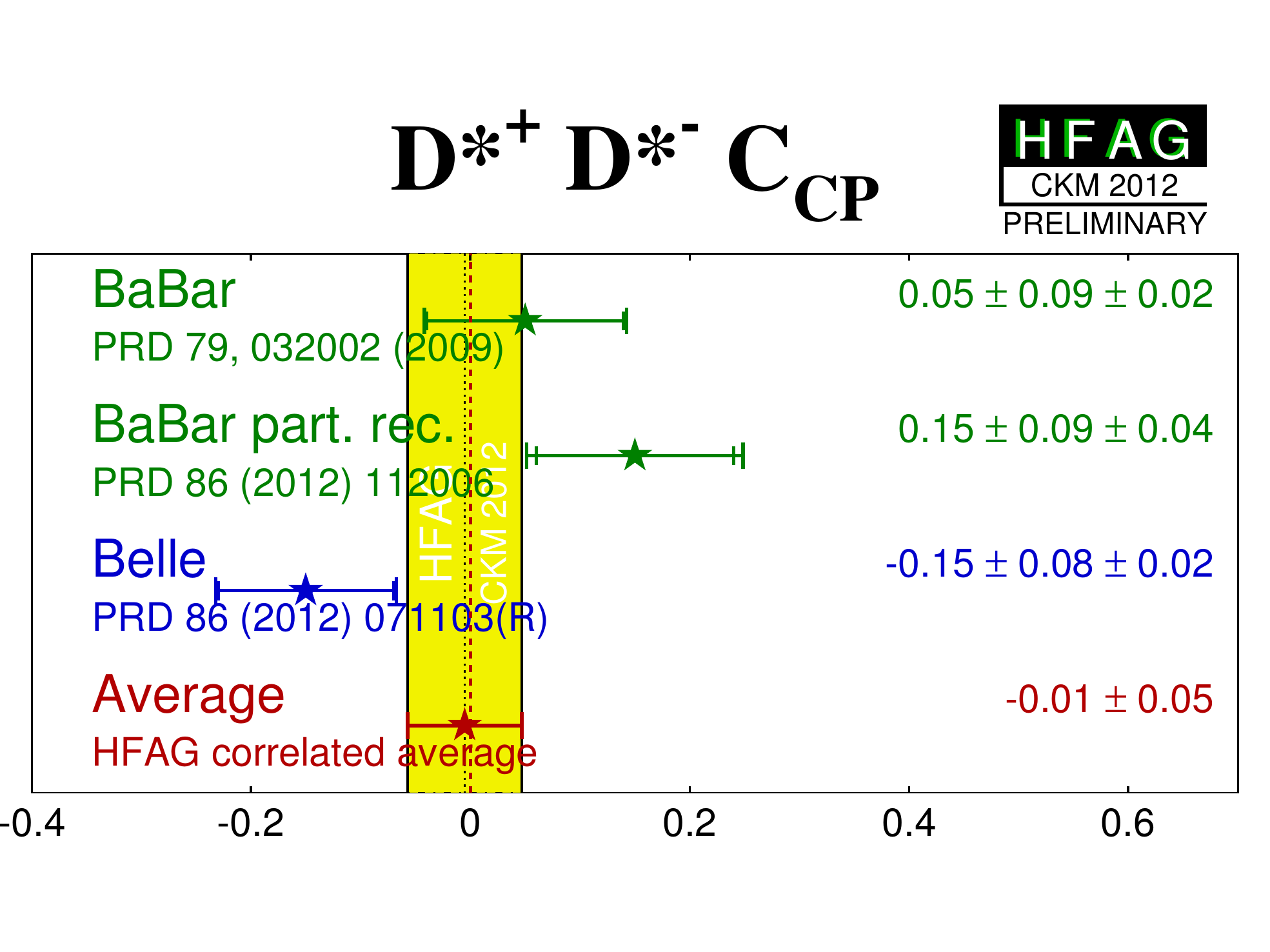}
      }
    \end{tabular}
  \end{center}
  \vspace{-0.8cm}
  \caption{
    Averages of 
    (left) $S_{b \to c\bar c d}$ and (right) $C_{b \to c\bar c d}$ 
    for the mode $\Bz \to D^{*+}D^{*-}$.
  }
  \label{fig:cp_uta:ccd:dstardstar}
\end{figure}

The vector-vector mode $D^{*+}D^{*-}$ 
is found to be dominated by the $\CP$-even longitudinally polarised component;
\babar\ measures a $\CP$-odd fraction of 
$0.158 \pm 0.028 \pm 0.006$~\cite{Aubert:2008ah} while
\belle\ measures a $\CP$-odd fraction of 
$0.125 \pm 0.043 \pm 0.023$~\cite{:2009za}.
These values, listed as $R_\perp$, are included in the averages which ensures
that the correlations are taken into account.\footnote{
  Note that the \babar\ value given in Table~\ref{tab:cp_uta:ccd2} differs from
  that given above, since that in the table is not corrected for efficiency.
}
\babar\ has also performed an additional fit in which the 
$\CP$-even and $\CP$-odd components are allowed to have different 
$\CP$ violation parameters $S$ and $C$.  
These results are included in Table~\ref{tab:cp_uta:ccd2}.
Results using $D^{*+}D^{*-}$ are shown in Fig.~\ref{fig:cp_uta:ccd:dstardstar}.


As discussed in Sec.~\ref{sec:cp_uta:notations:non_cp}, the most recent papers on the non-$\CP$ eigenstate mode $D^{*\pm}D^{\mp}$ use the ($A$, $S$, $\Delta S$, $C$, $\Delta C$) set of parameters, and we therefore perform the averages with this choice.

\begin{table}[htb]
	\begin{center}
		\caption{
     Averages for the $b \to c\bar{c}d$ modes,
     $\Bz \to J/\psi \pi^{0}$ and $D^+D^-$.
		}
		\vspace{0.2cm}
		\setlength{\tabcolsep}{0.0pc}
		\begin{tabular*}{\textwidth}{@{\extracolsep{\fill}}lrcccc} \hline
	\mc{2}{l}{Experiment} & $N(B\bar{B})$ & $S_{CP}$ & $C_{CP}$ & Correlation \\
	\hline
        \mc{6}{c}{$J/\psi \pi^{0}$} \\
	\babar & \cite{Aubert:2008bs} & 466M & $-1.23 \pm 0.21 \pm 0.04$ & $-0.20 \pm 0.19 \pm 0.03$ & $0.20$ \\
	\belle & \cite{:2007wd} & 535M & $-0.65 \pm 0.21 \pm 0.05$ & $-0.08 \pm 0.16 \pm 0.05$ & $-0.10$ \\
	\mc{3}{l}{\bf Average} & $-0.93 \pm 0.15$ & $-0.10 \pm 0.13$ & $0.04$ \\
	\mc{3}{l}{\small Confidence level} & \mc{2}{c}{\small $0.15~(1.4\sigma)$} & \\
		\hline

        \mc{6}{c}{$D^{+} D^{-}$} \\
	\babar & \cite{Aubert:2008ah} & 467M & $-0.65 \pm 0.36 \pm 0.05$ & $-0.07 \pm 0.23 \pm 0.03$ & $-0.01$ \\
	\belle & \cite{Rohrken:2012ta} & 772M & $-1.06 \,^{+0.21}_{-0.14} \pm 0.08$ & $-0.43 \pm 0.16 \pm 0.05$ & $-0.12$ \\
	\mc{3}{l}{\bf Average} & $-0.98 \pm 0.17$ & $-0.31 \pm 0.14$ & $-0.08$ \\
	\mc{3}{l}{\small Confidence level} & \mc{2}{c}{\small $0.26~(1.1\sigma)$} & \\
		\hline
 		\end{tabular*}
 		\label{tab:cp_uta:ccd1}
 	\end{center}
 \end{table}

\begin{sidewaystable}
 	\begin{center}
 		\caption{
      Averages for the $b \to c\bar{c}d$ modes,
      $D^{*+} D^{*-}$ and $D^{*\pm}D^\mp$.
 		}

 		\begin{tabular*}{\textwidth}{@{\extracolsep{\fill}}lrcccc} \hline
 		\mc{2}{l}{Experiment} & $N(B\bar{B})$ & $S_{CP}$ & $C_{CP}$ & $R_\perp$ \\
 		\hline
        \mc{6}{c}{$D^{*+} D^{*-}$} \\
	\babar & \cite{Aubert:2008ah} & 467M & $-0.70 \pm 0.16 \pm 0.03$ & $0.05 \pm 0.09 \pm 0.02$ & $0.17 \pm 0.03$ \\
	\babar part. rec. & \cite{Lees:2012px} & 471M & $-0.49 \pm 0.18 \pm 0.07 \pm 0.04$ & $0.15 \pm 0.09 \pm 0.04$ & $0.15 \pm 10.00$ \\
	\belle & \cite{Kronenbitter:2012ha} & 772M & $-0.79 \pm 0.13 \pm 0.03$ & $-0.15 \pm 0.08 \pm 0.02$ & $0.14 \pm 0.02 \pm 0.01$ \\
	\mc{3}{l}{\bf Average} & $-0.71 \pm 0.09$ & $-0.01 \pm 0.05$ & $0.15 \pm 0.02$ \\
	\mc{3}{l}{\small Confidence level} & \mc{3}{c}{\small $0.72~(0.4\sigma)$} \\
		\hline
		\end{tabular*}

                \vspace{2ex}

    \resizebox{\textwidth}{!}{
		\begin{tabular}{@{\extracolsep{2mm}}lrcccccc} \hline
	\mc{2}{l}{Experiment} & $N(B\bar{B})$ & $S_{CP+}$ & $C_{CP+}$ & $S_{CP-}$ & $C_{CP-}$ & $R_\perp$ \\
	\hline
        \mc{7}{c}{$D^{*+} D^{*-}$} \\
	\babar & \cite{Aubert:2008ah} & 467M & $-0.76 \pm 0.16 \pm 0.04$ & $0.02 \pm 0.12 \pm 0.02$ & $-1.81 \pm 0.71 \pm 0.16$ & $0.41 \pm 0.50 \pm 0.08$ & $0.15 \pm 0.03$ \\
		\hline
		\end{tabular}
    }

                \vspace{2ex}

    \resizebox{\textwidth}{!}{
		\begin{tabular}{@{\extracolsep{2mm}}lrcccccc} \hline
	\mc{2}{l}{Experiment} & $N(B\bar{B})$ & $S$ & $C$ & $\Delta S$ & $\Delta C$ & ${\cal A}$ \\
        \hline
        \mc{8}{c}{$D^{*\pm} D^{\mp}$} \\
	\babar & \cite{Aubert:2008ah} & 467M & $-0.68 \pm 0.15 \pm 0.04$ & $0.04 \pm 0.12 \pm 0.03$ & $0.05 \pm 0.15 \pm 0.02$ & $0.04 \pm 0.12 \pm 0.03$ & $0.01 \pm 0.05 \pm 0.01$ \\
	\belle & \cite{Rohrken:2012ta} & 772M & $-0.78 \pm 0.15 \pm 0.05$ & $-0.01 \pm 0.11 \pm 0.04$ & $-0.13 \pm 0.15 \pm 0.04$ & $0.12 \pm 0.11 \pm 0.03$ & $0.06 \pm 0.05 \pm 0.02$ \\
	\mc{3}{l}{\bf Average} & $-0.73 \pm 0.11$ & $0.01 \pm 0.09$ & $-0.04 \pm 0.11$ & $0.08 \pm 0.08$ & $0.03 \pm 0.04$ \\
	\mc{3}{l}{\small Confidence level} & {\small $0.65~(0.5\sigma)$} & {\small $0.77~(0.3\sigma)$} & {\small $0.41~(0.8\sigma)$} & {\small $0.63~(0.5\sigma)$} & {\small $0.48~(0.7\sigma)$} \\
        \hline
                \end{tabular}
    }
		\label{tab:cp_uta:ccd2}
	\end{center}
\end{sidewaystable}

In the absence of the penguin contribution (tree dominance),
the time-dependent parameters would be given by
$S_{b \to c\bar c d} = - \etacp \sin(2\beta)$,
$C_{b \to c\bar c d} = 0$,
$S_{+-} = \sin(2\beta + \delta)$,
$S_{-+} = \sin(2\beta - \delta)$,
$C_{+-} = - C_{-+}$ and 
${\cal A} = 0$,
where $\delta$ is the strong phase difference between the 
$D^{*+}D^-$ and $D^{*-}D^+$ decay amplitudes.
In the presence of the penguin contribution,
there is no clean interpretation in terms of CKM parameters,
however
direct $\CP$ violation may be observed as any of
$C_{b \to c\bar c d} \neq 0$, $C_{+-} \neq - C_{-+}$ or $A_{+-} \neq 0$.

\begin{figure}[htbp]
  \begin{center}
    \begin{tabular}{cc}
      \resizebox{0.46\textwidth}{!}{
        \includegraphics{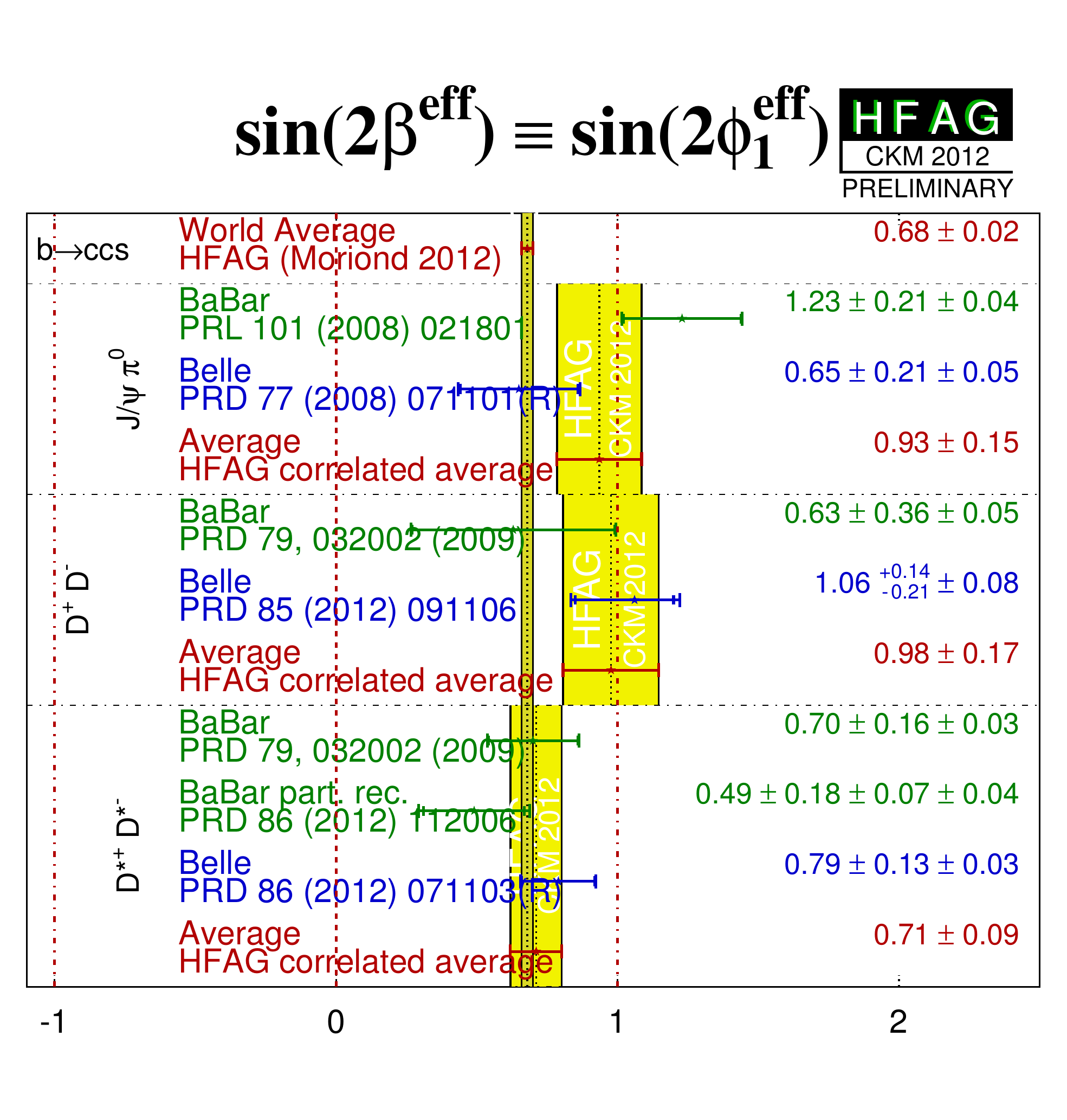}
      }
      &
      \resizebox{0.46\textwidth}{!}{
        \includegraphics{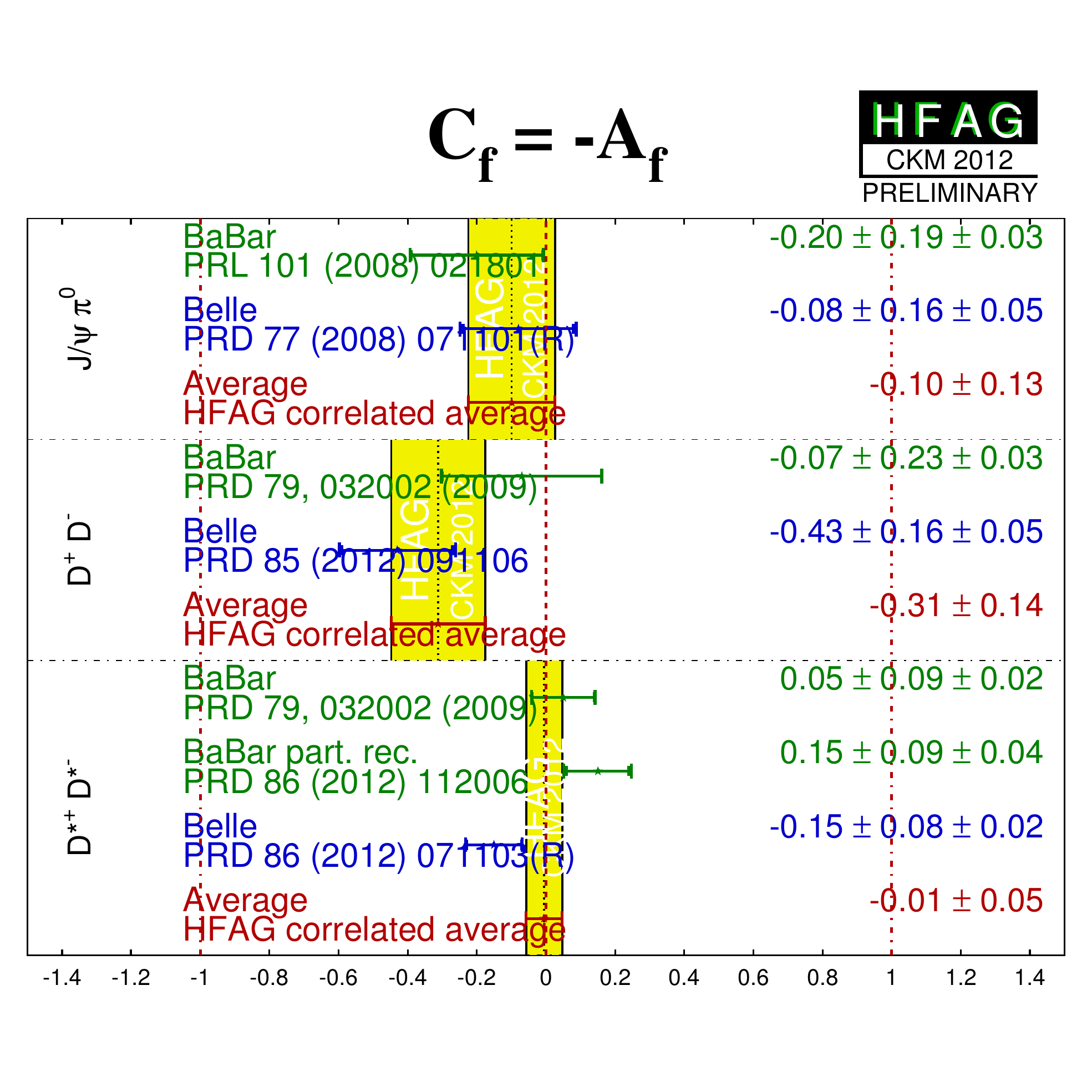}
      }
    \end{tabular}
  \end{center}
  \vspace{-0.8cm}
  \caption{
    Averages of 
    (left) $-\etacp S_{b \to c\bar c d}$ and (right) $C_{b \to c\bar c d}$.
    The $-\etacp S_{b \to q\bar q s}$ figure compares the results to 
    the world average 
    for $-\etacp S_{b \to c\bar c s}$ (see Sec.~\ref{sec:cp_uta:ccs:cp_eigen}).
  }
  \label{fig:cp_uta:ccd}
\end{figure}

\begin{figure}[htbp]
  \begin{center}
    \resizebox{0.66\textwidth}{!}{
      \includegraphics{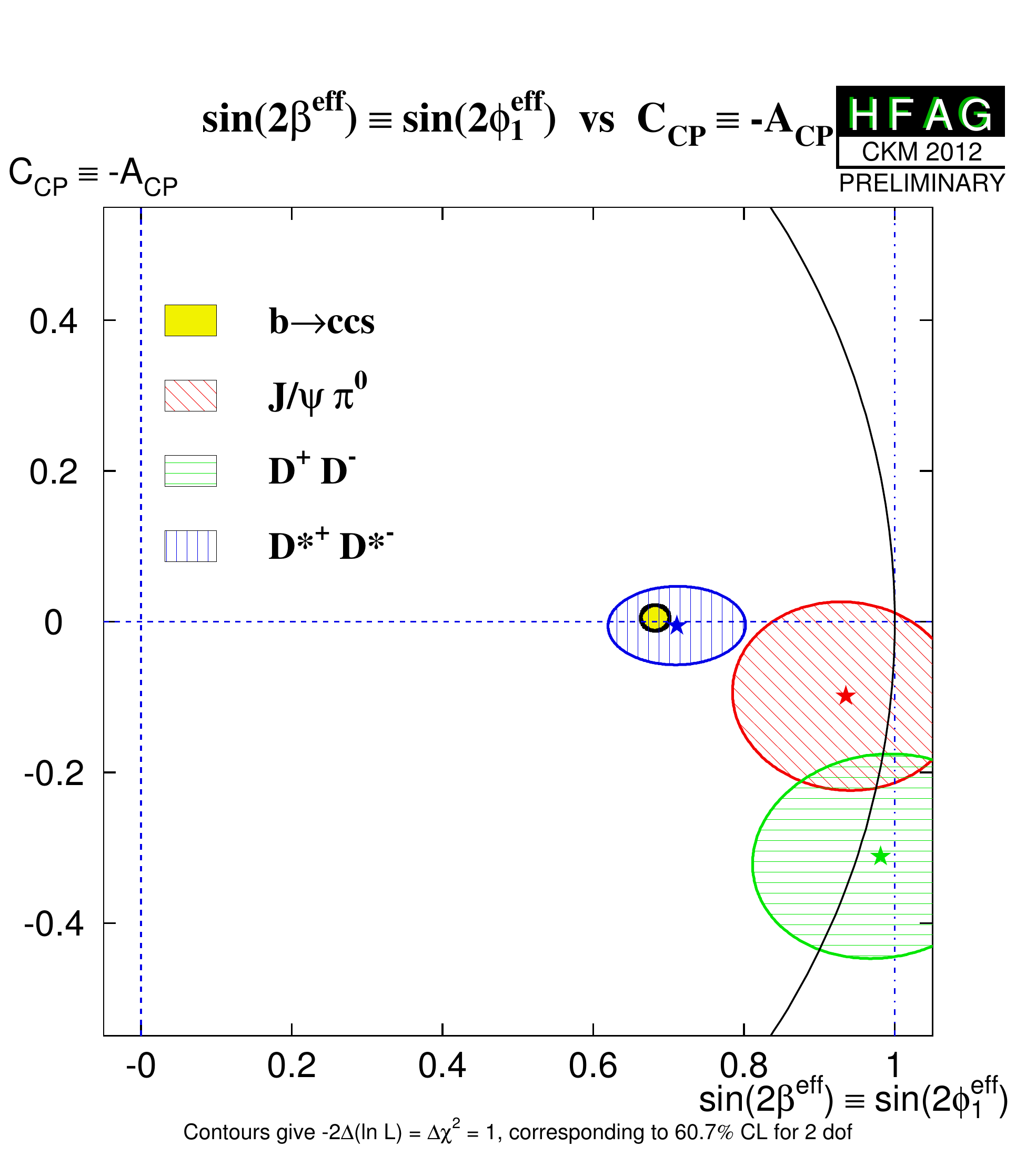}
    }
  \end{center}
  \vspace{-0.8cm}
  \caption{
    Compilation of constraints in the 
    $-\etacp S_{b \to c\bar c d}$ \vs\ $C_{b \to c\bar c d}$ plane.
  }
  \label{fig:cp_uta:ccd_SvsC-all}
\end{figure}

The averages for the $b \to c\bar c d$ modes 
are shown in Figs.~\ref{fig:cp_uta:ccd} and~\ref{fig:cp_uta:ccd_SvsC-all}.
Results are consistent with tree dominance,
and with the Standard Model,
though the \belle\ results in $\Bz \to D^+D^-$~\cite{Fratina:2007zk}
show an indication of $\CP$ violation in decay,
and hence a non-zero penguin contribution.
The average of $S_{b \to c\bar c d}$ in both $J/\psi \pi^{0}$ and
$D^{*+}D^{*-}$ final states is more than $5\sigma$ from zero, corresponding to
observations of \CP violation in these decay channels.
That in the $D^+D^-$ final state is more than $3\sigma$ from zero;
however, due to the large uncertainty and possible non-Gaussian effects,
any strong conclusion should be deferred.


\mysubsection{Time-dependent $\CP$ asymmetries in $b \to q\bar{q}d$ transitions
}
\label{sec:cp_uta:qqd}

Decays such as $\Bz\to\KS\KS$ are pure $b \to q\bar{q}d$ penguin transitions.
As shown in Eq.~(\ref{eq:cp_uta:b_to_d}),
this diagram has different contributing weak phases,
and therefore the observables are sensitive to the difference 
(which can be chosen to be either $\beta$ or $\gamma$).
Note that if the contribution with the top quark in the loop dominates,
the weak phase from the decay amplitudes should cancel that from mixing,
so that no $\CP$ violation (neither mixing-induced nor in decay) occurs.
Non-zero contributions from loops with intermediate up and charm quarks
can result in both types of effect 
(as usual, a strong phase difference is required for $\CP$ violation in decay
to occur).

\begin{figure}[htbp]
  \begin{center}
    \begin{tabular}{cc}
      \resizebox{0.46\textwidth}{!}{
        \includegraphics{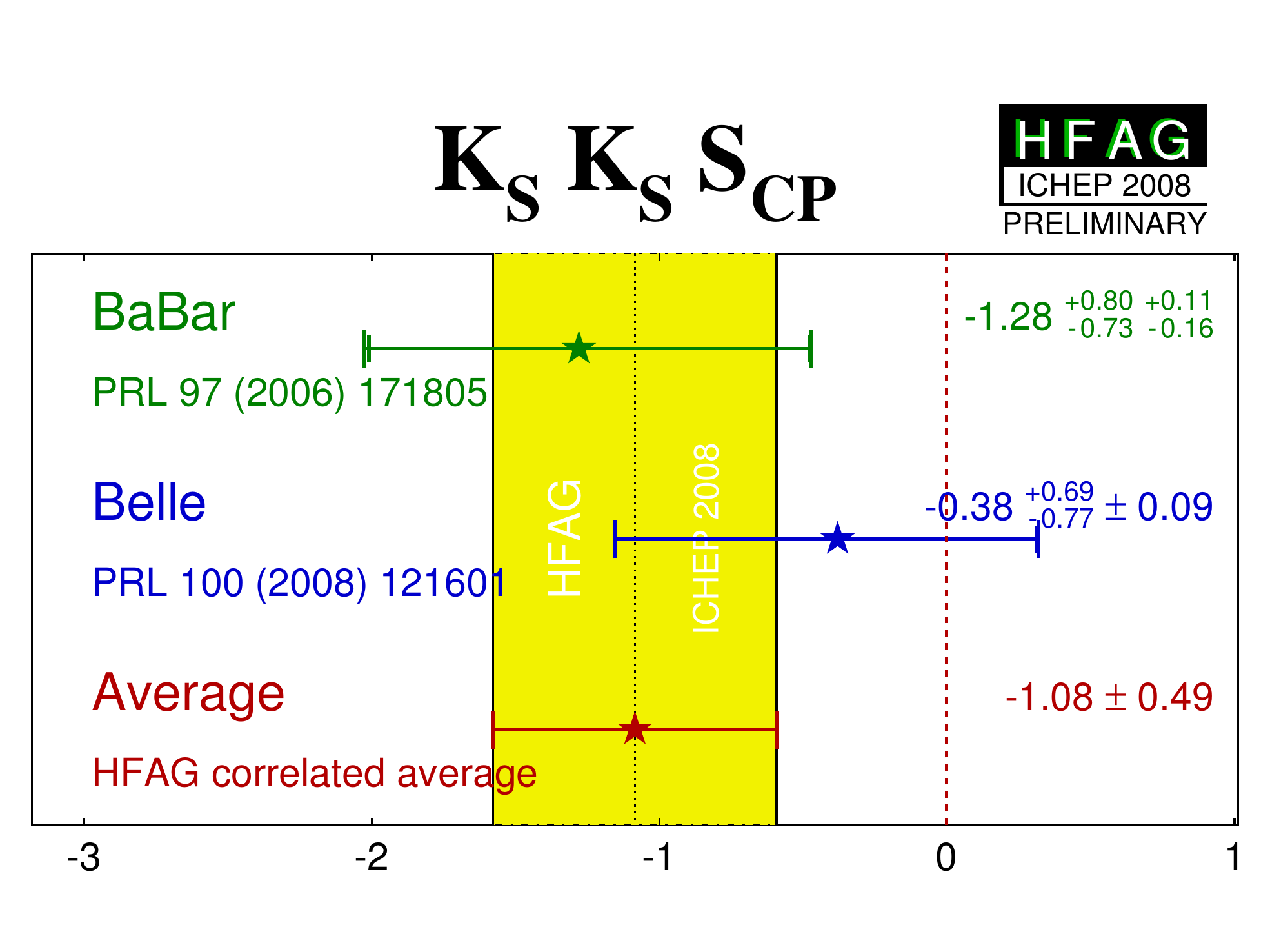}
      }
      &
      \resizebox{0.46\textwidth}{!}{
        \includegraphics{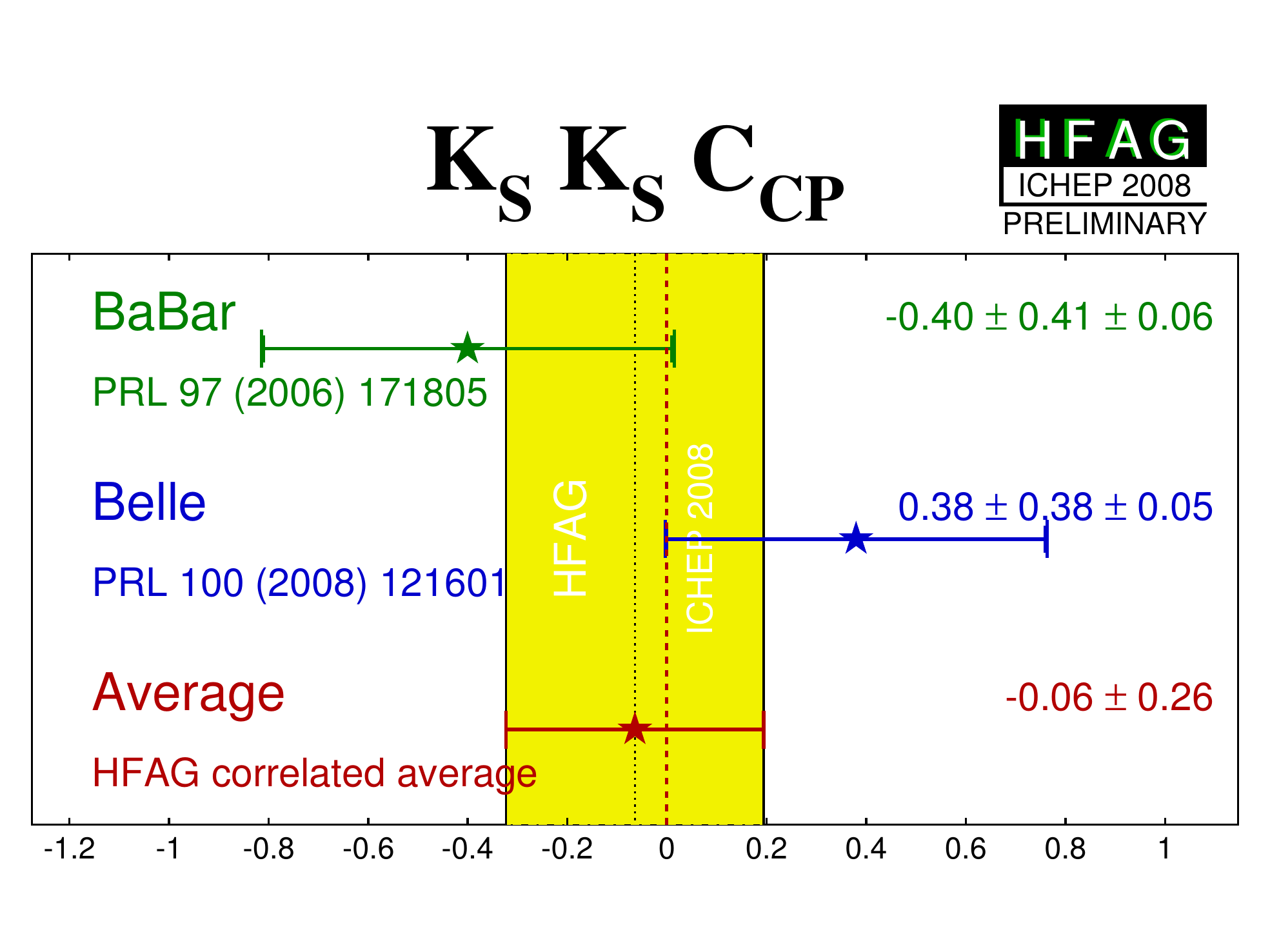}
      }
    \end{tabular}
  \end{center}
  \vspace{-0.8cm}
  \caption{
    Averages of 
    (left) $S_{b \to q\bar q d}$ and (right) $C_{b \to q\bar q d}$ 
    for the mode $\Bz \to \KS\KS$.
  }
  \label{fig:cp_uta:qqd:ksks}
\end{figure}

Both \babar~\cite{Aubert:2006gm} and \belle~\cite{Nakahama:2007dg}
have performed time-dependent analyses of $\Bz\to\KS\KS$.
The results are shown in Table~\ref{tab:cp_uta:qqd}
and Fig.~\ref{fig:cp_uta:qqd:ksks}.

\begin{table}[htb]
	\begin{center}
		\caption{
			Results for $\Bz \to \KS\KS$.
		}
		\vspace{0.2cm}
		\setlength{\tabcolsep}{0.0pc}
		\begin{tabular*}{\textwidth}{@{\extracolsep{\fill}}lrcccc} \hline
	\mc{2}{l}{Experiment} & $N(B\bar{B})$ & $S_{CP}$ & $C_{CP}$ & Correlation \\
	\hline
	\babar & \cite{Aubert:2006gm} & 350M & $-1.28 \,^{+0.80}_{-0.73} \,^{+0.11}_{-0.16}$ & $-0.40 \pm 0.41 \pm 0.06$ & $-0.32$ \\
	\belle & \cite{Nakahama:2007dg} & 657M & $-0.38 \,^{+0.69}_{-0.77} \pm 0.09$ & $0.38 \pm 0.38 \pm 0.05$ & $0.48$ \\
	\hline
	\mc{3}{l}{\bf Average} & $-1.08 \pm 0.49$ & $-0.06 \pm 0.26$ & $0.14$ \\
	\mc{3}{l}{\small Confidence level} & \mc{2}{c}{\small $0.29~(1.1\sigma)$} & \\
		\hline
		\end{tabular*}
		\label{tab:cp_uta:qqd}
	\end{center}
\end{table}

\mysubsection{Time-dependent asymmetries in $b \to s\gamma$ transitions
}
\label{sec:cp_uta:bsg}

The radiative decays $b \to s\gamma$ produce photons 
which are highly polarised in the Standard Model.
The decays $\Bz \to F \gamma$ and $\Bzb \to F \gamma$ 
produce photons with opposite helicities, 
and since the polarisation is, in principle, observable,
these final states cannot interfere.
The finite mass of the $s$ quark introduces small corrections
to the limit of maximum polarisation,
but any large mixing induced $\CP$ violation would be a signal for new physics.
Since a single weak phase dominates the $b \to s \gamma$ transition in the 
Standard Model, the cosine term is also expected to be small.

Atwood {\it et al.}~\cite{Atwood:2004jj} have shown that 
an inclusive analysis with respect to $\KS\pi^0\gamma$ can be performed,
since the properties of the decay amplitudes 
are independent of the angular momentum of the $\KS\pi^0$ system. 
However, if non-dipole operators contribute significantly to the amplitudes, 
then the Standard Model mixing-induced $\CP$ violation could be larger 
than the na\"\i ve expectation 
$S \simeq -2 (m_s/m_b) \sin \left(2\beta\right)$~\cite{Grinstein:2004uu,Grinstein:2005nu}.
In this case, 
the $\CP$ parameters may vary over the $\KS\pi^0\gamma$ Dalitz plot, 
for example as a function of the $\KS\pi^0$ invariant mass.
Explicit calculations indicate such corrections are small
for exclusive final states~\cite{Matsumori:2005ax,Ball:2006cva}.

With the above in mind, 
we quote two averages: one for $K^*(892)$ candidates only, 
and the other one for the inclusive $\KS\pi^0\gamma$ decay (including the $K^*(892)$).
If the Standard Model dipole operator is dominant, 
both should give the same quantities 
(the latter naturally with smaller statistical error). 
If not, care needs to be taken in interpretation of the inclusive parameters, 
while the results on the $K^*(892)$ resonance remain relatively clean.
Results from \babar\ and \belle\ are
used for both averages; both experiments use the invariant mass range 
$0.60 \ {\rm GeV}/c^2 < M_{\KS\pi^0} < 1.80 \ {\rm GeV}/c^2$
in the inclusive analysis.

In addition to the $\KS\pi^0\gamma$ decay, both \babar\ and \belle\ have presented results using $\KS\eta\gamma$ and $\KS\rho\gamma$, and \belle\ has in addition presented results using $\KS\phi\gamma$.
For the $\KS\rho\gamma$ case, due to the non-negligible width of the $\rho^0$ meson, decays selected as $\Bz \to \KS\rho^0\gamma$ can include a significant contribution from $K^{*\pm}\pi^\mp\gamma$ decays, which are flavour-specific and do not have the same oscillation phenomenology. 
Both \babar\ and \belle\ measure $S_{\rm eff}$ for all \B decay candidates with the $\rho^0$ selection being $0.6 < m(\pip\pim) < 0.9 \ {\rm GeV}/c^2$, obtaining $0.14 \pm 0.25 \,^{+0.04}_{-0.03}$ (\babar) and $0.09 \pm 0.27 \,^{+0.04}_{-0.07}$ (\belle). These values are then corrected for a ``dilution factor'', that is evaluated with different methods in the two experiments: \babar~\cite{Akar:2013ima} obtain $0.549 \,^{+0.096}_{-0.094}$ while \belle~\cite{Li:2008qma} obtain $0.83 \,^{+0.19}_{-0.03}$. Until the discrepancy between these values is understood, the average of the results should be treated with caution.

\begin{table}[htb]
	\begin{center}
		\caption{
      Averages for $b \to s \gamma$ modes.
		}
		\vspace{0.2cm}
		\setlength{\tabcolsep}{0.0pc}
		\begin{tabular*}{\textwidth}{@{\extracolsep{\fill}}lrcccc} \hline
	\mc{2}{l}{Experiment} & $N(B\bar{B})$ & $S_{CP} (b \to s \gamma)$ & $C_{CP} (b \to s \gamma)$ & Correlation \\
        \hline
        \mc{6}{c}{$\Kstar(892)\gamma$} \\
	\babar & \cite{Aubert:2008gy} & 467M & $-0.03 \pm 0.29 \pm 0.03$ & $-0.14 \pm 0.16 \pm 0.03$ & $0.05$ \\
	\belle & \cite{Ushiroda:2006fi} & 535M & $-0.32 \,^{+0.36}_{-0.33} \pm 0.05$ & $0.20 \pm 0.24 \pm 0.05$ & $0.08$ \\
	\mc{3}{l}{\bf Average} & $-0.16 \pm 0.22$ & $-0.04 \pm 0.14$ & $0.06$ \\
	\mc{3}{l}{\small Confidence level} & \mc{2}{c}{\small $0.40~(0.9\sigma)$} & \\
		\hline

        \mc{6}{c}{$\KS \pi^0 \gamma$ (including $\Kstar(892)\gamma$)} \\
	\babar & \cite{Aubert:2008gy} & 467M & $-0.17 \pm 0.26 \pm 0.03$ & $-0.19 \pm 0.14 \pm 0.03$ & $0.04$ \\
	\belle & \cite{Ushiroda:2006fi} & 535M & $-0.10 \pm 0.31 \pm 0.07$ & $0.20 \pm 0.20 \pm 0.06$ & $0.08$ \\
	\mc{3}{l}{\bf Average} & $-0.15 \pm 0.20$ & $-0.07 \pm 0.12$ & $0.05$ \\
        \mc{3}{l}{\small Confidence level} & \mc{2}{c}{\small $0.30~(1.0\sigma)$} & \\

		\hline

        \mc{6}{c}{$\KS \eta \gamma$} \\
	\babar & \cite{Aubert:2008js} & 465M & $-0.18 \,^{+0.49}_{-0.46} \pm 0.12$ & $-0.32 \,^{+0.40}_{-0.39} \pm 0.07$ & $-0.17$ \\
	\belle & \cite{Belle-KSetagamma} & 772M & $-1.32 \pm 0.77 \pm 0.36$ & $0.48 \pm 0.41 \pm 0.07$ & $-0.14$ \\
	\hline
	\mc{3}{l}{\bf Average} & $-0.49 \pm 0.42$ & $0.06 \pm 0.29$ & $-0.15$ \\
	\mc{3}{l}{\small Confidence level} & \mc{2}{c}{\small $0.24~(1.2\sigma)$} & \\

        \mc{6}{c}{$\KS \rho^0 \gamma$} \\
	\babar & \cite{Akar:2014hda} & 471M & $0.25 \pm 0.46 \,^{+0.08}_{-0.06}$ & $-0.39 \pm 0.20 \pm 0.05$ & $-0.09$ \\
	\belle & \cite{Li:2008qma} & 657M & $0.11 \pm 0.33 \,^{+0.05}_{-0.09}$ & $-0.05 \pm 0.18 \pm 0.06$ & $\phantom{-}0.04$ \\
	\hline
	\mc{3}{l}{\bf Average} & $0.14 \pm 0.27$ & $-0.20 \pm 0.14$ & $-0.01$ \\
	\mc{3}{l}{\small Confidence level} & \mc{2}{c}{\small $0.47~(0.7\sigma)$} & \\

        \mc{6}{c}{$\KS \phi \gamma$} \\
	\belle & \cite{Sahoo:2011zd} & 772M & $0.74 \,^{+0.72}_{-1.05} \,^{+0.10}_{-0.24}$ & $-0.35 \pm 0.58 \,^{+0.10}_{-0.23}$ & \textendash{} \\
	\hline
		\end{tabular*}
		\label{tab:cp_uta:bsg}
	\end{center}
\end{table}

\begin{figure}[htbp]
  \begin{center}
    \begin{tabular}{cc}
      \resizebox{0.46\textwidth}{!}{
        \includegraphics{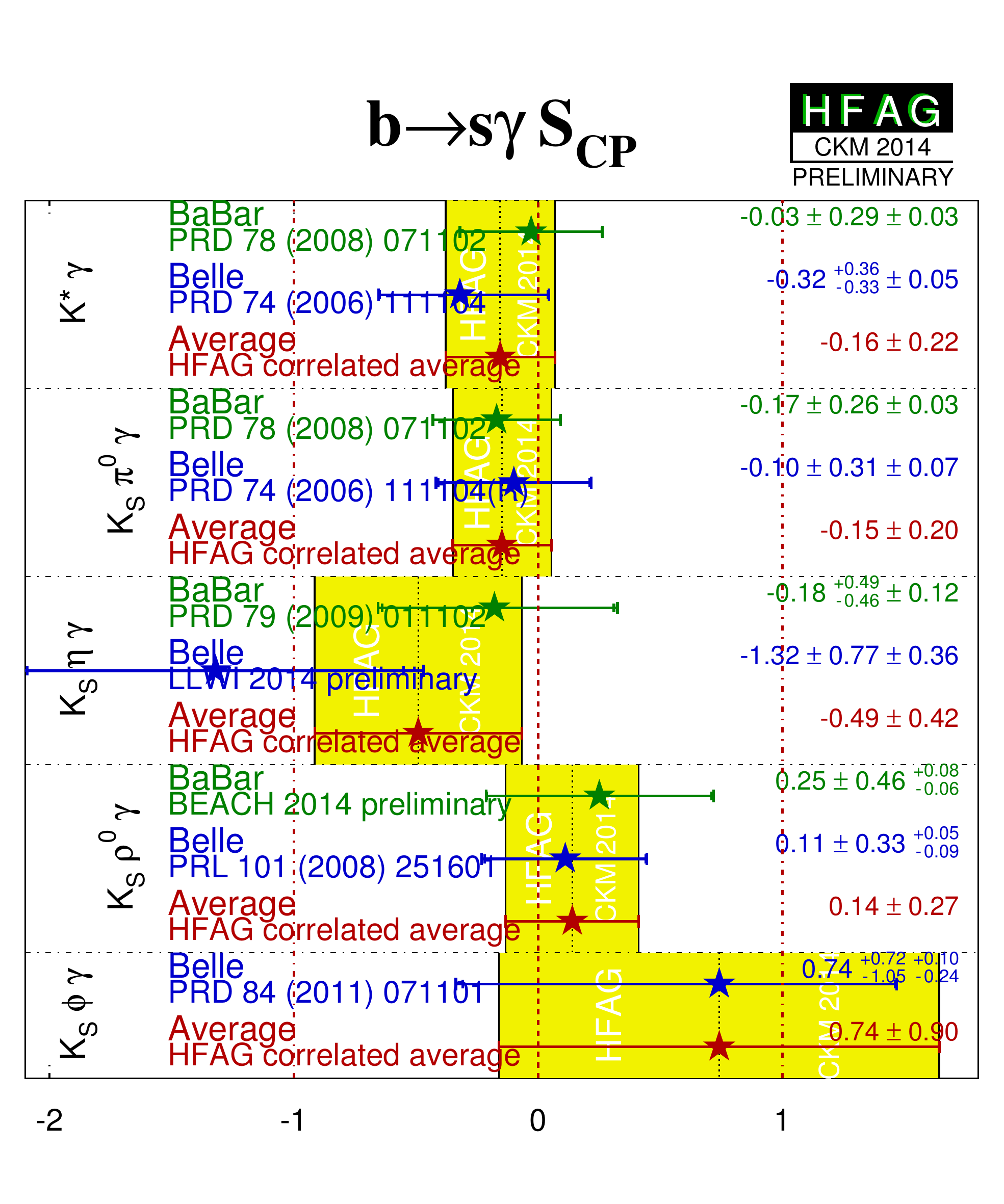}
      }
      &
      \resizebox{0.46\textwidth}{!}{
        \includegraphics{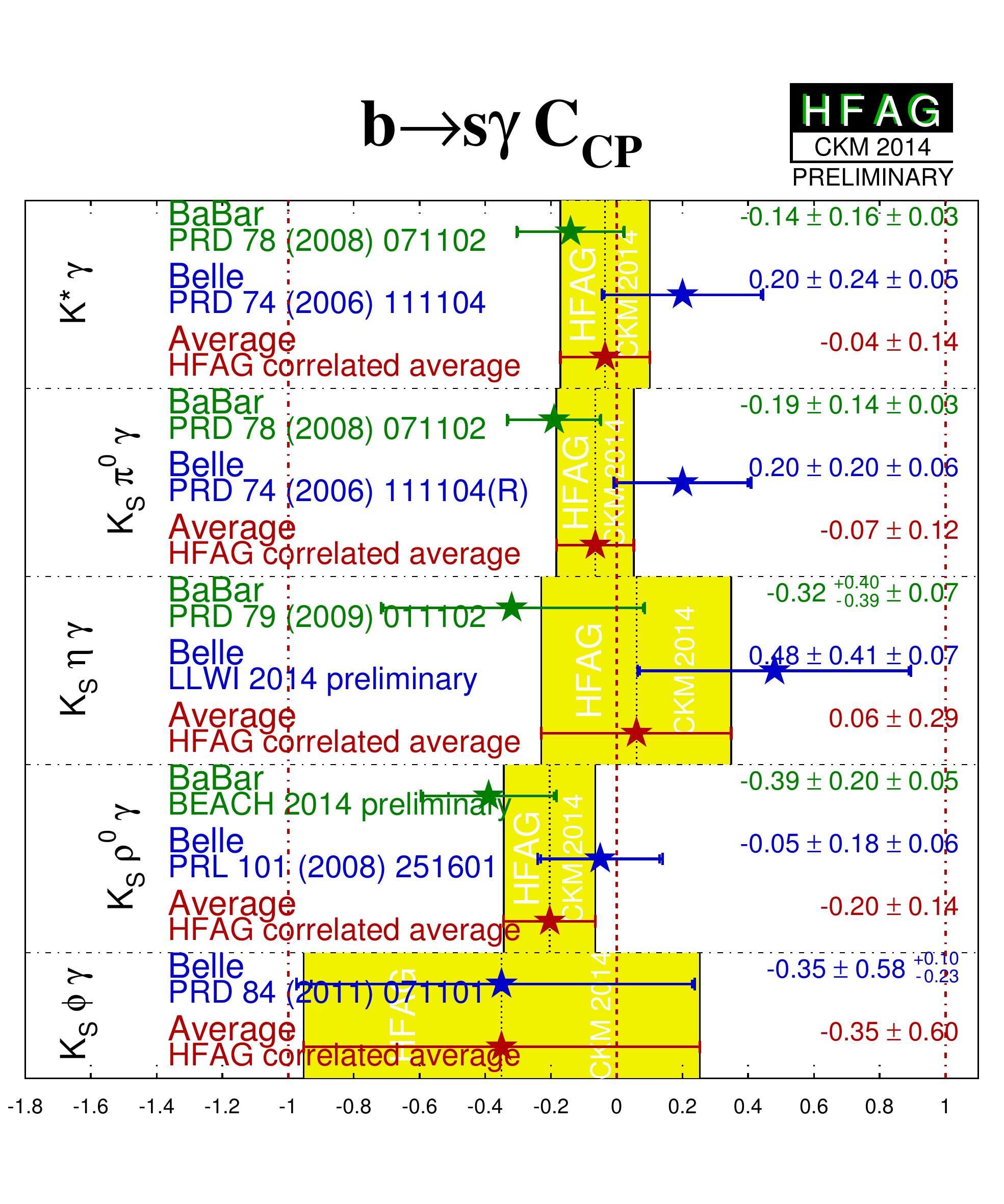}
      }
    \end{tabular}
  \end{center}
  \vspace{-0.8cm}
  \caption{
    Averages of (left) $S_{b \to s \gamma}$ and (right) $C_{b \to s \gamma}$.
    Recall that the data for $K^*\gamma$ is a subset of that for $\KS\pi^0\gamma$.
  }
  \label{fig:cp_uta:bsg}
\end{figure}

\begin{figure}[htbp]
  \begin{center}
    \resizebox{0.33\textwidth}{!}{
      \includegraphics{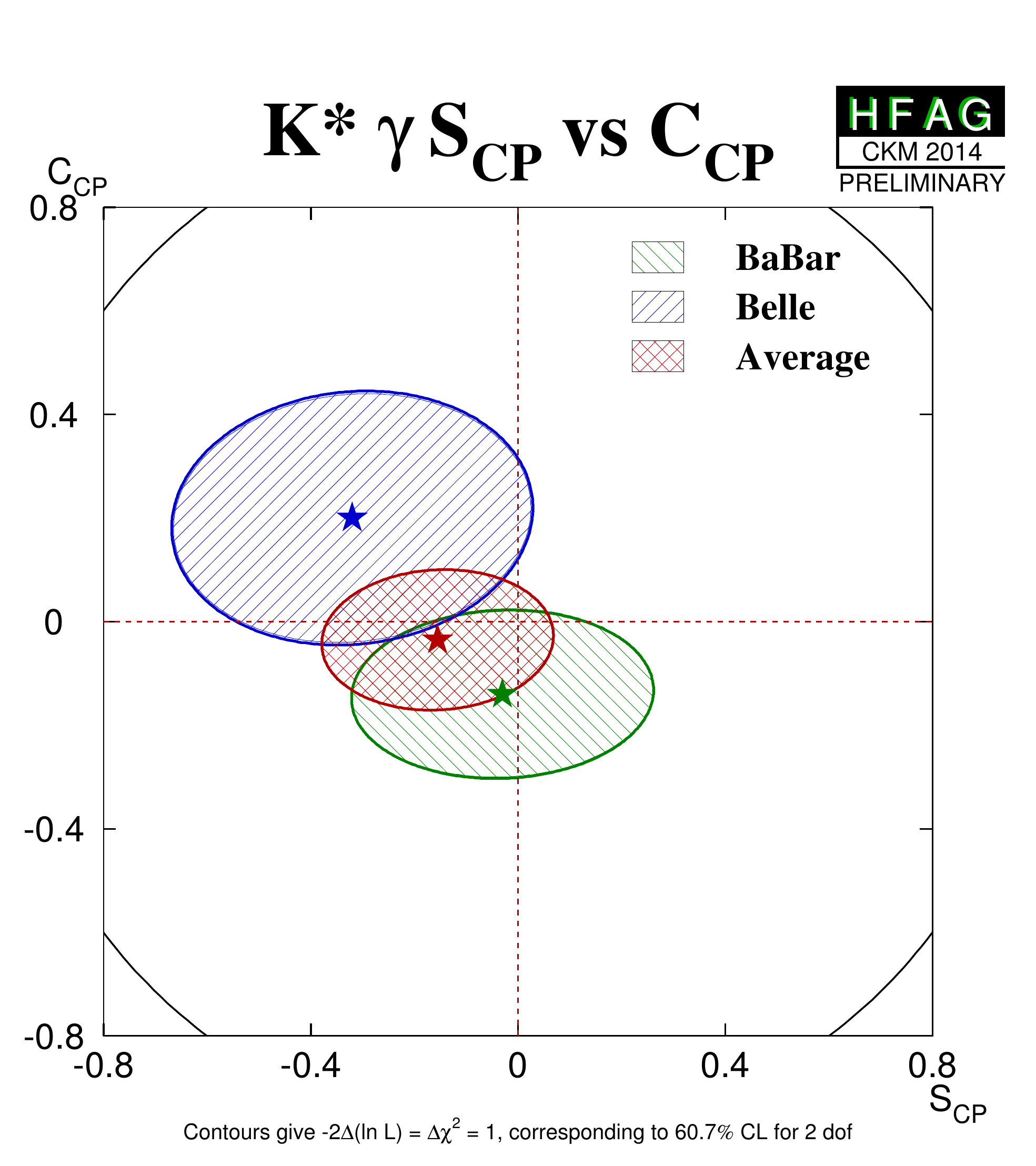}
    }
    \hspace{0.08\textwidth}
    \resizebox{0.33\textwidth}{!}{
      \includegraphics{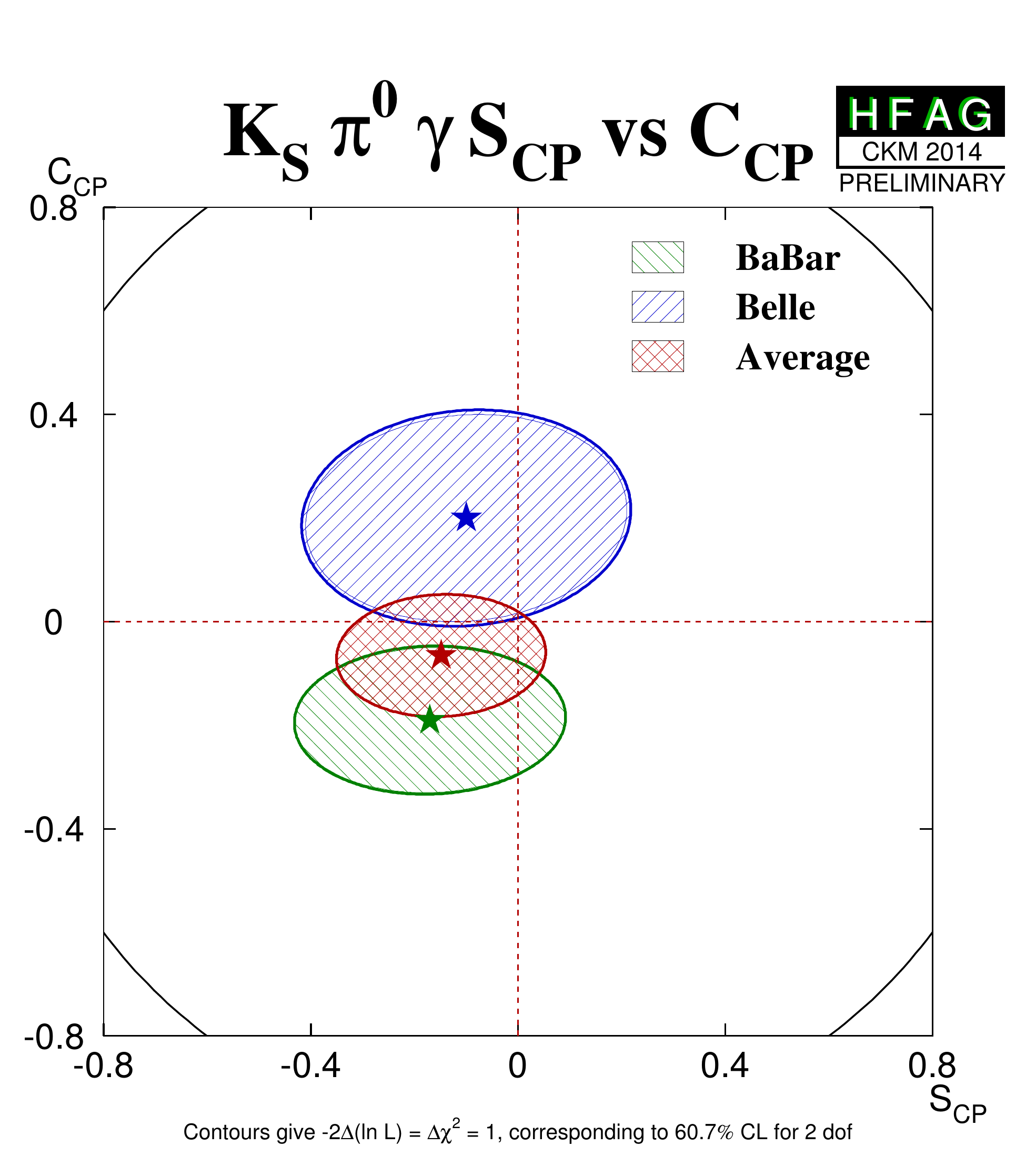}
    } \\
    \resizebox{0.33\textwidth}{!}{
      \includegraphics{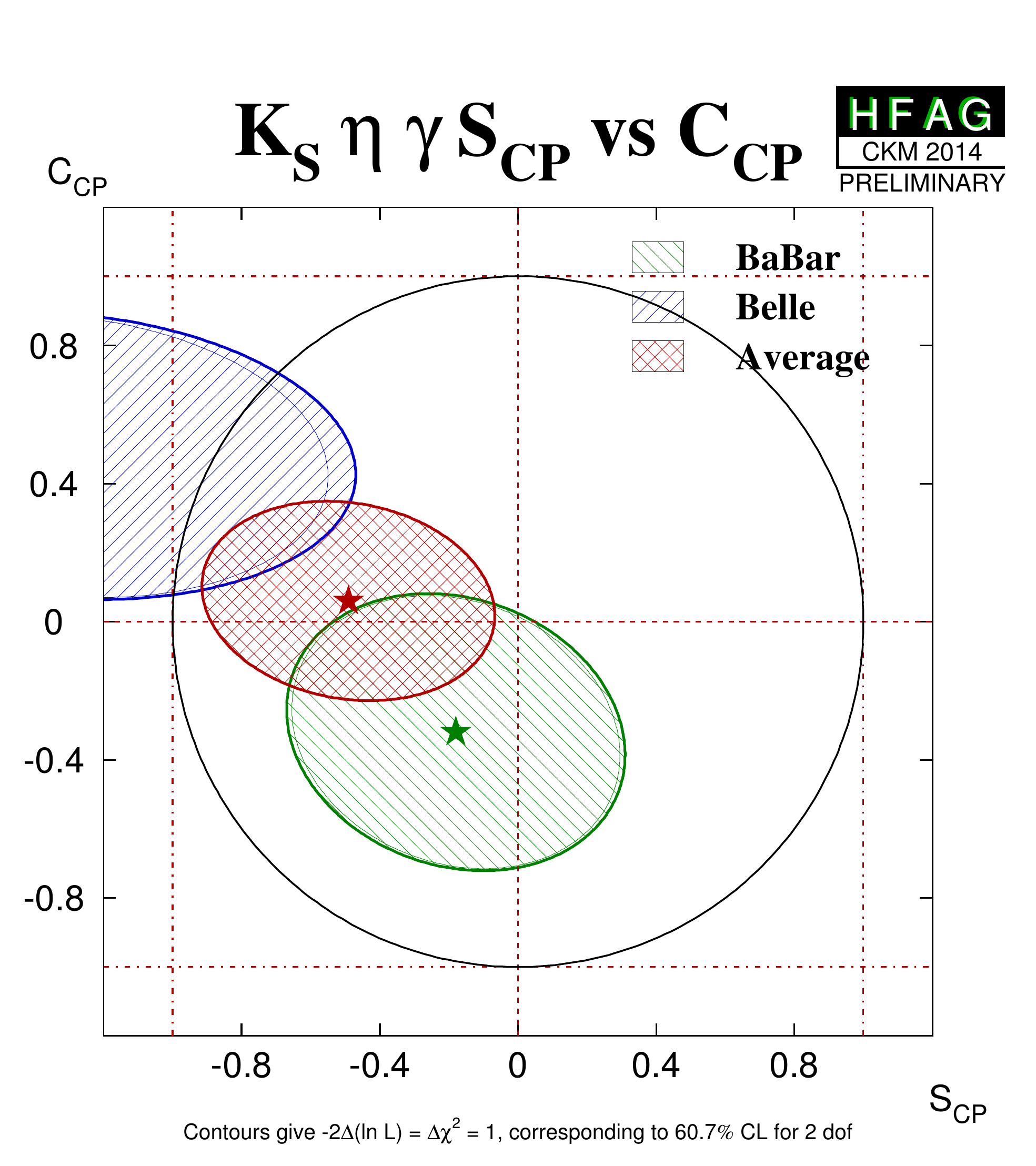}
    }
    \hspace{0.08\textwidth}
    \resizebox{0.33\textwidth}{!}{
      \includegraphics{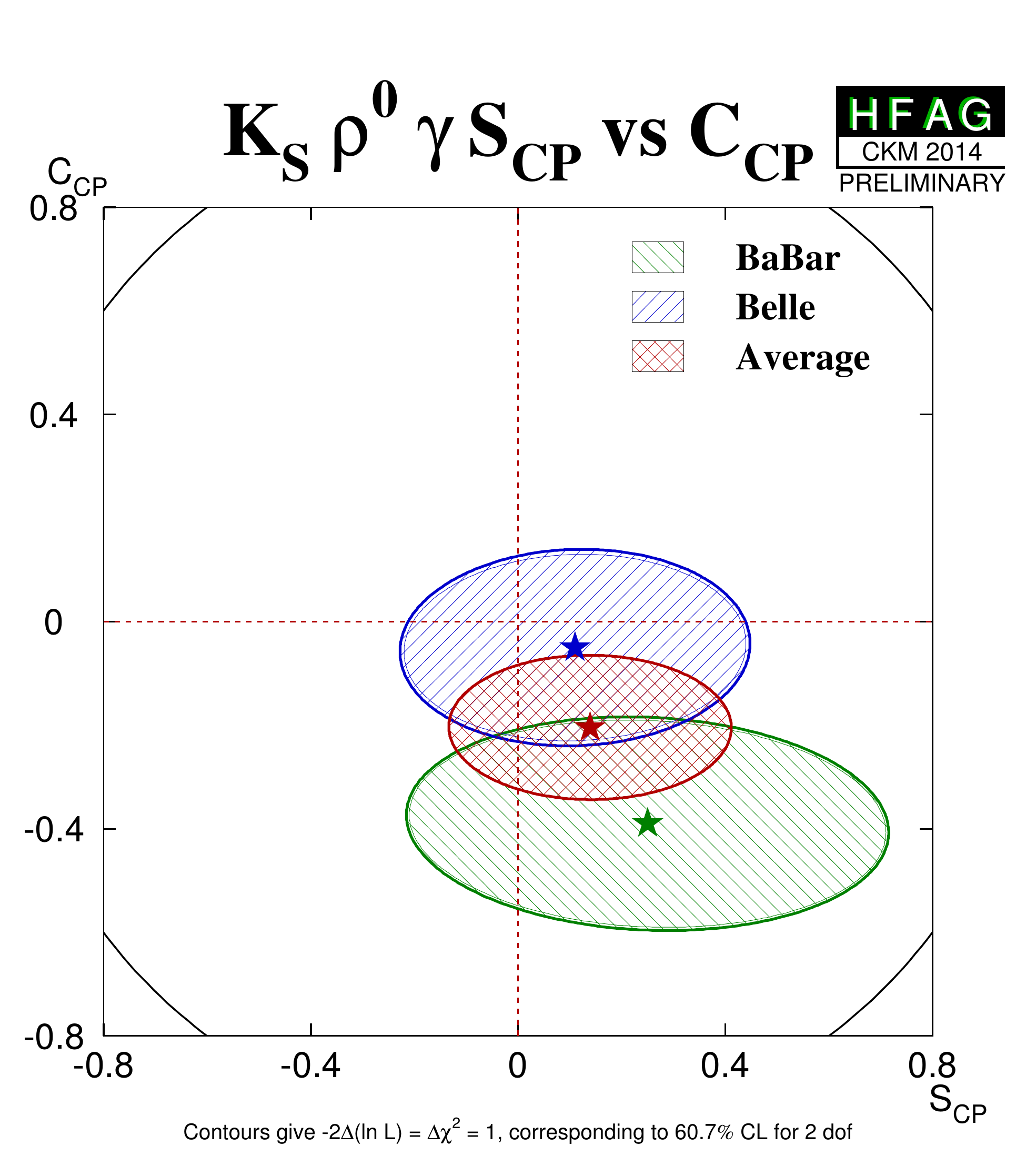}
    }
  \end{center}
  \vspace{-0.5cm}
  \caption{
    Averages of four $b \to s\gamma$ dominated channels,
    for which correlated averages are performed,
    in the $S_{\CP}$ \vs\ $C_{\CP}$ plane.
    (Top left) $\Bz \to K^*\gamma$, 
    (top right) $\Bz \to \KS\pi^0\gamma$ (including $K^*\gamma$),
    (bottom left) $\Bz \to \KS\eta\gamma$,
    (bottom right).
  }
  \label{fig:cp_uta:bsg_SvsC}
\end{figure}

The results are shown in Table~\ref{tab:cp_uta:bsg},
and in Figs.~\ref{fig:cp_uta:bsg} and~~\ref{fig:cp_uta:bsg_SvsC}.
No significant $\CP$ violation results are seen;
the results are consistent with the Standard Model
and with other measurements in the $b \to s\gamma$ system (see Sec.~\ref{sec:rare}).

\mysubsection{Time-dependent asymmetries in $b \to d\gamma$ transitions
}
\label{sec:cp_uta:bdg}

The formalism for the radiative decays $b \to d\gamma$ is much the same
as that for $b \to s\gamma$ discussed above.
Assuming dominance of the top quark in the loop,
the weak phase in decay should cancel with that from mixing,
so that the mixing-induced \CP\ violation parameter $S_{\CP}$ 
should be very small.
Corrections due to the finite light quark mass are smaller
compared to $b \to s\gamma$, since $m_d < m_s$,
and although QCD corrections may still play a role,
they cannot significantly affect the prediction $S_{b \to d \gamma} \simeq 0$.
Large \CP\ violation effects could, however, be seen through
a non-zero value of $C_{b \to d \gamma}$, 
since the top loop is not the only contribution.

Results using the mode $\Bz \to \rho^0\gamma$ are available from 
\belle\ and are shown in Table~\ref{tab:cp_uta:bdg}.

\begin{table}[htb]
	\begin{center}
		\caption{
			Averages for $\Bz \to \rho^{0} \gamma$.
		}
		\vspace{0.2cm}
		\setlength{\tabcolsep}{0.0pc}
		\begin{tabular*}{\textwidth}{@{\extracolsep{\fill}}lrcccc} \hline
	\mc{2}{l}{Experiment} & $N(B\bar{B})$ & $S_{CP}$ & $C_{CP}$ & Correlation \\
	\hline
	\belle & \cite{Ushiroda:2007jf} & 657M & $-0.83 \pm 0.65 \pm 0.18$ & $0.44 \pm 0.49 \pm 0.14$ & $-0.08$ \\
		\hline
		\end{tabular*}
		\label{tab:cp_uta:bdg}
	\end{center}
\end{table}

\mysubsection{Time-dependent $\CP$ asymmetries in $b \to u\bar{u}d$ transitions
}
\label{sec:cp_uta:uud}

The $b \to u \bar u d$ transition can be mediated by either 
a $b \to u$ tree amplitude or a $b \to d$ penguin amplitude.
These transitions can be investigated using 
the time dependence of $\Bz$ decays to final states containing light mesons.
Results are available from both \babar\ and \belle\ for the 
$\CP$ eigenstate ($\etacp = +1$) $\pi^+\pi^-$ final state
and for the vector-vector final state $\rho^+\rho^-$,
which is found to be dominated by the $\CP$-even
longitudinally polarised component
(\babar\ measure $f_{\rm long} = 
0.992 \pm 0.024 \, ^{+0.026}_{-0.013}$~\cite{Aubert:2007nua}
while \belle\ measure $f_{\rm long} = 
0.941 \, ^{+0.034}_{-0.040} \pm 0.030$~\cite{Somov:2006sg}).
\babar\ has also performed a time-dependent analysis of the 
vector-vector final state $\rho^0\rho^0$~\cite{:2008iha},
in which they measure  $f_{\rm long} = 0.70 \pm 0.14 \pm 0.05$;
\belle\ measure a smaller branching fraction than \babar\ for
$\Bz\to\rho^0\rho^0$~\cite{:2008et} with corresponding signal yields too small
to perform time-dependent or angular analyses.
\babar\ has furthermore performed a time-dependent analysis of the 
$\Bz \to a_1^\pm \pi^\mp$ decay~\cite{Aubert:2006gb}; further experimental
input for the extraction of $\alpha$ from this channel is reported in a later
publication~\cite{:2009ii}.

\begin{figure}[htbp]
  \begin{center}
    \begin{tabular}{cc}
      \resizebox{0.46\textwidth}{!}{
        \includegraphics{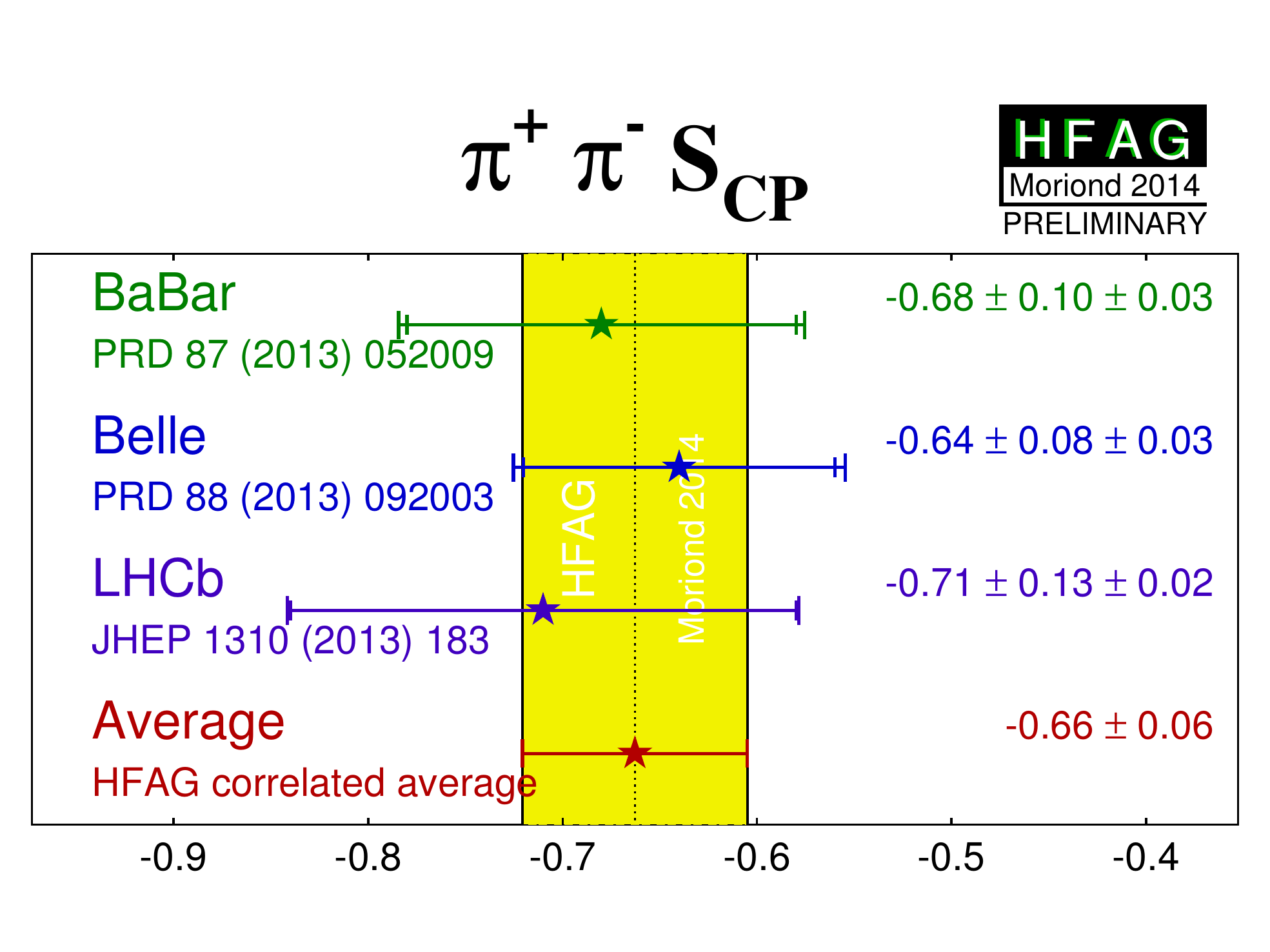}
      }
      &
      \resizebox{0.46\textwidth}{!}{
        \includegraphics{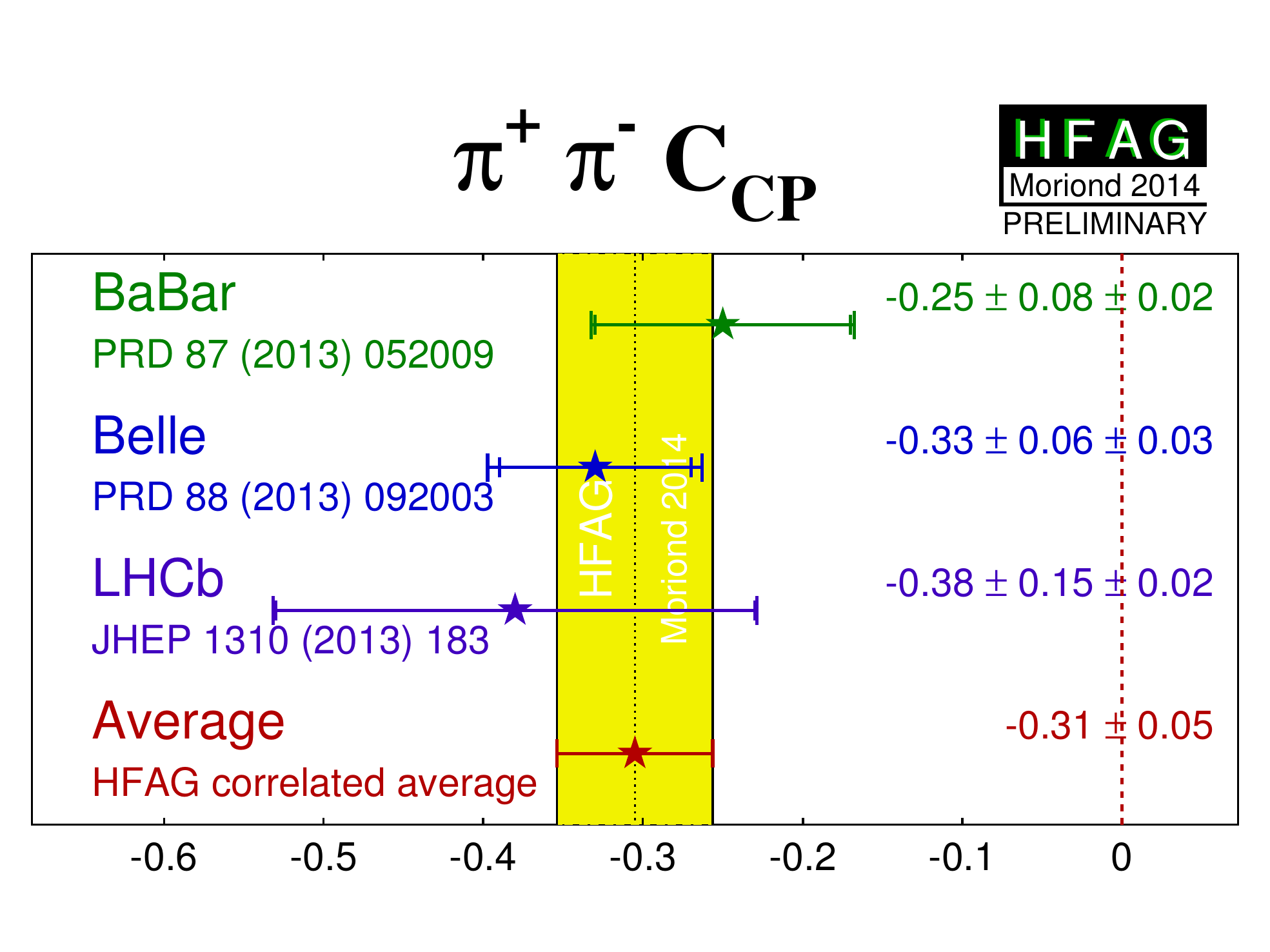}
      }
    \end{tabular}
  \end{center}
  \vspace{-0.8cm}
  \caption{
    Averages of (left) $S_{b \to u\bar u d}$ and (right) $C_{b \to u\bar u d}$
    for the mode $\Bz \to \pi^+\pi^-$.
  }
  \label{fig:cp_uta:uud:pipi}
\end{figure}

\begin{figure}[htbp]
  \begin{center}
    \begin{tabular}{cc}
      \resizebox{0.46\textwidth}{!}{
        \includegraphics{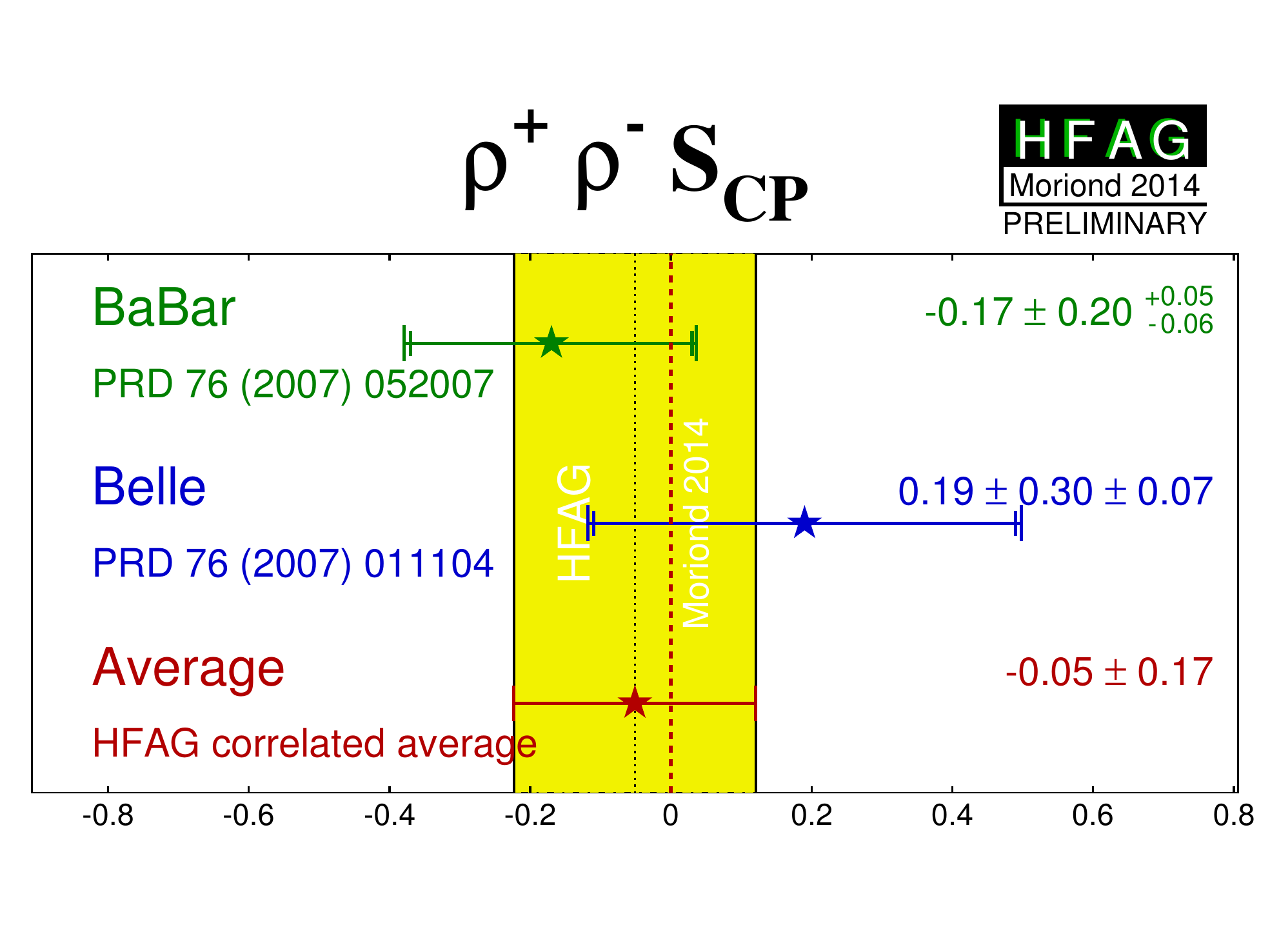}
      }
      &
      \resizebox{0.46\textwidth}{!}{
        \includegraphics{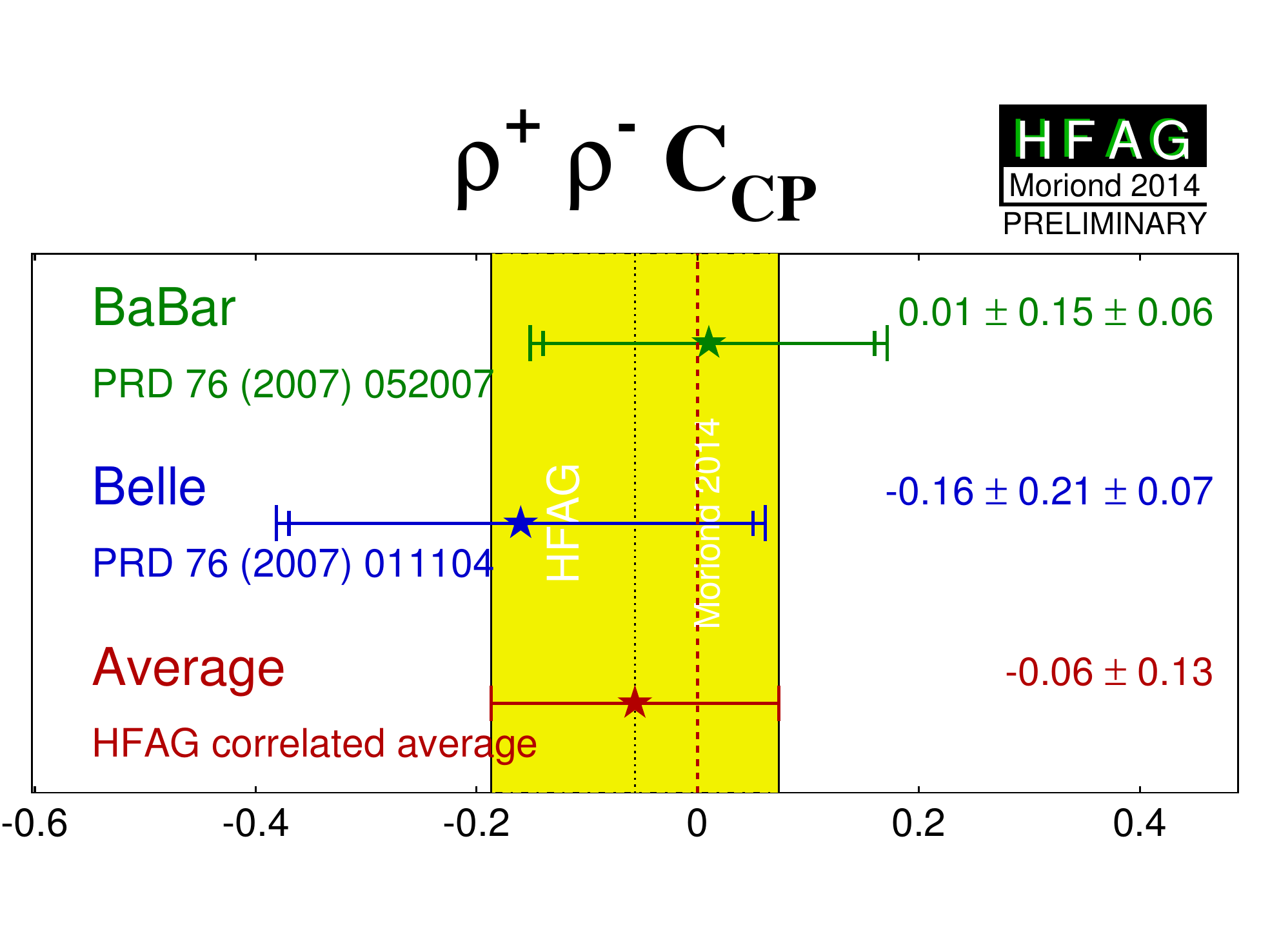}
      }
    \end{tabular}
  \end{center}
  \vspace{-0.8cm}
  \caption{
    Averages of (left) $S_{b \to u\bar u d}$ and (right) $C_{b \to u\bar u d}$
    for the mode $\Bz \to \rho^+\rho^-$.
  }
  \label{fig:cp_uta:uud:rhorho}
\end{figure}

\begin{figure}[htbp]
  \begin{center}
    \begin{tabular}{cc}
      \resizebox{0.46\textwidth}{!}{
        \includegraphics{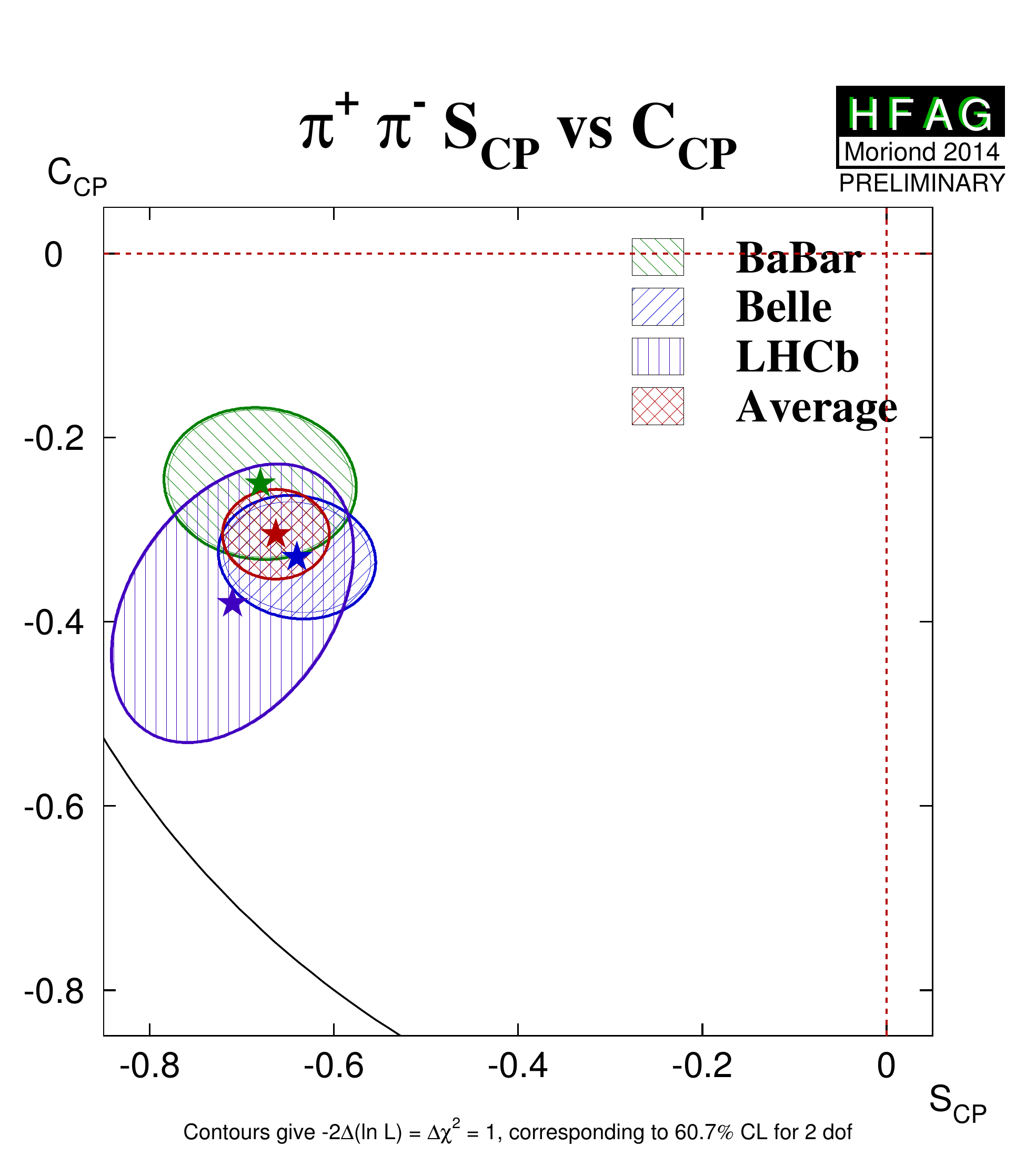}
      }      
      &
      \resizebox{0.46\textwidth}{!}{
        \includegraphics{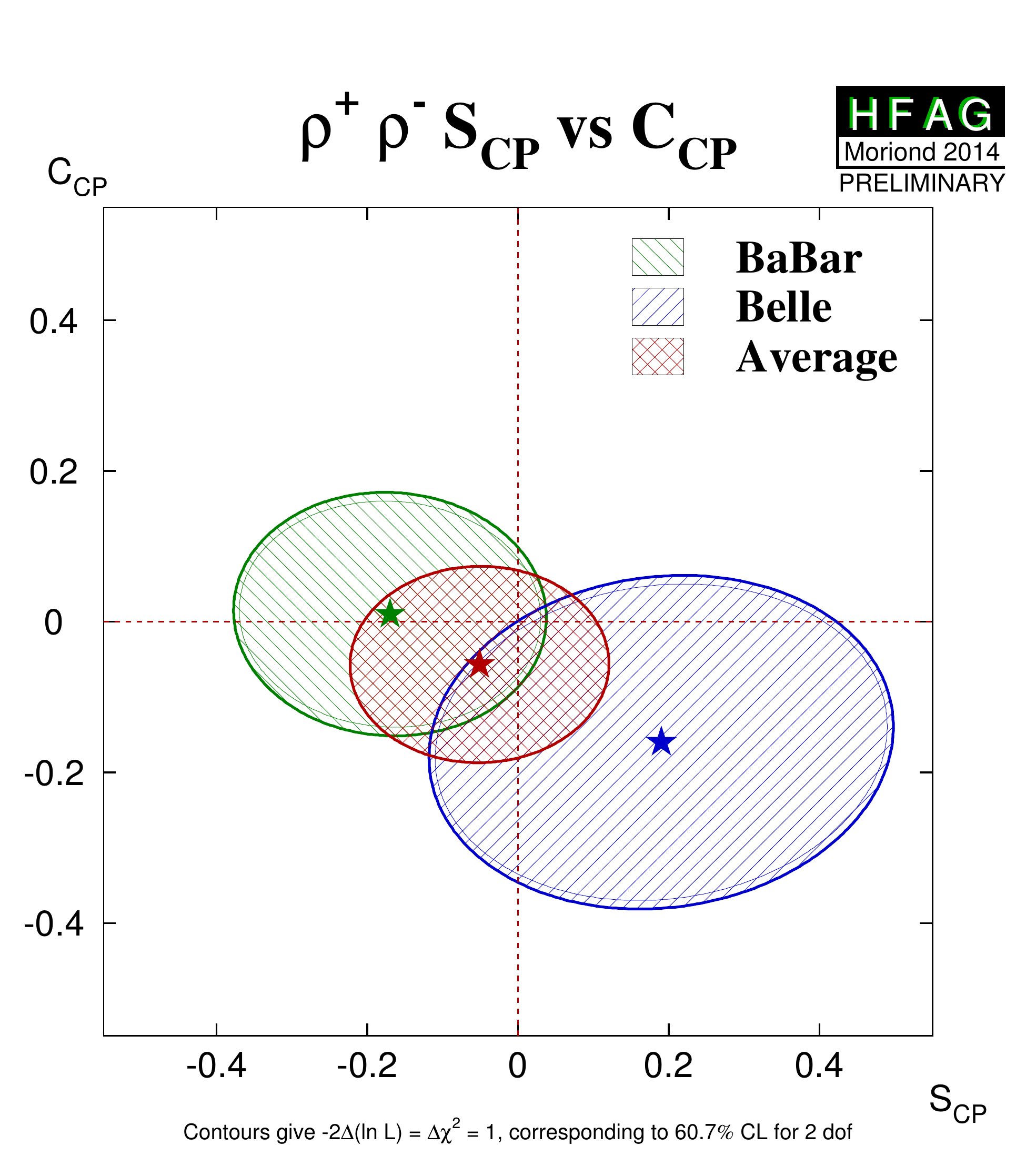}
      }
    \end{tabular}
  \end{center}
  \vspace{-0.8cm}
  \caption{
    Averages of $b \to u\bar u d$ dominated channels,
    for which correlated averages are performed,
    in the $S_{\CP}$ \vs\ $C_{\CP}$ plane.
    (Left) $\Bz \to \pi^+\pi^-$ and (right) $\Bz \to \rho^+\rho^-$.
  }
  \label{fig:cp_uta:uud_SvsC}
\end{figure}

\begin{figure}[htbp]
  \begin{center}
    \resizebox{0.46\textwidth}{!}{
      \includegraphics{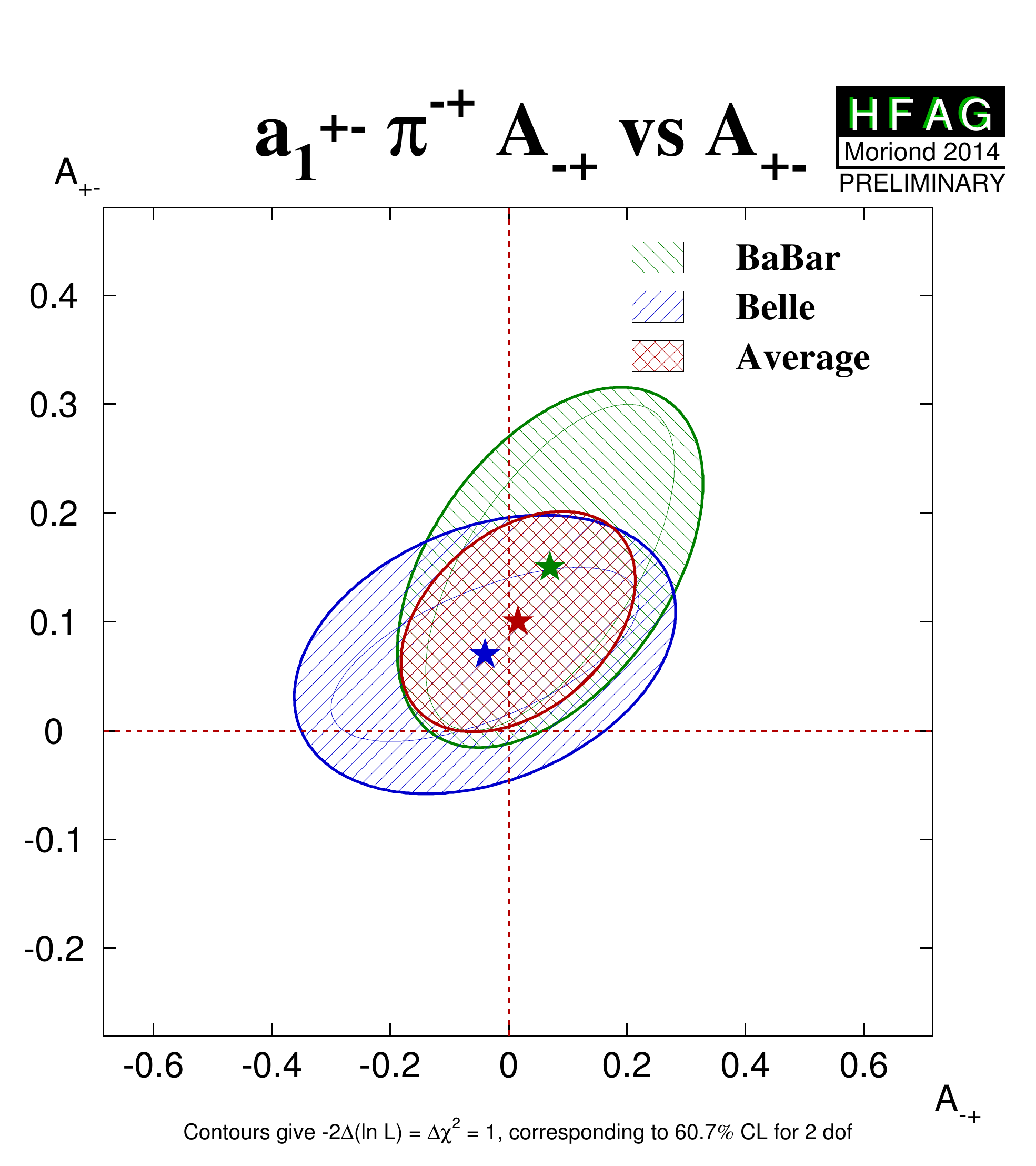}
    }
    \vspace{-0.3cm}
    \caption{
      Averages of \CP violation parameters in $\Bz \to a_1^\pm\pi^\mp$ in
      ${\cal A}^{-+}_{a_1\pi}$ \vs\ ${\cal A}^{+-}_{a_1\pi}$ space.
    }
    \label{fig:cp_uta:a1pi}
  \end{center}
\end{figure}

Results, and averages, of time-dependent \CP violation parameters in 
$b \to u \bar u d$ transitions are listed in Table~\ref{tab:cp_uta:uud}.
The averages for $\pi^+\pi^-$ are shown in Fig.~\ref{fig:cp_uta:uud:pipi},
and those for $\rho^+\rho^-$ are shown in Fig.~\ref{fig:cp_uta:uud:rhorho},
with the averages in the $S_{\CP}$ \vs\ $C_{\CP}$ plane 
shown in Fig.~\ref{fig:cp_uta:uud_SvsC} and
averages of \CP violation parameters in $\Bz \to a_1^\pm \pi^\mp$ decay shown in Fig.~\ref{fig:cp_uta:a1pi}.

\begin{sidewaystable}
	\begin{center}
		\caption{
      Averages for $b \to u \bar u d$ modes.
		}
		\vspace{0.2cm}
		\setlength{\tabcolsep}{0.0pc}
		\begin{tabular*}{\textwidth}{@{\extracolsep{\fill}}lrcccc} \hline
	\mc{2}{l}{Experiment} & Sample size & $S_{CP}$ & $C_{CP}$ & Correlation \\
	\hline
      \mc{6}{c}{$\pi^{+} \pi^{-}$} \\
	\babar & \cite{Lees:2012mma} & 467M & $-0.68 \pm 0.10 \pm 0.03$ & $-0.25 \pm 0.08 \pm 0.02$ & $-0.06$ \\
	\belle & \cite{Adachi:2013mae} & 772M & $-0.64 \pm 0.08 \pm 0.03$ & $-0.33 \pm 0.06 \pm 0.03$ & $-0.10$ \\
	LHCb & \cite{Aaij:2013tna} & $1.0 \ {\rm fb}^{-1}$ & $-0.71 \pm 0.13 \pm 0.02$ & $-0.38 \pm 0.15 \pm 0.02$ & $0.38$ \\
	\mc{3}{l}{\bf Average} & $-0.66 \pm 0.06$ & $-0.31 \pm 0.05$ & $0.00$ \\
	\mc{3}{l}{\small Confidence level} & \mc{2}{c}{\small $0.92~(0.1\sigma)$} & \\
		\hline

      \mc{6}{c}{$\rho^{+} \rho^{-}$} \\
	\babar & \cite{Aubert:2007nua} & $N(B\bar{B}) =$ 387M & $-0.17 \pm 0.20 \,^{+0.05}_{-0.06}$ & $0.01 \pm 0.15 \pm 0.06$ & $-0.04$ \\
	\belle & \cite{Abe:2007ez} & $N(B\bar{B}) =$ 535M & $0.19 \pm 0.30 \pm 0.07$ & $-0.16 \pm 0.21 \pm 0.07$ & $0.10$ \\
	\mc{3}{l}{\bf Average} & $-0.05 \pm 0.17$ & $-0.06 \pm 0.13$ & $0.01$ \\
	\mc{3}{l}{\small Confidence level} & \mc{2}{c}{\small $0.50~(0.7\sigma)$} & \\
		\hline

      \mc{6}{c}{$\rho^{0} \rho^{0}$} \\
	\babar & \cite{:2008iha} & $N(B\bar{B}) =$ 465M & $0.3 \pm 0.7 \pm 0.2$ & $0.2 \pm 0.8 \pm 0.3$ & $-0.04$ \\
 		\hline
 		\end{tabular*}

                \vspace{2ex}

    \resizebox{\textwidth}{!}{
 		\begin{tabular}{@{\extracolsep{2mm}}lrcccccc} \hline
 		\mc{2}{l}{Experiment} & $N(B\bar{B})$ & $A_{CP}^{a_1\pi}$ & $C_{a_1\pi}$ & $S_{a_1\pi}$ & $\Delta C_{a_1\pi}$ & $\Delta S_{a_1\pi}$ \\
 		\hline
      \mc{8}{c}{$a_1^{\pm} \pi^{\mp}$} \\
	\babar & \cite{Aubert:2006gb} & 384M & $-0.07 \pm 0.07 \pm 0.02$ & $-0.10 \pm 0.15 \pm 0.09$ & $0.37 \pm 0.21 \pm 0.07$ & $0.26 \pm 0.15 \pm 0.07$ & $-0.14 \pm 0.21 \pm 0.06$ \\
	\belle & \cite{Dalseno:2012hp} & 772M & $-0.06 \pm 0.05 \pm 0.07$ & $-0.01 \pm 0.11 \pm 0.09$ & $-0.51 \pm 0.14 \pm 0.08$ & $0.54 \pm 0.11 \pm 0.07$ & $-0.09 \pm 0.14 \pm 0.06$ \\ 
 	\hline
	\mc{3}{l}{\bf Average} & $-0.06 \pm 0.06$ & $-0.05 \pm 0.11$ & $-0.20 \pm 0.13$ & $0.43 \pm 0.10$ & $-0.10 \pm 0.12$ \\
	\mc{3}{l}{\small Confidence level} & \mc{5}{c}{\small $0.03~(2.1\sigma)$} \\
        \hline
		\end{tabular}
              }

                \vspace{2ex}

		\begin{tabular*}{\textwidth}{@{\extracolsep{\fill}}lrcccc} \hline
		\mc{2}{l}{Experiment} & $N(B\bar{B})$ & ${\cal A}^{-+}_{a_1\pi}$ & ${\cal A}^{+-}_{a_1\pi}$ & Correlation \\
		\hline
	\babar & \cite{Aubert:2006gb} & 384M & $0.07 \pm 0.21 \pm 0.15$ & $0.15 \pm 0.15 \pm 0.07$ & 0.63 \\
	\belle & \cite{Dalseno:2012hp} & 772M & $-0.04 \pm 0.26 \pm 0.19$ & $0.07 \pm 0.08 \pm 0.10$ & 0.61 \\
	\mc{3}{l}{\bf Average} & $0.02 \pm 0.20$ & $0.10 \pm 0.10$ & 0.38 \\
        \mc{3}{l}{\small Confidence level} & \mc{2}{c}{\small $0.92~(0.1\sigma)$} \\
		\hline
		\end{tabular*}

		\label{tab:cp_uta:uud}
	\end{center}
\end{sidewaystable}

If the penguin contribution is negligible, 
the time-dependent parameters for $\Bz \to \pi^+\pi^-$ 
and $\Bz \to \rho^+\rho^-$ are given by
$S_{b \to u\bar u d} = \etacp \sin(2\alpha)$ and
$C_{b \to u\bar u d} = 0$.
In the presence of the penguin contribution, 
$\CP$ violation in decay may arise, 
and there is no straightforward interpretation 
of $S_{b \to u\bar u d}$ and $C_{b \to u\bar u d}$.
An isospin analysis~\cite{Gronau:1990ka} 
can be used to disentangle the contributions and extract $\alpha$.

\begin{figure}[htbp]
  \begin{center}
    \begin{tabular}{cc}
      \resizebox{0.46\textwidth}{!}{
        \includegraphics{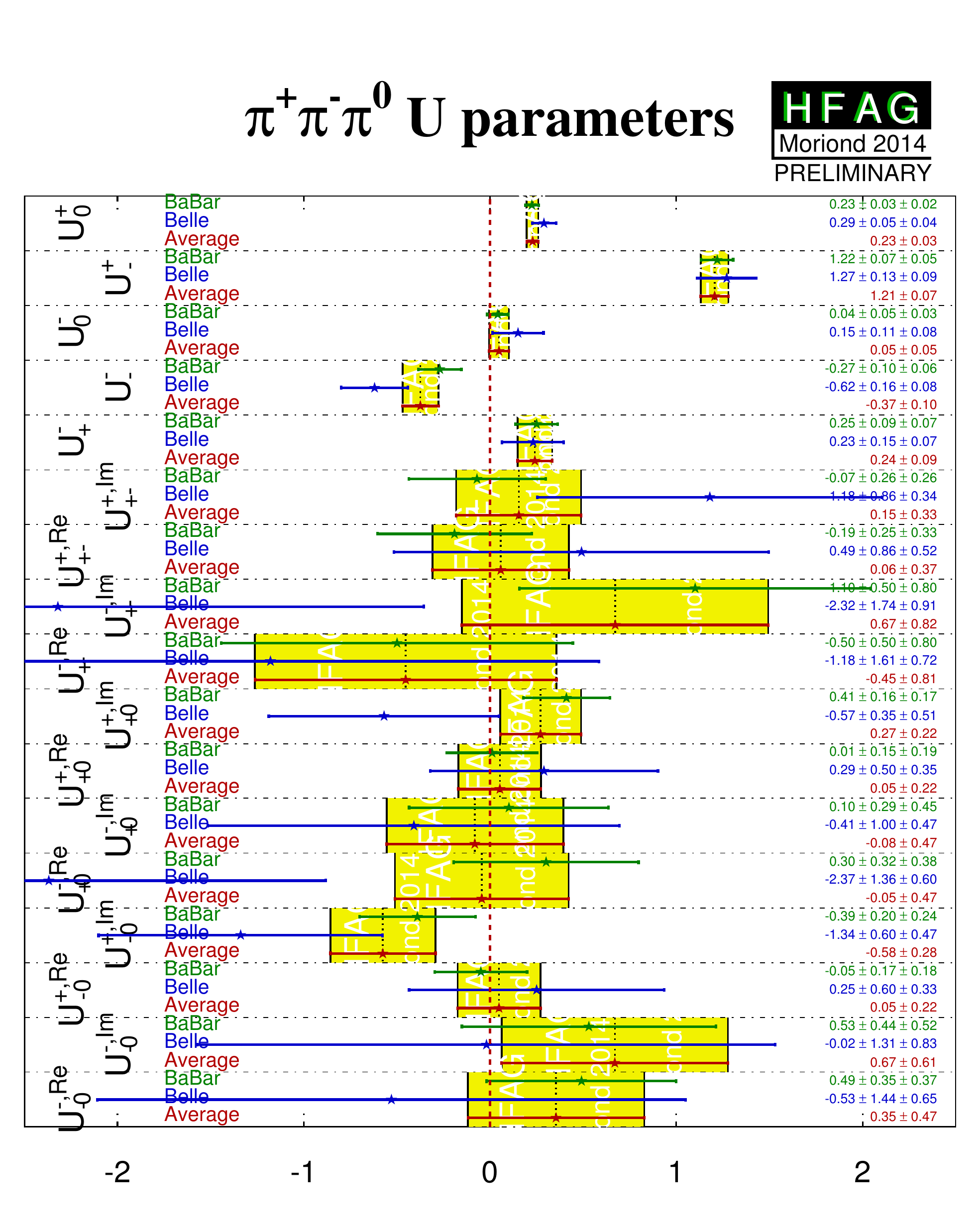}
      }
      &
      \resizebox{0.46\textwidth}{!}{
        \includegraphics{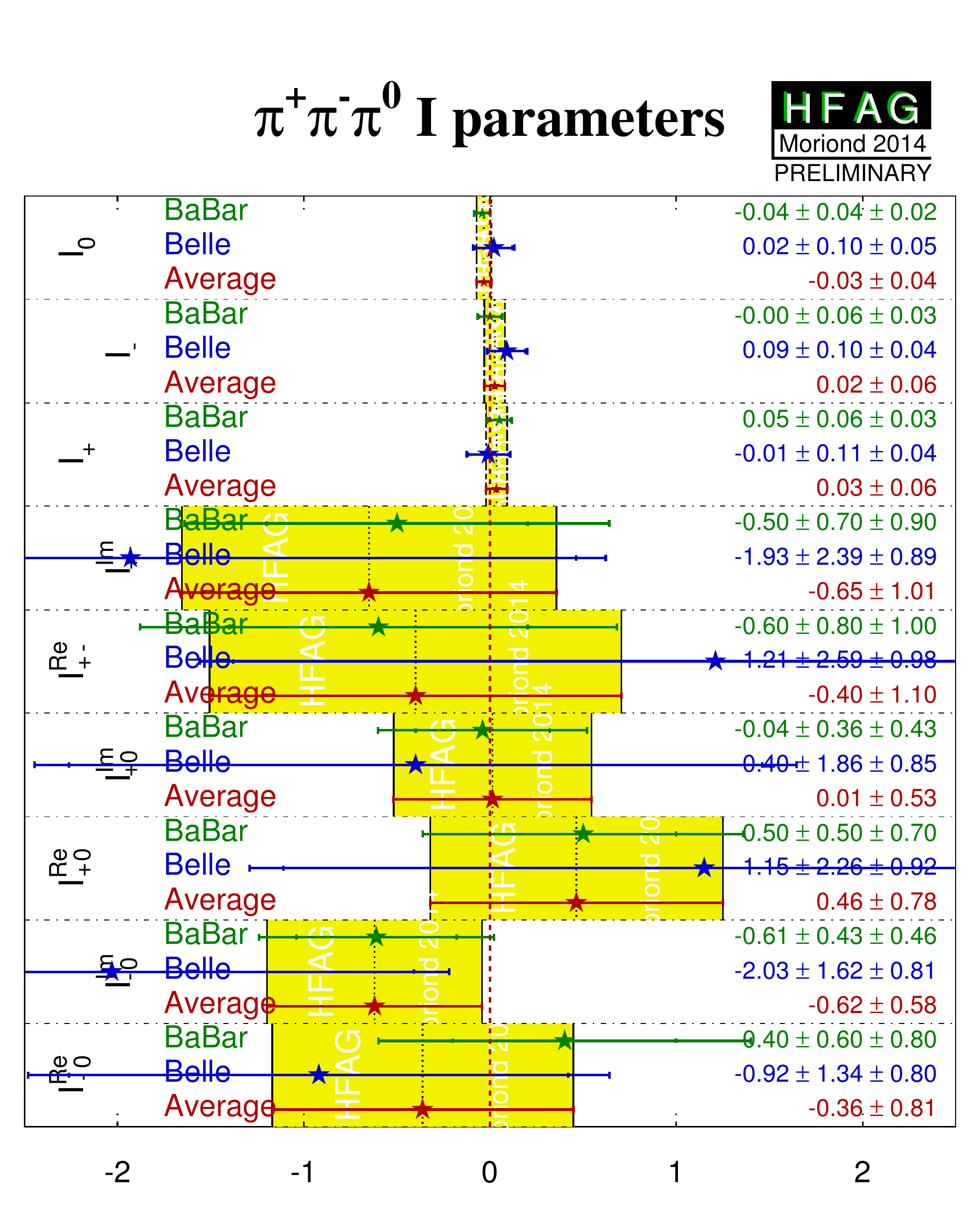}
      }
    \end{tabular}
  \end{center}
  \vspace{-0.8cm}
  \caption{
    Summary of the $U$ and $I$ parameters measured in the 
    time-dependent $\Bz \to \pi^+\pi^-\pi^0$ Dalitz plot analysis.
  }
  \label{fig:cp_uta:uud:uandi}
\end{figure}

For the non-$\CP$ eigenstate $\rho^{\pm}\pi^{\mp}$, 
both \babar~\cite{Aubert:2007jn} 
and \belle~\cite{Kusaka:2007dv,:2007mj} have performed 
time-dependent Dalitz plot (DP) analyses
of the $\pi^+\pi^-\pi^0$ final state~\cite{Snyder:1993mx};
such analyses allow direct measurements of the phases.
Both experiments have measured the $U$ and $I$ parameters discussed in 
Sec.~\ref{sec:cp_uta:notations:dalitz:pipipi0} and defined in 
Table~\ref{tab:cp_uta:pipipi0:uandi}.
We have performed a full correlated average of these parameters,
the results of which are summarised in Fig.~\ref{fig:cp_uta:uud:uandi}.

\begin{figure}[htbp]
  \begin{center}
    \begin{tabular}{cc}
      \resizebox{0.46\textwidth}{!}{
        \includegraphics{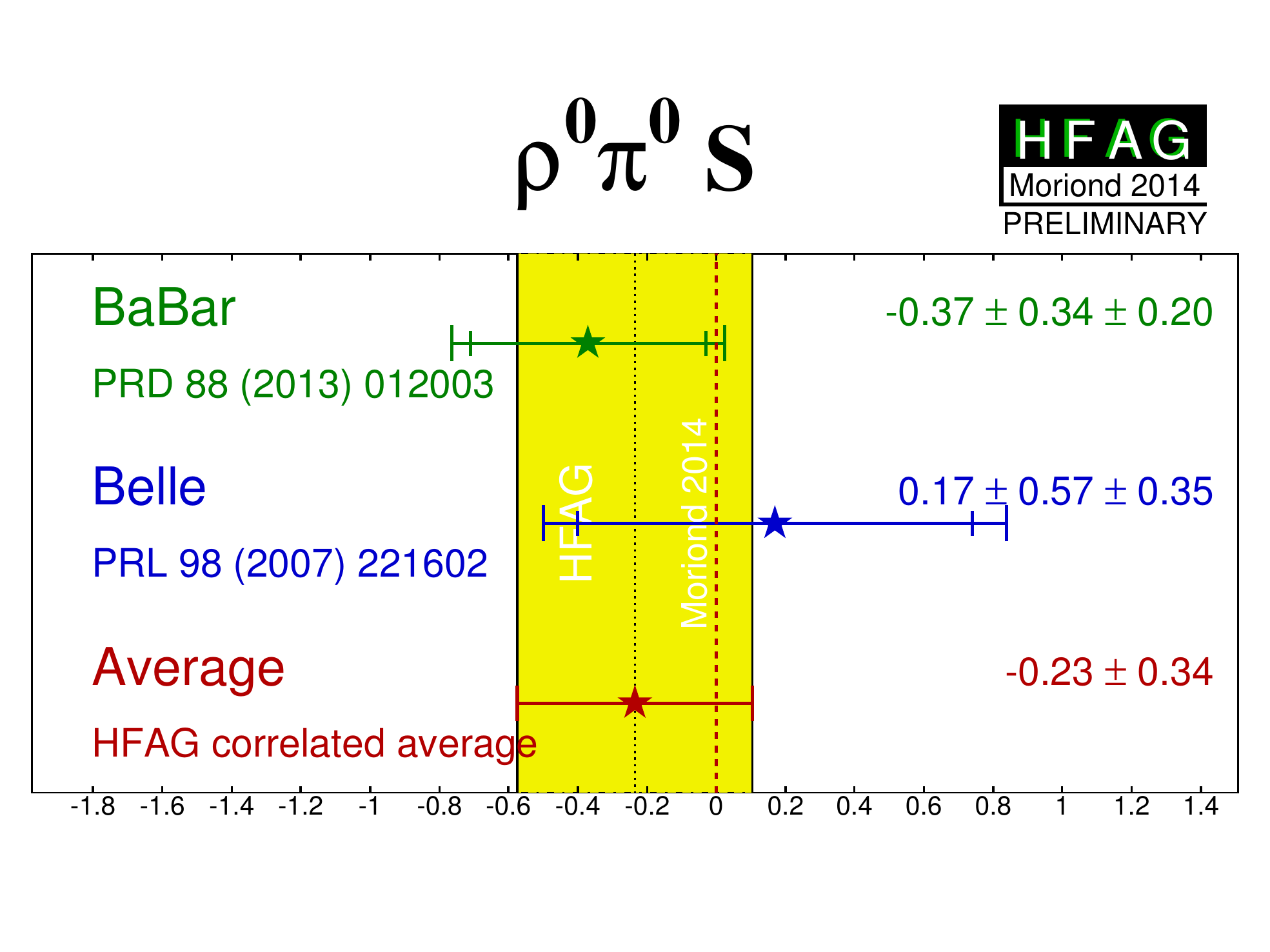}
      }
      &
      \resizebox{0.46\textwidth}{!}{
        \includegraphics{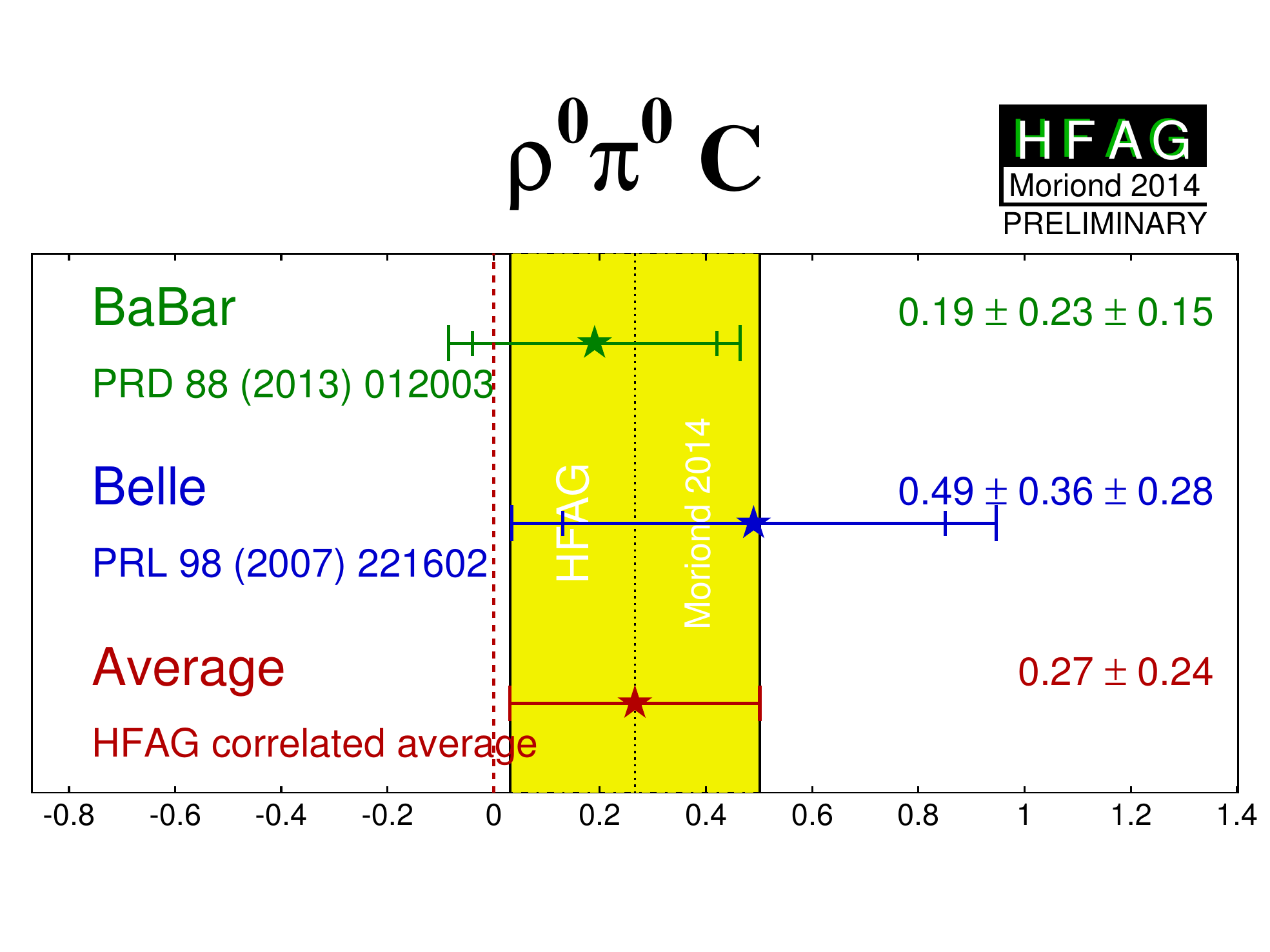}
      }
    \end{tabular}
  \end{center}
  \vspace{-0.8cm}
  \caption{
    Averages of (left) $S_{b \to u\bar u d}$ and (right) $C_{b \to u\bar u d}$
    for the mode $\Bz \to \rho^0\pi^0$.
  }
  \label{fig:cp_uta:uud:rho0pi0}
\end{figure}

\begin{figure}[htbp]
  \begin{center}
    \resizebox{0.46\textwidth}{!}{
      \includegraphics{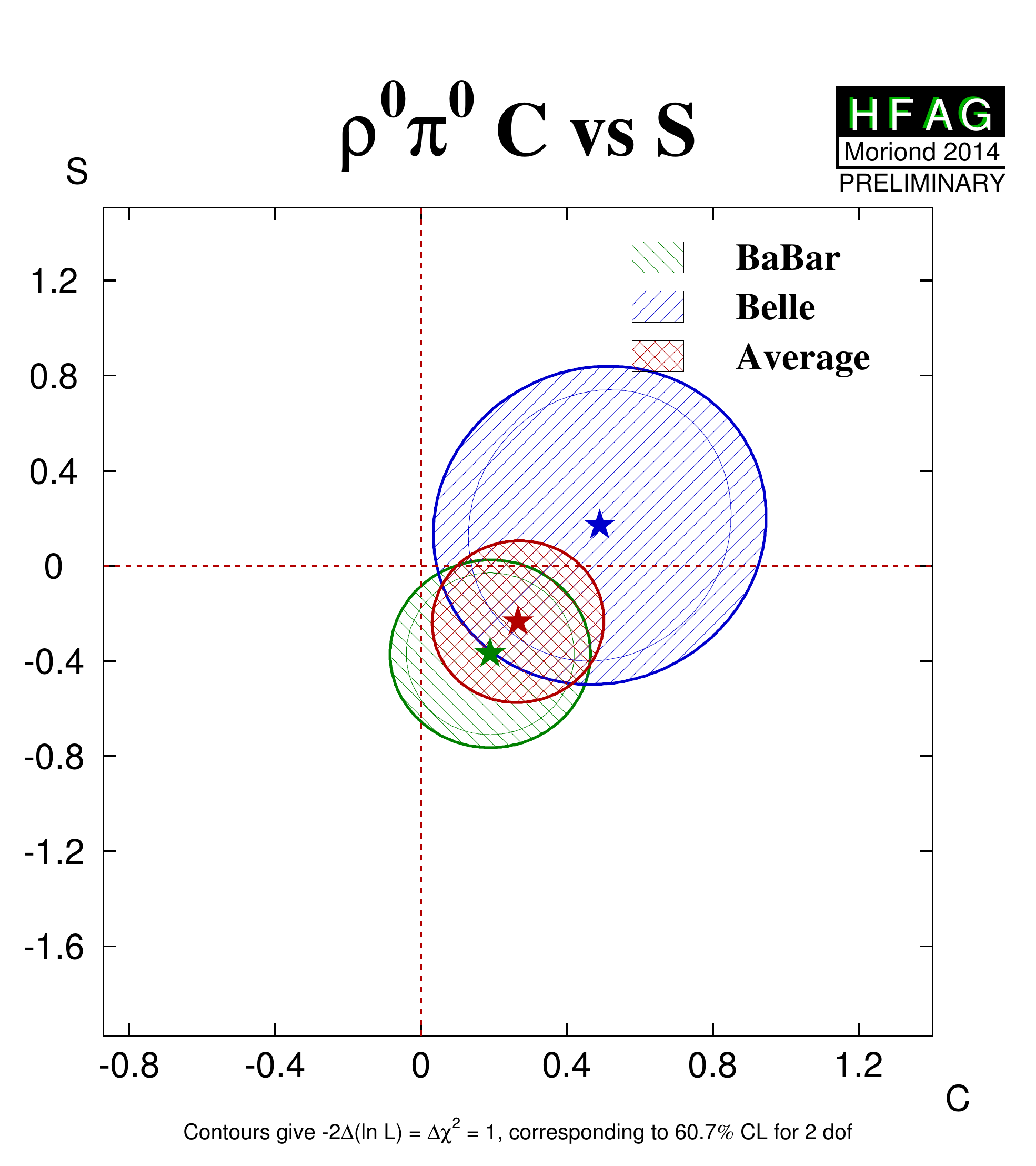}
    }      
  \end{center}
  \vspace{-0.8cm}
  \caption{
    Averages of $b \to u\bar u d$ dominated channels,
    for the mode $\Bz \to \rho^0\pi^0$
    in the $S_{\CP}$ \vs\ $C_{\CP}$ plane.
  }
  \label{fig:cp_uta:uud:rho0pi0_SvsC}
\end{figure}

Both experiments have also extracted the Q2B parameters.
We have performed a full correlated average of these parameters,
which is equivalent to determining the values from the 
averaged $U$ and $I$ parameters.
The results are shown in Table.~\ref{tab:cp_uta:uud:rhopi_q2b}.\footnote{
  The $\Bz \to \rho^\pm \pi^\mp$ Q2B parameters are equivalent to the
  parameters used for $\Bz \to a_1^\pm \pi^\mp$ decays, reported in
  Table~\ref{tab:cp_uta:uud}.   
  For the $\Bz \to a_1^\pm \pi^\mp$ case there has not yet been a full
  amplitude analysis of $\Bz \to \pi^+\pi^-\pi^+\pi^-$ and therefore only the
  Q2B parameters are available.
}
Averages of the $\Bz \to \rho^0\pi^0$ Q2B parameters are shown in 
Figs.~\ref{fig:cp_uta:uud:rho0pi0} and~\ref{fig:cp_uta:uud:rho0pi0_SvsC}.

\begin{sidewaystable}
	\begin{center}
		\caption{
                  Averages of quasi-two-body parameters extracted
                  from time-dependent Dalitz plot analysis of 
                  $\Bz \to \pi^+\pi^-\pi^0$.
		}
		\vspace{0.2cm}
		\setlength{\tabcolsep}{0.0pc}
    \resizebox{\textwidth}{!}{
		\begin{tabular}{@{\extracolsep{2mm}}lrcccccc} \hline
		\mc{2}{l}{Experiment} & $N(B\bar{B})$ & ${\cal A}_{CP}^{\rho\pi}$ & $C_{\rho\pi}$ & $S_{\rho\pi}$ & $\Delta C_{\rho\pi}$ & $\Delta S_{\rho\pi}$ \\
	\hline
	\babar & \cite{Lees:2013nwa} & 471M & $-0.10 \pm 0.03 \pm 0.02$ & $0.02 \pm 0.06 \pm 0.04$ & $0.05 \pm 0.08 \pm 0.03$ & $0.23 \pm 0.06 \pm 0.05$ & $0.05 \pm 0.08 \pm 0.04$ \\
	\belle & \cite{Kusaka:2007dv,:2007mj} & 449M & $-0.12 \pm 0.05 \pm 0.04$ & $-0.13 \pm 0.09 \pm 0.05$ & $0.06 \pm 0.13 \pm 0.05$ & $0.36 \pm 0.10 \pm 0.05$ & $-0.08 \pm 0.13 \pm 0.05$ \\
	\mc{3}{l}{\bf Average} & $-0.11 \pm 0.03$ & $-0.03 \pm 0.06$ & $0.06 \pm 0.07$ & $0.27 \pm 0.06$ & $0.01 \pm 0.08$ \\
	\mc{3}{l}{\small Confidence level} & \mc{5}{c}{\small $0.63~(0.5\sigma)$} \\
        \hline
		\end{tabular}
              }

                \vspace{2ex}

		\begin{tabular*}{\textwidth}{@{\extracolsep{\fill}}lrcccc} \hline
		\mc{2}{l}{Experiment} & $N(B\bar{B})$ & ${\cal A}^{-+}_{\rho\pi}$ & ${\cal A}^{+-}_{\rho\pi}$ & Correlation \\
		\hline
	\babar & \cite{Lees:2013nwa} & 471M & $-0.12 \pm 0.08 \,^{+0.04}_{-0.05}$ & $0.09 \,^{+0.05}_{-0.06} \pm 0.04$ & $0.55$ \\
	\belle & \cite{Kusaka:2007dv,:2007mj} & 449M & $0.08 \pm 0.16 \pm 0.11$ & $0.21 \pm 0.08 \pm 0.04$ & $0.47$ \\
	\mc{3}{l}{\bf Average} & $-0.08 \pm 0.08$ & $0.13 \pm 0.05$ & $0.37$ \\
        \mc{3}{l}{\small Confidence level} & \mc{2}{c}{\small $0.47~(0.7\sigma)$} \\
		\hline
		\end{tabular*}

                \vspace{2ex}

		\begin{tabular*}{\textwidth}{@{\extracolsep{\fill}}lrcccc} \hline
		\mc{2}{l}{Experiment} & $N(B\bar{B})$ & $C_{\rho^0\pi^0}$ & $S_{\rho^0\pi^0}$ & Correlation \\
		\hline
	\babar & \cite{Lees:2013nwa} & 471M & $0.19 \pm 0.23 \pm 0.15$ & $-0.37 \pm 0.34 \pm 0.20$ & $0.00$ \\
	\belle & \cite{Kusaka:2007dv,:2007mj} & 449M & $0.49 \pm 0.36 \pm 0.28$ & $0.17 \pm 0.57 \pm 0.35$ & $0.08$ \\
	\mc{3}{l}{\bf Average} & $0.27 \pm 0.24$ & $-0.23 \pm 0.34$ & $0.02$ \\
	\mc{3}{l}{\small Confidence level} & \mc{2}{c}{\small $0.68~(0.4\sigma)$} \\
		\hline
		\end{tabular*}
		\label{tab:cp_uta:uud:rhopi_q2b}
	\end{center}
\end{sidewaystable}

\begin{figure}[htbp]
  \begin{center}
    \resizebox{0.46\textwidth}{!}{
      \includegraphics{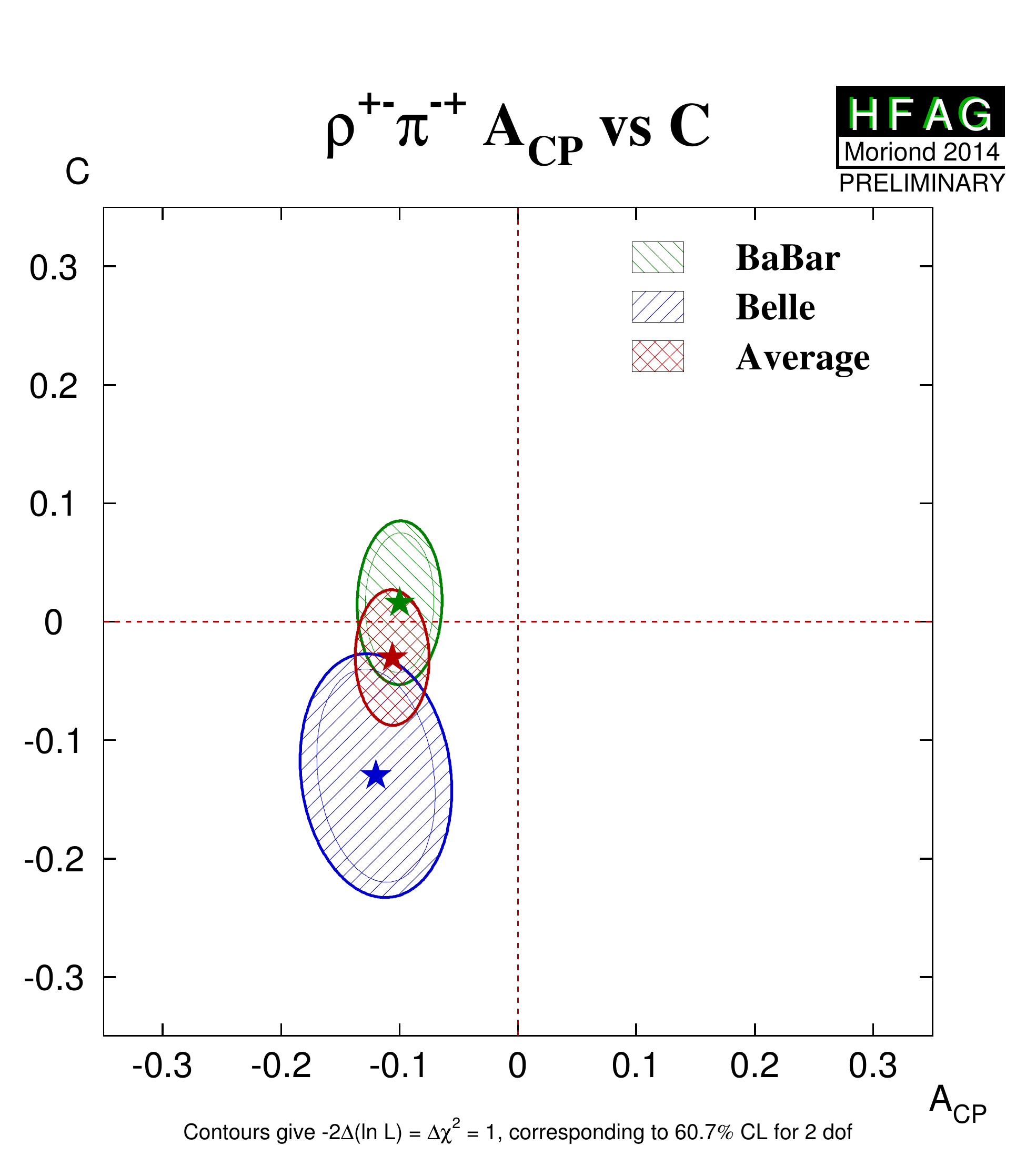}
    }
    \hfill
    \resizebox{0.46\textwidth}{!}{
      \includegraphics{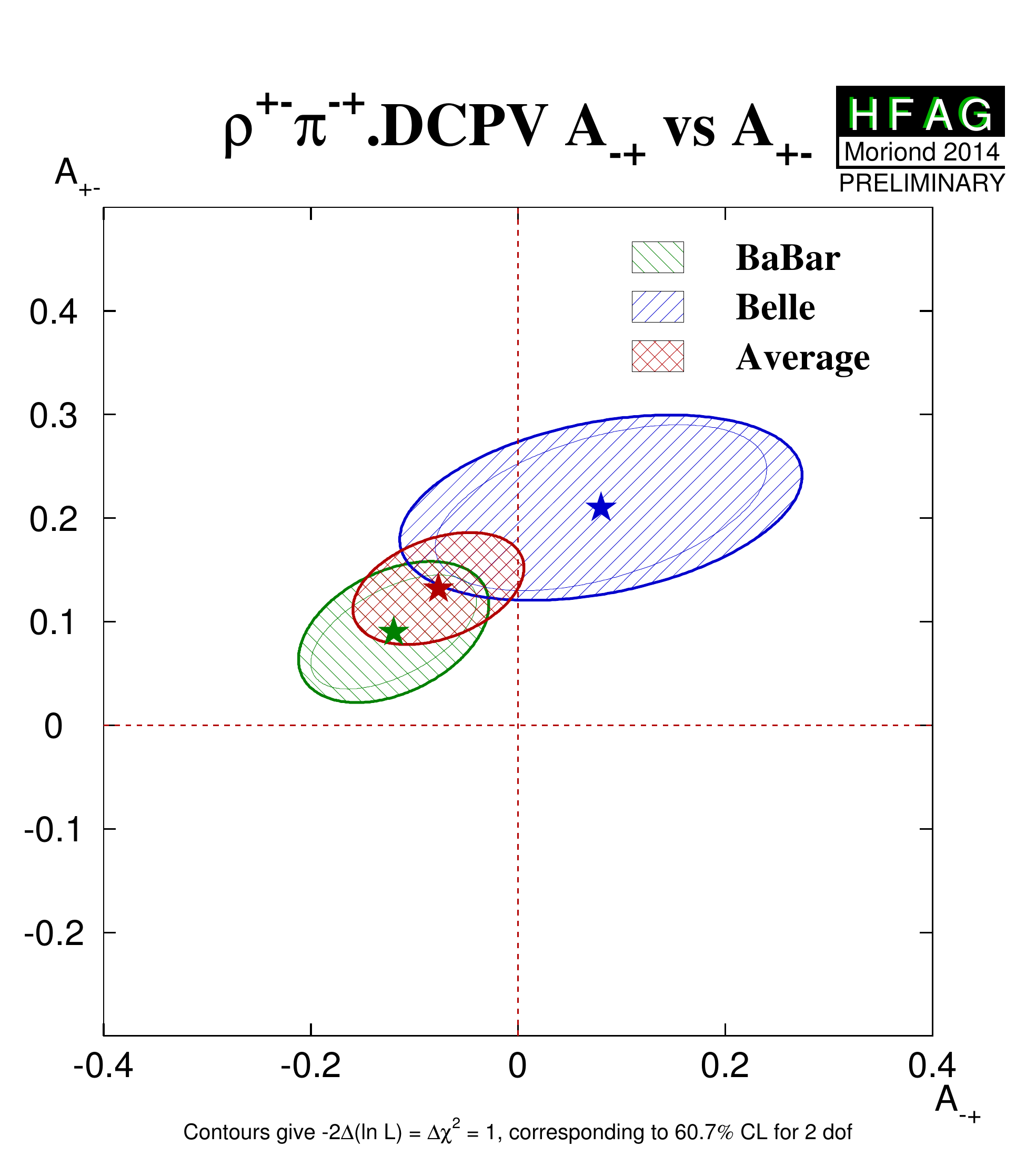}
    }
  \end{center}
  \vspace{-0.8cm}
  \caption{
    $\CP$ violation in $\Bz\to\rho^\pm\pi^\mp$ decays.
    (Left) ${\cal A}^{\rho\pi}_{\CP}$ \vs\ $C_{\rho\pi}$ space,
    (right) ${\cal A}^{-+}_{\rho\pi}$ \vs\ ${\cal A}^{+-}_{\rho\pi}$ space.
  }
  \label{fig:cp_uta:uud:rhopi-dircp}
\end{figure}

With the notation described in Sec.~\ref{sec:cp_uta:notations}
(Eq.~(\ref{eq:cp_uta:non-cp-s_and_deltas})), 
the time-dependent parameters for the Q2B $\Bz \to \rho^\pm\pi^\mp$ analysis are,
neglecting penguin contributions, given by
\begin{equation}
  S_{\rho\pi} = 
  \sqrt{1 - \left(\frac{\Delta C}{2}\right)^2}\sin(2\alpha)\cos(\delta)
  \ , \ \ \ 
  \Delta S_{\rho\pi} = 
  \sqrt{1 - \left(\frac{\Delta C}{2}\right)^2}\cos(2\alpha)\sin(\delta)
\end{equation} 
and $C_{\rho\pi} = {\cal A}_{\CP}^{\rho\pi} = 0$,
where $\delta=\arg(A_{-+}A^*_{+-})$ is the strong phase difference 
between the $\rho^-\pi^+$ and $\rho^+\pi^-$ decay amplitudes.
In the presence of the penguin contribution, there is no straightforward 
interpretation of the Q2B observables in the $\Bz \to \rho^\pm\pi^\mp$ system
in terms of CKM parameters.
However, $\CP$ violation in decay may arise,
resulting in either or both of $C_{\rho\pi} \neq 0$ and ${\cal A}_{\CP}^{\rho\pi} \neq 0$.
Equivalently,
$\CP$ violation in decay may be seen by either of
the decay-type-specific observables ${\cal A}^{+-}_{\rho\pi}$ 
and ${\cal A}^{-+}_{\rho\pi}$, defined in Eq.~(\ref{eq:cp_uta:non-cp-directcp}), 
deviating from zero.
Results and averages for these parameters
are also given in Table~\ref{tab:cp_uta:uud:rhopi_q2b}.
Averages of $\CP$ violation parameters in $\Bz \to \rho^\pm\pi^\mp$ decays
are shown in Fig.~\ref{fig:cp_uta:uud:rhopi-dircp},
both in 
${\cal A}^{\rho\pi}_{\CP}$ \vs\ $C_{\rho\pi}$ space and in 
${\cal A}^{-+}_{\rho\pi}$ \vs\ ${\cal A}^{+-}_{\rho\pi}$ space.

The averages for $S_{b \to u\bar u d}$ and $C_{b \to u\bar u d}$ 
in $\Bz \to \pi^+\pi^-$ are both more than $5\sigma$ away from zero,
suggesting that both mixing-induced and $\CP$ violation in decay
are well-established in this channel.
The discrepancy between results from \babar\ and Belle that used to exist in
this channel (see, for example, Ref.~\cite{Asner:2010qj}) is no longer
apparent, and the results from LHCb are also fully consistent with other
measurements.  
Some difference is, however, seen between the \babar\ and \belle\ measurements
in the $a_1^\pm\pi^\mp$ system. 
The confidence level of the five-dimensional average is $0.03$,
which corresponds to a $2.1\sigma$ discrepancy.  
As seen in Table~\ref{tab:cp_uta:uud}, this discrepancy is primarily in the
values of $S_{a_1\pi}$, and is not evident in the ${\cal A}^{-+}_{a_1\pi}$
\vs\ ${\cal A}^{+-}_{a_1\pi}$ projection shown in Fig.~\ref{fig:cp_uta:a1pi}.
Since there is no
evidence of systematic problems in either analysis,
we do not rescale the errors of the averages.

In $\Bz \to \rho^\pm\pi^\mp$ decays,
both experiments see an indication of $\CP$ violation in the 
${\cal A}^{\rho\pi}_{\CP}$ parameter 
(as seen in Fig.~\ref{fig:cp_uta:uud:rhopi-dircp}).
The average is more than $3\sigma$ from zero,
providing evidence of direct $\CP$ violation in this channel.
In $\Bz \to \rho^+\rho^-$ decays there is no evidence for $\CP$ violation,
either mixing-induced or in decay.
The absence of evidence of penguin contributions in this mode leads to
strong constraints on $\alpha \equiv \phi_2$.

\mysubsubsection{Constraints on $\alpha \equiv \phi_2$}
\label{sec:cp_uta:cus:alpha}

The precision of the measured $\CP$ violation parameters in
$b \to u\bar{u}d$ transitions allows 
constraints to be set on the UT angle $\alpha \equiv \phi_2$. 
Constraints have been obtained with various methods:
\begin{itemize}\setlength{\itemsep}{0.5ex}
\item 
  Both \babar~\cite{Lees:2012mma}
  and  \belle~\cite{Adachi:2013mae} have performed 
  isospin analyses in the $\pi\pi$ system.
  \belle\ exclude $23.8^\circ < \phi_2 < 66.8^\circ$ at the 68\%  C.L. while
  \babar\ give a confidence level interpretation for $\alpha$, and constrain
  $\alpha \in \left[ 71^\circ, 109^\circ \right]$ at the 68\%
  C.L. considering only the solution consistent with the Standard Model.
  Values in the range $\left[ 23^\circ, 67^\circ \right]$ at the 90\% C.L. are
  excluded.
  In both cases, only solutions in $0^\circ$--$180^\circ$ are considered.

\item
  Both experiments have also performed isospin analyses in the $\rho\rho$
  system. 
  The most recent result from \babar\ is given in an update of the
  measurements of the $B^+\to\rho^+\rho^0$ decay~\cite{Aubert:2009it}, and
  sets the constraint $\alpha = \left( 92.4 \,^{+6.0}_{-6.5}\right)^\circ$.
  The most recent result from \belle\ is given in an update of the
  search for the $\Bz \to \rho^0\rho^0$ decay and sets the constraint
  $\phi_2 = \left( 91.7 \pm 14.9 \right)^\circ$~\cite{:2008et}.

\item
  The time-dependent Dalitz plot analysis of the $\Bz \to \pi^+\pi^-\pi^0$
  decay allows a determination of $\alpha$ without input from any other 
  channels.
  \babar~\cite{Lees:2013nwa} present a scan, but not an interval, for $\alpha$, since
  their studies indicate that the scan is not statistically robust and cannot
  be interpreted as 1-C.L. 
  \belle~\cite{Kusaka:2007dv,:2007mj} have obtained a constraint on $\alpha$
  using additional information from the SU(2) partners of 
  $B \to \rho\pi$, which can be used to constrain $\alpha$
  via an isospin pentagon relation~\cite{Lipkin:1991st}. 
  With this analysis,
  \belle\ obtain the constraint $\phi_2 = (83 \, ^{+12}_{-23})^\circ$
  (where the errors correspond to $1\sigma$, \ie\ $68.3\%$ confidence level).

\item 
  The results from \babar\ on $\Bz \to a_1^\pm \pi^\mp$~\cite{Aubert:2006gb} can be
  combined with results from modes related by isospin ($a_1K$ and $K_1\pi$)~\cite{Gronau:2005kw}
  leading to the constraint
  $\alpha = \left( 79 \pm 7 \pm 11 \right)^\circ$~\cite{:2009ii}.


\item 
  The CKMfitter~\cite{Charles:2004jd} and 
  UTFit~\cite{Bona:2005vz} groups use the measurements 
  from \belle\ and \babar\ given above
  with other branching fractions and \CP asymmetries in 
  $\B\to\pi\pi$, $\rho\pi$ and $\rho\rho$ modes, 
  to perform isospin analyses for each system, 
  and to make combined constraints on $\alpha$.

\item
  The \babar\ and \belle\ collaborations have combined their results on $B \to \pi\pi$, $\pi\pi\pi^0$ and $\rho\rho$ to obtain~\cite{Bevan:2014iga}
  \begin{equation}
    \alpha \equiv \phi_2 = (88 \pm 5)^\circ \, .
  \end{equation}
  The above solution is that consistent with the Standard Model (an ambiguous solution shifted by $180^\circ$ exists). The strongest constraint currently comes from the $B \to \rho\rho$ system. The inclusion of results from $\Bz \to a_1^\pm \pi^\mp$ does not significantly affect the average. 

\end{itemize}

Note that methods based on isospin symmetry make extensive use of 
measurements of branching fractions and $\CP$ asymmetries,
as averaged by the HFAG Rare Decays subgroup (Sec.~\ref{sec:rare}).
Note also that each method suffers from discrete ambiguities in the solutions.
The model assumption in the $\Bz \to \pi^+\pi^-\pi^0$ analysis 
helps resolve some of the multiple solutions, 
and results in a single preferred value for $\alpha$ in $\left[ 0, \pi \right]$.
All the above measurements correspond to the choice
that is in agreement with the global CKM fit.

At present we make no attempt to provide an HFAG average for $\alpha \equiv \phi_2$.
More details on procedures to calculate a best fit value for $\alpha$ 
can be found in Refs.~\cite{Charles:2004jd,Bona:2005vz}.

\mysubsection{Time-dependent $\CP$ asymmetries in $b \to c\bar{u}d / u\bar{c}d$ transitions
}
\label{sec:cp_uta:cud}

Non-$\CP$ eigenstates such as $D^\mp\pi^\pm$, $D^{*\mp}\pi^\pm$ and $D^\mp\rho^\pm$ can be produced 
in decays of $\Bz$ mesons either via Cabibbo favoured ($b \to c$) or
doubly Cabibbo suppressed ($b \to u$) tree amplitudes. 
Since no penguin contribution is possible,
these modes are theoretically clean.
The ratio of the magnitudes of the suppressed and favoured amplitudes, $R$,
is sufficiently small (predicted to be about $0.02$),
that terms of ${\cal O}(R^2)$ can be neglected, 
and the sine terms give sensitivity to the combination of UT angles $2\beta+\gamma$.

\begin{figure}[htbp]
  \begin{center}
    \begin{tabular}{cc}
      \resizebox{0.46\textwidth}{!}{
        \includegraphics{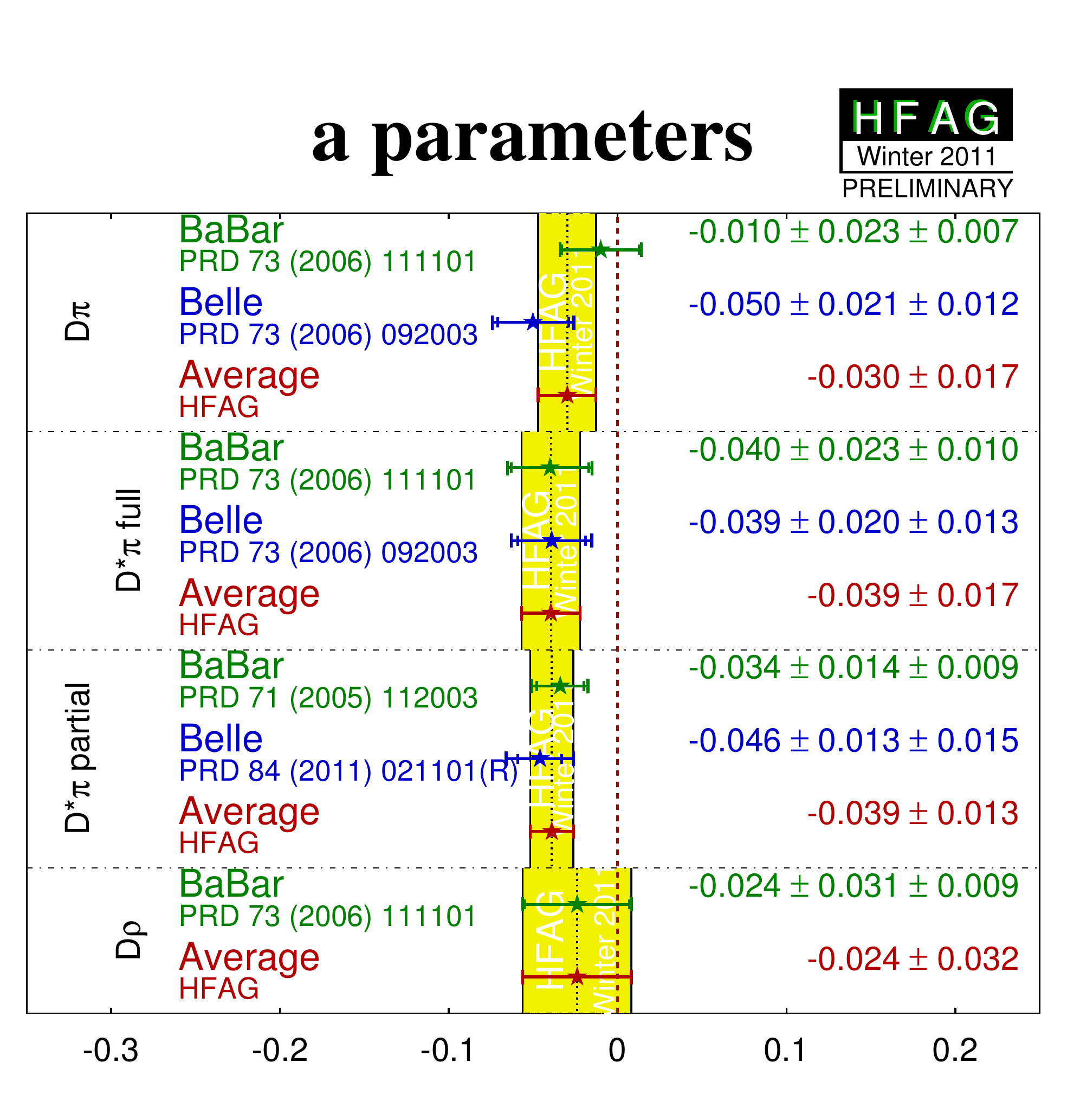}
      }
      &
      \resizebox{0.46\textwidth}{!}{
        \includegraphics{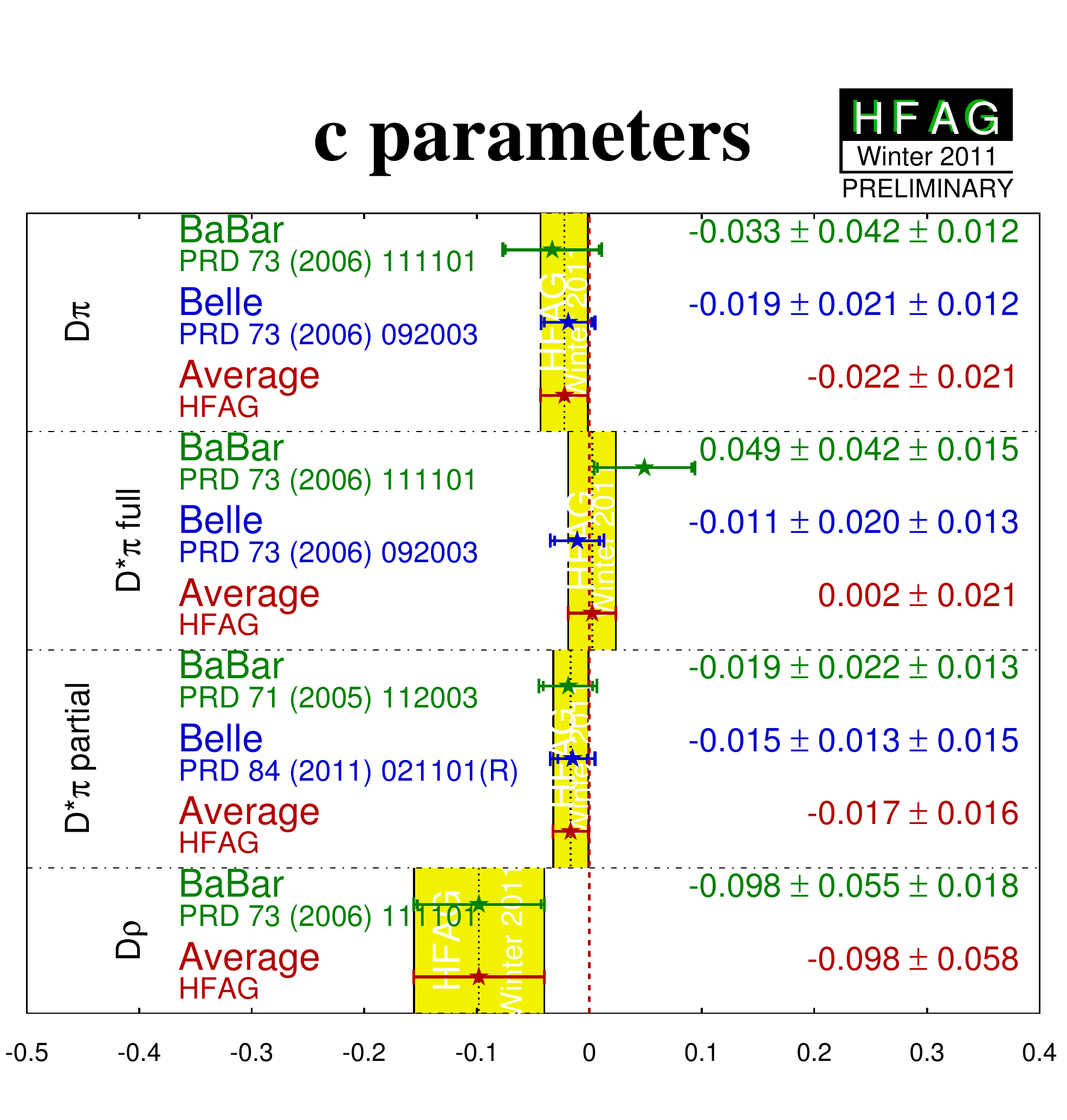}
      }
    \end{tabular}
  \end{center}
  \vspace{-0.8cm}
  \caption{
    Averages for $b \to c\bar{u}d / u\bar{c}d$ modes.
  }
  \label{fig:cp_uta:cud}
\end{figure}

\begin{figure}[htbp]
  \begin{center}
    \begin{tabular}{cc}
      \resizebox{0.46\textwidth}{!}{
        \includegraphics{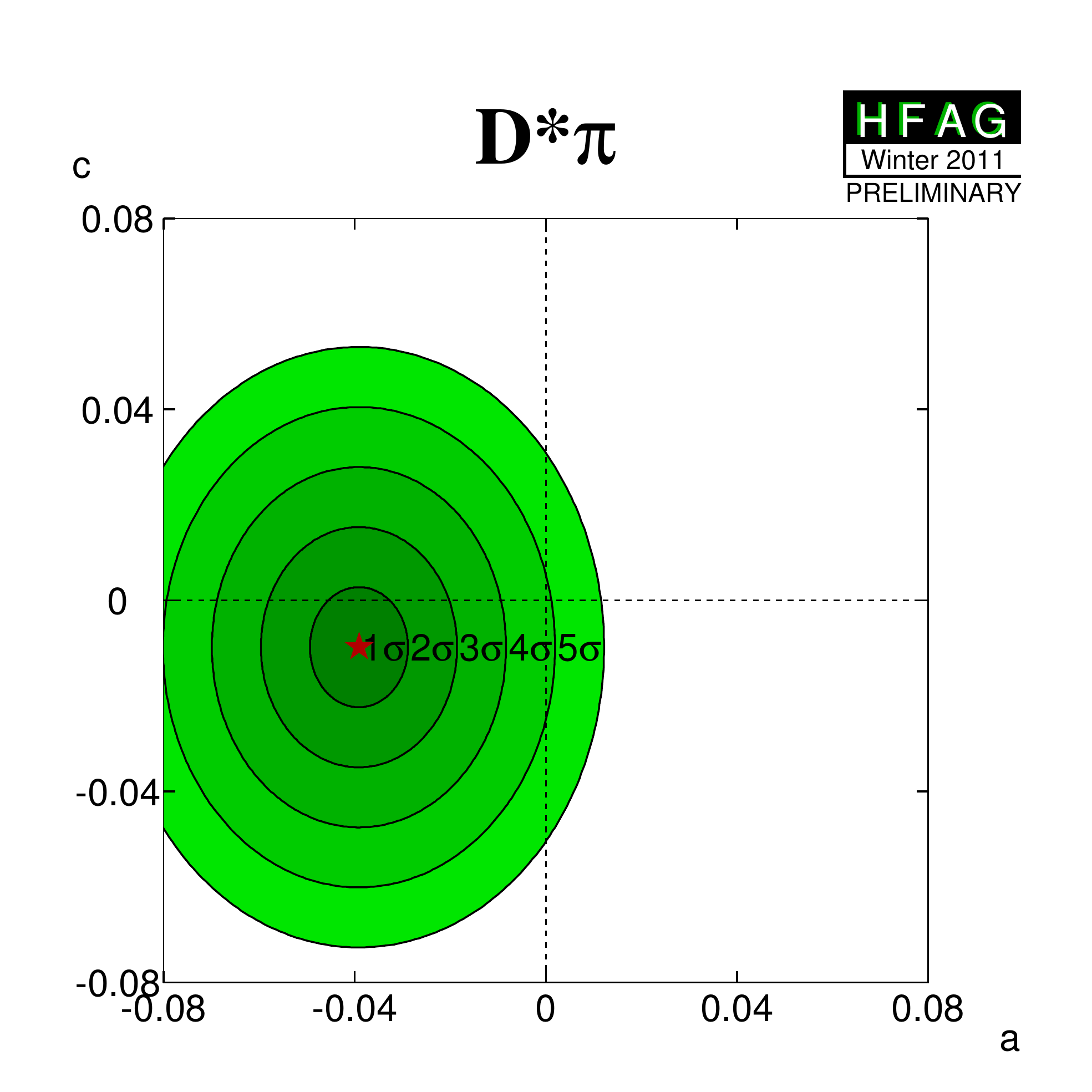}
      }
      &
      \resizebox{0.46\textwidth}{!}{
        \includegraphics{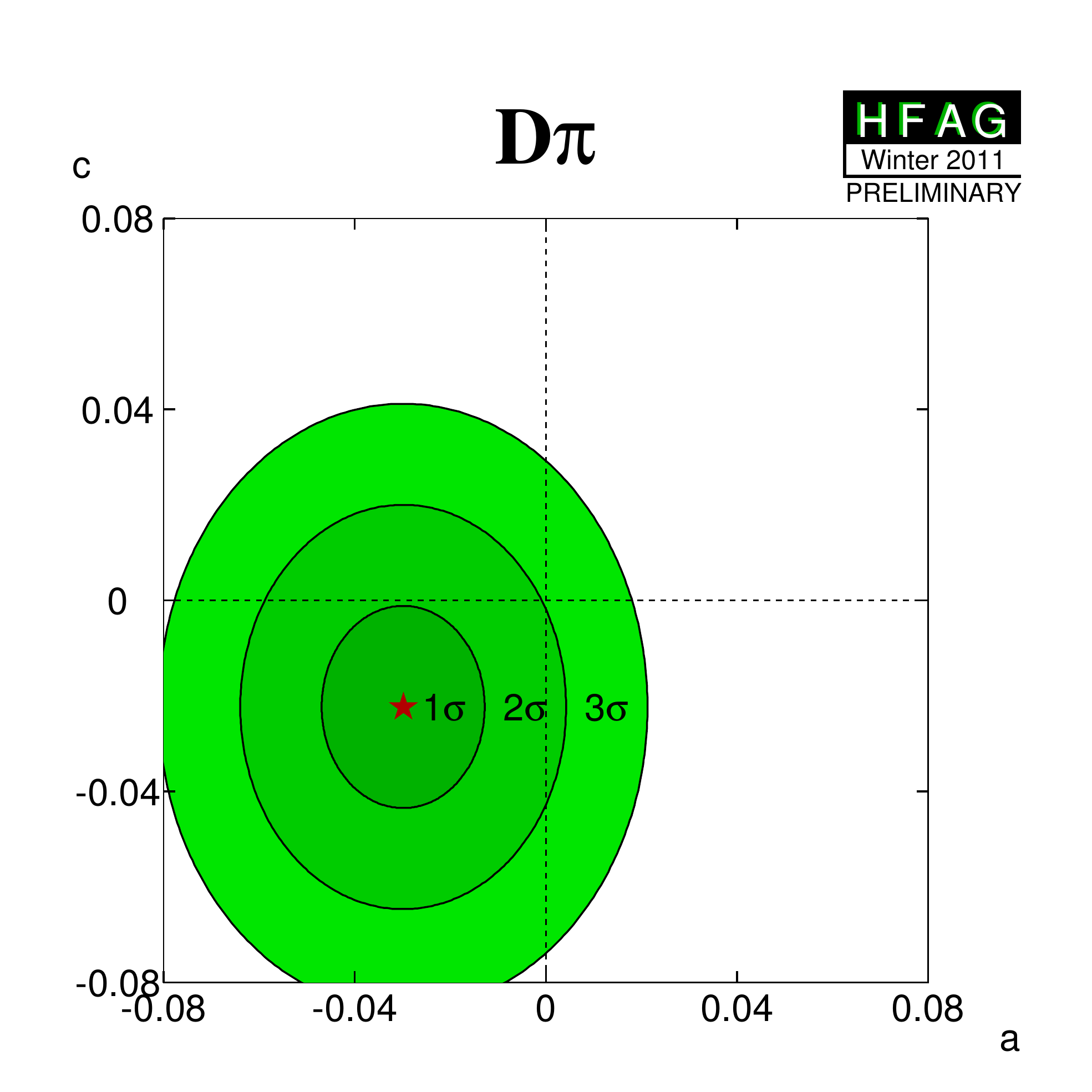}
      } \\
      \resizebox{0.46\textwidth}{!}{
        \includegraphics{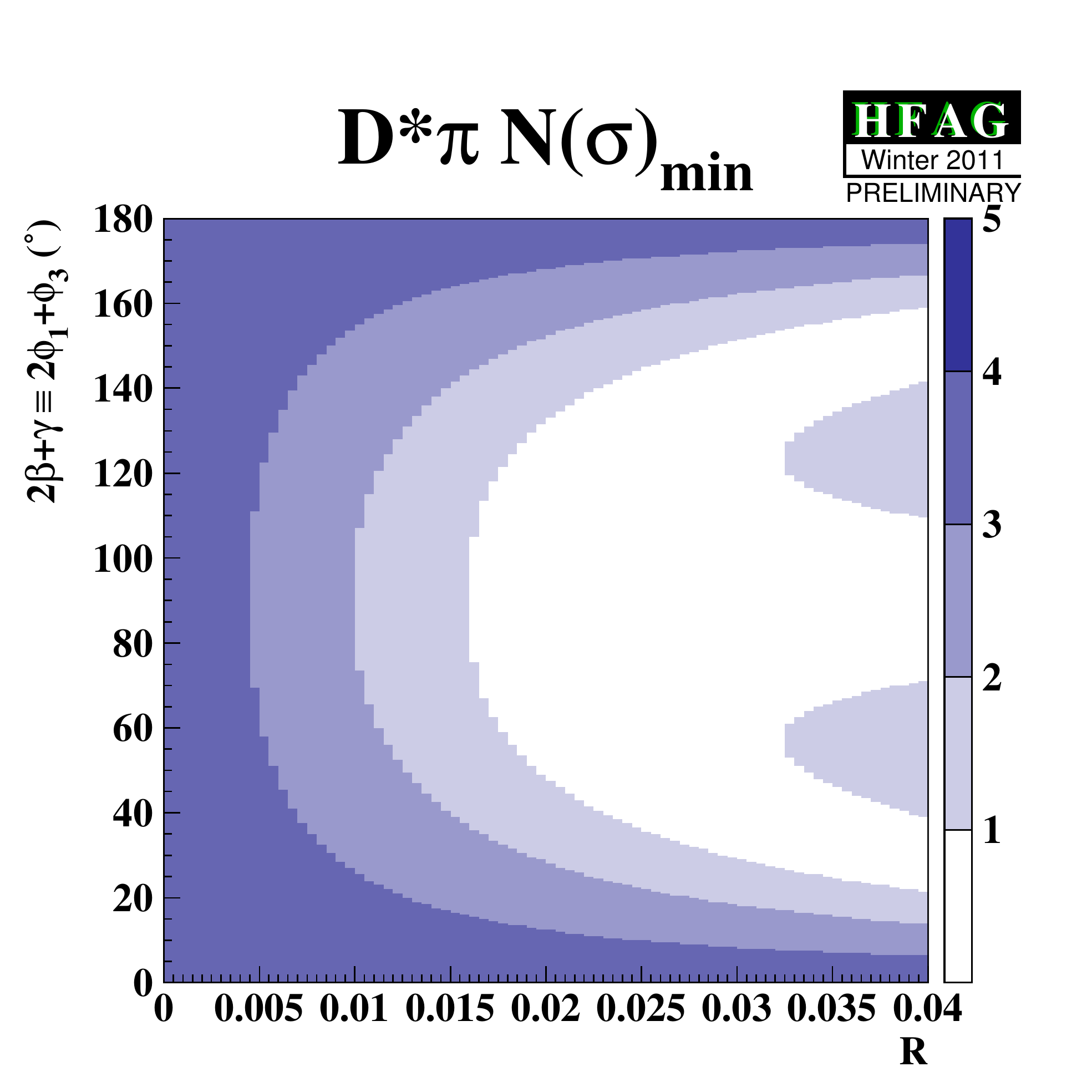}
      }
      &
      \resizebox{0.46\textwidth}{!}{
        \includegraphics{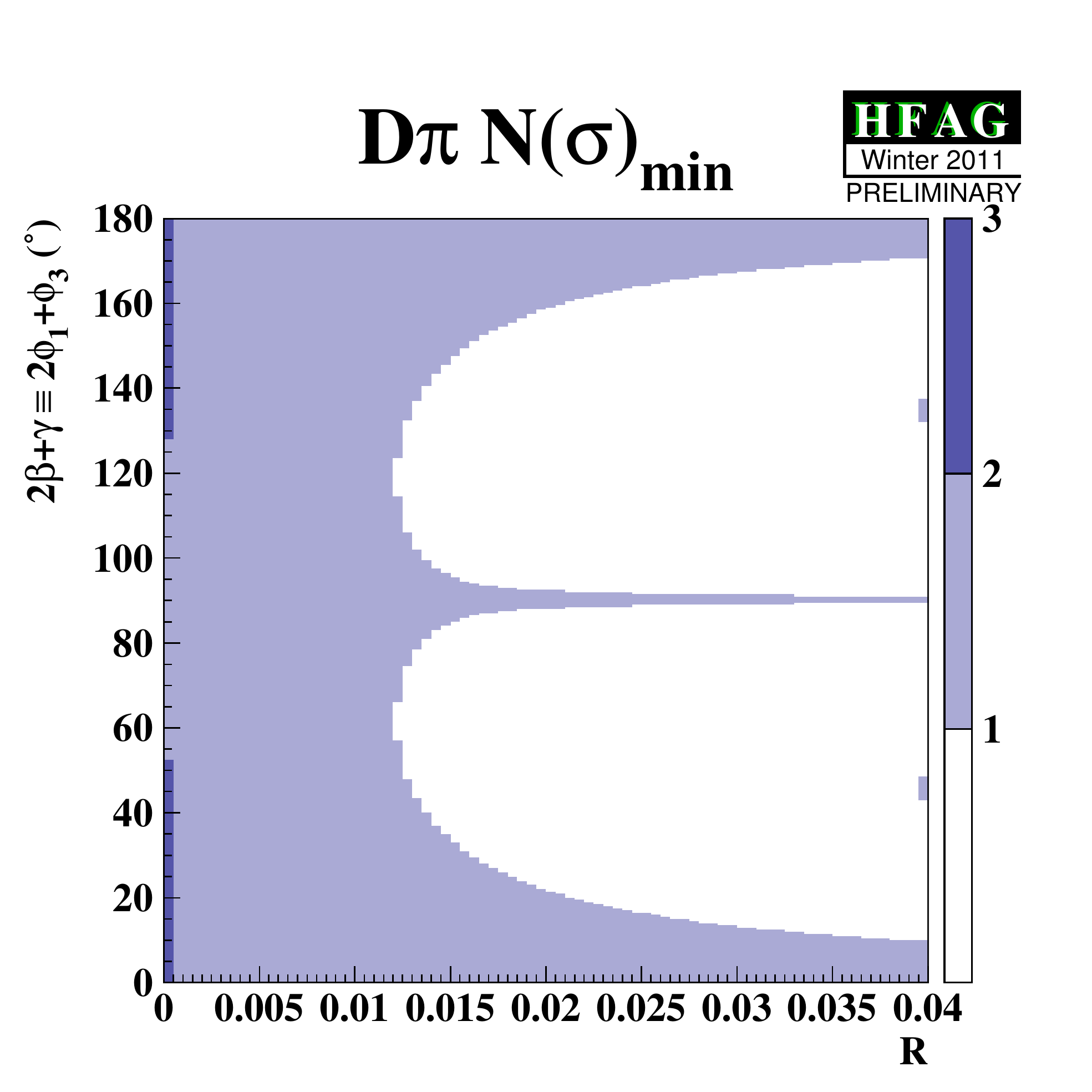}
      }          
    \end{tabular}
  \end{center}
  \vspace{-0.8cm}
  \caption{
    Results from $b \to c\bar{u}d / u\bar{c}d$ modes.
    (Top) Constraints in $c$ {\it vs.} $a$ space.
    (Bottom) Constraints in $2\beta+\gamma$ {\it vs.} $R$ space.
    (Left) $D^*\pi$ and (right) $D\pi$ modes.
  }
  \label{fig:cp_uta:cud_constraints}
\end{figure}

As described in Sec.~\ref{sec:cp_uta:notations:non_cp:dstarpi},
the averages are given in terms of parameters $a$ and $c$.
$\CP$ violation would appear as $a \neq 0$.
Results are available from both \babar\ and \belle\ in the modes
$D^\mp\pi^\pm$ and $D^{*\mp}\pi^\pm$; for the latter mode both experiments 
have used both full and partial reconstruction techniques.
Results are also available from \babar\ using $D^\mp\rho^\pm$.
These results, and their averages, are listed in Table~\ref{tab:cp_uta:cud},
and are shown in Fig.~\ref{fig:cp_uta:cud}.
The constraints in $c$ \vs\ $a$ space for the $D\pi$ and $D^*\pi$ modes
are shown in Fig.~\ref{fig:cp_uta:cud_constraints}.
It is notable that the average value of $a$ from $D^*\pi$ is more than
$3\sigma$ from zero, providing evidence of $\CP$ violation in this channel.

\begin{table}[htb]
	\begin{center}
		\caption{
      Averages for $b \to c\bar{u}d / u\bar{c}d$ modes.
                }
                \vspace{0.2cm}
                \setlength{\tabcolsep}{0.0pc}
                \begin{tabular*}{\textwidth}{@{\extracolsep{\fill}}lrccc} \hline 
	\mc{2}{l}{Experiment} & $N(B\bar{B})$ & $a$ & $c$ \\
	\hline
      \mc{5}{c}{$D^{\mp}\pi^{\pm}$} \\
	\babar (full rec.) & \cite{Aubert:2006tw} & 232M & $-0.010 \pm 0.023 \pm 0.007$ & $-0.033 \pm 0.042 \pm 0.012$ \\
	\belle (full rec.) & \cite{Ronga:2006hv} & 386M & $-0.050 \pm 0.021 \pm 0.012$ & $-0.019 \pm 0.021 \pm 0.012$ \\
        \mc{3}{l}{\bf Average} & $ -0.030 \pm 0.017$ & $ -0.022 \pm 0.021 $ \\
        \mc{3}{l}{\small Confidence level} & {\small $0.24~(1.2\sigma)$} & {\small $0.78~(0.3\sigma)$} \\
        \hline
      \mc{5}{c}{$D^{*\mp}\pi^{\pm}$} \\
      \babar (full rec.) & \cite{Aubert:2006tw} & 232M & $-0.040 \pm 0.023 \pm 0.010$ & $0.049 \pm 0.042 \pm 0.015$ \\
      \babar (partial rec.)  & \cite{Aubert:2005yf} & 232M & $-0.034 \pm 0.014 \pm 0.009$ & $-0.019 \pm 0.022 \pm 0.013$ \\
      \belle (full rec.) & \cite{Ronga:2006hv} & 386M & $-0.039 \pm 0.020 \pm 0.013$ & $-0.011 \pm 0.020 \pm 0.013$ \\
      \belle (partial rec.) & \cite{Bahinipati:2011yq} & 657M & $-0.046 \pm 0.013 \pm 0.015$ & $-0.015 \pm 0.013 \pm 0.015$ \\
	\mc{3}{l}{\bf Average} & $-0.039 \pm 0.010$ & $-0.010 \pm 0.013$ \\
      \mc{3}{l}{\small Confidence level} & {\small $0.97~(0.03\sigma)$} & {\small $0.59~(0.6\sigma)$} \\
      \hline
      \mc{5}{c}{$D^{\mp}\rho^{\pm}$} \\
      \babar (full rec.) & \cite{Aubert:2006tw} & 232M & $-0.024 \pm 0.031 \pm 0.009$ & $-0.098 \pm 0.055 \pm 0.018$ \\
      \hline 
    \end{tabular*}
    \label{tab:cp_uta:cud}
  \end{center}
\end{table}

For each mode, $D\pi$, $D^*\pi$ and $D\rho$, 
there are two measurements ($a$ and $c$, or $S^+$ and $S^-$) 
which depend on three unknowns ($R$, $\delta$ and $2\beta+\gamma$), 
of which two are different for each decay mode. 
Therefore, there is not enough information to solve directly for $2\beta+\gamma$. 
However, for each choice of $R$ and $2\beta+\gamma$, 
one can find the value of $\delta$ that allows $a$ and $c$ to be closest 
to their measured values, 
and calculate the distance in terms of numbers of standard deviations.
(We currently neglect experimental correlations in this analysis.) 
These values of $N(\sigma)_{\rm min}$ can then be plotted 
as a function of $R$ and $2\beta+\gamma$
(and can trivially be converted to confidence levels). 
These plots are given for the $D\pi$ and $D^*\pi$ modes 
in Fig.~\ref{fig:cp_uta:cud_constraints}; 
the uncertainties in the $D\rho$ mode are currently too large 
to give any meaningful constraint.

The constraints can be tightened if one is willing 
to use theoretical input on the values of $R$ and/or $\delta$. 
One popular choice is the use of SU(3) symmetry to obtain 
$R$ by relating the suppressed decay mode to $\B$ decays 
involving $D_s$ mesons. 
More details can be found 
in Refs.~\cite{Charles:2004jd,Bona:2005vz}.

\mysubsection{Time-dependent $\CP$ asymmetries in $b \to c\bar{u}s / u\bar{c}s$ transitions
}
\label{sec:cp_uta:cus-td}

\mysubsubsection{Time-dependent $\CP$ asymmetries in $\Bz \to D^\mp \KS \pi^\pm$}

Time-dependent analyses of transitions such as $\Bz \to D^\mp \KS \pi^\pm$ can
be used to probe $\sin(2\beta+\gamma)$ in a similar way to that discussed
above (Sec.~\ref{sec:cp_uta:cud}). Since the final state contains three
particles, a Dalitz plot analysis is necessary to maximise the
sensitivity. \babar~\cite{Aubert:2007qe} have carried out such an
analysis. They obtain $2\beta+\gamma = \left( 83 \pm 53 \pm 20 \right)^\circ$
(with an ambiguity $2\beta+\gamma \leftrightarrow 2\beta+\gamma+\pi$) assuming
the ratio of the $b \to u$ and $b \to c$ amplitude to be constant across the
Dalitz plot at 0.3.

\mysubsubsection{Time-dependent $\CP$ asymmetries in $\Bs \to D_s^\mp K^\pm$}

Time-dependent analysis of $\Bs \to D_s^\mp K^\pm$ decays can be used to determine $\gamma-2\beta_s$~\cite{Dunietz:1987bv,Aleksan:1991nh,Fleischer:2003yb}.
Compared to the situation for $\Bz \to D^{(*)\mp} \pi^\pm$ discussed above (Sec.~\ref{sec:cp_uta:cud}), the larger value of the ratio $R$ of the magnitudes of the suppressed and favoured amplitudes allows it to be determined from the data.  
Moreover, the non-zero value of $\Delta \Gamma_s$ allows the determination of additional terms, labelled $A^{\Delta\Gamma}$ and $\bar{A}{}^{\Delta\Gamma}$, that break ambiguities in the solutions for $\gamma-2\beta_s$.

LHCb~\cite{Aaij:2014fba} have measured the time-dependent \CP violation parameters in $\Bs \to D_s^\mp K^\pm$ decays, using $1.0 \ {\rm fb}^{-1}$ of data.  
The results are given in Table~\ref{tab:cp_uta:DsK}.
From these results, and a constraint on $2\beta_s$ from independent LHCb measurements, LHCb determine $\gamma = (115 \,^{+28}_{-43})^\circ$, $\delta_{D_sK} = (3 \,^{+19}_{-20})^\circ$ and $R_{D_sK} = 0.53 \,^{+0.17}_{-0.16}$. 

\begin{table}[!htb]
	\begin{center}
		\caption{
			Results for $\Bs \to D_s^\mp K^\pm$.
		}
    \resizebox{\textwidth}{!}{
		\begin{tabular}{@{\extracolsep{2mm}}lrcccccc} \hline
	\mc{2}{l}{Experiment} & Sample size & $C$ & $A^{\Delta\Gamma}$ & $\bar{A}{}^{\Delta\Gamma}$ & $S$ & $\bar{S}$ \\
	\hline
	LHCb & \cite{Aaij:2014fba} & 1 ${\rm fb}^{-1}$ & $0.53 \pm 0.25 \pm 0.04$ & $0.37 \pm 0.42 \pm 0.20$ & $0.20 \pm 0.41 \pm 0.20$ & $-1.09 \pm 0.33 \pm 0.08$ & $-0.36 \pm 0.34 \pm 0.08$ \\
		\hline
		\end{tabular}
    }
		\label{tab:cp_uta:DsK}
	\end{center}
\end{table}

\mysubsection{Rates and asymmetries in $\Bmp \to \DorDstar K^{(*)\mp}$ decays
}
\label{sec:cp_uta:cus}

As explained in Sec.~\ref{sec:cp_uta:notations:cus},
rates and asymmetries in $\Bmp \to \DorDstar K^{(*)\mp}$ decays
are sensitive to $\gamma$, and have negligible theoretical uncertainty~\cite{Brod:2013sga}.
Various methods using different $\DorDstar$ final states have been used.

\mysubsubsection{$D$ decays to $\CP$ eigenstates
}
\label{sec:cp_uta:cus:glw}

\begin{figure}[htbp]
  \begin{center}
    \begin{tabular}{cc}
      \resizebox{0.46\textwidth}{!}{
        \includegraphics{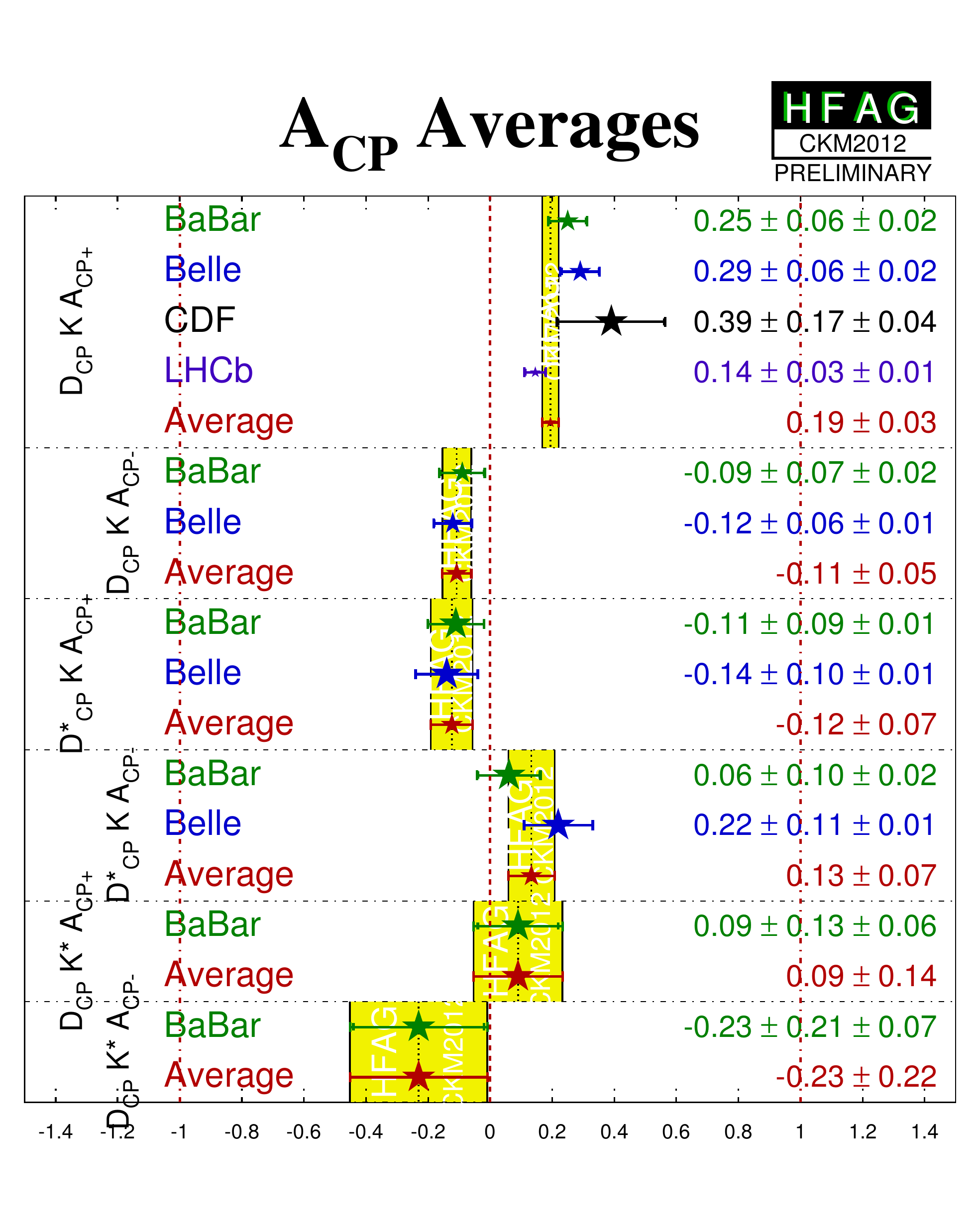}
      }
      &
      \resizebox{0.46\textwidth}{!}{
        \includegraphics{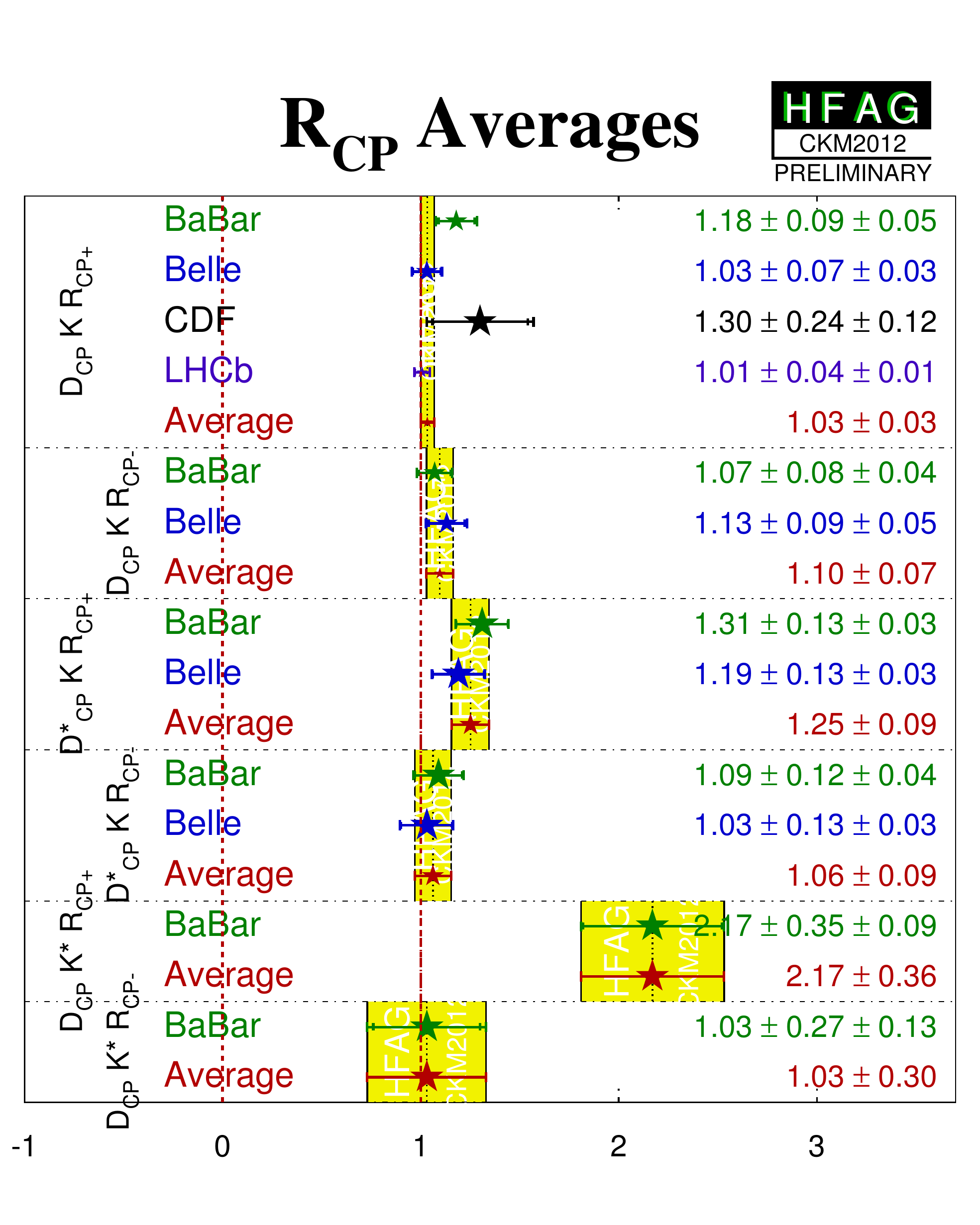}
      }
    \end{tabular}
 \end{center}
  \vspace{-0.8cm}
  \caption{
    Averages of $A_{\CP}$ and $R_{\CP}$ from GLW analyses.
  }
  \label{fig:cp_uta:cus:glw}
\end{figure}

Results are available from both \babar\ and \belle\ on GLW analyses in the
decay modes $\Bmp \to D\Kmp$, $\Bmp \to \Dstar\Kmp$ and 
$\Bmp \to D\Kstarmp$.
Both experiments use the $\CP$-even $D$ decay final states $K^+K^-$ and
$\pi^+\pi^-$ in all three modes; both experiments generally use the \CP-odd
decay modes $\KS\pi^0$, $\KS\omega$ and $\KS\phi$, though care is taken to
avoid statistical overlap with the $\KS K^+K^-$ sample used for Dalitz plot
analyses (see Sec.~\ref{sec:cp_uta:cus:dalitz}), 
and asymmetric systematic errors are assigned due to $\CP$-even pollution
under the $\KS\omega$ and $\KS\phi$ signals.
Both experiments use both the $\Dstar \to D\pi^0$ decay, 
which gives $\CP(\Dstar) = \CP(D)$,
and the $\Dstar \to D\gamma$ decays, 
which gives $\CP(\Dstar) = -\CP(D)$.
In addition, results from CDF and LHCb are available in the
decay mode $\Bmp \to D\Kmp$, 
and from LHCb in the decay mode $\Bmp \to D\Kmp\pi^+\pi^-$, 
for $\CP$-even final states ($K^+K^-$ and $\pipi$) only.
The results and averages are given in Table~\ref{tab:cp_uta:cus:glw}
and shown in Fig.~\ref{fig:cp_uta:cus:glw}.

\begin{table}[htb]
	\begin{center}
		\caption{
                        Averages from GLW analyses of $b \to c\bar{u}s / u\bar{c}s$ modes.
                }
                \vspace{0.2cm}
    \resizebox{\textwidth}{!}{
      \setlength{\tabcolsep}{0.0pc}
      \begin{tabular}{@{\extracolsep{2mm}}lrccccc} \hline 
        \mc{2}{l}{Experiment} & Sample size & $A_{\CP+}$ & $A_{\CP-}$ & $R_{\CP+}$ & $R_{\CP-}$ \\
        \hline
        \mc{7}{c}{$D_{\CP} K^-$} \\
	\babar & \cite{delAmoSanchez:2010ji} & $N(B\bar{B}) =$ 467M & $0.25 \pm 0.06 \pm 0.02$ & $-0.09 \pm 0.07 \pm 0.02$ & $1.18 \pm 0.09 \pm 0.05$ & $1.07 \pm 0.08 \pm 0.04$ \\
	\belle & \cite{belle:glwads:prelim} & $N(B\bar{B}) =$ 772M & $0.29 \pm 0.06 \pm 0.02$ & $-0.12 \pm 0.06 \pm 0.01$ & $1.03 \pm 0.07 \pm 0.03$ & $1.13 \pm 0.09 \pm 0.05$ \\
	CDF & \cite{Aaltonen:2009hz} & 1 ${\rm fb}^{-1}$ & $0.39 \pm 0.17 \pm 0.04$ & \textendash{} & $1.30 \pm 0.24 \pm 0.12$ &  \textendash{} \\
	LHCb & \cite{Aaij:2012kz} & 1 ${\rm fb}^{-1}$ & $0.14 \pm 0.03 \pm 0.01$ &  \textendash{} & $1.01 \pm 0.04 \pm 0.01$ &  \textendash{} \\
	\mc{3}{l}{\bf Average} & $0.19 \pm 0.03$ & $-0.11 \pm 0.05$ & $1.03 \pm 0.03$ & $1.10 \pm 0.07$ \\
	\mc{3}{l}{\small Confidence level} & {\small $0.09~(1.7\sigma)$} & {\small $0.75~(0.3\sigma)$} & {\small $0.33~(1.0\sigma)$} & {\small $0.66~(0.4\sigma)$} \\
		\hline

        \mc{7}{c}{$\Dstar_{\CP} K^-$} \\
	\babar & \cite{:2008jd} & $N(B\bar{B}) =$ 383M & $-0.11 \pm 0.09 \pm 0.01$ & $0.06 \pm 0.10 \pm 0.02$ & $1.31 \pm 0.13 \pm 0.03$ & $1.09 \pm 0.12 \pm 0.04$ \\
	\belle & \cite{Trabelsi:2013uj} & 772M & $-0.14 \pm 0.10 \pm 0.01$ & $0.22 \pm 0.11 \pm 0.01$ & $1.19 \pm 0.13 \pm 0.03$ & $1.03 \pm 0.13 \pm 0.03$ \\
	\mc{3}{l}{\bf Average} & $-0.12 \pm 0.07$ & $0.13 \pm 0.07$ & $1.25 \pm 0.09$ & $1.06 \pm 0.09$ \\
	\mc{3}{l}{\small Confidence level} & {\small $0.82~(0.2\sigma)$} & {\small $0.29~(1.1\sigma)$} & {\small $0.52~(0.6\sigma)$} & {\small $0.74~(0.3\sigma)$} \\
		\hline

        \mc{7}{c}{$D_{\CP} K^{*-}$} \\
	\babar & \cite{Aubert:2009yw} & $N(B\bar{B}) =$ 379M & $0.09 \pm 0.13 \pm 0.06$ & $-0.23 \pm 0.21 \pm 0.07$ & $2.17 \pm 0.35 \pm 0.09$ & $1.03 \pm 0.27 \pm 0.13$ \\
		\hline

        \mc{7}{c}{$D_{\CP} K^{-}\pi^+\pi^-$} \\
	LHCb & \cite{LHCb-CONF-2012-021} & 1 ${\rm fb}^{-1}$ & $-0.14 \pm 0.10 \pm 0.01$ & & $0.95 \pm 0.11 \pm 0.02$ & \\
	\hline

      \end{tabular}
    }
    \label{tab:cp_uta:cus:glw}
	\end{center}
\end{table}

LHCb have performed a GLW analysis using the $B^0 \to DK^{*0}$ decay with $D \to K^+K^-$ and $D \to \pi^+\pi^-$ channels~\cite{Aaij:2014eha}.
The results are presented separately to allow for possible \CP violation effects in the charm decays, which are, however, known to be small.
The results are shown in Table~\ref{tab:cp_uta:glw-DKstar} where an average is also given.

\begin{table}[!htb]
        \begin{center}
                \caption{
      Results from GLW analysis of $\Bz \to D\Kstarz$.
                }
                \vspace{0.2cm}
                \setlength{\tabcolsep}{0.0pc}
                \begin{tabular*}{\textwidth}{@{\extracolsep{\fill}}lrccc} \hline
        \mc{2}{l}{Experiment} & Sample size & $A_{\CP+}$ & $R_{\CP+}$ \\
        \hline
        LHCb ($D\to K^+K^-$) & \cite{Aaij:2014eha} & 3 ${\rm fb}^{-1}$ & $-0.20 \pm 0.15 \pm 0.02$ & $1.05 \,^{+0.17}_{-0.15} \pm 0.04$ \\
        LHCb ($D\to \pi^+\pi^-$) & \cite{Aaij:2014eha} & 3 ${\rm fb}^{-1}$ & $-0.09 \pm 0.22 \pm 0.02$ & $1.21 \,^{+0.28}_{-0.25} \pm 0.05$ \\
        \mc{3}{l}{\bf Average} & $-0.16 \pm 0.12$ & $1.10 \pm 0.14$ \\
                \hline
                \end{tabular*}
                \label{tab:cp_uta:glw-DKstar}
        \end{center}
\end{table}

\mysubsubsection{$D$ decays to suppressed final states}
\label{sec:cp_uta:cus:ads}

\begin{figure}[htbp]
  \begin{center}
    \begin{tabular}{cc}
      \resizebox{0.46\textwidth}{!}{
        \includegraphics{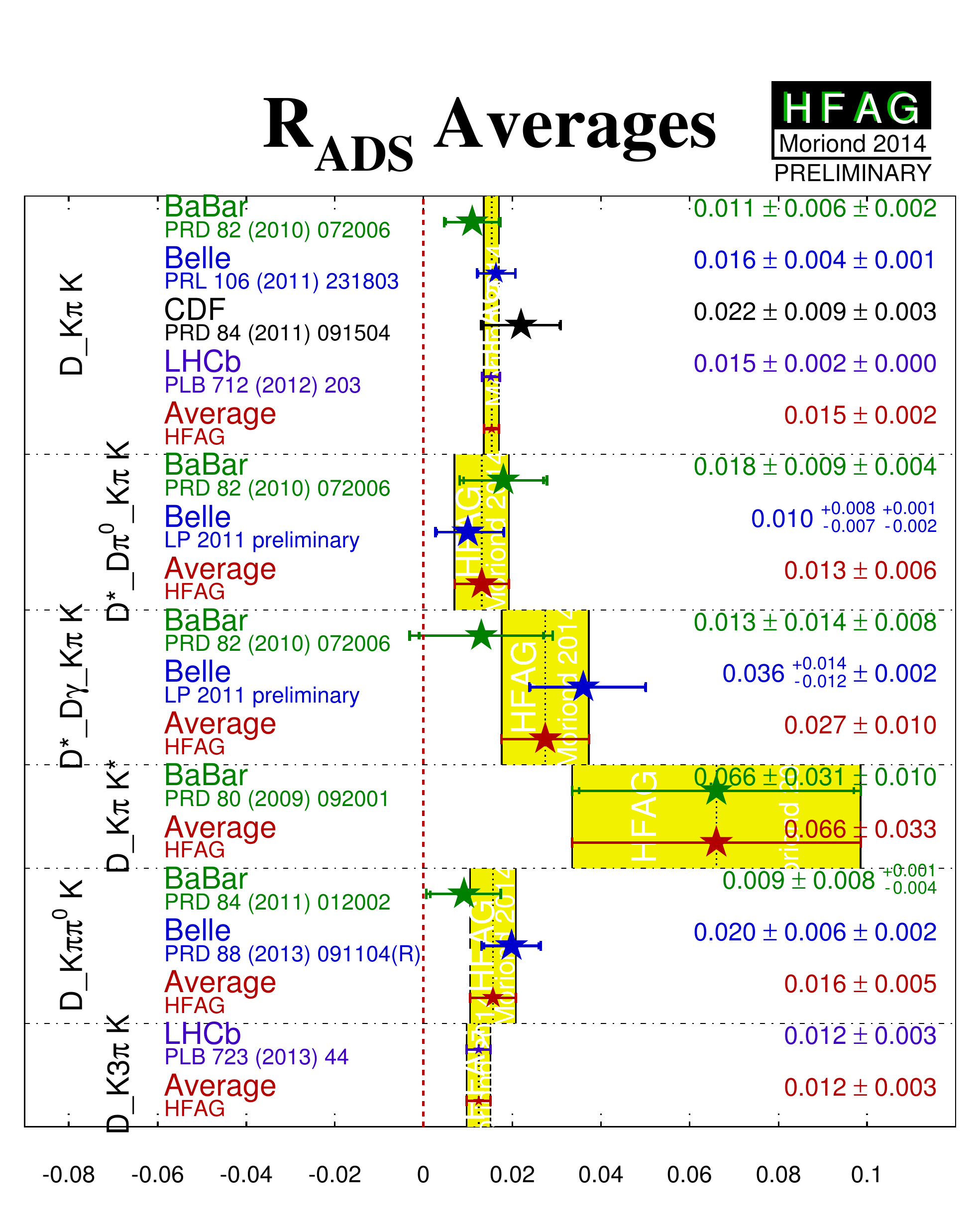}
      }
      &
      \resizebox{0.46\textwidth}{!}{
        \includegraphics{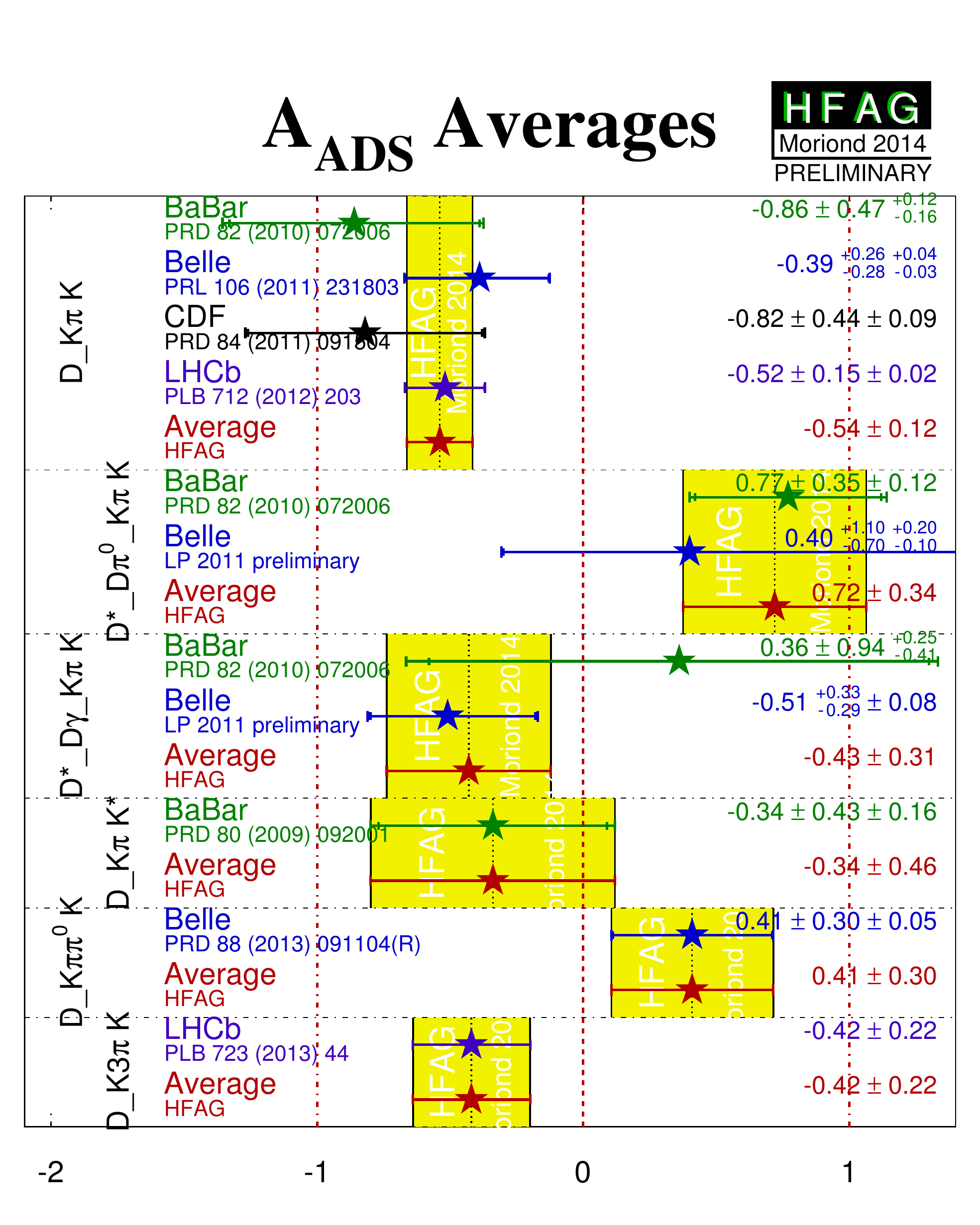}
      }
    \end{tabular}
  \end{center}
  \vspace{-0.8cm}
  \caption{
    Averages of $R_{\rm ADS}$ and $A_{\rm ADS}$ for $B \to D^{(*)}K^{(*)}$ decays.
  }
  \label{fig:cp_uta:cus:ads}
\end{figure}

For ADS analyses, all of \babar, \belle, CDF and LHCb have studied the modes 
$\Bmp \to D\Kmp$ and $\Bmp \to D\pi^\mp$. 
\babar\ and \belle\ have also analysed the $\Bmp \to \Dstar\Kmp$ mode.
There is an effective shift of $\pi$ in the strong phase difference between
the cases that the $\Dstar$ is reconstructed as $D\pi^0$ and
$D\gamma$~\cite{Bondar:2004bi}, therefore these modes are studied separately.
\babar\ has also studied the $\Bmp \to D\Kstarmp$ mode, 
where $\Kstarmp$ is reconstructed as $\KS\pi^\mp$.
In all cases the suppressed decay $D \to K^+\pi^-$ has been used.
\babar\ and \belle\ also have results using $\Bmp \to D\Kmp$ with $D \to K^+\pi^-\pi^0$,
while LHCb have results using $\Bmp \to D\Kmp$ with $D \to K^+\pi^-\pi^+\pi^-$.
The results and averages are given in Table~\ref{tab:cp_uta:cus:ads}
and shown in Fig.~\ref{fig:cp_uta:cus:ads}.

\begin{figure}[htbp]
  \begin{center}
    \begin{tabular}{cc}
      \resizebox{0.46\textwidth}{!}{
        \includegraphics{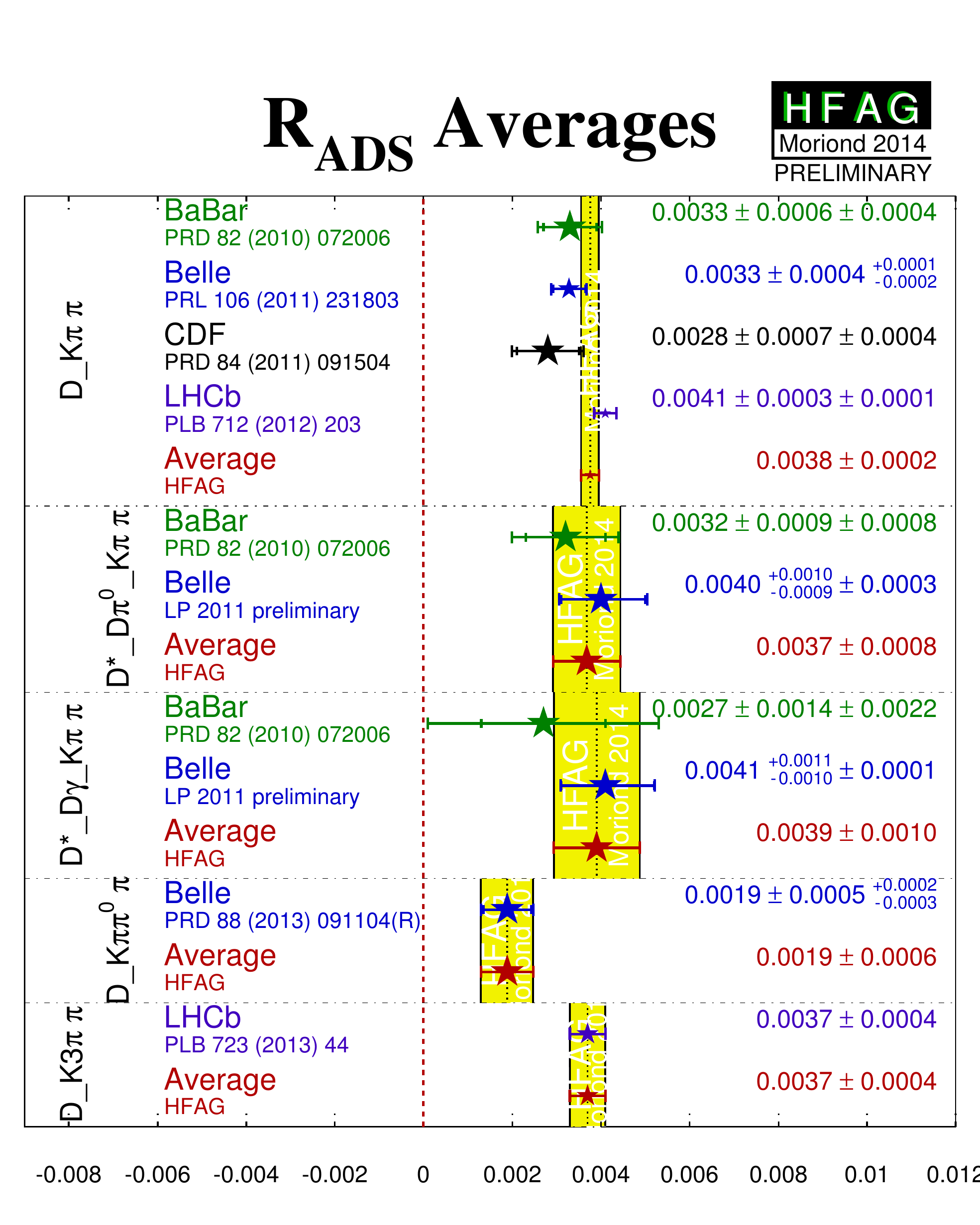}
      }
      &
      \resizebox{0.46\textwidth}{!}{
        \includegraphics{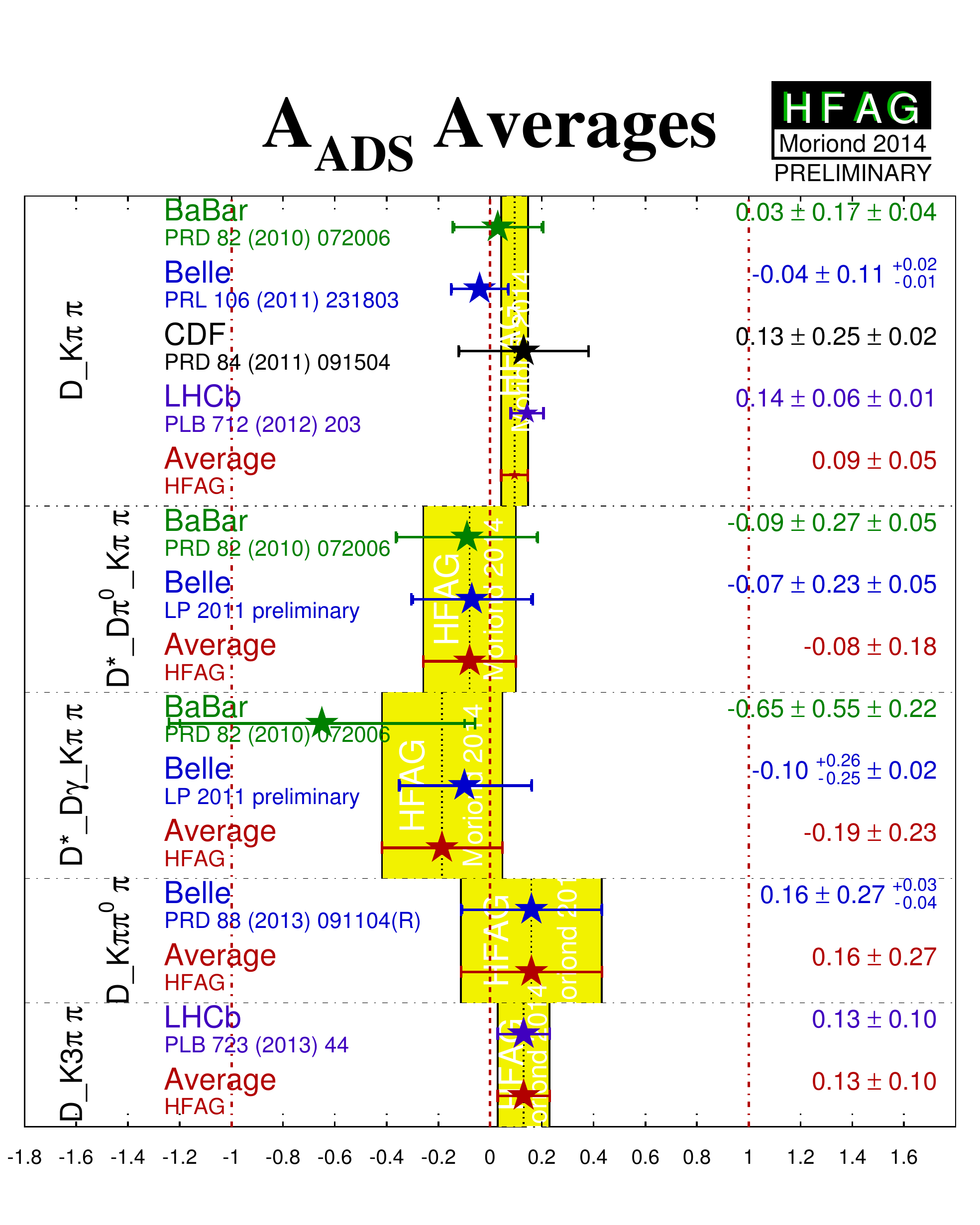}
      }
    \end{tabular}
  \end{center}
  \vspace{-0.8cm}
  \caption{
    Averages of $R_{\rm ADS}$ and $A_{\rm ADS}$ for $B \to D^{(*)}\pi$ decays.
  }
  \label{fig:cp_uta:cus:ads-Dpi}
\end{figure}

Similar phenomenology as for $B \to DK$ decays holds for $B \to D\pi$ decays, though in this case the interference is between $b \to c\bar{u}d$ and $b \to u\bar{c}d$ transitions, and the ratio of suppressed to favoured amplitudes is expected to be much smaller, ${\cal O}(1\%)$.
For most $D$ meson final states this implies that the interference effect is too small to be of interest, but in the case of ADS analysis it is possible that effects due to $\gamma$ may be observable.
Accordingly, the experiments now measure the corresponding observables in the $D\pi$ final states.
The results and averages are given in Table~\ref{tab:cp_uta:cus:ads2}
and shown in Fig.~\ref{fig:cp_uta:cus:ads-Dpi}.

\begin{table}[htb]
	\begin{center}
		\caption{
      Averages from ADS analyses of $b \to c\bar{u}s / u\bar{c}s$ modes.
                }
                \vspace{0.2cm}
                \setlength{\tabcolsep}{0.0pc}
                \begin{tabular*}{\textwidth}{@{\extracolsep{\fill}}lrccc} \hline 
        \mc{2}{l}{Experiment} & Sample size & $A_{\rm ADS}$ & $R_{\rm ADS}$ \\
	\hline
        \mc{5}{c}{$D K^-$, $D \to K^+\pi^-$} \\
	\babar & \cite{delAmoSanchez:2010dz} & $N(B\bar{B}) =$ 467M & $-0.86 \pm 0.47 \,^{+0.12}_{-0.16}$ & $0.011 \pm 0.006 \pm 0.002$ \\
	\belle & \cite{Belle:2011ac} & $N(B\bar{B}) =$ 772M & $-0.39 \,^{+0.26}_{-0.28} \,^{+0.04}_{-0.03}$ & $0.0163 \,^{+0.0044}_{-0.0041} \,^{+0.0007}_{-0.0013}$ \\
	CDF & \cite{Aaltonen:2011uu} & 7 ${\rm fb}^{-1}$ & $-0.82 \pm 0.44 \pm 0.09$ & $0.0220 \pm 0.0086 \pm 0.0026$ \\
	LHCb & \cite{Aaij:2012kz} & 1 ${\rm fb}^{-1}$ & $-0.52 \pm 0.15 \pm 0.02$ & $0.0152 \pm 0.0020 \pm 0.0004$ \\
	\mc{3}{l}{\bf Average} & $-0.54 \pm 0.12$ & $0.0153 \pm 0.0017$ \\
	\mc{3}{l}{\small Confidence level} & {\small $0.77~(0.3\sigma)$} & {\small $0.78~(0.3\sigma)$} \\
		\hline

        \mc{2}{l}{Experiment} & $N(B\bar{B})$ & $A_{\rm ADS}$ & $R_{\rm ADS}$ \\
        \mc{5}{c}{$\Dstar K^-$, $\Dstar \to D\pi^0$, $D \to K^+\pi^-$} \\
	\babar & \cite{delAmoSanchez:2010dz} & 467M & $0.77 \pm 0.35 \pm 0.12$ & $0.018 \pm 0.009 \pm 0.004$ \\
	\belle & \cite{belle:glwads:prelim} & 772M & $0.4 \,^{+1.1}_{-0.7} \,^{+0.2}_{-0.1}$ & $0.010 \,^{+0.008}_{-0.007} \,^{+0.001}_{-0.002}$ \\
	\mc{3}{l}{\bf Average} & $0.72 \pm 0.34$ & $0.013 \pm 0.006$ \\
	\mc{3}{l}{\small Confidence level} & {\small $0.71~(0.4\sigma)$} & {\small $0.52~(0.6\sigma)$} \\
 	\hline

        \mc{5}{c}{$\Dstar K^-$, $\Dstar \to D\gamma$, $D \to K^+\pi^-$} \\
	\babar & \cite{delAmoSanchez:2010dz} & 467M & $0.36 \pm 0.94 \,^{+0.25}_{-0.41}$ & $0.013 \pm 0.014 \pm 0.008$ \\
	\belle & \cite{belle:glwads:prelim} & 772M & $-0.51 \,^{+0.33}_{-0.29} \pm 0.08$ & $0.036 \,^{+0.014}_{-0.012} \pm 0.002$ \\
	\mc{3}{l}{\bf Average} & $-0.43 \pm 0.31$ & $0.027 \pm 0.010$ \\
	\mc{3}{l}{\small Confidence level} & {\small $0.42~(0.8\sigma)$} & {\small $0.26~(1.1\sigma)$} \\
		\hline

        \mc{5}{c}{$D K^{*-}$, $D \to K^+\pi^-$, $K^{*-} \to \KS \pi^-$} \\
	\babar & \cite{Aubert:2009yw} & 379M & $-0.34 \pm 0.43 \pm 0.16$ & $0.066 \pm 0.031 \pm 0.010$ \\
        \hline

        \mc{5}{c}{$D K^{-}$, $D \to K^+\pi^-\pi^0$} \\
	\babar & \cite{Lees:2011up} & 474M & \textendash{} & $0.0091 \,^{+0.0082}_{-0.0076} \,^{+0.0014}_{-0.0037}$ \\
	\belle & \cite{Nayak:2013tgg} & 772M & $0.41 \pm 0.30 \pm 0.05$ & $0.0198 \pm 0.0062 \pm 0.0024$ \\
	\mc{3}{l}{\bf Average} & \textendash{} & $0.0156 \pm 0.0052$ \\
 	\mc{3}{l}{\small Confidence level} & & {\small $0.32~(1.0\sigma)$} \\
 	\hline

        \mc{5}{c}{$D K^{-}$, $D \to K^+\pi^-\pi^+\pi^-$} \\
	LHCb & \cite{Aaij:2013mba} & 1 ${\rm fb}^{-1}$ & $-0.42 \pm 0.22$ & $0.0124 \pm 0.0027$ \\
        \hline
 		\end{tabular*}
                \label{tab:cp_uta:cus:ads}
 	\end{center}
 \end{table}

\begin{table}[htb]
	\begin{center}
		\caption{
      Averages from ADS analyses of $b \to c\bar{u}d / u\bar{c}d$ modes.
                }
                \vspace{0.2cm}
                \setlength{\tabcolsep}{0.0pc}
                \begin{tabular*}{\textwidth}{@{\extracolsep{\fill}}lrccc} \hline 
        \mc{2}{l}{Experiment} & Sample size & $A_{\rm ADS}$ & $R_{\rm ADS}$ \\
        \hline
       \mc{5}{c}{$D \pi^-$, $D \to K^+\pi^-$} \\
	\babar & \cite{delAmoSanchez:2010dz} & $N(B\bar{B}) =$ 467M & $0.03 \pm 0.17 \pm 0.04$ & $0.0033 \pm 0.0006 \pm 0.0004$ \\
	\belle & \cite{Belle:2011ac} & $N(B\bar{B}) =$ 772M & $-0.04 \pm 0.11 \,^{+0.02}_{-0.01}$ & $0.00328 \,^{+0.00038}_{-0.00036} \,^{+0.00012}_{-0.00018}$ \\
	CDF & \cite{Aaltonen:2011uu} & 7 ${\rm fb}^{-1}$ & $0.13 \pm 0.25 \pm 0.02$ & $0.0028 \pm 0.0007 \pm 0.0004$ \\
	LHCb & \cite{Aaij:2012kz} & 1 ${\rm fb}^{-1}$ & $0.143 \pm 0.062 \pm 0.011$ & $0.00410 \pm 0.00025 \pm 0.00005$ \\
	\mc{3}{l}{\bf Average} & $0.09 \pm 0.05$ & $0.00375 \pm 0.00020$ \\
	\mc{3}{l}{\small Confidence level} & {\small $0.53~(0.6\sigma)$} & {\small $0.17~(1.4\sigma)$} \\
       \hline 
       \mc{5}{c}{$\Dstar \pi^-$, $\Dstar \to D\pi^0$, $D \to K^+\pi^-$} \\
	\babar & \cite{delAmoSanchez:2010dz} & 467M & $-0.09 \pm 0.27 \pm 0.05$ & $0.0032 \pm 0.0009 \pm 0.0008$ \\
	\belle & \cite{belle:glwads:prelim} & 772M & $-0.07 \pm 0.23 \pm 0.05$ & $0.0040 \,^{+0.0010}_{-0.0009} \pm 0.0003$ \\
	\mc{3}{l}{\bf Average} & $-0.08 \pm 0.18$ & $0.0037 \pm 0.0008$ \\
	\mc{3}{l}{\small Confidence level} & {\small $0.96~(0.1\sigma)$} & {\small $0.61~(0.5\sigma)$} \\
       \hline 
       \mc{5}{c}{$\Dstar \pi^-$, $\Dstar \to D\gamma$, $D \to K^+\pi^-$} \\
	\babar & \cite{delAmoSanchez:2010dz} & 467M & $-0.65 \pm 0.55 \pm 0.22$ & $0.0027 \pm 0.0014 \pm 0.0022$ \\
	\belle & \cite{belle:glwads:prelim} & 772M & $-0.10 \,^{+0.26}_{-0.25} \pm 0.02$ & $0.0041 \,^{+0.0011}_{-0.0010} \pm 0.0001$ \\
	\mc{3}{l}{\bf Average} & $-0.19 \pm 0.23$ & $0.0039 \pm 0.0010$ \\
	\mc{3}{l}{\small Confidence level} & {\small $0.39~(0.9\sigma)$} & {\small $0.62~(0.5\sigma)$} \\
        \hline

        \mc{5}{c}{$D K^{-}$, $D \to K^+\pi^-\pi^0$} \\
	\belle & \cite{Nayak:2013tgg} & 772M & $0.16 \pm 0.27 \,^{+0.03}_{-0.04}$ & $0.00189 \pm 0.00054 \,^{+0.00022}_{-0.00025}$ \\
        \hline

        \mc{5}{c}{$D K^{-}$, $D \to K^+\pi^-\pi^+\pi^-$} \\
	LHCb & \cite{Aaij:2013mba} & 1 ${\rm fb}^{-1}$ & $0.13 \pm 0.10$ & $0.0037 \pm 0.0004$ \\
        \hline
 		\end{tabular*}
                \label{tab:cp_uta:cus:ads2}
 	\end{center}
 \end{table}

\babar~\cite{:2009au}, \belle~\cite{Negishi:2012uxa} and
LHCb~\cite{Aaij:2014eha} have also presented results from a similar analysis
method with self-tagging neutral $B$ decays: $\Bz \to DK^{*0}$ with $D \to K^-\pi^+$ (all), 
$D \to K^-\pi^+\pi^0$ and $D \to K^-\pi^+\pi^+\pi^-$ (\babar\ only)
(all with $K^{*0} \to K^+\pi^-$). 
Effects due to the natural width of the $K^{*0}$ are
handled using the parametrisation suggested by Gronau~\cite{Gronau:2002mu}. 

The following 95\% C.L. limits are set by \babar:
\begin{equation}
  R_{\rm ADS}(K\pi) < 0.244 \hspace{5mm}
  R_{\rm ADS}(K\pi\pi^0) < 0.181 \hspace{5mm}
  R_{\rm ADS}(K\pi\pi\pi) < 0.391 \, ,
\end{equation}
while \belle\ obtain
\begin{equation}
  R_{\rm ADS}(K\pi) < 0.16 \, .
\end{equation}
The results from LHCb, which are presented in terms of the parameters $R_+$ and $R_-$ instead of $R_{\rm ADS}$ and $A_{\rm ADS}$, are shown in Table~\ref{tab:cp_uta:ads-DKstar}.

\begin{table}[!htb]
        \begin{center}
                \caption{
      Results from ADS analysis of $\Bz \to D\Kstarz$, $D \to K^+\pi^-$.
                }
                \vspace{0.2cm}
                \setlength{\tabcolsep}{0.0pc}
                \begin{tabular*}{\textwidth}{@{\extracolsep{\fill}}lrccc} \hline
        \mc{2}{l}{Experiment} & Sample size & $R_{+}$ & $R_{-}$ \\
        \hline
        LHCb & \cite{Aaij:2014eha} & 3 ${\rm fb}^{-1}$ & $0.06 \pm 0.03 \pm 0.01$ & $0.06 \pm 0.03 \pm 0.01$ \\
                \hline
                \end{tabular*}
                \label{tab:cp_uta:ads-DKstar}
        \end{center}
\end{table}

Combining the results and using additional input from
CLEOc~\cite{Asner:2008ft,Lowery:2009id} a limit on the ratio between the 
$b \to u$ and $b \to c$ amplitudes of $r_s \in \left[ 0.07,0.41 \right]$ 
at 95\% C.L. limit is set by \babar.
Belle set a limit of $r_B(DK^{*0}) < 0.4$ at 95\% C.L. 
LHCb take input from Sec.~\ref{sec:charm_physics} and obtain $r_B(DK^{*0}) = 0.240 \,^{+0.055}_{-0.048}$ (different from zero with $2.7\sigma$ significance). 

\mysubsubsection{$D$ decays to multiparticle self-conjugate final states (model-dependent analysis)}
\label{sec:cp_uta:cus:dalitz}

\begin{figure}[htbp]
  \begin{center}
    \resizebox{0.30\textwidth}{!}{
      \includegraphics{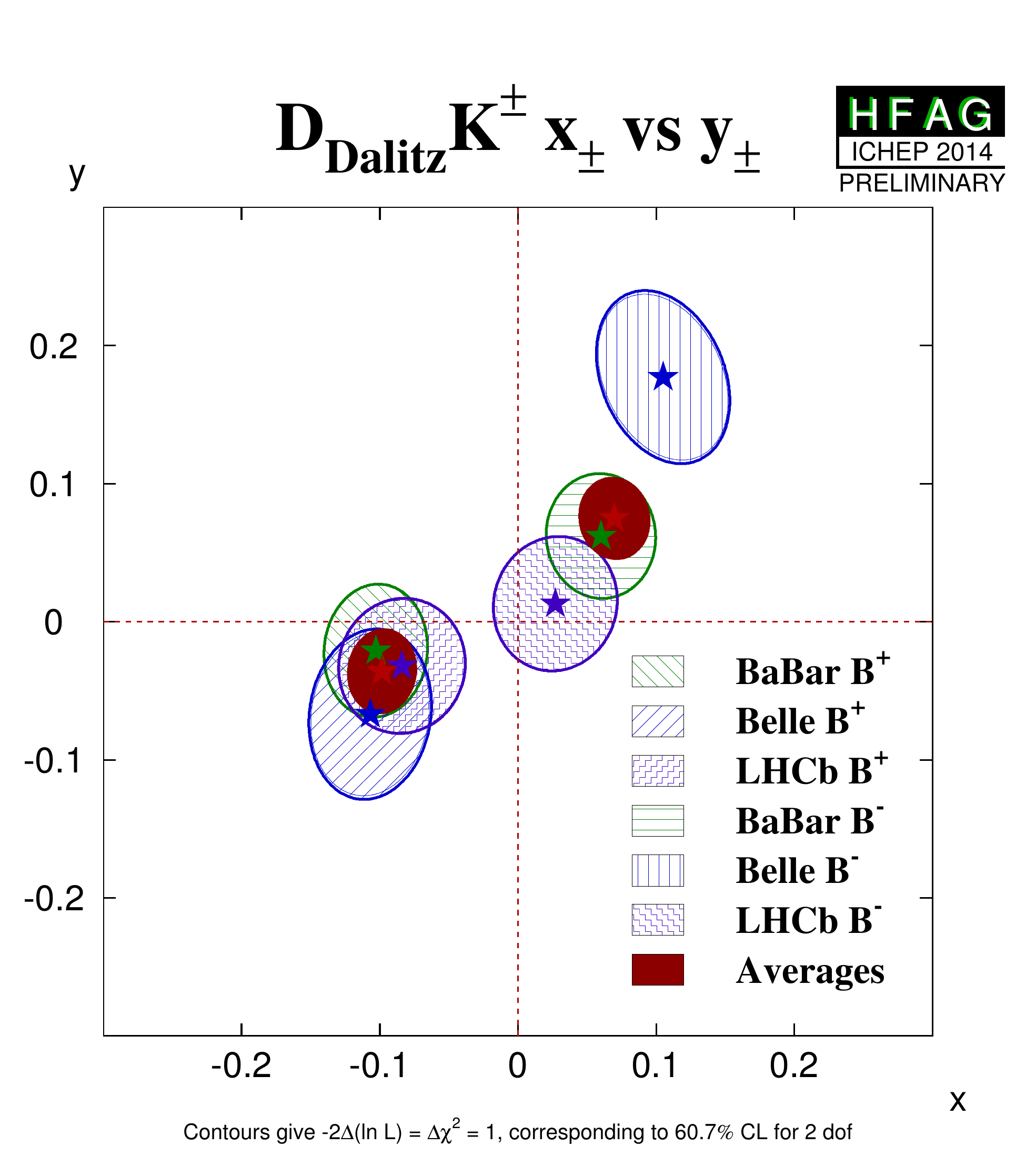}
    }
    \hfill
    \resizebox{0.30\textwidth}{!}{
      \includegraphics{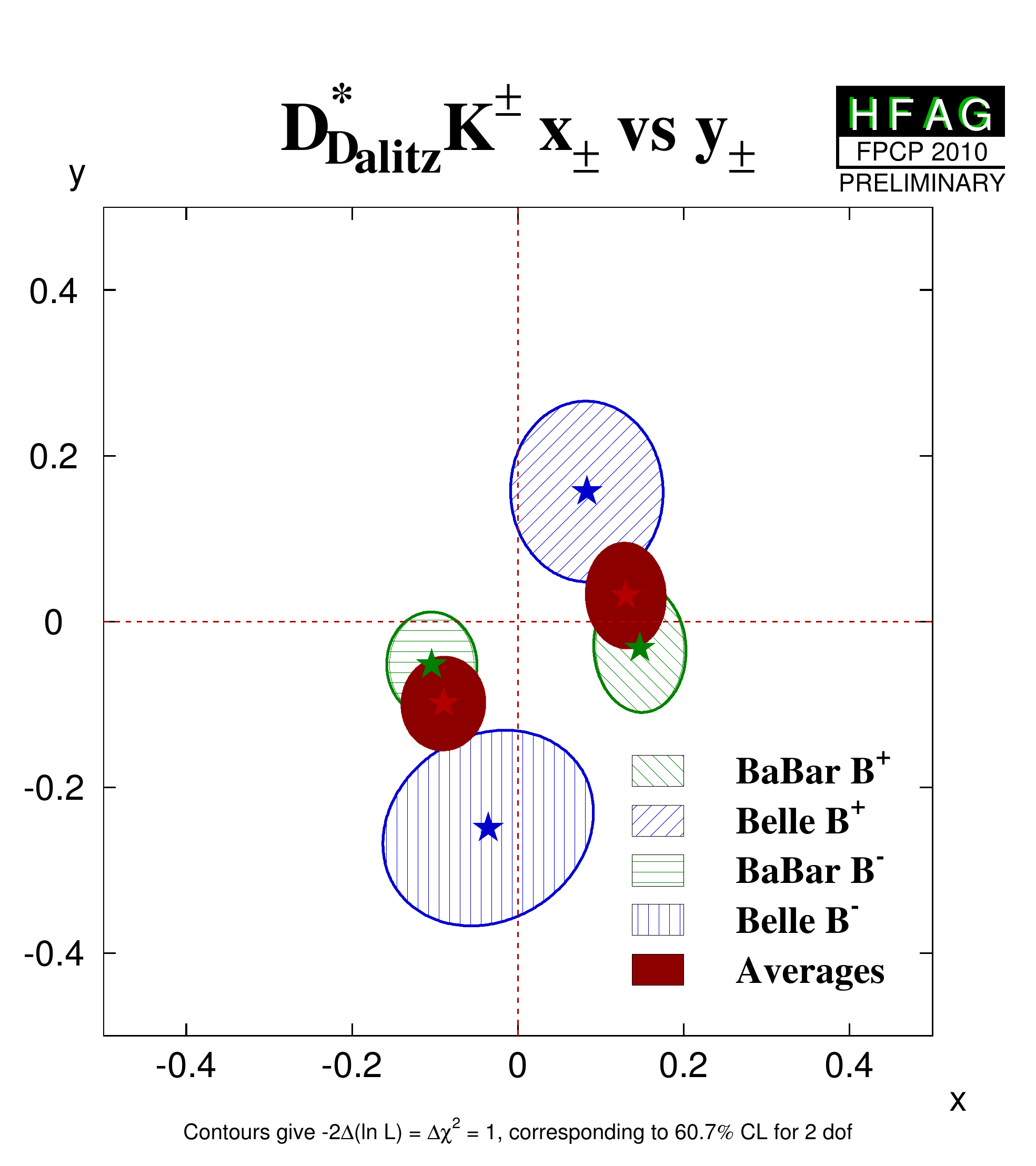}
    }
    \hfill
    \resizebox{0.30\textwidth}{!}{
      \includegraphics{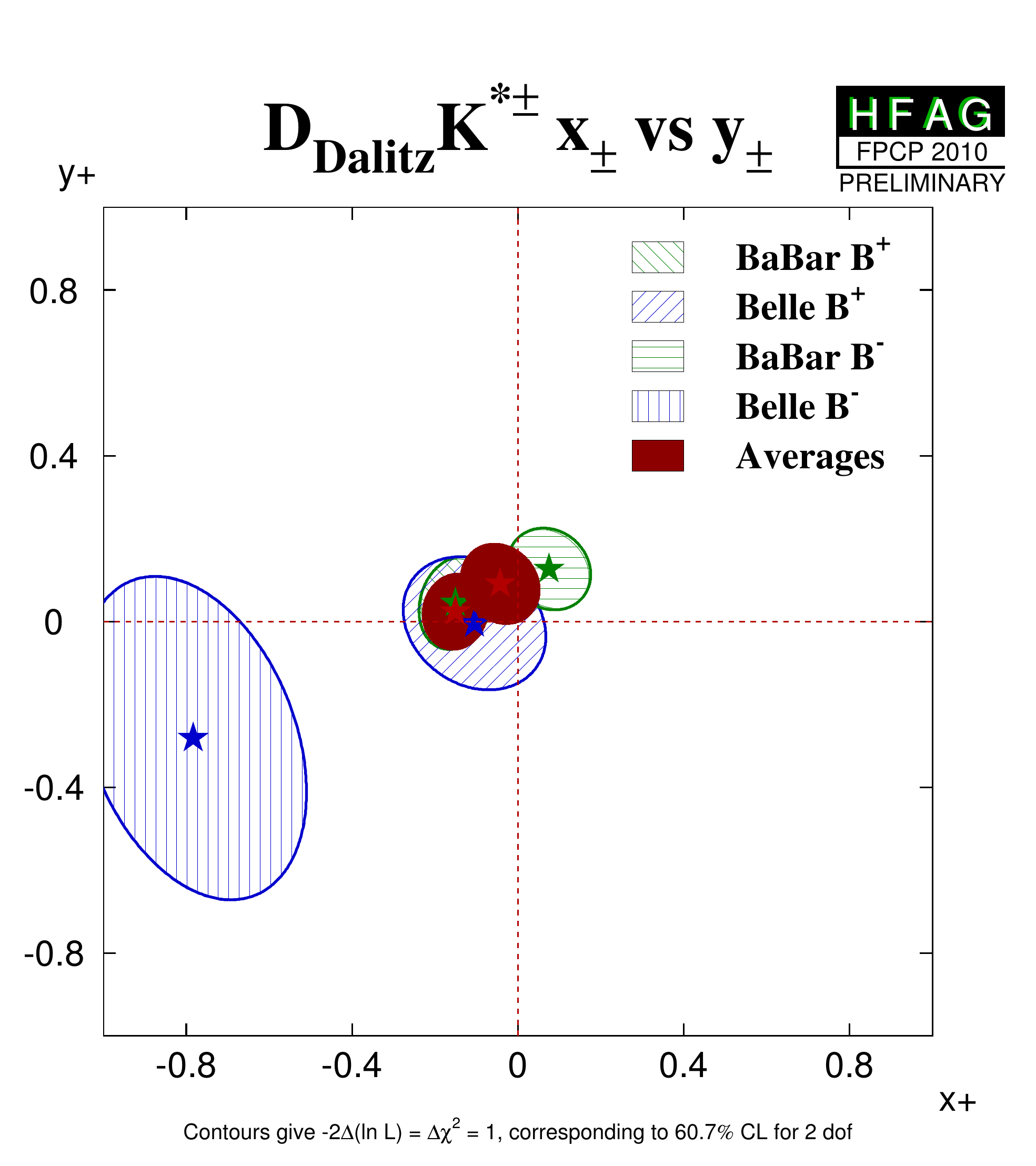}
    }
  \end{center}
  \vspace{-0.5cm}
  \caption{
    Contours in the $(x_\pm, y_\pm)$ from model-dependent analysis of $\Bmp \to D^{(*)}K^{(*)\pm}$, $D \to \KS h^+ h^-$ ($h = \pi,K$).
    (Left) $\Bmp \to D\Kmp$, 
    (middle) $\Bmp \to \Dstar\Kmp$,
    (right) $\Bmp \to D\Kstarmp$.
    Note that the uncertainties assigned to the averages given in these plots
    do not include model errors.        
  }
  \label{fig:cp_uta:cus:dalitz_2d}
\end{figure}

\begin{figure}[htbp]
  \begin{center}
    \resizebox{0.40\textwidth}{!}{
      \includegraphics{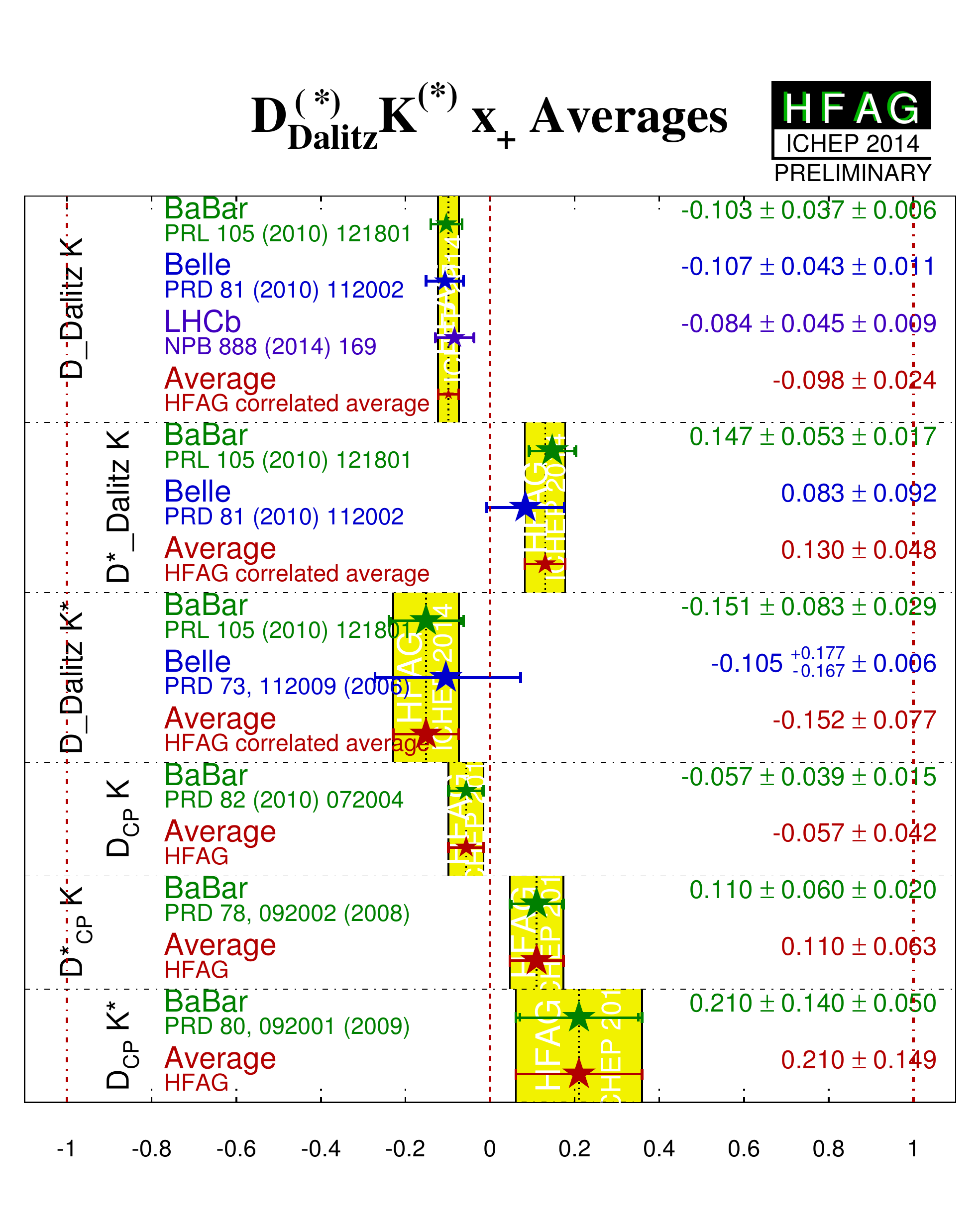}
    }
    \hspace{0.1\textwidth}
    \resizebox{0.40\textwidth}{!}{
      \includegraphics{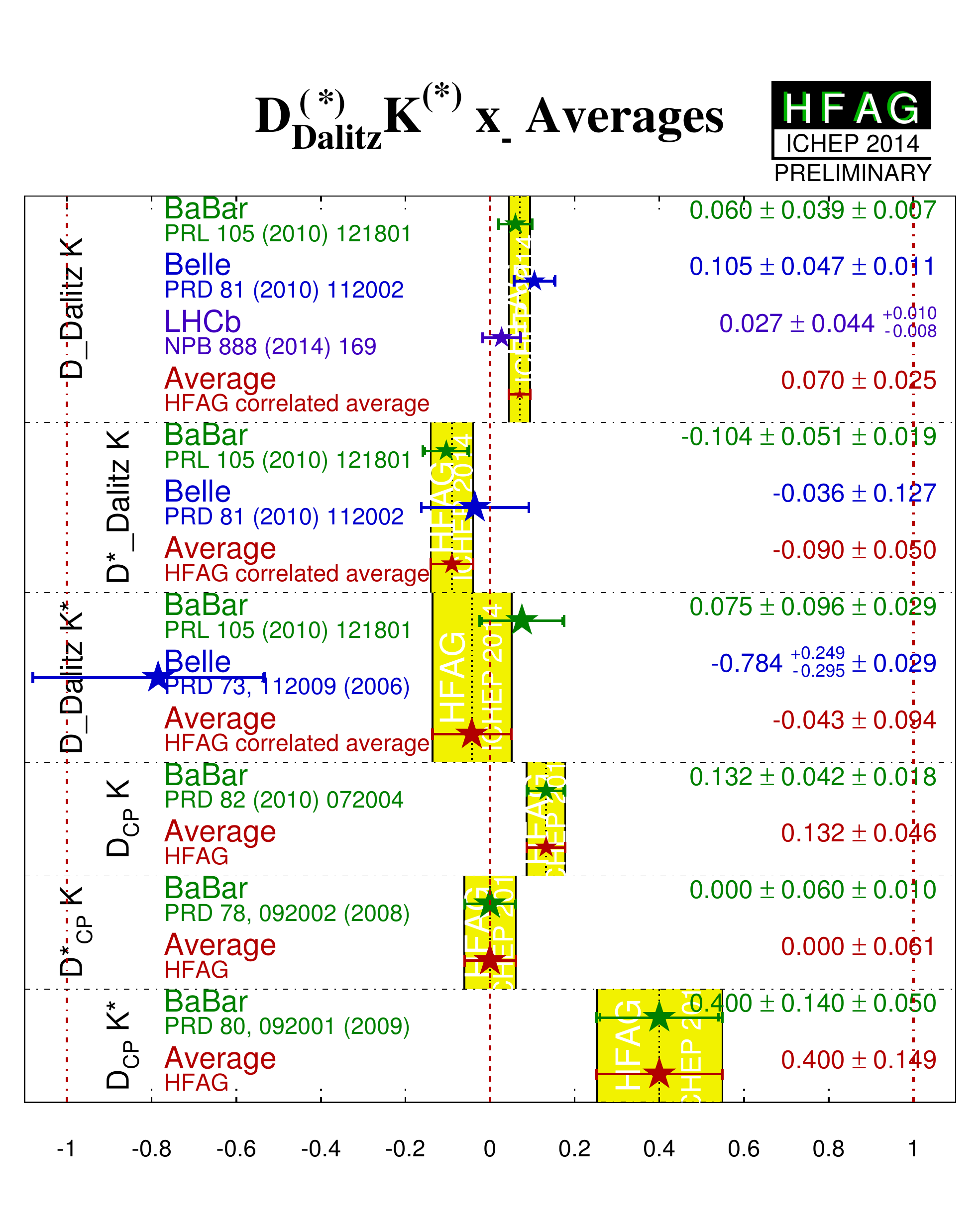}
    }
    \\
    \resizebox{0.40\textwidth}{!}{
      \includegraphics{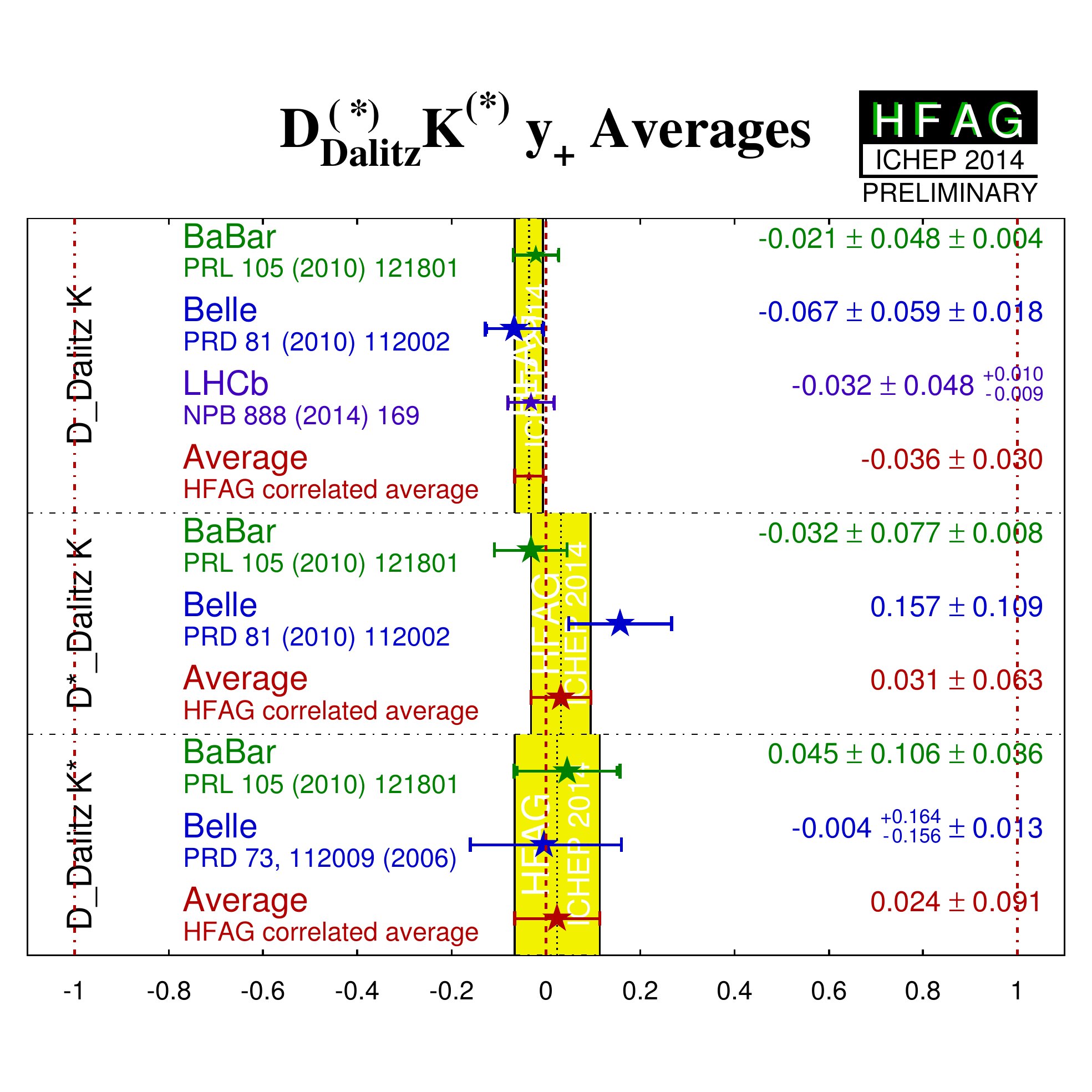}
    }
    \hspace{0.1\textwidth}
    \resizebox{0.40\textwidth}{!}{
      \includegraphics{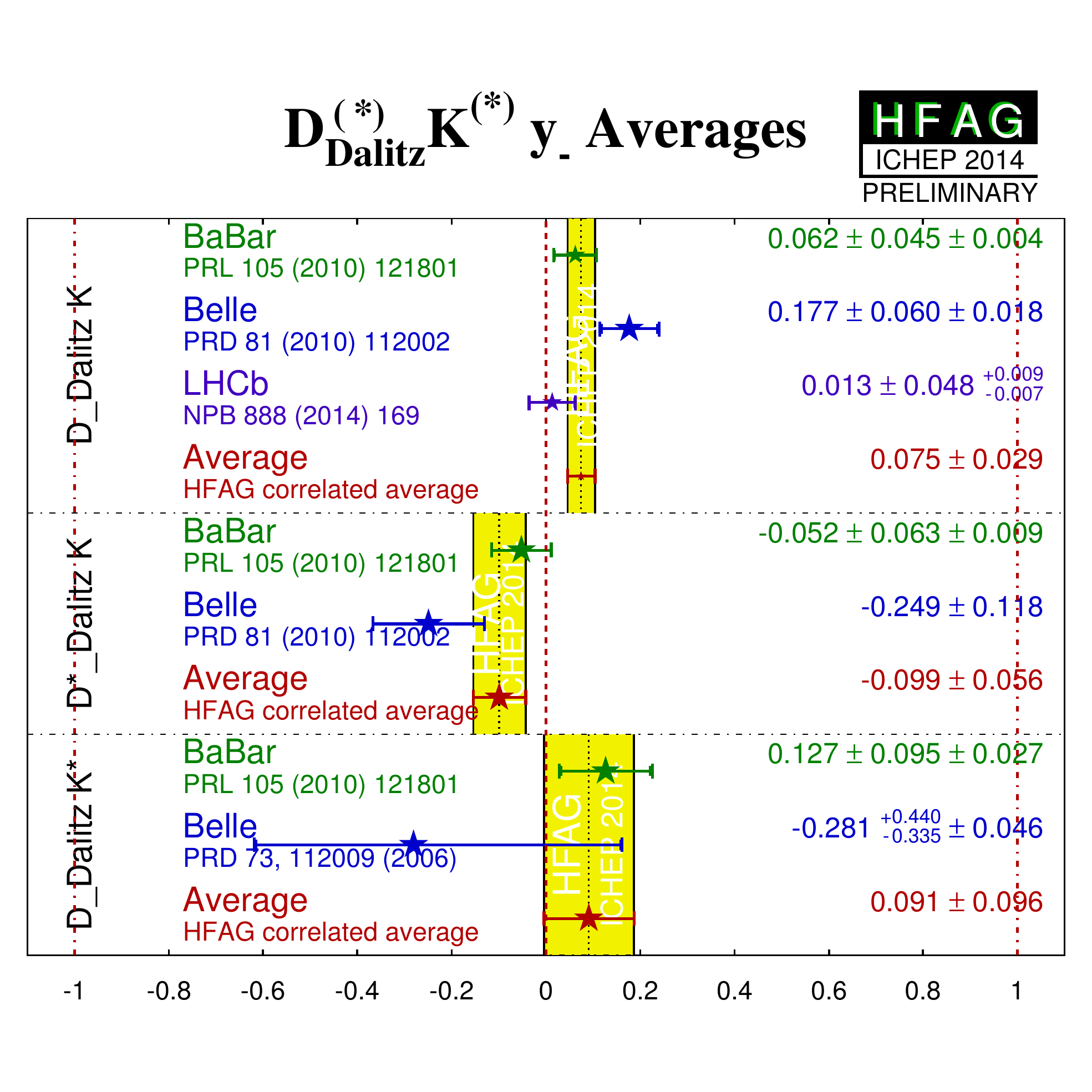}
    }
  \end{center}
  \vspace{-0.8cm}
  \caption{
    Averages of $(x_\pm, y_\pm)$ from model-dependent analyses of $\Bpm \to
    D^{(*)}K^{(*)\pm}$ with $D \to \KS h^+h^-$ ($h=\pi,K$).
    (Top left) $x_+$, (top right) $x_-$,
    (bottom left) $y_+$, (bottom right) $y_-$.
    The top plots include constraints on $x_{\pm}$ obtained from GLW analyses (see Sec.~\ref{sec:cp_uta:cus:glw}).
    Note that the uncertainties assigned to the averages given in these plots
    do not include model errors.        
  }
  \label{fig:cp_uta:cus:dalitz_1d}
\end{figure}

For the model-dependent Dalitz plot analysis, both 
\babar~\cite{Aubert:2008bd} and
\belle~\cite{Poluektov:2010wz,Poluektov:2006ia} have studied the modes 
$\Bmp \to D\Kmp$, $\Bmp \to \Dstar\Kmp$ and $\Bmp \to D\Kstarmp$.
For $\Bmp \to \Dstar\Kmp$,
both experiments have used both $\Dstar$ decay modes, $\Dstar \to D\pi^0$ and
$\Dstar \to D\gamma$, taking the effective shift in the strong phase
difference into account. 
In all cases the decay $D \to \KS\pi^+\pi^-$ has been used.
\babar\ also used the decay $D \to \KS K^+K^-$.
LHCb~\cite{Aaij:2014iba} has also studied $\Bmp \to D\Kmp$ decays with $D \to \KS\pi^+\pi^-$.
\babar\ has also performed an analysis of $\Bmp \to D\Kmp$ with 
$D \to \pi^+\pi^-\pi^0$~\cite{Aubert:2007ii}.
Results and averages are given in Table~\ref{tab:cp_uta:cus:dalitz}, and shown in Figs.~\ref{fig:cp_uta:cus:dalitz_2d} and~\ref{fig:cp_uta:cus:dalitz_1d}.
The third error on each measurement is due to $D$ decay model uncertainty.

The parameters measured in the analyses are explained in
Sec.~\ref{sec:cp_uta:notations:cus}.
Both \babar\ and \belle\ have measured the ``Cartesian''
$(x_\pm,y_\pm)$ variables, defined in Sec.~\ref{sec:cp_uta:notations:cus}, 
and perform frequentist statistical procedures,
to convert these into measurements of $\gamma$, $r_B$ and $\delta_B$.
In the $\Bmp \to D\Kmp$ with $D \to \pi^+\pi^-\pi^0$ analysis,
the parameters $(\rho^{\pm}, \theta^\pm)$ are used instead.

Both experiments reconstruct $\Kstarmp$ as $\KS\pi^\mp$,
but the treatment of possible nonresonant $\KS\pi^\mp$ differs:
\belle\ assign an additional model uncertainty,
while \babar\ use a parametrisation suggested by Gronau~\cite{Gronau:2002mu}.
The parameters $r_B$ and $\delta_B$ are replaced with 
effective parameters $\kappa r_s$ and $\delta_s$;
no attempt is made to extract the true hadronic parameters 
of the $\Bmp \to D\Kstarmp$ decay.

We perform averages using the following procedure, which is based on a set of
reasonable, though imperfect, assumptions. 

\begin{itemize}\setlength{\itemsep}{0.5ex}
\item 
  It is assumed that effects due to the different $D$ decay models 
  used by the two experiments are negligible. 
  Therefore, we do not rescale the results to a common model.
\item 
  It is further assumed that the model uncertainty is $100\%$ 
  correlated between experiments, 
  and therefore this source of error is not used in the averaging procedure.
  (This approximation is significantly less valid now that the \babar\ results
  include $D \to \KS K^+K^-$ decays in addition to $D \to \KS\pi^+\pi^-$.)
\item 
  We include in the average the effect of correlations 
  within each experiment's set of measurements.
\item 
  At present it is unclear how to assign an average model uncertainty. 
  We have not attempted to do so. 
  Our average includes only statistical and systematic errors. 
  An unknown amount of model uncertainty should be added to the final error.
\item 
  We follow the suggestion of Gronau~\cite{Gronau:2002mu} 
  in making the $DK^*$ averages. 
  Explicitly, we assume that the selection of $K^{*\pm} \to \KS\pi^\pm$
  is the same in both experiments 
  (so that $\kappa$, $r_s$ and $\delta_s$ are the same), 
  and drop the additional source of model uncertainty 
  assigned by Belle due to possible nonresonant decays.
\item 
  We do not consider common systematic errors, 
  other than the $D$ decay model. 
\end{itemize}

\begin{sidewaystable}
	\begin{center}
		\caption{
      Averages from Dalitz plot analyses of $b \to c\bar{u}s / u\bar{c}s$ modes.
      Note that the uncertainities assigned to the averages do not include model errors.	
		}
		\vspace{0.2cm}
		\setlength{\tabcolsep}{0.0pc}
    \resizebox{\textwidth}{!}{
		\begin{tabular}{@{\extracolsep{2mm}}lrccccc} \hline
	\mc{2}{l}{Experiment} & $N(B\bar{B})$ & $x_+$ & $y_+$ & $x_-$ & $y_-$ \\
	\hline
        \mc{7}{c}{$D K^-$, $D \to \KS \pi^+\pi^-$} \\
	\babar & \cite{delAmoSanchez:2010rq} & 468M & $-0.103 \pm 0.037 \pm 0.006 \pm 0.007$ & $-0.021 \pm 0.048 \pm 0.004 \pm 0.009$ & $0.060 \pm 0.039 \pm 0.007 \pm 0.006$ & $0.062 \pm 0.045 \pm 0.004 \pm 0.006$ \\
	\belle & \cite{Poluektov:2010wz} & 657M & $-0.107 \pm 0.043 \pm 0.011 \pm 0.055$ & $-0.067 \pm 0.059 \pm 0.018 \pm 0.063$ & $0.105 \pm 0.047 \pm 0.011 \pm 0.064$ & $0.177 \pm 0.060 \pm 0.018 \pm 0.054$ \\
	LHCb & \cite{Aaij:2014iba} & 1 ${\rm fb}^{-1}$ & $-0.084 \pm 0.045 \pm 0.009 \pm 0.005$ & $-0.032 \pm 0.048 \,^{+0.010}_{-0.009} \pm 0.008$ & $0.027 \pm 0.044 \,^{+0.010}_{-0.008} \pm 0.001$ & $0.013 \pm 0.048 \,^{+0.009}_{-0.007} \pm 0.003$ \\
	\mc{3}{l}{\bf Average} & $-0.098 \pm 0.024$ & $-0.036 \pm 0.030$ & $0.070 \pm 0.025$ & $0.075 \pm 0.029$ \\
        \mc{3}{l}{\small Confidence level} &  \mc{4}{c}{\small $0.52~(0.7\sigma)$} \\
 		\hline

                \mc{7}{c}{$\Dstar K^-$, $\Dstar \to D\pi^0$ or $D\gamma$, $D \to \KS \pi^+\pi^-$} \\
	\babar & \cite{delAmoSanchez:2010rq} & 468M & $0.147 \pm 0.053 \pm 0.017 \pm 0.003$ & $-0.032 \pm 0.077 \pm 0.008 \pm 0.006$ & $-0.104 \pm 0.051 \pm 0.019 \pm 0.002$ & $-0.052 \pm 0.063 \pm 0.009 \pm 0.007$ \\
	\belle & \cite{Poluektov:2010wz} & 657M & $0.083 \pm 0.092 \pm 0.081$ & $0.157 \pm 0.109 \pm 0.063$ & $-0.036 \pm 0.127 \pm 0.090$ & $-0.249 \pm 0.118 \pm 0.049$ \\
	\mc{3}{l}{\bf Average} & $0.130 \pm 0.048$ & $0.031 \pm 0.063$ & $-0.090 \pm 0.050$ & $-0.099 \pm 0.056$ \\
        \mc{3}{l}{\small Confidence level} & \mc{4}{c}{\small $0.29~(1.1\sigma)$} \\
 		\hline

                \mc{7}{c}{$D K^{*-}$, $D \to \KS \pi^+\pi^-$} \\
	\babar & \cite{delAmoSanchez:2010rq} & 468M & $-0.151 \pm 0.083 \pm 0.029 \pm 0.006$ & $0.045 \pm 0.106 \pm 0.036 \pm 0.008$ & $0.075 \pm 0.096 \pm 0.029 \pm 0.007$ & $0.127 \pm 0.095 \pm 0.027 \pm 0.006$ \\
 	\belle & \cite{Poluektov:2006ia} & 386M & $-0.105 \,^{+0.177}_{-0.167} \pm 0.006 \pm 0.088$ & $-0.004 \,^{+0.164}_{-0.156} \pm 0.013 \pm 0.095$ & $-0.784 \,^{+0.249}_{-0.295} \pm 0.029 \pm 0.097$ & $-0.281 \,^{+0.440}_{-0.335} \pm 0.046 \pm 0.086$ \\
	\mc{3}{l}{\bf Average} & $-0.152 \pm 0.077$ & $0.024 \pm 0.091$ & $-0.043 \pm 0.094$ & $0.091 \pm 0.096$ \\
        \mc{3}{l}{\small Confidence level} & \mc{4}{c}{\small $0.011~(2.5\sigma)$} \\
 		\hline

                \vspace{1ex} \\

	\hline
	\mc{2}{l}{Experiment} & $N(B\bar{B})$ & $\rho^{+}$ & $\theta^+$ & $\rho^{-}$ & $\theta^-$ \\
	\hline
        \mc{7}{c}{$D K^-$, $D \to \pi^+\pi^-\pi^0$} \\
	\babar & \cite{Aubert:2007ii} & 324M & $0.75 \pm 0.11 \pm 0.04$ & $147 \pm 23 \pm 1$ & $0.72 \pm 0.11 \pm 0.04$ & $173 \pm 42 \pm 2$ \\
	\hline
		\end{tabular}
              }
		\label{tab:cp_uta:cus:dalitz}
	\end{center}
\end{sidewaystable}

\vspace{3ex}

\noindent
\underline{Constraints on $\gamma$}

The measurements of $(x_\pm, y_\pm)$ can be used to obtain constraints on 
$\gamma$, as well as the hadronic parameters $r_B$ and $\delta_B$.
\babar~\cite{delAmoSanchez:2010rq},
\belle~\cite{Poluektov:2010wz,Poluektov:2006ia} and
LHCb~\cite{Aaij:2014iba}
have all done so using a frequentist procedure 
(there are some differences in the details of the techniques used).

\begin{itemize}\setlength{\itemsep}{0.5ex}

\item 
  \babar\ obtain $\gamma = (68 \,^{+15}_{-14} \pm 4 \pm 3)^\circ$
  from $D\Kpm$, $\Dstar\Kpm$ and $D\Kstarpm$.

\item
  \belle\ obtain $\phi_3 = (78 \,^{+11}_{-12} \pm 4 \pm 9)^\circ$
  from $D\Kpm$ and $\Dstar\Kpm$.

\item 
  LHCb obtain $\gamma = (84 \,^{+49}_{-42})^\circ$
  from $D\Kpm$ using 1 ${\rm fb}^{-1}$ of data (a more precise result using 3 ${\rm fb}^{-1}$ and the model-independent method is reported below).

\item
  The experiments also obtain values for the hadronic parameters as detailed
  in Table~\ref{tab:cp_uta:rBdeltaB_summary}.


\item 
  The CKMfitter~\cite{Charles:2004jd} and 
  UTFit~\cite{Bona:2005vz} groups use the measurements 
  from \belle\ and \babar\ given above
  to make combined constraints on $\gamma$.

\item 
  In the \babar\ analysis of $\Bmp \to D\Kmp$ with 
  $D \to \pi^+\pi^-\pi^0$~\cite{Aubert:2007ii},
  a constraint of $-30^\circ < \gamma < 76^\circ$ is obtained 
  at the 68\% confidence level.

\end{itemize}

\begin{table}
  \begin{center}
  \caption{
    Summary of constraints on hadronic parameters 
    in $\Bpm \to \DorDstar\KorKstarpm$ decays.
    Note the alternative parametrisation of the hadronic parameters used by
    \babar\ in the $D\Kstarpm$ mode.
  }
  \label{tab:cp_uta:rBdeltaB_summary}
  \begin{tabular}{lcc}
    \hline
    & $r_B$ & $\delta_B$ \\
    \hline
    \multicolumn{3}{c}{In $D\Kpm$} \\
    \babar & $0.096 \pm 0.029 \pm 0.005 \pm 0.004$ & $(119 \,^{+19}_{-20} \pm 3 \pm 3)^\circ$ \\
    \belle & $0.160 \,^{+0.040}_{-0.038} \pm 0.011 \,^{+0.05}_{-0.010}$ & 
    $(138 \,^{+13}_{-16} \pm 4 \pm 23)^\circ$ \\
    LHCb & $0.06 \pm 0.04$ & $(115 \,^{+41}_{-51})^\circ$ \\
    \hline
    \multicolumn{3}{c}{In $\Dstar\Kpm$} \\
    \babar & $0.133 \,^{+0.042}_{-0.039} \pm 0.014 \pm 0.003$ & $(-82 \pm 21 \pm 5 \pm 3)^\circ$ \\
    \belle & $0.196 \,^{+0.072}_{-0.069} \pm 0.012 \,^{+0.062}_{-0.012}$ &
    $(342 \,^{+19}_{-21} \pm 3 \pm 23)^\circ$ \\
    \hline
    \multicolumn{3}{c}{In $D\Kstarpm$} \\
    \babar & $\kappa r_S = 0.149 \,^{+0.066}_{-0.062} \pm 0.026 \pm 0.006$ &
    $\delta_S = (111 \pm 32 \pm 11 \pm 3)^\circ$ \\
    \belle & $0.56 \,^{+0.22}_{-0.16} \pm 0.04 \pm 0.08$ & 
    $(243 \,^{+20}_{-23} \pm 3 \pm 50)^\circ$ \\
    \hline
  \end{tabular}
  \end{center}
\end{table}

At present we make no attempt to provide an HFAG average for $\gamma$,
nor indeed for the hadronic parameters.
More details on procedures to calculate a best fit value for $\gamma$ 
can be found in Refs.~\cite{Charles:2004jd,Bona:2005vz}.

\babar~\cite{Aubert:2008yn} have also performed a similar Dalitz plot analysis
to that described above using the self-tagging neutral $B$ decay $\Bz \to
DK^{*0}$ (with $K^{*0} \to K^+\pi^-$). Effects due to the natural width of the
$K^{*0}$ are handled using the parametrisation suggested by
Gronau~\cite{Gronau:2002mu}.

\babar\ extract the three-dimensional likelihood for the parameters 
$\left( \gamma, \delta_S, r_S \right)$ and, combining with a separately
measured PDF for $r_S$ (using a Bayesian technique), obtain bounds on each of
the three parameters. 
\begin{equation}
  \gamma = (162 \pm 56)^\circ \hspace{5mm}
  \delta_S = (62 \pm 57)^\circ \hspace{5mm}
  r_S < 0.55  \, ,
\end{equation}
where the limit on $r_S$ is at 95\% probability.
Note that there is an ambiguity in the solutions 
$\left( \gamma, \delta_S \leftrightarrow \gamma+\pi, \delta_S+\pi \right)$.

\mysubsubsection{$D$ decays to multiparticle self-conjugate final states (model-independent analysis)}
\label{sec:cp_uta:cus:dalitz:modInd}

A model-independent approach to the analysis of $B^- \to \DorDstar K^-$ with multibody $D$ decays was proposed by Giri, Grossman, Soffer and Zupan~\cite{Giri:2003ty}, and further developed by Bondar and Poluektov~\cite{Bondar:2005ki,Bondar:2008hh}. 
The method relies on information on the average strong phase difference between $D^0$ and $\bar{D}{}^0$ decays in bins of Dalitz plot position that can be obtained from quantum-correlated $\psi(3770) \to D^0\bar{D}{}^0$ events. 
This information is measured in the form of parameters $c_i$ and $s_i$ that are the amplitude weighted averages of the cosine and sine of the strong phase difference in a Dalitz plot bin labelled by $i$, respectively. 
These quantities have been obtained for $D \to \KS \pi^+\pi^-$ (and $D \to \KS K^+K^-$) by CLEOc~\cite{Briere:2009aa,Libby:2010nu}.  
(Preliminary results from BESIII are also available.)

\belle~\cite{Aihara:2012aw} and LHCb~\cite{Aaij:2012hu,Aaij:2014uva} have 
used the model-independent Dalitz plot analysis approach to study the mode 
$\Bmp \to D\Kmp$ with $D \to \KS\pi^+\pi^-$.
LHCb have also included the $D \to \KS K^+K^-$ decay.
The variables $(x_\pm, y_\pm)$, defined in Sec.~\ref{sec:cp_uta:notations:cus}, are determined from the data. 
Note that due to the strong statistical and systematic correlations with the model-dependent results given in Sec.~\ref{sec:cp_uta:cus:dalitz}, these results cannot be combined. 

\begin{figure}[htbp]
  \begin{center}
    \resizebox{0.45\textwidth}{!}{
      \includegraphics{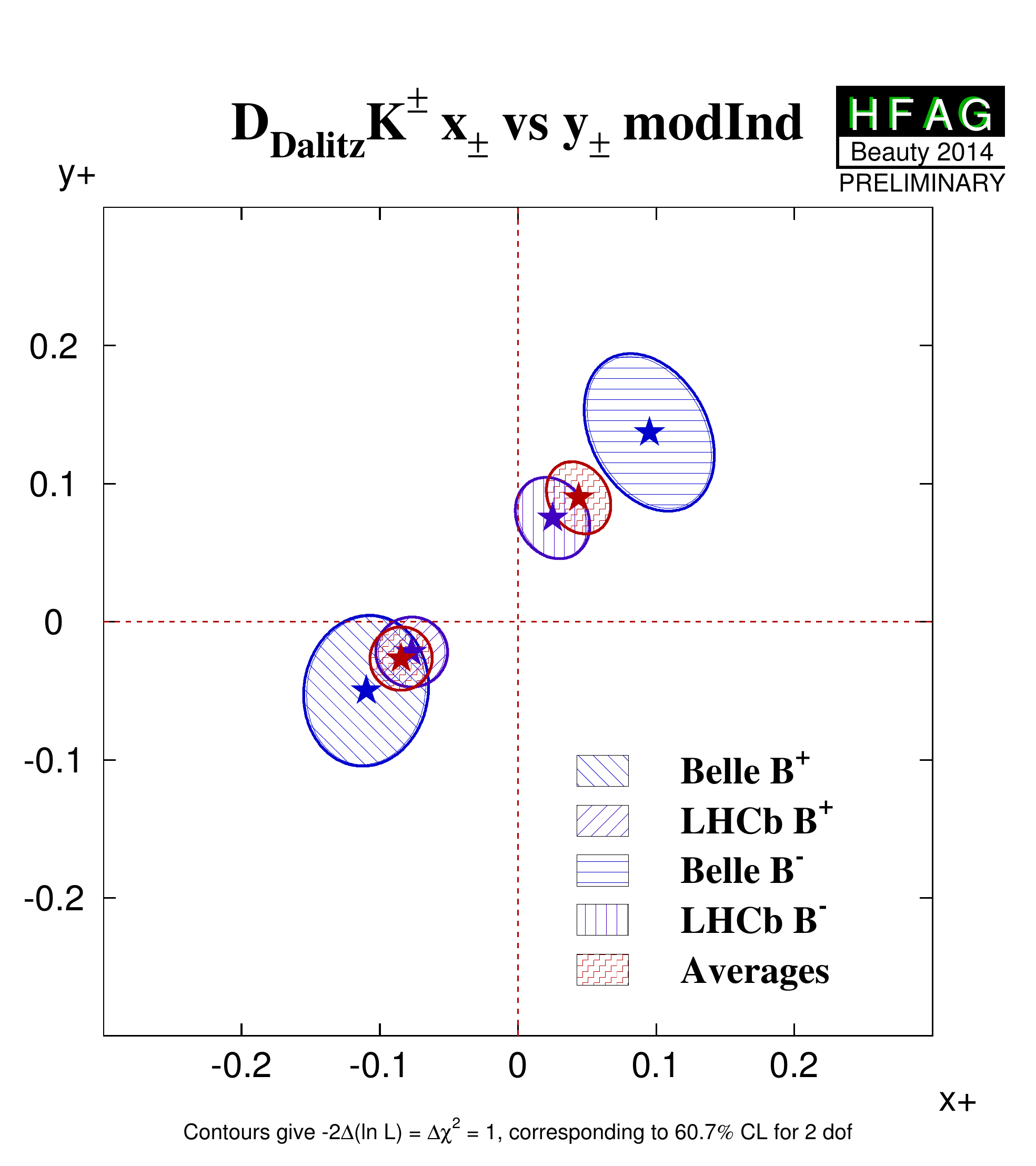}
    }
  \end{center}
  \vspace{-0.5cm}
  \caption{
    Contours in the $(x_\pm, y_\pm)$ plane from model-independent analysis of $\Bmp \to D\Kmp$, $D \to \KS h^+ h^-$ ($h = \pi,K$).
    Note that the uncertainties assigned to the averages given in these plots
    do not include model errors.        
  }
  \label{fig:cp_uta:cus:dalitz-modInd_2d}
\end{figure}

The results and averages are shown in Table~\ref{tab:cp_uta:cus:dalitz-modInd}, and shown in Figs.~\ref{fig:cp_uta:cus:dalitz-modInd_2d}.
The results have three sets of errors, which are statistical, systematic, and uncertainty coming from the knowledge of $c_i$ and $s_i$ respectively. 
To perform the average, we remove the last uncertainty, which should be 100\% correlated between the measurements. 
Since the size of the uncertainty from $c_i$ and $s_i$ is found to depend on the size of the $B \to DK$ data sample, we assign the LHCb uncertainties (which are mostly the smaller of the Belle and LHCb values) to the averaged result. 
This procedure should be conservative. 

\begin{sidewaystable}
	\begin{center}
		\caption{
      Averages from model-independent Dalitz plot analyses of $b \to c\bar{u}s / u\bar{c}s$ modes.
		}
		\vspace{0.2cm}
		\setlength{\tabcolsep}{0.0pc}
    \resizebox{\textwidth}{!}{
		\begin{tabular}{@{\extracolsep{2mm}}lrccccc} \hline
	\mc{2}{l}{Experiment} & Sample size & $x_+$ & $y_+$ & $x_-$ & $y_-$ \\
	\hline
        \mc{7}{c}{$D K^-$, $D \to \KS \pi^+\pi^-$} \\
	\belle & \cite{Aihara:2012aw} & 772M & $-0.110 \pm 0.043 \pm 0.014 \pm 0.007$ & $-0.050 \,^{+0.052}_{-0.055} \pm 0.011 \pm 0.007$ & $0.095 \pm 0.045 \pm 0.014 \pm 0.010$ & $0.137 \,^{+0.053}_{-0.057} \pm 0.015 \pm 0.023$ \\
	LHCb & \cite{Aaij:2014uva} & 3 ${\rm fb}^{-1}$ & $-0.077 \pm 0.024 \pm 0.010 \pm 0.004$ & $-0.022 \pm 0.025 \pm 0.004 \pm 0.010$ & $0.025 \pm 0.025 \pm 0.010 \pm 0.005$ & $0.075 \pm 0.029 \pm 0.005 \pm 0.014$ \\
	\mc{3}{l}{\bf Average} & $-0.085 \pm 0.023 \pm 0.04$ & $-0.027 \pm 0.023 \pm 0.010$ & $0.044 \pm 0.023 \pm 0.005$ & $0.090 \pm 0.026 \pm 0.014$ \\
        \mc{3}{l}{\small Confidence level} &  \mc{4}{c}{\small $0.39~(0.9\sigma)$} \\
 		\hline
		\end{tabular}
              }
		\label{tab:cp_uta:cus:dalitz-modInd}
	\end{center}
\end{sidewaystable}

\begin{sidewaystable}
	\begin{center}
		\caption{
      Results from model-independent Dalitz plot analysis of $B^- \to DK^-$, $D \to \KS\Kpm\pimp$.
		}
		\setlength{\tabcolsep}{0.0pc}
    \resizebox{\textwidth}{!}{
		\begin{tabular}{@{\extracolsep{2mm}}lrccccccc} \hline
	\mc{2}{l}{Experiment} & Sample size & $R_{\rm SS}$ & $R_{\rm OS}$ & $A_{{\rm SS},DK}$ & $A_{{\rm OS},DK}$ & $A_{{\rm SS},D\pi}$ & $A_{{\rm OS},D\pi}$ \\
	\hline
        \mc{9}{c}{$D \to \KS\Kpm\pimp$ (whole Dalitz plot)} \\
	LHCb & \cite{Aaij:2014dia} & 3 ${\rm fb}^{-1}$ & $0.092 \pm 0.009 \pm 0.004$ & $0.066 \pm 0.009 \pm 0.002$ & $0.040 \pm 0.091 \pm 0.018$ & $0.233 \pm 0.129 \pm 0.024$ & $-0.025 \pm 0.024 \pm 0.010$ & $-0.052 \pm 0.029 \pm 0.017$ \\
	\hline
        \mc{9}{c}{$D \to K^*(892)^\pm\pimp$} \\
	LHCb & \cite{Aaij:2014dia} & 3 ${\rm fb}^{-1}$ & $0.084 \pm 0.011 \pm 0.003$ & $0.056 \pm 0.013 \pm 0.002$ & $0.026 \pm 0.109 \pm 0.029$ & $0.336 \pm 0.208 \pm 0.026$ & $-0.012 \pm 0.028 \pm 0.010$ & $-0.054 \pm 0.043 \pm 0.017$ \\
        \hline
                \end{tabular}
    }
		\label{tab:cp_uta:cus:KSKpi-modInd}
	\end{center}
\end{sidewaystable}

\vspace{3ex}

\noindent
\underline{Constraints on $\gamma$}

The measurements of $(x_\pm, y_\pm)$ can be used to obtain constraints on 
$\gamma$, as well as the hadronic parameters $r_B$ and $\delta_B$.
\belle~\cite{Aihara:2012aw} and LHCb~\cite{Aaij:2012hu,Aaij:2014uva}
have done so using a frequentist procedure (there are some differences in the details of the techniques used).

\begin{itemize}\setlength{\itemsep}{0.5ex}

\item 
  \belle\ obtain
  $\phi_3 = (77.3 \,^{+15.1}_{-14.9} \pm 4.1 \pm 4.3)^\circ$.

\item
 LHCb obtain
 $\gamma = (62 \,^{+15}_{-14})^\circ$.

\item
  The experiments also obtain values for the hadronic parameters as detailed
  in Table~\ref{tab:cp_uta:rBdeltaB_summary-modInd}.

\end{itemize}

\begin{table}
  \begin{center}
  \caption{
    Summary of constraints on hadronic parameters  from model-independent analyses of $\Bpm \to D\Kpm$, $D \to \KS h^+h^-$ ($h=\pi,K$) decays.
  }
  \label{tab:cp_uta:rBdeltaB_summary-modInd}
  \begin{tabular}{lcc}
    \hline
    & $r_B$ & $\delta_B$ \\
    \hline
    \belle & $0.145 \pm 0.030 \pm 0.010 \pm 0.011$ & $(129.9 \pm 15.0 \pm 3.8 \pm 4.7)^\circ$ \\
    LHCb & $0.080 \,^{+0.019}_{-0.021}$ & $(134 \,^{+14}_{-15})^\circ$ \\
    \hline
  \end{tabular}
  \end{center}
\end{table}

At present we make no attempt to provide an HFAG average for $\gamma$,
nor indeed for the hadronic parameters.
More details on procedures to calculate a best fit value for $\gamma$ 
can be found in Refs.~\cite{Charles:2004jd,Bona:2005vz}.

\mysubsubsection{$D$ decays to multiparticle non-self-conjugate final states (model-independent analysis)}
\label{sec:cp_uta:cus:dalitz:KsKpi}

Following the original suggestion of Grossman, Ligeti and Soffer~\cite{Grossman:2002aq}, decays of $D$ mesons to $\KS\Kpm\pimp$ can be used in a similar approach to that discussed above to determine $\gamma \equiv \phi_3$. 
Since these decays are less abundant, the event samples available to date have not been sufficient for a fine binning of the Dalitz plots, but the analysis can be performed using only an overall coherence factor and related strong phase difference for the decay. 
These quantities have been determined by CLEOc~\cite{Insler:2012pm} both for the full Dalitz plots and in a restricted region $\pm 100 \ {\rm MeV}/c^2$ around the peak of the $K^*(892)^\pm$ resonance.

LHCb~\cite{Aaij:2014dia} have reported results of an analysis of $B^-\to D K^-$ and $B^- \to D \pi^-$ decays with $D \to \KS\Kpm\pimp$. 
The decays with different final states of the $D$ meson are distinguished by the charge of the kaon from the decay of the $D$ meson relative to the charge of the $B$ meson, and are labelled ``same sign'' (SS) and ``opposite sign'' (OS). 
Six observables potentially sensitive to $\gamma \equiv \phi_3$ are measured: two ratios of rates for $DK$ and $D\pi$ decays (one each for SS and OS) and four asymmetries (for $DK$ \& $D\pi$, SS \& OS). 
This is done both for the full Dalitz plot and for the $K^*(892)^\pm$-dominated region (with the same boundaries as used by CLEOc). 
Note that there is a significant overlap of events between the two samples. 
The results, shown in Table~\ref{tab:cp_uta:cus:KSKpi-modInd} do not yet have sufficient precision to set significant constraints on $\gamma \equiv \phi_3$. 

\mysubsubsection{Combinations of results on rates and asymmetries in $\Bmp \to \DorDstar K^{(*)\mp}$ decays to obtain constraints on $\gamma \equiv \phi_3$}
\label{sec:cp_uta:cus:gamma}

\babar, \belle\ and LHCb have all presented constraints on $\gamma \equiv \phi_3$ from combinations of their results on $B^- \to DK^-$ and related processes.
All use a frequentist procedure (there are some differences in the details of the techniques used).

\begin{itemize}\setlength{\itemsep}{0.5ex}

\item 
  \babar~\cite{Lees:2013nha} use results from $DK$, $D^*K$ and $DK^*$ modes with GLW, ADS and GGSZ analyses, to obtain $\gamma = (69 \,^{+17}_{-16})^\circ$.

\item 
  \belle~\cite{Trabelsi:2013uj} use results from $DK$ and $D^*K$ modes with GLW, ADS and GGSZ analyses, to obtain $\phi_3 = (68 \,^{+15}_{-14})^\circ$.

\item
  LHCb~\cite{LHCb-CONF-2014-004} use results from the $DK$ mode with GLW, ADS (both $K\pi$ and $K3\pi$), GGSZ ($\KS h^+h^-$) and GLS ($\KS\Kpm\pimp$) analyses, as well as $DK^{*0}$ with GLW and ADS analyses and $\Bs \to D_s^\mp\Kpm$ decays. 
  LHCb have in addition obtained a constraint (not quoted here) including results from $B \to D\pi$.
  The LHCb combination takes into account subleading effects due to charm mixing and \CP violation~\cite{Rama:2013voa}.  
  The result is $\gamma = (73 \,^{+9}_{-10})^\circ$.

\item
  All the combinations use inputs from CLEOc (and/or from the HFAG - Charm Physics global fits on charm mixing parameters, see Sec.~\ref{sec:charm:mixcpv}) to constrain the hadronic parameters in the charm system. 

\item 
  Constraints are also obtained on the hadronic parameters involved in the decays.
  A summary of these is given in Table~\ref{tab:cp_uta:rBdeltaB_combination}.

\item 
  The CKMfitter~\cite{Charles:2004jd} and 
  UTFit~\cite{Bona:2005vz} groups perform similar combinations of all available results to make combined constraints on $\gamma$.

\end{itemize}

\begin{table}
  \begin{center}
  \caption{
    Summary of constraints on hadronic parameters obtained from global combinations of results in $\Bpm \to \DorDstar\KorKstarpm$ decays.
  }
  \label{tab:cp_uta:rBdeltaB_combination}
  \begin{tabular}{l@{\hspace{5mm}}c@{\hspace{5mm}}c}
    \hline
    & $r_B$ & $\delta_B$ \\
    \hline
    \babar & $0.092 \,^{+0.013}_{-0.012}$ & $(105 \,^{+16}_{-17})^\circ$ \\
    \belle & $ 0.112 \,^{+0.014}_{-0.015}$ & $(116 \,^{+18}_{-21})^\circ$ \\
    LHCb & $0.0914 \,^{+0.0083}_{-0.0088}$ & $(127 \,^{+10}_{-12})^\circ$ \\
    \hline
  \end{tabular}
  \end{center}
\end{table}

\clearpage


\section{Semileptonic $B$ decays}
\label{sec:slbdecays}

This section contains our averages for semileptonic $B$~meson decays,
\ie\ decays of the type $B\to X\ell\nu_\ell$, where $X$ is a
hadronic system, $\ell$ a charged lepton and $\nu_\ell$ its corresponding
neutrino. Unless otherwise stated, $\ell$ stands for an electron
\emph{or} a muon, lepton universality is assumed, and both charge
conjugate states are included. Some averages assume isospin symmetry
and this will be explicitly mentioned at every instance.

The averages are organized by the flavour changing transition
(the CKM-favoured $b\to c$ transition and the CKM-suppressed $b\to u$
transition) and by the experimental definition of the hadronic
$X$~system. Measurements which are sensitive to only a specific
hadronic state ($X=D,D^*,\pi,\dots$) are called \emph{exclusive} while
analyses that measure all hadronic states within a given region of
phase space are \emph{inclusive}. The principal reason why semileptonic
$B$~decays are studied in experiments is the determination of the CKM
matrix element magnitudes $|V_{cb}|$ and $|V_{ub}|$. The averages in
the different subsections thus focus very much on these two
fundamental parameters of the Standard Model. In the last subsection,
we discuss semileptonic $B$~decays with a $\tau$-lepton. These
decays are relevant to the search for physics beyond the Standard
Model, \eg\ in the context of the type II
Two-Higgs-Doublet-Model (2HDM).

The technique for obtaining the averages follows the general HFAG
procedure (Sec.~\ref{sec:method}) unless otherwise stated. More
information on the averages, in particular the common input parameters
is available on the HFAG semileptonic webpage:

\centerline{\tt http://www.slac.stanford.edu/xorg/hfag/semi/pdg14/}





\subsection{Exclusive CKM-favoured decays}
\label{slbdecays_b2cexcl}
This section is organized as follows: First, we present averages for
the decays $\bar B\to D^*\ell^-\bar\nu_\ell$ and $\bar B\to
D\ell^-\bar\nu_\ell$. In addition to the branching fractions, the CKM
element $|V_{cb}|$ is extracted. We then provide
averages for the inclusive branching fractions $\cbf(\bar B\to
D^{(*)}\pi \ell^-\bar\nu_\ell)$ and for $B$ semileptonic decays into
orbitally-excited $P$-wave charm mesons ($D^{**}$). As the $D^{**}$
branching fraction is poorly known, we report the averages for the products 
$\cbf(B^-\to D^{**}(D^{(*)}\pi)\ell^-\bar\nu_\ell)\times
\cbf(D^{**}\to D^{(*)}\pi)$.


\mysubsubsection{$\bar B\to D^*\ell^-\bar\nu_\ell$}
\label{slbdecays_dstarlnu}

The kinematics of the decay $\bar B\to D^*\ell^-\bar\nu_\ell$ are
described by the form factor $\eta_\mathrm{EW}{\cal F}(w)$, where
$\eta_{EW}$ is a known electro-weak correction factor and $w$ is the
product of the $B$ and $D^*$ meson 4-velocities, $w=v_B\cdot
v_{D^*}$. Experiments measure the differential decay width as a
function of $w$ and determine the form factor $\eta_\mathrm{EW}{\cal
  F}(w)$ in the parameterization of Caprini, Lellouch and Neubert
(CLN), which describes the shape and normalization in terms of four
quantities: the normalization $\eta_\mathrm{EW}{\cal F}(1)$, the slope
$\rho^2$, and the amplitude ratios $R_1(1)$ and
$R_2(1)$~\cite{CLN}. Our main average and the determination of $\vcb$
are based on this parameterization.

We use the measurements of these form factor parameters shown in
Table~\ref{tab:vcbf1} and rescale them to the latest values of the
input parameters (mainly branching fractions of charmed
mesons)~\cite{HFAG_sl:inputparams}. Most of the measurements in
Table~\ref{tab:vcbf1} are based exclusively on the decay $\bar B^0\to
D^{*+}\ell^-\bar\nu_\ell$. Some
measurements~\cite{Adam:2002uw,Aubert:2009_1} are sensitive also to
$B^-\to D^{*0}\ell^-\bar\nu_\ell$ and one
measurement~\cite{Aubert:2009_3} is based exclusively on the decay
$B^-\to D^{*0}\ell^-\bar\nu_\ell$. Our analysis thus assumes isospin
symmetry.
\begin{table}[!htb]
\caption{Measurements of the Caprini, Lellouch and Neubert
  (CLN)~\cite{CLN} form factor parameters in $\bar B\to
  D^*\ell^-\bar\nu_\ell$ before and after rescaling. Most analyses
  (except \cite{Dungel:2010uk,Aubert:2006mb}) measure only
  $\eta_\mathrm{EW}{\cal F}(1)\vcb$, and $\rho^2$, so only these two
  parameters are shown here. The average is the result of a
  4-dimensional fit to the rescaled measurements of
  $\eta_\mathrm{EW}{\cal F}(1)\vcb$, $\rho^2$, $R_1(1)$ and
  $R_2(1)$ -- see the text for more details. The $\chi^2$~value of the
  combination is 30.0 for 23 degrees of freedom (CL=$15.0\%$).}
\begin{center}
\resizebox{0.99\textwidth}{!}{
\begin{tabular}{|l|c|c|}
  \hline
  Experiment
  & $\eta_\mathrm{EW}{\cal F}(1)\vcb [10^{-3}]$ (rescaled)
  & $\rho^2$ (rescaled)\\
  & $\eta_\mathrm{EW}{\cal F}(1)\vcb [10^{-3}]$ (published)
  & $\rho^2$ (published)\\
  \hline\hline
  ALEPH~\cite{Buskulic:1996yq}
  & $31.23\pm 1.80_{\rm stat}\pm 1.30_{\rm syst}$
  & $0.493\pm 0.228_{\rm stat}\pm 0.144_{\rm syst}$\\
  & $31.9\pm 1.8_{\rm stat}\pm 1.9_{\rm syst}$
  & $0.37\pm 0.26_{\rm stat}\pm 0.14_{\rm syst}$\\
  \hline
  CLEO~\cite{Adam:2002uw}
  & $39.94\pm 1.23_{\rm stat}\pm 1.62_{\rm syst}$
  & $1.367\pm 0.085_{\rm stat}\pm 0.086_{\rm syst}$\\
  & $43.1\pm 1.3_{\rm stat}\pm 1.8_{\rm syst}$
  & $1.61\pm 0.09_{\rm stat}\pm 0.21_{\rm syst}$\\
  \hline
  OPAL excl~\cite{Abbiendi:2000hk}
  & $36.50\pm 1.60_{\rm stat}\pm 1.49_{\rm syst}$
  & $1.234\pm 0.212_{\rm stat}\pm 0.145_{\rm syst}$\\
  & $36.8\pm 1.6_{\rm stat}\pm 2.0_{\rm syst}$
  & $1.31\pm 0.21_{\rm stat}\pm 0.16_{\rm syst}$\\
  \hline
  OPAL partial reco~\cite{Abbiendi:2000hk}
  & $37.14\pm 1.19_{\rm stat}\pm 2.36_{\rm syst}$
  & $1.152\pm 0.145_{\rm stat}\pm 0.294_{\rm syst}$\\
  & $37.5\pm 1.2_{\rm stat}\pm 2.5_{\rm syst}$
  & $1.12\pm 0.14_{\rm stat}\pm 0.29_{\rm syst}$\\
  \hline
  DELPHI partial reco~\cite{Abreu:2001ic}
  & $35.32\pm 1.40_{\rm stat}\pm 2.33_{\rm syst}$
  & $1.174\pm 0.126_{\rm stat} \pm 0.377_{\rm syst}$\\
  & $35.5\pm 1.4_{\rm stat}\ {}^{+2.3}_{-2.4}{}_{\rm syst}$
  & $1.34\pm 0.14_{\rm stat}\ {}^{+0.24}_{-0.22}{}_{\rm syst}$\\
  \hline
  DELPHI excl~\cite{Abdallah:2004rz}
  & $36.10\pm 1.70_{\rm stat}\pm 1.97_{\rm syst}$
  & $1.081\pm 0.142_{\rm stat} \pm 0.152_{\rm syst}$\\
  & $39.2\pm 1.8_{\rm stat}\pm 2.3_{\rm syst}$
  & $1.32\pm 0.15_{\rm stat}\pm 0.33_{\rm syst}$\\
  \hline
  \belle~\cite{Dungel:2010uk}
  & $34.60\pm 0.17_{\rm stat}\pm 1.02_{\rm syst}$
  & $1.212\pm 0.034_{\rm stat}\pm 0.009_{\rm syst}$\\
  & $34.6\pm 0.2_{\rm stat}\pm 1.0_{\rm syst}$
  & $1.214\pm 0.034_{\rm stat} \pm 0.009_{\rm syst}$\\
  \hline
  \babar\ excl~\cite{Aubert:2006mb}
  & $33.94\pm 0.30_{\rm stat}\pm 0.99_{\rm syst}$
  & $1.185\pm 0.048_{\rm stat}\pm 0.029_{\rm syst}$\\
  & $34.7\pm 0.3_{\rm stat}\pm 1.1_{\rm syst}$
  & $1.18\pm 0.05_{\rm stat}\pm 0.03_{\rm syst}$\\
  \hline
  \babar\ $D^{*0}$~\cite{Aubert:2009_3}
  & $35.22\pm 0.59_{\rm stat}\pm 1.33_{\rm syst}$
  & $1.128\pm 0.058_{\rm stat}\pm 0.055_{\rm syst}$\\
  & $35.9\pm 0.6_{\rm stat}\pm 1.4_{\rm syst}$
  & $1.16\pm 0.06_{\rm stat}\pm 0.08_{\rm syst}$\\
  \hline
  \babar\ global fit~\cite{Aubert:2009_1}
  & $35.76\pm 0.20_{\rm stat}\pm 1.10_{\rm syst}$
  & $1.193\pm 0.020_{\rm stat}\pm 0.061_{\rm syst}$\\
  & $35.7\pm 0.2_{\rm stat}\pm 1.2_{\rm syst}$
  & $1.21\pm 0.02_{\rm stat}\pm 0.07_{\rm syst}$\\
  \hline
  {\bf Average}
  & \mathversion{bold} $35.81\pm 0.11_{\rm stat}\pm 0.44_{\rm syst}$ &
  \mathversion{bold} $1.207\pm 0.015_{\rm stat}\pm 0.021_{\rm syst}$\\
  \hline 
\end{tabular}
}
\end{center}
\label{tab:vcbf1}
\end{table}

\begin{figure}[!ht]
  \begin{center}
  \unitlength 1.0cm 
  \begin{picture}(14.,11.0)
    \put(  7.5,-0.2){\includegraphics[width=9.5cm]{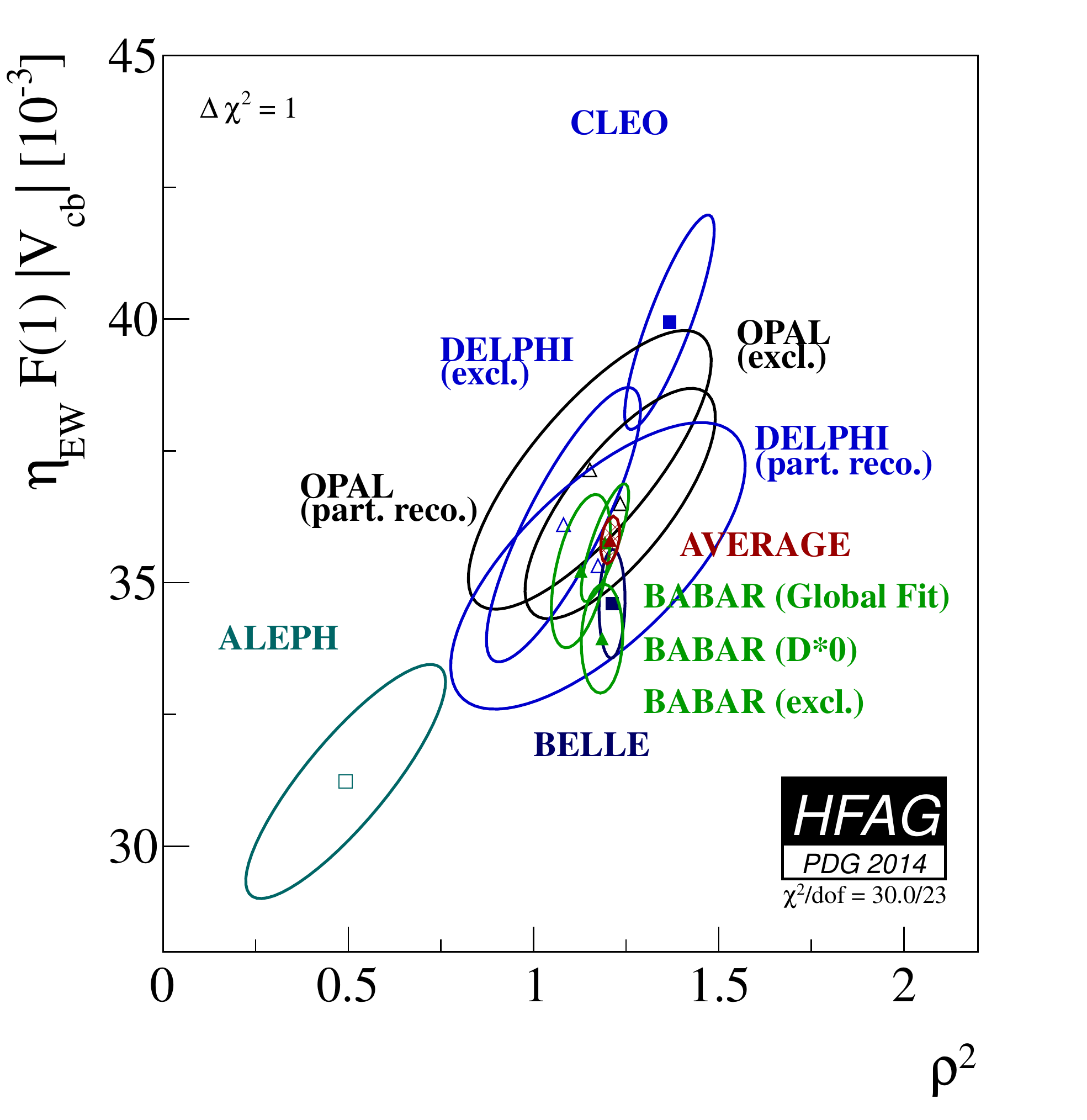}
    }
    \put( -1.5, 0.0){\includegraphics[width=9.0cm]{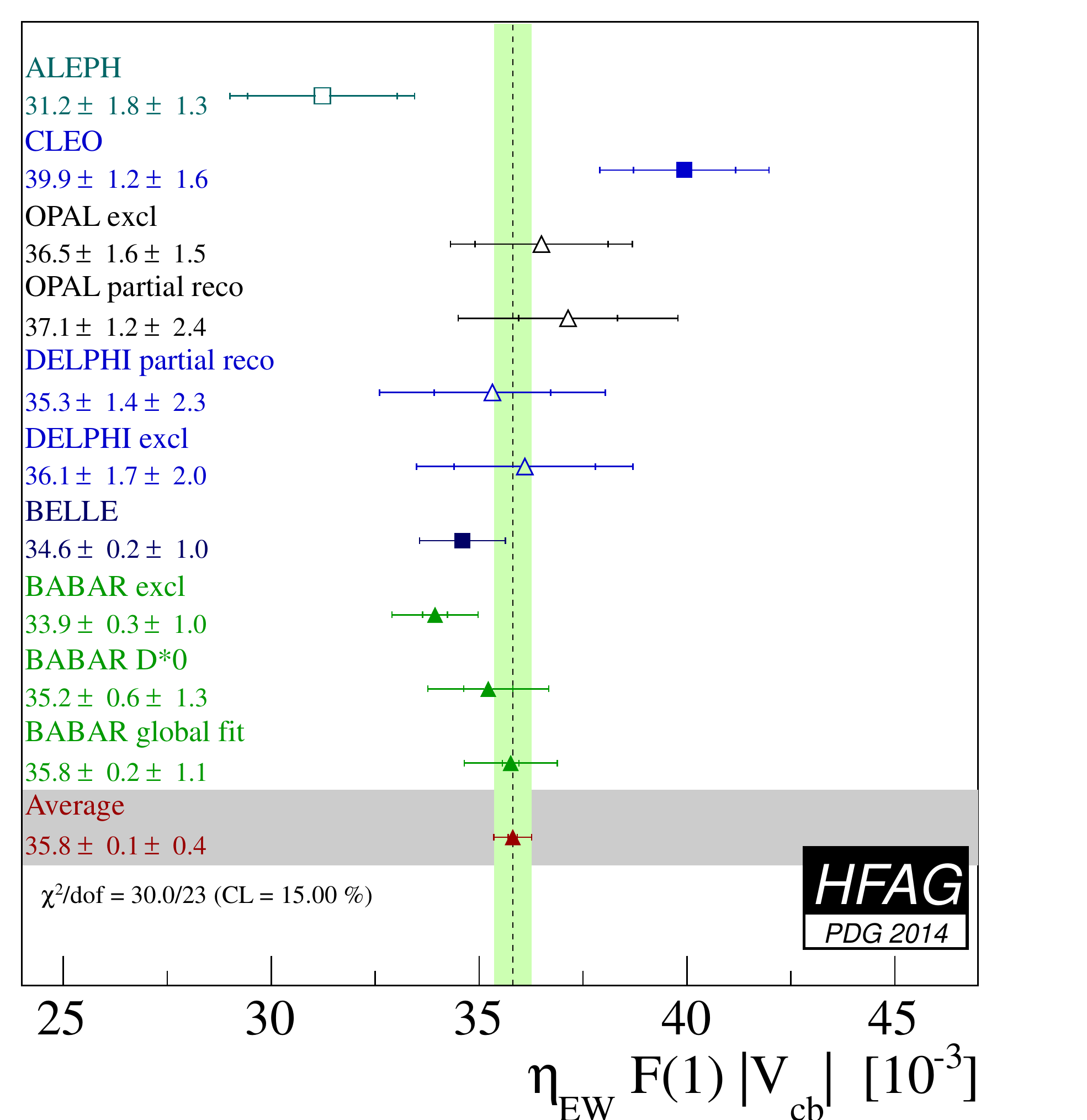}
    }
    \put(  5.8,  8.4){{\large\bf a)}}  
    \put( 14.7,  8.4){{\large\bf b)}}
  \end{picture}
  \caption{(a) Illustration of the $\eta_\mathrm{EW}{\cal F}(1)\vcb$
      average. (b) Illustration of the $\eta_\mathrm{EW}{\cal
        F}(1)\vcb$ vs.\ $\rho^2$ average. The error ellipses
      correspond  to $\Delta\chi^2 = 1$ (CL=39\%).} \label{fig:vcbf1}
  \end{center}
\end{figure}

In the next step, we perform a four-dimensional fit of the parameters
$\eta_\mathrm{EW}{\cal F}(1)\vcb$, $\rho^2$, $R_1(1)$ and $R_2(1)$
using the rescaled measurements and taking into account correlated
statistical and systematic uncertainties. Only two measurements
constrain all four parameters~\cite{Dungel:2010uk,Aubert:2006mb}, the remaining
measurements determine only the normalization $\eta_\mathrm{EW}{\cal
  F}(1)\vcb$ and the slope $\rho^2$. The result of the fit is
\begin{eqnarray}
  \eta_\mathrm{EW}{\cal F}(1)\vcb & = & (35.81\pm 0.45)\times
  10^{-3}~, \label{eq:vcbf1} \\
  \rho^2 & = & 1.207\pm 0.026~,\\
  R_1(1) & = & 1.406\pm 0.033~, \label{eq:r1} \\
  R_2(1) & = & 0.853\pm 0.020~, \label{eq:r2}
\end{eqnarray}
and the correlation coefficients are
\begin{eqnarray}
  \rho_{\eta_\mathrm{EW}{\cal F}(1)\vcb,\rho^2} & = & 0.323~,\\
  \rho_{\eta_\mathrm{EW}{\cal F}(1)\vcb,R_1(1)} & = & -0.108~,\\
  \rho_{\eta_\mathrm{EW}{\cal F}(1)\vcb,R_2(1)} & = & -0.063~,\\
  \rho_{\rho^2,R_1(1)} & = & 0.568~,\\
  \rho_{\rho^2,R_2(1)} & = & -0.809~,\\
  \rho_{R_1(1),R_2(1)} & = & -0.758~.
\end{eqnarray}
The uncertainties and correlations quoted here include both
statistical and systematic contributions. The $\chi^2$ of the fit is
30.0 for 23 degrees of freedom, which corresponds to a confidence
level of 15.0\%. An illustration of this fit result is given in
Fig.~\ref{fig:vcbf1}.

Using the lastest update from the FNAL/MILC
group~\cite{Bailey:2014tva}, the form factor normalization
$\eta_\mathrm{EW}{\cal F}(1)$ is
\begin{equation}
  \eta_\mathrm{EW}{\cal F}(1) = 0.920\pm 0.014~,
\end{equation}
which results in the following determination of $\vcb$ from
Eq.~\ref{eq:vcbf1},
\begin{equation}
  \vcb = (38.94\pm 0.49_{\rm exp}\pm 0.58_{\rm th})\times
  10^{-3}~, \label{eq:vcbdstar}
\end{equation}
where the first uncertainty is experimental and the second error is
theoretical (lattice QCD calculation and electro-weak correction).

From each rescaled measurement in Table~\ref{tab:vcbf1}, we
calculate the $\bar B\to D^*\ell^-\bar\nu_\ell$ form factor
$\eta_\mathrm{EW}{\cal F}(w)$ and, by numerical integration, the
branching ratio of the decay $\bar B^0\to
D^{*+}\ell^-\bar\nu_\ell$. For measurements which do not determine the
parameters $R_1(1)$ and $R_2(1)$ we assume the average
values~Eqs.~\ref{eq:r1} and \ref{eq:r2}. The results are quoted in
Table~\ref{tab:dstarlnu}. The branching ratio found for the average
values of $\eta_\mathrm{EW}{\cal F}(1)\vcb$, $\rho^2$, $R_1(1)$ and
$R_2(1)$ is
\begin{equation}
  \cbf(\BzbDstarlnu)=(4.93\pm 0.11)\%~.
\end{equation}

\begin{figure}[!ht]
  \begin{center}
  \unitlength1.0cm 
  \begin{picture}(14.,9.0)  
    \put( -1.5, 0.0){\includegraphics[width=9.0cm]{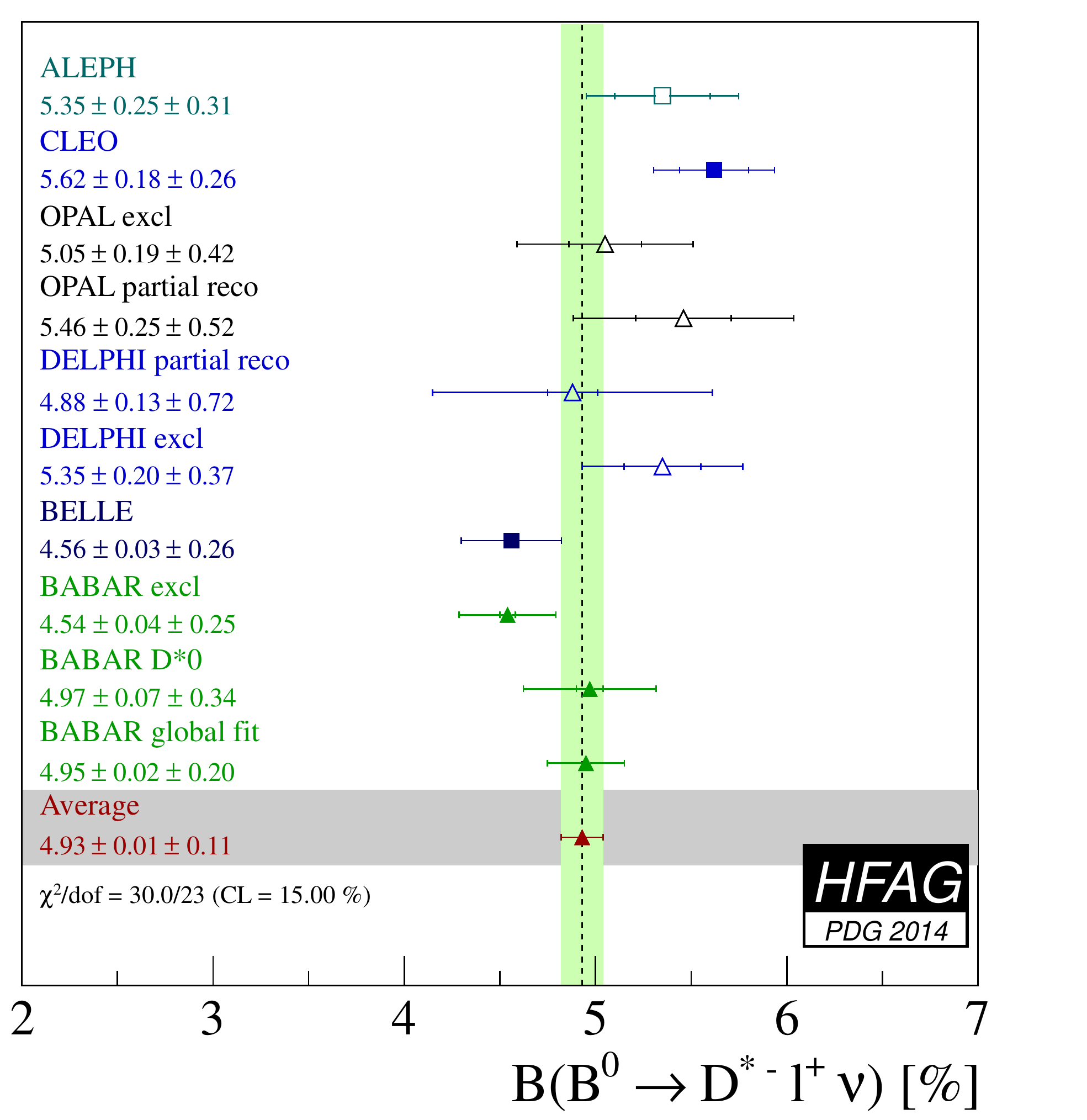}
    }
    \put(  7.5, 0.0){\includegraphics[width=9.0cm]{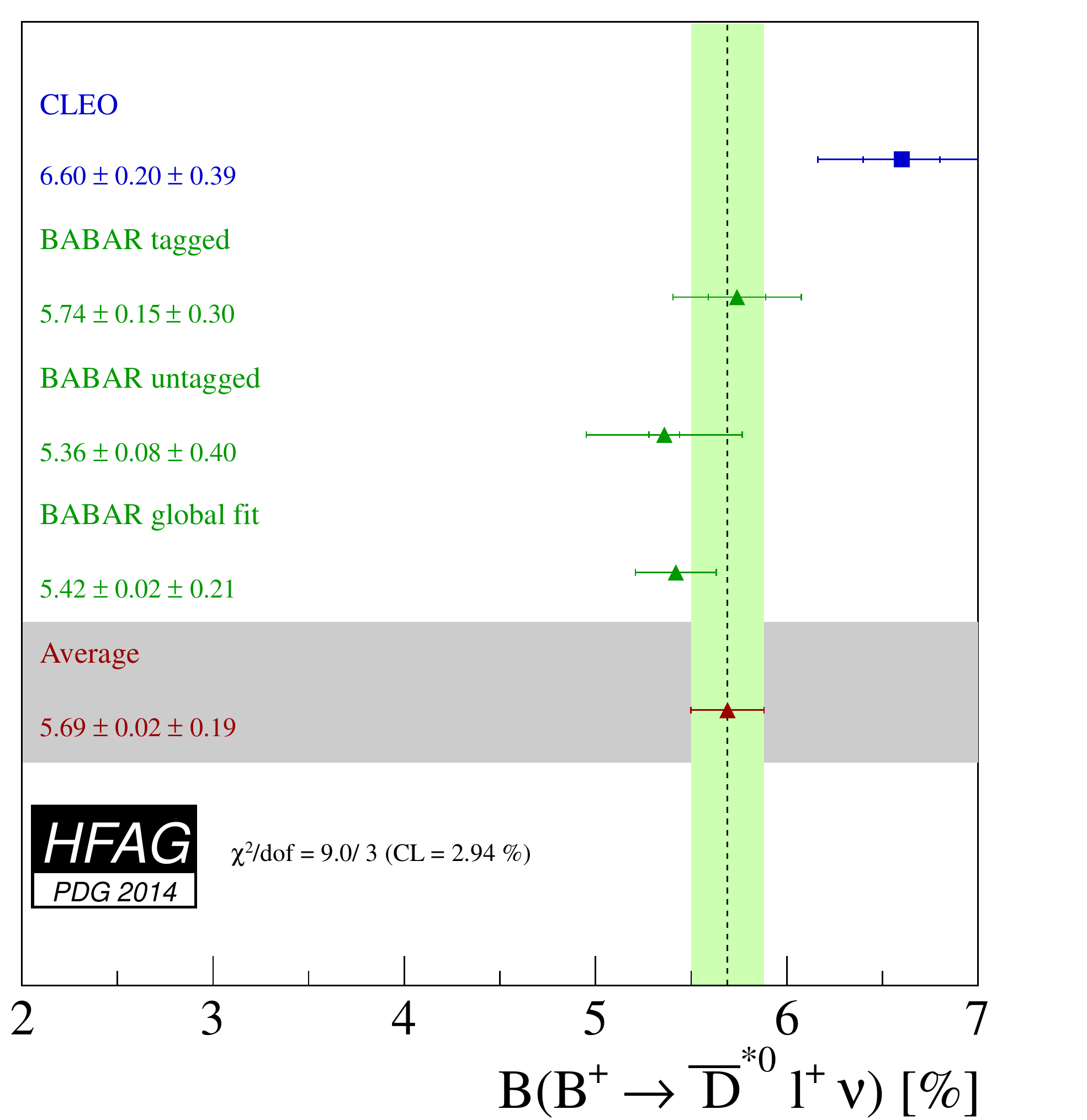}
    }
    \put(  5.8, 8.5){{\large\bf a)}}
    \put( 14.8, 8.5){{\large\bf b)}}
  \end{picture}
  \caption{(a) Average branching fractions of exclusive semileptonic
    $B$ decays $\bar B\to D^*\ell^-\bar\nu_\ell$: (a) $\BzbDstarlnu$
    (Table~\ref{tab:dstarlnu}) and (b) $B^-\to
    D^{*0}\ell^-\bar\nu_\ell$ (Table~\ref{tab:dstar0lnu}).} \label{fig:brdsl}
  \end{center}
\end{figure}

To test isospin symmetry, we have performed a simple 1-dimensional
average of measurements sensitive to the decay $B^-\to
D^{*0}\ell^-\bar\nu_\ell$ only, which is shown in
Table~\ref{tab:dstar0lnu}. Fig.~\ref{fig:brdsl} illustrates our two
averages of $\bar B\to D^*\ell^-\bar\nu_\ell$.
\begin{table}[!htb]
\caption{$\BzbDstarlnu$ branching fractions calculated from the
  rescaled CLN pameters in Table~\ref{tab:vcbf1}. For
  Ref.~\cite{Aubert:2009_3} the published value of ${\cal B}(B^-\to
  D^{*0}\ell^-\bar\nu_\ell)$ has been rescaled by the factor
  $\tau(B^0)/\tau(B^+)$ for comparison to the other measurements.}
\begin{center} 
\resizebox{0.99\textwidth}{!}{
\begin{tabular}{|l|c|c|}\hline
  Experiment & $\cbf(\BzbDstarlnu)$ [\%] (calculated) &
  $\cbf(\BzbDstarlnu)$ [\%] (published)\\
  \hline\hline
  ALEPH~\cite{Buskulic:1996yq}
  & $5.35\pm 0.25_{\rm stat} \pm 0.31_{\rm syst}$
  & $5.53\pm 0.26_{\rm stat} \pm 0.52_{\rm syst}$\\
  CLEO~\cite{Adam:2002uw}
  & $5.62\pm 0.18_{\rm stat} \pm 0.26_{\rm syst}$
  & $6.09\pm 0.19_{\rm stat} \pm 0.40_{\rm syst}$\\
  OPAL excl~\cite{Abbiendi:2000hk}
  & $5.05\pm 0.19_{\rm stat} \pm 0.42_{\rm syst}$
  & $5.11\pm 0.19_{\rm stat} \pm 0.49_{\rm syst}$\\
  OPAL partial reco~\cite{Abbiendi:2000hk}
  & $5.46\pm 0.25_{\rm stat} \pm 0.52_{\rm syst}$
  & $5.92\pm 0.27_{\rm stat} \pm 0.68_{\rm syst}$\\
  DELPHI partial reco~\cite{Abreu:2001ic}
  & $4.88\pm 0.13_{\rm stat} \pm 0.72_{\rm syst}$
  & $4.70\pm 0.13_{\rm stat} \ {}^{+0.36}_{-0.31}\ {}_{\rm syst}$\\
  DELPHI excl~\cite{Abdallah:2004rz}
  & $5.35\pm 0.20_{\rm stat} \pm 0.37_{\rm syst}$
  & $5.90\pm 0.22_{\rm stat} \pm 0.50_{\rm syst}$\\
  \belle~\cite{Dungel:2010uk}
  & $4.56\pm 0.03_{\rm stat} \pm 0.26_{\rm syst}$
  & $4.58\pm 0.03_{\rm stat} \pm 0.26_{\rm syst}$\\
  \babar\ excl~\cite{Aubert:2006mb}
  & $4.54\pm 0.04_{\rm stat}\pm 0.25_{\rm syst}$
  & $4.69\pm 0.04_{\rm stat} \pm 0.34_{\rm syst}$\\
  \babar\ $D^{*0}$~\cite{Aubert:2009_3}
  & $4.97\pm 0.07_{\rm stat}\pm 0.34_{\rm syst}$
  & $5.15\pm 0.07_{\rm stat} \pm 0.38_{\rm syst}$\\
  \babar\ global fit~\cite{Aubert:2009_1}
  & $4.95\pm 0.02_{\rm stat}\pm 0.20_{\rm syst}$
  & $5.00\pm 0.02_{\rm stat} \pm 0.19_{\rm syst}$\\
  \hline 
  {\bf Average} & \mathversion{bold}$4.93\pm 0.01_{\rm stat}\pm
  0.11_{\rm syst}$ & \mathversion{bold}$\chi^2/\dof = 30.0/23$ (CL=$15.0\%$)\\
  \hline 
\end{tabular}
}
\end{center}
\label{tab:dstarlnu}
\end{table}

\begin{table}[!htb]
\caption{Average of the $B^-\to D^{*0}\ell^-\bar\nu_\ell$ branching
  fraction measurements. This fit uses only measurements of $B^-\to
  D^{*0}\ell^-\bar\nu_\ell$.}
\begin{center}
\begin{tabular}{|l|c|c|}
  \hline
  Experiment & $\cbf(B^-\to D^{*0}\ell^-\bar\nu_\ell)$ [\%] (rescaled) &
  $\cbf(B^-\to D^{*0}\ell^-\bar\nu_\ell)$ [\%] (published)\\
  \hline \hline
  CLEO~\cite{Adam:2002uw}
  & $6.60\pm 0.20_{\rm stat}\pm 0.39_{\rm syst}$
  & $6.50\pm 0.20_{\rm stat}\pm 0.43_{\rm syst}$\\
  \babar tagged~\cite{Aubert:vcbExcl}
  & $5.74\pm 0.15_{\rm stat}\pm 0.30_{\rm syst}$
  & $5.83\pm 0.15_{\rm stat}\pm0.30_{\rm syst}$\\
  \babar~\cite{Aubert:2009_3}
  & $5.36\pm0.08_{\rm stat}\pm 0.40_{\rm syst}$
  & $5.56\pm0.08_{\rm stat} \pm0.41_{\rm syst}$\\
  \babar~\cite{Aubert:2009_1}
  & $5.42\pm 0.02_{\rm stat}\pm 0.21_{\rm syst}$
  & $5.40\pm 0.02_{\rm stat}\pm 0.21_{\rm syst}$\\
  \hline
  {\bf Average} & \mathversion{bold}$5.69\pm 0.02_{\rm stat}\pm
  0.19_{\rm syst}$ & \mathversion{bold}$\chi^2/\dof = 9.0/3$ (CL=$2.94\%$)\\
  \hline 
\end{tabular}
\end{center}
\label{tab:dstar0lnu}
\end{table}

\mysubsubsection{$\bar B\to D\ell^-\bar\nu_\ell$}
\label{slbdecays_dlnu}

The relevant form factor for the decay $\bar B\to D\ell^-\bar\nu_\ell$
is $\eta_\mathrm{EW}{\cal G}(w)$, which in CLN~\cite{CLN} is described
by only two parameters: the normalization $\eta_\mathrm{EW}{\cal
  G}(1)\vcb$ and the slope $\rho^2$.

Experiments measure the differential decay width as a
function of $w$ and determine these two form factor parameters. We use
the analyses shown in Table~\ref{tab:vcbg1} and correct them to match
the latest values of the input
parameters~\cite{HFAG_sl:inputparams}. These measurements are
sensitive to both isospin states ($\BzbDplnu$ and $B^-\to
D^0\ell^-\bar\nu_\ell$). So, isospin symmetry is assumed in the
analysis.
\begin{table}[!htb]
\caption{Measurements of the Caprini, Lellouch and Neubert
  (CLN)~\cite{CLN} form factor parameters in $\bar B\to
  D\ell^-\bar\nu_\ell$ before and after rescaling. The average is the
  result of a 2-dimensional fit to the rescaled measurements of
  $\eta_\mathrm{EW}{\cal G}(1)\vcb$ and $\rho^2$ -- see the text for
  more details. The $\chi^2$~value of the combination is 0.5 for 8
  degrees of freedom (CL=$100.0\%$).}
\begin{center}
\begin{tabular}{|l|c|c|}
  \hline
  Experiment
  & ${\cal G}(1)\vcb$ [10$^{-3}$] (rescaled)
  & $\rho^2$ (rescaled)\\
  & ${\cal G}(1)\vcb$ [10$^{-3}$] (published)
  & $\rho^2$ (published)\\
  \hline \hline
  ALEPH~\cite{Buskulic:1996yq}
  & $38.66\pm 11.80_{\rm stat}\pm 5.19_{\rm syst}$
  & $0.942\pm 0.980_{\rm stat}\pm 0.272_{\rm syst}$\\
  & $31.1\pm 9.9_{\rm stat}\pm 8.6_{\rm syst}$
  & $0.70\pm 0.98_{\rm stat}\pm 0.50_{\rm syst}$\\
  \hline
  CLEO~\cite{Bartelt:1998dq}
  & $44.88\pm 5.96_{\rm stat}\pm 3.25_{\rm syst}$
  & $1.270\pm 0.220_{\rm stat}\pm 0.119_{\rm syst}$\\
  & $44.8\pm 6.1_{\rm stat}\pm 3.7_{\rm syst}$
  & $1.30\pm 0.27_{\rm stat}\pm 0.14_{\rm syst}$\\
  \hline
  \belle~\cite{Abe:2001yf}
  & $40.96\pm 4.39_{\rm stat}\pm 5.03_{\rm syst}$
  & $1.120\pm 0.190_{\rm stat}\pm 0.111_{\rm syst}$\\
  & $41.1\pm 4.4_{\rm stat}\pm 5.1_{\rm syst}$
  & $1.12\pm 0.22_{\rm stat}\pm 0.14_{\rm syst}$\\
  \hline
  \babar global fit~\cite{Aubert:2009_1}
  & $43.25\pm 0.80_{\rm stat}\pm 2.07_{\rm syst}$
  & $1.201\pm 0.040_{\rm stat}\pm 0.057_{\rm syst}$\\
  & $43.1\pm 0.8_{\rm stat}\pm 2.3_{\rm syst}$
  & $1.20\pm 0.04_{\rm stat}\pm 0.07_{\rm syst}$\\
  \hline
  \babar tagged~\cite{Aubert:2009_2}
  & $42.54\pm 1.88_{\rm stat}\pm 1.02_{\rm syst}$
  & $1.177\pm 0.088_{\rm stat}\pm 0.057_{\rm syst}$\\
  & $42.3\pm 1.9_{\rm stat}\pm 1.0_{\rm syst}$
  & $1.20\pm 0.09_{\rm stat}\pm 0.04_{\rm syst}$\\
  \hline 
  {\bf Average }
  & \mathversion{bold}$42.65\pm 0.72_{\rm stat}\pm 1.35_{\rm syst}$
  & \mathversion{bold}$1.185\pm 0.035_{\rm stat}\pm 0.041_{\rm syst}$\\
  \hline 
\end{tabular}
\end{center}
\label{tab:vcbg1}
\end{table}

\begin{figure}[!ht]
  \begin{center}
  \unitlength1.0cm 
  \begin{picture}(14.,10.) 
    \put(  7.5, -0.2){\includegraphics[width=9.5cm]{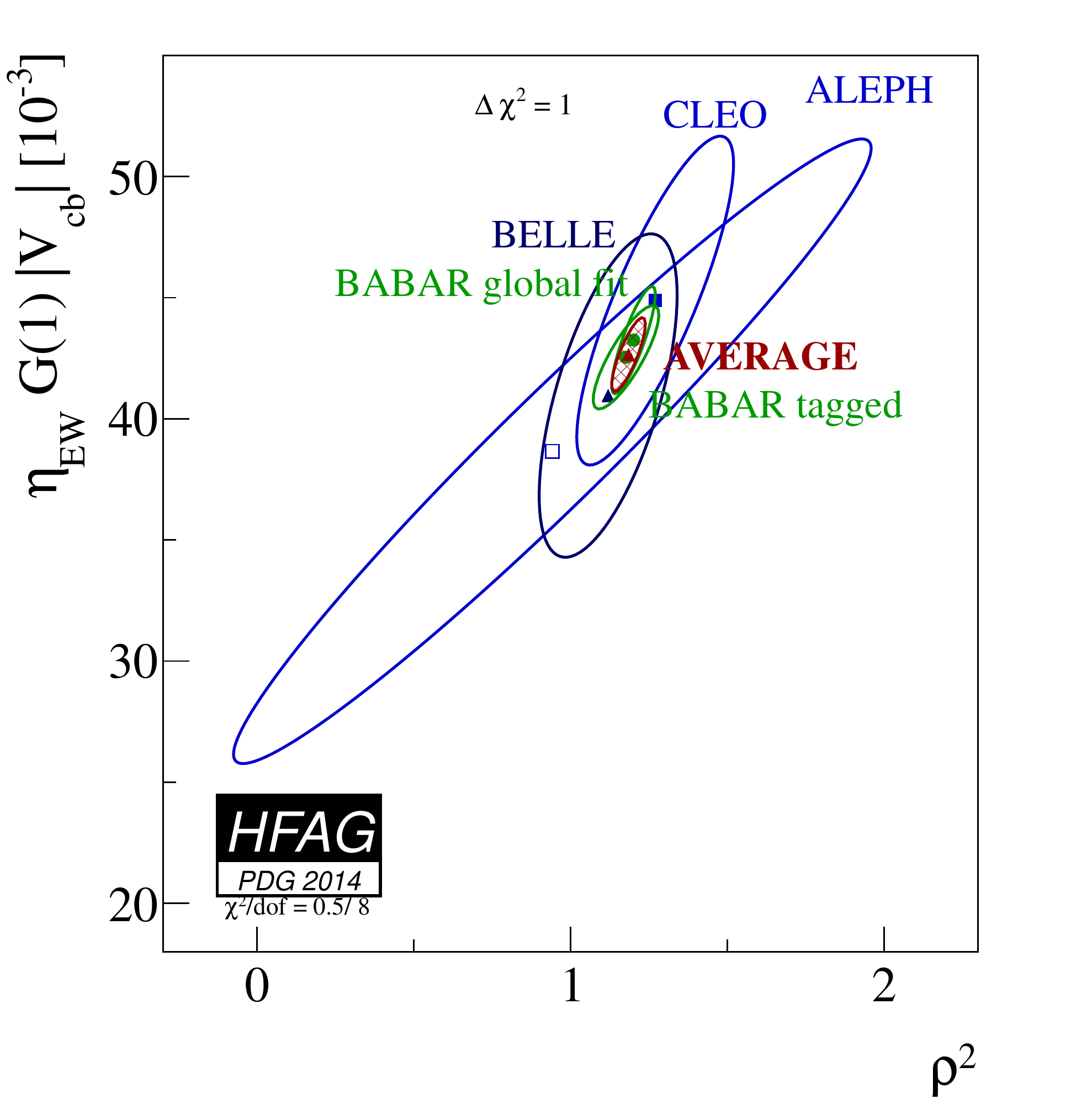}
    }
    \put( -1.5,  0.0){\includegraphics[width=9.0cm]{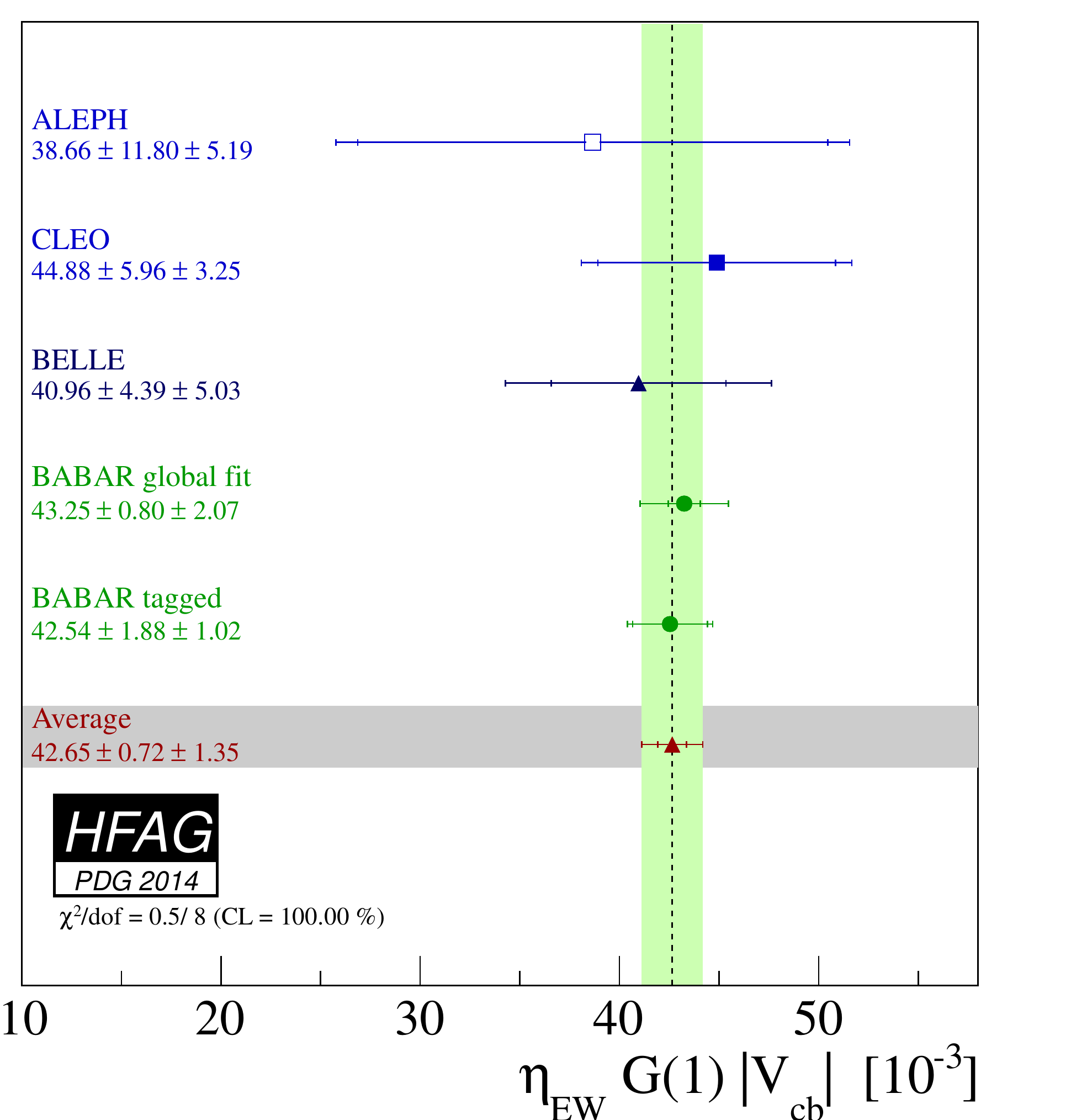}
    }
    \put(  5.8, 8.5){{\large\bf a)}}
    \put(  9.3, 8.5){{\large\bf b)}}
  \end{picture}
  \caption{(a) Illustration of the $\eta_\mathrm{EW}{\cal G}(1)\vcb$
    average. (b) Illustration of the $\eta_\mathrm{EW}{\cal G}(1)\vcb$
    vs.\ $\rho^2$ average. The error ellipses correspond  to
    $\Delta\chi^2 = 1$ (CL=39\%).}
  \label{fig:vcbg1}
  \end{center}
\end{figure}

The form factor parameters are extracted by a two-dimensional fit to
the rescaled measurements of $\eta_\mathrm{EW}{\cal G}(1)\vcb$ and
$\rho^2$ taking into account correlated statistical and systematic
uncertainties. The result of the fit reads
\begin{eqnarray}
  \eta_\mathrm{EW}{\cal G}(1)\vcb & = & (42.65\pm 1.53)\times
  10^{-3}~, \label{eq:vcbg1} \\
  \rho^2 & = & 1.185 \pm 0.054~,
\end{eqnarray}
with a correlation of
\begin{equation}
  \rho_{\eta_\mathrm{EW}{\cal G}(1)\vcb,\rho^2} = 0.824~.
\end{equation}
The uncertainties and the correlation coefficient include both
statistical and systematic contributions. The $\chi^2$ of the fit is
0.5 for 8 degrees of freedom, which corresponds to a confidence
level of 100.0\%. An illustration of this fit result is given in
Fig.~\ref{fig:vcbg1}.

The most recent lattice QCD result obtained for the form factor
normalization $\eta_\mathrm{EW}{\cal G}(1)$ is~\cite{Okamoto:2004xg}
\begin{equation}
  \eta_\mathrm{EW}{\cal G}(1) = 1.081\pm 0.024~,
\end{equation}
which can be used to turn Eq.~\ref{eq:vcbg1} into a determination of
$\vcb$,
\begin{equation}
  \vcb = (39.45\pm 1.42_{\rm exp}\pm 0.88_{\rm th})\times 10^{-3}~,
\end{equation}
where the first error is experimental and the second theoretical. This
number is in excellent agreement with $\vcb$ obtained from decays
$\bar B\to D^*\ell^-\bar\nu_\ell$ (Eq.~\ref{eq:vcbdstar}).

\begin{figure}[!ht]
  \begin{center}
    \includegraphics[width=9.5cm]{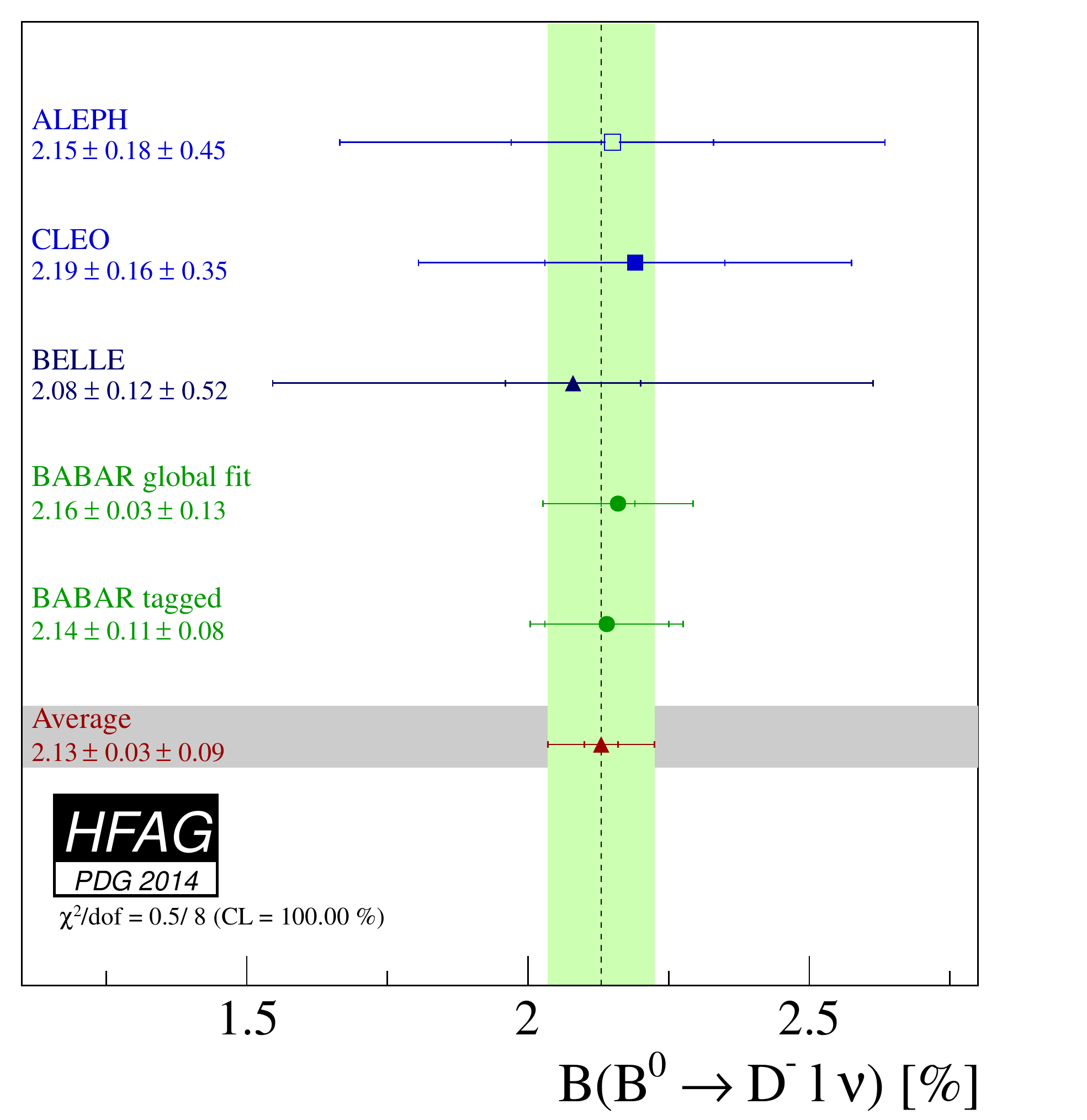}
    \caption{Illustration of Table~\ref{tab:dlnuIso}.} \label{fig:brdlIso}
  \end{center}
\end{figure}

From each rescaled measurement in Table~\ref{tab:vcbg1}, we have
calculated the $\bar B\to D\ell^-\bar\nu_\ell$ form factor ${\cal
  G}(w)$ and, by numerical integration, the branching ratio of the
decay $\BzbDplnu$. The results are quoted in Table~\ref{tab:dlnuIso} and
illustrated in Fig.~\ref{fig:brdlIso}. The branching ratio found for
the average values of $\eta_\mathrm{EW}{\cal G}(1)\vcb$ and $\rho^2$ is
\begin{equation}
  \cbf(\BzbDplnu)=(2.13\pm 0.09)\%~.
\end{equation}
\begin{table}[!htb]
\caption{$\bar B^0\to D^+\ell^-\bar\nu_\ell$ branching fractions
  calculated from the rescaled CLN pameters in Table~\ref{tab:vcbg1}.}
\begin{center}
\resizebox{0.99\textwidth}{!}{
\begin{tabular}{|l|c|c|}
  \hline
  Experiment
  & $\cbf(\bar B^0\to D^+\ell^-\bar\nu_\ell)$ [\%] (calculated)
  & $\cbf(\bar B^0\to D^+\ell^-\bar\nu_\ell)$ [\%] (published)\\
  \hline \hline
  ALEPH~\cite{Buskulic:1996yq}
  & $2.15\pm 0.18_{\rm stat}\pm 0.45_{\rm syst}$
  & $2.35\pm 0.20_{\rm stat}\pm 0.44_{\rm syst}$\\
  CLEO~\cite{Bartelt:1998dq}
  & $2.19\pm 0.16_{\rm stat}\pm 0.35_{\rm syst}$
  & $2.20\pm 0.16_{\rm stat}\pm 0.19_{\rm syst}$\\
  \belle~\cite{Abe:2001yf}
  & $2.08\pm 0.12_{\rm stat}\pm 0.52_{\rm syst}$
  & $2.13\pm 0.12_{\rm stat}\pm 0.39_{\rm syst}$\\
  \babar global fit~\cite{Aubert:2009_1}
  & $2.16\pm 0.03_{\rm stat}\pm 0.13_{\rm syst}$
  & $2.34\pm 0.03_{\rm stat}\pm 0.13_{\rm syst}$\\
  \babar tagged~\cite{Aubert:2009_2}
  & $2.14\pm 0.11_{\rm stat}\pm 0.08_{\rm syst}$
  & $2.23\pm 0.11_{\rm stat}\pm 0.11_{\rm syst}$\\
  \hline 
  {\bf Average}
  & \mathversion{bold}$2.13\pm 0.03_{\rm stat}\pm 0.09_{\rm syst}$
  & \mathversion{bold}$\chi^2/\dof = 0.5/8$ (CL=$100.0\%$)\\
  \hline 
\end{tabular}
}
\end{center}
\label{tab:dlnuIso}
\end{table}

We have also performed simple 1-dimensional averages of measurements
of $\BzbDplnu$ and $B^-\to D^0\ell^-\bar\nu_\ell$. These fits are
shown Tables~\ref{tab:dlnu} and \ref{tab:d0lnu}.
\begin{table}[!htb]
\caption{Average of $\BzbDplnu$ branching fraction
  measurements. This fit uses only measurements of the neutral mode.}
\begin{center}
\begin{tabular}{|l|c|c|}
  \hline
  Experiment
  & $\cbf(\BzbDplnu)$ [\%] (rescaled)
  & $\cbf(\BzbDplnu)$ [\%] (published)\\
  \hline \hline
  ALEPH~\cite{Buskulic:1996yq}
  & $2.28\pm 0.18_{\rm stat}\pm 0.35_{\rm syst}$
  & $2.35\pm 0.20_{\rm stat}\pm 0.44_{\rm syst}$\\
  CLEO~\cite{Bartelt:1998dq}
  & $2.13\pm 0.13_{\rm stat}\pm 0.15_{\rm syst}$
  & $2.20\pm 0.16_{\rm stat}\pm 0.19_{\rm syst}$\\
  \belle~\cite{Abe:2001yf}
  & $2.11\pm 0.12_{\rm stat}\pm 0.39_{\rm syst}$
  & $2.13\pm 0.12_{\rm stat}\pm 0.39_{\rm syst}$\\
  \babar~\cite{Aubert:vcbExcl}
  & $2.22\pm 0.11_{\rm stat}\pm 0.12_{\rm syst}$
  & $2.21\pm 0.11_{\rm stat}\pm 0.12_{\rm syst}$\\
  \hline 
  {\bf Average}
  & \mathversion{bold}$2.19\pm 0.06_{\rm stat}\pm 0.10_{\rm syst}$
  & \mathversion{bold}$\chi^2/\dof = 0.2/3$ (CL=$97.3\%$)\\
  \hline 
\end{tabular}
\end{center}
\label{tab:dlnu}
\end{table}

\begin{table}[!htb]
\caption{Average of $B^-\to D^0\ell^-\bar\nu_\ell$ branching fraction
  measurements. This fit uses only measurements of the charged mode.}
\begin{center}
\begin{tabular}{|l|c|c|}
  \hline
  Experiment
  & $\cbf(B^-\to D^0\ell^-\bar\nu_\ell)$ [\%] (rescaled)
  & $\cbf(B^-\to D^0\ell^-\bar\nu_\ell)$ [\%] (published)\\
  \hline \hline
  CLEO~\cite{Bartelt:1998dq}
  & $2.20\pm 0.13_{\rm stat}\pm 0.17_{\rm syst}$
  & $2.32\pm 0.17_{\rm stat}\pm 0.20_{\rm syst}$\\
  \babar~\cite{Aubert:vcbExcl}
  & $2.29\pm 0.09_{\rm stat}\pm 0.09_{\rm syst}$
  & $2.33\pm 0.09_{\rm stat}\pm 0.09_{\rm syst}$\\
  \hline
  {\bf Average}
  & \mathversion{bold}$2.27\pm 0.07_{\rm stat}\pm 0.08_{\rm syst}$
  & \mathversion{bold}$\chi^2/\dof = 0.1/1$ (CL=$72.1\%$)\\
  \hline
\end{tabular}
\end{center}
\label{tab:d0lnu}
\end{table}


\mysubsubsection{$\bar{B} \to D^{(*)}\pi \ell^-\bar{\nu}_{\ell}$}
\label{slbdecays_dpilnu}

\begin{figure}[!ht]
 \begin{center}
  \unitlength1.0cm 
  \begin{picture}(14.,9.5)  
   \put( -1.5,  0.0){\includegraphics[width=8.55cm]{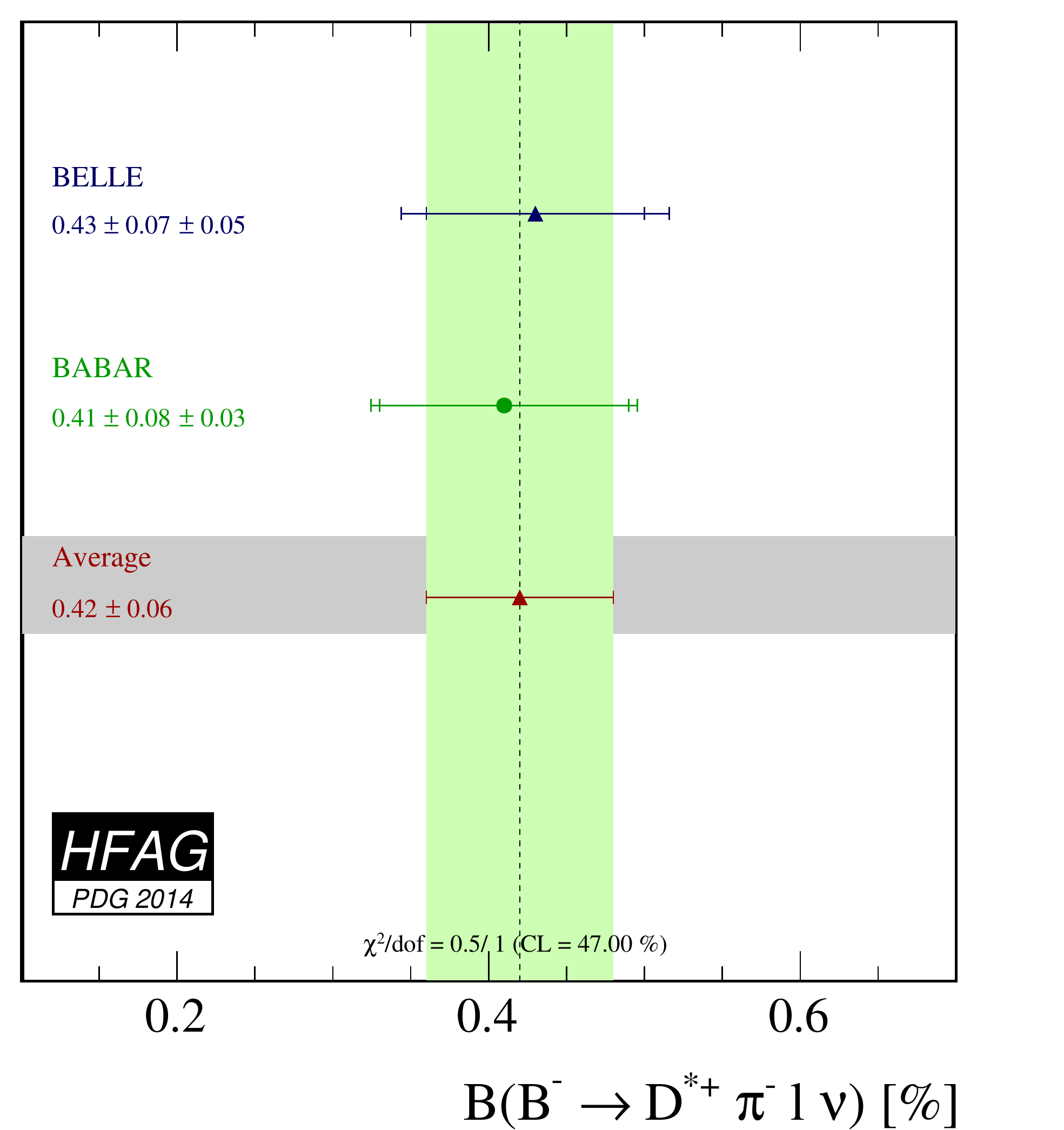}
   }
   \put(  7.5,  0.0){\includegraphics[width=8.8cm]{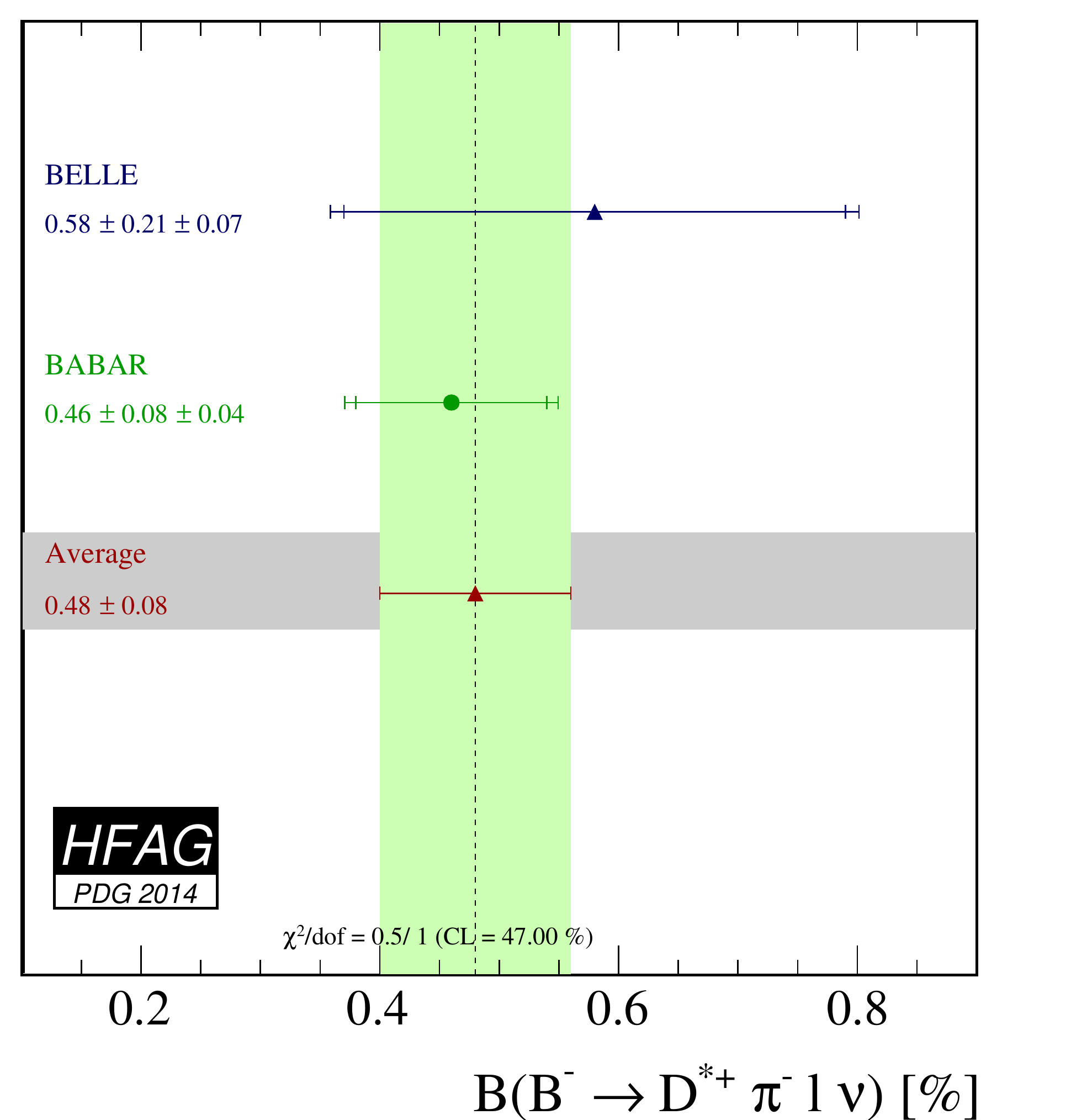}
   }
   \put(  5.5,  8.2){{\large\bf a)}}
   \put( 14.5,  8.2){{\large\bf b)}}
  \end{picture}
  \begin{picture}(14.,9.5)  
   \put( -1.5,  0.0){\includegraphics[width=8.55cm]{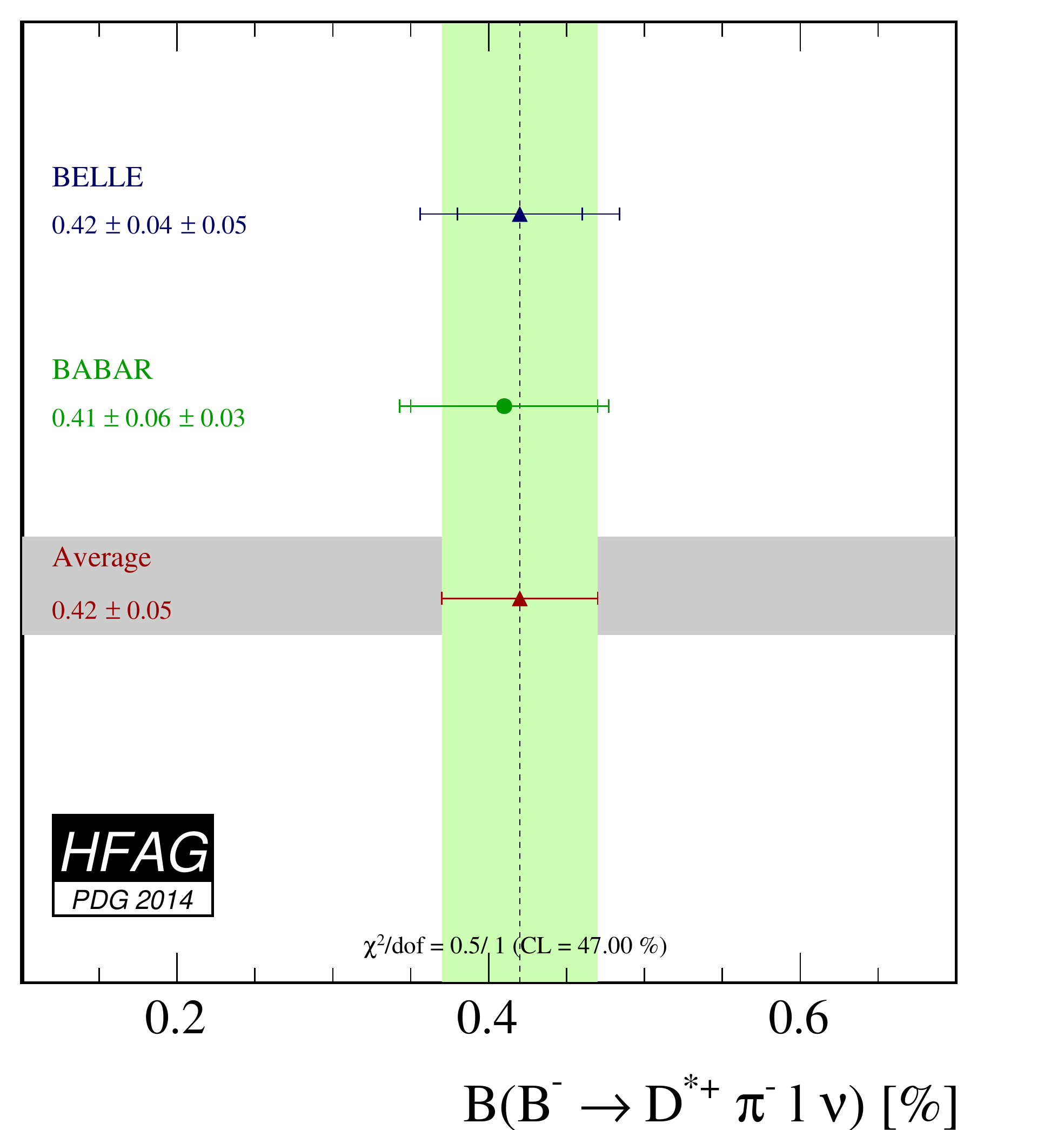}
   }
   \put(  7.5,  0.0){\includegraphics[width=8.8cm]{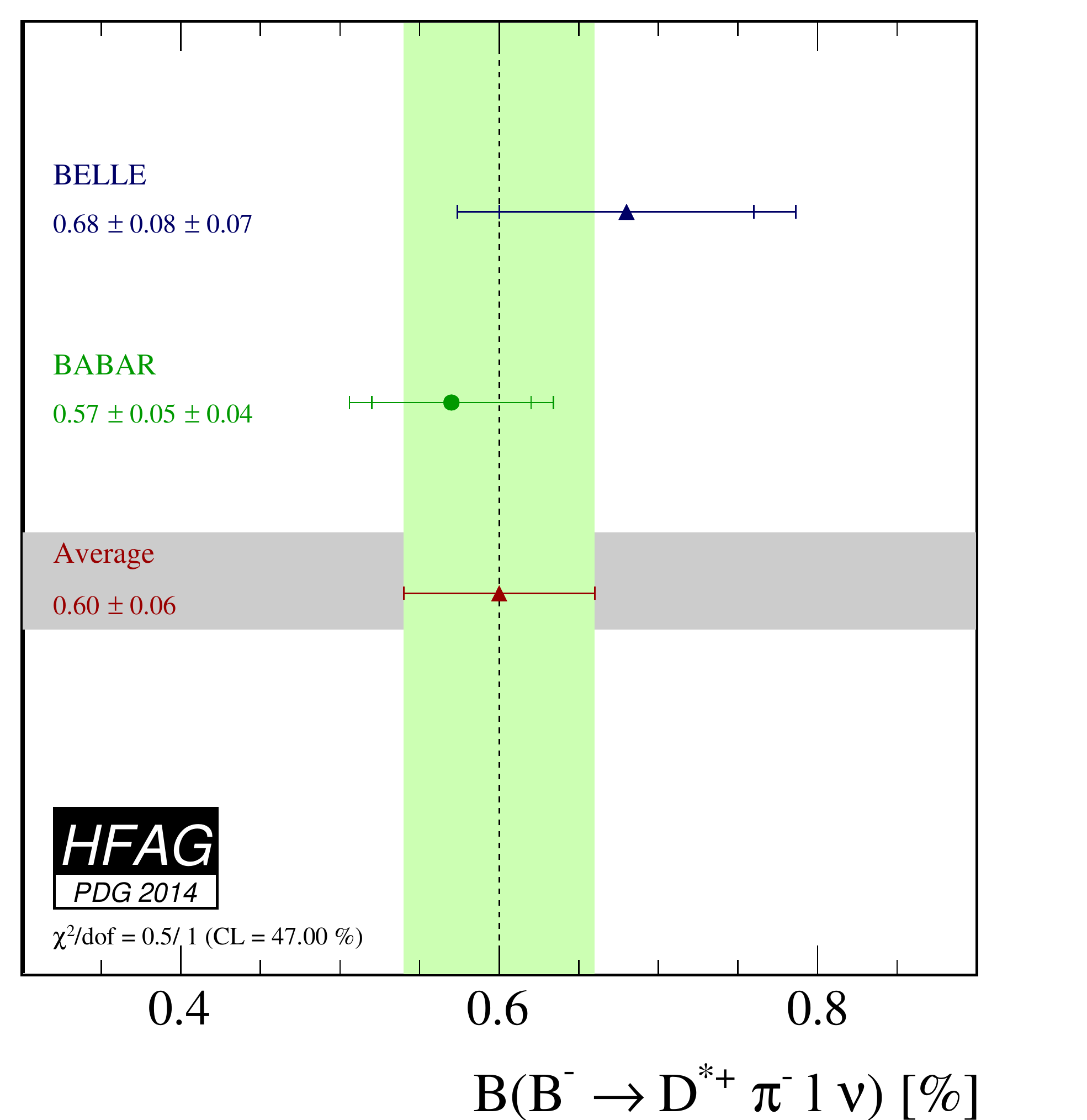}
   }
   \put(  5.5,  8.2){{\large\bf c)}}
   \put( 14.5,  8.2){{\large\bf d)}}
  \end{picture}
  \caption{Average branching fraction  of exclusive semileptonic $B$ decays
(a) $\bar{B}^0 \to D^0 \pi^+ \ell^-\bar{\nu}_{\ell}$, (b) $\bar{B}^0 \to D^{*0} \pi^+
\ell^-\bar{\nu}_{\ell}$, (c) $B^- \to D^+ \pi^-
\ell^-\bar{\nu}_{\ell}$, and (d) $B^- \to D^{*+} \pi^- \ell^-\bar{\nu}_{\ell}$.
The corresponding individual
  results are also shown.}
  \label{fig:brdpil}
 \end{center}
\end{figure}

The average inclusive branching fractions for $\bar{B} \to D^{*}\pi
\ell^-\bar{\nu}_{\ell}$ decays, where no constraint is applied to the
hadronic $D^{(*)}\pi$ system, are determined by the
combination of the results provided in Table~\ref{tab:dpilnu} for 
$\bar{B}^0 \to D^0 \pi^+ \ell^-\bar{\nu}_{\ell}$, $\bar{B}^0 \to D^{*0} \pi^+
\ell^-\bar{\nu}_{\ell}$, 
$B^- \to D^+ \pi^-
\ell^-\bar{\nu}_{\ell}$, and $B^- \to D^{*+} \pi^- \ell^-\bar{\nu}_{\ell}$.
The measurements included in the average 
are scaled to a consistent set of input
parameters and their errors~\cite{HFAG_sl:inputparams}.
For both the \babar\ and Belle results, the $B$ semileptonic signal yields are
 extracted from a fit to the missing mass squared in a sample of fully
 reconstructed \BB\ events.  
Figure~\ref{fig:brdpil} illustrates the measurements and the
resulting average.

\begin{table}[!htb]
\caption{Average of the branching fraction $B \to D^{(*)} \pi^- \ell^-\bar{\nu}_{\ell}$ and individual results.}
\begin{center}
\begin{tabular}{|l|c c|}\hline
Experiment                                 &$\cbf(B^- \to D^+ \pi^- \ell^-\bar{\nu}_{\ell}) [\%]$ (rescaled) & $\cbf(B^- \to D^+ \pi^- \ell^-\bar{\nu}_{\ell}) [\%]$ (published)\\
\hline
\belle  ~\cite{Live:Dss}             &$0.42 \pm0.04_{\rm stat} \pm0.05_{\rm syst}$  & $0.40 \pm0.04_{\rm stat} \pm0.06_{\rm syst}$\\
\babar  ~\cite{Aubert:vcbExcl}       &$0.41 \pm0.06_{\rm stat} \pm0.03_{\rm syst}$ & $0.42 \pm0.06_{\rm stat} \pm0.03_{\rm syst}$  \\
\hline 
{\bf Average}                              &\mathversion{bold}$0.42 \pm0.05$ &\mathversion{bold}$\chi^2/\dof = 0.01$ (CL=$90\%$)  \\
\hline\hline

Experiment                                 &$\cbf(B^- \to D^{*+} \pi^- \ell^-\bar{\nu}_{\ell}) [\%]$ (rescaled) & $\cbf(B^- \to D^{*+} \pi^- \ell^-\bar{\nu}_{\ell}) [\%]$ (published) \\
\hline\hline 
\belle  ~\cite{Live:Dss}           &$0.68 \pm0.08_{\rm stat} \pm0.07_{\rm syst}$   & $0.64 \pm0.08_{\rm stat} \pm0.09_{\rm syst}$  \\
\babar  ~\cite{Aubert:vcbExcl}       &$0.57 \pm0.05_{\rm stat} \pm0.04_{\rm syst}$   & $0.59 \pm0.05_{\rm stat} \pm0.04_{\rm syst}$ \\
\hline 
{\bf Average}                              &\mathversion{bold}$0.60 \pm0.06$   & \mathversion{bold}$\chi^2/\dof = 0.5$ (CL=$52\%$)    \\
\hline \hline

Experiment                               &$\cbf(\bar{B}^0 \to D^0 \pi^+ \ell^-\bar{\nu}_{\ell}) [\%]$ (rescaled) & $\cbf(\bar{B}^0 \to D^0 \pi^+ \ell^-\bar{\nu}_{\ell}) [\%]$ (published)\\
\hline\hline 
\belle  ~\cite{Live:Dss}           &$0.43 \pm0.07_{\rm stat} \pm0.05_{\rm syst}$ & $0.42 \pm0.07_{\rm stat} \pm0.06_{\rm syst}$ \\
\babar  ~\cite{Aubert:vcbExcl}     &$0.41 \pm0.08_{\rm stat} \pm0.03_{\rm syst}$ & $0.43 \pm0.08_{\rm stat} \pm0.03_{\rm syst}$ \\
\hline 
{\bf Average}                              &\mathversion{bold}$0.42 \pm0.06$  &\mathversion{bold}$\chi^2/\dof = 0.04$ (CL=$85\%$)  \\
\hline\hline

Experiment                                 &$\cbf(\bar{B}^0 \to D^{*0} \pi^+\ell^-\bar{\nu}_{\ell}) [\%]$ (rescaled) & $\cbf(\bar{B}^0 \to D^{*0} \pi^+\ell^-\bar{\nu}_{\ell}) [\%]$ (published) \\
\hline\hline 
\belle  ~\cite{Live:Dss}           &$0.58 \pm0.21_{\rm stat} \pm0.07_{\rm syst}$  & $0.56 \pm0.21_{\rm stat} \pm0.08_{\rm syst}$ \\
\babar  ~\cite{Aubert:vcbExcl}       &$0.46 \pm0.08_{\rm stat} \pm0.04_{\rm syst}$ &$0.48 \pm0.08_{\rm stat} \pm0.04_{\rm syst}$ \\ 
\hline 
{\bf Average}                              &\mathversion{bold}$0.48 \pm0.08$ &\mathversion{bold}$\chi^2/\dof = 0.25$ (CL=$62\%$)  \\
\hline\hline

\end{tabular}
\end{center}
\label{tab:dpilnu}
\end{table}

\mysubsubsection{$\bar{B} \to D^{**} \ell^-\bar{\nu}_{\ell}$}
\label{slbdecays_dsslnu}

The $D^{**}$ mesons contain one charm quark and one light quark with relative angular momentum $L=1$. According to Heavy Quark Symmetry (HQS)~\cite{Isgur:1991wq}, they form one doublet of states with angular momentum $j \equiv s_q + L= 3/2$  $\left[D_1(2420), D_2^*(2460)\right]$ and another doublet with $j=1/2$ $\left[D^*_0(2400), D_1'(2430)\right]$, where $s_q$ is the light quark spin. Parity and angular momentum conservation constrain the decays allowed for each state. The $D_1$ and $D_2^*$ states decay through a D-wave to $D^*\pi$ and $D^{(*)}\pi$, respectively, and have small decay widths, while the $D_0^*$ and $D_1'$  states decay through an S-wave to $D\pi$ and $D^*\pi$ and are very broad.
For the narrow states, the average  are determined by the
combination of the results provided in Table~\ref{tab:dss1lnu} and \ref{tab:dss2lnu} for 
$\cbf(B^- \to D_1^0\ell^-\bar{\nu}_{\ell})
\times \cbf(D_1^0 \to D^{*+}\pi^-)$ and $\cbf(B^- \to D_2^0\ell^-\bar{\nu}_{\ell})
\times \cbf(D_2^0 \to D^{*+}\pi^-)$. 
For the broad states, the average are determined by the
combination of the results provided in Table~\ref{tab:dss1plnu} and \ref{tab:dss0lnu} for 
$\cbf(B^- \to D_1'^0\ell^-\bar{\nu}_{\ell})
\times \cbf(D_1'^0 \to D^{*+}\pi^-)$ and $\cbf(B^- \to D_0^{*0}\ell^-\bar{\nu}_{\ell})
\times \cbf(D_0^{*0} \to D^{+}\pi^-)$. 
The measurements included in the average are scaled to a consistent set of input
parameters and their errors~\cite{HFAG_sl:inputparams}.  

\begin{figure}[!ht]
 \begin{center}
  \unitlength1.0cm 
  \begin{picture}(14.,9.0)  
   \put( -1.5,  0.0){\includegraphics[width=8.8cm]{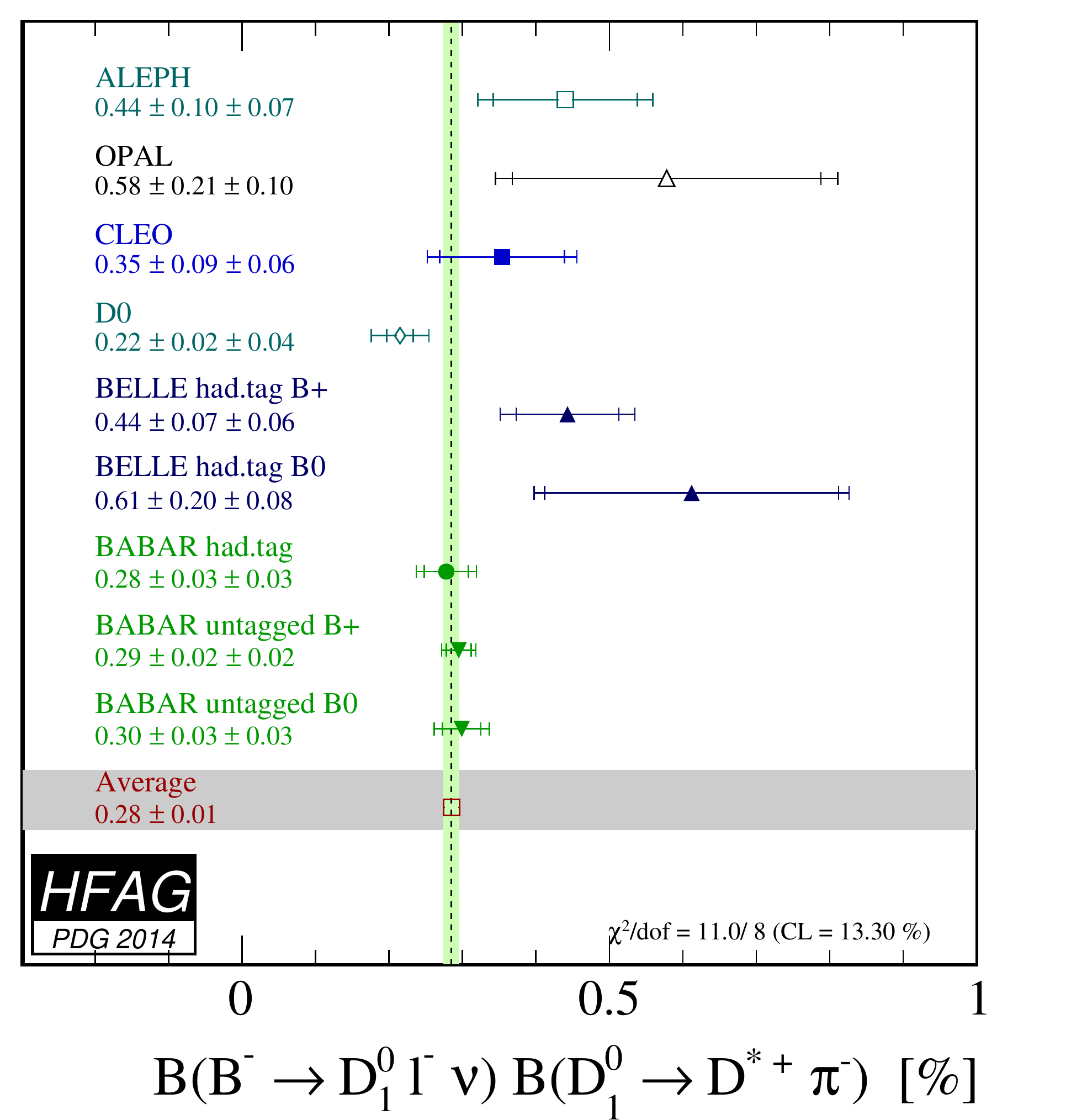}
   }
   \put(  7.5,  0.0){\includegraphics[width=8.55cm]{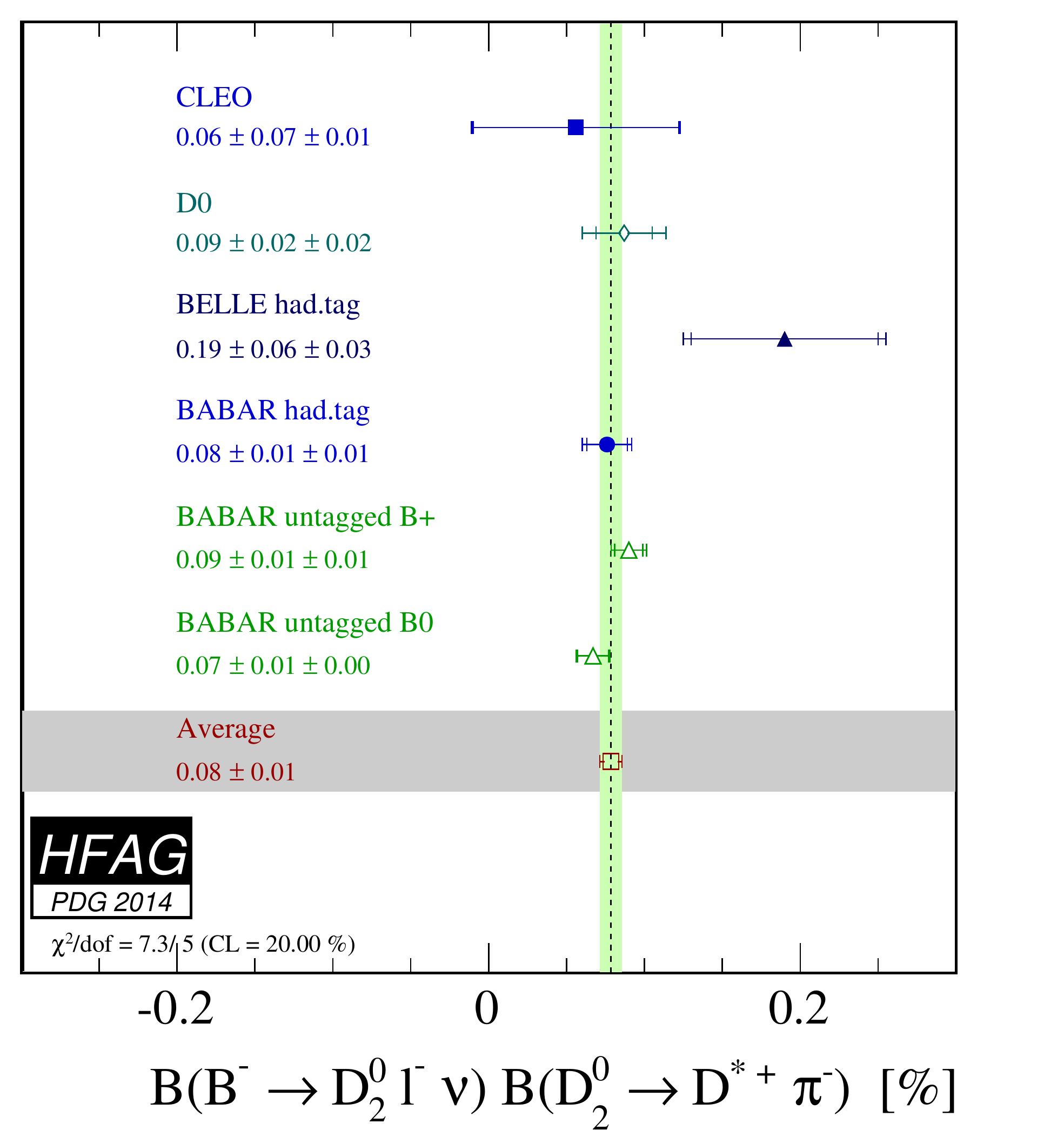}
   }
   \put(  5.6,  8.0){{\large\bf a)}}
   \put( 14.2,  8.0){{\large\bf b)}}
  \end{picture}
  \caption{Average of the product of branching fraction (a) 
  $\cbf(B^- \to D_1^0\ell^-\bar{\nu}_{\ell})
\times \cbf(D_1^0 \to D^{*+}\pi^-)$ and (b) $\cbf(B^- \to D_2^0\ell^-\bar{\nu}_{\ell})
\times \cbf(D_2^0 \to D^{*+}\pi^-)$. The corresponding individual results are also shown.}
  \label{fig:brdssl}
 \end{center}
\end{figure}

\begin{figure}[!ht]
 \begin{center}
  \unitlength1.0cm 
  \begin{picture}(14.,9.0)  
   \put( -1.5,  0.0){\includegraphics[width=8.8cm]{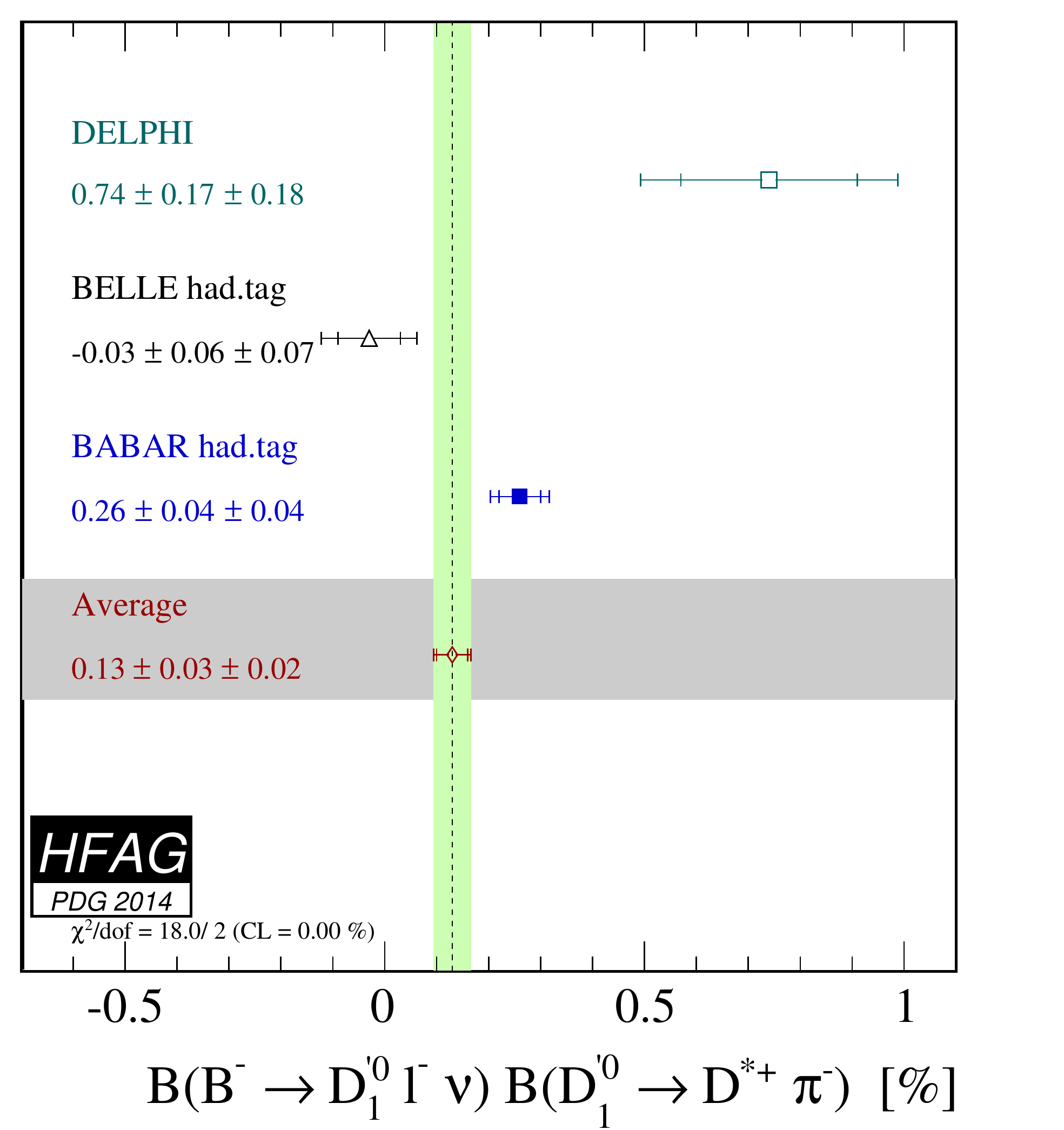}
   }
   \put(  7.5,  0.0){\includegraphics[width=8.8cm]{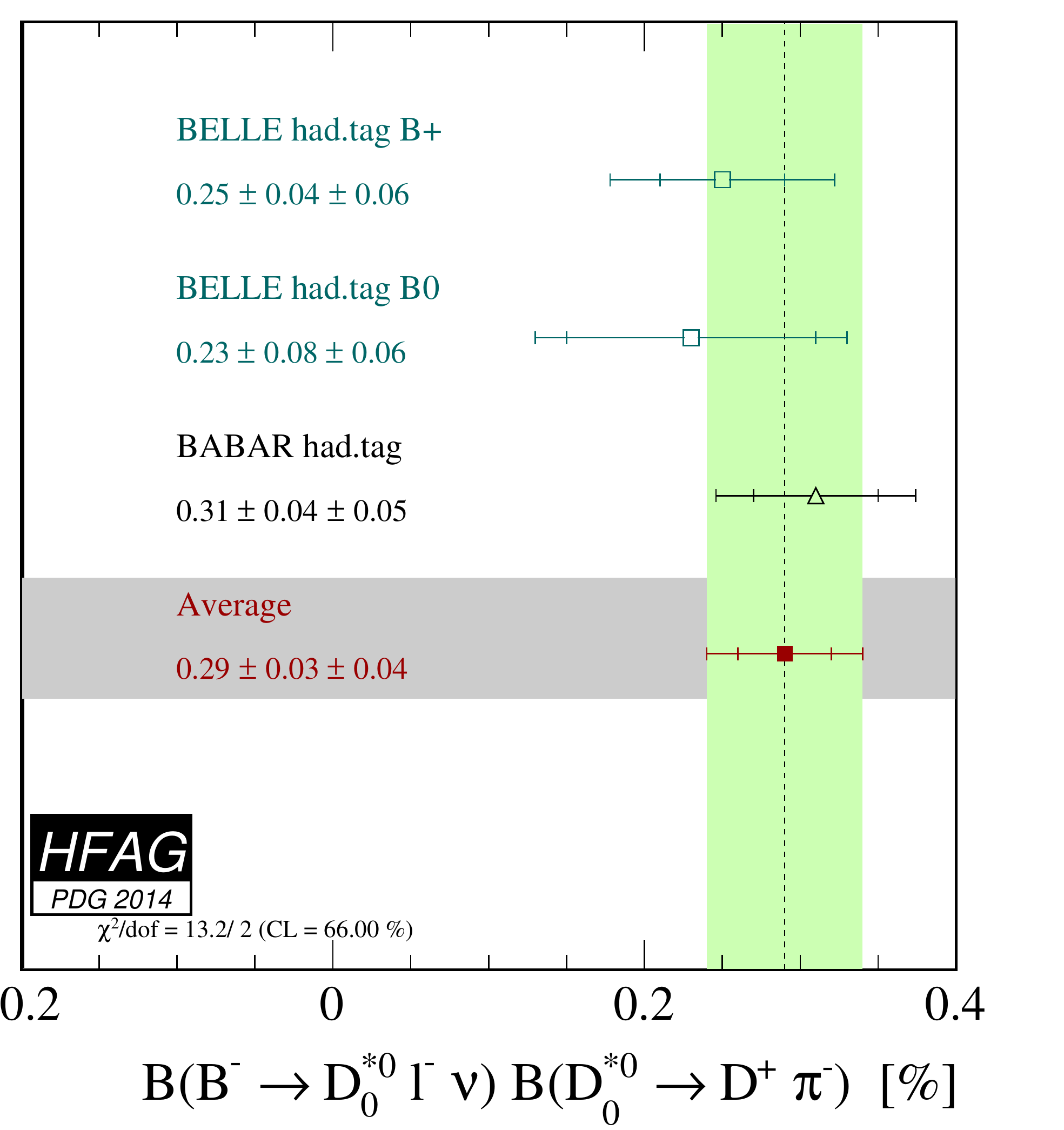}
   }
   \put(  5.8,  8.3){{\large\bf a)}}
   \put( 14.7,  8.3){{\large\bf b)}}
  \end{picture}
  \caption{Average of the product of branching fraction (a) 
  $\cbf(B^- \to D_1'^0\ell^-\bar{\nu}_{\ell})
\times \cbf(D_1'^0 \to D^{*+}\pi^-)$ and (b) $\cbf(B^- \to D_0^{*0}\ell^-\bar{\nu}_{\ell})
\times \cbf(D_0^{*0} \to D^{+}\pi^-)$. The corresponding individual
  results are also shown.}
  \label{fig:brdssl2}
 \end{center}
\end{figure}

For both the B-factory and the LEP and Tevatron results, the $B$ semileptonic 
signal yields are extracted from a fit to the invariant mass distribution of the $D^{(*)+}\pi^-$ system.
 Apart for the CLEO, \belle and \babar results, the other measurements 
 are for the $\bar{B} \to D^{**}(D^*\pi^-)X \ell^- \bar{\nu}_{\ell}$ final state and 
 we assume that no particles are left in the X system. The \babar tagged measurement~\cite{Aubert:2009_4} measures 
 $\bar{B} \to D_2^*(D\pi)X \ell^- \bar{\nu}_{\ell}$ and it has been translated in 
 a result on  $D_2^*\to D^*\pi$ decay mode, assuming 
 ${\cal B}(D_2^*\to D\pi)/{\cal B}(D_2^*\to D^*\pi)=1.54\pm 0.15$~\cite{PDG_2014}. 
Figure~\ref{fig:brdssl} and ~\ref{fig:brdssl2} illustrate the measurements and the
resulting average.

\begin{table}[!htb]
\caption{Average of the branching fraction $\cbf(B^- \to D_1^0\ell^-\bar{\nu}_{\ell})
\times \cbf(D_1^0 \to D^{*+}\pi^-)$ and individual results. The ALEPH, OPAL and D0 measurements are for the 
$D_1(D^*\pi)X$ final state and we assum that no particles are left in the X system.}
\begin{center}
\resizebox{0.99\textwidth}{!}{
\begin{tabular}{|l|c|c|}\hline
Experiment                                 &$\cbf(B^- \to D_1^0(D^{*+}\pi^-)\ell^-\bar{\nu}_{\ell})
 [\%]$  &$\cbf(B^- \to D_1^0(D^{*+}\pi^-)\ell^-\bar{\nu}_{\ell})
 [\%]$  \\
                                                & (rescaled) & (published) \\

\hline\hline 
ALEPH ~\cite{Aleph:Dss}        &$0.440 \pm0.098_{\rm stat} \pm0.068_{\rm syst}$ 
 &$0.47 \pm0.10_{\rm stat} \pm0.07_{\rm syst}$ \\
OPAL  ~\cite{opal:Dss}         &$0.578 \pm0.210_{\rm stat} \pm0.101_{\rm syst}$  
&$0.70 \pm0.21_{\rm stat} \pm0.10_{\rm syst}$ \\
CLEO  ~\cite{cleo:Dss}         &$0.354 \pm0.085_{\rm stat} \pm0.056_{\rm syst}$ 
 &$0.373 \pm0.085_{\rm stat} \pm0.057_{\rm syst}$ \\
D0  ~\cite{D0:Dss}         &$0.215 \pm0.018_{\rm stat} \pm0.035_{\rm syst}$  
&$0.219 \pm0.018_{\rm stat} \pm0.035_{\rm syst}$ \\
\belle Tagged $B^-$ ~\cite{Live:Dss}           &$0.443 \pm0.070_{\rm stat} \pm0.059_{\rm syst}$  
&$0.42 \pm0.07_{\rm stat} \pm0.07_{\rm syst}$ \\
\belle Tagged $B^0$ ~\cite{Live:Dss}           &$0.612 \pm0.200_{\rm stat} \pm0.077_{\rm syst}$  
&$0.42 \pm0.07_{\rm stat} \pm0.07_{\rm syst}$ \\ 
\babar Tagged ~\cite{Aubert:2009_4}           &$0.278 \pm0.030_{\rm stat} \pm0.028_{\rm syst}$
&$0.29 \pm0.03_{\rm stat} \pm0.03_{\rm syst}$ \\
\babar Untagged $B^-$ ~\cite{Aubert:2008zc}           &$0.295 \pm0.017_{\rm stat} \pm0.016_{\rm syst}$
&$0.30 \pm0.02_{\rm stat} \pm0.02_{\rm syst}$ \\
\babar Untagged $B^0$ ~\cite{Aubert:2008zc}           &$0.299 \pm0.026_{\rm stat} \pm0.027_{\rm syst}$
&$0.30 \pm0.02_{\rm stat} \pm0.02_{\rm syst}$ \\
\hline
{\bf Average}                              &\mathversion{bold}$0.285 \pm0.011 \pm 0.014$ 
    &\mathversion{bold}$\chi^2/\dof = 13.0/8$ (CL=$11.1\%$)  \\
\hline 
\end{tabular}
}
\end{center}
\label{tab:dss1lnu}
\end{table}

\begin{table}[!htb]
\caption{Average of the branching fraction $\cbf(B^- \to D_2^0\ell^-\bar{\nu}_{\ell})
\times \cbf(D_2^0 \to D^{*+}\pi^-)$ and individual results. The D0 measurement is for the $D_2^*(D^*\pi)X$
final state and we assume that no particles are left in the X system.
The \babar tagged measurement
has been translated in a result on $D_2^*\to D^*\pi$ decay mode, assuming 
${\cal B}(D_2^*\to D\pi)/{\cal B}(D_2^*\to D^*\pi)=1.54\pm
0.15$~\cite{PDG_2014}.}
\begin{center}
\resizebox{0.99\textwidth}{!}{
\begin{tabular}{|l|c|c|}\hline
Experiment                                 &$\cbf(B^- \to D_2^0(D^{*+}\pi^-)\ell^-\bar{\nu}_{\ell})
 [\%]$  &$\cbf(B^- \to D_2^0(D^{*+}\pi^-)\ell^-\bar{\nu}_{\ell})
 [\%]$  \\
                                                & (rescaled) & (published) \\
\hline\hline 
CLEO  ~\cite{cleo:Dss}         &$0.056 \pm0.066_{\rm stat} \pm0.011_{\rm syst}$ 
 &$0.059 \pm0.066_{\rm stat} \pm0.011_{\rm syst}$ \\
D0  ~\cite{D0:Dss}         &$0.087 \pm0.018_{\rm stat} \pm0.020_{\rm syst}$  
&$0.088 \pm0.018_{\rm stat} \pm0.020_{\rm syst}$ \\
\belle  ~\cite{Live:Dss}           &$0.190 \pm0.060_{\rm stat} \pm0.025_{\rm syst}$  
&$0.18 \pm0.06_{\rm stat} \pm0.03_{\rm syst}$ \\
\babar tagged ~\cite{Aubert:2009_4}           &$0.076 \pm0.013_{\rm stat} \pm0.009_{\rm syst}$
&$0.078 \pm0.013_{\rm stat} \pm0.010_{\rm syst}$ \\
\babar untagged $B^-$ ~\cite{Aubert:2008zc}           &$0.090 \pm0.009_{\rm stat} \pm0.007_{\rm syst}$
&$0.087 \pm0.013_{\rm stat} \pm0.007_{\rm syst}$ \\
\babar untagged $B^0$ ~\cite{Aubert:2008zc}           &$0.067 \pm0.010_{\rm stat} \pm0.004_{\rm syst}$
&$0.087 \pm0.013_{\rm stat} \pm0.007_{\rm syst}$ \\
\hline
{\bf Average}                              &\mathversion{bold}$0.078 \pm0.007 \pm 0.004$ 
    &\mathversion{bold}$\chi^2/\dof = 5.6/5$ (CL=$34.7\%$)  \\
\hline 
\end{tabular}
}
\end{center}
\label{tab:dss2lnu}
\end{table}

\begin{table}[!htb]
\caption{Average of the branching fraction $\cbf(B^- \to D_1^{'0}\ell^-\bar{\nu}_{\ell})
\times \cbf(D_1^{'0} \to D^{*+}\pi^-)$ and individual results. The DELPHI measurement 
is for the final state $D_1'(D^*\pi)X$ and we assume that no particles are left in the X system.}
\begin{center}
\begin{tabular}{|l|c|c|}\hline
Experiment                                 &$\cbf(B^- \to D_1^{'0}(D^{*+}\pi^-)\ell^-\bar{\nu}_{\ell})
 [\%]$  &$\cbf(B^- \to D_1^{'0}(D^{*+}\pi^-)\ell^-\bar{\nu}_{\ell})
 [\%]$  \\
                                                & (rescaled) & (published) \\
\hline\hline 
DELPHI ~\cite{Abdallah:2005cx}        &$0.74 \pm0.17_{\rm stat} \pm0.18_{\rm syst}$ 
 &$0.83 \pm0.17_{\rm stat} \pm0.18_{\rm syst}$ \\
\belle  ~\cite{Live:Dss}           &$-0.03 \pm0.06_{\rm stat} \pm0.07_{\rm syst}$  
&$-0.03 \pm0.06_{\rm stat} \pm0.07_{\rm syst}$ \\
\babar  ~\cite{Aubert:2009_4}           &$0.26 \pm0.04_{\rm stat} \pm0.04_{\rm syst}$
&$0.27 \pm0.04_{\rm stat} \pm0.05_{\rm syst}$ \\
\hline
{\bf Average}                              &\mathversion{bold}$0.13 \pm 0.03 \pm0.02$ 
    &\mathversion{bold}$\chi^2/\dof = 18./2$ (CL=$0.0001\%$)  \\
\hline 
\end{tabular}
\end{center}
\label{tab:dss1plnu}
\end{table}

\begin{table}[!htb]
\caption{Average of the branching fraction $\cbf(B^- \to D_0^{*0}\ell^-\bar{\nu}_{\ell})
\times \cbf(D_0^{*0} \to D^{+}\pi^-)$ and individual
results. }
\begin{center}
\begin{tabular}{|l|c|c|}\hline
Experiment                                 &$\cbf(B^- \to D_0^{*0}(D^{+}\pi^-)\ell^-\bar{\nu}_{\ell})
 [\%]$  &$\cbf(B^- \to D_0^{*0}(D^{+}\pi^-)\ell^-\bar{\nu}_{\ell})
 [\%]$ \\
						& (rescaled) & (published) \\
\hline\hline 
\belle Tagged $B^-$ ~\hfill\cite{Live:Dss}           &$0.25 \pm0.04_{\rm stat} \pm0.06_{\rm syst}$  
&$0.24 \pm0.04_{\rm stat} \pm0.06_{\rm syst}$ \\
\belle Tagged $B^0$ ~\hfill\cite{Live:Dss}           &$0.23 \pm0.08_{\rm stat} \pm0.06_{\rm syst}$  
&$0.24 \pm0.04_{\rm stat} \pm0.06_{\rm syst}$ \\
\babar Tagged ~\hfill\cite{Aubert:2009_4}            &$0.31 \pm0.04_{\rm stat} \pm0.05_{\rm syst}$
&$0.26 \pm0.05_{\rm stat} \pm0.04_{\rm syst}$ \\
\hline
{\bf Average}                              &\mathversion{bold}$0.29 \pm 0.03 \pm0.04$ 
    &\mathversion{bold}$\chi^2/\dof = 0.61/2$ (CL=$73.6\%$)  \\
\hline 
\end{tabular}
\end{center}
\label{tab:dss0lnu}
\end{table}

%
\subsection{Inclusive CKM-favored decays}
\label{slbdecays_b2cincl}

\subsubsection{Global analysis of $\bar B\to X_c\ell^-\bar\nu_\ell$}

The semileptonic width $\Gamma(\bar B\to X_c\ell^-\bar\nu_\ell)$ has
been calculated in the framework of the Operator Product
Expansion~\cite{Shifman:1986mx,*Chay:1990da,*Bigi:1992su,*Bigi:1992su_erratum}.
The result is a double-expansion in $\Lambda_{\rm QCD}/m_b$ and
$\alpha_s$, which depends on a number of non-perturbative
parameters. These parameters give information on the dynamics of the
$b$-quark inside the $B$~hadron and can be measured using other
observables in $\bar B\to X_c\ell^-\bar\nu_\ell$ decays, such as the
moments of the lepton energy and the hadronic mass spectrum.

Two independent sets of theoretical expressions, named after the
definition of the $b$-quark mass used, are available for this kind of
analysis: the kinetic~\cite{Benson:2003kp,Gambino:2004qm,Gambino:2011cq} and 1S
scheme expressions~\cite{Bauer:2004ve}. The non-perturbative
parameters in the kinetic scheme
are: the quark masses $m_b$ and $m_c$, $\mu^2_\pi$ and
$\mu^2_G$ at $O(1/m^2_b)$, and $\rho^3_D$ and $\rho^3_{LS}$ at
$O(1/m^3_b)$. In the 1S scheme, the parameters are: $m_b$, $\lambda_1$
at $O(1/m^2_b)$, and $\rho_1$, $\tau_1$, $\tau_2$ and $\tau_3$ at
$O(1/m^3_b)$. Note that due to the different definitions, the results
for the quark masses cannot be compared directly between the two
schemes.

Our analysis uses all available measurements of moments in $\bar B\to
X_c\ell^-\bar\nu_\ell$, excluding only points with too high
correlation to avoid numerical issues. The list of included
measurements is given in
Table~\ref{tab:gf_input}. The only external input is the average
lifetime~$\tau_B$ of neutral and charged $B$~mesons, taken to be
$(1.579\pm 0.005)$~ps (Sec.~\ref{sec:life_mix}).
\begin{table}[!htb]
\caption{Experimental inputs used in the global analysis of $\bar B\to
  X_c\ell^-\bar\nu_\ell$. $n$ is the order of the moment, $c$ is the
  threshold value of the lepton momentum in GeV. In total, there are
  23 measurements from \babar, 15 measurements from Belle and 12 from
  other experiments.} \label{tab:gf_input}
\begin{center}
\begin{tabular}{|l|l|l|}
  \hline
  Experiment
  & Hadron moments $\langle M^n_X\rangle$
  & Lepton moments $\langle E^n_\ell\rangle$\\
  \hline \hline
  \babar & $n=2$, $c=0.9,1.1,1.3,1.5$ & $n=0$, $c=0.6,1.2,1.5$\\
  & $n=4$, $c=0.8,1.0,1.2,1.4$ & $n=1$, $c=0.6,0.8,1.0,1.2,1.5$\\
  & $n=6$, $c=0.9,1.3$~\cite{Aubert:2009qda} & $n=2$, $c=0.6,1.0,1.5$\\
  & & $n=3$, $c=0.8,1.2$~\cite{Aubert:2009qda,Aubert:2004td}\\
  \hline
  Belle & $n=2$, $c=0.7,1.1,1.3,1.5$ & $n=0$, $c=0.6,1.4$\\
  & $n=4$, $c=0.7,0.9,1.3$~\cite{Schwanda:2006nf} & $n=1$,
  $c=1.0,1.4$\\
  & & $n=2$, $c=0.6,1.4$\\
  & & $n=3$, $c=0.8,1.2$~\cite{Urquijo:2006wd}\\
  \hline
  CDF & $n=2$, $c=0.7$ & \\
  & $n=4$, $c=0.7$~\cite{Acosta:2005qh} & \\
  \hline
  CLEO & $n=2$, $c=1.0,1.5$ & \\
  & $n=4$, $c=1.0,1.5$~\cite{Csorna:2004kp} & \\
  \hline
  DELPHI & $n=2$, $c=0.0$ & $n=1$, $c=0.0$ \\
  & $n=4$, $c=0.0$ & $n=2$, $c=0.0$ \\
  & $n=6$, $c=0.0$~\cite{Abdallah:2005cx} & $n=3$,
  $c=0.0$~\cite{Abdallah:2005cx}\\
  \hline
\end{tabular}
\end{center}
\end{table}

Both in the kinetic and 1S schemes, the moments in $\bar B\to
X_c\ell^-\bar\nu_\ell$ are not sufficient to determine the $b$-quark
mass precisely. In the kinetic scheme analysis we constrain the $c$-quark
mass (defined in the $\overline{\rm MS}$ scheme) to the value of
Ref.~\cite{Chetyrkin:2009fv},
\begin{equation}
  m_c^{\overline{\rm MS}}(3~{\rm GeV})=(0.986\pm 0.013)~{\rm GeV}~.
\end{equation}
In the 1S~scheme analysis, the $b$-quark mass is constrained with
measurements of the photon energy moments in $B\to
X_s\gamma$~\cite{Aubert:2005cua,Aubert:2006gg,Limosani:2009qg,Chen:2001fja}.

\subsubsection{Analysis in the kinetic scheme}
\label{globalfitsKinetic}

The fit relies on the calculations of the spectral moments in $\bar
B\to X_c\ell^-\bar\nu_\ell$~decays described in
Ref.~\cite{Gambino:2011cq} and closely follows the procedure of
Ref.~\cite{Gambino:2013rza}. The analysis determines $\vcb$ and the 6
non-perturbative parameters mentioned above.

\begin{figure}
\begin{center}
  \includegraphics[width=8.2cm]{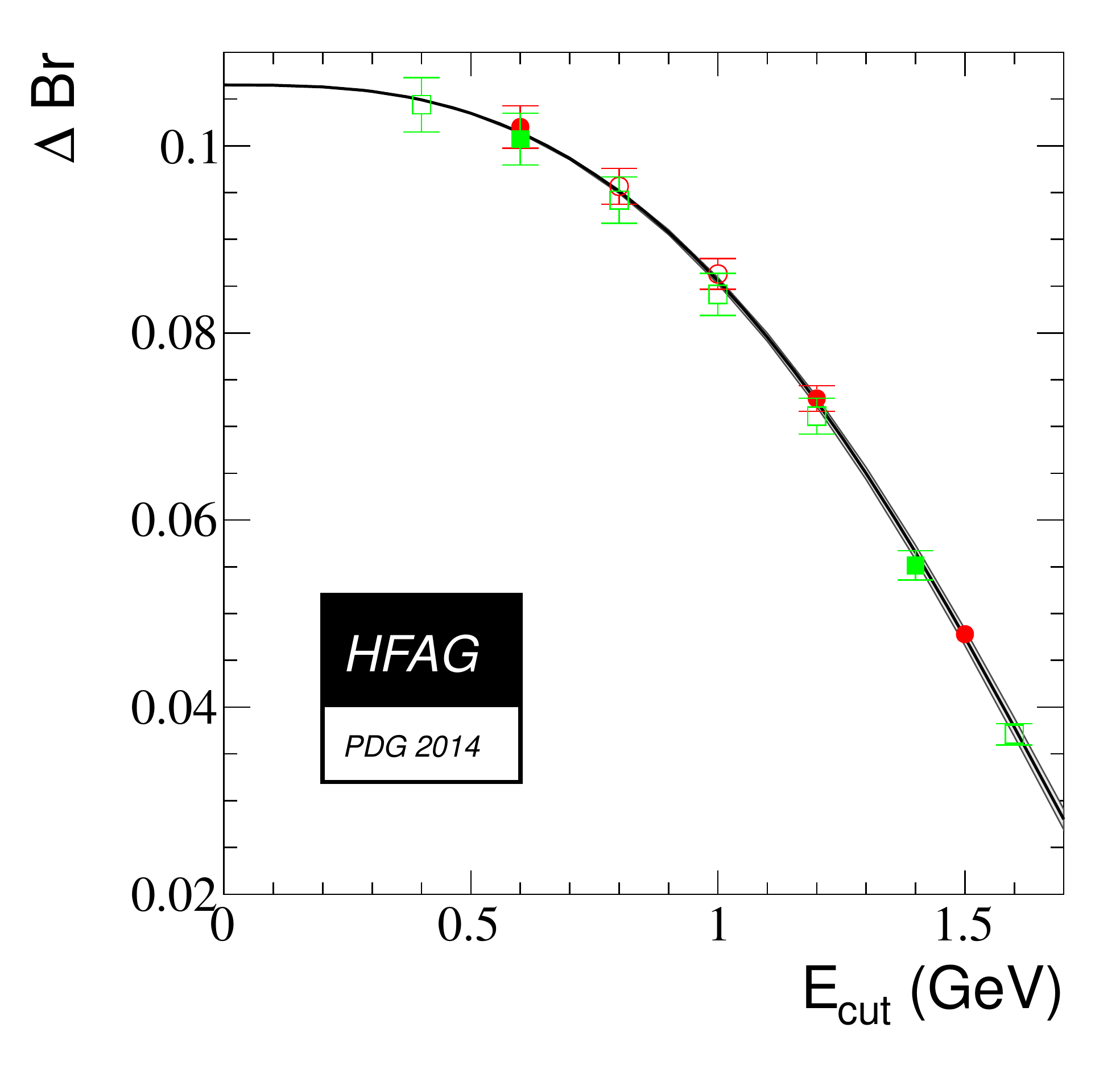}
  \includegraphics[width=8.2cm]{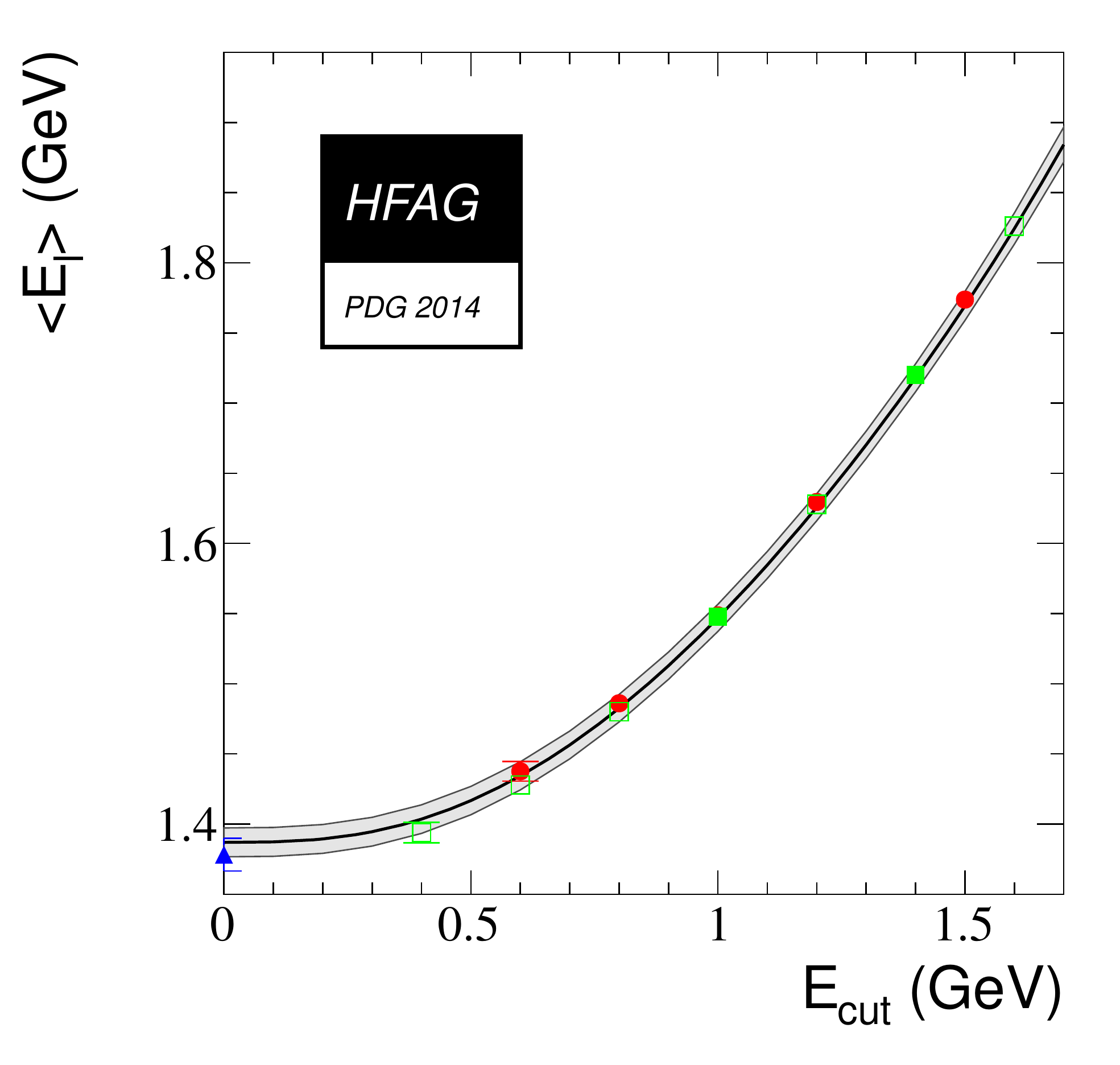}\\
  \includegraphics[width=8.2cm]{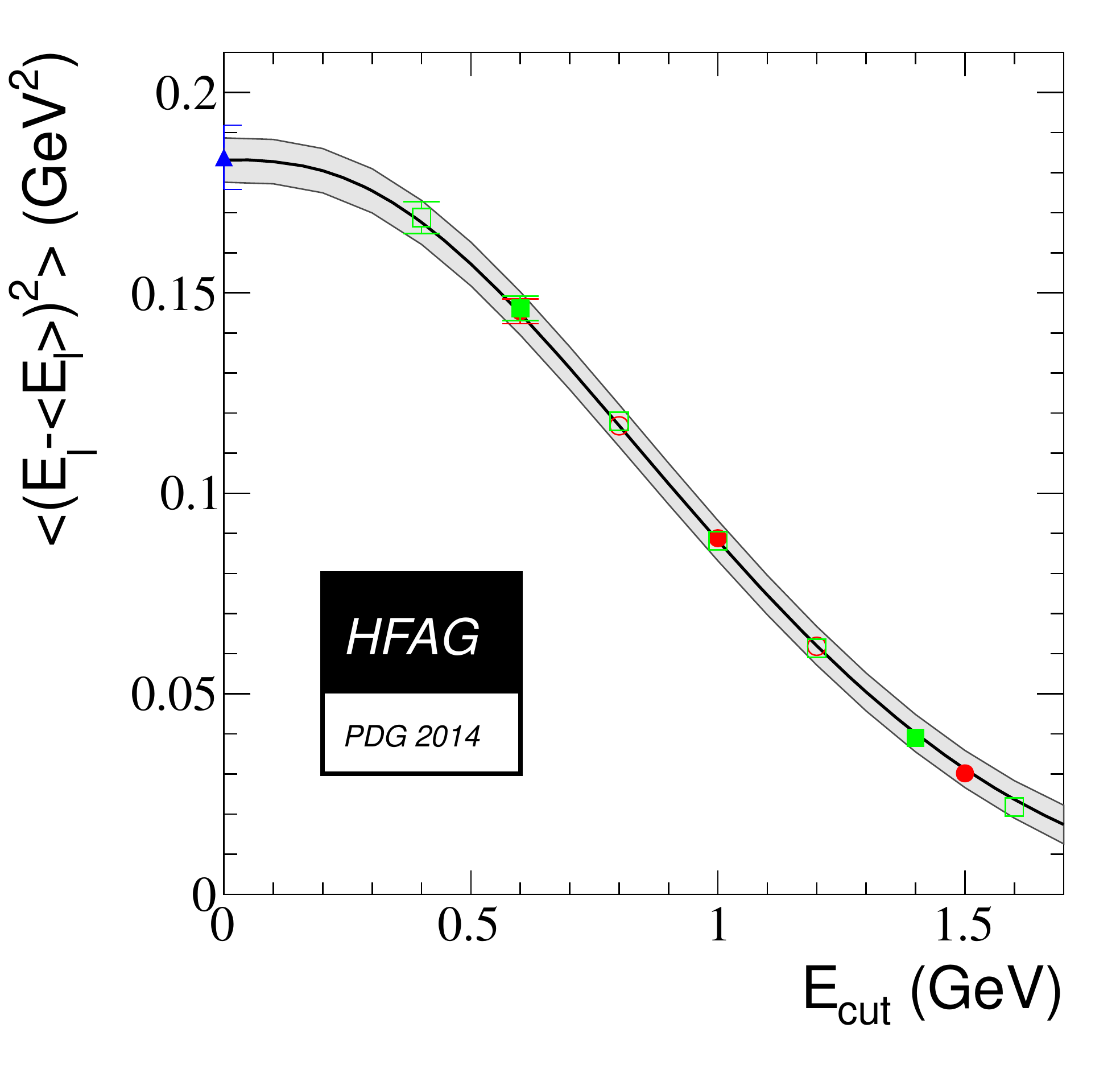}
  \includegraphics[width=8.2cm]{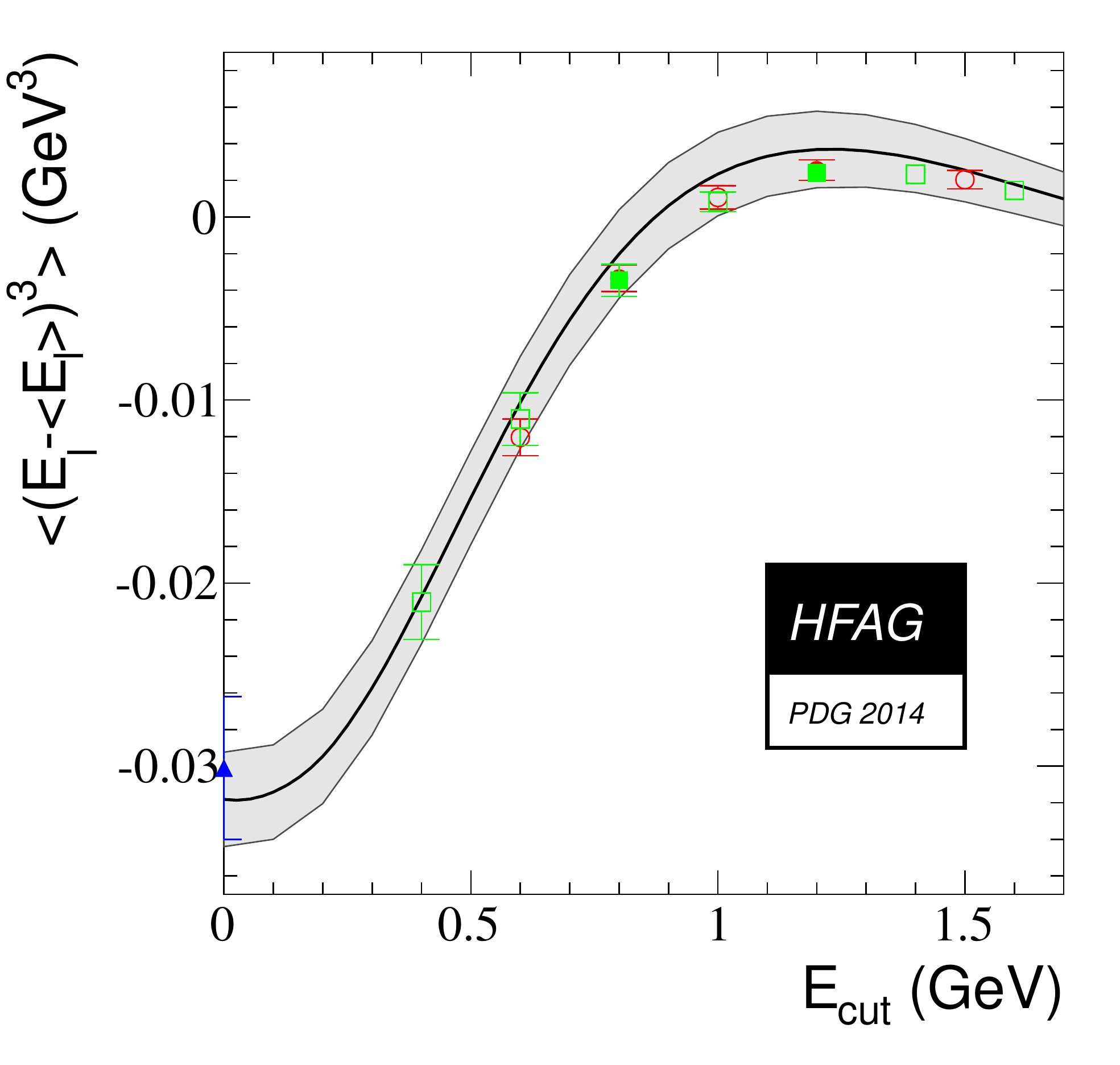}
\end{center}
\caption{Fit to the partial semileptonic branching ratios and to the
  lepton energy moments in the kinetic mass scheme. In all plots, the
  grey band is the theory prediction with total theory error. \babar
  data are shown by circles, Belle by squares and other experiments
  (DELPHI, CDF, CLEO) by triangles. Filled symbols mean that the point
  was used in the fit. Open symbols are measurements that were not
  used in the fit.} \label{fig:gf_res_kin_el}
\end{figure}

\begin{figure}
\begin{center}
  \includegraphics[width=8.2cm]{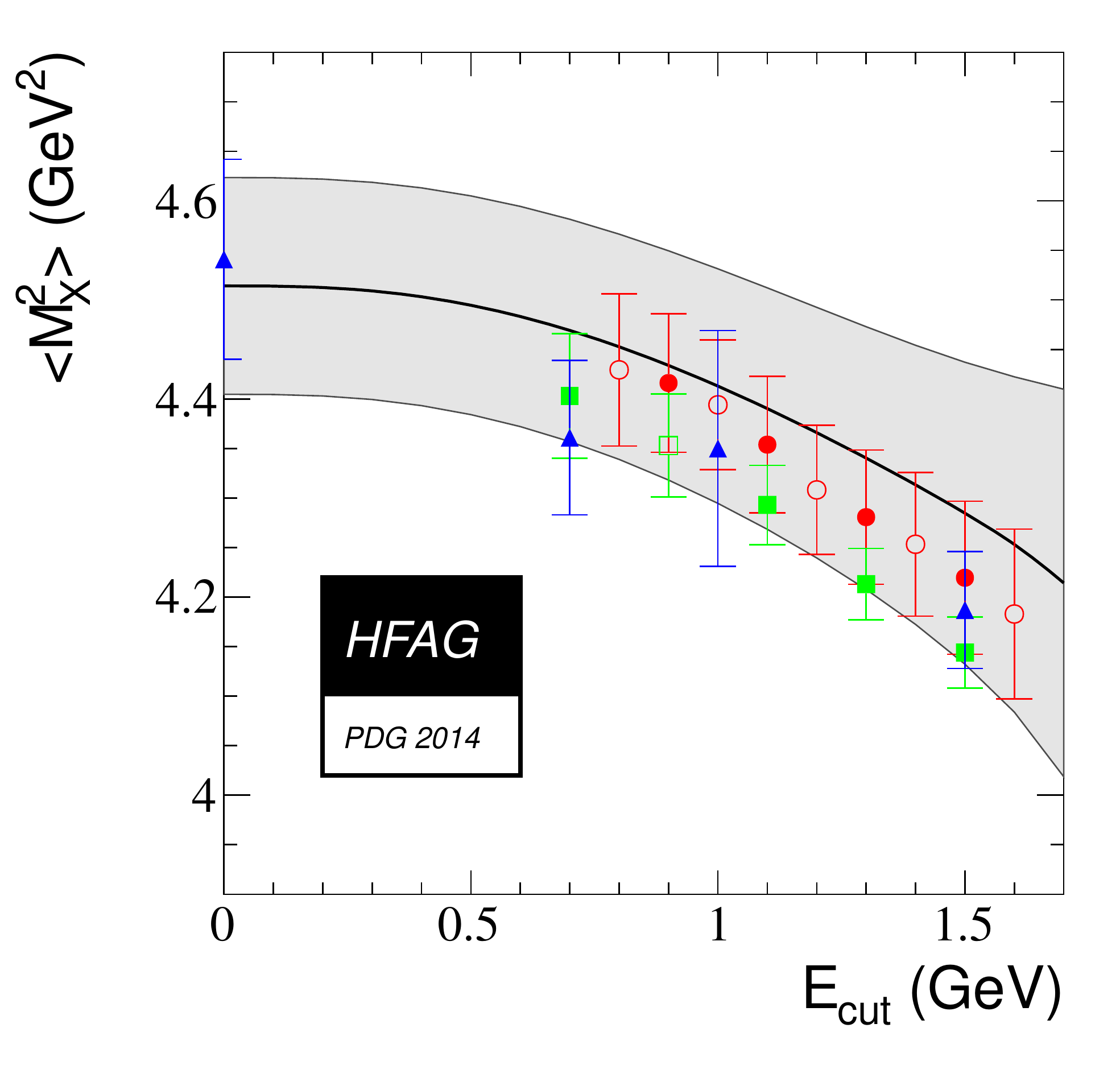}
  \includegraphics[width=8.2cm]{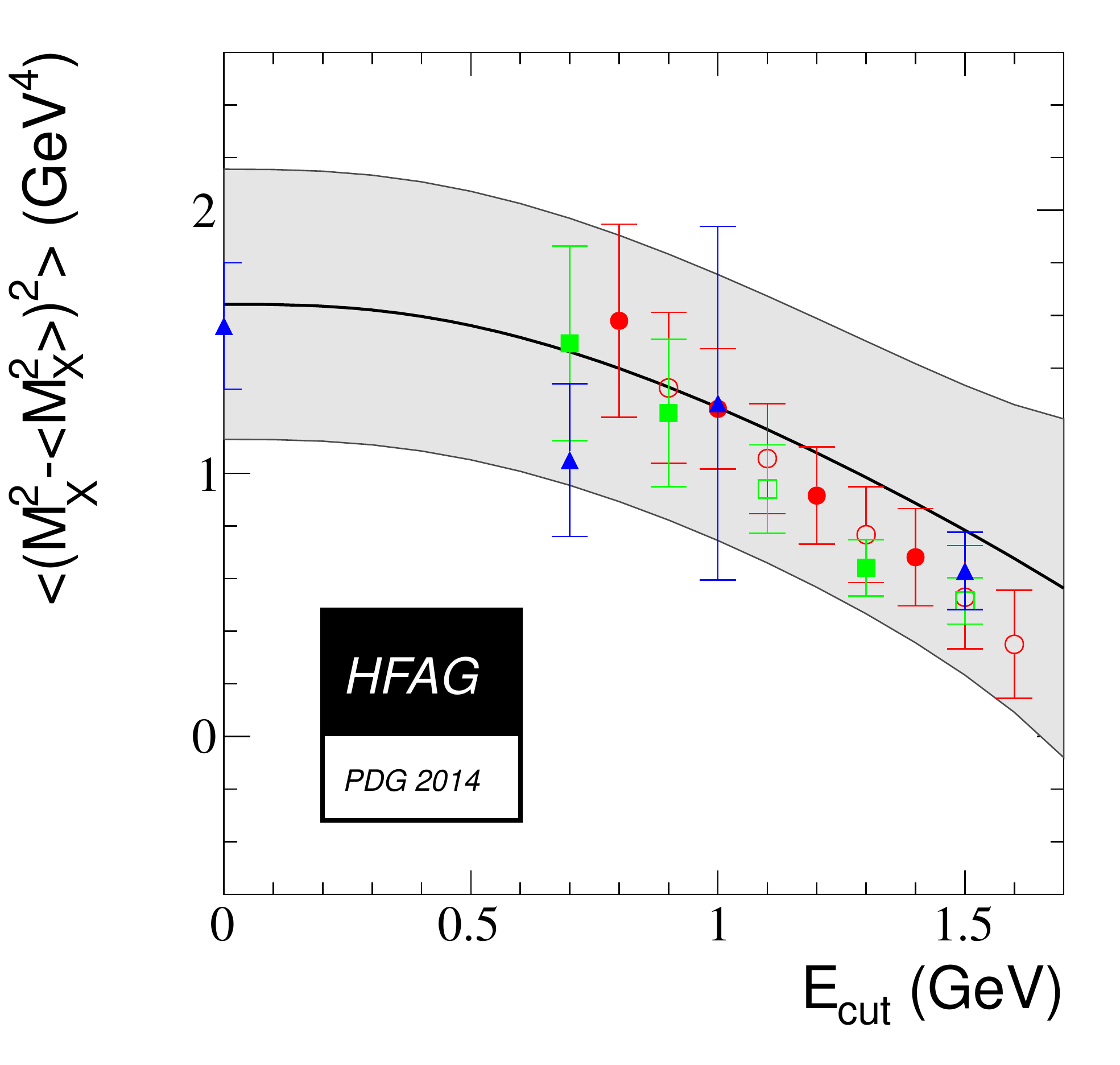}\\
  \includegraphics[width=8.2cm]{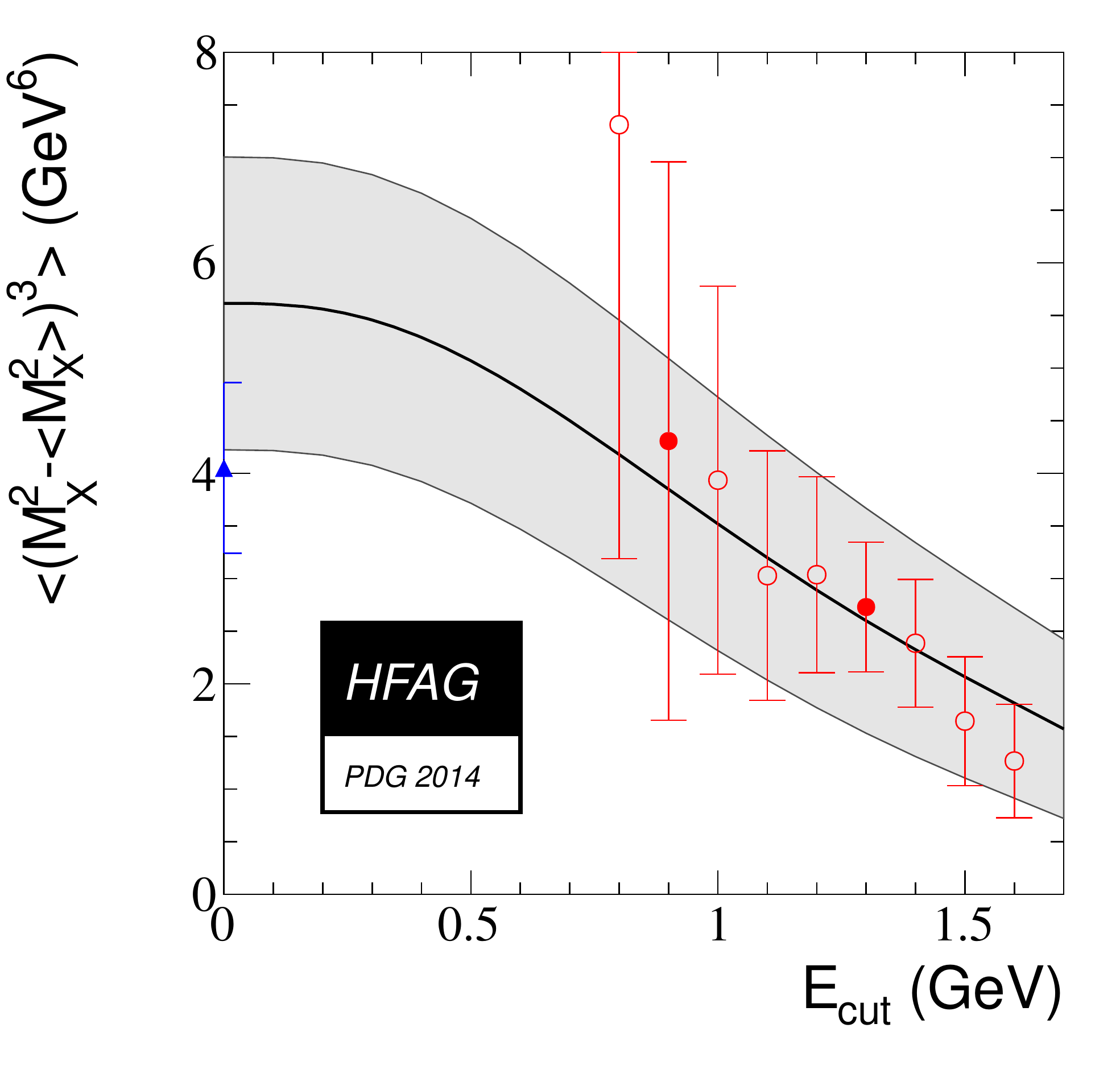}
\end{center}
\caption{Same as Fig.~\ref{fig:gf_res_kin_el} for the fit to the
  hadronic mass moments in the kinetic mass
  scheme.} \label{fig:gf_res_kin_mx}
\end{figure}

The result in terms of the main parameters is
\begin{eqnarray}
  \vcb & = & (42.46\pm 0.88)\times 10^{-3}~, \\
  m_b^{\rm kin} & = & 4.541\pm 0.023~{\rm GeV}~, \\
  \mu^2_\pi & = & 0.414\pm 0.078~{\rm GeV^2}~,
\end{eqnarray}
with a $\chi^2$ of 14.6 for $50-7$ degrees of freedom. The detailed
result and the matrix of the correlation coefficients is given in
Table~\ref{tab:gf_res_mc_kin}. The fit to the lepton energy and
hadronic mass moments is shown in Figs.~\ref{fig:gf_res_kin_el} and
\ref{fig:gf_res_kin_mx}, respectively.
\begin{table}[!htb]
\caption{Fit result in the kinetic scheme, using a precise $c$-quark
  mass constraint. The error matrix of the fit contains
  experimental and theoretical contributions. In the lower part of the
  table, the correlation matrix of the parameters is
  given.} \label{tab:gf_res_mc_kin}
\begin{center}
\resizebox{0.99\textwidth}{!}{
\begin{tabular}{|l|ccccccc|}
  \hline
  & \vcb\ [10$^{-3}$] & $m_b^{\rm kin}$ [GeV] &
  $m_c^{\overline{\rm MS}}$ [GeV] & $\mu^2_\pi$ [GeV$^2$]
  & $\rho^3_D$ [GeV$^3$] & $\mu^2_G$ [GeV$^2$] & $\rho^3_{LS}$ [GeV$^3$]\\
  \hline \hline
  value & 42.46 & \phantom{$-$}4.541 & \phantom{$-$}0.987 &
  \phantom{$-$}0.414 & \phantom{$-$}0.154 & \phantom{$-$}0.340 &
  $-$0.147\\
  error & 0.88 & \phantom{$-$}0.023 &
  \phantom{$-$}0.013 & \phantom{$-$}0.078 & \phantom{$-$}0.045 &
  \phantom{$-$}0.066 & \phantom{$-$}0.098\\
  \hline
  $|V_{cb}|$ & 1.000 & $-$0.466 & $-$0.049 &
  \phantom{$-$}0.344 & \phantom{$-$}0.161 & $-$0.190 &
  \phantom{$-$}0.019\\
  $m_b^{\rm kin}$ & & \phantom{$-$}1.000 & \phantom{$-$}0.506 &
  $-$0.113 & \phantom{$-$}0.219 & \phantom{$-$}0.487 & $-$0.156\\
  $m_c^{\overline{\rm MS}}$ & & & \phantom{$-$}1.000
  & $-$0.018 & \phantom{$-$}0.020 & \phantom{$-$}0.008 & $-$0.002\\
  $\mu^2_\pi$ & & & & \phantom{$-$}1.000 & \phantom{$-$}0.610 &
  \phantom{$-$}0.001 & \phantom{$-$}0.058\\
  $\rho^3_D$ & & & & & \phantom{$-$}1.000 & $-$0.038 & $-$0.126\\
  $\mu^2_G$ & & & & & & \phantom{$-$}1.000 & $-$0.014\\
  $\rho^3_{LS}$ & & & & & & & \phantom{$-$}1.000\\
  \hline
\end{tabular}
}
\end{center}
\end{table}

The inclusive $\bar B\to X_c\ell^-\bar\nu_\ell$ branching fraction
determined by this analysis is
\begin{equation}
  \cbf(\bar B\to X_c\ell^-\bar\nu_\ell)=(10.65\pm 0.16)\%~.
\end{equation}
Correcting for charmless semileptonic decays
(Sec.~\ref{slbdecays_b2uincl}), $\cbf(\bar B\to
X_u\ell^-\bar\nu_\ell)=(2.14\pm 0.31)\times 10^{-3}$, we obtain the
semileptonic branching fraction,
\begin{equation}
  \cbf(\bar B\to X\ell^-\bar\nu_\ell)=(10.86\pm 0.16)\%~.
\end{equation}

\subsubsection{Analysis in the 1S scheme}
\label{globalfits1S}

The fit relies on the calculations of the spectral moments described in
Ref.~\cite{Bauer:2004ve}. The theoretical uncertainties are estimated
as explained in Ref.~\cite{Schwanda:2008kw}. Only trivial theory
correlations, \ie\ between the same moment at the same
threshold are included in the analysis. The fit determines $\vcb$ and
the 6 non-perturbative parameters mentioned above.

The result of the fit using the $B\to X_s\gamma$ constraint is
\begin{eqnarray}
  \vcb & = & (41.98\pm 0.45)\times 10^{-3}~, \\
  m_b^{1S} & = & 4.691\pm 0.037~{\rm GeV}~, \\
  \lambda_1 & = & -0.362\pm 0.067~{\rm GeV^2}~,
\end{eqnarray}
with a $\chi^2$ of 23.0 for $66-7$ degrees of freedom. The detailed
result of the fit is given in Table~\ref{tab:gf_res_xsgamma_1s}.
\begin{table}[!htb]
\caption{Fit result in the 1S scheme, using $B\to X_s\gamma$~moments
  as a constraint. In the lower part of the table, the correlation
  matrix of the parameters is given.} \label{tab:gf_res_xsgamma_1s}
\begin{center}
\begin{tabular}{|l|ccccccc|}
  \hline
  & $m_b^{1S}$ [GeV] & $\lambda_1$ [GeV$^2$] & $\rho_1$ [GeV$^3$] &
  $\tau_1$ [GeV$^3$] & $\tau_2$ [GeV$^3$] & $\tau_3$ [GeV$^3$] &
  $\vcb$ [10$^{-3}$]\\
  \hline \hline
  value & 4.691 & $-0.362$ & \phantom{$-$}0.043 &
  \phantom{$-$}0.161 & $-0.017$ & \phantom{$-$}0.213 &
  \phantom{$-$}41.98\\
  error & 0.037 & \phantom{$-$}0.067 & \phantom{$-$}0.048 &
  \phantom{$-$}0.122 & \phantom{$-$}0.062 & \phantom{$-$}0.102 &
  \phantom{$-$}0.45\\
  \hline
  $m_b^{1S}$ & 1.000 & \phantom{$-$}0.434 & \phantom{$-$}0.213 &
  $-0.058$ & $-0.629$ & $-0.019$ & $-0.215$\\
  $\lambda_1$ & & \phantom{$-$}1.000 & $-0.467$ & $-0.602$ & $-0.239$
  & $-0.547$ & $-0.403$\\
  $\rho_1$ & & & \phantom{$-$}1.000 & \phantom{$-$}0.129 & $-0.624$ &
  \phantom{$-$}0.494 & \phantom{$-$}0.286\\
  $\tau_1$ & & & & \phantom{$-$}1.000 & \phantom{$-$}0.062 & $-0.148$ &
  \phantom{$-$}0.194\\
  $\tau_2$ & & & & & \phantom{$-$}1.000 & $-0.009$ & $-0.145$\\
  $\tau_3$ & & & & & & \phantom{$-$}1.000 & \phantom{$-$}0.376\\
  $\vcb$ & & & & & & & \phantom{$-$}1.000\\
  \hline
\end{tabular}
\end{center}
\end{table}

\subsection{Exclusive CKM-suppressed decays}
\label{slbdecays_b2uexcl}
In this section, we list results on exclusive charmless semileptonic branching fractions
and determinations of $\vub$ based on $\Bb\to\pi\ell\nub$ decays.
The measurements are based on two different event selections: tagged
events, in which case the second $B$ meson in the event is fully
reconstructed in either a hadronic decay (``had. tag'') or in a 
CKM-favored semileptonic decay (``sl. tag''); and untagged events, in which case the momentum
of the undetected neutrino is inferred from measurements of the total 
momentum sum of the detected particles and the knowledge of the initial state.
We also present averages for $\Bb\to\rho\ell\nub$, $\Bb\to\omega\ell\nub$, $\Bb\to\eta\ell\nub$ and
$\Bb\to\eta'\ell\nub$.

\begin{figure}[!ht]
 \begin{center}
  \unitlength1.0cm 
  \begin{picture}(13.,12.)  
   \put( 1.0, 0.0){\includegraphics[width=12.0cm]{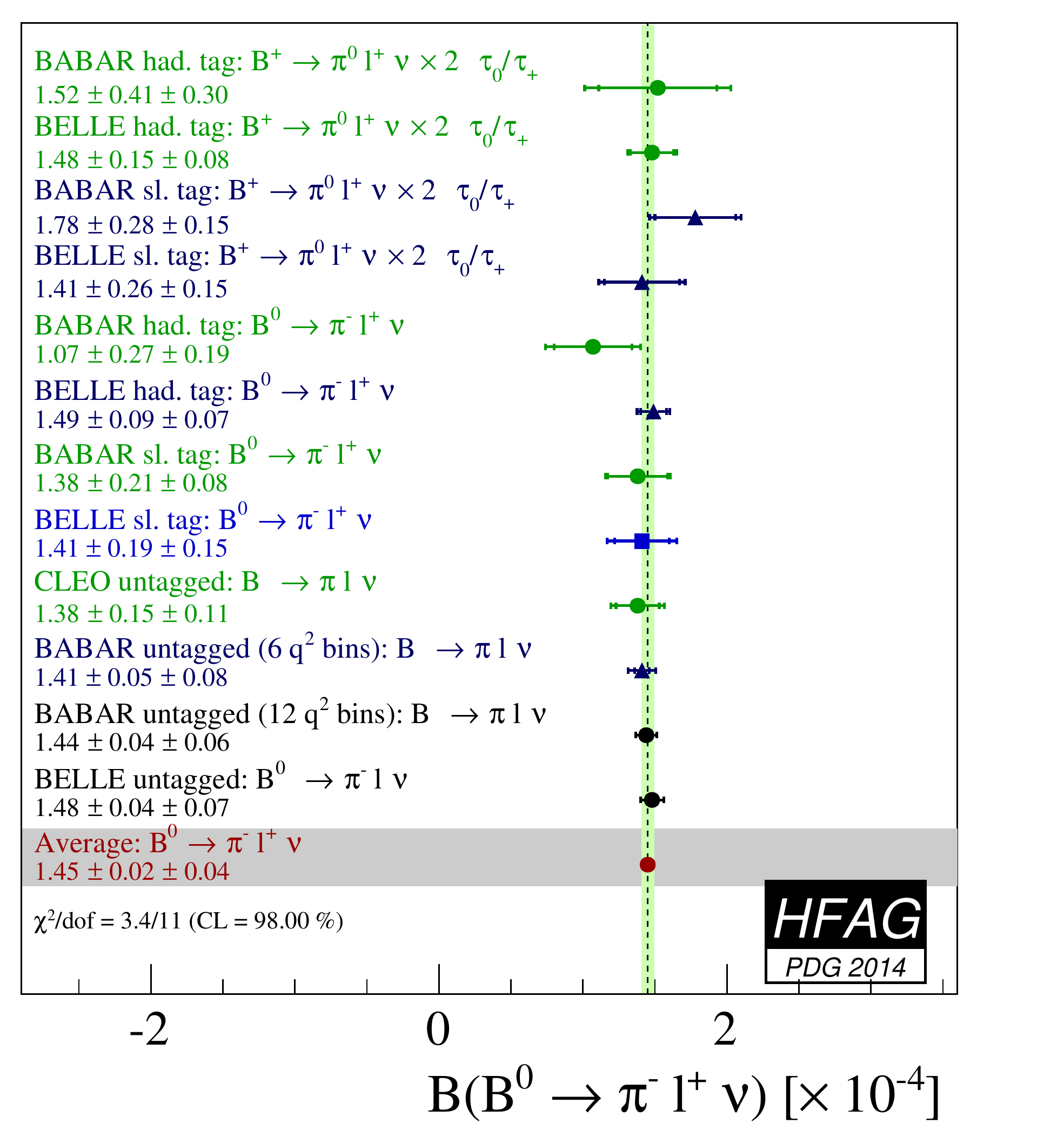}
   }
  \end{picture}\caption{
Summary of exclusive determinations of $\cbf(\Bb\to\pi\ell\nub)$ and their average.
Measured branching fractions for $B^+ \rightarrow \pi^0 l^+ \nu$ have been
multiplied by $2\times \tau_{B^0}/\tau_{B^+}$ in accordance with
isospin symmetry. The labels ``had. tag'', ``sl. tag'' and
``untagged'' refer to the constraint used for the second $B$~meson in
the event.
The results from the untagged measurements by CLEO and \babar are based on a combined analysis of
$B^0 \rightarrow \pi^- l^+ \nu$  and $B^+ \rightarrow \pi^0 l^+ \nu$ decays using isospin relations. 
}
\label{fig:xlnu}
\end{center}
\end{figure}

The results for the full and partial branching fractions for $\Bb\to\pi\ell\nub$ are given
in Table~\ref{tab:pilnubf} and shown in Figure~\ref{fig:xlnu}.   

When averaging these results, systematic uncertainties due to external
inputs, \eg\ form factor shapes and background estimates from the
modeling of $\Bb\to X_c\ell\nub$ and $\Bb\to X_u\ell\nub$ decays, are
treated as fully correlated (in the sense of Eq.~\ref{eq:correlrho} in
Sec.~\ref{sec:method}).
Uncertainties due to experimental reconstruction effects are treated
as fully correlated among measurements from a given experiment.

\begin{sidewaystable}[!htb]
\begin{center}
\caption{\label{tab:pilnubf}
Summary of exclusive determinations of $\cbf(\Bb\to\pi
\ell\nub)$. The errors quoted
correspond to statistical and systematic uncertainties, respectively.
Measured branching fractions for $B^+\rightarrow \pi^0 \ell^+ \nu$ have been
multiplied by $2\times \tau_{B^0}/\tau_{B^+}$ in accordance with
isospin symmetry. The labels ``had. tag'' and ``sl. tag'' refer to
the type of $B$ tag used in the measurement, and ``untagged'' refers to an untagged measurement.}
\begin{small}
\begin{tabular}{|lcccc|}
\hline
& $\cbf [10^{-4}]$
& $\cbf(q^2<12\,\gev^2/c^2) [10^{-4}]$
& $\cbf(q^2<16\,\gev^2/c^2) [10^{-4}]$
& $\cbf(q^2>16\,\gev^2/c^2) [10^{-4}]$
\\
\hline\hline
CLEO untagged $\pi^+,\pi^0$~\cite{Adam:2007pv}
& $1.38\pm 0.15\pm 0.11$ 
& $0.69\pm 0.12\pm 0.07$
& $0.97\pm 0.13\pm 0.09$
& $0.41\pm 0.08\pm 0.04$
\\ 
\babar untagged $\pi^+,\pi^0$~\cite{delAmoSanchez:2010af}
& $1.41\pm 0.05\pm 0.08$
& $0.88\pm 0.04\pm 0.05$
& $1.10\pm 0.04\pm 0.06$
& $0.32\pm 0.02\pm 0.03$
\\  
\babar untagged $\pi^+, \pi^0$~\cite{Lees:2012vv}
& $1.44\pm 0.04\pm 0.06$
& $0.83\pm 0.03\pm 0.04$
& $1.08\pm 0.03\pm 0.05$
& $0.40\pm 0.02\pm 0.02$
\\  
Belle untagged $\pi^+$~\cite{Ha:2010rf}
& $1.48\pm 0.04\pm 0.07$
& $0.82\pm 0.03\pm 0.04$
& $1.09\pm 0.03\pm 0.05$
& $0.40\pm 0.02\pm 0.02$
\\  
Belle sl. tag $\pi^+$~\cite{Hokuue:2006nr}
& $1.41\pm 0.19\pm 0.15$
& $0.80\pm 0.14\pm 0.08$
& $1.04\pm 0.16\pm 0.11$
& $0.37\pm 0.10\pm 0.04$
\\ 
Belle sl. tag $\pi^0$~\cite{Hokuue:2006nr}
& $1.41\pm 0.26\pm 0.15$
& $0.71\pm 0.17\pm 0.08$
& $1.05\pm 0.22\pm 0.12$
& $0.37\pm 0.15\pm 0.04$
\\ 
\babar sl. tag $\pi^+$~\cite{Aubert:2008bf}
& $1.38\pm 0.21\pm 0.08$
& $0.77\pm 0.14\pm 0.05$
& $0.92\pm 0.16\pm 0.05$
& $0.46\pm 0.13\pm 0.03$
\\ 
\babar sl. tag $\pi^0$~\cite{Aubert:2008bf}
& $1.78\pm 0.28\pm 0.15$
& $1.07\pm 0.20\pm 0.09$
& $1.35\pm 0.22\pm 0.11$
& $0.44\pm 0.17\pm 0.06$
\\ 
\babar had. tag $\pi^+$~\cite{Aubert:2006ry}
& $1.07\pm 0.27\pm 0.19$
& $0.26\pm 0.15\pm 0.04$
& $0.42\pm 0.18\pm 0.06$
& $0.65\pm 0.20\pm 0.13$
\\ 
\babar had. tag $\pi^0$~\cite{Aubert:2006ry}
& $1.52\pm 0.41 \pm0.30$
& $0.67\pm 0.30\pm 0.12$
& $1.04\pm 0.35\pm 0.18$
& $0.48\pm 0.22\pm 0.12$
\\ 
Belle had. tag $\pi^+$~\cite{Sibidanov:2013rkk}
& $1.49\pm 0.09\pm 0.07$
& $0.81\pm 0.06\pm 0.04$
& $1.06\pm 0.07\pm 0.05$
& $0.45\pm 0.05\pm 0.02$
\\ 
Belle had. tag $\pi^0$~\cite{Sibidanov:2013rkk}
& $1.48\pm 0.15\pm 0.08$
& $0.78\pm 0.11\pm 0.04$
& $1.09\pm 0.12\pm 0.06$
& $0.36\pm 0.07\pm 0.02$
\\  \hline
{\bf Average}
& \mathversion{bold}$1.45\pm 0.02\pm 0.04$
& \mathversion{bold}$0.81\pm 0.02\pm 0.02$
& \mathversion{bold}$1.06\pm 0.02\pm 0.03$
& \mathversion{bold}$0.38\pm 0.01\pm 0.01$
\\ 
\hline
\end{tabular}\\
\end{small}
\end{center}
\end{sidewaystable}

The determination of \vub\ from $\Bb\to\pi\ell\nub$ decays is shown in
Table~\ref{tab:pilnuvub}, and uses our averages for the partial
branching fractions given in Table~\ref{tab:pilnubf}, combined with
various form factor calculations. 
Two theoretical approaches are used: unquenched lattice QCD (LQCD) and
QCD light-cone sum rules (LCSR). The calculations make predictions for
different regions of $q^2$, where $q^2$ is the four-momentum
transfered to the lepton-neutrino system squared, $q^2=(p_\ell+p_\nu)^2$.
Lattice calculations of the form factors are limited to small hadron
momenta, \ie\ large $q^2$, while calculations based on light-cone sum
rules are restricted to small $q^2$.

\begin{table}[hbtf]
\caption{\label{tab:pilnuvub}
Determinations of \vub\ based on the average partial
$\Bb\to\pi\ell\nub$ decay branching fractions stated in
Table~\ref{tab:pilnubf}. 
The $q^2$ ranges for the partial branching fractions corresponding to the 
validity ranges of the form factor calculations are indicated. 
The first uncertainty is experimental and the second is from theory.  
}
\begin{center}
\renewcommand{\arraystretch}{1.2}
\begin{tabular}{|lcc|}
\hline
Method                                         & $q^2$ range [$\gev^2/c^2$] & $\Vub [10^{-3}]$ \\\hline\hline
Khodjamirian et al. (LCSR) ~\cite{Khodjamirian:2011ub} & 0 -- 12                    & $3.41\pm 0.06 {}^{+0.37}_{-0.32}$ \\ \hline
Ball \& Zwicky (LCSR)~\cite{Ball:2004ye}              & 0 -- 16                    & $3.58\pm 0.06 {}^{+0.59}_{-0.40}$ \\ \hline
HPQCD (LQCD)~\cite{Dalgic:2006dt}                     & 16 -- 26.4                 & $3.52\pm 0.08 {}^{+0.61}_{-0.40}$ \\  \hline
FNAL/MILC (LQCD)~\cite{Bailey:2008wp}                 & 16 -- 26.4                 & $3.36\pm 0.08 {}^{+0.37}_{-0.31}$ \\ 
\hline
\end{tabular}
\end{center}
\end{table}

An alternative method to determine \vub\ from $\Bb\to\pi\ell\nub$ decays that makes use
of the measurement over the full $q^2$ range is based on a simultaneous fit of a
$B\to \pi$ form factor parameterization to data and theory predictions. 
We choose the BCL (Bourrely, Caprini, Lellouch) parameterization~\cite{Bourrely:2008za} up to order $z^2$.
There are 3+1 fit parameters: the three coefficients of the BCL power series ($b_0$, $b_1$, $b_2$)
and \Vub, which is determined from the relative normalization between data and theory predictions. 
As the shape of the $q^2$ spectrum is determined by only two parameters ($b_1$ and $b_2$), we quote
the ratios $b_1/b_0$ and $b_2/b_0$ as results for the shape determined in the fit.

\begin{figure}[!ht]
 \begin{center}
  \unitlength1.0cm 

  \begin{picture}(16.,9.0)  
   \put( -1.0,  0.0){\includegraphics[width=9.5cm]{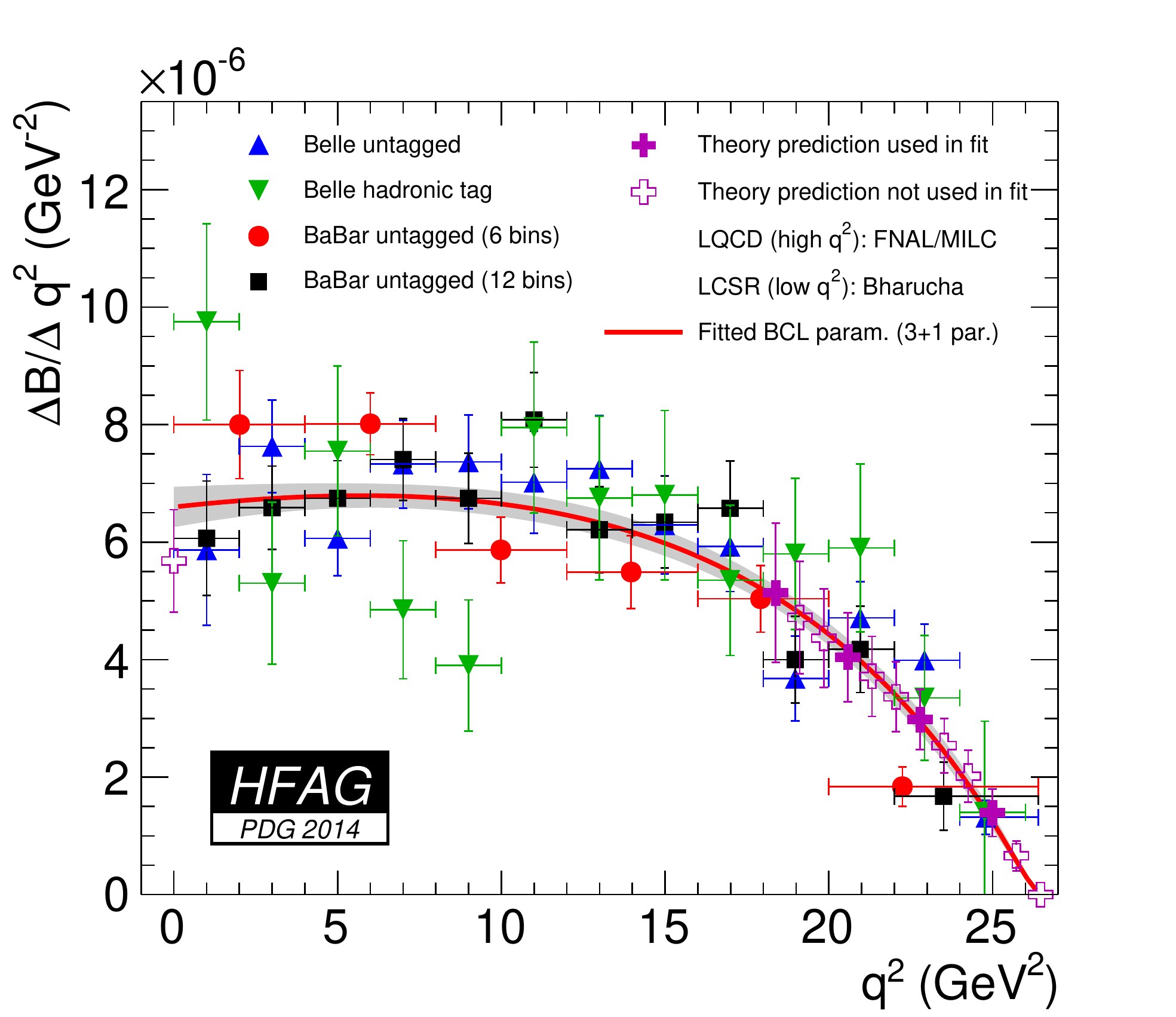}}
   \put( 8.0,  0.0){\includegraphics[width=9.5cm]{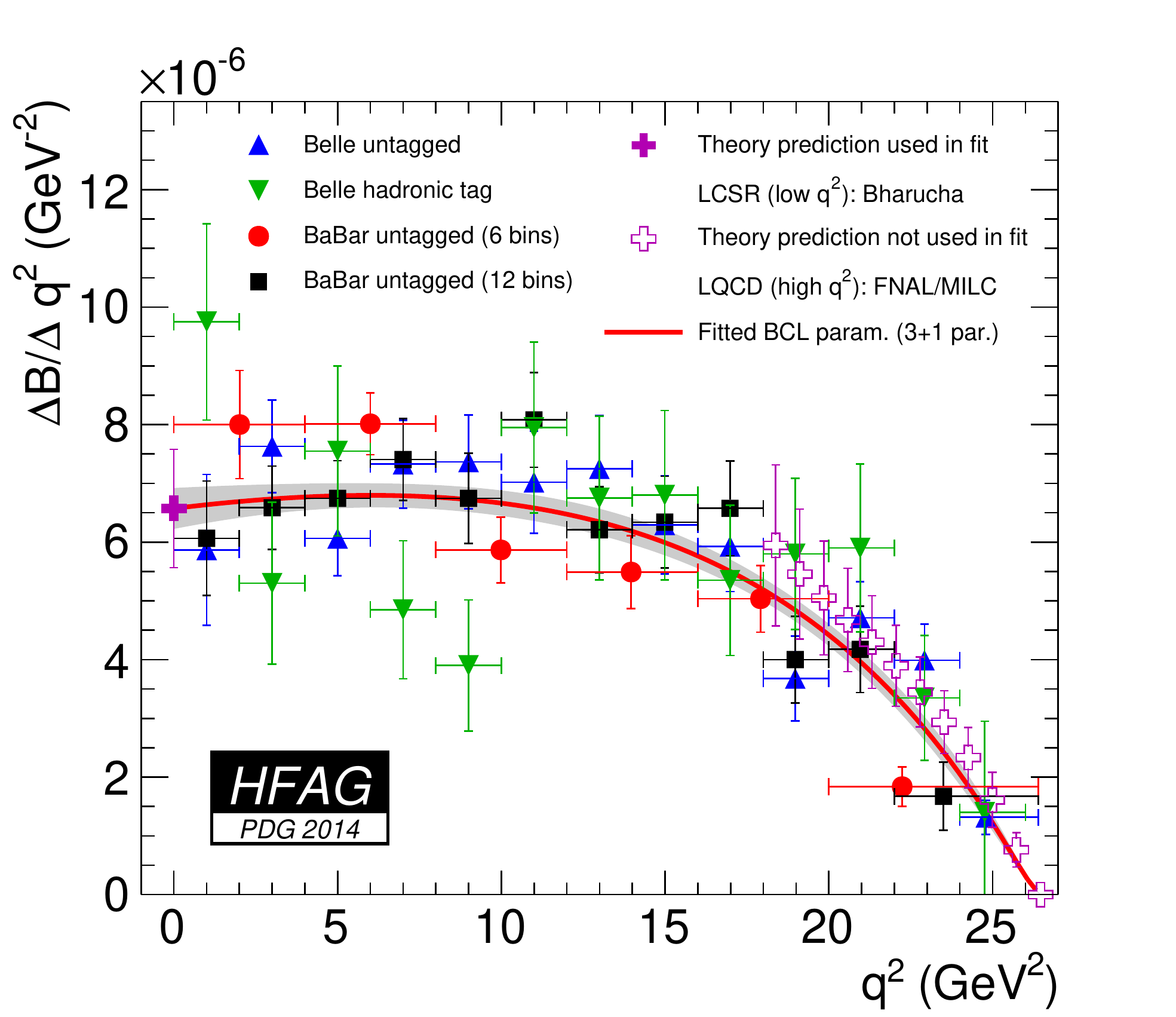}} 
   \put(  6.8,  4.6){{\large\bf a)}}     
   \put( 15.8,  4.6){{\large\bf b)}}
   \end{picture} 
   \caption{
    Simultaneous fit of the BCL parameterization with 3+1 parameters to \babar and Belle $\Bb\to\pi\ell\nub$ data
    and the theory prediction from (a) FNAL/MILC LQCD calculations~\cite{Bailey:2008wp} 
    (yielding $\Vub = (3.28 \pm 0.29) \times 10^{-3}$)
    and (b) LCSR~\cite{Bharucha:2012wy} at $q^2=0$ (yielding $\Vub = (3.53 \pm 0.29) \times 10^{-3}$).}
\label{fig:vub_pilnu_simultaneous}
\end{center}
\end{figure}

The result of the simultaneous fit to the four most precise measurements from \babar and Belle 
(\babar untagged 6 $q^2$ bins, \babar untagged 12 $q^2$ bins, Belle untagged, Belle had. tag)
and the FNAL/MILC LQCD calculations is shown in Figure~\ref{fig:vub_pilnu_simultaneous}~(a). 
The fit probability is 0.053 ($\chi^2/dof = 60.2/44$) and we obtain the following values:
\begin{eqnarray}
\Vub &=& (3.28 \pm 0.29) \times 10^{-3}, \\
b_1/b_0 &=& -1.02 \pm 0.18, \\
b_2/b_0 &=& -1.20 \pm 0.57.
\end{eqnarray}
The fit results correspond to a value of the product $f_+(0)\Vub$ of $(0.923 \pm 0.024) \times 10^{-3}$. 
The correlation matrix of the fit parameters is:
\begin{center}
\begin{tabular}{c|cccc}
      & $b_0$ & $b_1$ & $b_2$ & \Vub  \\
\hline
$b_0$ & 1.00  & -0.54 &  0.15 & -0.29 \\
$b_1$ &       &  1.00 & -0.89 &  0.24 \\ 
$b_2$ &       &       &  1.00 & -0.14 \\  
\Vub  &       &       &       &  1.00 \\
\end{tabular}
\end{center}

The simultaneous fit has also been performed using the most recent result for $f_+(0)$ from LCSR~\cite{Bharucha:2012wy}
($f_+(0) = 0.261^{+0.020}_{-0.023}$), \ie\ one point at $q^2=0$ is used as form factor normalization instead of the LQCD points at high $q^2$.
This fit has a probability of 0.029 ($\chi^2/dof = 59.8/41$) and yields consistent results (Fig.~\ref{fig:vub_pilnu_simultaneous}~(b)):
\begin{eqnarray}
\Vub &=& (3.53 \pm 0.29) \times 10^{-3}, \\
b_1/b_0 &=& -0.99 \pm 0.20, \\
b_2/b_0 &=& -1.28 \pm 0.61, \\
f_+(0)\Vub &=& (0.922 \pm 0.024) \times 10^{-3}. 
\end{eqnarray}\\

\begin{figure}[!ht]
 \begin{center}
  \unitlength1.0cm 
  \begin{picture}(14.,10.0)  
   \put( -1.5,  0.0){\includegraphics[width=9.0cm]{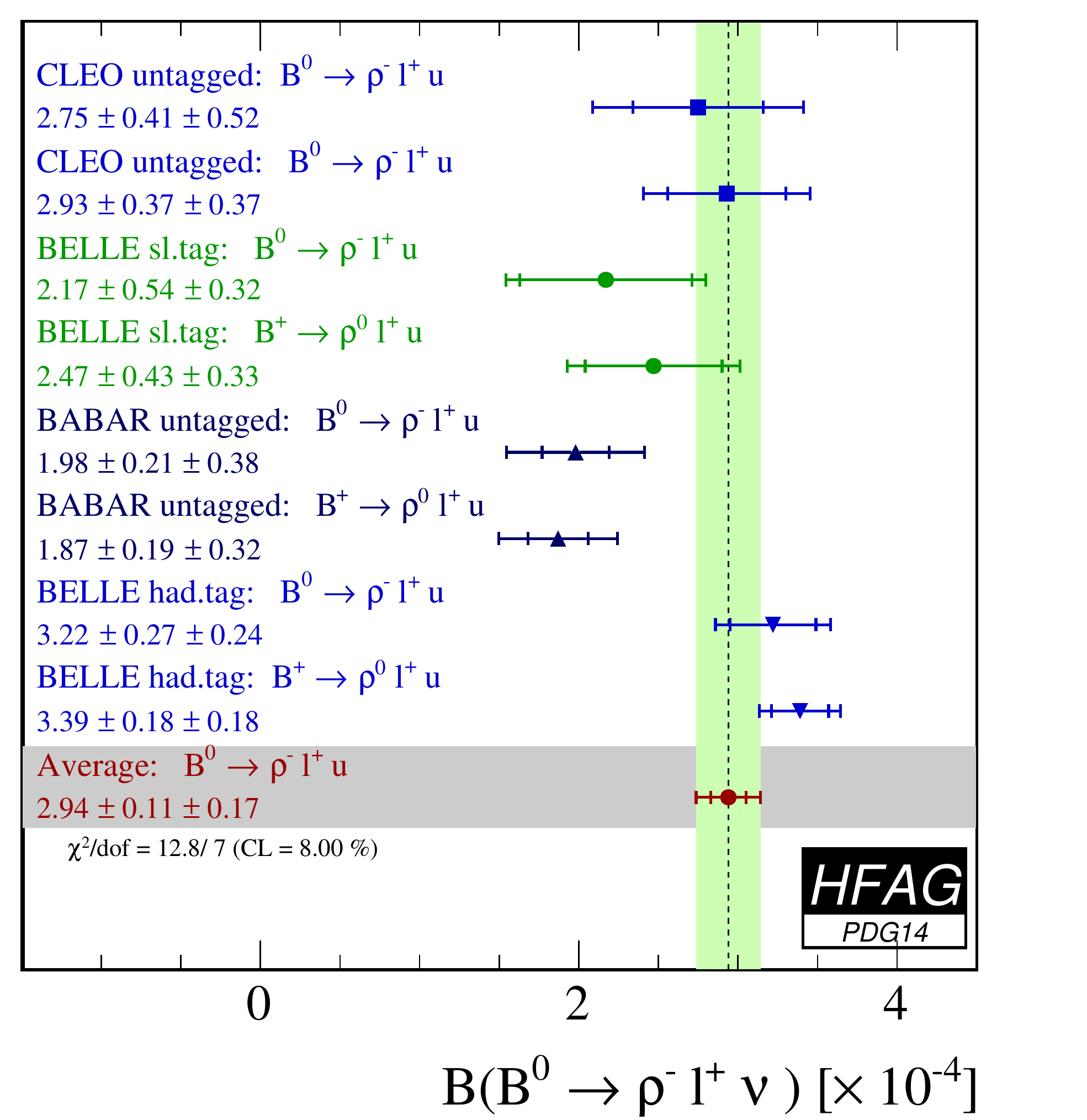}}
   \put( 7.5,  0.0){\includegraphics[width=9.0cm]{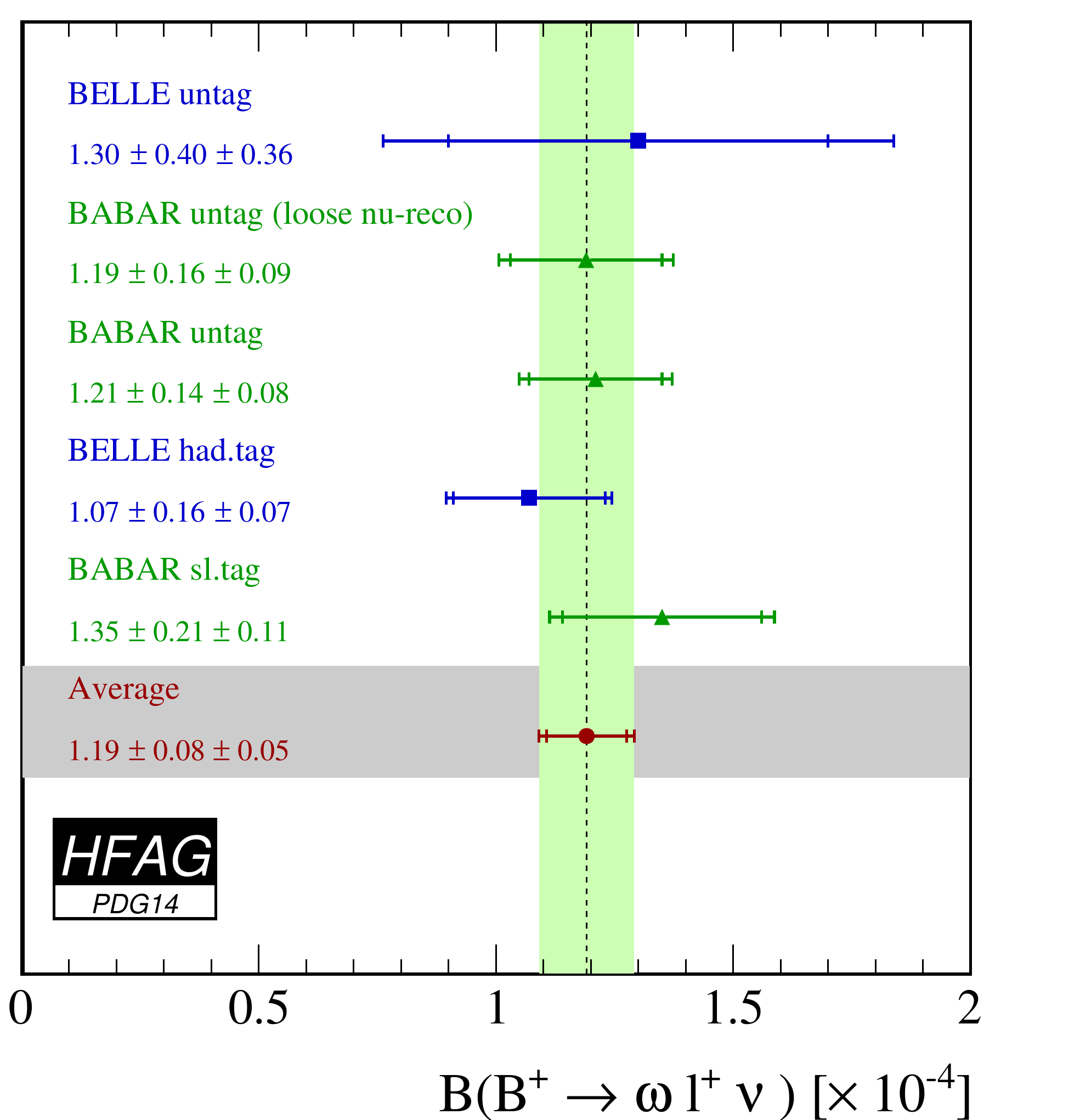}} 
   \put(  5.8,  8.5){{\large\bf a)}}  
   \put( 14.6,  8.5){{\large\bf b)}}
   \end{picture} \caption{
 (a) Summary of exclusive determinations of $\cbf(\Bb\to\rho\ell\nub)$ and their average. Measurements
 of $B^+ \to \rho^0\ell^+\nu$ branching fractions have been multiplied by $2\tau_{B^0}/\tau_{B^+}$ 
 in accordance with isospin symmetry.    
(b) Summary of exclusive determinations of $\cbf(\Bb\to\omega\ell\nub)$ and their average.
}
\label{fig:xulnu1}
\end{center}
\end{figure}

The branching fractions for 
$\Bb\to \rho\ell\nub$ decays is computed based on the measurements in
Table~\ref{tab:rholnu} and is shown in Figure~\ref{fig:xulnu1}. The determination of $\Vub$
from these other channels looks less promising than for $\Bb\to\pi\ell\nub$ and at the moment it is not extracted.

\begin{table}[!htb]
\begin{center}
\caption{Summary of exclusive determinations of $\cbf(\Bb\to\rho
\ell\nub)$. The errors quoted
correspond to statistical and systematic uncertainties, respectively.}
\label{tab:rholnu}
\begin{small}
\begin{tabular}{|lc|}
\hline
& $\cbf [10^{-4}]$
\\
\hline\hline
CLEO $\rho^+$~\cite{Behrens:1999vv}
& $2.75\pm 0.41\pm 0.52\ $ 
\\ 
CLEO $\rho^+$~\cite{Adam:2007pv}
& $2.93\pm 0.37\pm 0.37\ $ 
\\ 
Belle $\rho^+$~\cite{Sibidanov:2013rkk}
& $3.22\pm 0.27\pm 0.24\ $
\\
Belle $\rho^0$~\cite{Sibidanov:2013rkk}
& $3.39\pm 0.18\pm 0.18\ $
\\
Belle $\rho^+$~\cite{Hokuue:2006nr}
& $2.17\pm 0.54\pm 0.32\ $
\\
Belle $\rho^0$~\cite{Hokuue:2006nr}
& $2.47\pm 0.43\pm 0.33\ $
\\
\babar $\rho^+$~\cite{delAmoSanchez:2010af}
& $1.98\pm 0.21\pm 0.38\ $
\\
\babar $\rho^0$~\cite{delAmoSanchez:2010af}
& $1.87\pm 0.19\pm 0.32\ $

\\  \hline
{\bf Average}
& \mathversion{bold}$2.94 \pm 0.11\pm 0.17 $
\\ 
\hline
\end{tabular}\\
\end{small}
\end{center}
\end{table}

\begin{figure}[!ht]
 \begin{center}
  \unitlength1.0cm 
  \begin{picture}(14.,10.0)  
   \put( -1.5,  0.0){\includegraphics[width=9.0cm]{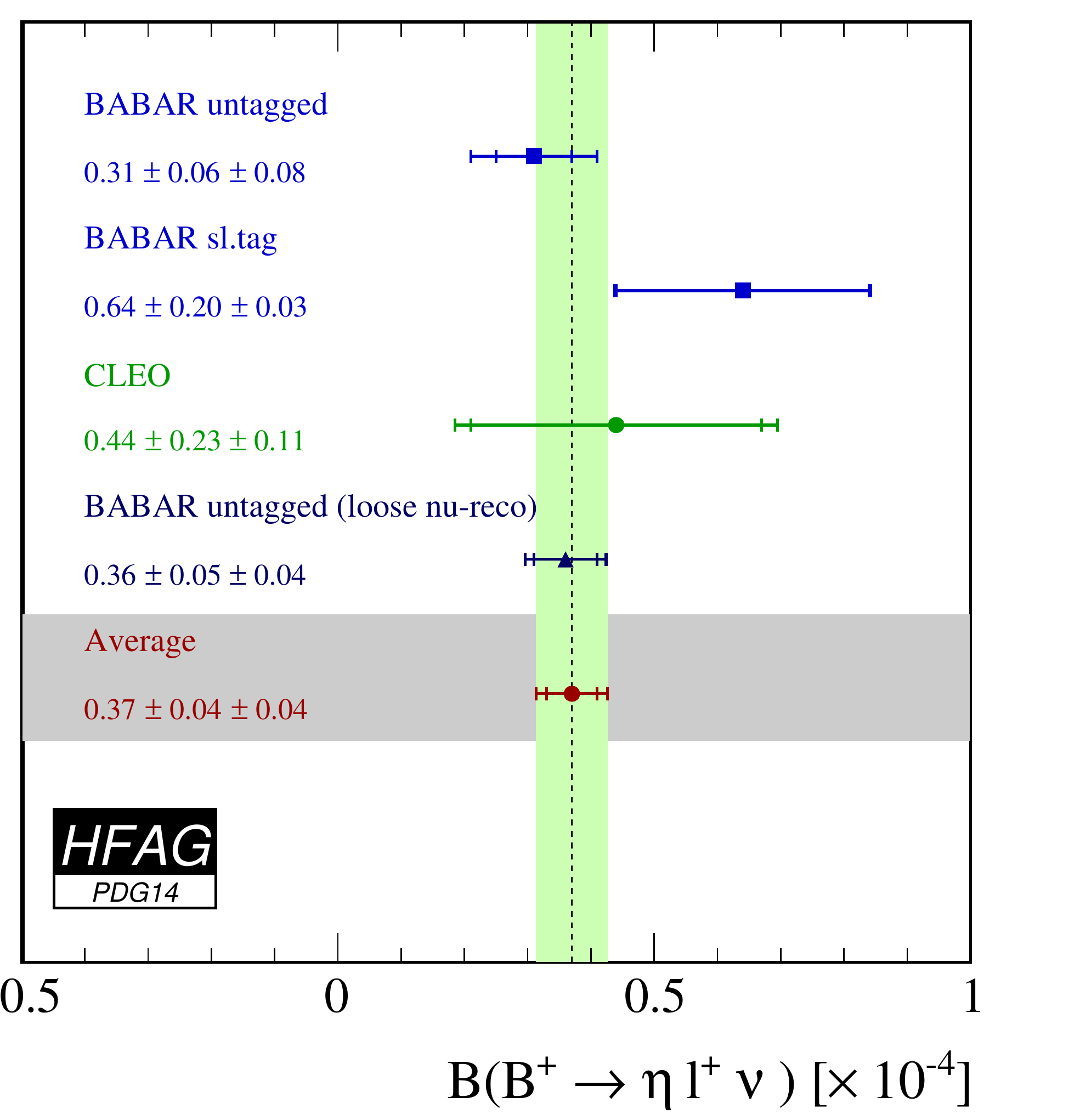}}
   \put( 7.5,  0.0){\includegraphics[width=9.0cm]{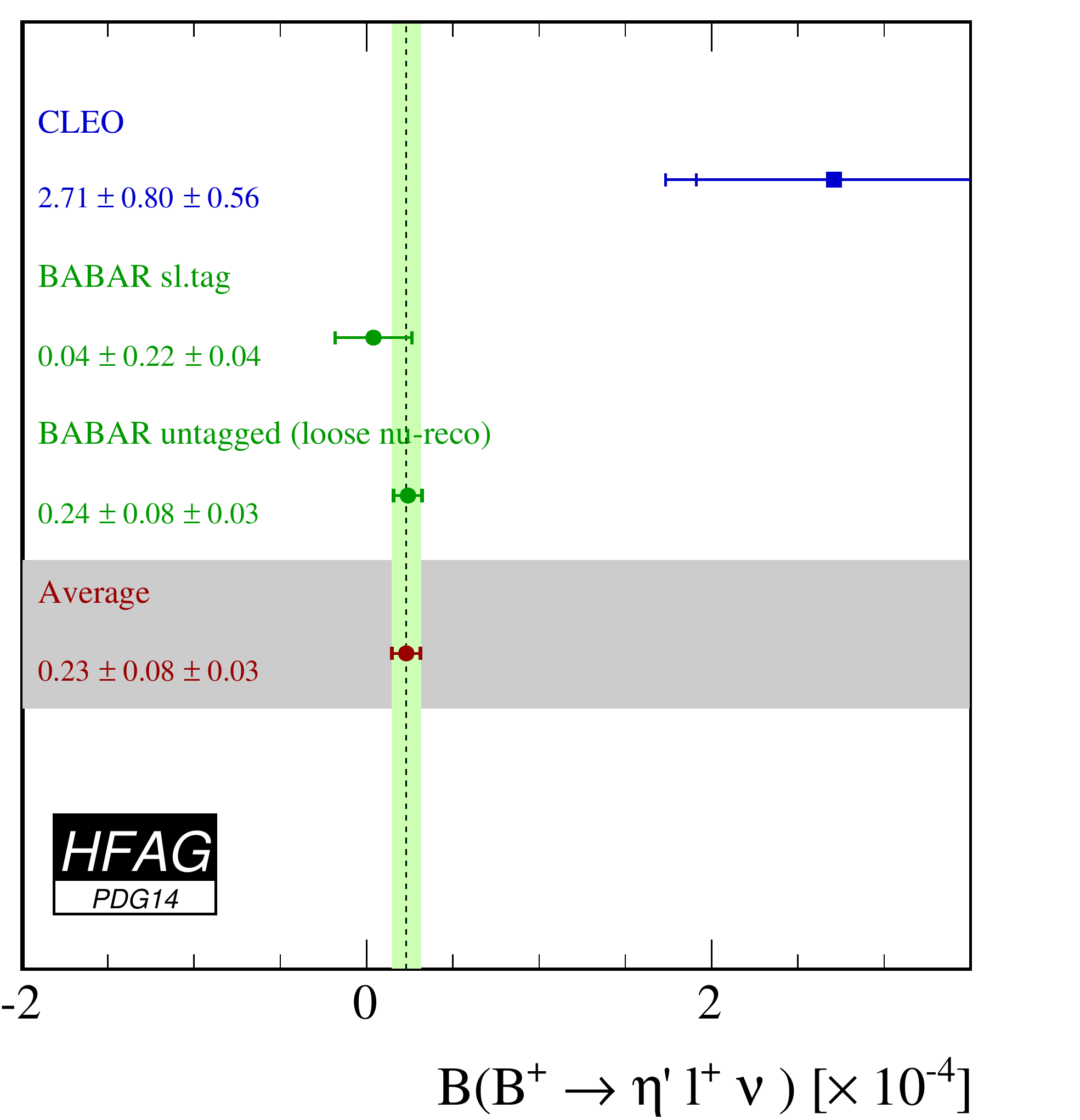}} 
   \put(  5.8,  8.5){{\large\bf a)}}     
   \put( 14.6,  8.5){{\large\bf b)}}
   
   \end{picture} \caption{
(a) Summary of exclusive determinations of $\cbf(\Bb\to\eta\ell\nub)$ and their average.
(b) Summary of exclusive determinations of $\cbf(\Bb\to\eta'\ell\nub)$ and their average.
}
\label{fig:xulnu2}
\end{center}
\end{figure}

We also report the branching fraction average for $\Bb\to\omega\ell\nub$, $\Bb\to\eta\ell\nub$ 
and $\Bb\to\eta'\ell\nub$. The measurements for $\Bb\to\omega\ell\nub$ are reported in Table~\ref{tab:omegalnu} 
and shown in Figure~\ref{fig:xulnu1}, while the ones for $\Bb\to\eta\ell\nub$ and  $\Bb\to\eta'\ell\nub$ are reported in Table~\ref{tab:etalnu} and~\ref{tab:etaprimelnu},  and are shown in Figure~\ref{fig:xulnu2}. 

\begin{table}[!htb]
\begin{center}
\caption{Summary of exclusive determinations of $\cbf(\Bb\to\omega
\ell\nub)$. The errors quoted
correspond to statistical and systematic uncertainties, respectively.}
\label{tab:omegalnu}
\begin{small}
\begin{tabular}{|lc|}
\hline
& $\cbf [10^{-4}]$
\\
\hline\hline
Belle $\omega$~\cite{Schwanda:2004fa}
& $1.30\pm 0.40\pm 0.36\ $
\\
\babar $\omega$~\cite{Lees:2012vv}
& $1.19\pm 0.16\pm 0.09\ $
\\  
\babar $\omega$~\cite{Lees:2012mq}
& $1.21\pm 0.14\pm 0.08\ $
\\  
Belle $\omega$~\cite{Sibidanov:2013rkk}
& $1.07\pm 0.16\pm 0.07 $
\\
\babar $\omega$~\cite{Lees:2013gja}
& $1.35\pm 0.21\pm 0.11\ $
\\  

\hline

{\bf Average}
& \mathversion{bold}$1.19 \pm 0.08 \pm 0.06\ $
\\ 
\hline
\end{tabular}\\
\end{small}
\end{center}
\end{table}

\begin{table}[!htb]
\begin{center}
\caption{Summary of exclusive determinations of $\cbf(\Bb\to\eta
\ell\nub)$. The errors quoted
correspond to statistical and systematic uncertainties, respectively.}
\label{tab:etalnu}
\begin{small}
\begin{tabular}{|lc|}
\hline
& $\cbf [10^{-4}]$
\\
\hline\hline
CLEO $\eta$~\cite{Gray:2007pw}
& $0.44\pm 0.23\pm 0.11\ $
\\
BABAR $\eta$~\cite{Aubert:2008ct}
& $0.31\pm 0.06\pm 0.08\ $
\\ 
BABAR $\eta$~\cite{Aubert:2008bf}
& $0.64\pm 0.20\pm 0.03\ $
\\
BABAR $\eta$~\cite{Lees:2012vv}
& $0.36\pm 0.05\pm 0.04\ $
\\  
 \hline
{\bf Average}
& \mathversion{bold}$0.37 \pm 0.04 \pm 0.04 $
\\ 
\hline
\end{tabular}\\
\end{small}
\end{center}
\end{table}

\begin{table}[!htb]
\begin{center}
\caption{Summary of exclusive determinations of $\cbf(\Bb\to\eta'
\ell\nub)$. The errors quoted
correspond to statistical and systematic uncertainties, respectively.}
\label{tab:etaprimelnu}
\begin{small}
\begin{tabular}{|lc|}
\hline
& $\cbf [10^{-4}]$
\\
\hline\hline
CLEO $\eta'$~\cite{Gray:2007pw}
& $2.71\pm 0.80\pm 0.56\ $
\\
BABAR $\eta'$~\cite{Aubert:2008bf}
& $0.04\pm 0.22\pm 0.04\ $
\\ 
BABAR $\eta'$~\cite{Lees:2012vv}
& $0.24\pm 0.08\pm 0.03\ $
\\  
 \hline
{\bf Average}
& \mathversion{bold}$0.23 \pm 0.08 \pm 0.03 $
\\ 
\hline
\end{tabular}\\
\end{small}
\end{center}
\end{table}

%
\subsection{Inclusive CKM-suppressed decays}
\label{slbdecays_b2uincl}
The large background from $\B\to X_c\ell^+\nul$ decays is the chief
experimental limitation in determinations of $\vub$.  Cuts designed to
reject this background limit the acceptance for $\B\to X_u\ell^+\nul$
decays. The calculation of partial rates for these restricted
acceptances is more complicated and requires substantial theoretical machinery.
In this update, we use several theoretical calculations
to extract \vub. We do not advocate the use of one method over another.
The authors for the different calculations have provided 
codes to compute the partial rates in limited regions of phase space covered by the measurements. 
Latest results by Belle~\cite{ref:belle-multivariate} and \babar~\cite{Lees:2011fv} 
explore bigger and bigger portions of phase space, with a consequent reduction of the theoretical 
uncertainties. 

For the averages we performed, the systematic errors associated with the
modeling of $\B\to X_c\ell^+\nul$ and $\B\to X_u\ell^+\nul$ decays and the theoretical
uncertainties are taken as fully correlated among all measurements.
Reconstruction-related uncertainties are taken as fully correlated within a given experiment.
We use all results published by \babar\ in Ref.~\cite{Lees:2011fv}, since the 
statistical correlations are given. 
To make use of the theoretical calculations of Ref.~\cite{ref:BLL}, we restrict the
kinematic range in $M_X$ and $q^2$, thereby reducing the size of the data
sample significantly, but also the theoretical uncertainty, as stated by the
authors~\cite{ref:BLL}.
The dependence of the quoted error on the measured value for each source of error
is taken into account in the calculation of the averages.
Measurements of partial branching fractions for $\B\to X_u\ell^+\nul$
transitions from $\Upsilon(4S)$ decays, together with the corresponding accepted region, 
are given in Table~\ref{tab:BFbulnu}.  
The signal yields for all the measurements shown in Table~\ref{tab:BFbulnu}
are not rescaled to common input values of the $B$ meson lifetime (see
Sec.~\ref{sec:life_mix}) and the semileptonic width~\cite{PDG_2014}.

It has been first suggested by Neubert~\cite{Neubert:1993um} and later detailed by Leibovich, 
Low, and Rothstein (LLR)~\cite{Leibovich:1999xf} and Lange, Neubert and Paz (LNP)~\cite{Lange:2005qn}, 
that the uncertainty of
the leading shape functions can be eliminated by comparing inclusive rates for
$\B\to X_u\ell^+\nul$ decays with the inclusive photon spectrum in $\B\to X_s\gamma$,
based on the assumption that the shape functions for transitions to light
quarks, $u$ or $s$, are the same to first order.
However, shape function uncertainties are only eliminated at the leading order
and they still enter via the signal models used for the determination of efficiency. 
For completeness, we provide a comparison of the results using 
calculations with reduced dependence on the shape function, as just
introduced, with our averages using different theoretical approaches.
Results are presented by \babar\ in Ref.\cite{Aubert:2006qi} using the LLR prescription. 
In another work (Ref.~\cite{Golubev:2007cs}), \vub\ was extracted from the 
endpoint spectrum of $\B\to X_u\ell^+\nul$ from \babar~\cite{ref:babar-endpoint}, 
using several theoretical approaches with reduced dependence on the shape function.
In both cases, the photon energy spectrum in the 
rest frame of the $B$-meson by \babar~\cite{Aubert:2005cua} has been used.

\begin{table}[!htb]
\caption{\label{tab:BFbulnu}
Summary of inclusive determinations of partial branching
fractions for $B\rightarrow X_u \ell^+ \nu_{\ell}$ decays.
The errors quoted on $\Delta\cbf$ correspond to
statistical and systematic uncertainties.
The $s_\mathrm{h}^{\mathrm{max}}$ variable is described in Refs.~\cite{ref:shmax,ref:babar-elq2}. }
\begin{center}
\begin{small}
\begin{tabular}{|llcl|}
\hline
Measurement & Accepted region &  $\Delta\cbf [10^{-4}]$ & Notes\\
\hline\hline
CLEO~\cite{ref:cleo-endpoint}
& $E_e>2.1\,\gev$ & $3.3\pm 0.2\pm 0.7$ &  \\ 
\babar~\cite{ref:babar-elq2}
& $E_e>2.0\,\gev$, $s_\mathrm{h}^{\mathrm{max}}<3.5\,\mathrm{GeV^2}$ & $4.0\pm 0.2\pm 0.3$ & \\
\babar~\cite{ref:babar-endpoint}
& $E_e>2.0\,\gev$  & $5.7\pm 0.4\pm 0.5$ & \\
Belle~\cite{ref:belle-endpoint}
& $E_e>1.9\,\gev$  & $8.5\pm 0.4\pm 1.5$ & \\
\babar~\cite{Lees:2011fv}
& $M_X<1.7\,\gev/c^2, q^2>8\,\gev^2/c^2$ & $6.9\pm 0.6\pm 0.4$ & 
\\
Belle~\cite{ref:belle-mxq2Anneal}
& $M_X<1.7\,\gev/c^2, q^2>8\,\gev^2/c^2$ & $7.4\pm 0.9\pm 1.3$ & \\
Belle~\cite{ref:belle-mx}
& $M_X<1.7\,\gev/c^2, q^2>8\,\gev^2/c^2$ & $8.5\pm 0.9\pm 1.0$ & used only in BLL average\\
\babar~\cite{Lees:2011fv}
& $P_+<0.66\,\gev$  & $9.9\pm 0.9\pm 0.8 $ & 
\\
\babar~\cite{Lees:2011fv}
& $M_X<1.7\,\gev/c^2$ & $11.6\pm 1.0\pm 0.8 $ &
\\ 
\babar~\cite{Lees:2011fv}
& $M_X<1.55\,\gev/c^2$ & $10.9\pm 0.8\pm 0.6 $ & 
\\ 
Belle~\cite{ref:belle-multivariate}
& $p^*_{\ell} > 1 \gev/c$ & $19.6\pm 1.7\pm 1.6$ & \\
\babar~\cite{Lees:2011fv}
& ($M_X, q^2$) fit, $p^*_{\ell} > 1 \gev/c$  & $18.2\pm 1.3\pm 1.5$ & 
\\ 
\babar~\cite{Lees:2011fv}
& $p^*_{\ell} > 1.3 \gev/c$  & $15.5\pm 1.3\pm 1.4$ & 
\\ \hline
\end{tabular}\\
\end{small}
\end{center}
\end{table}

\subsubsection{BLNP}
\begin{figure}[!ht]
 \begin{center}
  \unitlength1.0cm 
  \begin{picture}(14.,10.0)  
   \put( -1.5,  0.0){\includegraphics[width=9.4cm]{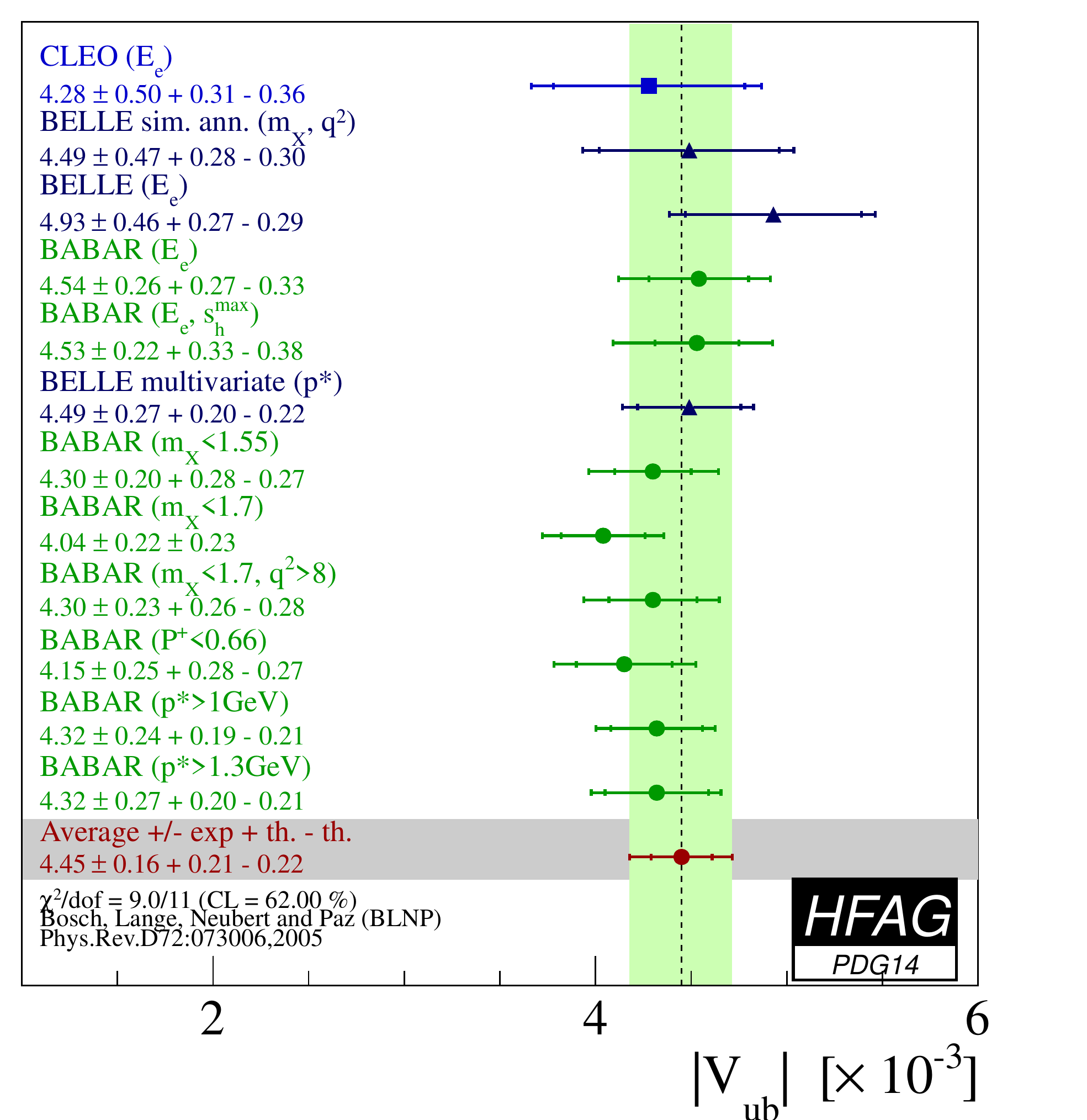}
   }
   \put(  7.4,  0.0){\includegraphics[width=9.2cm]{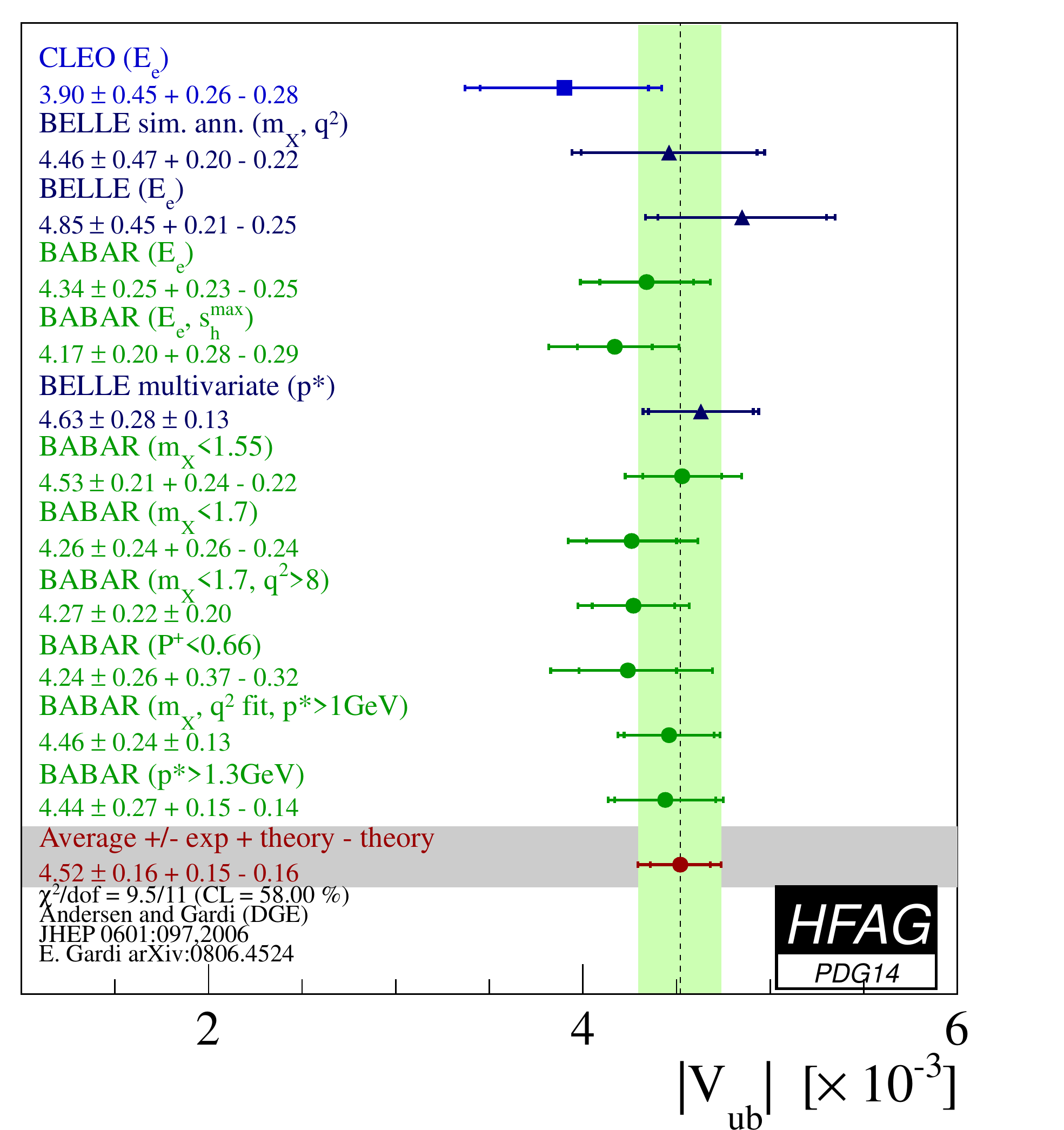}
   }
   \put(  6.0,  8.7){{\large\bf a)}}
   \put( 14.8,  8.7){{\large\bf b)}}
  \end{picture}
  \caption{Measurements of $\vub$ from inclusive semileptonic decays 
and their average based on the BLNP (a) and DGE (b) prescription. The
labels indicate the distributions and selections used to define the
signal regions in the different analyses, where $E_e$ is the electron
energy in the $B$~rest frame, $p^*$ the lepton momentum in the
$B$~frame and $m_X$ is the invariant mass of the hadronic system. The
light-cone momentum $P_+$ is defined in the $B$ rest frame as
$P_+=E_X-|\vec p_X|$.} \label{fig:BLNP_DGE}
 \end{center}
\end{figure}

Bosch, Lange, Neubert and Paz (BLNP)~\cite{ref:BLNP,
  ref:Neubert-new-1,ref:Neubert-new-2,ref:Neubert-new-3}
provide theoretical expressions for the triple
differential decay rate for $B\to X_u \ell^+ \nul$ events, incorporating all known
contributions, whilst smoothly interpolating between the 
``shape-function region'' of large hadronic
energy and small invariant mass, and the ``OPE region'' in which all
hadronic kinematical variables scale with the $b$-quark mass. BLNP assign
uncertainties to the $b$-quark mass which enters through the leading shape function, 
to sub-leading shape function forms, to possible weak annihilation
contribution, and to matching scales. 
The BLNP calculation uses the shape function renormalization scheme; the heavy quark parameters determined  
from the global fit in the kinetic scheme, described in \ref{globalfitsKinetic}, were therefore 
translated into the shape function scheme by using a prescription by Neubert 
\cite{Neubert:2004sp,Neubert:2005nt}. The resulting parameters are 
$m_b(SF)=(4.569 \pm 0.023 \pm 0.018)$ GeV, 
$\mu_\pi^2(SF)=(0.145 \pm 0.089 ^{+0.020}_{-0.040})$ GeV$^2$, 
where the second uncertainty is due to the scheme translation. 
The extracted values of \vub\, for each measurement along with their average are given in
Table~\ref{tab:bulnu} and illustrated in Figure~\ref{fig:BLNP_DGE}(a). 
The total uncertainty is $^{+5.8}_{-5.9}\%$ and is due to:
statistics ($^{+2.0}_{-2.1}\%$),
detector ($^{+1.7}_{-1.8}\%$),
$B\to X_c \ell^+ \nul$ model ($^{+1.2}_{-1.2}\%$),
$B\to X_u \ell^+ \nul$ model ($^{+1.9}_{-1.8}\%$),
heavy quark parameters ($^{+2.6}_{-2.7}\%$),
SF functional form ($^{+0.1}_{-0.3}\%$),
sub-leading shape functions ($^{+0.6}_{-0.6}\%$),
BLNP theory: matching scales $\mu,\mu_i,\mu_h$ ($^{+3.8}_{-3.8}\%$), and
weak annihilation ($^{+0.0}_{-1.4}\%$).
The error on the HQE parameters ($b$-quark mass and $\mu_\pi^2)$ 
is the source of the largest uncertainty, while the
uncertainty assigned for the matching scales is a close second. The uncertainty due to 
weak annihilation has been assumed to be asymmetric, \ie\ it only tends to decrease \vub.

\begin{table}[!htb]
\caption{\label{tab:bulnu}
Summary of input parameters used by the different theory calculations,
corresponding inclusive determinations of $\vub$ and their average.
The errors quoted on \vub\ correspond to
experimental and theoretical uncertainties, respectively.}
\begin{center}
\resizebox{0.99\textwidth}{!}{
\begin{tabular}{|lccccc|}
\hline
 & BLNP &DGE & GGOU & ADFR &BLL \\
\hline\hline
\multicolumn{6}{|c|}{Input parameters}\\ \hline
scheme & SF           & $\overline{MS}$ & kinetic &  $\overline{MS}$ & $1S$ \\ 
Ref.       & \cite{Neubert:2004sp,Neubert:2005nt} & Ref.~\cite{PDG_2010} & 
see Sec.~\ref{globalfitsKinetic}  & Ref.~\cite{PDG_2010} & Ref.~\cite{Barberio:2008fa} \\
$m_b$ (GeV)           & 4.569 $\pm$ 0.025 & 4.177 $\pm 0.043$ & 4.541 $\pm 0.023$ & 4.177 $\pm 0.043$ & 4.704 $\pm 0.029$ \\
$\mu_\pi^2$ (GeV$^2$) & 0.145 $^{+0.091}_{-0.097}$ & -                 & 0.414 $\pm 0.078$ & - &  - \\
\hline\hline
Ref. & \multicolumn{5}{c|}{$|V_{ub}|$ values}\\ 
\hline
$E_e$~\cite{ref:cleo-endpoint} &
$4.28\pm 0.50 ^{+0.31}_{-0.36}$ &
$3.90\pm 0.45 ^{+0.26}_{-0.28}$ &
$4.21\pm 0.49 ^{+0.23}_{-0.33}$ &
$3.44\pm 0.40 ^{+0.16}_{-0.16}$ &
- \\

$M_X, q^2$~\cite{ref:belle-mxq2Anneal}&
$4.49\pm 0.47 ^{+0.28}_{-0.30}$ &
$4.46\pm 0.47 ^{+0.20}_{-0.22}$ &
$4.50\pm 0.47 ^{+0.28}_{-0.31}$ &
$3.94\pm 0.41 ^{+0.17}_{-0.17}$ &
$4.68\pm 0.49 ^{+0.30}_{-0.30}$ \\

$E_e$~\cite{ref:belle-endpoint}&
$4.93\pm 0.46 ^{+0.27}_{-0.29}$ &
$4.85\pm 0.45 ^{+0.21}_{-0.25}$ &
$4.93\pm 0.46 ^{+0.17}_{-0.22}$ &
$4.50\pm 0.42 ^{+0.20}_{-0.20}$ &
-\\

$E_e$~\cite{ref:babar-endpoint}&
$4.54\pm 0.26 ^{+0.27}_{-0.33}$ &
$4.34\pm 0.25 ^{+0.23}_{-0.25}$ &
$4.50\pm 0.26 ^{+0.18}_{-0.25}$ &
$3.94\pm 0.22 ^{+0.20}_{-0.19}$ &
-\\

$E_e,s_\mathrm{h}^{\mathrm{max}}$~\cite{ref:babar-elq2}&
$4.53\pm 0.22 ^{+0.33}_{-0.38}$ &
$4.17\pm 0.20 ^{+0.28}_{-0.29}$ &
- &
$3.64\pm 0.18 ^{+0.17}_{-0.17}$ &
 \\ 

$p^*_{\ell}$~\cite{ref:belle-multivariate}&
$4.49\pm 0.27 ^{+0.20}_{-0.22}$ &
$4.63\pm 0.28 ^{+0.13}_{-0.13}$ &
$4.60\pm 0.27 ^{+0.10}_{-0.11}$ &
$4.52\pm 0.30 ^{+0.19}_{-0.19}$ &
- \\

$M_X$~\cite{Lees:2011fv}&
$4.30\pm 0.20 ^{+0.28}_{-0.27}$ &
$4.53\pm 0.21 ^{+0.24}_{-0.22}$ &
$4.29\pm 0.20 ^{+0.21}_{-0.22}$ &
$3.84\pm 0.18 ^{+0.19}_{-0.19}$ &
- \\
$M_X$~\cite{Lees:2011fv}&
$4.04\pm 0.22 ^{+0.23}_{-0.23}$ &
$4.26\pm 0.24 ^{+0.26}_{-0.24}$ &
$4.09\pm 0.23 ^{+0.18}_{-0.19}$ &
$3.76\pm 0.21 ^{+0.18}_{-0.17}$ &
- \\

$M_X,q^2$~\cite{Lees:2011fv}&
$4.30\pm 0.23 ^{+0.26}_{-0.28}$  &
$4.27\pm 0.22 ^{+0.20}_{-0.20}$  &
$4.32\pm 0.23 ^{+0.27}_{-0.30}$  &
$3.76\pm 0.20 ^{+0.17}_{-0.16}$  &
$4.50\pm 0.24 ^{+0.29}_{-0.29}$ \\

$P_+$~\cite{Lees:2011fv}&
$4.15\pm 0.25 ^{+0.28}_{-0.27}$  &
$4.24\pm 0.26 ^{+0.37}_{-0.32}$  &
$4.24\pm 0.26 ^{+0.32}_{-0.32}$  &
$3.59\pm 0.22 ^{+0.19}_{-0.18}$  &
- \\

$p^*_{\ell}$, $(M_X,q^2)$ fit~\cite{Lees:2011fv}&
$4.32\pm 0.24 ^{+0.19}_{-0.21}$  &
$4.46\pm 0.24 ^{+0.13}_{-0.13}$  &
$4.42\pm 0.24 ^{+0.09}_{-0.11}$  &
$4.35\pm 0.24 ^{+0.18}_{-0.18}$  &
- \\

$p^*_{\ell}$~\cite{Lees:2011fv}&
$4.32\pm 0.27 ^{+0.20}_{-0.21}$  &
$4.44\pm 0.27 ^{+0.15}_{-0.14}$  &
$4.41\pm 0.27 ^{+0.10}_{-0.12}$  &
$4.30\pm 0.27 ^{+0.19}_{-0.18}$  &
- \\

$M_X,q^2$~\cite{ref:belle-mx}&
- &
- &
- &
- &
$5.01\pm 0.39 ^{+0.32}_{-0.32}$ \\
\hline
Average &
$4.45\pm 0.16 ^{+0.21}_{-0.22}$ &
$4.52\pm 0.16 ^{+0.15}_{-0.16}$ &
$4.51\pm 0.16 ^{+0.12}_{-0.15}$ &
$4.05\pm 0.13 ^{+0.18}_{-0.11}$ &
$4.62\pm 0.20 ^{+0.29}_{-0.29}$ \\
\hline
\end{tabular}
}
\end{center}
\end{table}


\begin{figure}[!ht]
 \begin{center}
  \unitlength1.0cm 
  \begin{picture}(14.,10.0)  
   \put( -1.5,  0.0){\includegraphics[width=9.4cm]{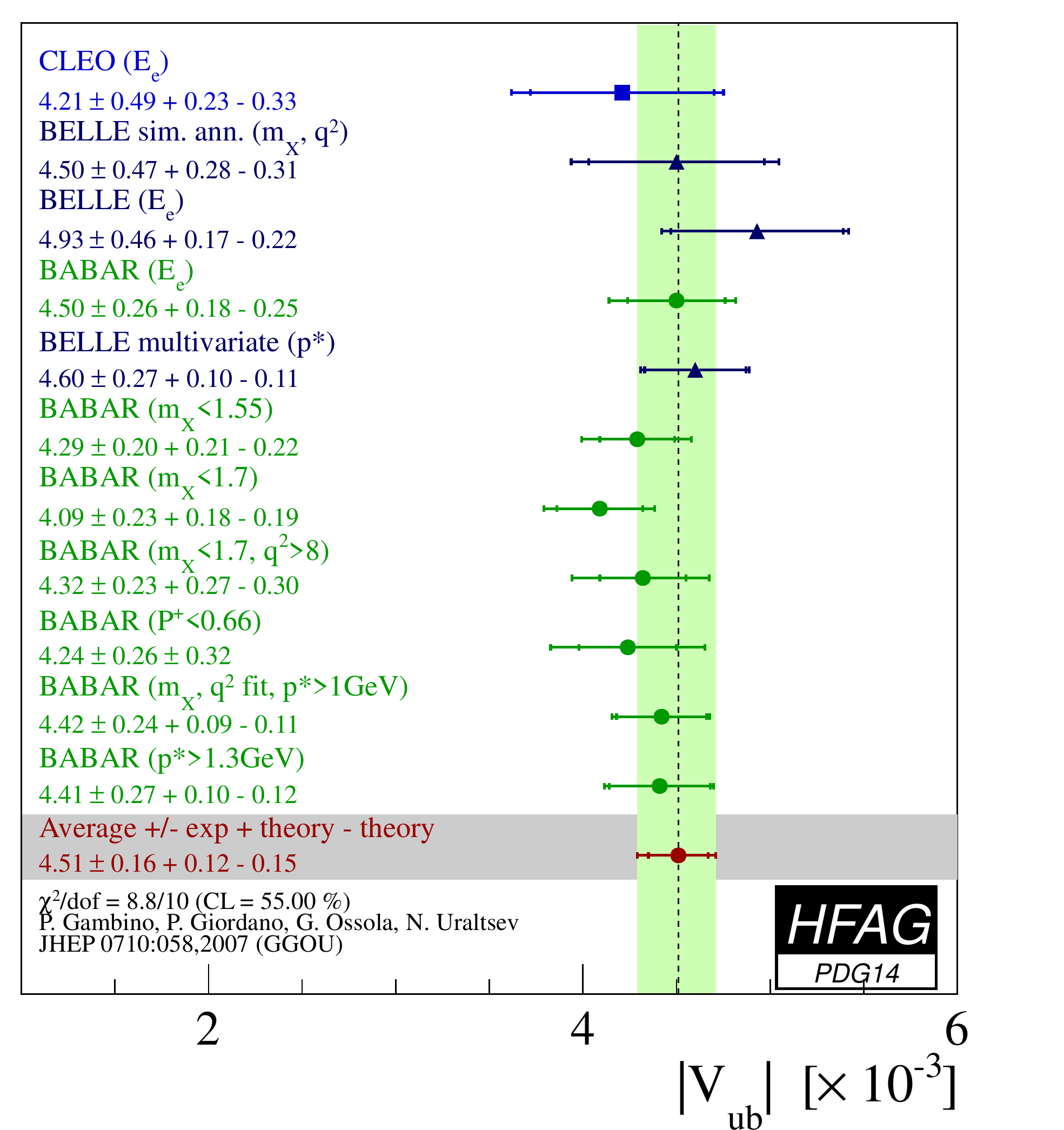}
   }
   \put(  7.4,  0.0){\includegraphics[width=9.4cm]{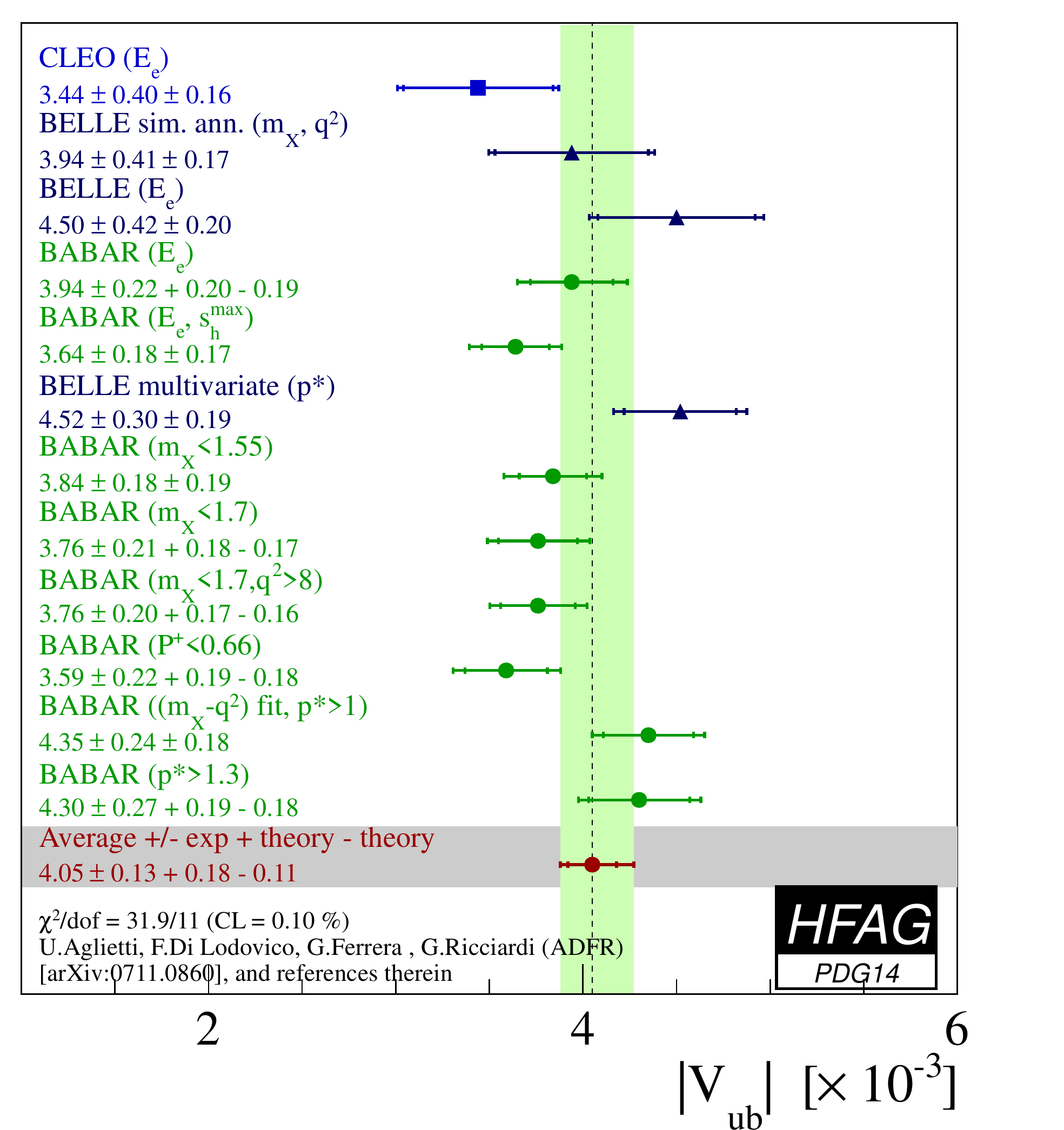}
   }
   \put(  6.0,  8.7){{\large\bf a)}}
   \put( 14.8,  8.7){{\large\bf b)}}
  \end{picture}
  \caption{Measurements of $\vub$ from inclusive semileptonic decays 
and their average based on the GGOU (a) and ADFR (b) prescription. The
labels indicate the distributions and selections used to define the
signal regions in the different analyses, where $E_e$ is the electron
energy in the $B$~rest frame, $p^*$ the lepton momentum in the
$B$~frame and $m_X$ is the invariant mass of the hadronic system. The
light-cone momentum $P_+$ is defined in the $B$ rest frame as
$P_+=E_X-|\vec p_X|$.} \label{fig:GGOU_ADFR}
 \end{center}
\end{figure}

\subsubsection{DGE}
J.R.~Andersen and E.~Gardi (Dressed Gluon Exponentiation, DGE)~\cite{ref:DGE} provide
a framework where the on-shell $b$-quark calculation, converted into hadronic variables, is
directly used as an approximation to the meson decay spectrum without
the use of a leading-power non-perturbative function (or, in other words,
a shape function). The on-shell mass of the $b$-quark within the $B$-meson ($m_b$) is
required as input. 
The DGE calculation uses the $\overline{MS}$ renormalization scheme; the heavy quark parameters determined  
from the global fit in the kinetic scheme, described in \ref{globalfitsKinetic}, were therefore 
translated into the $\overline{MS}$ scheme by using a calculation by Gardi, giving 
$m_b({\overline{MS}})=(4.177 \pm 0.043)$ GeV.
The extracted values
of \vub\, for each measurement along with their average are given in
Table~\ref{tab:bulnu} and illustrated in Figure~\ref{fig:BLNP_DGE}(b).
The total error is $^{+4.8}_{-5.0}\%$, whose breakdown is:
statistics ($^{+1.8}_{-1.8}\%$),
detector ($^{+1.7}_{-1.7}\%$),
$B\to X_c \ell^+ \nul$ model ($^{+1.3}_{-1.3}\%$),
$B\to X_u \ell^+ \nul$ model ($^{+2.1}_{-1.9}\%$),
strong coupling $\alpha_s$ ($^{+0.5}_{-0.5}\%$),
$m_b$ ($^{+3.2}_{-3.0}\%$),
weak annihilation ($^{+0.0}_{-1.9}\%$),
DGE theory: matching scales ($^{+0.5}_{-0.3}\%$).
The largest contribution to the total error is due to the effect of the uncertainty 
on $m_b$. 
The uncertainty due to 
weak annihilation has been assumed to be asymmetric, \ie\ it only tends to decrease \vub.


\subsubsection{GGOU}
Gambino, Giordano, Ossola and Uraltsev (GGOU)~\cite{Gambino:2007rp} 
compute the triple differential decay rates of $B \to X_u \ell^+ \nul$, 
including all perturbative and non--perturbative effects through $O(\alphas^2 \beta_0)$ 
and $O(1/m_b^3)$. 
The Fermi motion is parameterized in terms of a single light--cone function 
for each structure function and for any value of $q^2$, accounting for all subleading effects. 
The calculations are performed in the kinetic scheme, a framework characterized by a Wilsonian 
treatment with a hard cutoff $\mu \sim 1 $ GeV.
GGOU have not included calculations for the ``($E_e,s^{max}_h$)'' analysis. 
The heavy quark parameters determined  
from the global fit in the kinetic scheme, described in \ref{globalfitsKinetic}, are used as inputs: 
$m_b(kin)=(4.541 \pm 0.023)$ GeV, 
$\mu_\pi^2(kin)=(0.414 \pm 0.078)$ GeV$^2$. 
The extracted values
of \vub\, for each measurement along with their average are given in
Table~\ref{tab:bulnu} and illustrated in Figure~\ref{fig:GGOU_ADFR}(a).
The total error is $^{+4.3}_{-4.8}\%$ whose breakdown is:
statistics ($^{+1.9}_{-1.9}\%$),
detector ($^{+1.7}_{-1.7}\%$),
$B\to X_c \ell^+ \nul$ model ($^{+1.3}_{-1.3}\%$),
$B\to X_u \ell^+ \nul$ model ($^{+1.9}_{-1.9}\%$),
$\alpha_s$, $m_b$ and other non--perturbative parameters ($^{+1.6}_{-1.6}\%$), 
higher order perturbative and non--perturbative corrections ($^{+1.5}_{-1.5}\%$), 
modelling of the $q^2$ tail
($^{+1.4}_{-1.4}\%$), 
weak annihilations matrix element ($^{+0.0}_{-2.0}\%$), 
functional form of the distribution functions ($^{+0.2}_{-0.2}\%$), 
The leading uncertainties
on  \vub\ are both from theory, and are due to perturbative and non--perturbative
parameters and the modelling of the $q^2$ tail.
The uncertainty due to 
weak annihilation has been assumed to be asymmetric, \ie\ it only tends to decrease \vub.


\subsubsection{ADFR}
Aglietti, Di Lodovico, Ferrera and Ricciardi (ADFR)~\cite{Aglietti:2007ik}
use an approach to extract \vub, which makes use of the ratio
of the  $B \to X_c \ell^+ \nul$ and $B \to X_u \ell^+ \nul$ widths. 
The normalized triple differential decay rate for 
$B \to X_u \ell^+ \nul$~\cite{Aglietti:2006yb,Aglietti:2005mb, Aglietti:2005bm, Aglietti:2005eq}
is calculated with a model based on (i) soft--gluon resummation 
to next--to--next--leading order and (ii) an effective QCD coupling without
Landau pole. This coupling is constructed by means of an extrapolation to low
energy of the high--energy behaviour of the standard coupling. More technically,
an analyticity principle is used.
The lower cut on the electron energy for the endpoint analyses is 2.3~GeV~\cite{Aglietti:2006yb}.
The ADFR calculation uses the $\overline{MS}$ renormalization scheme; the heavy quark parameters determined  
from the global fit in the kinetic scheme, described in \ref{globalfitsKinetic}, were therefore 
translated into the $\overline{MS}$ scheme by using a calculation by Gardi, giving 
$m_b({\overline{MS}})=(4.177 \pm 0.043)$ GeV.
The extracted values
of \vub\, for each measurement along with their average are given in
Table~\ref{tab:bulnu} and illustrated in Figure~\ref{fig:GGOU_ADFR}(b).
The total error is $^{+5.4}_{-5.3}\%$ whose breakdown is:
statistics ($^{+1.8}_{-1.8}\%$),
detector ($^{+1.8}_{-2.0}\%$),
$B\to X_c \ell^+ \nul$ model ($^{+1.4}_{-1.4}\%$),
$B\to X_u \ell^+ \nul$ model ($^{+1.3}_{-1.3}\%$),
$\alpha_s$ ($^{+1.1}_{-1.0}\%$), 
$|V_{cb}|$ ($^{+2.0}_{-2.0}\%$), 
$m_b$ ($^{+0.7}_{-0.7}\%$), 
$m_c$ ($^{+0.4}_{-0.7}\%$), 
semileptonic branching fraction ($^{+0.8}_{-0.7}\%$), 
theory model ($^{+3.5}_{-3.5}\%$).
The leading
uncertainty, from theory, is due to the theory model.


\subsubsection{BLL}
\begin{figure}
\begin{center}
\includegraphics[width=0.52\textwidth]{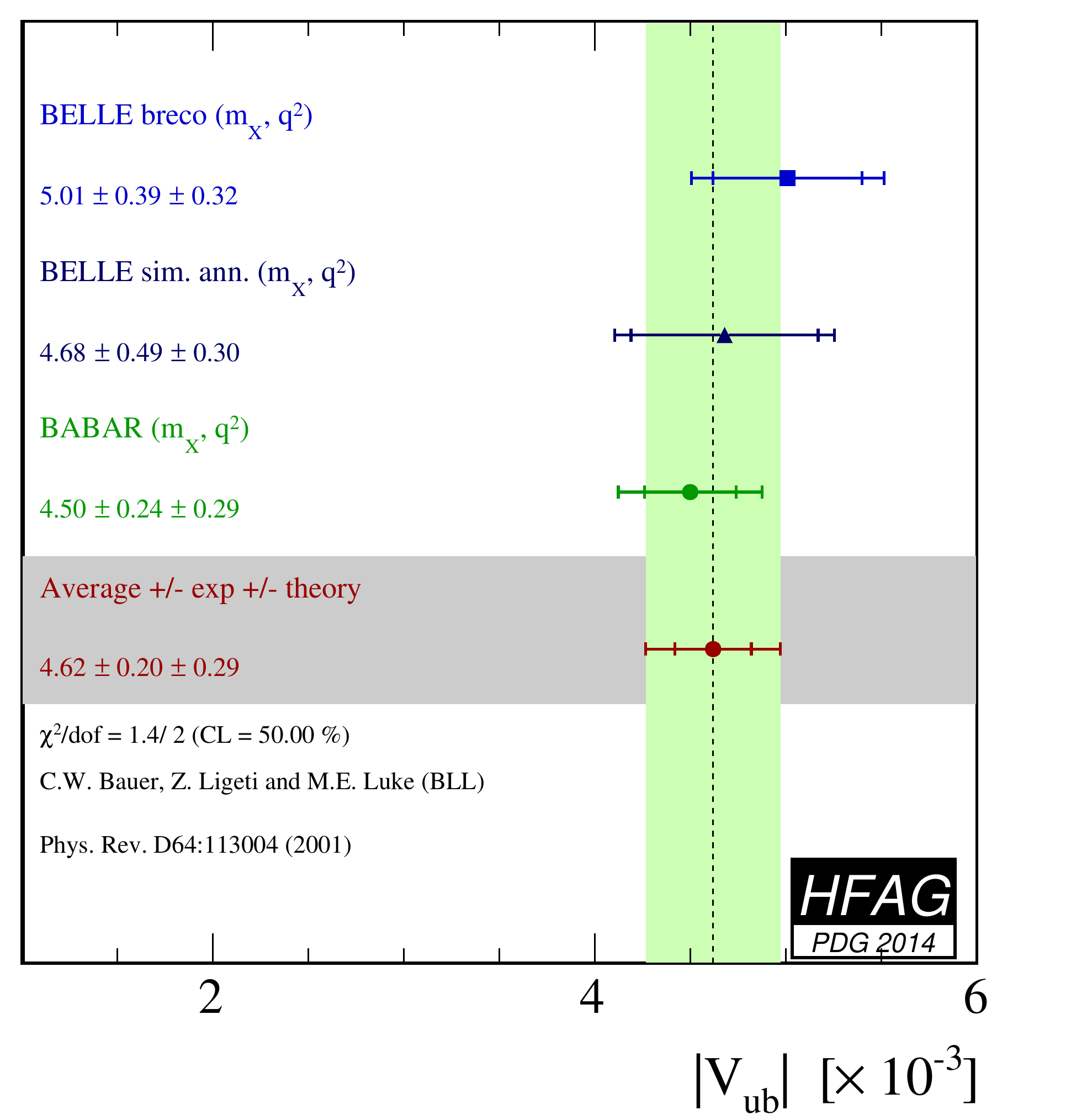}
\end{center}
\caption{Measurements of $\vub$ from inclusive semileptonic decays 
and their average in the BLL prescription.
``$(M_X, q^2)$'' indicates the analysis type.}
\label{fig:BLL}
\end{figure}

Bauer, Ligeti, and Luke (BLL)~\cite{ref:BLL} give a
HQET-based prescription that advocates combined cuts on the dilepton invariant mass, $q^2$,
and hadronic mass, $m_X$, to minimise the overall uncertainty on \vub.
In their reckoning a cut on $m_X$ only, although most efficient at
preserving phase space ($\sim$80\%), makes the calculation of the partial
rate untenable due to uncalculable corrections
to the $b$-quark distribution function or shape function. These corrections are
suppressed if events in the low $q^2$ region are removed. The cut combination used
in measurements is $M_x<1.7$ GeV/$c^2$ and $q^2 > 8$ GeV$^2$/$c^2$.  
The extracted values
of \vub\, for each measurement along with their average are given in
Table~\ref{tab:bulnu} and illustrated in Figure~\ref{fig:BLL}.
The total error is $^{+7.7}_{-7.7}\%$ whose breakdown is:
statistics ($^{+3.3}_{-3.3}\%$),
detector ($^{+3.0}_{-3.0}\%$),
$B\to X_c \ell^+ \nul$ model ($^{+1.6}_{-1.6}\%$),
$B\to X_u \ell^+ \nul$ model ($^{+1.1}_{-1.1}\%$),
spectral fraction ($m_b$) ($^{+3.0}_{-3.0}\%$),
perturbative : strong coupling $\alpha_s$ ($^{+3.0}_{-3.0}\%$),
residual shape function ($^{+2.5}_{-2.5}\%$),
third order terms in the OPE ($^{+4.0}_{-4.0}\%$),
The leading
uncertainties, both from theory, are due to residual shape function
effects and third order terms in the OPE expansion. The leading
experimental uncertainty is due to statistics. 

\subsubsection{Summary}
A summary of the averages presented in several different
frameworks and results by V.B.~Golubev, V.G.~Luth and Yu.I.~Skovpen~\cite{Golubev:2007cs},
based on prescriptions by LLR~\cite{Leibovich:1999xf} and LNP~\cite{Lange:2005qn} 
to reduce the leading shape function uncertainties are presented in 
Table~\ref{tab:vubcomparison}.
A value judgement based on a direct comparison should be
avoided at the moment, experimental and theoretical uncertainties play out
differently between the schemes and the theoretical assumptions for the
theory calculations are different.

\begin{table}[!htb]
\caption{\label{tab:vubcomparison}
Summary of inclusive determinations of $\vub$.
The errors quoted on \vub\ correspond to experimental and theoretical uncertainties, except for the last two 
measurements where the errors are due to the \babar\ endpoint analysis, the \babar $b\to s\gamma$ analysis~\cite{Aubert:2006qi}, 
the theoretical errors and $V_{ts}$ for the last averages. 
}
\begin{center}
\begin{small}
\begin{tabular}{|lc|}
\hline
Framework
&  $\Vub [10^{-3}]$\\
\hline\hline
BLNP
& $4.45 \pm 0.15 ^{+0.20}_{-0.21}$ \\ 
DGE
& $4.52 \pm 0.16 ^{+0.15}_{-0.16}$ \\
GGOU
& $4.51 \pm 0.16 ^{+0.12}_{-0.15}$ \\
ADFR
& $4.05 \pm 0.13 ^{+0.18}_{-0.11}$ \\
BLL ($m_X/q^2$ only)
& $4.62 \pm 0.20 \pm 0.29$ \\ 
LLR (\babar)~\cite{Aubert:2006qi}
& $4.43 \pm 0.45 \pm 0.29$ \\
LLR (\babar)~\cite{Golubev:2007cs}
& $4.28 \pm 0.29 \pm 0.29 \pm 0.26 \pm0.28$ \\
LNP (\babar)~\cite{Golubev:2007cs}
& $4.40 \pm 0.30 \pm 0.41 \pm 0.23$ \\
\hline
\end{tabular}\\
\end{small}
\end{center}
\end{table}

%
\subsection{Summary of the $B\to D^{(*)}\tau \nu_\tau$ decays}
\label{slbdecays_b2dtaunu}

The leptonic and semileptonic decays with $\tau$ in the final state 
are probes of physics beyond the SM. In the SM these decays proceed via the $W$ emission
diagrams. In models with extended Higgs sectors, such as the Two Higgs Doublet Models (2HDM)
or the MSSM, charged Higgs can contribute to the decay amplitude at the tree level. These further contributions
can affect the branching fraction. Compared to $B^+\to\tau\nu_\tau$, the $B\to D^{(*)}\tau \nu_\tau$ decay
has advantages: the branching fraction is relatively high, because it is not Cabibbo-suppressed, and because
it is a three-body decay, many observables beside the branching fraction can be studied, such as the $D^*$
polarisation, or the $q^2$ distribution (see Ref.~\cite{Duraisamy:2014sna} and reference therein for recent 
calculations).

The $B^0\to D^{*+}\tau\nu_\tau$ decay was first observed by Belle~\cite{Matyja:2007kt} performing 
an inclusive reconstruction of the $B_{tag}$ candidates using all the particles that remain after the selection
of the $B_{sig}$ decay products. Since than, both \babar and Belle have published improved measurements 
and have found evidence for the $B\to D\tau\nu_\tau$ decays~\cite{Aubert:2007dsa,Adachi:2009qg,Bozek:2010xy}. 
The most powerful way to study these decays is the full hadronic $B_{tag}$ technique widely used by both \babar and Belle.

Using the full dataset and an improved $B_{tag}$ selection, \babar measured~\cite{Lees:2012xj} the ratios:  
\begin{eqnarray}
{\cal R}(D)&=&\dfrac{ {\cal B}(B\to D\tau\nu_\tau) }{ {\cal B}(B\to D\ell\nu_\ell) }=0.440\pm 0.058\pm 0.042 \\
{\cal R}(D^*)&=&\dfrac{ {\cal B}(B\to D^*\tau\nu_\tau) }{ {\cal B}(B\to D^*\ell\nu_\ell) }=0.332\pm 0.024\pm 0.018 %
\end{eqnarray}
where $\ell=e,\mu$ and the $B^0$ and $B^+$ are combined in a isospin-constrained fit. The ratios ${\cal R}(D)$
and ${\cal R}(D^*)$ are independent of $|V_{cb}|$ and to a large extent of the parametrizations of the 
form factors, so the SM predictions for these ratios are quite precise, ${\cal R}(D)=0.297\pm 0.017$ and
 ${\cal R}(D)=0.252\pm 0.003$ (results obtained in Ref.\cite{Lees:2012xj,Lees:2013uzd} 
 updating the calculations in Ref.~\cite{Kamenik:2008tj,Fajfer:2012vx}
 with the recent $B\to D^{(*)}$ measurements from the B-Factories). The \babar result exceed SM predictions, in 
both $D$ and $D^*$ channels, by $2.0\sigma$ and $2.7\sigma$ respectively. 
The combined result disagree with the SM by $3.4\sigma$.  \babar also interpreted these measurements
in terms of the 2HDM type-II and found that their results are not compatible with this model for any value
of $\tan\beta$ and $m_H$.  
The result obtained by \babar and Belle on ${\cal R}(D)$ and ${\cal R}(D^*)$ are reported in Tab.\ref{tab:dtaunu}.
Before a published result by Belle using the new $B_{tag}$ reconstruction, we do not attempt to average the existing results.

\begin{table}[!htb]
\begin{center}
\caption{Summary of the results on ${\cal R}(D)$ and ${\cal R}(D^*)$. The errors quoted
correspond to statistical and systematic uncertainties, respectively.}
\label{tab:dtaunu}
\begin{small}
\begin{tabular}{|lcc|}
\hline
  & Belle~\cite{Adachi:2009qg}   & \babar~\cite{Lees:2012xj} \\
\hline\hline
${\cal R}(D^0) $    &  $0.70^{+0.19}_{-0.18}~^{+0.11}_{-0.09}$ & $0.99\pm 0.19\pm 0.13$\\
${\cal R}(D^{*0})$  &  $0.47^{+0.11}_{-0.10}~^{+0.06}_{-0.07}$ & $1.71\pm 0.17\pm 0.13$\\
${\cal R}(D^+) $    &  $0.48^{+0.22}_{-0.19}~^{+0.06}_{-0.05}$ & $1.01\pm 0.18\pm 0.12$\\
${\cal R}(D^{*+})$  &  $0.48^{+0.14}_{-0.12}~^{+0.06}_{-0.04}$ & $1.74\pm 0.19\pm 0.12$\\
\hline
\end{tabular}\\
\end{small}
\end{center}
\end{table}

%
%

\clearpage
\section{$b$-hadron decays to charmed hadrons}
\label{sec:b2c}
Ground state $B$ mesons and $b$-baryons dominantly decay to particles containing a charm quark via the $b \rightarrow c$ quark transition.
Therefore these decays are sensitive to the $|V_{cb}|$ CKM matrix element.
Usually semileptonic modes are used for $|V_{cb}|$ measurements which are discussed in Section~\ref{sec:slbdecays}.
Some $B$ meson decays to open charmed hadrons are fundamental decays for the measurements of $CP$-violation phases like $\phi_s^{c\bar{c}s}$ (Section~\ref{sec:life_mix}), $\beta=\phi_1$ and $\gamma=\phi_3$ (Section~\ref{sec:cp_uta}). 

The fact that decays to charmed hadrons are the dominant $b$-hadron decays
makes them a very important part of the experimental programme.
Understanding the rate of charm production in $b$-hadron decays is crucial to
validate the HQE that underpins much of the theoretical framework for $b$
physics (see, for example, Ref.~\cite{Lenz:2014nka} for a recent review).
Moreover, such decays are often used as the normalization mode for
measurements of rarer decays. 
In addition, they are the dominant background in many analyses.
To accurately model the background with simulated data it is essential to have a precise knowledge of the contributing decay modes.
In particular, with the expected increase in the data samples at LHCb and
Belle~II, the enhanced statistical sensitivity has to be matched by low
systematic uncertainties due to knowledge of the dominant $b$-hadron decay
modes. 
For multibody decays, knowledge of the distribution of decays across the
phase-space (\eg\ the Dalitz plot density for three-body decays or the
polarization amplitudes for vector-vector final states) is required in
addition to the total branching fraction.

The large yields of $b$-hadron decays to multibody final states containing
charm makes them ideal to study the spectroscopy of both open charm and
charmonia (or charmonia-like) mesons.  
In particular, they have been used to both discover and measure the properties
of exotic particles such as the $X(3872)$~\cite{Choi:2003ue,Aaij:2013zoa} and
$Z(4430)$~\cite{Choi:2007wga,Aaij:2014jqa} states.
The large yields available similarly make $b \to c$ decays very useful to
study baryon-antibaryon pair production.

In addition to the dominant $b$-hadron decays to final states containing
charmed hadrons, there are several decays in this category that are expected
to be highly suppressed in the Standard Model.
These are of interest to probe particular decay topologies (\eg\ the $B^- \to
\Dsm \phi$ decay, which is dominated by the so-called annihilation diagram)
and thereby constrain effects in other hadronic decays or to search for new physics.
There are also other $b \to c$ decays, such as $\Bzb \to \Dsm \pip$, that are
mediated by the $W$ emission involving the $|V_{ub}|$ CKM matrix element.
Finally, some $b \to c$ decays involving lepton flavour or number violation are
completely forbidden in the Standard Model, and therefore provide highly
sensitive null tests.

In this section, we give an exhaustive list of measured branching ratios of decay modes to charmed hadrons.
Compared to the previous version of $B$ to charm results the averaging procedure was updated to follow the methodology described in Section~\ref{sec:method}.
Where available, correlations between measurements 
are taken into account.
We provide averages of the polarization amplitudes of $B$ meson decays to
vector-vector states, but we do not currently provide detailed averages of
quantities obtained from Dalitz plot analyses, due to the complications
arising from the dependence on the model used.

The results are presented in subsections organized by the decaying bottom
hadron: $\Bzb$ (Sec.~\ref{sec:b2c:Bd}), $B^-$ (Sec.~\ref{sec:b2c:Bu}), $\Bzb/B^-$ admixture (Sec.~\ref{sec:b2c:B}), $\Bsb$ (Sec.~\ref{sec:b2c:Bs}), $B_c^-$ (Sec.~\ref{sec:b2c:Bc}), $b$ baryons (Sec.~\ref{sec:b2c:Bbaryon}).
For each subsection the measurements are arranged, considering the final state,
into the following groups: a single charmed meson, two charmed mesons, a
charmonium state, a charm baryon, or other states, like for example the
$X(3872)$ meson.
The individual measurements and averages are shown as numerical values in tables followed by a graphical representation of the averages.
The symbol $\mathcal{B}$ is used for branching ratios, $f$ for production fractions (see Section~\ref{sec:life_mix}), and $\sigma$ for cross sections.
The decay amplitudes for longitudinal, parallel, and perpendicular transverse polarization in pseudoscalar to vector-vector decays are denoted ${\cal{A}}_0$, ${\cal{A}}_\parallel$, and ${\cal{A}}_\perp$, respectively, and the definitions $\delta_\parallel = \arg({\cal{A}}_\parallel/{\cal{A}}_0)$ and $\delta_\perp = \arg({\cal{A}}_\perp/{\cal{A}}_0)$ are used for their relative phases.
The inclusion of charge conjugate modes is always implied.

Following the approach used by the PDG~\cite{PDG_2014}, for decays that involve
neutral kaons we mainly quote results in terms of final states including
either a $\Kz$ or $\Kzb$ meson (instead of a \KS or \KL).
In some cases where the decay is not flavour-specific and the final state is
not self-conjugate, the inclusion of the conjugate final state is implied --
in fact, the flavour of the neutral kaon is never determined experimentally,
and so the specification as \Kz or \Kzb simply follows the quark model
expectation for the dominant decay.
An exception occurs for some \Bs decays, specifically those to \CP
eigenstates, where the width difference between the mass eigenstates (see
Sec.~\ref{sec:life_mix}) means that the measured branching fraction,
integrated over decay time, is specific to the studied final
state~\cite{DeBruyn:2012wj}. 
Therefore it is appropriate to quote the branching fraction for, \eg, $\Bsb \to
\jpsi \KS$ instead of $\Bsb \to \jpsi \Kzb$.


\newenvironment{btocharmtab}[2]{\begin{table}[H]\begin{center}\caption{#2.}\label{tab:b2c:#1}\begin{tabular}{|l l l |}}{\end{tabular}\end{center}\end{table}}
\newcommand{\btocharmfig}[1]{\begin{figure}[H]\begin{center}\includegraphics[width=0.99\textwidth]{b2charm/figs/#1}\caption{Summary of the averages from Table~\ref{tab:b2c:#1}.}\label{fig:b2c:#1}\end{center}\end{figure}}
\newcommand{\btocharm}[1]{\input{b2charm/#1.tex}\btocharmfig{#1}}

\subsection{Decays of $\Bzb$ mesons}
\label{sec:b2c:Bd}
Measurements of $\Bzb$ decays to charmed hadrons are summarized in Sections \ref{sec:b2c:Bd_D} to \ref{sec:b2c:Bd_other}.

\subsubsection{Decays to a single open charm meson}
\label{sec:b2c:Bd_D}
Averages of $\Bzb$ decays to a single open charm meson are shown in Tables~\ref{tab:b2c:Bd_D_1}--\ref{tab:b2c:Bd_D_10} and Figs.~\ref{fig:b2c:Bd_D_1}--\ref{fig:b2c:Bd_D_10}.
\begin{btocharmtab}{Bd_D_1}{Decays to a $D^{(*)}$ meson and one or more pions $[10^{-3}]$}
\hline
\textbf{Parameter} & 
\begin{tabular}{l}\textbf{Measurements}\end{tabular} & \textbf{Average} \\
\hline
\hline
${\cal{B}} ( \bar{B}^0 \to D^0 \pi^0 )$ & \begin{tabular}{l} \babar \cite{Lees:2011gw}: $0.269 \pm 0.009 \pm 0.013$ \\ Belle \cite{Blyth:2006at}: $0.225 \pm 0.014 \pm 0.035$ \\ \end{tabular} & $0.262 \pm 0.015$ \\
\hline
${\cal{B}} ( \bar{B}^0 \to D^{*0}(2007) \pi^0 )$ & \begin{tabular}{l} \babar \cite{Lees:2011gw}: $0.305 \pm 0.014 \pm 0.028$ \\ Belle \cite{Blyth:2006at}: $0.139 \pm 0.018 \pm 0.026$ \\ \end{tabular} & $0.223 \pm 0.022$ \\
\hline
${\cal{B}} ( \bar{B}^0 \to D^+ \pi^- )$ & \begin{tabular}{l} \babar \cite{Aubert:2006jc}: $3.03 \pm 0.23 \pm 0.23$ \\ \babar \cite{Aubert:2006cd}: $2.55 \pm 0.05 \pm 0.16$ \\ \end{tabular} & $2.65 \pm 0.15$ \\
\hline
${\cal{B}} ( \bar{B}^0 \to D^{*+}(2010) \pi^- )$ & \begin{tabular}{l} \babar \cite{Aubert:2006jc}: $2.99 \pm 0.23 \pm 0.24$ \\ \babar \cite{Aubert:2006cd}: $2.79 \pm 0.08 \pm 0.17$ \\ Belle \cite{Abe:2004cw}: $2.30 \pm 0.06 \pm 0.19$ \\ \end{tabular} & $2.62 \pm 0.13$ \\
\hline
${\cal{B}} ( \bar{B}^0 \to D^0 \pi^+ \pi^- )$ & \begin{tabular}{l} Belle \cite{Abe:2004cw}: $1.07 \pm 0.06 \pm 0.10$ \\ Belle \cite{Satpathy:2002js}: $0.80 \pm 0.06 \pm 0.15$ \\ \end{tabular} & $0.98 \pm 0.09$ \\
\hline
${\cal{B}} ( \bar{B}^0 \to D^{*0}(2007) \pi^+ \pi^- )$ & \begin{tabular}{l} Belle \cite{Abe:2004cw}: $1.09 \pm 0.08 \pm 0.16$ \\ Belle \cite{Satpathy:2002js}: $0.62 \pm 0.12 \pm 0.18$ \\ \end{tabular} & $0.90 \pm 0.14$ \\
\hline
${\cal{B}} ( \bar{B}^0 \to D^{*+}(2010) \pi^- \pi^+ \pi^- )$ & \begin{tabular}{l} Belle \cite{Majumder:2004su}: $6.81 \pm 0.23 \pm 0.72$ \\ \end{tabular} & $6.81 \pm 0.76$ \\
\hline
${\cal{B}} ( \bar{B}^0 \to D^{*0}(2007) \pi^- \pi^+ \pi^- \pi^+ )$ & \begin{tabular}{l} Belle \cite{Majumder:2004su}: $2.60 \pm 0.47 \pm 0.37$ \\ \end{tabular} & $2.60 \pm 0.60$ \\
\hline
${\cal{B}} ( \bar{B}^0 \to D^{*+}(2010) \pi^- \pi^+ \pi^- \pi^+ \pi^- )$ & \begin{tabular}{l} Belle \cite{Majumder:2004su}: $4.72 \pm 0.59 \pm 0.71$ \\ \end{tabular} & $4.72 \pm 0.92$ \\
\hline
${\cal{B}} ( \bar{B}^0 \to D^{*+}(2010) \omega(782) \pi^- )$ & \begin{tabular}{l} \babar \cite{Aubert:2006zb}: $2.88 \pm 0.21 \pm 0.31$ \\ \end{tabular} & $2.88 \pm 0.38$ \\
\hline
\end{btocharmtab}
\btocharmfig{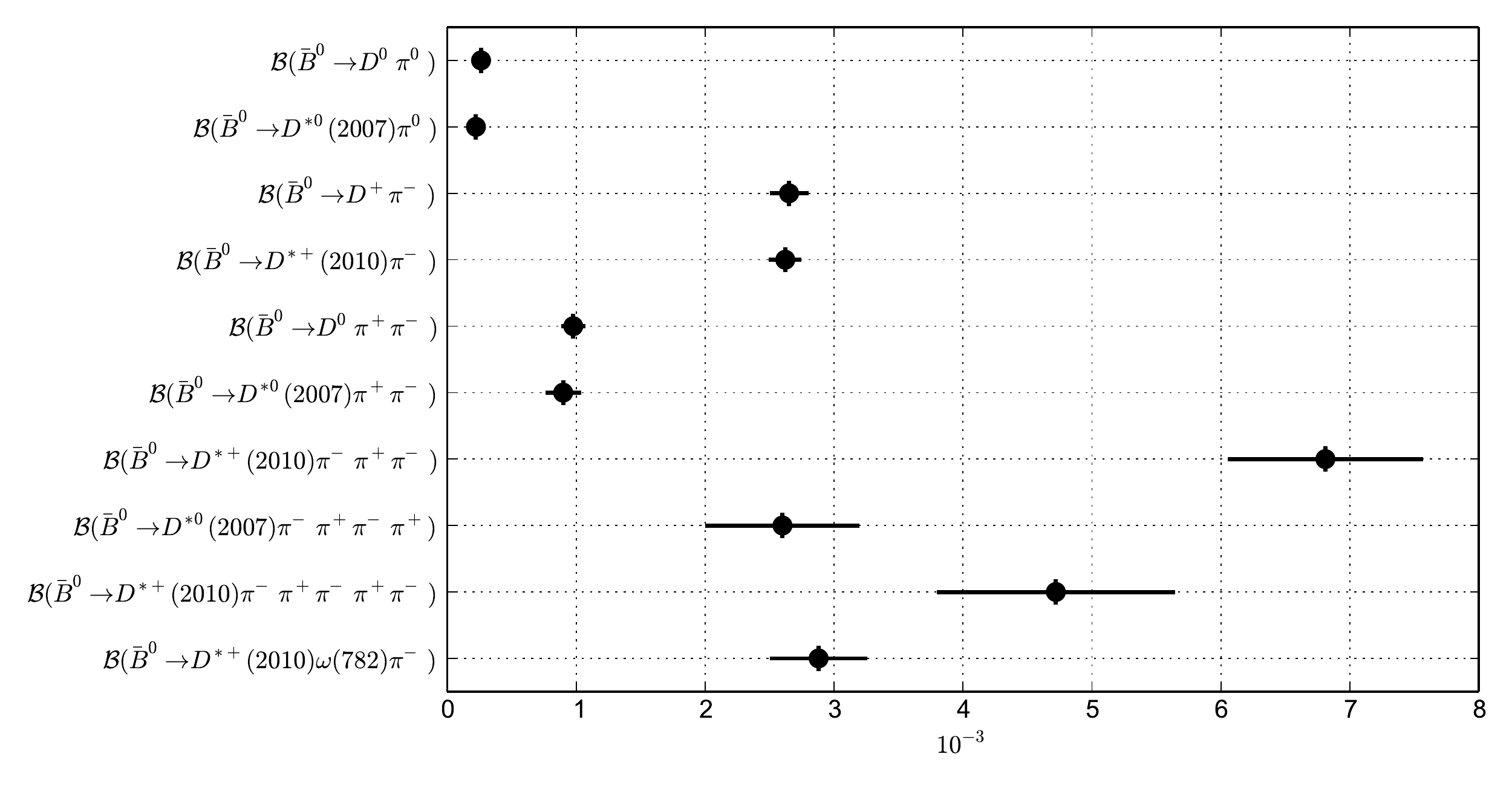}
\begin{btocharmtab}{Bd_D_2}{Decays to a $D^{(*)0}$ meson and a light meson $[10^{-4}]$}
\hline
\textbf{Parameter} & \begin{tabular}{l}\textbf{Measurements}\end{tabular} & \textbf{Average} \\
\hline
\hline
${\cal{B}} ( \bar{B}^0 \to D^0 \rho^0(770) )$ & \begin{tabular}{l} Belle \cite{Abe:2004cw}: $2.91 \pm 0.28 \,^{+0.34}_{-0.63}$ \\ Belle \cite{Satpathy:2002js}: $2.9 \pm 1.0 \pm 0.4$ \\ \end{tabular} & $2.91 \pm 0.51$ \\
\hline
${\cal{B}} ( \bar{B}^0 \to D^{*0}(2007) \rho^0(770) )$ & \begin{tabular}{l} Belle \cite{Abe:2004cw}: $3.73 \pm 0.87 \,^{+0.49}_{-0.47}$ \\ Belle \cite{Satpathy:2002js}: $< 5.1$ \\ \end{tabular} & $3.73 \,^{+1.00}_{-0.99}$ \\
\hline
${\cal{B}} ( \bar{B}^0 \to D^0 \eta )$ & \begin{tabular}{l} \babar \cite{Lees:2011gw}: $2.53 \pm 0.09 \pm 0.11$ \\ Belle \cite{Blyth:2006at}: $1.77 \pm 0.16 \pm 0.21$ \\ \end{tabular} & $2.36 \pm 0.13$ \\
\hline
${\cal{B}} ( \bar{B}^0 \to D^{*0}(2007) \eta )$ & \begin{tabular}{l} \babar \cite{Lees:2011gw}: $2.69 \pm 0.14 \pm 0.23$ \\ Belle \cite{Blyth:2006at}: $1.40 \pm 0.28 \pm 0.26$ \\ \end{tabular} & $2.26 \pm 0.22$ \\
\hline
${\cal{B}} ( \bar{B}^0 \to D^0 \eta^\prime(958) )$ & \begin{tabular}{l} \babar \cite{Lees:2011gw}: $1.48 \pm 0.13 \pm 0.07$ \\ Belle \cite{Schumann:2005ej}: $1.14 \pm 0.20 \,^{+0.10}_{-0.13}$ \\ \end{tabular} & $1.38 \pm 0.12$ \\
\hline
${\cal{B}} ( \bar{B}^0 \to D^{*0}(2007) \eta^\prime(958) )$ & \begin{tabular}{l} \babar \cite{Lees:2011gw}: $1.48 \pm 0.22 \pm 0.13$ \\ Belle \cite{Schumann:2005ej}: $1.21 \pm 0.34 \pm 0.22$ \\ \end{tabular} & $1.40 \pm 0.22$ \\
\hline
${\cal{B}} ( \bar{B}^0 \to D^0 \omega(782) )$ & \begin{tabular}{l} \babar \cite{Lees:2011gw}: $2.57 \pm 0.11 \pm 0.14$ \\ Belle \cite{Blyth:2006at}: $2.37 \pm 0.23 \pm 0.28$ \\ \end{tabular} & $2.53 \pm 0.16$ \\
\hline
${\cal{B}} ( \bar{B}^0 \to D^{*0}(2007) \omega(782) )$ & \begin{tabular}{l} \babar \cite{Lees:2011gw}: $4.55 \pm 0.24 \pm 0.39$ \\ Belle \cite{Blyth:2006at}: $2.29 \pm 0.39 \pm 0.40$ \\ \end{tabular} & $3.64 \pm 0.35$ \\
\hline
${\cal{B}} ( \bar{B}^0 \to f_2(1270) D^0 )$ & \begin{tabular}{l} Belle \cite{Abe:2004cw}: $1.95 \pm 0.34 \,^{+0.50}_{-0.38}$ \\ \end{tabular} & $1.95 \,^{+0.60}_{-0.51}$ \\
\hline
${\cal{B}} ( \bar{B}^0 \to f_2(1270) D^{*0}(2007) )$ & \begin{tabular}{l} Belle \cite{Abe:2004cw}: $1.86 \pm 0.65 \,^{+1.00}_{-0.79}$ \\ \end{tabular} & $1.86 \,^{+1.19}_{-1.03}$ \\
\hline
\end{btocharmtab}
\btocharmfig{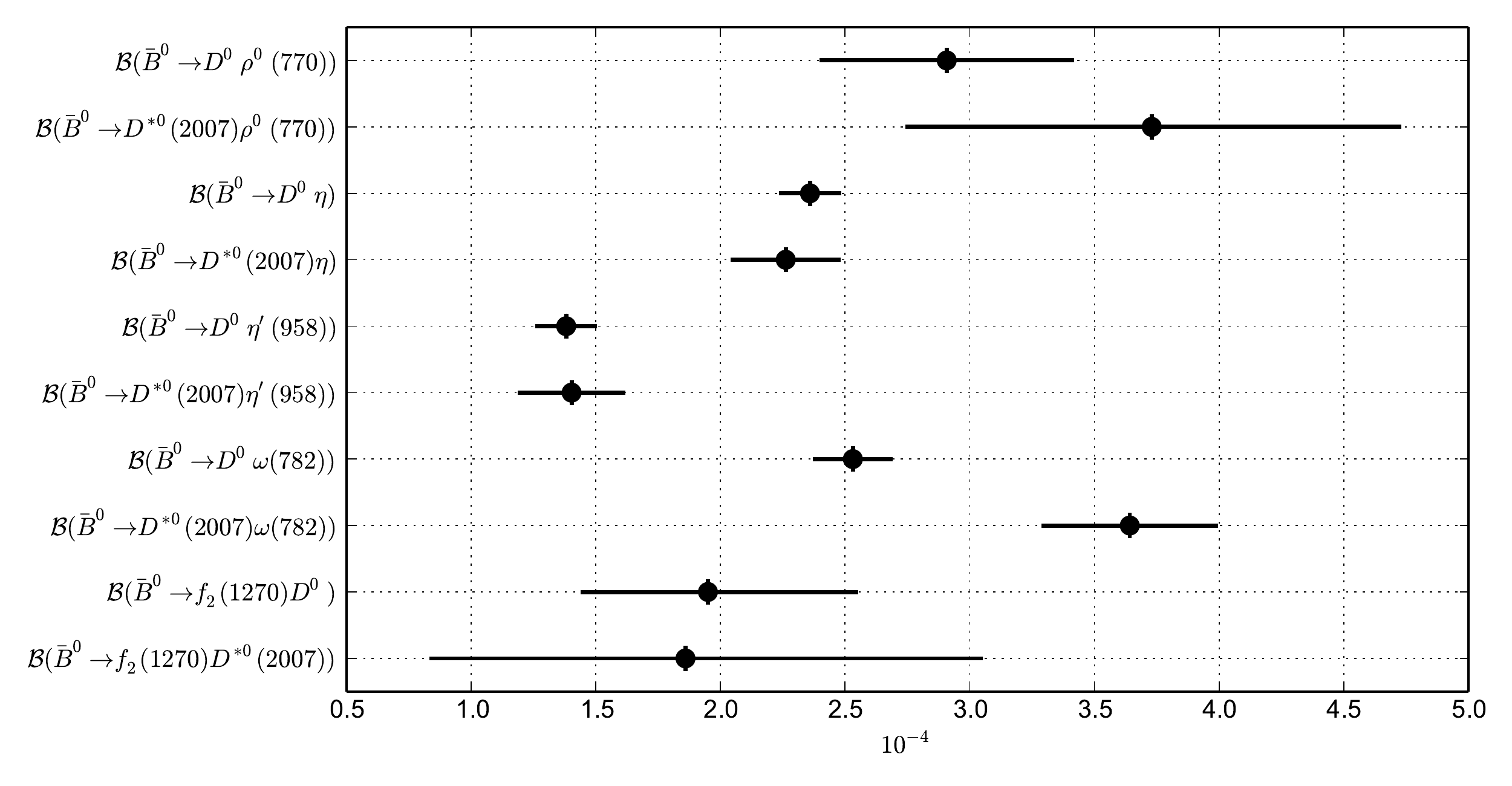}
\begin{btocharmtab}{Bd_D_3}{Decays to a $D^{(*)+}$ meson and one or more kaons $[10^{-3}]$}
\hline
\textbf{Parameter} & \begin{tabular}{l}\textbf{Measurements}\end{tabular} & \textbf{Average} \\
\hline
\hline
${\cal{B}} ( \bar{B}^0 \to D^+ K^- )$ & \begin{tabular}{l} LHCb \cite{Aaij:2013qqa}: $0.22 \pm 0.00 \pm 0.01$ \\ LHCb \cite{Aaij:2011hi}: $0.201 \pm 0.018 \pm 0.014$ \\ Belle \cite{Abe:2001waa}: $0.204 \pm 0.045 \pm 0.034$ \\ \end{tabular} & $0.215 \pm 0.011$ \\
\hline
${\cal{B}} ( \bar{B}^0 \to D^{*+}(2010) K^- )$ & \begin{tabular}{l} Belle \cite{Abe:2001waa}: $0.204 \pm 0.041 \pm 0.023$ \\ \end{tabular} & $0.204 \pm 0.047$ \\
\hline
${\cal{B}} ( \bar{B}^0 \to D^+ K^{*-}(892) )$ & \begin{tabular}{l} \babar \cite{Aubert:2004at}: $0.46 \pm 0.06 \pm 0.05$ \\ \end{tabular} & $0.46 \pm 0.08$ \\
\hline
${\cal{B}} ( \bar{B}^0 \to D^{*+}(2010) K^{*-}(892) )$ & \begin{tabular}{l} \babar \cite{Aubert:2004at}: $0.32 \pm 0.06 \pm 0.03$ \\ \end{tabular} & $0.32 \pm 0.07$ \\
\hline
${\cal{B}} ( \bar{B}^0 \to D^+ K^0 \pi^- )$ & \begin{tabular}{l} \babar \cite{Aubert:2004at}: $0.49 \pm 0.07 \pm 0.05$ \\ \end{tabular} & $0.49 \pm 0.09$ \\
\hline
${\cal{B}} ( \bar{B}^0 \to D^{*+}(2010) K^0 \pi^- )$ & \begin{tabular}{l} \babar \cite{Aubert:2004at}: $0.30 \pm 0.07 \pm 0.03$ \\ \end{tabular} & $0.30 \pm 0.08$ \\
\hline
${\cal{B}} ( \bar{B}^0 \to D^+ K^- K^0 )$ & \begin{tabular}{l} Belle \cite{Drutskoy:2002ib}: $< 0.31$ \\ \end{tabular} & $< 0.31$ \\
\hline
${\cal{B}} ( \bar{B}^0 \to D^{*+}(2010) K^- K^0 )$ & \begin{tabular}{l} Belle \cite{Drutskoy:2002ib}: $< 0.47$ \\ \end{tabular} & $< 0.47$ \\
\hline
${\cal{B}} ( \bar{B}^0 \to D^+ K^- K^{*0}(892) )$ & \begin{tabular}{l} Belle \cite{Drutskoy:2002ib}: $0.88 \pm 0.11 \pm 0.15$ \\ \end{tabular} & $0.88 \pm 0.19$ \\
\hline
${\cal{B}} ( \bar{B}^0 \to D^{*+}(2010) K^- K^{*0}(892) )$ & \begin{tabular}{l} Belle \cite{Drutskoy:2002ib}: $1.29 \pm 0.22 \pm 0.25$ \\ \end{tabular} & $1.29 \pm 0.33$ \\
\hline
\end{btocharmtab}
\btocharmfig{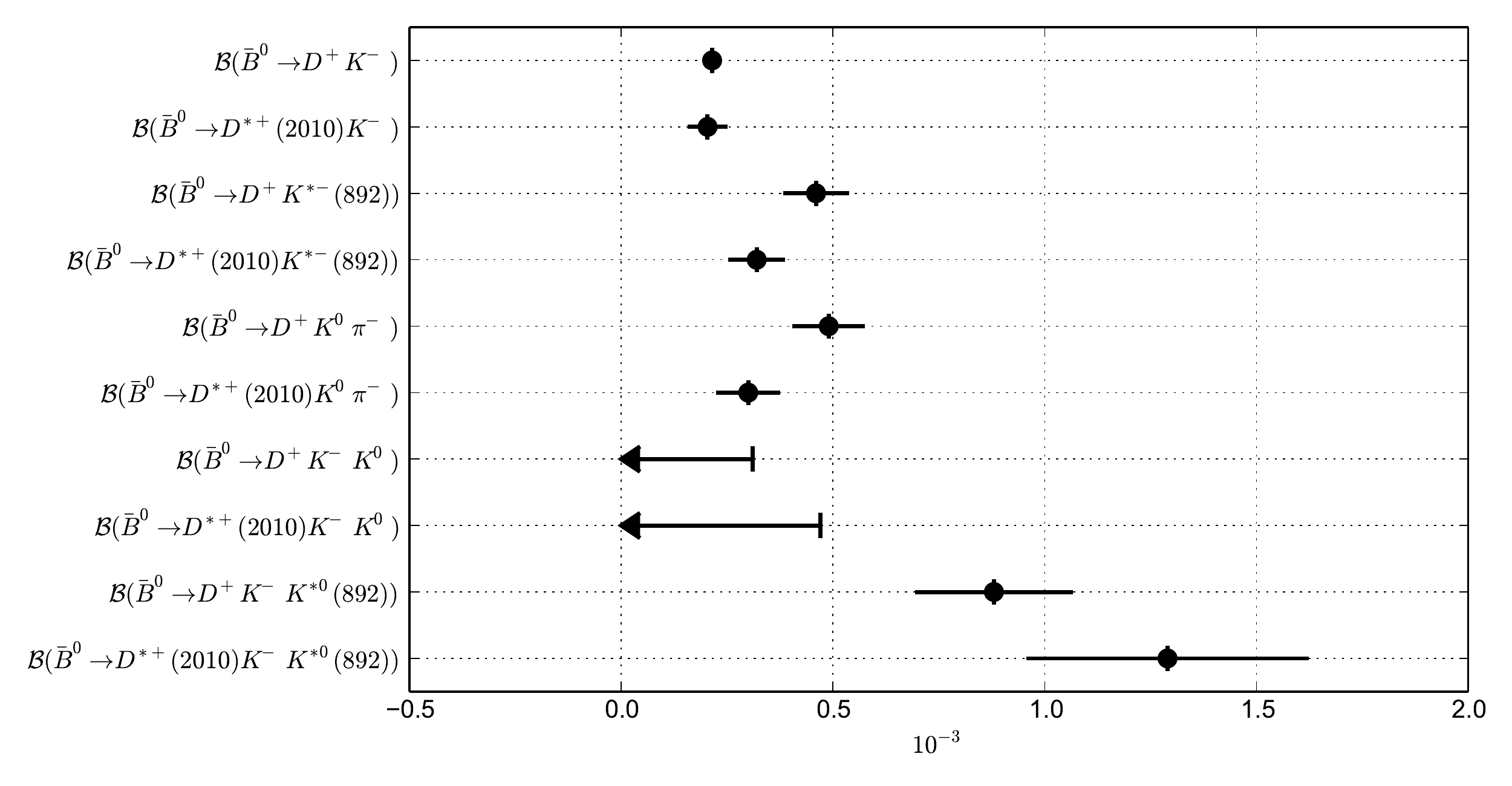}
\begin{btocharmtab}{Bd_D_4}{Decays to a $D^{(*)0}$ meson and a kaon $[10^{-4}]$}
\hline
\textbf{Parameter} & \begin{tabular}{l}\textbf{Measurements}\end{tabular} & \textbf{Average} \\
\hline
\hline
${\cal{B}} ( \bar{B}^0 \to D^0 \bar{K}^0 )$ & \begin{tabular}{l} \babar \cite{Aubert:2006qn}: $0.53 \pm 0.07 \pm 0.03$ \\ Belle \cite{Krokovny:2002ua}: $0.50 \,^{+0.13}_{-0.12} \pm 0.06$ \\ \end{tabular} & $0.52 \pm 0.07$ \\
\hline
${\cal{B}} ( \bar{B}^0 \to D^{*0}(2007) \bar{K}^0 )$ & \begin{tabular}{l} \babar \cite{Aubert:2006qn}: $0.36 \pm 0.12 \pm 0.03$ \\ Belle \cite{Krokovny:2002ua}: $< 0.66$ \\ \end{tabular} & $0.36 \pm 0.12$ \\
\hline
${\cal{B}} ( \bar{B}^0 \to D^0 K^- \pi^+ )$ & \begin{tabular}{l} \babar \cite{Aubert:2005yt}: $0.88 \pm 0.15 \pm 0.09$ \\ \end{tabular} & $0.88 \pm 0.17$ \\
\hline
${\cal{B}} ( \bar{B}^0 \to D^0 \bar{K}^{*0}(892) )$ & \begin{tabular}{l} \babar \cite{Aubert:2006qn}: $0.40 \pm 0.07 \pm 0.03$ \\ Belle \cite{Krokovny:2002ua}: $0.48 \,^{+0.11}_{-0.10} \pm 0.05$ \\ \end{tabular} & $0.42 \pm 0.06$ \\
\hline
\multicolumn{3}{|l|}{${{\cal{B}} ( \bar{B}^0 \to D^0 \bar{K}^{*0}(892) )\times {\cal{B}} ( \bar{K}^{*0}(892) \to K^- \pi^+ )}$}\\
 & \begin{tabular}{l} \babar \cite{Aubert:2005yt}: $0.38 \pm 0.06 \pm 0.04$ \\ \end{tabular} & $0.38 \pm 0.07$ \\
\hline
${\cal{B}} ( \bar{B}^0 \to D^{*0}(2007) \bar{K}^{*0}(892) )$ & \begin{tabular}{l} Belle \cite{Krokovny:2002ua}: $< 0.69$ \\ \end{tabular} & $< 0.69$ \\
\hline
${\cal{B}} ( \bar{B}^0 \to \bar{D}^{*0}(2007) \bar{K}^{*0}(892) )$ & \begin{tabular}{l} Belle \cite{Krokovny:2002ua}: $< 0.40$ \\ \end{tabular} & $< 0.40$ \\
\hline
${\cal{B}} ( \bar{B}^0 \to \bar{D}^0 K^- \pi^+ )$ & \begin{tabular}{l} \babar \cite{Aubert:2005yt}: $< 0.19$ \\ \end{tabular} & $< 0.19$ \\
\hline
${\cal{B}} ( \bar{B}^0 \to \bar{D}^0 \bar{K}^{*0}(892) )$ & \begin{tabular}{l} \babar \cite{Aubert:2006qn}: $0.00 \pm 0.05 \pm 0.03$ \\ Belle \cite{Krokovny:2002ua}: $< 0.18$ \\ \end{tabular} & $0.00 \pm 0.06$ \\
\hline
\end{btocharmtab}
\btocharmfig{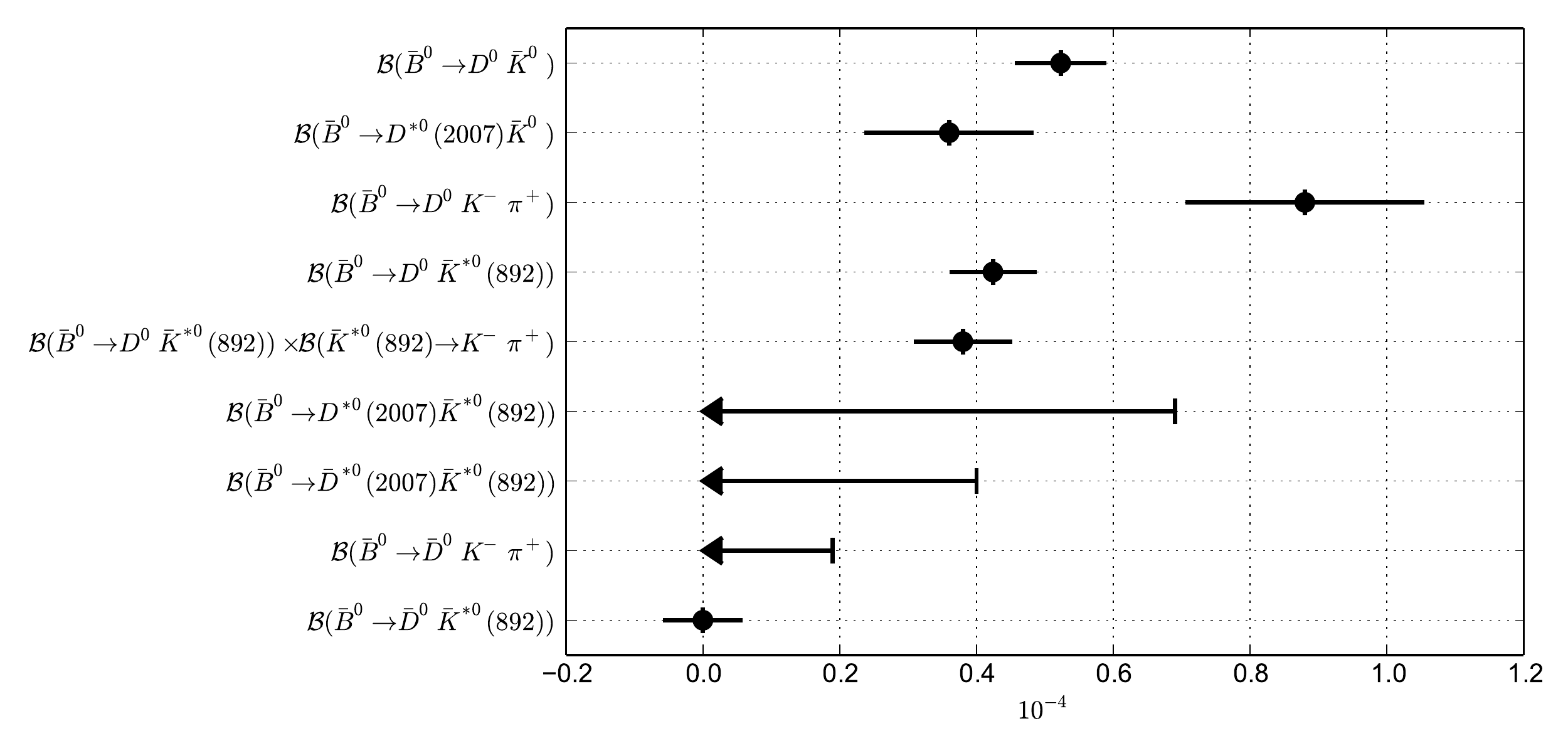}
\begin{btocharmtab}{Bd_D_5}{Decays to a $D_s^{(*)}$ meson $[10^{-4}]$}
\hline
\textbf{Parameter} & \begin{tabular}{l}\textbf{Measurements}\end{tabular} & \textbf{Average} \\
\hline
\hline
${\cal{B}} ( \bar{B}^0 \to D_s^- \pi^+ )$ & \begin{tabular}{l} \babar \cite{Aubert:2008zi}: $0.25 \pm 0.04 \pm 0.02$ \\ Belle \cite{Das:2010be}: $0.199 \pm 0.026 \pm 0.018$ \\ \end{tabular} & $0.216 \pm 0.026$ \\
\hline
${\cal{B}} ( \bar{B}^0 \to D_s^{*-} \pi^+ )$ & \begin{tabular}{l} \babar \cite{Aubert:2008zi}: $0.26 \,^{+0.05}_{-0.04} \pm 0.02$ \\ Belle \cite{Joshi:2009yv}: $0.175 \pm 0.034 \pm 0.020$ \\ \end{tabular} & $0.207 \pm 0.032$ \\
\hline
${\cal{B}} ( \bar{B}^0 \to D_s^- \rho^+(770) )$ & \begin{tabular}{l} \babar \cite{Aubert:2008zi}: $0.11 \,^{+0.09}_{-0.08} \pm 0.03$ \\ \end{tabular} & $0.11 \,^{+0.09}_{-0.09}$ \\
\hline
${\cal{B}} ( \bar{B}^0 \to D_s^{*-} \rho^+(770) )$ & \begin{tabular}{l} \babar \cite{Aubert:2008zi}: $0.41 \,^{+0.13}_{-0.12} \pm 0.04$ \\ \end{tabular} & $0.41 \,^{+0.14}_{-0.13}$ \\
\hline
${\cal{B}} ( \bar{B}^0 \to D_s^- a_0^+(980) )$ & \begin{tabular}{l} \babar \cite{Aubert:2005qt}: $0.06 \,^{+0.14}_{-0.11} \pm 0.01$ \\ \end{tabular} & $0.06 \,^{+0.14}_{-0.11}$ \\
\hline
${\cal{B}} ( \bar{B}^0 \to D_s^{*-} a_0^+(980) )$ & \begin{tabular}{l} \babar \cite{Aubert:2005qt}: $0.14 \,^{+0.21}_{-0.16} \pm 0.03$ \\ \end{tabular} & $0.14 \,^{+0.21}_{-0.16}$ \\
\hline
${\cal{B}} ( \bar{B}^0 \to D_s^- a_2^+(1320) )$ & \begin{tabular}{l} \babar \cite{Aubert:2005qt}: $0.64 \,^{+1.04}_{-0.57} \pm 0.15$ \\ \end{tabular} & $0.64 \,^{+1.05}_{-0.59}$ \\
\hline
${\cal{B}} ( \bar{B}^0 \to D_s^{*-} a_2^+(1320) )$ & \begin{tabular}{l} \babar \cite{Aubert:2005qt}: $< 2.0$ \\ \end{tabular} & $< 2.0$ \\
\hline
${\cal{B}} ( \bar{B}^0 \to D_s^+ K^- )$ & \begin{tabular}{l} \babar \cite{Aubert:2008zi}: $0.29 \pm 0.04 \pm 0.02$ \\ Belle \cite{Das:2010be}: $0.191 \pm 0.024 \pm 0.017$ \\ \end{tabular} & $0.221 \pm 0.025$ \\
\hline
${\cal{B}} ( \bar{B}^0 \to D_s^{*+} K^- )$ & \begin{tabular}{l} \babar \cite{Aubert:2008zi}: $0.24 \pm 0.04 \pm 0.02$ \\ Belle \cite{Joshi:2009yv}: $0.202 \pm 0.033 \pm 0.022$ \\ \end{tabular} & $0.219 \pm 0.031$ \\
\hline
${\cal{B}} ( \bar{B}^0 \to D_s^+ K^{*-}(892) )$ & \begin{tabular}{l} \babar \cite{Aubert:2008zi}: $0.35 \,^{+0.10}_{-0.09} \pm 0.04$ \\ \end{tabular} & $0.35 \,^{+0.11}_{-0.10}$ \\
\hline
${\cal{B}} ( \bar{B}^0 \to D_s^{*+} K^{*-}(892) )$ & \begin{tabular}{l} \babar \cite{Aubert:2008zi}: $0.32 \,^{+0.14}_{-0.12} \pm 0.04$ \\ \end{tabular} & $0.32 \,^{+0.15}_{-0.13}$ \\
\hline
${\cal{B}} ( \bar{B}^0 \to D_s^+ K_S^0 \pi^- )$ & \begin{tabular}{l} \babar \cite{Aubert:2007xma}: $0.55 \pm 0.13 \pm 0.10$ \\ \end{tabular} & $0.55 \pm 0.17$ \\
\hline
${\cal{B}} ( \bar{B}^0 \to D_s^{*+} K^0 \pi^- )$ & \begin{tabular}{l} \babar \cite{Aubert:2007xma}: $< 0.55$ \\ \end{tabular} & $< 0.55$ \\
\hline
\end{btocharmtab}
\btocharmfig{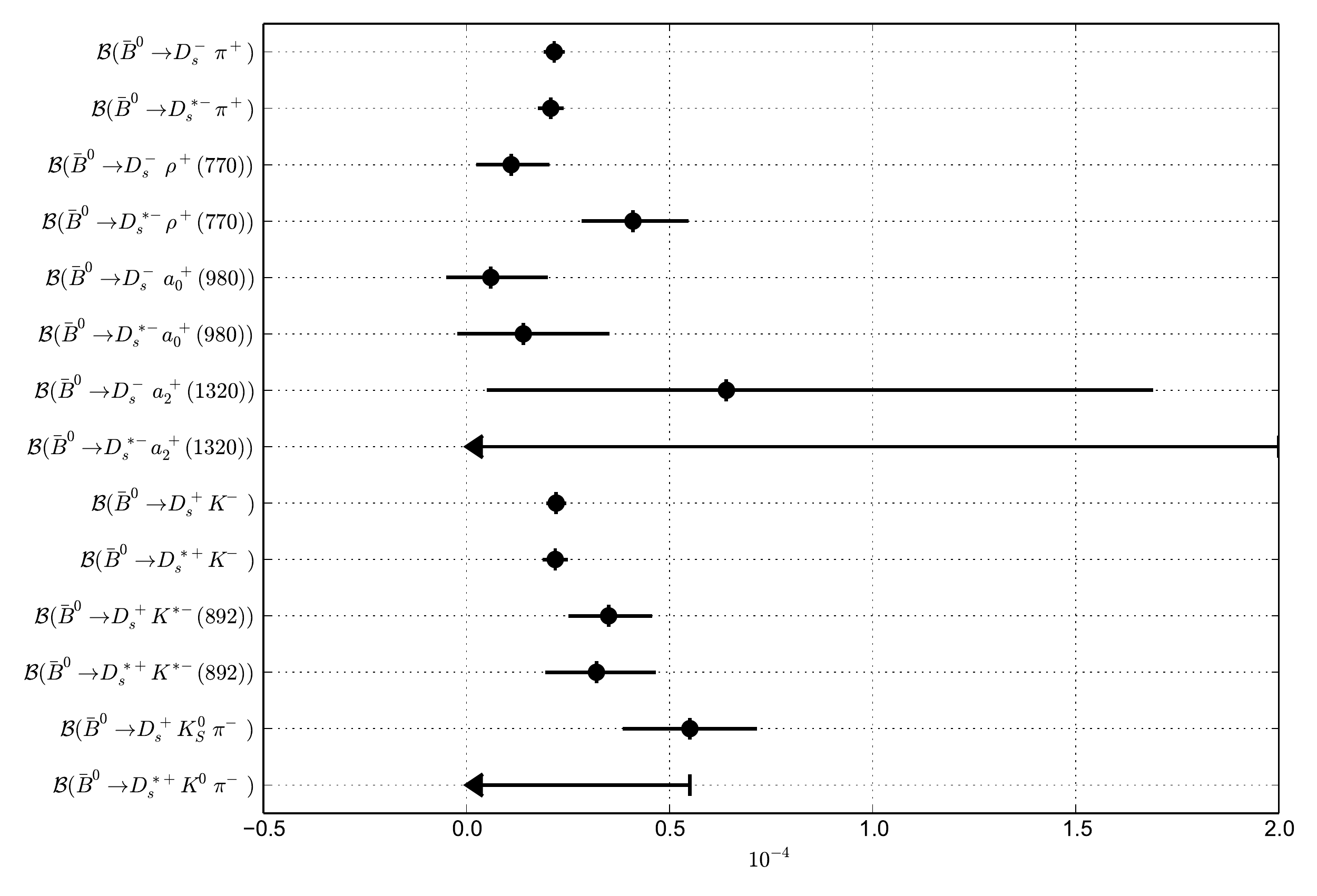}
\begin{btocharmtab}{Bd_D_6}{Relative decay rates}
\hline
\textbf{Parameter} & \begin{tabular}{l}\textbf{Measurements}\end{tabular} & \textbf{Average} \\
\hline
\hline
\multicolumn{3}{|l|}{${{\cal{B}} ( \bar{B}^0 \to D^0 \rho^0(770) )}/{{\cal{B}} ( \bar{B}^0 \to D^0 \omega(782) )}$}\\
 & \begin{tabular}{l} Belle \cite{Satpathy:2002js}: $1.6 \pm 0.8$ \\ \end{tabular} & $1.6 \pm 0.8$ \\
\hline
\multicolumn{3}{|l|}{${ {\cal{B}} ( \bar{B}^0 \to \bar{D}^0 K^- \pi^+ )}/  {{\cal{B}} ( \bar{B}^0 \to \bar{D}^0 \pi^- \pi^+ )}$}\\
 & \begin{tabular}{l} LHCb \cite{Aaij:2013pua}: $0.106 \pm 0.007 \pm 0.008$ \\ \end{tabular} & $0.106 \pm 0.011$ \\
\hline
\multicolumn{3}{|l|}{${ {\cal{B}} ( \bar{B}^0 \to \bar{D}^0 K^- K^+ )}/  {{\cal{B}} ( \bar{B}^0 \to \bar{D}^0 \pi^- \pi^+ )}$}\\
 & \begin{tabular}{l} LHCb \cite{Aaij:2012zka}: $0.056 \pm 0.110 \pm 0.007$ \\ \end{tabular} & $0.056 \pm 0.110$ \\
\hline
\multicolumn{3}{|l|}{${\cal{B}} ( \bar{B}^0 \to D^+ \pi^+ \pi^- \pi^- ) / {\cal{B}} ( \bar{B}^0 \to D^+ \pi^- )$}\\
 & \begin{tabular}{l} LHCb \cite{Aaij:2011rj}: $2.38 \pm 0.11 \pm 0.21$ \\ \end{tabular} & $2.38 \pm 0.24$ \\
\hline
$\frac{{\cal{B}} ( \bar{B}^0 \to D^+ K^- )}{{\cal{B}} ( \bar{B}^0 \to D^+ \pi^- )}$ & \begin{tabular}{l} LHCb \cite{Aaij:2013qqa}: $0.0822 \pm 0.0011 \pm 0.0025$ \\ Belle \cite{Abe:2001waa}: $0.068 \pm 0.015 \pm 0.007$ \\ \end{tabular} & $0.0818 \pm 0.0027$ \\
\hline
\multicolumn{3}{|l|}{${\cal{B}} ( \bar{B}^0 \to D^+ K^- \pi^+ \pi^- ) / {\cal{B}} ( \bar{B}^0 \to D^+ \pi^+ \pi^- \pi^- )$}\\
 & \begin{tabular}{l} LHCb \cite{Aaij:2012bw}: $0.059 \pm 0.011 \pm 0.005$ \\ \end{tabular} & $0.059 \pm 0.012$ \\
\hline
\multicolumn{3}{|l|}{${{\cal{B}} ( \bar{B}^0 \to D^{*+}(2010) \pi^- )}/{{\cal{B}} ( \bar{B}^0 \to D^+ \pi^- )}$}\\
 & \begin{tabular}{l} \babar \cite{Aubert:2006jc}: $0.99 \pm 0.11 \pm 0.08$ \\ \end{tabular} & $0.99 \pm 0.14$ \\
\hline
\multicolumn{3}{|l|}{${{\cal{B}} ( \bar{B}^0 \to D^{*+}(2010) K^- )}/{{\cal{B}} ( \bar{B}^0 \to D^{*+}(2010) \pi^- )}$}\\
 & \begin{tabular}{l} \babar \cite{Aubert:2005yt}: $0.0776 \pm 0.0034 \pm 0.0029$ \\ Belle \cite{Abe:2001waa}: $0.074 \pm 0.015 \pm 0.006$ \\ \end{tabular} & $0.0773 \pm 0.0043$ \\
\hline
\multicolumn{3}{|l|}{${{\cal{B}} ( \bar{B}^0 \to D^{**+} \pi^- )}/{{\cal{B}} ( \bar{B}^0 \to D^+ \pi^- )}$}\\
 & \begin{tabular}{l} \babar \cite{Aubert:2006jc}: $0.77 \pm 0.22 \pm 0.29$ \\ \end{tabular} & $0.77 \pm 0.36$ \\
\hline
\multicolumn{3}{|l|}{${{\cal{B}} ( \bar{B}^0 \to D_s^+ K^-  \pi^+  \pi^- )}/{{\cal{B}} ( \bar{B}_s^0 \to D_s^+ K^-  \pi^+  \pi^- )}$}\\
 & \begin{tabular}{l} LHCb \cite{Aaij:2012mra}: $0.54 \pm 0.07 \pm 0.07$ \\ \end{tabular} & $0.54 \pm 0.10$ \\
\hline
\end{btocharmtab}
\btocharmfig{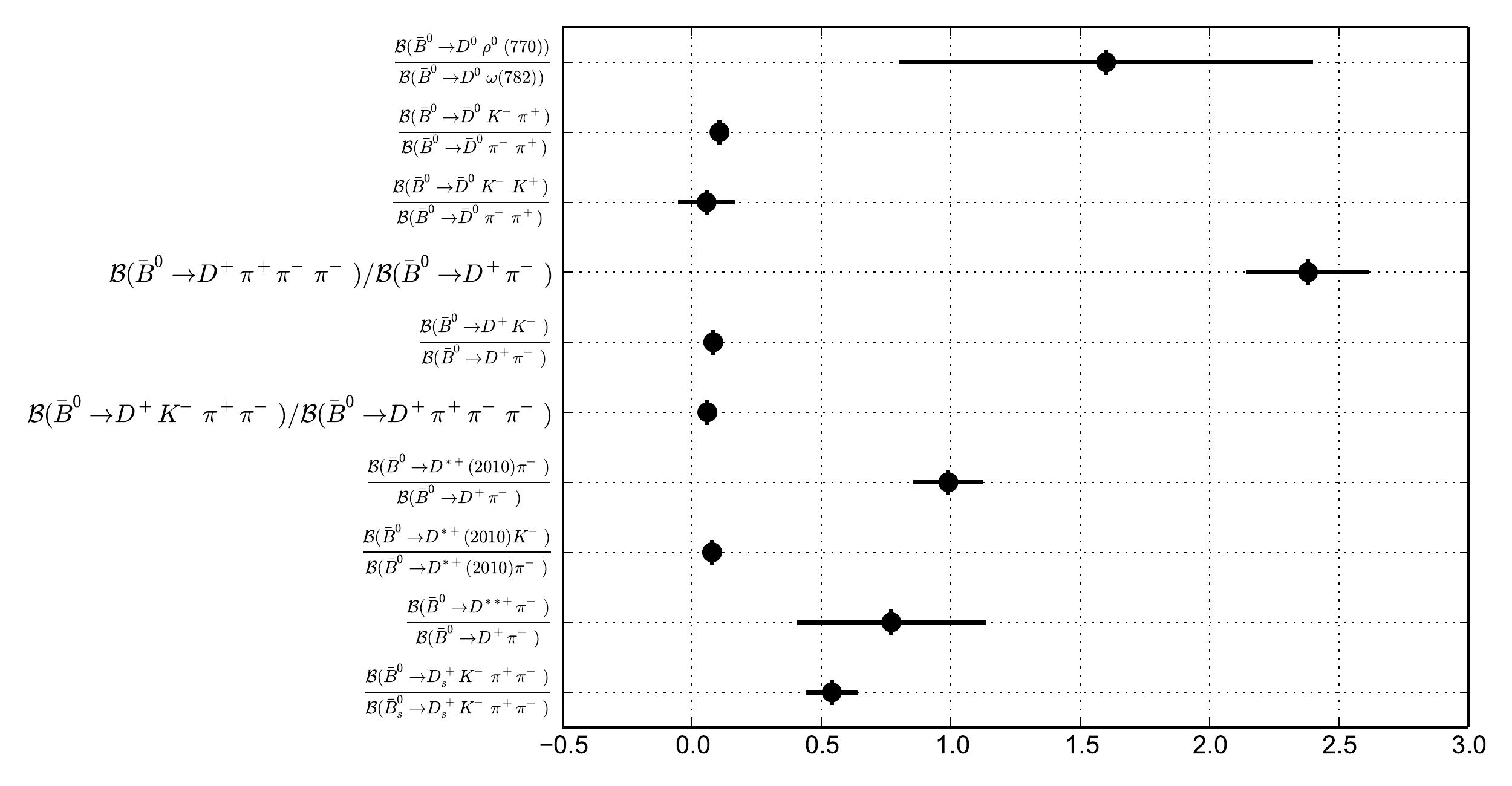}
\begin{btocharmtab}{Bd_D_7}{Absolute product decay rates to excited $D$ mesons $[10^{-3}]$}
\hline
\textbf{Parameter} & \begin{tabular}{l}\textbf{Measurements}\end{tabular} & \textbf{Average} \\
\hline
\hline
\multicolumn{3}{|l|}{${{\cal{B}} ( \bar{B}^0 \to D_1^+(2420) \pi^- )\times {\cal{B}} ( D_1^+(2420) \to D^{*0}(2007) \pi^+ )}$}\\
 & \begin{tabular}{l} Belle \cite{Abe:2004cw}: $0.368 \pm 0.060 \,^{+0.096}_{-0.050}$ \\ \end{tabular} & $0.368 \,^{+0.113}_{-0.078}$ \\
\hline
\multicolumn{3}{|l|}{${{\cal{B}} ( \bar{B}^0 \to D_1^+(2420) \pi^- )\times {\cal{B}} ( D_1^+(2420) \to D^+ \pi^- \pi^+ )}$}\\
 & \begin{tabular}{l} Belle \cite{Abe:2004sm}: $0.089 \pm 0.015 \,^{+0.017}_{-0.031}$ \\ \end{tabular} & $0.089 \,^{+0.023}_{-0.034}$ \\
\hline
\multicolumn{3}{|l|}{${{\cal{B}} ( \bar{B}^0 \to D_1^+(2420) \pi^- )\times {\cal{B}} ( D_1^+(2420) \to D^{*+}(2010) \pi^- \pi^+ )}$}\\
 & \begin{tabular}{l} Belle \cite{Abe:2004sm}: $< 0.033$ \\ \end{tabular} & $< 0.033$ \\
\hline
\multicolumn{3}{|l|}{${{\cal{B}} ( \bar{B}^0 \to D_2^{*+}(2460) \pi^- )\times {\cal{B}} ( D_2^{*+}(2460) \to D^0 \pi^+ )}$}\\
 & \begin{tabular}{l} Belle \cite{Abe:2004cw}: $0.308 \pm 0.033 \,^{+0.017}_{-0.009}$ \\ \end{tabular} & $0.308 \,^{+0.037}_{-0.034}$ \\
\hline
\multicolumn{3}{|l|}{${{\cal{B}} ( \bar{B}^0 \to D_2^{*+}(2460) \pi^- )\times {\cal{B}} ( D_2^{*+}(2460) \to D^{*0}(2007) \pi^+ )}$}\\
 & \begin{tabular}{l} Belle \cite{Abe:2004cw}: $0.245 \pm 0.042 \,^{+0.052}_{-0.048}$ \\ \end{tabular} & $0.245 \,^{+0.067}_{-0.064}$ \\
\hline
\multicolumn{3}{|l|}{${{\cal{B}} ( \bar{B}^0 \to D_2^{*+}(2460) \pi^- )\times {\cal{B}} ( D_2^{*+}(2460) \to D^{*+}(2010) \pi^- \pi^+ )}$}\\
 & \begin{tabular}{l} Belle \cite{Abe:2004sm}: $< 0.024$ \\ \end{tabular} & $< 0.024$ \\
\hline
\multicolumn{3}{|l|}{${{\cal{B}} ( \bar{B}^0 \to D_2^{*+}(2460) K^- )\times {\cal{B}} ( D_2^{*+}(2460) \to D^0 \pi^+ )}$}\\
 & \begin{tabular}{l} \babar \cite{Aubert:2005yt}: $0.0183 \pm 0.0040 \pm 0.0031$ \\ \end{tabular} & $0.0183 \pm 0.0051$ \\
\hline
\multicolumn{3}{|l|}{${{\cal{B}} ( \bar{B}^0 \to D_0^{*+} \pi^- )\times {\cal{B}} ( D_0^{*+} \to D^0 \pi^+ )}$}\\
 & \begin{tabular}{l} Belle \cite{Abe:2004cw}: $< 0.12$ \\ \end{tabular} & $< 0.12$ \\
\hline
\multicolumn{3}{|l|}{${{\cal{B}} ( \bar{B}^0 \to D_1^+(H) \pi^- )\times {\cal{B}} ( D_1^+(H) \to D^{*0}(2007) \pi^+ )}$}\\
 & \begin{tabular}{l} Belle \cite{Abe:2004cw}: $< 0.07$ \\ \end{tabular} & $< 0.07$ \\
\hline
\multicolumn{3}{|l|}{${{\cal{B}} ( \bar{B}^0 \to D_1^0(H) \omega(782) )\times {\cal{B}} ( D_1^0(H) \to D^{*+}(2010) \pi^- )}$}\\
 & \begin{tabular}{l} \babar \cite{Aubert:2006zb}: $0.41 \pm 0.12 \pm 0.11$ \\ \end{tabular} & $0.41 \pm 0.16$ \\
\hline
\multicolumn{3}{|l|}{${{\cal{B}} ( \bar{B}^0 \to D_{sJ}^{-}(2460) \pi^+ )\times {\cal{B}} ( D_{sJ}^{-}(2460) \to D_s^- \gamma )}$}\\
 & \begin{tabular}{l} Belle \cite{Abe:2004wz}: $< 0.0040$ \\ \end{tabular} & $< 0.0040$ \\
\hline
\multicolumn{3}{|l|}{${{\cal{B}} ( \bar{B}^0 \to D_{sJ}^{+}(2460) K^- )\times {\cal{B}} ( D_{sJ}^{+}(2460) \to D_s^+ \gamma )}$}\\
 & \begin{tabular}{l} Belle \cite{Abe:2005zy}: $0.0053 \pm 0.0020 \,^{+0.0016}_{-0.0015}$ \\ \end{tabular} & $0.0053 \,^{+0.0026}_{-0.0025}$ \\
\hline
\multicolumn{3}{|l|}{${{\cal{B}} ( \bar{B}^0 \to D_{sJ}^{*}(2317)^{-} \pi^+ )\times {\cal{B}} ( D_{sJ}^{*}(2317)^{-} \to D_s^- \pi^0 )}$}\\
 & \begin{tabular}{l} Belle \cite{Abe:2004wz}: $< 0.025$ \\ \end{tabular} & $< 0.025$ \\
\hline
\multicolumn{3}{|l|}{${{\cal{B}} ( \bar{B}^0 \to D_{sJ}^{*}(2317)^{+} K^- )\times {\cal{B}} ( D_{sJ}^{*}(2317)^{+} \to D_s^+ \pi^0 )}$}\\
 & \begin{tabular}{l} Belle \cite{Abe:2005zy}: $0.044 \pm 0.008 \pm 0.013$ \\ \end{tabular} & $0.044 \pm 0.015$ \\
\hline
\end{btocharmtab}
\btocharmfig{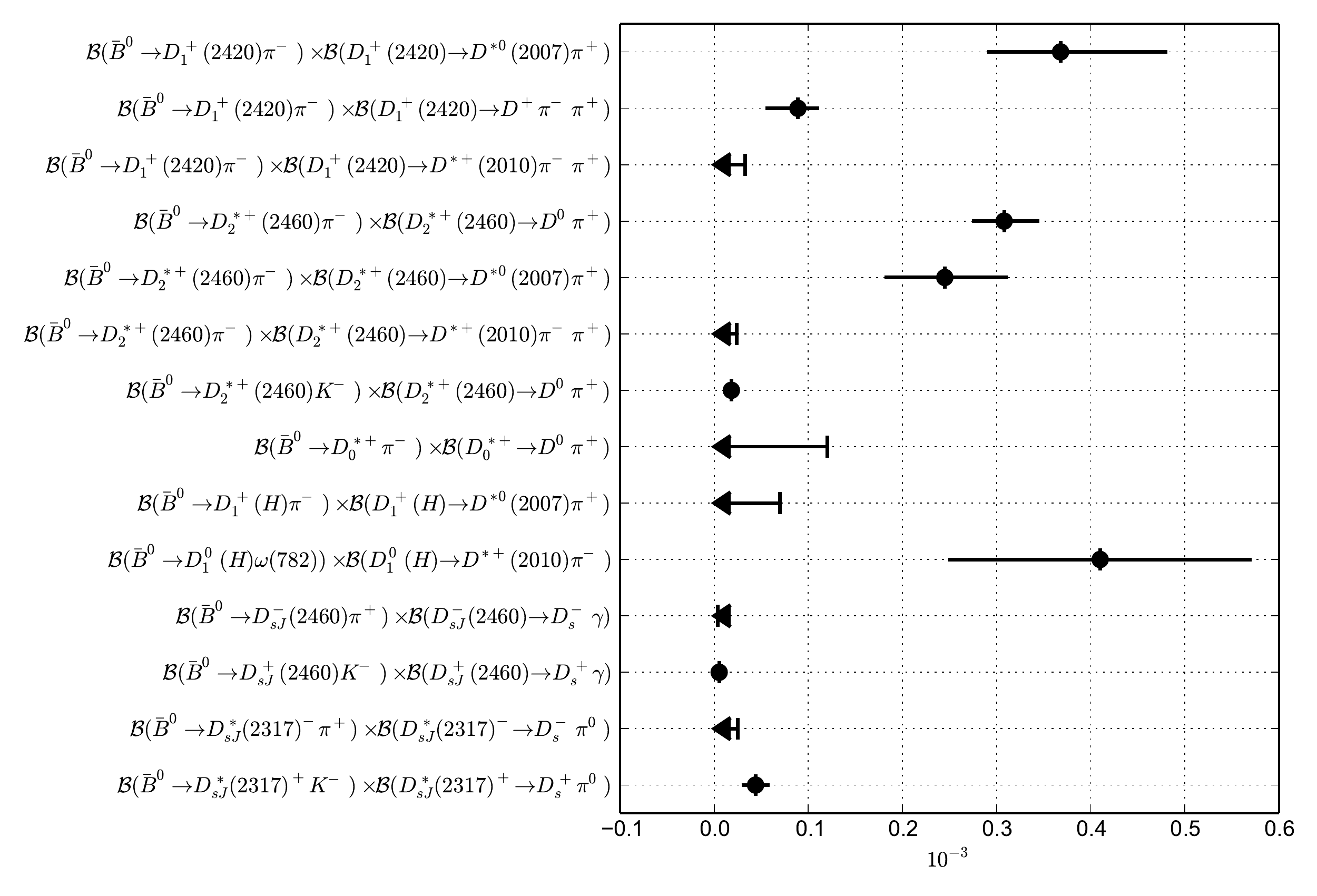}
\begin{btocharmtab}{Bd_D_8}{Absolute and relative decay rates to excited $D$ mesons $[10^{-2}]$}
\hline
\textbf{Parameter} & \begin{tabular}{l}\textbf{Measurements}\end{tabular} & \textbf{Average} \\
\hline
\hline
${\cal{B}} ( \bar{B}^0 \to D^{**+} \pi^- )$ & \begin{tabular}{l} \babar \cite{Aubert:2006jc}: $0.234 \pm 0.065 \pm 0.088$ \\ \end{tabular} & $0.234 \pm 0.109$ \\
\hline
\multicolumn{3}{|l|}{${\cal{B}} ( \bar{B}^0 \to D_1^+ \pi^- ) \times {\cal{B}} ( D_1^+ \to D^+ \pi^+ \pi^- ) / {\cal{B}} ( \bar{B}^0 \to D^+ \pi^+ \pi^- \pi^- )$}\\
 & \begin{tabular}{l} LHCb \cite{Aaij:2011rj}: $2.1 \pm 0.5 \,^{+0.3}_{-0.5}$ \\ \end{tabular} & $2.1 \,^{+0.6}_{-0.7}$ \\
\hline
\end{btocharmtab}
\btocharmfig{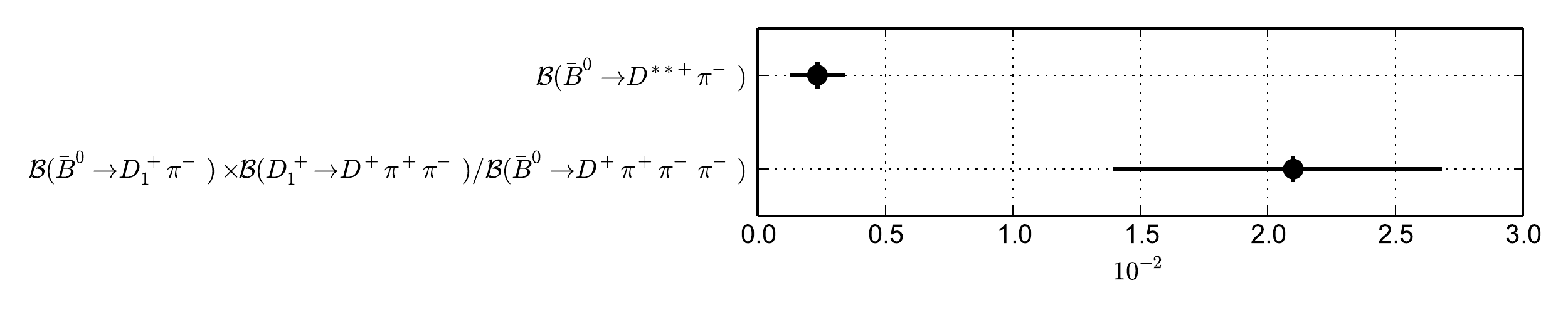}
\begin{btocharmtab}{Bd_D_9}{Baryonic decays $[10^{-4}]$}
\hline
\textbf{Parameter} & \begin{tabular}{l}\textbf{Measurements}\end{tabular} & \textbf{Average} \\
\hline
\hline
${\cal{B}} ( \bar{B}^0 \to D^0 p \bar{p} )$ & \begin{tabular}{l} \babar \cite{delAmoSanchez:2011gi}: $1.02 \pm 0.04 \pm 0.06$ \\ Belle \cite{Abe:2002tw}: $1.18 \pm 0.15 \pm 0.16$ \\ \end{tabular} & $1.04 \pm 0.07$ \\
\hline
${\cal{B}} ( \bar{B}^0 \to D^{*0}(2007) p \bar{p} )$ & \begin{tabular}{l} \babar \cite{delAmoSanchez:2011gi}: $0.97 \pm 0.07 \pm 0.09$ \\ Belle \cite{Abe:2002tw}: $1.20 \,^{+0.33}_{-0.29} \pm 0.21$ \\ \end{tabular} & $0.99 \pm 0.11$ \\
\hline
${\cal{B}} ( \bar{B}^0 \to D^+ p \bar{p} \pi^- )$ & \begin{tabular}{l} \babar \cite{delAmoSanchez:2011gi}: $3.32 \pm 0.10 \pm 0.29$ \\ \end{tabular} & $3.32 \pm 0.31$ \\
\hline
${\cal{B}} ( \bar{B}^0 \to D^{*+}(2010) p \bar{p} \pi^- )$ & \begin{tabular}{l} \babar \cite{delAmoSanchez:2011gi}: $4.55 \pm 0.16 \pm 0.39$ \\ \end{tabular} & $4.55 \pm 0.42$ \\
\hline
${\cal{B}} ( \bar{B}^0 \to D^0 p \bar{p} \pi^- \pi^+ )$ & \begin{tabular}{l} \babar \cite{delAmoSanchez:2011gi}: $2.99 \pm 0.21 \pm 0.45$ \\ \end{tabular} & $2.99 \pm 0.50$ \\
\hline
${\cal{B}} ( \bar{B}^0 \to D^{*0}(2007) p \bar{p} \pi^- \pi^+ )$ & \begin{tabular}{l} \babar \cite{delAmoSanchez:2011gi}: $1.91 \pm 0.36 \pm 0.29$ \\ \end{tabular} & $1.91 \pm 0.46$ \\
\hline
${\cal{B}} ( \bar{B}^0 \to D_s^+ \Lambda \bar{p} )$ & \begin{tabular}{l} Belle \cite{Medvedeva:2007af}: $0.29 \pm 0.07 \pm 0.06$ \\ \end{tabular} & $0.29 \pm 0.09$ \\
\hline
${\cal{B}} ( \bar{B}^0 \to D^0 \Lambda^0 \bar{\Lambda}^0 )$ & \begin{tabular}{l} \babar \cite{Lees:2014mka}: $0.098 \,^{+0.029}_{-0.026} \pm 0.019$ \\ Belle \cite{Chang:2008yw}: $0.105 \,^{+0.057}_{-0.044} \pm 0.014$ \\ \end{tabular} & $0.100 \pm 0.028$ \\
\hline
\multicolumn{3}{|l|}{${\cal{B}}(\bar{B}^0\to D^0\Sigma^0\bar\Lambda+\bar{B}^0\to D^0\Lambda\bar\Sigma^0) $}\\
 & \begin{tabular}{l} \babar \cite{Lees:2014mka}: $0.15 \,^{+0.09}_{-0.08} \pm 0.03$ \\ \end{tabular} & $0.15 \,^{+0.09}_{-0.09}$ \\
\hline
\end{btocharmtab}
\btocharmfig{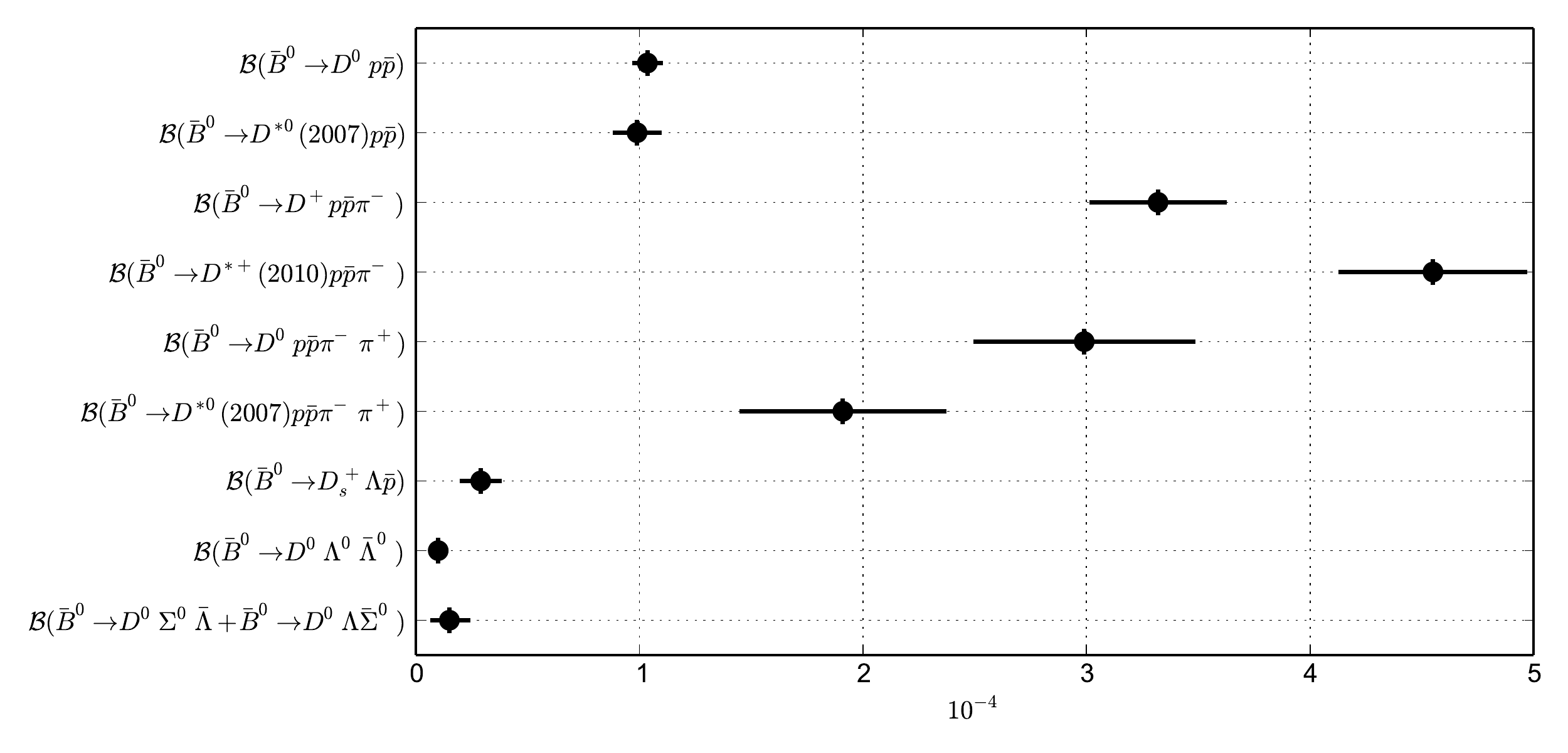}
\begin{btocharmtab}{Bd_D_10}{Semileptonic decays}
\hline
\textbf{Parameter} & \begin{tabular}{l}\textbf{Measurements}\end{tabular} & \textbf{Average} \\
\hline
\hline
\multicolumn{3}{|l|}{${{\cal{B}} ( \bar{B}^0 \to D^+ \mu^- \bar{\nu}_\mu )}/{{\cal{B}} ( \bar{B}^0 \to D^+ \pi^- )}$}\\
 & \begin{tabular}{l} CDF \cite{Aaltonen:2008eu}: $9.9 \pm 1.0 \pm 0.9$ \\ \end{tabular} & $9.9 \pm 1.3$ \\
\hline
\multicolumn{3}{|l|}{${{\cal{B}} ( \bar{B}^0 \to D^{*+}(2010) \mu^- \bar{\nu}_\mu )}/{{\cal{B}} ( \bar{B}^0 \to D^{*+}(2010) \pi^- )}$}\\
 & \begin{tabular}{l} CDF \cite{Aaltonen:2008eu}: $16.5 \pm 2.3 \pm 1.1$ \\ \end{tabular} & $16.5 \pm 2.6$ \\
\hline
\end{btocharmtab}
\btocharmfig{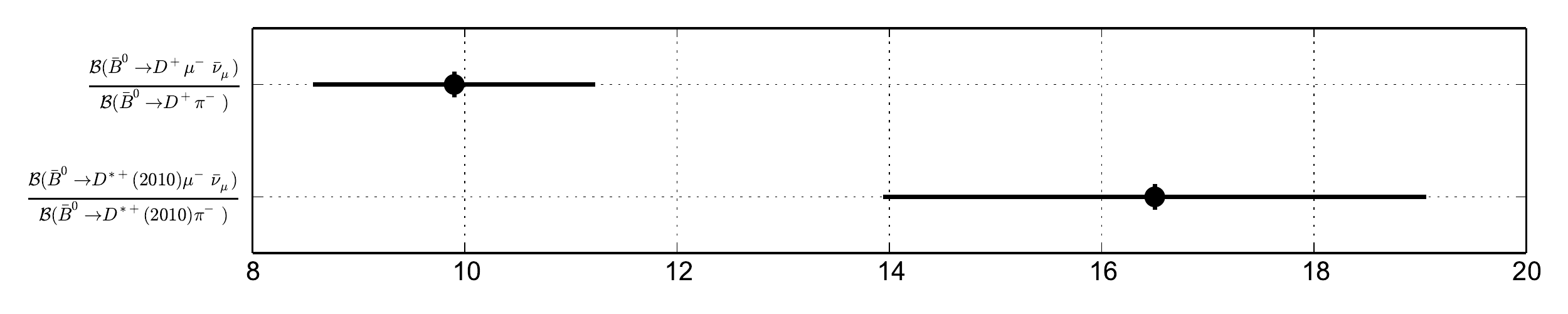}

\subsubsection{Decays to two open charm mesons}
\label{sec:b2c:Bd_DD}
Averages of $\Bzb$ decays to two open charm mesons are shown in Tables~\ref{tab:b2c:Bd_DD_1}--\ref{tab:b2c:Bd_DD_7} and Figs.~\ref{fig:b2c:Bd_DD_1}--\ref{fig:b2c:Bd_DD_7}.
\begin{btocharmtab}{Bd_DD_1}{Decays to $D^{(*)+}D^{(*)-}$ $[10^{-3}]$}
\hline
\textbf{Parameter} & \begin{tabular}{l}\textbf{Measurements}\end{tabular} & \textbf{Average} \\
\hline
\hline
${\cal{B}} ( \bar{B}^0 \to D^+ D^- )$ & \begin{tabular}{l} \babar \cite{Aubert:2006ia}: $0.28 \pm 0.04 \pm 0.05$ \\ Belle \cite{Rohrken:2012ta}: $0.212 \pm 0.016 \pm 0.018$ \\ \end{tabular} & $0.220 \pm 0.023$ \\
\hline
${\cal{B}} ( \bar{B}^0 \to D_s^- D^+ )$ & \begin{tabular}{l} \babar \cite{Aubert:2006ia}: $0.57 \pm 0.07 \pm 0.07$ \\ Belle \cite{Rohrken:2012ta}: $0.614 \pm 0.029 \pm 0.050$ \\ \end{tabular} & $0.603 \pm 0.050$ \\
\hline
${\cal{B}} ( \bar{B}^0 \to D^{*-} D^{*+}(2010) )$ & \begin{tabular}{l} \babar \cite{Aubert:2006ia}: $0.81 \pm 0.06 \pm 0.10$ \\ Belle \cite{Kronenbitter:2012ha}: $0.782 \pm 0.038 \pm 0.060$ \\ \end{tabular} & $0.790 \pm 0.061$ \\
\hline
${\cal{B}} ( \bar{B}^0 \to D^0 \bar{D}^0 )$ & \begin{tabular}{l} \babar \cite{Aubert:2006ia}: $< 0.06$ \\ Belle \cite{Adachi:2008cj}: $< 0.043$ \\ \end{tabular} & $< 0.043$ \\
\hline
${\cal{B}} ( \bar{B}^0 \to D^0 \bar{D}^{*0}(2007) )$ & \begin{tabular}{l} \babar \cite{Aubert:2006ia}: $< 0.29$ \\ \end{tabular} & $< 0.29$ \\
\hline
${\cal{B}} ( \bar{B}^0 \to D^{*0}(2007) \bar{D}^{*0}(2007) )$ & \begin{tabular}{l} \babar \cite{Aubert:2006ia}: $< 0.09$ \\ \end{tabular} & $< 0.09$ \\
\hline
\end{btocharmtab}
\btocharmfig{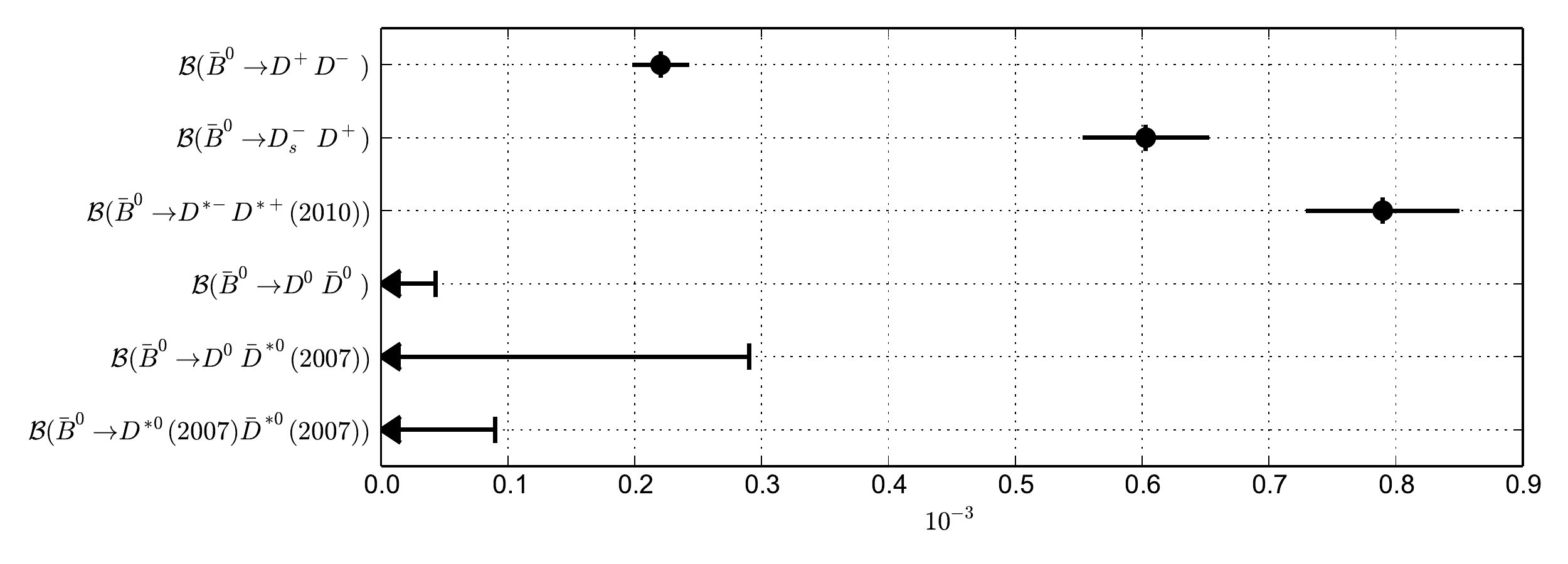}
\begin{btocharmtab}{Bd_DD_2}{Decays to two $D$ mesons and a kaon $[10^{-3}]$}
\hline
\textbf{Parameter} & \begin{tabular}{l}\textbf{Measurements}\end{tabular} & \textbf{Average} \\
\hline
\hline
${\cal{B}} ( \bar{B}^0 \to D^+ D^- \bar{K}^0 )$ & \begin{tabular}{l} \babar \cite{delAmoSanchez:2010pg}: $0.75 \pm 0.12 \pm 0.12$ \\ \end{tabular} & $0.75 \pm 0.17$ \\
\hline
${\cal{B}} ( \bar{B}^0 \to D^{*+}(2010) D^- \bar{K}^0 )$ & \begin{tabular}{l} \babar \cite{delAmoSanchez:2010pg}: $6.41 \pm 0.36 \pm 0.39$ \\ \end{tabular} & $6.41 \pm 0.53$ \\
\hline
${\cal{B}} ( \bar{B}^0 \to D^{*-}(2010) D^{*+}(2010) \bar{K}^0 )$ & \begin{tabular}{l} \babar \cite{delAmoSanchez:2010pg}: $8.26 \pm 0.43 \pm 0.67$ \\ \end{tabular} & $8.26 \pm 0.80$ \\
\hline
${\cal{B}} ( \bar{B}^0 \to D^{*+}(2010) D^{*-}(2010) K_S^0 )$ & \begin{tabular}{l} \babar \cite{Aubert:2006fh}: $4.4 \pm 0.4 \pm 0.7$ \\ Belle \cite{Dalseno:2007hx}: $3.4 \pm 0.4 \pm 0.7$ \\ \end{tabular} & $3.9 \pm 0.6$ \\
\hline
${\cal{B}} ( \bar{B}^0 \to D^+ \bar{D}^0 K^- )$ & \begin{tabular}{l} \babar \cite{delAmoSanchez:2010pg}: $1.07 \pm 0.07 \pm 0.09$ \\ \end{tabular} & $1.07 \pm 0.11$ \\
\hline
${\cal{B}} ( \bar{B}^0 \to D^{*+}(2010) \bar{D}^0 K^- )$ & \begin{tabular}{l} \babar \cite{delAmoSanchez:2010pg}: $2.47 \pm 0.10 \pm 0.18$ \\ \end{tabular} & $2.47 \pm 0.21$ \\
\hline
${\cal{B}} ( \bar{B}^0 \to D^+ \bar{D}^{*0}(2007) K^- )$ & \begin{tabular}{l} \babar \cite{delAmoSanchez:2010pg}: $3.46 \pm 0.18 \pm 0.37$ \\ \end{tabular} & $3.46 \pm 0.41$ \\
\hline
${\cal{B}} ( \bar{B}^0 \to D^{*+}(2010) \bar{D}^{*0}(2007) K^- )$ & \begin{tabular}{l} \babar \cite{delAmoSanchez:2010pg}: $10.6 \pm 0.3 \pm 0.9$ \\ \end{tabular} & $10.6 \pm 0.9$ \\
\hline
${\cal{B}} ( \bar{B}^0 \to D^0 \bar{D}^0 \bar{K}^0 )$ & \begin{tabular}{l} \babar \cite{delAmoSanchez:2010pg}: $0.27 \pm 0.10 \pm 0.05$ \\ \end{tabular} & $0.27 \pm 0.11$ \\
\hline
${\cal{B}} ( \bar{B}^0 \to D^0 \bar{D}^{*0}(2007) \bar{K}^0 )$ & \begin{tabular}{l} \babar \cite{delAmoSanchez:2010pg}: $1.08 \pm 0.32 \pm 0.36$ \\ \end{tabular} & $1.08 \pm 0.48$ \\
\hline
${\cal{B}} ( \bar{B}^0 \to D^{*0}(2007) \bar{D}^{*0}(2007) \bar{K}^0 )$ & \begin{tabular}{l} \babar \cite{delAmoSanchez:2010pg}: $2.40 \pm 0.55 \pm 0.67$ \\ \end{tabular} & $2.40 \pm 0.87$ \\
\hline
${\cal{B}} ( \bar{B}^0 \to D^0 \bar{D}^0 \pi^0 \bar{K}^0 )$ & \begin{tabular}{l} Belle \cite{Gokhroo:2006bt}: $0.173 \pm 0.070 \,^{+0.031}_{-0.053}$ \\ \end{tabular} & $0.173 \,^{+0.077}_{-0.088}$ \\
\hline
\end{btocharmtab}
\btocharmfig{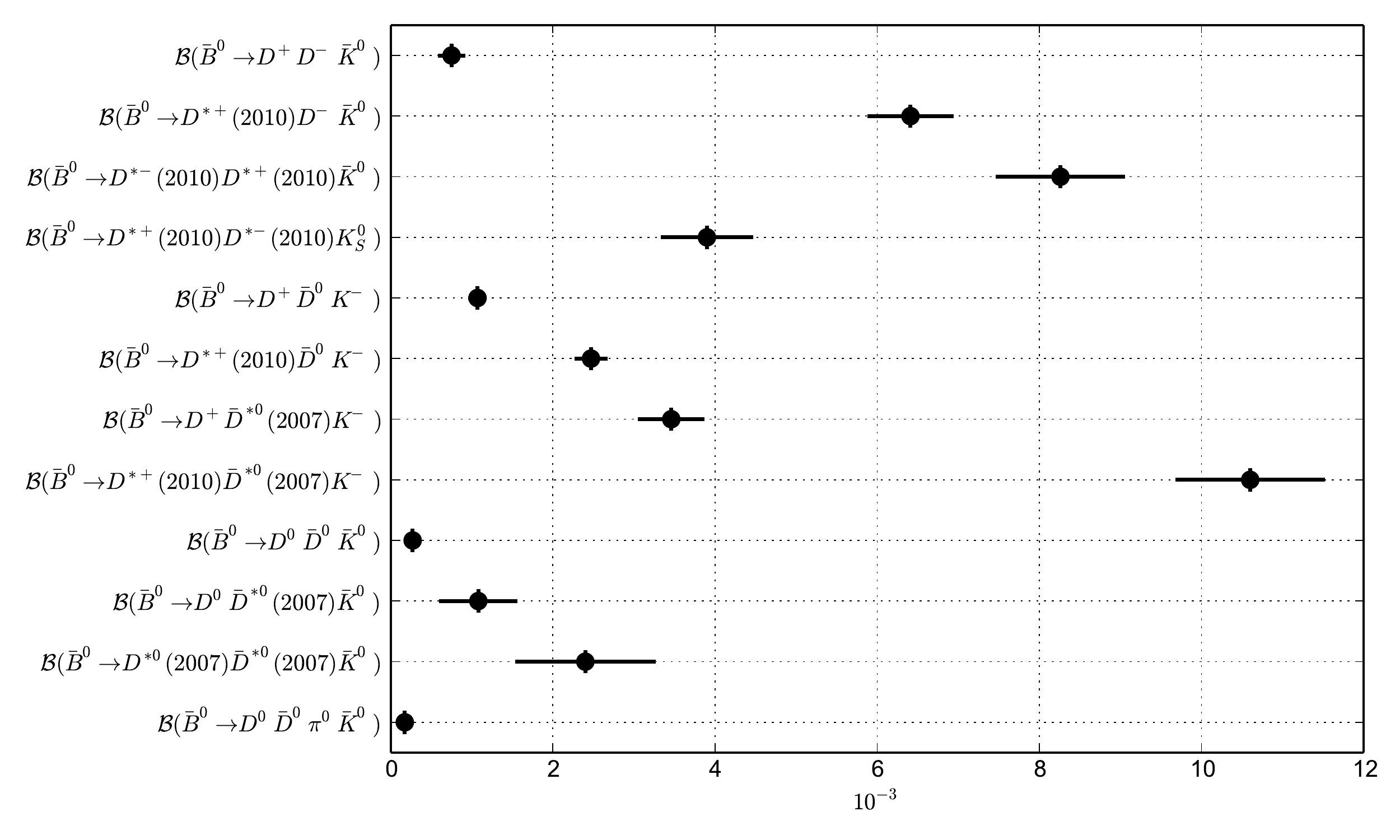}
\begin{btocharmtab}{Bd_DD_3}{Decays to $D_s^{(*)-}D^{(*)+}$ $[10^{-3}]$}
\hline
\textbf{Parameter} & \begin{tabular}{l}\textbf{Measurements}\end{tabular} & \textbf{Average} \\
\hline
\hline
${\cal{B}} ( \bar{B}^0 \to D_s^- D^+ )$ & \begin{tabular}{l} \babar \cite{Aubert:2006nm}: $9.0 \pm 1.8 \pm 1.4$ \\ Belle \cite{Zupanc:2007pu}: $7.5 \pm 0.2 \pm 1.1$ \\ \end{tabular} & $7.8 \pm 1.0$ \\
\hline
${\cal{B}} ( \bar{B}^0 \to D_s^{-} D^{*+}(2010) )$ & \begin{tabular}{l} \babar \cite{Aubert:2003jj}: $10.3 \pm 1.4 \pm 2.9$ \\ \babar \cite{Aubert:2006nm}: $5.7 \pm 1.6 \pm 0.9$ \\ \end{tabular} & $6.8 \pm 1.6$ \\
\hline
${\cal{B}} ( \bar{B}^0 \to D_s^{*-} D^{*+}(2010) )$ & \begin{tabular}{l} \babar \cite{Aubert:2003jj}: $19.7 \pm 1.5 \pm 5.7$ \\ \babar \cite{Aubert:2005xu}: $18.8 \pm 0.9 \pm 1.7$ \\ \babar \cite{Aubert:2006nm}: $16.5 \pm 2.3 \pm 1.9$ \\ \end{tabular} & $18.1 \pm 1.6$ \\
\hline
${\cal{B}} ( \bar{B}^0 \to D_s^{*-} D^+ )$ & \begin{tabular}{l} \babar \cite{Aubert:2006nm}: $6.7 \pm 2.0 \pm 1.1$ \\ \end{tabular} & $6.7 \pm 2.3$ \\
\hline
\multicolumn{3}{|l|}{${{\cal{B}} ( \bar{B}^0 \to D_s^- D^+ )\times {\cal{B}} ( D_s^- \to \phi(1020) \pi^- )\times {\cal{B}} ( \phi(1020) \to K^+ K^- )}$}\\
 & \begin{tabular}{l} Belle \cite{Abe:2005im}: $0.147 \pm 0.005 \pm 0.021$ \\ \end{tabular} & $0.147 \pm 0.022$ \\
\hline
\multicolumn{3}{|l|}{${{\cal{B}} ( \bar{B}^0 \to D_s^- D^+ )\times {\cal{B}} ( D_s^- \to \phi(1020) \pi^- )}$}\\
 & \begin{tabular}{l} \babar \cite{Aubert:2006nm}: $0.267 \pm 0.061 \pm 0.047$ \\ \end{tabular} & $0.267 \pm 0.077$ \\
\hline
\multicolumn{3}{|l|}{${{\cal{B}} ( \bar{B}^0 \to D_s^- D^{*+}(2010) )\times {\cal{B}} ( D_s^- \to \phi(1020) \pi^- )}$}\\
 & \begin{tabular}{l} \babar \cite{Aubert:2006nm}: $0.511 \pm 0.094 \pm 0.072$ \\ \end{tabular} & $0.511 \pm 0.118$ \\
\hline
\multicolumn{3}{|l|}{${{\cal{B}} ( \bar{B}^0 \to D_s^{*-} D^+ )\times {\cal{B}} ( D_s^- \to \phi(1020) \pi^- )}$}\\
 & \begin{tabular}{l} \babar \cite{Aubert:2006nm}: $0.414 \pm 0.119 \pm 0.094$ \\ \end{tabular} & $0.414 \pm 0.152$ \\
\hline
\multicolumn{3}{|l|}{${{\cal{B}} ( \bar{B}^0 \to D_s^{*-} D^{*+}(2010) )\times {\cal{B}} ( D_s^- \to \phi(1020) \pi^- )}$}\\
 & \begin{tabular}{l} \babar \cite{Aubert:2006nm}: $1.22 \pm 0.22 \pm 0.22$ \\ \end{tabular} & $1.22 \pm 0.31$ \\
\hline
\end{btocharmtab}
\btocharmfig{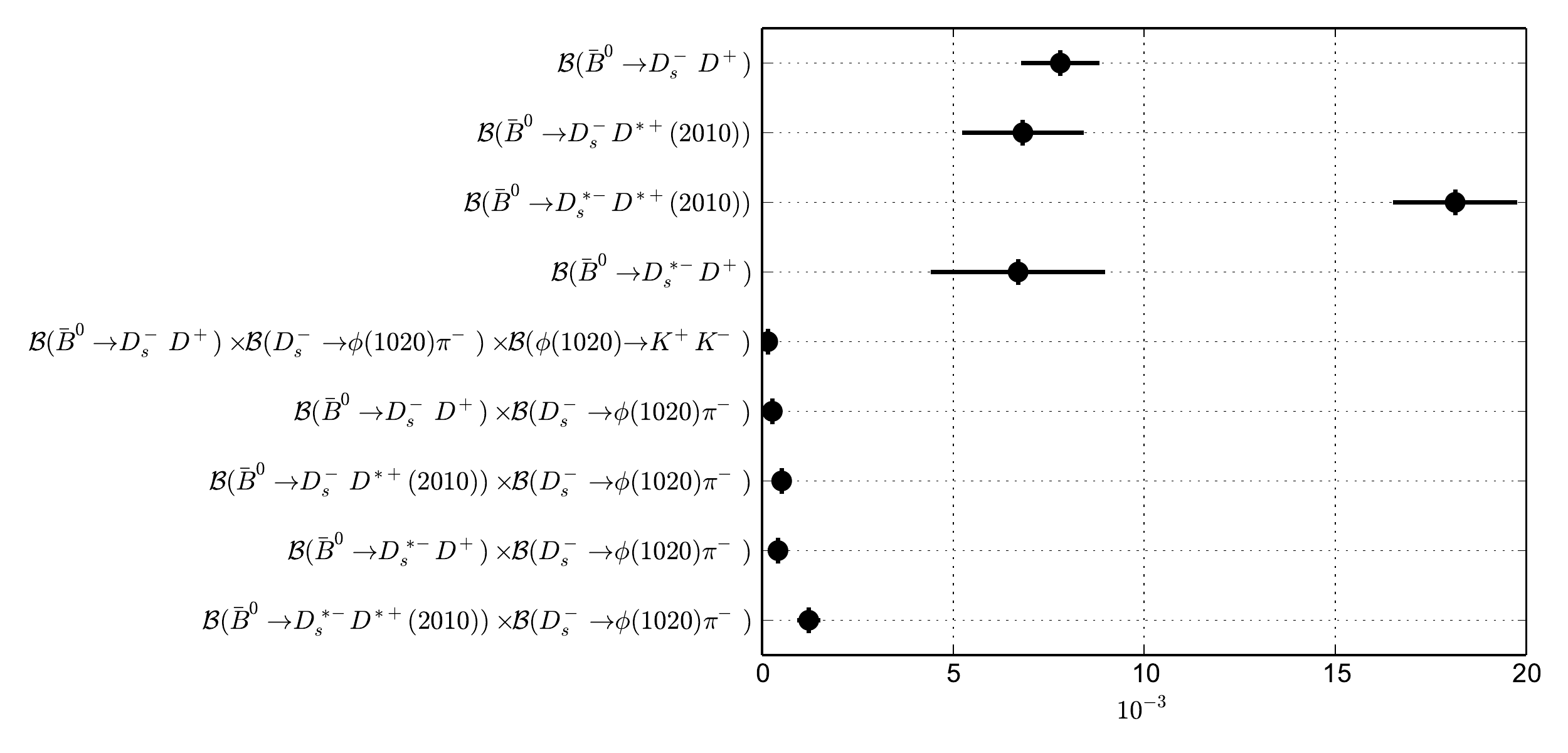}
\begin{btocharmtab}{Bd_DD_4}{Decays to $D_s^{(*)+}D_s^{(*)-}$ $[10^{-3}]$}
\hline
\textbf{Parameter} & \begin{tabular}{l}\textbf{Measurements}\end{tabular} & \textbf{Average} \\
\hline
\hline
${\cal{B}} ( \bar{B}^0 \to D_s^{-} D_s^{+} )$ & \begin{tabular}{l} \babar \cite{Aubert:2005jv}: $< 0.10$ \\ Belle \cite{Zupanc:2007pu}: $< 0.036$ \\ \end{tabular} & $< 0.036$ \\
\hline
${\cal{B}} ( \bar{B}^0 \to D_s^- D_s^{*+} )$ & \begin{tabular}{l} \babar \cite{Aubert:2005jv}: $< 0.13$ \\ \end{tabular} & $< 0.13$ \\
\hline
${\cal{B}} ( \bar{B}^0 \to D_s^{*+} D_s^{*-} )$ & \begin{tabular}{l} \babar \cite{Aubert:2005jv}: $< 0.24$ \\ \end{tabular} & $< 0.24$ \\
\hline
\end{btocharmtab}
\btocharmfig{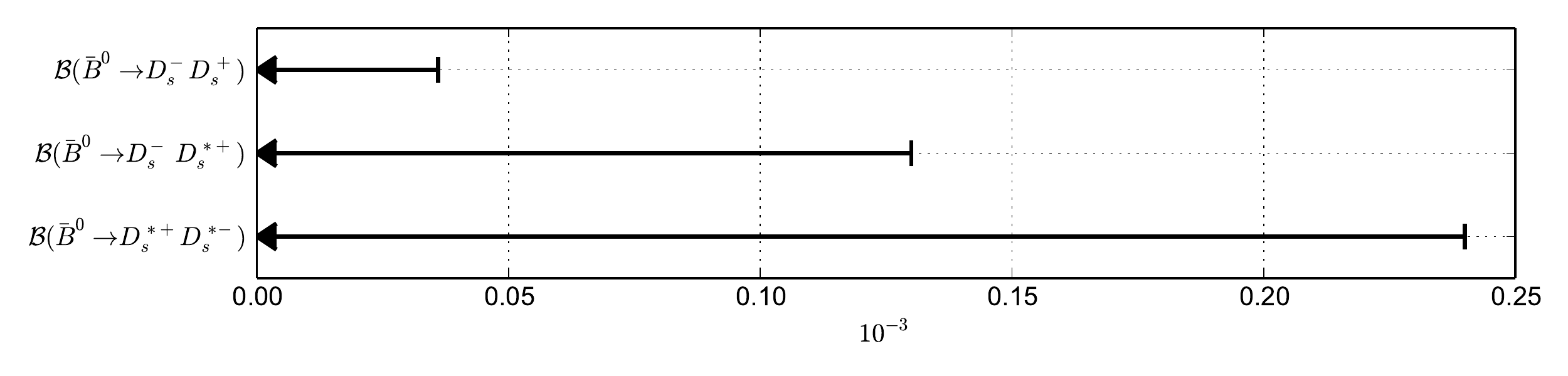}
\begin{btocharmtab}{Bd_DD_5}{Relative decay rates}
\hline
\textbf{Parameter} & \begin{tabular}{l}\textbf{Measurements}\end{tabular} & \textbf{Average} \\
\hline
\hline
\multicolumn{3}{|l|}{${\cal{B}} ( \bar{B}^0 \to D^0 \bar{D}^0 )/ {\cal{B}} ( \bar{B}^- \to D^0 \bar{D}_s^- )$}\\
 & \begin{tabular}{l} LHCb \cite{Aaij:2013fha}: $0.0014 \pm 0.0006 \pm 0.0002$ \\ \end{tabular} & $0.0014 \pm 0.0006$ \\
\hline
\multicolumn{3}{|l|}{${{\cal{B}} ( \bar{B}^0 \to D_s^- D^+ )}/{{\cal{B}} ( \bar{B}^0 \to D^+ \pi^+ \pi^- \pi^- )}$}\\
 & \begin{tabular}{l} CDF \cite{CDF:7925}: $1.99 \pm 0.13 \pm 0.46$ \\ \end{tabular} & $1.99 \pm 0.48$ \\
\hline
\multicolumn{3}{|l|}{${{\cal{B}} ( \bar{B}^0 \to D_s^- D^{*+}(2010) )}/{{\cal{B}} ( \bar{B}^0 \to D_s^- D^+ )}$}\\
 & \begin{tabular}{l} CDF \cite{CDF:7925}: $1.5 \pm 0.4 \pm 0.1$ \\ \end{tabular} & $1.5 \pm 0.4$ \\
\hline
\multicolumn{3}{|l|}{${{\cal{B}} ( \bar{B}^0 \to D_s^{*-} D^+ )}/{{\cal{B}} ( \bar{B}^0 \to D_s^- D^+ )}$}\\
 & \begin{tabular}{l} CDF \cite{CDF:7925}: $0.9 \pm 0.2 \pm 0.1$ \\ \end{tabular} & $0.9 \pm 0.2$ \\
\hline
\multicolumn{3}{|l|}{${{\cal{B}} ( \bar{B}^0 \to D_s^{*-} D^{*+}(2010) )}/{{\cal{B}} ( \bar{B}^0 \to D_s^- D^+ )}$}\\
 & \begin{tabular}{l} CDF \cite{CDF:7925}: $2.6 \pm 0.5 \pm 0.2$ \\ \end{tabular} & $2.6 \pm 0.5$ \\
\hline
\end{btocharmtab}
\btocharmfig{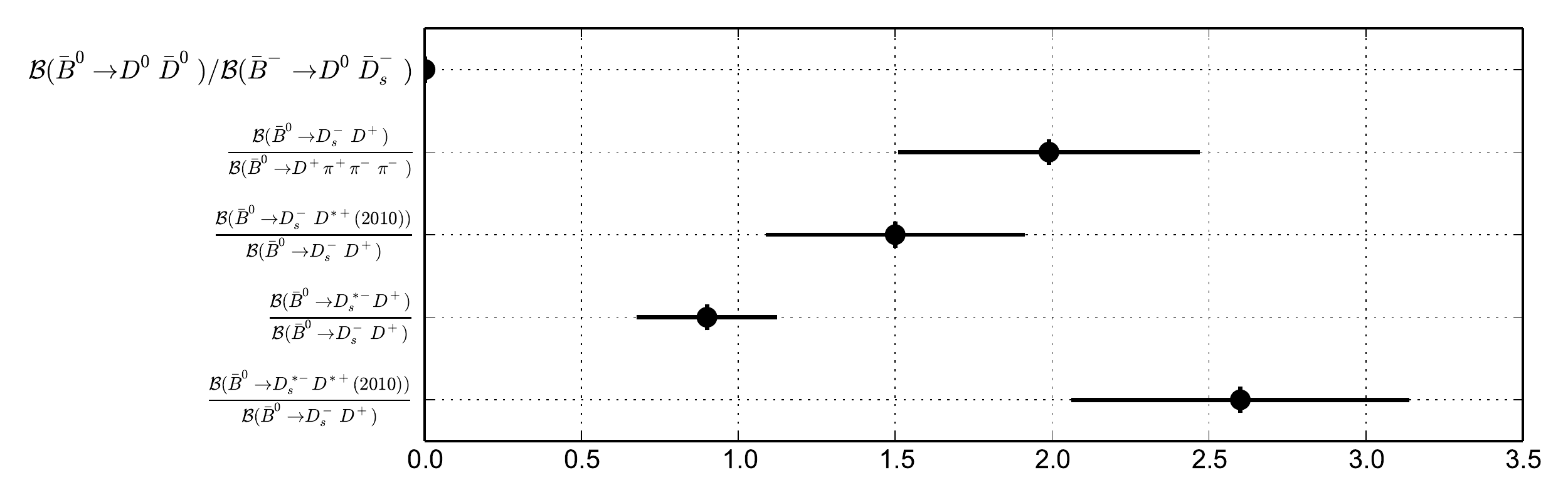}
\begin{btocharmtab}{Bd_DD_6}{Absolute decays rates to excited $D_s$ mesons $[10^{-3}]$}
\hline
\textbf{Parameter} & \begin{tabular}{l}\textbf{Measurements}\end{tabular} & \textbf{Average} \\
\hline
\hline
${\cal{B}} ( \bar{B}^0 \to D_{sJ}^{-}(2460) D^+ )$ & \begin{tabular}{l} \babar \cite{Aubert:2006nm}: $2.6 \pm 1.5 \pm 0.7$ \\ \end{tabular} & $2.6 \pm 1.7$ \\
\hline
${\cal{B}} ( \bar{B}^0 \to D_{sJ}^{-}(2460) D^{*+}(2010) )$ & \begin{tabular}{l} \babar \cite{Aubert:2006nm}: $8.8 \pm 2.0 \pm 1.4$ \\ \end{tabular} & $8.8 \pm 2.4$ \\
\hline
${\cal{B}} ( \bar{B}^0 \to D_{s1}^-(2536) D^{*+}(2010) )$ & \begin{tabular}{l} \babar \cite{Aubert:2006fh}: $92 \pm 24 \pm 1$ \\ \end{tabular} & $92 \pm 24$ \\
\hline
\end{btocharmtab}
\btocharmfig{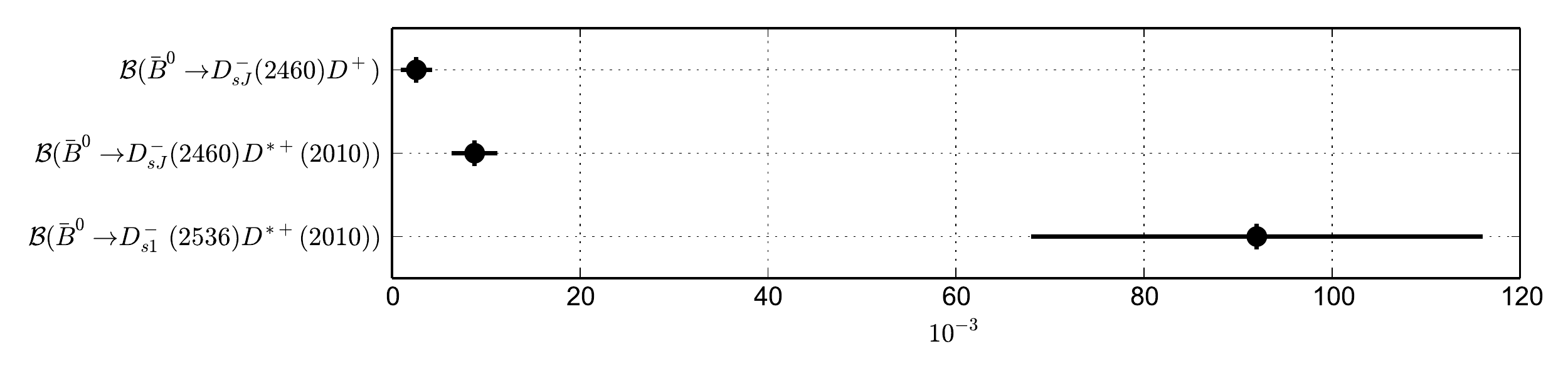}
\begin{btocharmtab}{Bd_DD_7}{Product decays rates to excited $D_s$ mesons $[10^{-3}]$}
\hline
\textbf{Parameter} & \begin{tabular}{l}\textbf{Measurements}\end{tabular} & \textbf{Average} \\
\hline
\hline
\multicolumn{3}{|l|}{${{\cal{B}} ( \bar{B}^0 \to D_{s1}^-(2536) D^+ )\times {\cal{B}} ( D_{s1}^-(2536) \to D^{*-}(2010) \bar{K}^0 )}$}\\
 & \begin{tabular}{l} \babar \cite{Aubert:2007rva}: $0.261 \pm 0.103 \pm 0.031$ \\ \end{tabular} & $0.261 \pm 0.108$ \\
\hline
\multicolumn{3}{|l|}{${{\cal{B}} ( \bar{B}^0 \to D_{s1}^-(2536) D^+ )\times {\cal{B}} ( D_{s1}^-(2536) \to K^- \bar{D}^{*0}(2007) )}$}\\
 & \begin{tabular}{l} \babar \cite{Aubert:2007rva}: $0.171 \pm 0.048 \pm 0.032$ \\ \end{tabular} & $0.171 \pm 0.058$ \\
\hline
\multicolumn{3}{|l|}{${{\cal{B}} ( \bar{B}^0 \to D_{s1}^-(2536) D^{*+}(2010) )\times {\cal{B}} ( D_{s1}^-(2536) \to D^{*-}(2010) \bar{K}^0 )}$}\\
 & \begin{tabular}{l} \babar \cite{Aubert:2007rva}: $0.500 \pm 0.151 \pm 0.067$ \\ \end{tabular} & $0.500 \pm 0.165$ \\
\hline
\multicolumn{3}{|l|}{${{\cal{B}} ( \bar{B}^0 \to D_{s1}^-(2536) D^{*+}(2010) )\times {\cal{B}} ( D_{s1}^-(2536) \to \bar{D}^{*0}(2007) K^+ )}$}\\
 & \begin{tabular}{l} \babar \cite{Aubert:2007rva}: $0.332 \pm 0.088 \pm 0.066$ \\ \end{tabular} & $0.332 \pm 0.110$ \\
\hline
\multicolumn{3}{|l|}{${{\cal{B}} ( \bar{B}^0 \to D_{s1}^+(2536) D^- )\times {\cal{B}} ( D_{s1}^+(2536) \to D^{*0}(2007) K^+ + D^{*+}(2010) K^0 )}$}\\
 & \begin{tabular}{l} Belle \cite{Belle:2011ad}: $0.275 \pm 0.062 \pm 0.036$ \\ \end{tabular} & $0.275 \pm 0.072$ \\
\hline
\multicolumn{3}{|l|}{${{\cal{B}} ( \bar{B}^0 \to D_{s1}^+(2536) D^{*-}(2010) )\times {\cal{B}} ( D_{s1}^+(2536) \to D^{*0}(2007) K^+ + D^{*+}(2010) K^0 )}$}\\
 & \begin{tabular}{l} Belle \cite{Belle:2011ad}: $0.501 \pm 0.121 \pm 0.070$ \\ \end{tabular} & $0.501 \pm 0.140$ \\
\hline
\multicolumn{3}{|l|}{${{\cal{B}} ( \bar{B}^0 \to D_{s1}^+(2536) D^{*-}(2010) )\times {\cal{B}} ( D_{s1}^+(2536) \to D^{*+}(2010) K_S^0 )}$}\\
 & \begin{tabular}{l} Belle \cite{Dalseno:2007hx}: $< 0.60$ \\ \end{tabular} & $< 0.60$ \\
\hline
\multicolumn{3}{|l|}{${{\cal{B}} ( \bar{B}^0 \to D^+ D_{sJ}^{-}(2460) )\times {\cal{B}} ( D_{sJ}^{-}(2460) \to D_s^- \pi^+ \pi^- )}$}\\
 & \begin{tabular}{l} Belle \cite{Krokovny:2003zq}: $< 0.20$ \\ \end{tabular} & $< 0.20$ \\
\hline
\multicolumn{3}{|l|}{${{\cal{B}} ( \bar{B}^0 \to D^+ D_{sJ}^{-}(2460) )\times {\cal{B}} ( D_{sJ}^{-}(2460) \to D_s^- \pi^0 )}$}\\
 & \begin{tabular}{l} Belle \cite{Krokovny:2003zq}: $< 0.36$ \\ \end{tabular} & $< 0.36$ \\
\hline
\multicolumn{3}{|l|}{${{\cal{B}} ( \bar{B}^0 \to D^+ D_{sJ}^{-}(2460) )\times {\cal{B}} ( D_{sJ}^{-}(2460) \to D_s^- \gamma )}$}\\
 & \begin{tabular}{l} \babar \cite{Aubert:2004pw}: $0.8 \pm 0.2 \,^{+0.3}_{-0.2}$ \\ Belle \cite{Krokovny:2003zq}: $0.82 \,^{+0.22}_{-0.19} \pm 0.25$ \\ \end{tabular} & $0.81 \pm 0.23$ \\
\hline
\multicolumn{3}{|l|}{${{\cal{B}} ( \bar{B}^0 \to D^+ D_{sJ}^{-}(2460) )\times {\cal{B}} ( D_{sJ}^{-}(2460) \to D_s^{*-} \pi^0 )}$}\\
 & \begin{tabular}{l} \babar \cite{Aubert:2004pw}: $2.8 \pm 0.8 \,^{+1.1}_{-0.8}$ \\ Belle \cite{Krokovny:2003zq}: $2.27 \,^{+0.73}_{-0.62} \pm 0.68$ \\ \end{tabular} & $2.47 \pm 0.76$ \\
\hline
\multicolumn{3}{|l|}{${{\cal{B}} ( \bar{B}^0 \to D^+ D_{sJ}^{-}(2460) )\times {\cal{B}} ( D_{sJ}^{-}(2460) \to D_s^{*-} \gamma )}$}\\
 & \begin{tabular}{l} Belle \cite{Krokovny:2003zq}: $0.13 \,^{+0.20}_{-0.14} \pm 0.00$ \\ \end{tabular} & $0.13 \,^{+0.20}_{-0.14}$ \\
\hline
\multicolumn{3}{|l|}{${{\cal{B}} ( \bar{B}^0 \to D_{sJ}^{-}(2460) D^{*+}(2010) )\times {\cal{B}} ( D_{sJ}^{-}(2460) \to D_s^{*-} \pi^0 )}$}\\
 & \begin{tabular}{l} \babar \cite{Aubert:2004pw}: $5.5 \pm 1.2 \,^{+2.1}_{-1.6}$ \\ \end{tabular} & $5.5 \,^{+2.5}_{-2.0}$ \\
\hline
\multicolumn{3}{|l|}{${{\cal{B}} ( \bar{B}^0 \to D_{sJ}^{-}(2460) D^{*+}(2010) )\times {\cal{B}} ( D_{sJ}^{-}(2460) \to D_s^- \gamma )}$}\\
 & \begin{tabular}{l} \babar \cite{Aubert:2004pw}: $2.3 \pm 0.3 \,^{+0.9}_{-0.6}$ \\ \end{tabular} & $2.3 \,^{+0.9}_{-0.7}$ \\
\hline
\multicolumn{3}{|l|}{${{\cal{B}} ( \bar{B}^0 \to D^+ D_{sJ}^{*}(2317)^{-} )\times {\cal{B}} ( D_{sJ}^{*}(2317)^{-} \to D_s^- \pi^0 )}$}\\
 & \begin{tabular}{l} \babar \cite{Aubert:2004pw}: $1.8 \pm 0.4 \,^{+0.7}_{-0.5}$ \\ Belle \cite{Krokovny:2003zq}: $0.86 \,^{+0.33}_{-0.26} \pm 0.26$ \\ \end{tabular} & $1.08 \pm 0.35$ \\
\hline
\multicolumn{3}{|l|}{${{\cal{B}} ( \bar{B}^0 \to D^+ D_{sJ}^{*}(2317)^{-} )\times {\cal{B}} ( D_{sJ}^{*}(2317)^{-} \to D_s^{*-} \gamma )}$}\\
 & \begin{tabular}{l} Belle \cite{Krokovny:2003zq}: $0.27 \,^{+0.29}_{-0.22} \pm 0.00$ \\ \end{tabular} & $0.27 \,^{+0.29}_{-0.22}$ \\
\hline
\multicolumn{3}{|l|}{${{\cal{B}} ( \bar{B}^0 \to D_{sJ}^{*}(2317)^{-} D^{*+}(2010) )\times {\cal{B}} ( D_{sJ}^{*}(2317)^{-} \to D_s^- \pi^0 )}$}\\
 & \begin{tabular}{l} \babar \cite{Aubert:2004pw}: $1.5 \pm 0.4 \,^{+0.5}_{-0.4}$ \\ \end{tabular} & $1.5 \,^{+0.7}_{-0.5}$ \\
\hline
\end{btocharmtab}
\btocharmfig{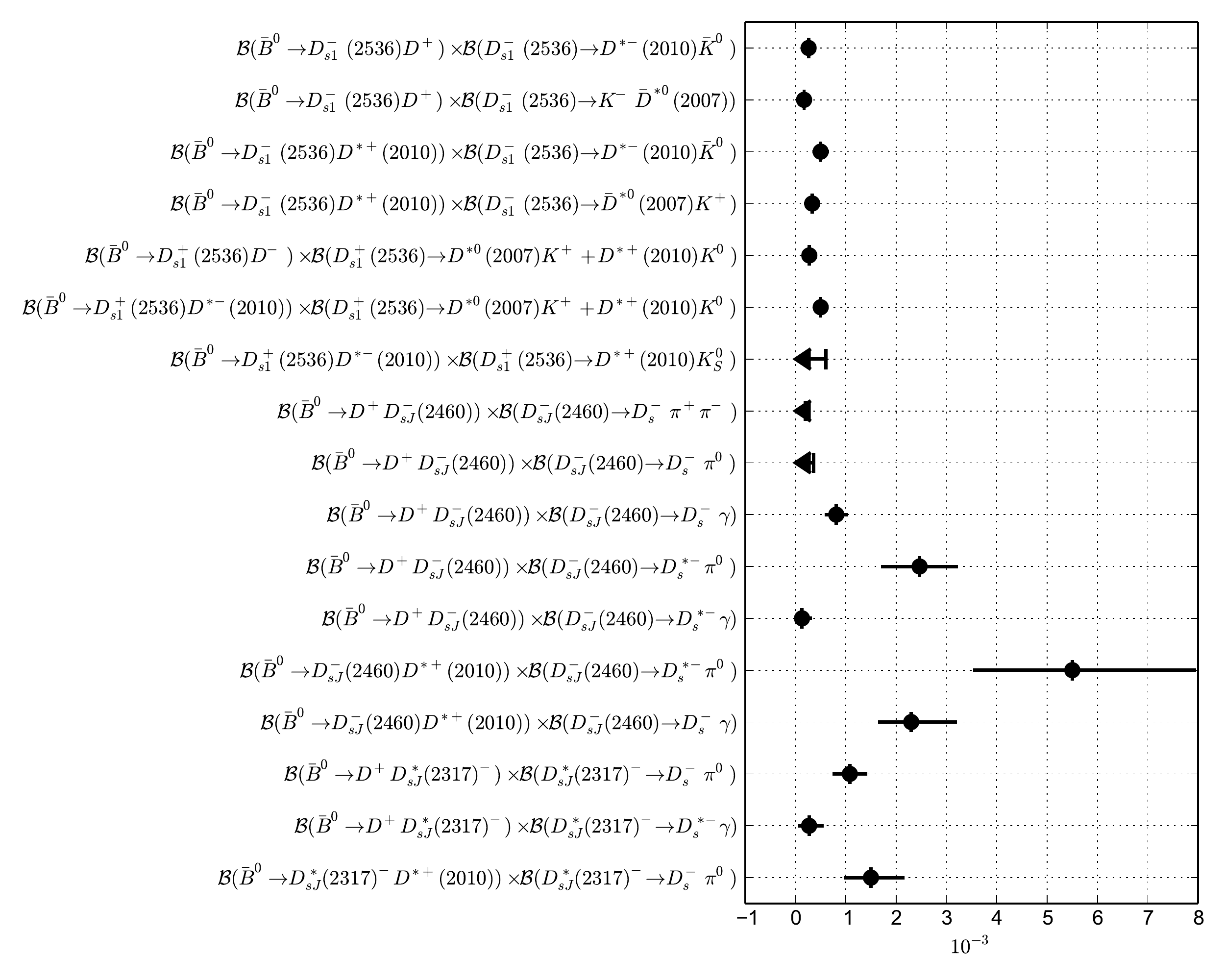}

\subsubsection{Decays to charmonium states}
\label{sec:b2c:Bd_cc}
Averages of $\Bzb$ decays to charmonium states are shown in Tables~\ref{tab:b2c:Bd_cc_1}--\ref{tab:b2c:Bd_cc_6} and Figs.~\ref{fig:b2c:Bd_cc_1}--\ref{fig:b2c:Bd_cc_6}.
\begin{btocharmtab}{Bd_cc_1}{Decays to $J/\psi$ and one kaon $[10^{-3}]$}
\hline
\textbf{Parameter} & \begin{tabular}{l}\textbf{Measurements}\end{tabular} & \textbf{Average} \\
\hline
\hline
${\cal{B}} ( \bar{B}^0 \to J/\psi(1S) \bar{K}^0 )$ & \begin{tabular}{l} CDF \cite{Abe:1995aw}: $1.15 \pm 0.23 \pm 0.17$ \\ \babar \cite{Aubert:2004rz}: $0.869 \pm 0.022 \pm 0.030$ \\ Belle \cite{Abe:2002rc}: $0.79 \pm 0.04 \pm 0.09$ \\ \end{tabular} & $0.863 \pm 0.035$ \\
\hline
${\cal{B}} ( \bar{B}^0 \to J/\psi(1S) K^- \pi^+  )$ & \begin{tabular}{l} Belle \cite{Chilikin:2014bkk}: $1.15 \pm 0.01 \pm 0.05$ \\ \end{tabular} & $1.15 \pm 0.05$ \\
\hline
${\cal{B}} ( \bar{B}^0 \to J/\psi(1S) \bar{K}^{*0}(892) )$ & \begin{tabular}{l} CDF \cite{Abe:1998yu}: $1.74 \pm 0.20 \pm 0.18$ \\ \babar \cite{Aubert:2004rz}: $1.309 \pm 0.026 \pm 0.077$ \\ Belle \cite{Chilikin:2014bkk}: $1.19 \pm 0.01 \pm 0.08$ \\ \end{tabular} & $1.270 \pm 0.056$ \\
\hline
${\cal{B}} ( \bar{B}^0 \to J/\psi(1S) \bar{K}^0 \pi^+ \pi^- )$ & \begin{tabular}{l} CDF \cite{Affolder:2001qi}: $1.03 \pm 0.33 \pm 0.15$ \\ \end{tabular} & $1.03 \pm 0.36$ \\
\hline
${\cal{B}} ( \bar{B}^0 \to J/\psi(1S) \bar{K}^0 \rho^0(770) )$ & \begin{tabular}{l} CDF \cite{Affolder:2001qi}: $0.54 \pm 0.29 \pm 0.09$ \\ \end{tabular} & $0.54 \pm 0.30$ \\
\hline
${\cal{B}} ( \bar{B}^0 \to J/\psi(1S) K^{*-}(892) \pi^+ )$ & \begin{tabular}{l} CDF \cite{Affolder:2001qi}: $0.77 \pm 0.41 \pm 0.13$ \\ \end{tabular} & $0.77 \pm 0.43$ \\
\hline
${\cal{B}} ( \bar{B}^0 \to J/\psi(1S) \omega(782) \bar{K}^0 )$ & \begin{tabular}{l} \babar \cite{delAmoSanchez:2010jr}: $0.23 \pm 0.03 \pm 0.03$ \\ \end{tabular} & $0.23 \pm 0.04$ \\
\hline
${\cal{B}} ( \bar{B}^0 \to J/\psi(1S) \phi(1020) \bar{K}^0 )$ & \begin{tabular}{l} \babar \cite{Aubert:2003ii}: $0.102 \pm 0.038 \pm 0.010$ \\ \end{tabular} & $0.102 \pm 0.039$ \\
\hline
${\cal{B}} ( \bar{B}^0 \to J/\psi(1S) K^{0} \pi^- \pi^+ )$ & \begin{tabular}{l} LHCb \cite{Aaij:2014naa}: $0.430 \pm 0.030 \pm 0.037$ \\ \end{tabular} & $0.430 \pm 0.047$ \\
\hline
${\cal{B}} ( \bar{B}^0 \to J/\psi(1S) \bar{K}_1^0(1270) )$ & \begin{tabular}{l} Belle \cite{Abe:2001wa}: $1.30 \pm 0.34 \pm 0.31$ \\ \end{tabular} & $1.30 \pm 0.46$ \\
\hline
${\cal{B}} ( \bar{B}^0 \to J/\psi(1S) \eta K_S^0 )$ & \begin{tabular}{l} \babar \cite{Aubert:2004fc}: $0.084 \pm 0.026 \pm 0.027$ \\ Belle \cite{Iwashita:2013wnn}: $0.0522 \pm 0.0078 \pm 0.0049$ \\ \end{tabular} & $0.0540 \pm 0.0089$ \\
\hline
${\cal{B}} ( \bar{B}^0 \to J/\psi(1S) \bar{K}^{*0}(892) \pi^+ \pi^- )$ & \begin{tabular}{l} CDF \cite{Affolder:2001qi}: $0.66 \pm 0.19 \pm 0.11$ \\ \end{tabular} & $0.66 \pm 0.22$ \\
\hline
\end{btocharmtab}
\btocharmfig{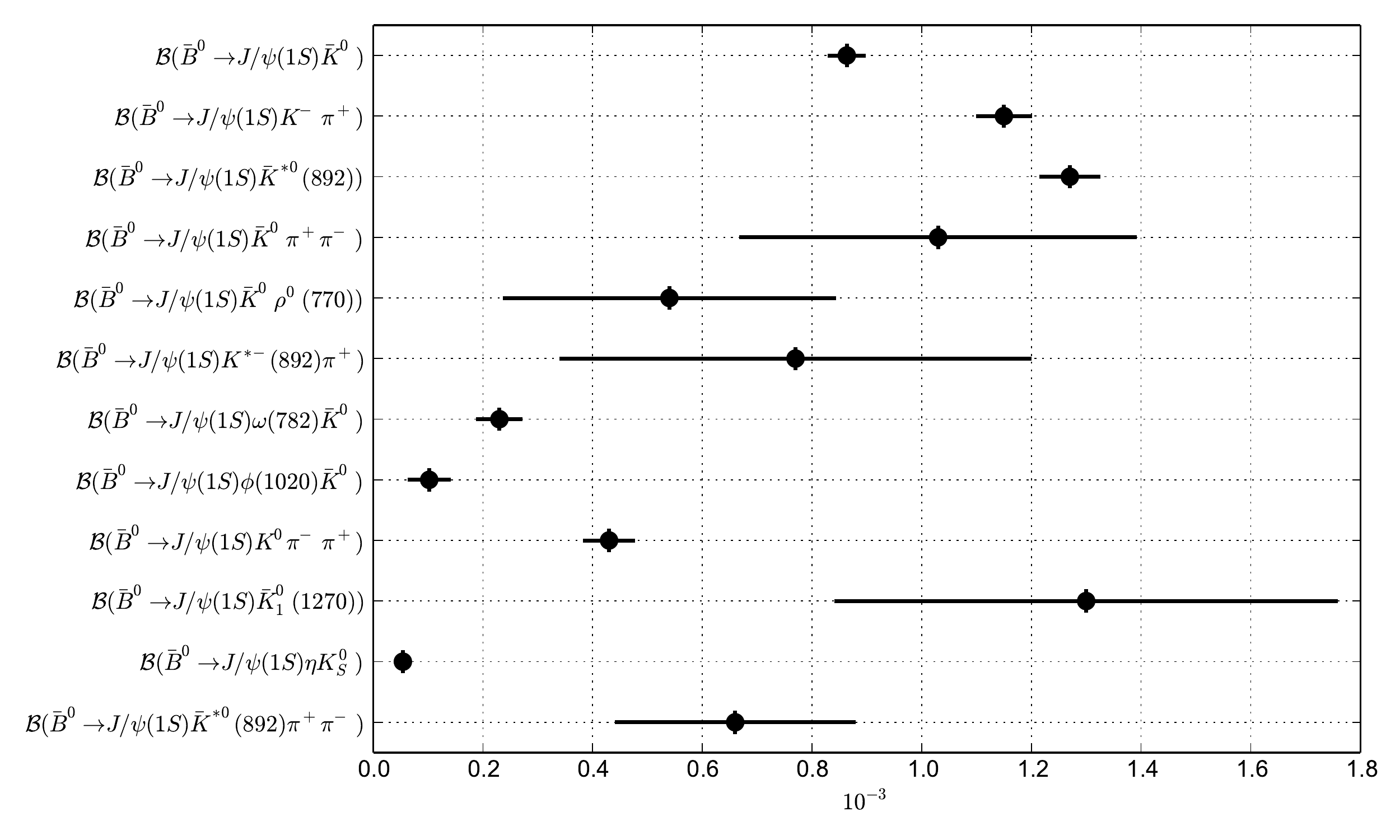}
\begin{btocharmtab}{Bd_cc_2}{Decays to charmonium other than $J/\psi$ and one kaon $[10^{-3}]$}
\hline
\textbf{Parameter} & \begin{tabular}{l}\textbf{Measurements}\end{tabular} & \textbf{Average} \\
\hline
\hline
${\cal{B}} ( \bar{B}^0 \to \psi(2S) \bar{K}^0 )$ & \begin{tabular}{l} \babar \cite{Aubert:2004rz}: $0.646 \pm 0.065 \pm 0.051$ \\ Belle \cite{Abe:2002rc}: $0.67 \pm 0.11$ \\ \end{tabular} & $0.655 \pm 0.066$ \\
\hline
${\cal{B}} ( \bar{B}^0 \to \psi(2S) \bar{K}^{*0}(892) )$ & \begin{tabular}{l} CDF \cite{Abe:1998yu}: $0.90 \pm 0.22 \pm 0.09$ \\ \babar \cite{Aubert:2004rz}: $0.649 \pm 0.059 \pm 0.097$ \\ Belle \cite{Abe:2003yx}: $0.720 \pm 0.043 \pm 0.065$ \\ \end{tabular} & $0.711 \pm 0.062$ \\
\hline
${\cal{B}} ( \bar{B}^0 \to \psi(2S) K^{0}  )$ & \begin{tabular}{l} LHCb \cite{Aaij:2014naa}: $0.47 \pm 0.07 \pm 0.07$ \\ \end{tabular} & $0.47 \pm 0.10$ \\
\hline
\multicolumn{3}{|l|}{${\cal{B}} ( \bar{B}^0 \to \psi(2S) K^0 )\times {\cal{B}} ( \psi(2S) \to \chi_{c1}(1P) \gamma )$}\\
 & \begin{tabular}{l} Belle \cite{Bhardwaj:2013rmw}: $0.68 \pm 0.10 \pm 0.07$ \\ \end{tabular} & $0.68 \pm 0.12$ \\
\hline
\multicolumn{3}{|l|}{${\cal{B}} ( \bar{B}^0 \to \psi(2S) K^0 )\times {\cal{B}} ( \psi(2S) \to \chi_{c2} \gamma )$}\\
 & \begin{tabular}{l} Belle \cite{Bhardwaj:2013rmw}: $0.47 \pm 0.16 \pm 0.08$ \\ \end{tabular} & $0.47 \pm 0.18$ \\
\hline
${\cal{B}} ( \bar{B}^0 \to K^{*0}(892) \psi(2S) )$ & \begin{tabular}{l} Belle \cite{Mizuk:2009da}: $0.552 \,^{+0.035}_{-0.032} \,^{+0.053}_{-0.058}$ \\ \end{tabular} & $0.552 \,^{+0.064}_{-0.066}$ \\
\hline
\multicolumn{3}{|l|}{${{\cal{B}} ( \bar{B}^0 \to \psi(3770) \bar{K}^0 )\times {\cal{B}} ( \psi(3770) \to D^0 \bar{D}^0 )}$}\\
 & \begin{tabular}{l} \babar \cite{Aubert:2007rva}: $< 0.123$ \\ \end{tabular} & $< 0.123$ \\
\hline
\multicolumn{3}{|l|}{${{\cal{B}} ( \bar{B}^0 \to \psi(3770) \bar{K}^0 )\times {\cal{B}} ( \psi(3770) \to D^+ D^- )}$}\\
 & \begin{tabular}{l} \babar \cite{Aubert:2007rva}: $< 0.188$ \\ \end{tabular} & $< 0.188$ \\
\hline
${\cal{B}} ( \bar{B}^0 \to \chi_{c0}(1P) \bar{K}^0 )$ & \begin{tabular}{l} \babar \cite{Aubert:2005vwa}: $< 1.24$ \\ \end{tabular} & $< 1.24$ \\
\hline
${\cal{B}} ( \bar{B}^0 \to \chi_{c0}(1P) \bar{K}^{*0}(892) )$ & \begin{tabular}{l} \babar \cite{Aubert:2005vwa}: $< 0.77$ \\ \babar \cite{Aubert:2008ak}: $0.17 \pm 0.03 \pm 0.02$ \\ \end{tabular} & $0.17 \pm 0.04$ \\
\hline
${\cal{B}} ( \bar{B}^0 \to \chi_{c1}(1P) \bar{K}^0 )$ & \begin{tabular}{l} \babar \cite{Aubert:2008ae}: $0.42 \pm 0.03 \pm 0.03$ \\ Belle \cite{Bhardwaj:2011dj}: $0.378 \,^{+0.017}_{-0.016} \pm 0.033$ \\ \end{tabular} & $0.396 \pm 0.028$ \\
\hline
${\cal{B}} ( \bar{B}^0 \to \chi_{c1}(1P) K^- \pi^+ )$ & \begin{tabular}{l} \babar \cite{Lees:2011ik}: $0.511 \pm 0.014 \pm 0.058$ \\ \end{tabular} & $0.511 \pm 0.060$ \\
\hline
${\cal{B}} ( \bar{B}^0 \to \chi_{c1}(1P) \bar{K}^{*0}(892) )$ & \begin{tabular}{l} \babar \cite{Aubert:2008ae}: $0.25 \pm 0.02 \pm 0.02$ \\ Belle \cite{Soni:2005fw}: $0.31 \pm 0.03 \pm 0.07$ \\ \end{tabular} & $0.26 \pm 0.03$ \\
\hline
${\cal{B}} ( \bar{B}^0 \to \chi_{c2}(1P) \bar{K}^0 )$ & \begin{tabular}{l} \babar \cite{Aubert:2008ae}: $0.015 \pm 0.009 \pm 0.003$ \\ Belle \cite{Bhardwaj:2011dj}: $< 0.015$ \\ \end{tabular} & $0.015 \pm 0.009$ \\
\hline
${\cal{B}} ( \bar{B}^0 \to \chi_{c2}(1P) \bar{K}^{*0}(892) )$ & \begin{tabular}{l} \babar \cite{Aubert:2008ae}: $0.066 \pm 0.018 \pm 0.005$ \\ \end{tabular} & $0.066 \pm 0.019$ \\
\hline
${\cal{B}} ( \bar{B}^0 \to \eta_c(1S) \bar{K}^0 )$ & \begin{tabular}{l} \babar \cite{Aubert:2004gc}: $1.14 \pm 0.15 \pm 0.34$ \\ \babar \cite{Aubert:2007qea}: $0.64 \,^{+0.22}_{-0.20} \,^{+0.28}_{-0.16}$ \\ Belle \cite{Fang:2002gi}: $1.23 \pm 0.23 \,^{+0.40}_{-0.41}$ \\ \end{tabular} & $0.85 \pm 0.24$ \\
\hline
${\cal{B}} ( \bar{B}^0 \to \eta_c(1S) \bar{K}^{*0}(892) )$ & \begin{tabular}{l} \babar \cite{Aubert:2008kp}: $0.57 \pm 0.06 \pm 0.09$ \\ \babar \cite{Aubert:2007qea}: $0.80 \,^{+0.21}_{-0.19} \,^{+0.37}_{-0.23}$ \\ Belle \cite{Fang:2002gi}: $1.62 \pm 0.32 \,^{+0.55}_{-0.60}$ \\ \end{tabular} & $0.61 \pm 0.10$ \\
\hline
${\cal{B}} ( \bar{B}^0 \to \eta_c(2S) \bar{K}^{*0}(892) )$ & \begin{tabular}{l} \babar \cite{Aubert:2008kp}: $< 0.39$ \\ \end{tabular} & $< 0.39$ \\
\hline
\multicolumn{3}{|l|}{${{\cal{B}} ( \bar{B}^0 \to h_c(1P) \bar{K}^{*0}(892) )\times {\cal{B}} ( h_c(1P) \to \eta_c(1S) \gamma )}$}\\
 & \begin{tabular}{l} \babar \cite{Aubert:2008kp}: $< 0.22$ \\ \end{tabular} & $< 0.22$ \\
\hline
\end{btocharmtab}
\btocharmfig{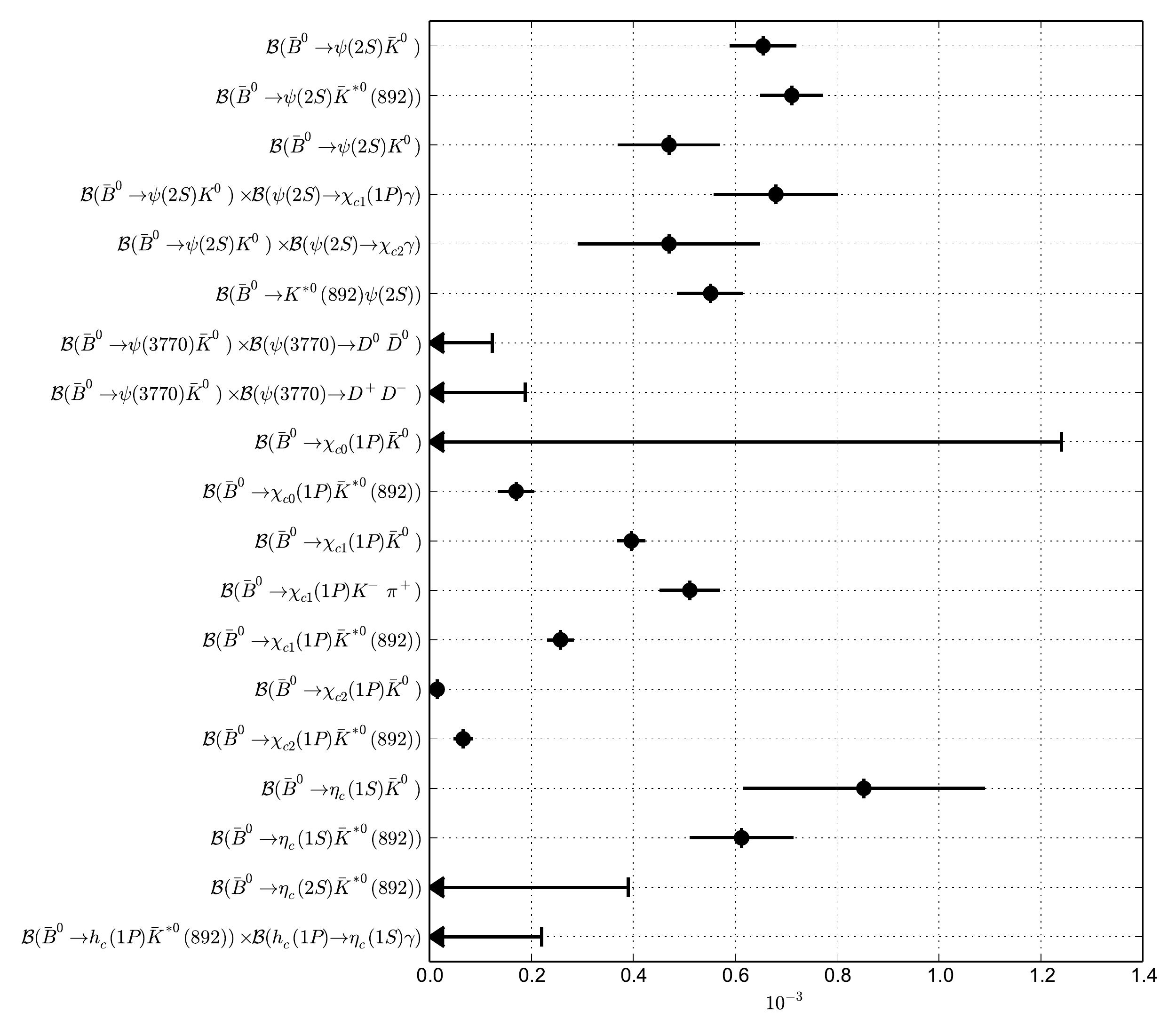}
\begin{btocharmtab}{Bd_cc_3}{Decays to charmonium and light mesons $[10^{-5}]$}
\hline
\textbf{Parameter} & \begin{tabular}{l}\textbf{Measurements}\end{tabular} & \textbf{Average} \\
\hline
\hline
${\cal{B}} ( \bar{B}^0 \to J/\psi(1S) \pi^0 )$ & \begin{tabular}{l} \babar \cite{Aubert:2008bs}: $1.69 \pm 0.14 \pm 0.07$ \\ Belle \cite{Abe:2002rc}: $2.3 \pm 0.5 \pm 0.2$ \\ \end{tabular} & $1.74 \pm 0.15$ \\
\hline
${\cal{B}} ( \bar{B}^0 \to J/\psi(1S) \pi^+ \pi^- )$ & \begin{tabular}{l} \babar \cite{Aubert:2007xw}: $< 1.2$ \\ Belle \cite{Abe:2004mv}: $< 1$ \\ \end{tabular} & $< 1$ \\
\hline
${\cal{B}} ( \bar{B}^0 \to J/\psi(1S) \rho^0(770) )$ & \begin{tabular}{l} \babar \cite{Aubert:2007xw}: $2.7 \pm 0.3 \pm 0.2$ \\ Belle \cite{Abe:2004mv}: $2.8 \pm 0.3 \pm 0.3$ \\ \end{tabular} & $2.7 \pm 0.3$ \\
\hline
${\cal{B}} ( \bar{B}^0 \to J/\psi(1S) \eta )$ & \begin{tabular}{l} \babar \cite{Aubert:2003ii}: $< 2.7$ \\ Belle \cite{Chang:2012gnb}: $1.23 \,^{+0.18}_{-0.17} \pm 0.07$ \\ \end{tabular} & $1.23 \,^{+0.19}_{-0.18}$ \\
\hline
${\cal{B}} ( \bar{B}^0 \to J/\psi(1S) \eta^\prime(958) )$ & \begin{tabular}{l} \babar \cite{Aubert:2003ii}: $< 6.3$ \\ Belle \cite{Chang:2012gnb}: $< 0.74$ \\ \end{tabular} & $< 0.74$ \\
\hline
${\cal{B}} ( \bar{B}^0 \to J/\psi(1S) f_2(1270) )$ & \begin{tabular}{l} \babar \cite{Aubert:2007xw}: $< 0.46$ \\ Belle \cite{Abe:2004mv}: $0.98 \pm 0.39 \pm 0.20$ \\ \end{tabular} & $0.98 \pm 0.44$ \\
\hline
\multicolumn{3}{|l|}{${{\cal{B}} ( \bar{B}^0 \to J/\psi(1S) a_0(980) )\times {\cal{B}} ( a_0(980) \to K^+ K^- )}$}\\
 & \begin{tabular}{l} LHCb \cite{Aaij:2013mtm}: $< 0.090$ \\ \end{tabular} & $< 0.090$ \\
\hline
\multicolumn{3}{|l|}{${{\cal{B}} ( \bar{B}^0 \to J/\psi(1S) f_0(980) )\times {\cal{B}} ( f_0(980) \to \pi^+ \pi^- )}$}\\
 & \begin{tabular}{l} LHCb \cite{Aaij:2013zpt}: $< 0.11$ \\ \end{tabular} & $< 0.11$ \\
\hline
${\cal{B}} ( \bar{B}^0 \to J/\psi(1S) f_1(1285) )$ & \begin{tabular}{l} LHCb \cite{Aaij:2013rja}: $0.837 \pm 0.195 \,^{+0.079}_{-0.075}$ \\ \end{tabular} & $0.837 \,^{+0.210}_{-0.209}$ \\
\hline
\multicolumn{3}{|l|}{${{\cal{B}} ( \bar{B}^0 \to J/\psi(1S) f_1(1285) )\times {\cal{B}} ( f_1(1285) \to \pi^+ \pi^- \pi^+ \pi^- )}$}\\
 & \begin{tabular}{l} LHCb \cite{Aaij:2013rja}: $0.0921 \pm 0.0214 \pm 0.0064$ \\ \end{tabular} & $0.0921 \pm 0.0223$ \\
\hline
${\cal{B}} ( \bar{B}^0 \to J/\psi(1S) K^+ K^- )$ & \begin{tabular}{l} LHCb \cite{Aaij:2013mtm}: $0.253 \pm 0.031 \pm 0.019$ \\ \end{tabular} & $0.253 \pm 0.036$ \\
\hline
${\cal{B}} ( \bar{B}^0 \to J/\psi(1S) \phi(1020) )$ & \begin{tabular}{l} LHCb \cite{Aaij:2013mtm}: $< 0.019$ \\ \babar \cite{Aubert:2003ii}: $< 0.9$ \\ Belle \cite{Liu:2008bta}: $< 0.094$ \\ \end{tabular} & $< 0.019$ \\
\hline
${\cal{B}} ( \bar{B}^0 \to J/\psi(1S) K^{0} K^{\pm} \pi^{\mp} )$ & \begin{tabular}{l} LHCb \cite{Aaij:2014naa}: $< 2.1$ \\ \end{tabular} & $< 2.1$ \\
\hline
${\cal{B}} ( \bar{B}^0 \to J/\psi(1S) K^{0} K^{+} K^{-} )$ & \begin{tabular}{l} LHCb \cite{Aaij:2014naa}: $2.02 \pm 0.43 \pm 0.19$ \\ \end{tabular} & $2.02 \pm 0.47$ \\
\hline
${\cal{B}} ( \bar{B}^0 \to \chi_{c1}(1P) \pi^0 )$ & \begin{tabular}{l} Belle \cite{Kumar:2008ir}: $1.12 \pm 0.25 \pm 0.12$ \\ \end{tabular} & $1.12 \pm 0.28$ \\
\hline
\end{btocharmtab}
\btocharmfig{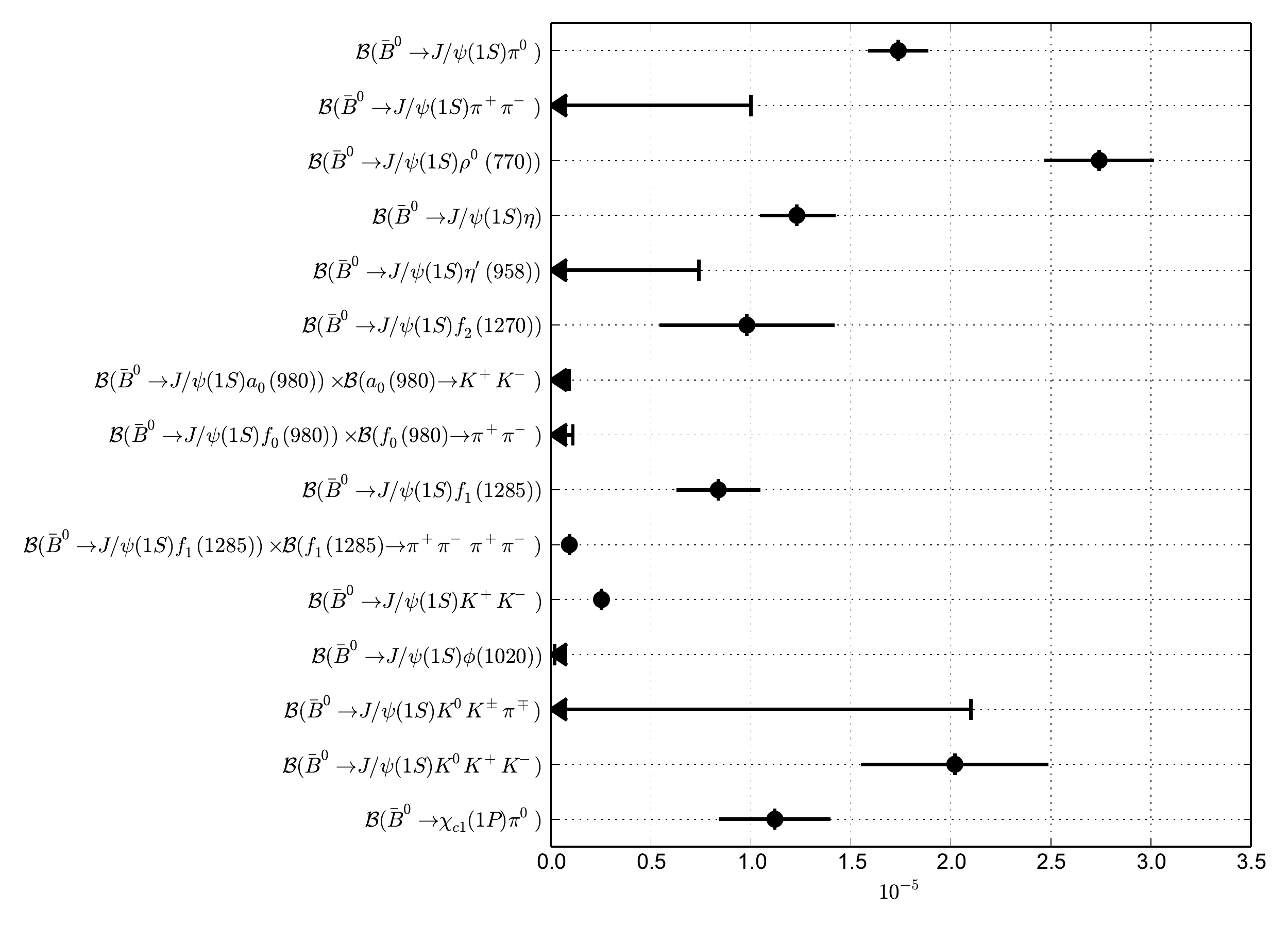}
\begin{btocharmtab}{Bd_cc_4}{Decays to  $J/\psi$ and photons, baryons, or heavy mesons $[10^{-5}]$}
\hline
\textbf{Parameter} & \begin{tabular}{l}\textbf{Measurements}\end{tabular} & \textbf{Average} \\
\hline
\hline
${\cal{B}} ( \bar{B}^0 \to J/\psi(1S) \gamma )$ & \begin{tabular}{l} \babar \cite{Aubert:2004xd}: $< 0.16$ \\ \end{tabular} & $< 0.16$ \\
\hline
${\cal{B}} ( \bar{B}^0 \to J/\psi(1S) \bar{p} p )$ & \begin{tabular}{l} LHCb \cite{Aaij:2013yba}: $< 0.052$ \\ \babar \cite{Aubert:2003ww}: $< 0.19$ \\ Belle \cite{Xie:2005tf}: $< 0.083$ \\ \end{tabular} & $< 0.052$ \\
\hline
${\cal{B}} ( \bar{B}^0 \to J/\psi(1S) D^0 )$ & \begin{tabular}{l} \babar \cite{Aubert:2005tr}: $< 1.3$ \\ Belle \cite{Zhang:2005bs}: $< 2.0$ \\ \end{tabular} & $< 1.3$ \\
\hline
\end{btocharmtab}
\btocharmfig{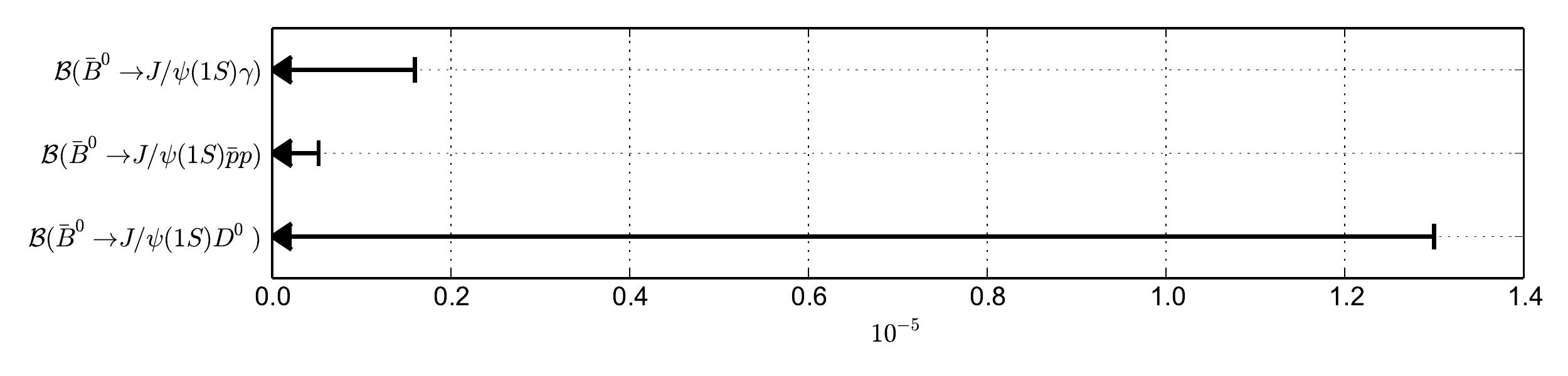}
\begin{btocharmtab}{Bd_cc_5}{Relative decay rates}
\hline
\textbf{Parameter} & \begin{tabular}{l}\textbf{Measurements}\end{tabular} & \textbf{Average} \\
\hline
\hline
\multicolumn{3}{|l|}{${{\cal{B}} ( \bar{B}^0 \to J/\psi(1S) \bar{K}_1^0(1270) )}/{{\cal{B}} ( B^- \to J/\psi(1S) K^- )}$}\\
 & \begin{tabular}{l} Belle \cite{Abe:2001wa}: $1.30 \pm 0.34 \pm 0.28$ \\ \end{tabular} & $1.30 \pm 0.44$ \\
\hline
\multicolumn{3}{|l|}{${{\cal{B}} ( \bar{B}^0 \to J/\psi(1S) \bar{K}^{*0}(892) )}/{{\cal{B}} ( \bar{B}^0 \to J/\psi(1S) \bar{K}^0 )}$}\\
 & \begin{tabular}{l} CDF \cite{Abe:1996kc}: $1.39 \pm 0.36 \pm 0.10$ \\ \babar \cite{Aubert:2004rz}: $1.51 \pm 0.05 \pm 0.08$ \\ \end{tabular} & $1.50 \pm 0.09$ \\
\hline
${\cal{B}} ( \bar{B}^0 \to J/\psi \omega) / {\cal{B}} ( \bar{B}^0 \to J/\psi \rho )$ & \begin{tabular}{l} LHCb \cite{LHCb:2012cw}: $0.89 \pm 0.19 \,^{+0.07}_{-0.13}$ \\ \end{tabular} & $0.89 \,^{+0.20}_{-0.23}$ \\
\hline
\multicolumn{3}{|l|}{${\cal{B}} ( \bar{B}^0 \to J/\psi \omega \bar{K}^0 ) / {\cal{B}} ( B^- \to J/\psi \omega K^- )$}\\
 & \begin{tabular}{l} \babar \cite{delAmoSanchez:2010jr}: $0.7 \pm 0.1 \pm 0.1$ \\ \end{tabular} & $0.7 \pm 0.1$ \\
\hline
\multicolumn{3}{|l|}{${{\cal{B}} ( \bar{B}^0 \to J/\psi(1S) K_S^{0} \pi^- \pi^+ )}/{{\cal{B}} ( \bar{B}^0 \to J/\psi(1S) K_S^0 )}$}\\
 & \begin{tabular}{l} LHCb \cite{Aaij:2014naa}: $0.493 \pm 0.034 \pm 0.027$ \\ \end{tabular} & $0.493 \pm 0.043$ \\
\hline
\multicolumn{3}{|l|}{${{\cal{B}} ( \bar{B}^0 \to J/\psi(1S) K_S^{0} K^{\pm} \pi^{\mp} )}/{{\cal{B}} ( \bar{B}^0 \to J/\psi(1S) \pi^+ \pi^- )}$}\\
 & \begin{tabular}{l} LHCb \cite{Aaij:2014naa}: $< 0.048$ \\ \end{tabular} & $< 0.048$ \\
\hline
\multicolumn{3}{|l|}{${{\cal{B}} ( \bar{B}^0 \to J/\psi(1S) K_S^{0} K^{+} K^{-} )}/{{\cal{B}} ( \bar{B}^0 \to J/\psi(1S) K_S^{0} \pi^+ \pi^- )}$}\\
 & \begin{tabular}{l} LHCb \cite{Aaij:2014naa}: $0.047 \pm 0.010 \pm 0.004$ \\ \end{tabular} & $0.047 \pm 0.011$ \\
\hline
\multicolumn{3}{|l|}{${   {\cal{B}}  ( \bar{B}^0 \to \psi(2S) K_S^{0} ) \times ( \psi(2S)\to  J/\psi(1S) \pi^- \pi^+)   }/ { {\cal{B}} ( \bar{B}^0 \to J/\psi(1S) K_S^0 )}$}\\
 & \begin{tabular}{l} LHCb \cite{Aaij:2014naa}: $0.183 \pm 0.027 \pm 0.015$ \\ \end{tabular} & $0.183 \pm 0.031$ \\
\hline
\multicolumn{3}{|l|}{${{\cal{B}} ( \bar{B}^0 \to \psi(2S) \bar{K}^{*0}(892) )}/{{\cal{B}} ( \bar{B}^0 \to \psi(2S) \bar{K}^0 )}$}\\
 & \begin{tabular}{l} \babar \cite{Aubert:2004rz}: $1.00 \pm 0.14 \pm 0.09$ \\ \end{tabular} & $1.00 \pm 0.17$ \\
\hline
\multicolumn{3}{|l|}{${{\cal{B}} ( \bar{B}^0 \to \psi(2S) K(892)^{*0} )}/{{\cal{B}} ( \bar{B}^0 \to J/\psi(1S)  K(892)^{*0} )}$}\\
 & \begin{tabular}{l} LHCb \cite{Aaij:2012dda}: $0.476 \pm 0.014 \pm 0.016$ \\ \end{tabular} & $0.476 \pm 0.021$ \\
\hline
\multicolumn{3}{|l|}{${{\cal{B}} \bar{B}^0 \to \psi(2S) \pi^+  \pi^- }/{{\cal{B}} \bar{B}^0 \to J/\psi \pi^+ \pi^-  }$}\\
 & \begin{tabular}{l} LHCb \cite{Aaij:2013cpa}: $0.56 \pm 0.07 \pm 0.05$ \\ \end{tabular} & $0.56 \pm 0.09$ \\
\hline
\multicolumn{3}{|l|}{${{\cal{B}} ( \bar{B}^0 \to \eta_c(1S) \bar{K}^0 )}/{{\cal{B}} ( B^- \to \eta_c(1S) K^- )}$}\\
 & \begin{tabular}{l} \babar \cite{Aubert:2004gc}: $0.87 \pm 0.13 \pm 0.07$ \\ \end{tabular} & $0.87 \pm 0.15$ \\
\hline
\multicolumn{3}{|l|}{${{\cal{B}} ( \bar{B}^0 \to \eta_c(1S) \bar{K}^0 )}/{{\cal{B}} ( \bar{B}^0 \to J/\psi(1S) \bar{K}^0 )}$}\\
 & \begin{tabular}{l} \babar \cite{Aubert:2004gc}: $1.34 \pm 0.19 \pm 0.40$ \\ \end{tabular} & $1.34 \pm 0.44$ \\
\hline
\multicolumn{3}{|l|}{${{\cal{B}} ( \bar{B}^0 \to \eta_c(1S) \bar{K}^{*0}(892) )}/{{\cal{B}} ( B^- \to \eta_c(1S) K^- )}$}\\
 & \begin{tabular}{l} \babar \cite{Aubert:2007ib}: $0.67 \pm 0.09 \pm 0.07$ \\ \end{tabular} & $0.67 \pm 0.11$ \\
\hline
\multicolumn{3}{|l|}{${{\cal{B}} ( \bar{B}^0 \to \eta_c(1S) \bar{K}^{*0}(892) )}/{{\cal{B}} ( \bar{B}^0 \to \eta_c(1S) \bar{K}^0 )}$}\\
 & \begin{tabular}{l} Belle \cite{Fang:2002gi}: $1.33 \pm 0.36 \,^{+0.24}_{-0.33}$ \\ \end{tabular} & $1.33 \,^{+0.43}_{-0.49}$ \\
\hline
\multicolumn{3}{|l|}{${{\cal{B}} ( \bar{B}^0 \to \chi_{c1}(1P) K^- \pi^+ )}/{{\cal{B}} ( \bar{B}^0 \to J/\psi(1S) K^- \pi^+ )}$}\\
 & \begin{tabular}{l} \babar \cite{Lees:2011ik}: $0.474 \pm 0.013 \pm 0.054$ \\ \end{tabular} & $0.474 \pm 0.056$ \\
\hline
\multicolumn{3}{|l|}{${{\cal{B}} ( \bar{B}^0 \to \chi_{c1}(1P) \bar{K}^{*0}(892) )}/{{\cal{B}} ( \bar{B}^0 \to \chi_{c1}(1P) \bar{K}^0 )}$}\\
 & \begin{tabular}{l} \babar \cite{Aubert:2004rz}: $0.72 \pm 0.11 \pm 0.12$ \\ \end{tabular} & $0.72 \pm 0.16$ \\
\hline
\multicolumn{3}{|l|}{${{\cal{B}} ( \bar{B}^0 \to h_c(1P) \bar{K}^{*0}(892) )\times {\cal{B}} ( h_c(1P) \to \eta_c(1S) \gamma )}/{{\cal{B}} ( B^- \to \eta_c(1S) K^- )}$}\\
 & \begin{tabular}{l} \babar \cite{Aubert:2007ib}: $< 0.26$ \\ \end{tabular} & $< 0.26$ \\
\hline
\multicolumn{3}{|l|}{${{\cal{B}} ( \bar{B}^0 \to h_c(1P) \bar{K}^{*0}(892) )\times {\cal{B}} ( h_c(1P) \to \eta_c(1S) \gamma )}/{{\cal{B}} ( \bar{B}^0 \to \eta_c(1S) \bar{K}^{*0}(892) )}$}\\
 & \begin{tabular}{l} \babar \cite{Aubert:2007ib}: $< 0.39$ \\ \end{tabular} & $< 0.39$ \\
\hline
\end{btocharmtab}
\btocharmfig{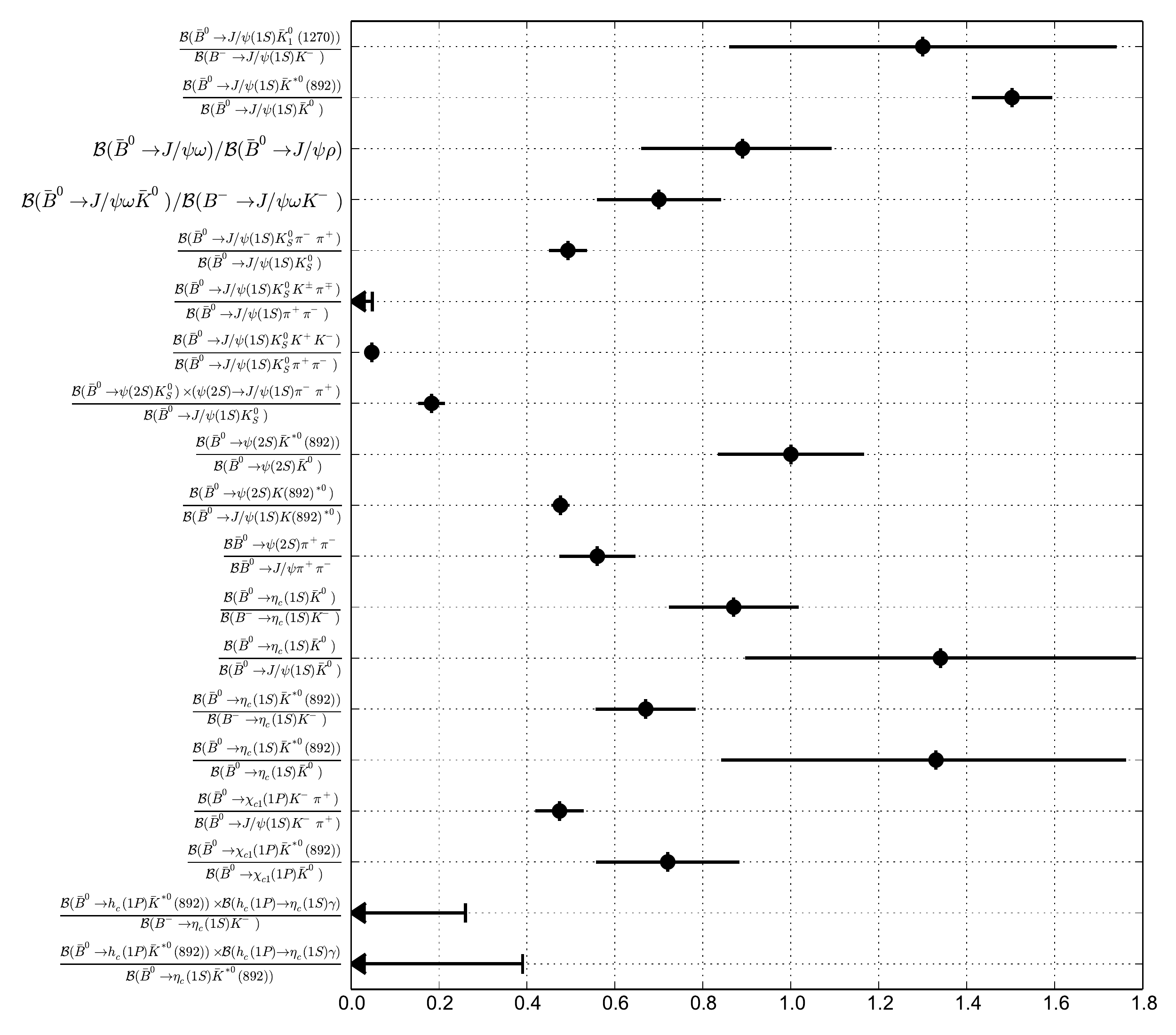}
\begin{btocharmtab}{Bd_cc_6}{Polarization fractions}
\hline
\textbf{Parameter} & \begin{tabular}{l}\textbf{Measurements}\end{tabular}%
 & $< 0.26$ \\
\hline
\end{btocharmtab}
\btocharmfig{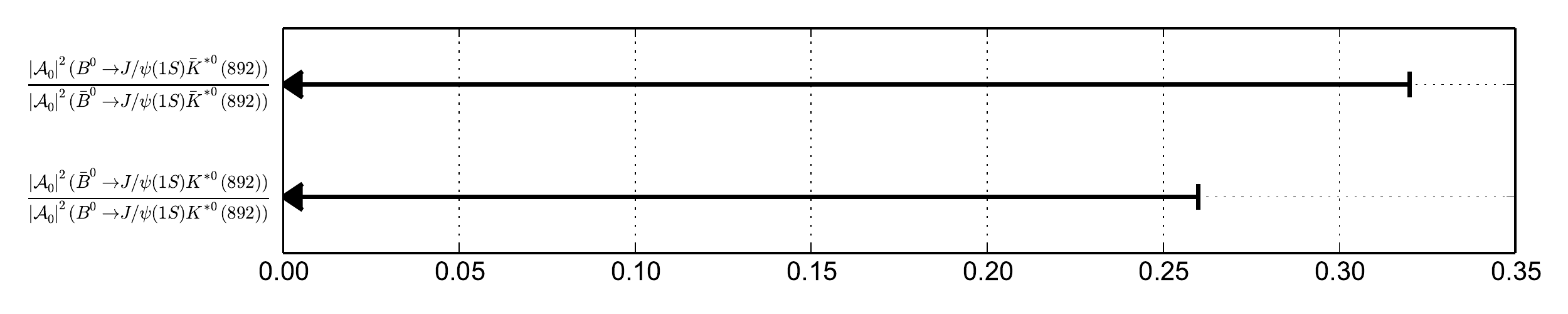}

\subsubsection{Decays to charm baryons}
\label{sec:b2c:Bd_baryon}
Averages of $\Bzb$ decays to charm baryons are shown in Tables~\ref{tab:b2c:Bd_baryon_1}--\ref{tab:b2c:Bd_baryon_2} and Figs.~\ref{fig:b2c:Bd_baryon_1}--\ref{fig:b2c:Bd_baryon_2}.
\begin{btocharmtab}{Bd_baryon_1}{Absolute decay rates to charm baryons $[10^{-4}]$}
\hline
\textbf{Parameter} & \begin{tabular}{l}\textbf{Measurements}\end{tabular} & \textbf{Average} \\
\hline
\hline
${\cal{B}} ( \bar{B}^0 \to \Lambda_c^+ \bar{p} \pi^0 )$ & \begin{tabular}{l} \babar \cite{Aubert:2010zv}: $1.94 \pm 0.17 \pm 0.52$ \\ \end{tabular}%
 & $< 0.028$ \\
\hline
\end{btocharmtab}
\btocharmfig{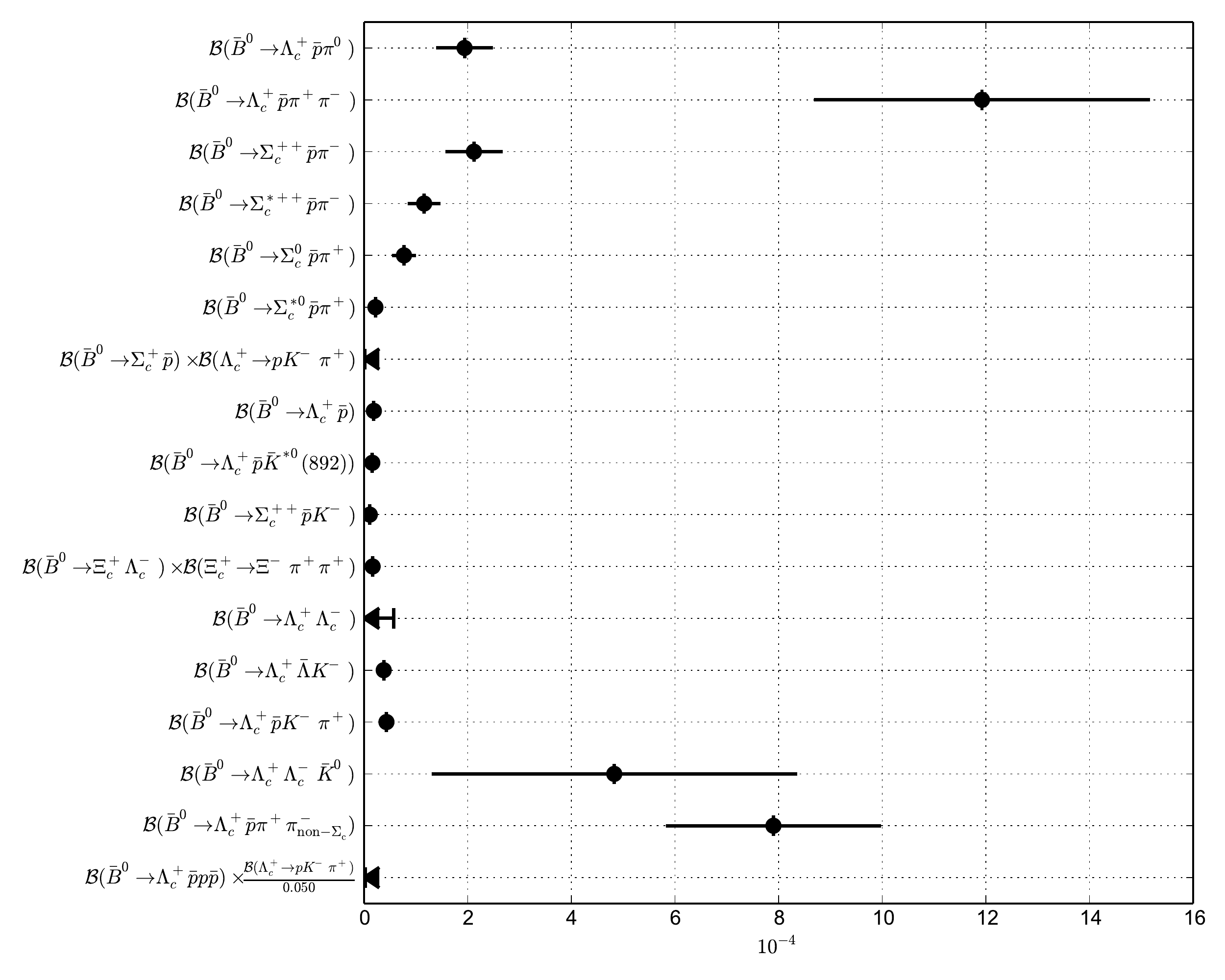}
\begin{btocharmtab}{Bd_baryon_2}{Relative decay rates to charm baryons $[10^{-3}]$}
\hline
\textbf{Parameter} & %
 & $< 2$ \\
\hline
\end{btocharmtab}
\btocharmfig{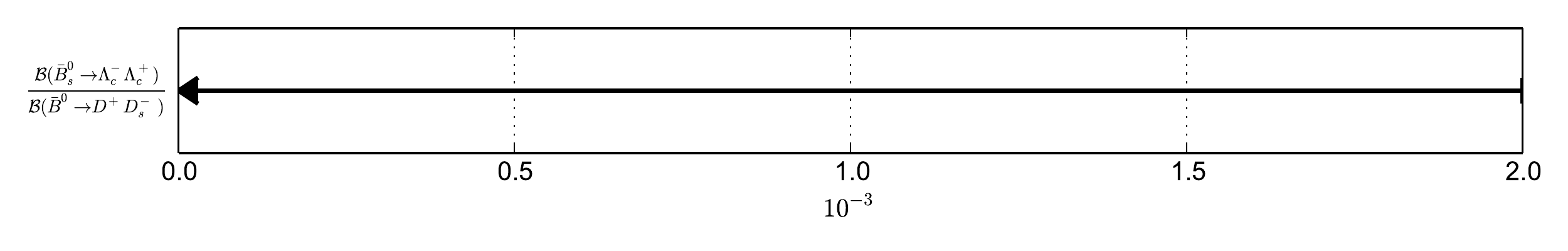}

\subsubsection{Decays to other ($XYZ$) states}
\label{sec:b2c:Bd_other}
Averages of $\Bzb$ decays to other ($XYZ$) states are shown in Tables~\ref{tab:b2c:Bd_other_1}--\ref{tab:b2c:Bd_other_5} and Figs.~\ref{fig:b2c:Bd_other_1}--\ref{fig:b2c:Bd_other_5}.
\begin{btocharmtab}{Bd_other_1}{Decays to $X(3872)$ $[10^{-5}]$}
\hline
\textbf{Parameter} & %
 & $< 4.37$ \\
\hline
\end{btocharmtab}
\btocharmfig{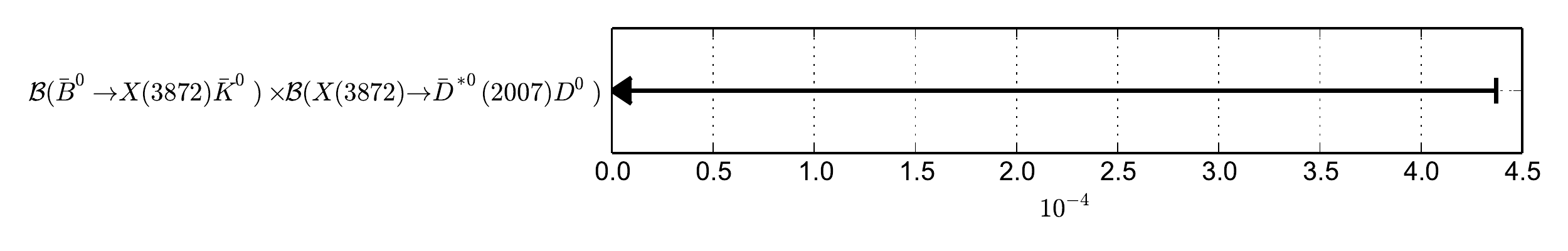}
\begin{btocharmtab}{Bd_other_3}{Decays to neutral states other than $X(3872)$ $[10^{-4}]$}
\hline
\textbf{Parameter} & %
 & $0.40 \,^{+1.98}_{-0.10}$ \\
\hline
\end{btocharmtab}
\btocharmfig{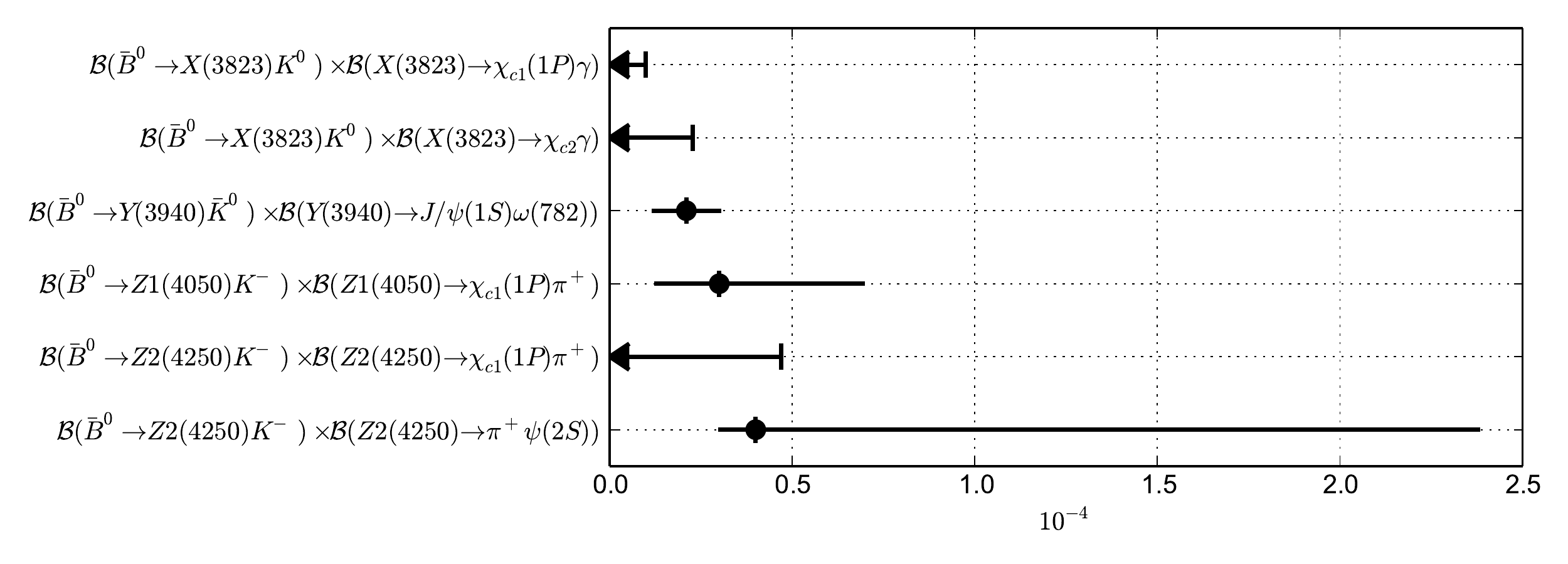}
\begin{btocharmtab}{Bd_other_4}{Decays to charged states $[10^{-4}]$}
\hline
\textbf{Parameter} & %
 & $0.22 \,^{+0.13}_{-0.08}$ \\
\hline
\end{btocharmtab}
\btocharmfig{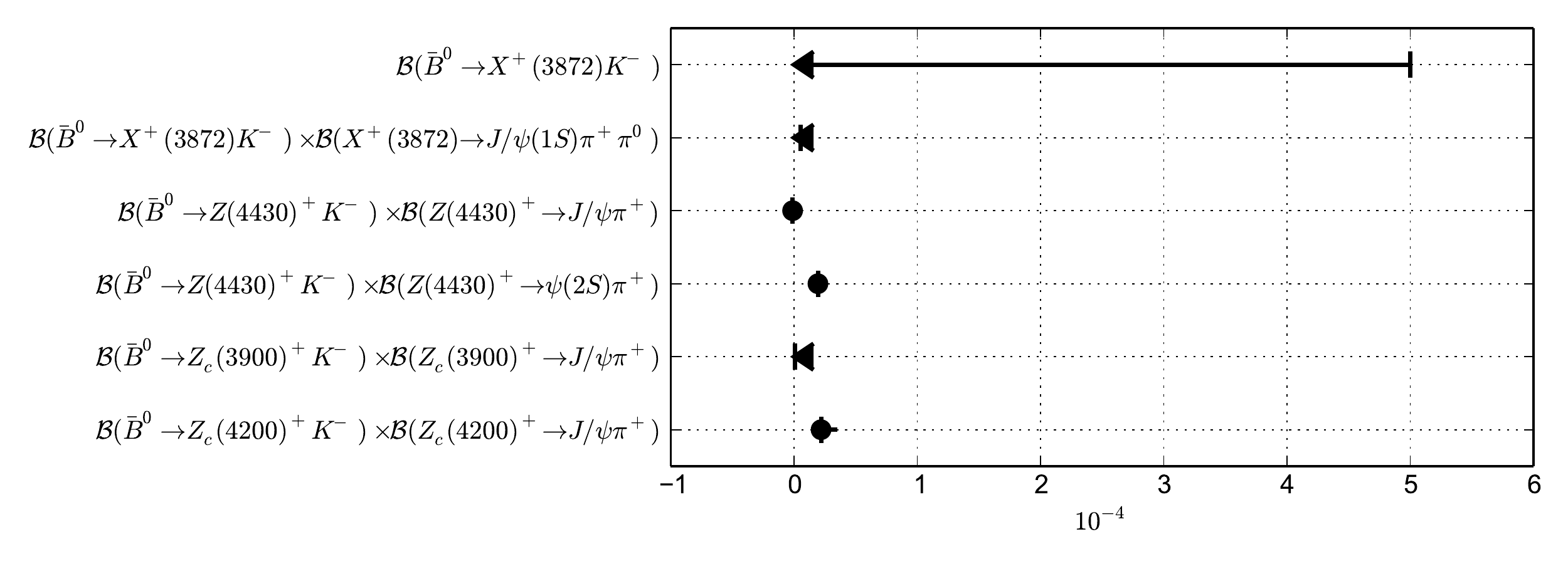}
\begin{btocharmtab}{Bd_other_5}{Relative decay rates}
\hline
\textbf{Parameter} & %
 & $0.7 \,^{+0.4}_{-0.3}$ \\
\hline
\end{btocharmtab}
\btocharmfig{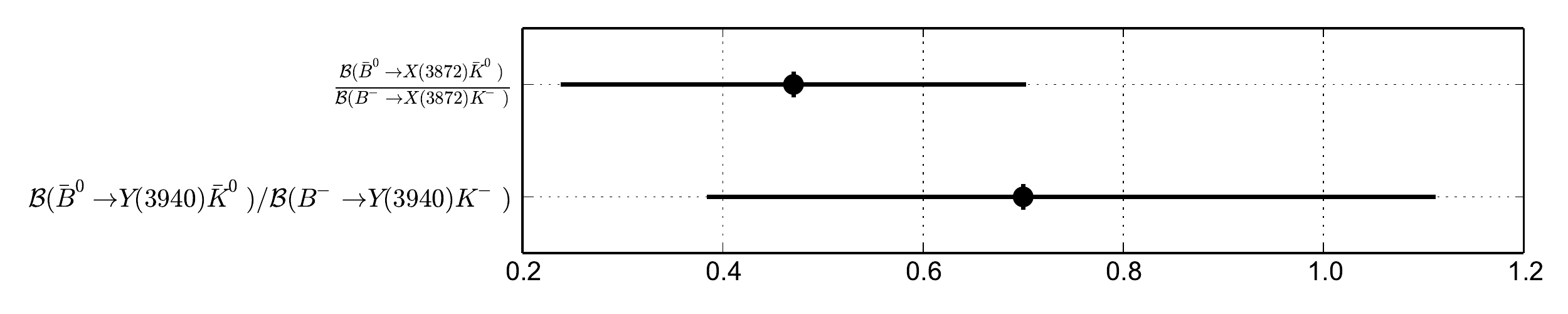}

\subsection{Decays of $B^-$ mesons}
\label{sec:b2c:Bu}
Measurements of $B^-$ decays to charmed hadrons are summarized in sections \ref{sec:b2c:Bu_D} to \ref{sec:b2c:Bu_other}.

\subsubsection{Decays to a single open charm meson}
\label{sec:b2c:Bu_D}
Averages of $B^-$ decays to a single open charm meson are shown in Tables~\ref{tab:b2c:Bu_D_1}--\ref{tab:b2c:Bu_D_11} and Figs.~\ref{fig:b2c:Bu_D_1}--\ref{fig:b2c:Bu_D_11}.
\begin{btocharmtab}{Bu_D_1}{Decays to a $D^{(*)}$ meson and one or more pions $[10^{-2}]$}
\hline
\textbf{Parameter} & 
 & $0.567 \pm 0.125$ \\
\hline
\end{btocharmtab}
\btocharmfig{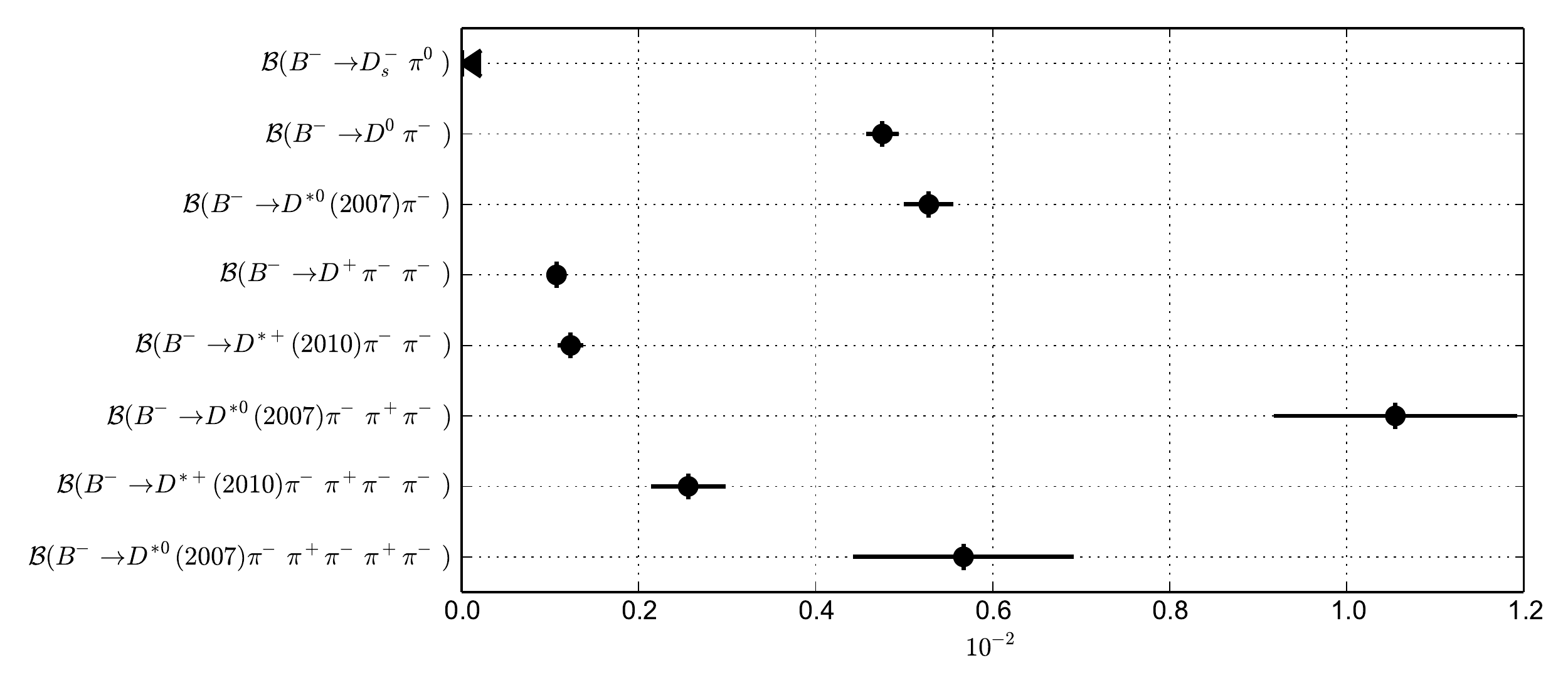}
\begin{btocharmtab}{Bu_D_2}{Decays to a $D^{(*)0}$ meson and one or more kaons $[10^{-3}]$}
\hline
\textbf{Parameter} & %
 & $< 0.09$ \\
\hline
\end{btocharmtab}
\btocharmfig{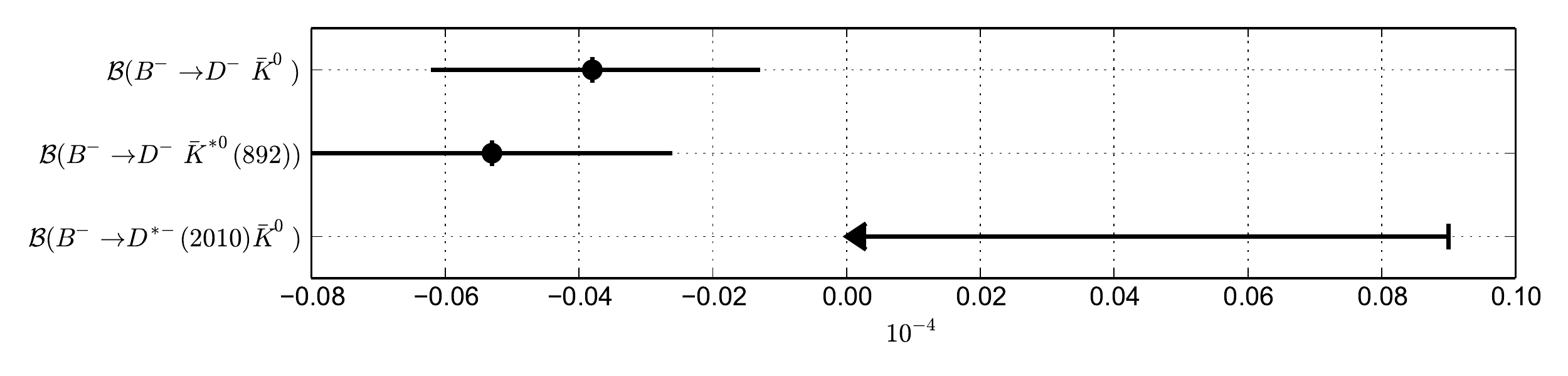}
\begin{btocharmtab}{Bu_D_4}{Relative decay rates to $D^0$ mesons}
\hline
\textbf{Parameter} & %
 & $0.094 \pm 0.016$ \\
\hline
\end{btocharmtab}
\btocharmfig{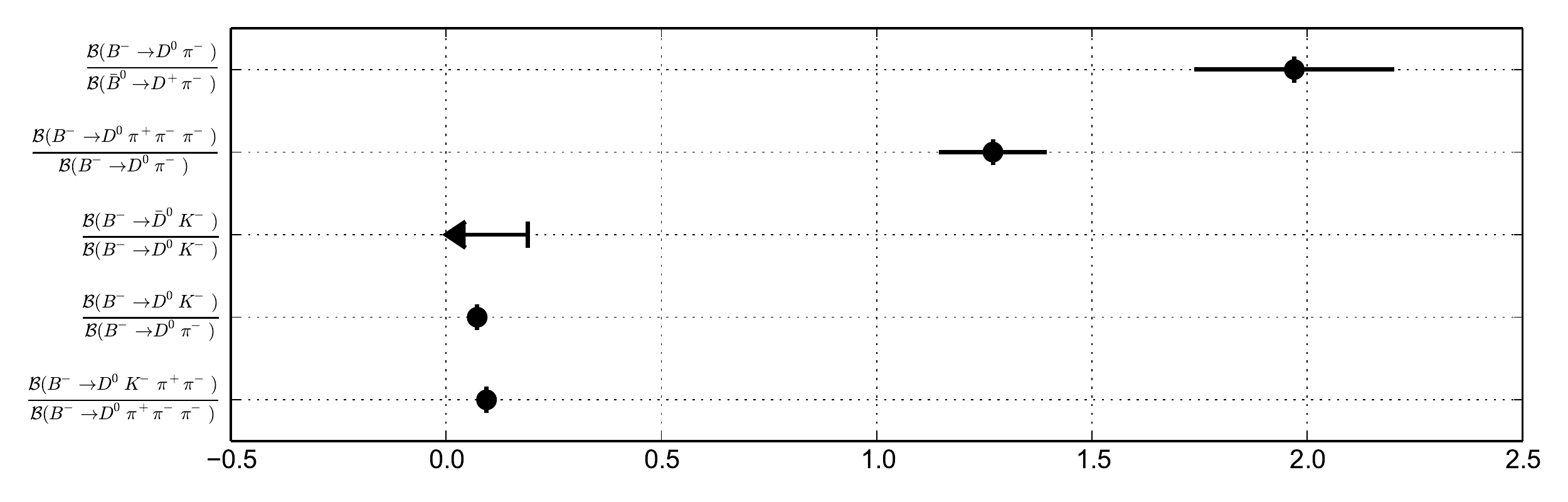}
\begin{btocharmtab}{Bu_D_5}{Absolute decay rates to excited $D$ mesons $[10^{-2}]$}
\hline
\textbf{Parameter} & %
 & $0.550 \pm 0.116$ \\
\hline
\end{btocharmtab}
\btocharmfig{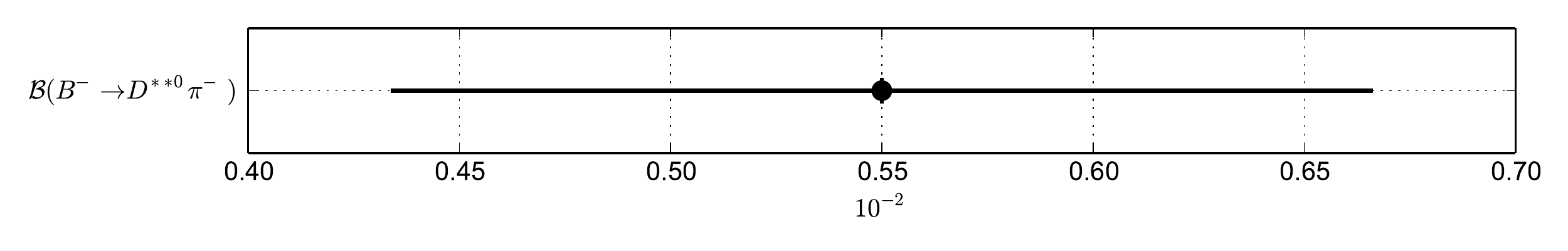}
\begin{btocharmtab}{Bu_D_6}{Absolute product decay rates to excited $D$ mesons $[10^{-3}]$}
\hline
\textbf{Parameter} & %
 & $< 0.022$ \\
\hline
\end{btocharmtab}
\btocharmfig{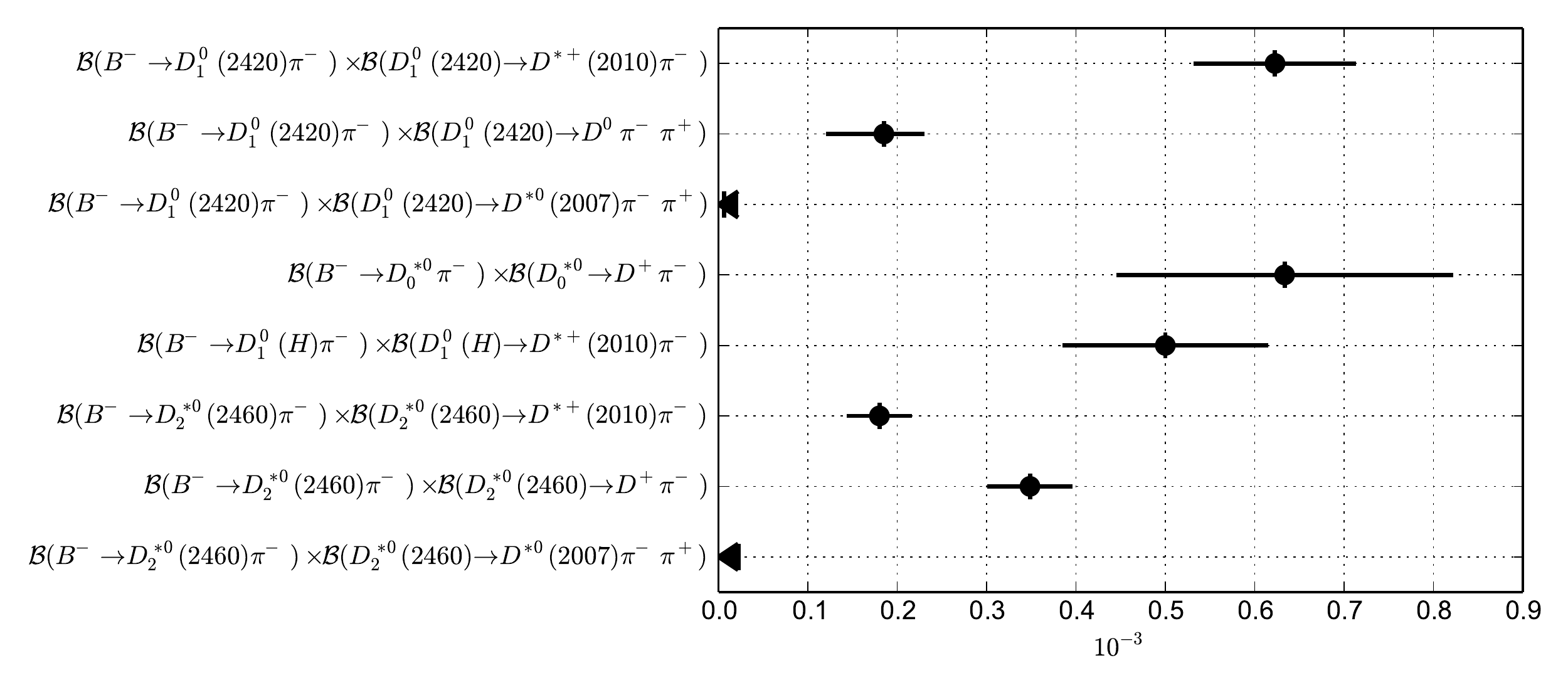}
\begin{btocharmtab}{Bu_D_7}{Relative decay rates to excited $D$ mesons}
\hline
\textbf{Parameter} & %
 & $0.0811 \pm 0.0053$ \\
\hline
\end{btocharmtab}
\btocharmfig{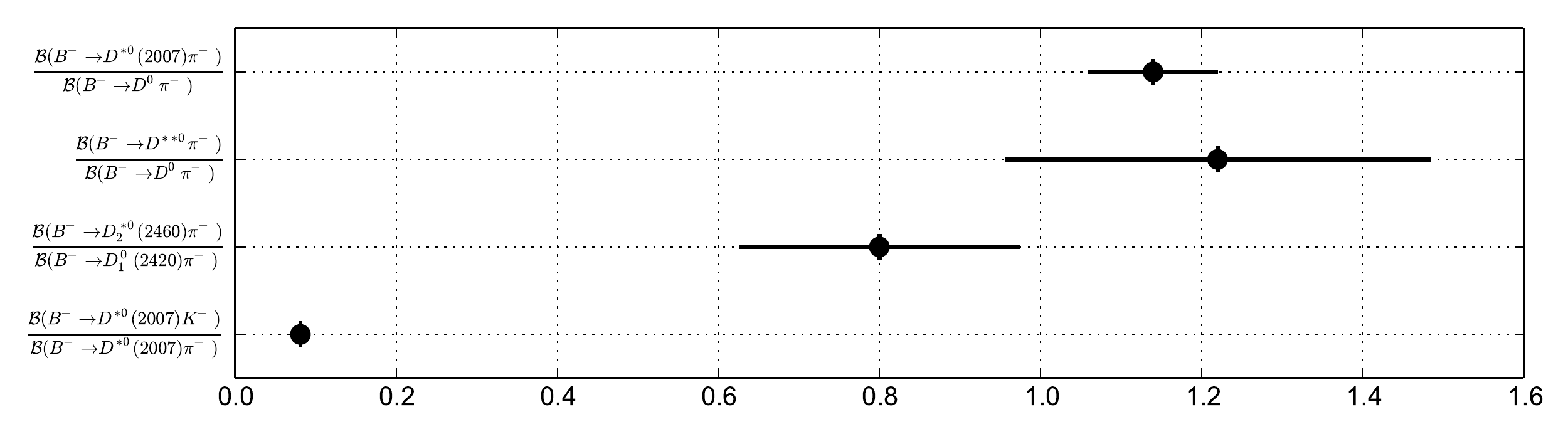}
\begin{btocharmtab}{Bu_D_8}{Relative product decay rates to excited $D$ mesons}
\hline
\textbf{Parameter} & %
 & $< 0.15$ \\
\hline
\end{btocharmtab}
\btocharmfig{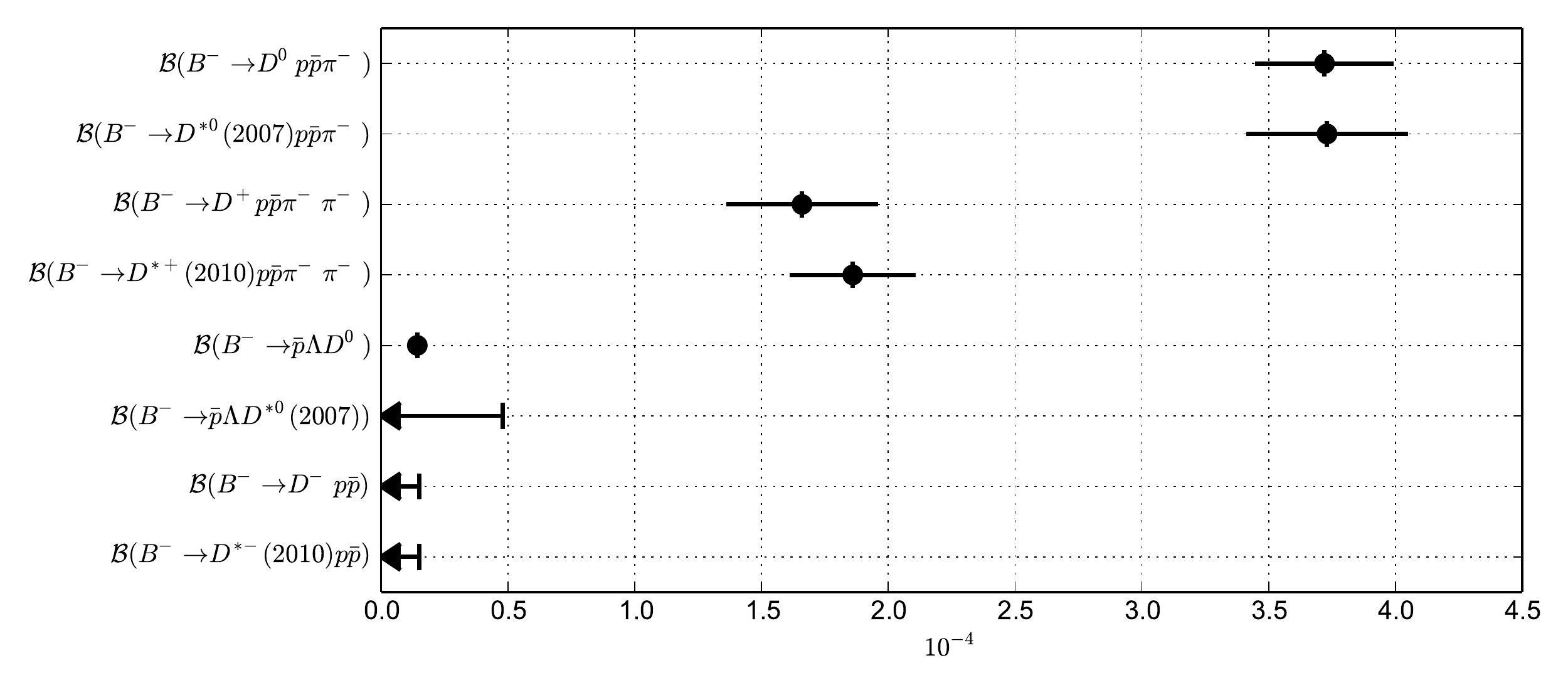}
\begin{btocharmtab}{Bu_D_11}{Lepton number violating decays $[10^{-6}]$}
\hline
\textbf{Parameter} & %
 & $< 1.0$ \\
\hline
\end{btocharmtab}
\btocharmfig{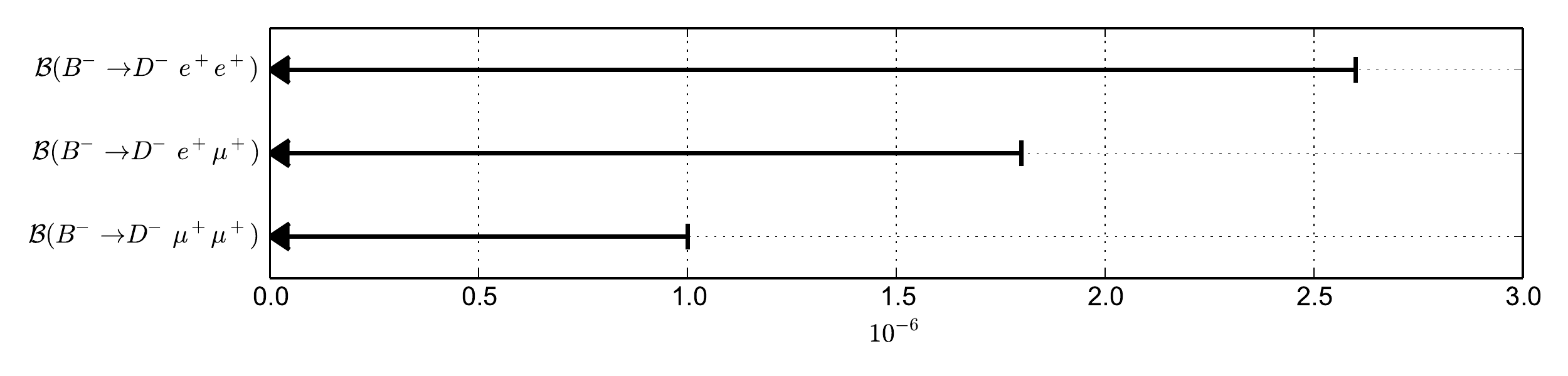}

\subsubsection{Decays to two open charm mesons}
\label{sec:b2c:Bu_DD}
Averages of $B^-$ decays to two open charm mesons are shown in Tables~\ref{tab:b2c:Bu_DD_1}--\ref{tab:b2c:Bu_DD_6} and Figs.~\ref{fig:b2c:Bu_DD_1}--\ref{fig:b2c:Bu_DD_6}.
\begin{btocharmtab}{Bu_DD_1}{Decays to $D^{(*)-}D^{(*)0}$ $[10^{-3}]$}
\hline
\textbf{Parameter} & %
 & $0.81 \pm 0.17$ \\
\hline
\end{btocharmtab}
\btocharmfig{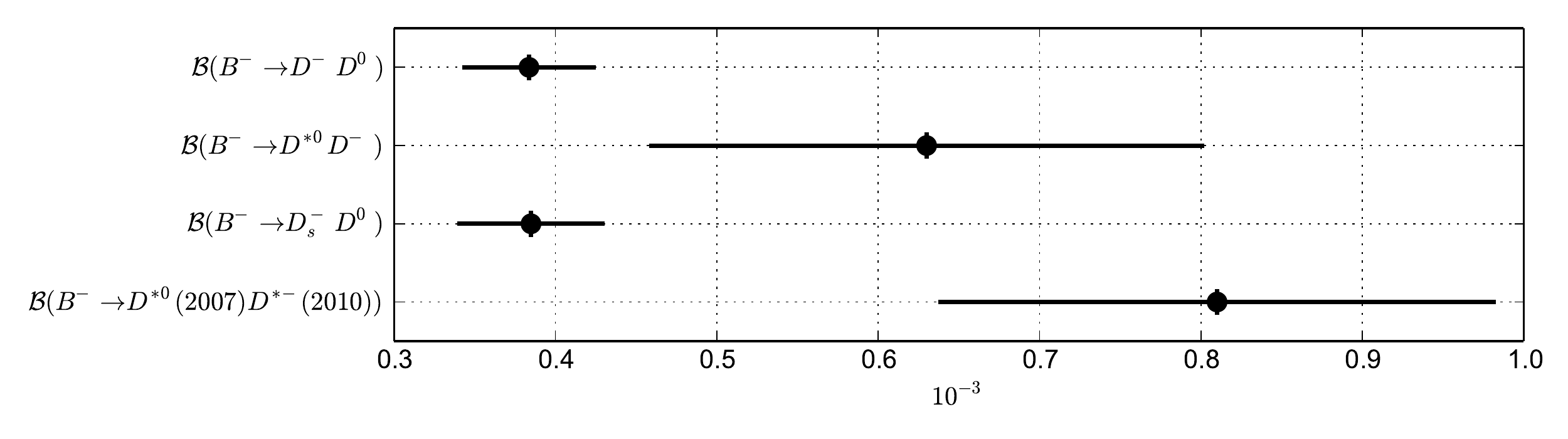}
\begin{btocharmtab}{Bu_DD_2}{Decays to two $D$ mesons and a kaon $[10^{-3}]$}
\hline
\textbf{Parameter} & %
 & $0.0857 \pm 0.0186$ \\
\hline
\end{btocharmtab}
\btocharmfig{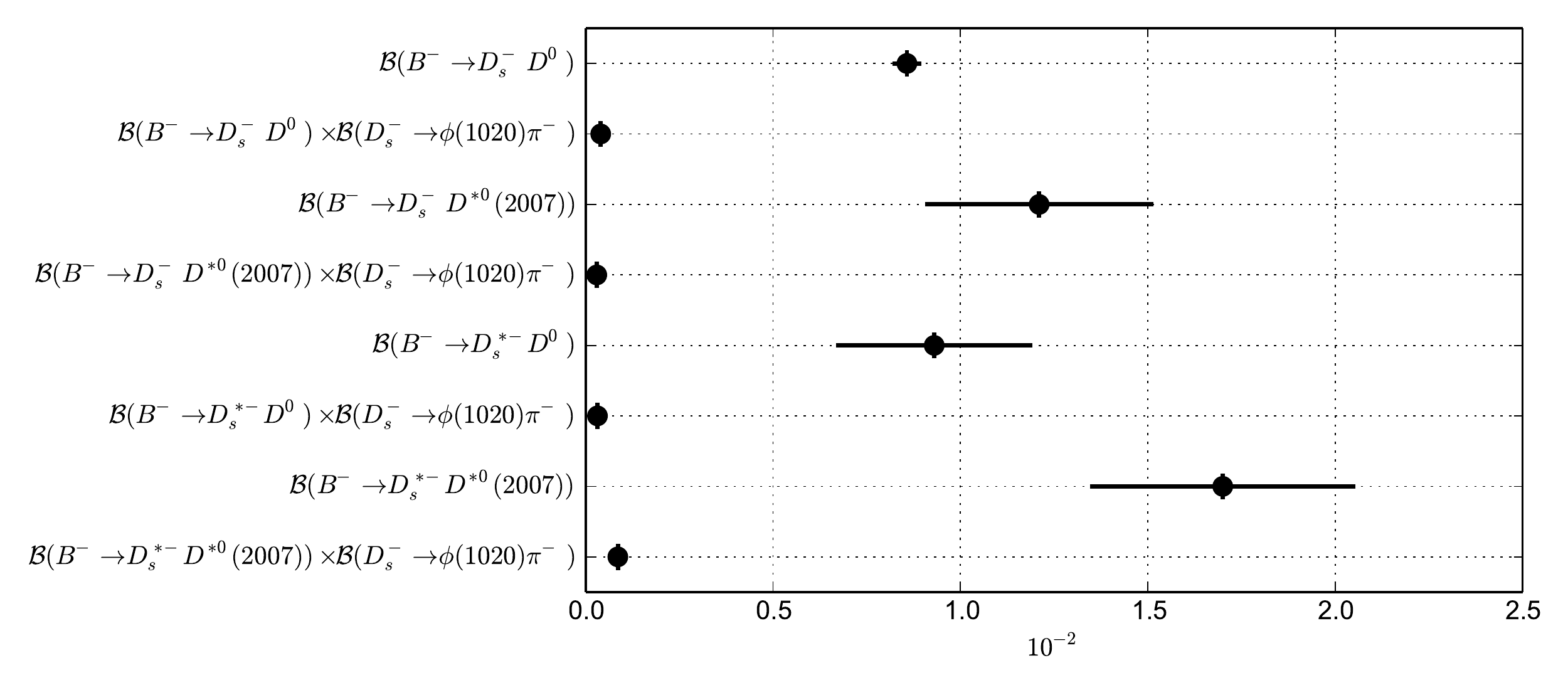}
\begin{btocharmtab}{Bu_DD_4}{Relative decay rates}
\hline
\textbf{Parameter} & %
 & $1.22 \pm 0.07$ \\
\hline
\end{btocharmtab}
\btocharmfig{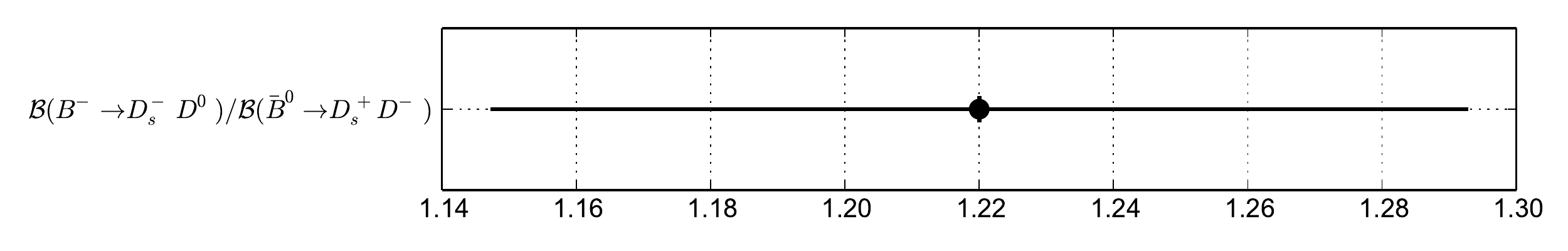}
\begin{btocharmtab}{Bu_DD_5}{Absolute decays rates to excited $D_s$ mesons $[10^{-3}]$}
\hline
\textbf{Parameter} & %
 & $11.2 \pm 3.3$ \\
\hline
\end{btocharmtab}
\btocharmfig{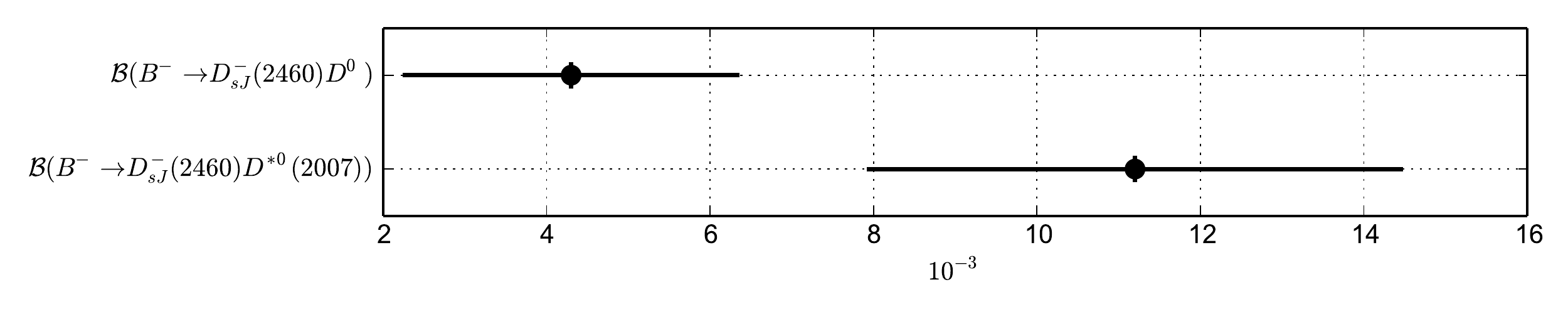}
\begin{btocharmtab}{Bu_DD_6}{Product decays rates to excited $D_s$ mesons $[10^{-3}]$}
\hline
\textbf{Parameter} & %
 & $< 1.069$ \\
\hline
\end{btocharmtab}
\btocharmfig{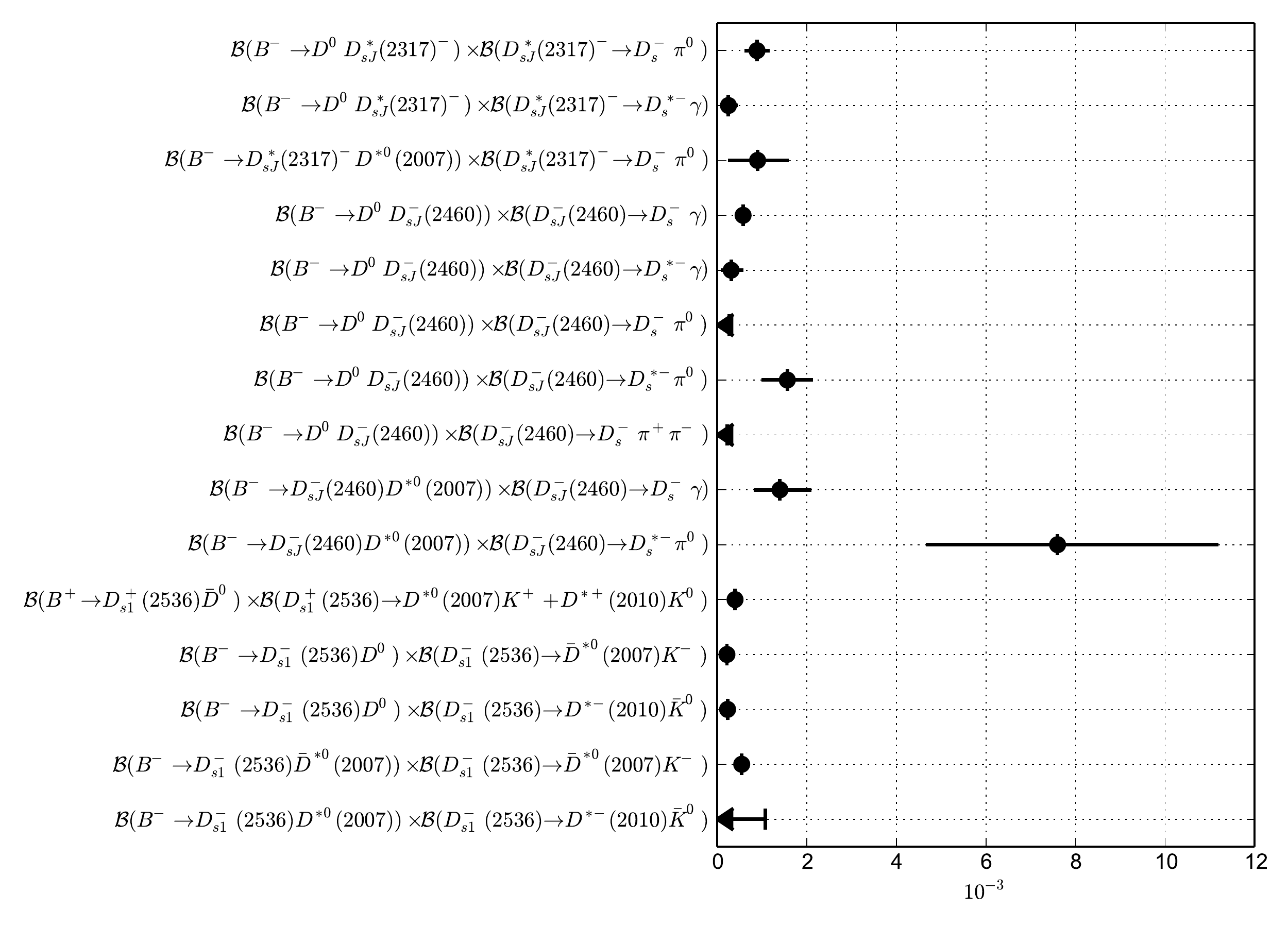}

\subsubsection{Decays to charmonium states}
\label{sec:b2c:Bu_cc}
Averages of $B^-$ decays to charmonium states are shown in Tables~\ref{tab:b2c:Bu_cc_1}--\ref{tab:b2c:Bu_cc_6} and Figs.~\ref{fig:b2c:Bu_cc_1}--\ref{fig:b2c:Bu_cc_6}.
\begin{btocharmtab}{Bu_cc_1}{Decays to $J/\psi$ and one kaon $[10^{-3}]$}
\hline
\textbf{Parameter} & %
 & $0.044 \pm 0.015$ \\
\hline
\end{btocharmtab}
\btocharmfig{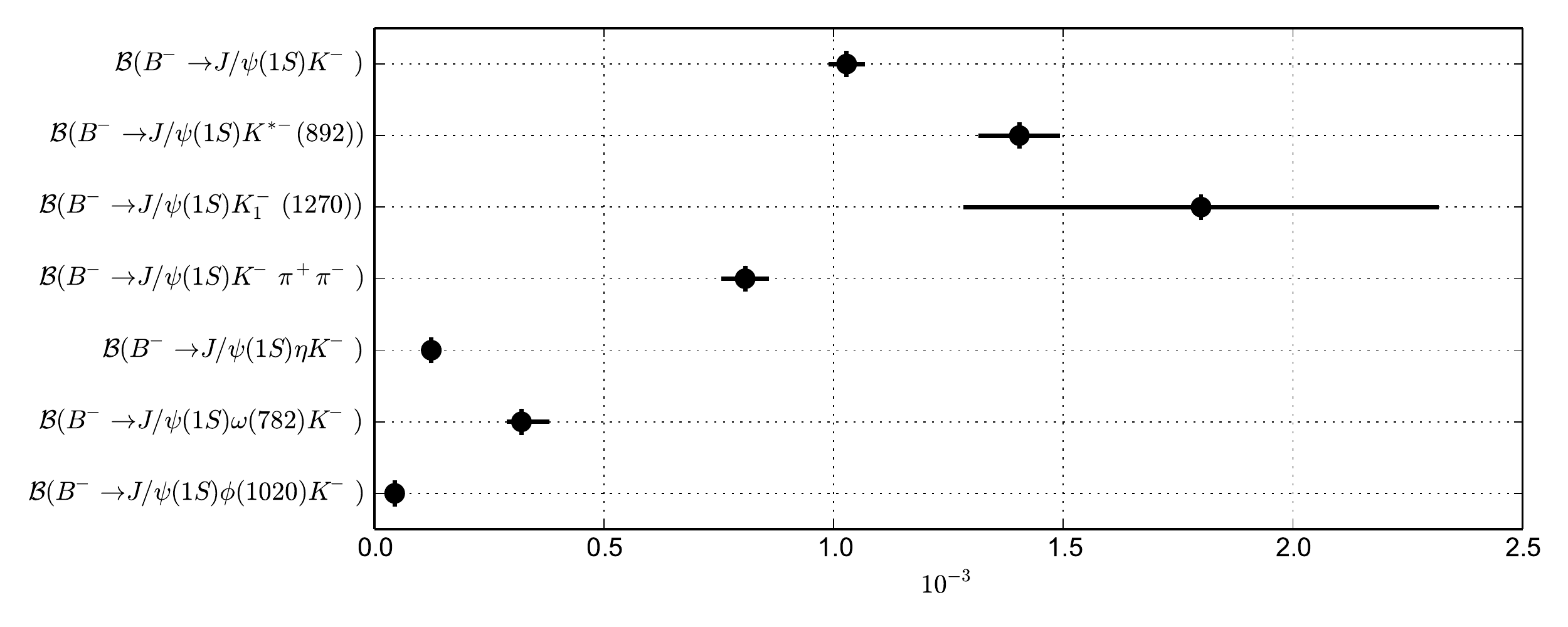}
{
\renewenvironment{btocharmtab}[2]{\begin{center}\begin{longtable}{| l l l |}\caption{#2}\label{tab:b2c:#1}\endfirsthead\multicolumn{3}{c}{continued from previous page}\endhead\endfoot\endlastfoot}{\end{longtable}\end{center}}
\begin{btocharmtab}{Bu_cc_2}{Decays to charmonium other than $J/\psi$ and one kaon $[10^{-3}]$}
\hline
\textbf{Parameter} & %
 & $< 0.0034$ \\
\hline
\end{btocharmtab}
}\btocharmfig{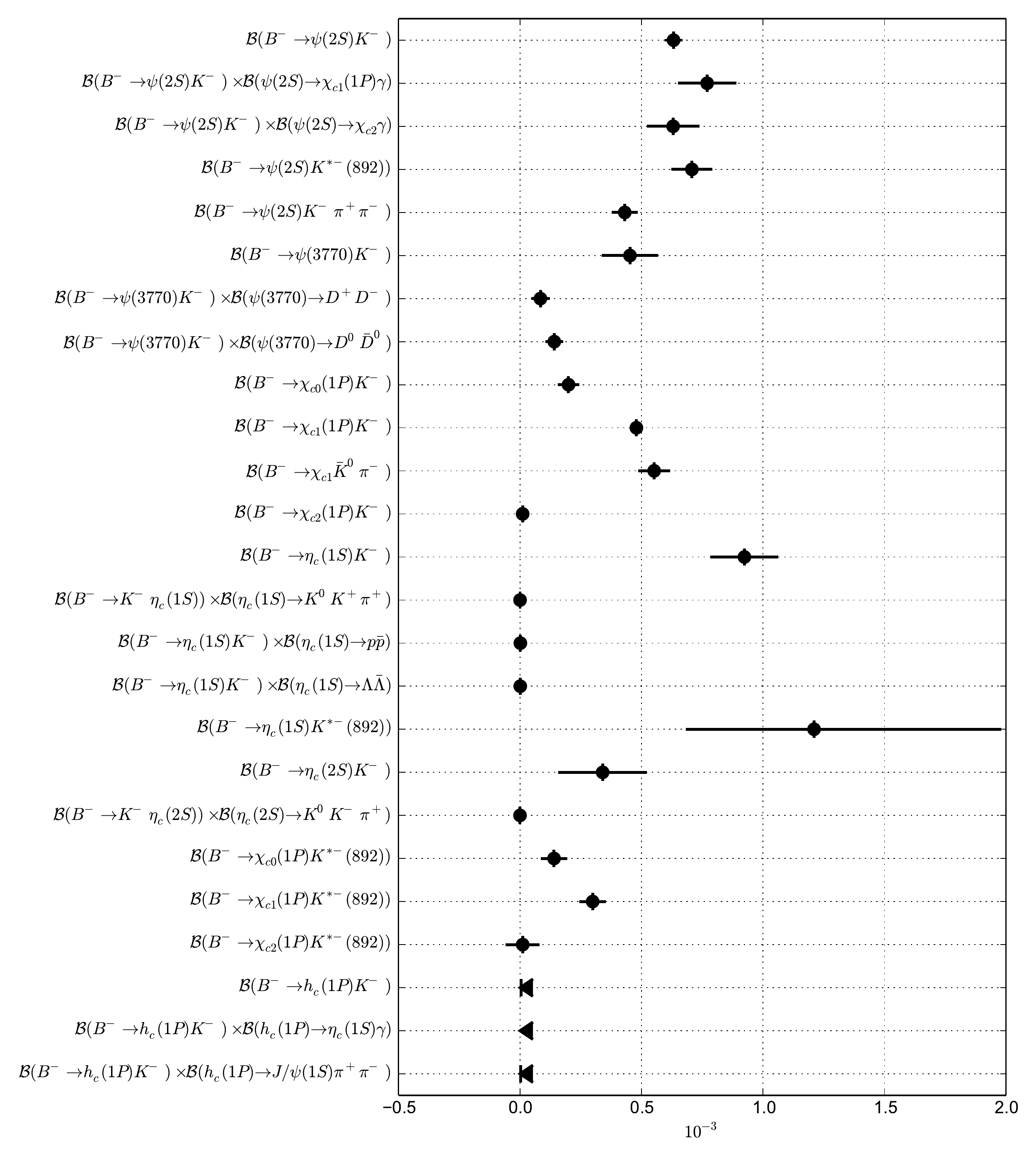}
\begin{btocharmtab}{Bu_cc_3}{Decays to charmonium and light mesons $[10^{-5}]$}
\hline
\textbf{Parameter} & %
 & $< 0.25$ \\
\hline
\end{btocharmtab}
\btocharmfig{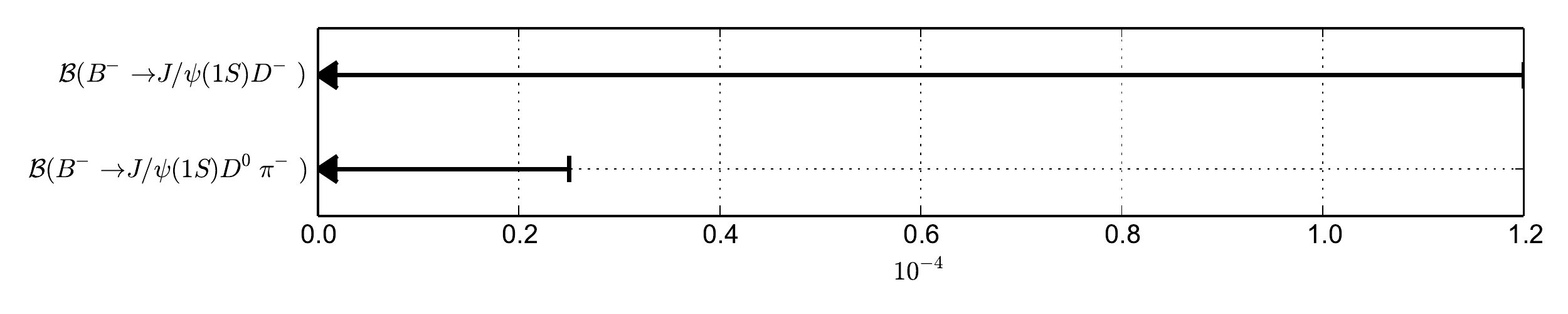}
\begin{btocharmtab}{Bu_cc_5}{Decays with baryons $[10^{-5}]$}
\hline
\textbf{Parameter} & %
 & $0.221 \pm 0.013$ \\
\hline
\end{btocharmtab}
\btocharmfig{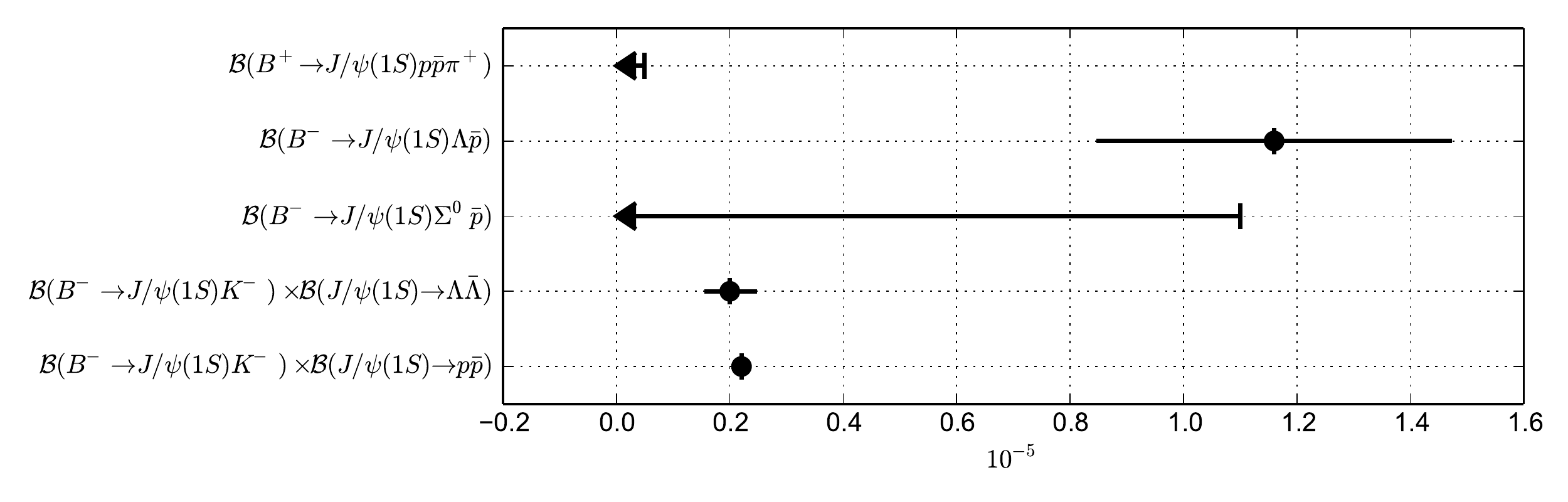}
\begin{btocharmtab}{Bu_cc_6}{Relative decay rates}
\hline
\textbf{Parameter} & %
 & $< 0.058$ \\
\hline
\end{btocharmtab}
\btocharmfig{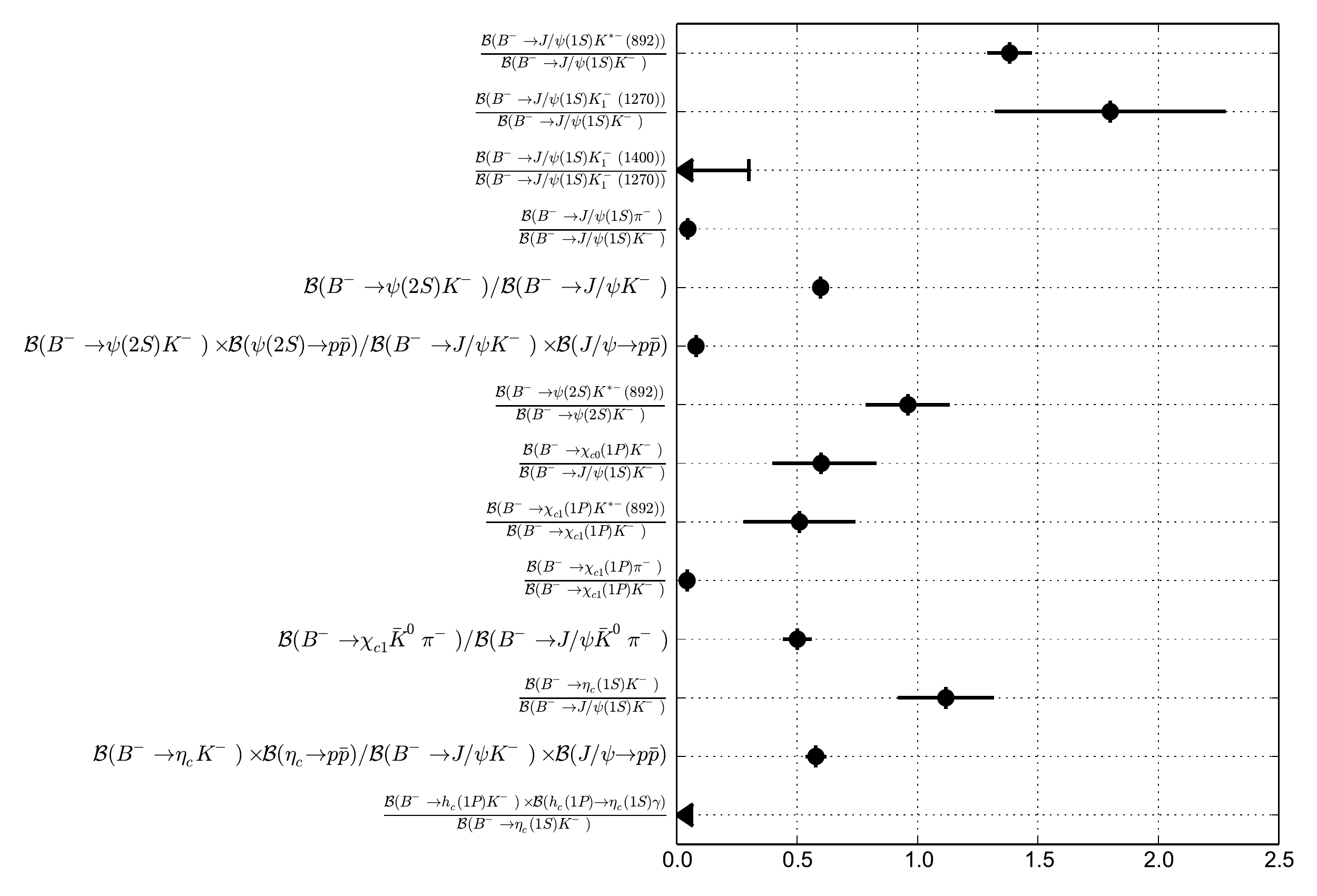}

\subsubsection{Decays to charm baryons}
\label{sec:b2c:Bu_baryon}
Averages of $B^-$ decays to charm baryons are shown in Tables~\ref{tab:b2c:Bu_baryon_1}--\ref{tab:b2c:Bu_baryon_2} and Figs.~\ref{fig:b2c:Bu_baryon_1}--\ref{fig:b2c:Bu_baryon_2}.
\begin{btocharmtab}{Bu_baryon_1}{Absolute (product) decay rates $[10^{-3}]$}
\hline
\textbf{Parameter} & %
 & $0.298 \pm 0.080$ \\
\hline
\end{btocharmtab}
\btocharmfig{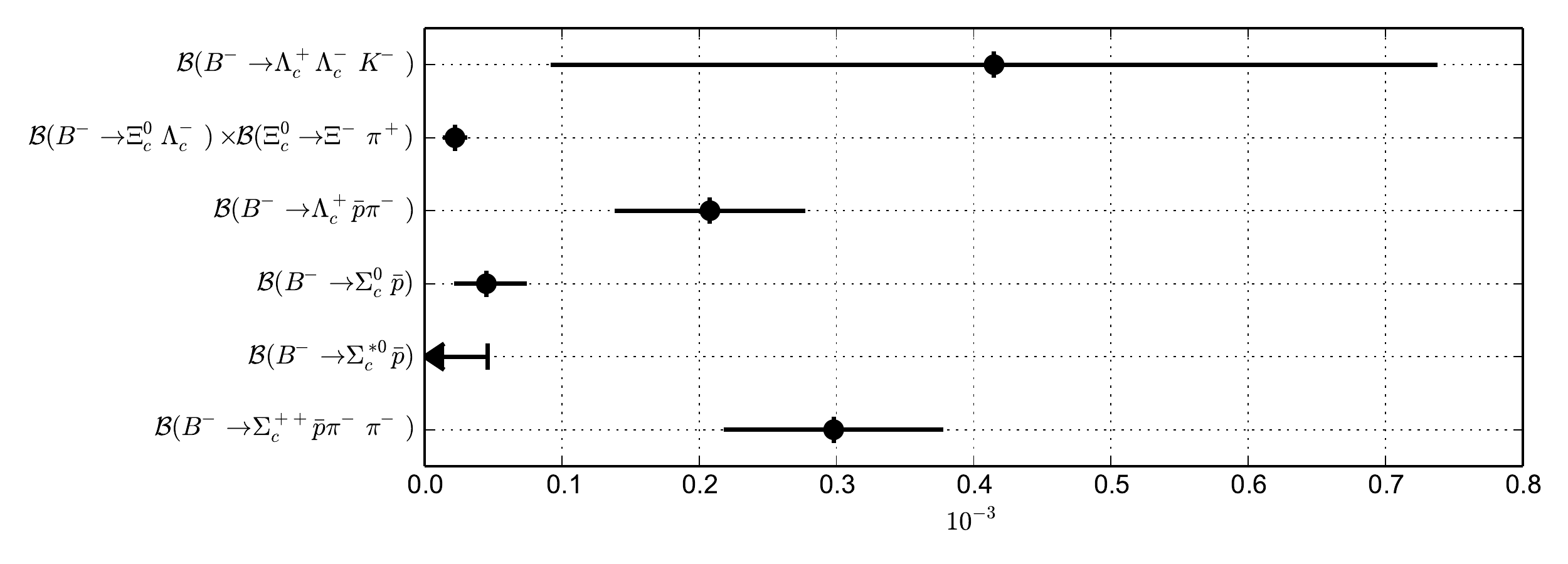}
\begin{btocharmtab}{Bu_baryon_2}{Relative decay rates}
\hline
\textbf{Parameter} & %
 & $0.117 \pm 0.033$ \\
\hline
\end{btocharmtab}
\btocharmfig{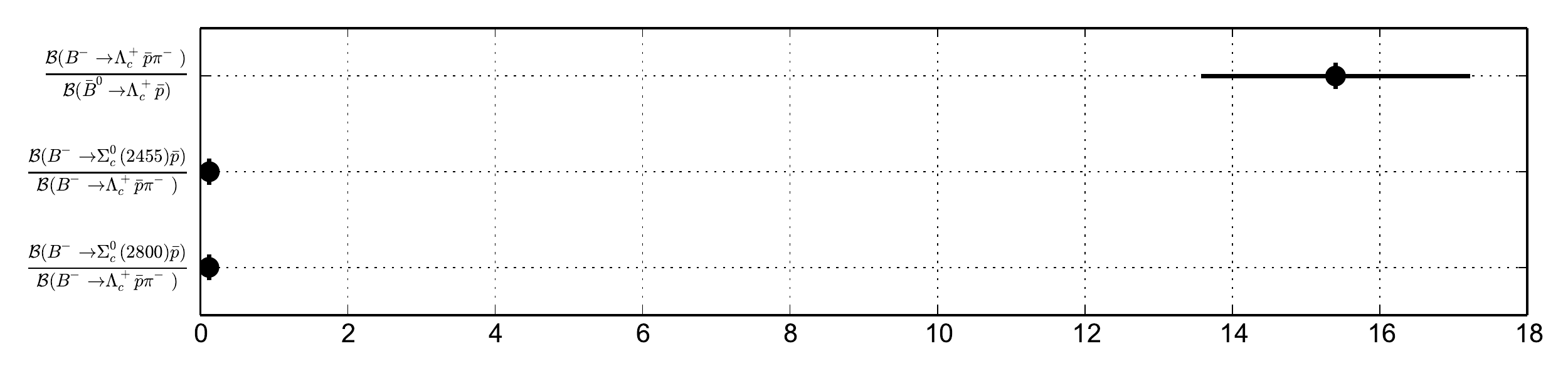}

\subsubsection{Decays to other ($XYZ$) states}
\label{sec:b2c:Bu_other}
Averages of $B^-$ decays to other ($XYZ$) states are shown in Tables~\ref{tab:b2c:Bu_other_1}--\ref{tab:b2c:Bu_other_5} and Figs.~\ref{fig:b2c:Bu_other_1}--\ref{fig:b2c:Bu_other_5}.
\begin{btocharmtab}{Bu_other_1}{Absolute decay rates $[10^{-4}]$}
\hline
\textbf{Parameter} & %
 & $< 3.2$ \\
\hline
\end{btocharmtab}
\btocharmfig{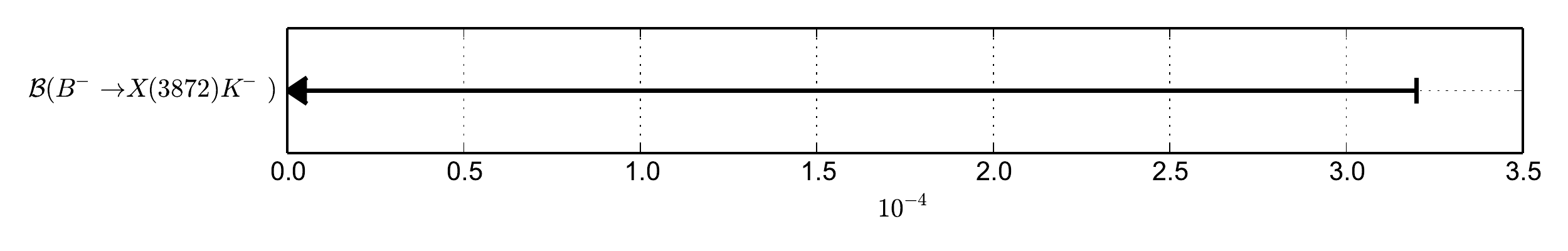}
\begin{btocharmtab}{Bu_other_2}{Product decay rates to $X(3872)$ $[10^{-4}]$}
\hline
\textbf{Parameter} & %
 & $< 0.067$ \\
\hline
\end{btocharmtab}
\btocharmfig{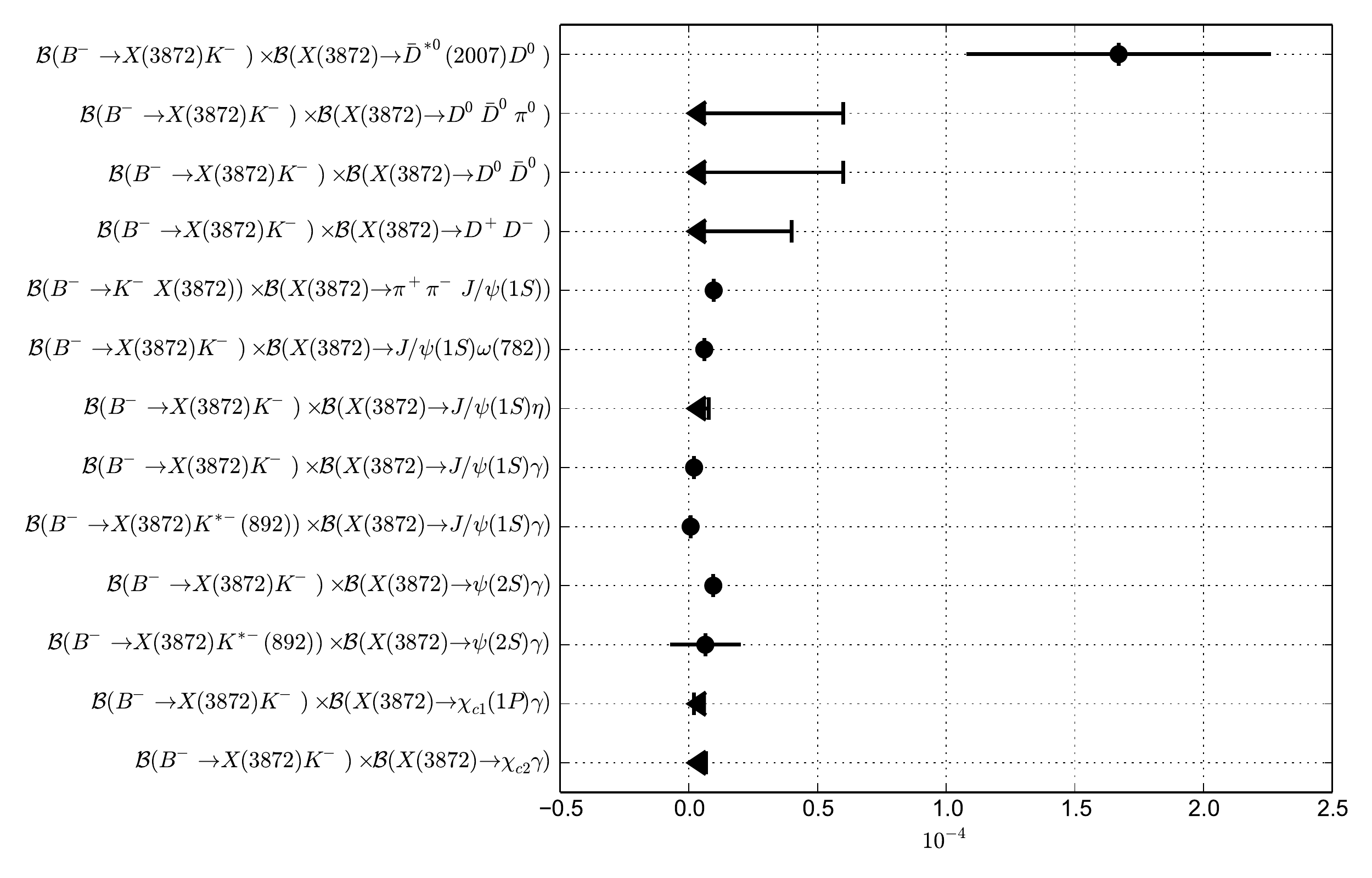}
\begin{btocharmtab}{Bu_other_3}{Product decay rates to neutral states other than $X(3872)$ $[10^{-5}]$}
\hline
\textbf{Parameter} & %
 & $2.0 \pm 0.7$ \\
\hline
\end{btocharmtab}
\btocharmfig{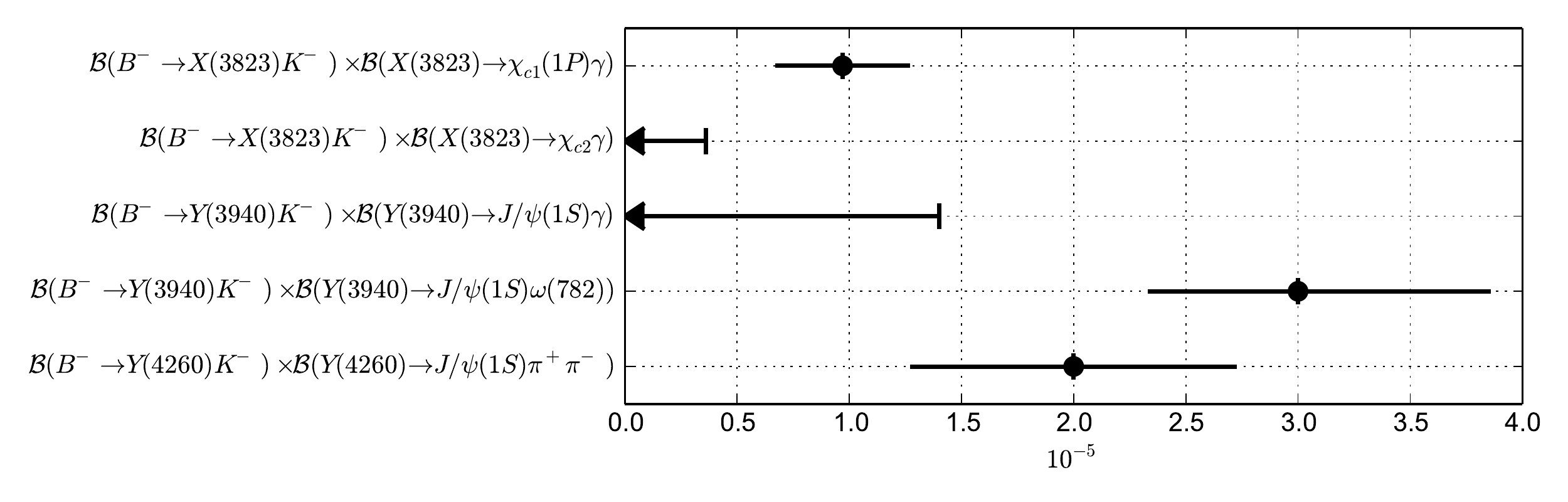}
\begin{btocharmtab}{Bu_other_4}{Relative product decay rates to states with $s\bar{s}$ component}
\hline
\textbf{Parameter} & %
 & $0.162 \pm 0.041$ \\
\hline
\end{btocharmtab}
\btocharmfig{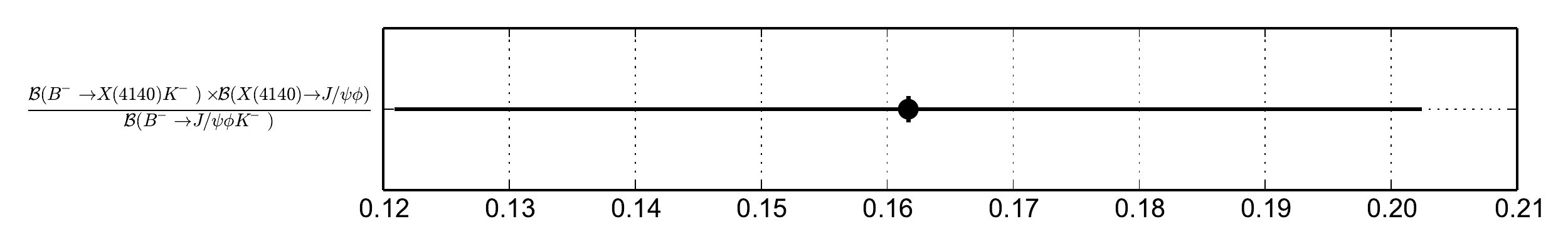}
\begin{btocharmtab}{Bu_other_5}{Product decay rates to charged states $[10^{-5}]$}
\hline
\textbf{Parameter} & %
 & $2.0 \pm 1.7$ \\
\hline
\end{btocharmtab}
\btocharmfig{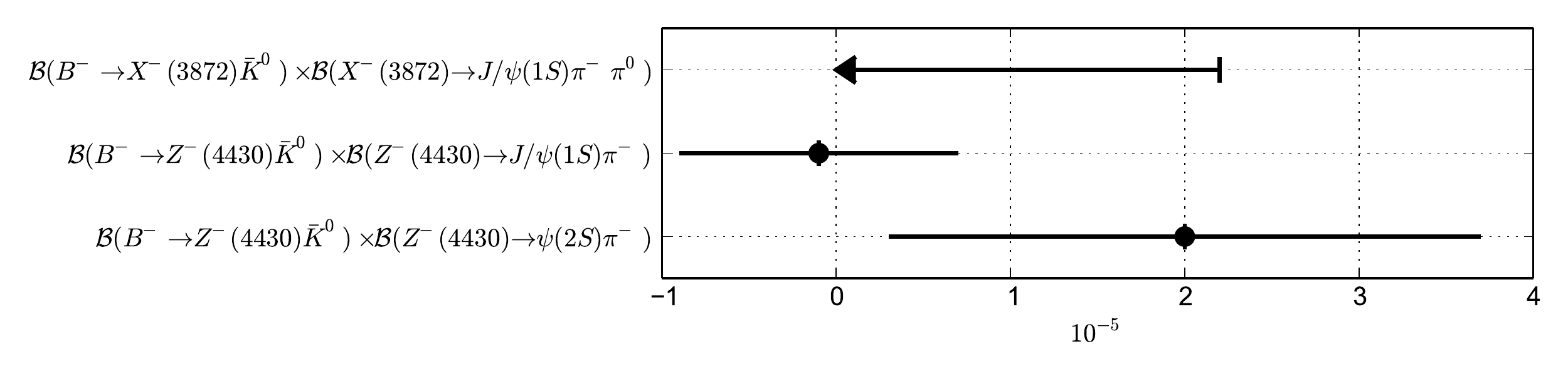}

\subsection{Decays of admixtures of $\Bzb$ / $B^-$ mesons}
\label{sec:b2c:B}
Measurements of $\Bzb$ / $B^-$ decays to charmed hadrons are summarized in sections \ref{sec:b2c:B_DD} to \ref{sec:b2c:B_other}.

\subsubsection{Decays to two open charm mesons}
\label{sec:b2c:B_DD}
Averages of $\Bzb$ / $B^-$ decays to two open charm mesons are shown in Table~\ref{tab:b2c:B_DD_1} and Fig.~\ref{fig:b2c:B_DD_1}.
\begin{btocharmtab}{B_DD_1}{ B decays to double charm $[10^{-4}]$}
\hline
\textbf{Parameter} & \begin{tabular}{l}\textbf{Measurements}\end{tabular} & \textbf{Average} \\
\hline
\hline
${\cal{B}} ( B \to D^0 \bar{D}^0 \pi^0 K )$ & \begin{tabular}{l} Belle \cite{Gokhroo:2006bt}: $1.27 \pm 0.31 \,^{+0.22}_{-0.39}$ \\ \end{tabular} & $1.27 \,^{+0.38}_{-0.50}$ \\
\hline
\end{btocharmtab}
\btocharmfig{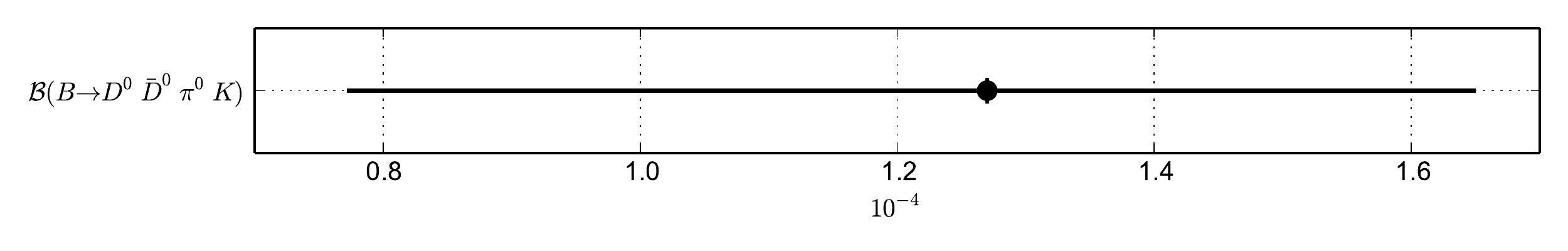}

\subsubsection{Decays to charmonium states}
\label{sec:b2c:B_cc}
Averages of $\Bzb$ / $B^-$ decays to charmonium states are shown in Tables~\ref{tab:b2c:B_cc_1}--\ref{tab:b2c:B_cc_5} and Figs.~\ref{fig:b2c:B_cc_1}--\ref{fig:b2c:B_cc_5}.
\begin{btocharmtab}{B_cc_1}{Decay amplitudes for parallel transverse polarization}
\hline
\textbf{Parameter} & \begin{tabular}{l}\textbf{Measurements}\end{tabular} & \textbf{Average} \\
\hline
\hline
$\vert{\cal{A}}_{\parallel}\vert^{2} ( B \to J/\psi(1S) K^{*} )$ & \begin{tabular}{l} \babar \cite{Aubert:2007hz}: $0.211 \pm 0.010 \pm 0.006$ \\ Belle \cite{Itoh:2005ks}: $0.231 \pm 0.012 \pm 0.008$ \\ \end{tabular} & $0.219 \pm 0.009$ \\
\hline
$\vert{\cal{A}}_{\parallel}\vert^{2} ( B \to \chi_{c1}(1P) K^{*} )$ & \begin{tabular}{l} \babar \cite{Aubert:2007hz}: $0.20 \pm 0.07 \pm 0.04$ \\ \end{tabular} & $0.20 \pm 0.08$ \\
\hline
$\vert{\cal{A}}_{\parallel}\vert^{2} ( B \to \psi(2S) K^{*} )$ & \begin{tabular}{l} \babar \cite{Aubert:2007hz}: $0.22 \pm 0.06 \pm 0.02$ \\ \end{tabular} & $0.22 \pm 0.06$ \\
\hline
\end{btocharmtab}
\btocharmfig{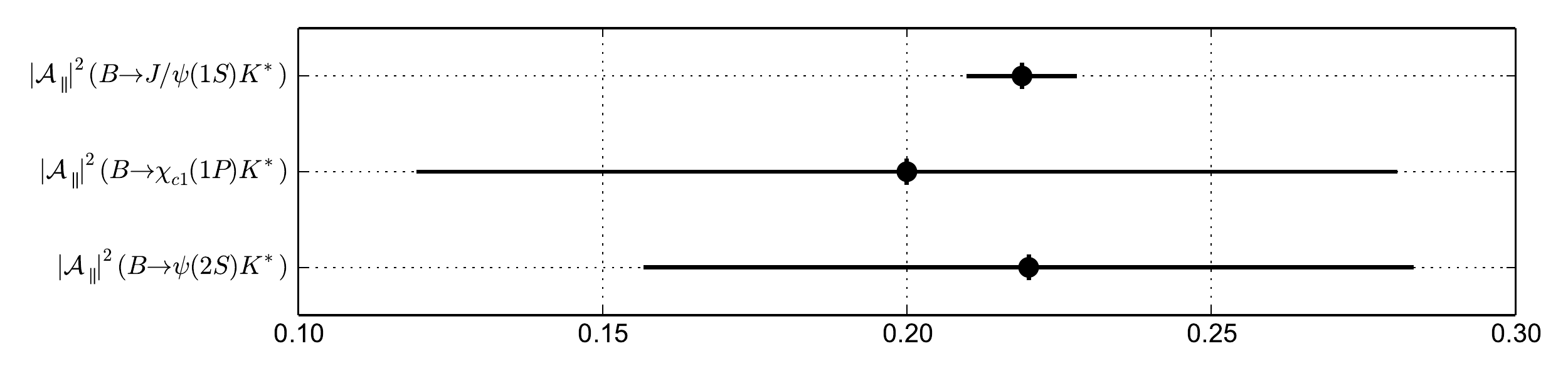}
\begin{btocharmtab}{B_cc_2}{Decay amplitudes for perpendicular transverse polarization}
\hline
\textbf{Parameter} & \begin{tabular}{l}\textbf{Measurements}\end{tabular} & \textbf{Average} \\
\hline
\hline
$\vert{\cal{A}}_{\perp}\vert^{2} ( B \to J/\psi(1S) K^{*} )$ & \begin{tabular}{l} \babar \cite{Aubert:2007hz}: $0.233 \pm 0.010 \pm 0.005$ \\ Belle \cite{Itoh:2005ks}: $0.195 \pm 0.012 \pm 0.008$ \\ \end{tabular} & $0.219 \pm 0.009$ \\
\hline
$\vert{\cal{A}}_{\perp}\vert^{2} ( B \to \chi_{c1}(1P) K^{*} )$ & \begin{tabular}{l} \babar \cite{Aubert:2007hz}: $0.03 \pm 0.04 \pm 0.02$ \\ \end{tabular} & $0.03 \pm 0.04$ \\
\hline
$\vert{\cal{A}}_{\perp}\vert^{2} ( B \to \psi(2S) K^{*} )$ & \begin{tabular}{l} \babar \cite{Aubert:2007hz}: $0.30 \pm 0.06 \pm 0.02$ \\ \end{tabular} & $0.30 \pm 0.06$ \\
\hline
\end{btocharmtab}
\btocharmfig{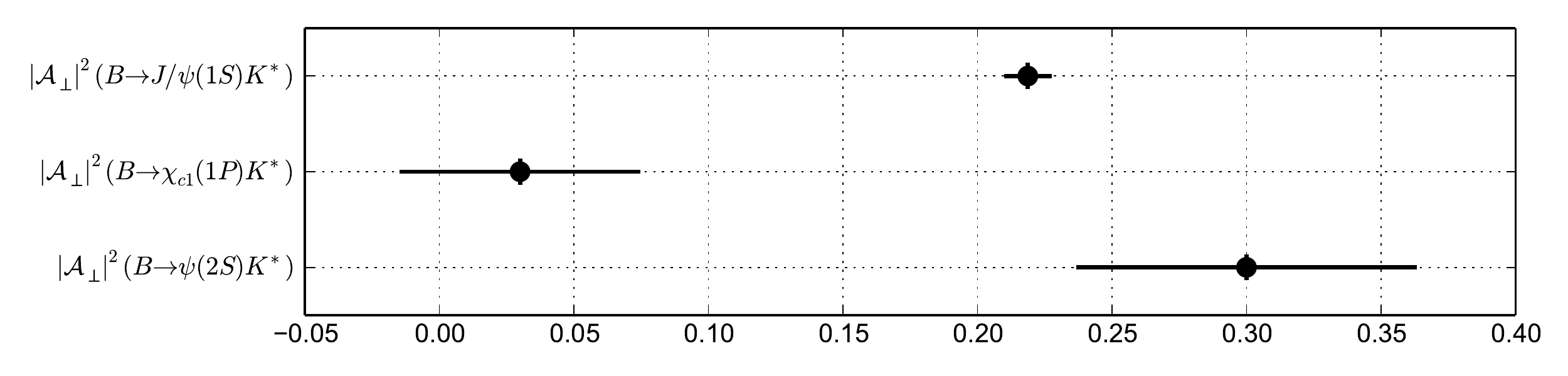}
\begin{btocharmtab}{B_cc_3}{Decay amplitudes for longitudinal polarization}
\hline
\textbf{Parameter} & \begin{tabular}{l}\textbf{Measurements}\end{tabular} & \textbf{Average} \\
\hline
\hline
$\vert{\cal{A}}_{0}\vert^{2} ( B \to J/\psi(1S) K^{*} )$ & \begin{tabular}{l} \babar \cite{Aubert:2007hz}: $0.556 \pm 0.009 \pm 0.010$ \\ Belle \cite{Itoh:2005ks}: $0.574 \pm 0.012 \pm 0.009$ \\ \end{tabular} & $0.564 \pm 0.010$ \\
\hline
$\vert{\cal{A}}_{0}\vert^{2} ( B \to \chi_{c1}(1P) K^{*} )$ & \begin{tabular}{l} \babar \cite{Aubert:2007hz}: $0.77 \pm 0.07 \pm 0.04$ \\ \end{tabular} & $0.77 \pm 0.08$ \\
\hline
$\vert{\cal{A}}_{0}\vert^{2} ( B \to \psi(2S) K^{*} )$ & \begin{tabular}{l} \babar \cite{Aubert:2007hz}: $0.48 \pm 0.05 \pm 0.02$ \\ \end{tabular} & $0.48 \pm 0.05$ \\
\hline
\end{btocharmtab}
\btocharmfig{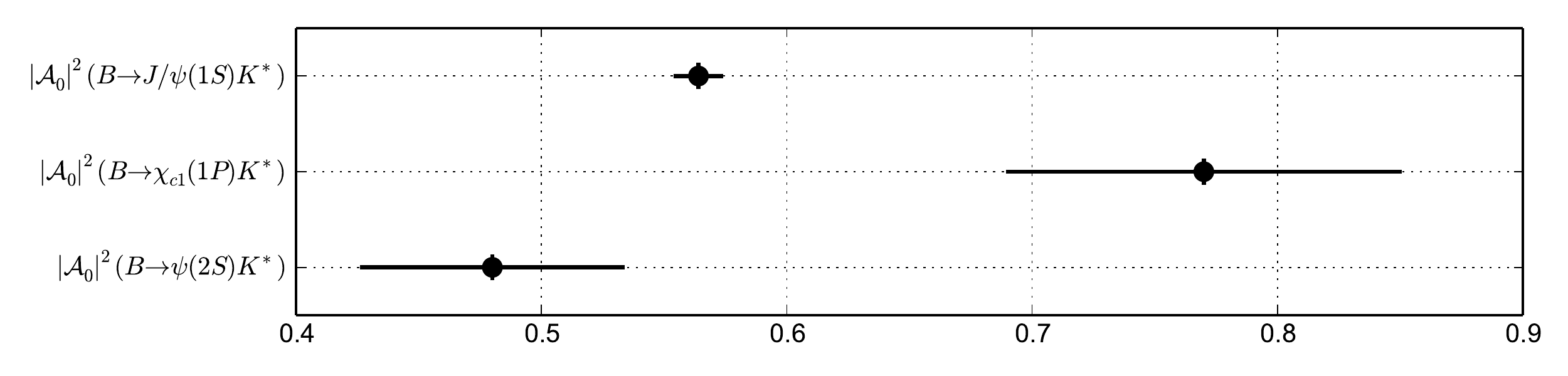}
\begin{btocharmtab}{B_cc_4}{Relative phases of parallel transverse polarization decay amplitudes}
\hline
\textbf{Parameter} & \begin{tabular}{l}\textbf{Measurements}\end{tabular} & \textbf{Average} \\
\hline
\hline
$\delta_{\parallel} ( B \to J/\psi(1S) K^{*} )$ & \begin{tabular}{l} \babar \cite{Aubert:2007hz}: $-2.93 \pm 0.08 \pm 0.04$ \\ Belle \cite{Itoh:2005ks}: $-2.887 \pm 0.090 \pm 0.008$ \\ \end{tabular} & $-2.909 \pm 0.064$ \\
\hline
$\delta_{\parallel} ( B \to \chi_{c1}(1P) K^{*} )$ & \begin{tabular}{l} \babar \cite{Aubert:2007hz}: $0.0 \pm 0.3 \pm 0.1$ \\ \end{tabular} & $0.0 \pm 0.3$ \\
\hline
$\delta_{\parallel} ( B \to \psi(2S) K^{*} )$ & \begin{tabular}{l} \babar \cite{Aubert:2007hz}: $-2.8 \pm 0.4 \pm 0.1$ \\ \end{tabular} & $-2.8 \pm 0.4$ \\
\hline
\end{btocharmtab}
\btocharmfig{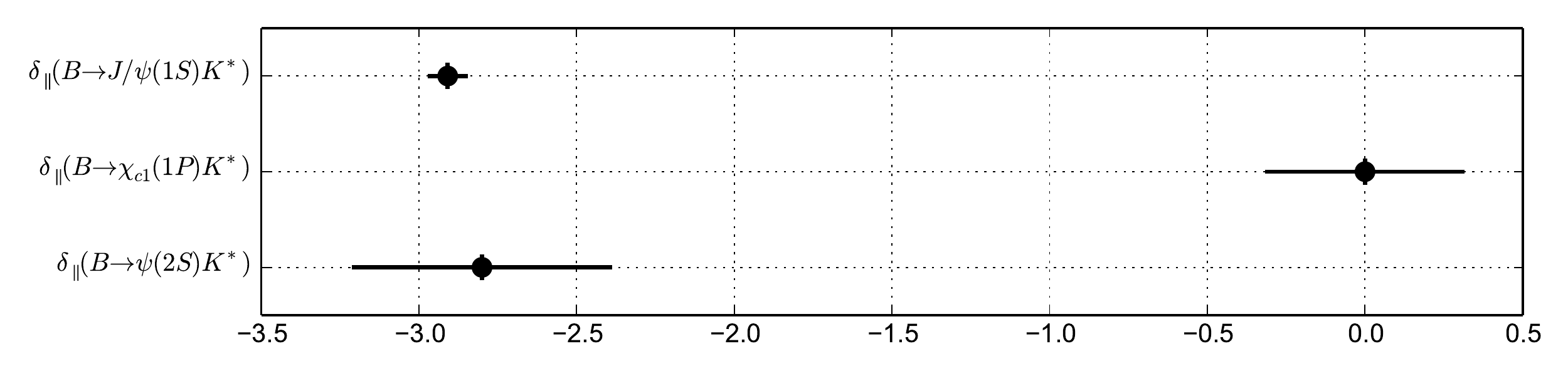}
\begin{btocharmtab}{B_cc_5}{Relative phases of perpendicular transverse polarization decay amplitudes}
\hline
\textbf{Parameter} & \begin{tabular}{l}\textbf{Measurements}\end{tabular} & \textbf{Average} \\
\hline
\hline
$\delta_{\perp} ( B \to J/\psi(1S) K^{*} )$ & \begin{tabular}{l} \babar \cite{Aubert:2007hz}: $2.91 \pm 0.05 \pm 0.03$ \\ Belle \cite{Itoh:2005ks}: $2.938 \pm 0.064 \pm 0.010$ \\ \end{tabular} & $2.923 \pm 0.043$ \\
\hline
$\delta_{\perp} ( B \to \psi(2S) K^{*} )$ & \begin{tabular}{l} \babar \cite{Aubert:2007hz}: $2.8 \pm 0.3 \pm 0.1$ \\ \end{tabular} & $2.8 \pm 0.3$ \\
\hline
\end{btocharmtab}
\btocharmfig{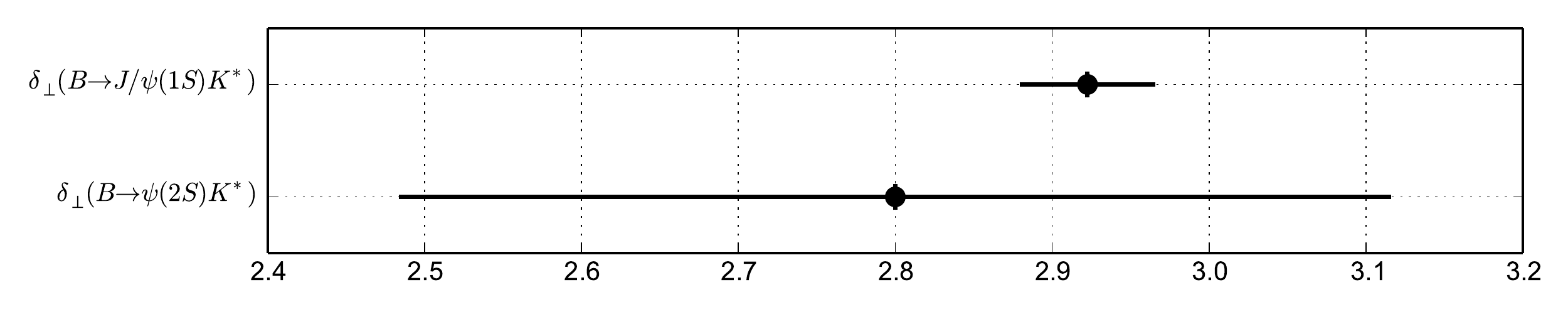}

\subsubsection{Decays to other ($XYZ$) states}
\label{sec:b2c:B_other}
Averages of $\Bzb$ / $B^-$ decays to other ($XYZ$) states are shown in Table~\ref{tab:b2c:B_other_1} and Fig.~\ref{fig:b2c:B_other_1}.
\begin{btocharmtab}{B_other_1}{Absolute decay rates to $X$/$Y$ states $[10^{-4}]$}
\hline
\textbf{Parameter} & \begin{tabular}{l}\textbf{Measurements}\end{tabular} & \textbf{Average} \\
\hline
\hline
\multicolumn{3}{|l|}{${{\cal{B}} ( B \to X(3872) K )\times {\cal{B}} ( X(3872) \to D^{*0}(2007) \bar{D}^0 )}$}\\
 & \begin{tabular}{l} Belle \cite{Adachi:2008sua}: $0.80 \pm 0.20 \pm 0.10$ \\ \end{tabular} & $0.80 \pm 0.22$ \\
\hline
\multicolumn{3}{|l|}{${{\cal{B}} ( B \to Y(3940) K )\times {\cal{B}} ( Y(3940) \to D^{*0}(2007) \bar{D}^0 )}$}\\
 & \begin{tabular}{l} Belle \cite{Adachi:2008sua}: $< 0.67$ \\ \end{tabular} & $< 0.67$ \\
\hline
\multicolumn{3}{|l|}{${{\cal{B}} ( B \to K Y(3940) )\times {\cal{B}} ( Y(3940) \to \omega(782) J/\psi(1S) )}$}\\
 & \begin{tabular}{l} Belle \cite{Abe:2004zs}: $0.71 \pm 0.13 \pm 0.31$ \\ \end{tabular} & $0.71 \pm 0.34$ \\
\hline
\end{btocharmtab}
\btocharmfig{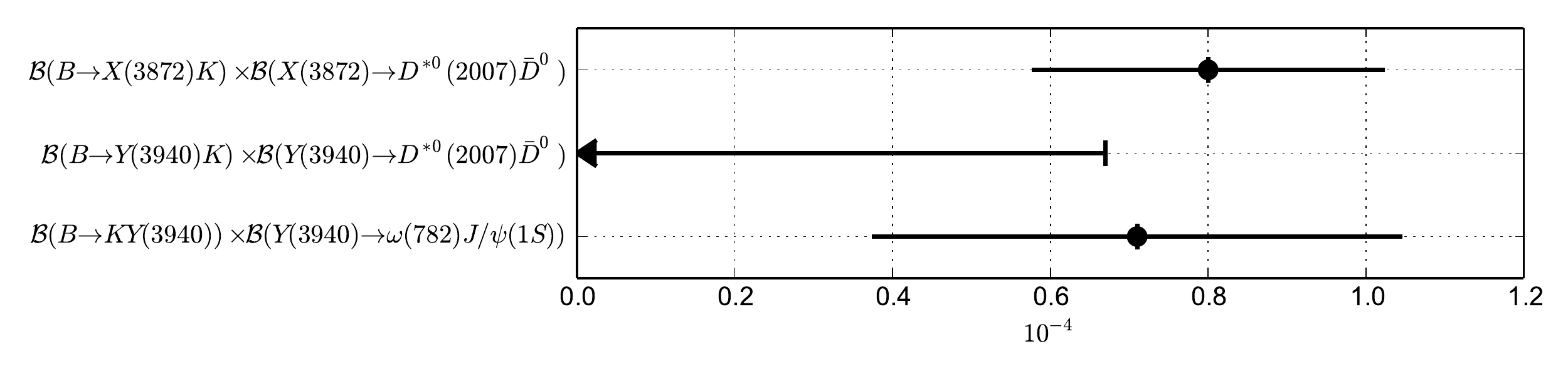}

\subsection{Decays of $\Bsb$ mesons}
\label{sec:b2c:Bs}
Measurements of $\Bsb$ decays to charmed hadrons are summarized in sections \ref{sec:b2c:Bs_D} to \ref{sec:b2c:Bs_baryon}.

\subsubsection{Decays to a single open charm meson}
\label{sec:b2c:Bs_D}
Averages of $\Bsb$ decays to a single open charm meson are shown in Tables~\ref{tab:b2c:Bs_D_1}--\ref{tab:b2c:Bs_D_4} and Figs.~\ref{fig:b2c:Bs_D_1}--\ref{fig:b2c:Bs_D_4}.
\begin{btocharmtab}{Bs_D_1}{Decays to a $D_s^{(*)}$ and a light meson $[10^{-3}]$}
\hline
\textbf{Parameter} & \begin{tabular}{l}\textbf{Measurements}\end{tabular} & \textbf{Average} \\
\hline
\hline
${\cal{B}} ( \bar{B}_s^0 \to D_s^+ \pi^- )$ & \begin{tabular}{l} LHCb \cite{Aaij:2012zz}: $2.95 \pm 0.05 \,^{+0.25}_{-0.28}$ \\ Belle \cite{Louvot:2008sc}: $3.67 \,^{+0.35}_{-0.33} \,^{+0.65}_{-0.65}$ \\ \end{tabular} & $3.03 \pm 0.25$ \\
\hline
${\cal{B}} ( \bar{B}_s^0 \to D_s^{*+} \pi^- )$ & \begin{tabular}{l} Belle \cite{Louvot:2010rd}: $2.4 \,^{+0.5}_{-0.4} \pm 0.4$ \\ \end{tabular} & $2.4 \,^{+0.7}_{-0.6}$ \\
\hline
${\cal{B}} ( \bar{B}_s^0 \to D_s^+ \rho^-(770) )$ & \begin{tabular}{l} Belle \cite{Louvot:2010rd}: $8.5 \,^{+1.3}_{-1.2} \pm 1.7$ \\ \end{tabular} & $8.5 \,^{+2.1}_{-2.1}$ \\
\hline
${\cal{B}} ( \bar{B}_s^0 \to D_s^{*+} \rho^-(770) )$ & \begin{tabular}{l} Belle \cite{Louvot:2010rd}: $11.8 \,^{+2.2}_{-2.0} \pm 2.5$ \\ \end{tabular} & $11.8 \,^{+3.3}_{-3.2}$ \\
\hline
${\cal{B}} ( \bar{B}_s^0 \to D_s^+ K^- )$ & \begin{tabular}{l} LHCb \cite{Aaij:2012zz}: $0.190 \pm 0.012 \,^{+0.018}_{-0.019}$ \\ Belle \cite{Louvot:2008sc}: $0.24 \,^{+0.12}_{-0.10} \pm 0.04$ \\ \end{tabular} & $0.192 \pm 0.022$ \\
\hline
\end{btocharmtab}
\btocharmfig{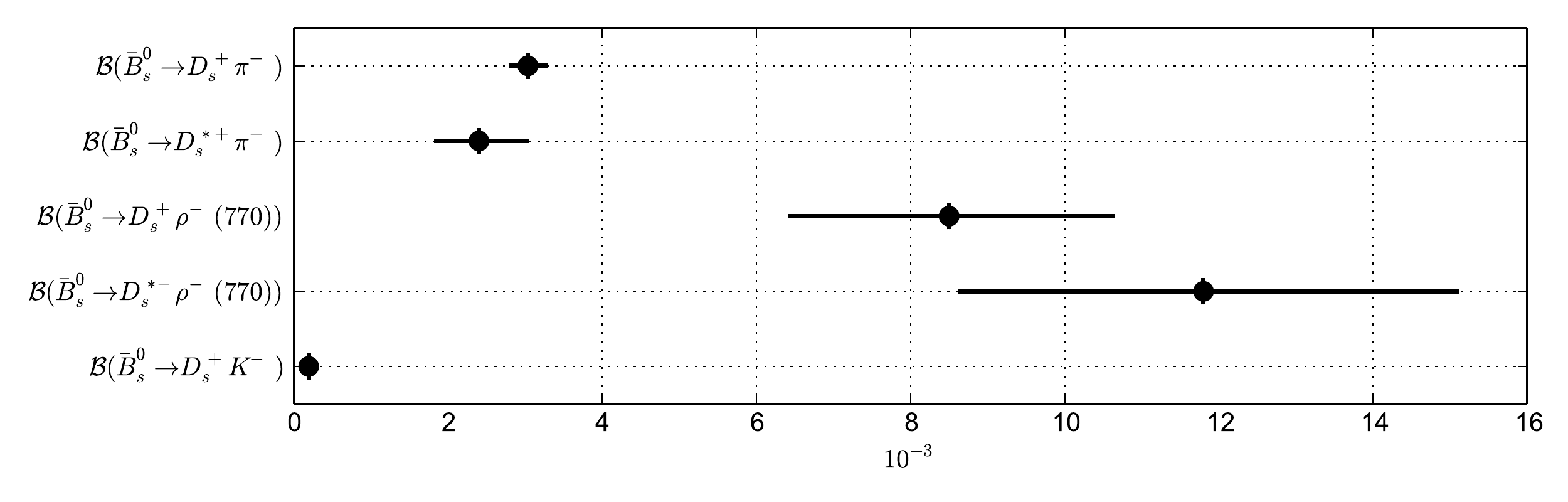}
\begin{btocharmtab}{Bs_D_2}{Decays to a $D^{(*)}$ and a light meson $[10^{-4}]$}
\hline
\textbf{Parameter} & \begin{tabular}{l}\textbf{Measurements}\end{tabular} & \textbf{Average} \\
\hline
\hline
${\cal{B}} ( \bar{B}_s^0 \to D_s^+ \pi^- )$ & \begin{tabular}{l} LHCb \cite{Aaij:2013fpa}: $< 0.061$ \\ \end{tabular} & $< 0.061$ \\
\hline
${\cal{B}} ( \bar{B}^0_s \to D^0 \bar{K}^{*0} )$ & \begin{tabular}{l} LHCb \cite{Aaij:2011tz}: $4.72 \pm 1.07 \pm 0.96$ \\ \end{tabular} & $4.72 \pm 1.44$ \\
\hline
\end{btocharmtab}
\btocharmfig{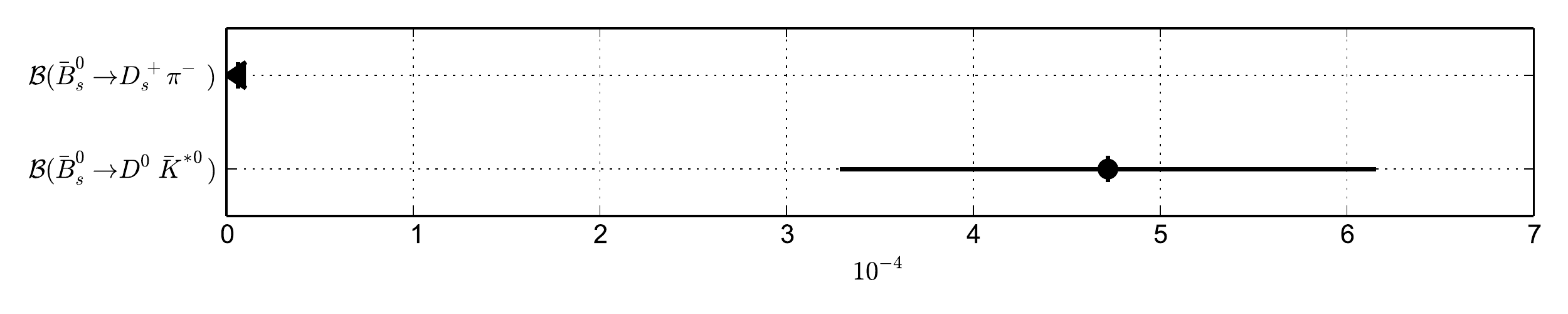}
\begin{btocharmtab}{Bs_D_3}{Relative decay rates}
\hline
\textbf{Parameter} & \begin{tabular}{l}\textbf{Measurements}\end{tabular} & \textbf{Average} \\
\hline
\hline
\multicolumn{3}{|l|}{${{\cal{B}} ( \bar{B}_s^0 \to D_s^+ \pi^- )}/{{\cal{B}} ( \bar{B}^0 \to D^+ \pi^- )}$}\\
 & \begin{tabular}{l} CDF \cite{Abulencia:2006aa}: $1.13 \pm 0.08 \pm 0.23$ \\ \end{tabular} & $1.13 \pm 0.25$ \\
\hline
\multicolumn{3}{|l|}{${\cal{B}} ( \bar{B}^0_s \to D_s^+ \pi^+ \pi^- \pi^- ) / {\cal{B}} ( \bar{B}^0_s \to D_s^+ \pi^- )$}\\
 & \begin{tabular}{l} LHCb \cite{Aaij:2011rj}: $2.01 \pm 0.37 \pm 0.20$ \\ \end{tabular} & $2.01 \pm 0.42$ \\
\hline
\multicolumn{3}{|l|}{${{\cal{B}} ( \bar{B}_s^0 \to D_s^+ \pi^+ \pi^- \pi^- )}/{{\cal{B}} ( \bar{B}^0 \to D^+ \pi^+ \pi^- \pi^- )}$}\\
 & \begin{tabular}{l} CDF \cite{Abulencia:2006aa}: $1.05 \pm 0.10 \pm 0.22$ \\ \end{tabular} & $1.05 \pm 0.24$ \\
\hline
\multicolumn{3}{|l|}{${{\cal{B}} ( \bar{B}_s^0 \to D_s^+ K^- )}/{{\cal{B}} ( \bar{B}_s^0 \to D_s^+ \pi^- )}$}\\
 & \begin{tabular}{l} CDF \cite{Aaltonen:2008ab}: $0.097 \pm 0.018 \pm 0.009$ \\ \end{tabular} & $0.097 \pm 0.020$ \\
\hline
\multicolumn{3}{|l|}{${{\cal{B}} ( \bar{B}_s^0 \to D_s^+ K^-  \pi^+  \pi^- )}/{{\cal{B}} ( \bar{B}^0 \to D_s^+ \pi^-  \pi^+  \pi^- )}$}\\
 & \begin{tabular}{l} LHCb \cite{Aaij:2012mra}: $0.052 \pm 0.005 \pm 0.003$ \\ \end{tabular} & $0.052 \pm 0.006$ \\
\hline
${\cal{B}} ( \bar{B}^0_s \to D^0 K^{*0} ) / {\cal{B}} ( B^- \to D^0 \rho^0 )$ & \begin{tabular}{l} LHCb \cite{Aaij:2011tz}: $1.48 \pm 0.34 \pm 0.19$ \\ \end{tabular} & $1.48 \pm 0.39$ \\
\hline
${\cal{B}} ( \bar{B}^0_s \to D^0 K^{*0} ) /{\cal{B}} ( \bar{B}^0 \to D^0 K^{*0} ) $ & \begin{tabular}{l} LHCb \cite{Aaij:2013dda}: $7.8 \pm 0.7 \pm 0.7$ \\ \end{tabular} & $7.8 \pm 1.0$ \\
\hline
\multicolumn{3}{|l|}{${\cal{B}} ( \bar{B}^0_s \to D^0 \phi(1020) ) /{\cal{B}} ( \bar{B}^0_s \to D^0 K^{*0} ) $}\\
 & \begin{tabular}{l} LHCb \cite{Aaij:2013dda}: $0.069 \pm 0.013 \pm 0.007$ \\ \end{tabular} & $0.069 \pm 0.015$ \\
\hline
\multicolumn{3}{|l|}{${ {\cal{B}} ( \bar{B}_s^0 \to \bar{D}^0 K^- \pi^+ )}/  {{\cal{B}} ( \bar{B}^0 \to \bar{D}^0 \pi^- \pi^+ )}$}\\
 & \begin{tabular}{l} LHCb \cite{Aaij:2013pua}: $1.18 \pm 0.05 \pm 0.12$ \\ \end{tabular} & $1.18 \pm 0.13$ \\
\hline
\multicolumn{3}{|l|}{${{\cal{B}} ( \bar{B}_s^0 \to D_{s1}^+ \pi^- ) \times (D_{s1}^+ \to  D_s^+ \pi^-  \pi^+)  }/{{\cal{B}} ( \bar{B}^0 \to D_s^+ \pi^-  \pi^+  \pi^- )}$}\\
 & \begin{tabular}{l} LHCb \cite{Aaij:2012mra}: $0.0040 \pm 0.0010 \pm 0.0004$ \\ \end{tabular} & $0.0040 \pm 0.0011$ \\
\hline
\end{btocharmtab}
\btocharmfig{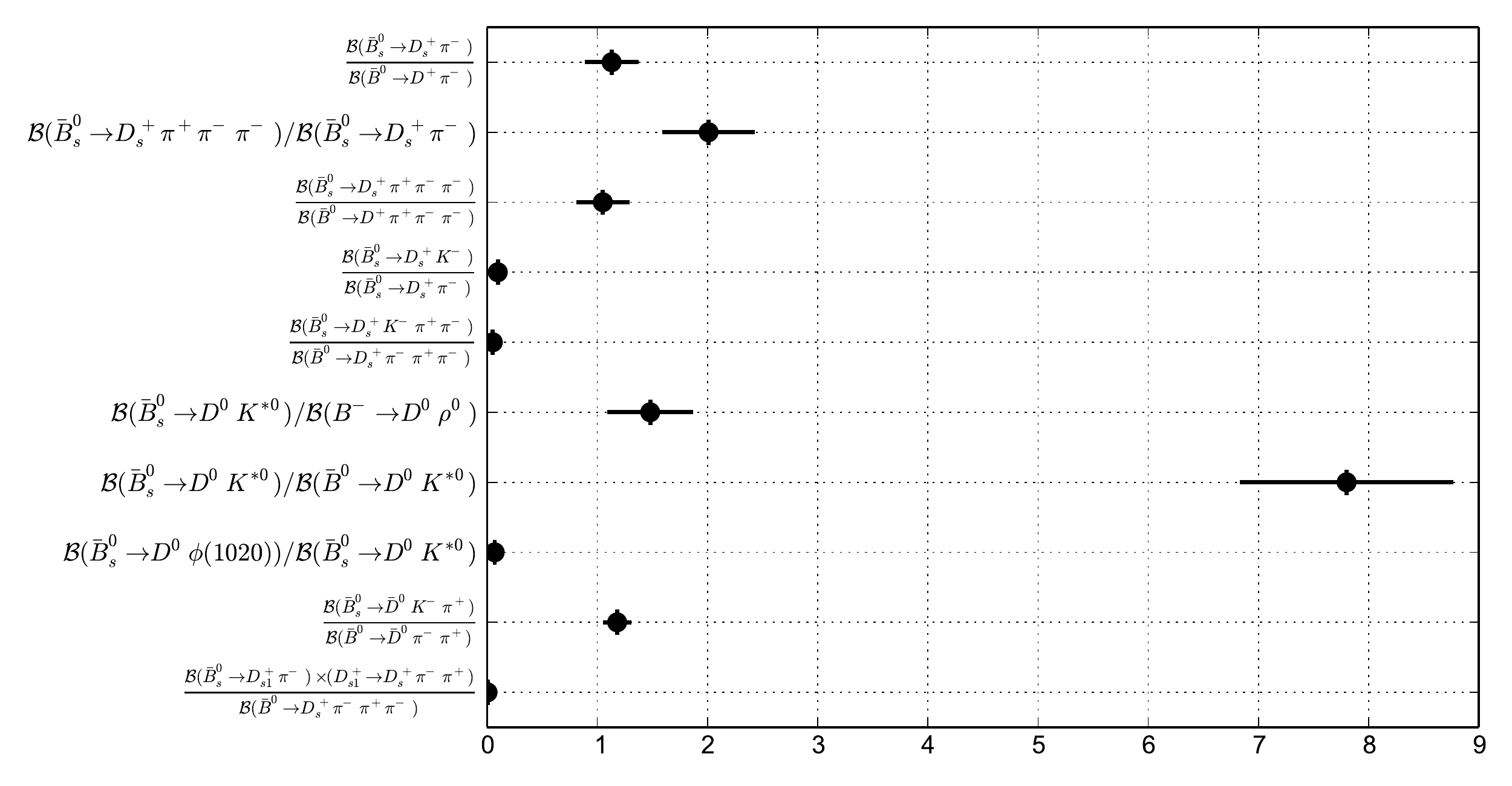}
\begin{btocharmtab}{Bs_D_4}{Semileptonic decays $[10^{-2}]$}
\hline
\textbf{Parameter} & \begin{tabular}{l}\textbf{Measurements}\end{tabular} & \textbf{Average} \\
\hline
\hline
\multicolumn{3}{|l|}{${\cal{B}} ( \bar{B}^0_s \to D_{s1}^+ \mu^- \bar{\nu}_\mu X ) / {\cal{B}} ( \bar{B}^0_s \to \mu^- \bar{\nu}_\mu X )$}\\
 & \begin{tabular}{l} LHCb \cite{Aaij:2011ju}: $5.4 \pm 1.2 \pm 0.5$ \\ \end{tabular} & $5.4 \pm 1.3$ \\
\hline
\multicolumn{3}{|l|}{${\cal{B}} ( \bar{B}^0_s \to D_{s2}^{*+} \mu^- \bar{\nu}_\mu X ) / {\cal{B}} ( \bar{B}^0_s \to \mu^- \bar{\nu}_\mu X )$}\\
 & \begin{tabular}{l} LHCb \cite{Aaij:2011ju}: $3.3 \pm 1.1 \pm 0.4$ \\ \end{tabular} & $3.3 \pm 1.2$ \\
\hline
\end{btocharmtab}
\btocharmfig{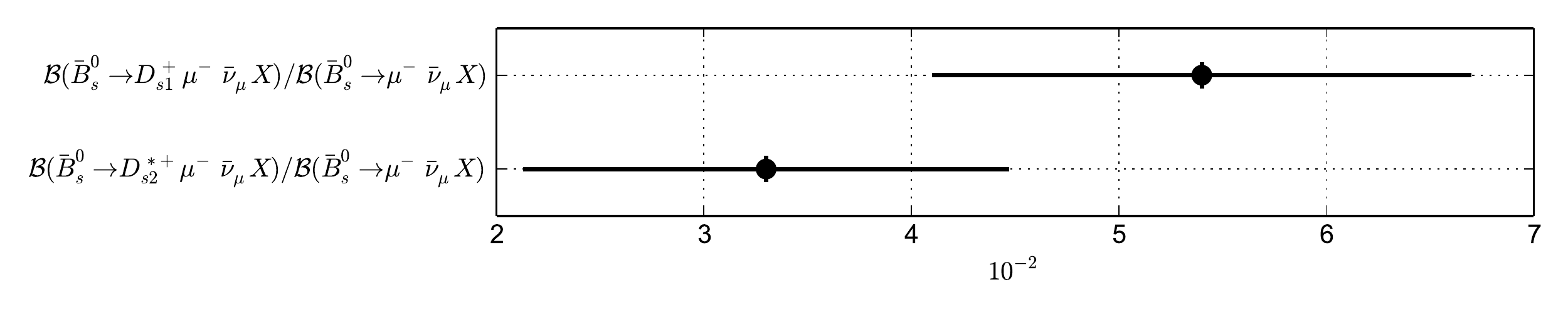}

\subsubsection{Decays to two open charm mesons}
\label{sec:b2c:Bs_DD}
Averages of $\Bsb$ decays to two open charm mesons are shown in Tables~\ref{tab:b2c:Bs_DD_1}--\ref{tab:b2c:Bs_DD_2} and Figs.~\ref{fig:b2c:Bs_DD_1}--\ref{fig:b2c:Bs_DD_2}.
\begin{btocharmtab}{Bs_DD_1}{Absolute branching fractions $[10^{-2}]$}
\hline
\textbf{Parameter} & \begin{tabular}{l}\textbf{Measurements}\end{tabular} & \textbf{Average} \\
\hline
\hline
${\cal{B}} ( \bar{B}_s^0 \to D_s^+ D_s^- )$ & \begin{tabular}{l} CDF \cite{Aaltonen:2012mg}: $0.49 \pm 0.06 \pm 0.09$ \\ Belle \cite{Esen:2012yz}: $0.58 \,^{+0.11}_{-0.09} \pm 0.13$ \\ \end{tabular} & $0.52 \pm 0.09$ \\
\hline
${\cal{B}} ( \bar{B}_s^0 \to D_s^+ D_s^{*-} )$ & \begin{tabular}{l} CDF \cite{Aaltonen:2012mg}: $1.13 \pm 0.12 \pm 0.21$ \\ Belle \cite{Esen:2012yz}: $1.76 \,^{+0.23}_{-0.22} \pm 0.40$ \\ \end{tabular} & $1.27 \pm 0.21$ \\
\hline
${\cal{B}} ( \bar{B}_s^0 \to D_s^{*+} D_s^{*-} )$ & \begin{tabular}{l} CDF \cite{Aaltonen:2012mg}: $1.75 \pm 0.19 \pm 0.34$ \\ Belle \cite{Esen:2012yz}: $1.98 \,^{+0.33}_{-0.31} \,^{+0.51}_{-0.50}$ \\ \end{tabular} & $1.82 \pm 0.32$ \\
\hline
${\cal{B}} ( \bar{B}_s^0 \to D_s^{(*)+} D_s^{(*)-} )$ & \begin{tabular}{l} \dzero \cite{Abazov:2008ig}: $3.5 \pm 1.0 \pm 1.1$ \\ CDF \cite{Aaltonen:2012mg}: $3.38 \pm 0.25 \pm 0.64$ \\ \end{tabular} & $3.40 \pm 0.62$ \\
\hline
\end{btocharmtab}
\btocharmfig{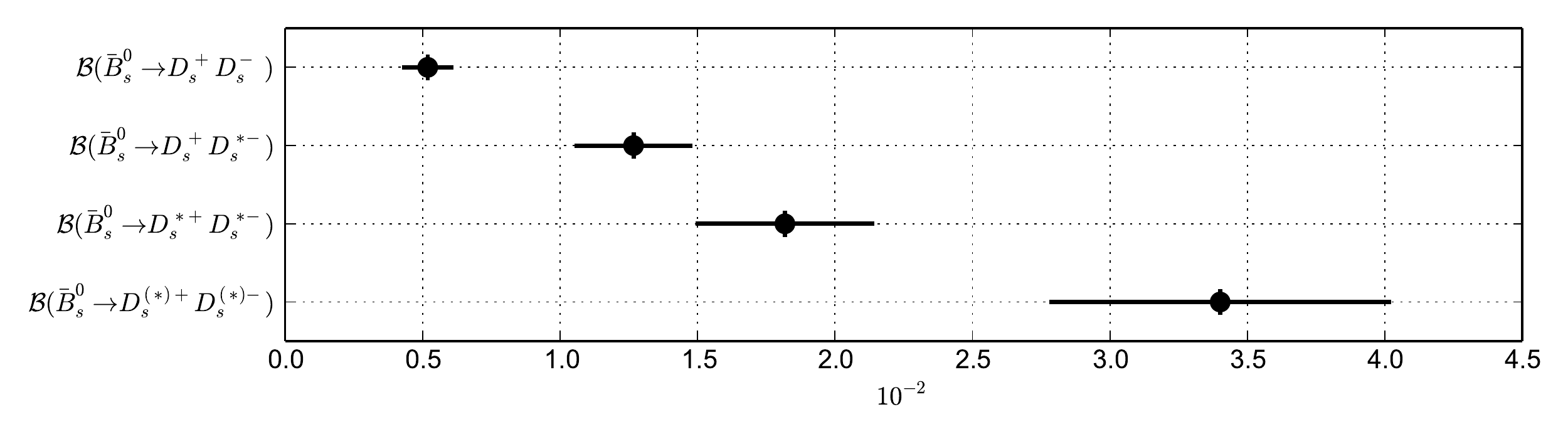}
\begin{btocharmtab}{Bs_DD_2}{Relative branching fractions}
\hline
\textbf{Parameter} & \begin{tabular}{l}\textbf{Measurements}\end{tabular} & \textbf{Average} \\
\hline
\hline
\multicolumn{3}{|l|}{${{\cal{B}} ( \bar{B}^0_s \to D^{-} D^{+} )}/{{\cal{B}} ( \bar{B}^0 \to D^- D^+ )}$}\\
 & \begin{tabular}{l} LHCb \cite{Aaij:2013fha}: $1.08 \pm 0.20 \pm 0.10$ \\ \end{tabular} & $1.08 \pm 0.22$ \\
\hline
\multicolumn{3}{|l|}{${{\cal{B}} ( \bar{B}_s^0 \to D_s^- D_s^+ )}/{{\cal{B}} ( \bar{B}^0 \to D_s^- D^+ )}$}\\
 & \begin{tabular}{l} LHCb \cite{Aaij:2013fha}: $0.56 \pm 0.03 \pm 0.04$ \\ \end{tabular} & $0.56 \pm 0.05$ \\
\hline
\multicolumn{3}{|l|}{${ {\cal{B}} ( \bar{B}^0_s \to D_s^{+} D^{-} )}/{ {\cal{B}} ( B^0 \to D_s^+ D^- )}$}\\
 & \begin{tabular}{l} LHCb \cite{Aaij:2013fha}: $0.050 \pm 0.008 \pm 0.004$ \\ \end{tabular} & $0.050 \pm 0.009$ \\
\hline
\multicolumn{3}{|l|}{${{\cal{B}} ( \bar{B}^0_s \to \bar{D}^{0} D^{0} )}/{{\cal{B}} ( B^- \to D^0 D_s^- )}$}\\
 & \begin{tabular}{l} LHCb \cite{Aaij:2013fha}: $0.019 \pm 0.003 \pm 0.003$ \\ \end{tabular} & $0.019 \pm 0.004$ \\
\hline
\end{btocharmtab}
\btocharmfig{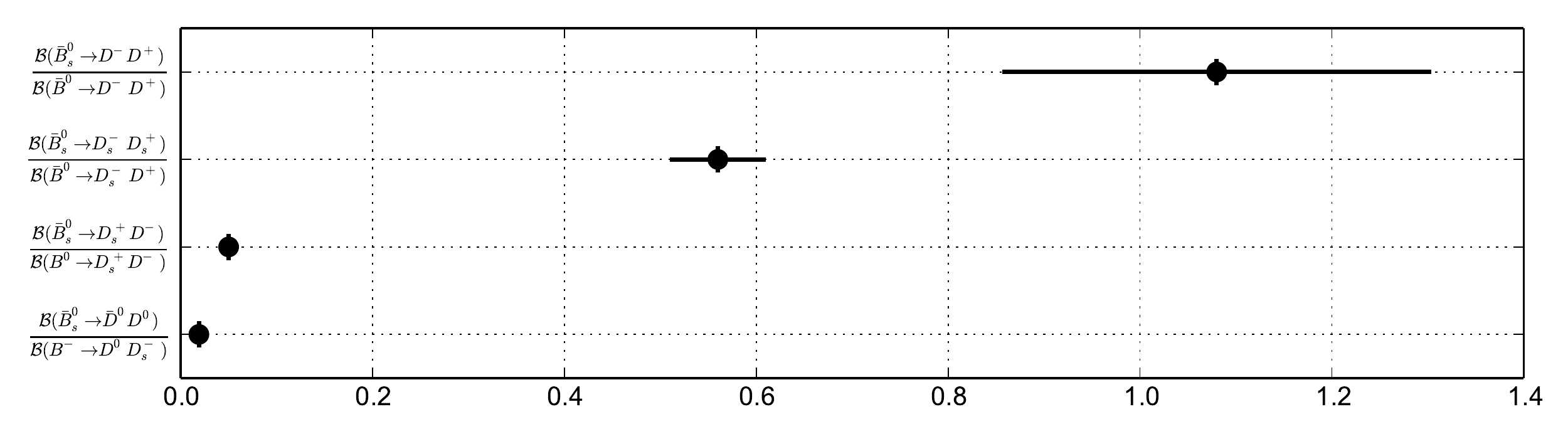}

\subsubsection{Decays to charmonium states}
\label{sec:b2c:Bs_cc}
Averages of $\Bsb$ decays to charmonium states are shown in Tables~\ref{tab:b2c:Bs_cc_1}--\ref{tab:b2c:Bs_cc_4} and Figs.~\ref{fig:b2c:Bs_cc_1}--\ref{fig:b2c:Bs_cc_4}.
\begin{btocharmtab}{Bs_cc_1}{Absolute decay rates I $[10^{-4}]$}
\hline
\textbf{Parameter} & \begin{tabular}{l}\textbf{Measurements}\end{tabular} & \textbf{Average} \\
\hline
\hline
${\cal{B}} ( \bar{B}_s^0 \to J/\psi(1S) \eta )$ & \begin{tabular}{l} Belle \cite{Belle:2012aa}: $5.10 \pm 0.50 \,^{+1.17}_{-0.83}$ \\ \end{tabular} & $5.10 \,^{+1.27}_{-0.97}$ \\
\hline
${\cal{B}} ( \bar{B}_s^0 \to J/\psi(1S) \eta^{\prime } )$ & \begin{tabular}{l} Belle \cite{Belle:2012aa}: $3.71 \pm 0.61 \,^{+0.85}_{-0.60}$ \\ \end{tabular} & $3.71 \,^{+1.05}_{-0.85}$ \\
\hline
${\cal{B}} ( \bar{B}_s^0 \to J/\psi(1S) \phi(1020) )$ & \begin{tabular}{l} CDF \cite{Abe:1996kc}: $9.3 \pm 2.8 \pm 1.7$ \\ \end{tabular} & $9.3 \pm 3.3$ \\
\hline
${\cal{B}} ( \bar{B}_s^0 \to J/\psi(1S) K^{0} K^{\pm} \pi^{\mp} )$ & \begin{tabular}{l} LHCb \cite{Aaij:2014naa}: $9.1 \pm 0.6 \pm 0.7$ \\ \end{tabular} & $9.1 \pm 0.9$ \\
\hline
\multicolumn{3}{|l|}{${{\cal{B}} ( \bar{B}_s^0 \to J/\psi(1S) f_0(980) )\times {\cal{B}} ( f_0(980) \to \pi^+ \pi^- )}$}\\
 & \begin{tabular}{l} Belle \cite{Li:2011pg}: $1.16 \,^{+0.31}_{-0.19} \,^{+0.30}_{-0.25}$ \\ \end{tabular} & $1.16 \,^{+0.43}_{-0.32}$ \\
\hline
\end{btocharmtab}
\btocharmfig{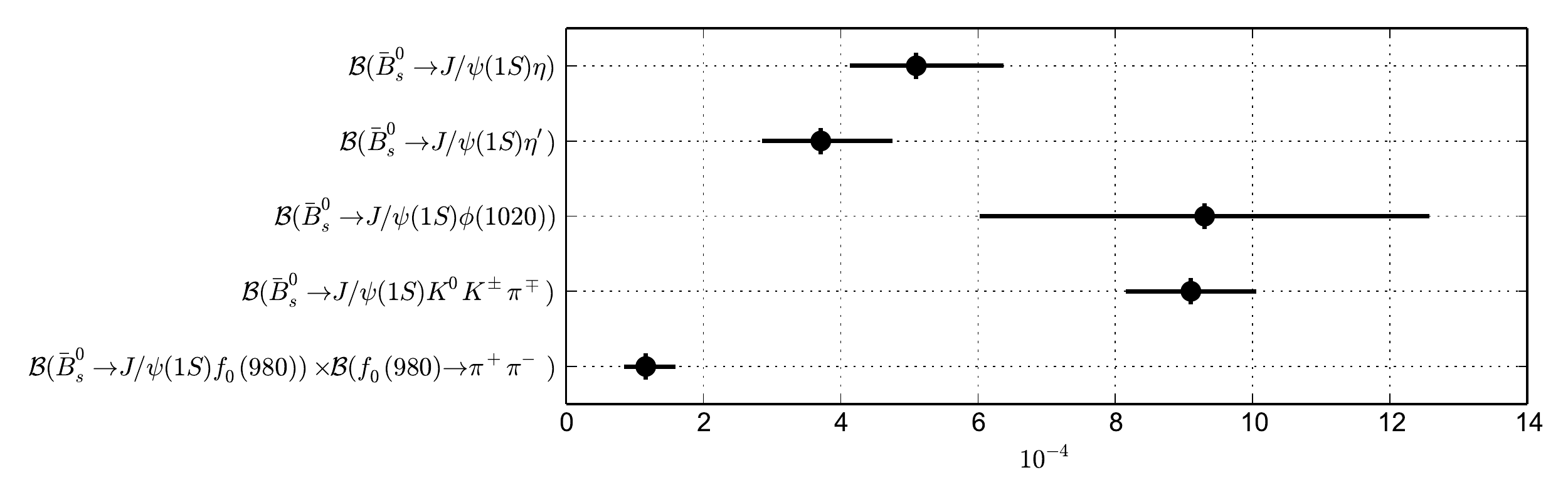}
\begin{btocharmtab}{Bs_cc_2}{Absolute decay rates II $[10^{-5}]$}
\hline
\textbf{Parameter} & \begin{tabular}{l}\textbf{Measurements}\end{tabular} & \textbf{Average} \\
\hline
\hline
${\cal{B}} ( \bar{B}^0_s \to J/\psi \bar{K}^0 )$ & \begin{tabular}{l} CDF \cite{Aaltonen:2011sy}: $3.5 \pm 0.6 \pm 0.6$ \\ \end{tabular} & $3.5 \pm 0.8$ \\
\hline
${\cal{B}} ( \bar{B}^0_s \to J/\psi K^{*0} )$ & \begin{tabular}{l} LHCb \cite{LHCb-CONF-2011-025}: $3.5 \,^{+1.1}_{-1.0} \pm 0.9$ \\ \end{tabular} & $3.5 \,^{+1.4}_{-1.3}$ \\
\hline
${\cal{B}} ( \bar{B}^0_s \to J/\psi \bar{K}^{*0} )$ & \begin{tabular}{l} LHCb \cite{Aaij:2012nh}: $4.4 \,^{+0.5}_{-0.4} \pm 0.8$ \\ CDF \cite{Aaltonen:2011sy}: $8.3 \pm 1.2 \pm 3.6$ \\ \end{tabular} & $4.6 \pm 0.9$ \\
\hline
${\cal{B}} ( \bar{B}^0_s \to J/\psi K^0_S ) $ & \begin{tabular}{l} LHCb \cite{Aaij:2012di}: $1.83 \pm 0.21 \pm 0.19$ \\ \end{tabular} & $1.83 \pm 0.28$ \\
\hline
${\cal{B}} ( \bar{B}_s^0 \to J/\psi(1S) p \bar{p} )$ & \begin{tabular}{l} LHCb \cite{Aaij:2013yba}: $< 0.48$ \\ \end{tabular} & $< 0.48$ \\
\hline
${\cal{B}} ( \bar{B}^0_s \to J/\psi K^+ \pi^- )$ & \begin{tabular}{l} LHCb \cite{LHCb-CONF-2011-025}: $3.94 \,^{+0.71}_{-0.62} \pm 0.66$ \\ \end{tabular} & $3.94 \,^{+0.97}_{-0.91}$ \\
\hline
${\cal{B}} ( \bar{B}_s^0 \to J/\psi(1S) f_1(1285) )$ & \begin{tabular}{l} LHCb \cite{Aaij:2013rja}: $7.14 \pm 0.99 \,^{+0.93}_{-1.00}$ \\ \end{tabular} & $7.14 \,^{+1.36}_{-1.41}$ \\
\hline
${\cal{B}} ( \bar{B}_s^0 \to J/\psi(1S) K^{0} \pi^{+} \pi^{-} )$ & \begin{tabular}{l} LHCb \cite{Aaij:2014naa}: $< 4.4$ \\ \end{tabular} & $< 4.4$ \\
\hline
${\cal{B}} ( \bar{B}_s^0 \to J/\psi(1S) K^{0} K^{+} K^{-} )$ & \begin{tabular}{l} LHCb \cite{Aaij:2014naa}: $< 1.2$ \\ \end{tabular} & $< 1.2$ \\
\hline
\multicolumn{3}{|l|}{${{\cal{B}} ( \bar{B}_s^0 \to J/\psi(1S) f_0(1370) )\times {\cal{B}} ( f_0(1370) \to \pi^+ \pi^- )}$}\\
 & \begin{tabular}{l} Belle \cite{Li:2011pg}: $3.4 \,^{+1.1}_{-1.4} \,^{+0.9}_{-0.5}$ \\ \end{tabular} & $3.4 \,^{+1.4}_{-1.5}$ \\
\hline
\multicolumn{3}{|l|}{${{\cal{B}} ( \bar{B}_s^0 \to J/\psi(1S) f_1(1285) )\times {\cal{B}} ( f_1(1285) \to  \pi^+ \pi^- \pi^+ \pi^- )}$}\\
 & \begin{tabular}{l} LHCb \cite{Aaij:2013rja}: $0.785 \pm 0.109 \,^{+0.089}_{-0.101}$ \\ \end{tabular} & $0.785 \,^{+0.141}_{-0.149}$ \\
\hline
\end{btocharmtab}
\btocharmfig{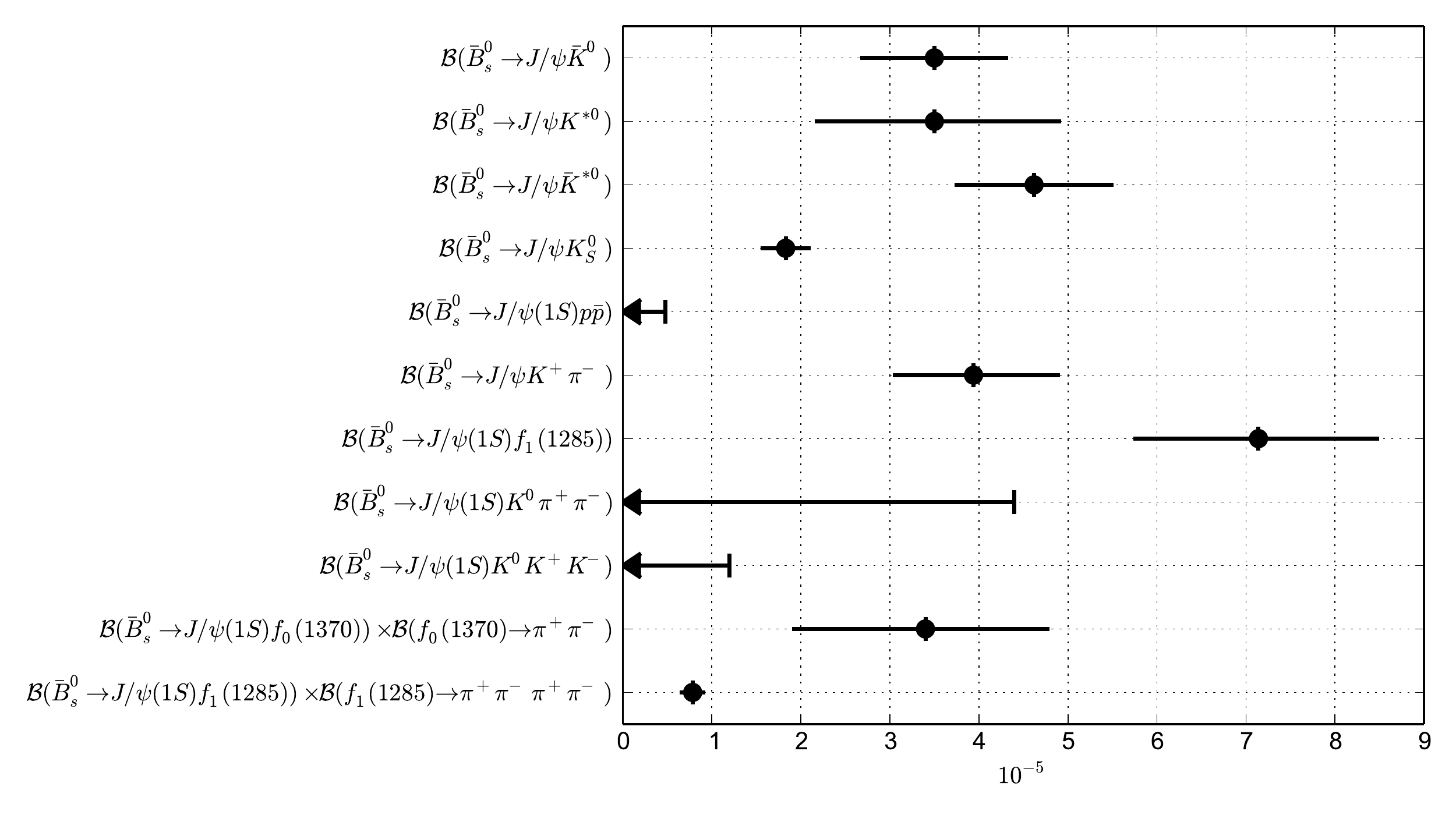}
\begin{btocharmtab}{Bs_cc_3}{Relative decay rates I}
\hline
\textbf{Parameter} & \begin{tabular}{l}\textbf{Measurements}\end{tabular} & \textbf{Average} \\
\hline
\hline
\multicolumn{3}{|l|}{${{\cal{B}} ( \bar{B}^0_s \to J/\psi \eta)}/ { {\cal{B}} ( \bar{B}^0 \to J/\psi \rho )}$}\\
 & \begin{tabular}{l} LHCb \cite{LHCb:2012cw}: $14.0 \pm 1.2 \,^{+1.6}_{-1.8}$ \\ \end{tabular} & $14.0 \,^{+2.0}_{-2.2}$ \\
\hline
\multicolumn{3}{|l|}{${ {\cal{B}} ( \bar{B}^0_s \to J/\psi \eta^{\prime })}/ {{\cal{B}} ( \bar{B}^0 \to J/\psi \rho )}$}\\
 & \begin{tabular}{l} LHCb \cite{LHCb:2012cw}: $12.7 \pm 1.1 \,^{+1.1}_{-0.9}$ \\ \end{tabular} & $12.7 \,^{+1.6}_{-1.4}$ \\
\hline
\multicolumn{3}{|l|}{${{\cal{B}} ( \bar{B}_s^0 \to J/\psi(1S) K_S^{0} K^{\pm} \pi^{\mp} )}/{{\cal{B}} ( \bar{B}^0 \to J/\psi(1S) \pi^+ \pi^- )}$}\\
 & \begin{tabular}{l} LHCb \cite{Aaij:2014naa}: $2.12 \pm 0.15 \pm 0.18$ \\ \end{tabular} & $2.12 \pm 0.23$ \\
\hline
\end{btocharmtab}
\btocharmfig{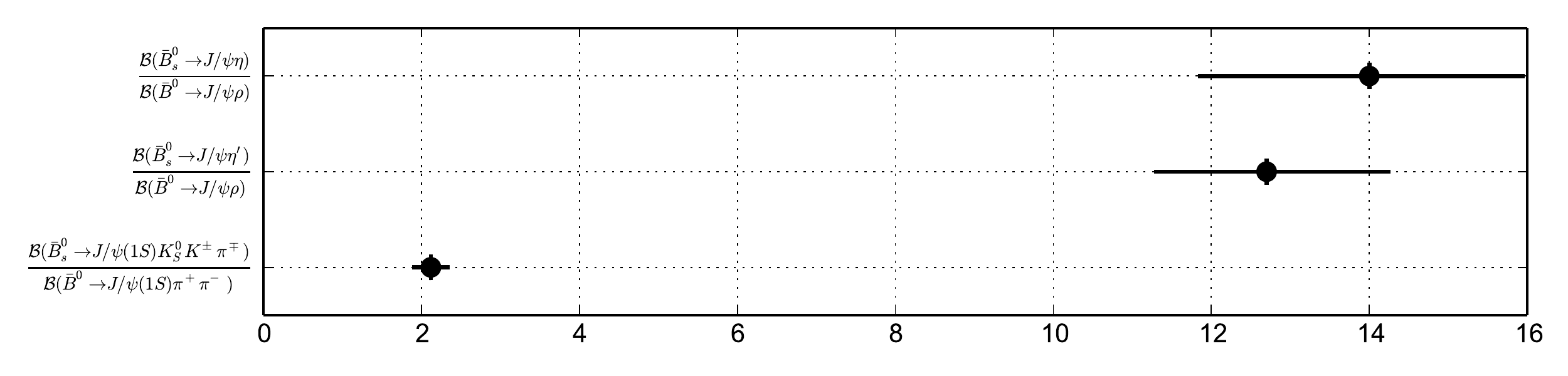}
\begin{btocharmtab}{Bs_cc_4}{Relative decay rates II}
\hline
\textbf{Parameter} & \begin{tabular}{l}\textbf{Measurements}\end{tabular} & \textbf{Average} \\
\hline
\hline
\multicolumn{3}{|l|}{${{\cal{B}} ( \bar{B}^0_s \to J/\psi K^0_S )}/ {{\cal{B}} ( \bar{B}^0 \to J/\psi K^0_S )}$}\\
 & \begin{tabular}{l} LHCb \cite{Aaij:2012di}: $0.0420 \pm 0.0049 \pm 0.0040$ \\ \end{tabular} & $0.0420 \pm 0.0063$ \\
\hline
\multicolumn{3}{|l|}{${ {\cal{B}} ( \bar{B}^0_s \to J/\psi \eta)}/ {{\cal{B}} ( \bar{B}^0 \to J/\psi \eta^{\prime} ) }$}\\
 & \begin{tabular}{l} Belle \cite{Belle:2012aa}: $0.73 \pm 0.14 \pm 0.02$ \\ \end{tabular} & $0.73 \pm 0.14$ \\
\hline
\multicolumn{3}{|l|}{$ \frac {{\cal{B}} ( \bar{B}^0_s \to J/\psi \eta^{\prime })}  {{\cal{B}} ( \bar{B}_{s}^0 \to J/\psi \eta )} $}\\
 & \begin{tabular}{l} LHCb \cite{LHCb:2012cw}: $0.90 \pm 0.09 \,^{+0.06}_{-0.02}$ \\ \end{tabular} & $0.90 \,^{+0.11}_{-0.09}$ \\
\hline
\multicolumn{3}{|l|}{${{\cal{B}} ( \bar{B}^0_s \to J/\psi f^\prime_2 )}/ { {\cal{B}} ( \bar{B}^0_s \to J/\psi \phi )}$}\\
 & \begin{tabular}{l} LHCb \cite{Aaij:2011ac}: $0.264 \pm 0.027 \pm 0.024$ \\ \dzero \cite{Abazov:2012dz}: $0.19 \pm 0.05 \pm 0.04$ \\ \end{tabular} & $0.246 \pm 0.031$ \\
\hline
\multicolumn{3}{|l|}{${{\cal{B}} ( \bar{B}^0_s \to J/\psi \phi )}/ { {\cal{B}} ( \bar{B}^0 \to J/\psi \bar{K}^{*0} )}$}\\
 & \begin{tabular}{l} CDF \cite{CDF:10795}: $0.89 \pm 0.01 \pm 0.13$ \\ \end{tabular} & $0.89 \pm 0.13$ \\
\hline
\multicolumn{3}{|l|}{${{\cal{B}} ( \bar{B}^0_s \to J/\psi f_0(980) )}/ { {\cal{B}} ( \bar{B}^0_s \to J/\psi \phi )}$}\\
 & \begin{tabular}{l} LHCb \cite{Aaij:2011fx}: $0.252 \,^{+0.046}_{-0.032} \,^{+0.027}_{-0.033}$ \\ \dzero \cite{Abazov:2011hv}: $0.275 \pm 0.041 \pm 0.061$ \\ \end{tabular} & $0.259 \pm 0.041$ \\
\hline
\multicolumn{3}{|l|}{${{\cal{B}} ( \bar{B}^0_s \to J/\psi \pi^+ \pi^- )}/ { {\cal{B}} ( \bar{B}^0_s \to J/\psi \phi )}$}\\
 & \begin{tabular}{l} LHCb \cite{Aaij:2011fx}: $0.162 \pm 0.022 \pm 0.016$ \\ \end{tabular} & $0.162 \pm 0.027$ \\
\hline
\multicolumn{3}{|l|}{${{\cal{B}} \bar{B}_s^0 \to \psi(2S) \pi^+  \pi^- }/{{\cal{B}} \bar{B}_s^0 \to J/\psi \pi^+ \pi^-  }$}\\
 & \begin{tabular}{l} LHCb \cite{Aaij:2013cpa}: $0.34 \pm 0.04 \pm 0.03$ \\ \end{tabular} & $0.34 \pm 0.05$ \\
\hline
\multicolumn{3}{|l|}{${{\cal{B}} ( \bar{B}_s^0 \to \psi(2S) \phi(1020) )}/{{\cal{B}} ( \bar{B}_s^0 \to J/\psi(1S) \phi(1020) )}$}\\
 & \begin{tabular}{l} LHCb \cite{Aaij:2012dda}: $0.489 \pm 0.026 \pm 0.024$ \\ \dzero \cite{Abazov:2008jk}: $0.55 \pm 0.11 \pm 0.09$ \\ CDF \cite{Abulencia:2006jp}: $0.52 \pm 0.13 \pm 0.07$ \\ \end{tabular} & $0.494 \pm 0.034$ \\
\hline
\multicolumn{3}{|l|}{${{\cal{B}} ( \bar{B}_s^0 \to J/\psi(1S) K_S^{0} \pi^{+} \pi^{-} )}/{{\cal{B}} ( \bar{B}^0 \to J/\psi(1S) \pi^+ \pi^- )}$}\\
 & \begin{tabular}{l} LHCb \cite{Aaij:2014naa}: $< 0.10$ \\ \end{tabular} & $< 0.10$ \\
\hline
\multicolumn{3}{|l|}{${{\cal{B}} ( \bar{B}_s^0 \to J/\psi(1S) K_S^{0} K^{+} K^{-} )}/{{\cal{B}} ( \bar{B}^0 \to J/\psi(1S) \pi^+ \pi^- )}$}\\
 & \begin{tabular}{l} LHCb \cite{Aaij:2014naa}: $< 0.027$ \\ \end{tabular} & $< 0.027$ \\
\hline
\multicolumn{3}{|l|}{${{\cal{B}} ( \bar{B}^0_s \to J/\psi f_0(980) ) \times {\cal{B}} ( f_0(980) \to \pi^+ \pi^- )}/ { {\cal{B}} ( \bar{B}^0_s \to J/\psi \phi )) \times {\cal{B}} ( \phi \to K^+ K^- )}$}\\
 & \begin{tabular}{l} CDF \cite{Aaltonen:2011nk}: $0.257 \pm 0.020 \pm 0.014$ \\ \end{tabular} & $0.257 \pm 0.024$ \\
\hline
\multicolumn{3}{|l|}{${{\cal{B}} ( \bar{B}^0_s \to J/\psi f_0(500) ) \times {\cal{B}} ( f_0(500) \to \pi^+ \pi^- )}/  { {\cal{B}} ( \bar{B}^0_s \to J/\psi f_0(980)  )) \times {\cal{B}} ( f_0(500)  \to \pi^+ \pi^- )}$}\\
 & \begin{tabular}{l} LHCb \cite{Aaij:2014emv}: $< 0.034$ \\ \end{tabular} & $< 0.034$ \\
\hline
\end{btocharmtab}
\btocharmfig{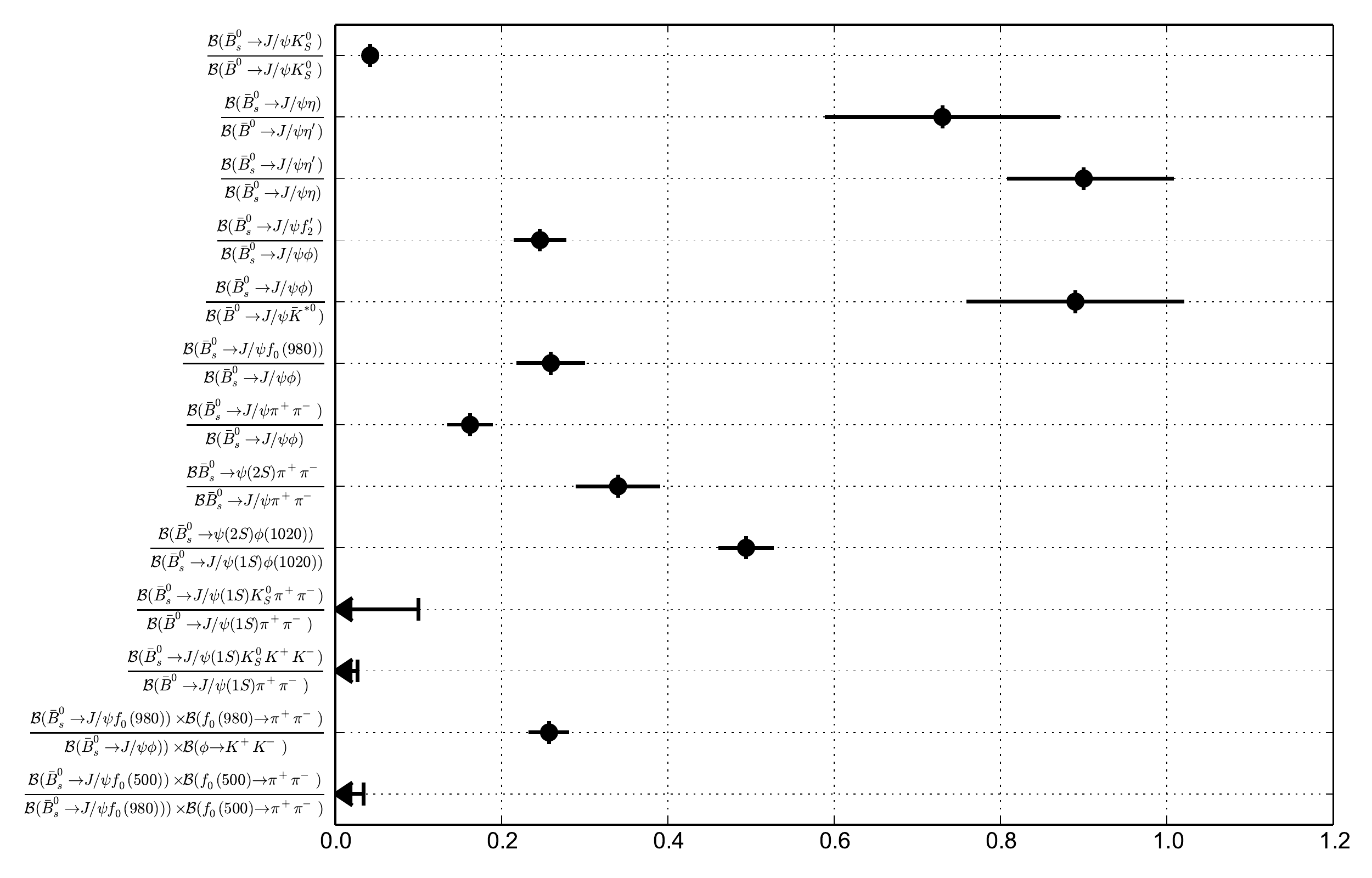}

\subsubsection{Decays to charm baryons}
\label{sec:b2c:Bs_baryon}
Averages of $\Bsb$ decays to charm baryons are shown in Tables~\ref{tab:b2c:Bs_baryon_1}--\ref{tab:b2c:Bs_baryon_2} and Figs.~\ref{fig:b2c:Bs_baryon_1}--\ref{fig:b2c:Bs_baryon_2}.
\begin{btocharmtab}{Bs_baryon_1}{Decays to one charm baryon $[10^{-4}]$}
\hline
\textbf{Parameter} & \begin{tabular}{l}\textbf{Measurements}\end{tabular} & \textbf{Average} \\
\hline
\hline
${\cal{B}} ( \bar{B}_s^0 \to \Lambda_c^+ \pi^- \bar{\Lambda} )$ & \begin{tabular}{l} Belle \cite{Solovieva:2013rhq}: $3.6 \pm 1.1 \,^{+1.2}_{-1.2}$ \\ \end{tabular} & $3.6 \,^{+1.6}_{-1.7}$ \\
\hline
\end{btocharmtab}
\btocharmfig{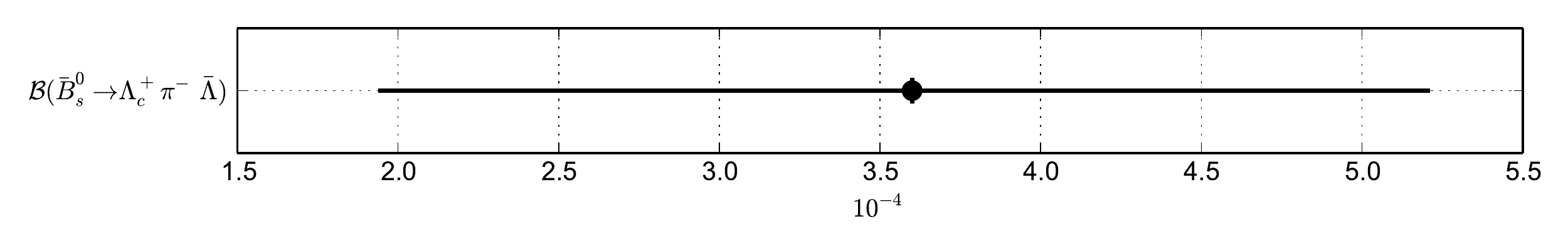}
\begin{btocharmtab}{Bs_baryon_2}{Decays to two charm baryons}
\hline
\textbf{Parameter} & \begin{tabular}{l}\textbf{Measurements}\end{tabular} & \textbf{Average} \\
\hline
\hline
\multicolumn{3}{|l|}{${{\cal{B}} ( \bar{B}^0_s \to \Lambda_c^{-} \Lambda_c^{+} )}/  { {\cal{B}} ( \bar{B}_s^0 \to D^- D_s^+ )}$}\\
 & \begin{tabular}{l} LHCb \cite{Aaij:2014pha}: $< 0.30$ \\ \end{tabular} & $< 0.30$ \\
\hline
\end{btocharmtab}
\btocharmfig{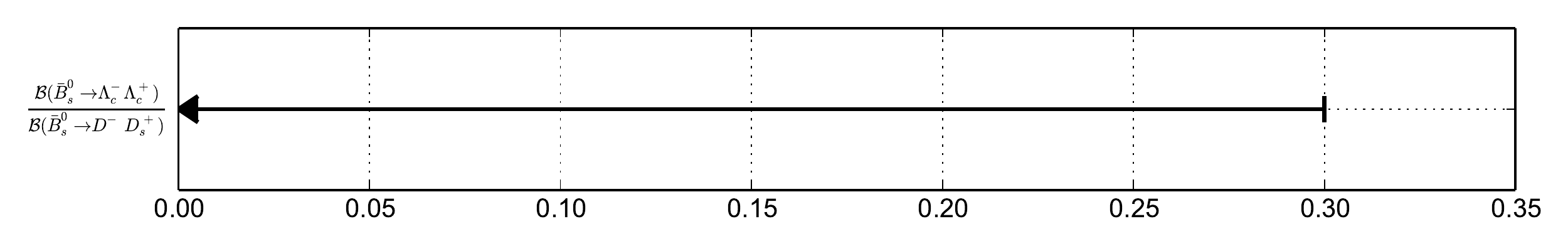}

\subsection{Decays of $B_c^-$ mesons}
\label{sec:b2c:Bc}
Measurements of $B_c^-$ decays to charmed hadrons are summarized in sections \ref{sec:b2c:Bc_cc} to \ref{sec:b2c:Bc_B}.

\subsubsection{Decays to charmonium states}
\label{sec:b2c:Bc_cc}
Averages of $B_c^-$ decays to charmonium states are shown in Tables~\ref{tab:b2c:Bc_cc_1}--\ref{tab:b2c:Bc_cc_2} and Figs.~\ref{fig:b2c:Bc_cc_1}--\ref{fig:b2c:Bc_cc_2}.
\begin{btocharmtab}{Bc_cc_1}{Relative decay rates}
\hline
\textbf{Parameter} & \begin{tabular}{l}\textbf{Measurements}\end{tabular} & \textbf{Average} \\
\hline
\hline
$\frac {{\cal{B}} (B_c^- \to J/\psi D_s^-)} {{\cal{B}} (B_c^- \to J/\psi \pi^-)}  $ & \begin{tabular}{l} LHCb \cite{Aaij:2013gia}: $2.90 \pm 0.57 \pm 0.24$ \\ \end{tabular} & $2.90 \pm 0.62$ \\
\hline
$ \frac{{\cal{B}}( B_c^- \to J/\psi K^-)} { {\cal{B}}( B_c^- \to J/\psi \pi^-)}$ & \begin{tabular}{l} LHCb \cite{Aaij:2013vcx}: $0.069 \pm 0.019 \pm 0.005$ \\ \end{tabular} & $0.069 \pm 0.020$ \\
\hline
\multicolumn{3}{|l|}{${{\cal{B}} ( B_c^- \to J/\psi K^- K^+ \pi^- )}/ { {\cal{B}} ( B_c^- \to J/\psi \pi^- )}$}\\
 & \begin{tabular}{l} LHCb \cite{Aaij:2013gxa}: $0.53 \pm 0.10 \pm 0.05$ \\ \end{tabular} & $0.53 \pm 0.11$ \\
\hline
\multicolumn{3}{|l|}{${{\cal{B}} ( B_c^- \to J/\psi \pi^+ \pi^- \pi^- )}/ {{\cal{B}} ( B_c^- \to J/\psi \pi^- )}$}\\
 & \begin{tabular}{l} LHCb \cite{LHCb-CONF-2011-040}: $3.0 \pm 0.6 \pm 0.4$ \\ LHCb \cite{LHCb:2012ag}: $2.41 \pm 0.30 \pm 0.33$ \\ CMS \cite{CMS-PAS-BPH-12-011}: $2.43 \pm 0.76 \,^{+0.46}_{-0.44}$ \\ \end{tabular} & $2.55 \pm 0.35$ \\
\hline
$\frac{{\cal{B}} ( B_c^- \to \psi(2S) \pi^- ) }  {{\cal{B}} ( B_c^- \to J/\psi \pi^- ) }$ & \begin{tabular}{l} LHCb \cite{Aaij:2013oya}: $0.25 \pm 0.07 \pm 0.02$ \\ \end{tabular} & $0.25 \pm 0.07$ \\
\hline
\end{btocharmtab}
\btocharmfig{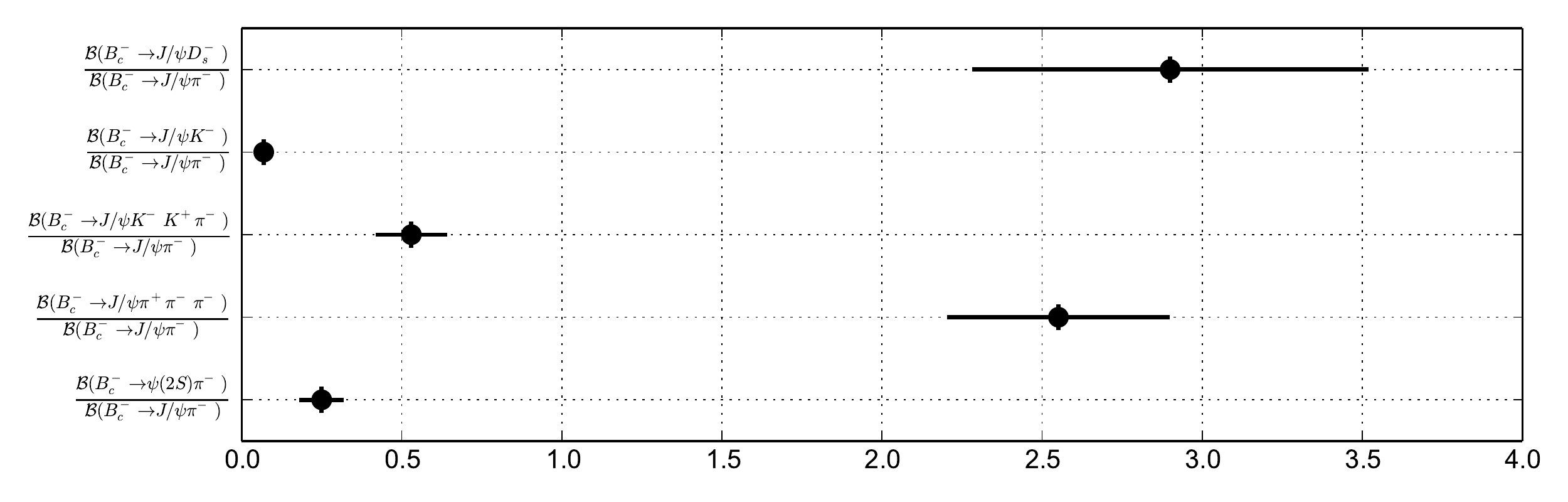}
\begin{btocharmtab}{Bc_cc_2}{Relative production times decay rates $[10^{-3}]$}
\hline
\textbf{Parameter} & \begin{tabular}{l}\textbf{Measurements}\end{tabular} & \textbf{Average} \\
\hline
\hline
\multicolumn{3}{|l|}{${{\sigma(B_c^-) \times \cal{B}} ( B_c^- \to J/\psi \pi^- )}/  {\sigma(B^-) \times {\cal{B}} ( B^- \to J/\psi K^- )}$}\\
 & \begin{tabular}{l} LHCb \cite{LHCb-CONF-2011-017}: $22 \pm 8 \pm 2$ \\ CMS \cite{CMS-PAS-BPH-12-011}: $4.8 \pm 0.5 \,^{+0.6}_{-0.5}$ \\ \end{tabular} & $4.9 \pm 0.8$ \\
\hline
\end{btocharmtab}
\btocharmfig{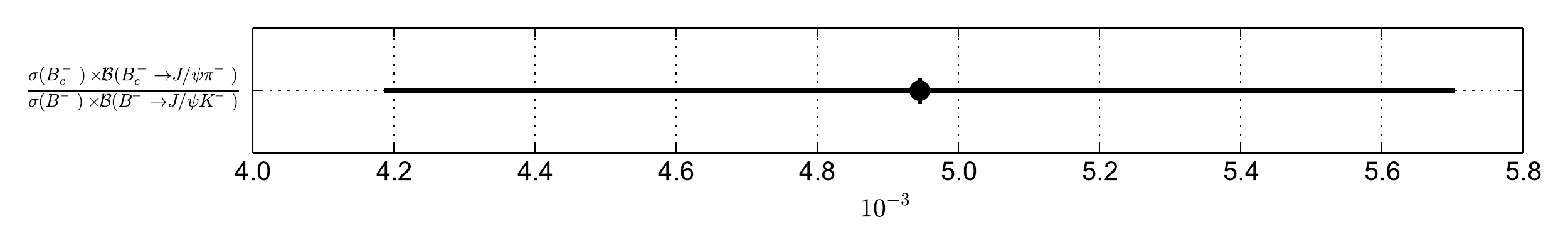}

\subsubsection{Decays to a $B$ meson}
\label{sec:b2c:Bc_B}
Averages of $B_c^-$ decays to a $B$ meson are shown in Table~\ref{tab:b2c:Bc_B_1} and Fig.~\ref{fig:b2c:Bc_B_1}.
\begin{btocharmtab}{Bc_B_1}{Decays to $B_s^0$ meson $[10^{-3}]$}
\hline
\textbf{Parameter} & \begin{tabular}{l}\textbf{Measurements}\end{tabular} & \textbf{Average} \\
\hline
\hline
\multicolumn{3}{|l|}{$ \frac{ \sigma(B_c^+)}{ \sigma(B_s^0)}  \times {\cal B} ( B_c^+ \to  B_s^0\pi^+ ) $}\\
 & \begin{tabular}{l} LHCb \cite{Aaij:2013cda}: $2.37 \pm 0.31 \,^{+0.20}_{-0.17}$ \\ \end{tabular} & $2.37 \,^{+0.37}_{-0.35}$ \\
\hline
\end{btocharmtab}
\btocharmfig{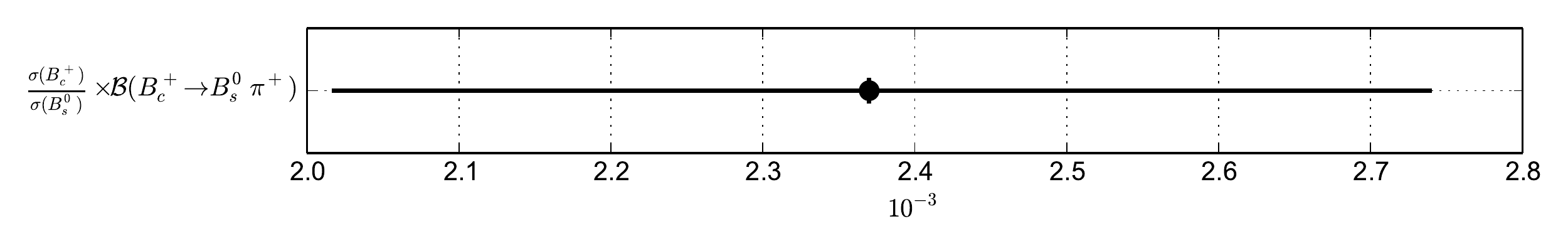}

\subsection{Decays of $b$ baryons}
\label{sec:b2c:Bbaryon}
Measurements of b baryons decays to charmed hadrons are summarized in sections \ref{sec:b2c:Bbaryon_D} to \ref{sec:b2c:Bbaryon_baryon}.

\subsubsection{Decays to a single open charm meson}
\label{sec:b2c:Bbaryon_D}
Averages of b baryons decays to a single open charm meson are shown in Table~\ref{tab:b2c:Bbaryon_D_1} and Fig.~\ref{fig:b2c:Bbaryon_D_1}.
\begin{btocharmtab}{Bbaryon_D_1}{Relative decay rates to $D^0$ mesons}
\hline
\textbf{Parameter} & \begin{tabular}{l}\textbf{Measurements}\end{tabular} & \textbf{Average} \\
\hline
\hline
\multicolumn{3}{|l|}{${\cal{B}} ( \Lambda_b^0 \to D^0 p K^- ) / {\cal{B}} ( \Lambda_b^ \to D^0 p \pi^- )$}\\
 & \begin{tabular}{l} LHCb \cite{Aaij:2013pka}: $0.073 \pm 0.008 \,^{+0.005}_{-0.006}$ \\ \end{tabular} & $0.073 \,^{+0.009}_{-0.010}$ \\
\hline
\multicolumn{3}{|l|}{$ \frac{ {\cal{B}} ( \Lambda_b^0 \to D^0 p \pi^-)  \times {\cal{B}} ( D^0 \to K^+ \pi^- )} {  {\cal{B}} ( \Lambda_b^0 \to \Lambda_c^+ \pi^- )   \times   {\cal{B}}  (\Lambda_c^+ \to  p K^- \pi^+) }$}\\
 & \begin{tabular}{l} LHCb \cite{Aaij:2013pka}: $0.0806 \pm 0.0023 \pm 0.0035$ \\ \end{tabular} & $0.0806 \pm 0.0042$ \\
\hline
\multicolumn{3}{|l|}{$  \frac{ f_{\Xi_b^0} \times{\cal{B}} ( \Xi_b^0 \to D^0 p K^- )   } { f_{\Lambda_b^0} \times{\cal{B}} ( \Lambda_b^0 \to D^0 p K^-)  } $}\\
 & \begin{tabular}{l} LHCb \cite{Aaij:2013pka}: $0.44 \pm 0.09 \pm 0.06$ \\ \end{tabular} & $0.44 \pm 0.11$ \\
\hline
\end{btocharmtab}
\btocharmfig{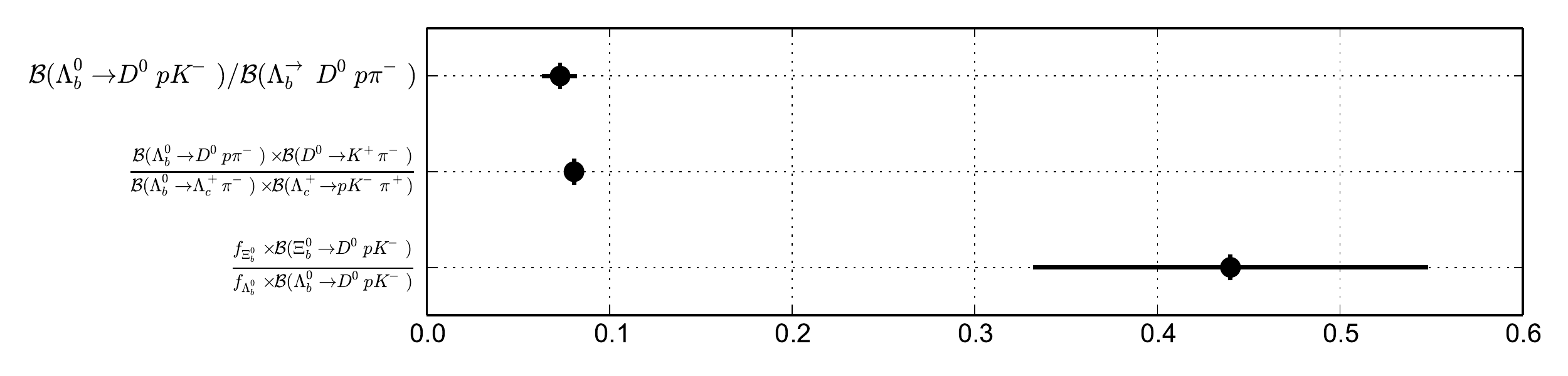}

\subsubsection{Decays to charmonium states}
\label{sec:b2c:Bbaryon_cc}
Averages of b baryons decays to charmonium states are shown in Tables~\ref{tab:b2c:Bbaryon_cc_1}--\ref{tab:b2c:Bbaryon_cc_3} and Figs.~\ref{fig:b2c:Bbaryon_cc_1}--\ref{fig:b2c:Bbaryon_cc_3}.
\begin{btocharmtab}{Bbaryon_cc_1}{$\Lambda_b^0$ decays to charmonium $[10^{-4}]$}
\hline
\textbf{Parameter} & \begin{tabular}{l}\textbf{Measurements}\end{tabular} & \textbf{Average} \\
\hline
\hline
${\cal{B}} ( \Lambda_b^0 \to J/\psi(1S) \Lambda )$ & \begin{tabular}{l} CDF \cite{Abe:1996tr}: $4.7 \pm 2.1 \pm 1.9$ \\ \end{tabular} & $4.7 \pm 2.8$ \\
\hline
\end{btocharmtab}
\btocharmfig{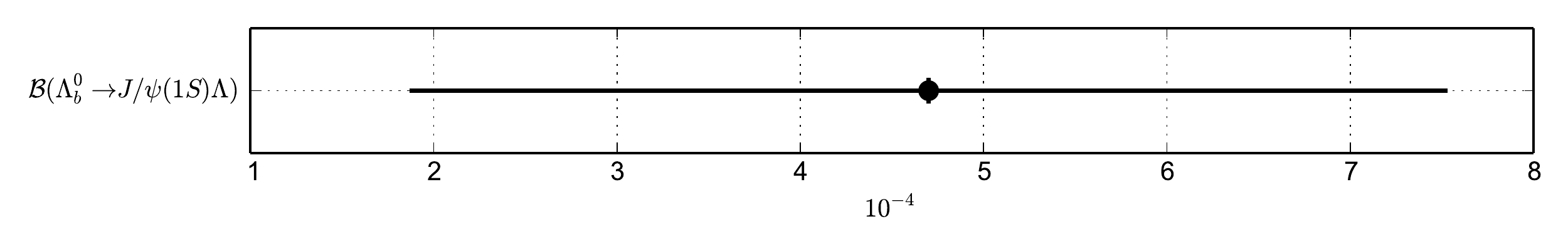}
\begin{btocharmtab}{Bbaryon_cc_2}{ $f_b$ times $\Lambda_b^0$ decay to charmonium $[10^{-5}]$}
\hline
\textbf{Parameter} & \begin{tabular}{l}\textbf{Measurements}\end{tabular} & \textbf{Average} \\
\hline
\hline
${f_b \times \cal{B}} ( \Lambda_b^0 \to J/\psi \Lambda )$ & \begin{tabular}{l} \dzero \cite{Abazov:2011wt}: $6.01 \pm 0.60 \pm 0.64$ \\ \end{tabular} & $6.01 \pm 0.88$ \\
\hline
\end{btocharmtab}
\btocharmfig{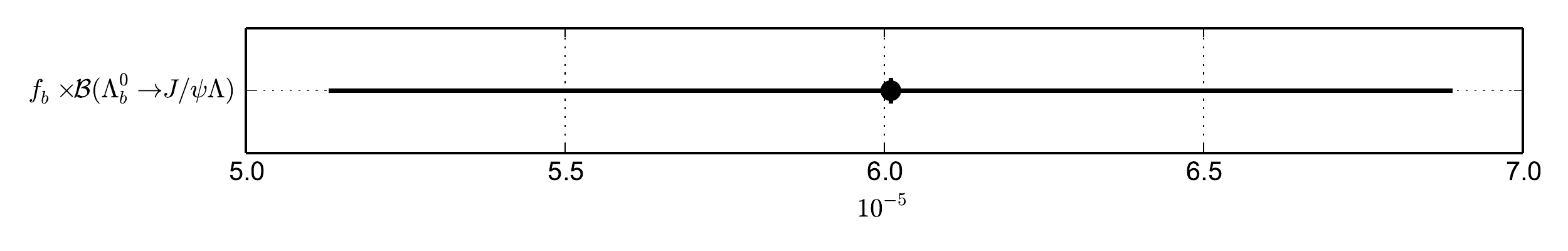}
\begin{btocharmtab}{Bbaryon_cc_3}{ $\Xi_b^-$ and $\Omega_b^-$ decays to charmonium}
\hline
\textbf{Parameter} & \begin{tabular}{l}\textbf{Measurements}\end{tabular} & \textbf{Average} \\
\hline
\hline
\multicolumn{3}{|l|}{$\sigma(\Xi_b^-) \times {\cal{B}} ( \Xi_b^- \to J/\psi \Xi^- ) / \sigma(\Lambda_b^0) \times {\cal{B}} ( \Lambda_b^0 \to J/\psi \Lambda )$}\\
 & \begin{tabular}{l} CDF \cite{Aaltonen:2009ny}: $0.167 \,^{+0.037}_{-0.025} \pm 0.012$ \\ \end{tabular} & $0.167 \,^{+0.039}_{-0.028}$ \\
\hline
\multicolumn{3}{|l|}{$\sigma(\Omega_b^-) \times {\cal{B}} ( \Omega_b^- \to J/\psi \Omega^- ) / \sigma(\Lambda_b^0)  \times   {\cal{B}} ( \Lambda_b^0 \to J/\psi \Lambda )$}\\
 & \begin{tabular}{l} CDF \cite{Aaltonen:2009ny}: $0.045 \,^{+0.017}_{-0.012} \pm 0.004$ \\ \end{tabular} & $0.045 \,^{+0.017}_{-0.013}$ \\
\hline
\end{btocharmtab}
\btocharmfig{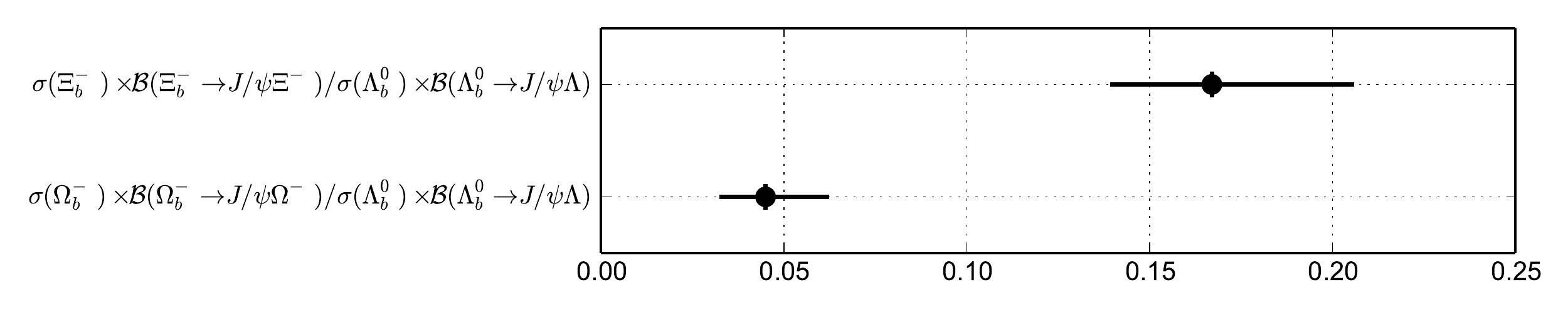}

\subsubsection{Decays to charm baryons}
\label{sec:b2c:Bbaryon_baryon}
Averages of b baryons decays to charm baryons are shown in Tables~\ref{tab:b2c:Bbaryon_baryon_1}--\ref{tab:b2c:Bbaryon_baryon_3} and Figs.~\ref{fig:b2c:Bbaryon_baryon_1}--\ref{fig:b2c:Bbaryon_baryon_3}.
\begin{btocharmtab}{Bbaryon_baryon_1}{Absolute decay rates $[10^{-2}]$}
\hline
\textbf{Parameter} & \begin{tabular}{l}\textbf{Measurements}\end{tabular} & \textbf{Average} \\
\hline
\hline
${\cal{B}} ( \Lambda_b^0 \to \Lambda_c^+ \pi^- )$ & \begin{tabular}{l} LHCb \cite{Aaij:2014jyk}: $0.430 \pm 0.003 \,^{+0.036}_{-0.035}$ \\ \end{tabular} & $0.430 \,^{+0.036}_{-0.035}$ \\
\hline
${\cal{B}} ( \Lambda_b^0 \to \Lambda_c^+ \pi^+ \pi^- \pi^- )$ & \begin{tabular}{l} CDF \cite{CDF:2011aa}: $2.68 \pm 0.29 \,^{+1.15}_{-1.09}$ \\ \end{tabular} & $2.68 \,^{+1.19}_{-1.12}$ \\
\hline
\end{btocharmtab}
\btocharmfig{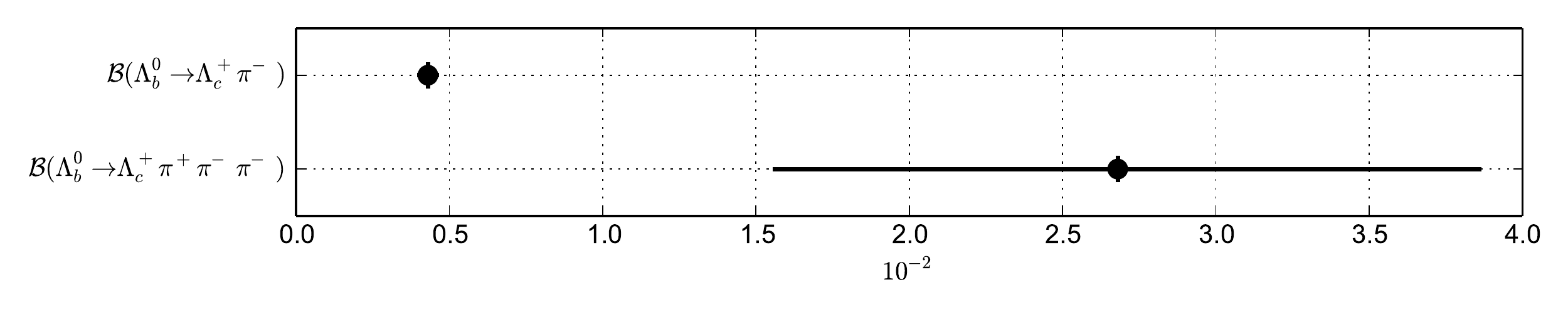}
\begin{btocharmtab}{Bbaryon_baryon_2}{Relative decay rates to $\Lambda_c$}
\hline
\textbf{Parameter} & \begin{tabular}{l}\textbf{Measurements}\end{tabular} & \textbf{Average} \\
\hline
\hline
\multicolumn{3}{|l|}{${{\cal{B}} ( \Lambda_b^0 \to \Lambda_c^+ \mu^- \bar{\nu}_\mu )}/{{\cal{B}} ( \Lambda_b^0 \to \Lambda_c^+ \pi^- )}$}\\
 & \begin{tabular}{l} CDF \cite{Aaltonen:2008eu}: $16.6 \pm 3.0 \,^{+2.8}_{-3.6}$ \\ \end{tabular} & $16.6 \,^{+4.1}_{-4.7}$ \\
\hline
\multicolumn{3}{|l|}{${{\cal{B}} ( \Lambda_b^0 \to \Lambda_c^+ \pi^- )}/{{\cal{B}} ( \bar{B}^0 \to D^+ \pi^- )}$}\\
 & \begin{tabular}{l} CDF \cite{Abulencia:2006df}: $3.3 \pm 0.3 \pm 1.2$ \\ \end{tabular} & $3.3 \pm 1.2$ \\
\hline
\multicolumn{3}{|l|}{${\cal{B}} ( \Lambda_b^0 \to \Lambda_c^+ \pi^+ \pi^- \pi^- ) / {\cal{B}} ( \Lambda_b^0 \to \Lambda_c^+ \pi^- )$}\\
 & \begin{tabular}{l} LHCb \cite{Aaij:2011rj}: $1.43 \pm 0.16 \pm 0.13$ \\ CDF \cite{CDF:2011aa}: $3.04 \pm 0.33 \,^{+0.70}_{-0.55}$ \\ \end{tabular} & $1.55 \pm 0.20$ \\
\hline
\multicolumn{3}{|l|}{$ \frac{ {\cal{B}} ( \Lambda_b^0 \to \Lambda_c^+ K^- )  }{ {\cal{B}} ( \Lambda_b^0 \to \Lambda_c^+ \pi^- )  }$}\\
 & \begin{tabular}{l} LHCb \cite{Aaij:2013pka}: $0.0731 \pm 0.0016 \pm 0.0016$ \\ \end{tabular} & $0.0731 \pm 0.0023$ \\
\hline
\multicolumn{3}{|l|}{$ \frac{ {\cal{B}} ( \Lambda_b^0 \to \Lambda_c^+ D^- )  }{ {\cal{B}} ( \Lambda_b^0 \to \Lambda_c^+ D_s^- )  }$}\\
 & \begin{tabular}{l} LHCb \cite{Aaij:2014pha}: $0.042 \pm 0.003 \pm 0.003$ \\ \end{tabular} & $0.042 \pm 0.004$ \\
\hline
\multicolumn{3}{|l|}{$ \frac{ {\cal{B}} ( \Xi_b^0 \to \Lambda_c^+ K^- )   \times   {\cal{B}}  (\Lambda_c^+ \to  p K^- \pi^+) } { {\cal{B}} ( \Xi_b^0 \to D^0 p K^-)  \times {\cal{B}} ( D^0 \to K^+ \pi^- ) }$}\\
 & \begin{tabular}{l} LHCb \cite{Aaij:2013pka}: $0.57 \pm 0.22 \pm 0.21$ \\ \end{tabular} & $0.57 \pm 0.30$ \\
\hline
\end{btocharmtab}
\btocharmfig{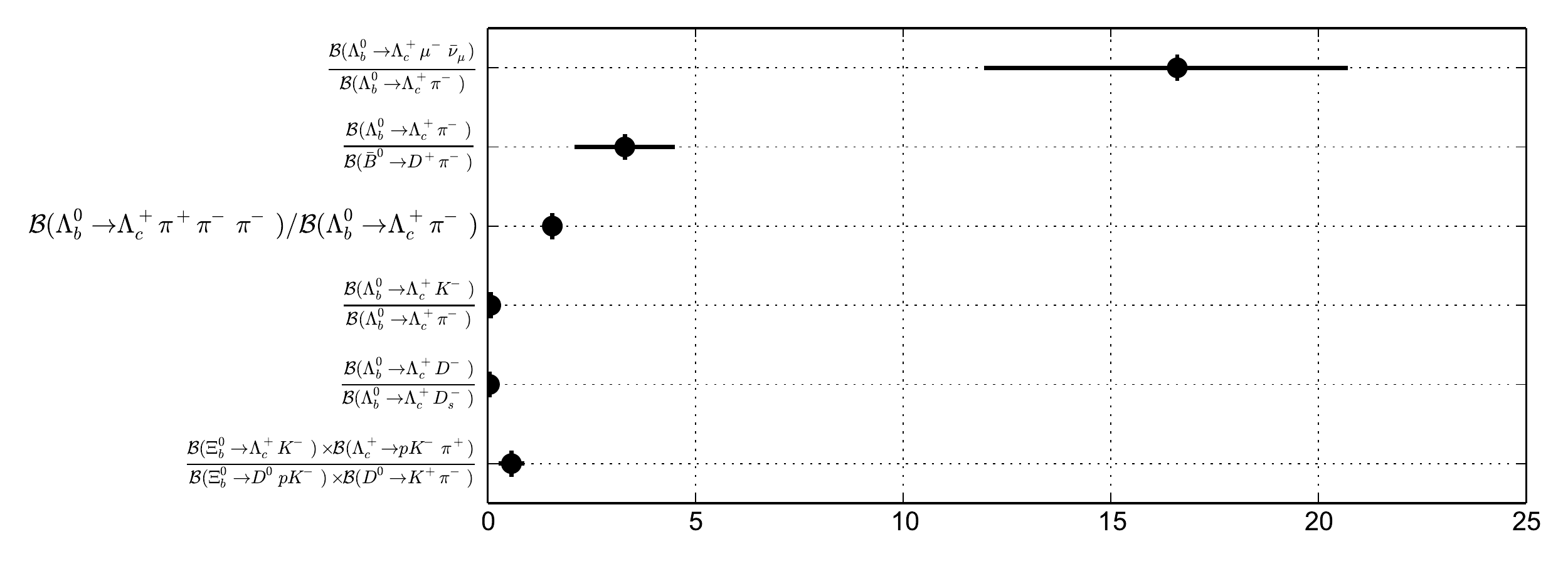}
\begin{btocharmtab}{Bbaryon_baryon_3}{Relative decay rates to excited or $\Sigma_c$ states}
\hline
\textbf{Parameter} & \begin{tabular}{l}\textbf{Measurements}\end{tabular} & \textbf{Average} \\
\hline
\hline
\multicolumn{3}{|l|}{${\cal{B}} ( \Lambda_b^0 \to \Lambda_c(2595)^+ \mu^- \bar{\nu}_\mu) / {\cal{B}} ( \Lambda_b^0 \to \Lambda_c^+ \mu^- \bar{\nu}_\mu )$}\\
 & \begin{tabular}{l} CDF \cite{Aaltonen:2008eu}: $0.126 \pm 0.033 \,^{+0.047}_{-0.038}$ \\ \end{tabular} & $0.126 \,^{+0.057}_{-0.050}$ \\
\hline
\multicolumn{3}{|l|}{${\cal{B}} ( \Lambda_b^0 \to \Lambda_c(2695)^+ \mu^- \bar{\nu}_\mu) / {\cal{B}} ( \Lambda_b^0 \to \Lambda_c^+ \mu^- \bar{\nu}_\mu )$}\\
 & \begin{tabular}{l} CDF \cite{Aaltonen:2008eu}: $0.210 \pm 0.042 \,^{+0.071}_{-0.050}$ \\ \end{tabular} & $0.210 \,^{+0.082}_{-0.065}$ \\
\hline
\multicolumn{3}{|l|}{${\cal{B}} ( \Lambda_b^0 \to \Sigma_c \pi \mu^- \bar{\nu}_\mu) / {\cal{B}} ( \Lambda_b^0 \to \Lambda_c^+ \mu^- \bar{\nu}_\mu )$}\\
 & \begin{tabular}{l} CDF \cite{Aaltonen:2008eu}: $0.108 \pm 0.044 \,^{+0.042}_{-0.036}$ \\ \end{tabular} & $0.108 \,^{+0.061}_{-0.057}$ \\
\hline
\multicolumn{3}{|l|}{${\cal{B}} ( \Lambda_b^0 \to \Lambda_c(2595)^+ \pi^- ) \times {\cal{B}} ( \Lambda_c(2595)^+ \to \Lambda_c^+ \pi^+ \pi^- ) / {\cal{B}} ( \Lambda_b^0 \to \Lambda_c^+ \pi^+ \pi^- \pi^- )$}\\
 & \begin{tabular}{l} LHCb \cite{Aaij:2011rj}: $0.044 \pm 0.017 \,^{+0.006}_{-0.004}$ \\ \end{tabular} & $0.044 \,^{+0.018}_{-0.017}$ \\
\hline
\multicolumn{3}{|l|}{${\cal{B}} ( \Lambda_b^0 \to \Lambda_c(2625)^+ \pi^- ) \times {\cal{B}} ( \Lambda_c(2625)^+ \to \Lambda_c^+ \pi^+ \pi^- ) / {\cal{B}} ( \Lambda_b^0 \to \Lambda_c^+ \pi^+ \pi^- \pi^- )$}\\
 & \begin{tabular}{l} LHCb \cite{Aaij:2011rj}: $0.043 \pm 0.015 \pm 0.004$ \\ \end{tabular} & $0.043 \pm 0.016$ \\
\hline
\multicolumn{3}{|l|}{${\cal{B}} ( \Lambda_b^0 \to \Sigma_c^0 \pi^+ \pi^- ) \times {\cal{B}} ( \Sigma_c^0 \to \Lambda_c^+ \pi^- ) / {\cal{B}} ( \Lambda_b^0 \to \Lambda_c^+ \pi^+ \pi^- \pi^- )$}\\
 & \begin{tabular}{l} LHCb \cite{Aaij:2011rj}: $0.074 \pm 0.024 \pm 0.012$ \\ \end{tabular} & $0.074 \pm 0.027$ \\
\hline
\multicolumn{3}{|l|}{${\cal{B}} ( \Lambda_b^0 \to \Sigma_c^{++} \pi^- \pi^- ) \times {\cal{B}} ( \Sigma_c^{++} \to \Lambda_c^+ \pi^+ ) / {\cal{B}} ( \Lambda_b^0 \to \Lambda_c^+ \pi^+ \pi^- \pi^- )$}\\
 & \begin{tabular}{l} LHCb \cite{Aaij:2011rj}: $0.042 \pm 0.018 \pm 0.007$ \\ \end{tabular} & $0.042 \pm 0.019$ \\
\hline
\end{btocharmtab}
\btocharmfig{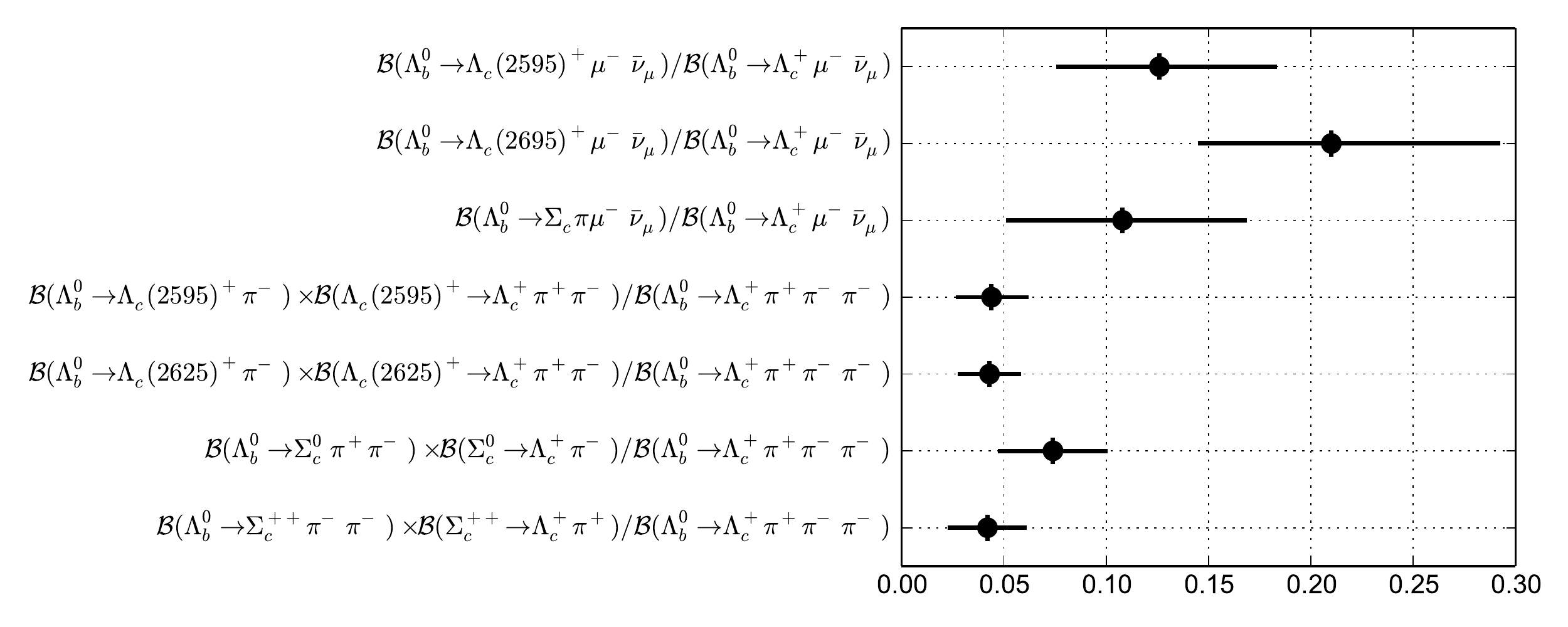}

\clearpage
\mysection{$B$ decays to charmless final states}

\label{sec:rare}

This section provides the branching fractions, polarization 
fractions, the partial rate asymmetries ($A_{\CP}$) and other observables of charmless 
$B$ decays.
The order of entries in the Tables correspond to that in the PDG, and the quoted RPP numbers correspond to the PDG numbers of the corresponding branching fractions.
The asymmetry is defined as 
$A_{\CP} = \frac{N_{\Bbar} -N_B}{N_{\Bbar} +N_B}$, where $N_{\Bbar}$ 
and $N_B$ are respectively number of $\Bzb/\Bm$ and $\Bz/\Bp$ decaying
into a specific final state.
Four different $B$ decay categories are considered: 
charmless mesonic, baryonic, radiative and leptonic. We also include
measurements of $B_s$ decays.
Measurements supported with  written documents are accepted in  
the averages; written documents include journal papers, 
conference contributed papers, preprints or conference proceedings.  
Results from  $A_{\CP}$ measurements  obtained from time-dependent analyses 
are listed and described in Sec.~\ref{sec:cp_uta}.

Most of the branching fractions from \babar\ and Belle assume equal production 
of charged and neutral $B$ pairs.  The best measurements to date show that this
is still a reasonable approximation (see Sec.~\ref{sec:life_mix}).
For branching fractions, we provide either averages or the most stringent
90\% confidence level (CL) upper limits.  If one or more experiments have
measurements with $>\!\!4 \sigma$ for a decay channel, all available central values
for that channel are used in the averaging.  We also give central values
and errors for cases where the significance of the average value is at
least $3 \sigma$, even if no single measurement is above $4 \sigma$. 
Since a few decay modes are sensitive to the contribution of
new physics and the current experimental upper limits are not far from the 
Standard Model expectation, we provide the combined upper limits or
averages in these cases.
Their upper limits can be estimated assuming that the errors are 
Gaussian.  For $A_{\CP}$ we provide averages in all cases.  

Our averaging is performed by maximizing the likelihood,
   $\displaystyle {\mathcal L} = \prod_i {\mathcal P}_i(x),$  
where ${\mathcal P_i}$ is the probability density function (PDF) of the
$i^{\rm th}$  measurement, and $x$ is the branching fraction or $A_{CP}$.
The PDF is modeled by an asymmetric Gaussian function with the measured
central value as its mean and the quadratic sum of the statistical
and systematic errors as the standard deviations. The experimental
uncertainties are considered to be uncorrelated with each other when the 
averaging is performed. No error scaling is applied when the fit $\chi^2$ is 
greater than 1 since we believe that tends to overestimate the errors
except in cases of extreme disagreement (we have no such cases).

At present, we have measurements of more than 500 decay modes, reported in
hundreds of 
papers. Because the number of references is so large, we do
not include them with the tables shown here but the full set of
references is available from active GIF files at the 
``2014'' link on 
the rare decays HFAG web page: {\tt http://www.slac.stanford.edu/xorg/hfag/rare/index.html}.
The largest improvement since the last report has come from the inclusion of a
variety of new measurements from the LHC, especially LHCb.  The
measurements of $B_s$ decays are particularly noteworthy.

Sections \ref{sec:rare-charmless} and \ref{sec:rare-bary} provide compilations of branching fractions of $\Bz$ and $\Bp$ to mesonic and baryonic charmless final states, respectively, 
while Sec.~\ref{sec:rare-lb} gives branching fractions of $\Lambda_b$ decays.
In Sec.~\ref{sec:rare-bs} and \ref{sec:rare-radll} different observables of interest are detailed in addition to branching fractions: in the former for \Bs-meson charmless decays, and in the latter for leptonic and radiative \Bz\ and \Bp\ meson decays, including processes in which the photon yields a pair of charged of neutral leptons. This section also contains limits from searches for lepton-number-violating decays.
Finally, Sec.~\ref{sec:rare-acp} and \ref{sec:rare-polar} give \CP asymmetries and results of polarization measurements, respectively, in different $b$-hadron charmless decays.

\mysubsection{Mesonic decays of \Bz and \Bp\ mesons}
\label{sec:rare-charmless}

This section provides branching fractions of charmless mesonic decays:
Tables~\ref{tab:charmless_BpFirst} to \ref{tab:charmless_BpLast} for \Bp\
and Tables~\ref{tab:charmless_BdFirst} to
\ref{tab:charmless_BdLast}
for \Bz\ mesons.
The tables are separated according to the presence or absence of kaons in the final state. 
Finally, Table~\ref{tab:charmless_Ratio} details several relative branching fractions of \Bz decays.

\begin{table}[!htb]
\begin{center}
\caption{Branching fractions (BF) of charmless mesonic  
$B^+$ decays with kaons (part 1) in units of $\times 10^{-6}$. Upper limits are
at 90\% CL. Values in \red{red} (\blue{blue}) are new \red{published}
(\blue{preliminary}) results since PDG2014.}
\label{tab:charmless_BpFirst}
\resizebox{\textwidth}{!}{

}
\end{center}

\vspace{-0.4cm}
\dag~Product BF - daughter BF taken to be 100\%.\\
\S~Product BF - $\times \BR{\eta(1295)\to\eta\pi\pi}$.\\
\ddag~Average of results in $\ks K^+K^-$, $\ks \ks K^+$~\cite{Lees:2012kx} and $K^+\pi^+\pi^-$~\cite{Aubert:2008bj}. Reference~\cite{Aubert:2008bj} includes an $f_X$ resonance with parameters that are compatible with $f_0(1500)$.
\end{table}


\begin{table}
\begin{center}
\vspace*{-1.2cm}
\caption{Branching fractions (BF) of charmless mesonic  
$B^+$ decays with kaons (part 2) in units of $\times 10^{-6}$. Upper limits are
at 90\% CL. Values in \red{red} (\blue{blue}) are new \red{published}
(\blue{preliminary}) results since PDG2014.}
\resizebox{\textwidth}{!}{

}
\end{center}

\vspace{-0.4cm}
\dag~Product BF - daughter BF taken to be 100\%. \\
~\S~$M_{\phi\phi}<2.85$ GeV/$c^2$.\\
\ddag~Average of results in $K^+K^-K^+$ and $\ks \ks K^+$.\\
$\diamond$ Result from ARGUS. Cited in the \babar column to avoid adding adding a column to the table. 
\end{table}


\begin{table}
\begin{center}
\vspace*{-0.5cm}
\caption{Branching fractions (BF) of charmless mesonic  
$B^+$ decays without kaons in units of $\times 10^{-6}$. Upper limits are
at 90\% CL. Values in \red{red} (\blue{blue}) are new \red{published}
(\blue{preliminary}) results since PDG2014.}
\label{tab:charmless_BpLast}
\resizebox{\textwidth}{!}{

}
\end{center}

\vspace{-0.4cm}
\dag~Product BF - daughter BF taken to be 100\%.
\end{table}
\clearpage


\begin{table}
\begin{center}
\caption{Branching fractions (BF) of charmless mesonic  
$\Bz$ decays with kaons (part 1) in units of $\times 10^{-6}$. Upper limits are
at 90\% CL. Values in \red{red} (\blue{blue}) are new \red{published}
(\blue{preliminary}) results since PDG2014.}
\label{tab:charmless_BdFirst}
\resizebox{\textwidth}{!}{

}
\end{center}

\vspace{-0.4cm}
\dag~Product BF - daughter BF taken to be 100\%.\\
\ddag~Relative BF converted to absolute BF.\\
$^{(1)}$~$0.755<M(K\pi)<1.250$ GeV/$c^2$.\\
$^{(2)}$~$K^{*0}_x$ stands for the possible candidates for $K^*(1410), K^*_0(1430), K^*_2(1430)$.\\
$^{(3)}$~Average of \babar results from $B^0 \to K^+ \pi^- \pi^0$ and $B^0 \to K^0 \pi^+ \pi^-$.\\
$\diamond$~Obtained from a fit to the ratios of BFs measured by LHCb (Ref.~\cite{Aaij:2013uta}) and to the averages of the BFs in their numerators, as measured by other experiments (RPP 292 and 298).
\end{table}


\begin{table}
\begin{center}
\caption{Branching fractions (BF) of charmless mesonic  
$\Bz$ decays with kaons (part 2) in units of $\times 10^{-6}$. Upper limits are
at 90\% CL. Values in \red{red} (\blue{blue}) are new \red{published}
(\blue{preliminary}) results since PDG2014.}
\resizebox{0.95\textwidth}{!}{

}
\end{center}
\vspace{-0.4cm}
{\footnotesize
\dag~Product BF - daughter BF taken to be 100\%;
~\S~$M_{\phi\phi}<2.85$ GeV/$c^2$;
\ddag~Relative BF converted to absolute BF;
$^{(1)}$~$0.55<M(\pi\pi)<1.42$ GeV/$c^2$;
$^{(2)}$~$0.75<M(K\pi)<1.20$ GeV/$c^2$;
$^{(3)}$~$0.70<M(K\pi)<1.70$ GeV/$c^2$;
$^{(4)}$~$1.60<M(K\pi)<2.15$ GeV/$c^2$;
$^{(5)}$~Result from ARGUS;
$\diamond$~Obtained from a fit to the ratios of BFs measured by LHCb (Ref.~\cite{Aaij:2013uta}) and to the averages of the BFs therein, as measured by other experiments (excluding the present line).
}
\end{table}
\clearpage


\begin{table}
\begin{center}
\caption{Branching fractions (BF) of charmless mesonic  
$\Bz$ decays without kaons in units of $\times 10^{-6}$. Upper limits are
at 90\% CL. Values in \red{red} (\blue{blue}) are new \red{published}
(\blue{preliminary}) results since PDG2014.}
\label{tab:charmless_BdLast}
\resizebox{\textwidth}{!}{

}
\end{center}
\vspace{-0.4cm}
\dag~Product BF - daughter BF taken to be 100\%.\\
\ddag~Relative BF converted to absolute BF.
\end{table}
\clearpage


\begin{table}
\begin{center}
\caption{Relative branching fractions (BF) of charmless mesonic  
$\Bz$ decays. Upper limits are
at 90\% CL. Values in \red{red} (\blue{blue}) are new \red{published}
(\blue{preliminary}) results since PDG2014.}
\label{tab:charmless_Ratio}
\resizebox{\textwidth}{!}{
\begin{tabular}{|lccccc|} \hline
RPP\# &Mode & PDG2014 Avg. &  CDF & LHCb & New avg.  \\ \sglinespb
273                                               & 
$\mathcal{B}(B^0\rightarrow K^+K^-)/\mathcal{B}(B^0\rightarrow K^+\pi^-)$& 
\nodata                                           & 
$0.012\pm0.005\pm0.005$                           & 
\nodata                                           & 
$0.012 \pm 0.007$                                 \\

356                                               & 
$\mathcal{B}(B^0\rightarrow\pi^+\pi^-)/\mathcal{B}(B^0\rightarrow K^+\pi^-)$& 
$0.261\pm 0.010$                                  & 
{$0.259\pm 0.017\pm 0.016$}                       & 
{$0.262\pm 0.009\pm 0.017$}                       & 
$0.261 \pm 0.015$                                 \\

$~-$                                              & 
$\BR(B^0 \to K^{*\mp}K^{\pm})/\BR(B^0 \to K^{*+}\pi^-)$ & 
New                                               & 
\nodata                                           & 
\blue{$<0.05$}                                    & 
$<0.05$                                           \\

\hline
\end{tabular}
}
\end{center}
\end{table}

\vspace{-0.8cm}

\mysubsection{Baryonic decays of \Bz and \Bp\ mesons}
\label{sec:rare-bary}

This section provides branching fractions of charmless baryonic decays of \Bz and \Bp\ mesons in Tables~\ref{tab:bary_Bp} and~\ref{tab:bary_Bz}, respectively. Relative branching fractions are given in Table~\ref{tab:bary_Ratio}.

\begin{table}[h!]
\begin{center}
\caption{Branching fractions (BF) of charmless baryonic  
$\Bp$ decays in units of $\times 10^{-6}$. Upper limits are
at 90\% CL. Values in \red{red} (\blue{blue}) are new \red{published}
(\blue{preliminary}) results since PDG2014.}
\label{tab:bary_Bp}
\resizebox{\textwidth}{!}{
\begin{tabular}{|lcccccc|}
\sgline
RPP\#   & Mode & PDG2014 Avg. & \babar & Belle & LHCb & New Avg. \\
\sgline

417                                               & 
$p \overline p \pi^+$                             & 
$1.62\pm0.20$                                     & 
{$\err{1.69}{0.29}{0.26}~\dag$}                   & 
{$\aerr{1.60}{0.22}{0.19}{0.12}$}                 & 
\nodata                                           & 
$\cerr{1.62}{0.21}{0.20}$                         \\

417                                               & 
$p \overline p \pi^+$~\S                          & 
\nodata                                           & 
\nodata                                           & 
\nodata                                           & 
\red{$\err{1.07}{0.11}{0.11}$}~\P                 & 
$1.07 \pm 0.16$                                   \\

420                                               & 
$p \overline p K^+$                               & 
$5.9\pm0.5$                                       & 
{$\err{6.7}{0.5}{0.4}~\dag$}                      & 
{$\aerr{5.54}{0.27}{0.25}{0.36}$}                 & 
\red{$\err{4.46}{0.21}{0.27}$}~\P~$\diamond$      & 
$5.14 \pm 0.25$                                   \\

421                                               & 
$\Theta^{++} \overline p$ $^{(1)}$                & 
$<0.091$                                          & 
{$<0.09$}                                         & 
{$<0.091$}                                        & 
\nodata                                           & 
{$<0.09$}                                         \\

422                                               & 
$f_J(2221) K^+$ $^{(2)}$                          & 
$<0.41$                                           & 
\nodata                                           & 
{$<0.41$}                                         & 
\nodata                                           & 
{$<0.41$}                                         \\

423                                               & 
$p \overline\Lambda(1520)$                        & 
$< 1.5$                                           & 
{$< 1.5$}                                         & 
\nodata                                           & 
\red{\err{0.315}{0.048}{0.027}}~\P                & 
$0.315 \pm 0.055$                                 \\

425                                               & 
$p \overline p K^{*+}$                            & 
$\cerr{3.6}{0.8}{0.7}$                            & 
{$\err{5.3}{1.5}{1.3}~\dag$}                      & 
{$\aerr{3.38}{0.73}{0.60}{0.39}~\ddag$}           & 
\nodata                                           & 
$\cerr{3.64}{0.79}{0.70}$                         \\

426                                               & 
$f_J(2221) K^{*+}$ $^{(2)}$                       & 
$<0.77$                                           & 
{$<0.77$}                                         & 
\nodata                                           & 
\nodata                                           & 
{$<0.77$}                                         \\

427                                               & 
$p \overline\Lambda$                              & 
$<0.32$                                           & 
\nodata                                           & 
{$< 0.32$}                                        & 
\nodata                                           & 
{$< 0.32$}                                        \\

429                                               & 
$p \overline\Lambda \pi^0$                        & 
$\cerr{3.00}{0.7}{0.6}$                           & 
\nodata                                           & 
{$\aerr{3.00}{0.61}{0.53}{0.33}$}                 & 
\nodata                                           & 
$\cerr{3.00}{0.69}{0.62}$                         \\

430                                               & 
$p \overline\Sigma(1385)^0$                       & 
$<0.47$                                           & 
\nodata                                           & 
$<0.47$                                           & 
\nodata                                           & 
$<0.47$                                           \\

431                                               & 
$\Delta^+\overline \Lambda$                       & 
$<0.82$                                           & 
\nodata                                           & 
$<0.82$                                           & 
\nodata                                           & 
$<0.82$                                           \\

433                                               & 
$p \overline{\Lambda} \pi^+\pi^-$ (NR)            & 
$5.9\pm1.1$                                       & 
\nodata                                           & 
$\aerr{5.92}{0.88}{0.84}{0.69}$                   & 
\nodata                                           & 
$\cerr{5.92}{1.12}{1.09}$                         \\

434                                               & 
$p \overline{\Lambda} \rho^0$                     & 
$4.8\pm0.9$                                       & 
\nodata                                           & 
$\aerr{4.78}{0.67}{0.64}{0.60}$                   & 
\nodata                                           & 
$\cerr{4.78}{0.90}{0.88}$                         \\

435                                               & 
$p \overline{\Lambda} f_2(1270)$                  & 
$2.0\pm0.8$                                       & 
\nodata                                           & 
$\aerr{2.03}{0.77}{0.72}{0.27}$                   & 
\nodata                                           & 
$\cerr{2.03}{0.82}{0.77}$                         \\

436                                               & 
$\Lambda \overline{\Lambda} \pi^+$                & 
$<0.94$                                           & 
\nodata                                           & 
$<0.94~\S$                                        & 
\nodata                                           & 
$<0.94~\S$                                        \\

437                                               & 
$\Lambda \overline{\Lambda} K^+$                  & 
$3.4\pm0.6$                                       & 
\nodata                                           & 
$\aerr{3.38}{0.41}{0.36}{0.41}~\ddag$             & 
\nodata                                           & 
$\cerr{3.38}{0.58}{0.55}$                         \\

438                                               & 
$\Lambda \overline{\Lambda} K^{*+}$               & 
$\cerr{2.2}{1.2}{0.9}$                            & 
\nodata                                           & 
$\aerr{2.19}{1.13}{0.88}{0.33}~\S$                & 
\nodata                                           & 
$\cerr{2.19}{1.18}{0.94}$                         \\

439                                               & 
$\overline{\Delta}^0 p$                           & 
$<1.38$                                           & 
\nodata                                           & 
{$<1.38$} $\S$                                    & 
\nodata                                           & 
{$<1.38$} $\S$                                    \\

440                                               & 
$\Delta^{++} \overline p$                         & 
$<0.14$                                           & 
\nodata                                           & 
{$<0.14$} $\S$                                    & 
\nodata                                           & 
{$<0.14$} $\S$                                    \\


\hline
\end{tabular}
}
\vspace*{0.01cm}
\end{center}
\S~Di-baryon mass is less than 2.85\gevcc.\\
$\dag$ Charmonium decays to $p\bar p$ have been statistically subtracted.\\
\P~Relative BF converted to absolute BF.\\
$\diamond$~Includes contribution where $p \bar{p}$ is produced in charmonia decays.\\
$^{(1)}~\Theta(1540)^{++}\to K^+p$ (pentaquark candidate). \\
$^{(2)}$ Product BF --- daughter BF taken to be 100\%. \\
$\ddag$ The charmonium mass region has been vetoed.\\
\end{table}


\begin{table}
\begin{center}
\caption{Branching fractions (BF) of charmless baryonic  
$\Bz$ decays in units of $\times 10^{-6}$. Upper limits are
at 90\% CL. Values in \red{red} (\blue{blue}) are new \red{published}
(\blue{preliminary}) results since PDG2014.}
\label{tab:bary_Bz}
\resizebox{\textwidth}{!}{
\begin{tabular}{|lcccccc|}
\sgline
RPP\#   & Mode & PDG2014 Avg. & \babar & Belle & LHCb & New Avg. \\
\sgline

407                                               & 
$p \overline{p}$                                  & 
$\cerr{0.015}{0.007}{0.005}$                      & 
{$<0.27$}                                         & 
{$<0.11$}                                         & 
$\aerrsy{0.0147}{0.0062}{0.0051}{0.0035}{0.0014}$ & 
$\cerr{0.0150}{0.0070}{0.0050}$                   \\

409                                               & 
$p \overline{p} K^0$                              & 
$2.66\pm0.32$                                     & 
{$\err{3.0}{0.5}{0.3}~\dag$}                      & 
$\aerr{2.51}{0.35}{0.29}{0.21}~\ddag$             & 
\nodata                                           & 
$\cerr{2.66}{0.34}{0.32}$                         \\

410                                               & 
$\Theta^+ \overline{p}$~$^{(1)}$                  & 
$<0.05$                                           & 
{$<0.05$}                                         & 
{$<0.23$}                                         & 
\nodata                                           & 
{$<0.05$}                                         \\

411                                               & 
$f_J(2221) K^0$ $^{(2)}$                          & 
$<0.45$                                           & 
{$<0.45$}                                         & 
\nodata                                           & 
\nodata                                           & 
{$<0.45$}                                         \\

412                                               & 
$p \overline{p} K^{*0}$                           & 
$\cerr{1.24}{0.28}{0.25}$                         & 
{$\err{1.47}{0.45}{0.40}~\dag$}                   & 
$\aerr{1.18}{0.29}{0.25}{0.11}~\ddag$             & 
\nodata                                           & 
$\cerr{1.24}{0.28}{0.25}$                         \\

413                                               & 
$f_J(2221) K^{*0}$ $^{(2)}$                       & 
$<0.15$                                           & 
{$<0.15$}                                         & 
\nodata                                           & 
\nodata                                           & 
{$<0.15$}                                         \\

414                                               & 
$p \overline\Lambda \pi^-$                        & 
$3.14\pm0.29$                                     & 
$\err{3.07}{0.31}{0.23}$                          & 
{$\aerr{3.23}{0.33}{0.29}{0.29}$}                 & 
\nodata                                           & 
$\cerr{3.14}{0.29}{0.28}$                         \\

415                                               & 
$p \overline\Sigma(1385)^-$                       & 
$<0.26$                                           & 
\nodata                                           & 
$<0.26$                                           & 
\nodata                                           & 
$<0.26$                                           \\

416                                               & 
$\Delta^0 \overline\Lambda$                       & 
$<0.93$                                           & 
\nodata                                           & 
$<0.93$                                           & 
\nodata                                           & 
$<0.93$                                           \\

417                                               & 
$p \overline\Lambda K^-$                          & 
$<0.82$                                           & 
\nodata                                           & 
$< 0.82$                                          & 
\nodata                                           & 
$< 0.82$                                          \\

418                                               & 
$p \overline\Sigma^0 \pi^-$                       & 
$<3.8$                                            & 
\nodata                                           & 
$< 3.8$                                           & 
\nodata                                           & 
$< 3.8$                                           \\

419                                               & 
$\overline\Lambda \Lambda$                        & 
$<0.32$                                           & 
\nodata                                           & 
{$<0.32$}                                         & 
\nodata                                           & 
{$<0.32$}                                         \\

420                                               & 
$\overline\Lambda \Lambda K^0$                    & 
$\cerr{4.8}{1.0}{0.9}$                            & 
\nodata                                           & 
$\aerr{4.76}{0.84}{0.68}{0.61}~\ddag$             & 
\nodata                                           & 
$\cerr{4.76}{1.04}{0.91}$                         \\

421                                               & 
$\Lambda \overline{\Lambda} K^{*0}$               & 
$\cerr{2.5}{0.9}{0.8}$                            & 
\nodata                                           & 
$\aerr{2.46}{0.87}{0.72}{0.34}~\ddag$             & 
\nodata                                           & 
$\cerr{2.46}{0.93}{0.80}$                         \\


\hline
\end{tabular}
}
\end{center}
$\dag$ Charmonium decays to $p\bar p$ have been statistically subtracted.\\
$\ddag$ The charmonium mass region has been vetoed.\\
$^{(1)}~\Theta(1540)^+\to p K^0$ (pentaquark candidate).\\
$^{(2)}$ Product BF --- daughter BF taken to be 100\%.
\end{table}


\begin{table}
\begin{center}
%
\caption{Relative branching fractions (BF) of charmless baryonic  
$\B$ decays. Values in \red{red} (\blue{blue}) are new \red{published}
(\blue{preliminary}) results since PDG2014.}
\label{tab:bary_Ratio}
\resizebox{\textwidth}{!}{
\begin{tabular}{|lcccc|} \hline
RPP\# & Mode & PDG2014 Avg. & LHCb & New Avg.  \\ \sglinespb
$417$                                             & 
$\mathcal{B}(B^+\rightarrow p \overline p \pi^+, m_{p \overline p}<2.85\gevcc)/\mathcal{B}(B^+\rightarrow J/\psi(\to p \bar{p})\pi^+)$& 
\nodata                                           & 
\red{$\err{12.0}{1.2}{0.3}$}                      & 
$12.0 \pm 1.2$                                    \\

$420$                                             & 
$\mathcal{B}(B^+\rightarrow p \overline p K^+)/\mathcal{B}(B^+\rightarrow J/\psi(\to p \bar{p})K^+)$& 
\nodata                                           & 
$\err{4.91}{0.19}{0.14}$~\dag                     & 
$4.91 \pm 0.24$                                   \\

$420$                                             & 
$\mathcal{B}(B^+\rightarrow p \overline p K^+)/\mathcal{B}(B^+\rightarrow J/\psi K^+)$& 
$\err{0.0104}{0.0005}{0.0001}$                    & 
$\err{0.0104}{0.0005}{0.0001}$~\dag~\S            & 
$0.0104 \pm 0.0005$                               \\

$423$                                             & 
$\mathcal{B}(B^+\rightarrow \overline\Lambda(1520)(\to K^+\bar{p}) p)/\mathcal{B}(B^+\rightarrow J/\psi(\to p \bar{p})\pi^+)$& 
\nodata                                           & 
\red{$\err{0.033}{0.005}{0.007}$}                 & 
$0.033 \pm 0.009$                                 \\

\sglinespt
\end{tabular}
}
\end{center}
\dag~Includes contribution where $p \bar{p}$ is produced in charmonia decays.\\
\S~Original experimental relative BF multiplied by the best values (PDG2014) of certain reference BFs. The first error is experimental, the second is from reference BF.\\
\end{table}

\clearpage

\mysubsection{Decays of \b baryons}
\label{sec:rare-lb}

A compilation of branching fractions of $\Lb$ baryon decays is given in Table~\ref{tab:bbaryons_Lb}. Table~\ref{tab:bbaryons_LbPartialBF} provides the partial branching fractions of $\Lb\to\Lambda\mup\mun$ decays.

\begin{table}[h!]
\begin{center}
%
%
\caption{Branching fractions (BF) of charmless $\Lb$ decays in units of
$\times 10^{-6}$. Values in \red{red} (\blue{blue}) are new \red{published}
(\blue{preliminary}) results since PDG2014.}
\label{tab:bbaryons_Lb}
\begin{tabular}{|lccccc|} \hline
RPP\# &Mode & PDG2012 Avg. & CDF & LHCb & New Avg.  \\ \sglinespb
$~19$                                             & 
$p\pi^-$                                          & 
$\err{3.5}{0.8}{0.6}$                             & 
$\err{3.5}{0.8}{0.6}$                             & 
\nodata                                           & 
$3.5 \pm 1.0$                                     \\

$~20$                                             & 
$p K^-$                                           & 
$\err{5.5}{1.0}{1.0}$                             & 
$\err{5.5}{1.0}{1.0}$                             & 
\nodata                                           & 
$5.5 \pm 1.4$                                     \\

$~21$                                             & 
$\Lambda \mu^+\mu^-$                              & 
$\err{17.2}{4.2}{5.5}$                            & 
$\err{17.2}{4.2}{5.5}$                            & 
\red{$\err{0.96}{0.16}{0.25}$}                    & 
$0.99 \pm 0.30$                                   \\

\hline
\end{tabular}
\end{center}
\end{table}


\begin{table}[h!]
\begin{center}
\caption{Partial branching fractions (BF) of $\Lb \to \mu^+\mu^-$
decays in intervals of $q^2=m^2_{\mu\mu}$ in units of
$\times 10^{-6}$. Values in \red{red} (\blue{blue}) are new \red{published} (\blue{preliminary}) results since PDG2014.}
\label{tab:bbaryons_LbPartialBF}
\resizebox{\textwidth}{!}{
\begin{tabular}{|llccccc|}
\sgline
RPP\# & Mode & $q^2~[(\mathrm{GeV}/c^2)^2]$~\dag &
PDG2012 Avg. & CDF & LHCb & New Avg. \\
\sglinespb
21                                                & 
$\Lambda\mu^+\mu^-$                               & 
$< 2.0$                                           & 
{$\err{0.15}{2.01}{0.05}$}                        & 
{$\err{0.15}{2.01}{0.05}$}                        & 
\red{$\err{0.56}{0.76}{0.80}$}                    & 
$0.41 \pm 0.87$                                   \\

\nodata                                           & 
$\Lambda\mu^+\mu^-$                               & 
$[2.0,4.3]$                                       & 
{$\err{1.8}{1.7}{0.6}$}                           & 
{$\err{1.8}{1.7}{0.6}$}                           & 
\red{$\err{0.71}{0.60}{0.10}$}                    & 
$0.91 \pm 0.55$                                   \\

\nodata                                           & 
$\Lambda\mu^+\mu^-$                               & 
$[4.3,8.68]$                                      & 
{$\err{-0.2}{1.6}{0.1}$}                          & 
{$\err{-0.2}{1.6}{0.1}$}                          & 
\red{$\err{0.66}{0.72}{0.16}$}                    & 
$0.40 \pm 0.62$                                   \\

\nodata                                           & 
$\Lambda\mu^+\mu^-$                               & 
$[10.09,12.86]$                                   & 
{$\err{3.0}{1.5}{1.0}$}                           & 
{$\err{3.0}{1.5}{1.0}$}                           & 
\red{$\err{1.55}{0.58}{0.55}$}                    & 
$1.96 \pm 0.68$                                   \\

\nodata                                           & 
$\Lambda\mu^+\mu^-$                               & 
$[14.18,16.00]$                                   & 
{$\err{1.0}{0.7}{0.3}$}                           & 
{$\err{1.0}{0.7}{0.3}$}                           & 
\red{$\err{1.44}{0.44}{0.42}$}                    & 
$1.19 \pm 0.40$                                   \\

\nodata                                           & 
$\Lambda\mu^+\mu^-$                               & 
$>16.00$                                          & 
{$\err{7.0}{1.9}{2.2}$}                           & 
{$\err{7.0}{1.9}{2.2}$}                           & 
\red{$\err{4.7}{0.8}{1.2}$}                       & 
$5.5 \pm 1.2$                                     \\

\sglinespt
\end{tabular}
}
\end{center}
\vspace{-0.3cm}
\dag~see the original papers for the exact $q^2$ selection.
\end{table}

\clearpage

\mysubsection{Decays of \Bs mesons}
\label{sec:rare-bs}

Tables~\ref{tab:Bs_BF} and~\ref{tab:Bs_BF_rel} detail branching fractions and relative branching fractions of \Bs meson decays, respectively. Tables~\ref{tab:Bs_PartBF} to~\ref{tab:Bs_A9} give different observables for $\Bs\to\phi\mup\mun$ decays in bins of the dimuon invariant mass.

\begin{table}[h!]
\begin{center}
%
\caption{Branching fractions (BF) of charmless 
$\Bs$ decays in units of $\times 10^{-6}$. Upper limits are
at 90\% CL. Values in \red{red} (\blue{blue}) are new \red{published}
(\blue{preliminary}) results since PDG2014.}
\label{tab:Bs_BF}
\resizebox{\textwidth}{!}{
\begin{tabular}{|lccccccccc|} \hline
RPP\# &Mode & PDG2014 Avg. &  Belle & CDF & D0 & LHCb & CMS & ATLAS & New Avg.  \\ \sglinespb
$45$                                              & 
$\pi^+\pi^-$                                      & 
$0.76\pm0.19$                                     & 
$<12$                                             & 
$\err{0.60}{0.17}{0.04}$~\ddag                    & 
\nodata                                           & 
$\aerr{0.98}{0.23}{0.19}{0.07}$~\ddag             & 
\nodata                                           & 
\nodata                                           & 
$0.76 \pm 0.13$                                   \\

$51$                                              & 
$\phi\phi$                                        & 
$19.1 \pm 3.1$                                    & 
\nodata                                           & 
$\err{19.1}{2.6}{1.6}$ $\ddag$                    & 
\nodata                                           & 
\nodata                                           & 
\nodata                                           & 
\nodata                                           & 
$19.1 \pm 3.1$                                    \\

$52$                                              & 
$\pi^+K^-$                                        & 
$5.5 \pm 0.6$                                     & 
$<26$                                             & 
$5.3 \pm 0.9 \pm 0.3$ $\ddag$                     & 
\nodata                                           & 
$\err{5.6}{0.6}{0.3}$ $\ddag$                     & 
\nodata                                           & 
\nodata                                           & 
$5.5 \pm 0.5$                                     \\

$53$                                              & 
$K^+K^-$                                          & 
$24.9 \pm 1.7$                                    & 
$\aerr{38}{10}{9}{7}$                             & 
$\err{25.9}{2.2}{1.7}$~\ddag                      & 
\nodata                                           & 
$\err{23.7}{1.6}{1.5}$~\ddag                      & 
\nodata                                           & 
\nodata                                           & 
$24.8 \pm 1.7$                                    \\

$54$                                              & 
$K^0\overline{K}^0$                               & 
$< 66$                                            & 
$< 66$                                            & 
\nodata                                           & 
\nodata                                           & 
\nodata                                           & 
\nodata                                           & 
\nodata                                           & 
$< 66$                                            \\

$55$                                              & 
$K^0\pi^+\pi^-$                                   & 
$19 \pm 5$                                        & 
\nodata                                           & 
\nodata                                           & 
\nodata                                           & 
$\err{19}{5}{2}$~\ddag                            & 
\nodata                                           & 
\nodata                                           & 
$19 \pm 5$                                        \\

$56$                                              & 
$K^0 K^- \pi^+$~\P                                & 
$97 \pm 17$                                       & 
\nodata                                           & 
\nodata                                           & 
\nodata                                           & 
$\err{97}{12}{12}$~\ddag                          & 
\nodata                                           & 
\nodata                                           & 
$97 \pm 16$                                       \\

$57$                                              & 
$K^0 K^+ K^-$                                     & 
$<4$                                              & 
\nodata                                           & 
\nodata                                           & 
\nodata                                           & 
$<4$~\ddag                                        & 
\nodata                                           & 
\nodata                                           & 
$<4$~\ddag                                        \\

$~-$                                              & 
$K^{*\pm}K^{\mp}$                                 & 
New                                               & 
\nodata                                           & 
\nodata                                           & 
\nodata                                           & 
\blue{$\err{12.7}{1.9}{1.9}$}~\dag                & 
\nodata                                           & 
\nodata                                           & 
$12.7 \pm 2.7$                                    \\

$~-$                                              & 
$K^{*-}\pi^+$                                     & 
New                                               & 
\nodata                                           & 
\nodata                                           & 
\nodata                                           & 
\blue{$\err{3.3}{1.1}{0.5}$}~\dag                 & 
\nodata                                           & 
\nodata                                           & 
$3.3 \pm 1.2$                                     \\

$59$                                              & 
$K^{*0}\overline{K}^{*0}$                         & 
$\err{28.1}{4.6}{5.6}$                            & 
\nodata                                           & 
\nodata                                           & 
\nodata                                           & 
$\err{28.1}{4.6}{5.6}$~\dag                       & 
\nodata                                           & 
\nodata                                           & 
$28.1 \pm 7.2$                                    \\

$60$                                              & 
$\phi \overline{K}^{*0}$                          & 
$1.13\pm0.3$                                      & 
\nodata                                           & 
\nodata                                           & 
\nodata                                           & 
$\err{1.13}{0.29}{0.06}$~\ddag                    & 
\nodata                                           & 
\nodata                                           & 
$1.13 \pm 0.30$                                   \\

$61$                                              & 
$p \overline{p}$                                  & 
$\cerr{0.028}{0.022}{0.017}$                      & 
\nodata                                           & 
\nodata                                           & 
\nodata                                           & 
\aerrsy{0.0284}{0.0203}{0.0168}{0.0085}{0.0018}~\dag& 
\nodata                                           & 
\nodata                                           & 
$\cerr{0.0280}{0.0220}{0.0170}$                   \\

$63$                                              & 
$\gamma\gamma$                                    & 
$<8.7 $                                           & 
$<8.7$                                            & 
\nodata                                           & 
\nodata                                           & 
\nodata                                           & 
\nodata                                           & 
\nodata                                           & 
$<8.7$                                            \\

$64$                                              & 
$\phi\gamma$                                      & 
$36 \pm 4$                                        & 
$\aerrsy{57}{18}{15}{12}{11}$                     & 
\nodata                                           & 
\nodata                                           & 
\err{35.1}{3.5}{1.2}~\ddag                        & 
\nodata                                           & 
\nodata                                           & 
$35.9 \pm 3.6$                                    \\

$65$                                              & 
$\mu^+\mu^-$                                      & 
$0.0031 \pm 0.0007$                               & 
\nodata                                           & 
$\cerr{0.013}{0.009}{0.007}~\dag$                 & 
$<0.012~\dag$                                     & 
$\aerrsy{0.0029}{0.0011}{0.0010}{0.0003}{0.0001}~\dag$& 
$\cerr{0.0030}{0.0010}{0.0009}~\dag$              & 
$<0.019~\dag$                                     & 
$0.0031 \pm 0.0007$                               \\

$65$                                              & 
$\mu^+\mu^-$                                      & 
CMS-LHCb comb.                                    & 
\nodata                                           & 
\nodata                                           & 
\nodata                                           & 
\blue{$\cerr{0.0028}{0.0007}{0.0006}$}            & 
\blue{$\cerr{0.0028}{0.0007}{0.0006}$}            & 
\nodata                                           & 
                                                  \\

$66$                                              & 
$e^+e^-$                                          & 
$<0.28$                                           & 
\nodata                                           & 
$<0.28$                                           & 
\nodata                                           & 
\nodata                                           & 
\nodata                                           & 
\nodata                                           & 
$<0.28$                                           \\

$67$                                              & 
$e^{\pm}\mu^{\mp}$                                & 
$<0.011$                                          & 
\nodata                                           & 
$<0.20$                                           & 
\nodata                                           & 
$<0.011$~\dag                                     & 
\nodata                                           & 
\nodata                                           & 
$<0.011$~\dag                                     \\

$68$                                              & 
$\mu^+ \mu^- \mu^+ \mu^-$                         & 
$<0.012$                                          & 
\nodata                                           & 
\nodata                                           & 
\nodata                                           & 
$<0.012$                                          & 
\nodata                                           & 
\nodata                                           & 
$<0.012$                                          \\

$70$                                              & 
$\phi\mu^+\mu^-$                                  & 
$0.76\pm0.15$                                     & 
\nodata                                           & 
\blue{$\err{1.17}{0.18}{0.37}~$}\dag              & 
$<3.2$ $\dag$                                     & 
$\aerr{0.707}{0.064}{0.059}{0.073}$~\dag          & 
\nodata                                           & 
\nodata                                           & 
$\cerr{0.731}{0.095}{0.092}$                      \\

\hline
\end{tabular}
}
\end{center}
\ddag~Original experimental relative BF multiplied by the best values (PDG2014) of reference BF. The first error is experimental, the second is from reference BF.\\[0.1cm]
$\dag$~Relative BF converted to absolute BF.\\
\P~Sum of charge conjugate states.
\end{table}


\begin{table}
\begin{center}
%
\caption{Relative branching fractions (BF) of charmless 
$\Bs$ decays. Upper limits are
at 90\% CL. Values in \red{red} (\blue{blue}) are new \red{published}
(\blue{preliminary}) results since PDG2014.}
\label{tab:Bs_BF_rel}
\resizebox{\textwidth}{!}{
\begin{tabular}{|lccccc|} \hline
RPP\# & Mode & PDG2014 Avg. & CDF & LHCb & New Avg.  \\ \sglinespb
$45$                                              & 
$\it{f}_s\mathcal{B}(B^0_s\rightarrow\pi^+\pi^-)/\it{f}_d\mathcal{B}(B^0\rightarrow K^+\pi^-)$& 
\nodata                                           & 
$0.008\pm0.002\pm0.001$                           & 
\nodata                                           & 
$0.008 \pm 0.002$                                 \\

$45$                                              & 
$\it{f}_s\mathcal{B}(B^0_s\rightarrow\pi^+\pi^-)/\it{f}_d\mathcal{B}(B^0\rightarrow \pi^+\pi^-)$& 
\nodata                                           & 
\nodata                                           & 
$\aerr{0.050}{0.011}{0.009}{0.004}$               & 
$\cerr{0.050}{0.012}{0.010}$                      \\

$51$                                              & 
$\mathcal{B}(B^0_s\rightarrow\phi\phi)/\mathcal{B}(B^0_s\rightarrow J/\psi\phi)$& 
\nodata                                           & 
$0.0178 \pm 0.0014 \pm 0.0020$                    & 
\nodata                                           & 
$0.0180 \pm 0.0020$                               \\

$52$                                              & 
$\it{f}_s\mathcal{B}(B^0_s\rightarrow K^+\pi^-)/\it{f}_d\mathcal{B}(B^0_d\rightarrow K^+\pi^-)$& 
\nodata                                           & 
$0.071\pm0.010\pm0.007 $                          & 
$0.074\pm0.006\pm0.006$                           & 
$0.073 \pm 0.007$                                 \\

$53$                                              & 
$\it{f}_s\mathcal{B}(B^0_s\rightarrow K^+K^-)/\it{f}_d\mathcal{B}(B^0_d\rightarrow K^+\pi^-)$& 
\nodata                                           & 
$0.347\pm0.020\pm0.021 $                          & 
\err{0.316}{0.009}{0.019}                         & 
$0.327 \pm 0.017$                                 \\

$55$                                              & 
$\it{f}_s\mathcal{B}(B^0_s\to K^0\pi^+\pi^-)/\it{f}_d\mathcal{B}(B^0 \to K^0\pi^+\pi^-)$& 
\nodata                                           & 
\nodata                                           & 
$\err{0.29}{0.06}{0.04}$                          & 
$0.29 \pm 0.07$                                   \\

$56$                                              & 
$\it{f}_s\mathcal{B}(B^0_s\to K^0 K^- \pi^+)/\it{f}_d\mathcal{B}(B^0 \to K^0 K^- \pi^+)$~\P & 
\nodata                                           & 
\nodata                                           & 
$\err{1.48}{0.12}{0.14}$                          & 
$1.48 \pm 0.18$                                   \\

$57$                                              & 
$\it{f}_s\mathcal{B}(B^0_s\to K^0 K^+ K^-)/\it{f}_d\mathcal{B}(B^0 \to K^0 K^+ K^-)$& 
\nodata                                           & 
\nodata                                           & 
$<0.068$                                          & 
$<0.068$                                          \\

$~-$                                              & 
$\BR(B^0_s\to K^{*-}K^+)/\BR(B^0 \to K^{*+}\pi^-)$& 
New                                               & 
\nodata                                           & 
\blue{$\err{1.49}{0.22}{0.18}$}                   & 
$1.49 \pm 0.28$                                   \\

$~-$                                              & 
$\BR(B^0_s\to K^{*-}\pi^+)/\BR(B^0 \to K^{*+}\pi^-)$& 
New                                               & 
\nodata                                           & 
\blue{$\err{0.39}{0.13}{0.05}$}                   & 
$0.39 \pm 0.14$                                   \\

$60$                                              & 
$\BR(B^0_s\to \phi \overline{K}^{*0})/\BR(B^0 \to \phi K^{*0})$& 
\nodata                                           & 
\nodata                                           & 
$\err{0.113}{0.024}{0.016}$                       & 
$0.113 \pm 0.029$                                 \\

$64$                                              & 
$\BR(B^0_s\to \phi \gamma)/\BR(B^0 \to K^{*0}\gamma)$& 
\nodata                                           & 
\nodata                                           & 
\err{0.81}{0.04}{0.07}                            & 
$0.81 \pm 0.08$                                   \\

$70$                                              & 
$\mathcal{B}(B^0_s\to \phi\mu^+\mu^-)/\mathcal{B}(B^0_s\to J/\psi\phi)\times10^3$& 
$0.71 \pm 0.13$                                   & 
\blue{$\err{0.90}{0.14}{0.07}$}                   & 
$\aerr{0.674}{0.061}{0.056}{0.016}$               & 
$\cerr{0.704}{0.060}{0.056}$                      \\

\sglinespt
\end{tabular}
}
\end{center}
\P~Sum of charge conjugate states in the numerator.
\end{table}


\begin{table}
\begin{center}
%
%
\caption{Partial branching fractions (BF) of
$\Bs\to \phi\mu^+\mu^-$ decays in units of $\times 10^{-7}$. Values in \red{red} (\blue{blue}) are new \red{published}
(\blue{preliminary}) results since PDG2014.}
\label{tab:Bs_PartBF}
\resizebox{\textwidth}{!}{
\begin{tabular}{|lccccc|}
\sgline
Mode & $q^2~[(\mathrm{GeV}/c^2)^2]$&
PDG2014 Avg. & CDF & LHCb & New Avg. \\
\sglinespb
$\phi\mu^+\mu^-$                                  & 
$< 2.0$~\dag                                      & 
$0.93 \pm 0.21$                                   & 
\blue{$\err{3.16}{0.92}{1.00}$}                   & 
$\aerr{0.90}{0.21}{0.19}{0.10}$                   & 
$\cerr{0.96}{0.23}{0.22}$                         \\

\nodata                                           & 
$[2.0,4.3]$                                       & 
$\cerr{0.55}{0.18}{0.16}$                         & 
\blue{$\err{0.27}{0.41}{0.09}$}                   & 
$\aerr{0.53}{0.18}{0.16}{0.06}$                   & 
$\cerr{0.49}{0.17}{0.16}$                         \\

\nodata                                           & 
$[4.3,8.68]$                                      & 
$1.40 \pm 0.26$                                   & 
\blue{$\err{0.64}{0.68}{0.20}$}                   & 
$\aerr{1.38}{0.25}{0.23}{0.15}$                   & 
$\cerr{1.28}{0.26}{0.25}$                         \\

\nodata                                           & 
$[10.09,12.86]$                                   & 
$1.22 \pm 0.25$                                   & 
\blue{$\err{2.25}{0.69}{0.71}$}                   & 
$\aerr{1.20}{0.23}{0.21}{0.14}$                   & 
$\cerr{1.27}{0.26}{0.25}$                         \\

\nodata                                           & 
$[14.18,16.00]$                                   & 
$0.80 \pm 0.20$                                   & 
\blue{$\err{1.11}{0.42}{0.35}$}                   & 
$\aerr{0.76}{0.19}{0.17}{0.09}$                   & 
$\cerr{0.80}{0.20}{0.18}$                         \\

\nodata                                           & 
$>16.00$~\dag                                     & 
$1.08 \pm 0.24$                                   & 
\blue{$\err{2.31}{0.59}{0.73}$}                   & 
$\aerr{1.06}{0.23}{0.21}{0.12}$                   & 
$\cerr{1.14}{0.25}{0.24}$                         \\

\nodata                                           & 
$[1.00,6.00]$                                     & 
{$1.15 \pm 0.25$}                                 & 
\blue{$\err{1.03}{0.70}{0.33}$}                   & 
$\aerr{1.14}{0.25}{0.23}{0.13}$                   & 
$\cerr{1.13}{0.26}{0.25}$                         \\

\sglinespt
\end{tabular}
}
\end{center}
\vspace{-0.3cm}
\dag~See the original papers for the exact $q^2$ interval, which very slightly differs between experiments.
\end{table}


\begin{table}
\begin{center}
%
\caption{Longitudinal polarization fraction ($F_L$) of
$\Bs\to \phi\mu^+\mu^-$ decays. Values in \red{red} (\blue{blue}) are new \red{published}
(\blue{preliminary}) results since PDG2014.}
\label{tab:Bs_FL}
\begin{tabular}{|lcccc|}
\sgline
Mode & $q^2~[(\mathrm{GeV}/c^2)^2]$ & PDG2014 Avg. & LHCb & New Avg. \\
\sglinespb
$\phi\mu^+\mu^-$                                  & 
$0.1-2.0$                                         & 
$\aerr{0.37}{0.19}{0.17}{0.07}$                   & 
$\aerr{0.37}{0.19}{0.17}{0.07}$                   & 
$\cerr{0.37}{0.20}{0.18}$                         \\

\nodata                                           & 
$[2.0,4.3]$                                       & 
$\aerr{0.53}{0.25}{0.23}{0.10}$                   & 
$\aerr{0.53}{0.25}{0.23}{0.10}$                   & 
$\cerr{0.53}{0.27}{0.25}$                         \\

\nodata                                           & 
$[4.3,8.68]$                                      & 
$\aerr{0.81}{0.11}{0.13}{0.05}$                   & 
$\aerr{0.81}{0.11}{0.13}{0.05}$                   & 
$\cerr{0.81}{0.12}{0.14}$                         \\

\nodata                                           & 
$[10.09,12.86]$                                   & 
$\aerr{0.33}{0.14}{0.12}{0.06}$                   & 
$\aerr{0.33}{0.14}{0.12}{0.06}$                   & 
$\cerr{0.33}{0.15}{0.13}$                         \\

\nodata                                           & 
$[14.18,16.00]$                                   & 
$\aerr{0.34}{0.18}{0.17}{0.07}$                   & 
$\aerr{0.34}{0.18}{0.17}{0.07}$                   & 
$\cerr{0.34}{0.19}{0.18}$                         \\

\nodata                                           & 
$16.00-19.00$                                     & 
$\aerr{0.16}{0.17}{0.10}{0.07}$                   & 
$\aerr{0.16}{0.17}{0.10}{0.07}$                   & 
$\cerr{0.16}{0.18}{0.12}$                         \\

\nodata                                           & 
$[1.00,6.00]$                                     & 
$\aerr{0.56}{0.17}{0.16}{0.09}$                   & 
$\aerr{0.56}{0.17}{0.16}{0.09}$                   & 
$\cerr{0.56}{0.19}{0.18}$                         \\

\sglinespt
\end{tabular}
\end{center}
\vspace{-0.3cm}
\end{table}


\begin{table}
\begin{center}
%
\caption{The parameter $S_3$ from the angular analysis of
$\Bs\to \phi\mu^+\mu^-$ decays (see reference for definition).
Values in \red{red} (\blue{blue}) are new \red{published}
(\blue{preliminary}) results since PDG2014.}
\label{tab:Bs_S3}
\begin{tabular}{|lcccc|}
\sgline
Mode & $q^2~[(\mathrm{GeV}/c^2)^2]$~\dag &
PDG2014 Avg. & LHCb & New Avg. \\
\sglinespb
$\phi\mu^+\mu^-$                                  & 
$0.1-2.0$                                         & 
\nodata                                           & 
$\aerr{-0.11}{0.28}{0.25}{0.05}$                  & 
$\cerr{-0.11}{0.28}{0.26}$                        \\

\nodata                                           & 
$[2.0,4.3]$                                       & 
\nodata                                           & 
$\aerr{-0.97}{0.53}{0.03}{0.17}$                  & 
$\cerr{-0.97}{0.56}{0.17}$                        \\

\nodata                                           & 
$[4.3,8.68]$                                      & 
\nodata                                           & 
$\aerr{0.25}{0.21}{0.24}{0.05}$                   & 
$\cerr{0.25}{0.22}{0.24}$                         \\

\nodata                                           & 
$[10.09,12.86]$                                   & 
\nodata                                           & 
$\aerr{0.24}{0.27}{0.25}{0.06}$                   & 
$\cerr{0.24}{0.28}{0.26}$                         \\

\nodata                                           & 
$[14.18,16.00]$                                   & 
\nodata                                           & 
$\aerr{-0.03}{0.29}{0.31}{0.06}$                  & 
$\cerr{-0.03}{0.30}{0.32}$                        \\

\nodata                                           & 
$16.00-19.00$                                     & 
\nodata                                           & 
$\aerr{0.19}{0.30}{0.31}{0.05}$                   & 
$\cerr{0.19}{0.30}{0.31}$                         \\

\nodata                                           & 
$[1.00,6.00]$                                     & 
\nodata                                           & 
$\aerr{-0.21}{0.24}{0.22}{0.08}$                  & 
$\cerr{-0.21}{0.25}{0.23}$                        \\

\sglinespt
\end{tabular}
\end{center}
\vspace{-0.3cm}
\end{table}


\begin{table}
\begin{center}
%
\caption{The parameter $A_6$ from the angular analysis of
$\Bs\to \phi\mu^+\mu^-$ decays (see reference for definition).
Values in \red{red} (\blue{blue}) are new \red{published}
(\blue{preliminary}) results since PDG2014.}
\label{tab:Bs_A6}
\begin{tabular}{|lcccc|}
\sgline
Mode & $q^2~[(\mathrm{GeV}/c^2)^2]$ &
PDG2014 Avg. & LHCb & New Avg. \\
\sglinespb
$\phi\mu^+\mu^-$                                  & 
$0.1-2.0$                                         & 
\nodata                                           & 
$\aerr{0.04}{0.27}{0.32}{0.12}$                   & 
$\cerr{0.04}{0.29}{0.34}$                         \\

\nodata                                           & 
$[2.0,4.3]$                                       & 
\nodata                                           & 
$\aerr{0.47}{0.39}{0.42}{0.14}$                   & 
$\cerr{0.47}{0.41}{0.44}$                         \\

\nodata                                           & 
$[4.3,8.68]$                                      & 
\nodata                                           & 
$\aerr{-0.02}{0.20}{0.21}{0.10}$                  & 
$\cerr{-0.02}{0.22}{0.23}$                        \\

\nodata                                           & 
$[10.09,12.86]$                                   & 
\nodata                                           & 
$\aerr{-0.06}{0.20}{0.20}{0.08}$                  & 
$-0.06 \pm 0.21$                                  \\

\nodata                                           & 
$[14.18,16.00]$                                   & 
\nodata                                           & 
$\aerr{-0.06}{0.30}{0.30}{0.08}$                  & 
$-0.06 \pm 0.31$                                  \\

\nodata                                           & 
$16.00-19.00$                                     & 
\nodata                                           & 
$\aerr{0.26}{0.22}{0.24}{0.08}$                   & 
$\cerr{0.26}{0.23}{0.25}$                         \\

\nodata                                           & 
$[1.00,6.00]$                                     & 
\nodata                                           & 
$\aerr{0.20}{0.29}{0.27}{0.07}$                   & 
$\cerr{0.20}{0.30}{0.28}$                         \\

\sglinespt
\end{tabular}
\end{center}
\vspace{-0.3cm}
\end{table}


\begin{table}
\begin{center}

\caption{The parameter $A_9$ from the angular analysis of
$\Bs\to \phi\mu^+\mu^-$ decays (see reference for definition).
Values in \red{red} (\blue{blue}) are new \red{published}
(\blue{preliminary}) results since PDG2014.}
\label{tab:Bs_A9}
\begin{tabular}{|lcccc|}
\sgline
Mode & $q^2~[(\mathrm{GeV}/c^2)^2]$ &
PDG2014 Avg. & LHCb & New Avg. \\
\sglinespb
$\phi\mu^+\mu^-$                                  & 
$0.1-2.0$                                         & 
\nodata                                           & 
$\aerr{-0.16}{0.30}{0.27}{0.09}$                  & 
$\cerr{-0.16}{0.31}{0.28}$                        \\

\nodata                                           & 
$[2.0,4.3]$                                       & 
\nodata                                           & 
$\aerr{-0.40}{0.52}{0.35}{0.11}$                  & 
$\cerr{-0.40}{0.53}{0.37}$                        \\

\nodata                                           & 
$[4.3,8.68]$                                      & 
\nodata                                           & 
$\aerr{-0.13}{0.27}{0.26}{0.10}$                  & 
$\cerr{-0.13}{0.29}{0.28}$                        \\

\nodata                                           & 
$[10.09,12.86]$                                   & 
\nodata                                           & 
$\aerr{0.29}{0.25}{0.26}{0.10}$                   & 
$\cerr{0.29}{0.27}{0.28}$                         \\

\nodata                                           & 
$[14.18,16.00]$                                   & 
\nodata                                           & 
$\aerr{0.24}{0.36}{0.35}{0.12}$                   & 
$\cerr{0.24}{0.38}{0.37}$                         \\

\nodata                                           & 
$16.00-19.00$                                     & 
\nodata                                           & 
$\aerr{0.27}{0.31}{0.28}{0.11}$                   & 
$\cerr{0.27}{0.33}{0.30}$                         \\

\nodata                                           & 
$[1.00,6.00]$                                     & 
\nodata                                           & 
$\aerr{-0.30}{0.30}{0.29}{0.11}$                  & 
$\cerr{-0.30}{0.32}{0.31}$                        \\

\sglinespt
\end{tabular}
\end{center}
\vspace{-0.3cm}
\end{table}

\clearpage

\mysubsection{Radiative and leptonic decays of \Bz and \Bp\ mesons}
\label{sec:rare-radll}

This section gives different observables for leptonic and radiative \Bz\ and \Bp\ meson decays, including processes in which the photon yields a pair of charged of neutral leptons. Tables~\ref{tab:radll_Bp}, \ref{tab:radll_Bz} and~\ref{tab:radll_B} provide compilations of branching fractions of \Bp, \Bz, and $\Bpm/\Bz$ admixture, respectively.
Table~\ref{tab:radll_lep} contains branching fractions of leptonic and radiative-leptonic $\Bp$ and $\Bz$ decays. It is followed by Tabs~\ref{tab:radll_rel} and~\ref{tab:radll_gluon}, which give relative branching fractions of $\Bp$ decays and a compilations of inclusive decays, respectively. The next two tables detail isospin asymmetries: Table~\ref{tab:radll_AI} contains overall measurements, and~\ref{tab:radll_AI_Kll} measurements in bins of dimuon invariant mass, $q^2$, bins for $\B\to K^{(\ast)}\ell^+\ell^-$ decays. Tables~\ref{tab:radll_AFB_Kll} and~\ref{tab:radll_FL_Kll} detail measurements of the forward-backward asymmetry and the fraction of longitudinal polarization, respectively, in $\B\to K^{(\ast)}\ell^+\ell^-$ decays in bins of $q^2$. Finally, Table~\ref{tab:radll_Aud} quotes the LHCb measurement of photon polarization via the Up-Down asymmetry in $B^+ \to K^+ \pi^- \pi^+ \gamma$ decays.

\begin{table}

\caption{Branching fractions (BF) of charmless semileptonic and radiative
$B^+$ decays in units of $\times 10^{-6}$. Upper limits are
at 90\% CL. Values in \red{red} (\blue{blue}) are new \red{published}
(\blue{preliminary}) results since PDG2014.}
\label{tab:radll_Bp}
%
\resizebox{\textwidth}{!}{

}
{
~\dag~$M_{K\pi\pi} < 1.8$ GeV/$c^2$;~\ddag~$1.0 < M_{K\pi\pi} < 2.0$ GeV/$c^2$;
~\S~$M_{K\pi\pi} < 2.4$ GeV/$c^2$.\\
\P~Relative BF converted to absolute BF.\\
$^{(1)}$~PDG2014 cites only the measurement: $\mathcal{B}(\pi^+\mu^+\mu^-)/\mathcal{B}(K^+\mu^+\mu^-)=\err{0.053}{0.014}{0.01}$.\\
$^{(2)}$~Differential BF in bins of $m(\mu\mu)$ is also available.\\
$^{(3)}$~At $95\%$ CL.\\
$^{(4)}$~PDG considers here the BF measured in $B^+ \to K^+\mu^+\mu^-$.
}
\end{table}
\clearpage


\begin{table}
%
\caption{Branching fractions (BF) of charmless semileptonic and radiative 
$\Bz$ decays in units of $\times 10^{-6}$. Upper limits are
at 90\% CL. Values in \red{red} (\blue{blue}) are new \red{published}
(\blue{preliminary}) results since PDG2014.}
\label{tab:radll_Bz}
\resizebox{\textwidth}{!}{

}
~\dag~$M_{K\pi\pi} < 1.8$ GeV/$c^2$.\\
~\ddag~$1.0 < M_{K\pi\pi} < 2.0$ GeV/$c^2$;
~\S~$1.25$ GeV/$c^2 < M_{K\pi} < 1.6$ GeV/$c^2$.
\end{table}


\begin{table}

\caption{Branching fractions (BF) of charmless semileptonic and radiative 
decays of $\Bpm/\Bz$ admixture in units of $\times 10^{-6}$. Upper limits are
at 90\% CL. Values in \red{red} (\blue{blue}) are new \red{published}
(\blue{preliminary}) results since PDG2014.}
\label{tab:radll_B}
\resizebox{\textwidth}{!}{
\begin{tabular}{|lccccccc|} \hline
 RPP\# &Mode & PDG2014 Avg. & \babar  & Belle & CLEO & CDF
& New Avg.  \\ \sglinespb
~66                                               & 
$K \eta \gamma$                                   & 
$\cerr{8.5}{1.8}{1.6}$                            & 
\nodata                                           & 
{$\aerr{8.5}{1.3}{1.2}{0.9}$}                     & 
\nodata                                           & 
\nodata                                           & 
$\cerr{8.5}{1.6}{1.5}$                            \\

~68                                               & 
$K_2^*(1430) \gamma $                             & 
$\cerr{17}{6}{5}$                                 & 
\nodata                                           & 
\nodata                                           & 
$\err{17}{6}{1}$                                  & 
\nodata                                           & 
$17 \pm 6$                                        \\

~70                                               & 
$K_3^*(1780) \gamma $                             & 
$<37$                                             & 
\nodata                                           & 
{$<2.8$}~\S                                       & 
\nodata                                           & 
\nodata                                           & 
{$<2.8$}~\S                                       \\

~77                                               & 
$s \gamma$                                        & 
$360\pm 23$                                       & 
$\err{300}{14}{20}$                               & 
$\err{345}{15}{40}$                               & 
$\berr{321}{43}{32}{29}$                          & 
\nodata                                           & 
$\err{343}{21}{7}$                                \\

~78                                               & 
$d \gamma$                                        & 
$9.2 \pm 3.0$                                     & 
{$\err{9.2}{2.0}{2.3}$}                           & 
\nodata                                           & 
\nodata                                           & 
\nodata                                           & 
$9.2 \pm 3.0$                                     \\

~84                                               & 
$\rho \gamma $                                    & 
$1.39\pm 0.25$                                    & 
{$\aerr{1.73}{0.34}{0.32}{0.17}$}                 & 
{$\aerr{1.21}{0.24}{0.22}{0.12}$}                 & 
$<14$                                             & 
\nodata                                           & 
$\cerr{1.39}{0.22}{0.21}$                         \\

~85                                               & 
$\rho/\omega \gamma $                             & 
$1.30\pm0.23$                                     & 
{$\aerr{1.63}{0.30}{0.28}{0.16}$}                 & 
{$\berr{1.14}{0.20}{0.10}{0.12}$}                 & 
$<14$                                             & 
\nodata                                           & 
$\cerr{1.30}{0.18}{0.19}$                         \\

119                                               & 
$se^+ e^-$~\ddag                                  & 
$4.7\pm 1.3$                                      & 
\red{$\aerrsy{7.69}{0.82}{0.77}{0.71}{0.60}$}     & 
\blue{$\berr{4.56}{1.15}{0.33}{0.40}$}            & 
$<57$                                             & 
\nodata                                           & 
$6.44 \pm 0.76$                                   \\

120                                               & 
$s\mu^+ \mu^-$~\ddag                              & 
$4.3\pm 1.2$                                      & 
\red{$\aerrsy{4.41}{1.31}{1.17}{0.63}{0.50}$}     & 
\blue{$\berr{1.91}{1.02}{0.16}{0.18}$}            & 
$<58$                                             & 
\nodata                                           & 
$2.90 \pm 0.80$                                   \\

121                                               & 
$s \ell^+ \ell^-$~\ddag                           & 
$4.5\pm 1.0$                                      & 
\red{$\aerrsy{6.73}{0.70}{0.64}{0.60}{0.56}$}     & 
\blue{$\berr{3.33}{0.80}{0.19}{0.24}$}            & 
$<42$                                             & 
\nodata                                           & 
$4.97 \pm 0.59$                                   \\

122                                               & 
$\pi \ell^+ \ell^-$                               & 
$<0.059$                                          & 
$<0.059$                                          & 
$<0.062$                                          & 
\nodata                                           & 
\nodata                                           & 
$<0.059$                                          \\

123                                               & 
$\pi e^+ e^-$                                     & 
$<0.110$                                          & 
$<0.110$                                          & 
\nodata                                           & 
\nodata                                           & 
\nodata                                           & 
$<0.110$                                          \\

124                                               & 
$\pi \mu^+ \mu^-$                                 & 
$<0.050$                                          & 
$<0.050$                                          & 
\nodata                                           & 
\nodata                                           & 
\nodata                                           & 
$<0.050$                                          \\

125                                               & 
$K e^+ e^-$                                       & 
$0.44 \pm 0.06$                                   & 
{$\aerr{0.39}{0.09}{0.08}{0.02}$}                 & 
{$\aerr{0.48}{0.08}{0.07}{0.03}$}                 & 
\nodata                                           & 
\nodata                                           & 
$0.44 \pm 0.06$                                   \\

126                                               & 
$K^* e^+ e^-$                                     & 
$1.19\pm0.20$                                     & 
{$\aerr{0.99}{0.23}{0.21}{0.06}$}                 & 
{$\aerr{1.39}{0.23}{0.20}{0.12}$}                 & 
\nodata                                           & 
\nodata                                           & 
$\cerr{1.19}{0.17}{0.16}$                         \\

127                                               & 
$K \mu^+ \mu^-$                                   & 
$0.44 \pm 0.04$                                   & 
{$\aerr{0.41}{0.13}{0.12}{0.02}$}                 & 
{$\err{0.50}{0.06}{0.03}$}                        & 
\nodata                                           & 
$\err{4.2}{0.4}{0.2}$                             & 
$0.55 \pm 0.06$                                   \\

128                                               & 
$K^* \mu^+ \mu^-$                                 & 
$1.06 \pm 0.09$                                   & 
{$\aerr{1.35}{0.35}{0.33}{0.10}$}                 & 
{$\aerr{1.10}{0.16}{0.14}{0.08}$}                 & 
\nodata                                           & 
$\err{10.1}{1.0}{0.5}$                            & 
$1.33 \pm 0.16$                                   \\

129                                               & 
$K \ell^+ \ell^-$                                 & 
$0.48 \pm 0.04$                                   & 
$\err{0.47}{0.06}{0.02}$                          & 
{$\aerr{0.48}{0.05}{0.04}{0.03}$}                 & 
$<1.7$                                            & 
\nodata                                           & 
$0.48 \pm 0.04$                                   \\

130                                               & 
$K^* \ell^+ \ell^-$                               & 
$1.05\pm 0.10$                                    & 
$\aerr{1.02}{0.14}{0.13}{0.05}$                   & 
{$\aerr{1.07}{0.11}{0.10}{0.09}$}                 & 
$<3.3$                                            & 
\nodata                                           & 
$1.05 \pm 0.10$                                   \\

131                                               & 
$K \nu \overline \nu$                             & 
$<17$                                             & 
{$<17$}                                           & 
\nodata                                           & 
\nodata                                           & 
\nodata                                           & 
{$<17$}                                           \\

132                                               & 
$K^* \nu \overline \nu$                           & 
$<76$                                             & 
$<76$                                             & 
\nodata                                           & 
\nodata                                           & 
\nodata                                           & 
$<76$                                             \\

134                                               & 
$\pi e^\pm \mu^\mp$                               & 
$<0.092$                                          & 
{$<0.092$}                                        & 
\nodata                                           & 
$<1.6$                                            & 
\nodata                                           & 
{$<0.092$}                                        \\

135                                               & 
$\rho e^\pm \mu^\mp$                              & 
$<3.2$                                            & 
\nodata                                           & 
\nodata                                           & 
$<3.2$                                            & 
\nodata                                           & 
$<3.2$                                            \\

136                                               & 
$K e^\pm \mu^\mp$                                 & 
$<0.038$                                          & 
{$<0.038$}                                        & 
\nodata                                           & 
$<1.6$                                            & 
\nodata                                           & 
{$<0.038$}                                        \\

137                                               & 
$K^* e^\pm \mu^\mp$                               & 
$<0.51$                                           & 
{$<0.51$}                                         & 
\nodata                                           & 
$<6.2$                                            & 
\nodata                                           & 
{$<0.51$}                                         \\

$~-$                                              & 
$s \gamma$ with baryons                           & 
$~-$                                              & 
\nodata                                           & 
\nodata                                           & 
$<38$~\dag                                        & 
\nodata                                           & 
$<38$~\dag                                        \\


\hline
\end{tabular}
}
\dag~$E_\gamma > 2.0$ GeV.\\
\ddag~Belle: $M(\ell^+\ell^-)>0.2$ GeV/$c^2$, \babar: $M^2(\ell^+\ell^-)>0.1$ GeV$^2$/$c^4$.\\
\S~Product BF ($\times \BR{K^*_3 \to K\eta}$). PDG gives the BF assuming $\BR{K^*_3 \to K\eta}=\cerr{11}{5}{4}$.
\end{table}


\begin{table}
\begin{center}
%
\caption{Branching fractions (BF) of leptonic and radiative-leptonic 
$\Bp$ and $\Bz$ decays in units of $\times 10^{-6}$. Upper limits are
at 90\% CL. Values in \red{red} (\blue{blue}) are new \red{published}
(\blue{preliminary}) results since PDG2014.}
\label{tab:radll_lep}
\resizebox{\textwidth}{!}{
\begin{tabular}{|lcccccccc|} \hline
 RPP\# &Mode & PDG2014 Avg. & \babar  & Belle & CDF & LHCb & CMS
& New Avg.  \\ \sglinespb
~29                                               & 
$e^+ \nu$                                         & 
$<0.98$                                           & 
{$<1.9$}                                          & 
$<0.98$~\dag                                      & 
\nodata                                           & 
\nodata                                           & 
\nodata                                           & 
$<0.98$~\dag                                      \\

~30                                               & 
$\mu^+ \nu$                                       & 
$<1.0$                                            & 
$<1.0$                                            & 
$<1.7$~\dag                                       & 
\nodata                                           & 
\nodata                                           & 
\nodata                                           & 
$<1.0$                                            \\

~31                                               & 
$\tau^+ \nu$                                      & 
$114\pm27$                                        & 
$179\pm48$~\ddag                                  & 
$96\pm26$~\ddag                                   & 
\nodata                                           & 
\nodata                                           & 
\nodata                                           & 
$114 \pm 22$                                      \\

~32                                               & 
$\ell^+ \nu_{\ell} \gamma$                        & 
$<15.6$                                           & 
$<15.6$                                           & 
\nodata                                           & 
\nodata                                           & 
\nodata                                           & 
\nodata                                           & 
$<15.6$                                           \\

~33                                               & 
$e^+ \nu_e \gamma$                                & 
$<17$                                             & 
{$<17$}                                           & 
\nodata                                           & 
\nodata                                           & 
\nodata                                           & 
\nodata                                           & 
{$<17$}                                           \\

~34                                               & 
$\mu^+ \nu_{\mu} \gamma$                          & 
$<24$                                             & 
{$<24$}                                           & 
\nodata                                           & 
\nodata                                           & 
\nodata                                           & 
\nodata                                           & 
{$<24$}                                           \\

457                                               & 
$\gamma \gamma$                                   & 
$<0.32$                                           & 
$<0.32$                                           & 
{$<0.62$}                                         & 
\nodata                                           & 
\nodata                                           & 
\nodata                                           & 
$<0.32$                                           \\

458                                               & 
$e^+ e^-$                                         & 
$<0.083$                                          & 
{$<0.113$}                                        & 
$<0.19$                                           & 
$<0.083$                                          & 
\nodata                                           & 
\nodata                                           & 
$<0.083$                                          \\

459                                               & 
$e^+ e^- \gamma$                                  & 
$<0.12$                                           & 
{$<0.12$}                                         & 
\nodata                                           & 
\nodata                                           & 
\nodata                                           & 
\nodata                                           & 
{$<0.12$}                                         \\

460                                               & 
$\mu^+ \mu^-$                                     & 
$<0.00063$                                        & 
{$<0.052$}                                        & 
$<0.16$                                           & 
$<0.0038$                                         & 
$<0.00063$                                        & 
$<0.00092$                                        & 
$<0.00063$                                        \\

460                                               & 
$\mu^+\mu^-$                                      & 
CMS-LHCb comb.                                    & 
\nodata                                           & 
\nodata                                           & 
\nodata                                           & 
\blue{$\cerr{0.00039}{0.00016}{0.00014}$}         & 
\blue{$\cerr{0.00039}{0.00016}{0.00014}$}         & 
                                                  \\

461                                               & 
$\mu^+ \mu^- \gamma$                              & 
$<0.16$                                           & 
{$<0.16$}                                         & 
\nodata                                           & 
\nodata                                           & 
\nodata                                           & 
\nodata                                           & 
{$<0.16$}                                         \\

462                                               & 
$\mu^+ \mu^- \mu^+ \mu^-$                         & 
$<0.0053$                                         & 
\nodata                                           & 
\nodata                                           & 
\nodata                                           & 
$<0.0053$                                         & 
\nodata                                           & 
$<0.0053$                                         \\

464                                               & 
$\tau^+ \tau^-$                                   & 
$<4100$                                           & 
{$<4100$}                                         & 
\nodata                                           & 
\nodata                                           & 
\nodata                                           & 
\nodata                                           & 
{$<4100$}                                         \\

482                                               & 
$e^\pm \mu^\mp$                                   & 
$<0.0028$                                         & 
{$<0.092$}                                        & 
$<0.17$                                           & 
$<0.064$                                          & 
$<0.0028$                                         & 
\nodata                                           & 
$<0.0028$                                         \\

488                                               & 
$e^\pm \tau^\mp$                                  & 
$<28$                                             & 
{$<28$}                                           & 
\nodata                                           & 
\nodata                                           & 
\nodata                                           & 
\nodata                                           & 
{$<28$}                                           \\

489                                               & 
$\mu^\pm \tau^\mp$                                & 
$<22$                                             & 
{$<22$}                                           & 
\nodata                                           & 
\nodata                                           & 
\nodata                                           & 
\nodata                                           & 
{$<22$}                                           \\

490                                               & 
$\nu \bar\nu$                                     & 
$<24$                                             & 
$<24$                                             & 
$<130$                                            & 
\nodata                                           & 
\nodata                                           & 
\nodata                                           & 
$<24$                                             \\

491                                               & 
$\nu \bar\nu \gamma$                              & 
$<17$                                             & 
$<17$                                             & 
\nodata                                           & 
\nodata                                           & 
\nodata                                           & 
\nodata                                           & 
$<17$                                             \\


\sglinespt
\end{tabular}
}
\end{center}
\dag~More recent results exist, with hadronic tagging (Ref.~\cite{Yook:2014kga}).
It does not improve the limits ({\blue{$<3.4$} and \blue{$<2.7$}} for $e^+\nu$ and $\mu^+\nu$, respectively).\\
\ddag~The authors average their results with earlier results
from \babar~\cite{Aubert:2009wt} and Belle~\cite{Hara:2010dk}.
\end{table}


\begin{table}
\begin{center}

\caption{Relative branching fractions (BF) of semileptonic and radiative 
$\Bp$ decays. Values in \red{red} (\blue{blue}) are new \red{published}
(\blue{preliminary}) results since PDG2014.}
\label{tab:radll_rel}
\resizebox{\textwidth}{!}{
\begin{tabular}{|lcccccc|} \hline
 RPP\# &Mode & PDG2012 Avg. &  CDF & D\O\ &  LHCb & New avg.  \\ \sglinespb
$~-$                                              & 
$10^4\times\mathcal{B}(B^+ \to K^+\pi^+\pi^-\mu^+\mu^-)/\mathcal{B}(B^+ \to \psi(2S)K^+)$& 
New                                               & 
\nodata                                           & 
\nodata                                           & 
\red{$\aerr{6.95}{0.46}{0.43}{0.34}$}             & 
$\cerr{6.95}{0.57}{0.55}$                         \\

$~-$                                              & 
$10^4\times\mathcal{B}(B^+ \to K^+\phi\mu^+\mu^-)/\mathcal{B}(B^+ \to \psi(2S)K^+)$& 
New                                               & 
\nodata                                           & 
\nodata                                           & 
\red{$\aerrsy{1.58}{0.36}{0.32}{0.19}{0.07}$}     & 
$\cerr{1.58}{0.41}{0.33}$                         \\

469                                               & 
$\mathcal{B}(\pi^+\mu^+\mu^-)/\mathcal{B}(K^+\mu^+\mu^-)$& 
$\err{0.053}{0.014}{0.01}$                        & 
\nodata                                           & 
\nodata                                           & 
$\err{0.053}{0.014}{0.01}$                        & 
$\err{0.053}{0.014}{0.01}$                        \\

\hline
\end{tabular}
}
\end{center}
\end{table}


\begin{table}
\begin{center}

\caption{Branching fractions (BF) of $B\to\bar{b}\to\bar{q}$ gluon decays in units of $\times 10^{-6}$. Upper limits are
at 90\% CL. Values in \red{red} (\blue{blue}) are new \red{published}
(\blue{preliminary}) results since PDG2014.}
\label{tab:radll_gluon}
\resizebox{\textwidth}{!}{
\begin{tabular}{|lcccccc|} \hline
 RPP\# & Mode & PDG2014 Avg. & \babar  & Belle & CLEO & New Avg.  \\
\sglinespb
$~80$                                             & 
$\eta X$                                          & 
$\cerr{260}{50}{80}$                              & 
\nodata                                           & 
{\berr{261}{30}{44}{74}}~\S                       & 
$<440$                                            & 
$\cerr{261}{53}{79}$                              \\

$~81$                                             & 
$\eta' X$                                         & 
$420 \pm 90$                                      & 
\err{390}{80}{90}~\dag                            & 
\nodata                                           & 
\err{460}{110}{60}~\dag                           & 
$423 \pm 86$                                      \\

$~82$                                             & 
$K^+ X$                                           & 
$<187$                                            & 
$<187$~\ddag                                      & 
\nodata                                           & 
\nodata                                           & 
$<187$~\ddag                                      \\

$~83$                                             & 
$K^0 X$                                           & 
\cerr{195}{71}{67}                                & 
\aerr{195}{51}{45}{50}~\ddag                      & 
\nodata                                           & 
\nodata                                           & 
$\cerr{195}{71}{67}$                              \\

$~94$                                             & 
$\pi^+ X$                                         & 
$370\pm80$                                        & 
{\aerr{372}{50}{47}{59}~\P}                       & 
\nodata                                           & 
\nodata                                           & 
$\cerr{372}{77}{75}$                              \\


\hline
\end{tabular}
}
\end{center}
~\S~$0.4 < m_{X} < 2.6$GeV/$c$;
~~~~\dag~$2.0 < p^*(\eta') < 2.7$GeV/$c$;\\
\ddag~$m_{X} < 1.69$GeV/$c$;
~~~~~~~~~~~\P~$m_{X} < 1.71$GeV/$c$.
\end{table}


\begin{table}
\begin{center}

\caption{Isospin asymmetry in radiative and semileptonic $B$ meson decays.
The notations are those adopted by the PDG.
Values in \red{red} (\blue{blue}) are new \red{published}
(\blue{preliminary}) results since PDG2014.}
\label{tab:radll_AI}
\resizebox{\textwidth}{!}{
\begin{tabular}{|cccccc|}
\sgline
 Parameter & PDG2014 Avg. & \babar & Belle & LHCb & New Avg. \\
\sglinespb
$\Delta_{0^-}(X_s\gamma)$                         & 
$-0.01 \pm 0.06$                                  & 
$-0.01 \pm 0.06$~\S                               & 
\nodata                                           & 
\nodata                                           & 
$-0.01 \pm 0.06$                                  \\

$\Delta_{0^+}(K^* \gamma)$                        & 
$0.052 \pm 0.026$                                 & 
{$\err{0.066}{0.021}{0.022}$}                     & 
{$\err{0.012}{0.044}{0.026}$}                     & 
\nodata                                           & 
$0.012 \pm 0.051$                                 \\

$\Delta_{\rho \gamma}$                            & 
$-0.46 \pm 0.17$                                  & 
{$\aerr{-0.43}{0.25}{0.22}{0.10}$}                & 
{$\aerrsy{-0.48}{0.21}{0.19}{0.08}{0.09}$}        & 
\nodata                                           & 
$\cerr{-0.48}{0.23}{0.21}$                        \\

$\Delta_{0-}(K\ell\ell)~\dag$                     & 
$-0.37 \pm 0.13$                                  & 
{$\aerr{-0.58}{0.29}{0.37}{0.02}$}                & 
{$\aerr{-0.31}{0.17}{0.14}{0.08}$}                & 
$\cerr{-0.35}{0.23}{0.27}$                        & 
$-0.32 \pm 0.14$                                  \\

$\Delta_{0-}(K^*\ell\ell)~\dag$                   & 
$-0.22 \pm 0.10$                                  & 
$\aerr{-0.25}{0.20}{0.17}{0.03}$                  & 
$\err{-0.29}{0.16}{0.09}$                         & 
$-0.15 \pm 0.16$                                  & 
$-0.21 \pm 0.12$                                  \\

$\Delta_{0-}(K^{(*)}\ell\ell)~\dag$               & 
$-0.45 \pm 0.17$                                  & 
{$\aerr{-0.64}{0.15}{0.14}{0.03}$}                & 
{$\aerr{-0.30}{0.12}{0.11}{0.08}$}                & 
\nodata                                           & 
$-0.30 \pm 0.14$                                  \\


\sglinespt
\end{tabular}
}
\end{center}
\vspace{-0.3cm}
\hspace*{0.1cm} \dag~See the references for precise $q^2=m^2_{\ell\ell}$ region. In all measurements  $m_{\ell\ell} < m_{J/\psi}$.\\
\hspace*{0.1cm} \S~Average of two independent measurements from \babar.
\end{table}


\begin{table}
\begin{center}

\caption{Isospin asymmetry in $K^{(*)}\ell^+\ell^-$ modes in bins of $q^2=m^2_{\ell\ell}$.
Values in \red{red} (\blue{blue}) are new \red{published}
(\blue{preliminary}) results since PDG2014.}
\label{tab:radll_AI_Kll}
\resizebox{\textwidth}{!}{
\begin{tabular}{|lccccccc|}
\sgline
 Mode &  $q^2~[(\mathrm{GeV}/c^2)^2]$~\dag &
PDG2014 Avg. & \babar & Belle & CDF~\ddag & LHCb~\ddag & New Avg. \\
\sglinespb
$K\ell^+\ell^-$                                   & 
$< 2.0$                                           & 
\nodata                                           & 
\cerr{-0.51}{0.49}{0.95}                          & 
{\cerr{-0.33}{0.34}{0.26}}                        & 
\blue{$\err{0.19}{0.34}{0.05}$}                   & 
\cerr{-0.55}{0.40}{0.56}                          & 
$\cerr{-0.24}{0.18}{0.19}$                        \\

\nodata                                           & 
$[2.0,4.3]$                                       & 
\nodata                                           & 
\cerr{-0.73}{0.48}{0.55}                          & 
{\cerr{-0.47}{0.50}{0.39}}                        & 
\blue{$\err{-0.07}{0.34}{0.07}$}                  & 
\cerr{-0.76}{0.45}{0.79}                          & 
$\cerr{-0.42}{0.20}{0.22}$                        \\

\nodata                                           & 
$[4.3,8.68]$                                      & 
\nodata                                           & 
\cerr{-0.32}{0.27}{0.30}                          & 
{\cerr{-0.19}{0.26}{0.22}}                        & 
\blue{$\err{-0.20}{0.26}{0.08}$}                  & 
\cerr{0.00}{0.14}{0.15}                           & 
$-0.11 \pm 0.11$                                  \\

\nodata                                           & 
$[10.09,12.86]$                                   & 
\nodata                                           & 
\cerr{-0.05}{0.25}{0.29}                          & 
{\cerr{-0.29}{0.37}{0.29}}                        & 
\blue{$\err{-0.27}{0.37}{0.08}$}                  & 
\cerr{-0.15}{0.19}{0.22}                          & 
$\cerr{-0.16}{0.14}{0.15}$                        \\

\nodata                                           & 
$[14.18,16.00]$                                   & 
\nodata                                           & 
\cerr{0.05}{0.32}{0.43}                           & 
{\cerr{-0.40}{0.61}{0.69}}                        & 
\blue{$\err{0.04}{0.23}{0.05}$}                   & 
$-0.40 \pm 0.22$                                  & 
$\cerr{-0.17}{0.14}{0.15}$                        \\

\nodata                                           & 
$>16.00$                                          & 
\nodata                                           & 
\cerr{-0.93}{0.83}{4.99}                          & 
{\cerr{0.11}{0.25}{0.22}}                         & 
\blue{$\err{-0.29}{0.28}{0.06}$}                  & 
\cerr{-0.52}{0.18}{0.22}                          & 
$\cerr{-0.28}{0.12}{0.13}$                        \\

\nodata                                           & 
$[1.00,6.00]$                                     & 
\nodata                                           & 
\cerr{-0.41}{0.25}{0.01}                          & 
{\cerr{-0.41}{0.26}{0.21}}                        & 
\blue{$\err{-0.06}{0.24}{0.07}$}                  & 
\cerr{-0.35}{0.23}{0.27}                          & 
$-0.30 \pm 0.12$                                  \\

$K\ell^+\ell^-$~\S                                & 
$0.1-2.0$                                         & 
\nodata                                           & 
\nodata                                           & 
\nodata                                           & 
\nodata                                           & 
\red{$\aerr{-0.37}{0.18}{0.21}{0.02}$}            & 
$\cerr{-0.37}{0.18}{0.21}$                        \\

\nodata                                           & 
$2.0-4.0$                                         & 
\nodata                                           & 
\nodata                                           & 
\nodata                                           & 
\nodata                                           & 
\red{$\aerr{-0.15}{0.13}{0.15}{0.02}$}            & 
$\cerr{-0.15}{0.13}{0.15}$                        \\

\nodata                                           & 
$4.0-6.0$                                         & 
\nodata                                           & 
\nodata                                           & 
\nodata                                           & 
\nodata                                           & 
\red{$\aerr{-0.10}{0.13}{0.16}{0.02}$}            & 
$\cerr{-0.10}{0.13}{0.16}$                        \\

\nodata                                           & 
$6.0-8.0$                                         & 
\nodata                                           & 
\nodata                                           & 
\nodata                                           & 
\nodata                                           & 
\red{$\aerr{0.09}{0.10}{0.11}{0.02}$}             & 
$\cerr{0.09}{0.10}{0.11}$                         \\

\nodata                                           & 
$11.0-12.5$                                       & 
\nodata                                           & 
\nodata                                           & 
\nodata                                           & 
\nodata                                           & 
\red{$\aerr{-0.16}{0.15}{0.18}{0.03}$}            & 
$\cerr{-0.16}{0.15}{0.18}$                        \\

\nodata                                           & 
$15.0-17.0$                                       & 
\nodata                                           & 
\nodata                                           & 
\nodata                                           & 
\nodata                                           & 
\red{$\aerr{-0.04}{0.11}{0.13}{0.02}$}            & 
$\cerr{-0.04}{0.11}{0.13}$                        \\

\nodata                                           & 
$17.0-22.0$                                       & 
\nodata                                           & 
\nodata                                           & 
\nodata                                           & 
\nodata                                           & 
\red{$\aerr{-0.12}{0.10}{0.11}{0.02}$}            & 
$\cerr{-0.12}{0.10}{0.11}$                        \\

\nodata                                           & 
$1.1-6.0$                                         & 
\nodata                                           & 
\nodata                                           & 
\nodata                                           & 
\nodata                                           & 
\red{$\aerr{-0.10}{0.08}{0.09}{0.02}$}            & 
$\cerr{-0.10}{0.08}{0.09}$                        \\

\nodata                                           & 
$15.0-22.0$                                       & 
\nodata                                           & 
\nodata                                           & 
\nodata                                           & 
\nodata                                           & 
\red{$\aerr{-0.09}{0.08}{0.08}{0.02}$}            & 
$-0.09 \pm 0.08$                                  \\

$K^{*}\ell^+\ell^-$                               & 
$< 2.0$                                           & 
\nodata                                           & 
\cerr{-0.17}{0.29}{0.24}                          & 
{\cerr{-0.67}{0.19}{0.17}}                        & 
\blue{$\err{0.15}{0.32}{0.06}$}                   & 
\cerr{0.05}{0.27}{0.21}                           & 
$\cerr{-0.25}{0.12}{0.11}$                        \\

\nodata                                           & 
$[2.0,4.3]$                                       & 
\nodata                                           & 
\cerr{-0.06}{0.56}{0.36}                          & 
{\cerr{1.45}{1.04}{1.15}}                         & 
\blue{$\err{0.00}{0.39}{0.07}$}                   & 
\cerr{-0.27}{0.29}{0.18}                          & 
$\cerr{-0.12}{0.23}{0.17}$                        \\

\nodata                                           & 
$[4.3,8.68]$                                      & 
\nodata                                           & 
\cerr{0.03}{0.43}{0.32}                           & 
{\cerr{-0.34}{0.32}{0.30}}                        & 
\blue{$\err{0.29}{0.41}{0.13}$}                   & 
\cerr{-0.06}{0.19}{0.14}                          & 
$\cerr{-0.06}{0.14}{0.11}$                        \\

\nodata                                           & 
$[10.09,12.86]$                                   & 
\nodata                                           & 
\cerr{-0.48}{0.23}{0.19}                          & 
{\cerr{0.00}{0.22}{0.23}}                         & 
\blue{$\err{0.43}{0.35}{0.10}$}                   & 
\cerr{-0.16}{0.17}{0.16}                          & 
$-0.14 \pm 0.11$                                  \\

\nodata                                           & 
$[14.18,16.00]$                                   & 
\nodata                                           & 
\cerr{0.24}{0.61}{0.39}                           & 
{\cerr{0.16}{0.31}{0.36}}                         & 
\blue{$\err{0.17}{0.29}{0.07}$}                   & 
\cerr{0.02}{0.23}{0.21}                           & 
$\cerr{0.11}{0.15}{0.14}$                         \\

\nodata                                           & 
$>16.00$                                          & 
\nodata                                           & 
\cerr{1.07}{4.28}{1.01}                           & 
{\cerr{-0.02}{0.22}{0.23}}                        & 
\blue{$\err{-0.23}{0.23}{0.06}$}                  & 
\cerr{0.02}{0.21}{0.20}                           & 
$-0.05 \pm 0.13$                                  \\

\nodata                                           & 
$[1.00,6.00]$                                     & 
\nodata                                           & 
\cerr{-0.20}{0.30}{0.23}                          & 
{\cerr{0.33}{0.38}{0.44}}                         & 
\blue{$\err{-0.26}{0.21}{0.07}$}                  & 
$-0.15 \pm 0.16$                                  & 
$\cerr{-0.16}{0.12}{0.11}$                        \\

$K^{*}\ell^+\ell^-$~\S                            & 
$0.1-2.0$                                         & 
\nodata                                           & 
\nodata                                           & 
\nodata                                           & 
\nodata                                           & 
\red{$\aerr{0.11}{0.12}{0.11}{0.02}$}             & 
$\cerr{0.11}{0.12}{0.11}$                         \\

\nodata                                           & 
$2.0-4.0$                                         & 
\nodata                                           & 
\nodata                                           & 
\nodata                                           & 
\nodata                                           & 
\red{$\aerr{-0.20}{0.15}{0.12}{0.03}$}            & 
$\cerr{-0.20}{0.15}{0.12}$                        \\

\nodata                                           & 
$4.0-6.0$                                         & 
\nodata                                           & 
\nodata                                           & 
\nodata                                           & 
\nodata                                           & 
\red{$\aerr{0.23}{0.21}{0.18}{0.02}$}             & 
$\cerr{0.23}{0.21}{0.18}$                         \\

\nodata                                           & 
$6.0-8.0$                                         & 
\nodata                                           & 
\nodata                                           & 
\nodata                                           & 
\nodata                                           & 
\red{$\aerr{0.19}{0.17}{0.15}{0.02}$}             & 
$\cerr{0.19}{0.17}{0.15}$                         \\

\nodata                                           & 
$11.0-12.5$                                       & 
\nodata                                           & 
\nodata                                           & 
\nodata                                           & 
\nodata                                           & 
\red{$\aerr{-0.25}{0.09}{0.08}{0.03}$}            & 
$\cerr{-0.25}{0.10}{0.09}$                        \\

\nodata                                           & 
$15.0-17.0$                                       & 
\nodata                                           & 
\nodata                                           & 
\nodata                                           & 
\nodata                                           & 
\red{$\aerr{-0.10}{0.10}{0.09}{0.03}$}            & 
$-0.10 \pm 0.10$                                  \\

\nodata                                           & 
$17.0-19.0$                                       & 
\nodata                                           & 
\nodata                                           & 
\nodata                                           & 
\nodata                                           & 
\red{$\aerr{0.51}{0.29}{0.24}{0.02}$}             & 
$\cerr{0.51}{0.29}{0.24}$                         \\

\nodata                                           & 
$1.1-6.0$                                         & 
\nodata                                           & 
\nodata                                           & 
\nodata                                           & 
\nodata                                           & 
\red{$\aerr{0.00}{0.12}{0.10}{0.02}$}             & 
$\cerr{0.00}{0.12}{0.10}$                         \\

\nodata                                           & 
$15.0-19.0$                                       & 
\nodata                                           & 
\nodata                                           & 
\nodata                                           & 
\nodata                                           & 
\red{$\aerr{0.06}{0.10}{0.09}{0.02}$}             & 
$\cerr{0.06}{0.10}{0.09}$                         \\


\sglinespt
\end{tabular}
}
\end{center}
\vspace{-0.3cm}
\dag~See the papers for the exact $q^2=M^2(\mu^+\mu^-)$ selection.\\
\ddag~Muon mode only ($\ell = \mu$).\\
\S~Results in two different sets of $q^2$ bins are available.
\end{table}


\begin{sidewaystable}
\begin{center}

\caption{Forward-backward asymmetry ($A_\mathrm{FB}$) in $K^{(*)}\ell^+\ell^-$ modes in bins of $q^2=m^2_{\ell\ell}$.
Values in \red{red} (\blue{blue}) are new \red{published}
(\blue{preliminary}) results since PDG2014.}
\label{tab:radll_AFB_Kll}

\hspace*{-1.5cm}
\tiny
\begin{tabular}{|lcccccccc|}
\sgline
 Mode & $q^2~[(\mathrm{GeV}/c^2)^2]$~\dag & PDG2014 Avg. & Belle & CDF~\ddag & LHCb~\ddag & CMS~\ddag & ATLAS~\ddag & New Avg. \\
\sglinespb
$K\ell^+\ell^-$                                   & 
$< 2.0$                                           & 
$\cerr{0.00}{0.06}{0.05}$                         & 
{$\aerr{0.06}{0.32}{0.35}{0.02}$}                 & 
\blue{$\aerr{-0.19}{0.37}{0.45}{0.09}$}           & 
$\aerrsy{0.00}{0.06}{0.05}{0.03}{0.01}$           & 
\nodata                                           & 
\nodata                                           & 
$\cerr{-0.00}{0.06}{0.05}$                        \\

\nodata                                           & 
$[2.0,4.3]$                                       & 
$\cerr{0.09}{0.10}{0.07}$                         & 
{$\aerr{-0.43}{0.38}{0.40}{0.09}$}                & 
\blue{$\aerr{0.32}{0.17}{0.13}{0.10}$}            & 
$\aerrsy{0.07}{0.08}{0.05}{0.02}{0.01}$           & 
\nodata                                           & 
\nodata                                           & 
$\cerr{0.09}{0.08}{0.06}$                         \\

\nodata                                           & 
$[4.3,8.68]$                                      & 
$\cerr{-0.04}{0.04}{0.05}$                        & 
{$\aerr{-0.20}{0.12}{0.14}{0.03}$}                & 
\blue{$\aerr{0.08}{0.08}{0.09}{0.01}$}            & 
$\aerr{-0.02}{0.03}{0.05}{0.03}$                  & 
\nodata                                           & 
\nodata                                           & 
$\cerr{-0.02}{0.04}{0.05}$                        \\

\nodata                                           & 
$[10.09,12.86]$                                   & 
$-0.05 \pm 0.06$                                  & 
{$\aerr{-0.21}{0.17}{0.15}{0.06}$}                & 
\blue{$\aerr{-0.04}{0.12}{0.10}{0.03}$}           & 
$\err{-0.03}{0.07}{0.01}$                         & 
\nodata                                           & 
\nodata                                           & 
$-0.05 \pm 0.06$                                  \\

\nodata                                           & 
$[14.18,16.00]$                                   & 
$\cerr{-0.02}{0.07}{0.05}$                        & 
{$\aerr{0.04}{0.13}{0.16}{0.05}$}                 & 
\blue{$\aerr{-0.07}{0.08}{0.08}{0.01}$}           & 
$\aerr{-0.01}{0.12}{0.06}{0.01}$                  & 
\nodata                                           & 
\nodata                                           & 
$\cerr{-0.03}{0.06}{0.04}$                        \\

\nodata                                           & 
$[16.0,18.0]$                                     & 
$\cerr{-0.09}{0.07}{0.09}$                        & 
\nodata                                           & 
\nodata                                           & 
$\aerrsy{-0.09}{0.07}{0.09}{0.02}{0.01}$          & 
\nodata                                           & 
\nodata                                           & 
$\cerr{-0.09}{0.07}{0.09}$                        \\

\nodata                                           & 
$[18.0,22.0]$                                     & 
$0.02 \pm 0.11$                                   & 
\nodata                                           & 
\nodata                                           & 
$\err{0.02}{0.11}{0.01}$                          & 
\nodata                                           & 
\nodata                                           & 
$0.02 \pm 0.11$                                   \\

\nodata                                           & 
$>16.00$                                          & 
$\cerr{0.04}{0.09}{0.07}$                         & 
{$\aerr{0.02}{0.11}{0.08}{0.02}$}                 & 
\blue{$\aerr{0.05}{0.18}{0.10}{0.05}$}            & 
\nodata                                           & 
\nodata                                           & 
\nodata                                           & 
$\cerr{0.03}{0.09}{0.07}$                         \\

\nodata                                           & 
$[1.00,6.00]$                                     & 
$\cerr{0.034}{0.040}{0.029}$                      & 
{$\aerr{-0.04}{0.13}{0.16}{0.05}$}                & 
\blue{$\aerr{0.13}{0.11}{0.10}{0.02}$}            & 
$\aerrsy{0.02}{0.05}{0.03}{0.02}{0.01}$           & 
\nodata                                           & 
\nodata                                           & 
$\cerr{0.03}{0.05}{0.03}$                         \\

$K^*\ell^+\ell^-$                                 & 
$< 2.0$                                           & 
$-0.01 \pm 0.14$                                  & 
{$\aerr{0.47}{0.26}{0.32}{0.03}$}                 & 
\blue{$\aerr{0.05}{0.28}{0.27}{0.10}$}            & 
$\err{-0.02}{0.12}{0.01}$                         & 
\nodata                                           & 
\nodata                                           & 
$0.04 \pm 0.11$                                   \\

\nodata                                           & 
$[1.0,2.0]$                                       & 
$\cerr{0.45}{0.26}{0.30}$                         & 
\nodata                                           & 
\nodata                                           & 
\nodata                                           & 
$\aerr{-0.29}{0.37}{0.00}{0.18}$                  & 
\nodata                                           & 
$\cerr{-0.29}{0.41}{0.18}$                        \\

\nodata                                           & 
$[2.0,4.3]$                                       & 
$-0.15 \pm 0.07$                                  & 
{$\aerr{0.11}{0.31}{0.36}{0.07}$}                 & 
\blue{$\aerr{-0.11}{0.34}{0.41}{0.16}$}           & 
$\err{-0.20}{0.08}{0.01}$                         & 
$\err{-0.07}{0.20}{0.02}$                         & 
\blue{$\err{0.22}{0.28}{0.14}$}                   & 
$-0.15 \pm 0.07$                                  \\

\nodata                                           & 
$[4.3,8.68]$                                      & 
$\cerr{0.13}{0.05}{0.06}$                         & 
{$\aerr{0.45}{0.15}{0.21}{0.15}$}                 & 
\blue{$\aerr{0.09}{0.14}{0.14}{0.04}$}            & 
$\aerr{0.16}{0.06}{0.05}{0.01}$                   & 
$\err{-0.01}{0.11}{0.03}$                         & 
\blue{$\err{0.24}{0.13}{0.01}$}                   & 
$0.15 \pm 0.04$                                   \\

\nodata                                           & 
$[10.09,12.86]$                                   & 
$0.34 \pm 0.05$                                   & 
{$\aerr{0.43}{0.18}{0.20}{0.03}$}                 & 
\blue{$\aerr{0.44}{0.12}{0.13}{0.08}$}            & 
$\aerr{0.28}{0.07}{0.06}{0.02}$                   & 
$\err{0.40}{0.08}{0.05}$                          & 
\blue{$\err{0.09}{0.09}{0.03}$}                   & 
$\cerr{0.29}{0.05}{0.04}$                         \\

\nodata                                           & 
$[14.18,16.00]$                                   & 
$\cerr{0.47}{0.07}{0.06}$                         & 
{$\aerr{0.70}{0.16}{0.22}{0.10}$}                 & 
\blue{$\aerr{0.53}{0.09}{0.09}{0.07}$}            & 
$\aerr{0.51}{0.07}{0.05}{0.02}$                   & 
$\err{0.29}{0.09}{0.05}$                          & 
\blue{$\err{0.48}{0.19}{0.05}$}                   & 
$0.48 \pm 0.04$                                   \\

\nodata                                           & 
$[16.0,19.0]$                                     & 
$0.40 \pm 0.06$                                   & 
{$\aerr{0.66}{0.11}{0.16}{0.04}$}                 & 
\blue{$\aerr{0.35}{0.17}{0.19}{0.06}$}            & 
$\berr{0.30}{0.08}{0.01}{0.02}$                   & 
$\err{0.41}{0.05}{0.03}$                          & 
\blue{$\err{0.16}{0.10}{0.03}$}                   & 
$0.36 \pm 0.04$                                   \\

\nodata                                           & 
$[1.00,6.00]$                                     & 
$-0.12 \pm 0.07$                                  & 
{$\aerr{0.26}{0.27}{0.30}{0.07}$}                 & 
\blue{$\aerr{0.19}{0.17}{0.21}{0.05}$}            & 
$\err{-0.17}{0.06}{0.01}$                         & 
$\err{-0.07}{0.12}{0.01}$                         & 
\blue{$\err{0.07}{0.20}{0.07}$}                   & 
$-0.11 \pm 0.05$                                  \\


\sglinespt
\end{tabular}
\end{center}
\vspace{-0.3cm}
\dag~see the original papers for the exact $q^2$ selection.\\
\ddag~muon mode only ($\ell = \mu$).
\end{sidewaystable}


\begin{sidewaystable}
\begin{center}

\caption{Fraction of the Longitudinal Polarization ($F_L$) in $K^{(*)}\ell^+\ell^-$ modes in bins of $q^2=m^2_{\ell\ell}$.
Values in \red{red} (\blue{blue}) are new \red{published}
(\blue{preliminary}) results since PDG2014.}
\label{tab:radll_FL_Kll}

\scriptsize
\begin{tabular}{|lcccccccc|}
\sgline
 Mode & $q^2~[(\mathrm{GeV}/c^2)^2]$~\dag &
PDG2014 Avg. & Belle & CDF~\ddag & LHCb~\ddag & CMS~\ddag & ATLAS~\ddag & New Avg. \\
\sglinespb

$K^*\ell^+\ell^-$                                 & 
$< 2.0$                                           & 
$\cerr{0.34}{0.08}{0.07}$                         & 
{$\aerr{0.29}{0.21}{0.18}{0.02}$}                 & 
\blue{$\aerr{0.25}{0.14}{0.13}{0.04}$}            & 
$\cerr{0.37}{0.11}{0.09}$                         & 
\nodata                                           & 
\nodata                                           & 
$\cerr{0.33}{0.08}{0.07}$                         \\

\nodata                                           & 
$[1.0,2.0]$                                       & 
$\aerr{0.60}{0.00}{0.28}{0.19}$                   & 
\nodata                                           & 
\nodata                                           & 
\nodata                                           & 
$\aerr{0.60}{0.00}{0.28}{0.19}$                   & 
\nodata                                           & 
$\cerr{0.60}{0.19}{0.34}$                         \\

\nodata                                           & 
$[2.0,4.3]$                                       & 
$0.69 \pm 0.08$                                   & 
{$0.71 \pm 0.24 \pm 0.05$}                        & 
\blue{$\aerr{0.71}{0.15}{0.17}{0.07}$}            & 
$\cerr{0.74}{0.10}{0.09}$                         & 
$\err{0.65}{0.17}{0.03}$                          & 
\blue{$\err{0.26}{0.18}{0.06}$}                   & 
$0.66 \pm 0.07$                                   \\

\nodata                                           & 
$[4.3,8.68]$                                      & 
$0.64 \pm 0.06$                                   & 
{$\aerr{0.64}{0.23}{0.24}{0.07}$}                 & 
\blue{$\aerr{0.72}{0.12}{0.13}{0.05}$}            & 
$\err{0.57}{0.07}{0.03}$                          & 
$\aerr{0.81}{0.13}{0.12}{0.05}$                   & 
\blue{$\err{0.37}{0.11}{0.02}$}                   & 
$0.59 \pm 0.05$                                   \\

\nodata                                           & 
$[10.09,12.86]$                                   & 
$0.43 \pm 0.06$                                   & 
{$\aerr{0.17}{0.17}{0.15}{0.03}$}                 & 
\blue{$\aerr{0.38}{0.11}{0.11}{0.04}$}            & 
$\aerr{0.48}{0.08}{0.09}{0.03}$                   & 
$\aerr{0.45}{0.10}{0.11}{0.04}$                   & 
\blue{$\err{0.50}{0.09}{0.04}$}                   & 
$0.44 \pm 0.05$                                   \\

\nodata                                           & 
$[14.18,16.00]$                                   & 
$0.35 \pm 0.08$                                   & 
{$\aerr{-0.15}{0.27}{0.23}{0.07}$}                & 
\blue{$\aerr{0.40}{0.11}{0.11}{0.04}$}            & 
$\aerr{0.33}{0.08}{0.07}{0.02}$                   & 
$\err{0.53}{0.12}{0.03}$                          & 
\blue{$\err{0.28}{0.16}{0.03}$}                   & 
$\cerr{0.36}{0.06}{0.05}$                         \\

\nodata                                           & 
$[16.0,19.0]$                                     & 
$0.37 \pm 0.06$                                   & 
{$\aerr{0.12}{0.15}{0.13}{0.02}$}                 & 
\blue{$\aerr{0.19}{0.12}{0.11}{0.07}$}            & 
$\aerr{0.38}{0.09}{0.08}{0.03}$                   & 
$\err{0.44}{0.07}{0.03}$                          & 
\blue{$\err{0.35}{0.08}{0.02}$}                   & 
$0.35 \pm 0.04$                                   \\

\nodata                                           & 
$[1.00,6.00]$                                     & 
$0.66 \pm 0.06$                                   & 
{$\err{0.67}{0.23}{0.05}$}                        & 
\blue{$\aerr{0.76}{0.12}{0.14}{0.07}$}            & 
$\aerr{0.65}{0.08}{0.07}{0.03}$                   & 
$\err{0.68}{0.10}{0.02}$                          & 
\blue{$\err{0.18}{0.15}{0.03}$}                   & 
$0.62 \pm 0.05$                                   \\


\sglinespt
\end{tabular}
\end{center}
\vspace{-0.3cm}
\dag~see the original paper for the exact $q^2$ selection.\\
\ddag~muon mode only ($\ell = \mu$).
\end{sidewaystable}


\begin{table}
\begin{center}

\caption{Up-down Asymmetry in $B^+ \to K^+ \pi^- \pi^+ \gamma$ decays in bins of $m_{K^+ \pi^- \pi^+}$.
Values in \red{red} (\blue{blue}) are new \red{published}
(\blue{preliminary}) results since PDG2014.}
\label{tab:radll_Aud}

\hspace*{-1.0cm}
\scriptsize
\begin{tabular}{|lcccc|}
\sgline
 Mode & $m_{K^+ \pi^- \pi^+}~[(\mathrm{GeV}/c^2)]$ & PDG2014 Avg. & LHCb &  New Avg. \\
\sglinespb
$K^+ \pi^- \pi^+ \gamma$                          & 
1.1-1.3                                           & 
\nodata                                           & 
\red{$6.9 \pm 1.7$}                               & 
$6.9 \pm 1.7$                                     \\

\nodata                                           & 
1.3-1.4                                           & 
\nodata                                           & 
\red{$4.9 \pm 2.0$}                               & 
$4.9 \pm 2.0$                                     \\

\nodata                                           & 
1.4-1.6                                           & 
\nodata                                           & 
\red{$5.6 \pm 1.8$}                               & 
$5.6 \pm 1.8$                                     \\

\nodata                                           & 
1.6-1.9                                           & 
\nodata                                           & 
\red{$-4.5 \pm 1.9$}                              & 
$-4.5 \pm 1.9$                                    \\

\nodata                                           & 
1.1-1.3                                           & 
\nodata                                           & 
\red{$-1.1 \pm 1.7$}                              & 
$-1.1 \pm 1.7$                                    \\

\nodata                                           & 
1.3-1.4                                           & 
\nodata                                           & 
\red{$7.2 \pm 2.0$}                               & 
$7.2 \pm 2.0$                                     \\

\nodata                                           & 
1.4-1.6                                           & 
\nodata                                           & 
\red{$6.4 \pm 1.8$}                               & 
$6.4 \pm 1.8$                                     \\

\nodata                                           & 
1.6-1.9                                           & 
\nodata                                           & 
\red{$-3.9 \pm 1.9$}                              & 
$-3.9 \pm 1.9$                                    \\

\sglinespt
\end{tabular}
\end{center}
\vspace{-0.3cm}
\end{table}

\clearpage

\mysubsection{Charge asymmetries in \b-hadron decays}
\label{sec:rare-acp}

This section contains, in Tables~\ref{tab:acp_Bp} to~\ref{tab:acp_Lb}
compilations of \CP\ asymmetries in decays of different \b-hadrons: \Bp, \Bz
mesons, $\Bpm/\Bz$ admixtures, \Bs mesons and finally \Lb baryons.
Measurements of time-dependent \CP\ asymmetries are not listed here but are
discussed in Sec.~\ref{sec:cp_uta}.
In addition, we do not attempt to include the model-independent studies of
$CP$ violation across the phase space of the three body decays $\Bp \to
\Kp\Km\Kp$, $\Kp\pim\pip$, $\Kp\Km\pip$ and $\pip\pim\pip$, where large
effects have been observed by LHCb~\cite{Aaij:2013sfa,Aaij:2013bla,Aaij:2014iva}.
The results, in terms of $CP$ asymmetries of quasi-two-body final states, of
model-dependent analyses, and of the $CP$ asymmetries for the inclusive
three-body final states, are however included.

\begin{table}
\begin{center}
%
%
\caption{\CP\ asymmetries of charmless hadronic $\Bp$ decays.
Values in \red{red} (\blue{blue}) are new \red{published}
(\blue{preliminary}) results since PDG2014.}
\label{tab:acp_Bp}
\resizebox{0.95\textwidth}{!}{

}
\end{center}
{\tiny 
\dag~PDG takes the value from the \babar amplitude analysis of $B^+ \to K^+ K^- K^+$ decays, while our numbers are from amplitude analyses of $B^+ \to K^+ \pi^- \pi^+$;~~~
\ddag~PDG swaps the BELLE results corresponding to $A_{C\!P}(p\bar{p}\pi^+)$ and $A_{C\!P}(p\bar{p}K^+)$;~~~
\S~PDG uses also a previous result from \babar (\cite{Aubert:2008ps});~~~
$^{(1)}$~PDG uses also a result from CLEO.
}
\end{table}


\begin{table}
\begin{center}
%

\caption{\CP\ asymmetries of charmless hadronic $\Bz$ decays.
Values in \red{red} (\blue{blue}) are new \red{published}
(\blue{preliminary}) results since PDG2014.}
\label{tab:acp_Bz}
\resizebox{\textwidth}{!}{

}
\vskip 0.5cm\end{center}

Measurements of time-dependent \CP\ asymmetries are listed in Sec.~\ref{sec:cp_uta}.
\\[0.3cm]
\normalsize
\dag~Extracted from measured $\Delta A_{CP}=A_{CP}
(\phi K^{*0})-A_{CP}(J/\psi K^{*0})=
\red{\err{0.015}{0.032}{0.005}}$.\\
$^{(1)}$~PDG uses also a result from CLEO.\\
$^{(2)}$ Average of \babar results from 
$B^0 \to K^+ \pi^- \pi^0$ and $B^0 \to K^0 \pi^+ \pi^-$.\\
$^{(3)}$ PDG quotes the opposite asymmetry.

\end{table}


\begin{table}

\caption{\CP\ asymmetries of charmless hadronic decays of $B^\pm/B^0$ admixture.
Values in \red{red} (\blue{blue}) are new \red{published}
(\blue{preliminary}) results since PDG2014.}
\label{tab:acp_B}
\resizebox{\textwidth}{!}{
\begin{tabular}{|lccccc|}
\hline
  RPP\#    & Mode & PDG2014 Avg. & \babar & Belle & New Avg. \\
\sglinespb
~65                                               & 
$K^* \gamma$                                      & 
$-0.003\pm 0.017$~\ddag                           & 
{$-0.003\pm 0.017\pm 0.007$}                      & 
{$-0.015\pm 0.044\pm 0.012$}                      & 
$-0.005 \pm 0.017$                                \\

~77                                               & 
$s \gamma$                                        & 
$-0.008\pm0.029$                                  & 
\blue{$\err{0.017}{0.019}{0.010}$}~\S             & 
$\err{0.002}{0.050}{0.030}$                       & 
$0.015 \pm 0.020$                                 \\

~80                                               & 
$s \eta$                                          & 
$\cerr{-0.13}{0.04}{0.05}$                        & 
\nodata                                           & 
{\berr{-0.13}{0.04}{0.02}{0.03}}                  & 
$\cerr{-0.13}{0.04}{0.05}$                        \\

~86                                               & 
$\pi^+ X$                                         & 
$0.10 \pm 0.17$                                   & 
{\err{0.10}{0.16}{0.05}}                          & 
\nodata                                           & 
$0.10 \pm 0.17$                                   \\

121                                               & 
$s \ell \ell$                                     & 
$-0.22\pm0.26$                                    & 
\red{$\err{0.04}{0.11}{0.01}$}                    & 
\nodata                                           & 
$0.04 \pm 0.11$                                   \\

126                                               & 
$K^*e^+e^-$                                       & 
$-0.18 \pm 0.15$                                  & 
\nodata                                           & 
$\err{-0.18}{0.15}{0.01}$                         & 
$-0.18 \pm 0.15$                                  \\

128                                               & 
$K^*\mu^+\mu^-$                                   & 
$-0.03 \pm 0.13$                                  & 
\nodata                                           & 
$\err{-0.03}{0.13}{0.02}$                         & 
$-0.03 \pm 0.13$                                  \\

129                                               & 
$K \ell \ell$                                     & 
New                                               & 
\red{$\err{-0.03}{0.14}{0.01}$}                   & 
\nodata                                           & 
$-0.03 \pm 0.14$                                  \\

130                                               & 
$K^* \ell \ell$                                   & 
$-0.04 \pm 0.07$                                  & 
$\err{0.03}{0.13}{0.01}$~\dag                     & 
$\err{-0.10}{0.10}{0.01}$                         & 
$-0.05 \pm 0.08$                                  \\

\hline
\end{tabular}
}
\S~\babar also measures the difference in direct $C\!P$ asymmetry for charged and neutral $B$ mesons: $\Delta A_{\CP}= +(5.0\pm3.9\pm1.5)\%$.\\
\dag~Previous \babar result is also included in the PDG Average.\\
\ddag~PDG include also a result from CLEO.
\end{table}


\begin{table}

\caption{\CP\ asymmetries of charmless hadronic $\Bs$ decays.
Values in \red{red} (\blue{blue}) are new \red{published}
(\blue{preliminary}) results since PDG2014.}
\label{tab:acp_Bs}
\resizebox{\textwidth}{!}{
\begin{tabular}{|lcccccc|}
\hline
  RPP\#    & Mode & PDG2014 Avg. & Belle & CDF & LHCb & New Avg. \\
\sglinespb
~52                                               & 
$\pi^+ K^-$                                       & 
$0.28 \pm 0.04$                                   & 
\nodata                                           & 
\blue{$\err{0.22}{0.07}{0.02}$}                   & 
$\err{0.27}{0.04}{0.01}$                          & 
$0.26 \pm 0.04$                                   \\

\hline
\end{tabular}
}
\end{table}


\begin{table}
\caption{\CP\ asymmetries of charmless hadronic $\Lb$ baryons decays.
Values in \red{red} (\blue{blue}) are new \red{published}
(\blue{preliminary}) results since PDG2014.}
\label{tab:acp_Lb}
\resizebox{\textwidth}{!}{
\begin{tabular}{|lccccc|}
\sgline
  RPP\#   & Mode & PDG2014 Avg. & CDF & LHCb & New Avg. \\
\hline
$~21$                                             & 
$p\pi^-$                                          & 
$0.03 \pm 0.18$                                   & 
\blue{$\err{0.06}{0.07}{0.03}$}                   & 
\nodata                                           & 
$0.06 \pm 0.08$                                   \\

$~22$                                             & 
$p K^-$                                           & 
$0.37 \pm 0.17$                                   & 
\blue{$\err{-0.10}{0.08}{0.04}$}                  & 
\nodata                                           & 
$-0.10 \pm 0.09$                                  \\

$~-$                                              & 
$\kzb p \pi^-$                                    & 
New                                               & 
\nodata                                           & 
\red{$\err{0.22}{0.13}{0.03}$}                    & 
$0.22 \pm 0.13$                                   \\

\sglinespt
\end{tabular}
}
\end{table}

\clearpage

\mysubsection{Polarization measurements in \b-hadron decays}
\label{sec:rare-polar}

In this section, compilations of polarization measurements in \b-hadron decays
are given. Tables~\ref{tab:polar_Bp} (\ref{tab:polar_Bz}) detail measurements
of the longitudinal fraction, $f_L$, in \Bp\ (\Bz) decays, and
Tables~\ref{tab:polar_BpAng} (\ref{tab:polar_BzAng}) the results of the full
angular analyses of $\Bp$ (\Bz) $\to\phi\Kstar$ decays. 
Table~\ref{tab:polar_BzAng1430} gives results of the full angular analysis of
$\Bz\to\phi K_2^{\ast 0}(1430)$ decays. 
Tables~\ref{tab:polar_Bs} to~\ref{tab:polar_BsAng2} detail quantities of \Bs decays: $f_L$ measurements, and observables from full angular analyses of decays to $\phi\phi$ and $\phi\Kstarzb$.

\begin{table}[h!]
\begin{center}
%
\caption{Longitudinal polarization fraction $f_L$ for $B^+$ decays.
Values in \red{red} (\blue{blue}) are new \red{published}
(\blue{preliminary}) results since PDG2014.}
\label{tab:polar_Bp}
\resizebox{\textwidth}{!}{
\begin{tabular}{|lccccc|}
\sgline
  RPP\#   & Mode & PDG2014 Avg. & \babar & Belle & New Avg. \\
\sglinespb
282                                               & 
$\omega K^{*+}$                                   & 
$\err{0.41}{0.18}{0.05}$                          & 
$\err{0.41}{0.18}{0.05}$                          & 
\nodata                                           & 
$0.41 \pm 0.19$                                   \\

285                                               & 
$\omega K_2^*(1430)^+$                            & 
$\err{0.56}{0.10}{0.04}$                          & 
$\err{0.56}{0.10}{0.04}$                          & 
\nodata                                           & 
$0.56 \pm 0.11$                                   \\

312                                               & 
$K^{*+} \rho^0$                                   & 
{$\err{0.78}{0.12}{0.03}$}                        & 
{$\err{0.78}{0.12}{0.03}$}                        & 
\nodata                                           & 
$0.78 \pm 0.12$                                   \\

316                                               & 
$K^{*0} \rho^+$                                   & 
$0.48\pm0.08$                                     & 
{$\err{0.52}{0.10}{0.04}$}                        & 
{$\berr{0.43}{0.11}{0.05}{0.02}$}                 & 
$0.48 \pm 0.08$                                   \\

338                                               & 
$K^{*+} \overline{K}^{*0}$                        & 
$\aerr{0.75}{0.16}{0.26}{0.03}$                   & 
$\aerr{0.75}{0.16}{0.26}{0.03}$                   & 
\nodata                                           & 
$\cerr{0.75}{0.16}{0.26}$                         \\

349                                               & 
$\phi K^{*+}$                                     & 
$0.50\pm0.05$                                     & 
{$\err{0.49}{0.05}{0.03}$}                        & 
{$\err{0.52}{0.08}{0.03}$}                        & 
$0.50 \pm 0.05$                                   \\

351                                               & 
$\phi K_1(1270)^+$                                & 
$\aerrsy{0.46}{0.12}{0.13}{0.06}{0.07}$           & 
$\aerrsy{0.46}{0.12}{0.13}{0.06}{0.07}$           & 
\nodata                                           & 
$\cerr{0.46}{0.13}{0.15}$                         \\

355                                               & 
$\phi K_2^*(1430)^+$                              & 
$\aerr{0.80}{0.09}{0.10}{0.03}$                   & 
$\aerr{0.80}{0.09}{0.10}{0.03}$                   & 
\nodata                                           & 
$0.80 \pm 0.10$                                   \\

391                                               & 
$\rho^+ \rho^0$                                   & 
$0.950\pm0.016$                                   & 
$\err{0.950}{0.015}{0.006}$                       & 
$\err{0.95}{0.11}{0.02}$                          & 
$0.950 \pm 0.016$                                 \\

396                                               & 
$\omega \rho^+$                                   & 
$\err{0.90}{0.05}{0.03}$                          & 
$\err{0.90}{0.05}{0.03}$                          & 
\nodata                                           & 
$0.90 \pm 0.06$                                   \\

\sglinespt
\end{tabular}
}
\end{center}
\end{table}


\begin{table}
\begin{center}

\caption{Longitudinal polarization fraction $f_L$ for $\Bz$ decays.
Values in \red{red} (\blue{blue}) are new \red{published}
(\blue{preliminary}) results since PDG2014.}
\label{tab:polar_Bz}
\resizebox{\textwidth}{!}{
\begin{tabular}{|lcccccc|}
\sgline
  RPP\#   & Mode & PDG2014 Avg. & \babar & Belle & LHCb & New Avg. \\
\hline
246                                               & 
$\omega K^{*0}$                                   & 
$0.69\pm0.13$                                     & 
$\err{0.72}{0.14}{0.02}$                          & 
$\berr{0.56}{0.29}{0.18}{0.08}$                   & 
\nodata                                           & 
$0.70 \pm 0.13$                                   \\

249                                               & 
$\omega K_2^*(1430)^0$                            & 
$\err{0.45}{0.12}{0.02}$                          & 
$\err{0.45}{0.12}{0.02}$                          & 
\nodata                                           & 
\nodata                                           & 
$0.45 \pm 0.12$                                   \\

279                                               & 
$K^{*0} \rho^0$                                   & 
$\err{0.40}{0.08}{0.11}$                          & 
$\err{0.40}{0.08}{0.11}$                          & 
\nodata                                           & 
\nodata                                           & 
$0.40 \pm 0.14$                                   \\

284                                               & 
$K^{*+} \rho^-$                                   & 
$\err{0.38}{0.13}{0.03}$                          & 
$\err{0.38}{0.13}{0.03}$                          & 
\nodata                                           & 
\nodata                                           & 
$0.38 \pm 0.13$                                   \\

312                                               & 
$\phi K^{*0}$                                     & 
$0.497\pm0.025$                                   & 
$\err{0.494}{0.034}{0.013}$                       & 
$\err{0.499}{0.030}{0.018}$                       & 
\red{$\err{0.497}{0.019}{0.015}$}                 & 
$0.497 \pm 0.017$                                 \\

315                                               & 
$K^{*0} \overline{K}^{*0}$                        & 
$\aerr{0.80}{0.10}{0.12}{0.06}$                   & 
{$\aerr{0.80}{0.10}{0.12}{0.06}$}                 & 
\nodata                                           & 
\nodata                                           & 
$\cerr{0.80}{0.12}{0.13}$                         \\

333                                               & 
$\phi K_2^*(1430)^0$                              & 
$\aerr{0.901}{0.046}{0.058}{0.037}$               & 
$\aerr{0.901}{0.046}{0.058}{0.037}$               & 
\nodata                                           & 
\nodata                                           & 
$\cerr{0.901}{0.059}{0.069}$                      \\

386                                               & 
$\rho^0 \rho^0$                                   & 
$\aerr{0.75}{0.11}{0.14}{0.05}$                   & 
{$\aerr{0.75}{0.11}{0.14}{0.05}$}                 & 
\red{$\aerr{0.21}{0.18}{0.22}{0.15}$}             & 
\nodata                                           & 
$0.59 \pm 0.13$                                   \\

394                                               & 
$\rho^+ \rho^-$                                   & 
$\cerr{0.977}{0.028}{0.024}$                      & 
{$\berr{0.992}{0.024}{0.026}{0.013}$}             & 
{$\aerr{0.941}{0.034}{0.040}{0.030}$}             & 
\nodata                                           & 
$\cerr{0.978}{0.025}{0.022}$                      \\

405                                               & 
$a_1^\pm a_1^\mp$                                 & 
$\err{0.31}{0.22}{0.10}$                          & 
{$\err{0.31}{0.22}{0.10}$}                        & 
\nodata                                           & 
\nodata                                           & 
$0.31 \pm 0.24$                                   \\

\sglinespt
\end{tabular}
}
\end{center}
\end{table}


\begin{table}

\caption{
Results of the full angular analyses of $B^+ \to \phi K^{*+}$.
Values in \red{red} (\blue{blue}) are new \red{published}
(\blue{preliminary}) results since PDG2014.}
\label{tab:polar_BpAng}
\begin{center}
\begin{tabular}{|lccccc|}
\sgline
   & Parameter & PDG2014 Avg. & \babar & Belle & New Avg. \\
\sglinespb
\nodata                                           & %
$f_\perp = \Lambda_{\perp\perp}$                  & 
$0.20\pm0.05$                                     & 
{$\err{0.21}{0.05}{0.02}$}                        & 
{$\err{0.19}{0.08}{0.02}$}                        & 
$0.20 \pm 0.05$                                   \\

\nodata                                           & %
$\phi_\parallel$                                  & 
$2.34\pm0.18$                                     & 
{$\err{2.47}{0.20}{0.07}$}                        & 
{$\err{2.10}{0.28}{0.04}$}                        & 
$2.34 \pm 0.17$                                   \\

\nodata                                           & %
$\phi_\perp$                                      & 
$2.58\pm0.17$                                     & 
{$\err{2.69}{0.20}{0.03}$}                        & 
{$\err{2.31}{0.30}{0.07}$}                        & 
$2.58 \pm 0.17$                                   \\

\nodata                                           & %
$\delta_0$                                        & 
{$\err{3.07}{0.18}{0.06}$}                        & 
{$\err{3.07}{0.18}{0.06}$}                        & 
\nodata                                           & 
$3.07 \pm 0.19$                                   \\

\nodata                                           & %
$A_{CP}^0$                                        & 
$\err{0.17}{0.11}{0.02}$                          & 
{$\err{0.17}{0.11}{0.02}$}                        & 
\nodata                                           & 
$0.17 \pm 0.11$                                   \\

\nodata                                           & %
$A_{CP}^\perp$                                    & 
$\err{0.22}{0.24}{0.08}$                          & 
{$\err{0.22}{0.24}{0.08}$}                        & 
\nodata                                           & 
$0.22 \pm 0.25$                                   \\

\nodata                                           & %
$\Delta\phi_\parallel$                            & 
$\err{0.07}{0.20}{0.05}$                          & 
{$\err{0.07}{0.20}{0.05}$}                        & 
\nodata                                           & 
$0.07 \pm 0.21$                                   \\

\nodata                                           & %
$\Delta\phi_\perp$                                & 
$\err{0.19}{0.20}{0.07}$                          & 
{$\err{0.19}{0.20}{0.07}$}                        & 
\nodata                                           & 
$0.19 \pm 0.21$                                   \\

\nodata                                           & %
$\Delta\delta_0$                                  & 
{$\err{0.20}{0.18}{0.03}$}                        & 
{$\err{0.20}{0.18}{0.03}$}                        & 
\nodata                                           & 
$0.20 \pm 0.18$                                   \\

\sglinespt
\end{tabular}
\end{center}
\hspace*{0.3cm}
{\small
Angles ($\phi$, $\delta$) are in radians. BF, $f_L$ and $A_{CP}$ are tabulated
separately.
}
\end{table}


\begin{table}

\caption{
Results of the full angular analyses of $B^0 \to \phi K^{*0}$.
Values in \red{red} (\blue{blue}) are new \red{published}
(\blue{preliminary}) results since PDG2014.}
\label{tab:polar_BzAng}
\resizebox{\textwidth}{!}{
\begin{tabular}{|lcccccc|}
\sgline
   & Parameter & PDG2014 Avg. & \babar & Belle & LHCb & New Avg. \\
\sglinespb
\nodata                                           & %
$f_\perp = \Lambda_{\perp\perp}$                  & 
$0.228 \pm 0.021$                                 & 
{$\err{0.212}{0.032}{0.013}$}                     & 
$\err{0.238}{0.026}{0.008}$                       & 
\red{$\err{0.221}{0.016}{0.013}$}                 & 
$0.225 \pm 0.015$                                 \\

\nodata                                           & %
$f_S(K\pi)$                                       & 
New                                               & 
\nodata                                           & 
\nodata                                           & 
\red{$\err{0.143}{0.013}{0.012}$}                 & 
$0.143 \pm 0.018$                                 \\

\nodata                                           & %
$f_S(KK)$                                         & 
New                                               & 
\nodata                                           & 
\nodata                                           & 
\red{$\err{0.122}{0.013}{0.008}$}                 & 
$0.122 \pm 0.015$                                 \\

\nodata                                           & %
$\phi_\parallel$                                  & 
$2.28 \pm 0.08$                                   & 
{$\err{2.40}{0.13}{0.08}$}                        & 
$\err{2.23}{0.10}{0.02}$                          & 
\red{$\err{2.562}{0.069}{0.040}$}                 & 
$2.430 \pm 0.058$                                 \\

\nodata                                           & %
$\phi_\perp$                                      & 
$2.36 \pm 0.09$                                   & 
{$\err{2.35}{0.13}{0.09}$}                        & 
$\err{2.37}{0.10}{0.04}$                          & 
\red{$\err{2.633}{0.062}{0.037}$}                 & 
$2.527 \pm 0.056$                                 \\

\nodata                                           & %
$\delta_0$                                        & 
$2.88 \pm 0.10$                                   & 
{$\err{2.82}{0.15}{0.09}$}                        & 
$\err{2.91}{0.10}{0.08}$                          & 
\nodata                                           & 
$2.88 \pm 0.10$                                   \\

\nodata                                           & %
$\phi_S(K\pi)$~\dag                               & 
New                                               & 
\nodata                                           & 
\nodata                                           & 
\red{$\err{2.222}{0.063}{0.081}$}                 & 
$2.222 \pm 0.103$                                 \\

\nodata                                           & %
$\phi_S(KK)$~\dag                                 & 
New                                               & 
\nodata                                           & 
\nodata                                           & 
\red{$\err{2.481}{0.072}{0.048}$}                 & 
$2.481 \pm 0.087$                                 \\

\nodata                                           & %
$A_{CP}^0$                                        & 
$-0.01 \pm 0.05$                                  & 
{$\err{0.01}{0.07}{0.02}$}                        & 
$\err{-0.03}{0.06}{0.01}$                         & 
\red{$\err{-0.003}{0.038}{0.005}$}                & 
$-0.007 \pm 0.030$                                \\

\nodata                                           & %
$A_{CP}^\perp$                                    & 
$-0.11 \pm 0.09$                                  & 
{$\err{-0.04}{0.15}{0.06}$}                       & 
$\err{-0.14}{0.11}{0.01}$                         & 
\red{$\err{0.047}{0.072}{0.009}$}                 & 
$-0.014 \pm 0.057$                                \\

\nodata                                           & %
${\cal A}_{CP}^S(K\pi)$                           & 
New                                               & 
\nodata                                           & 
\nodata                                           & 
\red{$\err{0.073}{0.091}{0.035}$}                 & 
$0.073 \pm 0.097$                                 \\

\nodata                                           & %
${\cal A}_{CP}^S(KK)$                             & 
New                                               & 
\nodata                                           & 
\nodata                                           & 
\red{$\err{-0.209}{0.105}{0.012}$}                & 
$-0.209 \pm 0.106$                                \\

\nodata                                           & %
$\Delta\phi_\parallel$                            & 
$0.06 \pm 0.11$                                   & 
{$\err{0.22}{0.12}{0.08}$}                        & 
$\err{-0.02}{0.10}{0.01}$                         & 
\red{$\err{0.045}{0.068}{0.015}$}                 & 
$0.051 \pm 0.053$                                 \\

\nodata                                           & %
$\Delta\phi_\perp$                                & 
$0.10 \pm 0.08$                                   & 
{$\err{0.21}{0.13}{0.08}$}                        & 
$\err{0.05}{0.10}{0.02}$                          & 
\red{$\err{0.062}{0.062}{0.006}$}                 & 
$0.075 \pm 0.050$                                 \\

\nodata                                           & %
$\Delta\delta_0$                                  & 
$0.13 \pm 0.09$                                   & 
{$\err{0.27}{0.14}{0.08}$}                        & 
$\err{0.08}{0.10}{0.01}$                          & 
\nodata                                           & 
$0.13 \pm 0.08$                                   \\

\nodata                                           & %
$\Delta \phi_S(K\pi)$~\dag                        & 
New                                               & 
\nodata                                           & 
\nodata                                           & 
\red{$\err{0.062}{0.062}{0.022}$}                 & 
$0.062 \pm 0.066$                                 \\

\nodata                                           & %
$\Delta \phi_S(KK)$~\dag                          & 
New                                               & 
\nodata                                           & 
\nodata                                           & 
\red{$\err{0.022}{0.072}{0.004}$}                 & 
$0.022 \pm 0.072$                                 \\

\sglinespt
\end{tabular}
}
\hspace*{0.3cm}
{\small
Angles ($\phi$, $\delta$) are in radians. BF, $f_L$ and $A_{CP}$ are tabulated
separately. \\
\dag~Original LHCb notation adapted to match similar existing quantities.
}
\end{table}


\begin{table}

\caption{
Results of the full angular analyses of $B^0 \to \phi K_2^{*0}(1430)$.
Values in \red{red} (\blue{blue}) are new \red{published}
(\blue{preliminary}) results since PDG2014.}
\label{tab:polar_BzAng1430}
\begin{center}
\begin{tabular}{|lccccc|}
\sgline
   & Parameter & PDG2014 Avg. & \babar & Belle & New Avg. \\
\sglinespb
\nodata                                           & %
$f_\perp = \Lambda_{\perp\perp}$                  & 
$\cerr{0.027}{0.031}{0.025}$                      & 
{$\aerr{0.002}{0.018}{0.002}{0.031}$}             & 
$\aerr{0.056}{0.050}{0.035}{0.009}$               & 
$\cerr{0.027}{0.027}{0.024}$                      \\

\nodata                                           & %
$\phi_\parallel$                                  & 
$4.0 \pm 0.4$                                     & 
{$\err{3.96}{0.38}{0.06}$}                        & 
$\err{3.76}{2.88}{1.32}$                          & 
$3.96 \pm 0.38$                                   \\

\nodata                                           & %
$\phi_\perp$                                      & 
$4.5 \pm 0.4$                                     & 
\nodata                                           & 
$\aerr{4.45}{0.43}{0.38}{0.13}$                   & 
$\cerr{4.45}{0.45}{0.40}$                         \\

\nodata                                           & %
$\delta_0$                                        & 
$3.46 \pm 0.14$                                   & 
{$\err{3.41}{0.13}{0.13}$}                        & 
$\err{3.53}{0.11}{0.19}$                          & 
$3.46 \pm 0.14$                                   \\

\nodata                                           & %
$A_{CP}^0$                                        & 
$-0.03 \pm 0.04$                                  & 
{$\err{-0.05}{0.06}{0.01}$}                       & 
$\aerr{-0.016}{0.066}{0.051}{0.008}$              & 
$\cerr{-0.032}{0.043}{0.038}$                     \\

\nodata                                           & %
$A_{CP}^\perp$                                    & 
$\cerr{0.0}{0.9}{0.7}$                            & 
\nodata                                           & 
$\aerr{-0.01}{0.85}{0.67}{0.09}$                  & 
$\cerr{-0.01}{0.85}{0.68}$                        \\

\nodata                                           & %
$\Delta\phi_\parallel$                            & 
$-0.9 \pm 0.4$                                    & 
{$\err{-1.00}{0.38}{0.09}$}                       & 
$\err{-0.02}{1.08}{1.01}$                         & 
$-0.94 \pm 0.38$                                  \\

\nodata                                           & %
$\Delta\phi_\perp$                                & 
$-0.2 \pm 0.4$                                    & 
\nodata                                           & 
$\err{-0.19}{0.42}{0.11}$                         & 
$-0.19 \pm 0.43$                                  \\

\nodata                                           & %
$\Delta\delta_0$                                  & 
$0.08 \pm 0.09$                                   & 
{$\err{0.11}{0.13}{0.06}$}                        & 
$\err{0.06}{0.11}{0.02}$                          & 
$0.08 \pm 0.09$                                   \\

\sglinespt
\end{tabular}
\end{center}
Angles ($\phi$, $\delta$) are in radians. BF, $f_L$ and $A_{CP}$ are tabulated separately.
\end{table}


\begin{table}
\begin{center}

\caption{Longitudinal polarization fraction $f_L$ for $\Bs$ decays.
Values in \red{red} (\blue{blue}) are new \red{published}
(\blue{preliminary}) results since PDG2014.}
\label{tab:polar_Bs}
\small
\begin{tabular}{|lccccc|}
\sgline
  RPP\#   & Mode & PDG2014 Avg. & CDF & LHCb & New Avg. \\
\hline
$51$                                              & 
$\phi\phi$                                        & 
$0.361 \pm 0.022$                                 & 
$\err{0.348}{0.041}{0.021}$                       & 
$\err{0.365}{0.022}{0.012}$                       & 
$0.361 \pm 0.022$                                 \\

$59$                                              & 
$K^{*0}\overline{K}^{*0}$                         & 
$0.31 \pm 0.13$                                   & 
\nodata                                           & 
$\err{0.31}{0.12}{0.04}$                          & 
$0.31 \pm 0.13$                                   \\

$60$                                              & 
$\phi\overline{K}^{*0}$                           & 
$0.51 \pm 0.17$                                   & 
\nodata                                           & 
$\err{0.51}{0.15}{0.07}$                          & 
$0.51 \pm 0.17$                                   \\

\sglinespt
\end{tabular}
\end{center}
\end{table}


\begin{table}
\begin{center}

\caption{
Results of the full angular analyses of $B_s \to \phi\phi$.
Values in \red{red} (\blue{blue}) are new \red{published}
(\blue{preliminary}) results since PDG2014.}
\label{tab:polar_BsAng1}
\resizebox{\textwidth}{!}{
\begin{tabular}{|lccccc|}
\sgline
   & Parameter & PDG2014 Avg. & CDF & LHCb & New Avg. \\
\sglinespb
\nodata                                           & %
$f_\perp = \Lambda_{\perp\perp}$                  & 
$0.306 \pm 0.030$                                 & 
$\err{0.365}{0.044}{0.027}$                       & 
$\err{0.291}{0.024}{0.010}$                       & 
$0.306 \pm 0.023$                                 \\

\nodata                                           & %
$\phi_\parallel$                                  & 
$2.59 \pm 0.15$                                   & 
$\aerr{2.71}{0.31}{0.36}{0.22}$                   & 
$\err{2.57}{0.15}{0.06}$                          & 
$2.59 \pm 0.15$                                   \\

\sglinespt
\end{tabular}
}
\end{center}
The parameter $\phi$ is in radians. BF, $f_L$ and $A_{CP}$ are tabulated separately.
\end{table}


\begin{table}
\begin{center}

\caption{
Results of the full angular analyses of $B_s \to \phi\overline{K}^{*0}$.
Values in \red{red} (\blue{blue}) are new \red{published}
(\blue{preliminary}) results since PDG2014.}
\label{tab:polar_BsAng2}
\begin{tabular}{|lcccc|}
\sgline
   & Parameter & PDG2014 Avg. & LHCb & New Avg. \\
\sglinespb
\nodata                                           & %
$f_\perp = \Lambda_{\perp\perp}$                  & 
\nodata                                           & 
$\err{0.28}{0.12}{0.03}$                          & 
$0.28 \pm 0.12$                                   \\

\nodata                                           & %
$f_0$                                             & 
\nodata                                           & 
$\err{0.51}{0.15}{0.07}$                          & 
$0.51 \pm 0.17$                                   \\

\nodata                                           & %
$f_\parallel$                                     & 
$0.21 \pm 0.11$                                   & 
$\err{0.21}{0.11}{0.02}$                          & 
$0.21 \pm 0.11$                                   \\

\nodata                                           & %
$\phi_\parallel$~\dag                             & 
$1.75 \pm 0.53 \pm 0.29$                          & 
$\aerrsy{1.75}{0.59}{0.53}{0.38}{0.30}$           & 
$\cerr{1.75}{0.70}{0.61}$                         \\

\sglinespt
\end{tabular}
\end{center}
The parameter $\phi$ is in radians. BF, $f_L$ and $A_{CP}$ are tabulated separately.\\
\dag~Converted from the measurement of $\cos(\phi_\parallel)$. PDG takes the smallest resulting asymmetric error as parabolic.
\end{table}

\clearpage
\section{$D$ decays}
\label{sec:charm_physics}

\def\kbar{\overline{K}{}^{\,0}}
\def\dbar{\overline{D}{}^{\,0}}
\def\bbar{\overline{B}{}^{\,0}}
\def\cp{$CP$}
\def\cpv{$CPV$}
\def\ra{\!\rightarrow\!}
\def\ddbar{$D^0$-$\dbar$}
\def\ycp{$y^{}_{\rm CP}$}

\def\dklnu{$D^0\ra K^+\ell^-\nu$}
\def\dkpi{$D^0\ra K^+\pi^-$}
\def\dkk{$D^0\ra K^+K^-$}
\def\dpipi{$D^0\ra\pi^+\pi^-$}
\def\dkkpp{$D^0\ra K^+K^-/\pi^+\pi^-$}
\def\dkspp{$D^0\ra K^0_S\,\pi^+\pi^-$}
\def\dkskk{$D^0\ra K^0_S\,K^+ K^-$}

\def\dsphipi{$D^+_s\ra\phi\,\pi^+$}

\def\gevm{~GeV/$c^2$}
\def\gevp{~GeV/$c$}
\def\geve{~GeV}
\def\mevm{~MeV/$c^2$}
\def\meve{~MeV}


\def\simge{\mathrel{%
   \rlap{\raise 0.511ex \hbox{$>$}}{\lower 0.511ex \hbox{$\sim$}}}}
\def\simle{\mathrel{
   \rlap{\raise 0.511ex \hbox{$<$}}{\lower 0.511ex \hbox{$\sim$}}}}

\newcommand{\Dnan}{\ensuremath{D_0^\ast(2400)^0}}
\newcommand{\Dtan}{\ensuremath{D_2^\ast(2460)^0}}
\newcommand{\Don}{\ensuremath{D_1(2420)^{0}}}
\newcommand{\Dopn}{\ensuremath{D_1(2430)^{0}}}
\newcommand{\Dnap}{\ensuremath{D_0^\ast(2400)^\pm}}
\newcommand{\Dtap}{\ensuremath{D_2^\ast(2460)^\pm}}
\newcommand{\Dop}{\ensuremath{D_1(2420)^{\pm}}}
\newcommand{\Dopp}{\ensuremath{D_1(2430)^{\pm}}}

\newcommand{\Dsa}{\ensuremath{D_s^{\ast\pm}}}
\newcommand{\Dsna}{\ensuremath{D_{s0}^\ast(2317)^{\pm}}}
\newcommand{\Dsop}{\ensuremath{D_{s1}(2460)^{\pm}}}
\newcommand{\Dso}{\ensuremath{D_{s1}(2536)^{\pm}}}
\newcommand{\Dst}{\ensuremath{D_{s2}(2573)^{\pm}}}
\newcommand{\Dsts}{\ensuremath{D_{sJ}(2700)^{\pm}}}
\newcommand{\Dste}{\ensuremath{D_{sJ}(2860)^{\pm}}}
\newcommand{\Dstsi}{\ensuremath{D_{sJ}(2632)^{\pm}}}

\newcommand{\citep}{\cite}

\newcommand{\kst}{K^*(892)^0}
\newcommand{\akst}{\overline{K}^*(892)^0}
\newcommand{\kstp}{K^*(1410)^0}
\newcommand{\akstp}{\overline{K}^*(1410)^0}
\newcommand{\kstd}{K^*_2(1430)^0}
\newcommand{\akstd}{\overline{K}^*_2(1430)^0}

\newcommand{\ksts}{K^*_0(1430)^0}
\newcommand{\aksts}{\overline{K}^*_0(1430)^0}

\newcommand{\ds}{D_{s}}
\newcommand{\dsp}{D_{s}^+}
\newcommand{\dsm}{D_{s}^-}
\newcommand{\dspm}{D_{s}^{\pm}}
\newcommand{\dsmunu}{\ds^+\to\mu^+\nu_{\mu}}
\newcommand{\dsellnu}{\ds^+\to\ell^+\nu_{\ell}}
\newcommand{\br}{{\cal B}}
\newcommand{\ellnu}{\ell^+\nu_{\ell}}
\newcommand{\enu}{e^+\nu_{e}}
\newcommand{\munu}{\mu^+\nu_{\mu}}
\newcommand{\taunu}{\tau^+\nu_{\tau}}
\newcommand{\taumunu}{\tau^+(\mu^+)\nu_{\tau}}
\newcommand{\tauenu}{\tau^+(e^+)\nu_{\tau}}
\newcommand{\taupinuCharm}{\tau^+(\pi^+)\nu_{\tau}}
\newcommand{\taurhonu}{\tau^+(\rho^+)\nu_{\tau}}

\subsection{\emph{$D^0$-$\dbar$} mixing and \emph{\cp}\ violation}
\label{sec:charm:mixcpv}

\subsubsection{Introduction}

In 2007 Belle~\cite{Staric:2007dt} and \babar~\cite{Aubert:2007wf} 
obtained the first evidence for $D^0$-$\dbar$ mixing, which 
had been searched for for more than two decades. 
These results were later confirmed by CDF~\cite{Aaltonen:2007uc},
and more recently by LHCb~\cite{Aaij:2013wda}.
There are now numerous measurements of $D^0$-$\dbar$ mixing 
with various levels of sensitivity. All the results are
input into a global fit to determine
world averages of mixing parameters, \cp-violation (\cpv) 
parameters, and strong phases.

Our notation is as follows.
The mass eigenstates are denoted 
$D^{}_1 = p|D^0\rangle-q|\dbar\rangle$ and
$D^{}_2 = p|D^0\rangle+q|\dbar\rangle$, 
where we use the convention 
$CP|D^0\rangle=-|\dbar\rangle$ and 
$CP|\dbar\rangle=-|D^0\rangle$. Thus in the absence of 
\cp\ violation, $D^{}_1$ is \cp-even and $D^{}_2$ is \cp-odd.
The weak phase $\phi$ is defined as ${\rm Arg}(q/p)$.
The mixing parameters are defined as 
$x\equiv(m^{}_1-m^{}_2)/\Gamma$ and 
$y\equiv (\Gamma^{}_1-\Gamma^{}_2)/(2\Gamma)$, where 
$m^{}_1,\,m^{}_2$ and $\Gamma^{}_1,\,\Gamma^{}_2$ are
the masses and decay widths for the mass eigenstates,
and $\Gamma\equiv (\Gamma^{}_1+\Gamma^{}_2)/2$.

The global fit determines central values and errors
for ten underlying parameters. These consist of the
mixing parameters $x$ and $y$; 
indirect \cpv\ parameters $|q/p|$ and $\phi$; 
the ratio of decay rates
$R^{}_D\equiv
[\Gamma(D^0\ra K^+\pi^-)+\Gamma(\dbar\ra K^-\pi^+)]/
[\Gamma(D^0\ra K^-\pi^+)+\Gamma(\dbar\ra K^+\pi^-)]$,
direct \cpv\ parameters $A^{}_K$, $A^{}_\pi$
(see Table~\ref{tab:relationships}), and
$A^{}_D =(R^+_D-R^-_D)/(R^+_D+R^-_D)$, where the $+\,(-)$
superscript corresponds to $D^0\,(\dbar)$ decays;
the strong phase difference
$\delta$ between $\dbar\ra K^-\pi^+$ and 
$D^0\ra K^-\pi^+$ amplitudes; and 
the strong phase difference $\delta^{}_{K\pi\pi}$ between 
$\dbar\ra K^-\rho^+$ and $D^0\ra K^-\rho^+$ amplitudes. 

The fit uses 45 observables taken from 
measurements of \dklnu, \dkk\ and \dpipi, \dkpi, 
$D^0\ra K^+\pi^-\pi^0$, 
\dkspp, and \dkskk\ decays,\footnote{Charge-conjugate modes
are implicitly included.} and from double-tagged branching 
fractions measured at the $\psi(3770)$ resonance. Correlations 
among observables are accounted for by using covariance matrices 
provided by the experimental collaborations. Errors are assumed
to be Gaussian, and systematic errors among different experiments 
are assumed uncorrelated unless specific correlations have been 
identified.
We have checked this method with a second method that adds
together three-dimensional log-likelihood functions 
for $x$, $y$, and $\delta$ obtained from several analyses;
this combination accounts for non-Gaussian errors.
When both methods are applied to the same set of 
measurements, equivalent results are obtained.

Mixing in heavy flavor systems such as those of $B^0$ and $B^0_s$ 
is governed by a short-distance box diagram. In the $D^0$ system,
however, this diagram is doubly-Cabibbo-suppressed relative to 
amplitudes dominating the decay width, and it is also GIM-suppressed.
Thus the short-distance mixing rate is tiny, and $D^0$-$\dbar$ 
mixing is expected to be dominated by long-distance processes. 
These are difficult to calculate reliably, and theoretical
estimates for $x$ and $y$ range by up to three orders of 
magnitude~\cite{Bigi:2000wn,Petrov:2003un,Petrov:2004rf,Falk:2004wg}.

Almost all methods besides that of the $\psi(3770)\ra DD$
measurements~\cite{Asner:2012xb} identify the flavor of the
$D^0$ or $\dbar$ when produced by reconstructing the decay
$D^{*+}\ra D^0\pi^+$ or $D^{*-}\ra\dbar\pi^-$. The charge
of the pion, which has low momentum and is usually 
referred to as the ``soft'' pion ($\pi^{}_s$),
identifies the $D$ flavor. For signal 
decays, $M^{}_{D^*}-M^{}_{D^0}-M^{}_{\pi^+}\equiv Q\approx 6$\meve, 
which is close to the threshold; thus analyses typically
require that the reconstructed $Q$ be small to suppress backgrounds. 
A recent LHCb measurement~\cite{Aaij:2014gsa} of the difference
between time-integrated \cp\ asymmetries
$A_{CP}(K^+K^-) - A_{CP}(\pi^+\pi^-)$ identifies the flavor of
the $D^0$ by partially reconstructing $\bbar\ra D^0\mu^- X$ 
decays (and charge-conjugates); in this case the charge of
the muon identifies the flavor of the $D^0$.

For time-dependent measurements, the $D^0$ decay time is 
calculated as 
$(\vec{\bf d}\cdot\vec{\bf p})/p^2\times M^{}_{D^0}$, 
where $\vec{\bf d}$ is the displacement vector between the
$D^*$ and $D^0$ decay vertices, and $\vec{\bf p}$ is the
$D^0$ momentum. The $D^*$ vertex position is 
taken to be at the primary vertex for $\bar{p}p$ and $pp$
experiments~\cite{Aaltonen:2007uc,Aaij:2013wda}, and at
the intersection of the $D^0$ momentum vector with the
beamspot profile for $e^+e^-$ experiments.

\subsubsection{Input observables}

The global fit determines central values and errors for
the underlying parameters using a $\chi^2$ statistic.
The fitted parameters are $x$, $y$, $R^{}_D$, $A^{}_D$,
$|q/p|$, $\phi$, $\delta$, $\delta^{}_{K\pi\pi}$,
$A^{}_K$ and $A^{}_\pi$.
In the $D\ra K^+\pi^-\pi^0$ 
Dalitz plot analysis that provides sensitivity to $x$ and $y$, 
the $\dbar\ra K^+\pi^-\pi^0$ isobar phases are determined 
relative to that for ${\cal A}(\dbar\ra K^+\rho^-)$, and 
the $D^0\ra K^+\pi^-\pi^0$ isobar phases are determined 
relative to that for ${\cal A}(D^0\ra K^+\rho^-)$. 
As the $\dbar$ and $D^0$ Dalitz plots are fit independently, 
the phase difference $\delta^{}_{K\pi\pi}$ between the
two ``normalizing amplitudes'' cannot be determined
from these fits.

\begin{figure}
\begin{center}
\includegraphics[width=4.2in]{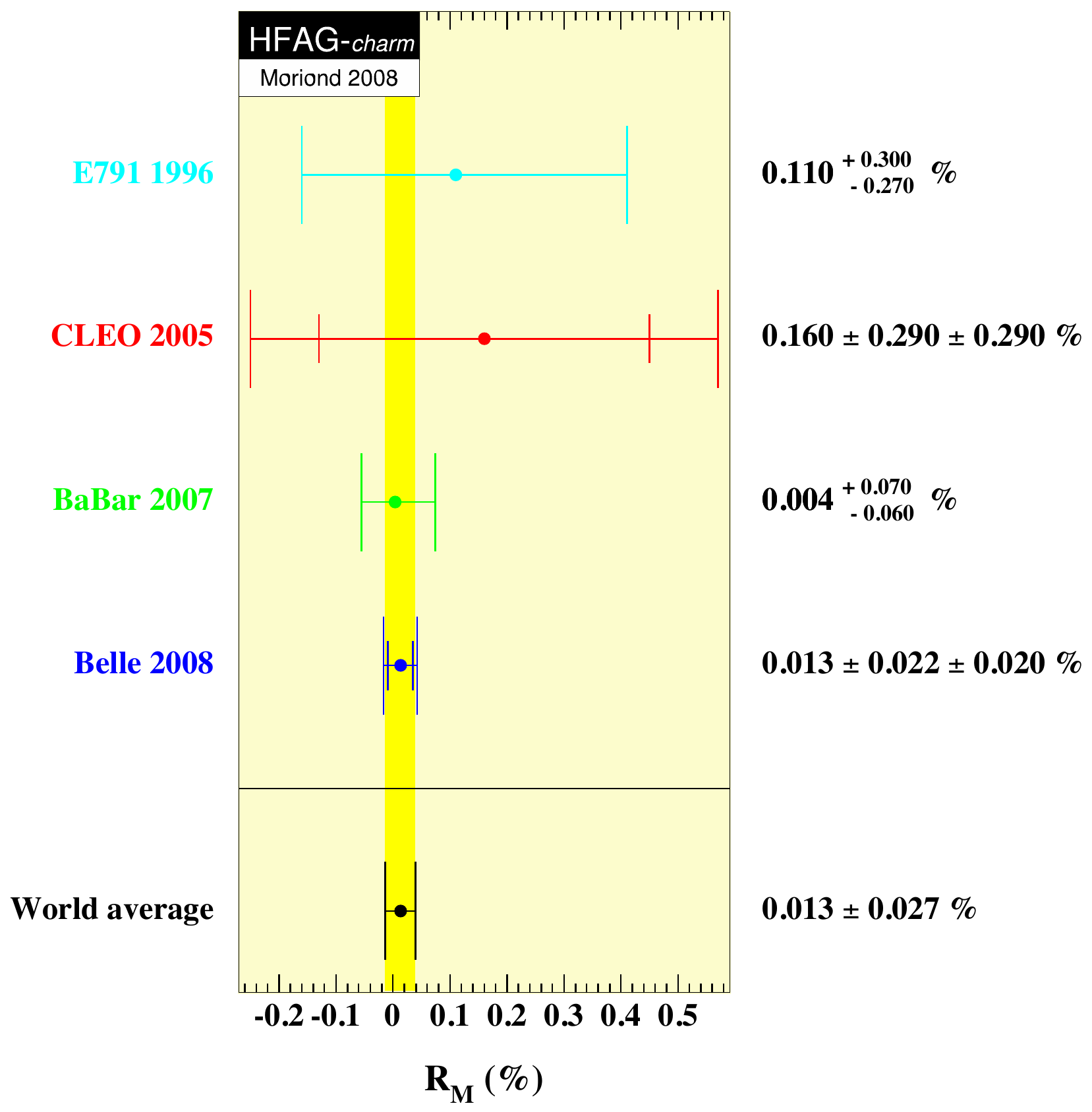}
\end{center}
\vskip-0.20in
\caption{\label{fig:rm_semi}
World average value of $R^{}_M$ from Ref.~\cite{HFAG_charm:webpage},
as calculated from $D^0\ra K^+\ell^-\nu$ 
measurements~\cite{Aitala:1996vz,Cawlfield:2005ze,Aubert:2007aa,Bitenc:2008bk}. }
\end{figure}

\begin{figure}
\begin{center}
\includegraphics[width=4.2in]{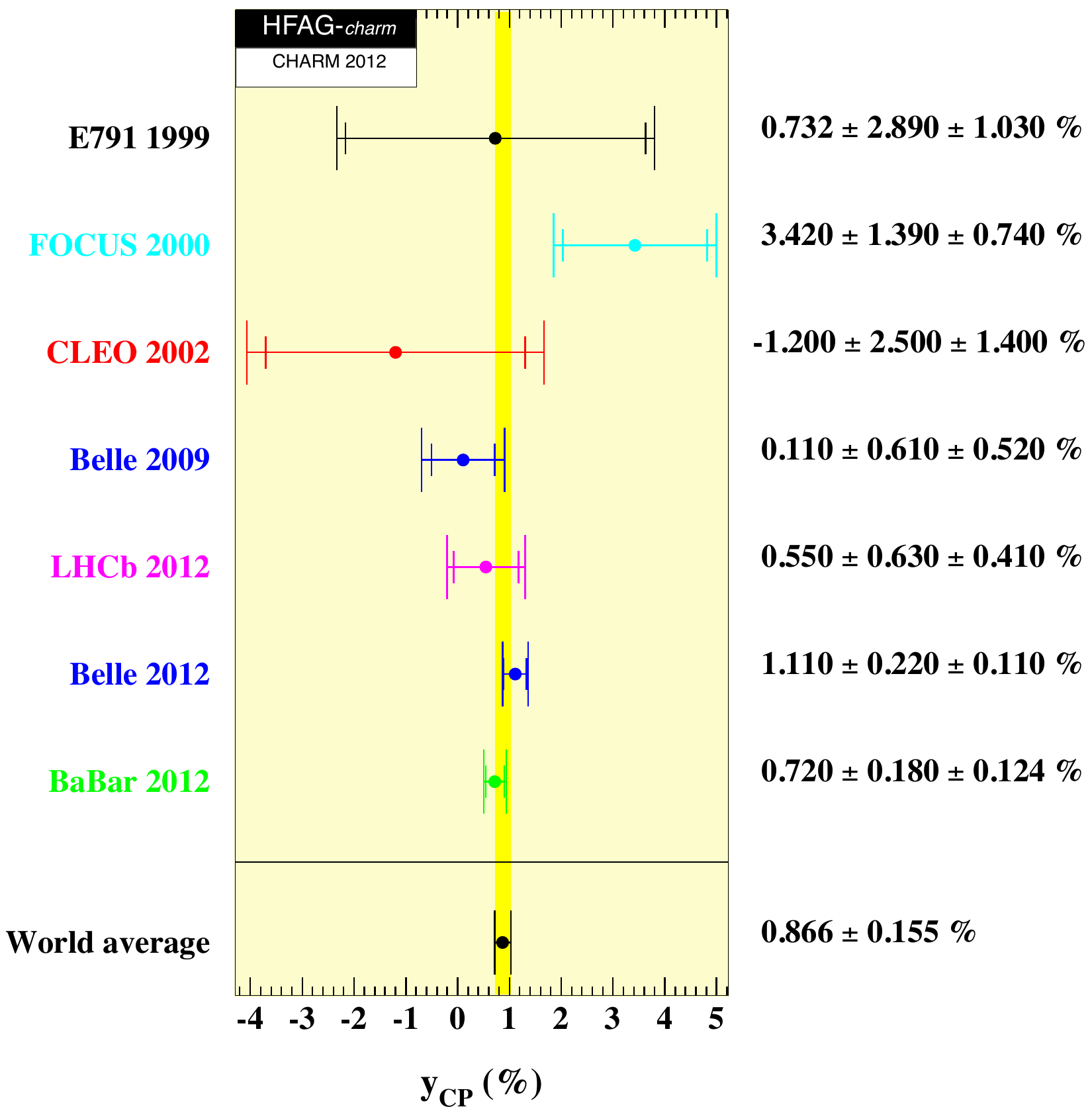}
\end{center}
\vskip-0.20in
\caption{\label{fig:ycp}
World average value of $y^{}_{CP}$ from 
Ref.~\cite{HFAG_charm:webpage}, as calculated from \dkkpp\ 
measurements~\cite{Aitala:1999dt,Link:2000cu,Csorna:2001ww,
Zupanc:2009sy,Staric:2012ta,Lees:2012qh,Aaij:2011ad}.  }
\end{figure}

\begin{figure}
\begin{center}
\includegraphics[width=4.2in]{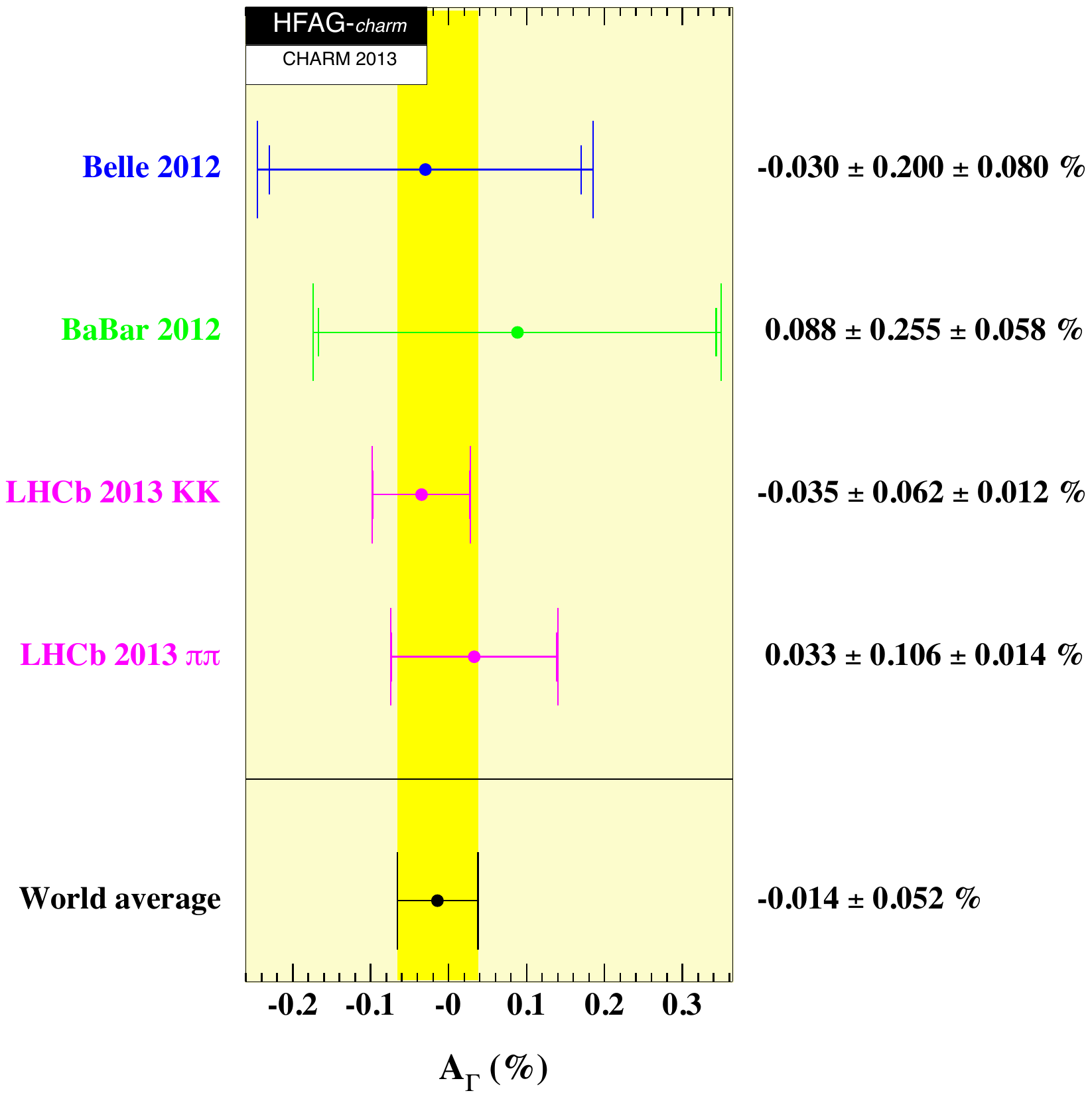}
\end{center}
\vskip-0.20in
\caption{\label{fig:Agamma}
World average value of $A^{}_\Gamma$ from 
Ref.~\cite{HFAG_charm:webpage}, as calculated from \dkkpp\ 
measurements~\cite{Staric:2012ta,Lees:2012qh,Aaij:2013ria}.}
\end{figure}

All input measurements are listed in 
Tables~\ref{tab:observables1}-\ref{tab:observables3}. 
The observable $R^{}_M=(x^2+y^2)/2$ is calculated from \dklnu\ 
decays~\cite{Aitala:1996vz,Cawlfield:2005ze,Aubert:2007aa,Bitenc:2008bk}
and is the world average (WA) value calculated by 
HFAG~\cite{HFAG_charm:webpage}. The inputs used for
this average are plotted in Fig.~\ref{fig:rm_semi}.
The observables 
$y^{}_{CP}= (1/2)(|q/p| + |p/q|)y\cos\phi - (1/2)(|q/p|-|p/q|)x\sin\phi$ 
and 
$A^{}_\Gamma= (1/2)(|q/p| - |p/q|)y\cos\phi - (1/2)(|q/p|+|p/q|)x\sin\phi$ 
are also HFAG WA values~\cite{HFAG_charm:webpage}; the inputs
used for these averages are plotted in
Figs.~\ref{fig:ycp} and \ref{fig:Agamma}, respectively.
The \dkpi\ observables used are from 
Belle~\cite{Zhang:2006dp,Ko:2014qvu}, 
\babar~\cite{Aubert:2007wf}, 
CDF~\cite{Aaltonen:2013pja}, and
LHCb~\cite{Aaij:2013wda};
earlier measurements have much less precision and are not used.
The observables from \dkspp\ decays are measured in two ways:
assuming \cp\ conservation ($D^0$ and $\dbar$ decays are combined),
and allowing for \cp\ violation ($D^0$ and $\dbar$ decays are
fitted separately). The no-\cpv\ measurements are from 
Belle~\cite{Peng:2014oda} and \babar~\cite{delAmoSanchez:2010xz}, 
but for the \cpv-allowed case only Belle 
measurements~\cite{Peng:2014oda} are available. The 
$D^0\ra K^+\pi^-\pi^0$ results are from \babar~\cite{Aubert:2008zh},
and the $\psi(3770)\ra\overline{D}D$ results are from 
CLEOc~\cite{Asner:2012xb}.

\begin{table}
\renewcommand{\arraystretch}{1.3}
\renewcommand{\arraycolsep}{0.02in}
\renewcommand{\tabcolsep}{0.05in}
\caption{\label{tab:observables1}
Observables used in the global fit except those from
\dkpi\ and those used for measuring direct \cpv. The 
$D^0\ra K^+\pi^-\pi^0$ observables are
$x'' \equiv x\cos\delta^{}_{K\pi\pi} + y\sin\delta^{}_{K\pi\pi}$ and 
$y'' \equiv -x\sin\delta^{}_{K\pi\pi} + y\cos\delta^{}_{K\pi\pi}$.}
\vspace*{6pt}
\footnotesize
\resizebox{0.99\textwidth}{!}{
\begin{tabular}{l|ccc}
\hline
{\bf Mode} & \textbf{Observable} & {\bf Values} & {\bf Correlation coefficients} \\
\hline
\begin{tabular}{l}  
$D^0\ra K^+K^-/\pi^+\pi^-$, \\
\hskip0.30in $\phi\,K^0_S$~\cite{HFAG_charm:webpage} 
\end{tabular}
&
\begin{tabular}{c}
 $y^{}_{CP}$  \\
 $A^{}_{\Gamma}$
\end{tabular} & 
$\begin{array}{c}
(0.866\pm 0.155)\% \\
(-0.014\pm 0.052)\% 
\end{array}$   & \\ 
\hline
\begin{tabular}{l}  
$D^0\ra K^0_S\,\pi^+\pi^-$~\cite{Peng:2014oda} \\
\ (Belle: no \cpv)
\end{tabular}
&
\begin{tabular}{c}
$x$ \\
$y$ 
\end{tabular} & 
\begin{tabular}{c}
 $(0.56\pm 0.19\,^{+0.067}_{-0.127})\%$ \\
 $(0.30\pm 0.15\,^{+0.050}_{-0.078})\%$ 
\end{tabular} & $+0.012$ \\ 
 & & \\
\begin{tabular}{l}  
$D^0\ra K^0_S\,\pi^+\pi^-$~\cite{Peng:2014oda} \\
\ (Belle: no direct \cpv)
\end{tabular}
&
\begin{tabular}{c}
$|q/p|$ \\
$\phi$  
\end{tabular} & 
\begin{tabular}{c}
 $0.90\,^{+0.16}_{-0.15}{}^{+0.078}_{-0.064}$ \\
 $(-6\pm 11\,^{+4.2}_{-5.0})$ degrees
\end{tabular} & \\
 & & \\
\begin{tabular}{l}  
$D^0\ra K^0_S\,\pi^+\pi^-$~\cite{Peng:2014oda} \\
\ (Belle: direct \cpv\ allowed)
\end{tabular}
&
\begin{tabular}{c}
$x$ \\
$y$ \\
$|q/p|$ \\
$\phi$  
\end{tabular} & 
\begin{tabular}{c}
 $(0.58\pm 0.19^{+0.0734}_{-0.1177})\%$ \\
 $(0.27\pm 0.16^{+0.0546}_{-0.0854})\%$ \\
 $0.82\,^{+0.20}_{-0.18}{}^{+0.0807}_{-0.0645}$ \\
 $(-13\,^{+12}_{-13}\,^{+4.15}_{-4.77})$ degrees
\end{tabular} & 
$\left\{ \begin{array}{cccc}
 1 &  0.054 & -0.074 & -0.031  \\
 0.054 &  1 & 0.034 & -0.019 \\
 -0.074 &  0.034 & 1 & 0.044  \\
 -0.031 &  -0.019 & 0.044 & 1 
\end{array} \right\}$  \\
 & & \\
\begin{tabular}{l}  
$D^0\ra K^0_S\,\pi^+\pi^-$~\cite{delAmoSanchez:2010xz} \\
\hskip0.30in $K^0_S\,K^+ K^-$ \\
\ (\babar: no \cpv) 
\end{tabular}
&
\begin{tabular}{c}
$x$ \\
$y$ 
\end{tabular} & 
\begin{tabular}{c}
 $(0.16\pm 0.23\pm 0.12\pm 0.08)\%$ \\
 $(0.57\pm 0.20\pm 0.13\pm 0.07)\%$ 
\end{tabular} &  $0.0615$ \\ 
\hline
\begin{tabular}{l}  
$D^0\ra K^+\ell^-\nu$~\cite{HFAG_charm:webpage}
\end{tabular} 
  & $R^{}_M$  & $(0.0130\pm 0.0269)\%$  &  \\ 
\hline
\begin{tabular}{l}  
$D^0\ra K^+\pi^-\pi^0$~\cite{Aubert:2008zh}
\end{tabular} 
&
\begin{tabular}{c}
$x''$ \\ 
$y''$ 
\end{tabular} &
\begin{tabular}{c}
$(2.61\,^{+0.57}_{-0.68}\,\pm 0.39)\%$ \\ 
$(-0.06\,^{+0.55}_{-0.64}\,\pm 0.34)\%$ 
\end{tabular} & $-0.75$ \\
\hline
\begin{tabular}{c}  
$\psi(3770)\ra\overline{D}D$~\cite{Asner:2012xb} \\
(CLEOc)
\end{tabular}
&
\begin{tabular}{c}
$R^{}_D$ \\
$x^2$ \\
$y$ \\
$\cos\delta$ \\
$\sin\delta$ 
\end{tabular} & 
\begin{tabular}{c}
$(0.533 \pm 0.107 \pm 0.045)\%$ \\
$(0.06 \pm 0.23 \pm 0.11)\%$ \\
$(4.2 \pm 2.0 \pm 1.0)\%$ \\
$0.81\,^{+0.22}_{-0.18}\,^{+0.07}_{-0.05}$ \\
$-0.01\pm 0.41\pm 0.04$
\end{tabular} &
$\left\{ \begin{array}{ccccc}
1 & 0 &  0    & -0.42 &  0.01 \\
  & 1 & -0.73 &  0.39 &  0.02 \\
  &   &  1    & -0.53 & -0.03 \\
  &   &       &  1    &  0.04 \\
  &   &       &       &  1    
\end{array} \right\}$ \\
\hline
\end{tabular}
}
\end{table}

\begin{table}
\renewcommand{\arraystretch}{1.3}
\renewcommand{\arraycolsep}{0.02in}
\caption{\label{tab:observables2}
\dkpi\ observables used for the global fit
($x'^{2\pm}$ and $y'^{\pm}$ are defined in Ref.~\cite{Aubert:2007wf}).}
\vspace*{6pt}
\footnotesize
\begin{center}
\begin{tabular}{l|ccc}
\hline
{\bf Mode} & \textbf{Observable} & {\bf Values} & {\bf Correlation coefficients} \\
\hline
\begin{tabular}{c}  
$D^0\ra K^+\pi^-$~\cite{Aubert:2007wf} \\
(\babar~384~fb$^{-1}$)
\end{tabular}
&
\begin{tabular}{c}
$R^{}_D$ \\
$x'^{2+}$ \\
$y'^+$ 
\end{tabular} & 
\begin{tabular}{c}
 $(0.303\pm 0.0189)\%$ \\
 $(-0.024\pm 0.052)\%$ \\
 $(0.98\pm 0.78)\%$ 
\end{tabular} &
$\left\{ \begin{array}{ccc}
 1 &  0.77 &  -0.87 \\
0.77 & 1 & -0.94 \\
-0.87 & -0.94 & 1 
\end{array} \right\}$ \\ \\
\begin{tabular}{c}  
$\dbar\ra K^-\pi^+$~\cite{Aubert:2007wf} \\
(\babar~384~fb$^{-1}$)
\end{tabular}
&
\begin{tabular}{c}
$A^{}_D$ \\
$x'^{2-}$ \\
$y'^-$ 
\end{tabular} & 
\begin{tabular}{c}
 $(-2.1\pm 5.4)\%$ \\
 $(-0.020\pm 0.050)\%$ \\
 $(0.96\pm 0.75)\%$ 
\end{tabular} & same as above \\
\hline
\begin{tabular}{c}  
$D^0\ra K^+\pi^-$~\cite{Ko:2014qvu} \\
(Belle 976~fb$^{-1}$ No \cpv)
\end{tabular}
&
\begin{tabular}{c}
$R^{}_D$ \\
$x'^{2}$ \\
$y'$ 
\end{tabular} & 
\begin{tabular}{c}
 $(0.353\pm 0.013)\%$ \\
 $(0.009\pm 0.022)\%$ \\
 $(0.46\pm 0.34)\%$ 
\end{tabular} &
$\left\{ \begin{array}{ccc}
 1 &  0.737 &  -0.865 \\
0.737 & 1 & -0.948 \\
-0.865 & -0.948 & 1 
\end{array} \right\}$ \\ \\
\begin{tabular}{c}  
$D^0\ra K^+\pi^-$~\cite{Zhang:2006dp} \\
(Belle 400~fb$^{-1}$ \cpv-allowed)
\end{tabular}
&
\begin{tabular}{c}
$R^{}_D$ \\
$x'^{2+}$ \\
$y'^+$ 
\end{tabular} & 
\begin{tabular}{c}
 $(0.364\pm 0.018)\%$ \\
 $(0.032\pm 0.037)\%$ \\
 $(-0.12\pm 0.58)\%$ 
\end{tabular} &
$\left\{ \begin{array}{ccc}
 1 &  0.655 &  -0.834 \\
0.655 & 1 & -0.909 \\
-0.834 & -0.909 & 1 
\end{array} \right\}$ \\ \\
\begin{tabular}{c}  
$\dbar\ra K^-\pi^+$~\cite{Zhang:2006dp} \\
(Belle 400~fb$^{-1}$ \cpv-allowed)
\end{tabular}
&
\begin{tabular}{c}
$A^{}_D$ \\
$x'^{2-}$ \\
$y'^-$ 
\end{tabular} & 
\begin{tabular}{c}
 $(2.3\pm 4.7)\%$ \\
 $(0.006\pm 0.034)\%$ \\
 $(0.20\pm 0.54)\%$ 
\end{tabular} & same as above \\
\hline
\begin{tabular}{c}  
$D^0\ra K^+\pi^-$~\cite{Aaltonen:2013pja} \\
(CDF 9.6~fb$^{-1}$ No \cpv)
\end{tabular}
&
\begin{tabular}{c}
$R^{}_D$ \\
$x'^{2}$ \\
$y'$ 
\end{tabular} & 
\begin{tabular}{c}
 $(0.351\pm 0.035)\%$ \\
 $(0.008\pm 0.018)\%$ \\
 $(0.43\pm 0.43)\%$ 
\end{tabular} & 
$\left\{ \begin{array}{ccc}
 1 &  0.90 &  -0.97 \\
0.90 & 1 & -0.98 \\
-0.97 & -0.98 & 1 
\end{array} \right\}$ \\ 
\hline
\begin{tabular}{c}  
$D^0\ra K^+\pi^-$~\cite{Aaij:2013wda} \\
(LHCb 3.0~fb$^{-1}$ \cpv-allowed)
\end{tabular}
&
\begin{tabular}{c}
$R^{+}_D$ \\
$x'^{2+}$ \\
$y'^+$ 
\end{tabular} & 
\begin{tabular}{c}
 $(0.3545\pm 0.0095)\%$ \\
 $(0.0049\pm 0.0070)\%$ \\
 $(0.51\pm 0.14)\%$ 
\end{tabular} &
$\left\{ \begin{array}{ccc}
 1 &  0.862 &  -0.942 \\
0.862 & 1 & -0.968 \\
-0.942 & -0.968 & 1 
\end{array} \right\}$ \\ \\
\begin{tabular}{c}  
$\dbar\ra K^-\pi^+$~\cite{Aaij:2013wda} \\
(LHCb 3.0~fb$^{-1}$ \cpv-allowed)
\end{tabular}
&
\begin{tabular}{c}
$R^{+}_D$ \\
$x'^{2-}$ \\
$y'^-$ 
\end{tabular} & 
\begin{tabular}{c}
 $(0.3591\pm 0.0094)\%$ \\
 $(0.0060\pm 0.0068)\%$ \\
 $(0.45\pm 0.14)\%$ 
\end{tabular} & 
$\left\{ \begin{array}{ccc}
 1 &  0.858 &  -0.941 \\
0.858 & 1 & -0.966 \\
-0.941 & -0.966 & 1 
\end{array} \right\}$ \\
\hline
\end{tabular}
\end{center}
\end{table}

\begin{table}
\renewcommand{\arraystretch}{1.3}
\renewcommand{\arraycolsep}{0.02in}
\caption{\label{tab:observables3}
Measurements of time-integrated \cp\ asymmetries. The observable 
$A^{}_{CP}(f)\equiv [\Gamma(D^0\ra f)-\Gamma(\dbar\ra f)]/
[\Gamma(D^0\ra f)+\Gamma(\dbar\ra f)]$, and
$\Delta\langle t\rangle$ is the difference
between mean lifetimes for $D^0\ra K^+K^-$ and
$D^0\ra\pi^+\pi^-$ decays (due to different
trigger/reconstruction efficiencies).}
\vspace*{6pt}
\footnotesize
\begin{center}
\resizebox{\textwidth}{!}{
\begin{tabular}{l|ccc}
\hline
{\bf Mode} & \textbf{Observable} & {\bf Values} & 
                  {\boldmath $\Delta\langle t\rangle/\tau^{}_D$} \\
\hline
\begin{tabular}{c}
$D^0\ra h^+ h^-$~\cite{Aubert:2007if} \\
(\babar\ 386 fb$^{-1}$)
\end{tabular} & 
\begin{tabular}{c}
$A^{}_{CP}(K^+K^-)$ \\
$A^{}_{CP}(\pi^+\pi^-)$ 
\end{tabular} & 
\begin{tabular}{c}
$(0.00 \pm 0.34 \pm 0.13)\%$ \\
$(-0.24 \pm 0.52 \pm 0.22)\%$ 
\end{tabular} &
0 \\
\hline
\begin{tabular}{c}
$D^0\ra h^+ h^-$~\cite{Ko:2012jh} \\
(Belle 976~fb$^{-1}$)
\end{tabular} & 
\begin{tabular}{c}
$A^{}_{CP}(K^+K^-)$ \\
$A^{}_{CP}(\pi^+\pi^-)$ 
\end{tabular} & 
\begin{tabular}{c}
$(-0.32 \pm 0.21 \pm 0.09)\%$ \\
$(0.55 \pm 0.36 \pm 0.09)\%$ 
\end{tabular} &
0 \\
\hline
\begin{tabular}{c}
$D^0\ra h^+ h^-$~\cite{CDF:webpage,Collaboration:2012qw} \\
(CDF 9.7~fb$^{-1}$)
\end{tabular} & 
\begin{tabular}{c}
$A^{}_{CP}(K^+K^-)-A^{}_{CP}(\pi^+\pi^-)$ \\
$A^{}_{CP}(K^+K^-)$ \\
$A^{}_{CP}(\pi^+\pi^-)$ 
\end{tabular} & 
\begin{tabular}{c}
$(-0.62 \pm 0.21 \pm 0.10)\%$ \\
$(-0.32 \pm 0.21)\%$ \\
$(0.31 \pm 0.22)\%$ 
\end{tabular} &
$0.27 \pm 0.01$ \\
\hline
\begin{tabular}{c}
$D^0\ra h^+ h^-$~\cite{LHCb-CONF-2013-003} \\
(LHCb 1.0~fb$^{-1}$, \\
$D^{*+}\ra D^0\pi^+$ tag)
\end{tabular} & 
$A^{}_{CP}(K^+K^-)-A^{}_{CP}(\pi^+\pi^-)$ &
$(-0.34 \pm 0.15 \pm 0.10)\%$ &
$0.1119 \pm 0.0013 \pm 0.0017$ \\
\hline
\begin{tabular}{c}
$D^0\ra h^+ h^-$~\cite{Aaij:2014gsa} \\
(LHCb 3~fb$^{-1}$, \\
 $B\ra D^0\mu^- X$ tag)
\end{tabular} & 
$A^{}_{CP}(K^+K^-)-A^{}_{CP}(\pi^+\pi^-)$ &
$(0.14 \pm 0.16 \pm 0.08)\%$ &
$0.014 \pm 0.004$ \\
\hline
\end{tabular}
}
\end{center}
\end{table}

The relationships between the observables and the fitted
parameters are listed in Table~\ref{tab:relationships}. 
For each set of correlated observables we construct a
difference vector $\vec{V}$ between measured values and
those calculated from fitted parameters using the
relations of Table~\ref{tab:relationships}; \eg\ for 
$D^0\ra K^0_S\,\pi^+\pi^-$ decays,
$\vec{V}=(\Delta x,\Delta y,\Delta |q/p|,\Delta \phi)$.
The contribution of a set of observables to the $\chi^2$ 
is calculated as $\vec{V}\cdot (M^{-1})\cdot\vec{V}^T$, 
where $M^{-1}$ is the inverse of the covariance matrix 
for the measurement. Covariance matrices are constructed 
from the correlation coefficients among the measured observables.
These coefficients (where applicable) are also listed in 
Tables~\ref{tab:observables1}-\ref{tab:observables3}. 

\begin{table}
\renewcommand{\arraycolsep}{0.02in}
\renewcommand{\arraystretch}{1.3}
\begin{center}
\caption{\label{tab:relationships}
Left: decay modes used to determine fitted parameters 
$x,\,y,\,\delta,\,\delta^{}_{K\pi\pi},\,R^{}_D,\,A^{}_D,\,|q/p|$, and $\phi$.
Middle: the observables measured for each decay mode. Right: the 
relationships between the observables measured and the fitted
parameters. $\langle t\rangle$ is the mean lifetime for 
$D^0\ra K^+K^-$ or $D^0\ra\pi^+\pi^-$ decays.}
\vspace*{6pt}
\footnotesize
\resizebox{0.99\textwidth}{!}{
\begin{tabular}{l|c|l}
\hline
\textbf{Decay Mode} & \textbf{Observables} & \textbf{Relationship} \\
\hline
$D^0\ra K^+K^-/\pi^+\pi^-$  & 
\begin{tabular}{c}
 $y^{}_{CP}$  \\
 $A^{}_{\Gamma}$
\end{tabular} & 
$\begin{array}{c}
2y^{}_{CP} = 
\left(\left|q/p\right|+\left|p/q\right|\right)y\cos\phi - \\
\hskip0.50in \left(\left|q/p\right|-\left|p/q\right|\right)x\sin\phi \\
2A^{}_\Gamma = 
\left(\left|q/p\right|-\left|p/q\right|\right)y\cos\phi - \\
\hskip0.50in \left(\left|q/p\right|+\left|p/q\right|\right)x\sin\phi
\end{array}$   \\
\hline
$D^0\ra K^0_S\,\pi^+\pi^-$ & 
$\begin{array}{c}
x \\ 
y \\ 
|q/p| \\ 
\phi
\end{array}$ &   \\ 
\hline
$D^0\ra K^+\ell^-\nu$ & $R^{}_M$  & $R^{}_M = (x^2 + y^2)/2$ \\
\hline
\begin{tabular}{l}
$D^0\ra K^+\pi^-\pi^0$ \\
(Dalitz plot analysis)
\end{tabular} & 
$\begin{array}{c}
x'' \\ 
y''
\end{array}$ &
$\begin{array}{l}
x'' = x\cos\delta^{}_{K\pi\pi} + y\sin\delta^{}_{K\pi\pi} \\ 
y'' = y\cos\delta^{}_{K\pi\pi} - x\sin\delta^{}_{K\pi\pi}
\end{array}$ \\
\hline
\begin{tabular}{l}
``Double-tagged'' \\
branching fractions \\
measured in \\
$\psi(3770)\ra DD$ decays
\end{tabular} & 
$\begin{array}{c}
R^{}_M \\
y \\
R^{}_D \\
\sqrt{R^{}_D}\cos\delta
\end{array}$ &   $R^{}_M = (x^2 + y^2)/2$ \\
\hline
$D^0\ra K^+\pi^-$ &
$\begin{array}{c}
x'^2,\ y' \\
x'^{2+},\ x'^{2-} \\
y'^+,\ y'^-
\end{array}$ & 
$\begin{array}{l}
x' = x\cos\delta + y\sin\delta \\ 
y' = y\cos\delta - x\sin\delta \\
A^{}_M\equiv (|q/p|^4-1)/(|q/p|^4+1) \\
x'^\pm = [(1\pm A^{}_M)/(1\mp A^{}_M)]^{1/4} \times \\
\hskip0.50in (x'\cos\phi\pm y'\sin\phi) \\
y'^\pm = [(1\pm A^{}_M)/(1\mp A^{}_M)]^{1/4} \times \\
\hskip0.50in (y'\cos\phi\mp x'\sin\phi) \\
\end{array}$ \\
\hline
\begin{tabular}{c}
$D^0\ra K^+\pi^-/K^-\pi^+$ \\
(time-integrated)
\end{tabular} & 
\begin{tabular}{c}
$\frac{\displaystyle \Gamma(D^0\ra K^+\pi^-)+\Gamma(\dbar\ra K^-\pi^+)}
{\displaystyle \Gamma(D^0\ra K^-\pi^+)+\Gamma(\dbar\ra K^+\pi^-)}$  \\ \\
$\frac{\displaystyle \Gamma(D^0\ra K^+\pi^-)-\Gamma(\dbar\ra K^-\pi^+)}
{\displaystyle \Gamma(D^0\ra K^+\pi^-)+\Gamma(\dbar\ra K^-\pi^+)}$ 
\end{tabular} & 
\begin{tabular}{c}
$R^{}_D$ \\ \\ \\
$A^{}_D$ 
\end{tabular} \\
\hline
\begin{tabular}{c}
$D^0\ra K^+K^-/\pi^+\pi^-$ \\
(time-integrated)
\end{tabular} & 
\begin{tabular}{c}
$\frac{\displaystyle \Gamma(D^0\ra K^+K^-)-\Gamma(\dbar\ra K^+K^-)}
{\displaystyle \Gamma(D^0\ra K^+K^-)+\Gamma(\dbar\ra K^+K^-)}$    \\ \\
$\frac{\displaystyle \Gamma(D^0\ra\pi^+\pi^-)-\Gamma(\dbar\ra\pi^+\pi^-)}
{\displaystyle \Gamma(D^0\ra\pi^+\pi^-)+\Gamma(\dbar\ra\pi^+\pi^-)}$ 
\end{tabular} & 
\begin{tabular}{c}
$A^{}_K  + \frac{\displaystyle \langle t\rangle}
{\displaystyle \tau^{}_D}\,{\cal A}_{CP}^{\rm indirect}$ 
\ \ (${\cal A}_{CP}^{\rm indirect}\approx -A^{}_\Gamma$)
\\ \\ \\
$A^{}_\pi + \frac{\displaystyle \langle t\rangle}
{\displaystyle \tau^{}_D}\,{\cal A}_{CP}^{\rm indirect}$ 
\ \ (${\cal A}_{CP}^{\rm indirect}\approx -A^{}_\Gamma$)
\end{tabular} \\
\hline
\end{tabular}
}
\end{center}
\end{table}

\subsubsection{Fit results}

The global fit uses MINUIT with the MIGRAD minimizer, 
and all errors are obtained from MINOS~\cite{MINUIT:webpage}. 
Four separate fits are performed: 
{\it (a)}\ assuming \cp\ conservation, \ie\ fixing
$A^{}_D\!=\!0$, $A_K\!=\!0$, $A^{}_\pi\!=\!0$, $\phi\!=\!0$, 
and $|q/p|\!=\!1$;
{\it (b)}\ assuming no direct \cpv\ in doubly Cabibbo-suppressed (DCS)
decays and fitting for parameters $(x,y,|q/p|)$ or $(x,y,\phi)$; 
{\it (c)}\ assuming no direct \cpv\ in DCS decays and fitting for
alternative parameters $x^{}_{12}= 2|M^{}_{12}|/\Gamma$, 
$y^{}_{12}= \Gamma^{}_{12}/\Gamma$, and 
$\phi^{}_{12}= {\rm Arg}(M^{}_{12}/\Gamma^{}_{12})$,
where $M^{}_{12}$ and $\Gamma^{}_{12}$ are the off-diagonal
elements of the $D^0$-$\dbar$ mass and decay matrices, respectively; and
{\it (d)}\ allowing full \cpv, \ie\ floating all parameters. 

For the fits assuming no-direct-\cpv\ in DCS decays, 
we set $A^{}_D\!=\!0$. In addition, for fit 
{\it (b)\/} we impose the relation~\cite{Ciuchini:2007cw,Kagan:2009gb}
$\tan\phi = (1-|q/p|^2)/(1+|q/p|^2)\times (x/y)$, which reduces 
four independent parameters to 
three.\footnote{One can also use Eq.~(15) of Ref.~\cite{Grossman:2009mn}
to reduce four parameters to three.} 
We impose this relationship in two ways. First we float parameters
$x$, $y$, and $\phi$ and from these derive $|q/p|$; subsequently we
repeat the fit floating $x$, $y$, and $|q/p|$ and from these derive 
$\phi$. The central values returned by the two fits are identical,
but the first fit yields MINOS errors for $\phi$, while the second
fit yields MINOS errors for $|q/p|$. For no-direct-\cpv\ fit 
{\it (c)}, we fit for the underlying parameters $x^{}_{12}$, $y^{}_{12}$, 
and $\phi^{}_{12}$, from which parameters $x$, $y$, $|q/p|$, and $\phi$ 
are derived. 

All fit results are listed in 
Table~\ref{tab:results}. For the \cpv-allowed fit,
individual contributions to the $\chi^2$ are listed 
in Table~\ref{tab:results_chi2}. The total $\chi^2$ 
is 66.8 for $45-10=35$ degrees of freedom; this 
corresponds to a confidence level of~0.001,
which is uncomfortably small.

\begin{figure}
\begin{center}
\includegraphics[width=4.2in]{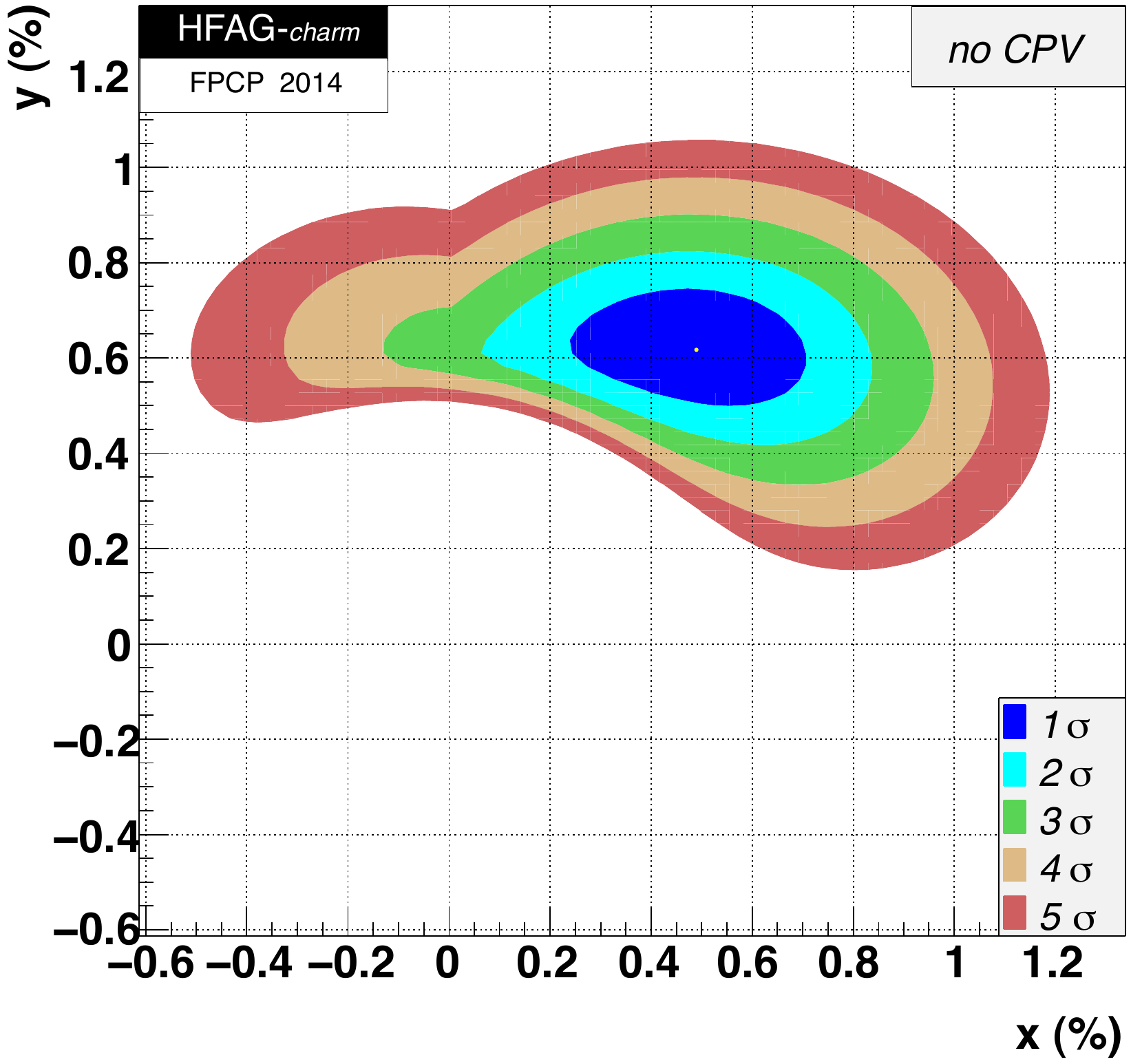}
\end{center}
\vskip-0.20in
\caption{\label{fig:contours_ncpv}
Two-dimensional contours for mixing parameters $(x,y)$, for no \cpv. }
\end{figure}

\begin{figure}
\begin{center}
\vbox{
\includegraphics[width=84mm]{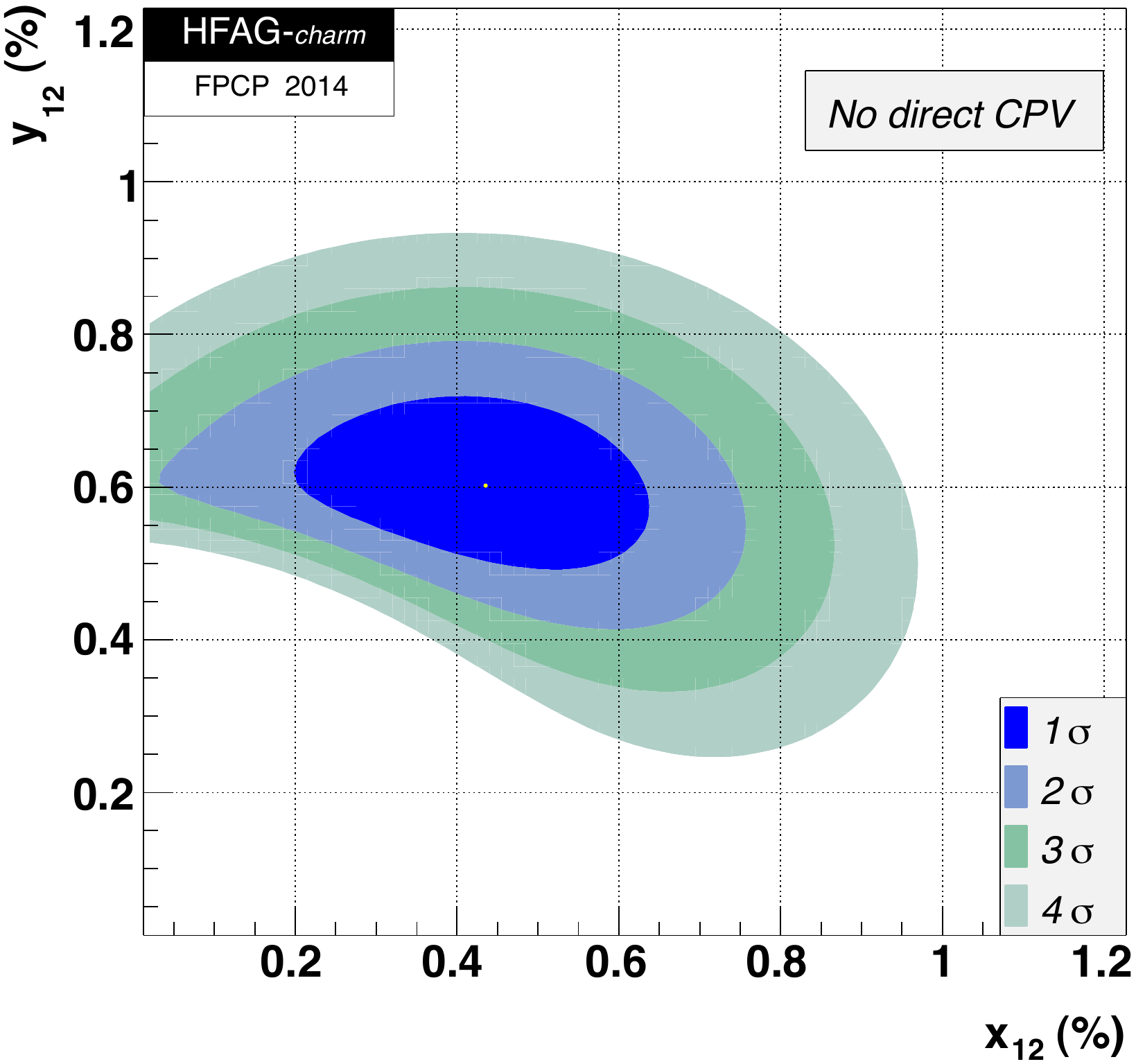}
\includegraphics[width=84mm]{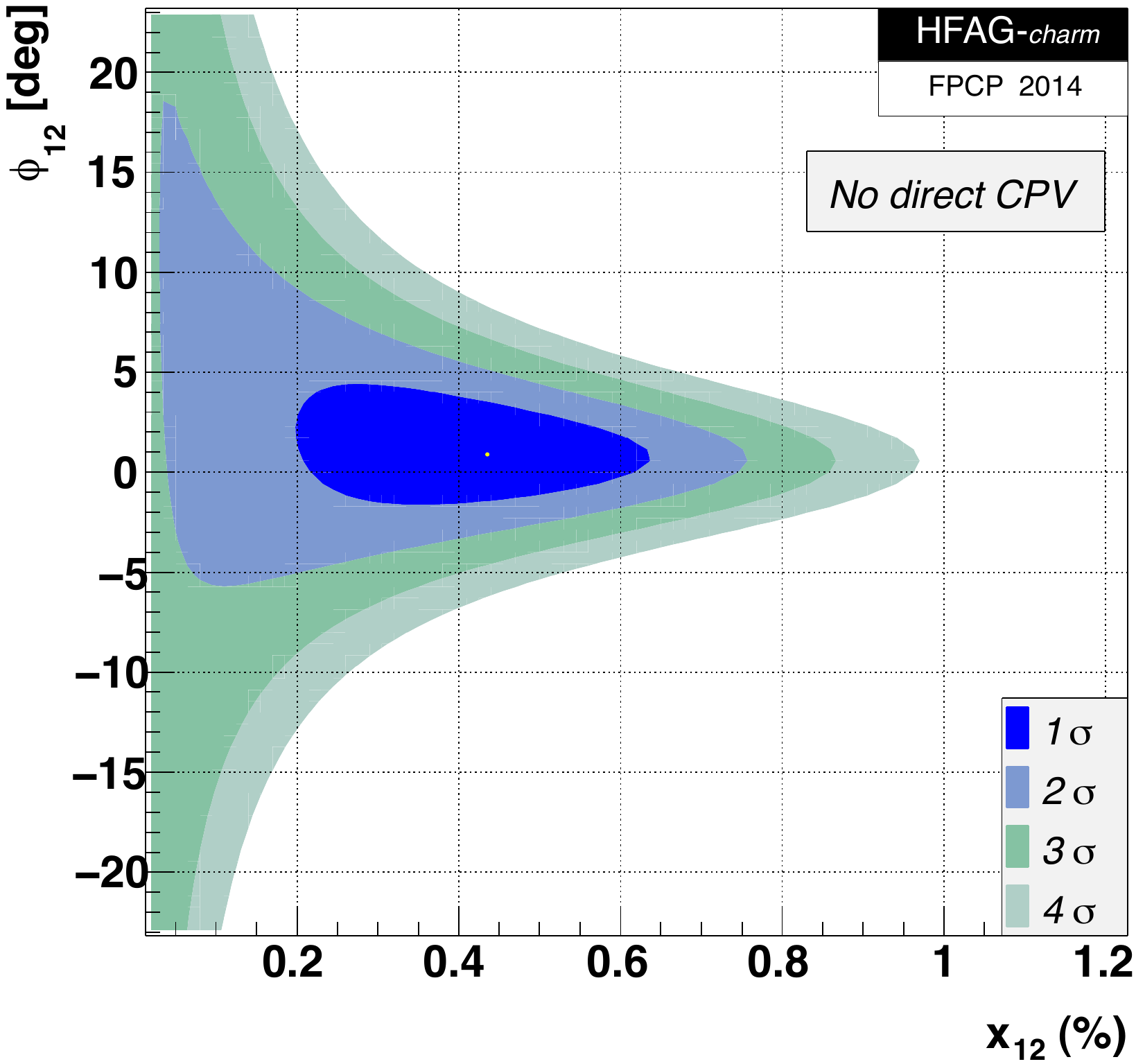}
\vskip0.30in
\includegraphics[width=84mm]{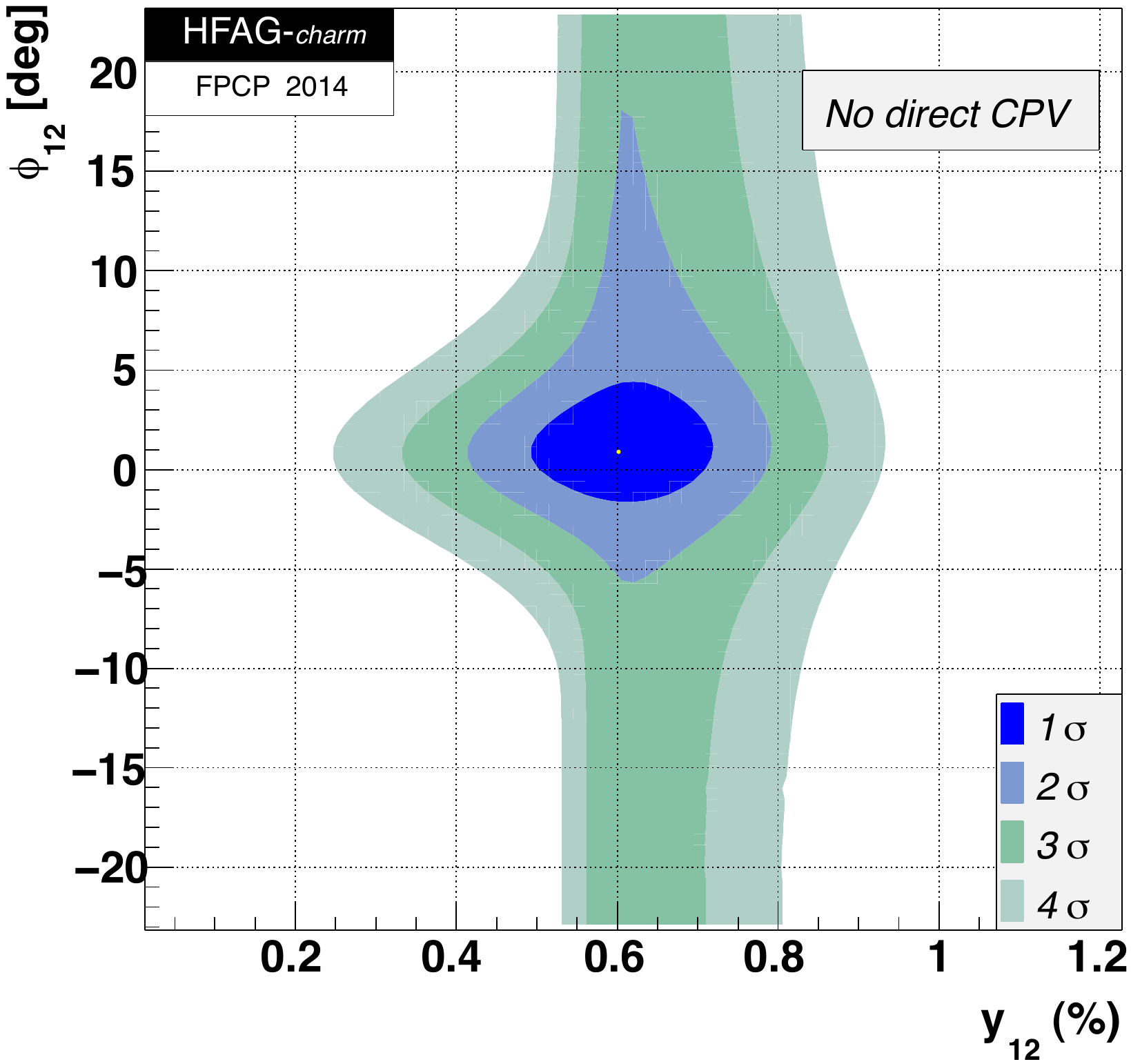}
}
\end{center}
\vskip-0.10in
\caption{\label{fig:contours_ndcpv}
Two-dimensional contours for theoretical parameters 
$(x^{}_{12},y^{}_{12})$ (top left), 
$(x^{}_{12},\phi^{}_{12})$ (top right), and 
$(y^{}_{12},\phi^{}_{12})$ (bottom), 
for no direct \cpv.}
\end{figure}

\begin{figure}
\begin{center}
\vbox{
\includegraphics[width=4.2in]{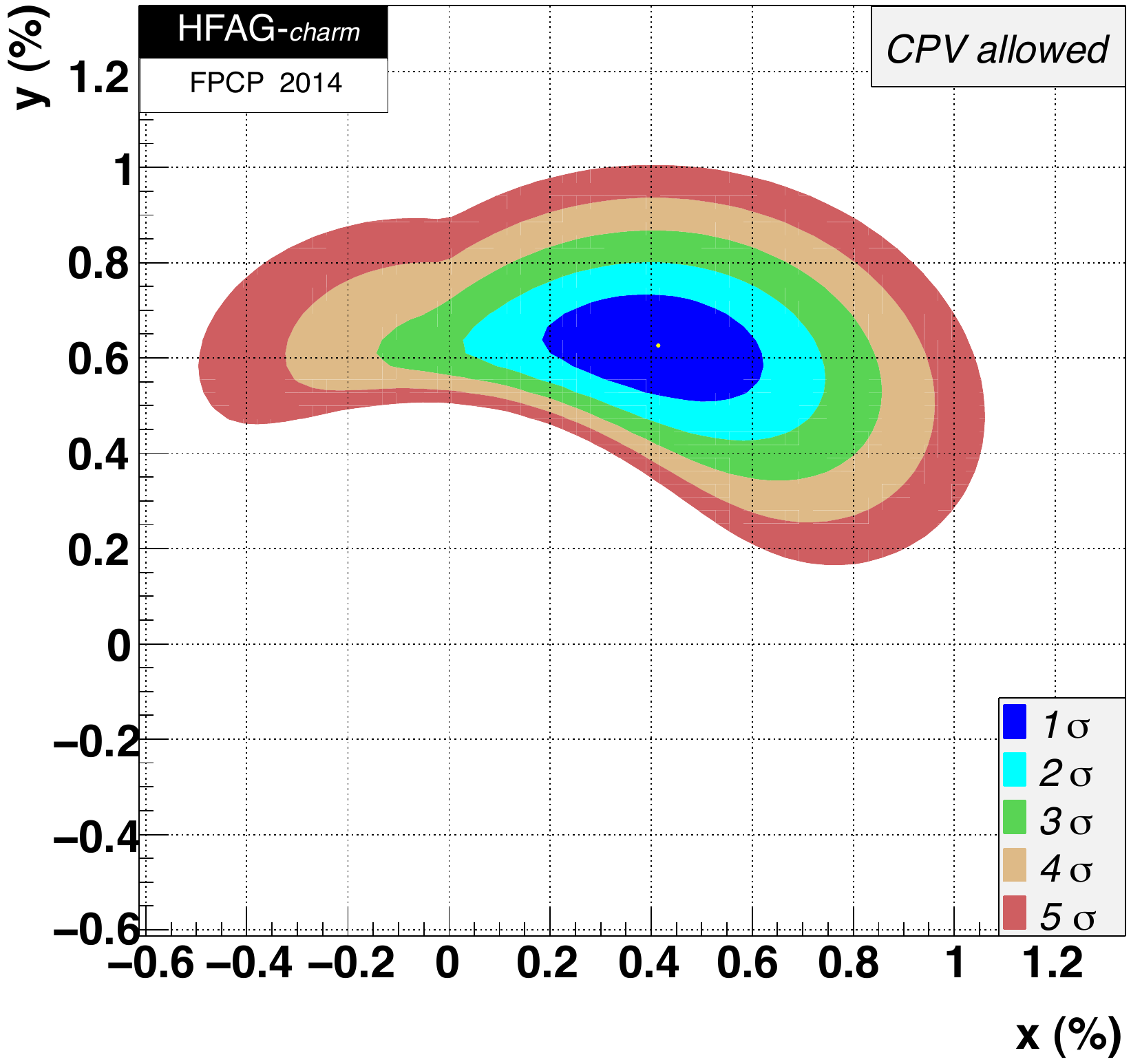}
\vskip0.10in
\includegraphics[width=4.2in]{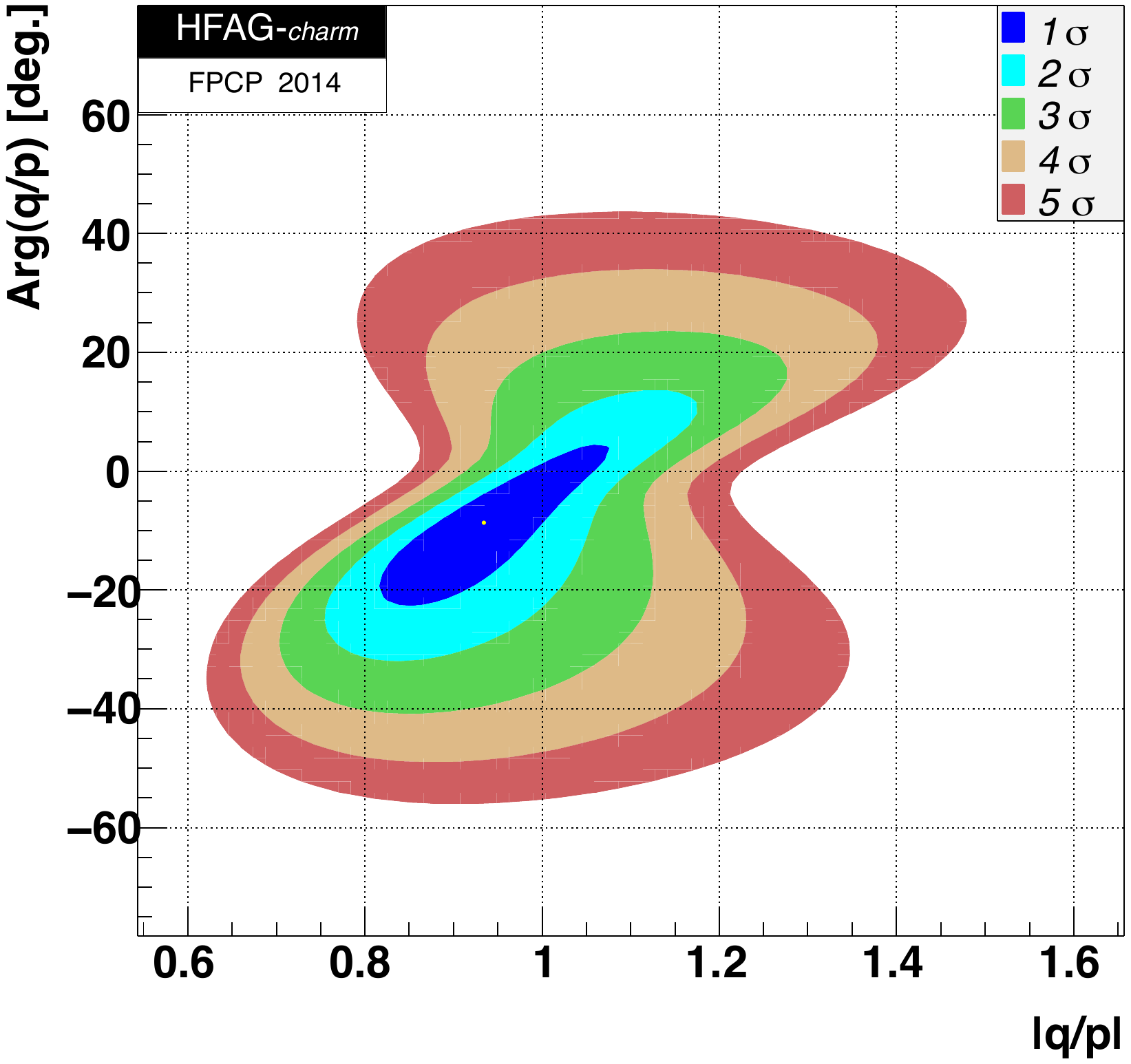}
}
\end{center}
\vskip-0.10in
\caption{\label{fig:contours_cpv}
Two-dimensional contours for parameters $(x,y)$ (top) 
and $(|q/p|,\phi)$ (bottom), allowing for \cpv.}
\end{figure}

Confidence contours in the two dimensions $(x,y)$ or 
in $(|q/p|,\phi)$ are obtained by allowing, for any point in the
two-dimensional plane, all other fitted parameters to take their 
preferred values. The resulting $1\sigma$-$5\sigma$ contours 
are shown 
in Fig.~\ref{fig:contours_ncpv} for the \cp-conserving case, 
in Fig.~\ref{fig:contours_ndcpv} for the no-direct-\cpv\ case, 
and in Fig.~\ref{fig:contours_cpv} for the \cpv-allowed 
case. The contours are determined from the increase of the
$\chi^2$ above the minimum value.
One observes that the $(x,y)$ contours for the no-\cpv\ fit 
are very similar to those for the \cpv-allowed fit. In the latter
fit, the $\chi^2$ at the no-mixing point $(x,y)\!=\!(0,0)$ is 421
units above the minimum value, which, for two degrees of freedom,
corresponds to a confidence level $>11.5\sigma$.\footnote{This is
the limit of the CERNLIB PROB routine used for this calculation.}
Thus, no mixing is excluded at this high level. In the $(|q/p|,\phi)$
plot, the point $(1,0)$ is within the $1\sigma$ contour; thus the
data is consistent with \cp\ conservation.

\begin{figure}
\begin{center}
\hbox{\hskip0.50in
\includegraphics[width=72mm]{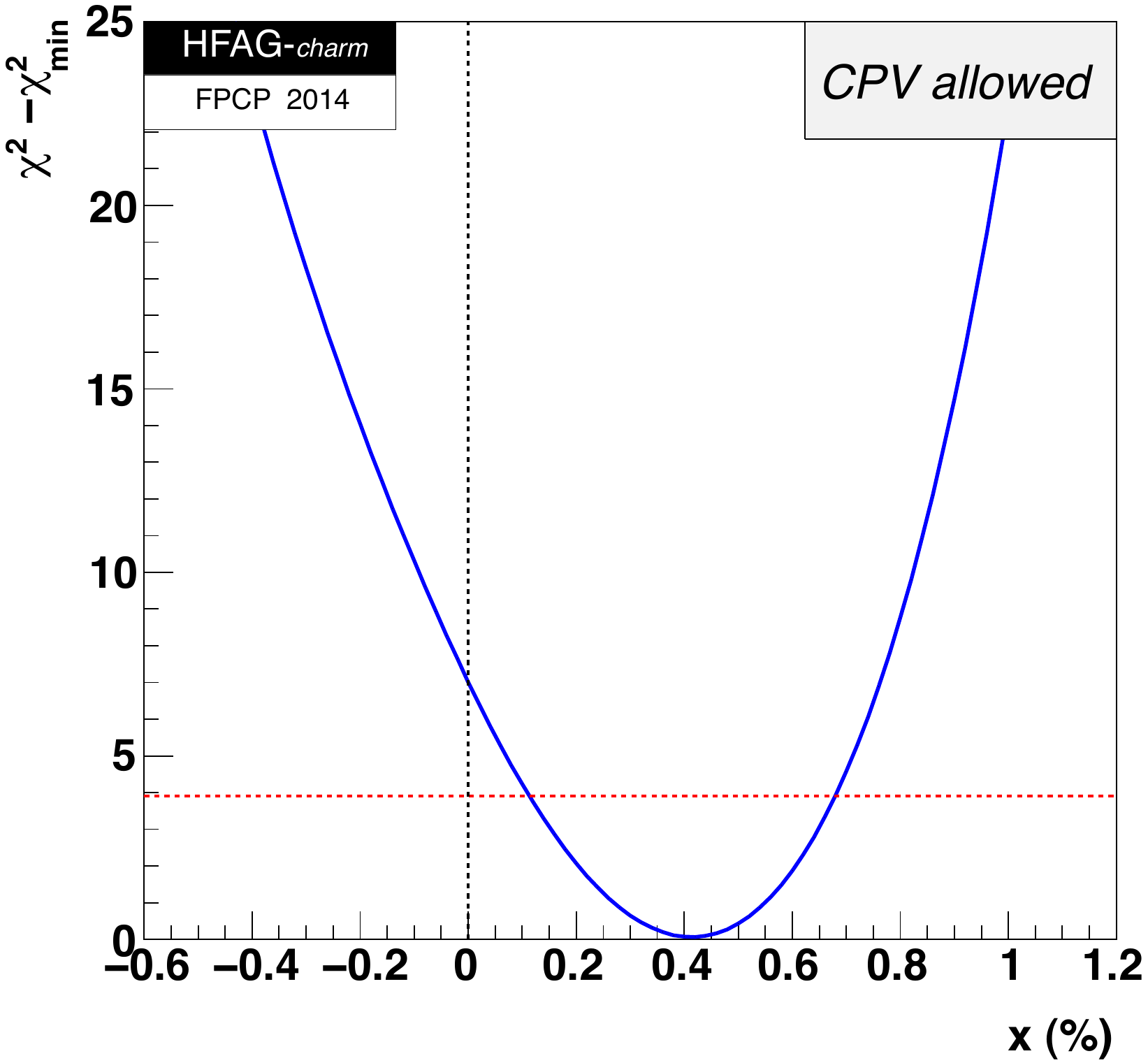}
\hskip0.20in
\includegraphics[width=72mm]{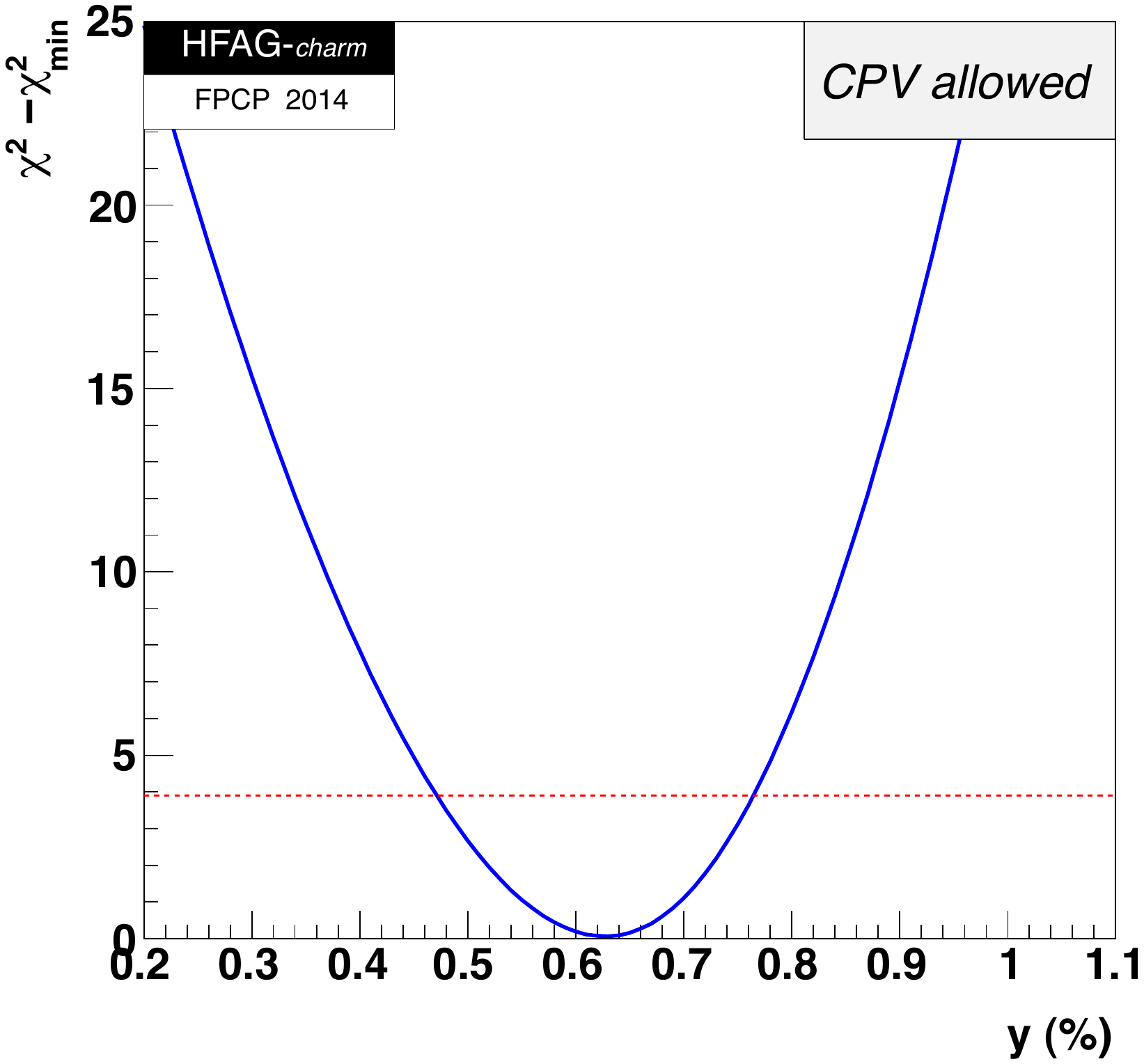}}
\hbox{\hskip0.50in
\includegraphics[width=72mm]{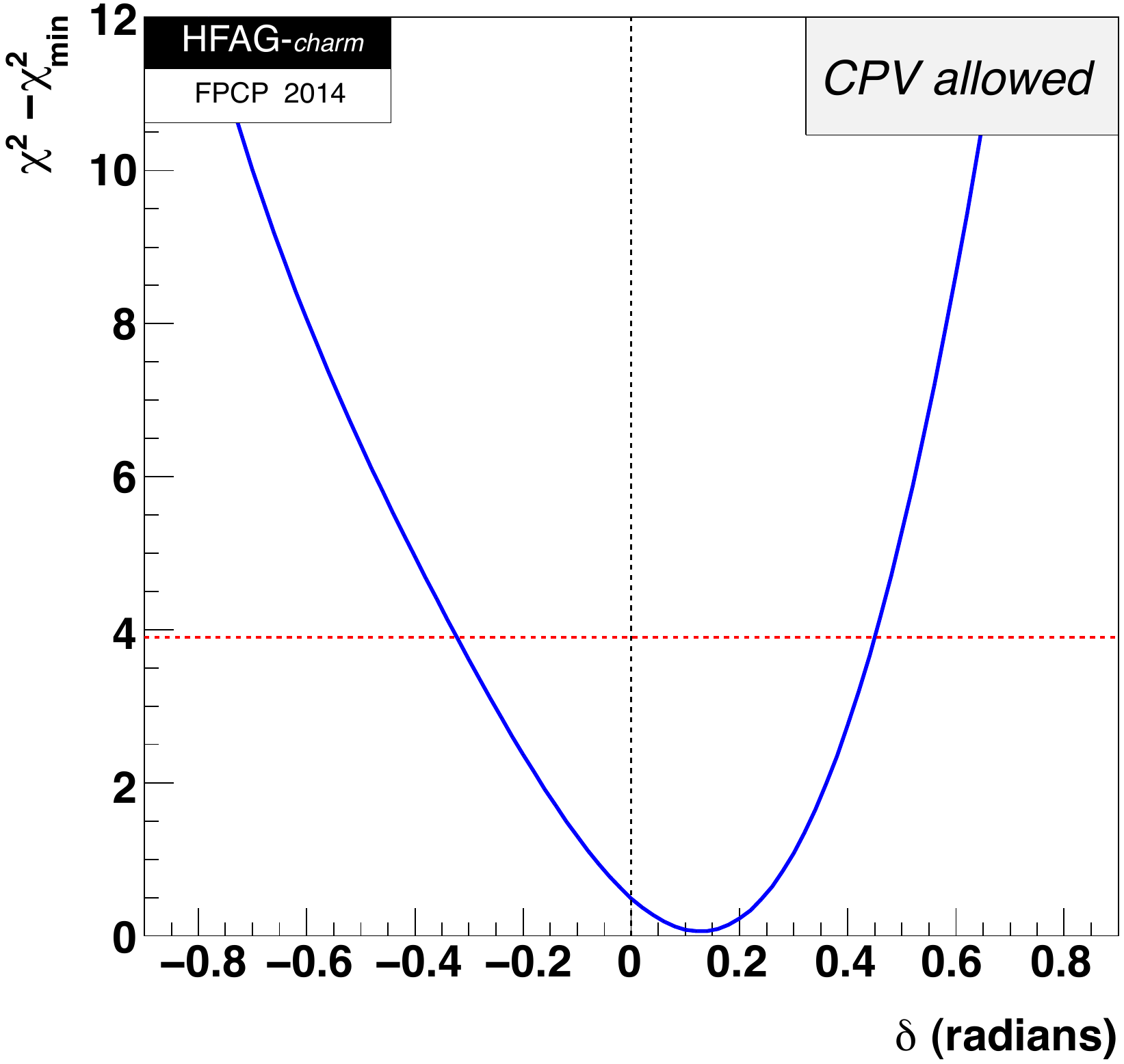}
\hskip0.20in
\includegraphics[width=72mm]{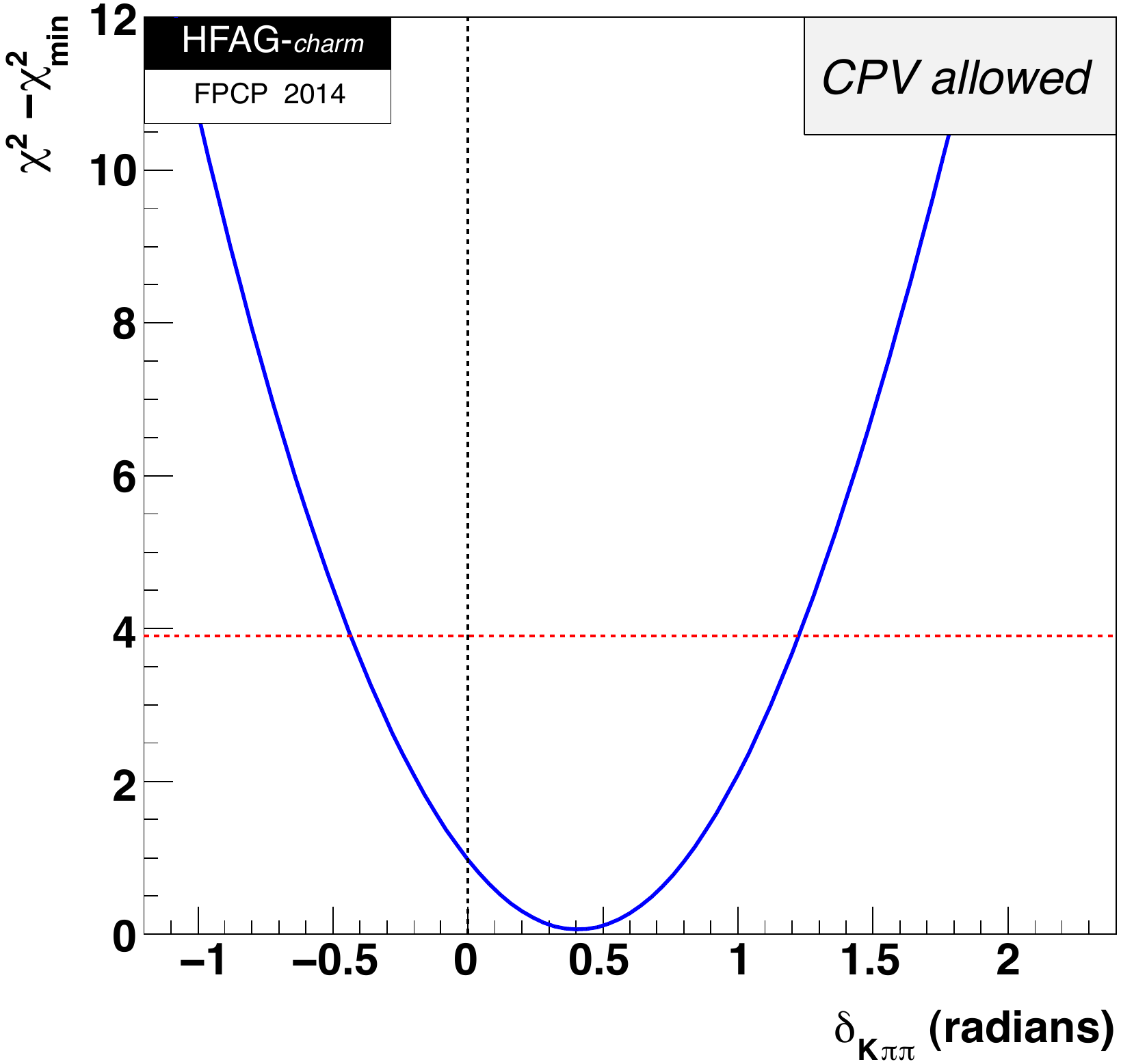}}
\hbox{\hskip0.50in
\includegraphics[width=72mm]{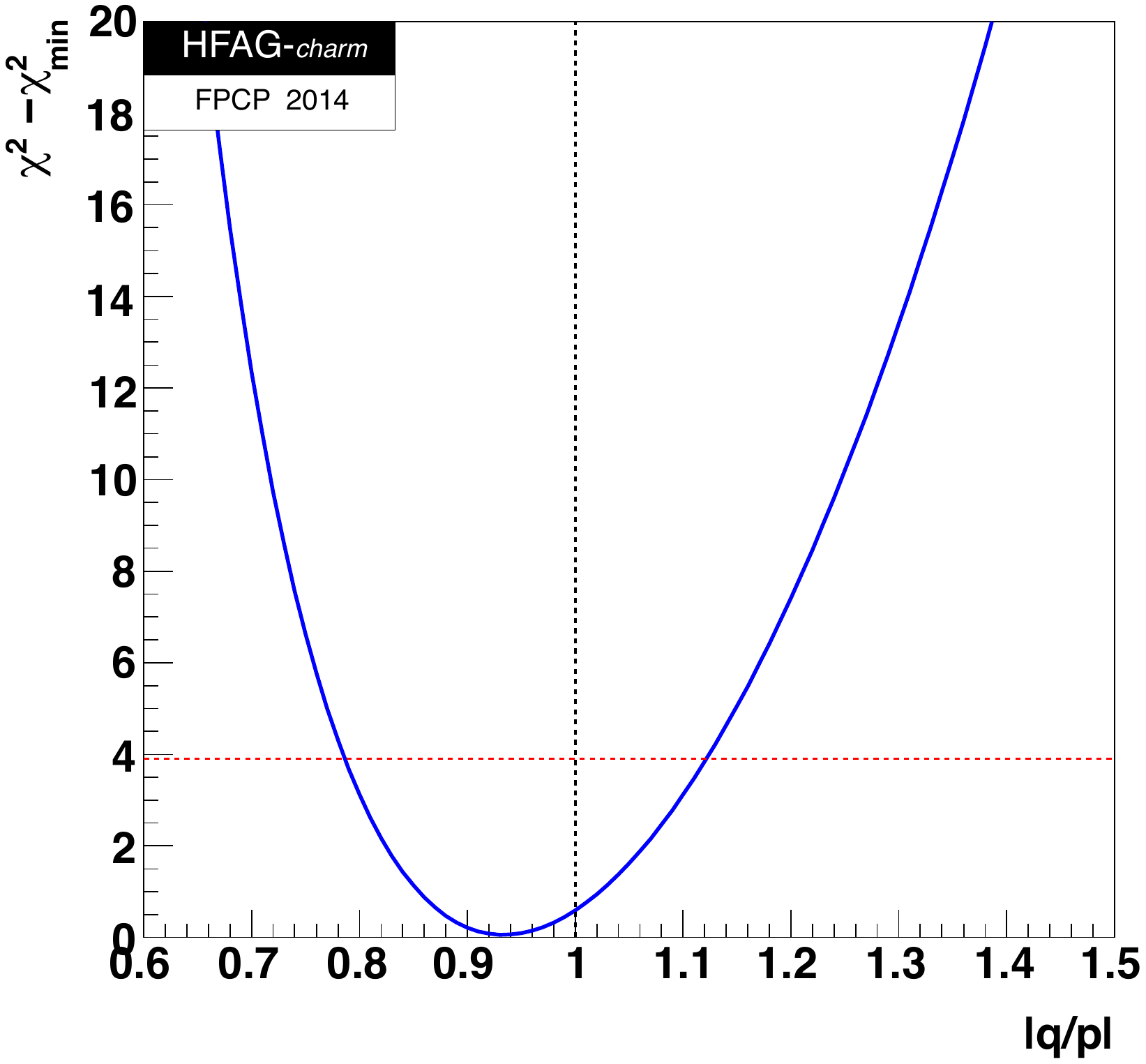}
\hskip0.20in
\includegraphics[width=72mm]{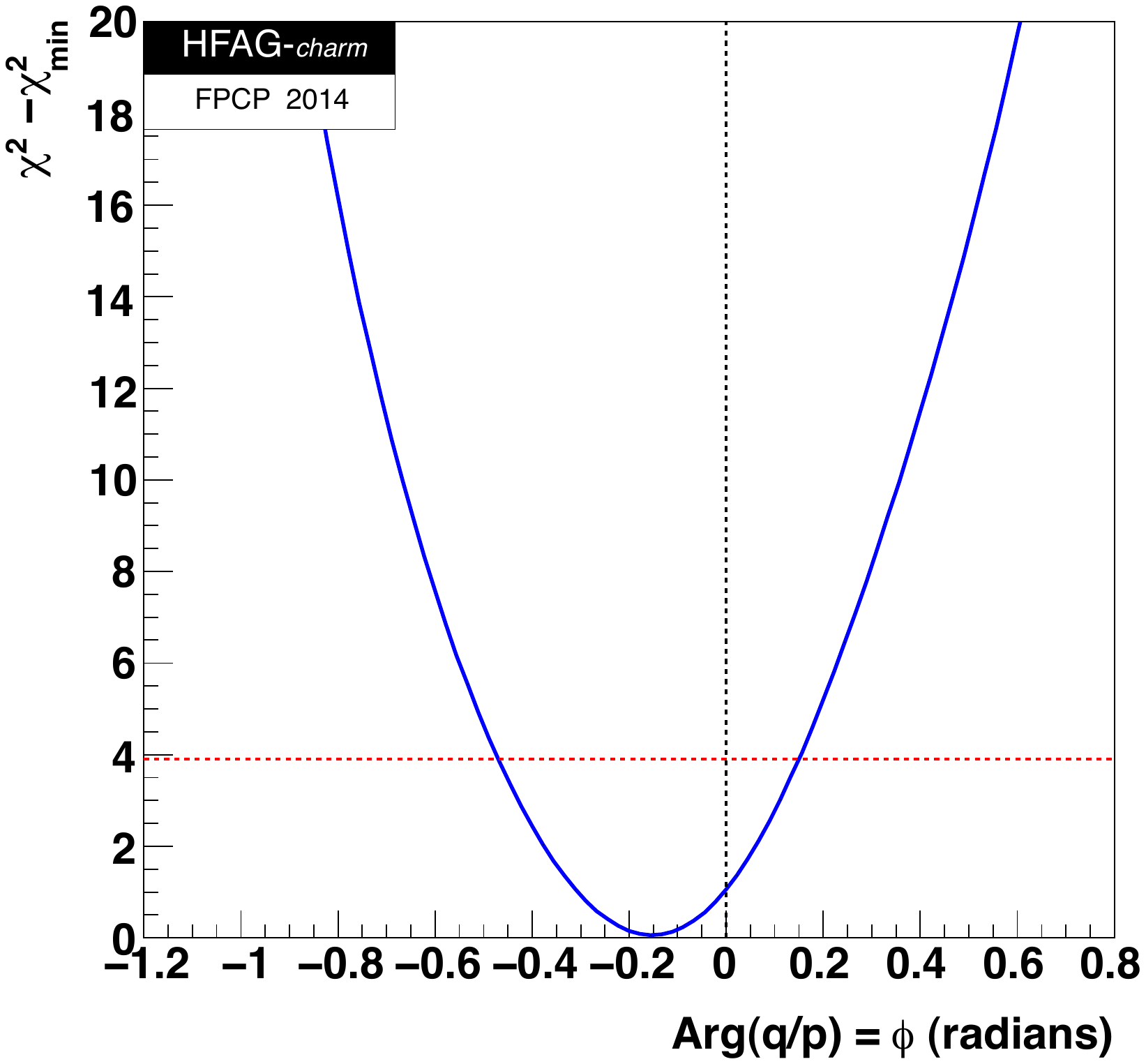}}
\end{center}
\vskip-0.30in
\caption{\label{fig:1dlikelihood}
The function $\Delta\chi^2=\chi^2-\chi^2_{\rm min}$ 
for fitted parameters
$x,\,y,\,\delta,\,\delta^{}_{K\pi\pi},\,|q/p|$, and $\phi$.
The points where $\Delta\chi^2=3.84$ (denoted by dashed 
horizontal lines) determine 95\% C.L. intervals. }
\end{figure}

One-dimensional confidence curves for individual parameters 
are obtained by allowing, for any value of the parameter, all other 
fitted parameters to take their preferred values. The resulting
functions $\Delta\chi^2=\chi^2-\chi^2_{\rm min}$ ($\chi^2_{\rm min}$
is the minimum value) are shown in Fig.~\ref{fig:1dlikelihood}.
The points where $\Delta\chi^2=3.84$ determine 95\% C.L. intervals 
for the parameters. These intervals are listed in Table~\ref{tab:results}.

\begin{table}
\renewcommand{\arraystretch}{1.4}
\begin{center}
\caption{\label{tab:results}
Results of the global fit for different assumptions concerning~\cpv.}
\vspace*{6pt}
\footnotesize
\begin{tabular}{c|cccc}
\hline
\textbf{Parameter} & \textbf{\boldmath No \cpv} & \textbf{\boldmath No direct \cpv} 
& \textbf{\boldmath \cpv-allowed} & \textbf{\boldmath \cpv-allowed} \\
 & & \textbf{\boldmath in DCS decays} & & \textbf{\boldmath 95\% CL Interval} \\
\hline
$\begin{array}{c}
x\ (\%) \\ 
y\ (\%) \\ 
\delta^{}_{K\pi}\ (^\circ) \\ 
R^{}_D\ (\%) \\ 
A^{}_D\ (\%) \\ 
|q/p| \\ 
\phi\ (^\circ) \\
\delta^{}_{K\pi\pi}\ (^\circ)  \\
A^{}_{\pi} (\%) \\
A^{}_K (\%) \\
x^{}_{12}\ (\%) \\ 
y^{}_{12}\ (\%) \\ 
\phi^{}_{12} (^\circ)
\end{array}$ & 
$\begin{array}{c}
0.49\,^{+0.14}_{-0.15} \\
0.62\,\pm 0.08 \\
7.8\,^{+9.6}_{-11.1} \\
0.350\,\pm 0.004 \\
- \\
- \\
- \\
18.7\,^{+23.2}_{-23.7} \\
- \\
- \\
- \\
- \\
- 
\end{array}$ &
$\begin{array}{c}
0.43\,^{+0.14}_{-0.15}\\
0.60\,\,\pm 0.07 \\
4.6\,^{+10.3}_{-12.0}\\
0.349\,\pm 0.004 \\
- \\
1.007\,^{+0.015}_{-0.014}\\
-0.30\,^{+0.58}_{-0.60} \\ 
20.8\,^{+23.9}_{-24.3} \\ 
0.11\,\pm 0.14 \\
-0.13\,\pm 0.13 \\
0.43\,^{+0.14}_{-0.15}\\
0.60\,\pm 0.07 \\
0.9\,^{+1.9}_{-1.7} 
\end{array}$ &
$\begin{array}{c}
0.41\,^{+0.14}_{-0.15}\\
0.63\,\,^{+0.07}_{-0.08}\\
7.3\,^{+9.8}_{-11.5} \\
0.349\,\pm 0.004 \\
-0.71\,^{+0.92}_{-0.95} \\
0.93\,^{+0.09}_{-0.08} \\ 
-8.7\,^{+8.7}_{-9.1} \\ 
23.3\,^{+23.9}_{-24.4} \\
0.14\,\pm 0.15 \\
-0.11\,^{+0.14}_{-0.13} \\
 \\
 \\
 \\
\end{array}$ &
$\begin{array}{c}
\left[ 0.11,\, 0.68\right] \\
\left[ 0.47,\, 0.76\right] \\
\left[ -18.5,\, 25.8\right] \\
\left[ 0.342,\, 0.356\right] \\
\left[ -2.6,\, 1.1\right] \\
\left[ 0.79,\, 1.12\right] \\\
\left[ -26.9,\, 8.6\right] \\
\left[ -24.8,\, 70.2\right] \\
\left[ -0.15,\, 0.42\right] \\
\left[ -0.37,\, 0.15\right] \\
\left[ 0.13,\, 0.69\right] \\
\left[ 0.45,\, 0.75\right] \\
\left[ -3.0,\, 6.1\right] \\
\end{array}$ \\
\hline
\end{tabular}
\end{center}
\end{table}

\begin{table}
\renewcommand{\arraystretch}{1.4}
\begin{center}
\caption{\label{tab:results_chi2}
Individual contributions to the $\chi^2$ for the \cpv-allowed fit.}
\vspace*{6pt}
\footnotesize
\begin{tabular}{l|rr}
\hline
\textbf{Observable} & \textbf{\boldmath $\chi^2$} & \textbf{\boldmath $\sum\chi^2$} \\
\hline
$y^{}_{CP}$                      & 2.57 & 2.57 \\
$A^{}_\Gamma$                    & 0.44 & 3.02 \\
\hline
$x^{}_{K^0\pi^+\pi^-}$ Belle       & 0.55 & 3.57 \\
$y^{}_{K^0\pi^+\pi^-}$ Belle       & 3.87 & 7.44 \\
$|q/p|^{}_{K^0\pi^+\pi^-}$ Belle   & 0.28 & 7.72 \\
$\phi^{}_{K^0\pi^+\pi^-}$  Belle   & 0.16 & 7.88 \\
\hline
$x^{}_{K^0 h^+ h^-}$ \babar        & 0.87 & 8.76 \\
$y^{}_{K^0 h^+ h^-}$ \babar        & 0.03 & 8.79 \\
\hline
$R^{}_M(K^+\ell^-\nu)$           & 0.14 & 8.93 \\
\hline
$x^{}_{K^+\pi^-\pi^0}$ \babar      & 7.15 & 16.08 \\
$y^{}_{K^+\pi^-\pi^0}$ \babar      & 3.99 & 20.08 \\
\hline
CLEOc                           &      &       \\
($x/y/R^{}_D/\cos\delta/\sin\delta$) 
                                & 10.08 & 30.16 \\
\hline
$R^+_D/x'{}^{2+}/y'{}^+$ \babar  & 11.66 & 41.82    \\
$R^-_D/x'{}^{2-}/y'{}^-$ \babar  &  6.02 & 47.83    \\
$R^+_D/x'{}^{2+}/y'{}^+$ Belle   &  2.13 & 49.96    \\
$R^-_D/x'{}^{2-}/y'{}^-$ Belle   &  3.18 & 53.14    \\
$R^{}_D/x'{}^{2}/y'$ CDF         &  1.15 & 54.29    \\
$R^+_D/x'{}^{2+}/y'{}^+$ LHCb    &  1.11 & 55.50    \\
$R^-_D/x'{}^{2-}/y'{}^-$ LHCb    &  1.27 & 56.67    \\
\hline
$A^{}_{KK}/A^{}_{\pi\pi}$  \babar & 0.53 & 57.19  \\
$A^{}_{KK}/A^{}_{\pi\pi}$  Belle  & 2.06 & 59.25  \\
$A^{}_{KK}/A^{}_{\pi\pi}$  CDF    & 2.64 & 61.89  \\
$A^{}_{KK}-A^{}_{\pi\pi}$  LHCb ($D^*$ tag)   
                                & 0.26 & 62.15  \\
$A^{}_{KK}-A^{}_{\pi\pi}$  LHCb ($B^0\ra D^0\mu X$ tag)
                                & 4.65 & 66.80  \\
\hline
\end{tabular}
\end{center}
\end{table}


\subsubsection{Conclusions}

From the fit results listed in Table~\ref{tab:results}
and shown in Figs.~\ref{fig:contours_cpv} and \ref{fig:1dlikelihood},
we conclude that:
\begin{itemize}
\item the experimental data consistently indicate that 
$D^0$ mesons undergo mixing. The no-mixing point $x=y=0$
is excluded at $>11.5\sigma$. The parameter $x$ differs
from zero by $2.4\sigma$, and $y$ differs from zero by
$9.4\sigma$. This mixing is presumably dominated 
by long-distance processes, which are difficult to calculate.
Thus, unless it turns out that $|x|\gg |y|$~\cite{Bigi:2000wn}
(which is not currently indicated), it will be difficult to
identify new physics from $(x,y)$ alone.
\item Since \ycp\ is positive, the \cp-even state is shorter-lived
as in the $K^0$-$\kbar$ system. However, since $x$ also appears
to be positive, the \cp-even state is heavier, 
unlike in the $K^0$-$\kbar$ system.
\item The LHCb and CDF experiments measured time-integrated
asymmetries that hint at {\it direct\/} \cpv\ in $D^0$ decays
(see Table~\ref{tab:observables3}). However, more statistics 
are needed to clarify this effect.
There is no evidence for \cpv\ arising from $D^0$-$\dbar$
mixing ($|q/p|\neq 1$) or from a phase difference between
the mixing amplitude and a direct decay amplitude ($\phi\neq 0$). 
\end{itemize}

\clearpage
\subsection{\emph{CP} asymmetries}\label{sec:cp_asym}

\emph{CP} violation occurs if the decay rate for a particle differs 
from that of its \emph{CP}-conjugate\cite{Bigi:2000yz}. 
In general there are two classes of \emph{CP} violation, termed
{\it indirect\/} and {\it direct\/}\cite{Nir:1999mg}. Indirect \emph{CP} 
violation refers to $\Delta C\!=\!2$ processes and 
arises in $D^0$ decays due to $D^0$-$\dbar$ mixing. 
It can occur as an asymmetry in the mixing itself, or it can 
result from interference between a decay 
amplitude arising via mixing and a non-mixed amplitude. 
Direct \emph{CP} violation refers to $\Delta C\!=\!1$
processes and can occur in both charged and neutral 
$D$ decays. It results from interference between two different decay
amplitudes (\eg\ a penguin and tree amplitude) that have
different weak (CKM) and strong phases\footnote{The weak 
phase difference will have opposite signs for $D\ra f$ and 
$\overline{D}\ra\bar{f}$ decays, while the strong phase difference 
will have the same sign. As a result, squaring the total amplitudes 
to obtain the decay rates gives interference terms having 
opposite sign, \ie\ non-identical decay rates.}.
A difference in strong phases typically arises due to 
final-state interactions (FSI)\cite{Buccella:1994nf}. A difference
in weak phases arises from different CKM vertex couplings, as 
is often the case for spectator and penguin diagrams.

\vspace{0.8cm}
The \emph{CP} asymmetry is defined as the difference between 
$D$ and $\overline{D}$ partial widths divided by their sum:
\begin{eqnarray}  
A_{CP} & = & \frac{\Gamma(D)-\Gamma(\overline{D})}
{\Gamma(D)+\Gamma(\overline{D})}\,.
\end{eqnarray}
However, to take into account differences in production rates between 
$D$ and $\overline{D}$ (which would affect the number of respective 
decays observed), some experiments normalize to a Cabibbo-favored 
mode. In this case there is the additional benefit that most corrections 
due to inefficiencies cancel out, reducing systematic uncertainties. An 
implicit assumption is that there is no measurable \emph{CP} 
violation in the Cabibbo-favored normalizing mode. 
The \emph{CP} asymmetry is calculated as
\begin{eqnarray}
A_{CP} & = & \frac{\eta(D)-\eta(\overline{D})}{\eta(D)+\eta(\overline{D})}\,,
\end{eqnarray}
where (considering, for example, $D^0 \to K^-K^+$)
\begin{eqnarray}
 \eta(D) & = & \frac{N(D^0 \rightarrow K^-K^+)}{N(D^0 \rightarrow K^-\pi^+)}\,, \\
 \eta(\overline{D}) & = & \frac{N(\dbar\rightarrow K^-K^+)}
{N(\dbar\rightarrow K^+\pi^-)}\,.
\end{eqnarray}
In the case of $D^+$ and $D^+_s$ decays, $A^{}_{CP}$ measures 
direct \emph{CP} violation; in the case of $D^0$ decays, $A^{}_{CP}$ 
measures direct and indirect \emph{CP} violation combined.
Values of $A^{}_{CP}$ for $D^+$, $D^0$ and $D_s^+$ decays are listed in
Tables~\ref{tab:cp_charged}, \ref{tab:cp_charged2}, \ref{tab:cp_neutral}, \ref{tab:cp_neutral2} and \ref{tab:cp_ds} respectively.
In these tables we report asymmetries for the actual
final state, \ie\ resonant substructure is implicitly 
included by not considered separately.

Overall, \emph{CP} asymmetry measurements have been
carried out for 47 charm decay modes, and in
several modes the sensitivity is below $5 \times 10^{-3}$. 
There is currently no evidence for \emph{CP} violation in
the charm sector, with one exception: the asymmetry
observed for $D^+ \to K^0_s\pi^+$. 
However, the \emph{CP} asymmetry observed is consistent
with that expected due to \emph{CP} violation in
$K^0$-$\kbar$ mixing~\cite{Grossman:2012aa},
and thus it is not attributed to charm.
Taken together, the limits obtained for \emph{CP} asymmetries 
in the charm sector pose tight constraints on new physics models.

\begin{table}[!htb]
\renewcommand{\arraystretch}{1.4}
\caption{\cp\ asymmetries 
$A^{}_{CP}= [\Gamma(D^+)-\Gamma(D^-)]/[\Gamma(D^+)+\Gamma(D^-)]$
for two-body $D^\pm$ decays.
\label{tab:cp_charged}}
\footnotesize
\begin{center}
\begin{tabular}{|l|c|c|c|} 
\hline
{\bf Mode} & {\bf Year} & {\bf Collaboration} & {\boldmath $A^{}_{CP}$} \\
\hline
{\boldmath $D^+ \to \mu^+ \nu$} &
  2008 & CLEOc~\cite{Eisenstein:2008aa} &  $ +0.08  \pm 0.08 $ \\
\hline
{\boldmath $D^+ \to \pi^+ \pi^0$} &
  2010 & CLEOc~\cite{Mendez:2009aa} &  $ +0.029  \pm 0.029 \pm 0.003 $ \\
\hline
{\boldmath $D^+ \to \pi^+ \eta$} &
  2011 & Belle~\cite{Won:2011ng}    &  $ +0.0174  \pm 0.0113 \pm 0.0019 $ \\
& 2010 & CLEOc~\cite{Mendez:2009aa} &  $ -0.020   \pm 0.023  \pm 0.003 $ \\
&      & COMBOS average             &  $ +0.010   \pm 0.010 $ \\
\hline
{\boldmath $D^+ \to \pi^+ \eta^\prime$} &
  2011 & Belle~\cite{Won:2011ng}    &  $ -0.0012  \pm 0.0112 \pm 0.0017 $ \\
& 2010 & CLEOc~\cite{Mendez:2009aa} &  $ -0.040   \pm 0.034  \pm 0.003  $ \\
&      & COMBOS average             &  $ -0.005   \pm 0.011 $ \\  
\hline
{\boldmath $D^+ \to K^+ \pi^0$} &
  2010 & CLEOc~\cite{Mendez:2009aa} &  $ -0.035  \pm 0.107 \pm 0.009 $ \\
\hline
{\boldmath $D^+ \to K^0_s\pi^+$}   &
   2014 & CLEOc~\cite{Bonvicini:2013vxi} &  $ -0.011   \pm 0.006   \pm 0.002   $ \\
&  2012 & Belle~\cite{Ko:2012aa}        &  $ -0.00363 \pm 0.00094 \pm 0.00067 $ \\
&  2011 & \babar~\cite{Amo:2011ab}      &  $ -0.0044  \pm 0.0013  \pm 0.0010  $ \\
&  2002 & FOCUS~\cite{Link:2001zj}      &  $ -0.016   \pm 0.015   \pm 0.009   $ \\
&       & COMBOS average                &  $ -0.0041  \pm 0.0009 $ \\
\hline
{\boldmath $D^+ \to K^0_sK^+$} &
  2014 & LHCb~\cite{Aaij:2014ac}        &  $ +0.0003 \pm 0.0017 \pm 0.0014 $ \\
& 2013 & \babar~\cite{Lees:2013aa}      &  $ +0.0013 \pm 0.0036 \pm 0.0025 $ \\
& 2013 & Belle~\cite{Ko:2013aa}         &  $ -0.0025 \pm 0.0028 \pm 0.0014 $ \\
& 2010 & CLEOc~\cite{Mendez:2009aa}     &  $ -0.002  \pm 0.015  \pm 0.009  $ \\  
& 2002 & FOCUS~\cite{Link:2001zj}       &  $ +0.071  \pm 0.061  \pm 0.012  $ \\
&      & COMBOS average                 &  $ -0.0003 \pm 0.0017 $ \\
\hline
\end{tabular}
\end{center} 
\end{table}

\begin{table}[!htb]
\renewcommand{\arraystretch}{1.4}
\caption{\cp\ asymmetries 
$A^{}_{CP}= [\Gamma(D^+)-\Gamma(D^-)]/[\Gamma(D^+)+\Gamma(D^-)]$
for three- and four-body $D^\pm$ decays.
\label{tab:cp_charged2}}
\footnotesize
\begin{center}
\begin{tabular}{|l|c|c|c|} 
\hline
{\bf Mode} & {\bf Year} & {\bf Collaboration} & {\boldmath $A^{}_{CP}$} \\
\hline
{\boldmath $D^+ \to \pi^+\pi^-\pi^+$} &
  2014 & LHCb~\cite{Aaij:2014aa}        &  Dalitz plot analysis, no evidence for CP violation \\
& 1997 & E791~\cite{Aitala:1996sh}      &  $ -0.017  \pm 0.042  $ (stat.) \\
\hline
{\boldmath $D^+ \to K^-\pi^+\pi^+$} &
  2014 & CLEOc~\cite{Bonvicini:2013vxi}  &  $ -0.003  \pm 0.002 \pm 0.004  $ \\
\hline
{\boldmath $D^+ \to K^0_s\pi^+\pi^0$} &
  2014 & CLEOc~\cite{Bonvicini:2013vxi} &  $ -0.001  \pm 0.007 \pm 0.002  $ \\
\hline
{\boldmath $D^+ \to K^+K^-\pi^+$} &
   2014 & CLEOc~\cite{Bonvicini:2013vxi} &  $ -0.001  \pm 0.009  \pm 0.004  $ \\
&  2013 & \babar~\cite{Lees:2013ab}      &  $ +0.0037 \pm 0.0030 \pm 0.0015 $ \\
&  2008 & CLEOc~\cite{Rubin:2008zi}     &  Dalitz plot analysis, no evidence for CP violation\\
&  2000 & FOCUS~\cite{Link:2000aw}       &  $ +0.006  \pm 0.011  \pm 0.005  $ \\
&  1997 & E791~\cite{Aitala:1996sh}      &  $ -0.014  \pm 0.029  $ (stat.)    \\
&       & COMBOS average                 &  $ +0.0032 \pm 0.0031 $            \\
\hline
{\boldmath $D^+ \to K^-\pi^+\pi^+\pi^0$} &
  2014 & CLEOc~\cite{Bonvicini:2013vxi}   &  $ -0.003  \pm 0.006  \pm 0.004  $ \\
\hline
{\boldmath $D^+ \to K^0_s\pi^+\pi^+\pi^-$} &
  2014 & CLEOc~\cite{Bonvicini:2013vxi}   &  $ +0.000  \pm 0.012  \pm 0.003  $ \\
\hline
{\boldmath $D^+ \to K^0_sK^+\pi^+\pi^-$} &
  2005 & FOCUS~\cite{Link:2005th}  &  $ -0.042  \pm 0.064  \pm 0.022  $ \\
\hline 
\end{tabular}
\end{center} 
\end{table}

\begin{table}[!htb]
\renewcommand{\arraystretch}{1.3}
\caption{\cp\ asymmetries 
$A^{}_{CP}=[\Gamma(D^0)-\Gamma(\dbar)]/[\Gamma(D^0)+\Gamma(\dbar)]$
for two-body $D^0,\dbar$ decays.
\label{tab:cp_neutral}}
\footnotesize
\begin{center}
\begin{tabular}{|l|c|c|c|} 
\hline
{\bf Mode} & {\bf Year} & {\bf Collaboration} & {\boldmath $A^{}_{CP}$} \\
\hline
{\boldmath $D^0 \to \pi^+\pi^-$} &
  2014 & LHCb~\cite{Aaij:2014gsa}     & $ -0.0020 \pm 0.0019 \pm 0.0010  $ \\
& 2012 & Belle~\cite{Ko:2012ab}      & $ +0.0055 \pm 0.0036 \pm 0.0009  $ \\
& 2012 & CDF~\cite{Aaltonen:2012ab}  & $ +0.0022 \pm 0.0024 \pm 0.0011  $ \\
& 2008 & \babar~\cite{Aubert:2007if} & $ -0.0024 \pm 0.0052 \pm 0.0022  $ \\
& 2002 & CLEO~\cite{Csorna:2001ww}   & $ +0.019  \pm 0.032  \pm 0.008   $ \\
& 2000 & FOCUS~\cite{Link:2000aw}    & $ +0.048  \pm 0.039  \pm 0.025   $ \\
& 1998 & E791~\cite{Aitala:1997ff}   & $ -0.049  \pm 0.078  \pm 0.030   $ \\
&      & COMBOS average              & $ +0.0005 \pm 0.0015 $ \\
\hline
{\boldmath $D^0 \to \pi^0\pi^0$} &
  2014 & Belle~\cite{Nisar:2014aa}     & $ -0.0003 \pm 0.0064 \pm 0.0010  $ \\
& 2001 & CLEO~\cite{Bonvicini:2000qm}  & $ +0.001  \pm 0.048 $ (stat. and syst. combined) \\
&      & COMBOS average                & $ -0.0003 \pm 0.0064 $ \\ 
\hline
{\boldmath $D^0 \to K_s^0\pi^0$} &
  2014 & Belle~\cite{Nisar:2014aa}     & $ -0.0021 \pm 0.0016 \pm 0.0007 $ \\
& 2001 & CLEO~\cite{Bonvicini:2000qm}  & $ +0.001  \pm 0.013 $ (stat. and syst. combined) \\
&      & COMBOS average                & $ -0.0020 \pm 0.0017 $ \\
\hline
{\boldmath $D^0 \to K_s^0\eta$} &
  2011 & Belle~\cite{Ko:2011ab}        & $ +0.0054 \pm 0.0051 \pm 0.0016 $ \\
\hline
{\boldmath $D^0 \to K_s^0\eta^\prime$} &
  2011 & Belle~\cite{Ko:2011ab}        & $ +0.0098 \pm 0.0067 \pm 0.0014 $ \\  
\hline
{\boldmath $D^0 \to K^0_sK^0_s$} &
 2001 & CLEO~\cite{Bonvicini:2000qm}   & $ -0.23  \pm 0.19  $ (stat. and syst. combined) \\
\hline
{\boldmath $D^0 \to K^+K^-$} &
  2014 & LHCb~\cite{Aaij:2014gsa}     & $ -0.0006 \pm 0.0015 \pm 0.0010 $ \\
& 2012 & Belle~\cite{Ko:2012ab}      & $ -0.0032 \pm 0.0021 \pm 0.0009 $ \\  
& 2012 & CDF~\cite{Aaltonen:2012ab}  & $ -0.0024 \pm 0.0022 \pm 0.0009 $ \\
& 2008 & \babar~\cite{Aubert:2007if} & $ +0.0000 \pm 0.0034 \pm 0.0013 $ \\
& 2002 & CLEO~\cite{Csorna:2001ww}   & $ +0.000  \pm 0.022  \pm 0.008  $ \\
& 2000 & FOCUS~\cite{Link:2000aw}    & $ -0.001  \pm 0.022  \pm 0.015  $ \\
& 1998 & E791~\cite{Aitala:1997ff}   & $ -0.010  \pm 0.049  \pm 0.012  $ \\
&      & COMBOS average              & $ -0.0016 \pm 0.0012            $ \\
\hline
\end{tabular}
\end{center} 
\end{table}

\begin{table}[!htb]
\renewcommand{\arraystretch}{1.3}
\caption{\cp\ asymmetries 
$A^{}_{CP}=[\Gamma(D^0)-\Gamma(\dbar)]/[\Gamma(D^0)+\Gamma(\dbar)]$
for three- and four-body $D^0,\dbar$ decays.
\label{tab:cp_neutral2}}
\footnotesize
\begin{center}
\begin{tabular}{|l|c|c|c|} 
\hline
{\bf Mode} & {\bf Year} & {\bf Collaboration} & {\boldmath $A^{}_{CP}$} \\
\hline
{\boldmath $D^0 \to \pi^+\pi^-\pi^0$} &
   2008 & \babar~\cite{Aubert:2008yd}       & $ -0.0031 \pm  0.0041 \pm  0.0017$ \\
&  2008 & Belle~\cite{Arinstein:2008zh}     & $ +0.0043 \pm  0.0130 $ (stat. and syst. combined) \\
&  2005 & CLEO~\cite{CroninHennessy:2005sy} & $ +0.01^{+0.09}_{-0.07} \pm  0.05 $ \\
&       & COMBOS average         & $ -0.0023 \pm 0.0042 $ \\
\hline
{\boldmath $D^0 \to K^-\pi^+\pi^0$} &
  2014 & CLEOc~\cite{Bonvicini:2013vxi} & $ +0.001  \pm 0.003 \pm 0.004   $ \\
& 2001 & CLEO~\cite{Kopp:2000gv}       & $ -0.031  \pm 0.086  $ (stat.) \\
&       & COMBOS average               & $ +0.0009 \pm 0.0050 $ \\
\hline   
{\boldmath $D^0 \to K^+\pi^-\pi^0$} &
  2005 & Belle~\cite{Tian:2005ik}        & $ -0.006  \pm 0.053  $ (stat.) \\
& 2001 & CLEO~\cite{Brandenburg:2001ze}  & $ +0.09^{+0.25}_{-0.22}  $ (stat.) \\
&      & COMBOS average                  & $ -0.0014 \pm 0.0517 $ \\
\hline
{\boldmath $D^0 \to K^0_s\pi^+\pi^-$} &
  2012 & CDF~\cite{Aaltonen:2012ac}  & $ -0.0005 \pm 0.0057 \pm 0.0054 $ \\
& 2004 & CLEO~\cite{Asner:2003uz}    & $ -0.009  \pm 0.021^{+0.016}_{-0.057} $ \\
&      & COMBOS average              & $ -0.0008 \pm 0.0077 $ \\
\hline
{\boldmath $D^0 \to K^+ K^-\pi^0$} &
   2008 & \babar~\cite{Aubert:2008yd} & $ 0.0100 \pm  0.0167 \pm  0.0025$ \\ 
\hline

{\boldmath $D^0 \to \pi^-\pi^-\pi^+\pi^+$} &
  2013 & LHCb~\cite{Aaij:2013aa}  & Amplitude analysis, no evidence for \CP violation\\
\hline
{\boldmath $D^0 \to K^+\pi^-\pi^+\pi^-$} &
  2005 & Belle~\cite{Tian:2005ik} & $ -0.018  \pm 0.044  $ (stat.) \\
\hline
{\boldmath $D^0 \to K^+K^-\pi^+\pi^-$} &
  2013 & LHCb~\cite{Aaij:2013aa}  & Amplitude analysis, no evidence for \CP violation \\
& 2005 & FOCUS~\cite{Link:2005th} & $ -0.082  \pm 0.056  \pm 0.047  $ \\
\hline                   
\end{tabular}
\end{center} 
\end{table}

\begin{table}[!htb]
\renewcommand{\arraystretch}{1.4}
\caption{\cp\ asymmetries 
$A^{}_{CP}= [\Gamma(D_s^+)-\Gamma(D_s^-)]/[\Gamma(D_s^+)+\Gamma(D_s^-)]$
for $D_s^\pm$ decays.
\label{tab:cp_ds}}
\footnotesize
\begin{center}
\begin{tabular}{|l|c|c|c|} 
\hline
{\bf Mode} & {\bf Year} & {\bf Collaboration} & {\boldmath $A^{}_{CP}$} \\
\hline
{\boldmath $D_s^+ \to \mu^+ \nu$} &
  2009 & CLEOc~\cite{Alexander:2009ux} & $ +0.048 \pm 0.061 $ \\
\hline
{\boldmath $D_s^+ \to \pi^+ \eta$} &
  2013 & CLEOc~\cite{Onyisi:2013bjt}     & $ +0.011 \pm 0.030 \pm 0.008 $ \\
\hline
{\boldmath $D_s^+ \to \pi^+ \eta^\prime$} &
  2013 & CLEOc~\cite{Onyisi:2013bjt}     & $ -0.022 \pm 0.022 \pm 0.006 $ \\
\hline
{\boldmath $D_s^+ \to K^0_s\pi^+$}  &
  2014 & LHCb~\cite{Aaij:2014ac}    & $ +0.0038 \pm 0.0046 \pm 0.0017 $ \\
& 2013 & \babar~\cite{Lees:2013aa}  & $ +0.006  \pm 0.020  \pm 0.003  $ \\  
& 2010 & Belle~\cite{Ko:2010ng}     & $ +0.0545 \pm 0.0250 \pm 0.0033 $ \\
& 2010 & CLEOc~\cite{Mendez:2009aa} & $ +0.163  \pm 0.073  \pm 0.003  $ \\
&      & COMBOS average             & $ +0.0063 \pm 0.0047 $            \\
\hline
{\boldmath $D_s^+ \to K^0_s K^+$}   &
  2013 & CLEOc~\cite{Onyisi:2013bjt}  & $ +0.026  \pm 0.015  \pm 0.006  $ \\
& 2013 & \babar~\cite{Lees:2013aa}  & $ -0.0005 \pm 0.0023 \pm 0.0024 $ \\  
& 2010 & Belle~\cite{Ko:2010ng}     & $ +0.0012 \pm 0.0036 \pm 0.0022 $ \\
&      & COMBOS average             & $ +0.0008 \pm 0.0026 $            \\
\hline
{\boldmath $D_s^+ \to K^+ \pi^0$}   &
  2010 & CLEOc~\cite{Mendez:2009aa} &  $ +0.266 \pm 0.228 \pm 0.009 $ \\
\hline
{\boldmath $D_s^+ \to K^+ \eta$}    &
  2010 & CLEOc~\cite{Mendez:2009aa} &  $ +0.093 \pm 0.152 \pm 0.009 $ \\
\hline
{\boldmath $D_s^+ \to K^+ \eta^\prime$}  &
  2010 & CLEOc~\cite{Mendez:2009aa}      &  $ +0.060 \pm 0.189 \pm 0.009 $ \\
\hline
{\boldmath $D_s^+ \to \pi^+ \pi^+ \pi^-$} &
  2013 & CLEOc~\cite{Onyisi:2013bjt}        & $ -0.007 \pm 0.030 \pm 0.006 $ \\
\hline
{\boldmath $D_s^+ \to \pi^+ \pi^0 \eta$}  &
  2013 & CLEOc~\cite{Onyisi:2013bjt}        & $ -0.005 \pm 0.039 \pm 0.020 $ \\
\hline
{\boldmath $D_s^+ \to \pi^+ \pi^0 \eta^\prime$} &
  2013 & CLEOc~\cite{Onyisi:2013bjt}        & $ -0.004 \pm 0.074 \pm 0.019 $ \\
\hline
{\boldmath $D_s^+ \to K^0_s K^+ \pi^0$}   &
  2013 & CLEOc~\cite{Onyisi:2013bjt}        & $ -0.016 \pm 0.060 \pm 0.011 $ \\
\hline
{\boldmath $D_s^+ \to K^0_s K^0_s \pi^+$} &
  2013 & CLEOc~\cite{Onyisi:2013bjt}        & $ +0.031 \pm 0.052 \pm 0.006 $ \\
\hline
{\boldmath $D_s^+ \to K^+ \pi^+ \pi^-$} &
  2013 & CLEOc~\cite{Onyisi:2013bjt}        & $ +0.045 \pm 0.048 \pm 0.006 $ \\
\hline
{\boldmath $D_s^+ \to K^+ K^- \pi^+$} &
  2013 & CLEOc~\cite{Onyisi:2013bjt}        & $ -0.005 \pm 0.008 \pm 0.004 $ \\
\hline
{\boldmath $D_s^+ \to K^0_s K^- \pi^+\pi^+$} &
  2013 & CLEOc~\cite{Onyisi:2013bjt}        & $ +0.041 \pm 0.027 \pm 0.009 $ \\
\hline
{\boldmath $D_s^+ \to K^0_s K^+ \pi^+\pi^-$} &
  2013 & CLEOc~\cite{Onyisi:2013bjt}        & $ -0.057 \pm 0.053 \pm 0.009 $ \\
\hline
{\boldmath $D_s^+ \to K^+ K^- \pi^+\pi^0$} &
  2013 & CLEOc~\cite{Onyisi:2013bjt}        & $ +0.000 \pm 0.027 \pm 0.012 $ \\ 
\hline 
\end{tabular}
\end{center} 
\end{table}

\clearpage
\subsection{\emph{$T$}-violating asymmetries}
                                               
$T$-violating asymmetries are measured using triple-product
correlations and assuming the validity of the $CPT$ theorem.
Triple-product correlations of the form 
$\vec{a}\cdot(\vec{b}\times\vec{c})$, 
where $a$, $b$, and $c$ are spins or momenta, are odd 
under time reversal~(\emph{T}).
For example, for $D^0 \to K^+K^-\pi^+\pi^-$ decays, 
$C_T \equiv \vec{p}^{}_{K^+}\cdot(\vec{p}_{\pi^+}\times \vec{p}_{\pi^-})$  
changes sign (\ie\ is odd) under a \emph{T} transformation.
The corresponding quantity for $\dbar$ is
$\overline{C}_T \equiv 
      \vec{p}^{}_{K^-}\cdot(\vec{p}_{\pi^-}\times \vec{p}_{\pi^+})$.
Defining  
\begin{eqnarray}
 A_{T} & = &
    \frac{\Gamma(C_T>0)-\Gamma(C_T<0)}{\Gamma(C_T>0)+\Gamma(C_T<0)}
\end{eqnarray}
for $D^0$ decays and
\begin{eqnarray}
\overline{A}_{T} & = & 
   \frac{\Gamma(-\overline{C}_T>0)-\Gamma(-\overline{C}_T<0)}
                        {\Gamma(-\overline{C}_T>0)+\Gamma(-\overline{C}_T<0)}
\end{eqnarray} 
for $\dbar$ decays, in the absence of strong phases
either $A^{}_T\neq 0$ or $\overline{A}^{}_T\neq 0$ indicates
$T$ violation. In these expressions the $\Gamma$'s are partial widths. 
The asymmetry
\begin{eqnarray}
A^{}_{T\,{\rm viol}} & \equiv & \frac{A_{T}-\overline{A}_{T}}{2}
\end{eqnarray}
tests for $T$ violation even with nonzero strong phases (see 
Refs.~\cite{Golowich:1988ig,Bigi:2001sg,Bensalem:2002ys,Bensalem:2000hq,Bensalem:2002pz}).
Values of $A_{T\,{\rm viol}}$ for some $D^+$, $D^+_s$, and
$D^0$ decay modes are listed in Table~\ref{tab:t_viol}.

$T$-violating asymmetry is a clean and alternative way to search for \emph{CP} violation
in the charm sector. Analogously to what is shown in Sec.~\ref{sec:cp_asym} 
there is no evidence of \emph{CP} violation.

\begin{table}[h]
\renewcommand{\arraystretch}{1.4}
\caption{$T$-violating asymmetries 
$A^{}_{T\,{\rm viol}} = (A_{T}-\overline{A}_{T})/2$.
\label{tab:t_viol}}
\footnotesize
\begin{center}
\begin{tabular}{|l|c|c|c|} 
\hline
{\bf Mode} & {\bf Year} & {\bf Collaboration} & {\boldmath $A^{}_{T\,{\rm viol}}$} \\
\hline
{\boldmath $D^0 \to K^+K^-\pi^+\pi^-$} &
   2014 & LHCb~\cite{Aaij:2014qwa}     &  $ +0.0018 \pm 0.0029 \pm 0.0004 $ \\
&  2010 & \babar~\cite{Sanchez:2010xj} &  $ +0.0010 \pm 0.0051 \pm 0.0044 $ \\
&  2005 & FOCUS~\cite{Link:2005th}     &  $ +0.010  \pm 0.057  \pm 0.037  $ \\
&       & COMBOS average               &  $ +0.0017 \pm 0.0027            $ \\  
\hline
{\boldmath $D^+ \to K^0_sK^+\pi^+\pi^-$} &
  2011 & \babar~\cite{Lees:2011ab} &  $ -0.0120 \pm 0.0100 \pm 0.0046 $ \\
& 2005 & FOCUS~\cite{Link:2005th}  &  $ +0.023  \pm 0.062  \pm 0.022  $ \\
&      & COMBOS average            &  $ -0.0110 \pm 0.0109            $ \\
\hline
{\boldmath $D^+_s \to K^0_sK^+\pi^+\pi^-$} &
  2011 & \babar~\cite{Lees:2011ab} &  $ -0.0136 \pm 0.0077 \pm 0.0034 $ \\
& 2005 & FOCUS~\cite{Link:2005th}  &  $ -0.036  \pm 0.067  \pm 0.023  $ \\
&      & COMBOS average            &  $ -0.0139 \pm 0.0084            $ \\
\hline                    
\end{tabular}
\end{center} 
\end{table}


\clearpage
\subsection{Interplay of direct and indirect \cp\ violation}
\label{sec:charm:cpvdir}

In decays of $D^0$ mesons, \cp\ asymmetry measurements have contributions from 
both direct and indirect \cp\ violation as discussed in Sec.~\ref{sec:charm:mixcpv}.
The contribution from indirect \cp\ violation depends on the decay-time distribution 
of the data sample~\cite{Kagan:2009gb}. This section describes a combination of 
measurements that allows the extraction of the individual contributions of the 
two types of \cp\ violation.
At the same time, the level of agreement for a no-\cp-violation hypothesis is 
tested. The observables are: 
\begin{equation}
A_{\Gamma} \equiv \frac{\tau(\dbar \ra h^+ h^-) - \tau(D^0 \ra h^+ h^- )}
{\tau(\dbar \ra h^+ h^-) + \tau(D^0 \ra h^+ h^- )},
\end{equation}
where $h^+ h^-$ can be $K^+ K^-$ or $\pi^+\pi^-$, and 
\begin{equation}
\Delta A_{\rm CP}   \equiv A_{\rm CP}(K^+K^-) - A_{\rm CP}(\pi^+\pi^-),
\end{equation}
where $A_{\rm CP}$ are time-integrated \cp\ asymmetries. The underlying 
theoretical parameters are: 
\begin{eqnarray}
a_{\rm CP}^{\rm dir} & \equiv & 
\frac{|{\cal A}_{D^0\rightarrow f} |^2 - |{\cal A}_{\dbar\rightarrow f} |^2} 
{|{\cal A}_{D^0\rightarrow f} |^2 + |{\cal A}_{\dbar\rightarrow f} |^2} ,\nonumber\\ 
a_{\rm CP}^{\rm ind}  & \equiv & \frac{1}{2} 
\left[ \left(\left|\frac{q}{p}\right| + \left|\frac{p}{q}\right|\right) x \sin \phi - 
\left(\left|\frac{q}{p}\right| - \left|\frac{p}{q}\right|\right) y \cos \phi \right] ,
\end{eqnarray}
where ${\cal A}_{D\rightarrow f}$ is the amplitude for $D\ra f$~\cite{Grossman:2006jg}. 
We use the following relations 
between the observables and the underlying parameters~\cite{Gersabeck:2011xj}: 
\begin{eqnarray}
A_{\Gamma} & = & - a_{\rm CP}^{\rm ind} - a_{\rm CP}^{\rm dir} y_{\rm CP},\nonumber\\ 
\Delta A_{\rm CP} & = &  \Delta a_{\rm CP}^{\rm dir} \left(1 + y_{\rm CP} 
\frac{\overline{\langle t\rangle}}{\tau} \right)   +   
   a_{\rm CP}^{\rm ind} \frac{\Delta\langle t\rangle}{\tau}   +   
  \overline{a_{\rm CP}^{\rm dir}} y_{\rm CP} \frac{\Delta\langle t\rangle}{\tau},\\ 
& \approx & \Delta a_{\rm CP}^{\rm dir} \left(1 + y_{\rm CP} 
\frac{\overline{\langle t\rangle}}{\tau} \right)   +   a_{\rm CP}^{\rm ind} 
\frac{\Delta\langle t\rangle}{\tau}.
\end{eqnarray}
The first relation constrains mostly indirect \cp\ violation, and the 
direct \cp\ violation contribution can differ for different final states. 
In the second relation, $\langle t\rangle/\tau$ denotes the mean decay 
time in units of the $D^0$ lifetime; $\Delta X$ denotes the difference 
in quantity $X$ between $K^+K^-$ and $\pi^+\pi^-$ final states; and $\overline{X}$ 
denotes the average for quantity $X$. 
We neglect the last term in this relation as all three factors are 
$\mathcal{O}(10^{-2})$ or smaller, and thus this term is negligible 
with respect to the other two terms. 
Note that $\Delta\langle t\rangle/\tau \ll\langle t\rangle/\tau$, and 
it is expected that $|a_{\rm CP}^{\rm dir}| < |\Delta a_{\rm CP}^{\rm dir}|$ 
because $a_{\rm CP}^{\rm dir}(K^+K^-)$ and $a_{\rm CP}^{\rm dir}(\pi^+\pi^-)$ 
are expected to have opposite signs~\cite{Grossman:2006jg}. 

A $\chi^2$ fit is performed in the plane $\Delta a_{\rm CP}^{\rm dir}$ 
vs. $a_{\rm CP}^{\rm ind}$. 
For the \babar result the difference of the quoted values for 
$A_{\rm CP}(K^+K^-)$ and $A_{\rm CP}(\pi^+\pi^-)$ is calculated, 
adding all uncertainties in quadrature. 
This may overestimate the systematic uncertainty for the difference 
as it neglects correlated errors; however, the result is conservative 
and the effect is small as all measurements are statistically limited. 
For all measurements, statistical and systematic uncertainties are added 
in quadrature when calculating the $\chi^2$. 
We use the current world average value $y_{\rm CP} = (0.866 \pm 0.155)\%$ 
(see Sec.~\ref{sec:charm:mixcpv}) and the measurements listed in 
Table~\ref{tab:charm:dir_indir_comb}. 

In this fit, $A_\Gamma(KK)$ and $A_\Gamma(\pi\pi)$ are assumed to be identical.
This assumption is supported by the most recent LHCb measurements~\cite{Aaij:2013ria}.
A significant relative shift due to final-state dependent $A_\Gamma$ values between $\Delta A_{\rm CP}$ measurements with different mean decay times is excluded by these measurements.

\begin{table}
\centering 
\caption{Inputs to the fit for direct and indirect \cp\ violation. 
The first uncertainty listed is statistical, and the second is systematic.}
\label{tab:charm:dir_indir_comb}
\vspace{3pt}
\begin{tabular}{ll|ccccc}
\hline \hline
Year & 	Experiment	& Results
& $\Delta \langle t\rangle/\tau$ & $\langle t\rangle/\tau$ & Reference\\
\hline
2012	& Belle	prel. & $A_\Gamma = (-0.03 \pm 0.20 \pm 0.08 )\%$ &	-&	-&	 
\cite{Staric:2012ta}\\
2012	& \babar	& $A_\Gamma = (0.09 \pm 0.26 \pm 0.06 )\%$ &	-&	-&	 
\cite{Lees:2012qh}\\
2013	& LHCb	& $A_\Gamma(KK) = (-0.035 \pm 0.062 \pm 0.012 )\%$ &	-&	-&	 
\cite{Aaij:2013ria}\\
    	&     	& $A_\Gamma(\pi\pi) = (0.033 \pm 0.106 \pm 0.014 )\%$ &	-&	-&	 
                   \\
2008	& \babar	& $A_{\rm CP}(KK) = (0.00 \pm 0.34 \pm 0.13 )\%$&&&\\ 
& & $A_{\rm CP}(\pi\pi) = (-0.24 \pm 0.52 \pm 0.22 )\%$ &	$0.00$ &	
$1.00$ &	 \cite{Aubert:2007if}\\
2012	& Belle	prel. & $\Delta A_{\rm CP} = (-0.87 \pm 0.41 \pm 0.06 )\%$ &	
$0.00$ &	$1.00$ &	 \cite{Ko:2012px}\\
2012	& CDF	 & $\Delta A_{\rm CP} = (-0.62 \pm 0.21 \pm 0.10 )\%$ &	
$0.25$ &	$2.58$ &	 \cite{Collaboration:2012qw}\\
2013	& LHCb	prel. & $\Delta A_{\rm CP} = (-0.34 \pm 0.15 \pm 0.10 )\%$ &	
$0.11$ &	$2.10$ &	 \cite{LHCb:2013dka}\\
2014	& LHCb	& $\Delta A_{\rm CP} = (0.14 \pm 0.16 \pm 0.08 )\%$ &	
$0.01$ &	$1.07$ &	 \cite{Aaij:2014gsa}\\
\hline
\end{tabular}
\end{table}

\begin{figure}
\begin{center}
\includegraphics[width=0.90\textwidth]{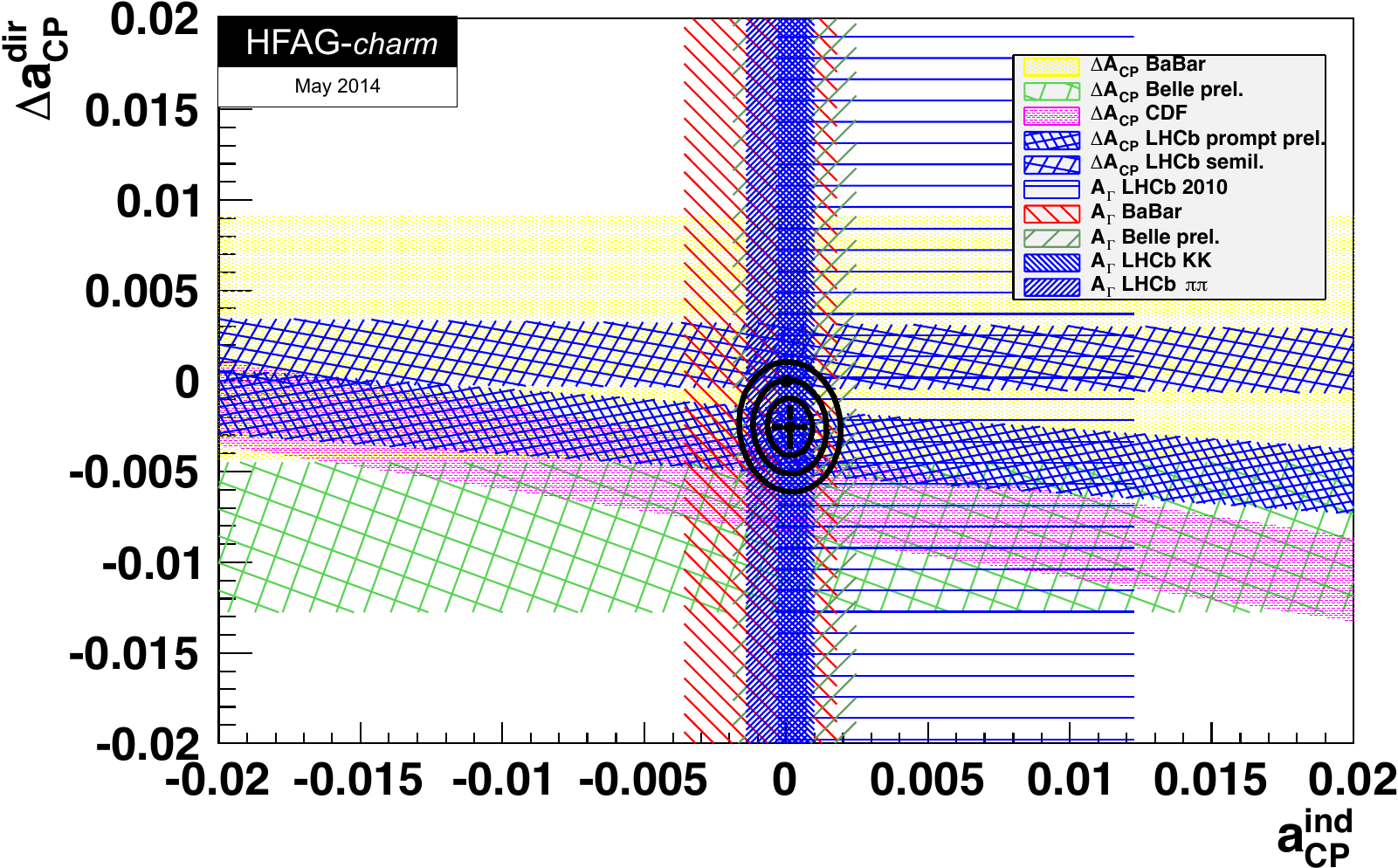}
\caption{Plot of all data and the fit result. Individual 
measurements are plotted as bands showing their $\pm1\sigma$ range. 
The no-\cpv\ point (0,0) is shown as a filled circle, and the best 
fit value is indicated by a cross showing the one-dimensional errors. 
Two-dimensional $68\%$ CL, $95\%$ CL, and $99.7\%$ CL regions are 
plotted as ellipses. }
\label{fig:charm:dir_indir_comb}
\end{center}
\end{figure}

The combination plot (see Fig.~\ref{fig:charm:dir_indir_comb}) shows the measurements listed in 
Table~\ref{tab:charm:dir_indir_comb} for
$\Delta A_{\rm CP}$ and $A_\Gamma$, where the bands represent $\pm1\sigma$ 
intervals.  The point of no \cp\ violation (0,0) is shown as a filled circle, 
and two-dimensional $68\%$ CL, $95\%$ CL, and $99.7\%$ CL regions are plotted 
as ellipses. The best fit value is indicated by a cross showing the
one-dimensional errors.

From the fit, the change in $\chi^2$ from the minimum value for the no-\cpv\ 
point (0,0) is $5.9$, which corresponds to a CL of $5.1\times 10^{-2}$ for 
two degrees of freedom. Thus the data are consistent with the no-\cp-violation 
hypothesis at $5.1\%$ CL. This $p$-value corresponds to $2.0\sigma$. The central
values and $\pm1\sigma$ errors for the individual parameters are
\begin{eqnarray}
a_{\rm CP}^{\rm ind} & = & (0.013 \pm 0.052 )\% \nonumber\\
\Delta a_{\rm CP}^{\rm dir} & = & (-0.253 \pm 0.104 )\%.
\end{eqnarray}
These values indicate that the present small deviation from 
no \cp\ violation is primarily due to a difference between direct
\cp\ violation in the two final states, rather than due to 
common indirect \cp\ violation.

\clearpage

\subsection{Semileptonic decays}
\label{sec:charm:semileptonic}

\subsubsection{Introduction}

Semileptonic decays of $D$ mesons involve the interaction of a leptonic
current with a hadronic current. The latter is nonperturbative
and cannot be calculated from first principles; thus it is usually
parameterized in terms of form factors. The transition matrix element 
is written
\begin{eqnarray}
  {\cal M} & = & -i\,\frac{G_F}{\sqrt{2}}\,V^{}_{cq}\,L^\mu H_\mu\,,
  \label{Melem}
\end{eqnarray}
where $G_F$ is the Fermi constant and $V^{}_{cq}$ is a CKM matrix element.
The leptonic current $L^\mu$ is evaluated directly from the lepton spinors 
and has a simple structure; this allows one to extract information about 
the form factors (in $H^{}_\mu$) from data on semileptonic decays~\cite{Becher:2005bg}.  
Conversely, because there are no final-state interactions between the
leptonic and hadronic systems, semileptonic decays for which the form 
factors can be calculated allow one to 
determine~$V^{}_{cq}$~\cite{Kobayashi:1973fv}.

\subsubsection{$D\ra P \overline \ell \nu_{\ell}$ decays}

When the final state hadron is a pseudoscalar, the hadronic 
current is given by
\begin{eqnarray}
\hspace{-1cm}
H_\mu & = & \left< P(p) | \bar{q}\gamma_\mu c | D(p') \right> \ =\  
f_+(q^2)\left[ (p' + p)_\mu -\frac{m_D^2-m_P^2}{q^2}q_\mu\right] + 
 f_0(q^2)\frac{m_D^2-m_P^2}{q^2}q_\mu\,,
\label{eq:hadronic}
\end{eqnarray}
where $m_D$ and $p'$ are the mass and four momentum of the 
parent $D$ meson, $m_P$ and $p$ are those of the daughter meson, 
$f_+(q^2)$ and $f_0(q^2)$ are form factors, and $q = p' - p$.  
Kinematics require that $f_+(0) = f_0(0)$.
The contraction $q_\mu L^\mu$ results in terms proportional 
to $m^{}_\ell$\cite{Gilman:1989uy}, and thus for $\ell=e $
the terms proportionals to $q_\mu$ in Eq.~(\ref{eq:hadronic}) are negligible. 
For light leptons only the $f_+(q^2)$ form factor 
is relevant and the differential partial width is
\begin{eqnarray}
\frac{d\Gamma(D \to P \bar \ell \nu_\ell)}{dq^2\, d\cos\theta_\ell} & = & 
   \frac{G_F^2|V_{cq}|^2}{32\pi^3} p^{*\,3}|f_{+}(q^2)|^2\sin\theta^2_\ell\,,
\label{eq:dGamma}
\end{eqnarray}
where ${p^*}$ is the magnitude of the momentum of the final state hadron
in the $D$ rest frame, and $\theta_\ell$ is the angle of the lepton in the 
$\ell\nu$ rest frame with respect to the direction of the pseudoscalar meson 
in the $D$ rest frame.


\subsubsection{Form factor parameterizations} 

The form factor is traditionally parameterized with an explicit pole 
and a sum of effective poles:
\begin{eqnarray}
f_+(q^2) & = & \frac{f_+(0)}{(1-\alpha)}\Bigg [
\left(\frac{1}{1- q^2/m^2_{\rm pole}}\right)\ +\ 
\sum_{k=1}^{N}\frac{\rho_k}{1- q^2/(\gamma_k\,m^2_{\rm pole})}\Bigg ]\,,
\label{eqn:expansion}
\end{eqnarray}
where $\rho_k$ and $\gamma_k$ are expansion parameters and $\alpha$ is such that provides the form factor normalization at $q^2=0$, $f_+(0)$. 
The parameter $m_{{\rm pole}}$ is the mass of the lowest-lying $c\bar{q}$ resonance
with the appropriate quantum numbers; this is expected to provide the
largest contribution to the form factor for the $c\ra q$ transition. The sum over $N$ gives the contribution of higher mass states.  
For example, for $D\to\pi$ transitions the dominant resonance is
expected to be $D^*(2010)$, and thus $m^{}_{\rm pole}=m^{}_{D^*(2010)}$.

\subsubsubsection{Simple pole}

Equation~(\ref{eqn:expansion}) can be simplified by neglecting the 
sum over effective poles, leaving only the explicit vector meson pole. 
This approximation is referred to as ``nearest pole dominance'' or 
``vector-meson dominance.''  The resulting parameterization is
\begin{eqnarray}
  f_+(q^2) & = & \frac{f_+(0)}{(1-q^2/m^2_{\rm pole})}\,. 
\label{SimplePole}
\end{eqnarray}
However, values of $m_{{\rm pole}}$ that give a good fit to the data 
do not agree with the expected vector meson masses~\cite{Hill:2006ub}. 
To address this problem, the ``modified pole'' or Becirevic-Kaidalov~(BK) 
parameterization~\cite{Becirevic:1999kt} was introduced.
$m_{\rm pole} /\sqrt{\alpha_{\rm BK}}$
is interpreted as the mass of an effective pole, higher than $m_{\rm pole}$, thus it is expected that $\alpha_{\rm BK}<1$.


The parameterization takes the form
\begin{eqnarray}
f_+(q^2) & = & \frac{f_+(0)}{(1-q^2/m^2_{\rm pole})}
\frac{1}{\left(1-\alpha^{}_{\rm BK}\frac{q^2}{m^2_{\rm pole}}\right)}\,.
\end{eqnarray}
These parameterizations have been used by several experiments to 
determine form factor parameters.
Measured values of $m^{}_{\rm pole}$ and $\alpha^{}_{\rm BK}$ are
listed in Tables~\ref{kPseudoPole} and~\ref{piPseudoPole} for
$D\to K\ell\nu_{\ell}$ and $D\to\pi\ell\nu_{\ell}$ decays, respectively.


\subsubsubsection{$z$ expansion}

An alternative series expansion around some value $q^2=t_0$ to parameterize 
$f^{}_+(q^2)$ can be used~\cite{Boyd:1994tt,Boyd:1997qw,Arnesen:2005ez,Becher:2005bg}. This parameterization is model 
independent and satisfies general QCD constraints, being suitable for fitting experimental data.
The expansion is given in terms of a complex parameter $z$,
which is the analytic continuation of $q^2$ into the complex plane:
\begin{eqnarray}
z(q^2,t_0) & = & \frac{\sqrt{t_+ - q^2} - \sqrt{t_+ - t_0}}{\sqrt{t_+ - q^2}
	  + \sqrt{t_+ - t_0}}\,, 
\end{eqnarray}
where $t_\pm \equiv (m_D \pm m_P)^2$ and $t_0$ is the (arbitrary) $q^2$ 
value corresponding to $z=0$. The physical region corresponds to $\pm|z|_{max} = \pm 0.051$ 
for $D\to K \ell \nu_\ell$ and $= \pm 0.17$ for  $D\to \pi \ell \nu_\ell$, 
using $t_{0}= t_{+} (1-\sqrt{1-t_{-}/t_{+}})$.  

The form factor is expressed as
\begin{eqnarray}
f_+(q^2) & = & \frac{1}{P(q^2)\,\phi(q^2,t_0)}\sum_{k=0}^\infty
a_k(t_0)[z(q^2,t_0)]^k\,,
\label{z_expansion}
\end{eqnarray}
where the $P(q^2)$ factor accommodates sub-threshold resonances via
\begin{eqnarray}
P(q^2) & \equiv & 
\begin{cases} 
1 & (D\to \pi) \\
z(q^2,M^2_{D^*_s}) & (D\to K)\,. 
\end{cases}
\end{eqnarray}
The ``outer'' function $\phi(t,t_0)$ can be any analytic function,
but a preferred choice (see, \eg\
Refs.~\cite{Boyd:1994tt,Boyd:1997qw,Bourrely:1980gp}) obtained
from the Operator Product Expansion (OPE) is
\begin{eqnarray}
\phi(q^2,t_0) & =  & \alpha 
\left(\sqrt{t_+ - q^2}+\sqrt{t_+ - t_0}\right) \times  \nonumber \\
 & & \hskip0.20in \frac{t_+ - q^2}{(t_+ - t_0)^{1/4}}\  
\frac{(\sqrt{t_+ - q^2}\ +\ \sqrt{t_+ - t_-})^{3/2}}
     {(\sqrt{t_+ - q^2}+\sqrt{t_+})^5}\,,
\label{eqn:outer}
\end{eqnarray}
with $\alpha = \sqrt{\pi m_c^2/3}$.
The OPE analysis provides a constraint upon the 
expansion coefficients, $\sum_{k=0}^{N}a_k^2 \leq 1$.
These coefficients receive $1/M_D$ corrections, and thus
the constraint is only approximate. However, the
expansion is expected to converge rapidly since 
$|z|<0.051\ (0.17)$ for $D\ra K$ ($D\ra\pi$) over 
the entire physical $q^2$ range, and Eq.~(\ref{z_expansion}) 
remains a useful parameterization. The main disadvantage as compared to 
phenomenological approaches is that there is no physical interpretation 
of the fitted coefficients $a_K$.

\subsubsubsection{Three-pole formalism}
A recent update of the vector pole dominance model has been developed for the $D \to \pi \ell \nu_\ell$ channel\cite{Becirevic:2014kaa}.
It uses information of the residues of the semileptonic form factor at its first two poles, the $D^\ast(2010)$ 
and $D^{\ast '}(2600)$ resonances.  
The form factor is expressed as an infinite sum of residues from $J^P =1^-$ states with masses $m_{D^\ast_n}$: 
\begin{eqnarray}
f_+(q^2) = \sum_{n=0}^\infty { \; \displaystyle{\underset{q^2= m_{D^\ast_n}^2}{\rm Res}} f_+(q^2)\;\over m_{D^\ast_n}^2-q^2} \,,
\label{ThreePole}
\end{eqnarray}
with the residues given by 
\begin{eqnarray}
\displaystyle{\underset{q^2=m_{D_n^\ast}^2}{\rm Res}} f_+(q^2)= {1\over 2} m_{D_n^\ast} f_{D_n^\ast} g_{D_n^\ast D\pi}\,. 
\label{Residua}
\end{eqnarray}
Values of the $f_{D^\ast}$ and $f_{D^{\ast '}}$ decay constants have been obtained by Lattice QCD calculations, 
relative to $f_{D}$, with 2$\%$ and 28$\%$ precision, respectively~\cite{Becirevic:2014kaa}. 
The couplings to the $D\pi$ state, $g_{D^\ast D\pi}$ and $g_{D^{\ast '} D\pi}$, are extracted from measurements of the $D^\ast(2010)$ and  $D^{\ast '}(2600)$ widths by \babar and LHCb experiments~\cite{Lees:2013uxa,delAmoSanchez:2010vq,Aaij:2013sza}. 
Thus the contribution from the first pole is known with a $3\%$ accuracy. 
The contribution from the $D^{\ast '}(2600)$ is determined with poorer accuracy, $\sim 30\%$, mainly due to lattice uncertainties.  A {\it superconvergence} condition is applied~\cite{Burdman:1996kr}: 
\begin{eqnarray}
\sum_{n=0}^\infty { \; \displaystyle{\underset{q^2=m_{D^\ast_n}^2}{\rm Res}} f_+(q^2) }= 0 \,,
\label{superconvergence}
\end{eqnarray}
protecting the form factor behavior at large $q^2$.
Within this model the first two poles are not sufficient to describe the data, and a third effective pole needs to be included. 

One of the advantages of this phenomenological model 
is that it can be extrapolated outside the charm physical region, providing a method to 
extract the CKM matrix element $V_{ub}$ using the ratio of the form factors of the 
$D\to \pi\ell \nu$ and $B\to \pi\ell \nu$ decay channels. It will be used once Lattice calculations provide
the form factor ratio $f^{+}_{B\pi}(q^2)/f^{+}_{D\pi}(q^2)$ at the same pion energy. 

\subsubsection{Experimental techniques and results}
 Different techniques by several experiments have been used to measure $D$ meson semileptonic decays with a 
pseudoscalar particle in the final state. The most recent results are provided by the \babar and BES III collaborations.
They have been presented at the ICHEP 2014 conference and are preliminary.  
Belle~\cite{Widhalm:2006wz}, \babar~\cite{Aubert:2007wg} and CLEO-c~\cite{Besson:2009uv,Dobbs:2007aa} 
collaborations have previously reported results. 
The Belle collaboration fully reconstructs the $D$ events from the continuum under the $\Upsilon(4S)$ resonance, 
achieving a very good $q^2$ resolution ($\Delta q^2 = 15 {\rm~MeV}^2$) and low background level, 
but having a low efficiency. Using 282~$\fb^{-1}$, about 1300 and 115 signal semileptonic decays are 
isolated for each lepton flavor ($e$ and $\mu$), respectively. 
The \babar experiment uses a partial reconstruction technique where the semileptonic decays are tagged 
through the $ D^{\ast +}\to D^0\pi^+$ decay. 
The $D$ direction and neutrino energy is obtained using information from the rest of the event. 
With 75~$\fb^{-1}$ 74000 signal events in the $D^0 \to {K}^- e^+ \nu$ mode are obtained. 
This technique provides larger statistics but a higher background level and poorer $q^2$ resolution ($\Delta q^2$ ranges from 66 to 219 MeV$^2$). In this case the measurement of the branching fraction is obtained by normalizing to the $D^0 \to K^- \pi^+$ decay channel and will benefit from future improvements in the determination of this reference channel. The measurement of the Cabibbo suppressed mode has been recently obtained 
using the same technique and 350~fb$^{-1}$ data. 5000 $D^0 \to {\pi}^- e^+ \nu$ signal events are reconstructed using this method~\cite{Lees:2014jka}.  
The CLEO-c experiment uses two different methods to measure charm semileptonic decays. 
Tagged analyses~\cite{Besson:2009uv} rely on the full reconstruction of $\Psi(3770)\to D {\overline D}$ events. One of the $D$ mesons is reconstructed in a hadronic decay mode, the other in the semileptonic channel. The only missing particle is the neutrino so the $q^2$ resolution is very good and the background level very low.   
With the entire CLEO-c data sample, 818 $\pb^{-1}$, 14123 and 1374 signal events are reconstructed for 
the $D^0 \to K^{-} e^+\nu$ and $D^0\to \pi^{-} e^+\nu$ channels, and 8467 and 838 for the 
$D^+\to {\overline K}^{0} e^+\nu$ and $D^+\to \pi^{0} e^+\nu$ decays, respectively. 
Another technique without tagging the $D$ meson in a hadronic mode (``untagged'' in the following) has been also 
used by CLEO-c~\cite{Dobbs:2007aa}. In this method, all missing energy and momentum in 
an event are associated with the neutrino four momentum, with the penalty of larger background as compared to the tagged method. 
Using the ``tagged'' method the BES III experiment has measured the $D^0 \to {K}^- e^+ \nu$ and 
$D^0 \to {\pi}^- e^+ \nu$ decay channels. With 2.9~fb$^{-1}$ they fully reconstruct 70700 and 6300 signal 
events for each channel, respectively. These results are preliminary. 

  Previous measurements were also performed by CLEO~III and FOCUS experiments. 
Events registered at the $\Upsilon (4S)$ energy corresponding to an integrated luminosity of 7 $\fb^{-1}$ were 
analyzed by CLEO~III~\cite{Huang:2004fra}. 
In the FOCUS fixed target photo-production experiment, $D^0$ semileptonic events were obtained from the decay of 
a $D^{\ast +}$, and the kaon or pion was reconstructed in the muon channel. 
 Results of the hadronic form factor parameters by the different groups are given in Tables
\ref{kPseudoPole} and \ref{piPseudoPole} for $m_{pole}$ and $\alpha_{BK}$. 
\begin{table}[htbp]
\caption{Results for $m_{\rm pole}$ and $\alpha_{\rm BK}$ from various
  experiments for $D^0\to K^-\ell^+\nu$ and $D^+\to K_S\ell^+\nu$
  decays. 
\label{kPseudoPole}}
\begin{center}
\begin{tabular}{cccc}
\hline
\vspace*{-10pt} & \\
 $D\to K\ell\nu_\ell$ Expt. & Ref.  & $m_{\rm pole}$ (GeV$/c^2$) 
& $\alpha^{}_{\rm BK}$       \\
\vspace*{-10pt} & \\
\hline
 \omit        & \omit                         & \omit                                  & \omit                  \\
 CLEO III     & \cite{Huang:2004fra}          & $1.89\pm0.05^{+0.04}_{-0.03}$          & $0.36\pm0.10^{+0.03}_{-0.07}$ \\
 FOCUS        & \cite{Link:2004dh}            & $1.93\pm0.05\pm0.03$                   & $0.28\pm0.08\pm0.07$     \\
 Belle        & \cite{Widhalm:2006wz}         & $1.82\pm0.04\pm0.03$                   & $0.52\pm0.08\pm0.06$     \\
 \babar        & \cite{Aubert:2007wg}          & $1.889\pm0.012\pm0.015$                & $0.366\pm0.023\pm0.029$  \\

 CLEO-c (tagged)   &\cite{Besson:2009uv}      & $1.93\pm0.02\pm0.01$                   & $0.30\pm0.03\pm0.01$     \\
 CLEO-c (untagged, $D^0$) &\cite{Dobbs:2007aa}       & $1.97 \pm0.03 \pm 0.01 $ & $0.21 \pm 0.05 \pm 0.03 $  \\
 CLEO-c (untagged, $D^+$) &\cite{Dobbs:2007aa}       & $1.96 \pm0.04 \pm 0.02 $ & $0.22 \pm 0.08 \pm 0.03$  \\
  BESIII (preliminary)     &\cite{BESIII-new}                & $1.921 \pm 0.010 \pm 0.007$ & $ 0.309 \pm 0.020 \pm 0.013$   \\ 
\vspace*{-10pt} & \\
\hline
\end{tabular}
\end{center}
\end{table}
\begin{table}[htbp]
\caption{Results for $m_{\rm pole}$ and
  $\alpha_{\rm BK}$ from various experiments for 
  $D^0\to \pi^-\ell^+\nu$ and $D^+\to \pi^0\ell^+\nu$ decays.  
\label{piPseudoPole}}
\begin{center}
\begin{tabular}{cccc}
\hline
\vspace*{-10pt} & \\
 $D\to \pi\ell\nu_\ell$ Expt. & Ref.               & $m_{\rm pole}$ (GeV$/c^2$) & $\alpha_{\rm BK}$ \\
\vspace*{-10pt} & \\
\hline
 \omit        & \omit                         & \omit                                  & \omit                  \\
 CLEO III     & \cite{Huang:2004fra}          & $1.86^{+0.10+0.07}_{-0.06-0.03}$       & $0.37^{+0.20}_{-0.31}\pm0.15$         \\
 FOCUS        & \cite{Link:2004dh}            & $1.91^{+0.30}_{-0.15}\pm0.07$          & --                                    \\
 Belle        & \cite{Widhalm:2006wz}         & $1.97\pm0.08\pm0.04$                   & $0.10\pm0.21\pm0.10$                  \\
 CLEO-c (tagged)   &\cite{Besson:2009uv}      & $1.91\pm0.02\pm0.01$                   & $0.21\pm0.07\pm0.02$     \\
 CLEO-c (untagged, $D^0$) &\cite{Dobbs:2007aa}       & $1.87 \pm0.03 \pm 0.01 $ & $0.37 \pm 0.08 \pm 0.03 $  \\
 CLEO-c (untagged, $D^+$) &\cite{Dobbs:2007aa}       & $1.97 \pm0.07 \pm 0.02 $ & $0.14 \pm 0.16 \pm 0.04$  \\
 BES III (preliminary)     &\cite{BESIII-new}                & $1.911 \pm 0.012 \pm 0.004$ & $ 0.279 \pm 0.035 \pm 0.011$   \\ 
  \babar (preliminary)     &\cite{Lees:2014jka}                & $1.906 \pm 0.029 \pm 0.023$ & $ 0.268 \pm 0.074 \pm 0.059$   \\
\vspace*{-10pt} & \\
\hline
\end{tabular}
\end{center}
\end{table}

The $z$-expansion formalism has been used by \babar~\cite{Aubert:2007wg,Lees:2014jka}, 
BES III\cite{BESIII-new} and CLEOc~\cite{Besson:2009uv},~\cite{Dobbs:2007aa}.
Their fits uses the first three terms of the expansion, 
and the results for the ratios $r_1\equiv a_1/a_0$ and $r_2\equiv a_2/a_0$ are 
listed in Tables~\ref{KPseudoZ} and~\ref{piPseudoZ}. 
The CLEO~III\cite{Huang:2004fra} and FOCUS\cite{Link:2004dh} results 
listed are obtained by refitting their data using the full
covariance matrix. The \babar correlation coefficient listed is 
obtained by refitting their published branching fraction using 
their published covariance matrix.  
These measurements correspond to using the standard 
outer function $\phi(q^2,t_0)$ of Eq.~(\ref{eqn:outer}) and 
$t_0=t_+\left(1-\sqrt{1-t_-/t_+}\right)$. This choice of $t^{}_0$
constrains $|z|$ to vary between $\pm z_{max.}$

\begin{table}[htbp]
\caption{Results for $r_1$ and $r_2$ from various experiments, for 
$D\to K\ell\nu_{\ell}$. The correlation coefficient listed is for the total uncertainties (statistical $\oplus$ systematic) on $r^{}_1$ and~$r^{}_2$.}
The combined result does not include the new BES III result presented at the ICHEP2014 conference~\cite{BESIII-new}, but the previous one with partial statistics~\cite{BESIII}. 
The fit is constrained by the branching fractions measured at Belle~\cite{Widhalm:2006wz}.
\label{KPseudoZ}
\begin{center}
\begin{tabular}{cccccc}
\hline
\vspace*{-10pt} & \\
Expt. $D\to K\ell\nu_{\ell}$     & mode &  Ref.                         & $r_1$               & $r_2$               & $\rho$        \\
\hline
 \omit    & \omit         & \omit                & \omit               & \omit               & \omit         \\
 CLEO III & \omit  & \cite{Huang:2004fra} & $0.2^{+3.6}_{-3.0}$ & $-89^{+104}_{-120}$ & -0.99         \\
 FOCUS    & \omit                & \cite{Link:2004dh}   & $-2.54\pm0.75$  & $7\pm 13$       & -0.97 \\
 \babar    & \omit        & \cite{Aubert:2007wg} & $-2.5\pm0.2\pm0.2$  & $2.5\pm6.0\pm5.0$     & -0.64         \\
 CLEO-c (tagged)     & $D^0\to K^-$    & \cite{Besson:2009uv}          & $-2.65\pm0.34\pm0.08$  & $13\pm9\pm1$       & -0.82 \\
 CLEO-c (tagged)     & $D^+\to \overline K^0$   & \cite{Besson:2009uv} & $-1.66\pm0.44\pm0.10$  & $-14\pm11\pm1$       & -0.82 \\
 CLEO-c (untagged)   & $D^0\to K^-$           &\cite{Dobbs:2007aa}     & $-2.4\pm0.4\pm0.1$  & $21\pm11\pm2$     & -0.81    \\
 CLEO-c (untagged)   & $D^+\to \overline K^0$ & \cite{Dobbs:2007aa}    & $-2.8\pm6\pm2$      & $32\pm18\pm4$       & -0.84         \\
  BES III (0.9 fb$^{-1}$)  & \omit         & \cite{BESIII}        & $-2.18\pm0.36\pm0.05$ & $5\pm 9\pm 1$  &            \\
  BES III (preliminary, 2.9 fb$^{-1}$) & \omit         & \cite{BESIII-new}        & $ -2.33\pm 0.16 \pm 0.08 $ & $ 3.4\pm 3.9 \pm 2.4 $  &            \\
\hline
\hline
 Combined (preliminary) & \omit         &  \omit               & $-2.39\pm0.17$       & $6.2\pm3.8$         & -0.82        \\ 
\hline
\end{tabular}
\end{center}
\end{table}

\begin{figure}[p]
\begin{center}
\includegraphics[width=0.47\textwidth]{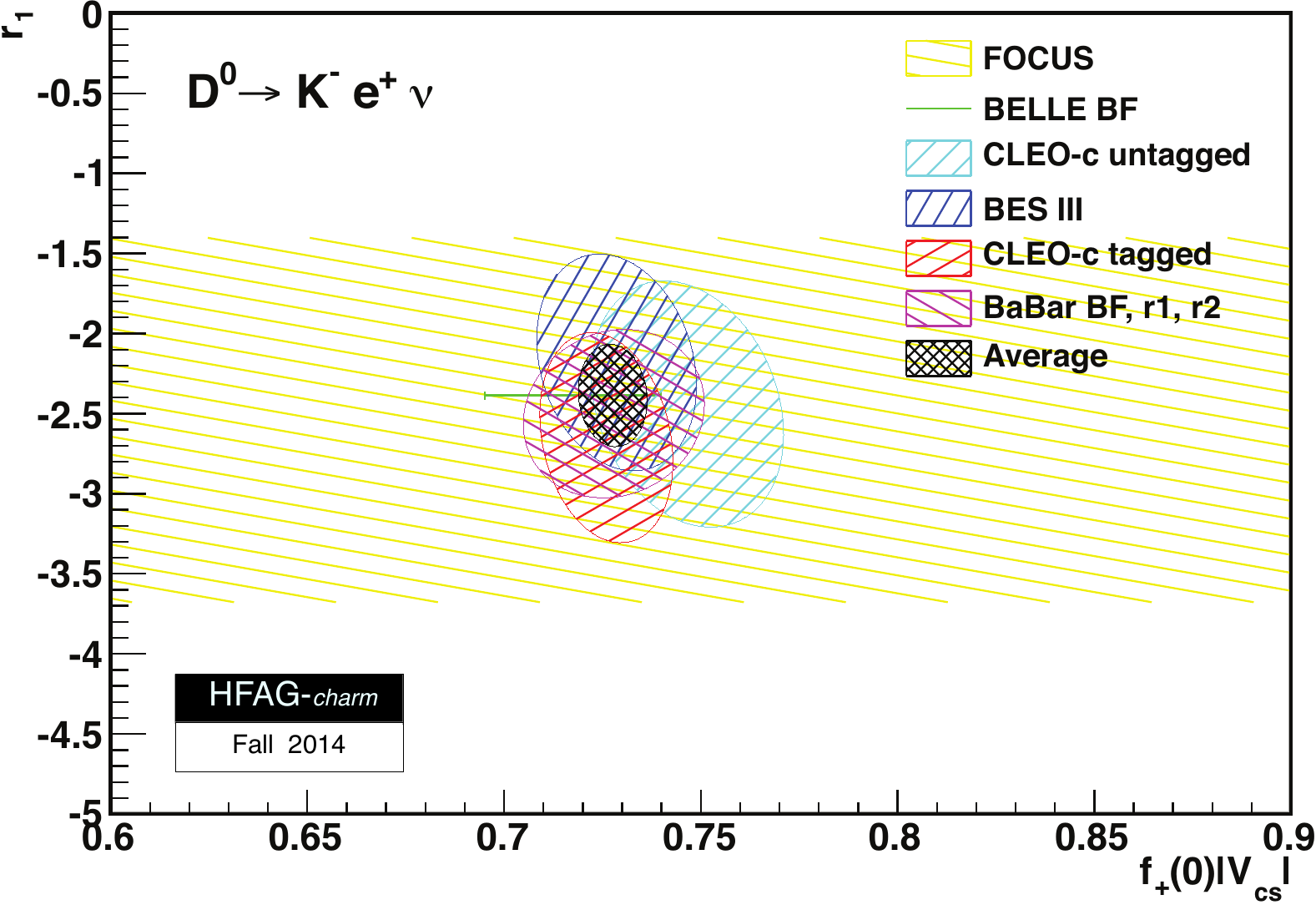}\hfill
\includegraphics[width=0.46\textwidth]{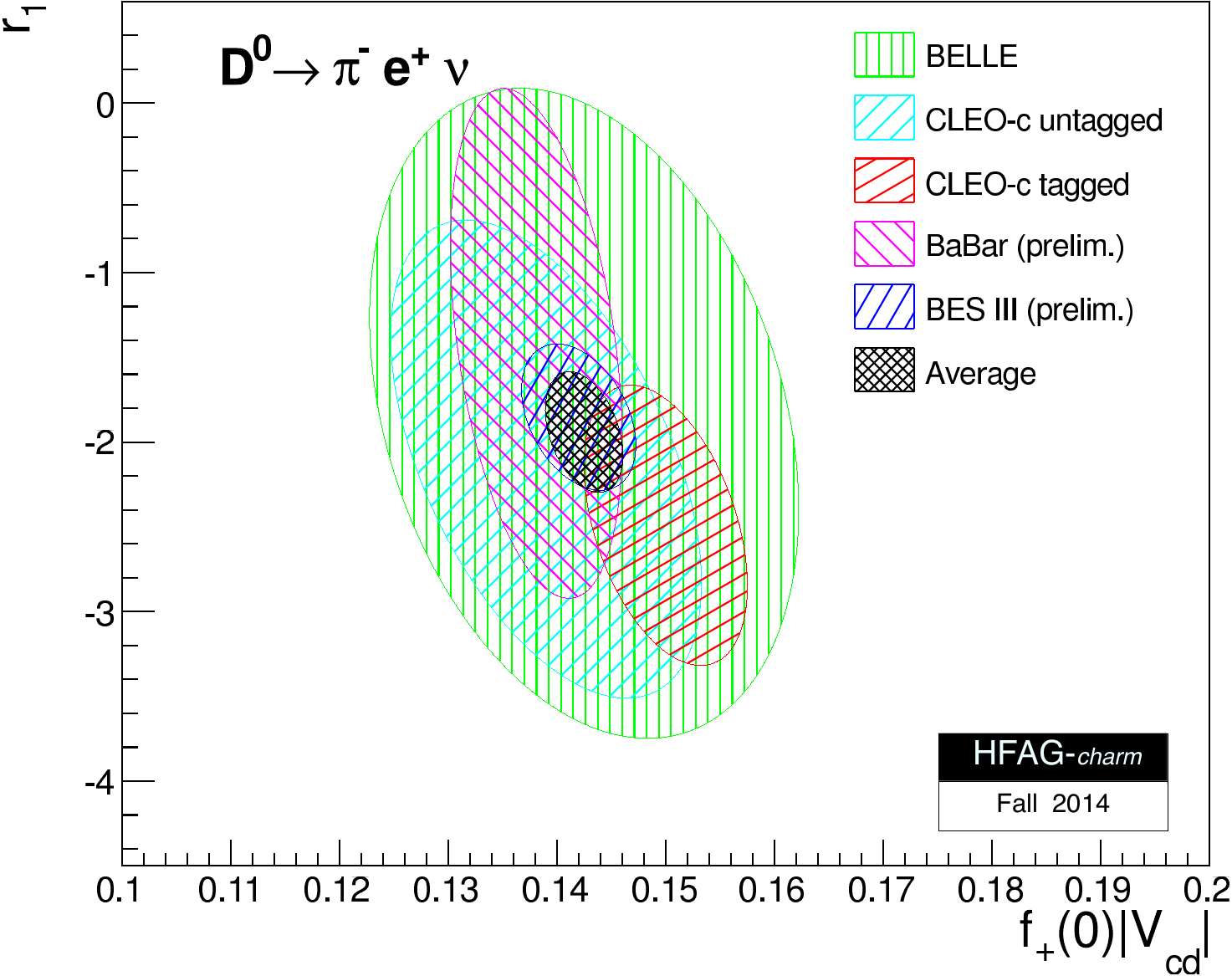}
\includegraphics[width=0.47\textwidth]{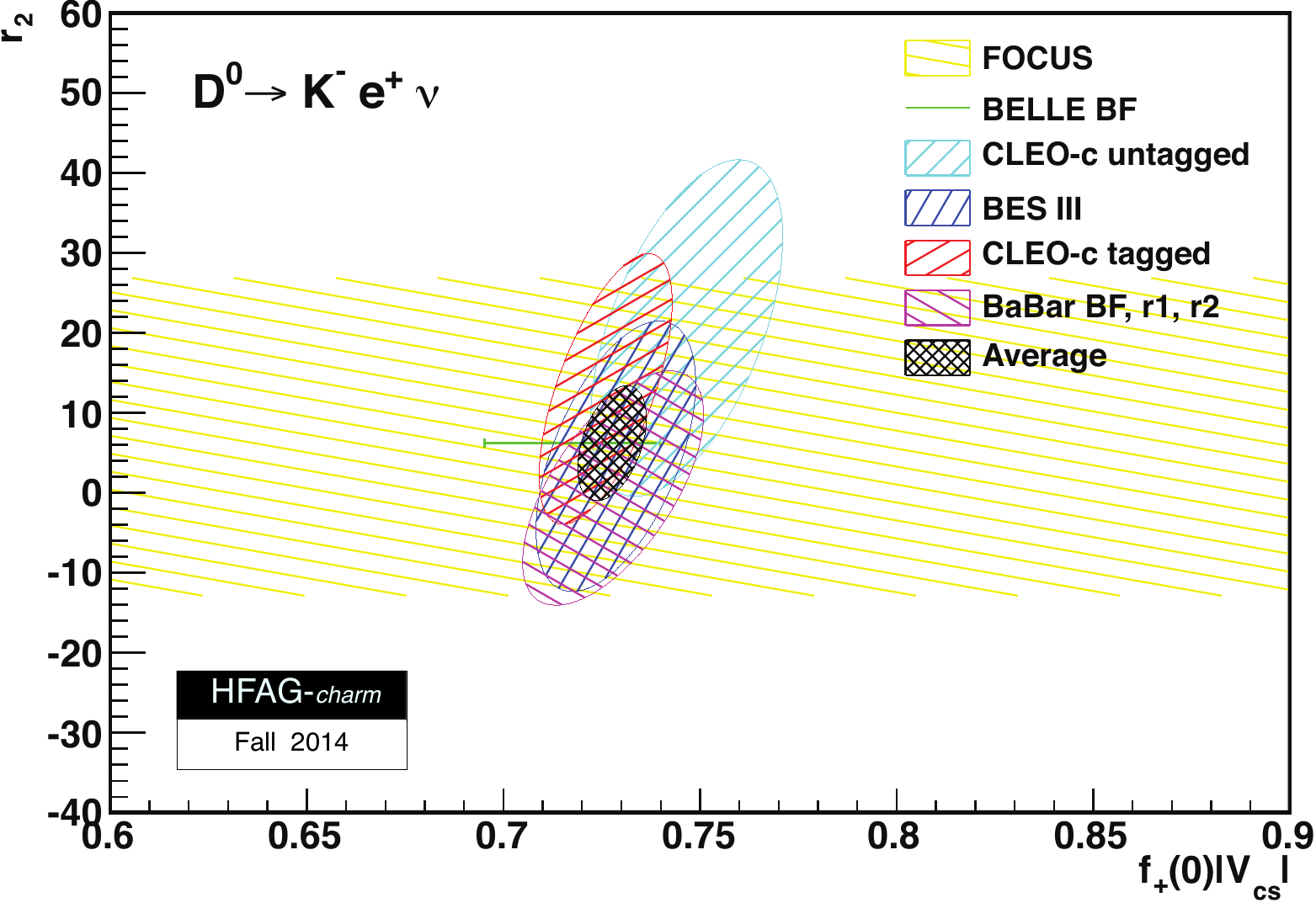}\hfill
\includegraphics[width=0.46\textwidth]{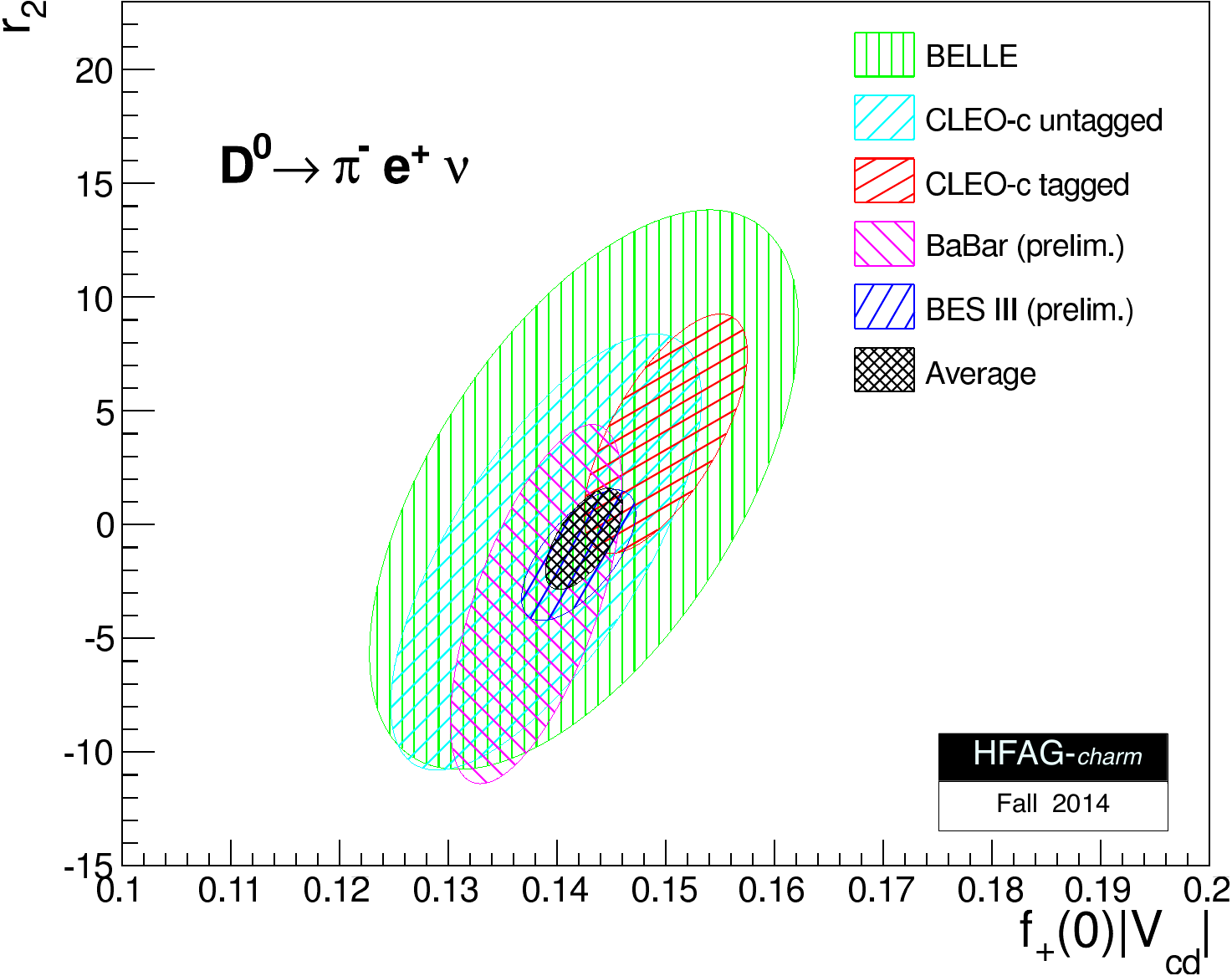}\hfill
\includegraphics[width=0.47\textwidth]{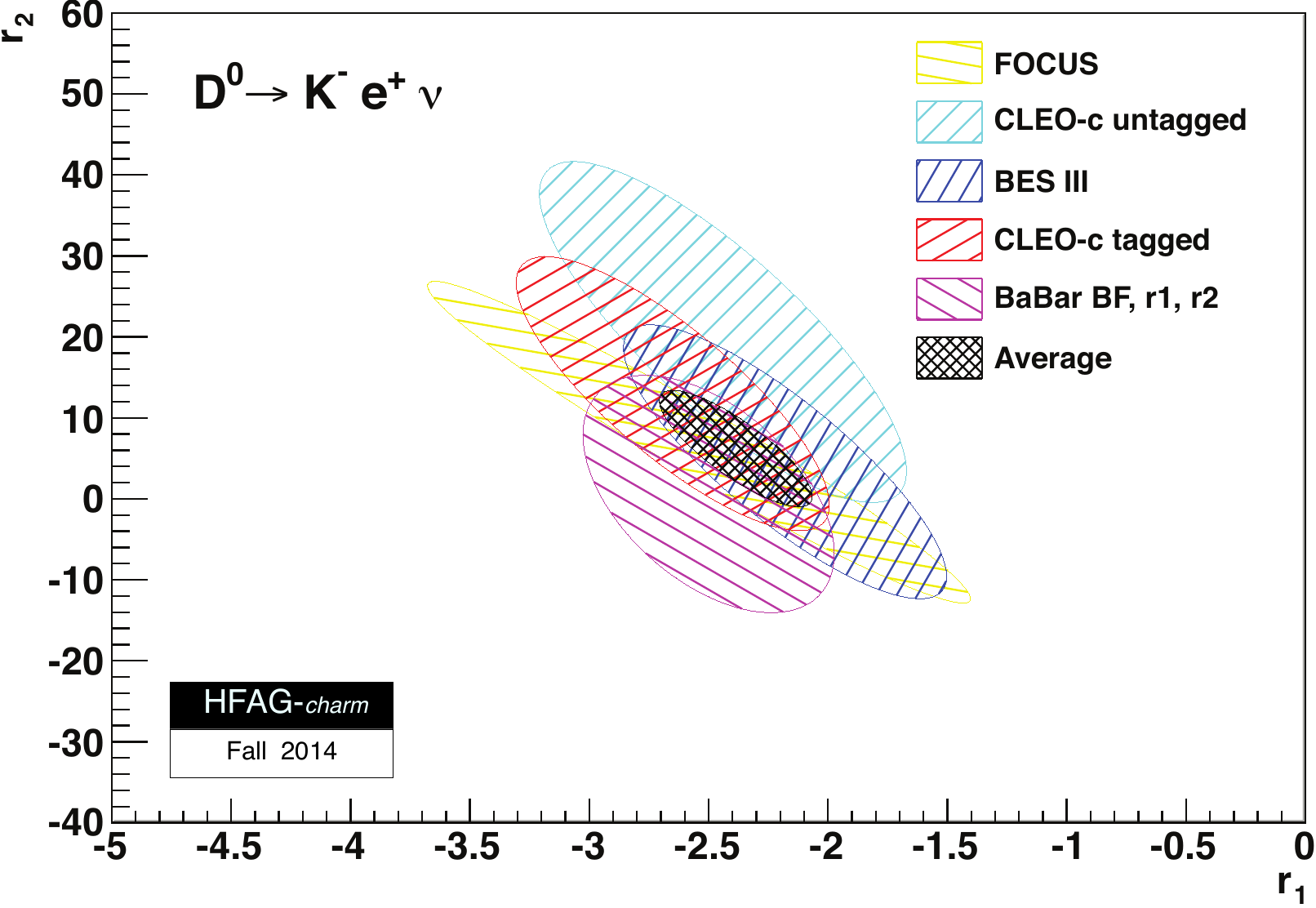}\hfill
\includegraphics[width=0.46\textwidth]{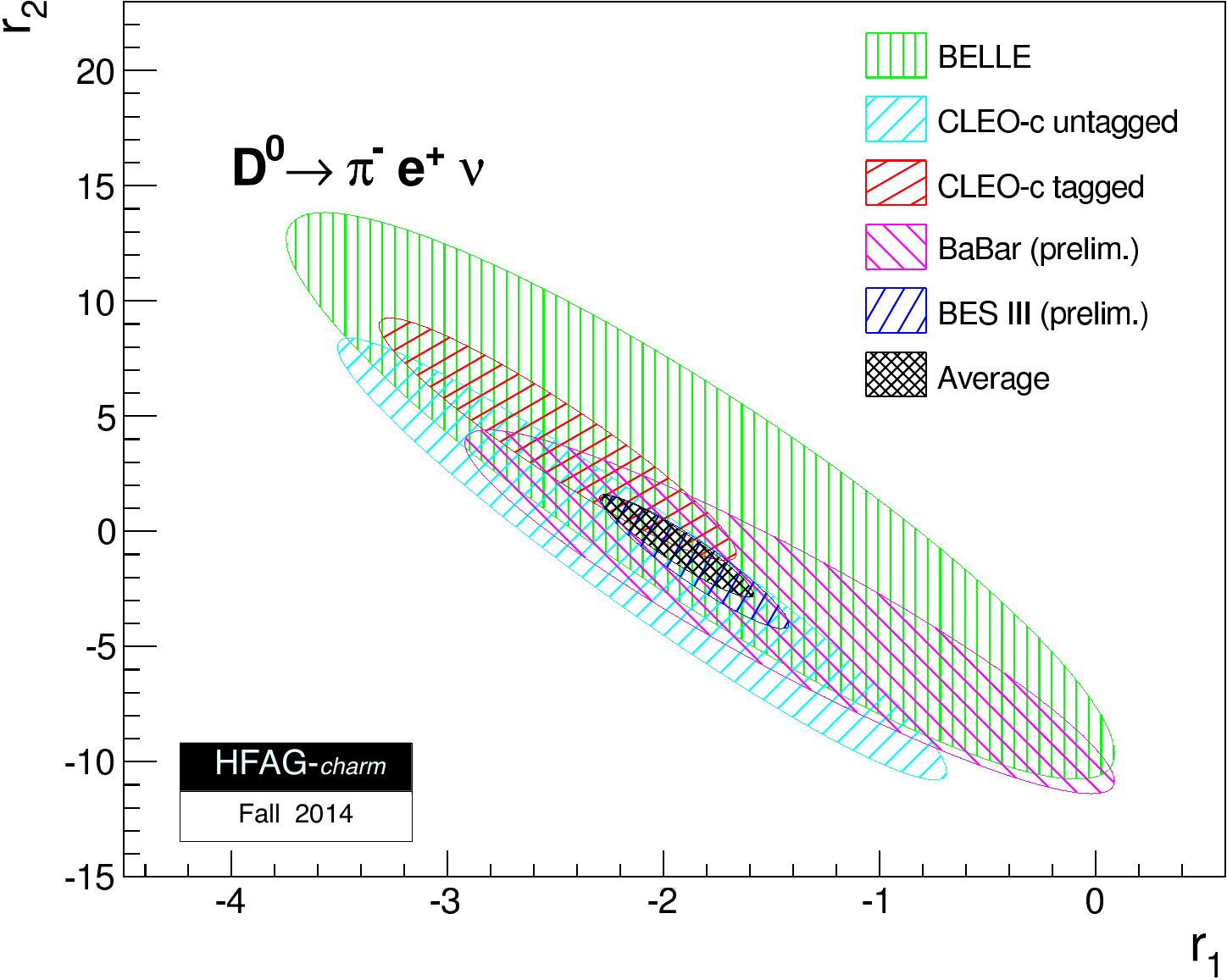}
\caption{The $D^0\to K^-e^+\nu$ (left) and $D^0\to \pi^-e^+\nu$ (right) 68\% C.L. error ellipses from the
average fit of the 3-parameter $z$-expansion results. 
\label{fig:fitellipse}}
\end{center}
\end{figure}

\begin{table}[htbp]
\caption{Results for $r_1$ and $r_2$ from various experiments, for $D\to \pi \ell\nu_{\ell}$. 
The correlation coefficient listed is for the total uncertainties (statistical $\oplus$ systematic) 
on $r^{}_1$ and~$r^{}_2$. 
The combined result includes preliminary results from \babar and BES III presented at ICHEP 2014. 
The Belle data is refitted in the $z$-expansion formalism using published values of $f^{D\pi}_{+} (q^2) \times |V_{cd}|$, 
and removing the uncertainty on $V_{cd}$ from the systematic error.}
\label{piPseudoZ}
\begin{center}
\begin{tabular}{cccccc}
\hline
\vspace*{-10pt} & \\
Expt. $D\to \pi\ell\nu_{\ell}$     & mode &  Ref.                         & $r_1$               & $r_2$               & $\rho$        \\
\hline
 \omit    & \omit         & \omit                & \omit               & \omit               & \omit         \\
\hline
\hline
 CLEO-c (tagged)     & $D^0\to\pi^+$ & \cite{Besson:2009uv}      &  $-2.80\pm0.49\pm0.04$ & $6\pm 3\pm$ 0 & -0.94 \\            
 CLEO-c (tagged)     & $D^+\to\pi^0$ & \cite{Besson:2009uv}      &  $-1.37\pm0.88\pm0.24$ & $-4\pm 5\pm$ 1 & -0.96 \\            
 CLEO-c  (untagged)  & $D^0\to\pi^+$ & \cite{Dobbs:2007aa}  & $-2.1\pm0.7\pm0.3$      & $-1.2\pm4.8\pm1.7$  & -0.96         \\
 CLEO-c   (untagged) & $D^+\to\pi^0$ & \cite{Dobbs:2007aa}  & $-0.2\pm1.5\pm0.4$    & $-9.8\pm9.1\pm2.1$  & -0.97         \\
 Belle  & \omit  & \cite{Widhalm:2006wz} & $-1.84\pm 1.02 $ & $1.69\pm 6.5$ &   -0.91\\
 BES III (preliminary) & \omit    & \cite{BESIII-new}                    & $-1.85 \pm 0.22 \pm 0.07$ & $-1.4 \pm 1.5 \pm 0.5$ & -0.93            \\
 \babar (preliminary) & \omit         &\cite{Lees:2014jka}                   & $ -1.31 \pm 0.70 \pm 0.43 $ & $-4.2 \pm 4.0 \pm 1.9$  & -0.97        \\
 \hline 
 \hline
  Combined (preliminary) & \omit         &  \omit               & $-1.94\pm 0.19$       & $ -0.62\pm 1.19$         & -0.94        \\ 
\vspace*{-10pt} & \\
\hline
\end{tabular}
\end{center}
\end{table}
The combined result for the $D\to K \ell\nu_{\ell}$ decay channel is obtained from a three-dimensional fit to \babar, BES III (preliminary), CLEO~III, CLEO-c and FOCUS data, taking the full correlations between $|V_{cq}|f_+(0)$, $r_1$ and $r_2$ into account. 
Data from each experiment is fitted with the $z$-expansion model and the combination is performed over the fitted results. The fit is constrained by the branching fraction measured at Belle~\cite{Widhalm:2006wz}. The normalization of the form factor is fixed for the FOCUS data.  
The effect of radiative events has been taken into account slightly modifying the values from \babar by 
correcting the numbers given in Tab.~III of Ref.~\cite{Aubert:2007wg} by the shifts quoted in the last 
column of Tab.~IV given in Ref.~\cite{Aubert:2007wg}. For this combination, the BES III (preliminary) 
result obtained with partial statistics is used~\cite{BESIII}.
Results of the combined fit are shown in Table~\ref{KPseudoZ}, Table~\ref{norma} and Figure \ref{fig:fitellipse}. The $\chi^2/d.o.f$ of the combined fit is $12/13$. The correlation matrix is given in Table \ref{tab:corrK}.

The combined result for the $D\to \pi \ell\nu_{\ell}$ decay channel is obtained from a fit 
to \babar (preliminary), Belle, BES III (preliminary), and CLEO-c data. The combination is performed in this case by fitting all the available measurements in bins of $q^2$ to the $z$-expansion model, instead of a combination fit to the individual fitted parameters in the $z$-expansion. 
Published values of $f^{D\pi}_{+} (q^2) \times |V_{cd}|$ Belle data~\cite{Widhalm:2006wz} are modified 
by subtracting the uncertainty on $V_{cd}$ from the systematic error. Since the experimental $q^2$ resolution is very high, measurements at different $q^2$ are assumed uncorrelated. 
Preliminary results obtained with the full BES III statistics~\cite{BESIII-new} are included 
in the combination of the $D\to \pi \ell\nu_{\ell}$ decay channel.
The new preliminary \babar results are already corrected for radiation effects~\cite{Lees:2014jka}.
Results of the combined fit are shown in Table~\ref{piPseudoZ}, Table~\ref{norma} and 
Figure~\ref{fig:fitellipse}. The $\chi^2/d.o.f$ of the combined fit is $51/55$. 
The correlation matrix is given in Table \ref{tab:corrpi}.


Results for the form factor normalization $f_+^K(0)|V_{cs}|$ and $f_+^{\pi}(0)|V_{cd}|$ for each individual measurement and for the combination is presented in Table \ref{norma}, obtained using the 
$z$-expansion formalism.
In the combination only the result for the $D \to \pi \ell \nu_\ell $ channel is obtained 
with the total BES III statistics~\cite{BESIII-new}, while the $D \to K \ell \nu_\ell $ 
channel includes results with partial statistics~\cite{BESIII}.\;

Assuming unitarity of the CKM matrix, the values of the CKM matrix elements entering in charm semileptonic decays are known 
from the $V_{ud}$, $V_{td}$ and $V_{cb}$ elements~\cite{PDG_2014}:
\begin{eqnarray}
\label{eq:ckm}
V_{cs} = 0.97343 \pm 0.00015  \nonumber \\
V_{cd} = 0.22521 \pm 0.00061 \, .
\end {eqnarray}
Using the combined values of $f_+^K(0)|V_{cs}|$ and $f_+^{\pi}(0)|V_{cd}|$ in Table \ref{norma}, this leads to the form factor values: 
\begin{eqnarray}
\label{ff_measured}
 f_+^K(0) = 0.7479 \pm 0.0051  \nonumber \\ 
 f_+^{\pi}(0) = 0.6327 \pm 0.0086 \, , 
\end {eqnarray}
which are in agreement with present Lattice QCD computations~\cite{FLAG}: $f_+^K(0) = 0.747 \pm 0.019$ and $f_+^\pi(0) = 0.666 \pm 0.029$.
If on the contrary one assumes the form factor values from Lattice, one obtains for the CKM matrix elements using the combined results in 
Table \ref{norma}:
\begin{eqnarray}
\label{ckm}
V_{cs} = 0.9746 \pm 0.0026  \nonumber \\ 
V_{cd} = 0.2140 \pm 0.0097 \, , 
\end {eqnarray} 
still compatible with unitarity of the CKM matrix.

\begin{table} 
\begin{center}
\caption{Correlation matrix for the combined fit for the $D^0\to K^-\ell^+\nu_\ell$ channel}
\label{tab:corrK}
\begin{tabular}{c  c c c }
\hline
 \omit & $|V_{cs}|f_{+}^{K}(0)$ & $r_1$ &  $r_2$ \\
\hline 
$|V_{cs}|f_{+}^{K}(0)$ & $1.000$ & $-0.088$ & $0.433$ \\
                 $r_1$ & $-0.088$ & $1.000$ &$-0.824$ \\
                 $r_2$ & $0.433$ & $-0.824$ & $1.000$ \\
\hline
\end{tabular}
\end{center}
\end{table}

\begin{table} 
\begin{center}
\caption{Correlation matrix for the combined fit for the $D^0\to \pi^-\ell^+\nu_\ell$ channel}
\label{tab:corrpi}
\begin{tabular}{c  c c c } \\
\hline
 \omit & $|V_{cd}|f_{+}^{\pi}(0)$ & $r_1$ &  $r_2$ \\
\hline
 $|V_{cd}|f_{+}^{\pi}(0)$ & $1.000$ &  $-0.379$ & $0.634$ \\
 $r_1$                 & $-0.379$ &  $1.000$ & $-0.936$ \\
 $r_2$                 &  $0.634$ & $-0.936$ & $1.000$ \\
\hline
\end{tabular}
\end{center}
\end{table}


\begin{table}[!htb]
\begin{center}
 \caption[]{Results for the form factor normalization
$f_+^K(0)|V_{cs}|$ and $f_+^{\pi}(0)|V_{cd}|$, obtained using the $z$-expansion formalism. 
Results of CLEO (2008) (untagged) shown in this table only refer to the $D^0$ channel.
In the combination only the result for the $D \to \pi \ell \nu_\ell $ channel is obtained 
with the total BES III statistics~\cite{BESIII-new}, while the $D \to K \ell \nu_\ell $ 
channel includes results with partial statistics~\cite{BESIII}.
 \label{norma}}
\begin{tabular}{c c c c}
\vspace*{-10pt} & \\
\hline
Experiment & Ref. & $f_+^K(0)|V_{cs}|$ & $f_+^{\pi}(0)|V_{cd}|$ \\
\hline\hline 
Belle (2006) & \cite{Widhalm:2006wz}& $0.692 \pm 0.007 \pm 0.022$ & $0.140\pm 0.004\pm0.007$ \\ 
\babar (2007), (preliminary 2014) & \cite{Aubert:2007wg,Lees:2014jka} & $0.720 \pm 0.007\pm 0.007$ & $0.137 \pm 0.004\pm 0.002$  \\
CLEO-c (2008)(untagged) &\cite{Dobbs:2007aa} & $0.747 \pm 0.009\pm 0.009$ & $0.139 \pm 0.007\pm 0.003$ \\
CLEO-c (2009) (tagged) &\cite{Besson:2009uv} & $0.719 \pm 0.006\pm 0.005$ & $0.150 \pm 0.004\pm 0.001$ \\
BESIII (2014) (preliminary) &\cite{BESIII-new} & $0.720 \pm 0.004\pm 0.004$ & $0.142 \pm 0.002\pm 0.001$ \\ 
\hline\hline
Combined fit (preliminary)                & \omit      & $ 0.728 \pm 0.005 $               &       $ 0.1425 \pm 0.0019$   \\
\hline
\hline
\end{tabular}
\end{center}
\end{table}

\vspace{0.5cm}
Results of the three-pole model~\cite{Becirevic:2014kaa} to \babar~\cite{Lees:2014jka}, Belle\cite{Widhalm:2006wz}, BES III\cite{BESIII-new} and 
CLEOc~\cite{Besson:2009uv},~\cite{Dobbs:2007aa} $D \to \pi \ell \nu_\ell $ data are shown in Table \ref{3Pole_pi}. 
Fitted parameters are the first two residues $\gamma_{0}=\underset{ q^2=m_{D^{\ast}}^2} {\rm Res} f_+(q^2) $ 
and $\gamma_{1}=\underset{q^2=m_{D^{\ast '}}^2}{\rm Res} f_+(q^2)$ (which are constrained using present measurements 
of masses and widths of the $D^\ast(2010)$ and $D^{\ast '}(2600)$ mesons, and lattice computations of decay constants, following~\cite{Becirevic:2014kaa}), 
and an effective mass, $m_{D^{\ast ''}_{eff}}$, accounting for higher mass hadronic contributions.
The $V_{cd}$ value enters in the fit; the value used is that prescribed by unitarity in Eq. \ref{eq:ckm}.
The $\chi^2/d.o.f$ of the combined fit is $57.5/57$. 

\begin{figure}[p]
\begin{center}
\includegraphics[width=0.7\textwidth]{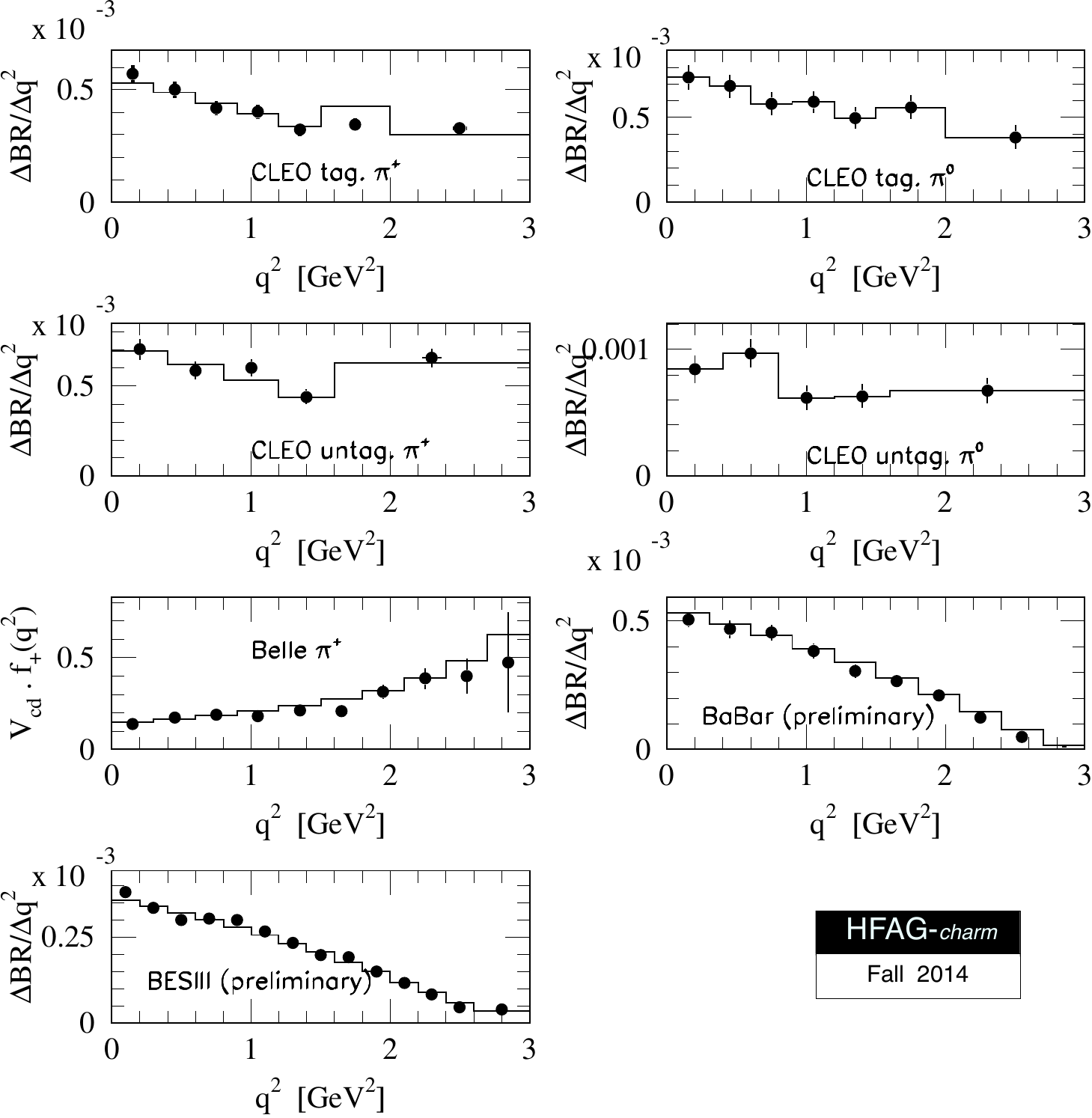}
\caption{Result of the three-pole model fit~\cite{Becirevic:2014kaa} to \babar~\cite{Lees:2014jka}, Belle\cite{Widhalm:2006wz}, 
BES III\cite{BESIII-new} and CLEOc~\cite{Besson:2009uv},~\cite{Dobbs:2007aa} $D \to \pi \ell \nu_\ell $ data. Points are the 
measured data in $q^2$ bins and the black line correspond to the result of the combined fit.
\label{plot3}
}
\end{center}
\end{figure}

The result for the effective mass $m_{D^{\ast ''}_{eff}}$ is larger than the mass of the 
second radially excited state with $J^P = 1^{-}$ ($\sim 3.11$~GeV), indicating that more contributions are needed to explain the form factor. 
Comparison of the combined fit with the individual data is shown in Figure \ref{plot3}.

\begin{table}[htbp]
\caption{Results of the three-pole model to \babar (preliminary), Belle, BES III (preliminary) and CLEOc (tagged and untagged) data. 
Fitted parameters are the first two residues 
$\gamma_0$  and $\gamma_{1}$, which are constrained using present measurements 
of masses and widths of the $D^\ast$ and $D^{\ast '}$ mesons, and lattice computations of decay constants, and the effective mass,  
$m_{D^{\ast ''}_{eff}}$, 
accounting for higher mass hadronic contributions. 
\label{3Pole_pi}}
\begin{center}
\begin{tabular}{cc}
\hline 
Parameter & Combined result ($D \to \pi \ell \nu_\ell $) \\ 
\hline 
\vspace*{-10pt} & \\
\hline
$\gamma_0$  & $3.878$ $\pm$ $0.090$  GeV$^2$ \\
$\gamma_1$  & $-1.18$ $\pm$ $0.30$  GeV$^2$  \\
$m_{D^{\ast ''}_{eff}}$  & $4.17$ $\pm$ $0.41$ GeV \\
\hline
\vspace*{-10pt} & \\
\hline
\end{tabular}
\end{center}
\end{table}

\subsubsection{$D\ra V\overline \ell \nu_\ell$ decays}

\begin{figure}[htbp]
  \begin{center}
\includegraphics[width=2.5in, bb=0 0 320 200]{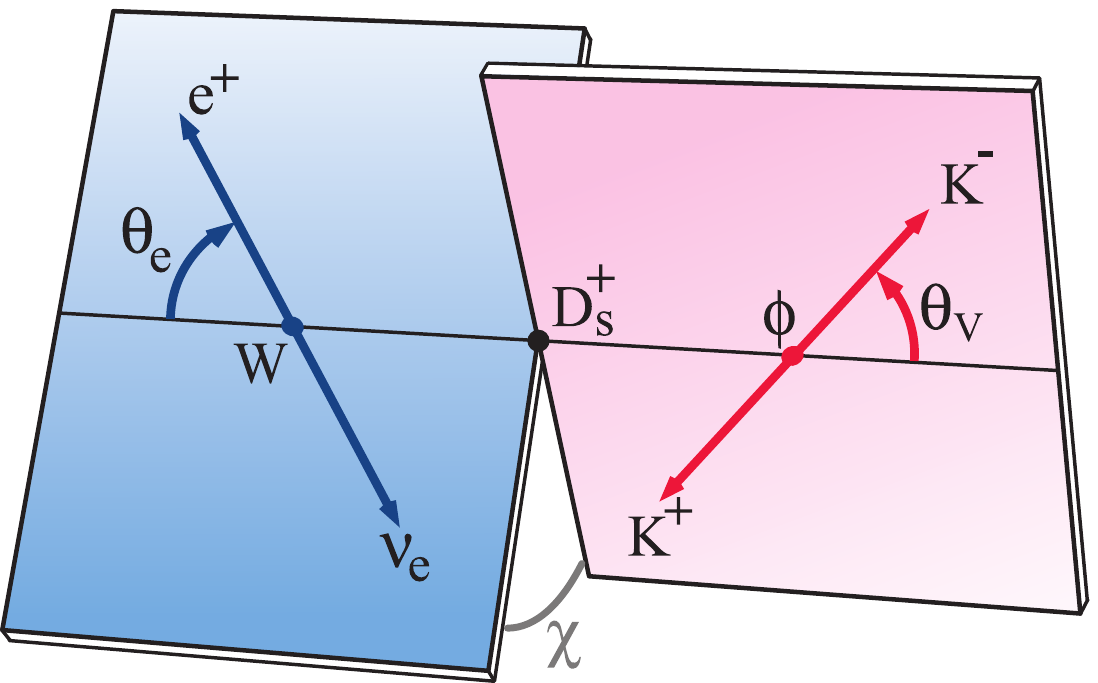}
  \end{center}
  \caption{
    Decay angles $\theta_V$, $\theta_\ell$ 
    and $\chi$. Note that the angle $\chi$ between the decay
    planes is defined in the $D$-meson reference frame, whereas
    the angles $\theta^{}_V$ and $\theta^{}_\ell$ are defined
    in the $V$ meson and $W$ reference frames, respectively.}
  \label{DecayAngles}
\end{figure}

When the final state hadron is a vector meson, the decay can proceed through
both vector and axial vector currents, and four form factors are needed.
The hadronic current is $H^{}_\mu = V^{}_\mu + A^{}_\mu$, 
where~\cite{Gilman:1989uy} 
\begin{eqnarray}
V_\mu & = & \left< V(p,\varepsilon) | \bar{q}\gamma_\mu c | D(p') \right> \ =\  
\frac{2V(q^2)}{m_D+m_V} 
\varepsilon_{\mu\nu\rho\sigma}\varepsilon^{*\nu}p^{\prime\rho}p^\sigma \\
 & & \nonumber\\
A_\mu & = & \left< V(p,\varepsilon) | -\bar{q}\gamma_\mu\gamma_5 c | D(p') \right> 
 \ =\  -i\,(m_D+m_V)A_1(q^2)\varepsilon^*_\mu \nonumber \\
 & & \hskip2.10in 
  +\ i \frac{A_2(q^2)}{m_D+m_V}(\varepsilon^*\cdot q)(p' + p)_\mu \nonumber \\
 & & \hskip2.30in 
+\ i\,\frac{2m_V}{q^2}\left(A_3(q^2)-A_0(q^2)\right)[\varepsilon^*\cdot (p' + p)] q_\mu\,.
\end{eqnarray}
In this expression, $m_V$ is the daughter meson mass and
\begin{eqnarray}A_3(q^2) & = & \frac{m_D + m_V}{2m_V}A_1(q^2)\ -\ \frac{m_D - m_V}{2m_V}A_2(q^2)\,.
\end{eqnarray}
Kinematics require that $A_3(0) = A_0(0)$. Terms proportional to $q_\mu$ are only important 
for the case of $\tau$ leptons. Thus, only three form factors are relevant in these decays: 
$A_1(q^2)$, $A_2(q^2)$ and $V(q^2)$. The differential partial width is
\begin{eqnarray}
\frac{d\Gamma(D \to V \overline \ell \nu_\ell)}{dq^2\, d\cos\theta_\ell} & = & 
  \frac{G_F^2\,|V_{cq}|^2}{128\pi^3m_D^2}\,p^*\,q^2 \times \nonumber \\
 & &  
\left[\frac{(1-\cos\theta_\ell)^2}{2}|H_-|^2\ +\  
\frac{(1+\cos\theta_\ell)^2}{2}|H_+|^2\ +\ \sin^2\theta_\ell|H_0|^2\right]\,,
\end{eqnarray}
where $H^{}_\pm$ and $H^{}_0$ are helicity amplitudes given by
\begin{eqnarray}
H_\pm & = & \frac{1}{m_D + m_V}\left[(m_D+m_V)^2A_1(q^2)\ \mp\ 
      2m^{}_D\,p^* V(q^2)\right] \\
 & & \nonumber \\
H_0 & = & \frac{1}{|q|}\frac{m_D^2}{2m_V(m_D + m_V)}\ \times\ \nonumber \\
 & & \hskip0.01in \left[
    \left(1- \frac{m_V^2 - q^2}{m_D^2}\right)(m_D + m_V)^2 A_1(q^2) 
    \ -\ 4{p^*}^2 A_2(q^2) \right]\,.
\label{HelDef}
\end{eqnarray}
$p^*$ is the magnitude of the three-momentum of the $V$ system, measured in the $D$ rest frame, and $\theta_\ell$ is 
defined in Figure \ref{DecayAngles} for the electron case ($\theta_e$).
The left-handed nature of the quark current manifests itself as
$|H_-|>|H_+|$. The differential decay rate for $D\ra V\ell\nu$ 
followed by the vector meson decaying into two pseudoscalars is

\begin{eqnarray}
\frac{d\Gamma(D\ra V\ell\nu, V\ra P_1P_2)}{dq^2 d\cos\theta_V d\cos\theta_\ell d\chi} 
 &  = & \frac{3G_F^2}{2048\pi^4}
       |V_{cq}|^2 \frac{p^*(q^2)q^2}{m_D^2} {\cal B}(V\to P_1P_2)\ \times \nonumber \\ 
 & & \hskip0.10in \Big\{ (1 + \cos\theta_\ell)^2 \sin^2\theta_V |H_+(q^2)|^2 \nonumber \\
 & & \hskip0.20in +\ (1 - \cos\theta_\ell)^2 \sin^2\theta_V |H_-(q^2)|^2 \nonumber \\
 & & \hskip0.30in +\ 4\sin^2\theta_\ell\cos^2\theta_V|H_0(q^2)|^2 \nonumber \\
 & & \hskip0.40in +\ 4\sin\theta_\ell (1 + \cos\theta_\ell) 
             \sin\theta_V \cos\theta_V \cos\chi H_+(q^2) H_0(q^2) \nonumber \\
 & & \hskip0.50in -\ 4\sin\theta_\ell (1 - \cos\theta_\ell) 
          \sin\theta_V \cos\theta_V \cos\chi H_-(q^2) H_0(q^2) \nonumber \\
 & & \hskip0.60in -\ 2\sin^2\theta_\ell \sin^2\theta_V 
                \cos 2\chi H_+(q^2) H_-(q^2) \Big\}\,,
\label{eq:dGammaVector}
\end{eqnarray}
where the angles $\theta^{}_\ell$, $\theta^{}_V$, and $\chi$ are defined
in Fig.~\ref{DecayAngles}. 

\begin{figure}[htbp]
  \begin{center}
    \includegraphics[width=0.7\textwidth]{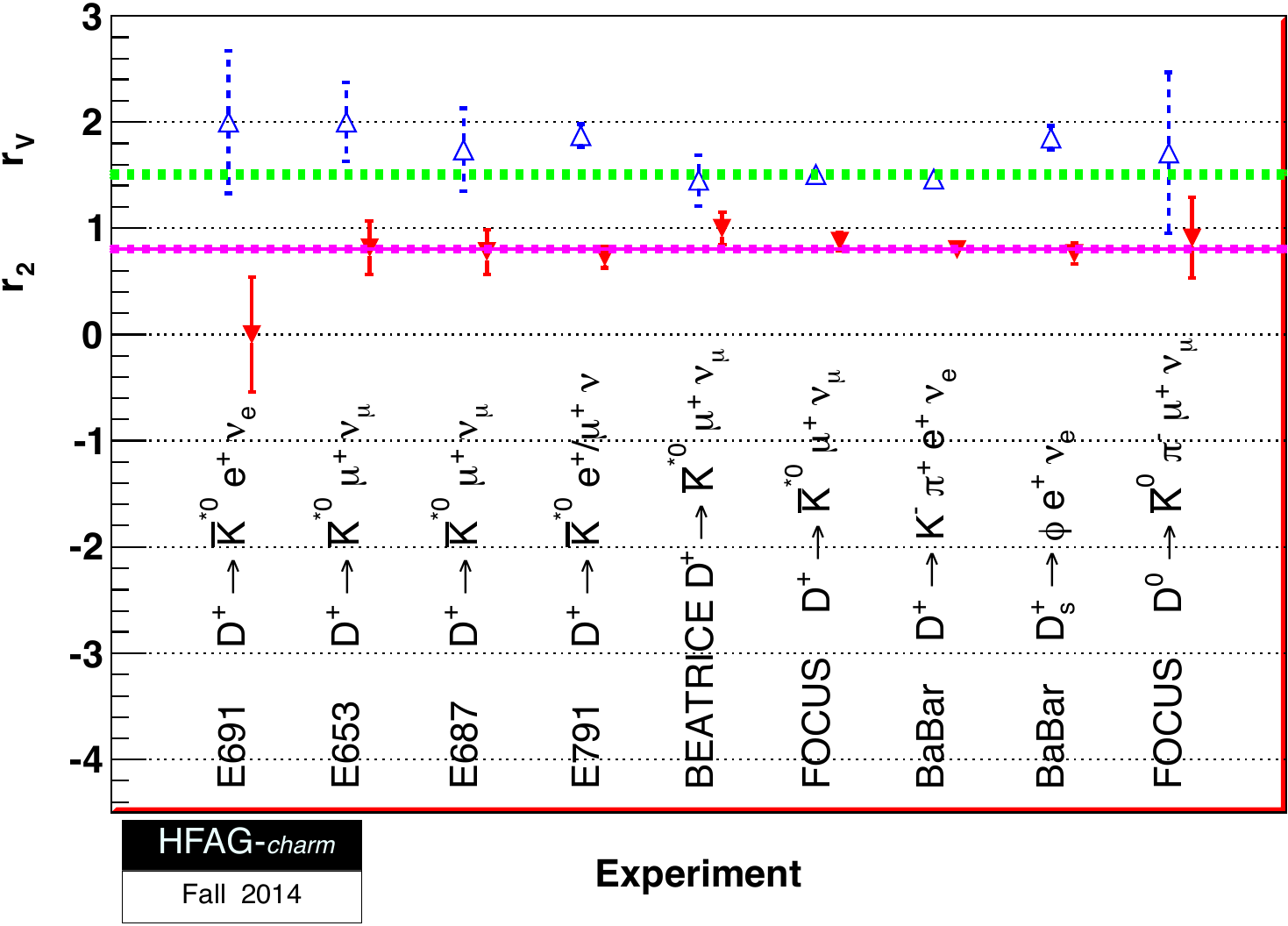}
  \end{center}
\vskip0.10in
  \caption{A comparison of $r_2$ and $r_V$ values 
    from various experiments. The first seven measurements are for $D^+
    \to K^-\pi^+ l^+\nu_l$ decays. Also shown as a line with
    1-$\sigma$ limits is the average of these. The last two points are
    $D_s^+$ decays and Cabibbo-suppressed $D$ decays.  
  \label{fig:r2rv}}
\end{figure}

Ratios between the values of the hadronic form factors expressed at $q^2=0$ are usually introduced:
\begin{eqnarray}
r_V \equiv V(0) / A_1(0), & &  r_2 \equiv A_2(0) / A_1(0) \label{rVr2_eq}\,.
\end{eqnarray}
Table \ref{Table1} lists measurements of $r_V$ and $r_2$ from several
experiments. Most of the measurements assume that the $q^2$ dependence of hadronic form factors 
is given by the simple pole ansatz. Some of these measurements do not consider a S-wave contribution and it is 
included in the measured values. 
The measurements are plotted in Fig.~\ref{fig:r2rv} which shows that they are all consistent.

\begin{table}[htbp]
\caption{Results for $r_V$ and $r_2$ from various experiments. 
\label{Table1}}
\begin{center}
\begin{tabular}{cccc}
\hline
\vspace*{-10pt} & \\
Experiment & Ref. & $r_V$ & $r_2$ \\
\vspace*{-10pt} & \\
\hline
\vspace*{-10pt} & \\
$D^+\to \overline{K}^{*0}l^+\nu$ & \omit & \omit & \omit         \\
E691         & \cite{Anjos:1990pn}     & 2.0$\pm$  0.6$\pm$  0.3  & 0.0$\pm$  0.5$\pm$  0.2    \\
E653         & \cite{Kodama:1992tn}     & 2.00$\pm$ 0.33$\pm$ 0.16 & 0.82$\pm$ 0.22$\pm$ 0.11   \\
E687         & \cite{Frabetti:1993jq}     & 1.74$\pm$ 0.27$\pm$ 0.28 & 0.78$\pm$ 0.18$\pm$ 0.11   \\
E791 (e)     & \cite{Aitala:1997cm}    & 1.90$\pm$ 0.11$\pm$ 0.09 & 0.71$\pm$ 0.08$\pm$ 0.09   \\
E791 ($\mu$) & \cite{Aitala:1998ey}    & 1.84$\pm$0.11$\pm$0.09   & 0.75$\pm$0.08$\pm$0.09     \\
Beatrice     & \cite{Adamovich:1998ia} & 1.45$\pm$ 0.23$\pm$ 0.07 & 1.00$\pm$ 0.15$\pm$ 0.03   \\
FOCUS        & \cite{Link:2002wg}   & 1.504$\pm$0.057$\pm$0.039& 0.875$\pm$0.049$\pm$0.064  \\
\hline
$D^0\to \overline{K}^0\pi^-\mu^+\nu$ & \omit & \omit & \omit         \\
FOCUS        & \cite{Link:2004uk}    & 1.706$\pm$0.677$\pm$0.342& 0.912$\pm$0.370$\pm$0.104 \\
\babar        & \cite{delAmoSanchez:2010fd} & $1.493 \pm 0.014 \pm 0.021$ & $0.775 \pm 0.011 \pm 0.011$ \\
\hline
$D_s^+ \to \phi\,e^+ \nu$ &\omit  &\omit     & \omit                  \\
\babar        & \cite{Aubert:2008rs}    & 1.849$\pm$0.060$\pm$0.095& 0.763$\pm$0.071$\pm$0.065\\
\hline
$D^0, D^+\to \rho\,e \nu$ & \omit  & \omit    & \omit                 \\
CLEO         & \cite{Mahlke:2007uf}    & 1.40$\pm$0.25$\pm$0.03   & 0.57$\pm$0.18$\pm$0.06    \\
\hline
\end{tabular}
\end{center}
\end{table}

\subsubsection{$S$-wave component}

In 2002 FOCUS reported~\cite{Link:2002ev} an asymmetry in
the observed $\cos(\theta_V)$ distribution. This is interpreted as
evidence for an $S$-wave component in the decay amplitude as follows. 
Since $H_0$ typically dominates over $H_{\pm}$, the distribution given 
by Eq.~(\ref{eq:dGammaVector}) is, after integration over $\chi$,
roughly proportional to $\cos^2\theta_V$. 
Inclusion of a constant $S$-wave amplitude of the form $A\,e^{i\delta}$ 
leads to an interference term proportional to 
$|A H_0 \sin\theta_\ell \cos\theta_V|$; this term causes an asymmetry 
in $\cos(\theta_V)$.
When FOCUS fit their data including this $S$-wave amplitude, 
they obtained $A = 0.330 \pm 0.022 \pm 0.015$ GeV$^{-1}$ and 
$\delta = 0.68 \pm 0.07 \pm 0.05$~\cite{Link:2002wg}. 

More recently, both \babar~\cite{Aubert:2008rs} and 
CLEO-c~\cite{Ecklund:2009fia} have also found evidence 
for an $f^{}_0$ component in semileptonic $D^{}_s$ decays.

\subsubsection{Model-independent form factor measurement}

\begin{figure}[htb]
  \begin{center}
    \vskip0.20in
    \includegraphics[width=0.55\textwidth]{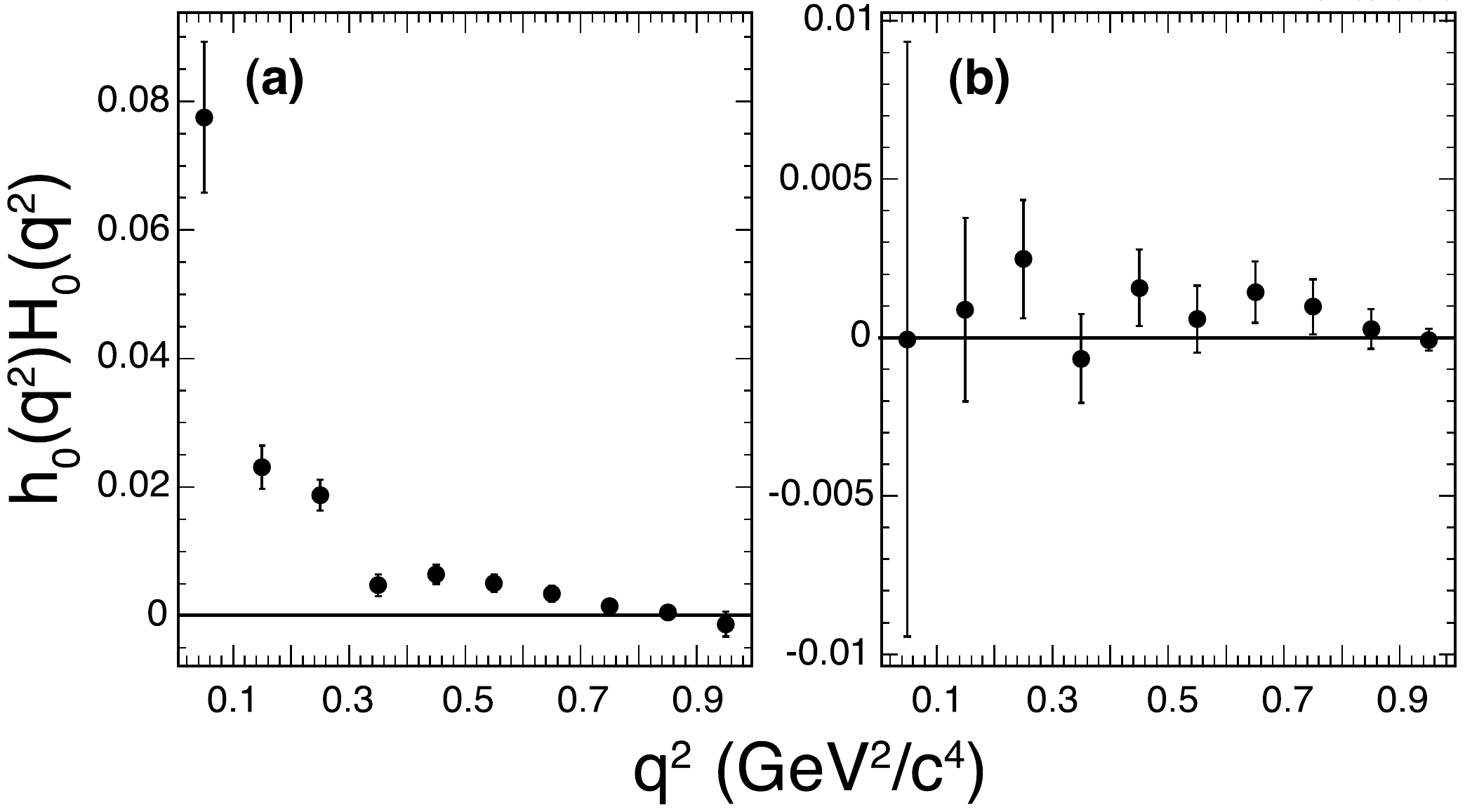}
  \end{center}
  \begin{center}
    \vskip0.20in
    \includegraphics[width=0.55\textwidth]{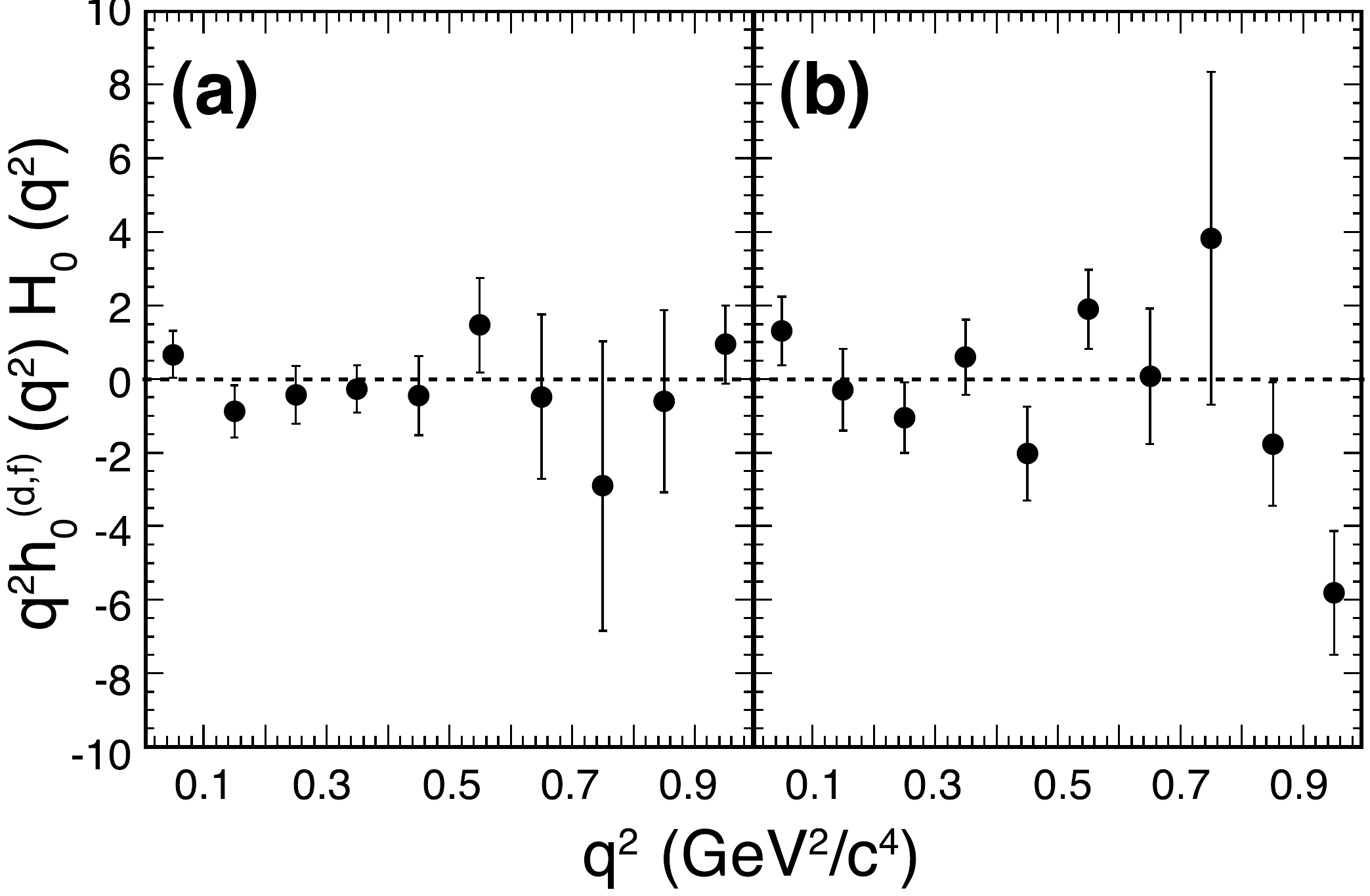}
  \end{center}
\vskip-0.20in
  \caption{Model-independent form factors $h_0(q^2)$ measured by 
    CLEO-c~\cite{Briere:2010zc}.
  \label{fig:cleoc_h0}}
\end{figure}

\begin{figure}[htb]
  \begin{center}
    \vskip0.20in
    \includegraphics[width=0.55\textwidth]{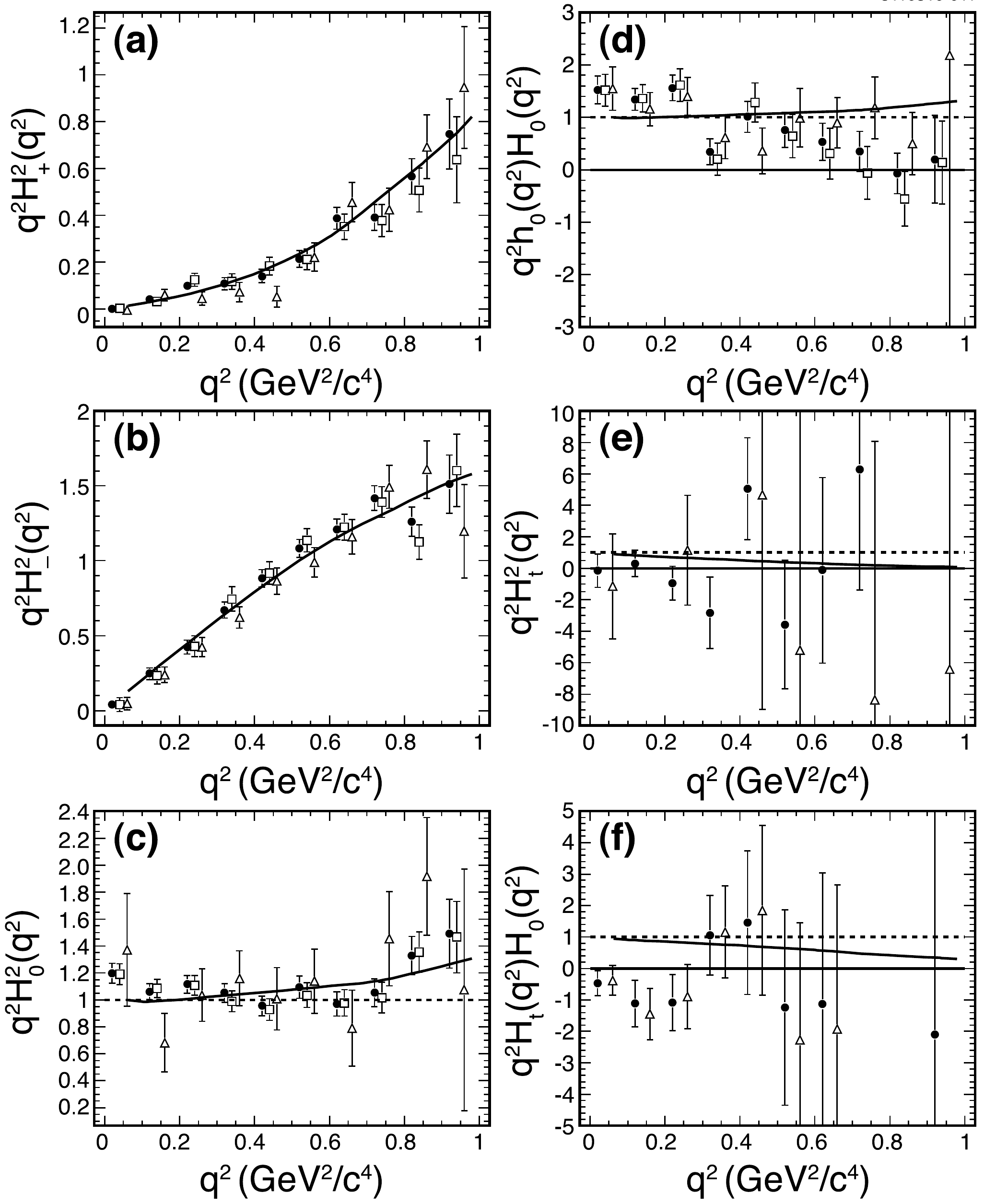}
  \end{center}
\vskip-0.20in
  \caption{Model-independent form factors $H(q^2)$ measured by 
    CLEO-c~\cite{Briere:2010zc}.
  \label{fig:cleoc_H}}
\end{figure}

Subsequently the CLEO-c collaboration extracted the form factors
$H_+(q^2)$, $H_-(q^2)$, and $H_0(q^2)$ in a model-independent fashion
directly as functions of $q^2$~\cite{Briere:2010zc} and also determined the
$S$-wave form factor $h_0(q^2)$ via the interference term, despite the
fact that the $K\pi$ mass distribution appears dominated by the vector
$K^*(892)$ state. Their results are shown in 
Figs.~\ref{fig:cleoc_h0} and \ref{fig:cleoc_H}.  Plots in Fig.~\ref{fig:cleoc_H} clearly show that
$H_0(q^2)$ dominates over essentially the full range of $q^2$, but
especially at low $q^2$. They also show that the transverse form factor
$H_t(q^2)$, which can be related to $A_3(q^2)$, is small compared to
Lattice Gauge Theory calculations and suggest that the form factor
ratio $r_3 \equiv A_3(0) / A_1(0)$ is large and negative.

The product $H_0(q^2)\times h_0(q^2)$ is shown in
Fig.~\ref{fig:cleoc_h0} and clearly indicates the existence of
$h_0(q^2)$, although it seems to fall faster with $q^2$ than $H_0(q^2)$.
The other plots in this figure show that $D$- and $F$-wave versions of
the $S$-wave $h_0(q^2)$ are not significant.

\subsubsection{Detailed measurements of the $D^+ \rightarrow K^- \pi^+ e^+ \nu_e$ 
decay channel}

\babar~\cite{delAmoSanchez:2010fd} has selected a large sample of $244\times 10^3$ signal events with a ratio $S/B\sim 2.3$ from an analyzed integrated
luminosity of $347~\fb^{-1}$. With four particles emitted in the final state, 
the differential decay rate depends on five variables.
In addition to the four variables defined in previous sections there is 
$m^2$, the mass squared of the $K\pi$ system.
Apart from this last variable, the reconstruction algorithm does not provide 
a high resolution on the other measured quantities 
and a multi-dimensional unfolding procedure
is not used to correct for efficiency and resolution effects. However, these
limitations still allow an essentially model independent measurement of
the differential decay rate. This is because, apart from the $q^2$
and mass dependence of the form factors, angular distributions are fixed by
kinematics. In addition, present accurate measurements of 
$D \rightarrow P \overline{\ell}\nu_{\ell}$ decays have shown that the 
$q^2$ dependence of the form factors can be well described by several models
as long as the corresponding model parameter(s) are fitted from data.
This is even more true in $D \rightarrow V \overline{\ell}\nu_{\ell}$ decays
because the $q^2$ range is reduced. To analyze the
$D^+ \rightarrow K^- \pi^+ e^+ \nu_e$ decay channel it is assumed
that all form factors have a $q^2$ variation given by
the simple pole model and the effective pole mass value,
$m_A=(2.63 \pm 0.10 \pm 0.13)~{\rm~GeV}/c^2$,
is fitted for the axial vector form factors. This value is compatible
with expectations when comparing with the mass
of $J^P=1^+$ charm mesons. Data are not sensitive to the effective mass
of the vector form factor for which $m_V=(2.1 \pm 0.1)~{\rm GeV}/c^2$ is used,
nor to the effective pole mass of the scalar component for which $m_A$ is used.
 For the mass dependence of the form factors,
a Breit-Wigner with a mass dependent width and a Blatt-Weisskopf damping factor
is used. For the S-wave amplitude, considering
what was measured in $D^+ \rightarrow K^- \pi^+\pi^+$ decays,
a polynomial variation below the $\overline{K}^*_0(1430)$
and a Breit-Wigner distribution, above are assumed. For the polynomial part, 
a linear term is sufficient to fit data.

\begin{figure}[!htb]
	\centering
\includegraphics[width=0.6\textwidth]{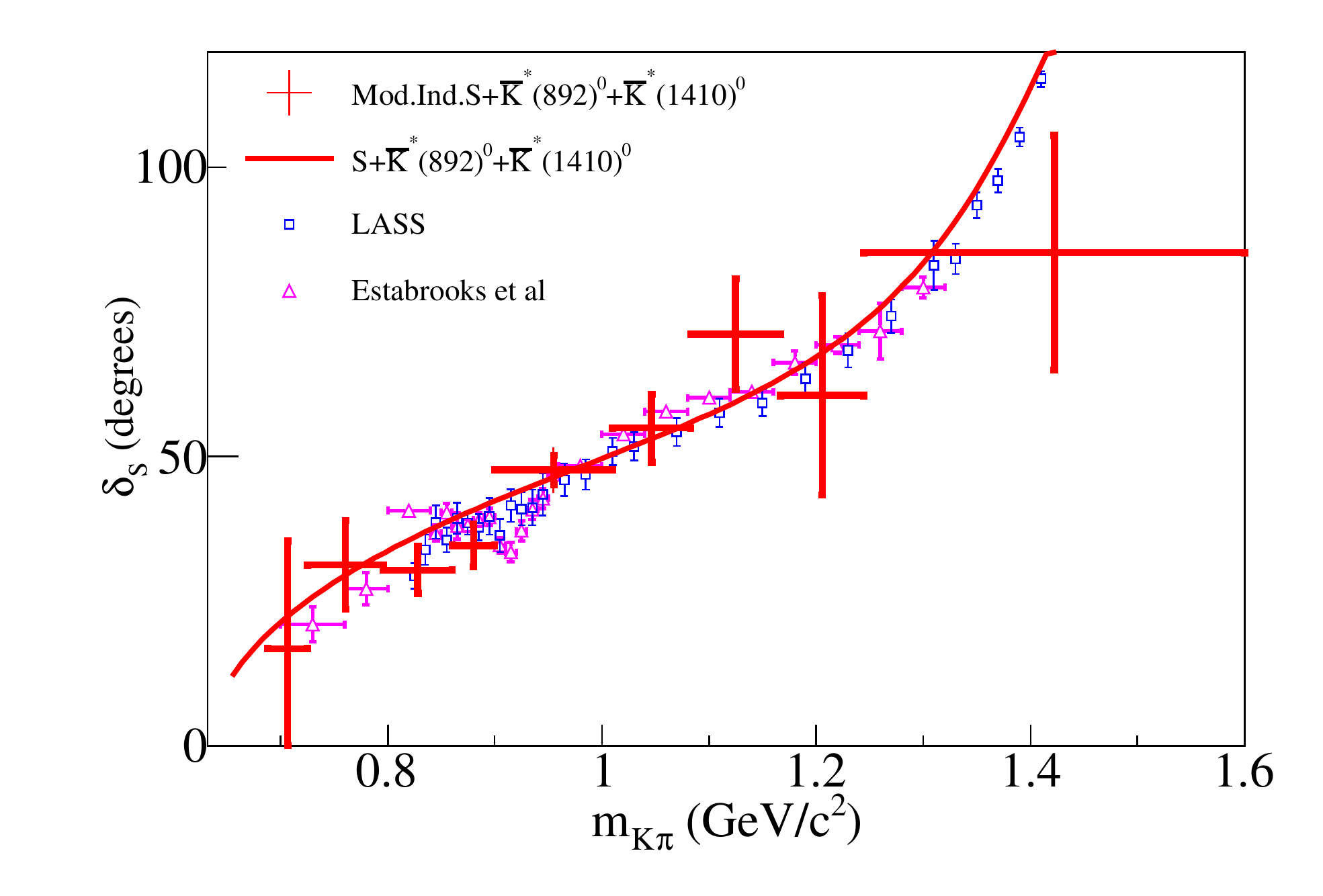}
\caption{Points (full circles) give the \babar $S$-wave phase variation 
assuming a signal containing $S$-wave, $\akst$ and $\akstp$ components~\cite{delAmoSanchez:2010fd}.
Error bars include systematic uncertainties.
The full line corresponds to a parameterized $S$-wave phase variation fitted on \babar data.
The phase variation measured in $K\pi$ scattering
by Ref.~\cite{Estabrooks:1977xe} (triangles) and LASS~\cite{Aston:1987ir} (squares), 
after correcting for $\delta^{3/2}$, are given.}
\label{fig:swave_phase}
\end{figure}

It is verified that the variation of the S-wave phase is compatible 
with expectations from elastic $K\pi$ scattering, according
to the Watson theorem. At variance with elastic scattering, 
a negative relative sign between the S- and P-waves is measured; 
this is compatible with 
the previous theorem. In Fig. \ref{fig:swave_phase}, the measured
S-wave phase is compared with the phase of the elastic, $I=1/2$,
$K\pi$ phase for different values of the $K\pi$ mass.

Contributions from other resonances decaying into $K^-\pi^+$ are considered.
A small signal from the $\overline{K}^*(1410)$ is observed, compatible
with expectations from $\tau$ decays and this component is included in the
nominal fit. In total, 11 parameters are fitted in addition to the total
number of signal events. They give a detailed description of the differential
decay rate versus the 5 variables and corresponding matrices for 
statistical and systematic uncertainties are provided allowing to 
evaluate the compatibility of data with future theoretical expectations.
Results of this analysis for the rates and few characteristics 
for S, P and D-waves are given in Table \ref{tab:comparison}.

\begin{figure}[htbp!]
 \begin{center}
\includegraphics[width=.50\textwidth]{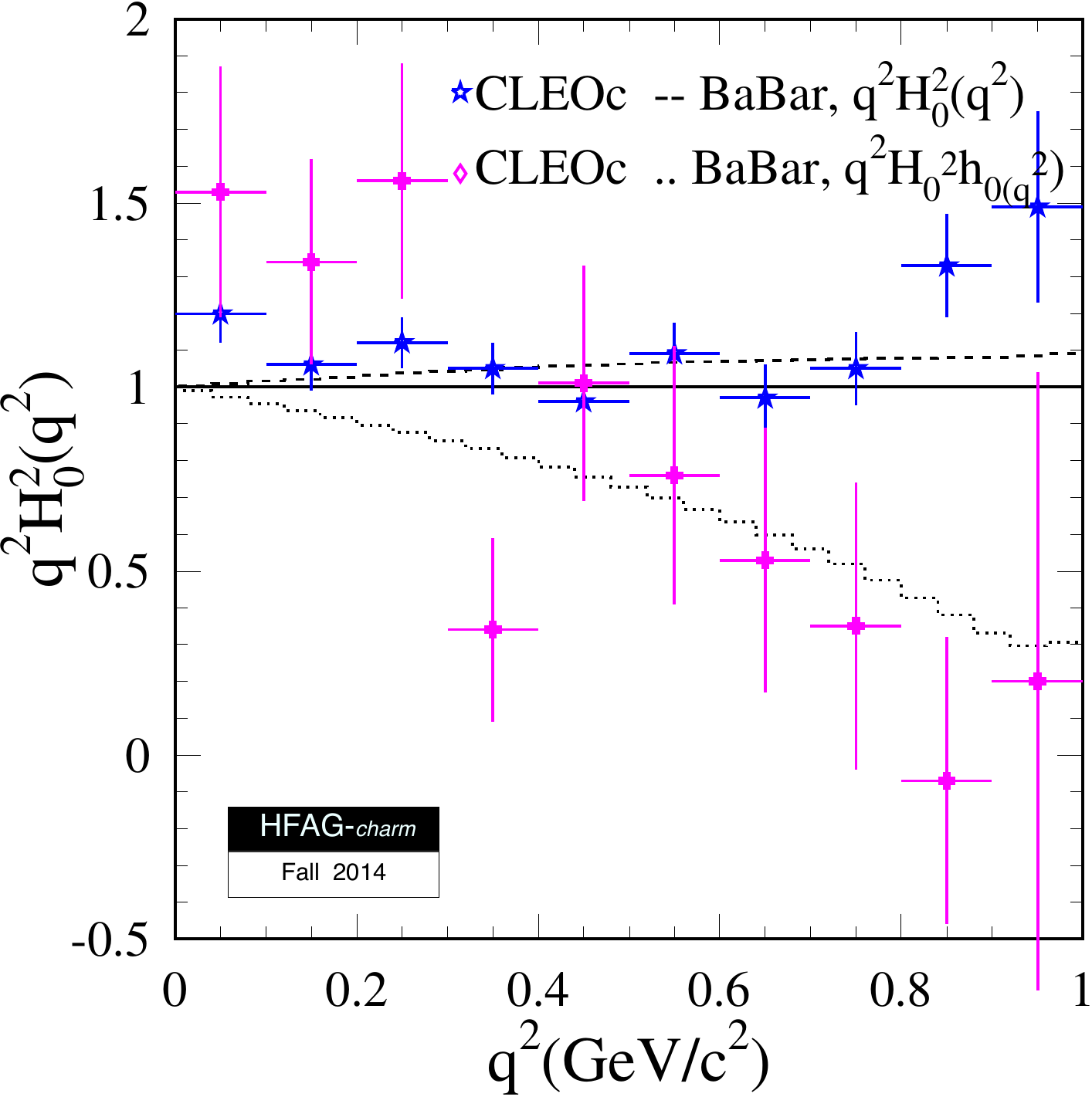}
 \end{center}
\caption[]{{Comparison between CLEO-c measurements and \babar results for the quantities $q^2H_0^2(q^2)$ and $q^2H_0(q^2t)h_0(q^2)$.}
   \label{fig:h0FF}}
\end{figure}

In Fig. \ref{fig:h0FF}, measured values from CLEO-c
of the products $q^2H_0^2(q^2)$ and $q^2h_0(q^2)H_0(q^2)$ are compared with 
corresponding results from \babar illustrating the difference in behavior
of the scalar $h_0$ component and the helicity zero $H_0$ P-wave form factor.
For this comparison, the plotted values from \babar for the two distributions
are fixed to 1 at $q^2=0$. The different behavior of $h_0(q^2)$
and $H_0(q^2)$ can be explained by they different dependence in the 
$p^*$ variable.

\begin{table}[!htb]
\begin{center}
 \caption[]{{Detailed determination of the properties of the 
$D^+ \rightarrow K^-\pi^+ e^+ \nu_e$ decay channel from \babar~\cite{delAmoSanchez:2010fd}. 
Values for ${\cal B}(D^+\rightarrow \akstp / \akstd e^+\nu_e)$ are corrected for their respectivebranching fractions into $K^-\pi^+$.}\hspace{1cm}
  \label{tab:comparison}}
\begin{tabular}{c c}
\vspace*{-10pt} & \\
\hline
\textbf{Measurement} & \textbf{\babar result} \\
\hline\hline
{ $m_{\kst}(~MeV/c^2)$} & {$895.4\pm{0.2}\pm0.2$}\\
{ $\Gamma^0_{\kst}(~MeV/c^2)$} & {$46.5\pm{0.3}\pm0.2$} \\
{$r_{BW}(~GeV/c)^{-1}$ }& {$2.1\pm{0.5}\pm 0.5$} \\
\hline
{$r_{V}$} & {$1.463\pm{0.017}\pm 0.031$} \\
{$r_{2}$} & {$0.801\pm{0.020}\pm 0.020$}  \\
{ $m_{A} (~GeV/c^2)$} & {$2.63\pm 0.10 \pm 0.13$}\\
\hline
${\cal B}(D^+ \rightarrow K^-\pi^+ e^+ \nu_e)(\%)$ &$4.04 \pm 0.03 \pm 0.04 \pm 0.09$ \\
${\cal B}(D^{+}\rightarrow K^- \pi^+ e^{+} \nu_{e})_{\overline{K}^{*0}}(\%)$ & $3.80\pm0.04\pm0.05 \pm0.09 $  \\ 
${\cal B}(D^+ \rightarrow K^-\pi^+ e^+ \nu_e)_{S-wave}(\%)$ &$0.234 \pm0.007  \pm0.007  \pm0.005  $ \\
${\cal B}(D^{+}\rightarrow \akstp e^{+} \nu_{e})(\%)$ & $0.30\pm0.12\pm0.18\pm0.06$ ($<0.6$ at 90$\%$ C.L.) \\ 
${\cal B}(D^{+}\rightarrow \akstd e^{+} \nu_{e})(\%)$ & $0.023\pm0.011\pm0.011\pm0.001$ ($<0.05$ at 90$\%$ C.L.)  \\ 
\hline\hline
\end{tabular}
\end{center}
\end{table}


\clearpage
\subsection{Leptonic decays}

Purely leptonic decays of $\Dp$ and $\dsp$ mesons are among the simplest and theoretically cleanest
probes of $c\to d$ and $c\to s$ quark flavor-changing transitions. The branching fraction of leptonic 
decays that proceed via the annihilation of the initial quark-antiquark pair ($c\overline{d}$ or 
$c\overline{s}$) into a virtual $W^+$ that finally materializes as an antilepton-neutrino pair ($\ellnu$) is 
given in the Standard Model by
\begin{equation}
 \br(D_{q}^+\to \ell^+\nu_{\ell})=\frac{G_F^2}{8\pi}\tau_{D_q}f_{D_{q}}^2|V_{cq}|^2m_{D_{q}}m_{\ell}^2\left(1-\frac{m_{\ell}^2}{m_{D_{q}}^2} \right)^2.
 \label{eq:brCharmLeptonicSM}
\end{equation}
Here, $m_{D_{q}}$ is the $D_{q}$ meson mass, $\tau_{D_q}$ is its lifetime, $m_{\ell}$ is the charged lepton mass, 
$|V_{cq}|$ is the magnitude of the relevant CKM matrix element, and $G_F$ is the Fermi coupling constant. The parameter 
$f_{D_{q}}$ is the $D_q$ meson decay constant and is related to the wave-function overlap of the meson's 
constituent quark and anti-quark. Within the SM, the decay constants have been predicted using several 
methods, the most precise being the lattice gauge theory (LQCD) calculations. The Flavor Lattice Averaging 
Group~\cite{FLAG} combines all LQCD calculations and provides averaged values for $f_D$ and $f_{D_s}$ (see 
Table~\ref{tab:Lattice}) that are used within this section to extract the magnitudes of the $V_{cd}$ and $V_{cs}$ CKM
matrix elements from experimentally measured branching fractions of $D^+\to \ell^+\nu_{\ell}$ and 
$D_s^+\to \ell^+\nu_{\ell}$ decays, respectively.
\begin{table}[b!]
\caption{The LQCD average for $D$ and $D_s$ meson decay constants and their ratio from the Flavor Lattice Averaging 
Group~\cite{FLAG}.
\label{tab:Lattice}}
\begin{center}
\begin{tabular}{ll}
\toprule
\rowcolor{Gray} Quantity & Value \\ 
\midrule
$f_D$ 		& $209.2\pm3.3$~MeV\\
$f_{D_s}$ 	& $248.6\pm2.7$~MeV\\
$f_{D_s}/f_D$	& $1.187\pm0.012$
\\ \bottomrule
\end{tabular}
\end{center}
\end{table}

The leptonic decays of pseudoscalar mesons 
are suppressed by helicity conservation and their decay rates are thus proportional to the square of 
the charged lepton mass. Leptonic decays into electrons with $\br\lesssim 10^{-7}$ are not experimentally 
observable yet whereas decays to taus are favored over decays to muons. In particular, the ratio of the 
latter decays is equal to 
$R^{D_q}_{\tau/\mu}\equiv \br(D^+_q\to\tau^+\nu_{\tau})/\br(D^+_q\to\mu^+\nu_{\mu})=m_{\tau}^2/m_{\mu}^2\cdot(1-m^2_{\tau}/m^2_{D_q})^2/(1-m^2_{\mu}/m^2_{D_q})^2=9.76\pm0.03$ 
in the case of $D_s^+$ decays and to $2.67\pm0.01$ in the case of $D^+$ decays based on the world average values of masses of the muon, tau and 
$D_q$ meson given in Ref.~\cite{PDG_2012}. 
Any deviation from this expectation could only be interpreted as violation of lepton universality in charged 
currents and would hence point to NP effects~\cite{Filipuzzi:2012mg}.

Averages presented within this subsection are weighted averages and correlations between measurements and dependencies on input parameters
are taken into account.

\subsubsection{$D^+\to \ell^+\nu_{\ell}$ decays and $|V_{cd}|$}

\begin{figure}[hbt!]
\centering
\includegraphics[width=0.7\textwidth]{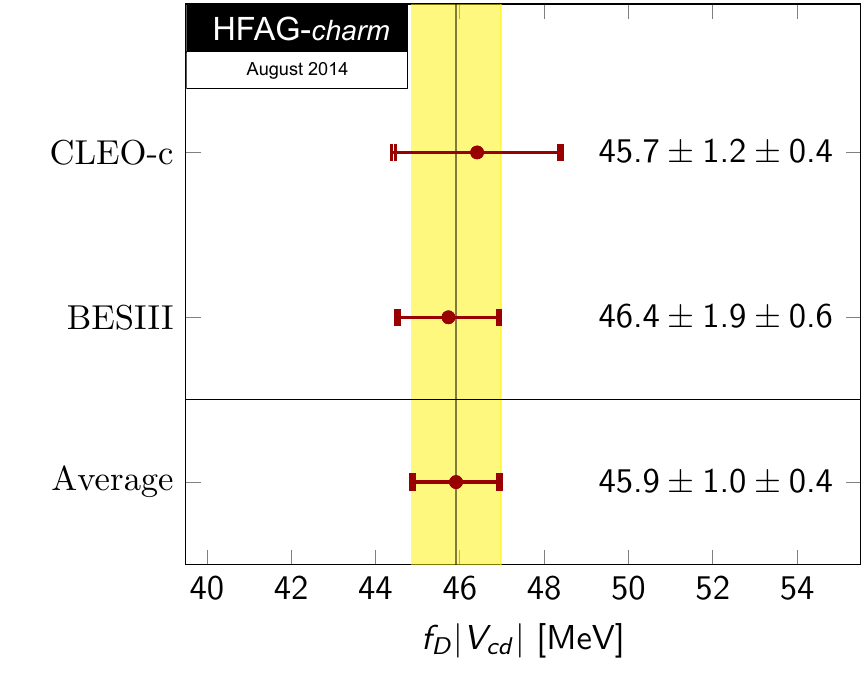}
\caption{
WA value for $f_{D}|V_{cd}|$. For each point, the first error listed is the statistical and the second error is the systematic error.
\label{fig:ExpDLeptonic}
}
\end{figure}

We use measurements of the branching fraction $\br(D^+\to\munu)$ from \mbox{CLEO-c}~\cite{Eisenstein:2008aa} and 
BESIII~\cite{Ablikim:2013uvu} to calculate its world average (WA) value. We obtain
\begin{equation}
 \br^{\rm WA}(D^+\to\munu) = (3.74\pm0.17)\times10^{-4},
 \label{eq:Br:WA:DtoMuNu}
\end{equation}
from which we determine the product of the decay constant and the CKM matrix element to be
\begin{equation}
 f_{D}|V_{cd}| = \left(45.9\pm1.1\right)~\mbox{MeV},
 \label{eq:expFDVCD}
\end{equation}
where the uncertainty includes the uncertainty on $\br^{\rm WA}(D^+\to\munu)$ and external inputs\footnote{These values (taken from the PDG~\cite{PDG_2012}) are
$m_{\mu} = (0.1056583715\pm0.0000000035)$~GeV/$c^2$, $m_D = (1.86962\pm0.00015)$~GeV/$c^2$ 
and $\tau_D = (1040\pm7)\times 10^{-15}$~s.} needed to extract $f_{D}|V_{cd}|$ from the 
measured branching fraction using Eq.~\ref{eq:brCharmLeptonicSM}. 
Using the LQCD value for $f_D$ from Table~\ref{tab:Lattice} we 
finally obtain the CKM matrix element $V_{cd}$ to be
\begin{equation}
 |V_{cd}| = 0.219\pm0.005(\rm exp.)\pm0.003(\rm LQCD),
 \label{eq:Vcd:WA:Leptonic}
\end{equation}
where the uncertainties are from the experiments and lattice calculations, respectively. All input values
and the resulting world averages are summarized in Table~\ref{tab:DExpLeptonic} and plotted in 
Fig.~\ref{fig:ExpDLeptonic}.
\begin{table}[t!]
\caption{Experimental results and world averages for ${\cal{B}}(D^+\to \ell^+\nu_{\ell})$ and $f_{D}|V_{cd}|$.
The first uncertainty is statistical and the second is experimental systematic. The third uncertainty in the case
of $f_{D^+}|V_{cd}|$ is due to external inputs (dominated by the uncertainty of $\tau_D$).
\label{tab:DExpLeptonic}}
\begin{center}
\begin{tabular}{lccll}
\toprule
\rowcolor{Gray} Mode 	& ${\cal{B}}$ ($10^{-4}$)	& $f_{D}|V_{cd}|$ (MeV)		& Reference & \\ 
\midrule
\multirow{2}{*}{$\munu$} & $3.82\pm0.32\pm 0.09$ 	& $46.4\pm1.9\pm0.5\pm0.2$	& CLEO-c & \cite{Eisenstein:2008aa}\\ 
			& $3.71\pm0.19\pm 0.06$ 	& $45.7\pm1.2\pm0.4\pm0.2$	& BESIII & \cite{Ablikim:2013uvu}\\
\midrule
\rowcolor{Gray}$\munu$ 	& $3.74\pm0.16\pm 0.05$		& $45.9\pm1.0\pm0.3\pm0.2$	& Average & \\
\midrule
$\enu$	 		& {$<0.088$ at 90\% C.L.}	&& CLEO-c & \cite{Eisenstein:2008aa}\\
\midrule
$\taunu$ 		& {$<12$ at 90\% C.L.}		&& CLEO-c & \cite{Eisenstein:2008aa}
\\ \bottomrule
\end{tabular}
\end{center}
\end{table}
 
The upper limit on the ratio of branching fractions is found to be $R_{\tau/\mu}^D<3.2$ at 90\%~C.L., which is just slightly above the SM expected value, $2.67\pm0.01$.

\subsubsection{$D_s^+\to \ell^+\nu_{\ell}$ decays and $|V_{cs}|$}

We use measurements of the absolute branching fraction $\br(D_s^+\to\munu)$ from CLEO-c~\cite{Alexander:2009ux}, \babar~\cite{delAmoSanchez:2010jg},
and Belle~\cite{Zupanc:2013byn} and obtain a WA value of
\begin{equation}
 \br^{\rm WA}(\dsmunu) = (5.57\pm0.24)\times10^{-3}.
 \label{eq:Br:WA:DstoMuNu}
\end{equation}
The WA value for $\br(D_s^+\to\taunu)$ is also calculated from CLEO-c, \babar, and Belle measurements. 
CLEO-c made separate measurements for $\tau^+\to e^+\nu_e\overline{\nu}{}_{\tau}$~\cite{Naik:2009tk},
$\tau^+\to\pi^+\overline{\nu}{}_{\tau}$~\cite{Alexander:2009ux}, and
$\tau^+\to\rho^+\overline{\nu}{}_{\tau}$~\cite{Onyisi:2009th};
\babar made separate measurements for 
$\tau^+\to e^+\nu_e\overline{\nu}{}_{\tau}$~\cite{delAmoSanchez:2010jg} and $\tau^+\to \mu^+\nu_{\mu}\overline{\nu}{}_{\tau}$; and
Belle made separate measurements for $\tau^+\to e^+\nu_e\overline{\nu}{}_{\tau}$, $\tau^+\to \mu^+\nu_{\mu}\overline{\nu}{}_{\tau}$, 
and $\tau^+\to\pi^+\overline{\nu}{}_{\tau}$~\cite{Zupanc:2013byn}.
Combining all of them we obtain the WA value of
\begin{equation}
 \br^{\rm WA}(\dsp\to\taunu) = (5.55\pm0.24)\times10^{-2}.
 \label{eq:Br:WA:DstoTauNu}
\end{equation}

The ratio of branching fractions is found to be
\begin{equation}
R_{\tau/\mu}^{\ds} = 9.96\pm0.57,
\label{eq:R:WA:Leptonic}
\end{equation}
and is consistent with the value expected in the SM, $9.76\pm0.03$.

\begin{figure}[hbt!]
\centering
\includegraphics[width=1\textwidth]{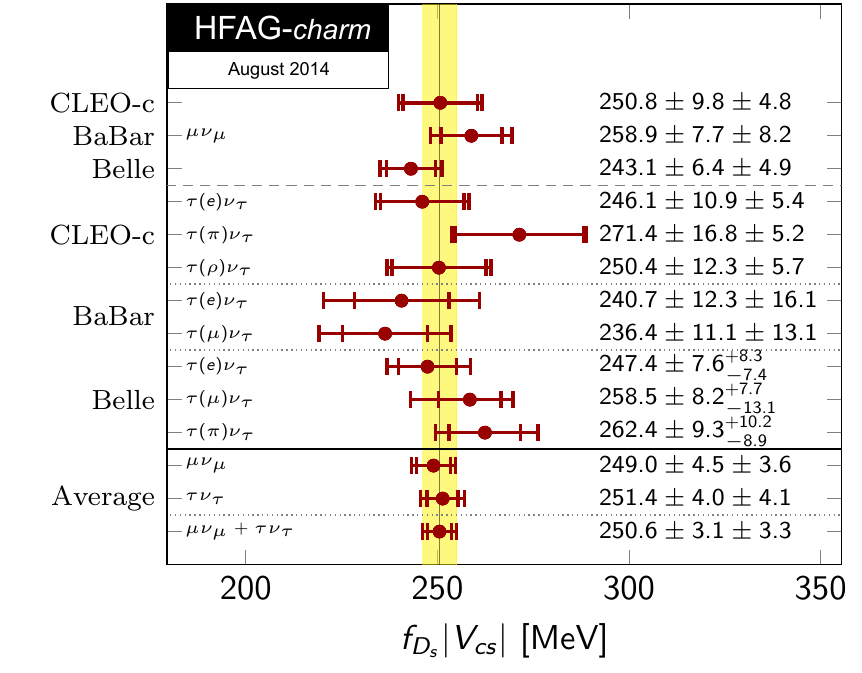}
\caption{
WA value for $f_{D_s}|V_{cs}|$. For each point, the first error listed is the statistical and the second error is the systematic error.
\label{fig:ExpDsLeptonic}
}
\end{figure}

From the average values of branching fractions of muonic and tauonic decays we determine\footnote{
We use the following values (taken from PDG~\cite{PDG_2012}) for external parameters entering 
Eq.~\ref{eq:brCharmLeptonicSM}: $m_{\tau} = (1.77682\pm0.00016)$~GeV/$c^2$, $m_{D_s} = (1.96850\pm0.00032)$~GeV/$c^2$ 
and $\tau_{D_s} = (500\pm7)\times 10^{-15}$~s.} the product of $D_s$ meson decay constant and 
the $|V_{cs}|$ CKM matrix element to be
\begin{equation}
 \fds|V_{cs}|=\left(250.6\pm4.5\right)~\mbox{MeV},
 \label{eq:expFDSVCS}
\end{equation}
where the uncertainty is due to the uncertainties on $\br^{\rm WA}(D_s^+\to\munu)$ and 
$\br^{\rm WA}(D_s^+\to\taunu)$ and the external inputs. All input values and the resulting world averages are 
summarized in Table~\ref{tab:DsLeptonic} and plotted in Fig.~\ref{fig:ExpDsLeptonic}. To obtain the 
averages given within this subsection and in Table~\ref{tab:DsLeptonic} we have taken into account
the correlations within each experiment\footnote{In the case of \babar we use the covariance matrix from 
the errata of~Ref.\cite{delAmoSanchez:2010jg}.} for the uncertainties related to: normalization, tracking, particle identification, 
signal and background parameterizations, and peaking background contributions.

Using the LQCD value for $\fds$ from Table~\ref{tab:Lattice} we 
finally obtain the CKM matrix element $V_{cs}$ to be
\begin{equation}
 |V_{cs}| = 1.008\pm0.018(\rm exp.)\pm0.011(\rm LQCD),
 \label{eq:Vcs:WA:Leptonic}
\end{equation}
where the uncertainties are from the experiments and lattice calculations, respectively.

\begin{table}[t!]
\caption{Experimental results and world averages for ${\cal{B}}(\dsellnu)$ and $f_{D_s}|V_{cs}|$.
The first uncertainty is statistical and the second is experimental systematic. The third uncertainty 
in the case of $f_{D_s}|V_{cs}|$ is due to external inputs (dominated by the uncertainty of $\tau_{D_s}$).
We have recalculated $\br(\dsp\to\taunu)$ quoted by CLEO-c and \babar using the latest 
values for branching fractions of $\tau$ decays to electron, muon, or pion and neutrinos~\cite{PDG_2012}.
CLEO-c and \babar include statistical uncertainty of number of $\ds$ tags (denominator in the calculation of 
branching fraction) in the statistical uncertainty of measured $\br$. We subtract this uncertainty from the
statistical one and add it to the systematic uncertainty. 
\label{tab:DsLeptonic}}
\begin{center}
\begin{tabular}{lccll}
\toprule
\rowcolor{Gray}
Mode 		& ${\cal{B}}$ ($10^{-2}$) 	& $f_{D_s}|V_{cs}|$ (MeV) 		& Reference & 
\\ \midrule
\multirow{3}{*}{$\munu$}	& $0.565\pm0.044\pm 0.020$ 	& $250.8 \pm 9.8 \pm 4.4 \pm 1.8$	& CLEO-c &\cite{Alexander:2009ux}\\		
				& $0.602\pm0.037\pm 0.032$ 	& $258.9 \pm 8.0 \pm 6.9 \pm 1.8$	& \babar  &\cite{delAmoSanchez:2010jg}\\
				& $0.531\pm0.028\pm 0.020$ 	& $243.1 \pm 6.4 \pm 4.6 \pm 1.7$ 	& Belle  &\cite{Zupanc:2013byn}\\
\midrule
\rowcolor{Gray}
$\munu$ 			& $0.557\pm0.020\pm0.014$ 		& $249.0 \pm 4.5 \pm 3.1 \pm 1.7$ 	& Average & \\
\midrule
$\tauenu$ 			& $5.31\pm0.47\pm0.22$ 		& $246.1 \pm 10.9 \pm 5.1 \pm 1.7$ 	& CLEO-c &\cite{Onyisi:2009th}\\
$\taupinuCharm$ 			& $6.46\pm0.80\pm0.23$ 		& $271.4 \pm 16.8 \pm 4.8 \pm 1.9$  & CLEO-c &\cite{Alexander:2009ux}\\
$\taurhonu$ 			& $5.50\pm0.54\pm0.24$ 		& $250.4 \pm 12.3 \pm 5.5 \pm 1.8$  & CLEO-c &\cite{Naik:2009tk}\\
\midrule
\rowcolor{LightGray}
$\taunu$			& $5.57\pm0.32\pm0.15$		& $252.0 \pm 7.2 \pm 3.4 \pm 1.8$   & CLEO-c & \\
\midrule
$\tauenu$ 			& $5.08\pm0.52\pm0.68$ 		& $240.7 \pm 12.3 \pm 16.1 \pm 1.7$	& \multirow{2}{*}{\babar} & \multirow{2}{*}{\cite{delAmoSanchez:2010jg}}\\
$\taumunu$ 			& $4.90\pm0.46\pm0.54$ 		& $236.4 \pm 11.1 \pm 13.0 \pm 1.7$	&  & \\
\midrule
\rowcolor{LightGray}
$\taunu$			& $4.95\pm0.36\pm0.58$		& $237.6 \pm 8.6 \pm 13.8 \pm 1.7$   & \babar &\\
\midrule
$\tauenu$  			& $5.37\pm0.33^{+0.35}_{-0.31}$ & $247.4 \pm 7.6^{+8.1}_{-7.1} \pm 1.7$  & \multirow{3}{*}{Belle} & \multirow{3}{*}{\cite{Zupanc:2013byn}} \\
$\taumunu$ 		 	& $5.86\pm0.37^{+0.34}_{-0.59}$ & $258.5 \pm 8.2^{+7.5}_{-13.0} \pm 1.8$  & &\\ 
$\taupinuCharm$  			& $6.04\pm0.43^{+0.46}_{-0.40}$ & $262.4 \pm 9.3^{+10.0}_{-8.7} \pm 1.8$  & &\\
\midrule
\rowcolor{LightGray}
$\taunu$			& $5.70\pm0.21\pm0.31$		& $254.9 \pm 4.7 \pm 6.9 \pm 1.8$   & Belle & \\
\midrule
\rowcolor{Gray}
$\taunu$ 			& $5.55\pm0.18\pm0.17$ 		& $251.5 \pm 4.1 \pm 3.9 \pm 1.8$	& Average & \\
\midrule
\rowcolor{Gray}
$\munu$  			&		 		& \multirow{2}{*}{$250.6\pm 3.1\pm 2.8\pm1.8$}	& \multirow{2}{*}{Average} & \\
\rowcolor{Gray}
$\taunu$ 			&		 		& \multirow{-2}{*}{$250.6\pm 3.1\pm 2.8\pm1.8$}	& \multirow{-2}{*}{Average} & \\
\midrule
$\enu$				& $<0.0083$ at 90\% C.L.	& 					& Belle & \cite{Zupanc:2013byn} \\
\bottomrule
\end{tabular}
\end{center}
\end{table}

\subsubsection{Comparison with other determinations of $|V_{cd}|$ and $|V_{cs}|$}

\begin{figure}[hbt!]
\centering
\includegraphics[width=0.49\textwidth]{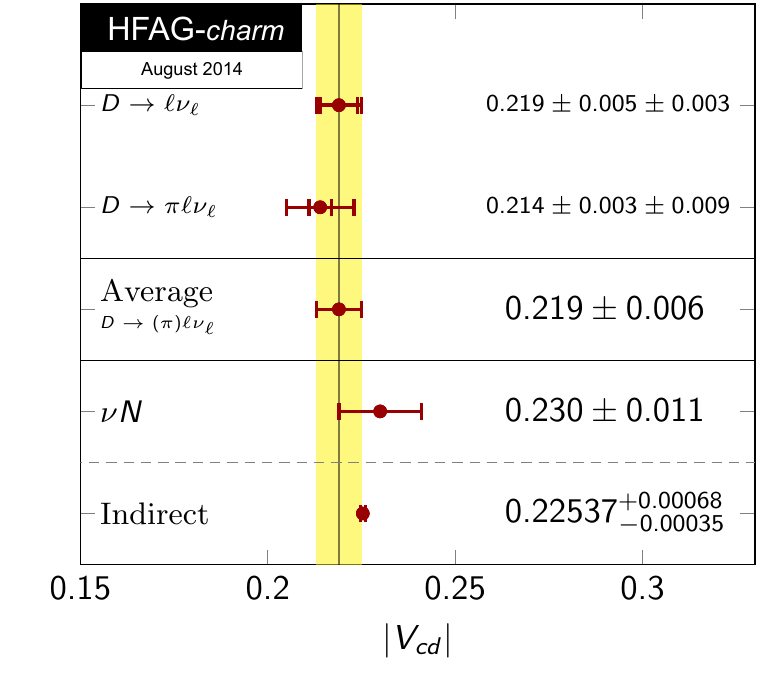}~
\includegraphics[width=0.49\textwidth]{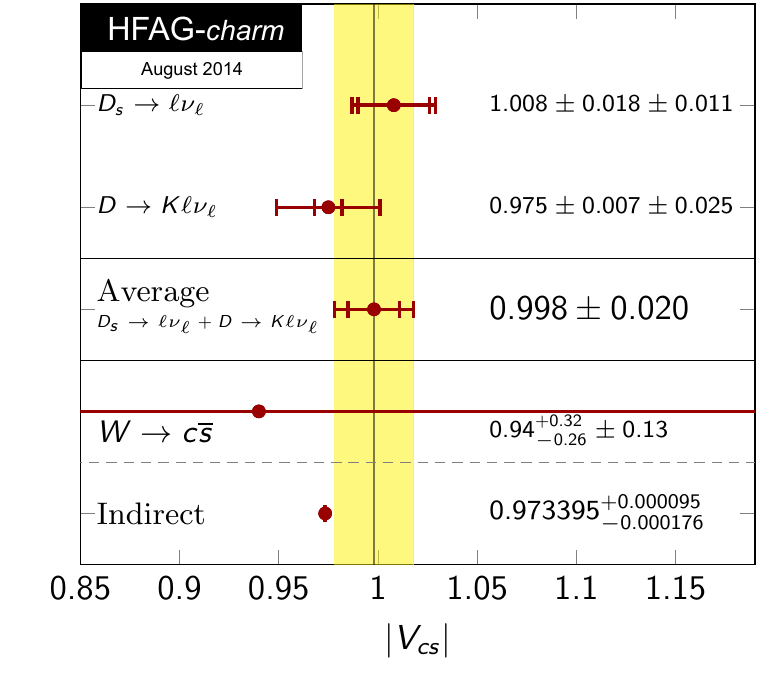}
\caption{
Comparison of magnitudes of the CKM matrix elements $|V_{cd}|$ (left) and $|V_{cs}|$ (right) determined from the (semi-)leptonic charm decays and from neutrino scattering data
or $W$ decays and indirect determination from the global fit assuming CKM unitarity~\cite{Charles:2004jd}.
\label{fig:VcdVcsComparions}
}
\end{figure}

Table~\ref{tab:CKMVcdVcs} summarizes and Fig.~\ref{fig:VcdVcsComparions} shows all determinations of the CKM matrix elements $|V_{cd}|$ and $|V_{cs}|$. As
can be seen, the most precise direct determinations of these CKM matrix elements are those from leptonic and semileptonic $D_{(s)}$ decays. The values are in agreement
within uncertainties with the values obtained from the global fit assuming CKM matrix unitarity.
\begin{table}[htb]
\caption{Average of the magnitudes of the CKM matrix elements $|V_{cd}|$ and $|V_{cs}|$ determined from the leptonic and semileptonic $D$ and $D_s$ decays.
In the calculation of average values we assume 100\% correlations in uncertainties due to LQCD.  The values determined from neutrino scattering 
or $W$ decays and indirect determination from the global fit to the CKM matrix are given for comparison as well.
\label{tab:CKMVcdVcs}}
\begin{center}
\begin{tabular}{lcc}
\toprule
\rowcolor{Gray} Method & Reference & Value \\ 
\midrule
&&{$|V_{cd}|$}\\
\cline{3-3}
$D\to\ell\nu_{\ell}$ 	 & This section			& $0.219\pm0.005(\rm exp.)\pm0.003(\rm LQCD)$\\
$D\to\pi\ell\nu_{\ell}$  & Section~\ref{sec:charm:semileptonic}		& $0.214\pm0.003(\rm exp.)\pm0.009(\rm LQCD)$\\
\midrule
\rowcolor{Gray} $D\to\ell\nu_{\ell}$ 	& \multirow{2}{*}{Average}	& \multirow{2}{*}{$0.219\pm0.006$}\\
\rowcolor{Gray} $D\to\pi\ell\nu_{\ell}$ & \multirow{-2}{*}{Average}	& \multirow{-2}{*}{$0.219\pm0.006$}\\
\midrule
$\nu N$			& PDG~\cite{PDG_2012}	& $0.230\pm0.011$\\
Indirect		& CKMFitter~\cite{Charles:2004jd}		& $0.22537^{+0.00068}_{-0.00035}$\\
\midrule
\midrule
&&{$|V_{cs}|$}\\
\cline{3-3}
$D_s\to\ell\nu_{\ell}$ 	 & This section			& $1.008\pm0.018(\rm exp.)\pm0.011(\rm LQCD)$\\
$D\to K\ell\nu_{\ell}$   & Section~\ref{sec:charm:semileptonic}		& $0.975\pm0.007(\rm exp.)\pm0.025(\rm LQCD)$\\
\midrule
\rowcolor{Gray} $D_s\to\ell\nu_{\ell}$ 	& \multirow{2}{*}{Average}	& \multirow{2}{*}{$0.998\pm0.020$}\\
\rowcolor{Gray} $D\to K\ell\nu_{\ell}$ & \multirow{-2}{*}{Average}	& \multirow{-2}{*}{$0.998\pm0.020$}\\
\midrule
$W\to c\overline{s}$	& PDG~\cite{PDG_2012}	& $0.94^{+0.32}_{-0.26}\pm0.13$\\
Indirect		& CKMFitter~\cite{Charles:2004jd}		& $0.973395^{+0.000095}_{-0.000176}$\\
\bottomrule
\end{tabular}
\end{center}
\end{table}

\subsubsection{Extraction of $D_{(s)}$ meson decay constants}

Assuming unitarity of the CKM matrix, the values of the elements relevant in the case of \mbox{(semi-)leptonic} charm decays are known from the global fit
of the CKM matrix, $|V_{cd}|=0.22537^{+0.00068}_{-0.00035}$, and
$|V_{cs}|=0.973395^{+0.000095}_{-0.000176}$~\cite{Charles:2004jd}. 
These values can be used to extract the $D$ and $D_s$ meson decay constants from the experimentally measured products $f_{D}|V_{cd}|$ (Eq.~\ref{eq:expFDVCD}) and $f_{D_s}|V_{cs}|$ (Eq.~\ref{eq:expFDSVCS}),
respectively. This leads to the experimentally measured $D_{(s)}$ meson decay constants to be:
\begin{eqnarray}
f_D^{\rm exp} & = & (203.7\pm4.9)~{\rm MeV,}\\ 
f_{D_s}^{\rm exp} & = & (257.4\pm4.6)~{\rm MeV,}
\end{eqnarray}
and the ratio of the constants is determined to be
\begin{equation}
f_{D_s}^{\rm exp}/f_D^{\rm exp} = 1.264\pm0.038.
\label{eq:fDsfDRatio:WA}
\end{equation}
The values are in agreement with the LQCD determinations given in Table~\ref{tab:Lattice} within the uncertainties. The largest discrepancy is in the determinations of 
the ratio of the decay constants where the agreement is only at the level of $1.9\sigma$.

\clearpage
\subsection{Hadronic decays of $D_s$ mesons}

\begin{figure}[hbt!]
\centering
\includegraphics[width=0.50\textwidth]{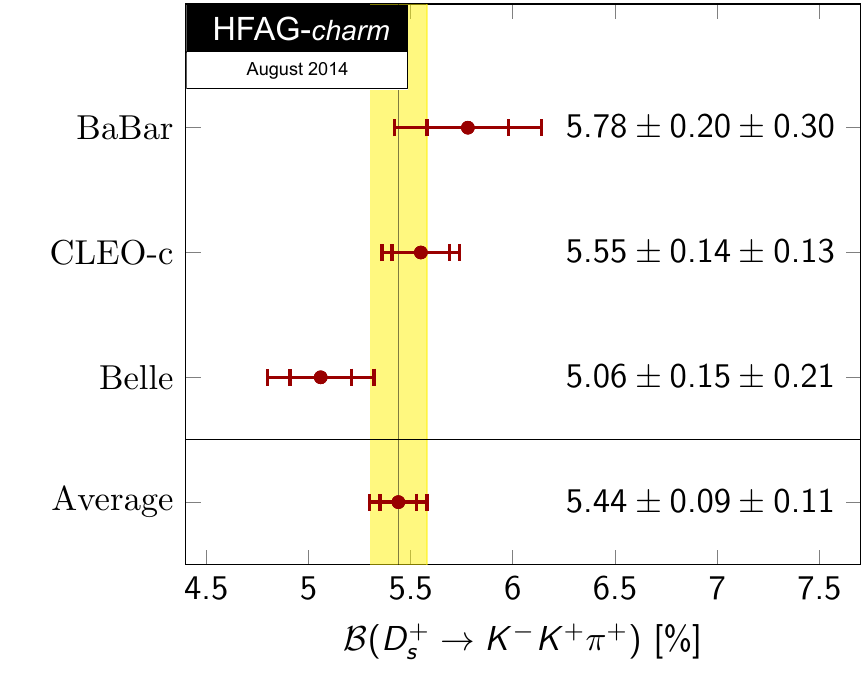}
\includegraphics[width=0.50\textwidth]{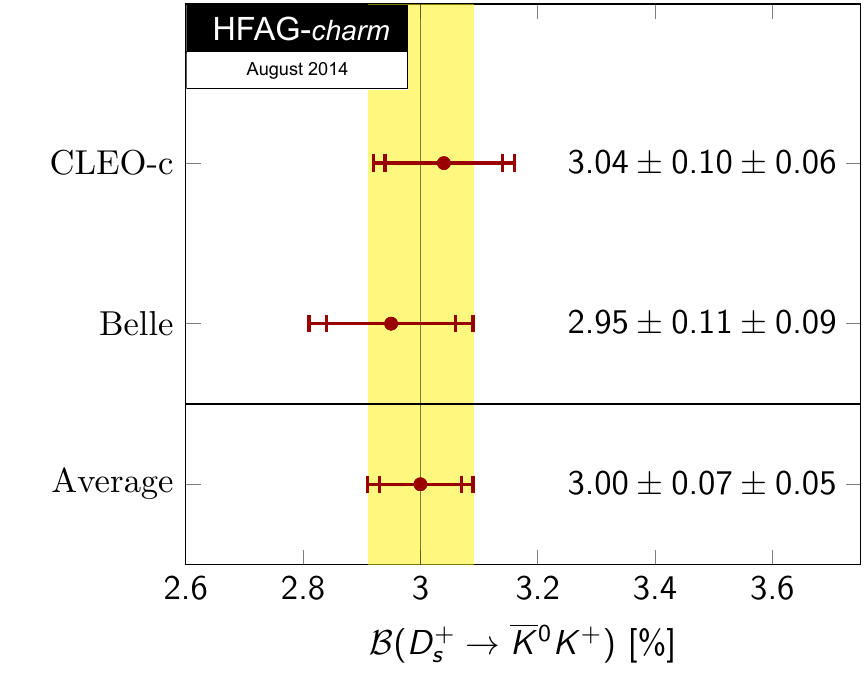}
\includegraphics[width=0.50\textwidth]{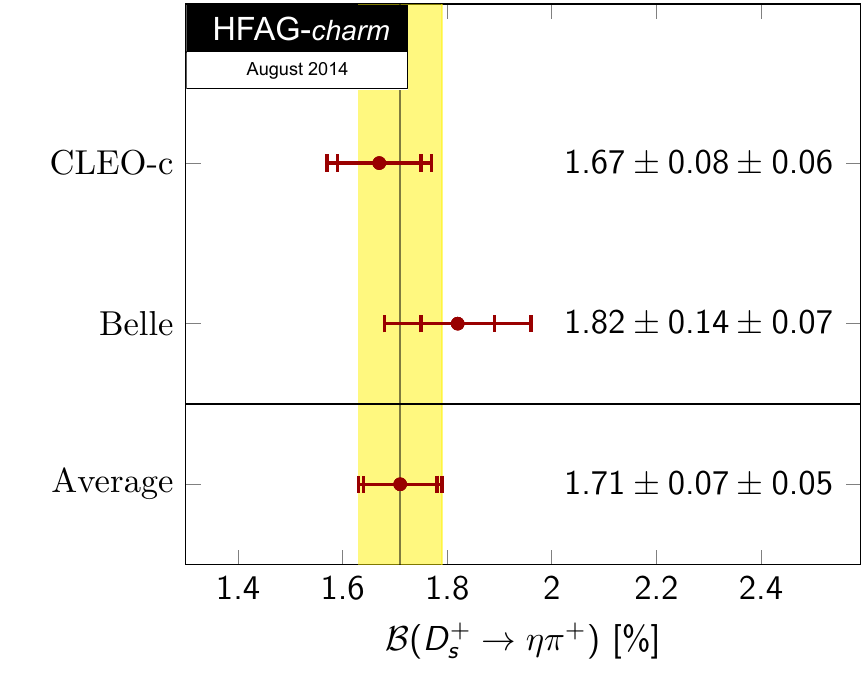}
\caption{
WA values for $\br(\dsp\to K^-K^+\pi^+)$ (top),
$\br(\dsp\to \overline{K}{}^0\pi^+)$ (middle), $\br(\dsp\to \eta\pi^+)$ (bottom).
For each point, the first error listed is the statistical and the second error
is the systematic error.
\label{fig:DSExpHadronic}
}
\end{figure}

\babar, CLEO-c and Belle collaborations have measured the absolute branching fractions of hadronic decays, $\dsp\to K^-K^+\pi^+$, $\dsp\to \overline{K}{}^0\pi^+$, and $\dsp \to \eta\pi^+$. The first two 
decay modes are the reference modes for the measurements of branching fractions of the $\dsp$ decays to any other final state. Table \ref{tab:DSExpHadronic} and 
Fig.~\ref{fig:DSExpHadronic} summarise the individual measurements and averaged values, which are found to be 
\begin{eqnarray}
\br^{\rm WA}(\dsp\to K^-K^+\pi^+) & = & (5.44\pm0.14)\%,\\
\br^{\rm WA}(\dsp\to \overline{K}{}^0\pi^+) & = & (3.00\pm0.09)\%,\\
\br^{\rm WA}(\dsp\to \eta\pi^+) & = & (1.71\pm0.08)\%,
\end{eqnarray}
where the uncertainties are total uncertainties. 

\begin{table}[t!]
\caption{Experimental results and world averages for branching fractions of $\dsp\to K^-K^+\pi^+$, $\dsp\to \overline{K}{}^0K^+$, and
$\dsp\to \eta\pi^+$ decays. The first uncertainty is statistical and the
second is experimental systematic. CLEO-c reports in Ref.~\cite{Onyisi:2013bjt}
$\br(\dsp\to K^0_SK^+)$. We include it in the average of $\br(\dsp\to \overline{K}{}^0K^+)$ by using the relation $\br(\dsp\to \overline{K}{}^0K^+)\equiv 2\br(\dsp\to K^0_SK^+)$.
\label{tab:DSExpHadronic}}
\begin{center}
\begin{tabular}{lcll}
\toprule
\rowcolor{Gray} Mode 	& ${\cal{B}}$ ($10^{-2}$)				& Reference 	& \\ 
\midrule
\multirow{3}{*}{$K^-K^+\pi^+$}  & $5.78\pm0.20\pm 0.30$ 		& \babar		& \cite{delAmoSanchez:2010jg}\\ 
						& $5.55\pm0.14\pm 0.13$ 		& CLEO-c		& \cite{Onyisi:2013bjt}\\ 
						& $5.06\pm0.15\pm 0.21$ 		& Belle   		& \cite{Zupanc:2013byn}\\
\midrule
\rowcolor{Gray}$K^-K^+\pi^+$	& $5.44\pm0.09\pm 0.11$			& Average & \\
\midrule
\multirow{2}{*}{$\overline{K}{}^0K^+$}		& $3.04\pm0.10\pm 0.06$ 		& CLEO-c		& \cite{Onyisi:2013bjt}\\ 
								& $2.95\pm0.11\pm 0.09$ 		& Belle   		& \cite{Zupanc:2013byn}\\
\midrule
\rowcolor{Gray}$\overline{K}{}^0K^+$		& $3.00\pm0.07\pm 0.05$			& Average & \\
\midrule
\multirow{2}{*}{$\eta\pi^+$}  	& $1.67\pm0.08\pm 0.06$ 		& CLEO-c		& \cite{Onyisi:2013bjt}\\ 
						& $1.82\pm0.14\pm 0.07$ 		& Belle   		& \cite{Zupanc:2013byn}\\
\midrule
\rowcolor{Gray}$\eta\pi^+$	& $1.71\pm0.07\pm 0.08$			& Average & 
\\ \bottomrule
\end{tabular}
\end{center}
\end{table}

\clearpage
\subsection{Two-body hadronic $D^0$ decays and final state radiation}

\begin{figure}[bh]
\begin{center}
\includegraphics[width=0.48\textwidth,angle=0.]{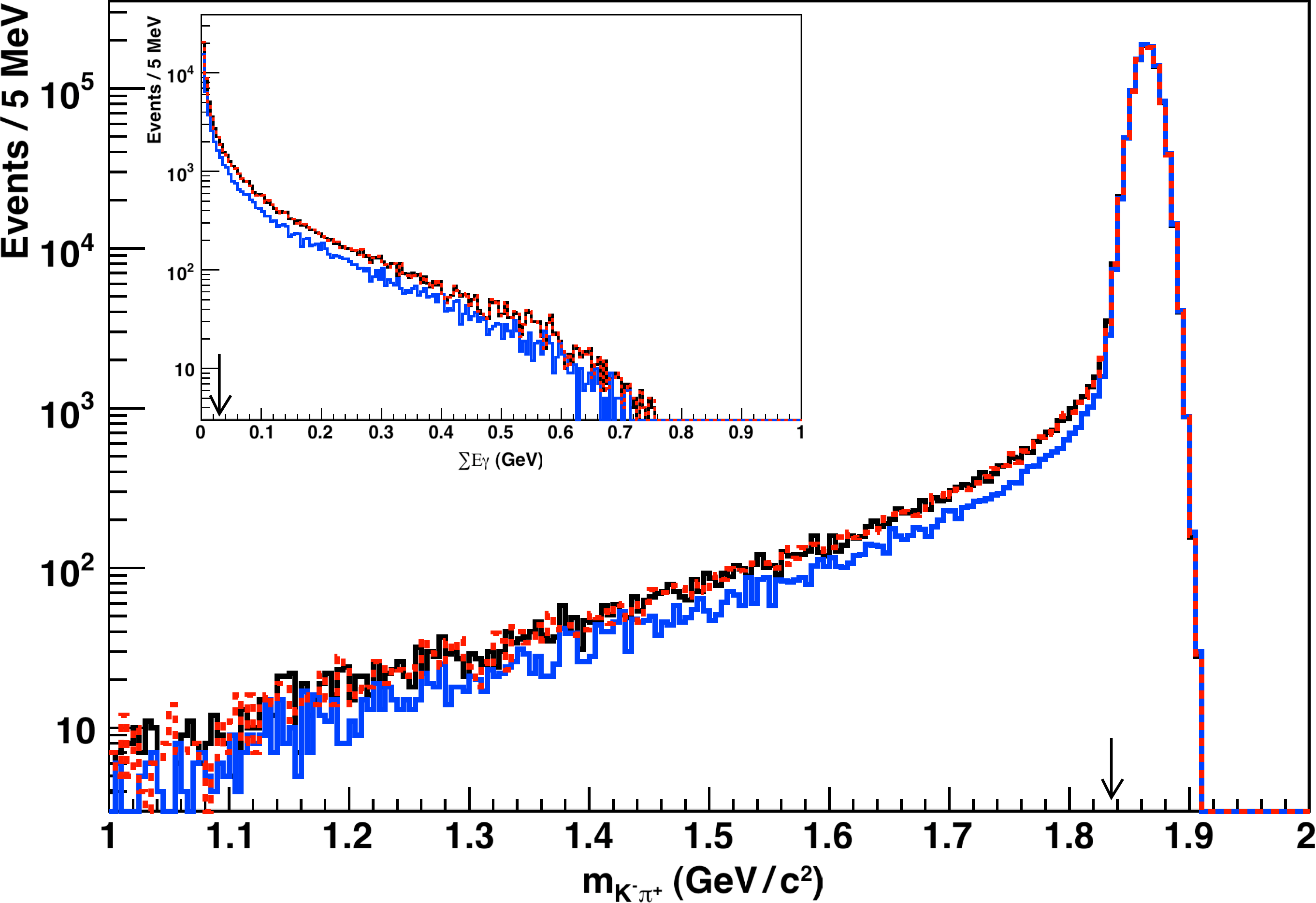}
\caption{The $K\pi$ invariant mass distribution for 
$D^0\to K^-\pi^+ (n\gamma)$ decays. The 3 curves correspond 
to three different configurations of PHOTOS for modeling FSR: 
version 2.02 without interference (blue/grey), version 2.02 with 
interference (red dashed) and version 2.15 with interference (black).  
The true invariant mass has been smeared with a typical experimental 
resolution of 10 MeV${}/c^2$.  Inset: The corresponding spectrum of 
total energy radiated per event.  The arrow indicates the $\sum E_\gamma$ 
value that begins to shift kinematic quantities outside of the range 
typically accepted in a measurement.}
\label{fig:FSR_mass_shift}
\end{center}
\end{figure}

Measurements of the branching fractions for the decays $D^0\to K^-\pi^+$,
$D^0\to \pi^+\pi^-$, and $D^0\to K^+ K^-$ have reached sufficient precision to
allow averages with ${\cal O}(1\%)$ relative uncertainties. 
At these precisions, Final 
State Radiation (FSR) must be treated correctly and consistently across 
the input measurements for the accuracy of the averages to match the 
precision.  The sensitivity of measurements to FSR arises because of 
a tail in the distribution of radiated energy that extends to the 
kinematic limit.  The tail beyond $\sum{E_\gamma} \approx 30$ MeV causes 
typical selection variables like the hadronic invariant mass to 
shift outside the selection range dictated by experimental 
resolution, as shown in Fig.~\ref{fig:FSR_mass_shift}.  While the 
differential rate for the tail is small, the integrated rate 
amounts to several percent of the total $h^+ h^-(n\gamma)$ 
rate because of the tail's extent.  The tail therefore 
translates directly into a several percent loss in 
experimental efficiency.

All measurements that include an FSR correction 
have a correction based on the use of 
PHOTOS~\cite{Barberio:1990ms,Barberio:1993qi,Golonka:2005pn,Golonka:2006tw} 
within the experiment's Monte Carlo simulation.  
PHOTOS itself, however, has evolved, over the period spanning the set of
measurements.  In particular, the incorporation of interference between
radiation off 
the two separate mesons has proceeded in stages: it was first
available for particle--antiparticle pairs in version 2.00 (1993), extended 
to any two-body, all-charged, final states in version 2.02 (1999), and 
further extended to multi-body final states in version 2.15 (2005).
The effects of interference are clearly visible, as shown in
Figure~\ref{fig:FSR_mass_shift}, and cause a 
roughly 30\% increase in the integrated rate into 
the high energy photon tail.  To evaluate the FSR 
correction incorporated into a given measurement, 
we must therefore note whether any correction was 
made, the version of PHOTOS used in correction, 
and whether the interference terms in PHOTOS were 
turned on.  

\subsubsection{Branching fraction corrections}

Before averaging the measured branching fractions, the published 
results are updated, as necessary, to the FSR prediction of 
PHOTOS~2.15 with interference included.  The correction will 
always shift a branching fraction to a higher value: with no 
FSR correction or with no interference term in the correction, 
the experimental efficiency determination will be biased high, 
and therefore the branching fraction will be biased low.

Most of the branching fraction analyses used the kinematic quantity 
sensitive to FSR in the candidate selection criteria.  For the 
analyses at the $\psi(3770)$, this variable was $\Delta E$, the 
difference between the candidate $D^0$ energy and the beam energy 
(\eg\ $E_K + E_\pi - E_{\rm beam}$ for $D^0\to K^-\pi^+$).  
In the remainder of the analyses, the relevant quantity was the 
reconstructed hadronic two-body mass $m_{h^+h^-}$.  To make the 
correction, 
we  only need to evaluate the fraction of decays that FSR moves 
outside of the range accepted for the analysis.  

The corrections were evaluated using an event generator (EvtGen 
\cite{Ryd:2005zz}) that incorporates PHOTOS to simulate the 
portions of the decay process most relevant to the correction.  
We compared corrections determined both with and without smearing 
to account for experimental resolution.  The differences were 
negligible, typically of ${\cal O}(1\%)$ of the correction itself.  
The immunity of the correction to resolution effects comes about because 
most of the long FSR-induced tail in, for example, the $m_{h^+h^-}$ 
distribution resides well away from the selection boundaries.  The 
smearing from resolution, on the other hand, mainly affects the 
distribution of events right at the boundary.

For measurements incorporating an FSR correction that did not 
include interference, we update by assessing the FSR-induced 
efficiency loss for both the PHOTOS version and configuration 
used in the analysis and our nominal version 2.15 with interference.  
For measurements that published their sensitivity to FSR, our 
generator-level predictions for the original efficiency loss 
agreed to within a few percent 
of the correction. 
This agreement 
lends additional credence to the procedure.

Once the event loss from FSR in the most sensitive kinematic 
quantity is accounted for, the event loss from other quantities 
is very small.  Analyses using $D^*$ tags, for example, showed 
little sensitivity to FSR in the reconstructed $D^*-D^0$ mass 
difference: for example, in $m_{K^-\pi^+\pi^+}-m_{K^-\pi^+}$. 
Because the effect of FSR tends to cancel in the difference of 
the reconstructed masses, this difference showed a much smaller 
sensitivity than the two-body mass even before a two-body mass 
requirement. In the $\psi(3770)$ analyses, the beam-constrained 
mass distributions 
($\sqrt{E_{\rm beam}^2 - |\vec{p}_K + \vec{p}_\pi|^2}$)  
also show much smaller sensitivity than the two-body mass.
\begin{figure}
\begin{center}
\includegraphics[width=1.00\textwidth]{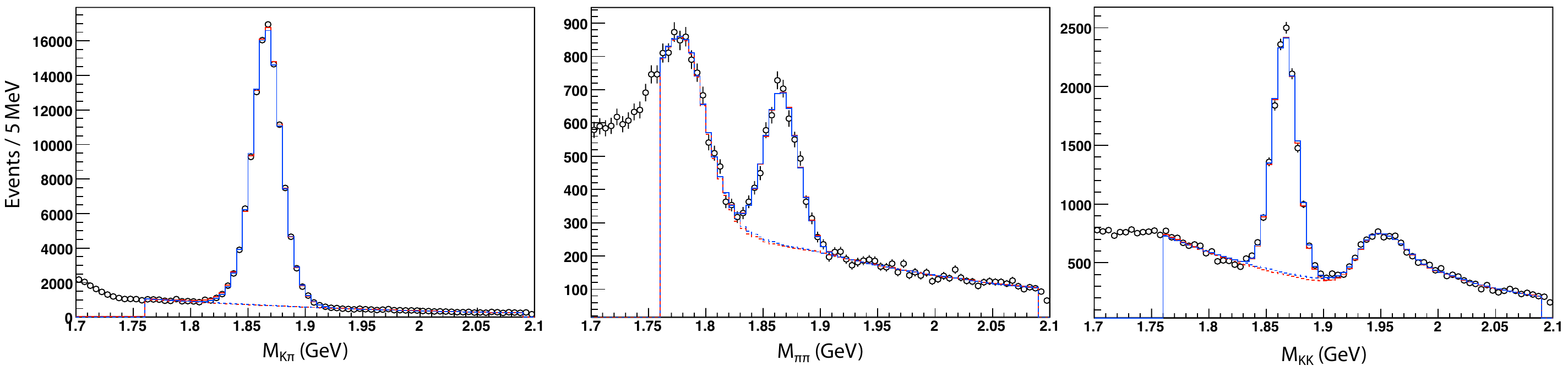}
\caption{FOCUS data (dots), original fits (blue) and 
toy MC parameterization (red) for $D^0\to K^-\pi^+$ (left), 
$D^0\to \pi^+\pi^-$ (center), and $D^0\to \pi^+\pi^-$ (right).}
\label{fig:FocusFits}
\end{center}
\end{figure}

The FOCUS~\cite{Link:2002hi} analysis of the branching ratios 
${\cal B}(D^0\to \pi^+\pi^-)/{\cal B}(D^0\to K^-\pi^+)$ and 
${\cal B}(D^0\to K^+ K^-)/{\cal B}(D^0\to K^-\pi^+)$ obtained 
yields using fits to the two-body mass distributions.  FSR will 
both distort the low end of the signal mass peak, and will 
contribute a signal component to the low side tail used to 
estimate the background.  The fitting procedure is not sensitive 
to signal events out in the FSR tail, which would be counted as 
part of the background.

A more complex toy Monte Carlo procedure was required to analyze 
the effect of FSR on the fitted yields, which were published with 
no FSR corrections applied.  A detailed description of the procedure 
and results is available on the HFAG web site, 
and a brief summary is provided
here.  Determining the correction involved an iterative procedure in which samples of similar size to the FOCUS sample were 
generated and then fit using the FOCUS signal and background 
parameterizations.  The MC parameterizations were tuned based 
on differences between the fits to the toy MC data and the FOCUS 
fits, and the procedure was repeated. These steps were iterated until 
the fit parameters matched the original FOCUS parameters.  

\begin{table}[!htb]
  \centering 
  \caption{The experimental measurements relating to ${\cal B}(D^0\to K^-\pi^+)$, ${\cal B}(D^0\to \pi^+\pi^-)$, and ${\cal B}(D^0\to K^+ K^-)$ after correcting to the common version and configuration of PHOTOS.  The uncertainties are statistical and total systematic, with the FSR-related systematic estimated in this procedure shown in parentheses.  Also listed are the percent shifts in the results from the correction, if any, applied here, as well as the original PHOTOS and interference configuration for each publication.}
  \label{tab:FSR_corrections}
\begin{tabular}{lccc}
\hline \hline
Experiment (acronym) & result (rescaled) & correction [\%] & PHOTOS \\ \hline
\multicolumn{4}{l}{$D^{0} \to K^{-} \pi^{+}$} \\
      CLEO-c 14  (CC14) \cite{Bonvicini:2013vxi} & $3.934 \pm 0.021 \pm 0.061(31)\%$ & --   & 2.15/Yes \\
      \babar 07  (BB07) \cite{Aubert:2007wn}   & $4.035 \pm 0.037 \pm 0.074(24)\%$ & 0.69 & 2.02/No \\
      CLEO II 98 (CL98) \cite{Artuso:1997mc}   & $3.920 \pm 0.154 \pm 0.168(32)\%$ & 2.80 & none \\
      ALEPH 97   (AL97) \cite{Barate:1997mm}   & $3.930 \pm 0.091 \pm 0.125(32)\%$ & 0.79 & 2.0/No \\
      ARGUS 94   (AR94) \cite{Albrecht:1994nb} & $3.490 \pm 0.123 \pm 0.288(24)\%$ & 2.33 & none \\
      CLEO II 93 (CL93) \cite{Akerib:1993pm}   & $3.960 \pm 0.080 \pm 0.171(15)\%$ & 0.38 & 2.0/No \\
      ALEPH 91   (AL91) \cite{Decamp:1991jw}   & $3.730 \pm 0.351 \pm 0.455(34)\%$ & 3.12 & none \\
\multicolumn{4}{l}{$D^{0} \to \pi^{+}\pi^{-} / D^{0} \to K^{-} \pi^{+}$} \\
      CLEO-c 10  (CC10) \cite{Mendez:2009aa}   & $0.0370  \pm 0.0006  \pm 0.0009(02)$  & --   & 2.15/Yes \\
      CDF 05     (CD05) \cite{Acosta:2004ts}   & $0.03594 \pm 0.00054 \pm 0.00043(15)$ & --   & 2.15/Yes \\
      FOCUS 02   (FO02) \cite{Link:2002hi}     & $0.0364  \pm 0.0012  \pm 0.0006(02)$  & 3.10 & none \\
\multicolumn{4}{l}{$D^{0} \to K^{+}K^{-} / D^{0} \to K^{-} \pi^{+}$} \\
      CLEO-c 10   \cite{Mendez:2009aa}         & $0.1041 \pm 0.0011 \pm 0.0012(03)$ & --    & 2.15/Yes \\ 
      CDF 05      \cite{Acosta:2004ts}         & $0.0992 \pm 0.0011 \pm 0.0012(01)$ & --    & 2.15/Yes \\
      FOCUS 02    \cite{Link:2002hi}           & $0.0982 \pm 0.0014 \pm 0.0014(01)$ & -1.12 & none \\ \hline
\end{tabular}
\end{table}

The toy MC samples for the first iteration were based on the generator-level 
distribution of $m_{K^-\pi^+}$, $m_{\pi^+\pi^-}$, and $m_{K^+K^-}$, including 
the effects of FSR, smeared according to the original FOCUS resolution 
function,  and on backgrounds 
generated 
using the parameterization from the final
FOCUS fits.  For each iteration, 400 to 1600 individual 
data-sized samples were 
generated 
and fit. The means of the parameters from these fits determined the 
corrections to the generator parameters for the following iteration.  The 
ratio between the number of signal events generated and the final signal 
yield provides the required FSR correction in the final iteration.  Only a 
few iterations were required in each mode.  Figure ~\ref{fig:FocusFits} 
shows the FOCUS data, the published FOCUS fits, and the final toy MC 
parameterizations.  The toy MC provides an excellent description of the 
data.

The corrections obtained to the individual FOCUS yields were 
$1.0298\pm 0.0001$ for $K^-\pi^+$, $1.062 \pm 0.001$ for $\pi^+\pi^-$, 
and $1.0183 \pm 0.0003$ for $K^+K^-$.  These corrections tend to cancel 
in the branching ratios, leading to corrections of 1.031 to  
${\cal B}(D^0\to \pi^+\pi^-)/{\cal B}(D^0\to K^-\pi^+)$, and 0.9888 for 
${\cal B}(D^0\to K^+ K^-)/{\cal B}(D^0\to K^-\pi^+)$.

Table~\ref{tab:FSR_corrections} summarizes the corrected branching fractions. 
The published FSR-related modeling uncertainties have been replaced by with a
new, common, estimate based on the assumption that the dominant uncertainty 
in the FSR corrections comes from the fact that the mesons are treated like 
structureless particles. No contributions from structure-dependent terms in 
the decay process (\eg\ radiation off individual quarks) are included in PHOTOS. 
Internal studies done by various experiments have indicated that 
in $K\pi$ decays, 
the PHOTOS corrections agree with data at the 20-30\% level. 
We therefore attribute a 25\% uncertainty to the FSR prediction from potential 
structure-dependent contributions. For the other two modes, the only difference 
in structure is the final state valence quark content. While radiative corrections 
typically come in with a $1/M$ dependence, one would expect the additional 
contribution from the structure terms to come in on time scales shorter than 
the hadronization time scale. In this case, you might expect
$\rm \Lambda_{\rm QCD}$ to be the relevant scale, rather than the quark masses,
and therefore that the amplitude is the same for the three modes. In treating
the correlations among the measurements this is what we assume. We also assume
that the PHOTOS amplitudes and any missing structure amplitudes are relatively 
real with constructive interference.  The uncertainties largely cancel 
in the branching fraction ratios. For the final average branching 
fractions, the FSR uncertainty on $K\pi$ dominates. Note that because 
of the relative sizes of FSR in the different modes, the $\pi\pi/K\pi$ 
branching ratio uncertainty from FSR is 
positively correlated with that 
for the $K\pi$ branching fraction, while the $KK/K\pi$ branching ratio FSR
uncertainty is negatively correlated.

The ${\cal B}(D^0\to K^-\pi^+)$ measurement of reference~\cite{Coan:1997ye}, the  
${\cal B}(D^0\to \pi^+\pi^-)/{\cal B}(D^0\to K^-\pi^+)$ measurements of 
references~\cite{Aitala:1997ff} 
and~\cite{Csorna:2001ww}, and the 
${\cal B}(D^0\to K^+ K^-)/{\cal B}(D^0\to K^-\pi^+)$ measurement
of reference~\cite{Csorna:2001ww} are excluded from the branching 
fraction averages presented here.
These measurements appear not to have incorporated any FSR corrections, 
and insufficient information
is available to determine the 2-3\% corrections that would be required.

\begin{sidewaystable}[p]
  \centering 
  \caption{The correlation matrix corresponding to the full covariance matrix. 
  Subscripts $h$ denote which of the $D^0 \to h^+ h^-$ decay results from a single experiment
  is represented in that row or column.}\label{tab:correlations}
  \small
\begin{tabular}{lr@{.}lr@{.}lr@{.}lr@{.}lr@{.}lr@{.}lr@{.}lr@{.}lr@{.}lr@{.}lr@{.}lr@{.}lr@{.}l}
\hline\hline
           & \multicolumn{2}{c}{CC14}
                   & \multicolumn{2}{c}{BB07}
                           & \multicolumn{2}{c}{CL98}
                                   & \multicolumn{2}{c}{AL97}
                                           & \multicolumn{2}{c}{AR94} 
                                                   & \multicolumn{2}{c}{CL93} 
                                                           & \multicolumn{2}{c}{AL91} 
                                                                   & \multicolumn{2}{c}{FO02$_\pi$} 
                                                                           & \multicolumn{2}{c}{CD05$_\pi$} 
                                                                                   & \multicolumn{2}{c}{CC10$_\pi$} 
                                                                                           & \multicolumn{2}{c}{FO02$_K$}
                                                                                                    & \multicolumn{2}{c}{CD05$_K$} 
                                                                                                            & \multicolumn{2}{c}{CC10$_K$} \\ \hline
CC14 & 1&000 & 0&139 & 0&057 & 0&084 & 0&031 & 0&033 & 0&023 & 0&070 & 0&103 & 0&068 &-0&019 &-0&032 &-0&085 \\
BB07 & 0&139 & 1&000 & 0&035 & 0&051 & 0&019 & 0&020 & 0&014 & 0&042 & 0&062 & 0&041 &-0&012 &-0&019 &-0&051 \\
CL98 & 0&057 & 0&035 & 1&000 & 0&021 & 0&008 & 0&298 & 0&006 & 0&017 & 0&026 & 0&017 &-0&005 &-0&008 &-0&021 \\
AL97 & 0&084 & 0&051 & 0&021 & 1&000 & 0&011 & 0&012 & 0&116 & 0&025 & 0&038 & 0&025 &-0&007 &-0&012 &-0&031 \\
AR94 & 0&031 & 0&019 & 0&008 & 0&011 & 1&000 & 0&004 & 0&003 & 0&009 & 0&014 & 0&009 &-0&003 &-0&004 &-0&011 \\
CL93 & 0&033 & 0&020 & 0&298 & 0&012 & 0&004 & 1&000 & 0&003 & 0&010 & 0&015 & 0&010 &-0&003 &-0&005 &-0&012 \\
AL91 & 0&023 & 0&014 & 0&006 & 0&116 & 0&003 & 0&003 & 1&000 & 0&007 & 0&010 & 0&007 &-0&002 &-0&003 &-0&009 \\
FO02$_\pi$ & 0&070 & 0&042 & 0&017 & 0&025 & 0&009 & 0&010 & 0&007 & 1&000 & 0&031 & 0&021 &-0&006 &-0&010 &-0&026 \\
CD05$_\pi$ & 0&103 & 0&062 & 0&026 & 0&038 & 0&014 & 0&015 & 0&010 & 0&031 & 1&000 & 0&031 &-0&009 &-0&014 &-0&038 \\
CC10$_\pi$ & 0&068 & 0&041 & 0&017 & 0&025 & 0&009 & 0&010 & 0&007 & 0&021 & 0&031 & 1&000 &-0&006 &-0&010 &-0&025 \\
FO02$_K$ &-0&019 &-0&012 &-0&005 &-0&007 &-0&003 &-0&003 &-0&002 &-0&006 &-0&009 &-0&006 & 1&000 & 0&003 & 0&007 \\
CD05$_K$ &-0&032 &-0&019 &-0&008 &-0&012 &-0&004 &-0&005 &-0&003 &-0&010 &-0&014 &-0&010 & 0&003 & 1&000 & 0&012 \\
CC10$_K$ &-0&085 &-0&051 &-0&021 &-0&031 &-0&011 &-0&012 &-0&009 &-0&026 &-0&038 &-0&025 & 0&007 & 0&012 & 1&000 \\
\hline
\end{tabular}
\end{sidewaystable}

\subsubsection{Average branching fractions}
\begin{figure}
\begin{center}
\includegraphics[width=0.6\textwidth,angle=0.]{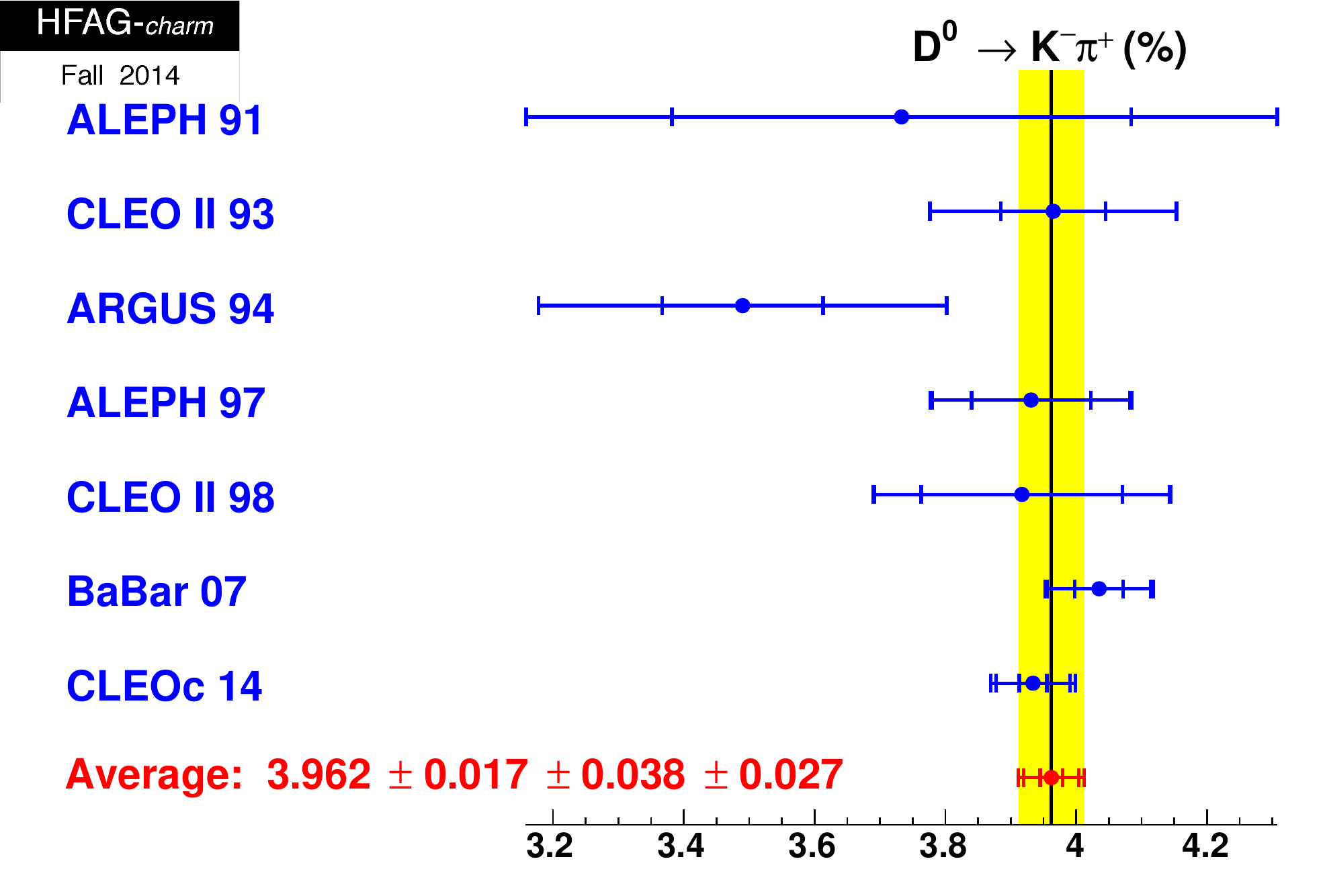}
\caption{Comparison of measurements of 
${\cal B}(D^0\to K^-\pi^+)$ (blue) with the average 
branching fraction obtained here (red, and yellow band).}
\label{D0bfs}
\end{center}
\end{figure}


The average branching fractions for 
$D^0\to K^-\pi^+$, $D^0\to \pi^+\pi^-$ and $D^0\to K^+ K^-$ 
are obtained from a single $\chi^2$ minimization procedure, 
in which the three branching fractions are floating parameters. 
The central values derive from a fit in which the full covariance matrix, 
accounting for all statistical, systematic (excluding FSR), and FSR measurement uncertainties, is used.  
Table~\ref{tab:correlations} presents the correlation matrix for 
this nominal fit. 
We then obtain the three reported uncertainties on those central values as follows:
The statistical uncertainties are obtained from a fit using only the statistical covariance matrix.  
The systematic uncertainties are obtained by subtracting (in quadrature) the statistical uncertainties 
from the uncertainties determined via a fit using a covariance matrix that accounts for both statistical and systematic measurement uncertainties.  
The FSR uncertainties are obtained by subtracting (in quadrature)
the uncertainties determined via a fit using a covariance matrix that accounts for both statistical and systematic measurement uncertainties
from the uncertainties determined via the fit using the full covariance matrix.

In forming the full covariance matrix, the FSR
uncertainties are treated as fully correlated (or anti-correlated) as 
described above.  
For the covariance matrices involving systematic measurement uncertainties, ALEPH's systematic 
uncertainties in the $\theta_{D^*}$ parameter are treated
as fully correlated between the ALEPH 97 and ALEPH 91 measurements.  Similarly,
the tracking efficiency uncertainties in the CLEO II 98 and the
CLEO II 93 measurements are treated as fully correlated.  

The averaging procedure results in a 
final $\chi^2$ of $11.0$ for $10$ ($13-3$) degrees 
of freedom.  The branching
fractions obtained are
\begin{eqnarray*}
  {\cal B}(D^0\to K^-\pi^+)   & = & ( 3.962 \pm 0.017 \pm 0.038 \pm 0.027 )\,\% \\
  {\cal B}(D^0\to \pi^+\pi^-) & = & ( 0.144 \pm 0.002 \pm 0.002 \pm 0.002 )\,\% \\
  {\cal B}(D^0\to K^+ K^-)    & = & ( 0.399 \pm 0.003 \pm 0.005 \pm 0.002 )\,\%\,. 
\end{eqnarray*}The uncertainties, estimated as described above, are statistical, 
systematic (excluding FSR), and
FSR modeling.  The correlation coefficients from the fit using the 
total uncertainties are
\begin{center}
\begin{tabular}{llll}
               & $K^-\pi^+$ & $\pi^+\pi^-$ & $K^+ K^-$ \\
$K^-\pi^+$     &  1.00 & 0.71 & 0.76  \\
$\pi^+\pi^-$   &  0.71 & 1.00 & 0.53  \\
$K^+ K^-$      &  0.76 & 0.53 & 1.00  \\
\end{tabular}
\end{center}

\begin{table}[!htb]
  \centering 
  \caption{Evolution of the $D^0\to K^-\pi^+$ branching fraction from a fit with
  no FSR corrections or correlations (similar to the average in the 
  PDG 2014 update~\cite{PDG_2014}) to the nominal fit presented
here.}\label{tab:fit_evolution}
\begin{tabular}{cccll}
\hline\hline
Modes &  description                       & ${\cal B}(D^0\to K^-\pi^+)$ (\%)           & $\chi^2$ / (d.o.f.) \\
fit        &                               &                                       & \\ \hline
$K^-\pi^+$ & PDG summer 2014 equivalent    & $3.913 \pm 0.022 \pm 0.043 $ & 6.0 / (8-1)\\
$K^-\pi^+$ & drop Ref.~\cite{Coan:1997ye}  & $3.921 \pm 0.023 \pm 0.044$           & 4.8 / (7-1)\\
$K^-\pi^+$ & use Ref.~\cite{Bonvicini:2013vxi} instead of Ref.\cite{Dobbs:2007zt}  & $3.938 \pm 0.017 \pm 0.042$ & 4.5 / (7-1)\\
$K^-\pi^+$ & add FSR corrections           & $3.955 \pm 0.017 \pm 0.038 \pm 0.018$ & 3.5 / (7-1)\\
$K^-\pi^+$ & add FSR correlations          & $3.956 \pm 0.017 \pm 0.038 \pm 0.027$ & 3.6 / (7-1)\\
all        & --   & $3.962 \pm 0.017 \pm 0.038 \pm 0.027$ &11.0 /(13-3) \\
\hline
\end{tabular}
\end{table}

\begin{figure}
\begin{center}
\includegraphics[width=0.47\textwidth,angle=0.]{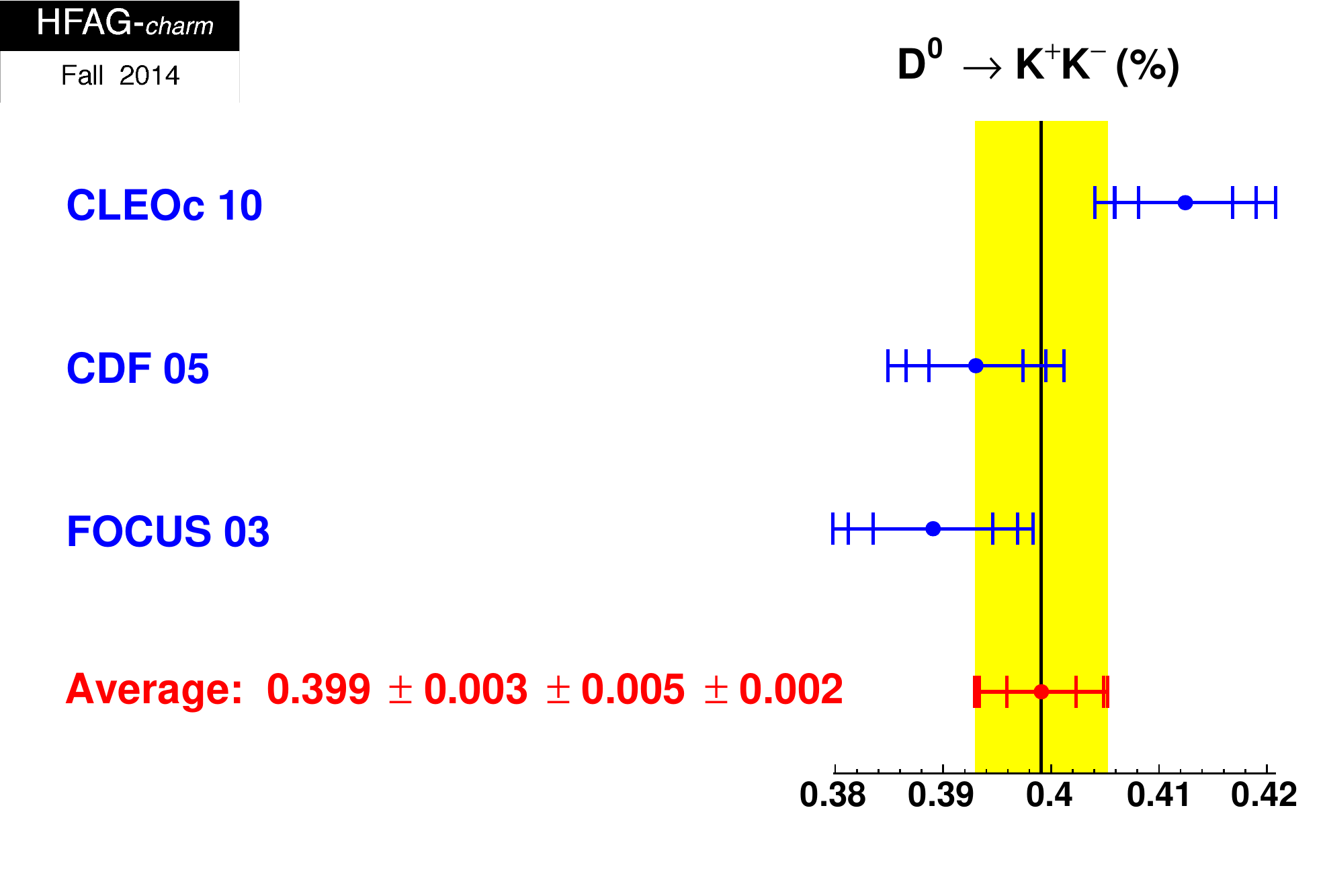}\hfill
\includegraphics[width=0.47\textwidth,angle=0.]{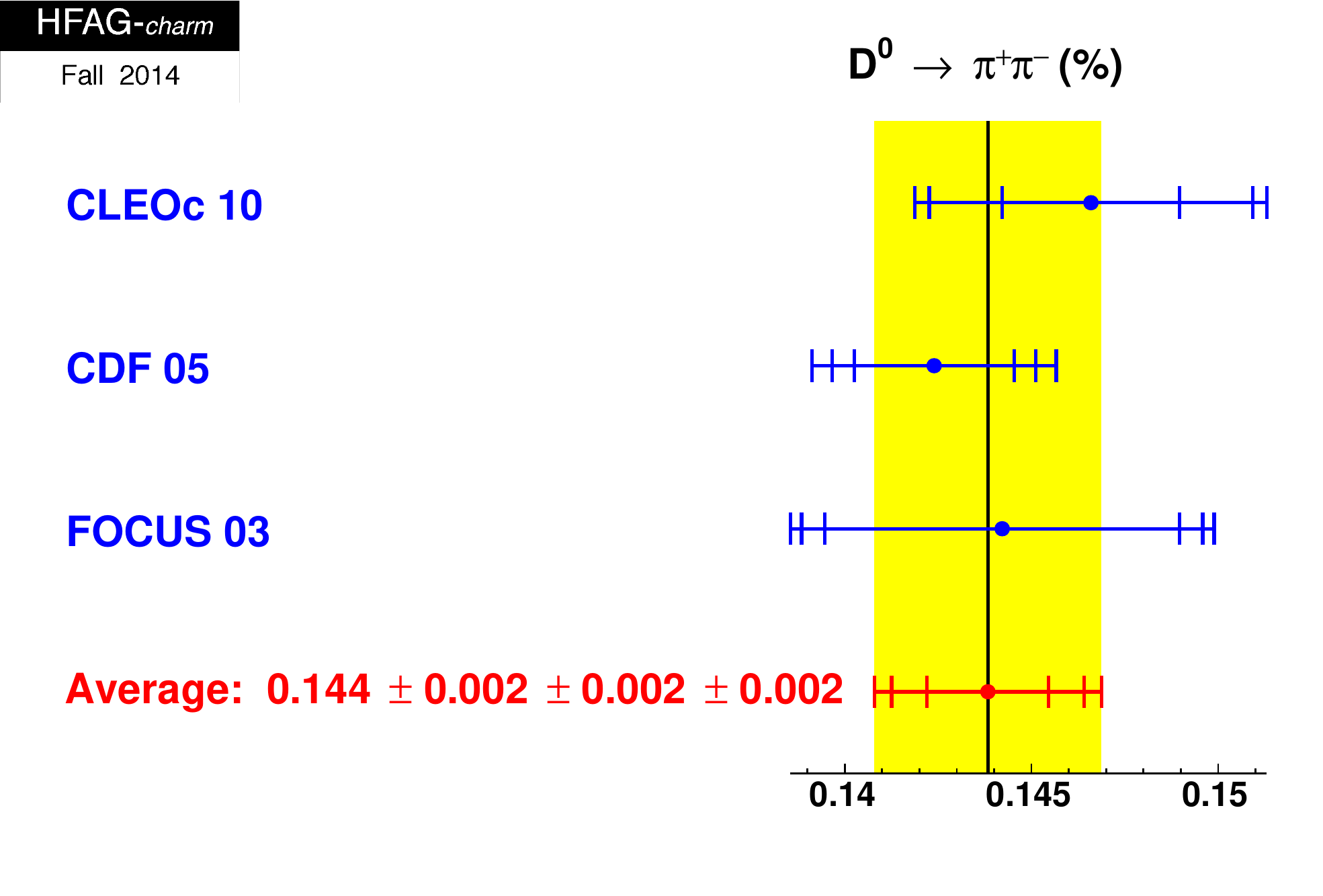}
\caption{The ${\cal B}(D^0\to K^+K^-)$ (left) and ${\cal B}(D^0\to \pi^+\pi^-)$ (right) 
values obtained by scaling the measured branching ratios with the ${\cal B}(D^0\to K^-\pi^+)$ branching fraction
average obtained here.  For the measurements (blue points), the error bars correspond to the statistical, systematic
and $K\pi$ normalization uncertainties.  The average obtained here (red point, yellow band) lists the statistical,
systematics excluding FSR, and the FSR systematic.
\label{fig:kkpipi}}
\end{center}
\end{figure}

As the $\chi^2$ would suggest and Fig.~\ref{D0bfs} shows, the average 
value for ${\cal B}(D^0\to K^-\pi^+)$ and
the input branching fractions agree very well.  With the estimated 
uncertainty in the FSR modeling used here,
the FSR uncertainty dominates the statistical uncertainty 
in the average, suggesting that experimental
work in the near future should focus on verification of FSR with 
$\sum E_\gamma \simge 100$ MeV.  Note that the systematic uncertainty 
excluding FSR
is still larger than the FSR uncertainty; in the most 
precise measurements of these branching fractions, the 
largest systematic
uncertainty is the uncertainty on the tracking efficiency. The ${\cal B}(D^0\to
K^+K^-)$ and ${\cal B}(D^0\to \pi^+\pi^-)$ measurements inferred
from the branching ratio measurements also agree well 
(Fig.~\ref{fig:kkpipi}). 

The ${\cal B}(D^0\to K^-\pi^+)$ average obtained here is 
approximately two statistical standard deviations higher than the
2014 PDG update average~\cite{PDG_2014}. 
Table~\ref{tab:fit_evolution} shows the evolution from a
fit similar to the PDG's (no FSR corrections or correlations, 
reference~\cite{Coan:1997ye}
included, uses reference~\cite{Dobbs:2007zt} instead of reference~\cite{Bonvicini:2013vxi} [the latter being a recent, superseding result]) to the average presented here.
There are three main contributions to the difference. The branching 
fraction in reference~\cite{Coan:1997ye} is
low, and its exclusion shifts the result upwards. The dominant shifts
($+0.017\%$ each) are due to the FSR corrections, which as
expected shift the result upwards, and the more precise result from
reference~\cite{Bonvicini:2013vxi}.

\clearpage
\subsection{Excited \emph{$D_{(s)}$} mesons}
\begin{figure}[htb!]
\begin{centering}
\includegraphics[width=0.45\textwidth]{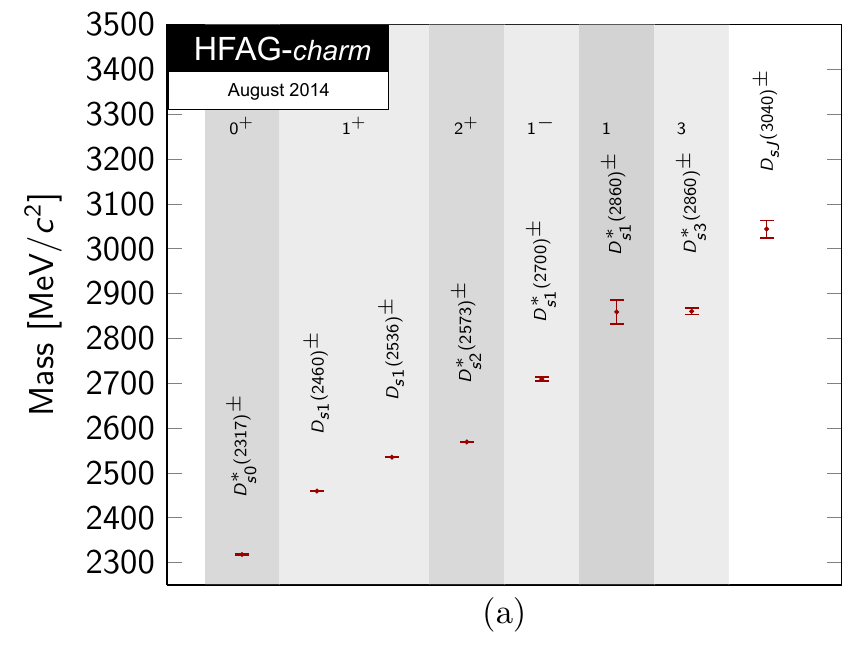}
\quad
\includegraphics[width=0.45\textwidth]{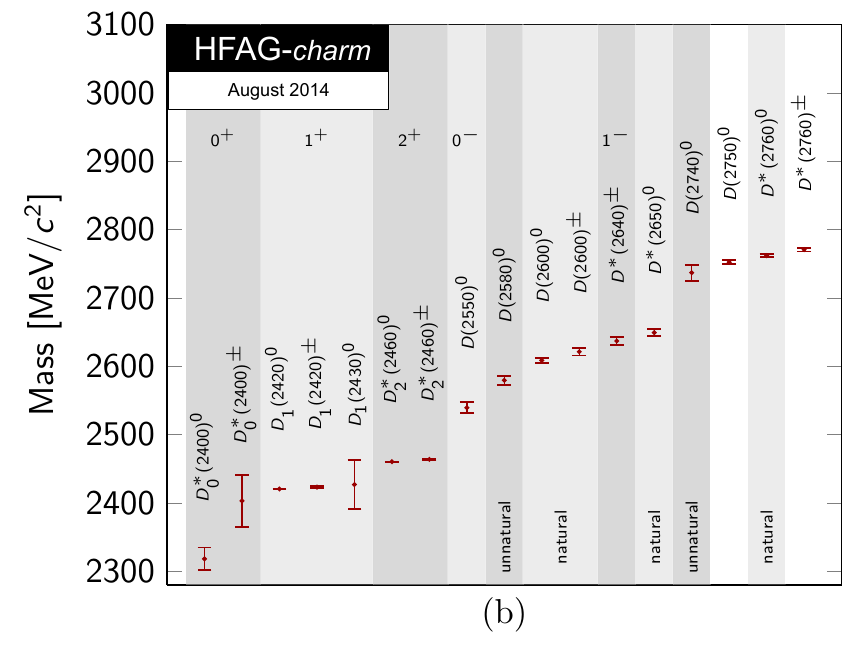}\\
\includegraphics[width=0.45\textwidth]{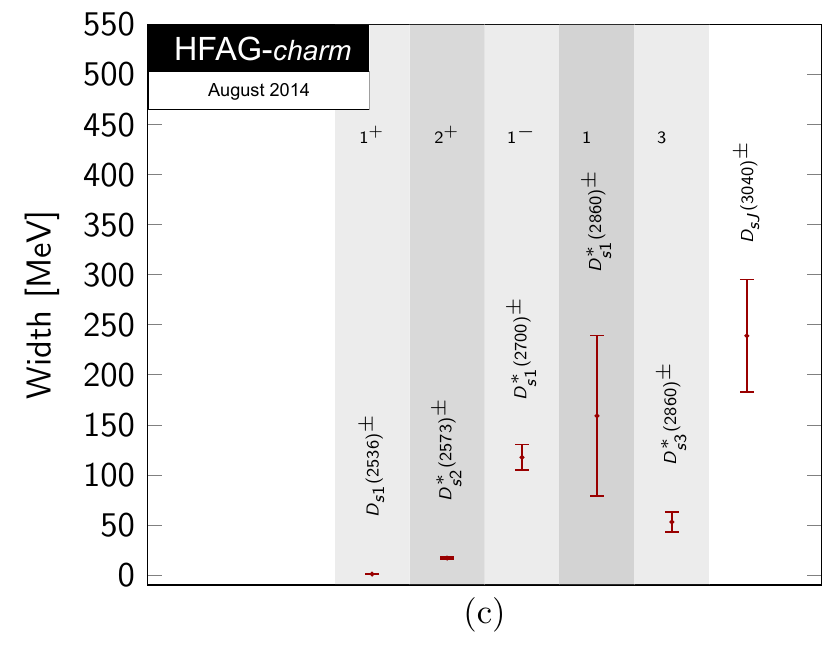}
\quad
\includegraphics[width=0.45\textwidth]{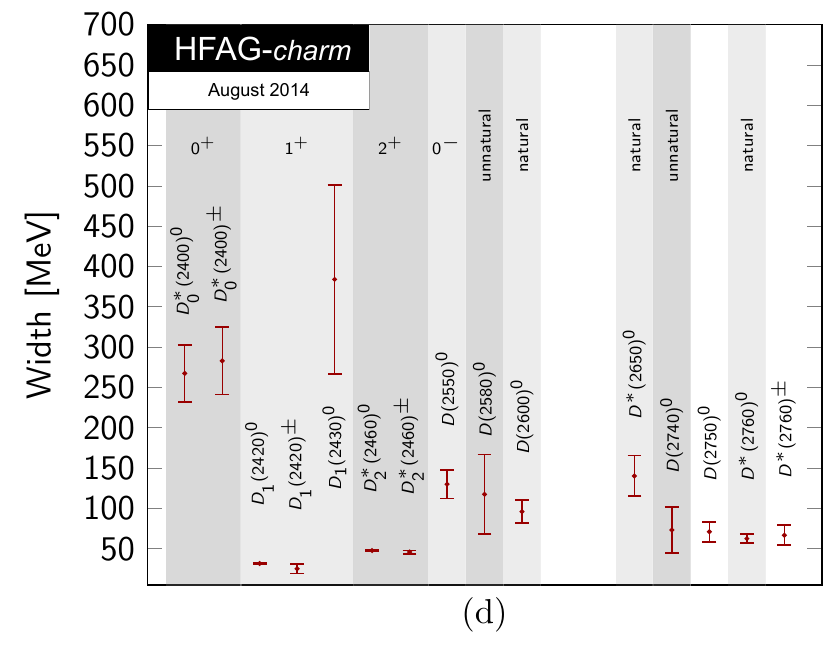}

\caption{\label{fig:charm:spect:1}  Averaged masses for $D_{s}$ mesons are shown in subfigure (a) and for $D$ mesons in subfigure (b). The average widths for $D_{s}$ mesons are shown in subfigure (c) and for $D$ mesons in subfigure (d). The vertical shaded regions distinguish between different spin parity states.}
\end{centering}
\end{figure}

Excited charm meson states have received increased attention since the first observation of states that could not be accommodated by QCD predictions~\cite{Aubert:2003fg,
Besson:2003cp,Abe:2003jk,Aubert:2003pe}. Tables \ref{table:charm:spect:1} and \ref{tabel:charm:spect:2} summarize recent measurements of the masses and widths of excited $D$ and $D_{s}$ mesons, respectively. If a preferred assignment of spin and parity was measured it is listed in the column $J^{P}$, where the label natural denotes $J^{P}=0^{-},1^{+},2^{-}\ldots$ and unnatural $J^{P}=0^{+},1^{-},2^{+}\ldots$. If possible an average mass and width was calculated, which is listed in the gray shaded row. The calculation of the averages assumes no correlation between individual measurements. A summary of the averaged masses and widths is shown in Figure~\ref{fig:charm:spect:1}. 

In the study of $B_{s}^{0}\to \overline{D}{}^{0}K^{-}\pi^{+}$ decays the LHCb collaboration searched for excited $D_{s}$ mesons~\cite{Aaij:2014xza}.  Previous measurements by \babar{}~\cite{Aubert:2009ah} and LHCb~\cite{Aaij:2012pc} indicated the existence of a strange-charm $D^{*}_{sJ}(2860)^{-}$ meson. The new measurement of LHCb showed with $10\sigma$ significance that this state is comprised of two different particles,  one of spin 1 and one of spin 3. This represents the first measurement of a heavy flavored spin-3 particle, and the first observation $B$ mesons decays to spin 3 particles.

The masses and widths of narrow ($\Gamma<50$~MeV) orbitally excited $D$ mesons (denoted $D^{\ast\ast}$), both neutral and charged, are well established. Measurements of broad states ($\Gamma\sim$ 200--400~MeV) are less abundant, as identifying the signal is more challenging. There is a slight discrepancy between the 
$D_0^\ast(2400)^0$ masses measured by the Belle~\cite{Abe:2003zm} and  FOCUS~\cite{Link:2003bd} experiments. No data exist yet for the $D_1(2430)^{\pm}$ state. Dalitz plot analyses of $B\to D^{(\ast)}\pi\pi$ decays strongly favor the assignments $0^+$ and $1^+$ for the spin-parity quantum numbers of the $D_0^\ast(2400)^0/D_0^\ast(2400)^\pm$ and $D_1(2430)^{0}$ states, respectively. The measured masses and widths, as well as the $J^P$ values, are in agreement with theoretical predictions based on potential models~\cite{Godfrey:1985xj, Godfrey:1986wj, Isgur:1991wq, Schweitzer:2002nm}.

Tables~\ref{table:charm:spect:3} and \ref{table:charm:spect:4} summarizes the branching fractions of $B$ mesons decays to excited $D$ and $D_{s}$ states, respectively. It can be noted that the branching fractions for $B$ mesons decaying to a narrow $D^{\ast\ast}$ state and a pion are similar for charged and neutral $B$ initial states, the branching fractions to a broad $D^{\ast\ast}$ state and $\pi^+$ are much larger for $B^+$ than for $B^0$. This may be due to the fact that color-suppressed amplitudes contribute only to the $B^+$ decay and not to the $B^0$ decay (for a theoretical discussion, see Ref.~\citep{Jugeau:2005yr,Colangelo:2004vu}). Measurements of individual branching fractions of $D$ mesons are difficult due to the unknown fragmentation of $c\bar c \to D^{\ast\ast}$ or due to the unknown $B \to D^{\ast\ast} X$ branching fractions.

The discoveries of the $D_{s0}^\ast(2317)^{\pm}$ and $D_{s1}(2460)^{\pm}$ have triggered increased interest in properties of, and searches for, excited $D_s$ mesons (here generically denoted $D_s^{\ast\ast}$). While the masses and widths of $D_{s1}(2536)^{\pm}$ and $D_{s2}(2573)^{\pm}$ states are in relatively good agreement with potential model predictions, the masses of $D_{s0}^\ast(2317)^{\pm}$ and $D_{s1}(2460)^{\pm}$ states are significantly lower than expected (see Ref.~\cite{Cahn:2003cw} for a discussion of $c\bar{s}$ models). Moreover, the mass splitting between these two states greatly exceeds that between the $D_{s1}(2536)^{\pm}$ and $D_{s2}(2573)^{\pm}$. These unexpected properties have led to interpretations of the $D_{s0}^\ast(2317)^{\pm}$ and $D_{s1}(2460)^{\pm}$ as exotic four-quark states~\cite{Barnes:2003dj,Lipkin:2003zk}.

While there are few measurements of the $J^P$ values of $D_{s0}^\ast(2317)^{\pm}$ and $D_{s1}(2460)^{\pm}$, the available data favor $0^+$ and $1^+$, respectively. A molecule-like ($DK$) interpretation of the $D_{s0}^\ast(2317)^{\pm}$ and $D_{s1}(2460)^{\pm}$~\cite{Barnes:2003dj,Lipkin:2003zk} that can account for their low masses and isospin-breaking decay modes is tested by searching for charged and neutral isospin partners of these states; thus far such searches have yielded negative results. Therefore the subset of models that predict equal production rates for different charged states is excluded. The molecular picture can also be tested by measuring the rates for the radiative processes $D_{s0}^\ast(2317)^{\pm}/D_{s1}(2460)^{\pm}\to D_s^{(\ast)}\gamma$ and comparing to theoretical predictions. The predicted rates, however, are below the sensitivity of current experiments. 

Another model successful in explaining the total widths and the $D_{s0}^\ast(2317)^{\pm}$ -- $D_{s1}(2460)^{\pm}$ mass splitting is based on the assumption that these states are chiral partners of the ground states $D_{s}^{+}$ and~$D_{s}^{*}$~\cite{Bardeen:2003kt}. While some measured branching fraction ratios agree with predicted
values, further experimental tests with better sensitivity are needed to confirm or refute this scenario. A summary of the mass difference measurements is given in Table~\ref{table:charm:spect:5}.

In addition to the $D_{s0}^\ast(2317)^{\pm}$ and $D_{s1}(2460)^{\pm}$ states, other excited $D_s$ states may have been observed. SELEX has reported a $D_{sJ}(2632)^{\pm}$ candidate~\cite{Evdokimov:2004iy}, but this has not been confirmed by other experiments. Recently, Belle, \babar{} and LHCb have observed $D_{s1}(2700)^{\pm}$ which may be radial excitations of the $D_s^{\ast\pm}$. Equally the $D_{s1}(2860)^{\pm}$ measured by LHCb and $D_{sJ}(3040)^{\pm}$ measured by \babar{} could be excitations of $D_{s0}^\ast(2317)^{\pm}$ and $D_{s1}(2460)^{\pm}$ or $D_{s1}(2536)^{\pm}$, respectively (for a theoretical discussion, see Ref~\cite{Matsuki:2006rz}).

Table~\ref{table:charm:spect:6} summarizes measurements of the polarization amplitude $A_{D}$ (sometimes also referred as helicity parameter), which describes the initial polarization of the $D$ meson. In $D^{\ast\ast}$ meson decay the helicity distribution varies like $1 + A_{D}\cos^{2}\theta_{H}$, where $\theta_{H}$ is the angle in the $D^{\ast}$ rest frame between the two pions emitted by decay $D^{\ast\ast} \to D^{\ast}\pi$ and the $D^{\ast} \to D \pi$. The parameter is sensitive to possible S-wave contributions in the decay. In the case of an unpolarized $D$ meson decay decaying purely via D-wave the polarization amplitude is predicted to give $A_{D}=3$. 
Studies of the $D_{1}(2420)^{0}$ meson by the ZUES and \babar{} collaborations suggest that there is an S-wave admixture in the decay, which is  contrary to Heavy Quark Effective Theory calculations~\cite{Isgur:1989vq,Neubert:1993mb}.

\begin{table}[htb!]
\caption{\label{table:charm:spect:1} Recent measurements of mass and width for different excited $D_{s}$ mesons. The column $J^{P}$ list the most significant assignment of spin and parity. If possible an average mass or width is calculated.}
\begin{adjustbox}{width=\textwidth,center}
{\setlength\tabcolsep{0pt}

}
\end{adjustbox}

\end{table} 

\begin{table}[htb!]
\caption{\label{tabel:charm:spect:2} Recent measurements of mass and width for different excited $D$ mesons. The column $J^{P}$ list the most significant assignment of spin and parity. If possible an average mass or width is calculated.}
\begin{adjustbox}{width=0.76\textwidth,center}
{\setlength\tabcolsep{0pt}
	%
}
\end{adjustbox}

\end{table}

\begin{table}[htb!]
\begin{center}
\caption{\label{table:charm:spect:3} 
Product of $B$ meson branching fraction and 
(daughter) excited $D$ meson branching fraction. }
\begin{adjustbox}{width=\textwidth,center}
{\setlength\tabcolsep{0pt}
	%
}
\end{adjustbox}

\end{center}
\end{table} 

\begin{table}[htb!]
\begin{center}
\caption{\label{table:charm:spect:4} 
Product of $B$ meson branching fraction and 
(daughter) excited $D^{}_s$ meson branching fraction. }
\begin{adjustbox}{width=\textwidth,center}
{\setlength\tabcolsep{0pt}
	%
}
\end{adjustbox}

\end{center}
\end{table}

\begin{table}[htb!]
\begin{center}
\caption{\label{table:charm:spect:5} 
Mass difference measurements for excited $D$ mesons.}
{\setlength\tabcolsep{0pt}
	%
}

\end{center}
\end{table} 

\begin{table}[htb!]
\begin{center}
\caption{\label{table:charm:spect:6}
Measurements of polarization amplitudes for excited $D$ mesons.}
{\setlength\tabcolsep{0pt}
	%
}

\end{center}
\end{table}

\clearpage
\subsection{Charm baryons}

In this section we summarize the present status of excited charm
baryons, decaying strongly or electromagnetically. We list their
masses (or the mass difference between the excited baryon and the
corresponding ground state), natural widths, decay modes, and 
assigned quantum numbers. 
Table~\ref{sumtable1} summarizes the excited $\Lambda_c^+$ baryons.  
The first two states, $\Lambda_c(2595)^+$ and $\Lambda_c(2625)^+$,
are well-established. 
Based on the measured masses, it is believed they are orbitally 
excited $\Lambda_c^+$ baryons with total angular momentum of the
light quarks $L=1$. Thus their quantum numbers are assigned to be 
$J^P=(\frac{1}{2})^-$ and $J^P=(\frac{3}{2})^-$, respectively. 
Recently, their masses 
were precisely measured by CDF~\cite{Aaltonen:2011sf}: 
$M(\Lambda_c(2595)^+)=2592.25\pm 0.24\pm 0.14$~MeV/c$^2$ and
$M(\Lambda_c(2625)^+)=2628.11\pm 0.13\pm 0.14$~MeV/c$^2$. 

The next two states, $\Lambda_c(2765)^+$ and $\Lambda_c(2880)^+$, 
were discovered by CLEO~\cite{Artuso:2000xy} in the $\Lambda_c^+\pi^+\pi^-$ 
final state. CLEO found that $\Lambda_c(2880)^+$ decays also through
the $\Sigma_c(2445)^{++/0}\pi^{-/+}$ mode. 
Later, \babar~\cite{Aubert:2006sp} 
observed that this state has also a $D^0 p$ decay mode. It is the 
first example of an excited charm baryon decaying into a charm meson 
and a light baryon. (Excited charm baryons typically decay into charm 
baryons and light mesons.) In that analysis, \babar observed for the
first time an additional state, $\Lambda_c(2940)^+$, 
which decays into $D^0 p$. Looking for the $D^+ p$ final state,
\babar found no signal; this implies that the $\Lambda_c(2880)^+$ 
and $\Lambda_c(2940)^+$ are really $\Lambda_c^+$ excited states
rather than $\Sigma_c$ excitations. 
Belle reported the result of an angular analysis that favors
$5/2$ for the $\Lambda_c(2880)^+$ spin hypothesis. 
Moreover, the measured ratio of branching fractions 
${\cal B}(\Lambda_c(2880)^+\rightarrow \Sigma_c(2520)\pi^{\pm})/{\cal B}(\Lambda_c(2880)^+\rightarrow \Sigma_c(2455)\pi^{\pm})=(0.225\pm 0.062\pm 0.025)$, combined 
with theoretical predictions based on HQS~\cite{Isgur:1991wq,Cheng:2006dk}, 
favor even parity.     
The current open questions in the excited $\Lambda_c^+$ family are
the determination of quantum numbers for almost all states, and 
the nature of the $\Lambda_c(2765)^+$ state, \ie\ whether it is
an excited $\Sigma_c^+$ or $\Lambda_c^+$.

\begin{table}[t]
\caption{Summary of excited $\Lambda_c^+$ baryons family.} 
\vskip0.15in
\resizebox{0.99\textwidth}{!}{
\begin{tabular}{c|c|c|c|c}
\hline
Charmed Baryon   & Mode  & Mass & Natural Width  & $J^P$  \\
Excited State &  &  (MeV/c$^2$) & (MeV/c$^2$)  \\
\hline
$\Lambda_c(2595)^+$ & $\Lambda_c^+\pi^+\pi^-$, $\Sigma_c\pi$ &  $2595.4\pm 0.6$ & $3.6^{+2.0}_{-1.3}$  & $1/2^-$  \\
\hline
$\Lambda_c(2625)^+$ & $\Lambda_c^+\pi^+\pi^-$, $\Sigma_c\pi$ & $2628.1\pm 0.6$ & $<1.9$ & $3/2^-$  \\
\hline
$\Lambda_c(2765)^+$ & $\Lambda_c^+\pi^+\pi^-$, $\Sigma_c\pi$ & $2766.6\pm 2.4$ & $50$ & ??  \\
\hline
$\Lambda_c(2880)^+$ & $\Lambda_c^+\pi^+\pi^-$, $\Sigma_c\pi$,  &$2881.53\pm 0.35$ & $5.8\pm 1.1$ & $5/2^+$ \\
 &  $\Sigma_c(2520)\pi$, $D^0p$     & & & (experimental evidence) \\
\hline
$\Lambda_c(2940)^+$ & $D^0p$, $\Sigma_c\pi$ & $2939.3^{+1.4}_{-1.5}$ & $17^{+8}_{-6}$  & ??  \\
\hline 
\end{tabular}
}
\label{sumtable1} 
\end{table}

Table~\ref{sumtable2} summarizes the excited $\Sigma_c^{++,+,0}$ baryons.
The triplet of $\Sigma_c(2520)^{++,+,0}$ baryons is well-established. 
Recently Belle~\cite{SHLee:2014} 
precisely measured the mass differences (see above for the definition)  
and widths of the double charged and neutral members of this triplet.
The results are
\begin{eqnarray}
\Delta M(\Sigma_c(2520)^{++}) & = & (231.99\pm 0.10\pm 0.02){\rm\ MeV}/c^2 \\ 
\Gamma(\Sigma_c(2520)^{++}) & = & (14.77\pm 0.25^{+0.18}_{-0.30}){\rm\ MeV} \\ 
\Delta M(\Sigma_c(2520)^{0}) & = & (231.98\pm 0.11\pm 0.04){\rm\ MeV}/c^2 \\ 
\Gamma(\Sigma_c(2520)^{0}) & = & (15.41\pm 0.41^{+0.20}_{-0.32}){\rm\ MeV}\,.
\end{eqnarray} 
This short list of excited $\Sigma_c$ baryons completes the triplet 
of $\Sigma_c(2800)$ states observed by Belle~\cite{Mizuk:2004yu}. Based 
on the measured masses and theoretical predictions~\cite{Copley:1979wj,Pirjol:1997nh}, 
these states are identified as members of the predicted $\Sigma_{c2}$ $3/2^-$
triplet. From a study of resonant substructure 
in $B^-\rightarrow \Lambda_c^+\bar{p}\pi^-$ decays, \babar found 
a significant signal for $\Lambda_c^+\pi^-$ with a mean value 
higher than that measured by Belle by about $3\sigma$
(Table~\ref{sumtable2}). The decay widths measured by
Belle and \babar are consistent.

\begin{table}[!htb]
\caption{Summary of excited $\Sigma_c^{++,+,0}$ baryons family.} 
\vskip0.15in
\resizebox{\textwidth}{!}{
 \begin{tabular}{c|c|c|c|c}
\hline
Charmed Baryon   & Mode  & $\Delta M$ & Natural Width  & $J^P$  \\
Excited State &  &  (MeV/c$^2$) & (MeV/c$^2$)  \\
\hline
$\Sigma_c(2520)^{++}$ &$\Lambda_c^+\pi^+$  & $231.99\pm 0.10\pm 0.02$ & $14.77\pm 0.25^{+0.18}_{-0.30}$ & $3/2^+$   \\
$\Sigma_c(2520)^{+}$ &$\Lambda_c^+\pi^+$  & $231.0\pm 2.3$ & $<17$~@~90$\%$~CL & $3/2^+$ \\
$\Sigma_c(2520)^{0}$ &$\Lambda_c^+\pi^+$  & $231.98\pm 0.11\pm 0.04$ & $15.41\pm 0.41^{+0.20}_{-0.32}$ & $3/2^+$    \\
\hline
$\Sigma_c(2800)^{++}$ & $\Lambda_c^+\pi^{+}$ & $514.5^{+3.4+2.8}_{-3.1-4.9}$ & $75^{+18+12}_{-13-11}$ & tentatively identified      \\
$\Sigma_c(2800)^{+}$ & $\Lambda_c^+\pi^{0}$&$505.4^{+5.8+12.4}_{-4.6-2.0}$ &$62^{+37+52}_{-23-38}$ & as members of the predicted  \\
$\Sigma_c(2800)^{0}$ & $\Lambda_c^+\pi^{-}$&$515.4^{+3.2+2.1}_{-3.1-6.0}$ & $61^{+18+22}_{-13-13}$ &$\Sigma_{c2}$ $3/2^-$ isospin triplet  \\
 & $\Lambda_c^+\pi^{-}$ & $560\pm 8\pm 10$ & $86^{+33}_{-22}$  \\

\hline 
\end{tabular}
}
\label{sumtable2} 
\end{table}

Table~\ref{sumtable3} summarizes the excited $\Xi_c^{+,0}$ and $\Omega_c^0$ 
baryons. Recently, the list of excited $\Xi_c$ baryons has increased,
with several states having masses above 2900~MeV/c$^2$ and decaying into 
$\Lambda_c^+ K^-$, $\Lambda_c^+ K^{-/0}\pi^{+/-}$ and $\Sigma_c(2455/2520) K$. 
Some of these states ($\Xi_c(2980)^+$ and $\Xi_c(3080)^{+,0}$) are seen by
both Belle~\cite{Chistov:2006zj,YKato:2014} 
and \babar~\cite{Aubert:2007dt} and are considered well-established.
The $\Xi_c(2930)^0$ state decaying into $\Lambda_c^+ K^-$ is seen
by \babar~\cite{Aubert:2007eb} but still needs confirmation.  
The $\Xi_c(3055)^+$ found in the $\Sigma_c(2455)^{++}\pi^-$ final
state by \babar~\cite{Aubert:2007dt} was recently confirmed by
Belle~\cite{YKato:2014}. 
The width and mass measurements for the $\Xi_c(3055)^+$
listed in Table~\ref{sumtable3} 
 are not averaged but quoted separately for \babar and Belle. 
However, the $\Xi_c(3123)^+$ observed by \babar~\cite{Aubert:2007dt}
in the $\Sigma_c(2520)^{++}\pi^-$ final state has not been
confirmed by Belle~\cite{YKato:2014} with twice as much statistics.   
We note that Belle~\cite{YKato:2014} recently performed the first
significant measurement of the $\Xi_c(2645)^+$ width, as listed
in Table~\ref{sumtable3}.

The excited $\Omega_c^0$ double charm baryon is seen by both 
\babar~\cite{Aubert:2006je} and Belle~\cite{Solovieva:2008fw};
the mass differences $\delta M=M(\Omega_c^{*0})-M(\Omega_c^0)$ 
measured by the experiments are in good agreement
and are also consistent with most theoretical 
predictions~\cite{Rosner:1995yu,Glozman:1995xy,Jenkins:1996de,
Burakovsky:1997vm}.

\begin{table}[b]
\caption{Summary of excited $\Xi_c^{+,0}$ and $\Omega_c^0$ baryons families.} 
\vskip0.15in
\resizebox{\textwidth}{!}{
\begin{tabular}{c|c|c|c|c}
\hline
Charmed Baryon   & Mode  & Mass & Natural Width  & $J^P$  \\
Excited State &  &  (MeV/c$^2$) & (MeV/c$^2$)  \\
\hline
$\Xi_c'^+$ & $\Xi_c^+\gamma$ & $2575.6\pm 3.1$  &  & $1/2^+$    \\
$\Xi_c'^0$ & $\Xi_c^0\gamma$ & $2577.9\pm 2.9$  &  & $1/2^+$   \\
\hline
$\Xi_c(2645)^+$ & $\Xi_c^0\pi^+$ & $2645.9^{+0.6}_{-0.5}$  & $2.6\pm 0.2\pm 0.4$ & $3/2^+$   \\
$\Xi_c(2645)^0$ & $\Xi_c^+\pi^-$ & $2645.9\pm 0.5$  & $<5.5$ & $3/2^+$   \\
\hline
$\Xi_c(2790)^+$ &$\Xi_c'^0\pi^+$ & $2789.1\pm 3.2$ & $<15$ & $1/2^-$   \\
$\Xi_c(2790)^0$ &$\Xi_c'^+\pi^-$ & $2791.8\pm 3.3$ & $<12$ & $1/2^-$   \\
\hline
$\Xi_c(2815)^+$ &$\Xi_c^+\pi^+\pi^-$, $\Xi_c(2645)^0\pi^+$ & $2816.6\pm 0.9$ & $<3.5$ &  $3/2^-$  \\
$\Xi_c(2815)^0$ & $\Xi_c^0\pi^+\pi^-$, $\Xi_c(2645)^+\pi^-$& $2819.6\pm 1.2$  & $<6.5$  & $3/2^-$   \\
\hline
$\Xi_c(2930)^0$ & $\Lambda_c^+ K^-$ & $2931.6\pm 6$ & $36\pm 13$ & ??     \\
\hline
$\Xi_c(2980)^+$ & $\Lambda_c^+K^-\pi^+$, $\Sigma_c^{++}K^-$, $\Xi_c(2645)^0\pi^+$
 &  $2971.4\pm 3.3$  & $26\pm 7$ & ??     \\
$\Xi_c(2980)^0$ & $\Xi_c(2645)^+\pi^-$
&  $2968.0\pm 2.6$ &$20\pm 7$ & ??       \\
\hline
$\Xi_c(3055)^+$ & $\Sigma_c^{++}K^-$ & 	$3054.2\pm 1.3$ (\babar) &     	$17\pm 13$ (\babar) & ??   \\
       &        &   $3058.1\pm 1.0\pm 2.1$   (Belle)   &   $9.7\pm 3.4\pm 3.3$  (Belle)           \\
\hline
$\Xi_c(3080)^+$ & $\Lambda_c^+K^-\pi^+$, $\Sigma_c^{++}K^-$, $\Sigma_c(2520)^{++}K^-$ & $3077.0\pm 0.4$ & $5.8\pm 1.0$ & ??   \\
$\Xi_c(3080)^0$ &$\Lambda_c^+ K^0_S\pi^-$, $\Sigma_c^0K^0_S$, $\Sigma_c(2520)^{0}K^0_S$ & $3079.9\pm 1.4$ & $5.6\pm 2.2$ & ??   \\
\hline
\\
\hline
$\Omega_c(2770)^0$ & $\Omega_c^0\gamma$& $2765.9\pm 2.0$  & $70.7^{+0.8}_{-0.9}$ & $3/2^+$  \\
\hline 
\end{tabular}
}
\label{sumtable3} 
\end{table}

\begin{figure}[!htb]
\includegraphics[width=1.0\textwidth]{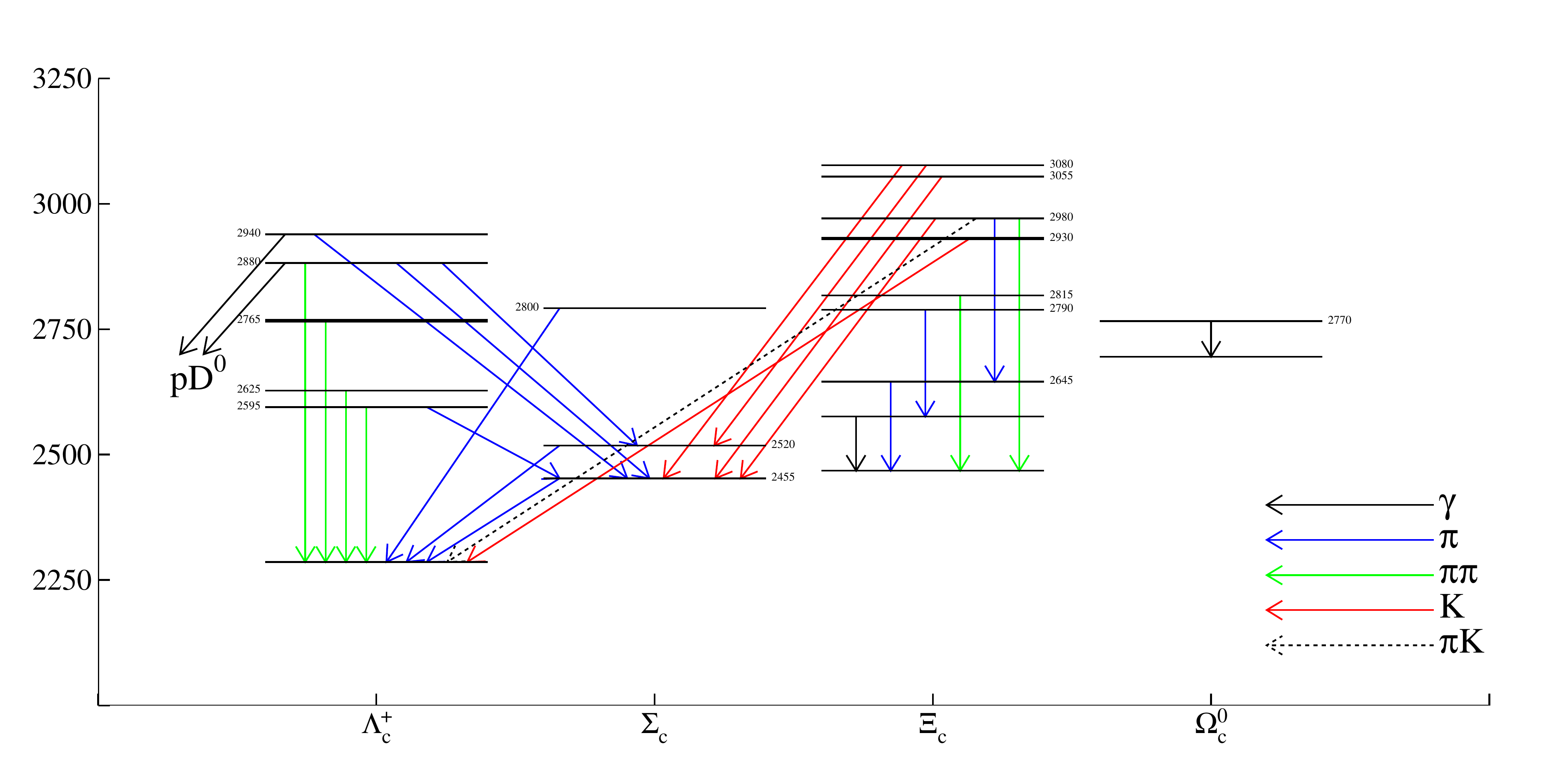}
\caption{Level diagram for excited charm baryons.}
\label{charm:leveldiagram}
\end{figure}

Figure~\ref{charm:leveldiagram} shows the levels of excited charm
baryons along with corresponding transitions between them, and
also transitions to the ground states.
In summary, we note that Belle and \babar  recently discovered
that transitions between families are possible, \ie\ between the 
$\Xi_c$ and $\Lambda_c^+$ families of excited charm baryons. Also,
highly excited $\Lambda_c^+$ baryons are found to decay into 
a charm meson and a proton.

\clearpage
\subsection{Rare and forbidden decays}
\label{sec:charm:rare}

\begin{figure}
\begin{center}
\includegraphics[width=6.0in]{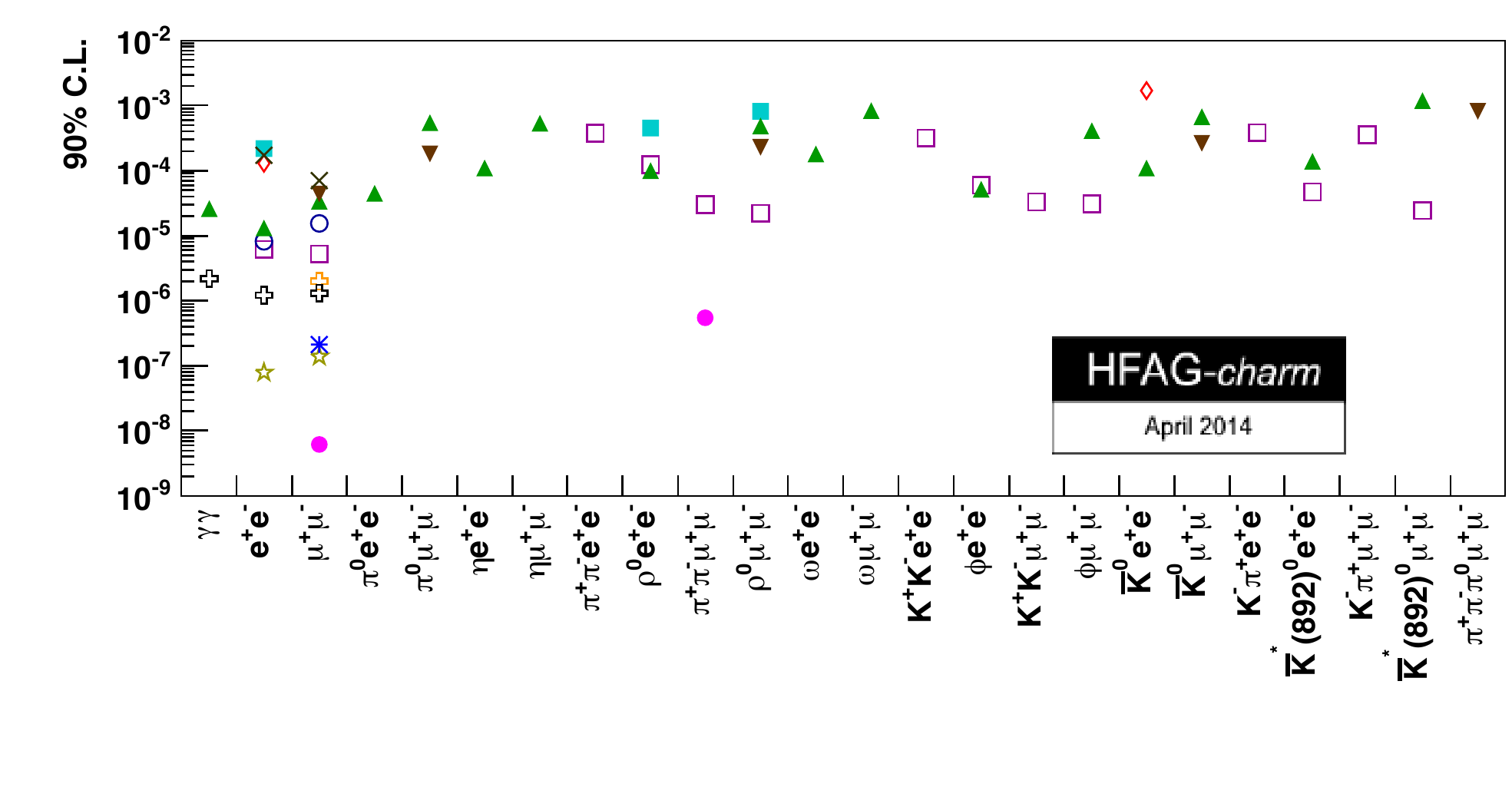}
\vskip-0.10in
\includegraphics[width=6.0in]{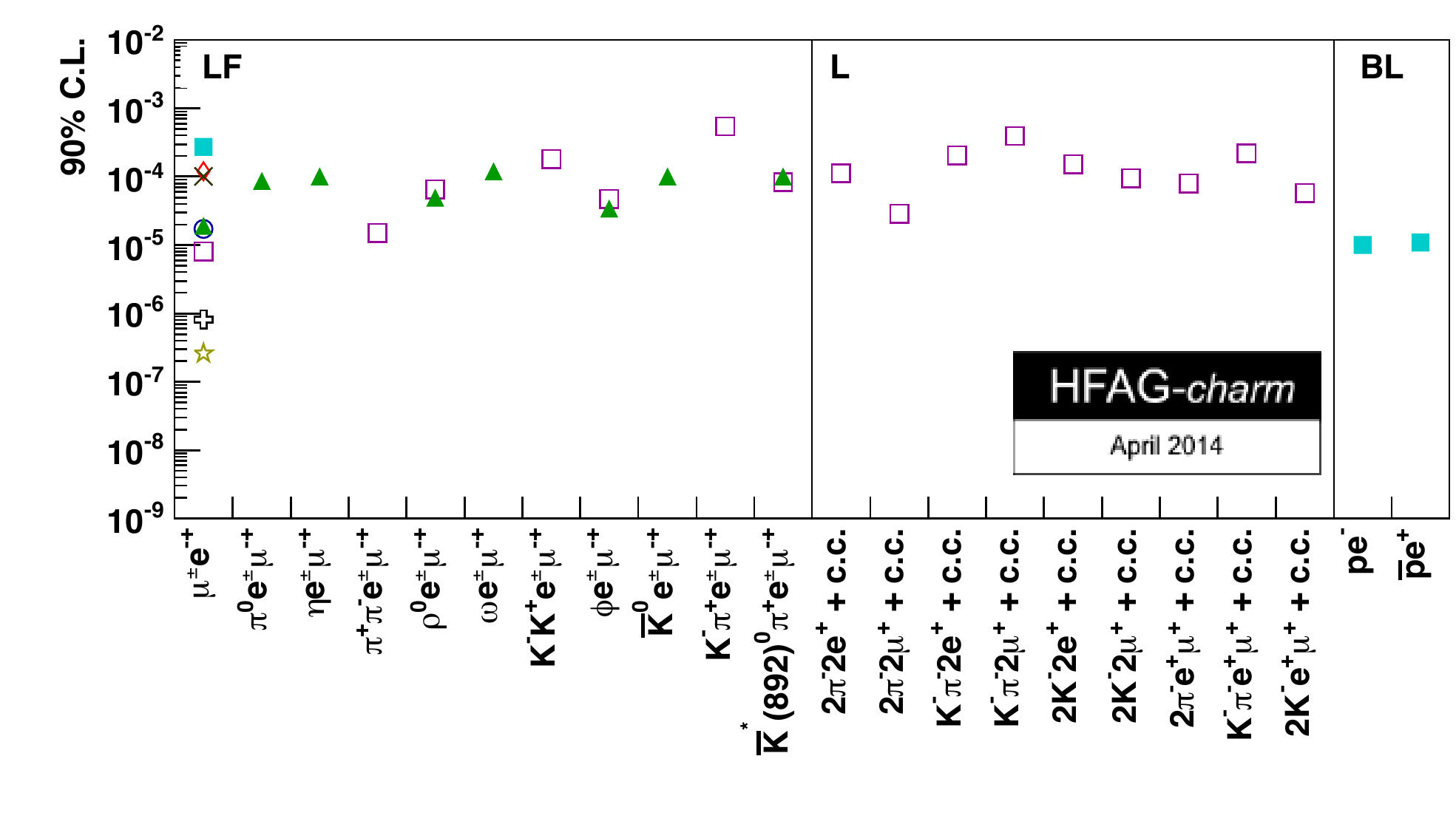}
\caption{Upper limits at $90\%$ CL for $D^0$ decays. The top plot
shows flavor-changing neutral current decays, and the bottom plot
shows lepton-flavor-changing (LF), lepton-number-changing (L), and 
both baryon- and lepton-number-changing (BL) decays.
The legend is given in Fig.~\ref{fig:charm:lambdac}.}
\label{fig:charm:rare_d0}
\end{center}
\end{figure}

\begin{figure}
\begin{center}
\includegraphics[width=5.0in]{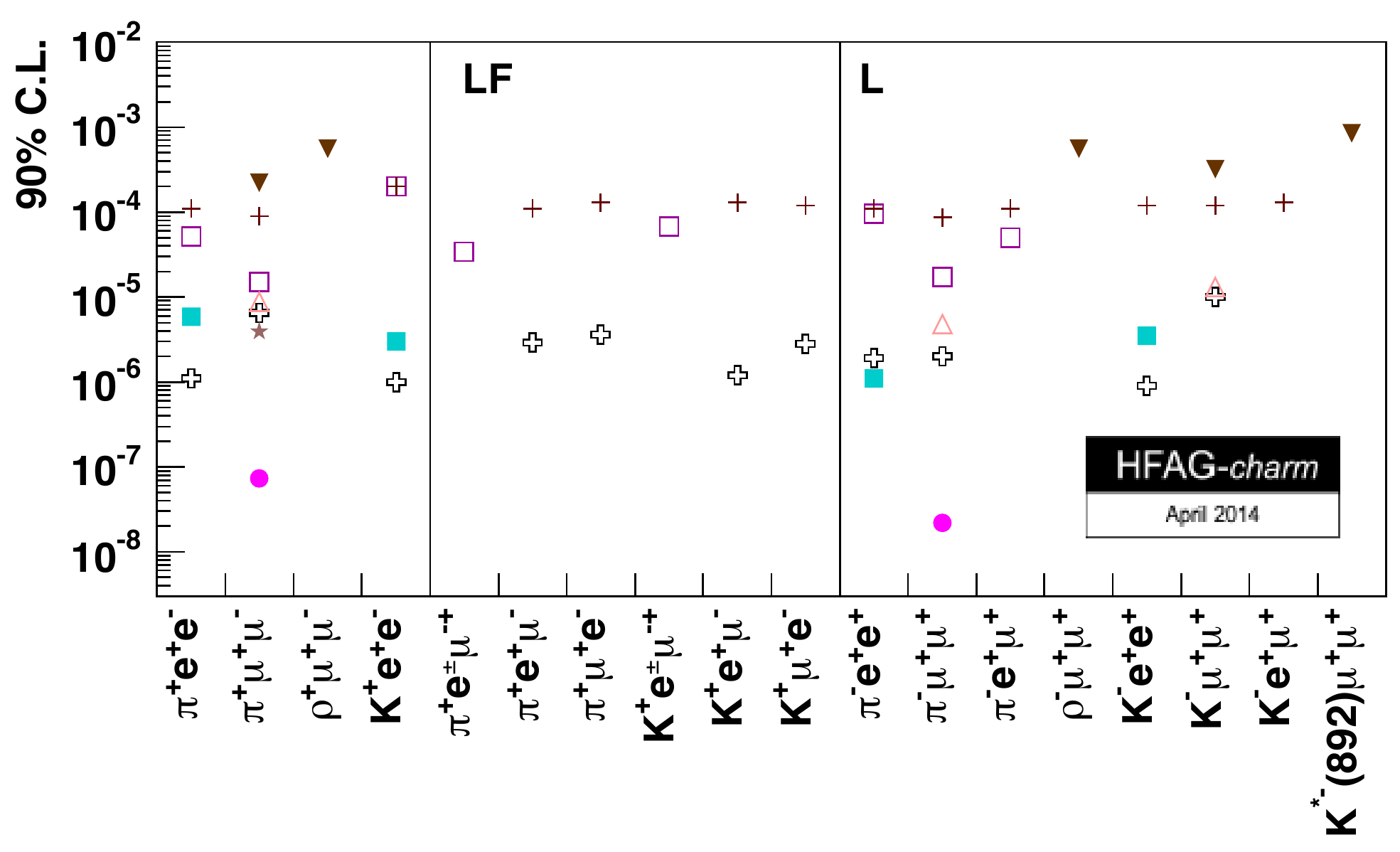}
\vskip0.10in
\includegraphics[width=5.0in]{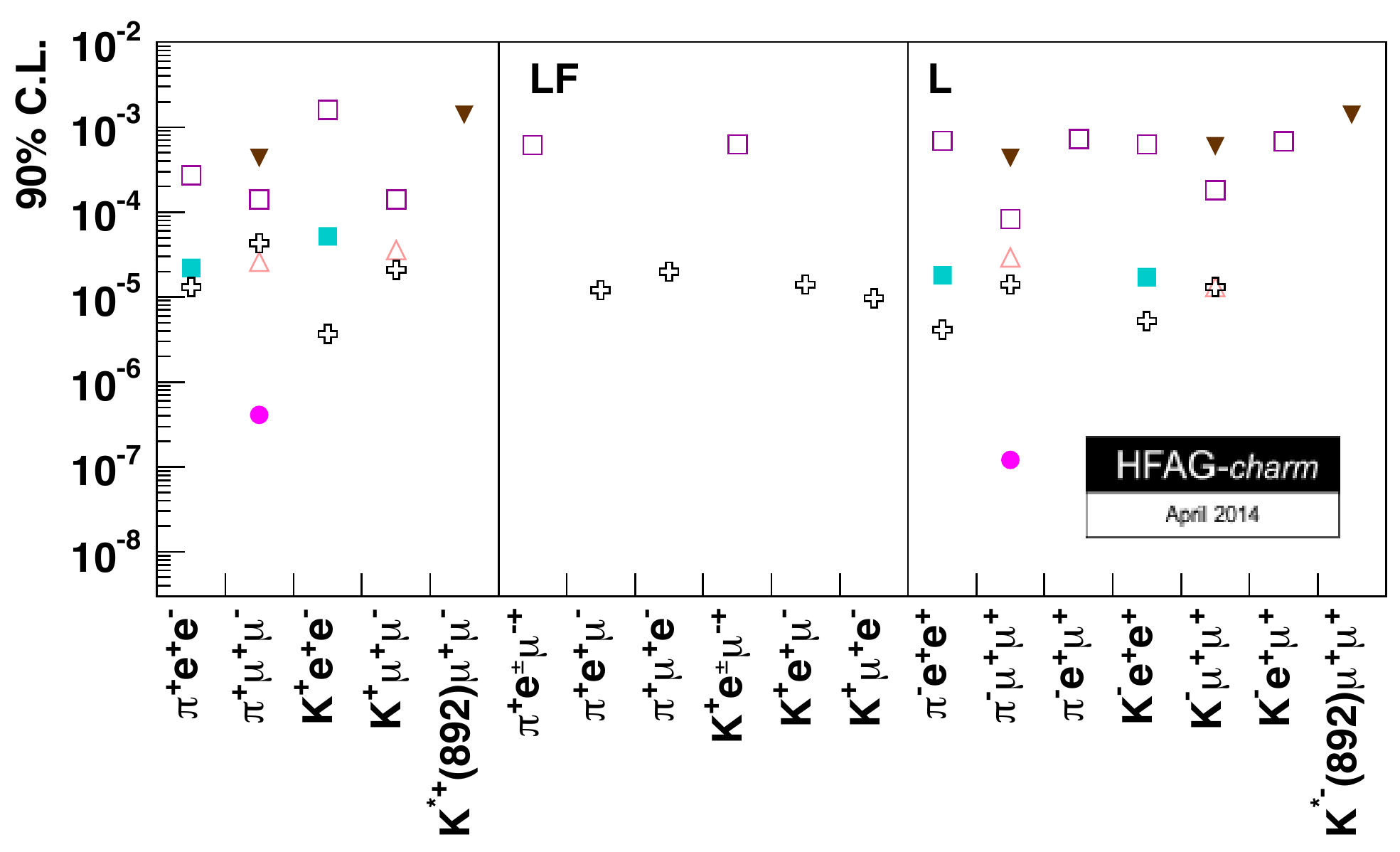}
\caption{Upper limits at $90\%$ CL for $D^+$ (top) and $D_s^+$ (bottom) 
decays. Each plot shows flavor-changing neutral current decays, 
lepton-flavor-changing decays (LF), and lepton-number-changing (L) decays. 
The legend is given in Fig.~\ref{fig:charm:lambdac}.}
\label{fig:charm:rare_charged}
\end{center}
\end{figure}

\begin{figure}
\begin{center}
\hbox{
\includegraphics[width=3.0in]{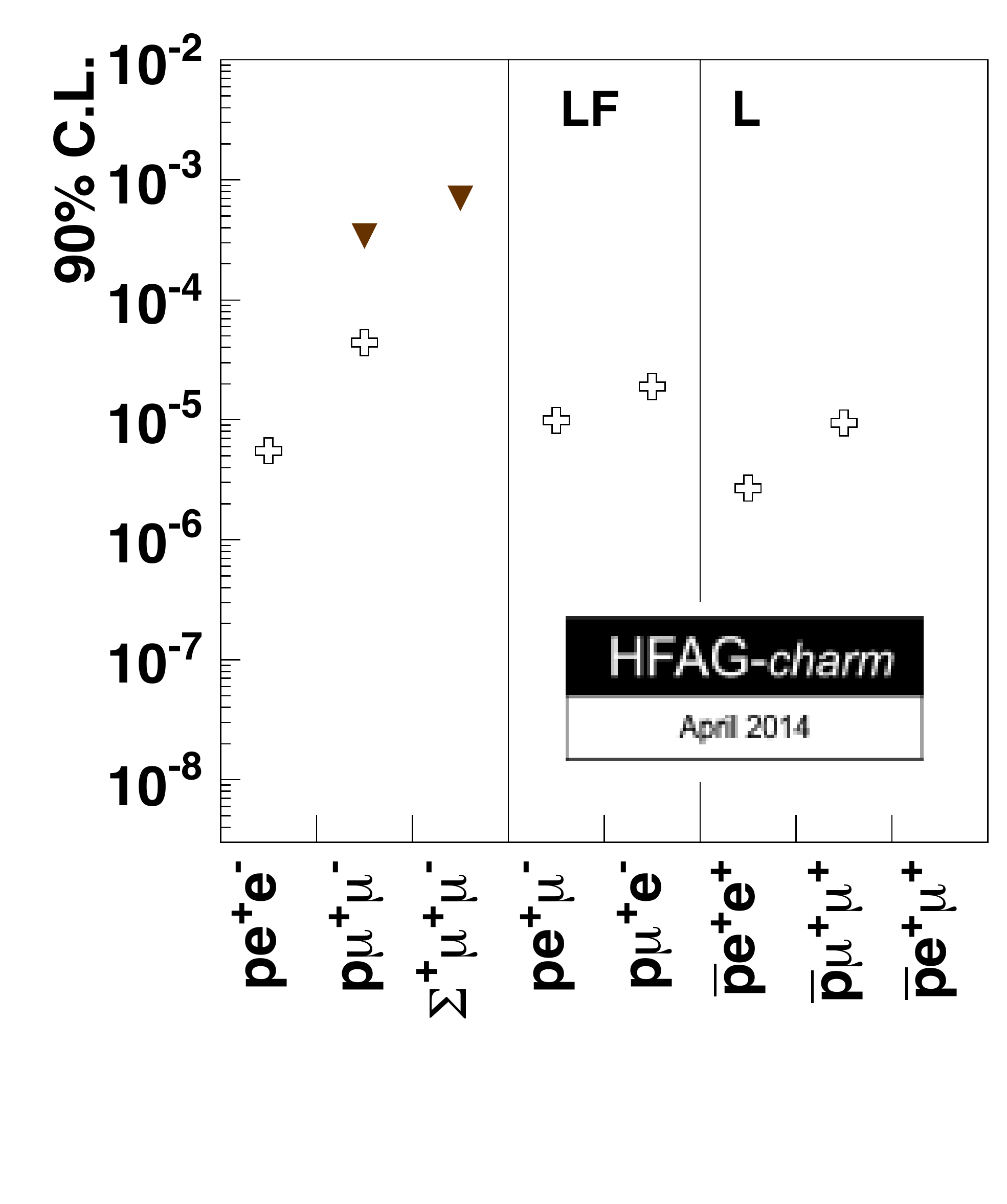}
\hskip-1.80in
\vbox{
\includegraphics[width=2.5in]{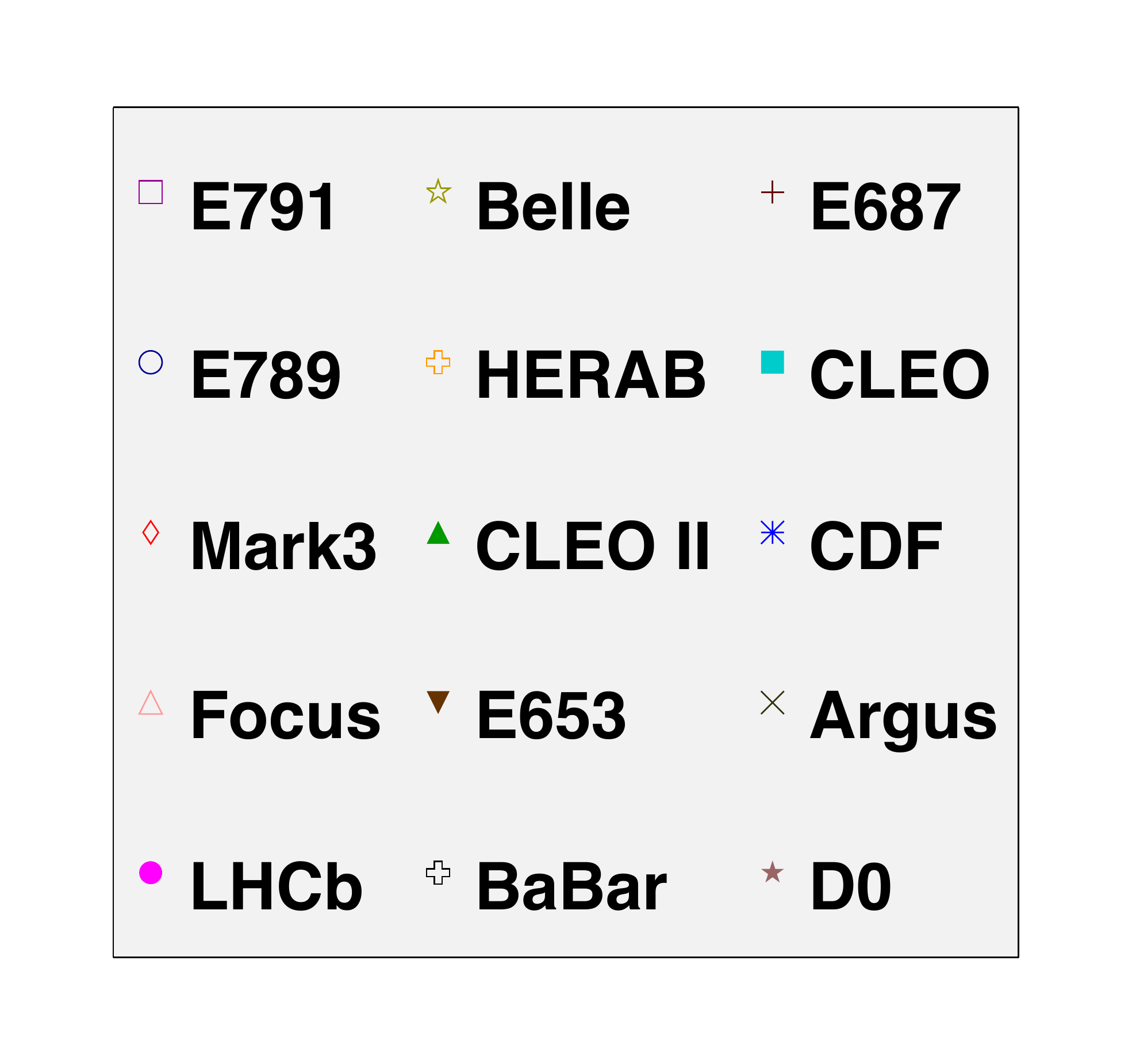}
\vskip1.2in
}}
\vskip-0.20in
\caption{Upper limits at $90\%$ CL for $\Lambda_c^+$ decays. Shown are 
flavor-changing neutral current decays, lepton-flavor-changing (LF) 
decays, and lepton-number-changing (L) decays. }
\label{fig:charm:lambdac}
\end{center}
\end{figure}

This section provides a summary of rare and forbidden charm decays
in tabular form. The decay modes can be categorized as 
flavor-changing neutral currents, lepton-flavor-violating, 
lepton-number-violating, and both baryon- and lepton-number-violating decays.
Figures~\ref{fig:charm:rare_d0}-\ref{fig:charm:lambdac} plot the 
upper limits for $D^0$, $D^+$, $D_s^+$, and $\Lambda_c^+$ decays. 
Tables~\ref{tab:charm:rare_d0}-\ref{tab:charm:rare_lambdac} give the 
corresponding numerical results. Some theoretical predictions are given in 
Refs.~\cite{Burdman:2001tf,Fajfer:2002bu,Fajfer:2007dy,Golowich:2009ii,Paul:2010pq,Borisov:2011aa}.

In several cases the rare-decay final states have been observed with the di-lepton pair being the decay product of a hadronic resonance.
For these measurements the quoted limits are those expected for the non-resonant di-lepton spectrum.
For the extrapolation to the full spectrum a phase-space distribution of the non-resonant component has been assumed.
This applies to the CLEO measurement of the decays $D_{(s)}^+\to(K^+\pi^+)e^+e^-$~\cite{Rubin:2010cq}, to the D0 measurements of the decays $D_{(s)}^+\to\pi^+\mu^+\mu^-$~\cite{Abazov:2007aj}, and to the \babar measurements of the decays $D_{(s)}^+\to(K^+\pi^+)e^+e^-$ and $D_{(s)}^+\to(K^+\pi^+)\mu^+\mu^-$, where the contribution from $\phi\to l^+l^-$ ($l=e,\mu$) has been excluded.
In the case of the LHCb measurements of the decays $D^0\to\pi^+\pi^-\mu^+\mu^-$~\cite{Aaij:2013uoa} as well as the decays $D_{(s)}^+\to\pi^+\mu^+\mu^-$~\cite{Aaij:2013sua} the contributions from $\phi\to l^+l^-$ as well as from $\rho,\omega\to l^+l^-$ ($l=e,\mu$) have been excluded. 

\begin{longtable}{l|ccc}
\caption{Upper limits at $90\%$ CL for $D^0$ decays.
\label{tab:charm:rare_d0}
}\\

\hline\hline
Decay & Limit $\times10^6$ & Experiment & Reference\\
\endfirsthead
\multicolumn{4}{c}{\tablename\ \thetable{} -- continued from previous page} \\ \hline
Decay & Limit $\times10^6$ & Experiment & Reference\\
\endhead
\hline
$\gamma{}\gamma{}$ & 26.0 & CLEO II & \cite{Coan:2002te}\\
& 2.2 & \babar Preliminary & \cite{Lees:2011qz}\\
\hline
$e^+e^-$ & 220.0 & CLEO & \cite{Haas:1988bh}\\
& 170.0 & ARGUS & \cite{Albrecht:1988ge}\\
& 130.0 & Mark3 & \cite{Adler:1987cp}\\
& 13.0 & CLEO II & \cite{Freyberger:1996it}\\
& 8.19 & E789 & \cite{Pripstein:1999tq}\\
& 6.2 & E791 & \cite{Aitala:1999db}\\
& 1.2 & \babar & \cite{Aubert:2004bs}\\
& 0.079 & Belle & \cite{Petric:2010yt}\\
\hline
$\mu{}^+\mu{}^-$ & 70.0 & ARGUS & \cite{Albrecht:1988ge}\\
& 44.0 & E653 & \cite{Kodama:1995ia}\\
& 34.0 & CLEO II & \cite{Freyberger:1996it}\\
& 15.6 & E789 & \cite{Pripstein:1999tq}\\
& 5.2 & E791 & \cite{Aitala:1999db}\\
& 2.0 & HERAb & \cite{Abt:2004hn}\\
& 1.3 & \babar & \cite{Aubert:2004bs}\\
& 0.21 & CDF & \cite{Aaltonen:2010hz}\\
& 0.14 & Belle & \cite{Petric:2010yt}\\
& 0.0062 & LHCb & \cite{Aaij:2013cza}\\
\hline
$\pi{}^0e^+e^-$ & 45.0 & CLEO II & \cite{Freyberger:1996it}\\
\hline
$\pi{}^0\mu{}^+\mu{}^-$ & 540.0 & CLEO II & \cite{Freyberger:1996it}\\
& 180.0 & E653 & \cite{Kodama:1995ia}\\
\hline
$\eta{}e^+e^-$ & 110.0 & CLEO II & \cite{Freyberger:1996it}\\
\hline
$\eta{}\mu{}^+\mu{}^-$ & 530.0 & CLEO II & \cite{Freyberger:1996it}\\
\hline
$\pi{}^+\pi{}^-e^+e^-$ & 370.0 & E791 & \cite{Aitala:2000kk}\\
\hline
$\rho{}e^+e^-$ & 450.0 & CLEO & \cite{Haas:1988bh}\\
& 124.0 & E791 & \cite{Aitala:2000kk}\\
& 100.0 & CLEO II & \cite{Freyberger:1996it}\\
\hline
$\pi{}^+\pi{}^-\mu{}^+\mu{}^-$ & 30.0 & E791 & \cite{Aitala:2000kk}\\
& 0.55 & LHCb & \cite{Aaij:2013uoa}\\
\hline
$\rho{}\mu{}^+\mu{}^-$ & 810.0 & CLEO & \cite{Haas:1988bh}\\
& 490.0 & CLEO II & \cite{Freyberger:1996it}\\
& 230.0 & E653 & \cite{Kodama:1995ia}\\
& 22.0 & E791 & \cite{Aitala:2000kk}\\
\hline
$\omega{}e^+e^-$ & 180.0 & CLEO II & \cite{Freyberger:1996it}\\
\hline
$\omega{}\mu{}^+\mu{}^-$ & 830.0 & CLEO II & \cite{Freyberger:1996it}\\
\hline
$K^+K^-e^+e^-$ & 315.0 & E791 & \cite{Aitala:2000kk}\\
\hline
$\phi{}e^+e^-$ & 59.0 & E791 & \cite{Aitala:2000kk}\\
& 52.0 & CLEO II & \cite{Freyberger:1996it}\\
\hline
$K^+K^-\mu{}^+\mu{}^-$ & 33.0 & E791 & \cite{Aitala:2000kk}\\
\hline
$\phi{}\mu{}^+\mu{}^-$ & 410.0 & CLEO II & \cite{Freyberger:1996it}\\
& 31.0 & E791 & \cite{Aitala:2000kk}\\
\hline
$\overline{K}^0e^+e^-$ & 1700.0 & Mark3 & \cite{Adler:1988es}\\
& 110.0 & CLEO II & \cite{Freyberger:1996it}\\
\hline
$\overline{K}^0\mu{}^+\mu{}^-$ & 670.0 & CLEO II & \cite{Freyberger:1996it}\\
& 260.0 & E653 & \cite{Kodama:1995ia}\\
\hline
$K^-\pi{}^+e^+e^-$ & 385.0 & E791 & \cite{Aitala:2000kk}\\
\hline
$\overline{K}^{*0}(892)e^+e^-$ & 140.0 & CLEO II & \cite{Freyberger:1996it}\\
& 47.0 & E791 & \cite{Aitala:2000kk}\\
\hline
$K^-\pi{}^+\mu{}^+\mu{}^-$ & 360.0 & E791 & \cite{Aitala:2000kk}\\
\hline
$\overline{K}^{*0}(892)\mu{}^+\mu{}^-$ & 1180.0 & CLEO II & \cite{Freyberger:1996it}\\
& 24.0 & E791 & \cite{Aitala:2000kk}\\
\hline
$\pi{}^+\pi{}^-\pi{}^0\mu{}^+\mu{}^-$ & 810.0 & E653 & \cite{Kodama:1995ia}\\
\hline
$\mu{}^{\pm}e^{\mp}$ & 270.0 & CLEO & \cite{Haas:1988bh}\\
& 120.0 & Mark3 & \cite{Becker:1987mu}\\
& 100.0 & ARGUS & \cite{Albrecht:1988ge}\\
& 19.0 & CLEO II & \cite{Freyberger:1996it}\\
& 17.2 & E789 & \cite{Pripstein:1999tq}\\
& 8.1 & E791 & \cite{Aitala:1999db}\\
& 0.81 & \babar & \cite{Aubert:2004bs}\\
& 0.26 & Belle & \cite{Petric:2010yt}\\
\hline
$\pi{}^0e^{\pm}\mu{}^{\mp}$ & 86.0 & CLEO II & \cite{Freyberger:1996it}\\
\hline
$\eta{}e^{\pm}\mu{}^{\mp}$ & 100.0 & CLEO II & \cite{Freyberger:1996it}\\
\hline
$\pi{}^+\pi{}^-e^{\pm}\mu{}^{\mp}$ & 15.0 & E791 & \cite{Aitala:2000kk}\\
\hline
$\rho{}e^{\pm}\mu{}^{\mp}$ & 66.0 & E791 & \cite{Aitala:2000kk}\\
& 49.0 & CLEO II & \cite{Freyberger:1996it}\\
\hline
$\omega{}e^{\pm}\mu{}^{\mp}$ & 120.0 & CLEO II & \cite{Freyberger:1996it}\\
\hline
$K^+K^-e^{\pm}\mu{}^{\mp}$ & 180.0 & E791 & \cite{Aitala:2000kk}\\
\hline
$\phi{}e^{\pm}\mu{}^{\mp}$ & 47.0 & E791 & \cite{Aitala:2000kk}\\
& 34.0 & CLEO II & \cite{Freyberger:1996it}\\
\hline
$\overline{K}^0e^{\pm}\mu{}^{\mp}$ & 100.0 & CLEO II & \cite{Freyberger:1996it}\\
\hline
$K^-\pi{}^+e^{\pm}\mu{}^{\mp}$ & 550.0 & E791 & \cite{Aitala:2000kk}\\
\hline
$K^{*0}(892)e^{\pm}\mu{}^{\mp}$ & 100.0 & CLEO II & \cite{Freyberger:1996it}\\
& 83.0 & E791 & \cite{Aitala:2000kk}\\
\hline
$\pi{}^{\mp}\pi{}^{\mp}e^{\pm}e^{\pm}$ & 112.0 & E791 & \cite{Aitala:2000kk}\\
\hline
$\pi{}^{\mp}\pi{}^{\mp}\mu{}^{\pm}\mu{}^{\pm}$ & 29.0 & E791 & \cite{Aitala:2000kk}\\
\hline
$K^{\mp}\pi{}^{\mp}e^{\pm}e^{\pm}$ & 206.0 & E791 & \cite{Aitala:2000kk}\\
\hline
$K^{\mp}\pi{}^{\mp}\mu{}^{\pm}\mu{}^{\pm}$ & 390.0 & E791 & \cite{Aitala:2000kk}\\
\hline
$K^{\mp}K^{\mp}e^{\pm}e^{\pm}$ & 152.0 & E791 & \cite{Aitala:2000kk}\\
\hline
$K^{\mp}K^{\mp}\mu{}^{\pm}\mu{}^{\pm}$ & 94.0 & E791 & \cite{Aitala:2000kk}\\
\hline
$\pi{}^{\mp}\pi{}^{\mp}e^{\pm}\mu{}^{\pm}$ & 79.0 & E791 & \cite{Aitala:2000kk}\\
\hline
$K^{\mp}\pi{}^{\mp}e^{\pm}\mu{}^{\pm}$ & 218.0 & E791 & \cite{Aitala:2000kk}\\
\hline
$K^{\mp}K^{\mp}e^{\pm}\mu{}^{\pm}$ & 57.0 & E791 & \cite{Aitala:2000kk}\\
\hline
$pe^-$ & 10.0 & CLEO & \cite{Rubin:2009aa}\\
\hline
$\overline{p}e^+$ & 11.0 & CLEO & \cite{Rubin:2009aa}\\
\hline
\end{longtable}

\pagebreak

\begin{longtable}{l|ccc}
\caption{Upper limits at $90\%$ CL for $D^+$ decays.\label{tab:charm:rare_dplus}}\\
\hline\hline
Decay & Limit $\times10^6$ & Experiment & Reference\\
\endfirsthead
\multicolumn{4}{c}{\tablename\ \thetable{} -- continued from previous page} \\ \hline
Decay & Limit $\times10^6$ & Experiment & Reference\\
\endhead

\hline
$\pi{}^+e^+e^-$ & 110.0 & E687 & \cite{Frabetti:1997wp}\\
& 52.0 & E791 & \cite{Aitala:1999db}\\
& 5.9 & CLEO & \cite{Rubin:2010cq}\\
& 1.1 & \babar & \cite{Lees:2011hb}\\
\hline
$\pi{}^+\mu{}^+\mu{}^-$ & 220.0 & E653 & \cite{Kodama:1995ia}\\
& 89.0 & E687 & \cite{Frabetti:1997wp}\\
& 15.0 & E791 & \cite{Aitala:1999db}\\
& 8.8 & Focus & \cite{Link:2003qp}\\
& 6.5 & \babar & \cite{Lees:2011hb}\\
& 3.9 & D0 & \cite{Abazov:2007aj}\\
& 0.073 & LHCb & \cite{Aaij:2013sua}\\
\hline
$\rho{}^+\mu{}^+\mu{}^-$ & 560.0 & E653 & \cite{Kodama:1995ia}\\
\hline
$K^+e^+e^-$ & 200.0 & E687 & \cite{Frabetti:1997wp}\\
& 3.0 & CLEO & \cite{Rubin:2010cq}\\
& 1.0 & \babar & \cite{Lees:2011hb}\\
\hline
$\pi{}^+e^{\pm}\mu{}^{\mp}$ & 34.0 & E791 & \cite{Aitala:1999db}\\
\hline
$\pi{}^+e^+\mu{}^-$ & 110.0 & E687 & \cite{Frabetti:1997wp}\\
& 2.9 & \babar & \cite{Lees:2011hb}\\
\hline
$\pi{}^+\mu{}^+e^-$ & 130.0 & E687 & \cite{Frabetti:1997wp}\\
& 3.6 & \babar & \cite{Lees:2011hb}\\
\hline
$K^+e^{\pm}\mu{}^{\mp}$ & 68.0 & E791 & \cite{Aitala:1999db}\\
\hline
$K^+e^+\mu{}^-$ & 130.0 & E687 & \cite{Frabetti:1997wp}\\
& 1.2 & \babar & \cite{Lees:2011hb}\\
\hline
$K^+\mu{}^+e^-$ & 120.0 & E687 & \cite{Frabetti:1997wp}\\
& 2.8 & \babar & \cite{Lees:2011hb}\\
\hline
$\pi{}^-e^+e^+$ & 110.0 & E687 & \cite{Frabetti:1997wp}\\
& 96.0 & E791 & \cite{Aitala:1999db}\\
& 1.9 & \babar & \cite{Lees:2011hb}\\
& 1.1 & CLEO & \cite{Rubin:2010cq}\\
\hline
$\pi{}^-\mu{}^+\mu{}^+$ & 87.0 & E687 & \cite{Frabetti:1997wp}\\
& 17.0 & E791 & \cite{Aitala:1999db}\\
& 4.8 & Focus & \cite{Link:2003qp}\\
& 2.0 & \babar & \cite{Lees:2011hb}\\
& 0.022 & LHCb & \cite{Aaij:2013sua}\\
\hline
$\pi{}^-e^+\mu{}^+$ & 110.0 & E687 & \cite{Frabetti:1997wp}\\
& 50.0 & E791 & \cite{Aitala:1999db}\\
\hline
$\rho{}^-\mu{}^+\mu{}^+$ & 560.0 & E653 & \cite{Kodama:1995ia}\\
\hline
$K^-e^+e^+$ & 120.0 & E687 & \cite{Frabetti:1997wp}\\
& 3.5 & CLEO & \cite{Rubin:2010cq}\\
& 0.9 & \babar & \cite{Lees:2011hb}\\
\hline
$K^-\mu{}^+\mu{}^+$ & 320.0 & E653 & \cite{Kodama:1995ia}\\
& 120.0 & E687 & \cite{Frabetti:1997wp}\\
& 13.0 & Focus & \cite{Link:2003qp}\\
& 10.0 & \babar & \cite{Lees:2011hb}\\
\hline
$K^-e^+\mu{}^+$ & 130.0 & E687 & \cite{Frabetti:1997wp}\\
\hline
$K^{*-}(892)\mu{}^+\mu{}^+$ & 850.0 & E653 & \cite{Kodama:1995ia}\\
\hline
\end{longtable}

\begin{longtable}{l|ccc}
\caption{Upper limits at $90\%$ CL for $D_s^+$ decays.\label{tab:charm:rare_dsplus}}\\
\hline\hline
Decay & Limit $\times10^6$ & Experiment & Reference\\
\endfirsthead
\multicolumn{4}{c}{\tablename\ \thetable{} -- continued from previous page} \\ \hline
Decay & Limit $\times10^6$ & Experiment & Reference\\
\endhead

\hline
$\pi{}^+e^+e^-$ & 270.0 & E791 & \cite{Aitala:1999db}\\
& 22.0 & CLEO & \cite{Rubin:2010cq}\\
& 13.0 & \babar & \cite{Lees:2011hb}\\
\hline
$\pi{}^+\mu{}^+\mu{}^-$ & 430.0 & E653 & \cite{Kodama:1995ia}\\
& 140.0 & E791 & \cite{Aitala:1999db}\\
& 43.0 & \babar & \cite{Lees:2011hb}\\
& 26.0 & Focus & \cite{Link:2003qp}\\
& 0.41 & LHCb & \cite{Aaij:2013sua}\\
\hline
$K^+e^+e^-$ & 1600.0 & E791 & \cite{Aitala:1999db}\\
& 52.0 & CLEO & \cite{Rubin:2010cq}\\
& 3.7 & \babar & \cite{Lees:2011hb}\\
\hline
$K^+\mu{}^+\mu{}^-$ & 140.0 & E791 & \cite{Aitala:1999db}\\
& 36.0 & Focus & \cite{Link:2003qp}\\
& 21.0 & \babar & \cite{Lees:2011hb}\\
\hline
$K^{*+}(892)\mu{}^+\mu{}^-$ & 1400.0 & E653 & \cite{Kodama:1995ia}\\
\hline
$\pi{}^+e^{\pm}\mu{}^{\mp}$ & 610.0 & E791 & \cite{Aitala:1999db}\\
\hline
$\pi{}^+e^+\mu{}^-$ & 12.0 & \babar & \cite{Lees:2011hb}\\
\hline
$\pi{}^+\mu{}^+e^-$ & 20.0 & \babar & \cite{Lees:2011hb}\\
\hline
$K^+e^{\pm}\mu{}^{\mp}$ & 630.0 & E791 & \cite{Aitala:1999db}\\
\hline
$K^+e^+\mu{}^-$ & 14.0 & \babar & \cite{Lees:2011hb}\\
\hline
$K^+\mu{}^+e^-$ & 9.7 & \babar & \cite{Lees:2011hb}\\
\hline
$\pi{}^-e^+e^+$ & 690.0 & E791 & \cite{Aitala:1999db}\\
& 18.0 & CLEO & \cite{Rubin:2010cq}\\
& 4.1 & \babar & \cite{Lees:2011hb}\\
\hline
$\pi{}^-\mu{}^+\mu{}^+$ & 430.0 & E653 & \cite{Kodama:1995ia}\\
& 82.0 & E791 & \cite{Aitala:1999db}\\
& 29.0 & Focus & \cite{Link:2003qp}\\
& 14.0 & \babar & \cite{Lees:2011hb}\\
& 0.12 & LHCb & \cite{Aaij:2013sua}\\
\hline
$\pi{}^-e^+\mu{}^+$ & 730.0 & E791 & \cite{Aitala:1999db}\\
\hline
$K^-e^+e^+$ & 630.0 & E791 & \cite{Aitala:1999db}\\
& 17.0 & CLEO & \cite{Rubin:2010cq}\\
& 5.2 & \babar & \cite{Lees:2011hb}\\
\hline
$K^-\mu{}^+\mu{}^+$ & 590.0 & E653 & \cite{Kodama:1995ia}\\
& 180.0 & E791 & \cite{Aitala:1999db}\\
& 13.0 & \babar & \cite{Lees:2011hb}\\
\hline
$K^-e^+\mu{}^+$ & 680.0 & E791 & \cite{Aitala:1999db}\\
\hline
$K^{*-}(892)\mu{}^+\mu{}^+$ & 1400.0 & E653 & \cite{Kodama:1995ia}\\
\hline
\end{longtable}

\begin{longtable}{l|ccc}
\caption{Upper limits at $90\%$ CL for $\Lambda_c^+$ decays.\label{tab:charm:rare_lambdac}}\\
\hline\hline
Decay & Limit $\times10^6$ & Experiment & Reference\\
\endfirsthead
\multicolumn{4}{c}{\tablename\ \thetable{} -- continued from previous page} \\ \hline
Decay & Limit $\times10^6$ & Experiment & Reference\\
\endhead

\hline
$pe^+e^-$ & 5.5 & \babar & \cite{Lees:2011hb}\\
\hline
$p\mu{}^+\mu{}^-$ & 340.0 & E653 & \cite{Kodama:1995ia}\\
& 44.0 & \babar & \cite{Lees:2011hb}\\
\hline
$\sigma{}^+\mu{}^+\mu{}^-$ & 700.0 & E653 & \cite{Kodama:1995ia}\\
\hline
$pe^+\mu{}^-$ & 9.9 & \babar & \cite{Lees:2011hb}\\
\hline
$p\mu{}^+e^-$ & 19.0 & \babar & \cite{Lees:2011hb}\\
\hline
$\overline{p}e^+e^+$ & 2.7 & \babar & \cite{Lees:2011hb}\\
\hline
$\overline{p}\mu{}^+\mu{}^+$ & 9.4 & \babar & \cite{Lees:2011hb}\\
\hline
\end{longtable}

\clearpage
\newcommand{\hfagTauTag}{Summer 2014\xspace}
\newcommand{\hfagFitLabel}{HFAG \hfagTauTag fit}

\newif\ifhevea\heveafalse
\providecommand{\cutname}[1]{}

\makeatletter
\newenvironment*{fleqn}[1][\leftmargini minus\leftmargini]{\@fleqntrue
  \setlength\@mathmargin{#1}\ignorespaces
}{%
  \ignorespacesafterend
}

\newenvironment{ceqn}{\@fleqnfalse
  \@mathmargin\@centering \ignorespaces
}{%
  \ignorespacesafterend
}

\newenvironment{citenoleadsp}{\@fleqnfalse
  \let\cite@adjust\@empty
}{%
  \ignorespacesafterend
}
\makeatother

\let\citeOld\cite
\renewcommand{\cite}[1]{\ifmmode\text{\citeOld{#1}}\else\citeOld{#1}\fi}

\newenvironment{ensuredisplaymath}
  {\(\displaystyle}
  {\)}

\makeatletter
\DeclareRobustCommand\bfseries{%
  \not@math@alphabet\bfseries\mathbf
  \fontseries\bfdefault\selectfont\boldmath}
\DeclareRobustCommand*{\bm}[1]{%
    \mathchoice{\bmbox{\displaystyle}{#1}}%
               {\bmbox{\textstyle}{#1}}%
               {\bmbox{\scriptstyle}{#1}}%
               {\bmbox{\scriptscriptstyle}{#1}}}
\DeclareRobustCommand*{\bmbox}[2]{\mbox{\bfseries$#1 #2$}}
\makeatother

\makeatletter
\def\LT@makecaption#1#2#3{%
  \LT@mcol\LT@cols c{\hbox to\z@{\hss\parbox[t]\LTcapwidth{%
    \sbox\@tempboxa{\normalsize#1{#2: }#3}%
    \ifdim\wd\@tempboxa>\hsize
      \normalsize#1{#2: }#3%
    \else
      \hbox to\hsize{\hfil\box\@tempboxa\hfil}%
    \fi
    \endgraf\vskip\baselineskip}%
  \hss}}}
\makeatother

\makeatletter

\newcommand{\htdef}[2]{%
  \@namedef{hfagtau@#1}{#2}%
}

\iffalse
\newcommand{\htuse}[1]{%
  \@nameuse{hfagtau@#1}}%
\else
\newcommand{\htuse}[1]{%
  \ifcsname hfagtau@#1\endcsname
  \@nameuse{hfagtau@#1}%
  \else
  \@latex@error{Undefined name hfagtau@#1}\@eha
  \fi}
\fi

\newcommand{\htconstrdef}[4]{%
  \@namedef{hfagtau@#1.left}{\ensuremath{#2}}%
  \@namedef{hfagtau@#1.right}{\ensuremath{#3}}%
  \@namedef{hfagtau@#1.right.split}{\ensuremath{#4}}%
  \@namedef{hfagtau@#1.constr.eq}{\htuse{#1.left} ={}& \htuse{#1.right}}%
}

\iffalse
\newcommand{\htquantdef}[6]{%
  \@namedef{hfagtau@#1.gn}{\ensuremath{#2}}%
  \@namedef{hfagtau@#1.td}{\ensuremath{#3}}%
  \@namedef{hfagtau@#1}{\ensuremath{#4}}%
  \@namedef{hfagtau@#1.v}{\ensuremath{#5}}%
  \@namedef{hfagtau@#1.e}{\ensuremath{#6}}%
}
\else
\newcommand{\htquantdef}[6]{%
  \ifx&#2&\else
  \@namedef{hfagtau@#1.gn}{\ensuremath{#2}}%
  \fi
  \ifx&#3&\else
  \@namedef{hfagtau@#1.td}{\ensuremath{#3}}%
  \fi
  \ifx&#6&%
    \@namedef{hfagtau@#1}{\ensuremath{#5}}%
  \else
    \ifthenelse{\equal{#6}{0}}{%
      \@namedef{hfagtau@#1}{\ensuremath{#5}}%
    }{%
      \@namedef{hfagtau@#1}{\ensuremath{#4}}%
      \@namedef{hfagtau@#1.v}{\ensuremath{#5}}%
      \@namedef{hfagtau@#1.e}{\ensuremath{#6}}%
    }%
  \fi
}
\fi

\newcommand{\htmeasdef}[8]{%
  \@namedef{hfagtau@#1,quant}{\ensuremath{#2}}%
  \@namedef{hfagtau@#1,exp}{#3}%
  \@namedef{hfagtau@#1,ref}{\cite{#4}}%
  \@namedef{hfagtau@#1}{\ensuremath{#5}}%
  \@namedef{hfagtau@#1,val}{\ensuremath{#6}}%
  \@namedef{hfagtau@#1,stat}{\ensuremath{#7}}%
  \@namedef{hfagtau@#1,syst}{\ensuremath{#8}}%
}
\makeatother

\newcommand{\BRF}[2]{#2}
\renewcommand{\BR}{\ensuremath{B}\xspace}
\ifhevea
\renewcommand{\babar}{BaBar\xspace}
\else
\renewcommand{\babar}{\mbox{\slshape B{\smaller A}B{\smaller AR}}\xspace}
\fi
\newcommand{\lepth}{\ensuremath{L}{}\xspace}%
\newcommand{\leptl}{\ensuremath{\ell}\xspace}%
\newcommand{\opdelta}{\ensuremath{r}}%
\newcommand{\Rhad}{\ensuremath{R_{\text{had}}}\xspace}
\newcommand{\BRhad}{\ensuremath{\BR_{\text{had}}}\xspace}
\newcommand{\Gammahad}{\ensuremath{\Gamma_{\text{had}}}\xspace}
\newcommand{\Rstrange}{\ensuremath{R_s}\xspace}
\newcommand{\BRstrange}{\ensuremath{\BR_s}\xspace}
\newcommand{\Gammastrange}{\ensuremath{\Gamma_s}\xspace}
\newcommand{\Rnonstrange}{\ensuremath{R_{\text{VA}}}\xspace}
\newcommand{\BRnonstrange}{\ensuremath{\BR_{\text{VA}}}\xspace}
\newcommand{\Gammanonstrange}{\ensuremath{\Gamma_{\text{VA}}}\xspace}
\newcommand{\tauknu}{\ensuremath{\tau^{-} \to K^{-} \nut}\xspace}
\newcommand{\taupinu}{\ensuremath{\tau^{-} \to \pi^{-} \nut}\xspace}
\newcommand{\BFtautoknu}{\ensuremath{\BR(\tauknu)}\xspace}
\newcommand{\BFtautopinu}{\ensuremath{\BR(\taupinu)}\xspace}
\newcommand{\VusUni}{\ensuremath{\Vus_{\text{uni}}}\xspace}
\newcommand{\VusTauIncl}{\ensuremath{\Vus_{\tau s}}\xspace}
\newcommand{\VusTauKpi}{\ensuremath{\Vus_{\tau K/\pi}}\xspace}
\newcommand{\VusTauKnu}{\ensuremath{\Vus_{\tau K}}\xspace}
\newcommand{\hfagtau}{HFAG-Tau\xspace}

\htdef{UnitarityResid}{\ensuremath{(9.902 \pm 9.850) \cdot 10^{-4}}}%
\htdef{MeasNum}{174}%
\htdef{QuantNum}{103}%
\htdef{BaseQuantNum}{47}%
\htdef{ConstrNum}{56}%
\htdef{Chisq}{142.5}%
\htdef{Dof}{127}%
\htdef{ChisqProb}{16.45\%}%
\htmeasdef{ALEPH.Gamma10.pub.BARATE.99K}{Gamma10}{ALEPH}{Barate:1999hi}{0.00696 \pm 0.0002865}{0.00696}{\pm 0.0002865}{0}%
\htmeasdef{ALEPH.Gamma103.pub.SCHAEL.05C}{Gamma103}{ALEPH}{Schael:2005am}{0.00072 \pm 0.00015}{0.00072}{\pm 0.00015}{0}%
\htmeasdef{ALEPH.Gamma104.pub.SCHAEL.05C}{Gamma104}{ALEPH}{Schael:2005am}{0.00021 \pm 9.21954\cdot 10^{-5}}{0.00021}{\pm 9.21954\cdot 10^{-5}}{0}%
\htmeasdef{ALEPH.Gamma126.pub.BUSKULIC.97C}{Gamma126}{ALEPH}{Buskulic:1996qs}{0.0018 \pm 0.0004472}{0.0018}{\pm 0.0004472}{0}%
\htmeasdef{ALEPH.Gamma128.pub.BUSKULIC.97C}{Gamma128}{ALEPH}{Buskulic:1996qs}{( 2.9 {}^{+1.3\cdot 10^{-4}}_{-1.2} \pm 0.7 ) \cdot 10^{ -4 }}{2.9\cdot 10^{-4}}{{}^{+1.3\cdot 10^{-4}}_{-1.2\cdot 10^{-4}}}{0.7\cdot 10^{-4}}%
\htmeasdef{ALEPH.Gamma13.pub.SCHAEL.05C}{Gamma13}{ALEPH}{Schael:2005am}{0.25924 \pm 0.00128973}{0.25924}{\pm 0.00128973}{0}%
\htmeasdef{ALEPH.Gamma150.pub.BUSKULIC.97C}{Gamma150}{ALEPH}{Buskulic:1996qs}{0.0191 \pm 0.000922}{0.0191}{\pm 0.000922}{0}%
\htmeasdef{ALEPH.Gamma150by66.pub.BUSKULIC.96}{Gamma150by66}{ALEPH}{Buskulic:1995ty}{0.431 \pm 0.033}{0.431}{\pm 0.033}{0}%
\htmeasdef{ALEPH.Gamma152.pub.BUSKULIC.97C}{Gamma152}{ALEPH}{Buskulic:1996qs}{0.0043 \pm 0.000781}{0.0043}{\pm 0.000781}{0}%
\htmeasdef{ALEPH.Gamma16.pub.BARATE.99K}{Gamma16}{ALEPH}{Barate:1999hi}{0.00444 \pm 0.0003538}{0.00444}{\pm 0.0003538}{0}%
\htmeasdef{ALEPH.Gamma19.pub.SCHAEL.05C}{Gamma19}{ALEPH}{Schael:2005am}{0.09295 \pm 0.00121655}{0.09295}{\pm 0.00121655}{0}%
\htmeasdef{ALEPH.Gamma23.pub.BARATE.99K}{Gamma23}{ALEPH}{Barate:1999hi}{0.00056 \pm 0.00025}{0.00056}{\pm 0.00025}{0}%
\htmeasdef{ALEPH.Gamma26.pub.SCHAEL.05C}{Gamma26}{ALEPH}{Schael:2005am}{0.01082 \pm 0.000925581}{0.01082}{\pm 0.000925581}{0}%
\htmeasdef{ALEPH.Gamma28.pub.BARATE.99K}{Gamma28}{ALEPH}{Barate:1999hi}{0.00037 \pm 0.0002371}{0.00037}{\pm 0.0002371}{0}%
\htmeasdef{ALEPH.Gamma3.pub.SCHAEL.05C}{Gamma3}{ALEPH}{Schael:2005am}{0.17319 \pm 0.000769675}{0.17319}{\pm 0.000769675}{0}%
\htmeasdef{ALEPH.Gamma30.pub.SCHAEL.05C}{Gamma30}{ALEPH}{Schael:2005am}{0.00112 \pm 0.000509313}{0.00112}{\pm 0.000509313}{0}%
\htmeasdef{ALEPH.Gamma33.pub.BARATE.98E}{Gamma33}{ALEPH}{Barate:1997tt}{0.0097 \pm 0.000849}{0.0097}{\pm 0.000849}{0}%
\htmeasdef{ALEPH.Gamma35.pub.BARATE.99K}{Gamma35}{ALEPH}{Barate:1999hi}{0.00928 \pm 0.000564}{0.00928}{\pm 0.000564}{0}%
\htmeasdef{ALEPH.Gamma37.pub.BARATE.98E}{Gamma37}{ALEPH}{Barate:1997tt}{0.00158 \pm 0.0004531}{0.00158}{\pm 0.0004531}{0}%
\htmeasdef{ALEPH.Gamma37.pub.BARATE.99K}{Gamma37}{ALEPH}{Barate:1999hi}{0.00162 \pm 0.0002371}{0.00162}{\pm 0.0002371}{0}%
\htmeasdef{ALEPH.Gamma40.pub.BARATE.98E}{Gamma40}{ALEPH}{Barate:1997tt}{0.00294 \pm 0.0008184}{0.00294}{\pm 0.0008184}{0}%
\htmeasdef{ALEPH.Gamma40.pub.BARATE.99K}{Gamma40}{ALEPH}{Barate:1999hi}{0.00347 \pm 0.0006464}{0.00347}{\pm 0.0006464}{0}%
\htmeasdef{ALEPH.Gamma42.pub.BARATE.98E}{Gamma42}{ALEPH}{Barate:1997tt}{0.00152 \pm 0.0007885}{0.00152}{\pm 0.0007885}{0}%
\htmeasdef{ALEPH.Gamma42.pub.BARATE.99K}{Gamma42}{ALEPH}{Barate:1999hi}{0.00143 \pm 0.0002915}{0.00143}{\pm 0.0002915}{0}%
\htmeasdef{ALEPH.Gamma44.pub.BARATE.99R}{Gamma44}{ALEPH}{Barate:1999hj}{0.00026 \pm 0.00024}{0.00026}{\pm 0.00024}{0}%
\htmeasdef{ALEPH.Gamma46.pub.BARATE.98E}{Gamma46}{ALEPH}{Barate:1997tt}{0.00153 \pm 0.00034}{0.00153}{\pm 0.00034}{0}%
\htmeasdef{ALEPH.Gamma47.pub.BARATE.98E}{Gamma47}{ALEPH}{Barate:1997tt}{0.00026 \pm 0.0001118}{0.00026}{\pm 0.0001118}{0}%
\htmeasdef{ALEPH.Gamma48.pub.BARATE.98E}{Gamma48}{ALEPH}{Barate:1997tt}{0.00101 \pm 0.0002642}{0.00101}{\pm 0.0002642}{0}%
\htmeasdef{ALEPH.Gamma5.pub.SCHAEL.05C}{Gamma5}{ALEPH}{Schael:2005am}{0.17837 \pm 0.000804984}{0.17837}{\pm 0.000804984}{0}%
\htmeasdef{ALEPH.Gamma51.pub.BARATE.98E}{Gamma51}{ALEPH}{Barate:1997tt}{( 3.1 \pm 1.1 \pm 0.5 ) \cdot 10^{ -4 }}{3.1\cdot 10^{-4}}{\pm 1.1\cdot 10^{-4}}{0.5\cdot 10^{-4}}%
\htmeasdef{ALEPH.Gamma53.pub.BARATE.98E}{Gamma53}{ALEPH}{Barate:1997tt}{0.00023 \pm 0.000202485}{0.00023}{\pm 0.000202485}{0}%
\htmeasdef{ALEPH.Gamma58.pub.SCHAEL.05C}{Gamma58}{ALEPH}{Schael:2005am}{0.09469 \pm 0.000957758}{0.09469}{\pm 0.000957758}{0}%
\htmeasdef{ALEPH.Gamma66.pub.SCHAEL.05C}{Gamma66}{ALEPH}{Schael:2005am}{0.04734 \pm 0.000766942}{0.04734}{\pm 0.000766942}{0}%
\htmeasdef{ALEPH.Gamma76.pub.SCHAEL.05C}{Gamma76}{ALEPH}{Schael:2005am}{0.00435 \pm 0.000460977}{0.00435}{\pm 0.000460977}{0}%
\htmeasdef{ALEPH.Gamma8.pub.SCHAEL.05C}{Gamma8}{ALEPH}{Schael:2005am}{0.11524 \pm 0.00104805}{0.11524}{\pm 0.00104805}{0}%
\htmeasdef{ALEPH.Gamma805.pub.SCHAEL.05C}{Gamma805}{ALEPH}{Schael:2005am}{( 4 \pm 2 ) \cdot 10^{ -4 }}{4\cdot 10^{-4}}{\pm 2\cdot 10^{-4}}{0}%
\htmeasdef{ALEPH.Gamma85.pub.BARATE.98}{Gamma85}{ALEPH}{Barate:1997ma}{0.00214 \pm 0.0004701}{0.00214}{\pm 0.0004701}{0}%
\htmeasdef{ALEPH.Gamma88.pub.BARATE.98}{Gamma88}{ALEPH}{Barate:1997ma}{0.00061 \pm 0.0004295}{0.00061}{\pm 0.0004295}{0}%
\htmeasdef{ALEPH.Gamma93.pub.BARATE.98}{Gamma93}{ALEPH}{Barate:1997ma}{0.00163 \pm 0.0002702}{0.00163}{\pm 0.0002702}{0}%
\htmeasdef{ALEPH.Gamma94.pub.BARATE.98}{Gamma94}{ALEPH}{Barate:1997ma}{0.00075 \pm 0.0003265}{0.00075}{\pm 0.0003265}{0}%
\htmeasdef{ARGUS.Gamma103.pub.ALBRECHT.88B}{Gamma103}{ARGUS}{Albrecht:1987zf}{0.00064 \pm 0.0002508}{0.00064}{\pm 0.0002508}{0}%
\htmeasdef{ARGUS.Gamma3by5.pub.ALBRECHT.92D}{Gamma3by5}{ARGUS}{Albrecht:1991rh}{0.997 \pm 0.05315}{0.997}{\pm 0.05315}{0}%
\htmeasdef{BaBar.Gamma10by5.pub.AUBERT.10F}{Gamma10by5}{BaBar}{Aubert:2009qj}{0.03882 \pm 0.000630207 \pm 0.000173608}{0.03882}{\pm 0.000630207}{0.000173608}%
\htmeasdef{BaBar.Gamma128.pub.DEL-AMO-SANCHEZ.11E}{Gamma128}{BaBar}{delAmoSanchez:2010pc}{0.000142 \pm 1.1\cdot 10^{-5} \pm 7\cdot 10^{-6}}{0.000142}{\pm 1.1\cdot 10^{-5}}{7\cdot 10^{-6}}%
\htmeasdef{BaBar.Gamma16.pub.AUBERT.07AP}{Gamma16}{BaBar}{Aubert:2007jh}{0.00416 \pm 3\cdot 10^{-5} \pm 0.00018}{0.00416}{\pm 3\cdot 10^{-5}}{0.00018}%
\htmeasdef{BaBar.Gamma35.prelim.ICHEP08}{Gamma35}{BaBar}{Aubert:2008an}{0.0084 \pm 4\cdot 10^{-5} \pm 0.00023}{0.0084}{\pm 4\cdot 10^{-5}}{0.00023}%
\htmeasdef{BaBar.Gamma3by5.pub.AUBERT.10F}{Gamma3by5}{BaBar}{Aubert:2009qj}{0.9796 \pm 0.00390406 \pm 0.00052753}{0.9796}{\pm 0.00390406}{0.00052753}%
\htmeasdef{BaBar.Gamma40.prelim.DPF09}{Gamma40}{BaBar}{Paramesvaran:2009ec}{0.00342 \pm 6\cdot 10^{-5} \pm 0.00015}{0.00342}{\pm 6\cdot 10^{-5}}{0.00015}%
\htmeasdef{BaBar.Gamma47.pub.LEES.2012Y}{Gamma47}{BaBar}{Lees:2012de}{( 2.31 \pm 0.04 \pm 0.08 ) \cdot 10^{ -4 }}{2.31\cdot 10^{-4}}{\pm 0.04\cdot 10^{-4}}{0.08\cdot 10^{-4}}%
\htmeasdef{BaBar.Gamma50.pub.LEES.2012Y}{Gamma50}{BaBar}{Lees:2012de}{( 1.60 \pm 0.20 \pm 0.22 ) \cdot 10^{ -5 }}{1.60\cdot 10^{-5}}{\pm 0.20\cdot 10^{-5}}{0.22\cdot 10^{-5}}%
\htmeasdef{BaBar.Gamma60.pub.AUBERT.08}{Gamma60}{BaBar}{Aubert:2007mh}{0.088337 \pm 7.4\cdot 10^{-5} \pm 0.00126724}{0.088337}{\pm 7.4\cdot 10^{-5}}{0.00126724}%
\htmeasdef{BaBar.Gamma811.pub.LEES.2012X}{Gamma811}{BaBar}{Lees:2012ks}{( 7.3 \pm 1.2 \pm 1.2 ) \cdot 10^{ -5 }}{7.3\cdot 10^{-5}}{\pm 1.2\cdot 10^{-5}}{1.2\cdot 10^{-5}}%
\htmeasdef{BaBar.Gamma812.pub.LEES.2012X}{Gamma812}{BaBar}{Lees:2012ks}{( 0.1 \pm 0.08 \pm 0.30 ) \cdot 10^{ -4 }}{0.1\cdot 10^{-4}}{\pm 0.08\cdot 10^{-4}}{0.30\cdot 10^{-4}}%
\htmeasdef{BaBar.Gamma821.pub.LEES.2012X}{Gamma821}{BaBar}{Lees:2012ks}{( 7.68 \pm 0.04 \pm 0.40 ) \cdot 10^{ -4 }}{7.68\cdot 10^{-4}}{\pm 0.04\cdot 10^{-4}}{0.40\cdot 10^{-4}}%
\htmeasdef{BaBar.Gamma822.pub.LEES.2012X}{Gamma822}{BaBar}{Lees:2012ks}{( 0.6 \pm 0.5 \pm 1.1 ) \cdot 10^{ -6 }}{0.6\cdot 10^{-6}}{\pm 0.5\cdot 10^{-6}}{1.1\cdot 10^{-6}}%
\htmeasdef{BaBar.Gamma831.pub.LEES.2012X}{Gamma831}{BaBar}{Lees:2012ks}{( 8.4 \pm 0.4 \pm 0.6 ) \cdot 10^{ -5 }}{8.4\cdot 10^{-5}}{\pm 0.4\cdot 10^{-5}}{0.6\cdot 10^{-5}}%
\htmeasdef{BaBar.Gamma832.pub.LEES.2012X}{Gamma832}{BaBar}{Lees:2012ks}{( 0.36 \pm 0.03 \pm 0.09 ) \cdot 10^{ -4 }}{0.36\cdot 10^{-4}}{\pm 0.03\cdot 10^{-4}}{0.09\cdot 10^{-4}}%
\htmeasdef{BaBar.Gamma833.pub.LEES.2012X}{Gamma833}{BaBar}{Lees:2012ks}{( 1.1 \pm 0.4 \pm 0.4 ) \cdot 10^{ -6 }}{1.1\cdot 10^{-6}}{\pm 0.4\cdot 10^{-6}}{0.4\cdot 10^{-6}}%
\htmeasdef{BaBar.Gamma85.pub.AUBERT.08}{Gamma85}{BaBar}{Aubert:2007mh}{0.0027257 \pm 1.8\cdot 10^{-5} \pm 9.2441\cdot 10^{-5}}{0.0027257}{\pm 1.8\cdot 10^{-5}}{9.2441\cdot 10^{-5}}%
\htmeasdef{BaBar.Gamma910.pub.LEES.2012X}{Gamma910}{BaBar}{Lees:2012ks}{( 8.27 \pm 0.88 \pm 0.81 ) \cdot 10^{ -5 }}{8.27\cdot 10^{-5}}{\pm 0.88\cdot 10^{-5}}{0.81\cdot 10^{-5}}%
\htmeasdef{BaBar.Gamma911.pub.LEES.2012X}{Gamma911}{BaBar}{Lees:2012ks}{( 4.57 \pm 0.77 \pm 0.50 ) \cdot 10^{ -5 }}{4.57\cdot 10^{-5}}{\pm 0.77\cdot 10^{-5}}{0.50\cdot 10^{-5}}%
\htmeasdef{BaBar.Gamma920.pub.LEES.2012X}{Gamma920}{BaBar}{Lees:2012ks}{( 5.20 \pm 0.31 \pm 0.37 ) \cdot 10^{ -5 }}{5.20\cdot 10^{-5}}{\pm 0.31\cdot 10^{-5}}{0.37\cdot 10^{-5}}%
\htmeasdef{BaBar.Gamma93.pub.AUBERT.08}{Gamma93}{BaBar}{Aubert:2007mh}{0.0013461 \pm 1\cdot 10^{-5} \pm 3.6413\cdot 10^{-5}}{0.0013461}{\pm 1\cdot 10^{-5}}{3.6413\cdot 10^{-5}}%
\htmeasdef{BaBar.Gamma930.pub.LEES.2012X}{Gamma930}{BaBar}{Lees:2012ks}{( 5.39 \pm 0.27 \pm 0.41 ) \cdot 10^{ -5 }}{5.39\cdot 10^{-5}}{\pm 0.27\cdot 10^{-5}}{0.41\cdot 10^{-5}}%
\htmeasdef{BaBar.Gamma944.pub.LEES.2012X}{Gamma944}{BaBar}{Lees:2012ks}{( 8.26 \pm 0.35 \pm 0.51 ) \cdot 10^{ -5 }}{8.26\cdot 10^{-5}}{\pm 0.35\cdot 10^{-5}}{0.51\cdot 10^{-5}}%
\htmeasdef{BaBar.Gamma96.pub.AUBERT.08}{Gamma96}{BaBar}{Aubert:2007mh}{1.5777\cdot 10^{-5} \pm 1.3\cdot 10^{-6} \pm 1.2308\cdot 10^{-6}}{1.5777\cdot 10^{-5}}{\pm 1.3\cdot 10^{-6}}{1.2308\cdot 10^{-6}}%
\htmeasdef{BaBar.Gamma9by5.pub.AUBERT.10F}{Gamma9by5}{BaBar}{Aubert:2009qj}{0.5945 \pm 0.00574448 \pm 0.00248413}{0.5945}{\pm 0.00574448}{0.00248413}%
\htmeasdef{Belle.Gamma126.pub.INAMI.09}{Gamma126}{Belle}{Inami:2008ar}{0.00135 \pm 3\cdot 10^{-5} \pm 7\cdot 10^{-5}}{0.00135}{\pm 3\cdot 10^{-5}}{7\cdot 10^{-5}}%
\htmeasdef{Belle.Gamma128.pub.INAMI.09}{Gamma128}{Belle}{Inami:2008ar}{0.000158 \pm 5\cdot 10^{-6} \pm 9\cdot 10^{-6}}{0.000158}{\pm 5\cdot 10^{-6}}{9\cdot 10^{-6}}%
\htmeasdef{Belle.Gamma13.pub.FUJIKAWA.08}{Gamma13}{Belle}{Fujikawa:2008ma}{0.2567 \pm 1\cdot 10^{-4} \pm 0.0039}{0.2567}{\pm 1\cdot 10^{-4}}{0.0039}%
\htmeasdef{Belle.Gamma130.pub.INAMI.09}{Gamma130}{Belle}{Inami:2008ar}{4.6\cdot 10^{-5} \pm 1.1\cdot 10^{-5} \pm 4\cdot 10^{-6}}{4.6\cdot 10^{-5}}{\pm 1.1\cdot 10^{-5}}{4\cdot 10^{-6}}%
\htmeasdef{Belle.Gamma132.pub.INAMI.09}{Gamma132}{Belle}{Inami:2008ar}{8.8\cdot 10^{-5} \pm 1.4\cdot 10^{-5} \pm 6\cdot 10^{-6}}{8.8\cdot 10^{-5}}{\pm 1.4\cdot 10^{-5}}{6\cdot 10^{-6}}%
\htmeasdef{Belle.Gamma33.pub.Ryu:2014vpc}{Gamma33}{Belle}{Ryu:2014vpc}{( 9.15 \pm 0.01 \pm 0.15 ) \cdot 10^{ -3 }}{9.15\cdot 10^{-3}}{\pm 0.01\cdot 10^{-3}}{0.15\cdot 10^{-3}}%
\htmeasdef{Belle.Gamma35.pub.Ryu:2014vpc}{Gamma35}{Belle}{Ryu:2014vpc}{( 8.32 \pm 0.02 \pm 0.16 ) \cdot 10^{ -3 }}{8.32\cdot 10^{-3}}{\pm 0.02\cdot 10^{-3}}{0.16\cdot 10^{-3}}%
\htmeasdef{Belle.Gamma37.pub.Ryu:2014vpc}{Gamma37}{Belle}{Ryu:2014vpc}{( 14.8 \pm 0.14 \pm 0.54 ) \cdot 10^{ -4 }}{14.8\cdot 10^{-4}}{\pm 0.14\cdot 10^{-4}}{0.54\cdot 10^{-4}}%
\htmeasdef{Belle.Gamma40.pub.Ryu:2014vpc}{Gamma40}{Belle}{Ryu:2014vpc}{( 3.86 \pm 0.04 \pm 0.14 ) \cdot 10^{ -3 }}{3.86\cdot 10^{-3}}{\pm 0.04\cdot 10^{-3}}{0.14\cdot 10^{-3}}%
\htmeasdef{Belle.Gamma42.pub.Ryu:2014vpc}{Gamma42}{Belle}{Ryu:2014vpc}{( 14.96 \pm 0.20 \pm 0.74 ) \cdot 10^{ -4 }}{14.96\cdot 10^{-4}}{\pm 0.20\cdot 10^{-4}}{0.74\cdot 10^{-4}}%
\htmeasdef{Belle.Gamma47.pub.Ryu:2014vpc}{Gamma47}{Belle}{Ryu:2014vpc}{( 2.33 \pm 0.03 \pm 0.09 ) \cdot 10^{ -4 }}{2.33\cdot 10^{-4}}{\pm 0.03\cdot 10^{-4}}{0.09\cdot 10^{-4}}%
\htmeasdef{Belle.Gamma50.pub.Ryu:2014vpc}{Gamma50}{Belle}{Ryu:2014vpc}{( 2.00 \pm 0.22 \pm 0.20 ) \cdot 10^{ -5 }}{2.00\cdot 10^{-5}}{\pm 0.22\cdot 10^{-5}}{0.20\cdot 10^{-5}}%
\htmeasdef{Belle.Gamma60.pub.LEE.10}{Gamma60}{Belle}{Lee:2010tc}{0.0842 \pm 3.3211\cdot 10^{-5} \pm 0.0025879}{0.0842}{\pm 3.3211\cdot 10^{-5}}{0.0025879}%
\htmeasdef{Belle.Gamma85.pub.LEE.10}{Gamma85}{Belle}{Lee:2010tc}{0.0033 \pm 1.274\cdot 10^{-5} \pm 0.00016625}{0.0033}{\pm 1.274\cdot 10^{-5}}{0.00016625}%
\htmeasdef{Belle.Gamma93.pub.LEE.10}{Gamma93}{Belle}{Lee:2010tc}{0.00155 \pm 6.575\cdot 10^{-6} \pm 5.5579\cdot 10^{-5}}{0.00155}{\pm 6.575\cdot 10^{-6}}{5.5579\cdot 10^{-5}}%
\htmeasdef{Belle.Gamma96.pub.LEE.10}{Gamma96}{Belle}{Lee:2010tc}{3.29\cdot 10^{-5} \pm 1.6941\cdot 10^{-6} \pm 1.9621\cdot 10^{-6}}{3.29\cdot 10^{-5}}{\pm 1.6941\cdot 10^{-6}}{1.9621\cdot 10^{-6}}%
\htmeasdef{CELLO.Gamma54.pub.BEHREND.89B}{Gamma54}{CELLO}{Behrend:1989wc}{0.15 \pm 0.005}{0.15}{\pm 0.005}{0}%
\htmeasdef{CLEO.Gamma10.pub.BATTLE.94}{Gamma10}{CLEO}{Battle:1994by}{0.0066 \pm 0.00114}{0.0066}{\pm 0.00114}{0}%
\htmeasdef{CLEO.Gamma102.pub.GIBAUT.94B}{Gamma102}{CLEO}{Gibaut:1994ik}{0.00097 \pm 0.0001208}{0.00097}{\pm 0.0001208}{0}%
\htmeasdef{CLEO.Gamma103.pub.GIBAUT.94B}{Gamma103}{CLEO}{Gibaut:1994ik}{0.00077 \pm 0.000103}{0.00077}{\pm 0.000103}{0}%
\htmeasdef{CLEO.Gamma104.pub.ANASTASSOV.01}{Gamma104}{CLEO}{Anastassov:2000xu}{0.00017 \pm 2.828\cdot 10^{-5}}{0.00017}{\pm 2.828\cdot 10^{-5}}{0}%
\htmeasdef{CLEO.Gamma126.pub.ARTUSO.92}{Gamma126}{CLEO}{Artuso:1992qu}{0.0017 \pm 0.0002828}{0.0017}{\pm 0.0002828}{0}%
\htmeasdef{CLEO.Gamma128.pub.BARTELT.96}{Gamma128}{CLEO}{Bartelt:1996iv}{( 2.6 \pm 0.5 \pm 0.5 ) \cdot 10^{ -4 }}{2.6\cdot 10^{-4}}{\pm 0.5\cdot 10^{-4}}{0.5\cdot 10^{-4}}%
\htmeasdef{CLEO.Gamma13.pub.ARTUSO.94}{Gamma13}{CLEO}{Artuso:1994ii}{0.2587 \pm 0.004368}{0.2587}{\pm 0.004368}{0}%
\htmeasdef{CLEO.Gamma130.pub.BISHAI.99}{Gamma130}{CLEO}{Bishai:1998gf}{0.000177 \pm 9.04268\cdot 10^{-5}}{0.000177}{\pm 9.04268\cdot 10^{-5}}{0}%
\htmeasdef{CLEO.Gamma132.pub.BISHAI.99}{Gamma132}{CLEO}{Bishai:1998gf}{0.00022 \pm 7.33757\cdot 10^{-5}}{0.00022}{\pm 7.33757\cdot 10^{-5}}{0}%
\htmeasdef{CLEO.Gamma150.pub.BARINGER.87}{Gamma150}{CLEO}{Baringer:1987tr}{0.016 \pm 0.004909}{0.016}{\pm 0.004909}{0}%
\htmeasdef{CLEO.Gamma150by66.pub.BALEST.95C}{Gamma150by66}{CLEO}{Balest:1995kq}{0.464 \pm 0.02335}{0.464}{\pm 0.02335}{0}%
\htmeasdef{CLEO.Gamma152by76.pub.BORTOLETTO.93}{Gamma152by76}{CLEO}{Bortoletto:1993px}{0.81 \pm 0.08485}{0.81}{\pm 0.08485}{0}%
\htmeasdef{CLEO.Gamma16.pub.BATTLE.94}{Gamma16}{CLEO}{Battle:1994by}{0.0051 \pm 0.001221}{0.0051}{\pm 0.001221}{0}%
\htmeasdef{CLEO.Gamma19by13.pub.PROCARIO.93}{Gamma19by13}{CLEO}{Procario:1992hd}{0.342 \pm 0.01709}{0.342}{\pm 0.01709}{0}%
\htmeasdef{CLEO.Gamma23.pub.BATTLE.94}{Gamma23}{CLEO}{Battle:1994by}{9\cdot 10^{-4} \pm 0.001044}{9\cdot 10^{-4}}{\pm 0.001044}{0}%
\htmeasdef{CLEO.Gamma26by13.pub.PROCARIO.93}{Gamma26by13}{CLEO}{Procario:1992hd}{0.044 \pm 0.005831}{0.044}{\pm 0.005831}{0}%
\htmeasdef{CLEO.Gamma29.pub.PROCARIO.93}{Gamma29}{CLEO}{Procario:1992hd}{0.0016 \pm 0.0007071}{0.0016}{\pm 0.0007071}{0}%
\htmeasdef{CLEO.Gamma31.pub.BATTLE.94}{Gamma31}{CLEO}{Battle:1994by}{0.017 \pm 0.002247}{0.017}{\pm 0.002247}{0}%
\htmeasdef{CLEO.Gamma34.pub.COAN.96}{Gamma34}{CLEO}{Coan:1996iu}{0.00855 \pm 0.0008139}{0.00855}{\pm 0.0008139}{0}%
\htmeasdef{CLEO.Gamma37.pub.COAN.96}{Gamma37}{CLEO}{Coan:1996iu}{0.00151 \pm 0.0003041}{0.00151}{\pm 0.0003041}{0}%
\htmeasdef{CLEO.Gamma39.pub.COAN.96}{Gamma39}{CLEO}{Coan:1996iu}{0.00562 \pm 0.0006931}{0.00562}{\pm 0.0006931}{0}%
\htmeasdef{CLEO.Gamma3by5.pub.ANASTASSOV.97}{Gamma3by5}{CLEO}{Anastassov:1996tc}{0.9777 \pm 0.01074}{0.9777}{\pm 0.01074}{0}%
\htmeasdef{CLEO.Gamma42.pub.COAN.96}{Gamma42}{CLEO}{Coan:1996iu}{0.00145 \pm 0.0004118}{0.00145}{\pm 0.0004118}{0}%
\htmeasdef{CLEO.Gamma47.pub.COAN.96}{Gamma47}{CLEO}{Coan:1996iu}{0.00023 \pm 5.831\cdot 10^{-5}}{0.00023}{\pm 5.831\cdot 10^{-5}}{0}%
\htmeasdef{CLEO.Gamma5.pub.ANASTASSOV.97}{Gamma5}{CLEO}{Anastassov:1996tc}{0.1776 \pm 0.001803}{0.1776}{\pm 0.001803}{0}%
\htmeasdef{CLEO.Gamma57.pub.BALEST.95C}{Gamma57}{CLEO}{Balest:1995kq}{0.0951 \pm 0.002119}{0.0951}{\pm 0.002119}{0}%
\htmeasdef{CLEO.Gamma66.pub.BALEST.95C}{Gamma66}{CLEO}{Balest:1995kq}{0.0423 \pm 0.00228}{0.0423}{\pm 0.00228}{0}%
\htmeasdef{CLEO.Gamma69.pub.EDWARDS.00A}{Gamma69}{CLEO}{Edwards:1999fj}{0.0419 \pm 0.002326}{0.0419}{\pm 0.002326}{0}%
\htmeasdef{CLEO.Gamma76by54.pub.BORTOLETTO.93}{Gamma76by54}{CLEO}{Bortoletto:1993px}{0.034 \pm 0.003606}{0.034}{\pm 0.003606}{0}%
\htmeasdef{CLEO.Gamma78.pub.ANASTASSOV.01}{Gamma78}{CLEO}{Anastassov:2000xu}{0.00022 \pm 5\cdot 10^{-5}}{0.00022}{\pm 5\cdot 10^{-5}}{0}%
\htmeasdef{CLEO.Gamma8.pub.ANASTASSOV.97}{Gamma8}{CLEO}{Anastassov:1996tc}{0.1152 \pm 0.0013}{0.1152}{\pm 0.0013}{0}%
\htmeasdef{CLEO.Gamma80by60.pub.RICHICHI.99}{Gamma80by60}{CLEO}{Richichi:1998bc}{0.0544 \pm 0.005701}{0.0544}{\pm 0.005701}{0}%
\htmeasdef{CLEO.Gamma81by69.pub.RICHICHI.99}{Gamma81by69}{CLEO}{Richichi:1998bc}{0.0261 \pm 0.006155}{0.0261}{\pm 0.006155}{0}%
\htmeasdef{CLEO.Gamma93by60.pub.RICHICHI.99}{Gamma93by60}{CLEO}{Richichi:1998bc}{0.016 \pm 0.003354}{0.016}{\pm 0.003354}{0}%
\htmeasdef{CLEO.Gamma94by69.pub.RICHICHI.99}{Gamma94by69}{CLEO}{Richichi:1998bc}{0.0079 \pm 0.004682}{0.0079}{\pm 0.004682}{0}%
\htmeasdef{CLEO3.Gamma151.pub.ARMS.05}{Gamma151}{CLEO3}{Arms:2005qg}{0.00041 \pm 9.21954\cdot 10^{-5}}{0.00041}{\pm 9.21954\cdot 10^{-5}}{0}%
\htmeasdef{CLEO3.Gamma60.pub.BRIERE.03}{Gamma60}{CLEO3}{Briere:2003fr}{0.0913 \pm 0.004627}{0.0913}{\pm 0.004627}{0}%
\htmeasdef{CLEO3.Gamma85.pub.BRIERE.03}{Gamma85}{CLEO3}{Briere:2003fr}{0.00384 \pm 0.000405}{0.00384}{\pm 0.000405}{0}%
\htmeasdef{CLEO3.Gamma88.pub.ARMS.05}{Gamma88}{CLEO3}{Arms:2005qg}{0.00074 \pm 0.000136}{0.00074}{\pm 0.000136}{0}%
\htmeasdef{CLEO3.Gamma93.pub.BRIERE.03}{Gamma93}{CLEO3}{Briere:2003fr}{0.00155 \pm 0.0001082}{0.00155}{\pm 0.0001082}{0}%
\htmeasdef{CLEO3.Gamma94.pub.ARMS.05}{Gamma94}{CLEO3}{Arms:2005qg}{( 5.5 \pm 1.844 ) \cdot 10^{ -5 }}{5.5\cdot 10^{-5}}{\pm 1.844\cdot 10^{-5}}{0}%
\htmeasdef{DELPHI.Gamma10.pub.ABREU.94K}{Gamma10}{DELPHI}{Abreu:1994fi}{0.0085 \pm 0.0018}{0.0085}{\pm 0.0018}{0}%
\htmeasdef{DELPHI.Gamma103.pub.ABDALLAH.06A}{Gamma103}{DELPHI}{Abdallah:2003cw}{0.00097 \pm 0.0001581}{0.00097}{\pm 0.0001581}{0}%
\htmeasdef{DELPHI.Gamma104.pub.ABDALLAH.06A}{Gamma104}{DELPHI}{Abdallah:2003cw}{0.00016 \pm 0.0001342}{0.00016}{\pm 0.0001342}{0}%
\htmeasdef{DELPHI.Gamma13.pub.ABDALLAH.06A}{Gamma13}{DELPHI}{Abdallah:2003cw}{0.2574 \pm 0.002438}{0.2574}{\pm 0.002438}{0}%
\htmeasdef{DELPHI.Gamma19.pub.ABDALLAH.06A}{Gamma19}{DELPHI}{Abdallah:2003cw}{0.09498 \pm 0.004219}{0.09498}{\pm 0.004219}{0}%
\htmeasdef{DELPHI.Gamma25.pub.ABDALLAH.06A}{Gamma25}{DELPHI}{Abdallah:2003cw}{0.01403 \pm 0.003098}{0.01403}{\pm 0.003098}{0}%
\htmeasdef{DELPHI.Gamma3.pub.ABREU.99X}{Gamma3}{DELPHI}{Abreu:1999rb}{0.17325 \pm 0.001223}{0.17325}{\pm 0.001223}{0}%
\htmeasdef{DELPHI.Gamma31.pub.ABREU.94K}{Gamma31}{DELPHI}{Abreu:1994fi}{0.0154 \pm 0.0024}{0.0154}{\pm 0.0024}{0}%
\htmeasdef{DELPHI.Gamma5.pub.ABREU.99X}{Gamma5}{DELPHI}{Abreu:1999rb}{0.17877 \pm 0.001549}{0.17877}{\pm 0.001549}{0}%
\htmeasdef{DELPHI.Gamma57.pub.ABDALLAH.06A}{Gamma57}{DELPHI}{Abdallah:2003cw}{0.09317 \pm 0.001218}{0.09317}{\pm 0.001218}{0}%
\htmeasdef{DELPHI.Gamma66.pub.ABDALLAH.06A}{Gamma66}{DELPHI}{Abdallah:2003cw}{0.04545 \pm 0.001478}{0.04545}{\pm 0.001478}{0}%
\htmeasdef{DELPHI.Gamma7.pub.ABREU.92N}{Gamma7}{DELPHI}{Abreu:1992gn}{0.124 \pm 0.009899}{0.124}{\pm 0.009899}{0}%
\htmeasdef{DELPHI.Gamma74.pub.ABDALLAH.06A}{Gamma74}{DELPHI}{Abdallah:2003cw}{0.00561 \pm 0.001168}{0.00561}{\pm 0.001168}{0}%
\htmeasdef{DELPHI.Gamma8.pub.ABDALLAH.06A}{Gamma8}{DELPHI}{Abdallah:2003cw}{0.11571 \pm 0.001655}{0.11571}{\pm 0.001655}{0}%
\htmeasdef{HRS.Gamma102.pub.BYLSMA.87}{Gamma102}{HRS}{Bylsma:1986zy}{0.00102 \pm 0.00029}{0.00102}{\pm 0.00029}{0}%
\htmeasdef{HRS.Gamma103.pub.BYLSMA.87}{Gamma103}{HRS}{Bylsma:1986zy}{0.00051 \pm 2\cdot 10^{-4}}{0.00051}{\pm 2\cdot 10^{-4}}{0}%
\htmeasdef{L3.Gamma102.pub.ACHARD.01D}{Gamma102}{L3}{Achard:2001pk}{0.0017 \pm 0.0003406}{0.0017}{\pm 0.0003406}{0}%
\htmeasdef{L3.Gamma13.pub.ACCIARRI.95}{Gamma13}{L3}{Acciarri:1994vr}{0.2505 \pm 0.006103}{0.2505}{\pm 0.006103}{0}%
\htmeasdef{L3.Gamma19.pub.ACCIARRI.95}{Gamma19}{L3}{Acciarri:1994vr}{0.0888 \pm 0.005597}{0.0888}{\pm 0.005597}{0}%
\htmeasdef{L3.Gamma26.pub.ACCIARRI.95}{Gamma26}{L3}{Acciarri:1994vr}{0.017 \pm 0.004494}{0.017}{\pm 0.004494}{0}%
\htmeasdef{L3.Gamma3.pub.ACCIARRI.01F}{Gamma3}{L3}{Acciarri:2001sg}{0.17342 \pm 0.001288}{0.17342}{\pm 0.001288}{0}%
\htmeasdef{L3.Gamma35.pub.ACCIARRI.95F}{Gamma35}{L3}{Acciarri:1995kx}{0.0095 \pm 0.001616}{0.0095}{\pm 0.001616}{0}%
\htmeasdef{L3.Gamma40.pub.ACCIARRI.95F}{Gamma40}{L3}{Acciarri:1995kx}{0.0041 \pm 0.001237}{0.0041}{\pm 0.001237}{0}%
\htmeasdef{L3.Gamma5.pub.ACCIARRI.01F}{Gamma5}{L3}{Acciarri:2001sg}{0.17806 \pm 0.001288}{0.17806}{\pm 0.001288}{0}%
\htmeasdef{L3.Gamma54.pub.ADEVA.91F}{Gamma54}{L3}{Adeva:1991qq}{0.144 \pm 0.006708}{0.144}{\pm 0.006708}{0}%
\htmeasdef{L3.Gamma55.pub.ACHARD.01D}{Gamma55}{L3}{Achard:2001pk}{0.14556 \pm 0.001296}{0.14556}{\pm 0.001296}{0}%
\htmeasdef{L3.Gamma7.pub.ACCIARRI.95}{Gamma7}{L3}{Acciarri:1994vr}{0.1247 \pm 0.005025}{0.1247}{\pm 0.005025}{0}%
\htmeasdef{OPAL.Gamma10.pub.ABBIENDI.01J}{Gamma10}{OPAL}{Abbiendi:2000ee}{0.00658 \pm 0.0003962}{0.00658}{\pm 0.0003962}{0}%
\htmeasdef{OPAL.Gamma103.pub.ACKERSTAFF.99E}{Gamma103}{OPAL}{Ackerstaff:1998ia}{0.00091 \pm 0.0001523}{0.00091}{\pm 0.0001523}{0}%
\htmeasdef{OPAL.Gamma104.pub.ACKERSTAFF.99E}{Gamma104}{OPAL}{Ackerstaff:1998ia}{0.00027 \pm 0.0002012}{0.00027}{\pm 0.0002012}{0}%
\htmeasdef{OPAL.Gamma13.pub.ACKERSTAFF.98M}{Gamma13}{OPAL}{Ackerstaff:1997tx}{0.2589 \pm 0.003362}{0.2589}{\pm 0.003362}{0}%
\htmeasdef{OPAL.Gamma16.pub.ABBIENDI.04J}{Gamma16}{OPAL}{Abbiendi:2004xa}{0.00471 \pm 0.0006332}{0.00471}{\pm 0.0006332}{0}%
\htmeasdef{OPAL.Gamma17.pub.ACKERSTAFF.98M}{Gamma17}{OPAL}{Ackerstaff:1997tx}{0.0991 \pm 0.004111}{0.0991}{\pm 0.004111}{0}%
\htmeasdef{OPAL.Gamma3.pub.ABBIENDI.03}{Gamma3}{OPAL}{Abbiendi:2002jw}{0.1734 \pm 0.001082}{0.1734}{\pm 0.001082}{0}%
\htmeasdef{OPAL.Gamma31.pub.ABBIENDI.01J}{Gamma31}{OPAL}{Abbiendi:2000ee}{0.01528 \pm 0.0005587}{0.01528}{\pm 0.0005587}{0}%
\htmeasdef{OPAL.Gamma33.pub.AKERS.94G}{Gamma33}{OPAL}{Akers:1994td}{0.0097 \pm 0.001082}{0.0097}{\pm 0.001082}{0}%
\htmeasdef{OPAL.Gamma35.pub.ABBIENDI.00C}{Gamma35}{OPAL}{Abbiendi:1999pm}{0.00933 \pm 0.0008382}{0.00933}{\pm 0.0008382}{0}%
\htmeasdef{OPAL.Gamma38.pub.ABBIENDI.00C}{Gamma38}{OPAL}{Abbiendi:1999pm}{0.0033 \pm 0.0006742}{0.0033}{\pm 0.0006742}{0}%
\htmeasdef{OPAL.Gamma43.pub.ABBIENDI.00C}{Gamma43}{OPAL}{Abbiendi:1999pm}{0.00324 \pm 0.000991564}{0.00324}{\pm 0.000991564}{0}%
\htmeasdef{OPAL.Gamma5.pub.ABBIENDI.99H}{Gamma5}{OPAL}{Abbiendi:1998cx}{0.1781 \pm 0.001082}{0.1781}{\pm 0.001082}{0}%
\htmeasdef{OPAL.Gamma55.pub.AKERS.95Y}{Gamma55}{OPAL}{Akers:1995ry}{0.1496 \pm 0.002377}{0.1496}{\pm 0.002377}{0}%
\htmeasdef{OPAL.Gamma57by55.pub.AKERS.95Y}{Gamma57by55}{OPAL}{Akers:1995ry}{0.66 \pm 0.01456}{0.66}{\pm 0.01456}{0}%
\htmeasdef{OPAL.Gamma7.pub.ALEXANDER.91D}{Gamma7}{OPAL}{Alexander:1991am}{0.121 \pm 0.008602}{0.121}{\pm 0.008602}{0}%
\htmeasdef{OPAL.Gamma8.pub.ACKERSTAFF.98M}{Gamma8}{OPAL}{Ackerstaff:1997tx}{0.1198 \pm 0.002062}{0.1198}{\pm 0.002062}{0}%
\htmeasdef{OPAL.Gamma85.pub.ABBIENDI.04J}{Gamma85}{OPAL}{Abbiendi:2004xa}{0.00415 \pm 0.000664}{0.00415}{\pm 0.000664}{0}%
\htmeasdef{OPAL.Gamma92.pub.ABBIENDI.00D}{Gamma92}{OPAL}{Abbiendi:1999cq}{0.00159 \pm 0.0005665}{0.00159}{\pm 0.0005665}{0}%
\htmeasdef{TPC.Gamma54.pub.AIHARA.87B}{Gamma54}{TPC}{Aihara:1986mw}{0.151 \pm 0.01}{0.151}{\pm 0.01}{0}%
\htmeasdef{TPC.Gamma82.pub.BAUER.94}{Gamma82}{TPC}{Bauer:1993wn}{0.0058 \pm 0.001845}{0.0058}{\pm 0.001845}{0}%
\htmeasdef{TPC.Gamma92.pub.BAUER.94}{Gamma92}{TPC}{Bauer:1993wn}{0.0015 \pm 0.00085515}{0.0015}{\pm 0.00085515}{0}%
\htdef{Gamma3.qt}{\ensuremath{0.17391 \pm 0.00040}}%
\htdef{ALEPH.Gamma3.pub.SCHAEL.05C,qt}{\ensuremath{0.17319 \pm 0.00077 \pm 0.00000}}%
\htdef{DELPHI.Gamma3.pub.ABREU.99X,qt}{\ensuremath{0.17325 \pm 0.00122 \pm 0.00000}}%
\htdef{L3.Gamma3.pub.ACCIARRI.01F,qt}{\ensuremath{0.17342 \pm 0.00129 \pm 0.00000}}%
\htdef{OPAL.Gamma3.pub.ABBIENDI.03,qt}{\ensuremath{0.17340 \pm 0.00108 \pm 0.00000}}%
\htdef{Gamma3by5.qt}{\ensuremath{0.97610 \pm 0.00278}}%
\htdef{ARGUS.Gamma3by5.pub.ALBRECHT.92D,qt}{\ensuremath{0.99700 \pm 0.05315 \pm 0.00000}}%
\htdef{BaBar.Gamma3by5.pub.AUBERT.10F,qt}{\ensuremath{0.97960 \pm 0.00390 \pm 0.00053}}%
\htdef{CLEO.Gamma3by5.pub.ANASTASSOV.97,qt}{\ensuremath{0.97770 \pm 0.01074 \pm 0.00000}}%
\htdef{Gamma5.qt}{\ensuremath{0.17817 \pm 0.00041}}%
\htdef{ALEPH.Gamma5.pub.SCHAEL.05C,qt}{\ensuremath{0.17837 \pm 0.00080 \pm 0.00000}}%
\htdef{CLEO.Gamma5.pub.ANASTASSOV.97,qt}{\ensuremath{0.17760 \pm 0.00180 \pm 0.00000}}%
\htdef{DELPHI.Gamma5.pub.ABREU.99X,qt}{\ensuremath{0.17877 \pm 0.00155 \pm 0.00000}}%
\htdef{L3.Gamma5.pub.ACCIARRI.01F,qt}{\ensuremath{0.17806 \pm 0.00129 \pm 0.00000}}%
\htdef{OPAL.Gamma5.pub.ABBIENDI.99H,qt}{\ensuremath{0.17810 \pm 0.00108 \pm 0.00000}}%
\htdef{Gamma7.qt}{\ensuremath{0.12026 \pm 0.00054}}%
\htdef{DELPHI.Gamma7.pub.ABREU.92N,qt}{\ensuremath{0.12400 \pm 0.00990 \pm 0.00000}}%
\htdef{L3.Gamma7.pub.ACCIARRI.95,qt}{\ensuremath{0.12470 \pm 0.00502 \pm 0.00000}}%
\htdef{OPAL.Gamma7.pub.ALEXANDER.91D,qt}{\ensuremath{0.12100 \pm 0.00860 \pm 0.00000}}%
\htdef{Gamma8.qt}{\ensuremath{0.11509 \pm 0.00054}}%
\htdef{ALEPH.Gamma8.pub.SCHAEL.05C,qt}{\ensuremath{0.11524 \pm 0.00105 \pm 0.00000}}%
\htdef{CLEO.Gamma8.pub.ANASTASSOV.97,qt}{\ensuremath{0.11520 \pm 0.00130 \pm 0.00000}}%
\htdef{DELPHI.Gamma8.pub.ABDALLAH.06A,qt}{\ensuremath{0.11571 \pm 0.00166 \pm 0.00000}}%
\htdef{OPAL.Gamma8.pub.ACKERSTAFF.98M,qt}{\ensuremath{0.11980 \pm 0.00206 \pm 0.00000}}%
\htdef{Gamma9.qt}{\ensuremath{0.10813 \pm 0.00053}}%
\htdef{Gamma9by5.qt}{\ensuremath{0.6069 \pm 0.0032}}%
\htdef{BaBar.Gamma9by5.pub.AUBERT.10F,qt}{\ensuremath{0.5945 \pm 0.0057 \pm 0.0025}}%
\htdef{Gamma10.qt}{\ensuremath{(0.6955 \pm 0.0096) \cdot 10^{-2}}}%
\htdef{ALEPH.Gamma10.pub.BARATE.99K,qt}{\ensuremath{(0.6960 \pm 0.0287 \pm 0.0000) \cdot 10^{-2} }}%
\htdef{CLEO.Gamma10.pub.BATTLE.94,qt}{\ensuremath{(0.6600 \pm 0.1140 \pm 0.0000) \cdot 10^{-2} }}%
\htdef{DELPHI.Gamma10.pub.ABREU.94K,qt}{\ensuremath{(0.8500 \pm 0.1800 \pm 0.0000) \cdot 10^{-2} }}%
\htdef{OPAL.Gamma10.pub.ABBIENDI.01J,qt}{\ensuremath{(0.6580 \pm 0.0396 \pm 0.0000) \cdot 10^{-2} }}%
\htdef{Gamma10by5.qt}{\ensuremath{(3.903 \pm 0.054) \cdot 10^{-2}}}%
\htdef{BaBar.Gamma10by5.pub.AUBERT.10F,qt}{\ensuremath{(3.882 \pm 0.063 \pm 0.017) \cdot 10^{-2} }}%
\htdef{Gamma13.qt}{\ensuremath{0.25936 \pm 0.00090}}%
\htdef{ALEPH.Gamma13.pub.SCHAEL.05C,qt}{\ensuremath{0.25924 \pm 0.00129 \pm 0.00000}}%
\htdef{Belle.Gamma13.pub.FUJIKAWA.08,qt}{\ensuremath{0.25670 \pm 0.00010 \pm 0.00390}}%
\htdef{CLEO.Gamma13.pub.ARTUSO.94,qt}{\ensuremath{0.25870 \pm 0.00437 \pm 0.00000}}%
\htdef{DELPHI.Gamma13.pub.ABDALLAH.06A,qt}{\ensuremath{0.25740 \pm 0.00244 \pm 0.00000}}%
\htdef{L3.Gamma13.pub.ACCIARRI.95,qt}{\ensuremath{0.25050 \pm 0.00610 \pm 0.00000}}%
\htdef{OPAL.Gamma13.pub.ACKERSTAFF.98M,qt}{\ensuremath{0.25890 \pm 0.00336 \pm 0.00000}}%
\htdef{Gamma14.qt}{\ensuremath{0.25502 \pm 0.00092}}%
\htdef{Gamma16.qt}{\ensuremath{(0.4331 \pm 0.0149) \cdot 10^{-2}}}%
\htdef{ALEPH.Gamma16.pub.BARATE.99K,qt}{\ensuremath{(0.4440 \pm 0.0354 \pm 0.0000) \cdot 10^{-2} }}%
\htdef{BaBar.Gamma16.pub.AUBERT.07AP,qt}{\ensuremath{(0.4160 \pm 0.0030 \pm 0.0180) \cdot 10^{-2} }}%
\htdef{CLEO.Gamma16.pub.BATTLE.94,qt}{\ensuremath{(0.5100 \pm 0.1221 \pm 0.0000) \cdot 10^{-2} }}%
\htdef{OPAL.Gamma16.pub.ABBIENDI.04J,qt}{\ensuremath{(0.4710 \pm 0.0633 \pm 0.0000) \cdot 10^{-2} }}%
\htdef{Gamma17.qt}{\ensuremath{0.10804 \pm 0.00095}}%
\htdef{OPAL.Gamma17.pub.ACKERSTAFF.98M,qt}{\ensuremath{0.09910 \pm 0.00411 \pm 0.00000}}%
\htdef{Gamma19.qt}{\ensuremath{(9.303 \pm 0.097) \cdot 10^{-2}}}%
\htdef{ALEPH.Gamma19.pub.SCHAEL.05C,qt}{\ensuremath{(9.295 \pm 0.122 \pm 0.000) \cdot 10^{-2} }}%
\htdef{DELPHI.Gamma19.pub.ABDALLAH.06A,qt}{\ensuremath{(9.498 \pm 0.422 \pm 0.000) \cdot 10^{-2} }}%
\htdef{L3.Gamma19.pub.ACCIARRI.95,qt}{\ensuremath{(8.880 \pm 0.560 \pm 0.000) \cdot 10^{-2} }}%
\htdef{Gamma19by13.qt}{\ensuremath{0.3587 \pm 0.0044}}%
\htdef{CLEO.Gamma19by13.pub.PROCARIO.93,qt}{\ensuremath{0.3420 \pm 0.0171 \pm 0.0000}}%
\htdef{Gamma20.qt}{\ensuremath{(9.240 \pm 0.100) \cdot 10^{-2}}}%
\htdef{Gamma23.qt}{\ensuremath{(0.0630 \pm 0.0220) \cdot 10^{-2}}}%
\htdef{ALEPH.Gamma23.pub.BARATE.99K,qt}{\ensuremath{(0.0560 \pm 0.0250 \pm 0.0000) \cdot 10^{-2} }}%
\htdef{CLEO.Gamma23.pub.BATTLE.94,qt}{\ensuremath{(0.0900 \pm 0.1044 \pm 0.0000) \cdot 10^{-2} }}%
\htdef{Gamma25.qt}{\ensuremath{(1.233 \pm 0.065) \cdot 10^{-2}}}%
\htdef{DELPHI.Gamma25.pub.ABDALLAH.06A,qt}{\ensuremath{(1.403 \pm 0.310 \pm 0.000) \cdot 10^{-2} }}%
\htdef{Gamma26.qt}{\ensuremath{(1.157 \pm 0.072) \cdot 10^{-2}}}%
\htdef{ALEPH.Gamma26.pub.SCHAEL.05C,qt}{\ensuremath{(1.082 \pm 0.093 \pm 0.000) \cdot 10^{-2} }}%
\htdef{L3.Gamma26.pub.ACCIARRI.95,qt}{\ensuremath{(1.700 \pm 0.449 \pm 0.000) \cdot 10^{-2} }}%
\htdef{Gamma26by13.qt}{\ensuremath{(4.460 \pm 0.277) \cdot 10^{-2}}}%
\htdef{CLEO.Gamma26by13.pub.PROCARIO.93,qt}{\ensuremath{(4.400 \pm 0.583 \pm 0.000) \cdot 10^{-2} }}%
\htdef{Gamma27.qt}{\ensuremath{(1.030 \pm 0.075) \cdot 10^{-2}}}%
\htdef{Gamma28.qt}{\ensuremath{(4.190 \pm 2.160) \cdot 10^{-4}}}%
\htdef{ALEPH.Gamma28.pub.BARATE.99K,qt}{\ensuremath{(3.700 \pm 2.371 \pm 0.000) \cdot 10^{-4} }}%
\htdef{Gamma29.qt}{\ensuremath{(0.1566 \pm 0.0391) \cdot 10^{-2}}}%
\htdef{CLEO.Gamma29.pub.PROCARIO.93,qt}{\ensuremath{(0.1600 \pm 0.0707 \pm 0.0000) \cdot 10^{-2} }}%
\htdef{Gamma30.qt}{\ensuremath{(0.1097 \pm 0.0391) \cdot 10^{-2}}}%
\htdef{ALEPH.Gamma30.pub.SCHAEL.05C,qt}{\ensuremath{(0.1120 \pm 0.0509 \pm 0.0000) \cdot 10^{-2} }}%
\htdef{Gamma31.qt}{\ensuremath{(1.548 \pm 0.030) \cdot 10^{-2}}}%
\htdef{CLEO.Gamma31.pub.BATTLE.94,qt}{\ensuremath{(1.700 \pm 0.225 \pm 0.000) \cdot 10^{-2} }}%
\htdef{DELPHI.Gamma31.pub.ABREU.94K,qt}{\ensuremath{(1.540 \pm 0.240 \pm 0.000) \cdot 10^{-2} }}%
\htdef{OPAL.Gamma31.pub.ABBIENDI.01J,qt}{\ensuremath{(1.528 \pm 0.056 \pm 0.000) \cdot 10^{-2} }}%
\htdef{Gamma33.qt}{\ensuremath{(0.9019 \pm 0.0081) \cdot 10^{-2}}}%
\htdef{ALEPH.Gamma33.pub.BARATE.98E,qt}{\ensuremath{(0.9700 \pm 0.0849 \pm 0.0000) \cdot 10^{-2} }}%
\htdef{Belle.Gamma33.pub.Ryu:2014vpc,qt}{\ensuremath{(0.9150 \pm 0.0010 \pm 0.0150) \cdot 10^{-2} }}%
\htdef{OPAL.Gamma33.pub.AKERS.94G,qt}{\ensuremath{(0.9700 \pm 0.1082 \pm 0.0000) \cdot 10^{-2} }}%
\htdef{Gamma34.qt}{\ensuremath{(0.9878 \pm 0.0119) \cdot 10^{-2}}}%
\htdef{CLEO.Gamma34.pub.COAN.96,qt}{\ensuremath{(0.8550 \pm 0.0814 \pm 0.0000) \cdot 10^{-2} }}%
\htdef{Gamma35.qt}{\ensuremath{(0.8378 \pm 0.0123) \cdot 10^{-2}}}%
\htdef{ALEPH.Gamma35.pub.BARATE.99K,qt}{\ensuremath{(0.9280 \pm 0.0564 \pm 0.0000) \cdot 10^{-2} }}%
\htdef{BaBar.Gamma35.prelim.ICHEP08,qt}{\ensuremath{(0.8400 \pm 0.0040 \pm 0.0230) \cdot 10^{-2} }}%
\htdef{Belle.Gamma35.pub.Ryu:2014vpc,qt}{\ensuremath{(0.8320 \pm 0.0020 \pm 0.0160) \cdot 10^{-2} }}%
\htdef{L3.Gamma35.pub.ACCIARRI.95F,qt}{\ensuremath{(0.9500 \pm 0.1616 \pm 0.0000) \cdot 10^{-2} }}%
\htdef{OPAL.Gamma35.pub.ABBIENDI.00C,qt}{\ensuremath{(0.9330 \pm 0.0838 \pm 0.0000) \cdot 10^{-2} }}%
\htdef{Gamma37.qt}{\ensuremath{(0.1500 \pm 0.0050) \cdot 10^{-2}}}%
\htdef{ALEPH.Gamma37.pub.BARATE.98E,qt}{\ensuremath{(0.1580 \pm 0.0453 \pm 0.0000) \cdot 10^{-2} }}%
\htdef{ALEPH.Gamma37.pub.BARATE.99K,qt}{\ensuremath{(0.1620 \pm 0.0237 \pm 0.0000) \cdot 10^{-2} }}%
\htdef{Belle.Gamma37.pub.Ryu:2014vpc,qt}{\ensuremath{(0.1480 \pm 0.0014 \pm 0.0054) \cdot 10^{-2} }}%
\htdef{CLEO.Gamma37.pub.COAN.96,qt}{\ensuremath{(0.1510 \pm 0.0304 \pm 0.0000) \cdot 10^{-2} }}%
\htdef{Gamma38.qt}{\ensuremath{(0.3029 \pm 0.0074) \cdot 10^{-2}}}%
\htdef{OPAL.Gamma38.pub.ABBIENDI.00C,qt}{\ensuremath{(0.3300 \pm 0.0674 \pm 0.0000) \cdot 10^{-2} }}%
\htdef{Gamma39.qt}{\ensuremath{(0.5209 \pm 0.0114) \cdot 10^{-2}}}%
\htdef{CLEO.Gamma39.pub.COAN.96,qt}{\ensuremath{(0.5620 \pm 0.0693 \pm 0.0000) \cdot 10^{-2} }}%
\htdef{Gamma40.qt}{\ensuremath{(0.3680 \pm 0.0103) \cdot 10^{-2}}}%
\htdef{ALEPH.Gamma40.pub.BARATE.98E,qt}{\ensuremath{(0.2940 \pm 0.0818 \pm 0.0000) \cdot 10^{-2} }}%
\htdef{ALEPH.Gamma40.pub.BARATE.99K,qt}{\ensuremath{(0.3470 \pm 0.0646 \pm 0.0000) \cdot 10^{-2} }}%
\htdef{BaBar.Gamma40.prelim.DPF09,qt}{\ensuremath{(0.3420 \pm 0.0060 \pm 0.0150) \cdot 10^{-2} }}%
\htdef{Belle.Gamma40.pub.Ryu:2014vpc,qt}{\ensuremath{(0.3860 \pm 0.0040 \pm 0.0140) \cdot 10^{-2} }}%
\htdef{L3.Gamma40.pub.ACCIARRI.95F,qt}{\ensuremath{(0.4100 \pm 0.1237 \pm 0.0000) \cdot 10^{-2} }}%
\htdef{Gamma42.qt}{\ensuremath{(0.1528 \pm 0.0070) \cdot 10^{-2}}}%
\htdef{ALEPH.Gamma42.pub.BARATE.98E,qt}{\ensuremath{(0.1520 \pm 0.0789 \pm 0.0000) \cdot 10^{-2} }}%
\htdef{ALEPH.Gamma42.pub.BARATE.99K,qt}{\ensuremath{(0.1430 \pm 0.0291 \pm 0.0000) \cdot 10^{-2} }}%
\htdef{Belle.Gamma42.pub.Ryu:2014vpc,qt}{\ensuremath{(0.1496 \pm 0.0020 \pm 0.0074) \cdot 10^{-2} }}%
\htdef{CLEO.Gamma42.pub.COAN.96,qt}{\ensuremath{(0.1450 \pm 0.0412 \pm 0.0000) \cdot 10^{-2} }}%
\htdef{Gamma43.qt}{\ensuremath{(0.3805 \pm 0.0229) \cdot 10^{-2}}}%
\htdef{OPAL.Gamma43.pub.ABBIENDI.00C,qt}{\ensuremath{(0.3240 \pm 0.0992 \pm 0.0000) \cdot 10^{-2} }}%
\htdef{Gamma44.qt}{\ensuremath{(1.245 \pm 2.043) \cdot 10^{-4}}}%
\htdef{ALEPH.Gamma44.pub.BARATE.99R,qt}{\ensuremath{(2.600 \pm 2.400 \pm 0.000) \cdot 10^{-4} }}%
\htdef{Gamma46.qt}{\ensuremath{(0.1329 \pm 0.0110) \cdot 10^{-2}}}%
\htdef{ALEPH.Gamma46.pub.BARATE.98E,qt}{\ensuremath{(0.1530 \pm 0.0340 \pm 0.0000) \cdot 10^{-2} }}%
\htdef{Gamma47.qt}{\ensuremath{(2.359 \pm 0.061) \cdot 10^{-4}}}%
\htdef{ALEPH.Gamma47.pub.BARATE.98E,qt}{\ensuremath{(2.600 \pm 1.118 \pm 0.000) \cdot 10^{-4} }}%
\htdef{BaBar.Gamma47.pub.LEES.2012Y,qt}{\ensuremath{(2.310 \pm 0.040 \pm 0.080) \cdot 10^{-4} }}%
\htdef{Belle.Gamma47.pub.Ryu:2014vpc,qt}{\ensuremath{(2.330 \pm 0.030 \pm 0.090) \cdot 10^{-4} }}%
\htdef{CLEO.Gamma47.pub.COAN.96,qt}{\ensuremath{(2.300 \pm 0.583 \pm 0.000) \cdot 10^{-4} }}%
\htdef{Gamma48.qt}{\ensuremath{(0.0857 \pm 0.0104) \cdot 10^{-2}}}%
\htdef{ALEPH.Gamma48.pub.BARATE.98E,qt}{\ensuremath{(0.1010 \pm 0.0264 \pm 0.0000) \cdot 10^{-2} }}%
\htdef{Gamma49.qt}{\ensuremath{(2.896 \pm 1.051) \cdot 10^{-4}}}%
\htdef{Gamma50.qt}{\ensuremath{(1.845 \pm 0.206) \cdot 10^{-5}}}%
\htdef{BaBar.Gamma50.pub.LEES.2012Y,qt}{\ensuremath{(1.600 \pm 0.200 \pm 0.220) \cdot 10^{-5} }}%
\htdef{Belle.Gamma50.pub.Ryu:2014vpc,qt}{\ensuremath{(2.000 \pm 0.220 \pm 0.200) \cdot 10^{-5} }}%
\htdef{Gamma51.qt}{\ensuremath{(2.527 \pm 1.047) \cdot 10^{-4}}}%
\htdef{ALEPH.Gamma51.pub.BARATE.98E,qt}{\ensuremath{(3.100 \pm 1.100 \pm 0.500) \cdot 10^{-4} }}%
\htdef{Gamma53.qt}{\ensuremath{(2.221 \pm 2.024) \cdot 10^{-4}}}%
\htdef{ALEPH.Gamma53.pub.BARATE.98E,qt}{\ensuremath{(2.300 \pm 2.025 \pm 0.000) \cdot 10^{-4} }}%
\htdef{Gamma54.qt}{\ensuremath{0.15201 \pm 0.00059}}%
\htdef{CELLO.Gamma54.pub.BEHREND.89B,qt}{\ensuremath{0.15000 \pm 0.00500 \pm 0.00000}}%
\htdef{L3.Gamma54.pub.ADEVA.91F,qt}{\ensuremath{0.14400 \pm 0.00671 \pm 0.00000}}%
\htdef{TPC.Gamma54.pub.AIHARA.87B,qt}{\ensuremath{0.15100 \pm 0.01000 \pm 0.00000}}%
\htdef{Gamma55.qt}{\ensuremath{0.14573 \pm 0.00056}}%
\htdef{L3.Gamma55.pub.ACHARD.01D,qt}{\ensuremath{0.14556 \pm 0.00130 \pm 0.00000}}%
\htdef{OPAL.Gamma55.pub.AKERS.95Y,qt}{\ensuremath{0.14960 \pm 0.00238 \pm 0.00000}}%
\htdef{Gamma57.qt}{\ensuremath{(9.448 \pm 0.053) \cdot 10^{-2}}}%
\htdef{CLEO.Gamma57.pub.BALEST.95C,qt}{\ensuremath{(9.510 \pm 0.212 \pm 0.000) \cdot 10^{-2} }}%
\htdef{DELPHI.Gamma57.pub.ABDALLAH.06A,qt}{\ensuremath{(9.317 \pm 0.122 \pm 0.000) \cdot 10^{-2} }}%
\htdef{Gamma57by55.qt}{\ensuremath{0.6483 \pm 0.0029}}%
\htdef{OPAL.Gamma57by55.pub.AKERS.95Y,qt}{\ensuremath{0.6600 \pm 0.0146 \pm 0.0000}}%
\htdef{Gamma58.qt}{\ensuremath{(9.418 \pm 0.053) \cdot 10^{-2}}}%
\htdef{ALEPH.Gamma58.pub.SCHAEL.05C,qt}{\ensuremath{(9.469 \pm 0.096 \pm 0.000) \cdot 10^{-2} }}%
\htdef{Gamma60.qt}{\ensuremath{(9.0097 \pm 0.0510) \cdot 10^{-2}}}%
\htdef{BaBar.Gamma60.pub.AUBERT.08,qt}{\ensuremath{(8.8337 \pm 0.0074 \pm 0.1267) \cdot 10^{-2} }}%
\htdef{Belle.Gamma60.pub.LEE.10,qt}{\ensuremath{(8.4200 \pm 0.0033 \pm 0.2588) \cdot 10^{-2} }}%
\htdef{CLEO3.Gamma60.pub.BRIERE.03,qt}{\ensuremath{(9.1300 \pm 0.4627 \pm 0.0000) \cdot 10^{-2} }}%
\htdef{Gamma62.qt}{\ensuremath{(8.980 \pm 0.051) \cdot 10^{-2}}}%
\htdef{Gamma66.qt}{\ensuremath{(4.603 \pm 0.051) \cdot 10^{-2}}}%
\htdef{ALEPH.Gamma66.pub.SCHAEL.05C,qt}{\ensuremath{(4.734 \pm 0.077 \pm 0.000) \cdot 10^{-2} }}%
\htdef{CLEO.Gamma66.pub.BALEST.95C,qt}{\ensuremath{(4.230 \pm 0.228 \pm 0.000) \cdot 10^{-2} }}%
\htdef{DELPHI.Gamma66.pub.ABDALLAH.06A,qt}{\ensuremath{(4.545 \pm 0.148 \pm 0.000) \cdot 10^{-2} }}%
\htdef{Gamma69.qt}{\ensuremath{(4.516 \pm 0.052) \cdot 10^{-2}}}%
\htdef{CLEO.Gamma69.pub.EDWARDS.00A,qt}{\ensuremath{(4.190 \pm 0.233 \pm 0.000) \cdot 10^{-2} }}%
\htdef{Gamma70.qt}{\ensuremath{(2.767 \pm 0.071) \cdot 10^{-2}}}%
\htdef{Gamma74.qt}{\ensuremath{(0.5130 \pm 0.0311) \cdot 10^{-2}}}%
\htdef{DELPHI.Gamma74.pub.ABDALLAH.06A,qt}{\ensuremath{(0.5610 \pm 0.1168 \pm 0.0000) \cdot 10^{-2} }}%
\htdef{Gamma76.qt}{\ensuremath{(0.4919 \pm 0.0310) \cdot 10^{-2}}}%
\htdef{ALEPH.Gamma76.pub.SCHAEL.05C,qt}{\ensuremath{(0.4350 \pm 0.0461 \pm 0.0000) \cdot 10^{-2} }}%
\htdef{Gamma76by54.qt}{\ensuremath{(3.236 \pm 0.202) \cdot 10^{-2}}}%
\htdef{CLEO.Gamma76by54.pub.BORTOLETTO.93,qt}{\ensuremath{(3.400 \pm 0.361 \pm 0.000) \cdot 10^{-2} }}%
\htdef{Gamma77.qt}{\ensuremath{(9.734 \pm 3.546) \cdot 10^{-4}}}%
\htdef{Gamma78.qt}{\ensuremath{(2.109 \pm 0.299) \cdot 10^{-4}}}%
\htdef{CLEO.Gamma78.pub.ANASTASSOV.01,qt}{\ensuremath{(2.200 \pm 0.500 \pm 0.000) \cdot 10^{-4} }}%
\htdef{Gamma80by60.qt}{\ensuremath{(4.845 \pm 0.081) \cdot 10^{-2}}}%
\htdef{CLEO.Gamma80by60.pub.RICHICHI.99,qt}{\ensuremath{(5.440 \pm 0.570 \pm 0.000) \cdot 10^{-2} }}%
\htdef{Gamma81by69.qt}{\ensuremath{(1.932 \pm 0.266) \cdot 10^{-2}}}%
\htdef{CLEO.Gamma81by69.pub.RICHICHI.99,qt}{\ensuremath{(2.610 \pm 0.615 \pm 0.000) \cdot 10^{-2} }}%
\htdef{Gamma82.qt}{\ensuremath{(0.4796 \pm 0.0138) \cdot 10^{-2}}}%
\htdef{TPC.Gamma82.pub.BAUER.94,qt}{\ensuremath{(0.5800 \pm 0.1845 \pm 0.0000) \cdot 10^{-2} }}%
\htdef{Gamma85.qt}{\ensuremath{(0.2929 \pm 0.0068) \cdot 10^{-2}}}%
\htdef{ALEPH.Gamma85.pub.BARATE.98,qt}{\ensuremath{(0.2140 \pm 0.0470 \pm 0.0000) \cdot 10^{-2} }}%
\htdef{BaBar.Gamma85.pub.AUBERT.08,qt}{\ensuremath{(0.2726 \pm 0.0018 \pm 0.0092) \cdot 10^{-2} }}%
\htdef{Belle.Gamma85.pub.LEE.10,qt}{\ensuremath{(0.3300 \pm 0.0013 \pm 0.0166) \cdot 10^{-2} }}%
\htdef{CLEO3.Gamma85.pub.BRIERE.03,qt}{\ensuremath{(0.3840 \pm 0.0405 \pm 0.0000) \cdot 10^{-2} }}%
\htdef{OPAL.Gamma85.pub.ABBIENDI.04J,qt}{\ensuremath{(0.4150 \pm 0.0664 \pm 0.0000) \cdot 10^{-2} }}%
\htdef{Gamma88.qt}{\ensuremath{(8.113 \pm 1.168) \cdot 10^{-4}}}%
\htdef{ALEPH.Gamma88.pub.BARATE.98,qt}{\ensuremath{(6.100 \pm 4.295 \pm 0.000) \cdot 10^{-4} }}%
\htdef{CLEO3.Gamma88.pub.ARMS.05,qt}{\ensuremath{(7.400 \pm 1.360 \pm 0.000) \cdot 10^{-4} }}%
\htdef{Gamma92.qt}{\ensuremath{(0.1497 \pm 0.0033) \cdot 10^{-2}}}%
\htdef{OPAL.Gamma92.pub.ABBIENDI.00D,qt}{\ensuremath{(0.1590 \pm 0.0566 \pm 0.0000) \cdot 10^{-2} }}%
\htdef{TPC.Gamma92.pub.BAUER.94,qt}{\ensuremath{(0.1500 \pm 0.0855 \pm 0.0000) \cdot 10^{-2} }}%
\htdef{Gamma93.qt}{\ensuremath{(0.14363 \pm 0.00274) \cdot 10^{-2}}}%
\htdef{ALEPH.Gamma93.pub.BARATE.98,qt}{\ensuremath{(0.16300 \pm 0.02702 \pm 0.00000) \cdot 10^{-2} }}%
\htdef{BaBar.Gamma93.pub.AUBERT.08,qt}{\ensuremath{(0.13461 \pm 0.00100 \pm 0.00364) \cdot 10^{-2} }}%
\htdef{Belle.Gamma93.pub.LEE.10,qt}{\ensuremath{(0.15500 \pm 0.00066 \pm 0.00556) \cdot 10^{-2} }}%
\htdef{CLEO3.Gamma93.pub.BRIERE.03,qt}{\ensuremath{(0.15500 \pm 0.01082 \pm 0.00000) \cdot 10^{-2} }}%
\htdef{Gamma93by60.qt}{\ensuremath{(1.594 \pm 0.030) \cdot 10^{-2}}}%
\htdef{CLEO.Gamma93by60.pub.RICHICHI.99,qt}{\ensuremath{(1.600 \pm 0.335 \pm 0.000) \cdot 10^{-2} }}%
\htdef{Gamma94.qt}{\ensuremath{(0.611 \pm 0.183) \cdot 10^{-4}}}%
\htdef{ALEPH.Gamma94.pub.BARATE.98,qt}{\ensuremath{(7.500 \pm 3.265 \pm 0.000) \cdot 10^{-4} }}%
\htdef{CLEO3.Gamma94.pub.ARMS.05,qt}{\ensuremath{(0.550 \pm 0.184 \pm 0.000) \cdot 10^{-4} }}%
\htdef{Gamma94by69.qt}{\ensuremath{(0.1353 \pm 0.0406) \cdot 10^{-2}}}%
\htdef{CLEO.Gamma94by69.pub.RICHICHI.99,qt}{\ensuremath{(0.7900 \pm 0.4682 \pm 0.0000) \cdot 10^{-2} }}%
\htdef{Gamma96.qt}{\ensuremath{(2.156 \pm 0.800) \cdot 10^{-5}}}%
\htdef{BaBar.Gamma96.pub.AUBERT.08,qt}{\ensuremath{(1.578 \pm 0.130 \pm 0.123) \cdot 10^{-5} }}%
\htdef{Belle.Gamma96.pub.LEE.10,qt}{\ensuremath{(3.290 \pm 0.169 \pm 0.196) \cdot 10^{-5} }}%
\htdef{Gamma102.qt}{\ensuremath{(0.0986 \pm 0.0037) \cdot 10^{-2}}}%
\htdef{CLEO.Gamma102.pub.GIBAUT.94B,qt}{\ensuremath{(0.0970 \pm 0.0121 \pm 0.0000) \cdot 10^{-2} }}%
\htdef{HRS.Gamma102.pub.BYLSMA.87,qt}{\ensuremath{(0.1020 \pm 0.0290 \pm 0.0000) \cdot 10^{-2} }}%
\htdef{L3.Gamma102.pub.ACHARD.01D,qt}{\ensuremath{(0.1700 \pm 0.0341 \pm 0.0000) \cdot 10^{-2} }}%
\htdef{Gamma103.qt}{\ensuremath{(8.224 \pm 0.315) \cdot 10^{-4}}}%
\htdef{ALEPH.Gamma103.pub.SCHAEL.05C,qt}{\ensuremath{(7.200 \pm 1.500 \pm 0.000) \cdot 10^{-4} }}%
\htdef{ARGUS.Gamma103.pub.ALBRECHT.88B,qt}{\ensuremath{(6.400 \pm 2.508 \pm 0.000) \cdot 10^{-4} }}%
\htdef{CLEO.Gamma103.pub.GIBAUT.94B,qt}{\ensuremath{(7.700 \pm 1.030 \pm 0.000) \cdot 10^{-4} }}%
\htdef{DELPHI.Gamma103.pub.ABDALLAH.06A,qt}{\ensuremath{(9.700 \pm 1.581 \pm 0.000) \cdot 10^{-4} }}%
\htdef{HRS.Gamma103.pub.BYLSMA.87,qt}{\ensuremath{(5.100 \pm 2.000 \pm 0.000) \cdot 10^{-4} }}%
\htdef{OPAL.Gamma103.pub.ACKERSTAFF.99E,qt}{\ensuremath{(9.100 \pm 1.523 \pm 0.000) \cdot 10^{-4} }}%
\htdef{Gamma104.qt}{\ensuremath{(1.637 \pm 0.113) \cdot 10^{-4}}}%
\htdef{ALEPH.Gamma104.pub.SCHAEL.05C,qt}{\ensuremath{(2.100 \pm 0.922 \pm 0.000) \cdot 10^{-4} }}%
\htdef{CLEO.Gamma104.pub.ANASTASSOV.01,qt}{\ensuremath{(1.700 \pm 0.283 \pm 0.000) \cdot 10^{-4} }}%
\htdef{DELPHI.Gamma104.pub.ABDALLAH.06A,qt}{\ensuremath{(1.600 \pm 1.342 \pm 0.000) \cdot 10^{-4} }}%
\htdef{OPAL.Gamma104.pub.ACKERSTAFF.99E,qt}{\ensuremath{(2.700 \pm 2.012 \pm 0.000) \cdot 10^{-4} }}%
\htdef{Gamma110.qt}{\ensuremath{(2.882 \pm 0.047) \cdot 10^{-2}}}%
\htdef{Gamma126.qt}{\ensuremath{(0.1387 \pm 0.0072) \cdot 10^{-2}}}%
\htdef{ALEPH.Gamma126.pub.BUSKULIC.97C,qt}{\ensuremath{(0.1800 \pm 0.0447 \pm 0.0000) \cdot 10^{-2} }}%
\htdef{Belle.Gamma126.pub.INAMI.09,qt}{\ensuremath{(0.1350 \pm 0.0030 \pm 0.0070) \cdot 10^{-2} }}%
\htdef{CLEO.Gamma126.pub.ARTUSO.92,qt}{\ensuremath{(0.1700 \pm 0.0283 \pm 0.0000) \cdot 10^{-2} }}%
\htdef{Gamma128.qt}{\ensuremath{(1.548 \pm 0.080) \cdot 10^{-4}}}%
\htdef{ALEPH.Gamma128.pub.BUSKULIC.97C,qt}{\ensuremath{(2.900 {}^{+1.300}_{-1.200} \pm 0.700) \cdot 10^{-4} }}%
\htdef{BaBar.Gamma128.pub.DEL-AMO-SANCHEZ.11E,qt}{\ensuremath{(1.420 \pm 0.110 \pm 0.070) \cdot 10^{-4} }}%
\htdef{Belle.Gamma128.pub.INAMI.09,qt}{\ensuremath{(1.580 \pm 0.050 \pm 0.090) \cdot 10^{-4} }}%
\htdef{CLEO.Gamma128.pub.BARTELT.96,qt}{\ensuremath{(2.600 \pm 0.500 \pm 0.500) \cdot 10^{-4} }}%
\htdef{Gamma130.qt}{\ensuremath{(0.483 \pm 0.116) \cdot 10^{-4}}}%
\htdef{Belle.Gamma130.pub.INAMI.09,qt}{\ensuremath{(0.460 \pm 0.110 \pm 0.040) \cdot 10^{-4} }}%
\htdef{CLEO.Gamma130.pub.BISHAI.99,qt}{\ensuremath{(1.770 \pm 0.904 \pm 0.000) \cdot 10^{-4} }}%
\htdef{Gamma132.qt}{\ensuremath{(0.934 \pm 0.149) \cdot 10^{-4}}}%
\htdef{Belle.Gamma132.pub.INAMI.09,qt}{\ensuremath{(0.880 \pm 0.140 \pm 0.060) \cdot 10^{-4} }}%
\htdef{CLEO.Gamma132.pub.BISHAI.99,qt}{\ensuremath{(2.200 \pm 0.734 \pm 0.000) \cdot 10^{-4} }}%
\htdef{Gamma136.qt}{\ensuremath{(2.186 \pm 0.129) \cdot 10^{-4}}}%
\htdef{Gamma150.qt}{\ensuremath{(1.995 \pm 0.064) \cdot 10^{-2}}}%
\htdef{ALEPH.Gamma150.pub.BUSKULIC.97C,qt}{\ensuremath{(1.910 \pm 0.092 \pm 0.000) \cdot 10^{-2} }}%
\htdef{CLEO.Gamma150.pub.BARINGER.87,qt}{\ensuremath{(1.600 \pm 0.491 \pm 0.000) \cdot 10^{-2} }}%
\htdef{Gamma150by66.qt}{\ensuremath{0.4333 \pm 0.0139}}%
\htdef{ALEPH.Gamma150by66.pub.BUSKULIC.96,qt}{\ensuremath{0.4310 \pm 0.0330 \pm 0.0000}}%
\htdef{CLEO.Gamma150by66.pub.BALEST.95C,qt}{\ensuremath{0.4640 \pm 0.0233 \pm 0.0000}}%
\htdef{Gamma151.qt}{\ensuremath{(4.100 \pm 0.922) \cdot 10^{-4}}}%
\htdef{CLEO3.Gamma151.pub.ARMS.05,qt}{\ensuremath{(4.100 \pm 0.922 \pm 0.000) \cdot 10^{-4} }}%
\htdef{Gamma152.qt}{\ensuremath{(0.4054 \pm 0.0418) \cdot 10^{-2}}}%
\htdef{ALEPH.Gamma152.pub.BUSKULIC.97C,qt}{\ensuremath{(0.4300 \pm 0.0781 \pm 0.0000) \cdot 10^{-2} }}%
\htdef{Gamma152by76.qt}{\ensuremath{0.8243 \pm 0.0757}}%
\htdef{CLEO.Gamma152by76.pub.BORTOLETTO.93,qt}{\ensuremath{0.8100 \pm 0.0848 \pm 0.0000}}%
\htdef{Gamma800.qt}{\ensuremath{(1.954 \pm 0.065) \cdot 10^{-2}}}%
\htdef{Gamma801.qt}{\ensuremath{(3.664 \pm 1.360) \cdot 10^{-5}}}%
\htdef{Gamma802.qt}{\ensuremath{(0.2922 \pm 0.0068) \cdot 10^{-2}}}%
\htdef{Gamma803.qt}{\ensuremath{(4.101 \pm 1.429) \cdot 10^{-4}}}%
\htdef{Gamma804.qt}{\ensuremath{(2.359 \pm 0.061) \cdot 10^{-4}}}%
\htdef{Gamma805.qt}{\ensuremath{(4.000 \pm 2.000) \cdot 10^{-4}}}%
\htdef{ALEPH.Gamma805.pub.SCHAEL.05C,qt}{\ensuremath{(4.000 \pm 2.000 \pm 0.000) \cdot 10^{-4} }}%
\htdef{Gamma806.qt}{\ensuremath{(1.845 \pm 0.206) \cdot 10^{-5}}}%
\htdef{Gamma810.qt}{\ensuremath{(1.925 \pm 0.298) \cdot 10^{-4}}}%
\htdef{Gamma811.qt}{\ensuremath{(7.110 \pm 1.586) \cdot 10^{-5}}}%
\htdef{BaBar.Gamma811.pub.LEES.2012X,qt}{\ensuremath{(7.300 \pm 1.200 \pm 1.200) \cdot 10^{-5} }}%
\htdef{Gamma812.qt}{\ensuremath{(1.336 \pm 2.682) \cdot 10^{-5}}}%
\htdef{BaBar.Gamma812.pub.LEES.2012X,qt}{\ensuremath{(1.000 \pm 0.800 \pm 3.000) \cdot 10^{-5} }}%
\htdef{Gamma820.qt}{\ensuremath{(8.205 \pm 0.315) \cdot 10^{-4}}}%
\htdef{Gamma821.qt}{\ensuremath{(7.685 \pm 0.296) \cdot 10^{-4}}}%
\htdef{BaBar.Gamma821.pub.LEES.2012X,qt}{\ensuremath{(7.680 \pm 0.040 \pm 0.400) \cdot 10^{-4} }}%
\htdef{Gamma822.qt}{\ensuremath{(0.596 \pm 1.208) \cdot 10^{-6}}}%
\htdef{BaBar.Gamma822.pub.LEES.2012X,qt}{\ensuremath{(0.600 \pm 0.500 \pm 1.100) \cdot 10^{-6} }}%
\htdef{Gamma830.qt}{\ensuremath{(1.626 \pm 0.113) \cdot 10^{-4}}}%
\htdef{Gamma831.qt}{\ensuremath{(8.370 \pm 0.624) \cdot 10^{-5}}}%
\htdef{BaBar.Gamma831.pub.LEES.2012X,qt}{\ensuremath{(8.400 \pm 0.400 \pm 0.600) \cdot 10^{-5} }}%
\htdef{Gamma832.qt}{\ensuremath{(3.783 \pm 0.873) \cdot 10^{-5}}}%
\htdef{BaBar.Gamma832.pub.LEES.2012X,qt}{\ensuremath{(3.600 \pm 0.300 \pm 0.900) \cdot 10^{-5} }}%
\htdef{Gamma833.qt}{\ensuremath{(1.108 \pm 0.566) \cdot 10^{-6}}}%
\htdef{BaBar.Gamma833.pub.LEES.2012X,qt}{\ensuremath{(1.100 \pm 0.400 \pm 0.400) \cdot 10^{-6} }}%
\htdef{Gamma910.qt}{\ensuremath{(7.144 \pm 0.423) \cdot 10^{-5}}}%
\htdef{BaBar.Gamma910.pub.LEES.2012X,qt}{\ensuremath{(8.270 \pm 0.880 \pm 0.810) \cdot 10^{-5} }}%
\htdef{Gamma911.qt}{\ensuremath{(4.424 \pm 0.867) \cdot 10^{-5}}}%
\htdef{BaBar.Gamma911.pub.LEES.2012X,qt}{\ensuremath{(4.570 \pm 0.770 \pm 0.500) \cdot 10^{-5} }}%
\htdef{Gamma920.qt}{\ensuremath{(5.202 \pm 0.444) \cdot 10^{-5}}}%
\htdef{BaBar.Gamma920.pub.LEES.2012X,qt}{\ensuremath{(5.200 \pm 0.310 \pm 0.370) \cdot 10^{-5} }}%
\htdef{Gamma930.qt}{\ensuremath{(5.010 \pm 0.297) \cdot 10^{-5}}}%
\htdef{BaBar.Gamma930.pub.LEES.2012X,qt}{\ensuremath{(5.390 \pm 0.270 \pm 0.410) \cdot 10^{-5} }}%
\htdef{Gamma944.qt}{\ensuremath{(8.615 \pm 0.510) \cdot 10^{-5}}}%
\htdef{BaBar.Gamma944.pub.LEES.2012X,qt}{\ensuremath{(8.260 \pm 0.350 \pm 0.510) \cdot 10^{-5} }}%
\htdef{Gamma998.qt}{\ensuremath{(9.902 \pm 9.850) \cdot 10^{-4}}}%
\htdef{Gamma3.qm}{%
\begin{ensuredisplaymath}
\htuse{Gamma3.gn} = \htuse{Gamma3.td}
\end{ensuredisplaymath}
 & \htuse{Gamma3.qt} & \hfagFitLabel\\
\htuse{ALEPH.Gamma3.pub.SCHAEL.05C,qt} & \htuse{ALEPH.Gamma3.pub.SCHAEL.05C,exp} & \htuse{ALEPH.Gamma3.pub.SCHAEL.05C,ref} \\
\htuse{DELPHI.Gamma3.pub.ABREU.99X,qt} & \htuse{DELPHI.Gamma3.pub.ABREU.99X,exp} & \htuse{DELPHI.Gamma3.pub.ABREU.99X,ref} \\
\htuse{L3.Gamma3.pub.ACCIARRI.01F,qt} & \htuse{L3.Gamma3.pub.ACCIARRI.01F,exp} & \htuse{L3.Gamma3.pub.ACCIARRI.01F,ref} \\
\htuse{OPAL.Gamma3.pub.ABBIENDI.03,qt} & \htuse{OPAL.Gamma3.pub.ABBIENDI.03,exp} & \htuse{OPAL.Gamma3.pub.ABBIENDI.03,ref}
}%
\htdef{Gamma3by5.qm}{%
\begin{ensuredisplaymath}
\htuse{Gamma3by5.gn} = \htuse{Gamma3by5.td}
\end{ensuredisplaymath}
 & \htuse{Gamma3by5.qt} & \hfagFitLabel\\
\htuse{ARGUS.Gamma3by5.pub.ALBRECHT.92D,qt} & \htuse{ARGUS.Gamma3by5.pub.ALBRECHT.92D,exp} & \htuse{ARGUS.Gamma3by5.pub.ALBRECHT.92D,ref} \\
\htuse{BaBar.Gamma3by5.pub.AUBERT.10F,qt} & \htuse{BaBar.Gamma3by5.pub.AUBERT.10F,exp} & \htuse{BaBar.Gamma3by5.pub.AUBERT.10F,ref} \\
\htuse{CLEO.Gamma3by5.pub.ANASTASSOV.97,qt} & \htuse{CLEO.Gamma3by5.pub.ANASTASSOV.97,exp} & \htuse{CLEO.Gamma3by5.pub.ANASTASSOV.97,ref}
}%
\htdef{Gamma5.qm}{%
\begin{ensuredisplaymath}
\htuse{Gamma5.gn} = \htuse{Gamma5.td}
\end{ensuredisplaymath}
 & \htuse{Gamma5.qt} & \hfagFitLabel\\
\htuse{ALEPH.Gamma5.pub.SCHAEL.05C,qt} & \htuse{ALEPH.Gamma5.pub.SCHAEL.05C,exp} & \htuse{ALEPH.Gamma5.pub.SCHAEL.05C,ref} \\
\htuse{CLEO.Gamma5.pub.ANASTASSOV.97,qt} & \htuse{CLEO.Gamma5.pub.ANASTASSOV.97,exp} & \htuse{CLEO.Gamma5.pub.ANASTASSOV.97,ref} \\
\htuse{DELPHI.Gamma5.pub.ABREU.99X,qt} & \htuse{DELPHI.Gamma5.pub.ABREU.99X,exp} & \htuse{DELPHI.Gamma5.pub.ABREU.99X,ref} \\
\htuse{L3.Gamma5.pub.ACCIARRI.01F,qt} & \htuse{L3.Gamma5.pub.ACCIARRI.01F,exp} & \htuse{L3.Gamma5.pub.ACCIARRI.01F,ref} \\
\htuse{OPAL.Gamma5.pub.ABBIENDI.99H,qt} & \htuse{OPAL.Gamma5.pub.ABBIENDI.99H,exp} & \htuse{OPAL.Gamma5.pub.ABBIENDI.99H,ref}
}%
\htdef{Gamma7.qm}{%
\begin{ensuredisplaymath}
\htuse{Gamma7.gn} = \htuse{Gamma7.td}
\end{ensuredisplaymath}
 & \htuse{Gamma7.qt} & \hfagFitLabel\\
\htuse{DELPHI.Gamma7.pub.ABREU.92N,qt} & \htuse{DELPHI.Gamma7.pub.ABREU.92N,exp} & \htuse{DELPHI.Gamma7.pub.ABREU.92N,ref} \\
\htuse{L3.Gamma7.pub.ACCIARRI.95,qt} & \htuse{L3.Gamma7.pub.ACCIARRI.95,exp} & \htuse{L3.Gamma7.pub.ACCIARRI.95,ref} \\
\htuse{OPAL.Gamma7.pub.ALEXANDER.91D,qt} & \htuse{OPAL.Gamma7.pub.ALEXANDER.91D,exp} & \htuse{OPAL.Gamma7.pub.ALEXANDER.91D,ref}
}%
\htdef{Gamma8.qm}{%
\begin{ensuredisplaymath}
\htuse{Gamma8.gn} = \htuse{Gamma8.td}
\end{ensuredisplaymath}
 & \htuse{Gamma8.qt} & \hfagFitLabel\\
\htuse{ALEPH.Gamma8.pub.SCHAEL.05C,qt} & \htuse{ALEPH.Gamma8.pub.SCHAEL.05C,exp} & \htuse{ALEPH.Gamma8.pub.SCHAEL.05C,ref} \\
\htuse{CLEO.Gamma8.pub.ANASTASSOV.97,qt} & \htuse{CLEO.Gamma8.pub.ANASTASSOV.97,exp} & \htuse{CLEO.Gamma8.pub.ANASTASSOV.97,ref} \\
\htuse{DELPHI.Gamma8.pub.ABDALLAH.06A,qt} & \htuse{DELPHI.Gamma8.pub.ABDALLAH.06A,exp} & \htuse{DELPHI.Gamma8.pub.ABDALLAH.06A,ref} \\
\htuse{OPAL.Gamma8.pub.ACKERSTAFF.98M,qt} & \htuse{OPAL.Gamma8.pub.ACKERSTAFF.98M,exp} & \htuse{OPAL.Gamma8.pub.ACKERSTAFF.98M,ref}
}%
\htdef{Gamma9.qm}{%
\begin{ensuredisplaymath}
\htuse{Gamma9.gn} = \htuse{Gamma9.td}
\end{ensuredisplaymath}
 & \htuse{Gamma9.qt} & \hfagFitLabel}%
\htdef{Gamma9by5.qm}{%
\begin{ensuredisplaymath}
\htuse{Gamma9by5.gn} = \htuse{Gamma9by5.td}
\end{ensuredisplaymath}
 & \htuse{Gamma9by5.qt} & \hfagFitLabel\\
\htuse{BaBar.Gamma9by5.pub.AUBERT.10F,qt} & \htuse{BaBar.Gamma9by5.pub.AUBERT.10F,exp} & \htuse{BaBar.Gamma9by5.pub.AUBERT.10F,ref}
}%
\htdef{Gamma10.qm}{%
\begin{ensuredisplaymath}
\htuse{Gamma10.gn} = \htuse{Gamma10.td}
\end{ensuredisplaymath}
 & \htuse{Gamma10.qt} & \hfagFitLabel\\
\htuse{ALEPH.Gamma10.pub.BARATE.99K,qt} & \htuse{ALEPH.Gamma10.pub.BARATE.99K,exp} & \htuse{ALEPH.Gamma10.pub.BARATE.99K,ref} \\
\htuse{CLEO.Gamma10.pub.BATTLE.94,qt} & \htuse{CLEO.Gamma10.pub.BATTLE.94,exp} & \htuse{CLEO.Gamma10.pub.BATTLE.94,ref} \\
\htuse{DELPHI.Gamma10.pub.ABREU.94K,qt} & \htuse{DELPHI.Gamma10.pub.ABREU.94K,exp} & \htuse{DELPHI.Gamma10.pub.ABREU.94K,ref} \\
\htuse{OPAL.Gamma10.pub.ABBIENDI.01J,qt} & \htuse{OPAL.Gamma10.pub.ABBIENDI.01J,exp} & \htuse{OPAL.Gamma10.pub.ABBIENDI.01J,ref}
}%
\htdef{Gamma10by5.qm}{%
\begin{ensuredisplaymath}
\htuse{Gamma10by5.gn} = \htuse{Gamma10by5.td}
\end{ensuredisplaymath}
 & \htuse{Gamma10by5.qt} & \hfagFitLabel\\
\htuse{BaBar.Gamma10by5.pub.AUBERT.10F,qt} & \htuse{BaBar.Gamma10by5.pub.AUBERT.10F,exp} & \htuse{BaBar.Gamma10by5.pub.AUBERT.10F,ref}
}%
\htdef{Gamma13.qm}{%
\begin{ensuredisplaymath}
\htuse{Gamma13.gn} = \htuse{Gamma13.td}
\end{ensuredisplaymath}
 & \htuse{Gamma13.qt} & \hfagFitLabel\\
\htuse{ALEPH.Gamma13.pub.SCHAEL.05C,qt} & \htuse{ALEPH.Gamma13.pub.SCHAEL.05C,exp} & \htuse{ALEPH.Gamma13.pub.SCHAEL.05C,ref} \\
\htuse{Belle.Gamma13.pub.FUJIKAWA.08,qt} & \htuse{Belle.Gamma13.pub.FUJIKAWA.08,exp} & \htuse{Belle.Gamma13.pub.FUJIKAWA.08,ref} \\
\htuse{CLEO.Gamma13.pub.ARTUSO.94,qt} & \htuse{CLEO.Gamma13.pub.ARTUSO.94,exp} & \htuse{CLEO.Gamma13.pub.ARTUSO.94,ref} \\
\htuse{DELPHI.Gamma13.pub.ABDALLAH.06A,qt} & \htuse{DELPHI.Gamma13.pub.ABDALLAH.06A,exp} & \htuse{DELPHI.Gamma13.pub.ABDALLAH.06A,ref} \\
\htuse{L3.Gamma13.pub.ACCIARRI.95,qt} & \htuse{L3.Gamma13.pub.ACCIARRI.95,exp} & \htuse{L3.Gamma13.pub.ACCIARRI.95,ref} \\
\htuse{OPAL.Gamma13.pub.ACKERSTAFF.98M,qt} & \htuse{OPAL.Gamma13.pub.ACKERSTAFF.98M,exp} & \htuse{OPAL.Gamma13.pub.ACKERSTAFF.98M,ref}
}%
\htdef{Gamma14.qm}{%
\begin{ensuredisplaymath}
\htuse{Gamma14.gn} = \htuse{Gamma14.td}
\end{ensuredisplaymath}
 & \htuse{Gamma14.qt} & \hfagFitLabel}%
\htdef{Gamma16.qm}{%
\begin{ensuredisplaymath}
\htuse{Gamma16.gn} = \htuse{Gamma16.td}
\end{ensuredisplaymath}
 & \htuse{Gamma16.qt} & \hfagFitLabel\\
\htuse{ALEPH.Gamma16.pub.BARATE.99K,qt} & \htuse{ALEPH.Gamma16.pub.BARATE.99K,exp} & \htuse{ALEPH.Gamma16.pub.BARATE.99K,ref} \\
\htuse{BaBar.Gamma16.pub.AUBERT.07AP,qt} & \htuse{BaBar.Gamma16.pub.AUBERT.07AP,exp} & \htuse{BaBar.Gamma16.pub.AUBERT.07AP,ref} \\
\htuse{CLEO.Gamma16.pub.BATTLE.94,qt} & \htuse{CLEO.Gamma16.pub.BATTLE.94,exp} & \htuse{CLEO.Gamma16.pub.BATTLE.94,ref} \\
\htuse{OPAL.Gamma16.pub.ABBIENDI.04J,qt} & \htuse{OPAL.Gamma16.pub.ABBIENDI.04J,exp} & \htuse{OPAL.Gamma16.pub.ABBIENDI.04J,ref}
}%
\htdef{Gamma17.qm}{%
\begin{ensuredisplaymath}
\htuse{Gamma17.gn} = \htuse{Gamma17.td}
\end{ensuredisplaymath}
 & \htuse{Gamma17.qt} & \hfagFitLabel\\
\htuse{OPAL.Gamma17.pub.ACKERSTAFF.98M,qt} & \htuse{OPAL.Gamma17.pub.ACKERSTAFF.98M,exp} & \htuse{OPAL.Gamma17.pub.ACKERSTAFF.98M,ref}
}%
\htdef{Gamma19.qm}{%
\begin{ensuredisplaymath}
\htuse{Gamma19.gn} = \htuse{Gamma19.td}
\end{ensuredisplaymath}
 & \htuse{Gamma19.qt} & \hfagFitLabel\\
\htuse{ALEPH.Gamma19.pub.SCHAEL.05C,qt} & \htuse{ALEPH.Gamma19.pub.SCHAEL.05C,exp} & \htuse{ALEPH.Gamma19.pub.SCHAEL.05C,ref} \\
\htuse{DELPHI.Gamma19.pub.ABDALLAH.06A,qt} & \htuse{DELPHI.Gamma19.pub.ABDALLAH.06A,exp} & \htuse{DELPHI.Gamma19.pub.ABDALLAH.06A,ref} \\
\htuse{L3.Gamma19.pub.ACCIARRI.95,qt} & \htuse{L3.Gamma19.pub.ACCIARRI.95,exp} & \htuse{L3.Gamma19.pub.ACCIARRI.95,ref}
}%
\htdef{Gamma19by13.qm}{%
\begin{ensuredisplaymath}
\htuse{Gamma19by13.gn} = \htuse{Gamma19by13.td}
\end{ensuredisplaymath}
 & \htuse{Gamma19by13.qt} & \hfagFitLabel\\
\htuse{CLEO.Gamma19by13.pub.PROCARIO.93,qt} & \htuse{CLEO.Gamma19by13.pub.PROCARIO.93,exp} & \htuse{CLEO.Gamma19by13.pub.PROCARIO.93,ref}
}%
\htdef{Gamma20.qm}{%
\begin{ensuredisplaymath}
\htuse{Gamma20.gn} = \htuse{Gamma20.td}
\end{ensuredisplaymath}
 & \htuse{Gamma20.qt} & \hfagFitLabel}%
\htdef{Gamma23.qm}{%
\begin{ensuredisplaymath}
\htuse{Gamma23.gn} = \htuse{Gamma23.td}
\end{ensuredisplaymath}
 & \htuse{Gamma23.qt} & \hfagFitLabel\\
\htuse{ALEPH.Gamma23.pub.BARATE.99K,qt} & \htuse{ALEPH.Gamma23.pub.BARATE.99K,exp} & \htuse{ALEPH.Gamma23.pub.BARATE.99K,ref} \\
\htuse{CLEO.Gamma23.pub.BATTLE.94,qt} & \htuse{CLEO.Gamma23.pub.BATTLE.94,exp} & \htuse{CLEO.Gamma23.pub.BATTLE.94,ref}
}%
\htdef{Gamma25.qm}{%
\begin{ensuredisplaymath}
\htuse{Gamma25.gn} = \htuse{Gamma25.td}
\end{ensuredisplaymath}
 & \htuse{Gamma25.qt} & \hfagFitLabel\\
\htuse{DELPHI.Gamma25.pub.ABDALLAH.06A,qt} & \htuse{DELPHI.Gamma25.pub.ABDALLAH.06A,exp} & \htuse{DELPHI.Gamma25.pub.ABDALLAH.06A,ref}
}%
\htdef{Gamma26.qm}{%
\begin{ensuredisplaymath}
\htuse{Gamma26.gn} = \htuse{Gamma26.td}
\end{ensuredisplaymath}
 & \htuse{Gamma26.qt} & \hfagFitLabel\\
\htuse{ALEPH.Gamma26.pub.SCHAEL.05C,qt} & \htuse{ALEPH.Gamma26.pub.SCHAEL.05C,exp} & \htuse{ALEPH.Gamma26.pub.SCHAEL.05C,ref} \\
\htuse{L3.Gamma26.pub.ACCIARRI.95,qt} & \htuse{L3.Gamma26.pub.ACCIARRI.95,exp} & \htuse{L3.Gamma26.pub.ACCIARRI.95,ref}
}%
\htdef{Gamma26by13.qm}{%
\begin{ensuredisplaymath}
\htuse{Gamma26by13.gn} = \htuse{Gamma26by13.td}
\end{ensuredisplaymath}
 & \htuse{Gamma26by13.qt} & \hfagFitLabel\\
\htuse{CLEO.Gamma26by13.pub.PROCARIO.93,qt} & \htuse{CLEO.Gamma26by13.pub.PROCARIO.93,exp} & \htuse{CLEO.Gamma26by13.pub.PROCARIO.93,ref}
}%
\htdef{Gamma27.qm}{%
\begin{ensuredisplaymath}
\htuse{Gamma27.gn} = \htuse{Gamma27.td}
\end{ensuredisplaymath}
 & \htuse{Gamma27.qt} & \hfagFitLabel}%
\htdef{Gamma28.qm}{%
\begin{ensuredisplaymath}
\htuse{Gamma28.gn} = \htuse{Gamma28.td}
\end{ensuredisplaymath}
 & \htuse{Gamma28.qt} & \hfagFitLabel\\
\htuse{ALEPH.Gamma28.pub.BARATE.99K,qt} & \htuse{ALEPH.Gamma28.pub.BARATE.99K,exp} & \htuse{ALEPH.Gamma28.pub.BARATE.99K,ref}
}%
\htdef{Gamma29.qm}{%
\begin{ensuredisplaymath}
\htuse{Gamma29.gn} = \htuse{Gamma29.td}
\end{ensuredisplaymath}
 & \htuse{Gamma29.qt} & \hfagFitLabel\\
\htuse{CLEO.Gamma29.pub.PROCARIO.93,qt} & \htuse{CLEO.Gamma29.pub.PROCARIO.93,exp} & \htuse{CLEO.Gamma29.pub.PROCARIO.93,ref}
}%
\htdef{Gamma30.qm}{%
\begin{ensuredisplaymath}
\htuse{Gamma30.gn} = \htuse{Gamma30.td}
\end{ensuredisplaymath}
 & \htuse{Gamma30.qt} & \hfagFitLabel\\
\htuse{ALEPH.Gamma30.pub.SCHAEL.05C,qt} & \htuse{ALEPH.Gamma30.pub.SCHAEL.05C,exp} & \htuse{ALEPH.Gamma30.pub.SCHAEL.05C,ref}
}%
\htdef{Gamma31.qm}{%
\begin{ensuredisplaymath}
\htuse{Gamma31.gn} = \htuse{Gamma31.td}
\end{ensuredisplaymath}
 & \htuse{Gamma31.qt} & \hfagFitLabel\\
\htuse{CLEO.Gamma31.pub.BATTLE.94,qt} & \htuse{CLEO.Gamma31.pub.BATTLE.94,exp} & \htuse{CLEO.Gamma31.pub.BATTLE.94,ref} \\
\htuse{DELPHI.Gamma31.pub.ABREU.94K,qt} & \htuse{DELPHI.Gamma31.pub.ABREU.94K,exp} & \htuse{DELPHI.Gamma31.pub.ABREU.94K,ref} \\
\htuse{OPAL.Gamma31.pub.ABBIENDI.01J,qt} & \htuse{OPAL.Gamma31.pub.ABBIENDI.01J,exp} & \htuse{OPAL.Gamma31.pub.ABBIENDI.01J,ref}
}%
\htdef{Gamma33.qm}{%
\begin{ensuredisplaymath}
\htuse{Gamma33.gn} = \htuse{Gamma33.td}
\end{ensuredisplaymath}
 & \htuse{Gamma33.qt} & \hfagFitLabel\\
\htuse{ALEPH.Gamma33.pub.BARATE.98E,qt} & \htuse{ALEPH.Gamma33.pub.BARATE.98E,exp} & \htuse{ALEPH.Gamma33.pub.BARATE.98E,ref} \\
\htuse{Belle.Gamma33.pub.Ryu:2014vpc,qt} & \htuse{Belle.Gamma33.pub.Ryu:2014vpc,exp} & \htuse{Belle.Gamma33.pub.Ryu:2014vpc,ref} \\
\htuse{OPAL.Gamma33.pub.AKERS.94G,qt} & \htuse{OPAL.Gamma33.pub.AKERS.94G,exp} & \htuse{OPAL.Gamma33.pub.AKERS.94G,ref}
}%
\htdef{Gamma34.qm}{%
\begin{ensuredisplaymath}
\htuse{Gamma34.gn} = \htuse{Gamma34.td}
\end{ensuredisplaymath}
 & \htuse{Gamma34.qt} & \hfagFitLabel\\
\htuse{CLEO.Gamma34.pub.COAN.96,qt} & \htuse{CLEO.Gamma34.pub.COAN.96,exp} & \htuse{CLEO.Gamma34.pub.COAN.96,ref}
}%
\htdef{Gamma35.qm}{%
\begin{ensuredisplaymath}
\htuse{Gamma35.gn} = \htuse{Gamma35.td}
\end{ensuredisplaymath}
 & \htuse{Gamma35.qt} & \hfagFitLabel\\
\htuse{ALEPH.Gamma35.pub.BARATE.99K,qt} & \htuse{ALEPH.Gamma35.pub.BARATE.99K,exp} & \htuse{ALEPH.Gamma35.pub.BARATE.99K,ref} \\
\htuse{BaBar.Gamma35.prelim.ICHEP08,qt} & \htuse{BaBar.Gamma35.prelim.ICHEP08,exp} & \htuse{BaBar.Gamma35.prelim.ICHEP08,ref} \\
\htuse{Belle.Gamma35.pub.Ryu:2014vpc,qt} & \htuse{Belle.Gamma35.pub.Ryu:2014vpc,exp} & \htuse{Belle.Gamma35.pub.Ryu:2014vpc,ref} \\
\htuse{L3.Gamma35.pub.ACCIARRI.95F,qt} & \htuse{L3.Gamma35.pub.ACCIARRI.95F,exp} & \htuse{L3.Gamma35.pub.ACCIARRI.95F,ref} \\
\htuse{OPAL.Gamma35.pub.ABBIENDI.00C,qt} & \htuse{OPAL.Gamma35.pub.ABBIENDI.00C,exp} & \htuse{OPAL.Gamma35.pub.ABBIENDI.00C,ref}
}%
\htdef{Gamma37.qm}{%
\begin{ensuredisplaymath}
\htuse{Gamma37.gn} = \htuse{Gamma37.td}
\end{ensuredisplaymath}
 & \htuse{Gamma37.qt} & \hfagFitLabel\\
\htuse{ALEPH.Gamma37.pub.BARATE.98E,qt} & \htuse{ALEPH.Gamma37.pub.BARATE.98E,exp} & \htuse{ALEPH.Gamma37.pub.BARATE.98E,ref} \\
\htuse{ALEPH.Gamma37.pub.BARATE.99K,qt} & \htuse{ALEPH.Gamma37.pub.BARATE.99K,exp} & \htuse{ALEPH.Gamma37.pub.BARATE.99K,ref} \\
\htuse{Belle.Gamma37.pub.Ryu:2014vpc,qt} & \htuse{Belle.Gamma37.pub.Ryu:2014vpc,exp} & \htuse{Belle.Gamma37.pub.Ryu:2014vpc,ref} \\
\htuse{CLEO.Gamma37.pub.COAN.96,qt} & \htuse{CLEO.Gamma37.pub.COAN.96,exp} & \htuse{CLEO.Gamma37.pub.COAN.96,ref}
}%
\htdef{Gamma38.qm}{%
\begin{ensuredisplaymath}
\htuse{Gamma38.gn} = \htuse{Gamma38.td}
\end{ensuredisplaymath}
 & \htuse{Gamma38.qt} & \hfagFitLabel\\
\htuse{OPAL.Gamma38.pub.ABBIENDI.00C,qt} & \htuse{OPAL.Gamma38.pub.ABBIENDI.00C,exp} & \htuse{OPAL.Gamma38.pub.ABBIENDI.00C,ref}
}%
\htdef{Gamma39.qm}{%
\begin{ensuredisplaymath}
\htuse{Gamma39.gn} = \htuse{Gamma39.td}
\end{ensuredisplaymath}
 & \htuse{Gamma39.qt} & \hfagFitLabel\\
\htuse{CLEO.Gamma39.pub.COAN.96,qt} & \htuse{CLEO.Gamma39.pub.COAN.96,exp} & \htuse{CLEO.Gamma39.pub.COAN.96,ref}
}%
\htdef{Gamma40.qm}{%
\begin{ensuredisplaymath}
\htuse{Gamma40.gn} = \htuse{Gamma40.td}
\end{ensuredisplaymath}
 & \htuse{Gamma40.qt} & \hfagFitLabel\\
\htuse{ALEPH.Gamma40.pub.BARATE.98E,qt} & \htuse{ALEPH.Gamma40.pub.BARATE.98E,exp} & \htuse{ALEPH.Gamma40.pub.BARATE.98E,ref} \\
\htuse{ALEPH.Gamma40.pub.BARATE.99K,qt} & \htuse{ALEPH.Gamma40.pub.BARATE.99K,exp} & \htuse{ALEPH.Gamma40.pub.BARATE.99K,ref} \\
\htuse{BaBar.Gamma40.prelim.DPF09,qt} & \htuse{BaBar.Gamma40.prelim.DPF09,exp} & \htuse{BaBar.Gamma40.prelim.DPF09,ref} \\
\htuse{Belle.Gamma40.pub.Ryu:2014vpc,qt} & \htuse{Belle.Gamma40.pub.Ryu:2014vpc,exp} & \htuse{Belle.Gamma40.pub.Ryu:2014vpc,ref} \\
\htuse{L3.Gamma40.pub.ACCIARRI.95F,qt} & \htuse{L3.Gamma40.pub.ACCIARRI.95F,exp} & \htuse{L3.Gamma40.pub.ACCIARRI.95F,ref}
}%
\htdef{Gamma42.qm}{%
\begin{ensuredisplaymath}
\htuse{Gamma42.gn} = \htuse{Gamma42.td}
\end{ensuredisplaymath}
 & \htuse{Gamma42.qt} & \hfagFitLabel\\
\htuse{ALEPH.Gamma42.pub.BARATE.98E,qt} & \htuse{ALEPH.Gamma42.pub.BARATE.98E,exp} & \htuse{ALEPH.Gamma42.pub.BARATE.98E,ref} \\
\htuse{ALEPH.Gamma42.pub.BARATE.99K,qt} & \htuse{ALEPH.Gamma42.pub.BARATE.99K,exp} & \htuse{ALEPH.Gamma42.pub.BARATE.99K,ref} \\
\htuse{Belle.Gamma42.pub.Ryu:2014vpc,qt} & \htuse{Belle.Gamma42.pub.Ryu:2014vpc,exp} & \htuse{Belle.Gamma42.pub.Ryu:2014vpc,ref} \\
\htuse{CLEO.Gamma42.pub.COAN.96,qt} & \htuse{CLEO.Gamma42.pub.COAN.96,exp} & \htuse{CLEO.Gamma42.pub.COAN.96,ref}
}%
\htdef{Gamma43.qm}{%
\begin{ensuredisplaymath}
\htuse{Gamma43.gn} = \htuse{Gamma43.td}
\end{ensuredisplaymath}
 & \htuse{Gamma43.qt} & \hfagFitLabel\\
\htuse{OPAL.Gamma43.pub.ABBIENDI.00C,qt} & \htuse{OPAL.Gamma43.pub.ABBIENDI.00C,exp} & \htuse{OPAL.Gamma43.pub.ABBIENDI.00C,ref}
}%
\htdef{Gamma44.qm}{%
\begin{ensuredisplaymath}
\htuse{Gamma44.gn} = \htuse{Gamma44.td}
\end{ensuredisplaymath}
 & \htuse{Gamma44.qt} & \hfagFitLabel\\
\htuse{ALEPH.Gamma44.pub.BARATE.99R,qt} & \htuse{ALEPH.Gamma44.pub.BARATE.99R,exp} & \htuse{ALEPH.Gamma44.pub.BARATE.99R,ref}
}%
\htdef{Gamma46.qm}{%
\begin{ensuredisplaymath}
\htuse{Gamma46.gn} = \htuse{Gamma46.td}
\end{ensuredisplaymath}
 & \htuse{Gamma46.qt} & \hfagFitLabel\\
\htuse{ALEPH.Gamma46.pub.BARATE.98E,qt} & \htuse{ALEPH.Gamma46.pub.BARATE.98E,exp} & \htuse{ALEPH.Gamma46.pub.BARATE.98E,ref}
}%
\htdef{Gamma47.qm}{%
\begin{ensuredisplaymath}
\htuse{Gamma47.gn} = \htuse{Gamma47.td}
\end{ensuredisplaymath}
 & \htuse{Gamma47.qt} & \hfagFitLabel\\
\htuse{ALEPH.Gamma47.pub.BARATE.98E,qt} & \htuse{ALEPH.Gamma47.pub.BARATE.98E,exp} & \htuse{ALEPH.Gamma47.pub.BARATE.98E,ref} \\
\htuse{BaBar.Gamma47.pub.LEES.2012Y,qt} & \htuse{BaBar.Gamma47.pub.LEES.2012Y,exp} & \htuse{BaBar.Gamma47.pub.LEES.2012Y,ref} \\
\htuse{Belle.Gamma47.pub.Ryu:2014vpc,qt} & \htuse{Belle.Gamma47.pub.Ryu:2014vpc,exp} & \htuse{Belle.Gamma47.pub.Ryu:2014vpc,ref} \\
\htuse{CLEO.Gamma47.pub.COAN.96,qt} & \htuse{CLEO.Gamma47.pub.COAN.96,exp} & \htuse{CLEO.Gamma47.pub.COAN.96,ref}
}%
\htdef{Gamma48.qm}{%
\begin{ensuredisplaymath}
\htuse{Gamma48.gn} = \htuse{Gamma48.td}
\end{ensuredisplaymath}
 & \htuse{Gamma48.qt} & \hfagFitLabel\\
\htuse{ALEPH.Gamma48.pub.BARATE.98E,qt} & \htuse{ALEPH.Gamma48.pub.BARATE.98E,exp} & \htuse{ALEPH.Gamma48.pub.BARATE.98E,ref}
}%
\htdef{Gamma49.qm}{%
\begin{ensuredisplaymath}
\htuse{Gamma49.gn} = \htuse{Gamma49.td}
\end{ensuredisplaymath}
 & \htuse{Gamma49.qt} & \hfagFitLabel}%
\htdef{Gamma50.qm}{%
\begin{ensuredisplaymath}
\htuse{Gamma50.gn} = \htuse{Gamma50.td}
\end{ensuredisplaymath}
 & \htuse{Gamma50.qt} & \hfagFitLabel\\
\htuse{BaBar.Gamma50.pub.LEES.2012Y,qt} & \htuse{BaBar.Gamma50.pub.LEES.2012Y,exp} & \htuse{BaBar.Gamma50.pub.LEES.2012Y,ref} \\
\htuse{Belle.Gamma50.pub.Ryu:2014vpc,qt} & \htuse{Belle.Gamma50.pub.Ryu:2014vpc,exp} & \htuse{Belle.Gamma50.pub.Ryu:2014vpc,ref}
}%
\htdef{Gamma51.qm}{%
\begin{ensuredisplaymath}
\htuse{Gamma51.gn} = \htuse{Gamma51.td}
\end{ensuredisplaymath}
 & \htuse{Gamma51.qt} & \hfagFitLabel\\
\htuse{ALEPH.Gamma51.pub.BARATE.98E,qt} & \htuse{ALEPH.Gamma51.pub.BARATE.98E,exp} & \htuse{ALEPH.Gamma51.pub.BARATE.98E,ref}
}%
\htdef{Gamma53.qm}{%
\begin{ensuredisplaymath}
\htuse{Gamma53.gn} = \htuse{Gamma53.td}
\end{ensuredisplaymath}
 & \htuse{Gamma53.qt} & \hfagFitLabel\\
\htuse{ALEPH.Gamma53.pub.BARATE.98E,qt} & \htuse{ALEPH.Gamma53.pub.BARATE.98E,exp} & \htuse{ALEPH.Gamma53.pub.BARATE.98E,ref}
}%
\htdef{Gamma54.qm}{%
\begin{ensuredisplaymath}
\htuse{Gamma54.gn} = \htuse{Gamma54.td}
\end{ensuredisplaymath}
 & \htuse{Gamma54.qt} & \hfagFitLabel\\
\htuse{CELLO.Gamma54.pub.BEHREND.89B,qt} & \htuse{CELLO.Gamma54.pub.BEHREND.89B,exp} & \htuse{CELLO.Gamma54.pub.BEHREND.89B,ref} \\
\htuse{L3.Gamma54.pub.ADEVA.91F,qt} & \htuse{L3.Gamma54.pub.ADEVA.91F,exp} & \htuse{L3.Gamma54.pub.ADEVA.91F,ref} \\
\htuse{TPC.Gamma54.pub.AIHARA.87B,qt} & \htuse{TPC.Gamma54.pub.AIHARA.87B,exp} & \htuse{TPC.Gamma54.pub.AIHARA.87B,ref}
}%
\htdef{Gamma55.qm}{%
\begin{ensuredisplaymath}
\htuse{Gamma55.gn} = \htuse{Gamma55.td}
\end{ensuredisplaymath}
 & \htuse{Gamma55.qt} & \hfagFitLabel\\
\htuse{L3.Gamma55.pub.ACHARD.01D,qt} & \htuse{L3.Gamma55.pub.ACHARD.01D,exp} & \htuse{L3.Gamma55.pub.ACHARD.01D,ref} \\
\htuse{OPAL.Gamma55.pub.AKERS.95Y,qt} & \htuse{OPAL.Gamma55.pub.AKERS.95Y,exp} & \htuse{OPAL.Gamma55.pub.AKERS.95Y,ref}
}%
\htdef{Gamma57.qm}{%
\begin{ensuredisplaymath}
\htuse{Gamma57.gn} = \htuse{Gamma57.td}
\end{ensuredisplaymath}
 & \htuse{Gamma57.qt} & \hfagFitLabel\\
\htuse{CLEO.Gamma57.pub.BALEST.95C,qt} & \htuse{CLEO.Gamma57.pub.BALEST.95C,exp} & \htuse{CLEO.Gamma57.pub.BALEST.95C,ref} \\
\htuse{DELPHI.Gamma57.pub.ABDALLAH.06A,qt} & \htuse{DELPHI.Gamma57.pub.ABDALLAH.06A,exp} & \htuse{DELPHI.Gamma57.pub.ABDALLAH.06A,ref}
}%
\htdef{Gamma57by55.qm}{%
\begin{ensuredisplaymath}
\htuse{Gamma57by55.gn} = \htuse{Gamma57by55.td}
\end{ensuredisplaymath}
 & \htuse{Gamma57by55.qt} & \hfagFitLabel\\
\htuse{OPAL.Gamma57by55.pub.AKERS.95Y,qt} & \htuse{OPAL.Gamma57by55.pub.AKERS.95Y,exp} & \htuse{OPAL.Gamma57by55.pub.AKERS.95Y,ref}
}%
\htdef{Gamma58.qm}{%
\begin{ensuredisplaymath}
\htuse{Gamma58.gn} = \htuse{Gamma58.td}
\end{ensuredisplaymath}
 & \htuse{Gamma58.qt} & \hfagFitLabel\\
\htuse{ALEPH.Gamma58.pub.SCHAEL.05C,qt} & \htuse{ALEPH.Gamma58.pub.SCHAEL.05C,exp} & \htuse{ALEPH.Gamma58.pub.SCHAEL.05C,ref}
}%
\htdef{Gamma60.qm}{%
\begin{ensuredisplaymath}
\htuse{Gamma60.gn} = \htuse{Gamma60.td}
\end{ensuredisplaymath}
 & \htuse{Gamma60.qt} & \hfagFitLabel\\
\htuse{BaBar.Gamma60.pub.AUBERT.08,qt} & \htuse{BaBar.Gamma60.pub.AUBERT.08,exp} & \htuse{BaBar.Gamma60.pub.AUBERT.08,ref} \\
\htuse{Belle.Gamma60.pub.LEE.10,qt} & \htuse{Belle.Gamma60.pub.LEE.10,exp} & \htuse{Belle.Gamma60.pub.LEE.10,ref} \\
\htuse{CLEO3.Gamma60.pub.BRIERE.03,qt} & \htuse{CLEO3.Gamma60.pub.BRIERE.03,exp} & \htuse{CLEO3.Gamma60.pub.BRIERE.03,ref}
}%
\htdef{Gamma62.qm}{%
\begin{ensuredisplaymath}
\htuse{Gamma62.gn} = \htuse{Gamma62.td}
\end{ensuredisplaymath}
 & \htuse{Gamma62.qt} & \hfagFitLabel}%
\htdef{Gamma66.qm}{%
\begin{ensuredisplaymath}
\htuse{Gamma66.gn} = \htuse{Gamma66.td}
\end{ensuredisplaymath}
 & \htuse{Gamma66.qt} & \hfagFitLabel\\
\htuse{ALEPH.Gamma66.pub.SCHAEL.05C,qt} & \htuse{ALEPH.Gamma66.pub.SCHAEL.05C,exp} & \htuse{ALEPH.Gamma66.pub.SCHAEL.05C,ref} \\
\htuse{CLEO.Gamma66.pub.BALEST.95C,qt} & \htuse{CLEO.Gamma66.pub.BALEST.95C,exp} & \htuse{CLEO.Gamma66.pub.BALEST.95C,ref} \\
\htuse{DELPHI.Gamma66.pub.ABDALLAH.06A,qt} & \htuse{DELPHI.Gamma66.pub.ABDALLAH.06A,exp} & \htuse{DELPHI.Gamma66.pub.ABDALLAH.06A,ref}
}%
\htdef{Gamma69.qm}{%
\begin{ensuredisplaymath}
\htuse{Gamma69.gn} = \htuse{Gamma69.td}
\end{ensuredisplaymath}
 & \htuse{Gamma69.qt} & \hfagFitLabel\\
\htuse{CLEO.Gamma69.pub.EDWARDS.00A,qt} & \htuse{CLEO.Gamma69.pub.EDWARDS.00A,exp} & \htuse{CLEO.Gamma69.pub.EDWARDS.00A,ref}
}%
\htdef{Gamma70.qm}{%
\begin{ensuredisplaymath}
\htuse{Gamma70.gn} = \htuse{Gamma70.td}
\end{ensuredisplaymath}
 & \htuse{Gamma70.qt} & \hfagFitLabel}%
\htdef{Gamma74.qm}{%
\begin{ensuredisplaymath}
\htuse{Gamma74.gn} = \htuse{Gamma74.td}
\end{ensuredisplaymath}
 & \htuse{Gamma74.qt} & \hfagFitLabel\\
\htuse{DELPHI.Gamma74.pub.ABDALLAH.06A,qt} & \htuse{DELPHI.Gamma74.pub.ABDALLAH.06A,exp} & \htuse{DELPHI.Gamma74.pub.ABDALLAH.06A,ref}
}%
\htdef{Gamma76.qm}{%
\begin{ensuredisplaymath}
\htuse{Gamma76.gn} = \htuse{Gamma76.td}
\end{ensuredisplaymath}
 & \htuse{Gamma76.qt} & \hfagFitLabel\\
\htuse{ALEPH.Gamma76.pub.SCHAEL.05C,qt} & \htuse{ALEPH.Gamma76.pub.SCHAEL.05C,exp} & \htuse{ALEPH.Gamma76.pub.SCHAEL.05C,ref}
}%
\htdef{Gamma76by54.qm}{%
\begin{ensuredisplaymath}
\htuse{Gamma76by54.gn} = \htuse{Gamma76by54.td}
\end{ensuredisplaymath}
 & \htuse{Gamma76by54.qt} & \hfagFitLabel\\
\htuse{CLEO.Gamma76by54.pub.BORTOLETTO.93,qt} & \htuse{CLEO.Gamma76by54.pub.BORTOLETTO.93,exp} & \htuse{CLEO.Gamma76by54.pub.BORTOLETTO.93,ref}
}%
\htdef{Gamma77.qm}{%
\begin{ensuredisplaymath}
\htuse{Gamma77.gn} = \htuse{Gamma77.td}
\end{ensuredisplaymath}
 & \htuse{Gamma77.qt} & \hfagFitLabel}%
\htdef{Gamma78.qm}{%
\begin{ensuredisplaymath}
\htuse{Gamma78.gn} = \htuse{Gamma78.td}
\end{ensuredisplaymath}
 & \htuse{Gamma78.qt} & \hfagFitLabel\\
\htuse{CLEO.Gamma78.pub.ANASTASSOV.01,qt} & \htuse{CLEO.Gamma78.pub.ANASTASSOV.01,exp} & \htuse{CLEO.Gamma78.pub.ANASTASSOV.01,ref}
}%
\htdef{Gamma80by60.qm}{%
\begin{ensuredisplaymath}
\htuse{Gamma80by60.gn} = \htuse{Gamma80by60.td}
\end{ensuredisplaymath}
 & \htuse{Gamma80by60.qt} & \hfagFitLabel\\
\htuse{CLEO.Gamma80by60.pub.RICHICHI.99,qt} & \htuse{CLEO.Gamma80by60.pub.RICHICHI.99,exp} & \htuse{CLEO.Gamma80by60.pub.RICHICHI.99,ref}
}%
\htdef{Gamma81by69.qm}{%
\begin{ensuredisplaymath}
\htuse{Gamma81by69.gn} = \htuse{Gamma81by69.td}
\end{ensuredisplaymath}
 & \htuse{Gamma81by69.qt} & \hfagFitLabel\\
\htuse{CLEO.Gamma81by69.pub.RICHICHI.99,qt} & \htuse{CLEO.Gamma81by69.pub.RICHICHI.99,exp} & \htuse{CLEO.Gamma81by69.pub.RICHICHI.99,ref}
}%
\htdef{Gamma82.qm}{%
\begin{ensuredisplaymath}
\htuse{Gamma82.gn} = \htuse{Gamma82.td}
\end{ensuredisplaymath}
 & \htuse{Gamma82.qt} & \hfagFitLabel\\
\htuse{TPC.Gamma82.pub.BAUER.94,qt} & \htuse{TPC.Gamma82.pub.BAUER.94,exp} & \htuse{TPC.Gamma82.pub.BAUER.94,ref}
}%
\htdef{Gamma85.qm}{%
\begin{ensuredisplaymath}
\htuse{Gamma85.gn} = \htuse{Gamma85.td}
\end{ensuredisplaymath}
 & \htuse{Gamma85.qt} & \hfagFitLabel\\
\htuse{ALEPH.Gamma85.pub.BARATE.98,qt} & \htuse{ALEPH.Gamma85.pub.BARATE.98,exp} & \htuse{ALEPH.Gamma85.pub.BARATE.98,ref} \\
\htuse{BaBar.Gamma85.pub.AUBERT.08,qt} & \htuse{BaBar.Gamma85.pub.AUBERT.08,exp} & \htuse{BaBar.Gamma85.pub.AUBERT.08,ref} \\
\htuse{Belle.Gamma85.pub.LEE.10,qt} & \htuse{Belle.Gamma85.pub.LEE.10,exp} & \htuse{Belle.Gamma85.pub.LEE.10,ref} \\
\htuse{CLEO3.Gamma85.pub.BRIERE.03,qt} & \htuse{CLEO3.Gamma85.pub.BRIERE.03,exp} & \htuse{CLEO3.Gamma85.pub.BRIERE.03,ref} \\
\htuse{OPAL.Gamma85.pub.ABBIENDI.04J,qt} & \htuse{OPAL.Gamma85.pub.ABBIENDI.04J,exp} & \htuse{OPAL.Gamma85.pub.ABBIENDI.04J,ref}
}%
\htdef{Gamma88.qm}{%
\begin{ensuredisplaymath}
\htuse{Gamma88.gn} = \htuse{Gamma88.td}
\end{ensuredisplaymath}
 & \htuse{Gamma88.qt} & \hfagFitLabel\\
\htuse{ALEPH.Gamma88.pub.BARATE.98,qt} & \htuse{ALEPH.Gamma88.pub.BARATE.98,exp} & \htuse{ALEPH.Gamma88.pub.BARATE.98,ref} \\
\htuse{CLEO3.Gamma88.pub.ARMS.05,qt} & \htuse{CLEO3.Gamma88.pub.ARMS.05,exp} & \htuse{CLEO3.Gamma88.pub.ARMS.05,ref}
}%
\htdef{Gamma92.qm}{%
\begin{ensuredisplaymath}
\htuse{Gamma92.gn} = \htuse{Gamma92.td}
\end{ensuredisplaymath}
 & \htuse{Gamma92.qt} & \hfagFitLabel\\
\htuse{OPAL.Gamma92.pub.ABBIENDI.00D,qt} & \htuse{OPAL.Gamma92.pub.ABBIENDI.00D,exp} & \htuse{OPAL.Gamma92.pub.ABBIENDI.00D,ref} \\
\htuse{TPC.Gamma92.pub.BAUER.94,qt} & \htuse{TPC.Gamma92.pub.BAUER.94,exp} & \htuse{TPC.Gamma92.pub.BAUER.94,ref}
}%
\htdef{Gamma93.qm}{%
\begin{ensuredisplaymath}
\htuse{Gamma93.gn} = \htuse{Gamma93.td}
\end{ensuredisplaymath}
 & \htuse{Gamma93.qt} & \hfagFitLabel\\
\htuse{ALEPH.Gamma93.pub.BARATE.98,qt} & \htuse{ALEPH.Gamma93.pub.BARATE.98,exp} & \htuse{ALEPH.Gamma93.pub.BARATE.98,ref} \\
\htuse{BaBar.Gamma93.pub.AUBERT.08,qt} & \htuse{BaBar.Gamma93.pub.AUBERT.08,exp} & \htuse{BaBar.Gamma93.pub.AUBERT.08,ref} \\
\htuse{Belle.Gamma93.pub.LEE.10,qt} & \htuse{Belle.Gamma93.pub.LEE.10,exp} & \htuse{Belle.Gamma93.pub.LEE.10,ref} \\
\htuse{CLEO3.Gamma93.pub.BRIERE.03,qt} & \htuse{CLEO3.Gamma93.pub.BRIERE.03,exp} & \htuse{CLEO3.Gamma93.pub.BRIERE.03,ref}
}%
\htdef{Gamma93by60.qm}{%
\begin{ensuredisplaymath}
\htuse{Gamma93by60.gn} = \htuse{Gamma93by60.td}
\end{ensuredisplaymath}
 & \htuse{Gamma93by60.qt} & \hfagFitLabel\\
\htuse{CLEO.Gamma93by60.pub.RICHICHI.99,qt} & \htuse{CLEO.Gamma93by60.pub.RICHICHI.99,exp} & \htuse{CLEO.Gamma93by60.pub.RICHICHI.99,ref}
}%
\htdef{Gamma94.qm}{%
\begin{ensuredisplaymath}
\htuse{Gamma94.gn} = \htuse{Gamma94.td}
\end{ensuredisplaymath}
 & \htuse{Gamma94.qt} & \hfagFitLabel\\
\htuse{ALEPH.Gamma94.pub.BARATE.98,qt} & \htuse{ALEPH.Gamma94.pub.BARATE.98,exp} & \htuse{ALEPH.Gamma94.pub.BARATE.98,ref} \\
\htuse{CLEO3.Gamma94.pub.ARMS.05,qt} & \htuse{CLEO3.Gamma94.pub.ARMS.05,exp} & \htuse{CLEO3.Gamma94.pub.ARMS.05,ref}
}%
\htdef{Gamma94by69.qm}{%
\begin{ensuredisplaymath}
\htuse{Gamma94by69.gn} = \htuse{Gamma94by69.td}
\end{ensuredisplaymath}
 & \htuse{Gamma94by69.qt} & \hfagFitLabel\\
\htuse{CLEO.Gamma94by69.pub.RICHICHI.99,qt} & \htuse{CLEO.Gamma94by69.pub.RICHICHI.99,exp} & \htuse{CLEO.Gamma94by69.pub.RICHICHI.99,ref}
}%
\htdef{Gamma96.qm}{%
\begin{ensuredisplaymath}
\htuse{Gamma96.gn} = \htuse{Gamma96.td}
\end{ensuredisplaymath}
 & \htuse{Gamma96.qt} & \hfagFitLabel\\
\htuse{BaBar.Gamma96.pub.AUBERT.08,qt} & \htuse{BaBar.Gamma96.pub.AUBERT.08,exp} & \htuse{BaBar.Gamma96.pub.AUBERT.08,ref} \\
\htuse{Belle.Gamma96.pub.LEE.10,qt} & \htuse{Belle.Gamma96.pub.LEE.10,exp} & \htuse{Belle.Gamma96.pub.LEE.10,ref}
}%
\htdef{Gamma102.qm}{%
\begin{ensuredisplaymath}
\htuse{Gamma102.gn} = \htuse{Gamma102.td}
\end{ensuredisplaymath}
 & \htuse{Gamma102.qt} & \hfagFitLabel\\
\htuse{CLEO.Gamma102.pub.GIBAUT.94B,qt} & \htuse{CLEO.Gamma102.pub.GIBAUT.94B,exp} & \htuse{CLEO.Gamma102.pub.GIBAUT.94B,ref} \\
\htuse{HRS.Gamma102.pub.BYLSMA.87,qt} & \htuse{HRS.Gamma102.pub.BYLSMA.87,exp} & \htuse{HRS.Gamma102.pub.BYLSMA.87,ref} \\
\htuse{L3.Gamma102.pub.ACHARD.01D,qt} & \htuse{L3.Gamma102.pub.ACHARD.01D,exp} & \htuse{L3.Gamma102.pub.ACHARD.01D,ref}
}%
\htdef{Gamma103.qm}{%
\begin{ensuredisplaymath}
\htuse{Gamma103.gn} = \htuse{Gamma103.td}
\end{ensuredisplaymath}
 & \htuse{Gamma103.qt} & \hfagFitLabel\\
\htuse{ALEPH.Gamma103.pub.SCHAEL.05C,qt} & \htuse{ALEPH.Gamma103.pub.SCHAEL.05C,exp} & \htuse{ALEPH.Gamma103.pub.SCHAEL.05C,ref} \\
\htuse{ARGUS.Gamma103.pub.ALBRECHT.88B,qt} & \htuse{ARGUS.Gamma103.pub.ALBRECHT.88B,exp} & \htuse{ARGUS.Gamma103.pub.ALBRECHT.88B,ref} \\
\htuse{CLEO.Gamma103.pub.GIBAUT.94B,qt} & \htuse{CLEO.Gamma103.pub.GIBAUT.94B,exp} & \htuse{CLEO.Gamma103.pub.GIBAUT.94B,ref} \\
\htuse{DELPHI.Gamma103.pub.ABDALLAH.06A,qt} & \htuse{DELPHI.Gamma103.pub.ABDALLAH.06A,exp} & \htuse{DELPHI.Gamma103.pub.ABDALLAH.06A,ref} \\
\htuse{HRS.Gamma103.pub.BYLSMA.87,qt} & \htuse{HRS.Gamma103.pub.BYLSMA.87,exp} & \htuse{HRS.Gamma103.pub.BYLSMA.87,ref} \\
\htuse{OPAL.Gamma103.pub.ACKERSTAFF.99E,qt} & \htuse{OPAL.Gamma103.pub.ACKERSTAFF.99E,exp} & \htuse{OPAL.Gamma103.pub.ACKERSTAFF.99E,ref}
}%
\htdef{Gamma104.qm}{%
\begin{ensuredisplaymath}
\htuse{Gamma104.gn} = \htuse{Gamma104.td}
\end{ensuredisplaymath}
 & \htuse{Gamma104.qt} & \hfagFitLabel\\
\htuse{ALEPH.Gamma104.pub.SCHAEL.05C,qt} & \htuse{ALEPH.Gamma104.pub.SCHAEL.05C,exp} & \htuse{ALEPH.Gamma104.pub.SCHAEL.05C,ref} \\
\htuse{CLEO.Gamma104.pub.ANASTASSOV.01,qt} & \htuse{CLEO.Gamma104.pub.ANASTASSOV.01,exp} & \htuse{CLEO.Gamma104.pub.ANASTASSOV.01,ref} \\
\htuse{DELPHI.Gamma104.pub.ABDALLAH.06A,qt} & \htuse{DELPHI.Gamma104.pub.ABDALLAH.06A,exp} & \htuse{DELPHI.Gamma104.pub.ABDALLAH.06A,ref} \\
\htuse{OPAL.Gamma104.pub.ACKERSTAFF.99E,qt} & \htuse{OPAL.Gamma104.pub.ACKERSTAFF.99E,exp} & \htuse{OPAL.Gamma104.pub.ACKERSTAFF.99E,ref}
}%
\htdef{Gamma110.qm}{%
\begin{ensuredisplaymath}
\htuse{Gamma110.gn} = \htuse{Gamma110.td}
\end{ensuredisplaymath}
 & \htuse{Gamma110.qt} & \hfagFitLabel}%
\htdef{Gamma126.qm}{%
\begin{ensuredisplaymath}
\htuse{Gamma126.gn} = \htuse{Gamma126.td}
\end{ensuredisplaymath}
 & \htuse{Gamma126.qt} & \hfagFitLabel\\
\htuse{ALEPH.Gamma126.pub.BUSKULIC.97C,qt} & \htuse{ALEPH.Gamma126.pub.BUSKULIC.97C,exp} & \htuse{ALEPH.Gamma126.pub.BUSKULIC.97C,ref} \\
\htuse{Belle.Gamma126.pub.INAMI.09,qt} & \htuse{Belle.Gamma126.pub.INAMI.09,exp} & \htuse{Belle.Gamma126.pub.INAMI.09,ref} \\
\htuse{CLEO.Gamma126.pub.ARTUSO.92,qt} & \htuse{CLEO.Gamma126.pub.ARTUSO.92,exp} & \htuse{CLEO.Gamma126.pub.ARTUSO.92,ref}
}%
\htdef{Gamma128.qm}{%
\begin{ensuredisplaymath}
\htuse{Gamma128.gn} = \htuse{Gamma128.td}
\end{ensuredisplaymath}
 & \htuse{Gamma128.qt} & \hfagFitLabel\\
\htuse{ALEPH.Gamma128.pub.BUSKULIC.97C,qt} & \htuse{ALEPH.Gamma128.pub.BUSKULIC.97C,exp} & \htuse{ALEPH.Gamma128.pub.BUSKULIC.97C,ref} \\
\htuse{BaBar.Gamma128.pub.DEL-AMO-SANCHEZ.11E,qt} & \htuse{BaBar.Gamma128.pub.DEL-AMO-SANCHEZ.11E,exp} & \htuse{BaBar.Gamma128.pub.DEL-AMO-SANCHEZ.11E,ref} \\
\htuse{Belle.Gamma128.pub.INAMI.09,qt} & \htuse{Belle.Gamma128.pub.INAMI.09,exp} & \htuse{Belle.Gamma128.pub.INAMI.09,ref} \\
\htuse{CLEO.Gamma128.pub.BARTELT.96,qt} & \htuse{CLEO.Gamma128.pub.BARTELT.96,exp} & \htuse{CLEO.Gamma128.pub.BARTELT.96,ref}
}%
\htdef{Gamma130.qm}{%
\begin{ensuredisplaymath}
\htuse{Gamma130.gn} = \htuse{Gamma130.td}
\end{ensuredisplaymath}
 & \htuse{Gamma130.qt} & \hfagFitLabel\\
\htuse{Belle.Gamma130.pub.INAMI.09,qt} & \htuse{Belle.Gamma130.pub.INAMI.09,exp} & \htuse{Belle.Gamma130.pub.INAMI.09,ref} \\
\htuse{CLEO.Gamma130.pub.BISHAI.99,qt} & \htuse{CLEO.Gamma130.pub.BISHAI.99,exp} & \htuse{CLEO.Gamma130.pub.BISHAI.99,ref}
}%
\htdef{Gamma132.qm}{%
\begin{ensuredisplaymath}
\htuse{Gamma132.gn} = \htuse{Gamma132.td}
\end{ensuredisplaymath}
 & \htuse{Gamma132.qt} & \hfagFitLabel\\
\htuse{Belle.Gamma132.pub.INAMI.09,qt} & \htuse{Belle.Gamma132.pub.INAMI.09,exp} & \htuse{Belle.Gamma132.pub.INAMI.09,ref} \\
\htuse{CLEO.Gamma132.pub.BISHAI.99,qt} & \htuse{CLEO.Gamma132.pub.BISHAI.99,exp} & \htuse{CLEO.Gamma132.pub.BISHAI.99,ref}
}%
\htdef{Gamma136.qm}{%
\begin{ensuredisplaymath}
\htuse{Gamma136.gn} = \htuse{Gamma136.td}
\end{ensuredisplaymath}
 & \htuse{Gamma136.qt} & \hfagFitLabel}%
\htdef{Gamma150.qm}{%
\begin{ensuredisplaymath}
\htuse{Gamma150.gn} = \htuse{Gamma150.td}
\end{ensuredisplaymath}
 & \htuse{Gamma150.qt} & \hfagFitLabel\\
\htuse{ALEPH.Gamma150.pub.BUSKULIC.97C,qt} & \htuse{ALEPH.Gamma150.pub.BUSKULIC.97C,exp} & \htuse{ALEPH.Gamma150.pub.BUSKULIC.97C,ref} \\
\htuse{CLEO.Gamma150.pub.BARINGER.87,qt} & \htuse{CLEO.Gamma150.pub.BARINGER.87,exp} & \htuse{CLEO.Gamma150.pub.BARINGER.87,ref}
}%
\htdef{Gamma150by66.qm}{%
\begin{ensuredisplaymath}
\htuse{Gamma150by66.gn} = \htuse{Gamma150by66.td}
\end{ensuredisplaymath}
 & \htuse{Gamma150by66.qt} & \hfagFitLabel\\
\htuse{ALEPH.Gamma150by66.pub.BUSKULIC.96,qt} & \htuse{ALEPH.Gamma150by66.pub.BUSKULIC.96,exp} & \htuse{ALEPH.Gamma150by66.pub.BUSKULIC.96,ref} \\
\htuse{CLEO.Gamma150by66.pub.BALEST.95C,qt} & \htuse{CLEO.Gamma150by66.pub.BALEST.95C,exp} & \htuse{CLEO.Gamma150by66.pub.BALEST.95C,ref}
}%
\htdef{Gamma151.qm}{%
\begin{ensuredisplaymath}
\htuse{Gamma151.gn} = \htuse{Gamma151.td}
\end{ensuredisplaymath}
 & \htuse{Gamma151.qt} & \hfagFitLabel\\
\htuse{CLEO3.Gamma151.pub.ARMS.05,qt} & \htuse{CLEO3.Gamma151.pub.ARMS.05,exp} & \htuse{CLEO3.Gamma151.pub.ARMS.05,ref}
}%
\htdef{Gamma152.qm}{%
\begin{ensuredisplaymath}
\htuse{Gamma152.gn} = \htuse{Gamma152.td}
\end{ensuredisplaymath}
 & \htuse{Gamma152.qt} & \hfagFitLabel\\
\htuse{ALEPH.Gamma152.pub.BUSKULIC.97C,qt} & \htuse{ALEPH.Gamma152.pub.BUSKULIC.97C,exp} & \htuse{ALEPH.Gamma152.pub.BUSKULIC.97C,ref}
}%
\htdef{Gamma152by76.qm}{%
\begin{ensuredisplaymath}
\htuse{Gamma152by76.gn} = \htuse{Gamma152by76.td}
\end{ensuredisplaymath}
 & \htuse{Gamma152by76.qt} & \hfagFitLabel\\
\htuse{CLEO.Gamma152by76.pub.BORTOLETTO.93,qt} & \htuse{CLEO.Gamma152by76.pub.BORTOLETTO.93,exp} & \htuse{CLEO.Gamma152by76.pub.BORTOLETTO.93,ref}
}%
\htdef{Gamma800.qm}{%
\begin{ensuredisplaymath}
\htuse{Gamma800.gn} = \htuse{Gamma800.td}
\end{ensuredisplaymath}
 & \htuse{Gamma800.qt} & \hfagFitLabel}%
\htdef{Gamma801.qm}{%
\begin{ensuredisplaymath}
\htuse{Gamma801.gn} = \htuse{Gamma801.td}
\end{ensuredisplaymath}
 & \htuse{Gamma801.qt} & \hfagFitLabel}%
\htdef{Gamma802.qm}{%
\begin{ensuredisplaymath}
\htuse{Gamma802.gn} = \htuse{Gamma802.td}
\end{ensuredisplaymath}
 & \htuse{Gamma802.qt} & \hfagFitLabel}%
\htdef{Gamma803.qm}{%
\begin{ensuredisplaymath}
\htuse{Gamma803.gn} = \htuse{Gamma803.td}
\end{ensuredisplaymath}
 & \htuse{Gamma803.qt} & \hfagFitLabel}%
\htdef{Gamma804.qm}{%
\begin{ensuredisplaymath}
\htuse{Gamma804.gn} = \htuse{Gamma804.td}
\end{ensuredisplaymath}
 & \htuse{Gamma804.qt} & \hfagFitLabel}%
\htdef{Gamma805.qm}{%
\begin{ensuredisplaymath}
\htuse{Gamma805.gn} = \htuse{Gamma805.td}
\end{ensuredisplaymath}
 & \htuse{Gamma805.qt} & \hfagFitLabel\\
\htuse{ALEPH.Gamma805.pub.SCHAEL.05C,qt} & \htuse{ALEPH.Gamma805.pub.SCHAEL.05C,exp} & \htuse{ALEPH.Gamma805.pub.SCHAEL.05C,ref}
}%
\htdef{Gamma806.qm}{%
\begin{ensuredisplaymath}
\htuse{Gamma806.gn} = \htuse{Gamma806.td}
\end{ensuredisplaymath}
 & \htuse{Gamma806.qt} & \hfagFitLabel}%
\htdef{Gamma810.qm}{%
\begin{ensuredisplaymath}
\htuse{Gamma810.gn} = \htuse{Gamma810.td}
\end{ensuredisplaymath}
 & \htuse{Gamma810.qt} & \hfagFitLabel}%
\htdef{Gamma811.qm}{%
\begin{ensuredisplaymath}
\htuse{Gamma811.gn} = \htuse{Gamma811.td}
\end{ensuredisplaymath}
 & \htuse{Gamma811.qt} & \hfagFitLabel\\
\htuse{BaBar.Gamma811.pub.LEES.2012X,qt} & \htuse{BaBar.Gamma811.pub.LEES.2012X,exp} & \htuse{BaBar.Gamma811.pub.LEES.2012X,ref}
}%
\htdef{Gamma812.qm}{%
\begin{ensuredisplaymath}
\htuse{Gamma812.gn} = \htuse{Gamma812.td}
\end{ensuredisplaymath}
 & \htuse{Gamma812.qt} & \hfagFitLabel\\
\htuse{BaBar.Gamma812.pub.LEES.2012X,qt} & \htuse{BaBar.Gamma812.pub.LEES.2012X,exp} & \htuse{BaBar.Gamma812.pub.LEES.2012X,ref}
}%
\htdef{Gamma820.qm}{%
\begin{ensuredisplaymath}
\htuse{Gamma820.gn} = \htuse{Gamma820.td}
\end{ensuredisplaymath}
 & \htuse{Gamma820.qt} & \hfagFitLabel}%
\htdef{Gamma821.qm}{%
\begin{ensuredisplaymath}
\htuse{Gamma821.gn} = \htuse{Gamma821.td}
\end{ensuredisplaymath}
 & \htuse{Gamma821.qt} & \hfagFitLabel\\
\htuse{BaBar.Gamma821.pub.LEES.2012X,qt} & \htuse{BaBar.Gamma821.pub.LEES.2012X,exp} & \htuse{BaBar.Gamma821.pub.LEES.2012X,ref}
}%
\htdef{Gamma822.qm}{%
\begin{ensuredisplaymath}
\htuse{Gamma822.gn} = \htuse{Gamma822.td}
\end{ensuredisplaymath}
 & \htuse{Gamma822.qt} & \hfagFitLabel\\
\htuse{BaBar.Gamma822.pub.LEES.2012X,qt} & \htuse{BaBar.Gamma822.pub.LEES.2012X,exp} & \htuse{BaBar.Gamma822.pub.LEES.2012X,ref}
}%
\htdef{Gamma830.qm}{%
\begin{ensuredisplaymath}
\htuse{Gamma830.gn} = \htuse{Gamma830.td}
\end{ensuredisplaymath}
 & \htuse{Gamma830.qt} & \hfagFitLabel}%
\htdef{Gamma831.qm}{%
\begin{ensuredisplaymath}
\htuse{Gamma831.gn} = \htuse{Gamma831.td}
\end{ensuredisplaymath}
 & \htuse{Gamma831.qt} & \hfagFitLabel\\
\htuse{BaBar.Gamma831.pub.LEES.2012X,qt} & \htuse{BaBar.Gamma831.pub.LEES.2012X,exp} & \htuse{BaBar.Gamma831.pub.LEES.2012X,ref}
}%
\htdef{Gamma832.qm}{%
\begin{ensuredisplaymath}
\htuse{Gamma832.gn} = \htuse{Gamma832.td}
\end{ensuredisplaymath}
 & \htuse{Gamma832.qt} & \hfagFitLabel\\
\htuse{BaBar.Gamma832.pub.LEES.2012X,qt} & \htuse{BaBar.Gamma832.pub.LEES.2012X,exp} & \htuse{BaBar.Gamma832.pub.LEES.2012X,ref}
}%
\htdef{Gamma833.qm}{%
\begin{ensuredisplaymath}
\htuse{Gamma833.gn} = \htuse{Gamma833.td}
\end{ensuredisplaymath}
 & \htuse{Gamma833.qt} & \hfagFitLabel\\
\htuse{BaBar.Gamma833.pub.LEES.2012X,qt} & \htuse{BaBar.Gamma833.pub.LEES.2012X,exp} & \htuse{BaBar.Gamma833.pub.LEES.2012X,ref}
}%
\htdef{Gamma910.qm}{%
\begin{ensuredisplaymath}
\htuse{Gamma910.gn} = \htuse{Gamma910.td}
\end{ensuredisplaymath}
 & \htuse{Gamma910.qt} & \hfagFitLabel\\
\htuse{BaBar.Gamma910.pub.LEES.2012X,qt} & \htuse{BaBar.Gamma910.pub.LEES.2012X,exp} & \htuse{BaBar.Gamma910.pub.LEES.2012X,ref}
}%
\htdef{Gamma911.qm}{%
\begin{ensuredisplaymath}
\htuse{Gamma911.gn} = \htuse{Gamma911.td}
\end{ensuredisplaymath}
 & \htuse{Gamma911.qt} & \hfagFitLabel\\
\htuse{BaBar.Gamma911.pub.LEES.2012X,qt} & \htuse{BaBar.Gamma911.pub.LEES.2012X,exp} & \htuse{BaBar.Gamma911.pub.LEES.2012X,ref}
}%
\htdef{Gamma920.qm}{%
\begin{ensuredisplaymath}
\htuse{Gamma920.gn} = \htuse{Gamma920.td}
\end{ensuredisplaymath}
 & \htuse{Gamma920.qt} & \hfagFitLabel\\
\htuse{BaBar.Gamma920.pub.LEES.2012X,qt} & \htuse{BaBar.Gamma920.pub.LEES.2012X,exp} & \htuse{BaBar.Gamma920.pub.LEES.2012X,ref}
}%
\htdef{Gamma930.qm}{%
\begin{ensuredisplaymath}
\htuse{Gamma930.gn} = \htuse{Gamma930.td}
\end{ensuredisplaymath}
 & \htuse{Gamma930.qt} & \hfagFitLabel\\
\htuse{BaBar.Gamma930.pub.LEES.2012X,qt} & \htuse{BaBar.Gamma930.pub.LEES.2012X,exp} & \htuse{BaBar.Gamma930.pub.LEES.2012X,ref}
}%
\htdef{Gamma944.qm}{%
\begin{ensuredisplaymath}
\htuse{Gamma944.gn} = \htuse{Gamma944.td}
\end{ensuredisplaymath}
 & \htuse{Gamma944.qt} & \hfagFitLabel\\
\htuse{BaBar.Gamma944.pub.LEES.2012X,qt} & \htuse{BaBar.Gamma944.pub.LEES.2012X,exp} & \htuse{BaBar.Gamma944.pub.LEES.2012X,ref}
}%
\htdef{Gamma998.qm}{%
\begin{ensuredisplaymath}
\htuse{Gamma998.gn} = \htuse{Gamma998.td}
\end{ensuredisplaymath}
 & \htuse{Gamma998.qt} & \hfagFitLabel}%
\htdef{BrVal}{%
\htuse{Gamma3.qm} \\
\hline
\htuse{Gamma3by5.qm} \\
\hline
\htuse{Gamma5.qm} \\
\hline
\htuse{Gamma7.qm} \\
\hline
\htuse{Gamma8.qm} \\
\hline
\htuse{Gamma9.qm} \\
\hline
\htuse{Gamma9by5.qm} \\
\hline
\htuse{Gamma10.qm} \\
\hline
\htuse{Gamma10by5.qm} \\
\hline
\htuse{Gamma13.qm} \\
\hline
\htuse{Gamma14.qm} \\
\hline
\htuse{Gamma16.qm} \\
\hline
\htuse{Gamma17.qm} \\
\hline
\htuse{Gamma19.qm} \\
\hline
\htuse{Gamma19by13.qm} \\
\hline
\htuse{Gamma20.qm} \\
\hline
\htuse{Gamma23.qm} \\
\hline
\htuse{Gamma25.qm} \\
\hline
\htuse{Gamma26.qm} \\
\hline
\htuse{Gamma26by13.qm} \\
\hline
\htuse{Gamma27.qm} \\
\hline
\htuse{Gamma28.qm} \\
\hline
\htuse{Gamma29.qm} \\
\hline
\htuse{Gamma30.qm} \\
\hline
\htuse{Gamma31.qm} \\
\hline
\htuse{Gamma33.qm} \\
\hline
\htuse{Gamma34.qm} \\
\hline
\htuse{Gamma35.qm} \\
\hline
\htuse{Gamma37.qm} \\
\hline
\htuse{Gamma38.qm} \\
\hline
\htuse{Gamma39.qm} \\
\hline
\htuse{Gamma40.qm} \\
\hline
\htuse{Gamma42.qm} \\
\hline
\htuse{Gamma43.qm} \\
\hline
\htuse{Gamma44.qm} \\
\hline
\htuse{Gamma46.qm} \\
\hline
\htuse{Gamma47.qm} \\
\hline
\htuse{Gamma48.qm} \\
\hline
\htuse{Gamma49.qm} \\
\hline
\htuse{Gamma50.qm} \\
\hline
\htuse{Gamma51.qm} \\
\hline
\htuse{Gamma53.qm} \\
\hline
\htuse{Gamma54.qm} \\
\hline
\htuse{Gamma55.qm} \\
\hline
\htuse{Gamma57.qm} \\
\hline
\htuse{Gamma57by55.qm} \\
\hline
\htuse{Gamma58.qm} \\
\hline
\htuse{Gamma60.qm} \\
\hline
\htuse{Gamma62.qm} \\
\hline
\htuse{Gamma66.qm} \\
\hline
\htuse{Gamma69.qm} \\
\hline
\htuse{Gamma70.qm} \\
\hline
\htuse{Gamma74.qm} \\
\hline
\htuse{Gamma76.qm} \\
\hline
\htuse{Gamma76by54.qm} \\
\hline
\htuse{Gamma77.qm} \\
\hline
\htuse{Gamma78.qm} \\
\hline
\htuse{Gamma80by60.qm} \\
\hline
\htuse{Gamma81by69.qm} \\
\hline
\htuse{Gamma82.qm} \\
\hline
\htuse{Gamma85.qm} \\
\hline
\htuse{Gamma88.qm} \\
\hline
\htuse{Gamma92.qm} \\
\hline
\htuse{Gamma93.qm} \\
\hline
\htuse{Gamma93by60.qm} \\
\hline
\htuse{Gamma94.qm} \\
\hline
\htuse{Gamma94by69.qm} \\
\hline
\htuse{Gamma96.qm} \\
\hline
\htuse{Gamma102.qm} \\
\hline
\htuse{Gamma103.qm} \\
\hline
\htuse{Gamma104.qm} \\
\hline
\htuse{Gamma110.qm} \\
\hline
\htuse{Gamma126.qm} \\
\hline
\htuse{Gamma128.qm} \\
\hline
\htuse{Gamma130.qm} \\
\hline
\htuse{Gamma132.qm} \\
\hline
\htuse{Gamma136.qm} \\
\hline
\htuse{Gamma150.qm} \\
\hline
\htuse{Gamma150by66.qm} \\
\hline
\htuse{Gamma151.qm} \\
\hline
\htuse{Gamma152.qm} \\
\hline
\htuse{Gamma152by76.qm} \\
\hline
\htuse{Gamma800.qm} \\
\hline
\htuse{Gamma801.qm} \\
\hline
\htuse{Gamma802.qm} \\
\hline
\htuse{Gamma803.qm} \\
\hline
\htuse{Gamma804.qm} \\
\hline
\htuse{Gamma805.qm} \\
\hline
\htuse{Gamma806.qm} \\
\hline
\htuse{Gamma810.qm} \\
\hline
\htuse{Gamma811.qm} \\
\hline
\htuse{Gamma812.qm} \\
\hline
\htuse{Gamma820.qm} \\
\hline
\htuse{Gamma821.qm} \\
\hline
\htuse{Gamma822.qm} \\
\hline
\htuse{Gamma830.qm} \\
\hline
\htuse{Gamma831.qm} \\
\hline
\htuse{Gamma832.qm} \\
\hline
\htuse{Gamma833.qm} \\
\hline
\htuse{Gamma910.qm} \\
\hline
\htuse{Gamma911.qm} \\
\hline
\htuse{Gamma920.qm} \\
\hline
\htuse{Gamma930.qm} \\
\hline
\htuse{Gamma944.qm} \\
\hline
\htuse{Gamma998.qm}}%
\htdef{BARATE 98.cite}{\cite{Barate:1997ma}}%
\htdef{BARATE 98.collab}{ALEPH}%
\htdef{BARATE 98.ref}{BARATE 98 (ALEPH) \cite{Barate:1997ma}}%
\htdef{BARATE 98.meas}{%
\begin{ensuredisplaymath}
\htuse{Gamma85.gn} = \htuse{Gamma85.td}
\end{ensuredisplaymath} & \htuse{ALEPH.Gamma85.pub.BARATE.98}
\\
\begin{ensuredisplaymath}
\htuse{Gamma88.gn} = \htuse{Gamma88.td}
\end{ensuredisplaymath} & \htuse{ALEPH.Gamma88.pub.BARATE.98}
\\
\begin{ensuredisplaymath}
\htuse{Gamma93.gn} = \htuse{Gamma93.td}
\end{ensuredisplaymath} & \htuse{ALEPH.Gamma93.pub.BARATE.98}
\\
\begin{ensuredisplaymath}
\htuse{Gamma94.gn} = \htuse{Gamma94.td}
\end{ensuredisplaymath} & \htuse{ALEPH.Gamma94.pub.BARATE.98}}%
\htdef{BARATE 98E.cite}{\cite{Barate:1997tt}}%
\htdef{BARATE 98E.collab}{ALEPH}%
\htdef{BARATE 98E.ref}{BARATE 98E (ALEPH) \cite{Barate:1997tt}}%
\htdef{BARATE 98E.meas}{%
\begin{ensuredisplaymath}
\htuse{Gamma33.gn} = \htuse{Gamma33.td}
\end{ensuredisplaymath} & \htuse{ALEPH.Gamma33.pub.BARATE.98E}
\\
\begin{ensuredisplaymath}
\htuse{Gamma37.gn} = \htuse{Gamma37.td}
\end{ensuredisplaymath} & \htuse{ALEPH.Gamma37.pub.BARATE.98E}
\\
\begin{ensuredisplaymath}
\htuse{Gamma40.gn} = \htuse{Gamma40.td}
\end{ensuredisplaymath} & \htuse{ALEPH.Gamma40.pub.BARATE.98E}
\\
\begin{ensuredisplaymath}
\htuse{Gamma42.gn} = \htuse{Gamma42.td}
\end{ensuredisplaymath} & \htuse{ALEPH.Gamma42.pub.BARATE.98E}
\\
\begin{ensuredisplaymath}
\htuse{Gamma46.gn} = \htuse{Gamma46.td}
\end{ensuredisplaymath} & \htuse{ALEPH.Gamma46.pub.BARATE.98E}
\\
\begin{ensuredisplaymath}
\htuse{Gamma47.gn} = \htuse{Gamma47.td}
\end{ensuredisplaymath} & \htuse{ALEPH.Gamma47.pub.BARATE.98E}
\\
\begin{ensuredisplaymath}
\htuse{Gamma48.gn} = \htuse{Gamma48.td}
\end{ensuredisplaymath} & \htuse{ALEPH.Gamma48.pub.BARATE.98E}
\\
\begin{ensuredisplaymath}
\htuse{Gamma51.gn} = \htuse{Gamma51.td}
\end{ensuredisplaymath} & \htuse{ALEPH.Gamma51.pub.BARATE.98E}
\\
\begin{ensuredisplaymath}
\htuse{Gamma53.gn} = \htuse{Gamma53.td}
\end{ensuredisplaymath} & \htuse{ALEPH.Gamma53.pub.BARATE.98E}}%
\htdef{BARATE 99K.cite}{\cite{Barate:1999hi}}%
\htdef{BARATE 99K.collab}{ALEPH}%
\htdef{BARATE 99K.ref}{BARATE 99K (ALEPH) \cite{Barate:1999hi}}%
\htdef{BARATE 99K.meas}{%
\begin{ensuredisplaymath}
\htuse{Gamma10.gn} = \htuse{Gamma10.td}
\end{ensuredisplaymath} & \htuse{ALEPH.Gamma10.pub.BARATE.99K}
\\
\begin{ensuredisplaymath}
\htuse{Gamma16.gn} = \htuse{Gamma16.td}
\end{ensuredisplaymath} & \htuse{ALEPH.Gamma16.pub.BARATE.99K}
\\
\begin{ensuredisplaymath}
\htuse{Gamma23.gn} = \htuse{Gamma23.td}
\end{ensuredisplaymath} & \htuse{ALEPH.Gamma23.pub.BARATE.99K}
\\
\begin{ensuredisplaymath}
\htuse{Gamma28.gn} = \htuse{Gamma28.td}
\end{ensuredisplaymath} & \htuse{ALEPH.Gamma28.pub.BARATE.99K}
\\
\begin{ensuredisplaymath}
\htuse{Gamma35.gn} = \htuse{Gamma35.td}
\end{ensuredisplaymath} & \htuse{ALEPH.Gamma35.pub.BARATE.99K}
\\
\begin{ensuredisplaymath}
\htuse{Gamma37.gn} = \htuse{Gamma37.td}
\end{ensuredisplaymath} & \htuse{ALEPH.Gamma37.pub.BARATE.99K}
\\
\begin{ensuredisplaymath}
\htuse{Gamma40.gn} = \htuse{Gamma40.td}
\end{ensuredisplaymath} & \htuse{ALEPH.Gamma40.pub.BARATE.99K}
\\
\begin{ensuredisplaymath}
\htuse{Gamma42.gn} = \htuse{Gamma42.td}
\end{ensuredisplaymath} & \htuse{ALEPH.Gamma42.pub.BARATE.99K}}%
\htdef{BARATE 99R.cite}{\cite{Barate:1999hj}}%
\htdef{BARATE 99R.collab}{ALEPH}%
\htdef{BARATE 99R.ref}{BARATE 99R (ALEPH) \cite{Barate:1999hj}}%
\htdef{BARATE 99R.meas}{%
\begin{ensuredisplaymath}
\htuse{Gamma44.gn} = \htuse{Gamma44.td}
\end{ensuredisplaymath} & \htuse{ALEPH.Gamma44.pub.BARATE.99R}}%
\htdef{BUSKULIC 96.cite}{\cite{Buskulic:1995ty}}%
\htdef{BUSKULIC 96.collab}{ALEPH}%
\htdef{BUSKULIC 96.ref}{BUSKULIC 96 (ALEPH) \cite{Buskulic:1995ty}}%
\htdef{BUSKULIC 96.meas}{%
\begin{ensuredisplaymath}
\htuse{Gamma150by66.gn} = \htuse{Gamma150by66.td}
\end{ensuredisplaymath} & \htuse{ALEPH.Gamma150by66.pub.BUSKULIC.96}}%
\htdef{BUSKULIC 97C.cite}{\cite{Buskulic:1996qs}}%
\htdef{BUSKULIC 97C.collab}{ALEPH}%
\htdef{BUSKULIC 97C.ref}{BUSKULIC 97C (ALEPH) \cite{Buskulic:1996qs}}%
\htdef{BUSKULIC 97C.meas}{%
\begin{ensuredisplaymath}
\htuse{Gamma126.gn} = \htuse{Gamma126.td}
\end{ensuredisplaymath} & \htuse{ALEPH.Gamma126.pub.BUSKULIC.97C}
\\
\begin{ensuredisplaymath}
\htuse{Gamma128.gn} = \htuse{Gamma128.td}
\end{ensuredisplaymath} & \htuse{ALEPH.Gamma128.pub.BUSKULIC.97C}
\\
\begin{ensuredisplaymath}
\htuse{Gamma150.gn} = \htuse{Gamma150.td}
\end{ensuredisplaymath} & \htuse{ALEPH.Gamma150.pub.BUSKULIC.97C}
\\
\begin{ensuredisplaymath}
\htuse{Gamma152.gn} = \htuse{Gamma152.td}
\end{ensuredisplaymath} & \htuse{ALEPH.Gamma152.pub.BUSKULIC.97C}}%
\htdef{SCHAEL 05C.cite}{\cite{Schael:2005am}}%
\htdef{SCHAEL 05C.collab}{ALEPH}%
\htdef{SCHAEL 05C.ref}{SCHAEL 05C (ALEPH) \cite{Schael:2005am}}%
\htdef{SCHAEL 05C.meas}{%
\begin{ensuredisplaymath}
\htuse{Gamma3.gn} = \htuse{Gamma3.td}
\end{ensuredisplaymath} & \htuse{ALEPH.Gamma3.pub.SCHAEL.05C}
\\
\begin{ensuredisplaymath}
\htuse{Gamma5.gn} = \htuse{Gamma5.td}
\end{ensuredisplaymath} & \htuse{ALEPH.Gamma5.pub.SCHAEL.05C}
\\
\begin{ensuredisplaymath}
\htuse{Gamma8.gn} = \htuse{Gamma8.td}
\end{ensuredisplaymath} & \htuse{ALEPH.Gamma8.pub.SCHAEL.05C}
\\
\begin{ensuredisplaymath}
\htuse{Gamma13.gn} = \htuse{Gamma13.td}
\end{ensuredisplaymath} & \htuse{ALEPH.Gamma13.pub.SCHAEL.05C}
\\
\begin{ensuredisplaymath}
\htuse{Gamma19.gn} = \htuse{Gamma19.td}
\end{ensuredisplaymath} & \htuse{ALEPH.Gamma19.pub.SCHAEL.05C}
\\
\begin{ensuredisplaymath}
\htuse{Gamma26.gn} = \htuse{Gamma26.td}
\end{ensuredisplaymath} & \htuse{ALEPH.Gamma26.pub.SCHAEL.05C}
\\
\begin{ensuredisplaymath}
\htuse{Gamma30.gn} = \htuse{Gamma30.td}
\end{ensuredisplaymath} & \htuse{ALEPH.Gamma30.pub.SCHAEL.05C}
\\
\begin{ensuredisplaymath}
\htuse{Gamma58.gn} = \htuse{Gamma58.td}
\end{ensuredisplaymath} & \htuse{ALEPH.Gamma58.pub.SCHAEL.05C}
\\
\begin{ensuredisplaymath}
\htuse{Gamma66.gn} = \htuse{Gamma66.td}
\end{ensuredisplaymath} & \htuse{ALEPH.Gamma66.pub.SCHAEL.05C}
\\
\begin{ensuredisplaymath}
\htuse{Gamma76.gn} = \htuse{Gamma76.td}
\end{ensuredisplaymath} & \htuse{ALEPH.Gamma76.pub.SCHAEL.05C}
\\
\begin{ensuredisplaymath}
\htuse{Gamma103.gn} = \htuse{Gamma103.td}
\end{ensuredisplaymath} & \htuse{ALEPH.Gamma103.pub.SCHAEL.05C}
\\
\begin{ensuredisplaymath}
\htuse{Gamma104.gn} = \htuse{Gamma104.td}
\end{ensuredisplaymath} & \htuse{ALEPH.Gamma104.pub.SCHAEL.05C}
\\
\begin{ensuredisplaymath}
\htuse{Gamma805.gn} = \htuse{Gamma805.td}
\end{ensuredisplaymath} & \htuse{ALEPH.Gamma805.pub.SCHAEL.05C}}%
\htdef{ALBRECHT 88B.cite}{\cite{Albrecht:1987zf}}%
\htdef{ALBRECHT 88B.collab}{ARGUS}%
\htdef{ALBRECHT 88B.ref}{ALBRECHT 88B (ARGUS) \cite{Albrecht:1987zf}}%
\htdef{ALBRECHT 88B.meas}{%
\begin{ensuredisplaymath}
\htuse{Gamma103.gn} = \htuse{Gamma103.td}
\end{ensuredisplaymath} & \htuse{ARGUS.Gamma103.pub.ALBRECHT.88B}}%
\htdef{ALBRECHT 92D.cite}{\cite{Albrecht:1991rh}}%
\htdef{ALBRECHT 92D.collab}{ARGUS}%
\htdef{ALBRECHT 92D.ref}{ALBRECHT 92D (ARGUS) \cite{Albrecht:1991rh}}%
\htdef{ALBRECHT 92D.meas}{%
\begin{ensuredisplaymath}
\htuse{Gamma3by5.gn} = \htuse{Gamma3by5.td}
\end{ensuredisplaymath} & \htuse{ARGUS.Gamma3by5.pub.ALBRECHT.92D}}%
\htdef{AUBERT 07AP.cite}{\cite{Aubert:2007jh}}%
\htdef{AUBERT 07AP.collab}{\babar}%
\htdef{AUBERT 07AP.ref}{AUBERT 07AP (\babar) \cite{Aubert:2007jh}}%
\htdef{AUBERT 07AP.meas}{%
\begin{ensuredisplaymath}
\htuse{Gamma16.gn} = \htuse{Gamma16.td}
\end{ensuredisplaymath} & \htuse{BaBar.Gamma16.pub.AUBERT.07AP}}%
\htdef{AUBERT 08.cite}{\cite{Aubert:2007mh}}%
\htdef{AUBERT 08.collab}{\babar}%
\htdef{AUBERT 08.ref}{AUBERT 08 (\babar) \cite{Aubert:2007mh}}%
\htdef{AUBERT 08.meas}{%
\begin{ensuredisplaymath}
\htuse{Gamma60.gn} = \htuse{Gamma60.td}
\end{ensuredisplaymath} & \htuse{BaBar.Gamma60.pub.AUBERT.08}
\\
\begin{ensuredisplaymath}
\htuse{Gamma85.gn} = \htuse{Gamma85.td}
\end{ensuredisplaymath} & \htuse{BaBar.Gamma85.pub.AUBERT.08}
\\
\begin{ensuredisplaymath}
\htuse{Gamma93.gn} = \htuse{Gamma93.td}
\end{ensuredisplaymath} & \htuse{BaBar.Gamma93.pub.AUBERT.08}
\\
\begin{ensuredisplaymath}
\htuse{Gamma96.gn} = \htuse{Gamma96.td}
\end{ensuredisplaymath} & \htuse{BaBar.Gamma96.pub.AUBERT.08}}%
\htdef{AUBERT 10F.cite}{\cite{Aubert:2009qj}}%
\htdef{AUBERT 10F.collab}{\babar}%
\htdef{AUBERT 10F.ref}{AUBERT 10F (\babar) \cite{Aubert:2009qj}}%
\htdef{AUBERT 10F.meas}{%
\begin{ensuredisplaymath}
\htuse{Gamma3by5.gn} = \htuse{Gamma3by5.td}
\end{ensuredisplaymath} & \htuse{BaBar.Gamma3by5.pub.AUBERT.10F}
\\
\begin{ensuredisplaymath}
\htuse{Gamma9by5.gn} = \htuse{Gamma9by5.td}
\end{ensuredisplaymath} & \htuse{BaBar.Gamma9by5.pub.AUBERT.10F}
\\
\begin{ensuredisplaymath}
\htuse{Gamma10by5.gn} = \htuse{Gamma10by5.td}
\end{ensuredisplaymath} & \htuse{BaBar.Gamma10by5.pub.AUBERT.10F}}%
\htdef{DEL-AMO-SANCHEZ 11E.cite}{\cite{delAmoSanchez:2010pc}}%
\htdef{DEL-AMO-SANCHEZ 11E.collab}{\babar}%
\htdef{DEL-AMO-SANCHEZ 11E.ref}{DEL-AMO-SANCHEZ 11E (\babar) \cite{delAmoSanchez:2010pc}}%
\htdef{DEL-AMO-SANCHEZ 11E.meas}{%
\begin{ensuredisplaymath}
\htuse{Gamma128.gn} = \htuse{Gamma128.td}
\end{ensuredisplaymath} & \htuse{BaBar.Gamma128.pub.DEL-AMO-SANCHEZ.11E}}%
\htdef{LEES 2012X.cite}{\cite{Lees:2012ks}}%
\htdef{LEES 2012X.collab}{\babar}%
\htdef{LEES 2012X.ref}{LEES 2012X (\babar) \cite{Lees:2012ks}}%
\htdef{LEES 2012X.meas}{%
\begin{ensuredisplaymath}
\htuse{Gamma811.gn} = \htuse{Gamma811.td}
\end{ensuredisplaymath} & \htuse{BaBar.Gamma811.pub.LEES.2012X}
\\
\begin{ensuredisplaymath}
\htuse{Gamma812.gn} = \htuse{Gamma812.td}
\end{ensuredisplaymath} & \htuse{BaBar.Gamma812.pub.LEES.2012X}
\\
\begin{ensuredisplaymath}
\htuse{Gamma821.gn} = \htuse{Gamma821.td}
\end{ensuredisplaymath} & \htuse{BaBar.Gamma821.pub.LEES.2012X}
\\
\begin{ensuredisplaymath}
\htuse{Gamma822.gn} = \htuse{Gamma822.td}
\end{ensuredisplaymath} & \htuse{BaBar.Gamma822.pub.LEES.2012X}
\\
\begin{ensuredisplaymath}
\htuse{Gamma831.gn} = \htuse{Gamma831.td}
\end{ensuredisplaymath} & \htuse{BaBar.Gamma831.pub.LEES.2012X}
\\
\begin{ensuredisplaymath}
\htuse{Gamma832.gn} = \htuse{Gamma832.td}
\end{ensuredisplaymath} & \htuse{BaBar.Gamma832.pub.LEES.2012X}
\\
\begin{ensuredisplaymath}
\htuse{Gamma833.gn} = \htuse{Gamma833.td}
\end{ensuredisplaymath} & \htuse{BaBar.Gamma833.pub.LEES.2012X}
\\
\begin{ensuredisplaymath}
\htuse{Gamma910.gn} = \htuse{Gamma910.td}
\end{ensuredisplaymath} & \htuse{BaBar.Gamma910.pub.LEES.2012X}
\\
\begin{ensuredisplaymath}
\htuse{Gamma911.gn} = \htuse{Gamma911.td}
\end{ensuredisplaymath} & \htuse{BaBar.Gamma911.pub.LEES.2012X}
\\
\begin{ensuredisplaymath}
\htuse{Gamma920.gn} = \htuse{Gamma920.td}
\end{ensuredisplaymath} & \htuse{BaBar.Gamma920.pub.LEES.2012X}
\\
\begin{ensuredisplaymath}
\htuse{Gamma930.gn} = \htuse{Gamma930.td}
\end{ensuredisplaymath} & \htuse{BaBar.Gamma930.pub.LEES.2012X}
\\
\begin{ensuredisplaymath}
\htuse{Gamma944.gn} = \htuse{Gamma944.td}
\end{ensuredisplaymath} & \htuse{BaBar.Gamma944.pub.LEES.2012X}}%
\htdef{LEES 2012Y.cite}{\cite{Lees:2012de}}%
\htdef{LEES 2012Y.collab}{\babar}%
\htdef{LEES 2012Y.ref}{LEES 2012Y (\babar) \cite{Lees:2012de}}%
\htdef{LEES 2012Y.meas}{%
\begin{ensuredisplaymath}
\htuse{Gamma47.gn} = \htuse{Gamma47.td}
\end{ensuredisplaymath} & \htuse{BaBar.Gamma47.pub.LEES.2012Y}
\\
\begin{ensuredisplaymath}
\htuse{Gamma50.gn} = \htuse{Gamma50.td}
\end{ensuredisplaymath} & \htuse{BaBar.Gamma50.pub.LEES.2012Y}}%
\htdef{BaBar prelim. DPF09.cite}{\cite{Paramesvaran:2009ec}}%
\htdef{BaBar prelim. DPF09.collab}{BaBar}%
\htdef{BaBar prelim. DPF09.ref}{\babar prelim. DPF09 \cite{Paramesvaran:2009ec}}%
\htdef{BaBar prelim. DPF09.meas}{%
\begin{ensuredisplaymath}
\htuse{Gamma40.gn} = \htuse{Gamma40.td}
\end{ensuredisplaymath} & \htuse{BaBar.Gamma40.prelim.DPF09}}%
\htdef{BaBar prelim. ICHEP08.cite}{\cite{Aubert:2008an}}%
\htdef{BaBar prelim. ICHEP08.collab}{BaBar}%
\htdef{BaBar prelim. ICHEP08.ref}{\babar prelim. ICHEP08 \cite{Aubert:2008an}}%
\htdef{BaBar prelim. ICHEP08.meas}{%
\begin{ensuredisplaymath}
\htuse{Gamma35.gn} = \htuse{Gamma35.td}
\end{ensuredisplaymath} & \htuse{BaBar.Gamma35.prelim.ICHEP08}}%
\htdef{FUJIKAWA 08.cite}{\cite{Fujikawa:2008ma}}%
\htdef{FUJIKAWA 08.collab}{Belle}%
\htdef{FUJIKAWA 08.ref}{FUJIKAWA 08 (Belle) \cite{Fujikawa:2008ma}}%
\htdef{FUJIKAWA 08.meas}{%
\begin{ensuredisplaymath}
\htuse{Gamma13.gn} = \htuse{Gamma13.td}
\end{ensuredisplaymath} & \htuse{Belle.Gamma13.pub.FUJIKAWA.08}}%
\htdef{INAMI 09.cite}{\cite{Inami:2008ar}}%
\htdef{INAMI 09.collab}{Belle}%
\htdef{INAMI 09.ref}{INAMI 09 (Belle) \cite{Inami:2008ar}}%
\htdef{INAMI 09.meas}{%
\begin{ensuredisplaymath}
\htuse{Gamma126.gn} = \htuse{Gamma126.td}
\end{ensuredisplaymath} & \htuse{Belle.Gamma126.pub.INAMI.09}
\\
\begin{ensuredisplaymath}
\htuse{Gamma128.gn} = \htuse{Gamma128.td}
\end{ensuredisplaymath} & \htuse{Belle.Gamma128.pub.INAMI.09}
\\
\begin{ensuredisplaymath}
\htuse{Gamma130.gn} = \htuse{Gamma130.td}
\end{ensuredisplaymath} & \htuse{Belle.Gamma130.pub.INAMI.09}
\\
\begin{ensuredisplaymath}
\htuse{Gamma132.gn} = \htuse{Gamma132.td}
\end{ensuredisplaymath} & \htuse{Belle.Gamma132.pub.INAMI.09}}%
\htdef{LEE 10.cite}{\cite{Lee:2010tc}}%
\htdef{LEE 10.collab}{Belle}%
\htdef{LEE 10.ref}{LEE 10 (Belle) \cite{Lee:2010tc}}%
\htdef{LEE 10.meas}{%
\begin{ensuredisplaymath}
\htuse{Gamma60.gn} = \htuse{Gamma60.td}
\end{ensuredisplaymath} & \htuse{Belle.Gamma60.pub.LEE.10}
\\
\begin{ensuredisplaymath}
\htuse{Gamma85.gn} = \htuse{Gamma85.td}
\end{ensuredisplaymath} & \htuse{Belle.Gamma85.pub.LEE.10}
\\
\begin{ensuredisplaymath}
\htuse{Gamma93.gn} = \htuse{Gamma93.td}
\end{ensuredisplaymath} & \htuse{Belle.Gamma93.pub.LEE.10}
\\
\begin{ensuredisplaymath}
\htuse{Gamma96.gn} = \htuse{Gamma96.td}
\end{ensuredisplaymath} & \htuse{Belle.Gamma96.pub.LEE.10}}%
\htdef{Ryu:2014vpc.cite}{\cite{Ryu:2014vpc}}%
\htdef{Ryu:2014vpc.collab}{Belle}%
\htdef{Ryu:2014vpc.ref}{Ryu:2014vpc (Belle) \cite{Ryu:2014vpc}}%
\htdef{Ryu:2014vpc.meas}{%
\begin{ensuredisplaymath}
\htuse{Gamma33.gn} = \htuse{Gamma33.td}
\end{ensuredisplaymath} & \htuse{Belle.Gamma33.pub.Ryu:2014vpc}
\\
\begin{ensuredisplaymath}
\htuse{Gamma35.gn} = \htuse{Gamma35.td}
\end{ensuredisplaymath} & \htuse{Belle.Gamma35.pub.Ryu:2014vpc}
\\
\begin{ensuredisplaymath}
\htuse{Gamma37.gn} = \htuse{Gamma37.td}
\end{ensuredisplaymath} & \htuse{Belle.Gamma37.pub.Ryu:2014vpc}
\\
\begin{ensuredisplaymath}
\htuse{Gamma40.gn} = \htuse{Gamma40.td}
\end{ensuredisplaymath} & \htuse{Belle.Gamma40.pub.Ryu:2014vpc}
\\
\begin{ensuredisplaymath}
\htuse{Gamma42.gn} = \htuse{Gamma42.td}
\end{ensuredisplaymath} & \htuse{Belle.Gamma42.pub.Ryu:2014vpc}
\\
\begin{ensuredisplaymath}
\htuse{Gamma47.gn} = \htuse{Gamma47.td}
\end{ensuredisplaymath} & \htuse{Belle.Gamma47.pub.Ryu:2014vpc}
\\
\begin{ensuredisplaymath}
\htuse{Gamma50.gn} = \htuse{Gamma50.td}
\end{ensuredisplaymath} & \htuse{Belle.Gamma50.pub.Ryu:2014vpc}}%
\htdef{BEHREND 89B.cite}{\cite{Behrend:1989wc}}%
\htdef{BEHREND 89B.collab}{CELLO}%
\htdef{BEHREND 89B.ref}{BEHREND 89B (CELLO) \cite{Behrend:1989wc}}%
\htdef{BEHREND 89B.meas}{%
\begin{ensuredisplaymath}
\htuse{Gamma54.gn} = \htuse{Gamma54.td}
\end{ensuredisplaymath} & \htuse{CELLO.Gamma54.pub.BEHREND.89B}}%
\htdef{ANASTASSOV 01.cite}{\cite{Anastassov:2000xu}}%
\htdef{ANASTASSOV 01.collab}{CLEO}%
\htdef{ANASTASSOV 01.ref}{ANASTASSOV 01 (CLEO) \cite{Anastassov:2000xu}}%
\htdef{ANASTASSOV 01.meas}{%
\begin{ensuredisplaymath}
\htuse{Gamma78.gn} = \htuse{Gamma78.td}
\end{ensuredisplaymath} & \htuse{CLEO.Gamma78.pub.ANASTASSOV.01}
\\
\begin{ensuredisplaymath}
\htuse{Gamma104.gn} = \htuse{Gamma104.td}
\end{ensuredisplaymath} & \htuse{CLEO.Gamma104.pub.ANASTASSOV.01}}%
\htdef{ANASTASSOV 97.cite}{\cite{Anastassov:1996tc}}%
\htdef{ANASTASSOV 97.collab}{CLEO}%
\htdef{ANASTASSOV 97.ref}{ANASTASSOV 97 (CLEO) \cite{Anastassov:1996tc}}%
\htdef{ANASTASSOV 97.meas}{%
\begin{ensuredisplaymath}
\htuse{Gamma3by5.gn} = \htuse{Gamma3by5.td}
\end{ensuredisplaymath} & \htuse{CLEO.Gamma3by5.pub.ANASTASSOV.97}
\\
\begin{ensuredisplaymath}
\htuse{Gamma5.gn} = \htuse{Gamma5.td}
\end{ensuredisplaymath} & \htuse{CLEO.Gamma5.pub.ANASTASSOV.97}
\\
\begin{ensuredisplaymath}
\htuse{Gamma8.gn} = \htuse{Gamma8.td}
\end{ensuredisplaymath} & \htuse{CLEO.Gamma8.pub.ANASTASSOV.97}}%
\htdef{ARTUSO 92.cite}{\cite{Artuso:1992qu}}%
\htdef{ARTUSO 92.collab}{CLEO}%
\htdef{ARTUSO 92.ref}{ARTUSO 92 (CLEO) \cite{Artuso:1992qu}}%
\htdef{ARTUSO 92.meas}{%
\begin{ensuredisplaymath}
\htuse{Gamma126.gn} = \htuse{Gamma126.td}
\end{ensuredisplaymath} & \htuse{CLEO.Gamma126.pub.ARTUSO.92}}%
\htdef{ARTUSO 94.cite}{\cite{Artuso:1994ii}}%
\htdef{ARTUSO 94.collab}{CLEO}%
\htdef{ARTUSO 94.ref}{ARTUSO 94 (CLEO) \cite{Artuso:1994ii}}%
\htdef{ARTUSO 94.meas}{%
\begin{ensuredisplaymath}
\htuse{Gamma13.gn} = \htuse{Gamma13.td}
\end{ensuredisplaymath} & \htuse{CLEO.Gamma13.pub.ARTUSO.94}}%
\htdef{BALEST 95C.cite}{\cite{Balest:1995kq}}%
\htdef{BALEST 95C.collab}{CLEO}%
\htdef{BALEST 95C.ref}{BALEST 95C (CLEO) \cite{Balest:1995kq}}%
\htdef{BALEST 95C.meas}{%
\begin{ensuredisplaymath}
\htuse{Gamma57.gn} = \htuse{Gamma57.td}
\end{ensuredisplaymath} & \htuse{CLEO.Gamma57.pub.BALEST.95C}
\\
\begin{ensuredisplaymath}
\htuse{Gamma66.gn} = \htuse{Gamma66.td}
\end{ensuredisplaymath} & \htuse{CLEO.Gamma66.pub.BALEST.95C}
\\
\begin{ensuredisplaymath}
\htuse{Gamma150by66.gn} = \htuse{Gamma150by66.td}
\end{ensuredisplaymath} & \htuse{CLEO.Gamma150by66.pub.BALEST.95C}}%
\htdef{BARINGER 87.cite}{\cite{Baringer:1987tr}}%
\htdef{BARINGER 87.collab}{CLEO}%
\htdef{BARINGER 87.ref}{BARINGER 87 (CLEO) \cite{Baringer:1987tr}}%
\htdef{BARINGER 87.meas}{%
\begin{ensuredisplaymath}
\htuse{Gamma150.gn} = \htuse{Gamma150.td}
\end{ensuredisplaymath} & \htuse{CLEO.Gamma150.pub.BARINGER.87}}%
\htdef{BARTELT 96.cite}{\cite{Bartelt:1996iv}}%
\htdef{BARTELT 96.collab}{CLEO}%
\htdef{BARTELT 96.ref}{BARTELT 96 (CLEO) \cite{Bartelt:1996iv}}%
\htdef{BARTELT 96.meas}{%
\begin{ensuredisplaymath}
\htuse{Gamma128.gn} = \htuse{Gamma128.td}
\end{ensuredisplaymath} & \htuse{CLEO.Gamma128.pub.BARTELT.96}}%
\htdef{BATTLE 94.cite}{\cite{Battle:1994by}}%
\htdef{BATTLE 94.collab}{CLEO}%
\htdef{BATTLE 94.ref}{BATTLE 94 (CLEO) \cite{Battle:1994by}}%
\htdef{BATTLE 94.meas}{%
\begin{ensuredisplaymath}
\htuse{Gamma10.gn} = \htuse{Gamma10.td}
\end{ensuredisplaymath} & \htuse{CLEO.Gamma10.pub.BATTLE.94}
\\
\begin{ensuredisplaymath}
\htuse{Gamma16.gn} = \htuse{Gamma16.td}
\end{ensuredisplaymath} & \htuse{CLEO.Gamma16.pub.BATTLE.94}
\\
\begin{ensuredisplaymath}
\htuse{Gamma23.gn} = \htuse{Gamma23.td}
\end{ensuredisplaymath} & \htuse{CLEO.Gamma23.pub.BATTLE.94}
\\
\begin{ensuredisplaymath}
\htuse{Gamma31.gn} = \htuse{Gamma31.td}
\end{ensuredisplaymath} & \htuse{CLEO.Gamma31.pub.BATTLE.94}}%
\htdef{BISHAI 99.cite}{\cite{Bishai:1998gf}}%
\htdef{BISHAI 99.collab}{CLEO}%
\htdef{BISHAI 99.ref}{BISHAI 99 (CLEO) \cite{Bishai:1998gf}}%
\htdef{BISHAI 99.meas}{%
\begin{ensuredisplaymath}
\htuse{Gamma130.gn} = \htuse{Gamma130.td}
\end{ensuredisplaymath} & \htuse{CLEO.Gamma130.pub.BISHAI.99}
\\
\begin{ensuredisplaymath}
\htuse{Gamma132.gn} = \htuse{Gamma132.td}
\end{ensuredisplaymath} & \htuse{CLEO.Gamma132.pub.BISHAI.99}}%
\htdef{BORTOLETTO 93.cite}{\cite{Bortoletto:1993px}}%
\htdef{BORTOLETTO 93.collab}{CLEO}%
\htdef{BORTOLETTO 93.ref}{BORTOLETTO 93 (CLEO) \cite{Bortoletto:1993px}}%
\htdef{BORTOLETTO 93.meas}{%
\begin{ensuredisplaymath}
\htuse{Gamma76by54.gn} = \htuse{Gamma76by54.td}
\end{ensuredisplaymath} & \htuse{CLEO.Gamma76by54.pub.BORTOLETTO.93}
\\
\begin{ensuredisplaymath}
\htuse{Gamma152by76.gn} = \htuse{Gamma152by76.td}
\end{ensuredisplaymath} & \htuse{CLEO.Gamma152by76.pub.BORTOLETTO.93}}%
\htdef{COAN 96.cite}{\cite{Coan:1996iu}}%
\htdef{COAN 96.collab}{CLEO}%
\htdef{COAN 96.ref}{COAN 96 (CLEO) \cite{Coan:1996iu}}%
\htdef{COAN 96.meas}{%
\begin{ensuredisplaymath}
\htuse{Gamma34.gn} = \htuse{Gamma34.td}
\end{ensuredisplaymath} & \htuse{CLEO.Gamma34.pub.COAN.96}
\\
\begin{ensuredisplaymath}
\htuse{Gamma37.gn} = \htuse{Gamma37.td}
\end{ensuredisplaymath} & \htuse{CLEO.Gamma37.pub.COAN.96}
\\
\begin{ensuredisplaymath}
\htuse{Gamma39.gn} = \htuse{Gamma39.td}
\end{ensuredisplaymath} & \htuse{CLEO.Gamma39.pub.COAN.96}
\\
\begin{ensuredisplaymath}
\htuse{Gamma42.gn} = \htuse{Gamma42.td}
\end{ensuredisplaymath} & \htuse{CLEO.Gamma42.pub.COAN.96}
\\
\begin{ensuredisplaymath}
\htuse{Gamma47.gn} = \htuse{Gamma47.td}
\end{ensuredisplaymath} & \htuse{CLEO.Gamma47.pub.COAN.96}}%
\htdef{EDWARDS 00A.cite}{\cite{Edwards:1999fj}}%
\htdef{EDWARDS 00A.collab}{CLEO}%
\htdef{EDWARDS 00A.ref}{EDWARDS 00A (CLEO) \cite{Edwards:1999fj}}%
\htdef{EDWARDS 00A.meas}{%
\begin{ensuredisplaymath}
\htuse{Gamma69.gn} = \htuse{Gamma69.td}
\end{ensuredisplaymath} & \htuse{CLEO.Gamma69.pub.EDWARDS.00A}}%
\htdef{GIBAUT 94B.cite}{\cite{Gibaut:1994ik}}%
\htdef{GIBAUT 94B.collab}{CLEO}%
\htdef{GIBAUT 94B.ref}{GIBAUT 94B (CLEO) \cite{Gibaut:1994ik}}%
\htdef{GIBAUT 94B.meas}{%
\begin{ensuredisplaymath}
\htuse{Gamma102.gn} = \htuse{Gamma102.td}
\end{ensuredisplaymath} & \htuse{CLEO.Gamma102.pub.GIBAUT.94B}
\\
\begin{ensuredisplaymath}
\htuse{Gamma103.gn} = \htuse{Gamma103.td}
\end{ensuredisplaymath} & \htuse{CLEO.Gamma103.pub.GIBAUT.94B}}%
\htdef{PROCARIO 93.cite}{\cite{Procario:1992hd}}%
\htdef{PROCARIO 93.collab}{CLEO}%
\htdef{PROCARIO 93.ref}{PROCARIO 93 (CLEO) \cite{Procario:1992hd}}%
\htdef{PROCARIO 93.meas}{%
\begin{ensuredisplaymath}
\htuse{Gamma19by13.gn} = \htuse{Gamma19by13.td}
\end{ensuredisplaymath} & \htuse{CLEO.Gamma19by13.pub.PROCARIO.93}
\\
\begin{ensuredisplaymath}
\htuse{Gamma26by13.gn} = \htuse{Gamma26by13.td}
\end{ensuredisplaymath} & \htuse{CLEO.Gamma26by13.pub.PROCARIO.93}
\\
\begin{ensuredisplaymath}
\htuse{Gamma29.gn} = \htuse{Gamma29.td}
\end{ensuredisplaymath} & \htuse{CLEO.Gamma29.pub.PROCARIO.93}}%
\htdef{RICHICHI 99.cite}{\cite{Richichi:1998bc}}%
\htdef{RICHICHI 99.collab}{CLEO}%
\htdef{RICHICHI 99.ref}{RICHICHI 99 (CLEO) \cite{Richichi:1998bc}}%
\htdef{RICHICHI 99.meas}{%
\begin{ensuredisplaymath}
\htuse{Gamma80by60.gn} = \htuse{Gamma80by60.td}
\end{ensuredisplaymath} & \htuse{CLEO.Gamma80by60.pub.RICHICHI.99}
\\
\begin{ensuredisplaymath}
\htuse{Gamma81by69.gn} = \htuse{Gamma81by69.td}
\end{ensuredisplaymath} & \htuse{CLEO.Gamma81by69.pub.RICHICHI.99}
\\
\begin{ensuredisplaymath}
\htuse{Gamma93by60.gn} = \htuse{Gamma93by60.td}
\end{ensuredisplaymath} & \htuse{CLEO.Gamma93by60.pub.RICHICHI.99}
\\
\begin{ensuredisplaymath}
\htuse{Gamma94by69.gn} = \htuse{Gamma94by69.td}
\end{ensuredisplaymath} & \htuse{CLEO.Gamma94by69.pub.RICHICHI.99}}%
\htdef{ARMS 05.cite}{\cite{Arms:2005qg}}%
\htdef{ARMS 05.collab}{CLEO3}%
\htdef{ARMS 05.ref}{ARMS 05 (CLEO3) \cite{Arms:2005qg}}%
\htdef{ARMS 05.meas}{%
\begin{ensuredisplaymath}
\htuse{Gamma88.gn} = \htuse{Gamma88.td}
\end{ensuredisplaymath} & \htuse{CLEO3.Gamma88.pub.ARMS.05}
\\
\begin{ensuredisplaymath}
\htuse{Gamma94.gn} = \htuse{Gamma94.td}
\end{ensuredisplaymath} & \htuse{CLEO3.Gamma94.pub.ARMS.05}
\\
\begin{ensuredisplaymath}
\htuse{Gamma151.gn} = \htuse{Gamma151.td}
\end{ensuredisplaymath} & \htuse{CLEO3.Gamma151.pub.ARMS.05}}%
\htdef{BRIERE 03.cite}{\cite{Briere:2003fr}}%
\htdef{BRIERE 03.collab}{CLEO3}%
\htdef{BRIERE 03.ref}{BRIERE 03 (CLEO3) \cite{Briere:2003fr}}%
\htdef{BRIERE 03.meas}{%
\begin{ensuredisplaymath}
\htuse{Gamma60.gn} = \htuse{Gamma60.td}
\end{ensuredisplaymath} & \htuse{CLEO3.Gamma60.pub.BRIERE.03}
\\
\begin{ensuredisplaymath}
\htuse{Gamma85.gn} = \htuse{Gamma85.td}
\end{ensuredisplaymath} & \htuse{CLEO3.Gamma85.pub.BRIERE.03}
\\
\begin{ensuredisplaymath}
\htuse{Gamma93.gn} = \htuse{Gamma93.td}
\end{ensuredisplaymath} & \htuse{CLEO3.Gamma93.pub.BRIERE.03}}%
\htdef{ABDALLAH 06A.cite}{\cite{Abdallah:2003cw}}%
\htdef{ABDALLAH 06A.collab}{DELPHI}%
\htdef{ABDALLAH 06A.ref}{ABDALLAH 06A (DELPHI) \cite{Abdallah:2003cw}}%
\htdef{ABDALLAH 06A.meas}{%
\begin{ensuredisplaymath}
\htuse{Gamma8.gn} = \htuse{Gamma8.td}
\end{ensuredisplaymath} & \htuse{DELPHI.Gamma8.pub.ABDALLAH.06A}
\\
\begin{ensuredisplaymath}
\htuse{Gamma13.gn} = \htuse{Gamma13.td}
\end{ensuredisplaymath} & \htuse{DELPHI.Gamma13.pub.ABDALLAH.06A}
\\
\begin{ensuredisplaymath}
\htuse{Gamma19.gn} = \htuse{Gamma19.td}
\end{ensuredisplaymath} & \htuse{DELPHI.Gamma19.pub.ABDALLAH.06A}
\\
\begin{ensuredisplaymath}
\htuse{Gamma25.gn} = \htuse{Gamma25.td}
\end{ensuredisplaymath} & \htuse{DELPHI.Gamma25.pub.ABDALLAH.06A}
\\
\begin{ensuredisplaymath}
\htuse{Gamma57.gn} = \htuse{Gamma57.td}
\end{ensuredisplaymath} & \htuse{DELPHI.Gamma57.pub.ABDALLAH.06A}
\\
\begin{ensuredisplaymath}
\htuse{Gamma66.gn} = \htuse{Gamma66.td}
\end{ensuredisplaymath} & \htuse{DELPHI.Gamma66.pub.ABDALLAH.06A}
\\
\begin{ensuredisplaymath}
\htuse{Gamma74.gn} = \htuse{Gamma74.td}
\end{ensuredisplaymath} & \htuse{DELPHI.Gamma74.pub.ABDALLAH.06A}
\\
\begin{ensuredisplaymath}
\htuse{Gamma103.gn} = \htuse{Gamma103.td}
\end{ensuredisplaymath} & \htuse{DELPHI.Gamma103.pub.ABDALLAH.06A}
\\
\begin{ensuredisplaymath}
\htuse{Gamma104.gn} = \htuse{Gamma104.td}
\end{ensuredisplaymath} & \htuse{DELPHI.Gamma104.pub.ABDALLAH.06A}}%
\htdef{ABREU 92N.cite}{\cite{Abreu:1992gn}}%
\htdef{ABREU 92N.collab}{DELPHI}%
\htdef{ABREU 92N.ref}{ABREU 92N (DELPHI) \cite{Abreu:1992gn}}%
\htdef{ABREU 92N.meas}{%
\begin{ensuredisplaymath}
\htuse{Gamma7.gn} = \htuse{Gamma7.td}
\end{ensuredisplaymath} & \htuse{DELPHI.Gamma7.pub.ABREU.92N}}%
\htdef{ABREU 94K.cite}{\cite{Abreu:1994fi}}%
\htdef{ABREU 94K.collab}{DELPHI}%
\htdef{ABREU 94K.ref}{ABREU 94K (DELPHI) \cite{Abreu:1994fi}}%
\htdef{ABREU 94K.meas}{%
\begin{ensuredisplaymath}
\htuse{Gamma10.gn} = \htuse{Gamma10.td}
\end{ensuredisplaymath} & \htuse{DELPHI.Gamma10.pub.ABREU.94K}
\\
\begin{ensuredisplaymath}
\htuse{Gamma31.gn} = \htuse{Gamma31.td}
\end{ensuredisplaymath} & \htuse{DELPHI.Gamma31.pub.ABREU.94K}}%
\htdef{ABREU 99X.cite}{\cite{Abreu:1999rb}}%
\htdef{ABREU 99X.collab}{DELPHI}%
\htdef{ABREU 99X.ref}{ABREU 99X (DELPHI) \cite{Abreu:1999rb}}%
\htdef{ABREU 99X.meas}{%
\begin{ensuredisplaymath}
\htuse{Gamma3.gn} = \htuse{Gamma3.td}
\end{ensuredisplaymath} & \htuse{DELPHI.Gamma3.pub.ABREU.99X}
\\
\begin{ensuredisplaymath}
\htuse{Gamma5.gn} = \htuse{Gamma5.td}
\end{ensuredisplaymath} & \htuse{DELPHI.Gamma5.pub.ABREU.99X}}%
\htdef{BYLSMA 87.cite}{\cite{Bylsma:1986zy}}%
\htdef{BYLSMA 87.collab}{HRS}%
\htdef{BYLSMA 87.ref}{BYLSMA 87 (HRS) \cite{Bylsma:1986zy}}%
\htdef{BYLSMA 87.meas}{%
\begin{ensuredisplaymath}
\htuse{Gamma102.gn} = \htuse{Gamma102.td}
\end{ensuredisplaymath} & \htuse{HRS.Gamma102.pub.BYLSMA.87}
\\
\begin{ensuredisplaymath}
\htuse{Gamma103.gn} = \htuse{Gamma103.td}
\end{ensuredisplaymath} & \htuse{HRS.Gamma103.pub.BYLSMA.87}}%
\htdef{ACCIARRI 01F.cite}{\cite{Acciarri:2001sg}}%
\htdef{ACCIARRI 01F.collab}{L3}%
\htdef{ACCIARRI 01F.ref}{ACCIARRI 01F (L3) \cite{Acciarri:2001sg}}%
\htdef{ACCIARRI 01F.meas}{%
\begin{ensuredisplaymath}
\htuse{Gamma3.gn} = \htuse{Gamma3.td}
\end{ensuredisplaymath} & \htuse{L3.Gamma3.pub.ACCIARRI.01F}
\\
\begin{ensuredisplaymath}
\htuse{Gamma5.gn} = \htuse{Gamma5.td}
\end{ensuredisplaymath} & \htuse{L3.Gamma5.pub.ACCIARRI.01F}}%
\htdef{ACCIARRI 95.cite}{\cite{Acciarri:1994vr}}%
\htdef{ACCIARRI 95.collab}{L3}%
\htdef{ACCIARRI 95.ref}{ACCIARRI 95 (L3) \cite{Acciarri:1994vr}}%
\htdef{ACCIARRI 95.meas}{%
\begin{ensuredisplaymath}
\htuse{Gamma7.gn} = \htuse{Gamma7.td}
\end{ensuredisplaymath} & \htuse{L3.Gamma7.pub.ACCIARRI.95}
\\
\begin{ensuredisplaymath}
\htuse{Gamma13.gn} = \htuse{Gamma13.td}
\end{ensuredisplaymath} & \htuse{L3.Gamma13.pub.ACCIARRI.95}
\\
\begin{ensuredisplaymath}
\htuse{Gamma19.gn} = \htuse{Gamma19.td}
\end{ensuredisplaymath} & \htuse{L3.Gamma19.pub.ACCIARRI.95}
\\
\begin{ensuredisplaymath}
\htuse{Gamma26.gn} = \htuse{Gamma26.td}
\end{ensuredisplaymath} & \htuse{L3.Gamma26.pub.ACCIARRI.95}}%
\htdef{ACCIARRI 95F.cite}{\cite{Acciarri:1995kx}}%
\htdef{ACCIARRI 95F.collab}{L3}%
\htdef{ACCIARRI 95F.ref}{ACCIARRI 95F (L3) \cite{Acciarri:1995kx}}%
\htdef{ACCIARRI 95F.meas}{%
\begin{ensuredisplaymath}
\htuse{Gamma35.gn} = \htuse{Gamma35.td}
\end{ensuredisplaymath} & \htuse{L3.Gamma35.pub.ACCIARRI.95F}
\\
\begin{ensuredisplaymath}
\htuse{Gamma40.gn} = \htuse{Gamma40.td}
\end{ensuredisplaymath} & \htuse{L3.Gamma40.pub.ACCIARRI.95F}}%
\htdef{ACHARD 01D.cite}{\cite{Achard:2001pk}}%
\htdef{ACHARD 01D.collab}{L3}%
\htdef{ACHARD 01D.ref}{ACHARD 01D (L3) \cite{Achard:2001pk}}%
\htdef{ACHARD 01D.meas}{%
\begin{ensuredisplaymath}
\htuse{Gamma55.gn} = \htuse{Gamma55.td}
\end{ensuredisplaymath} & \htuse{L3.Gamma55.pub.ACHARD.01D}
\\
\begin{ensuredisplaymath}
\htuse{Gamma102.gn} = \htuse{Gamma102.td}
\end{ensuredisplaymath} & \htuse{L3.Gamma102.pub.ACHARD.01D}}%
\htdef{ADEVA 91F.cite}{\cite{Adeva:1991qq}}%
\htdef{ADEVA 91F.collab}{L3}%
\htdef{ADEVA 91F.ref}{ADEVA 91F (L3) \cite{Adeva:1991qq}}%
\htdef{ADEVA 91F.meas}{%
\begin{ensuredisplaymath}
\htuse{Gamma54.gn} = \htuse{Gamma54.td}
\end{ensuredisplaymath} & \htuse{L3.Gamma54.pub.ADEVA.91F}}%
\htdef{ABBIENDI 00C.cite}{\cite{Abbiendi:1999pm}}%
\htdef{ABBIENDI 00C.collab}{OPAL}%
\htdef{ABBIENDI 00C.ref}{ABBIENDI 00C (OPAL) \cite{Abbiendi:1999pm}}%
\htdef{ABBIENDI 00C.meas}{%
\begin{ensuredisplaymath}
\htuse{Gamma35.gn} = \htuse{Gamma35.td}
\end{ensuredisplaymath} & \htuse{OPAL.Gamma35.pub.ABBIENDI.00C}
\\
\begin{ensuredisplaymath}
\htuse{Gamma38.gn} = \htuse{Gamma38.td}
\end{ensuredisplaymath} & \htuse{OPAL.Gamma38.pub.ABBIENDI.00C}
\\
\begin{ensuredisplaymath}
\htuse{Gamma43.gn} = \htuse{Gamma43.td}
\end{ensuredisplaymath} & \htuse{OPAL.Gamma43.pub.ABBIENDI.00C}}%
\htdef{ABBIENDI 00D.cite}{\cite{Abbiendi:1999cq}}%
\htdef{ABBIENDI 00D.collab}{OPAL}%
\htdef{ABBIENDI 00D.ref}{ABBIENDI 00D (OPAL) \cite{Abbiendi:1999cq}}%
\htdef{ABBIENDI 00D.meas}{%
\begin{ensuredisplaymath}
\htuse{Gamma92.gn} = \htuse{Gamma92.td}
\end{ensuredisplaymath} & \htuse{OPAL.Gamma92.pub.ABBIENDI.00D}}%
\htdef{ABBIENDI 01J.cite}{\cite{Abbiendi:2000ee}}%
\htdef{ABBIENDI 01J.collab}{OPAL}%
\htdef{ABBIENDI 01J.ref}{ABBIENDI 01J (OPAL) \cite{Abbiendi:2000ee}}%
\htdef{ABBIENDI 01J.meas}{%
\begin{ensuredisplaymath}
\htuse{Gamma10.gn} = \htuse{Gamma10.td}
\end{ensuredisplaymath} & \htuse{OPAL.Gamma10.pub.ABBIENDI.01J}
\\
\begin{ensuredisplaymath}
\htuse{Gamma31.gn} = \htuse{Gamma31.td}
\end{ensuredisplaymath} & \htuse{OPAL.Gamma31.pub.ABBIENDI.01J}}%
\htdef{ABBIENDI 03.cite}{\cite{Abbiendi:2002jw}}%
\htdef{ABBIENDI 03.collab}{OPAL}%
\htdef{ABBIENDI 03.ref}{ABBIENDI 03 (OPAL) \cite{Abbiendi:2002jw}}%
\htdef{ABBIENDI 03.meas}{%
\begin{ensuredisplaymath}
\htuse{Gamma3.gn} = \htuse{Gamma3.td}
\end{ensuredisplaymath} & \htuse{OPAL.Gamma3.pub.ABBIENDI.03}}%
\htdef{ABBIENDI 04J.cite}{\cite{Abbiendi:2004xa}}%
\htdef{ABBIENDI 04J.collab}{OPAL}%
\htdef{ABBIENDI 04J.ref}{ABBIENDI 04J (OPAL) \cite{Abbiendi:2004xa}}%
\htdef{ABBIENDI 04J.meas}{%
\begin{ensuredisplaymath}
\htuse{Gamma16.gn} = \htuse{Gamma16.td}
\end{ensuredisplaymath} & \htuse{OPAL.Gamma16.pub.ABBIENDI.04J}
\\
\begin{ensuredisplaymath}
\htuse{Gamma85.gn} = \htuse{Gamma85.td}
\end{ensuredisplaymath} & \htuse{OPAL.Gamma85.pub.ABBIENDI.04J}}%
\htdef{ABBIENDI 99H.cite}{\cite{Abbiendi:1998cx}}%
\htdef{ABBIENDI 99H.collab}{OPAL}%
\htdef{ABBIENDI 99H.ref}{ABBIENDI 99H (OPAL) \cite{Abbiendi:1998cx}}%
\htdef{ABBIENDI 99H.meas}{%
\begin{ensuredisplaymath}
\htuse{Gamma5.gn} = \htuse{Gamma5.td}
\end{ensuredisplaymath} & \htuse{OPAL.Gamma5.pub.ABBIENDI.99H}}%
\htdef{ACKERSTAFF 98M.cite}{\cite{Ackerstaff:1997tx}}%
\htdef{ACKERSTAFF 98M.collab}{OPAL}%
\htdef{ACKERSTAFF 98M.ref}{ACKERSTAFF 98M (OPAL) \cite{Ackerstaff:1997tx}}%
\htdef{ACKERSTAFF 98M.meas}{%
\begin{ensuredisplaymath}
\htuse{Gamma8.gn} = \htuse{Gamma8.td}
\end{ensuredisplaymath} & \htuse{OPAL.Gamma8.pub.ACKERSTAFF.98M}
\\
\begin{ensuredisplaymath}
\htuse{Gamma13.gn} = \htuse{Gamma13.td}
\end{ensuredisplaymath} & \htuse{OPAL.Gamma13.pub.ACKERSTAFF.98M}
\\
\begin{ensuredisplaymath}
\htuse{Gamma17.gn} = \htuse{Gamma17.td}
\end{ensuredisplaymath} & \htuse{OPAL.Gamma17.pub.ACKERSTAFF.98M}}%
\htdef{ACKERSTAFF 99E.cite}{\cite{Ackerstaff:1998ia}}%
\htdef{ACKERSTAFF 99E.collab}{OPAL}%
\htdef{ACKERSTAFF 99E.ref}{ACKERSTAFF 99E (OPAL) \cite{Ackerstaff:1998ia}}%
\htdef{ACKERSTAFF 99E.meas}{%
\begin{ensuredisplaymath}
\htuse{Gamma103.gn} = \htuse{Gamma103.td}
\end{ensuredisplaymath} & \htuse{OPAL.Gamma103.pub.ACKERSTAFF.99E}
\\
\begin{ensuredisplaymath}
\htuse{Gamma104.gn} = \htuse{Gamma104.td}
\end{ensuredisplaymath} & \htuse{OPAL.Gamma104.pub.ACKERSTAFF.99E}}%
\htdef{AKERS 94G.cite}{\cite{Akers:1994td}}%
\htdef{AKERS 94G.collab}{OPAL}%
\htdef{AKERS 94G.ref}{AKERS 94G (OPAL) \cite{Akers:1994td}}%
\htdef{AKERS 94G.meas}{%
\begin{ensuredisplaymath}
\htuse{Gamma33.gn} = \htuse{Gamma33.td}
\end{ensuredisplaymath} & \htuse{OPAL.Gamma33.pub.AKERS.94G}}%
\htdef{AKERS 95Y.cite}{\cite{Akers:1995ry}}%
\htdef{AKERS 95Y.collab}{OPAL}%
\htdef{AKERS 95Y.ref}{AKERS 95Y (OPAL) \cite{Akers:1995ry}}%
\htdef{AKERS 95Y.meas}{%
\begin{ensuredisplaymath}
\htuse{Gamma55.gn} = \htuse{Gamma55.td}
\end{ensuredisplaymath} & \htuse{OPAL.Gamma55.pub.AKERS.95Y}
\\
\begin{ensuredisplaymath}
\htuse{Gamma57by55.gn} = \htuse{Gamma57by55.td}
\end{ensuredisplaymath} & \htuse{OPAL.Gamma57by55.pub.AKERS.95Y}}%
\htdef{ALEXANDER 91D.cite}{\cite{Alexander:1991am}}%
\htdef{ALEXANDER 91D.collab}{OPAL}%
\htdef{ALEXANDER 91D.ref}{ALEXANDER 91D (OPAL) \cite{Alexander:1991am}}%
\htdef{ALEXANDER 91D.meas}{%
\begin{ensuredisplaymath}
\htuse{Gamma7.gn} = \htuse{Gamma7.td}
\end{ensuredisplaymath} & \htuse{OPAL.Gamma7.pub.ALEXANDER.91D}}%
\htdef{AIHARA 87B.cite}{\cite{Aihara:1986mw}}%
\htdef{AIHARA 87B.collab}{TPC}%
\htdef{AIHARA 87B.ref}{AIHARA 87B (TPC) \cite{Aihara:1986mw}}%
\htdef{AIHARA 87B.meas}{%
\begin{ensuredisplaymath}
\htuse{Gamma54.gn} = \htuse{Gamma54.td}
\end{ensuredisplaymath} & \htuse{TPC.Gamma54.pub.AIHARA.87B}}%
\htdef{BAUER 94.cite}{\cite{Bauer:1993wn}}%
\htdef{BAUER 94.collab}{TPC}%
\htdef{BAUER 94.ref}{BAUER 94 (TPC) \cite{Bauer:1993wn}}%
\htdef{BAUER 94.meas}{%
\begin{ensuredisplaymath}
\htuse{Gamma82.gn} = \htuse{Gamma82.td}
\end{ensuredisplaymath} & \htuse{TPC.Gamma82.pub.BAUER.94}
\\
\begin{ensuredisplaymath}
\htuse{Gamma92.gn} = \htuse{Gamma92.td}
\end{ensuredisplaymath} & \htuse{TPC.Gamma92.pub.BAUER.94}}%
\htdef{MeasPaper}{%
\multicolumn{2}{l}{\htuse{BARATE 98.ref}} \\
\htuse{BARATE 98.meas} \\\hline
\multicolumn{2}{l}{\htuse{BARATE 98E.ref}} \\
\htuse{BARATE 98E.meas} \\\hline
\multicolumn{2}{l}{\htuse{BARATE 99K.ref}} \\
\htuse{BARATE 99K.meas} \\\hline
\multicolumn{2}{l}{\htuse{BARATE 99R.ref}} \\
\htuse{BARATE 99R.meas} \\\hline
\multicolumn{2}{l}{\htuse{BUSKULIC 96.ref}} \\
\htuse{BUSKULIC 96.meas} \\\hline
\multicolumn{2}{l}{\htuse{BUSKULIC 97C.ref}} \\
\htuse{BUSKULIC 97C.meas} \\\hline
\multicolumn{2}{l}{\htuse{SCHAEL 05C.ref}} \\
\htuse{SCHAEL 05C.meas} \\\hline
\multicolumn{2}{l}{\htuse{ALBRECHT 88B.ref}} \\
\htuse{ALBRECHT 88B.meas} \\\hline
\multicolumn{2}{l}{\htuse{ALBRECHT 92D.ref}} \\
\htuse{ALBRECHT 92D.meas} \\\hline
\multicolumn{2}{l}{\htuse{AUBERT 07AP.ref}} \\
\htuse{AUBERT 07AP.meas} \\\hline
\multicolumn{2}{l}{\htuse{AUBERT 08.ref}} \\
\htuse{AUBERT 08.meas} \\\hline
\multicolumn{2}{l}{\htuse{AUBERT 10F.ref}} \\
\htuse{AUBERT 10F.meas} \\\hline
\multicolumn{2}{l}{\htuse{DEL-AMO-SANCHEZ 11E.ref}} \\
\htuse{DEL-AMO-SANCHEZ 11E.meas} \\\hline
\multicolumn{2}{l}{\htuse{LEES 2012X.ref}} \\
\htuse{LEES 2012X.meas} \\\hline
\multicolumn{2}{l}{\htuse{LEES 2012Y.ref}} \\
\htuse{LEES 2012Y.meas} \\\hline
\multicolumn{2}{l}{\htuse{BaBar prelim. DPF09.ref}} \\
\htuse{BaBar prelim. DPF09.meas} \\\hline
\multicolumn{2}{l}{\htuse{BaBar prelim. ICHEP08.ref}} \\
\htuse{BaBar prelim. ICHEP08.meas} \\\hline
\multicolumn{2}{l}{\htuse{FUJIKAWA 08.ref}} \\
\htuse{FUJIKAWA 08.meas} \\\hline
\multicolumn{2}{l}{\htuse{INAMI 09.ref}} \\
\htuse{INAMI 09.meas} \\\hline
\multicolumn{2}{l}{\htuse{LEE 10.ref}} \\
\htuse{LEE 10.meas} \\\hline
\multicolumn{2}{l}{\htuse{Ryu:2014vpc.ref}} \\
\htuse{Ryu:2014vpc.meas} \\\hline
\multicolumn{2}{l}{\htuse{BEHREND 89B.ref}} \\
\htuse{BEHREND 89B.meas} \\\hline
\multicolumn{2}{l}{\htuse{ANASTASSOV 01.ref}} \\
\htuse{ANASTASSOV 01.meas} \\\hline
\multicolumn{2}{l}{\htuse{ANASTASSOV 97.ref}} \\
\htuse{ANASTASSOV 97.meas} \\\hline
\multicolumn{2}{l}{\htuse{ARTUSO 92.ref}} \\
\htuse{ARTUSO 92.meas} \\\hline
\multicolumn{2}{l}{\htuse{ARTUSO 94.ref}} \\
\htuse{ARTUSO 94.meas} \\\hline
\multicolumn{2}{l}{\htuse{BALEST 95C.ref}} \\
\htuse{BALEST 95C.meas} \\\hline
\multicolumn{2}{l}{\htuse{BARINGER 87.ref}} \\
\htuse{BARINGER 87.meas} \\\hline
\multicolumn{2}{l}{\htuse{BARTELT 96.ref}} \\
\htuse{BARTELT 96.meas} \\\hline
\multicolumn{2}{l}{\htuse{BATTLE 94.ref}} \\
\htuse{BATTLE 94.meas} \\\hline
\multicolumn{2}{l}{\htuse{BISHAI 99.ref}} \\
\htuse{BISHAI 99.meas} \\\hline
\multicolumn{2}{l}{\htuse{BORTOLETTO 93.ref}} \\
\htuse{BORTOLETTO 93.meas} \\\hline
\multicolumn{2}{l}{\htuse{COAN 96.ref}} \\
\htuse{COAN 96.meas} \\\hline
\multicolumn{2}{l}{\htuse{EDWARDS 00A.ref}} \\
\htuse{EDWARDS 00A.meas} \\\hline
\multicolumn{2}{l}{\htuse{GIBAUT 94B.ref}} \\
\htuse{GIBAUT 94B.meas} \\\hline
\multicolumn{2}{l}{\htuse{PROCARIO 93.ref}} \\
\htuse{PROCARIO 93.meas} \\\hline
\multicolumn{2}{l}{\htuse{RICHICHI 99.ref}} \\
\htuse{RICHICHI 99.meas} \\\hline
\multicolumn{2}{l}{\htuse{ARMS 05.ref}} \\
\htuse{ARMS 05.meas} \\\hline
\multicolumn{2}{l}{\htuse{BRIERE 03.ref}} \\
\htuse{BRIERE 03.meas} \\\hline
\multicolumn{2}{l}{\htuse{ABDALLAH 06A.ref}} \\
\htuse{ABDALLAH 06A.meas} \\\hline
\multicolumn{2}{l}{\htuse{ABREU 92N.ref}} \\
\htuse{ABREU 92N.meas} \\\hline
\multicolumn{2}{l}{\htuse{ABREU 94K.ref}} \\
\htuse{ABREU 94K.meas} \\\hline
\multicolumn{2}{l}{\htuse{ABREU 99X.ref}} \\
\htuse{ABREU 99X.meas} \\\hline
\multicolumn{2}{l}{\htuse{BYLSMA 87.ref}} \\
\htuse{BYLSMA 87.meas} \\\hline
\multicolumn{2}{l}{\htuse{ACCIARRI 01F.ref}} \\
\htuse{ACCIARRI 01F.meas} \\\hline
\multicolumn{2}{l}{\htuse{ACCIARRI 95.ref}} \\
\htuse{ACCIARRI 95.meas} \\\hline
\multicolumn{2}{l}{\htuse{ACCIARRI 95F.ref}} \\
\htuse{ACCIARRI 95F.meas} \\\hline
\multicolumn{2}{l}{\htuse{ACHARD 01D.ref}} \\
\htuse{ACHARD 01D.meas} \\\hline
\multicolumn{2}{l}{\htuse{ADEVA 91F.ref}} \\
\htuse{ADEVA 91F.meas} \\\hline
\multicolumn{2}{l}{\htuse{ABBIENDI 00C.ref}} \\
\htuse{ABBIENDI 00C.meas} \\\hline
\multicolumn{2}{l}{\htuse{ABBIENDI 00D.ref}} \\
\htuse{ABBIENDI 00D.meas} \\\hline
\multicolumn{2}{l}{\htuse{ABBIENDI 01J.ref}} \\
\htuse{ABBIENDI 01J.meas} \\\hline
\multicolumn{2}{l}{\htuse{ABBIENDI 03.ref}} \\
\htuse{ABBIENDI 03.meas} \\\hline
\multicolumn{2}{l}{\htuse{ABBIENDI 04J.ref}} \\
\htuse{ABBIENDI 04J.meas} \\\hline
\multicolumn{2}{l}{\htuse{ABBIENDI 99H.ref}} \\
\htuse{ABBIENDI 99H.meas} \\\hline
\multicolumn{2}{l}{\htuse{ACKERSTAFF 98M.ref}} \\
\htuse{ACKERSTAFF 98M.meas} \\\hline
\multicolumn{2}{l}{\htuse{ACKERSTAFF 99E.ref}} \\
\htuse{ACKERSTAFF 99E.meas} \\\hline
\multicolumn{2}{l}{\htuse{AKERS 94G.ref}} \\
\htuse{AKERS 94G.meas} \\\hline
\multicolumn{2}{l}{\htuse{AKERS 95Y.ref}} \\
\htuse{AKERS 95Y.meas} \\\hline
\multicolumn{2}{l}{\htuse{ALEXANDER 91D.ref}} \\
\htuse{ALEXANDER 91D.meas} \\\hline
\multicolumn{2}{l}{\htuse{AIHARA 87B.ref}} \\
\htuse{AIHARA 87B.meas} \\\hline
\multicolumn{2}{l}{\htuse{BAUER 94.ref}} \\
\htuse{BAUER 94.meas}}%
\htdef{BrStrangeVal}{%
\begin{ensuredisplaymath}
\htuse{Gamma10.gn} = \htuse{Gamma10.td}
\end{ensuredisplaymath}
 & \ensuremath{(0.6955 \pm 0.0096) \cdot 10^{-2}} \\
\begin{ensuredisplaymath}
\htuse{Gamma16.gn} = \htuse{Gamma16.td}
\end{ensuredisplaymath}
 & \ensuremath{(0.4331 \pm 0.0149) \cdot 10^{-2}} \\
\begin{ensuredisplaymath}
\htuse{Gamma23.gn} = \htuse{Gamma23.td}
\end{ensuredisplaymath}
 & \ensuremath{(0.0630 \pm 0.0220) \cdot 10^{-2}} \\
\begin{ensuredisplaymath}
\htuse{Gamma28.gn} = \htuse{Gamma28.td}
\end{ensuredisplaymath}
 & \ensuremath{(0.0419 \pm 0.0216) \cdot 10^{-2}} \\
\begin{ensuredisplaymath}
\htuse{Gamma35.gn} = \htuse{Gamma35.td}
\end{ensuredisplaymath}
 & \ensuremath{(0.8378 \pm 0.0123) \cdot 10^{-2}} \\
\begin{ensuredisplaymath}
\htuse{Gamma40.gn} = \htuse{Gamma40.td}
\end{ensuredisplaymath}
 & \ensuremath{(0.3680 \pm 0.0103) \cdot 10^{-2}} \\
\begin{ensuredisplaymath}
\htuse{Gamma44.gn} = \htuse{Gamma44.td}
\end{ensuredisplaymath}
 & \ensuremath{(0.0124 \pm 0.0204) \cdot 10^{-2}} \\
\begin{ensuredisplaymath}
\htuse{Gamma53.gn} = \htuse{Gamma53.td}
\end{ensuredisplaymath}
 & \ensuremath{(0.0222 \pm 0.0202) \cdot 10^{-2}} \\
\begin{ensuredisplaymath}
\htuse{Gamma128.gn} = \htuse{Gamma128.td}
\end{ensuredisplaymath}
 & \ensuremath{(0.0155 \pm 0.0008) \cdot 10^{-2}} \\
\begin{ensuredisplaymath}
\htuse{Gamma130.gn} = \htuse{Gamma130.td}
\end{ensuredisplaymath}
 & \ensuremath{(0.0048 \pm 0.0012) \cdot 10^{-2}} \\
\begin{ensuredisplaymath}
\htuse{Gamma132.gn} = \htuse{Gamma132.td}
\end{ensuredisplaymath}
 & \ensuremath{(0.0093 \pm 0.0015) \cdot 10^{-2}} \\
\begin{ensuredisplaymath}
\htuse{Gamma151.gn} = \htuse{Gamma151.td}
\end{ensuredisplaymath}
 & \ensuremath{(0.0410 \pm 0.0092) \cdot 10^{-2}} \\
\begin{ensuredisplaymath}
\htuse{Gamma801.gn} = \htuse{Gamma801.td}
\end{ensuredisplaymath}
 & \ensuremath{(0.0037 \pm 0.0014) \cdot 10^{-2}} \\
\begin{ensuredisplaymath}
\htuse{Gamma802.gn} = \htuse{Gamma802.td}
\end{ensuredisplaymath}
 & \ensuremath{(0.2922 \pm 0.0068) \cdot 10^{-2}} \\
\begin{ensuredisplaymath}
\htuse{Gamma803.gn} = \htuse{Gamma803.td}
\end{ensuredisplaymath}
 & \ensuremath{(0.0410 \pm 0.0143) \cdot 10^{-2}} \\
\begin{ensuredisplaymath}
\htuse{Gamma822.gn} = \htuse{Gamma822.td}
\end{ensuredisplaymath}
 & \ensuremath{(0.0001 \pm 0.0001) \cdot 10^{-2}} \\
\begin{ensuredisplaymath}
\htuse{Gamma833.gn} = \htuse{Gamma833.td}
\end{ensuredisplaymath}
 & \ensuremath{(0.0001 \pm 0.0001) \cdot 10^{-2}}}%
\htdef{BrStrangeTotVal}{%
\begin{ensuredisplaymath}
\htuse{Gamma110.gn} = \htuse{Gamma110.td}
\end{ensuredisplaymath}
 & \ensuremath{(2.8817 \pm 0.0470) \cdot 10^{-2}}}%
\htdef{UnitarityQuants}{%
\begin{ensuredisplaymath}
\htuse{Gamma3.gn} = \htuse{Gamma3.td}
\end{ensuredisplaymath}
 & \ensuremath{(17.3909 \pm 0.0396) \cdot 10^{-2}} \\
\begin{ensuredisplaymath}
\htuse{Gamma5.gn} = \htuse{Gamma5.td}
\end{ensuredisplaymath}
 & \ensuremath{(17.8167 \pm 0.0409) \cdot 10^{-2}} \\
\begin{ensuredisplaymath}
\htuse{Gamma9.gn} = \htuse{Gamma9.td}
\end{ensuredisplaymath}
 & \ensuremath{(10.8135 \pm 0.0527) \cdot 10^{-2}} \\
\begin{ensuredisplaymath}
\htuse{Gamma10.gn} = \htuse{Gamma10.td}
\end{ensuredisplaymath}
 & \ensuremath{(0.6955 \pm 0.0096) \cdot 10^{-2}} \\
\begin{ensuredisplaymath}
\htuse{Gamma14.gn} = \htuse{Gamma14.td}
\end{ensuredisplaymath}
 & \ensuremath{(25.5024 \pm 0.0917) \cdot 10^{-2}} \\
\begin{ensuredisplaymath}
\htuse{Gamma16.gn} = \htuse{Gamma16.td}
\end{ensuredisplaymath}
 & \ensuremath{(0.4331 \pm 0.0149) \cdot 10^{-2}} \\
\begin{ensuredisplaymath}
\htuse{Gamma20.gn} = \htuse{Gamma20.td}
\end{ensuredisplaymath}
 & \ensuremath{(9.2398 \pm 0.0997) \cdot 10^{-2}} \\
\begin{ensuredisplaymath}
\htuse{Gamma23.gn} = \htuse{Gamma23.td}
\end{ensuredisplaymath}
 & \ensuremath{(0.0630 \pm 0.0220) \cdot 10^{-2}} \\
\begin{ensuredisplaymath}
\htuse{Gamma27.gn} = \htuse{Gamma27.td}
\end{ensuredisplaymath}
 & \ensuremath{(1.0299 \pm 0.0749) \cdot 10^{-2}} \\
\begin{ensuredisplaymath}
\htuse{Gamma28.gn} = \htuse{Gamma28.td}
\end{ensuredisplaymath}
 & \ensuremath{(0.0419 \pm 0.0216) \cdot 10^{-2}} \\
\begin{ensuredisplaymath}
\htuse{Gamma30.gn} = \htuse{Gamma30.td}
\end{ensuredisplaymath}
 & \ensuremath{(0.1097 \pm 0.0391) \cdot 10^{-2}} \\
\begin{ensuredisplaymath}
\htuse{Gamma35.gn} = \htuse{Gamma35.td}
\end{ensuredisplaymath}
 & \ensuremath{(0.8378 \pm 0.0123) \cdot 10^{-2}} \\
\begin{ensuredisplaymath}
\htuse{Gamma37.gn} = \htuse{Gamma37.td}
\end{ensuredisplaymath}
 & \ensuremath{(0.1500 \pm 0.0050) \cdot 10^{-2}} \\
\begin{ensuredisplaymath}
\htuse{Gamma40.gn} = \htuse{Gamma40.td}
\end{ensuredisplaymath}
 & \ensuremath{(0.3680 \pm 0.0103) \cdot 10^{-2}} \\
\begin{ensuredisplaymath}
\htuse{Gamma42.gn} = \htuse{Gamma42.td}
\end{ensuredisplaymath}
 & \ensuremath{(0.1528 \pm 0.0070) \cdot 10^{-2}} \\
\begin{ensuredisplaymath}
\htuse{Gamma44.gn} = \htuse{Gamma44.td}
\end{ensuredisplaymath}
 & \ensuremath{(0.0124 \pm 0.0204) \cdot 10^{-2}} \\
\begin{ensuredisplaymath}
\htuse{Gamma47.gn} = \htuse{Gamma47.td}
\end{ensuredisplaymath}
 & \ensuremath{(0.0236 \pm 0.0006) \cdot 10^{-2}} \\
\begin{ensuredisplaymath}
\htuse{Gamma48.gn} = \htuse{Gamma48.td}
\end{ensuredisplaymath}
 & \ensuremath{(0.0857 \pm 0.0104) \cdot 10^{-2}} \\
\begin{ensuredisplaymath}
\htuse{Gamma50.gn} = \htuse{Gamma50.td}
\end{ensuredisplaymath}
 & \ensuremath{(0.0018 \pm 0.0002) \cdot 10^{-2}} \\
\begin{ensuredisplaymath}
\htuse{Gamma51.gn} = \htuse{Gamma51.td}
\end{ensuredisplaymath}
 & \ensuremath{(0.0253 \pm 0.0105) \cdot 10^{-2}} \\
\begin{ensuredisplaymath}
\htuse{Gamma53.gn} = \htuse{Gamma53.td}
\end{ensuredisplaymath}
 & \ensuremath{(0.0222 \pm 0.0202) \cdot 10^{-2}} \\
\begin{ensuredisplaymath}
\htuse{Gamma62.gn} = \htuse{Gamma62.td}
\end{ensuredisplaymath}
 & \ensuremath{(8.9798 \pm 0.0511) \cdot 10^{-2}} \\
\begin{ensuredisplaymath}
\htuse{Gamma70.gn} = \htuse{Gamma70.td}
\end{ensuredisplaymath}
 & \ensuremath{(2.7672 \pm 0.0710) \cdot 10^{-2}} \\
\begin{ensuredisplaymath}
\htuse{Gamma77.gn} = \htuse{Gamma77.td}
\end{ensuredisplaymath}
 & \ensuremath{(0.0973 \pm 0.0355) \cdot 10^{-2}} \\
\begin{ensuredisplaymath}
\htuse{Gamma78.gn} = \htuse{Gamma78.td}
\end{ensuredisplaymath}
 & \ensuremath{(0.0211 \pm 0.0030) \cdot 10^{-2}} \\
\begin{ensuredisplaymath}
\htuse{Gamma93.gn} = \htuse{Gamma93.td}
\end{ensuredisplaymath}
 & \ensuremath{(0.1436 \pm 0.0027) \cdot 10^{-2}} \\
\begin{ensuredisplaymath}
\htuse{Gamma94.gn} = \htuse{Gamma94.td}
\end{ensuredisplaymath}
 & \ensuremath{(0.0061 \pm 0.0018) \cdot 10^{-2}} \\
\begin{ensuredisplaymath}
\htuse{Gamma103.gn} = \htuse{Gamma103.td}
\end{ensuredisplaymath}
 & \ensuremath{(0.0822 \pm 0.0031) \cdot 10^{-2}} \\
\begin{ensuredisplaymath}
\htuse{Gamma104.gn} = \htuse{Gamma104.td}
\end{ensuredisplaymath}
 & \ensuremath{(0.0164 \pm 0.0011) \cdot 10^{-2}} \\
\begin{ensuredisplaymath}
\htuse{Gamma126.gn} = \htuse{Gamma126.td}
\end{ensuredisplaymath}
 & \ensuremath{(0.1387 \pm 0.0072) \cdot 10^{-2}} \\
\begin{ensuredisplaymath}
\htuse{Gamma128.gn} = \htuse{Gamma128.td}
\end{ensuredisplaymath}
 & \ensuremath{(0.0155 \pm 0.0008) \cdot 10^{-2}} \\
\begin{ensuredisplaymath}
\htuse{Gamma130.gn} = \htuse{Gamma130.td}
\end{ensuredisplaymath}
 & \ensuremath{(0.0048 \pm 0.0012) \cdot 10^{-2}} \\
\begin{ensuredisplaymath}
\htuse{Gamma132.gn} = \htuse{Gamma132.td}
\end{ensuredisplaymath}
 & \ensuremath{(0.0093 \pm 0.0015) \cdot 10^{-2}} \\
\begin{ensuredisplaymath}
\htuse{Gamma151.gn} = \htuse{Gamma151.td}
\end{ensuredisplaymath}
 & \ensuremath{(0.0410 \pm 0.0092) \cdot 10^{-2}} \\
\begin{ensuredisplaymath}
\htuse{Gamma152.gn} = \htuse{Gamma152.td}
\end{ensuredisplaymath}
 & \ensuremath{(0.4054 \pm 0.0418) \cdot 10^{-2}} \\
\begin{ensuredisplaymath}
\htuse{Gamma800.gn} = \htuse{Gamma800.td}
\end{ensuredisplaymath}
 & \ensuremath{(1.9539 \pm 0.0647) \cdot 10^{-2}} \\
\begin{ensuredisplaymath}
\htuse{Gamma801.gn} = \htuse{Gamma801.td}
\end{ensuredisplaymath}
 & \ensuremath{(0.0037 \pm 0.0014) \cdot 10^{-2}} \\
\begin{ensuredisplaymath}
\htuse{Gamma802.gn} = \htuse{Gamma802.td}
\end{ensuredisplaymath}
 & \ensuremath{(0.2922 \pm 0.0068) \cdot 10^{-2}} \\
\begin{ensuredisplaymath}
\htuse{Gamma803.gn} = \htuse{Gamma803.td}
\end{ensuredisplaymath}
 & \ensuremath{(0.0410 \pm 0.0143) \cdot 10^{-2}} \\
\begin{ensuredisplaymath}
\htuse{Gamma804.gn} = \htuse{Gamma804.td}
\end{ensuredisplaymath}
 & \ensuremath{(0.0236 \pm 0.0006) \cdot 10^{-2}} \\
\begin{ensuredisplaymath}
\htuse{Gamma805.gn} = \htuse{Gamma805.td}
\end{ensuredisplaymath}
 & \ensuremath{(0.0400 \pm 0.0200) \cdot 10^{-2}} \\
\begin{ensuredisplaymath}
\htuse{Gamma806.gn} = \htuse{Gamma806.td}
\end{ensuredisplaymath}
 & \ensuremath{(0.0018 \pm 0.0002) \cdot 10^{-2}} \\
\begin{ensuredisplaymath}
\htuse{Gamma998.gn} = \htuse{Gamma998.td}
\end{ensuredisplaymath}
 & \ensuremath{(0.0990 \pm 0.0985) \cdot 10^{-2}}}%
\htdef{BrCorr}{%
\ifhevea\begin{table}\fi
\begin{center}
\ifhevea
\caption{Base nodes correlation coefficients in percent, section 1.\label{tab:tau:br-fit-corr1}}%
\else
\begin{minipage}{\linewidth}
\begin{center}
\captionof{table}{Base nodes correlation coefficients in percent, section 1.}\label{tab:tau:br-fit-corr1}%
\fi
\begin{envsmall}
\begin{center}
\renewcommand*{\arraystretch}{1.1}%
\begin{tabular}{rrrrrrrrrrrrrrr}
\hline
\( \Gamma_{5} \) &   23 &  &  &  &  &  &  &  &  &  &  &  &  &  \\
\( \Gamma_{9} \) &    7 &    5 &  &  &  &  &  &  &  &  &  &  &  &  \\
\( \Gamma_{10} \) &    3 &    6 &    1 &  &  &  &  &  &  &  &  &  &  &  \\
\( \Gamma_{14} \) &  -13 &  -14 &  -13 &   -3 &  &  &  &  &  &  &  &  &  &  \\
\( \Gamma_{16} \) &   -0 &   -1 &    2 &   -1 &  -16 &  &  &  &  &  &  &  &  &  \\
\( \Gamma_{20} \) &   -5 &   -5 &   -7 &   -1 &  -40 &    2 &  &  &  &  &  &  &  &  \\
\( \Gamma_{23} \) &    0 &    0 &   -0 &   -2 &    2 &  -13 &  -22 &  &  &  &  &  &  &  \\
\( \Gamma_{27} \) &   -4 &   -3 &   -8 &   -1 &    0 &    3 &  -36 &    6 &  &  &  &  &  &  \\
\( \Gamma_{28} \) &    0 &    0 &   -0 &   -2 &    2 &  -13 &    5 &  -21 &  -29 &  &  &  &  &  \\
\( \Gamma_{30} \) &   -5 &   -4 &  -11 &   -2 &   -9 &   -0 &    6 &    0 &  -42 &    0 &  &  &  &  \\
\( \Gamma_{35} \) &   -0 &   -0 &    1 &    0 &   -0 &    2 &   -1 &    1 &   -0 &    1 &   -0 &  &  &  \\
\( \Gamma_{37} \) &    0 &    0 &   -0 &   -0 &    0 &   -2 &    1 &   -2 &    1 &   -2 &    0 &  -29 &  &  \\
\( \Gamma_{40} \) &   -0 &   -1 &    1 &    0 &   -0 &    2 &   -0 &    1 &   -2 &    1 &   -0 &  -10 &   -1 &  \\
 & \( \Gamma_{3} \) & \( \Gamma_{5} \) & \( \Gamma_{9} \) & \( \Gamma_{10} \) & \( \Gamma_{14} \) & \( \Gamma_{16} \) & \( \Gamma_{20} \) & \( \Gamma_{23} \) & \( \Gamma_{27} \) & \( \Gamma_{28} \) & \( \Gamma_{30} \) & \( \Gamma_{35} \) & \( \Gamma_{37} \) & \( \Gamma_{40} \)
\\\hline
\end{tabular}
\end{center}
\end{envsmall}
\ifhevea\else
\end{center}
\end{minipage}
\fi
\end{center}
\ifhevea\end{table}\fi
\ifhevea\begin{table}\fi
\begin{center}
\ifhevea
\caption{Base nodes correlation coefficients in percent, section 2.\label{tab:tau:br-fit-corr2}}%
\else
\begin{minipage}{\linewidth}
\begin{center}
\captionof{table}{Base nodes correlation coefficients in percent, section 2.}\label{tab:tau:br-fit-corr2}%
\fi
\begin{envsmall}
\begin{center}
\renewcommand*{\arraystretch}{1.1}%
\begin{tabular}{rrrrrrrrrrrrrrr}
\hline
\( \Gamma_{42} \) &   -0 &    0 &   -0 &   -0 &    1 &   -3 &    1 &   -5 &    0 &   -5 &    0 &   -3 &  -27 &  -17 \\
\( \Gamma_{44} \) &   -0 &   -0 &    0 &   -0 &   -0 &   -0 &    0 &   -1 &    0 &   -1 &    0 &   13 &   16 &    1 \\
\( \Gamma_{47} \) &    0 &    0 &   -0 &    0 &   -0 &    1 &   -0 &    1 &   -0 &    1 &   -0 &   -9 &   -9 &   -6 \\
\( \Gamma_{48} \) &   -0 &   -0 &    1 &   -0 &    0 &   -1 &    1 &   -3 &    0 &   -3 &   -0 &   40 &   51 &   12 \\
\( \Gamma_{50} \) &    0 &    0 &   -0 &   -0 &    0 &   -0 &    0 &   -0 &    0 &   -0 &   -0 &   -3 &    3 &   -1 \\
\( \Gamma_{51} \) &   -0 &   -0 &    0 &   -0 &    0 &   -0 &    0 &   -1 &    0 &   -1 &   -0 &   13 &   17 &    4 \\
\( \Gamma_{53} \) &    0 &    0 &    0 &    0 &    0 &   -0 &    0 &    0 &    0 &    0 &    0 &   -0 &   -0 &   -0 \\
\( \Gamma_{62} \) &   -3 &   -5 &    8 &    0 &   -4 &    5 &   -7 &   -1 &   -5 &   -1 &   -5 &    3 &   -1 &    2 \\
\( \Gamma_{70} \) &   -6 &   -6 &   -7 &   -1 &   -9 &   -1 &   -1 &    0 &   -1 &    0 &    3 &   -1 &    0 &   -1 \\
\( \Gamma_{77} \) &   -1 &   -0 &   -3 &   -1 &   -2 &   -0 &   -0 &    0 &    2 &    0 &    2 &   -0 &    0 &   -0 \\
\( \Gamma_{93} \) &   -1 &   -1 &    2 &    0 &   -1 &    2 &   -1 &   -0 &   -1 &   -0 &   -1 &    1 &   -0 &    1 \\
\( \Gamma_{94} \) &   -0 &   -0 &   -0 &   -0 &   -0 &   -0 &   -0 &    0 &   -0 &    0 &    0 &   -0 &    0 &   -0 \\
\( \Gamma_{126} \) &    0 &    0 &    0 &    0 &    0 &    0 &   -1 &   -0 &    0 &   -0 &   -2 &    0 &   -0 &    0 \\
\( \Gamma_{128} \) &   -0 &   -0 &    1 &   -0 &   -0 &    1 &   -0 &   -1 &   -0 &   -1 &   -0 &    1 &   -0 &    0 \\
 & \( \Gamma_{3} \) & \( \Gamma_{5} \) & \( \Gamma_{9} \) & \( \Gamma_{10} \) & \( \Gamma_{14} \) & \( \Gamma_{16} \) & \( \Gamma_{20} \) & \( \Gamma_{23} \) & \( \Gamma_{27} \) & \( \Gamma_{28} \) & \( \Gamma_{30} \) & \( \Gamma_{35} \) & \( \Gamma_{37} \) & \( \Gamma_{40} \)
\\\hline
\end{tabular}
\end{center}
\end{envsmall}
\ifhevea\else
\end{center}
\end{minipage}
\fi
\end{center}
\ifhevea\end{table}\fi
\ifhevea\begin{table}\fi
\begin{center}
\ifhevea
\caption{Base nodes correlation coefficients in percent, section 3.\label{tab:tau:br-fit-corr3}}%
\else
\begin{minipage}{\linewidth}
\begin{center}
\captionof{table}{Base nodes correlation coefficients in percent, section 3.}\label{tab:tau:br-fit-corr3}%
\fi
\begin{envsmall}
\begin{center}
\renewcommand*{\arraystretch}{1.1}%
\begin{tabular}{rrrrrrrrrrrrrrr}
\hline
\( \Gamma_{130} \) &    0 &    0 &    0 &    0 &    0 &    0 &   -0 &   -0 &    0 &   -0 &   -0 &    0 &   -0 &    0 \\
\( \Gamma_{132} \) &    0 &    0 &    0 &    0 &    0 &    0 &   -0 &   -0 &    0 &   -0 &   -0 &    1 &    1 &    0 \\
\( \Gamma_{136} \) &    0 &    0 &    0 &    0 &   -0 &    0 &   -0 &   -0 &   -0 &   -0 &   -1 &    0 &   -0 &    0 \\
\( \Gamma_{151} \) &    0 &    0 &    0 &   -0 &    0 &    0 &    0 &   -0 &    0 &   -0 &    0 &   -0 &   -0 &   -0 \\
\( \Gamma_{152} \) &   -1 &   -0 &   -3 &   -1 &   -2 &   -0 &   -1 &    0 &    2 &    0 &    2 &    0 &    0 &    0 \\
\( \Gamma_{800} \) &   -2 &   -2 &   -2 &   -0 &   -3 &   -0 &   -0 &    0 &   -0 &    0 &    1 &   -0 &    0 &   -0 \\
\( \Gamma_{801} \) &   -0 &   -0 &    0 &   -0 &   -0 &   -0 &    0 &   -0 &    0 &   -0 &   -0 &    1 &    1 &    0 \\
\( \Gamma_{802} \) &   -1 &   -1 &    0 &    0 &   -1 &   -1 &   -2 &    0 &   -2 &    0 &   -1 &   -0 &    0 &   -0 \\
\( \Gamma_{803} \) &   -0 &   -0 &   -0 &   -0 &   -0 &   -0 &   -0 &    0 &   -0 &    0 &    0 &   -0 &   -0 &   -0 \\
\( \Gamma_{805} \) &    0 &    0 &    0 &    0 &    0 &    0 &    0 &    0 &    0 &    0 &    0 &    0 &    0 &    0 \\
\( \Gamma_{811} \) &    0 &    0 &    0 &    0 &    0 &    0 &   -0 &   -0 &   -0 &   -0 &   -0 &    0 &   -0 &    0 \\
\( \Gamma_{812} \) &    0 &    1 &    0 &    0 &    0 &   -0 &    0 &    0 &   -0 &    0 &   -0 &   -0 &   -0 &   -0 \\
\( \Gamma_{821} \) &    0 &    0 &    0 &    0 &    0 &    0 &   -1 &   -0 &   -0 &   -0 &   -1 &    0 &   -0 &    0 \\
\( \Gamma_{822} \) &    0 &    0 &    0 &    0 &    0 &   -0 &   -0 &    0 &   -0 &    0 &   -0 &   -0 &    0 &   -0 \\
 & \( \Gamma_{3} \) & \( \Gamma_{5} \) & \( \Gamma_{9} \) & \( \Gamma_{10} \) & \( \Gamma_{14} \) & \( \Gamma_{16} \) & \( \Gamma_{20} \) & \( \Gamma_{23} \) & \( \Gamma_{27} \) & \( \Gamma_{28} \) & \( \Gamma_{30} \) & \( \Gamma_{35} \) & \( \Gamma_{37} \) & \( \Gamma_{40} \)
\\\hline
\end{tabular}
\end{center}
\end{envsmall}
\ifhevea\else
\end{center}
\end{minipage}
\fi
\end{center}
\ifhevea\end{table}\fi
\ifhevea\begin{table}\fi
\begin{center}
\ifhevea
\caption{Base nodes correlation coefficients in percent, section 4.\label{tab:tau:br-fit-corr4}}%
\else
\begin{minipage}{\linewidth}
\begin{center}
\captionof{table}{Base nodes correlation coefficients in percent, section 4.}\label{tab:tau:br-fit-corr4}%
\fi
\begin{envsmall}
\begin{center}
\renewcommand*{\arraystretch}{1.1}%
\begin{tabular}{rrrrrrrrrrrrrrr}
\hline
\( \Gamma_{831} \) &   -0 &    0 &    0 &    0 &   -0 &    0 &   -0 &   -0 &   -0 &   -0 &   -1 &    0 &   -0 &    0 \\
\( \Gamma_{832} \) &   -0 &   -0 &   -0 &   -0 &   -0 &   -0 &    0 &    0 &    0 &    0 &   -0 &    0 &    0 &   -0 \\
\( \Gamma_{833} \) &   -0 &   -0 &   -0 &   -0 &   -0 &   -0 &    0 &    0 &    0 &    0 &   -0 &   -0 &    0 &   -0 \\
\( \Gamma_{920} \) &    0 &    0 &    0 &    0 &    0 &    0 &   -0 &   -0 &   -0 &   -0 &   -0 &    0 &   -0 &    0 \\
 & \( \Gamma_{3} \) & \( \Gamma_{5} \) & \( \Gamma_{9} \) & \( \Gamma_{10} \) & \( \Gamma_{14} \) & \( \Gamma_{16} \) & \( \Gamma_{20} \) & \( \Gamma_{23} \) & \( \Gamma_{27} \) & \( \Gamma_{28} \) & \( \Gamma_{30} \) & \( \Gamma_{35} \) & \( \Gamma_{37} \) & \( \Gamma_{40} \)
\\\hline
\end{tabular}
\end{center}
\end{envsmall}
\ifhevea\else
\end{center}
\end{minipage}
\fi
\end{center}
\ifhevea\end{table}\fi
\ifhevea\begin{table}\fi
\begin{center}
\ifhevea
\caption{Base nodes correlation coefficients in percent, section 5.\label{tab:tau:br-fit-corr5}}%
\else
\begin{minipage}{\linewidth}
\begin{center}
\captionof{table}{Base nodes correlation coefficients in percent, section 5.}\label{tab:tau:br-fit-corr5}%
\fi
\begin{envsmall}
\begin{center}
\renewcommand*{\arraystretch}{1.1}%
\begin{tabular}{rrrrrrrrrrrrrrr}
\hline
\( \Gamma_{44} \) &    7 &  &  &  &  &  &  &  &  &  &  &  &  &  \\
\( \Gamma_{47} \) &   -4 &   17 &  &  &  &  &  &  &  &  &  &  &  &  \\
\( \Gamma_{48} \) &   20 &  -96 &   48 &  &  &  &  &  &  &  &  &  &  &  \\
\( \Gamma_{50} \) &    3 &    6 &  -12 &   20 &  &  &  &  &  &  &  &  &  &  \\
\( \Gamma_{51} \) &    7 &  -32 &   17 & -100 &    7 &  &  &  &  &  &  &  &  &  \\
\( \Gamma_{53} \) &   -0 &    0 &   -0 &   -0 &   -0 &   -0 &  &  &  &  &  &  &  &  \\
\( \Gamma_{62} \) &   -1 &    1 &    0 &    1 &    0 &    0 &   -0 &  &  &  &  &  &  &  \\
\( \Gamma_{70} \) &    0 &    0 &   -0 &   -0 &   -0 &   -0 &   -0 &  -19 &  &  &  &  &  &  \\
\( \Gamma_{77} \) &    0 &   -0 &    0 &    0 &    0 &    0 &    0 &   -1 &   -7 &  &  &  &  &  \\
\( \Gamma_{93} \) &   -0 &    0 &    0 &    1 &    0 &    0 &   -0 &   14 &   -4 &   -0 &  &  &  &  \\
\( \Gamma_{94} \) &    0 &    0 &   -0 &   -0 &   -0 &   -0 &   -0 &   -0 &   -2 &   -0 &   -0 &  &  &  \\
\( \Gamma_{126} \) &   -0 &    0 &    1 &    0 &    0 &    0 &   -0 &    1 &   -0 &   -5 &    0 &   -0 &  &  \\
\( \Gamma_{128} \) &   -0 &    0 &    0 &    0 &    0 &    0 &   -0 &    2 &   -0 &   -0 &    1 &   -0 &    4 &  \\
 & \( \Gamma_{42} \) & \( \Gamma_{44} \) & \( \Gamma_{47} \) & \( \Gamma_{48} \) & \( \Gamma_{50} \) & \( \Gamma_{51} \) & \( \Gamma_{53} \) & \( \Gamma_{62} \) & \( \Gamma_{70} \) & \( \Gamma_{77} \) & \( \Gamma_{93} \) & \( \Gamma_{94} \) & \( \Gamma_{126} \) & \( \Gamma_{128} \)
\\\hline
\end{tabular}
\end{center}
\end{envsmall}
\ifhevea\else
\end{center}
\end{minipage}
\fi
\end{center}
\ifhevea\end{table}\fi
\ifhevea\begin{table}\fi
\begin{center}
\ifhevea
\caption{Base nodes correlation coefficients in percent, section 6.\label{tab:tau:br-fit-corr6}}%
\else
\begin{minipage}{\linewidth}
\begin{center}
\captionof{table}{Base nodes correlation coefficients in percent, section 6.}\label{tab:tau:br-fit-corr6}%
\fi
\begin{envsmall}
\begin{center}
\renewcommand*{\arraystretch}{1.1}%
\begin{tabular}{rrrrrrrrrrrrrrr}
\hline
\( \Gamma_{130} \) &   -0 &    0 &    0 &    0 &    0 &    0 &   -0 &    0 &   -0 &   -1 &    0 &   -0 &    1 &    1 \\
\( \Gamma_{132} \) &    0 &   -1 &    1 &   -4 &    0 &   -1 &   -0 &    0 &   -0 &   -0 &    0 &   -0 &    2 &    1 \\
\( \Gamma_{136} \) &   -0 &    0 &    0 &    0 &   -0 &    0 &   -0 &   -0 &   -1 &    0 &    0 &   -0 &    0 &    0 \\
\( \Gamma_{151} \) &   -0 &    0 &   -0 &    0 &   -0 &   -0 &   -0 &    0 &   12 &   -0 &    0 &    0 &   -0 &   -0 \\
\( \Gamma_{152} \) &    0 &   -0 &    0 &    0 &    0 &    0 &    0 &   -1 &  -11 &  -64 &   -0 &   -0 &   -0 &   -0 \\
\( \Gamma_{800} \) &    0 &    0 &   -0 &   -0 &   -0 &   -0 &   -0 &   -8 &  -69 &   -2 &   -1 &    0 &   -0 &   -0 \\
\( \Gamma_{801} \) &    0 &   -1 &    1 &   -4 &    0 &   -1 &   -0 &   -1 &   -0 &   -0 &    1 &   -0 &    0 &    0 \\
\( \Gamma_{802} \) &    0 &   -0 &   -0 &   -0 &   -0 &   -0 &   -0 &   17 &   -6 &   -0 &   -0 &   -0 &   -0 &   -0 \\
\( \Gamma_{803} \) &   -0 &   -0 &   -0 &   -0 &   -0 &   -0 &   -0 &   -1 &  -19 &   -0 &   -0 &   -2 &   -0 &   -1 \\
\( \Gamma_{805} \) &    0 &    0 &    0 &    0 &    0 &    0 &    0 &    0 &    0 &    0 &    0 &    0 &    0 &    0 \\
\( \Gamma_{811} \) &   -0 &   -0 &    0 &   -0 &   -0 &   -0 &   -0 &   -0 &   -0 &   -0 &   -0 &   -0 &    0 &    0 \\
\( \Gamma_{812} \) &   -0 &   -0 &    0 &   -0 &   -1 &   -0 &   -0 &   -0 &   -1 &   -0 &   -0 &   -0 &   -0 &   -0 \\
\( \Gamma_{821} \) &   -0 &    0 &    0 &    0 &    0 &    0 &    0 &    0 &   -1 &    0 &    0 &   -0 &    0 &    0 \\
\( \Gamma_{822} \) &    0 &   -0 &   -0 &   -0 &    0 &   -0 &    0 &   -0 &   -0 &    0 &   -0 &   -0 &   -0 &   -0 \\
 & \( \Gamma_{42} \) & \( \Gamma_{44} \) & \( \Gamma_{47} \) & \( \Gamma_{48} \) & \( \Gamma_{50} \) & \( \Gamma_{51} \) & \( \Gamma_{53} \) & \( \Gamma_{62} \) & \( \Gamma_{70} \) & \( \Gamma_{77} \) & \( \Gamma_{93} \) & \( \Gamma_{94} \) & \( \Gamma_{126} \) & \( \Gamma_{128} \)
\\\hline
\end{tabular}
\end{center}
\end{envsmall}
\ifhevea\else
\end{center}
\end{minipage}
\fi
\end{center}
\ifhevea\end{table}\fi
\ifhevea\begin{table}\fi
\begin{center}
\ifhevea
\caption{Base nodes correlation coefficients in percent, section 7.\label{tab:tau:br-fit-corr7}}%
\else
\begin{minipage}{\linewidth}
\begin{center}
\captionof{table}{Base nodes correlation coefficients in percent, section 7.}\label{tab:tau:br-fit-corr7}%
\fi
\begin{envsmall}
\begin{center}
\renewcommand*{\arraystretch}{1.1}%
\begin{tabular}{rrrrrrrrrrrrrrr}
\hline
\( \Gamma_{831} \) &   -0 &    0 &    0 &    0 &   -0 &    0 &    0 &   -0 &   -0 &    0 &    0 &   -0 &    0 &    0 \\
\( \Gamma_{832} \) &   -0 &   -0 &    0 &   -0 &   -0 &   -0 &    0 &   -0 &   -0 &    0 &   -0 &   -0 &    0 &    0 \\
\( \Gamma_{833} \) &    0 &   -0 &   -0 &   -0 &    0 &   -0 &    0 &   -0 &    0 &    0 &   -0 &    0 &   -0 &   -0 \\
\( \Gamma_{920} \) &   -0 &    0 &    0 &    0 &    0 &    0 &    0 &    0 &   -0 &    0 &    0 &   -0 &    0 &    0 \\
 & \( \Gamma_{42} \) & \( \Gamma_{44} \) & \( \Gamma_{47} \) & \( \Gamma_{48} \) & \( \Gamma_{50} \) & \( \Gamma_{51} \) & \( \Gamma_{53} \) & \( \Gamma_{62} \) & \( \Gamma_{70} \) & \( \Gamma_{77} \) & \( \Gamma_{93} \) & \( \Gamma_{94} \) & \( \Gamma_{126} \) & \( \Gamma_{128} \)
\\\hline
\end{tabular}
\end{center}
\end{envsmall}
\ifhevea\else
\end{center}
\end{minipage}
\fi
\end{center}
\ifhevea\end{table}\fi
\ifhevea\begin{table}\fi
\begin{center}
\ifhevea
\caption{Base nodes correlation coefficients in percent, section 8.\label{tab:tau:br-fit-corr8}}%
\else
\begin{minipage}{\linewidth}
\begin{center}
\captionof{table}{Base nodes correlation coefficients in percent, section 8.}\label{tab:tau:br-fit-corr8}%
\fi
\begin{envsmall}
\begin{center}
\renewcommand*{\arraystretch}{1.1}%
\begin{tabular}{rrrrrrrrrrrrrrr}
\hline
\( \Gamma_{132} \) &    0 &  &  &  &  &  &  &  &  &  &  &  &  &  \\
\( \Gamma_{136} \) &    0 &   -0 &  &  &  &  &  &  &  &  &  &  &  &  \\
\( \Gamma_{151} \) &   -0 &   -0 &   -0 &  &  &  &  &  &  &  &  &  &  &  \\
\( \Gamma_{152} \) &   -0 &   -0 &    0 &   -0 &  &  &  &  &  &  &  &  &  &  \\
\( \Gamma_{800} \) &   -0 &   -0 &   -0 &  -14 &   -3 &  &  &  &  &  &  &  &  &  \\
\( \Gamma_{801} \) &    0 &   -0 &    0 &   -0 &   -0 &   -0 &  &  &  &  &  &  &  &  \\
\( \Gamma_{802} \) &   -0 &   -0 &   -0 &   -2 &   -0 &   -1 &    1 &  &  &  &  &  &  &  \\
\( \Gamma_{803} \) &   -0 &   -0 &   -0 &  -58 &   -0 &    9 &   -0 &    1 &  &  &  &  &  &  \\
\( \Gamma_{805} \) &    0 &    0 &    0 &    0 &    0 &    0 &    0 &    0 &    0 &  &  &  &  &  \\
\( \Gamma_{811} \) &    0 &   -1 &   20 &   -0 &   -0 &   -0 &   -0 &   -0 &   -0 &    0 &  &  &  &  \\
\( \Gamma_{812} \) &   -0 &   -2 &   -8 &   -0 &   -0 &   -0 &   -0 &   -0 &   -0 &    0 &  -16 &  &  &  \\
\( \Gamma_{821} \) &    0 &   -0 &   47 &   -0 &    0 &   -0 &    0 &   -0 &   -0 &    0 &    8 &   -4 &  &  \\
\( \Gamma_{822} \) &   -0 &    0 &   -1 &   -0 &    0 &   -0 &   -0 &    0 &   -0 &    0 &   -0 &    0 &   -1 &  \\
 & \( \Gamma_{130} \) & \( \Gamma_{132} \) & \( \Gamma_{136} \) & \( \Gamma_{151} \) & \( \Gamma_{152} \) & \( \Gamma_{800} \) & \( \Gamma_{801} \) & \( \Gamma_{802} \) & \( \Gamma_{803} \) & \( \Gamma_{805} \) & \( \Gamma_{811} \) & \( \Gamma_{812} \) & \( \Gamma_{821} \) & \( \Gamma_{822} \)
\\\hline
\end{tabular}
\end{center}
\end{envsmall}
\ifhevea\else
\end{center}
\end{minipage}
\fi
\end{center}
\ifhevea\end{table}\fi
\ifhevea\begin{table}\fi
\begin{center}
\ifhevea
\caption{Base nodes correlation coefficients in percent, section 9.\label{tab:tau:br-fit-corr9}}%
\else
\begin{minipage}{\linewidth}
\begin{center}
\captionof{table}{Base nodes correlation coefficients in percent, section 9.}\label{tab:tau:br-fit-corr9}%
\fi
\begin{envsmall}
\begin{center}
\renewcommand*{\arraystretch}{1.1}%
\begin{tabular}{rrrrrrrrrrrrrrr}
\hline
\( \Gamma_{831} \) &    0 &   -0 &   39 &   -0 &    0 &   -0 &    0 &   -0 &   -0 &    0 &   14 &   -4 &   39 &   -1 \\
\( \Gamma_{832} \) &    0 &   -0 &    3 &   -0 &    0 &   -0 &   -0 &   -0 &   -0 &    0 &    2 &   -0 &    3 &   -0 \\
\( \Gamma_{833} \) &   -0 &    0 &   -1 &   -0 &    0 &    0 &   -0 &   -0 &    0 &    0 &   -0 &    0 &   -1 &    0 \\
\( \Gamma_{920} \) &    0 &   -0 &   20 &   -0 &    0 &   -0 &    0 &   -0 &   -0 &    0 &    3 &   -2 &   35 &   -1 \\
 & \( \Gamma_{130} \) & \( \Gamma_{132} \) & \( \Gamma_{136} \) & \( \Gamma_{151} \) & \( \Gamma_{152} \) & \( \Gamma_{800} \) & \( \Gamma_{801} \) & \( \Gamma_{802} \) & \( \Gamma_{803} \) & \( \Gamma_{805} \) & \( \Gamma_{811} \) & \( \Gamma_{812} \) & \( \Gamma_{821} \) & \( \Gamma_{822} \)
\\\hline
\end{tabular}
\end{center}
\end{envsmall}
\ifhevea\else
\end{center}
\end{minipage}
\fi
\end{center}
\ifhevea\end{table}\fi
\ifhevea\begin{table}\fi
\begin{center}
\ifhevea
\caption{Base nodes correlation coefficients in percent, section 10.\label{tab:tau:br-fit-corr10}}%
\else
\begin{minipage}{\linewidth}
\begin{center}
\captionof{table}{Base nodes correlation coefficients in percent, section 10.}\label{tab:tau:br-fit-corr10}%
\fi
\begin{envsmall}
\begin{center}
\renewcommand*{\arraystretch}{1.1}%
\begin{tabular}{rrrrr}
\hline
\( \Gamma_{832} \) &   -2 &  &  &  \\
\( \Gamma_{833} \) &   -1 &   -1 &  &  \\
\( \Gamma_{920} \) &   17 &    1 &   -0 &  \\
 & \( \Gamma_{831} \) & \( \Gamma_{832} \) & \( \Gamma_{833} \) & \( \Gamma_{920} \)
\\\hline
\end{tabular}
\end{center}
\end{envsmall}
\ifhevea\else
\end{center}
\end{minipage}
\fi
\end{center}
\ifhevea\end{table}\fi}%
\htconstrdef{Gamma3by5.c}{\frac{\Gamma_{3}}{\Gamma_{5}}}{\frac{\Gamma_{3}}{\Gamma_{5}}}{\frac{\Gamma_{3}}{\Gamma_{5}}}%
\htconstrdef{Gamma7.c}{\Gamma_{7}}{\Gamma_{35}\cdot{}\Gamma_{<\bar{K}^0|K_L>} + \Gamma_{9} + \Gamma_{804} + \Gamma_{37}\cdot{}\Gamma_{<K^0|K_L>} + \Gamma_{10}}{\Gamma_{35}\cdot{}\Gamma_{<\bar{K}^0|K_L>} + \Gamma_{9} + \Gamma_{804} + \Gamma_{37}\cdot{}\Gamma_{<K^0|K_L>} + \Gamma_{10}}%
\htconstrdef{Gamma8.c}{\Gamma_{8}}{\Gamma_{9} + \Gamma_{10}}{\Gamma_{9} + \Gamma_{10}}%
\htconstrdef{Gamma9by5.c}{\frac{\Gamma_{9}}{\Gamma_{5}}}{\frac{\Gamma_{9}}{\Gamma_{5}}}{\frac{\Gamma_{9}}{\Gamma_{5}}}%
\htconstrdef{Gamma10by5.c}{\frac{\Gamma_{10}}{\Gamma_{5}}}{\frac{\Gamma_{10}}{\Gamma_{5}}}{\frac{\Gamma_{10}}{\Gamma_{5}}}%
\htconstrdef{Gamma13.c}{\Gamma_{13}}{\Gamma_{14} + \Gamma_{16}}{\Gamma_{14} + \Gamma_{16}}%
\htconstrdef{Gamma17.c}{\Gamma_{17}}{\Gamma_{128}\cdot{}\Gamma_{\eta\to3\pi^0} + \Gamma_{30} + \Gamma_{23} + \Gamma_{28} + \Gamma_{35}\cdot{}(\Gamma_{<K^0|K_S>}\cdot{}\Gamma_{K_S\to\pi^0\pi^0}) + \Gamma_{40}\cdot{}(\Gamma_{<K^0|K_S>}\cdot{}\Gamma_{K_S\to\pi^0\pi^0}) + \Gamma_{42}\cdot{}(\Gamma_{<K^0|K_S>}\cdot{}\Gamma_{K_S\to\pi^0\pi^0}) + \Gamma_{20} + \Gamma_{27} + \Gamma_{47}\cdot{}(\Gamma_{K_S\to\pi^0\pi^0}\cdot{}\Gamma_{K_S\to\pi^0\pi^0}) + \Gamma_{48}\cdot{}\Gamma_{K_S\to\pi^0\pi^0} + \Gamma_{50}\cdot{}(\Gamma_{K_S\to\pi^0\pi^0}\cdot{}\Gamma_{K_S\to\pi^0\pi^0}) + \Gamma_{51}\cdot{}\Gamma_{K_S\to\pi^0\pi^0} + \Gamma_{126}\cdot{}\Gamma_{\eta\to3\pi^0} + \Gamma_{37}\cdot{}(\Gamma_{<K^0|K_S>}\cdot{}\Gamma_{K_S\to\pi^0\pi^0}) + \Gamma_{130}\cdot{}\Gamma_{\eta\to3\pi^0}}{\Gamma_{128}\cdot{}\Gamma_{\eta\to3\pi^0} + \Gamma_{30} + \Gamma_{23} + \Gamma_{28} + \Gamma_{35}\cdot{}(\Gamma_{<K^0|K_S>}\cdot{}\Gamma_{K_S\to\pi^0\pi^0})  \\ 
  {}& + \Gamma_{40}\cdot{}(\Gamma_{<K^0|K_S>}\cdot{}\Gamma_{K_S\to\pi^0\pi^0}) + \Gamma_{42}\cdot{}(\Gamma_{<K^0|K_S>}\cdot{}\Gamma_{K_S\to\pi^0\pi^0}) + \Gamma_{20}  \\ 
  {}& + \Gamma_{27} + \Gamma_{47}\cdot{}(\Gamma_{K_S\to\pi^0\pi^0}\cdot{}\Gamma_{K_S\to\pi^0\pi^0}) + \Gamma_{48}\cdot{}\Gamma_{K_S\to\pi^0\pi^0} + \Gamma_{50}\cdot{}(\Gamma_{K_S\to\pi^0\pi^0}\cdot{}\Gamma_{K_S\to\pi^0\pi^0})  \\ 
  {}& + \Gamma_{51}\cdot{}\Gamma_{K_S\to\pi^0\pi^0} + \Gamma_{126}\cdot{}\Gamma_{\eta\to3\pi^0} + \Gamma_{37}\cdot{}(\Gamma_{<K^0|K_S>}\cdot{}\Gamma_{K_S\to\pi^0\pi^0})  \\ 
  {}& + \Gamma_{130}\cdot{}\Gamma_{\eta\to3\pi^0}}%
\htconstrdef{Gamma19.c}{\Gamma_{19}}{\Gamma_{23} + \Gamma_{20}}{\Gamma_{23} + \Gamma_{20}}%
\htconstrdef{Gamma19by13.c}{\frac{\Gamma_{19}}{\Gamma_{13}}}{\frac{\Gamma_{19}}{\Gamma_{13}}}{\frac{\Gamma_{19}}{\Gamma_{13}}}%
\htconstrdef{Gamma25.c}{\Gamma_{25}}{\Gamma_{128}\cdot{}\Gamma_{\eta\to3\pi^0} + \Gamma_{30} + \Gamma_{28} + \Gamma_{27} + \Gamma_{126}\cdot{}\Gamma_{\eta\to3\pi^0} + \Gamma_{130}\cdot{}\Gamma_{\eta\to3\pi^0}}{\Gamma_{128}\cdot{}\Gamma_{\eta\to3\pi^0} + \Gamma_{30} + \Gamma_{28} + \Gamma_{27} + \Gamma_{126}\cdot{}\Gamma_{\eta\to3\pi^0}  \\ 
  {}& + \Gamma_{130}\cdot{}\Gamma_{\eta\to3\pi^0}}%
\htconstrdef{Gamma26.c}{\Gamma_{26}}{\Gamma_{128}\cdot{}\Gamma_{\eta\to3\pi^0} + \Gamma_{28} + \Gamma_{40}\cdot{}(\Gamma_{<K^0|K_S>}\cdot{}\Gamma_{K_S\to\pi^0\pi^0}) + \Gamma_{42}\cdot{}(\Gamma_{<K^0|K_S>}\cdot{}\Gamma_{K_S\to\pi^0\pi^0}) + \Gamma_{27}}{\Gamma_{128}\cdot{}\Gamma_{\eta\to3\pi^0} + \Gamma_{28} + \Gamma_{40}\cdot{}(\Gamma_{<K^0|K_S>}\cdot{}\Gamma_{K_S\to\pi^0\pi^0}) + \Gamma_{42}\cdot{}(\Gamma_{<K^0|K_S>}\cdot{}\Gamma_{K_S\to\pi^0\pi^0})  \\ 
  {}& + \Gamma_{27}}%
\htconstrdef{Gamma26by13.c}{\frac{\Gamma_{26}}{\Gamma_{13}}}{\frac{\Gamma_{26}}{\Gamma_{13}}}{\frac{\Gamma_{26}}{\Gamma_{13}}}%
\htconstrdef{Gamma29.c}{\Gamma_{29}}{\Gamma_{30} + \Gamma_{126}\cdot{}\Gamma_{\eta\to3\pi^0} + \Gamma_{130}\cdot{}\Gamma_{\eta\to3\pi^0}}{\Gamma_{30} + \Gamma_{126}\cdot{}\Gamma_{\eta\to3\pi^0} + \Gamma_{130}\cdot{}\Gamma_{\eta\to3\pi^0}}%
\htconstrdef{Gamma31.c}{\Gamma_{31}}{\Gamma_{128}\cdot{}\Gamma_{\eta\to\text{neutral}} + \Gamma_{23} + \Gamma_{28} + \Gamma_{42} + \Gamma_{16} + \Gamma_{37} + \Gamma_{10} + \Gamma_{801}\cdot{}(\Gamma_{\phi\to K_S K_L}\cdot{}\Gamma_{K_S\to\pi^0\pi^0})/(\Gamma_{\phi\to K^+K^-}+\Gamma_{\phi\to K_S K_L})}{\Gamma_{128}\cdot{}\Gamma_{\eta\to\text{neutral}} + \Gamma_{23} + \Gamma_{28} + \Gamma_{42} + \Gamma_{16} + \Gamma_{37}  \\ 
  {}& + \Gamma_{10} + \Gamma_{801}\cdot{}(\Gamma_{\phi\to K_S K_L}\cdot{}\Gamma_{K_S\to\pi^0\pi^0})/(\Gamma_{\phi\to K^+K^-}+\Gamma_{\phi\to K_S K_L})}%
\htconstrdef{Gamma33.c}{\Gamma_{33}}{\Gamma_{35}\cdot{}\Gamma_{<\bar{K}^0|K_S>} + \Gamma_{40}\cdot{}\Gamma_{<\bar{K}^0|K_S>} + \Gamma_{42}\cdot{}\Gamma_{<K^0|K_S>} + \Gamma_{47} + \Gamma_{48} + \Gamma_{50} + \Gamma_{51} + \Gamma_{37}\cdot{}\Gamma_{<K^0|K_S>} + \Gamma_{132}\cdot{}(\Gamma_{<\bar{K}^0|K_S>}\cdot{}\Gamma_{\eta\to\text{neutral}}) + \Gamma_{44}\cdot{}\Gamma_{<\bar{K}^0|K_S>} + \Gamma_{801}\cdot{}\Gamma_{\phi\to K_S K_L}/(\Gamma_{\phi\to K^+K^-}+\Gamma_{\phi\to K_S K_L})}{\Gamma_{35}\cdot{}\Gamma_{<\bar{K}^0|K_S>} + \Gamma_{40}\cdot{}\Gamma_{<\bar{K}^0|K_S>} + \Gamma_{42}\cdot{}\Gamma_{<K^0|K_S>} + \Gamma_{47}  \\ 
  {}& + \Gamma_{48} + \Gamma_{50} + \Gamma_{51} + \Gamma_{37}\cdot{}\Gamma_{<K^0|K_S>} + \Gamma_{132}\cdot{}(\Gamma_{<\bar{K}^0|K_S>}\cdot{}\Gamma_{\eta\to\text{neutral}})  \\ 
  {}& + \Gamma_{44}\cdot{}\Gamma_{<\bar{K}^0|K_S>} + \Gamma_{801}\cdot{}\Gamma_{\phi\to K_S K_L}/(\Gamma_{\phi\to K^+K^-}+\Gamma_{\phi\to K_S K_L})}%
\htconstrdef{Gamma34.c}{\Gamma_{34}}{\Gamma_{35} + \Gamma_{37}}{\Gamma_{35} + \Gamma_{37}}%
\htconstrdef{Gamma38.c}{\Gamma_{38}}{\Gamma_{42} + \Gamma_{37}}{\Gamma_{42} + \Gamma_{37}}%
\htconstrdef{Gamma39.c}{\Gamma_{39}}{\Gamma_{40} + \Gamma_{42}}{\Gamma_{40} + \Gamma_{42}}%
\htconstrdef{Gamma43.c}{\Gamma_{43}}{\Gamma_{40} + \Gamma_{44}}{\Gamma_{40} + \Gamma_{44}}%
\htconstrdef{Gamma46.c}{\Gamma_{46}}{\Gamma_{48} + \Gamma_{47} + \Gamma_{804}}{\Gamma_{48} + \Gamma_{47} + \Gamma_{804}}%
\htconstrdef{Gamma49.c}{\Gamma_{49}}{\Gamma_{50} + \Gamma_{51} + \Gamma_{806}}{\Gamma_{50} + \Gamma_{51} + \Gamma_{806}}%
\htconstrdef{Gamma54.c}{\Gamma_{54}}{\Gamma_{128}\cdot{}\Gamma_{\eta\to\text{charged}} + \Gamma_{152}\cdot{}(\Gamma_{\omega\to\pi^+\pi^-\pi^0}+\Gamma_{\omega\to\pi^+\pi^-}) + \Gamma_{35}\cdot{}(\Gamma_{<K^0|K_S>}\cdot{}\Gamma_{K_S\to\pi^+\pi^-}) + \Gamma_{40}\cdot{}(\Gamma_{<K^0|K_S>}\cdot{}\Gamma_{K_S\to\pi^+\pi^-}) + \Gamma_{42}\cdot{}(\Gamma_{<K^0|K_S>}\cdot{}\Gamma_{K_S\to\pi^+\pi^-}) + \Gamma_{78} + \Gamma_{47}\cdot{}(2\cdot{}\Gamma_{K_S\to\pi^+\pi^-}\cdot{}\Gamma_{K_S\to\pi^0\pi^0}) + \Gamma_{77} + \Gamma_{48}\cdot{}\Gamma_{K_S\to\pi^+\pi^-} + \Gamma_{50}\cdot{}(2\cdot{}\Gamma_{K_S\to\pi^+\pi^-}\cdot{}\Gamma_{K_S\to\pi^0\pi^0}) + \Gamma_{51}\cdot{}\Gamma_{K_S\to\pi^+\pi^-} + \Gamma_{94} + \Gamma_{62} + \Gamma_{70} + \Gamma_{93} + \Gamma_{126}\cdot{}\Gamma_{\eta\to\text{charged}} + \Gamma_{37}\cdot{}(\Gamma_{<K^0|K_S>}\cdot{}\Gamma_{K_S\to\pi^+\pi^-}) + \Gamma_{802} + \Gamma_{803} + \Gamma_{800}\cdot{}(\Gamma_{\omega\to\pi^+\pi^-\pi^0}+\Gamma_{\omega\to\pi^+\pi^-}) + \Gamma_{151}\cdot{}(\Gamma_{\omega\to\pi^+\pi^-\pi^0}+\Gamma_{\omega\to\pi^+\pi^-}) + \Gamma_{130}\cdot{}\Gamma_{\eta\to\text{charged}} + \Gamma_{132}\cdot{}(\Gamma_{<\bar{K}^0|K_L>}\cdot{}\Gamma_{\eta\to\pi^+\pi^-\pi^0} + \Gamma_{<\bar{K}^0|K_S>}\cdot{}\Gamma_{K_S\to\pi^0\pi^0}\cdot{}\Gamma_{\eta\to\pi^+\pi^-\pi^0} + \Gamma_{<\bar{K}^0|K_S>}\cdot{}\Gamma_{K_S\to\pi^+\pi^-}\cdot{}\Gamma_{\eta\to3\pi^0}) + \Gamma_{53}\cdot{}(\Gamma_{<\bar{K}^0|K_S>}\cdot{}\Gamma_{K_S\to\pi^0\pi^0}+\Gamma_{<\bar{K}^0|K_L>}) + \Gamma_{801}\cdot{}(\Gamma_{\phi\to K^+K^-} + \Gamma_{\phi\to K_S K_L}\cdot{}\Gamma_{K_S\to\pi^+\pi^-})/(\Gamma_{\phi\to K^+K^-}+\Gamma_{\phi\to K_S K_L})}{\Gamma_{128}\cdot{}\Gamma_{\eta\to\text{charged}} + \Gamma_{152}\cdot{}(\Gamma_{\omega\to\pi^+\pi^-\pi^0}+\Gamma_{\omega\to\pi^+\pi^-}) + \Gamma_{35}\cdot{}(\Gamma_{<K^0|K_S>}\cdot{}\Gamma_{K_S\to\pi^+\pi^-})  \\ 
  {}& + \Gamma_{40}\cdot{}(\Gamma_{<K^0|K_S>}\cdot{}\Gamma_{K_S\to\pi^+\pi^-}) + \Gamma_{42}\cdot{}(\Gamma_{<K^0|K_S>}\cdot{}\Gamma_{K_S\to\pi^+\pi^-}) + \Gamma_{78}  \\ 
  {}& + \Gamma_{47}\cdot{}(2\cdot{}\Gamma_{K_S\to\pi^+\pi^-}\cdot{}\Gamma_{K_S\to\pi^0\pi^0}) + \Gamma_{77} + \Gamma_{48}\cdot{}\Gamma_{K_S\to\pi^+\pi^-}  \\ 
  {}& + \Gamma_{50}\cdot{}(2\cdot{}\Gamma_{K_S\to\pi^+\pi^-}\cdot{}\Gamma_{K_S\to\pi^0\pi^0}) + \Gamma_{51}\cdot{}\Gamma_{K_S\to\pi^+\pi^-} + \Gamma_{94}  \\ 
  {}& + \Gamma_{62} + \Gamma_{70} + \Gamma_{93} + \Gamma_{126}\cdot{}\Gamma_{\eta\to\text{charged}} + \Gamma_{37}\cdot{}(\Gamma_{<K^0|K_S>}\cdot{}\Gamma_{K_S\to\pi^+\pi^-})  \\ 
  {}& + \Gamma_{802} + \Gamma_{803} + \Gamma_{800}\cdot{}(\Gamma_{\omega\to\pi^+\pi^-\pi^0}+\Gamma_{\omega\to\pi^+\pi^-}) + \Gamma_{151}\cdot{}(\Gamma_{\omega\to\pi^+\pi^-\pi^0} \\ 
  {}& +\Gamma_{\omega\to\pi^+\pi^-}) + \Gamma_{130}\cdot{}\Gamma_{\eta\to\text{charged}} + \Gamma_{132}\cdot{}(\Gamma_{<\bar{K}^0|K_L>}\cdot{}\Gamma_{\eta\to\pi^+\pi^-\pi^0} + \Gamma_{<\bar{K}^0|K_S>}\cdot{}\Gamma_{K_S\to\pi^0\pi^0}\cdot{}\Gamma_{\eta\to\pi^+\pi^-\pi^0}  \\ 
  {}& + \Gamma_{<\bar{K}^0|K_S>}\cdot{}\Gamma_{K_S\to\pi^+\pi^-}\cdot{}\Gamma_{\eta\to3\pi^0}) + \Gamma_{53}\cdot{}(\Gamma_{<\bar{K}^0|K_S>}\cdot{}\Gamma_{K_S\to\pi^0\pi^0}+\Gamma_{<\bar{K}^0|K_L>})  \\ 
  {}& + \Gamma_{801}\cdot{}(\Gamma_{\phi\to K^+K^-} + \Gamma_{\phi\to K_S K_L}\cdot{}\Gamma_{K_S\to\pi^+\pi^-})/(\Gamma_{\phi\to K^+K^-}+\Gamma_{\phi\to K_S K_L})}%
\htconstrdef{Gamma55.c}{\Gamma_{55}}{\Gamma_{128}\cdot{}\Gamma_{\eta\to\text{charged}} + \Gamma_{152}\cdot{}(\Gamma_{\omega\to\pi^+\pi^-\pi^0}+\Gamma_{\omega\to\pi^+\pi^-}) + \Gamma_{78} + \Gamma_{77} + \Gamma_{94} + \Gamma_{62} + \Gamma_{70} + \Gamma_{93} + \Gamma_{126}\cdot{}\Gamma_{\eta\to\text{charged}} + \Gamma_{802} + \Gamma_{803} + \Gamma_{800}\cdot{}(\Gamma_{\omega\to\pi^+\pi^-\pi^0}+\Gamma_{\omega\to\pi^+\pi^-}) + \Gamma_{151}\cdot{}(\Gamma_{\omega\to\pi^+\pi^-\pi^0}+\Gamma_{\omega\to\pi^+\pi^-}) + \Gamma_{130}\cdot{}\Gamma_{\eta\to\text{charged}} + \Gamma_{801}\cdot{}\Gamma_{\phi\to K^+K^-}/(\Gamma_{\phi\to K^+K^-}+\Gamma_{\phi\to K_S K_L})}{\Gamma_{128}\cdot{}\Gamma_{\eta\to\text{charged}} + \Gamma_{152}\cdot{}(\Gamma_{\omega\to\pi^+\pi^-\pi^0}+\Gamma_{\omega\to\pi^+\pi^-}) + \Gamma_{78} + \Gamma_{77}  \\ 
  {}& + \Gamma_{94} + \Gamma_{62} + \Gamma_{70} + \Gamma_{93} + \Gamma_{126}\cdot{}\Gamma_{\eta\to\text{charged}} + \Gamma_{802}  \\ 
  {}& + \Gamma_{803} + \Gamma_{800}\cdot{}(\Gamma_{\omega\to\pi^+\pi^-\pi^0}+\Gamma_{\omega\to\pi^+\pi^-}) + \Gamma_{151}\cdot{}(\Gamma_{\omega\to\pi^+\pi^-\pi^0}+\Gamma_{\omega\to\pi^+\pi^-})  \\ 
  {}& + \Gamma_{130}\cdot{}\Gamma_{\eta\to\text{charged}} + \Gamma_{801}\cdot{}\Gamma_{\phi\to K^+K^-}/(\Gamma_{\phi\to K^+K^-}+\Gamma_{\phi\to K_S K_L})}%
\htconstrdef{Gamma57.c}{\Gamma_{57}}{\Gamma_{62} + \Gamma_{93} + \Gamma_{802} + \Gamma_{800}\cdot{}\Gamma_{\omega\to\pi^+\pi^-} + \Gamma_{151}\cdot{}\Gamma_{\omega\to\pi^+\pi^-} + \Gamma_{801}\cdot{}\Gamma_{\phi\to K^+K^-}/(\Gamma_{\phi\to K^+K^-}+\Gamma_{\phi\to K_S K_L})}{\Gamma_{62} + \Gamma_{93} + \Gamma_{802} + \Gamma_{800}\cdot{}\Gamma_{\omega\to\pi^+\pi^-} + \Gamma_{151}\cdot{}\Gamma_{\omega\to\pi^+\pi^-}  \\ 
  {}& + \Gamma_{801}\cdot{}\Gamma_{\phi\to K^+K^-}/(\Gamma_{\phi\to K^+K^-}+\Gamma_{\phi\to K_S K_L})}%
\htconstrdef{Gamma57by55.c}{\frac{\Gamma_{57}}{\Gamma_{55}}}{\frac{\Gamma_{57}}{\Gamma_{55}}}{\frac{\Gamma_{57}}{\Gamma_{55}}}%
\htconstrdef{Gamma58.c}{\Gamma_{58}}{\Gamma_{62} + \Gamma_{93} + \Gamma_{802} + \Gamma_{801}\cdot{}\Gamma_{\phi\to K^+K^-}/(\Gamma_{\phi\to K^+K^-}+\Gamma_{\phi\to K_S K_L})}{\Gamma_{62} + \Gamma_{93} + \Gamma_{802} + \Gamma_{801}\cdot{}\Gamma_{\phi\to K^+K^-}/(\Gamma_{\phi\to K^+K^-}+\Gamma_{\phi\to K_S K_L})}%
\htconstrdef{Gamma60.c}{\Gamma_{60}}{\Gamma_{62} + \Gamma_{800}\cdot{}\Gamma_{\omega\to\pi^+\pi^-}}{\Gamma_{62} + \Gamma_{800}\cdot{}\Gamma_{\omega\to\pi^+\pi^-}}%
\htconstrdef{Gamma66.c}{\Gamma_{66}}{\Gamma_{128}\cdot{}\Gamma_{\eta\to\pi^+\pi^-\pi^0} + \Gamma_{152}\cdot{}\Gamma_{\omega\to\pi^+\pi^-} + \Gamma_{94} + \Gamma_{70} + \Gamma_{803} + \Gamma_{800}\cdot{}\Gamma_{\omega\to\pi^+\pi^-\pi^0} + \Gamma_{151}\cdot{}\Gamma_{\omega\to\pi^+\pi^-\pi^0}}{\Gamma_{128}\cdot{}\Gamma_{\eta\to\pi^+\pi^-\pi^0} + \Gamma_{152}\cdot{}\Gamma_{\omega\to\pi^+\pi^-} + \Gamma_{94} + \Gamma_{70} + \Gamma_{803}  \\ 
  {}& + \Gamma_{800}\cdot{}\Gamma_{\omega\to\pi^+\pi^-\pi^0} + \Gamma_{151}\cdot{}\Gamma_{\omega\to\pi^+\pi^-\pi^0}}%
\htconstrdef{Gamma69.c}{\Gamma_{69}}{\Gamma_{152}\cdot{}\Gamma_{\omega\to\pi^+\pi^-} + \Gamma_{70} + \Gamma_{800}\cdot{}\Gamma_{\omega\to\pi^+\pi^-\pi^0}}{\Gamma_{152}\cdot{}\Gamma_{\omega\to\pi^+\pi^-} + \Gamma_{70} + \Gamma_{800}\cdot{}\Gamma_{\omega\to\pi^+\pi^-\pi^0}}%
\htconstrdef{Gamma74.c}{\Gamma_{74}}{\Gamma_{152}\cdot{}\Gamma_{\omega\to\pi^+\pi^-\pi^0} + \Gamma_{78} + \Gamma_{77} + \Gamma_{126}\cdot{}\Gamma_{\eta\to\pi^+\pi^-\pi^0} + \Gamma_{130}\cdot{}\Gamma_{\eta\to\pi^+\pi^-\pi^0}}{\Gamma_{152}\cdot{}\Gamma_{\omega\to\pi^+\pi^-\pi^0} + \Gamma_{78} + \Gamma_{77} + \Gamma_{126}\cdot{}\Gamma_{\eta\to\pi^+\pi^-\pi^0} + \Gamma_{130}\cdot{}\Gamma_{\eta\to\pi^+\pi^-\pi^0}}%
\htconstrdef{Gamma76.c}{\Gamma_{76}}{\Gamma_{152}\cdot{}\Gamma_{\omega\to\pi^+\pi^-\pi^0} + \Gamma_{77} + \Gamma_{126}\cdot{}\Gamma_{\eta\to\pi^+\pi^-\pi^0} + \Gamma_{130}\cdot{}\Gamma_{\eta\to\pi^+\pi^-\pi^0}}{\Gamma_{152}\cdot{}\Gamma_{\omega\to\pi^+\pi^-\pi^0} + \Gamma_{77} + \Gamma_{126}\cdot{}\Gamma_{\eta\to\pi^+\pi^-\pi^0} + \Gamma_{130}\cdot{}\Gamma_{\eta\to\pi^+\pi^-\pi^0}}%
\htconstrdef{Gamma76by54.c}{\frac{\Gamma_{76}}{\Gamma_{54}}}{\frac{\Gamma_{76}}{\Gamma_{54}}}{\frac{\Gamma_{76}}{\Gamma_{54}}}%
\htconstrdef{Gamma78.c}{\Gamma_{78}}{\Gamma_{810} + \Gamma_{50}\cdot{}2\cdot{}\Gamma_{K_S\to\pi^+\pi^-}\cdot{}\Gamma_{K_S\to\pi^0\pi^0} + \Gamma_{132}\cdot{}(\Gamma_{<\bar{K}^0|K_S>}\cdot{}\Gamma_{K_S\to\pi^+\pi^-}\cdot{}\Gamma_{\eta\to3\pi^0})}{\Gamma_{810} + \Gamma_{50}\cdot{}2\cdot{}\Gamma_{K_S\to\pi^+\pi^-}\cdot{}\Gamma_{K_S\to\pi^0\pi^0} + \Gamma_{132}\cdot{}(\Gamma_{<\bar{K}^0|K_S>}\cdot{}\Gamma_{K_S\to\pi^+\pi^-}\cdot{}\Gamma_{\eta\to3\pi^0})}%
\htconstrdef{Gamma80by60.c}{\frac{\Gamma_{80}}{\Gamma_{60}}}{\frac{\Gamma_{80}}{\Gamma_{60}}}{\frac{\Gamma_{80}}{\Gamma_{60}}}%
\htconstrdef{Gamma81by69.c}{\frac{\Gamma_{81}}{\Gamma_{69}}}{\frac{\Gamma_{81}}{\Gamma_{69}}}{\frac{\Gamma_{81}}{\Gamma_{69}}}%
\htconstrdef{Gamma82.c}{\Gamma_{82}}{\Gamma_{128}\cdot{}\Gamma_{\eta\to\text{charged}} + \Gamma_{42}\cdot{}(\Gamma_{<K^0|K_S>}\cdot{}\Gamma_{K_S\to\pi^+\pi^-}) + \Gamma_{802} + \Gamma_{803} + \Gamma_{151}\cdot{}(\Gamma_{\omega\to\pi^+\pi^-\pi^0}+\Gamma_{\omega\to\pi^+\pi^-}) + \Gamma_{37}\cdot{}(\Gamma_{<K^0|K_S>}\cdot{}\Gamma_{K_S\to\pi^+\pi^-})}{\Gamma_{128}\cdot{}\Gamma_{\eta\to\text{charged}} + \Gamma_{42}\cdot{}(\Gamma_{<K^0|K_S>}\cdot{}\Gamma_{K_S\to\pi^+\pi^-}) + \Gamma_{802} + \Gamma_{803}  \\ 
  {}& + \Gamma_{151}\cdot{}(\Gamma_{\omega\to\pi^+\pi^-\pi^0}+\Gamma_{\omega\to\pi^+\pi^-}) + \Gamma_{37}\cdot{}(\Gamma_{<K^0|K_S>}\cdot{}\Gamma_{K_S\to\pi^+\pi^-})}%
\htconstrdef{Gamma85.c}{\Gamma_{85}}{\Gamma_{802} + \Gamma_{151}\cdot{}\Gamma_{\omega\to\pi^+\pi^-}}{\Gamma_{802} + \Gamma_{151}\cdot{}\Gamma_{\omega\to\pi^+\pi^-}}%
\htconstrdef{Gamma88.c}{\Gamma_{88}}{\Gamma_{128}\cdot{}\Gamma_{\eta\to\pi^+\pi^-\pi^0} + \Gamma_{803} + \Gamma_{151}\cdot{}\Gamma_{\omega\to\pi^+\pi^-\pi^0}}{\Gamma_{128}\cdot{}\Gamma_{\eta\to\pi^+\pi^-\pi^0} + \Gamma_{803} + \Gamma_{151}\cdot{}\Gamma_{\omega\to\pi^+\pi^-\pi^0}}%
\htconstrdef{Gamma92.c}{\Gamma_{92}}{\Gamma_{94} + \Gamma_{93}}{\Gamma_{94} + \Gamma_{93}}%
\htconstrdef{Gamma93by60.c}{\frac{\Gamma_{93}}{\Gamma_{60}}}{\frac{\Gamma_{93}}{\Gamma_{60}}}{\frac{\Gamma_{93}}{\Gamma_{60}}}%
\htconstrdef{Gamma94by69.c}{\frac{\Gamma_{94}}{\Gamma_{69}}}{\frac{\Gamma_{94}}{\Gamma_{69}}}{\frac{\Gamma_{94}}{\Gamma_{69}}}%
\htconstrdef{Gamma96.c}{\Gamma_{96}}{\Gamma_{801}\cdot{}\Gamma_{\phi\to K^+K^-}/(\Gamma_{\phi\to K^+K^-}+\Gamma_{\phi\to K_S K_L})}{\Gamma_{801}\cdot{}\Gamma_{\phi\to K^+K^-}/(\Gamma_{\phi\to K^+K^-}+\Gamma_{\phi\to K_S K_L})}%
\htconstrdef{Gamma102.c}{\Gamma_{102}}{\Gamma_{103} + \Gamma_{104}}{\Gamma_{103} + \Gamma_{104}}%
\htconstrdef{Gamma103.c}{\Gamma_{103}}{\Gamma_{820} + \Gamma_{822} + \Gamma_{831}\cdot{}\Gamma_{\omega\to\pi^+\pi^-}}{\Gamma_{820} + \Gamma_{822} + \Gamma_{831}\cdot{}\Gamma_{\omega\to\pi^+\pi^-}}%
\htconstrdef{Gamma104.c}{\Gamma_{104}}{\Gamma_{830} + \Gamma_{833}}{\Gamma_{830} + \Gamma_{833}}%
\htconstrdef{Gamma110.c}{\Gamma_{110}}{\Gamma_{10} + \Gamma_{16} + \Gamma_{23} + \Gamma_{28} + \Gamma_{35} + \Gamma_{40} + \Gamma_{128} + \Gamma_{802} + \Gamma_{803} + \Gamma_{151} + \Gamma_{130} + \Gamma_{132} + \Gamma_{44} + \Gamma_{53} + \Gamma_{801} + \Gamma_{822} + \Gamma_{833}}{\Gamma_{10} + \Gamma_{16} + \Gamma_{23} + \Gamma_{28} + \Gamma_{35} + \Gamma_{40} + \Gamma_{128}  \\ 
  {}& + \Gamma_{802} + \Gamma_{803} + \Gamma_{151} + \Gamma_{130} + \Gamma_{132} + \Gamma_{44} + \Gamma_{53}  \\ 
  {}& + \Gamma_{801} + \Gamma_{822} + \Gamma_{833}}%
\htconstrdef{Gamma150.c}{\Gamma_{150}}{\Gamma_{800} + \Gamma_{151}}{\Gamma_{800} + \Gamma_{151}}%
\htconstrdef{Gamma150by66.c}{\frac{\Gamma_{150}}{\Gamma_{66}}}{\frac{\Gamma_{150}}{\Gamma_{66}}}{\frac{\Gamma_{150}}{\Gamma_{66}}}%
\htconstrdef{Gamma152by76.c}{\frac{\Gamma_{152}}{\Gamma_{76}}}{\frac{\Gamma_{152}}{\Gamma_{76}}}{\frac{\Gamma_{152}}{\Gamma_{76}}}%
\htconstrdef{Gamma804.c}{\Gamma_{804}}{\Gamma_{47} \cdot{} (\Gamma_{<K^0|K_L>}\cdot{}\Gamma_{<\bar{K}^0|K_L>}) / (\Gamma_{<K^0|K_S>}\cdot{}\Gamma_{<\bar{K}^0|K_S>})}{\Gamma_{47} \cdot{} (\Gamma_{<K^0|K_L>}\cdot{}\Gamma_{<\bar{K}^0|K_L>}) / (\Gamma_{<K^0|K_S>}\cdot{}\Gamma_{<\bar{K}^0|K_S>})}%
\htconstrdef{Gamma806.c}{\Gamma_{806}}{\Gamma_{50} \cdot{} (\Gamma_{<K^0|K_L>}\cdot{}\Gamma_{<\bar{K}^0|K_L>}) / (\Gamma_{<K^0|K_S>}\cdot{}\Gamma_{<\bar{K}^0|K_S>})}{\Gamma_{50} \cdot{} (\Gamma_{<K^0|K_L>}\cdot{}\Gamma_{<\bar{K}^0|K_L>}) / (\Gamma_{<K^0|K_S>}\cdot{}\Gamma_{<\bar{K}^0|K_S>})}%
\htconstrdef{Gamma810.c}{\Gamma_{810}}{\Gamma_{910} + \Gamma_{911} + \Gamma_{811}\cdot{}\Gamma_{\omega\to\pi^+\pi^-\pi^0} + \Gamma_{812}}{\Gamma_{910} + \Gamma_{911} + \Gamma_{811}\cdot{}\Gamma_{\omega\to\pi^+\pi^-\pi^0} + \Gamma_{812}}%
\htconstrdef{Gamma820.c}{\Gamma_{820}}{\Gamma_{920} + \Gamma_{821}}{\Gamma_{920} + \Gamma_{821}}%
\htconstrdef{Gamma830.c}{\Gamma_{830}}{\Gamma_{930} + \Gamma_{831}\cdot{}\Gamma_{\omega\to\pi^+\pi^-\pi^0} + \Gamma_{832}}{\Gamma_{930} + \Gamma_{831}\cdot{}\Gamma_{\omega\to\pi^+\pi^-\pi^0} + \Gamma_{832}}%
\htconstrdef{Gamma910.c}{\Gamma_{910}}{\Gamma_{136}\cdot{}\Gamma_{\eta\to3\pi^0}}{\Gamma_{136}\cdot{}\Gamma_{\eta\to3\pi^0}}%
\htconstrdef{Gamma930.c}{\Gamma_{930}}{\Gamma_{136}\cdot{}\Gamma_{\eta\to\pi^+\pi^-\pi^0}}{\Gamma_{136}\cdot{}\Gamma_{\eta\to\pi^+\pi^-\pi^0}}%
\htconstrdef{Gamma944.c}{\Gamma_{944}}{\Gamma_{136}\cdot{}\Gamma_{\eta\to\gamma\gamma}}{\Gamma_{136}\cdot{}\Gamma_{\eta\to\gamma\gamma}}%
\htconstrdef{GammaAll.c}{\Gamma_{\text{All}}}{\Gamma_{3} + \Gamma_{5} + \Gamma_{9} + \Gamma_{10} + \Gamma_{14} + \Gamma_{16} + \Gamma_{20} + \Gamma_{23} + \Gamma_{27} + \Gamma_{28} + \Gamma_{30} + \Gamma_{35} + \Gamma_{37} + \Gamma_{40} + \Gamma_{42} + \Gamma_{47} + \Gamma_{48} + \Gamma_{804} + \Gamma_{62} + \Gamma_{70} + \Gamma_{77} + \Gamma_{78} + \Gamma_{93} + \Gamma_{94} + \Gamma_{104} + \Gamma_{126} + \Gamma_{128} + \Gamma_{802} + \Gamma_{803} + \Gamma_{800} + \Gamma_{151} + \Gamma_{130} + \Gamma_{132} + \Gamma_{44} + \Gamma_{53} + \Gamma_{50} + \Gamma_{51} + \Gamma_{806} + \Gamma_{805} + \Gamma_{801} + \Gamma_{152} + \Gamma_{103}}{\Gamma_{3} + \Gamma_{5} + \Gamma_{9} + \Gamma_{10} + \Gamma_{14} + \Gamma_{16} + \Gamma_{20}  \\ 
  {}& + \Gamma_{23} + \Gamma_{27} + \Gamma_{28} + \Gamma_{30} + \Gamma_{35} + \Gamma_{37} + \Gamma_{40}  \\ 
  {}& + \Gamma_{42} + \Gamma_{47} + \Gamma_{48} + \Gamma_{804} + \Gamma_{62} + \Gamma_{70} + \Gamma_{77}  \\ 
  {}& + \Gamma_{78} + \Gamma_{93} + \Gamma_{94} + \Gamma_{104} + \Gamma_{126} + \Gamma_{128} + \Gamma_{802}  \\ 
  {}& + \Gamma_{803} + \Gamma_{800} + \Gamma_{151} + \Gamma_{130} + \Gamma_{132} + \Gamma_{44} + \Gamma_{53}  \\ 
  {}& + \Gamma_{50} + \Gamma_{51} + \Gamma_{806} + \Gamma_{805} + \Gamma_{801} + \Gamma_{152} + \Gamma_{103}}%
\htconstrdef{Unitarity}{1}{\Gamma_{\text{All}} + \Gamma_{998}}{\Gamma_{\text{All}} + \Gamma_{998}}%
\htdef{ConstrEqs}{%
\begin{align*}
\htuse{Gamma7.c.left} ={}& \htuse{Gamma7.c.right.split}
\end{align*}
\begin{align*}
\htuse{Gamma8.c.left} ={}& \htuse{Gamma8.c.right.split}
\end{align*}
\begin{align*}
\htuse{Gamma13.c.left} ={}& \htuse{Gamma13.c.right.split}
\end{align*}
\begin{align*}
\htuse{Gamma17.c.left} ={}& \htuse{Gamma17.c.right.split}
\end{align*}
\begin{align*}
\htuse{Gamma19.c.left} ={}& \htuse{Gamma19.c.right.split}
\end{align*}
\begin{align*}
\htuse{Gamma25.c.left} ={}& \htuse{Gamma25.c.right.split}
\end{align*}
\begin{align*}
\htuse{Gamma26.c.left} ={}& \htuse{Gamma26.c.right.split}
\end{align*}
\begin{align*}
\htuse{Gamma29.c.left} ={}& \htuse{Gamma29.c.right.split}
\end{align*}
\begin{align*}
\htuse{Gamma31.c.left} ={}& \htuse{Gamma31.c.right.split}
\end{align*}
\begin{align*}
\htuse{Gamma33.c.left} ={}& \htuse{Gamma33.c.right.split}
\end{align*}
\begin{align*}
\htuse{Gamma34.c.left} ={}& \htuse{Gamma34.c.right.split}
\end{align*}
\begin{align*}
\htuse{Gamma38.c.left} ={}& \htuse{Gamma38.c.right.split}
\end{align*}
\begin{align*}
\htuse{Gamma39.c.left} ={}& \htuse{Gamma39.c.right.split}
\end{align*}
\begin{align*}
\htuse{Gamma43.c.left} ={}& \htuse{Gamma43.c.right.split}
\end{align*}
\begin{align*}
\htuse{Gamma46.c.left} ={}& \htuse{Gamma46.c.right.split}
\end{align*}
\begin{align*}
\htuse{Gamma49.c.left} ={}& \htuse{Gamma49.c.right.split}
\end{align*}
\begin{align*}
\htuse{Gamma54.c.left} ={}& \htuse{Gamma54.c.right.split}
\end{align*}
\begin{align*}
\htuse{Gamma55.c.left} ={}& \htuse{Gamma55.c.right.split}
\end{align*}
\begin{align*}
\htuse{Gamma57.c.left} ={}& \htuse{Gamma57.c.right.split}
\end{align*}
\begin{align*}
\htuse{Gamma58.c.left} ={}& \htuse{Gamma58.c.right.split}
\end{align*}
\begin{align*}
\htuse{Gamma60.c.left} ={}& \htuse{Gamma60.c.right.split}
\end{align*}
\begin{align*}
\htuse{Gamma66.c.left} ={}& \htuse{Gamma66.c.right.split}
\end{align*}
\begin{align*}
\htuse{Gamma69.c.left} ={}& \htuse{Gamma69.c.right.split}
\end{align*}
\begin{align*}
\htuse{Gamma74.c.left} ={}& \htuse{Gamma74.c.right.split}
\end{align*}
\begin{align*}
\htuse{Gamma76.c.left} ={}& \htuse{Gamma76.c.right.split}
\end{align*}
\begin{align*}
\htuse{Gamma78.c.left} ={}& \htuse{Gamma78.c.right.split}
\end{align*}
\begin{align*}
\htuse{Gamma82.c.left} ={}& \htuse{Gamma82.c.right.split}
\end{align*}
\begin{align*}
\htuse{Gamma85.c.left} ={}& \htuse{Gamma85.c.right.split}
\end{align*}
\begin{align*}
\htuse{Gamma88.c.left} ={}& \htuse{Gamma88.c.right.split}
\end{align*}
\begin{align*}
\htuse{Gamma92.c.left} ={}& \htuse{Gamma92.c.right.split}
\end{align*}
\begin{align*}
\htuse{Gamma96.c.left} ={}& \htuse{Gamma96.c.right.split}
\end{align*}
\begin{align*}
\htuse{Gamma102.c.left} ={}& \htuse{Gamma102.c.right.split}
\end{align*}
\begin{align*}
\htuse{Gamma103.c.left} ={}& \htuse{Gamma103.c.right.split}
\end{align*}
\begin{align*}
\htuse{Gamma104.c.left} ={}& \htuse{Gamma104.c.right.split}
\end{align*}
\begin{align*}
\htuse{Gamma110.c.left} ={}& \htuse{Gamma110.c.right.split}
\end{align*}
\begin{align*}
\htuse{Gamma150.c.left} ={}& \htuse{Gamma150.c.right.split}
\end{align*}
\begin{align*}
\htuse{Gamma804.c.left} ={}& \htuse{Gamma804.c.right.split}
\end{align*}
\begin{align*}
\htuse{Gamma806.c.left} ={}& \htuse{Gamma806.c.right.split}
\end{align*}
\begin{align*}
\htuse{Gamma810.c.left} ={}& \htuse{Gamma810.c.right.split}
\end{align*}
\begin{align*}
\htuse{Gamma820.c.left} ={}& \htuse{Gamma820.c.right.split}
\end{align*}
\begin{align*}
\htuse{Gamma830.c.left} ={}& \htuse{Gamma830.c.right.split}
\end{align*}
\begin{align*}
\htuse{Gamma910.c.left} ={}& \htuse{Gamma910.c.right.split}
\end{align*}
\begin{align*}
\htuse{Gamma930.c.left} ={}& \htuse{Gamma930.c.right.split}
\end{align*}
\begin{align*}
\htuse{Gamma944.c.left} ={}& \htuse{Gamma944.c.right.split}
\end{align*}
\begin{align*}
\htuse{GammaAll.c.left} ={}& \htuse{GammaAll.c.right.split}
\end{align*}}%
\htdef{NumMeasALEPH}{40}%
\htdef{NumMeasARGUS}{2}%
\htdef{NumMeasBaBar}{25}%
\htdef{NumMeasBelle}{16}%
\htdef{NumMeasCELLO}{1}%
\htdef{NumMeasCLEO}{35}%
\htdef{NumMeasCLEO3}{6}%
\htdef{NumMeasDELPHI}{14}%
\htdef{NumMeasHRS}{2}%
\htdef{NumMeasL3}{11}%
\htdef{NumMeasOPAL}{19}%
\htdef{NumMeasTPC}{3}%

\htquantdef{B_tau_had_fit}{B_tau_had_fit}{}{64.69 \pm 0.10}{64.69}{0.10}%
\htquantdef{B_tau_s_fit}{B_tau_s_fit}{}{2.882 \pm 0.047}{2.882}{0.047}%
\htquantdef{B_tau_s_unitarity}{B_tau_s_unitarity}{}{\ensuremath{(2.981 \pm 0.100) \cdot 10^{-2}}}{2.981\cdot 10^{-2}}{0.100\cdot 10^{-2}}%
\htquantdef{B_tau_VA}{B_tau_VA}{}{\ensuremath{0.6181 \pm 0.0010}}{0.6181}{0.0010}%
\htquantdef{B_tau_VA_fit}{B_tau_VA_fit}{}{61.81 \pm 0.10}{61.81}{0.10}%
\htquantdef{B_tau_VA_unitarity}{B_tau_VA_unitarity}{}{\ensuremath{0.61911 \pm 0.00079}}{0.61911}{0.00079}%
\htquantdef{Be_fit}{Be_fit}{}{\ensuremath{0.17817 \pm 0.00041}}{0.17817}{0.00041}%
\htquantdef{Be_from_Bmu}{Be_from_Bmu}{}{\ensuremath{0.17882 \pm 0.00041}}{0.17882}{0.00041}%
\htquantdef{Be_from_taulife}{Be_from_taulife}{}{\ensuremath{0.17778 \pm 0.00033}}{0.17778}{0.00033}%
\htquantdef{Be_unitarity}{Be_unitarity}{}{\ensuremath{0.17916 \pm 0.00097}}{0.17916}{0.00097}%
\htquantdef{Be_univ}{Be_univ}{}{17.814 \pm 0.023}{17.814}{0.023}%
\htquantdef{Bmu_by_Be_th}{Bmu_by_Be_th}{}{\ensuremath{0.9725594 \pm 0.0000048}}{0.9725594}{0.0000048}%
\htquantdef{Bmu_fit}{Bmu_fit}{}{\ensuremath{0.17391 \pm 0.00040}}{0.17391}{0.00040}%
\htquantdef{Bmu_from_taulife}{Bmu_from_taulife}{}{\ensuremath{0.17290 \pm 0.00032}}{0.17290}{0.00032}%
\htquantdef{Bmu_unitarity}{Bmu_unitarity}{}{\ensuremath{0.17490 \pm 0.00097}}{0.17490}{0.00097}%
\htquantdef{BR_eta_2gam}{BR_eta_2gam}{}{0.3941}{0.3941}{0}%
\htquantdef{BR_eta_3piz}{BR_eta_3piz}{}{0.3268}{0.3268}{0}%
\htquantdef{BR_eta_charged}{BR_eta_charged}{}{0.2810}{0.2810}{0}%
\htquantdef{BR_eta_neutral}{BR_eta_neutral}{}{0.7212}{0.7212}{0}%
\htquantdef{BR_eta_pimpippiz}{BR_eta_pimpippiz}{}{0.2292}{0.2292}{0}%
\htquantdef{BR_f1_2pip2pim}{BR_f1_2pip2pim}{}{0.1100}{0.1100}{0}%
\htquantdef{BR_f1_2pizpippim}{BR_f1_2pizpippim}{}{0.2200}{0.2200}{0}%
\htquantdef{BR_KS_2piz}{BR_KS_2piz}{}{0.3069}{0.3069}{0}%
\htquantdef{BR_KS_pimpip}{BR_KS_pimpip}{}{0.6920}{0.6920}{0}%
\htquantdef{BR_om_pimpip}{BR_om_pimpip}{}{1.530\cdot 10^{-2}}{1.530\cdot 10^{-2}}{0}%
\htquantdef{BR_om_pimpippiz}{BR_om_pimpippiz}{}{0.8920}{0.8920}{0}%
\htquantdef{BR_om_pizgamma}{BR_om_pizgamma}{}{8.280\cdot 10^{-2}}{8.280\cdot 10^{-2}}{0}%
\htquantdef{BR_phi_KmKp}{BR_phi_KmKp}{}{0.4890}{0.4890}{0}%
\htquantdef{BR_phi_KSKL}{BR_phi_KSKL}{}{0.3420}{0.3420}{0}%
\htquantdef{BRA_Kz_KL_KET}{BRA_Kz_KL_KET}{}{0.5000}{0.5000}{0}%
\htquantdef{BRA_Kz_KS_KET}{BRA_Kz_KS_KET}{}{0.5000}{0.5000}{0}%
\htquantdef{BRA_Kzbar_KL_KET}{BRA_Kzbar_KL_KET}{}{0.5000}{0.5000}{0}%
\htquantdef{BRA_Kzbar_KS_KET}{BRA_Kzbar_KS_KET}{}{0.5000}{0.5000}{0}%
\htquantdef{delta_K}{delta_K}{}{\ensuremath{(0.9000 \pm 0.2200) \cdot 10^{-2}}}{0.9000\cdot 10^{-2}}{0.2200\cdot 10^{-2}}%
\htquantdef{delta_LD_tauK_Kmu}{delta_LD_tauK_Kmu}{}{0.90 \pm 0.22}{0.90}{0.22}%
\htquantdef{delta_LD_taupi_pimu}{delta_LD_taupi_pimu}{}{0.16 \pm 0.14}{0.16}{0.14}%
\htquantdef{delta_mu_gamma}{delta_mu_gamma}{}{0.9958}{0.9958}{0}%
\htquantdef{delta_mu_W}{delta_mu_W}{}{\ensuremath{1.00000103622 \pm 0.00000000039}}{1.00000103622}{0.00000000039}%
\htquantdef{delta_pi}{delta_pi}{}{\ensuremath{(0.1600 \pm 0.1400) \cdot 10^{-2}}}{0.1600\cdot 10^{-2}}{0.1400\cdot 10^{-2}}%
\htquantdef{delta_tau_gamma}{delta_tau_gamma}{}{0.9957}{0.9957}{0}%
\htquantdef{delta_tau_W}{delta_tau_W}{}{\ensuremath{1.00029304 \pm 0.00000012}}{1.00029304}{0.00000012}%
\htquantdef{deltaR_su3break}{deltaR_su3break}{}{0.239 \pm 0.030}{0.239}{0.030}%
\htquantdef{deltaR_su3break_d2pert}{deltaR_su3break_d2pert}{}{\ensuremath{9.300 \pm 3.400}}{9.300}{3.400}%
\htquantdef{deltaR_su3break_pheno}{deltaR_su3break_pheno}{}{\ensuremath{0.1544 \pm 0.0037}}{0.1544}{0.0037}%
\htquantdef{deltaR_su3break_remain}{deltaR_su3break_remain}{}{\ensuremath{(0.3400 \pm 0.2800) \cdot 10^{-2}}}{0.3400\cdot 10^{-2}}{0.2800\cdot 10^{-2}}%
\htquantdef{f_K}{f_K}{}{156.3 \pm 0.9}{156.3}{0.9}%
\htquantdef{f_K_by_f_pi}{f_K_by_f_pi}{}{\ensuremath{1.1920 \pm 0.0050}}{1.1920}{0.0050}%
\htquantdef{fp0_Kpi}{fp0_Kpi}{}{\ensuremath{0.9661 \pm 0.0032}}{0.9661}{0.0032}%
\htquantdef{G_F_by_hcut3_c3}{G_F_by_hcut3_c3}{}{\ensuremath{(1.16637870 \pm 0.00000060) \cdot 10^{-11}}}{1.16637870\cdot 10^{-11}}{0.00000060\cdot 10^{-11}}%
\htquantdef{Gamma10}{\Gamma_{10}}{\BRF{\tau^-}{K^- \nu_\tau}}{\ensuremath{(0.6955 \pm 0.0096) \cdot 10^{-2}}}{0.6955\cdot 10^{-2}}{0.0096\cdot 10^{-2}}%
\htquantdef{Gamma102}{\Gamma_{102}}{\BRF{\tau^-}{3h^- 2h^+ \ge{}0 \text{neutrals} \nu_\tau\;(\text{ex.~} K^0)}}{\ensuremath{(9.861 \pm 0.369) \cdot 10^{-4}}}{9.861\cdot 10^{-4}}{0.369\cdot 10^{-4}}%
\htquantdef{Gamma103}{\Gamma_{103}}{\BRF{\tau^-}{3h^- 2h^+ \nu_\tau ~(\text{ex.~}K^0)}}{\ensuremath{(8.224 \pm 0.315) \cdot 10^{-4}}}{8.224\cdot 10^{-4}}{0.315\cdot 10^{-4}}%
\htquantdef{Gamma104}{\Gamma_{104}}{\BRF{\tau^-}{3h^- 2h^+ \pi^0 \nu_\tau ~(\text{ex.~}K^0)}}{\ensuremath{(1.637 \pm 0.113) \cdot 10^{-4}}}{1.637\cdot 10^{-4}}{0.113\cdot 10^{-4}}%
\htquantdef{Gamma10by5}{\frac{\Gamma_{10}}{\Gamma_{5}}}{\frac{\BRF{\tau^-}{K^- \nu_\tau}}{\BRF{\tau^-}{e^- \bar{\nu}_e \nu_\tau}}}{\ensuremath{(3.903 \pm 0.054) \cdot 10^{-2}}}{3.903\cdot 10^{-2}}{0.054\cdot 10^{-2}}%
\htquantdef{Gamma10by9}{\frac{\Gamma_{10}}{\Gamma_{9}}}{\frac{\BRF{\tau^-}{K^- \nu_\tau}}{\BRF{\tau^-}{\pi^- \nu_\tau}}}{\ensuremath{(6.431 \pm 0.094) \cdot 10^{-2}}}{6.431\cdot 10^{-2}}{0.094\cdot 10^{-2}}%
\htquantdef{Gamma110}{\Gamma_{110}}{\BRF{\tau^-}{X_s^- \nu_\tau}}{\ensuremath{(2.882 \pm 0.047) \cdot 10^{-2}}}{2.882\cdot 10^{-2}}{0.047\cdot 10^{-2}}%
\htquantdef{Gamma110_pdg09}{\Gamma_{110}_pdg09}{}{\ensuremath{(2.825 \pm 0.036) \cdot 10^{-2}}}{2.825\cdot 10^{-2}}{0.036\cdot 10^{-2}}%
\htquantdef{Gamma126}{\Gamma_{126}}{\BRF{\tau^-}{\pi^- \pi^0 \eta \nu_\tau}}{\ensuremath{(0.1387 \pm 0.0072) \cdot 10^{-2}}}{0.1387\cdot 10^{-2}}{0.0072\cdot 10^{-2}}%
\htquantdef{Gamma128}{\Gamma_{128}}{\BRF{\tau^-}{K^- \eta \nu_\tau}}{\ensuremath{(1.548 \pm 0.080) \cdot 10^{-4}}}{1.548\cdot 10^{-4}}{0.080\cdot 10^{-4}}%
\htquantdef{Gamma13}{\Gamma_{13}}{\BRF{\tau^-}{h^- \pi^0 \nu_\tau}}{\ensuremath{0.25936 \pm 0.00090}}{0.25936}{0.00090}%
\htquantdef{Gamma130}{\Gamma_{130}}{\BRF{\tau^-}{K^- \pi^0 \eta \nu_\tau}}{\ensuremath{(4.829 \pm 1.161) \cdot 10^{-5}}}{4.829\cdot 10^{-5}}{1.161\cdot 10^{-5}}%
\htquantdef{Gamma132}{\Gamma_{132}}{\BRF{\tau^-}{\pi^- \bar{K}^0 \eta \nu_\tau}}{\ensuremath{(9.341 \pm 1.490) \cdot 10^{-5}}}{9.341\cdot 10^{-5}}{1.490\cdot 10^{-5}}%
\htquantdef{Gamma136}{\Gamma_{136}}{\BRF{\tau^-}{\pi^- \pi^- \pi^+ \eta \nu_\tau\;(\text{ex.~} K^0)}}{\ensuremath{(2.186 \pm 0.129) \cdot 10^{-4}}}{2.186\cdot 10^{-4}}{0.129\cdot 10^{-4}}%
\htquantdef{Gamma14}{\Gamma_{14}}{\BRF{\tau^-}{\pi^- \pi^0 \nu_\tau}}{\ensuremath{0.25502 \pm 0.00092}}{0.25502}{0.00092}%
\htquantdef{Gamma150}{\Gamma_{150}}{\BRF{\tau^-}{h^- \omega \nu_\tau}}{\ensuremath{(1.995 \pm 0.064) \cdot 10^{-2}}}{1.995\cdot 10^{-2}}{0.064\cdot 10^{-2}}%
\htquantdef{Gamma150by66}{\frac{\Gamma_{150}}{\Gamma_{66}}}{\frac{\BRF{\tau^-}{h^- \omega \nu_\tau}}{\BRF{\tau^-}{h^- h^- h^+ \pi^0 \nu_\tau\;(\text{ex.~} K^0)}}}{\ensuremath{0.4333 \pm 0.0139}}{0.4333}{0.0139}%
\htquantdef{Gamma151}{\Gamma_{151}}{\BRF{\tau^-}{K^- \omega \nu_\tau}}{\ensuremath{(4.100 \pm 0.922) \cdot 10^{-4}}}{4.100\cdot 10^{-4}}{0.922\cdot 10^{-4}}%
\htquantdef{Gamma152}{\Gamma_{152}}{\BRF{\tau^-}{h^- \pi^0 \omega \nu_\tau}}{\ensuremath{(0.4054 \pm 0.0418) \cdot 10^{-2}}}{0.4054\cdot 10^{-2}}{0.0418\cdot 10^{-2}}%
\htquantdef{Gamma152by76}{\frac{\Gamma_{152}}{\Gamma_{76}}}{\frac{\BRF{\tau^-}{h^- \omega \pi^0 \nu_\tau}}{\BRF{\tau^-}{h^- h^- h^+ 2\pi^0 \nu_\tau\;(\text{ex.~} K^0)}}}{\ensuremath{0.8243 \pm 0.0757}}{0.8243}{0.0757}%
\htquantdef{Gamma16}{\Gamma_{16}}{\BRF{\tau^-}{K^- \pi^0 \nu_\tau}}{\ensuremath{(0.4331 \pm 0.0149) \cdot 10^{-2}}}{0.4331\cdot 10^{-2}}{0.0149\cdot 10^{-2}}%
\htquantdef{Gamma17}{\Gamma_{17}}{\BRF{\tau^-}{h^- \ge{}2 \pi^0 \nu_\tau}}{\ensuremath{0.10804 \pm 0.00095}}{0.10804}{0.00095}%
\htquantdef{Gamma19}{\Gamma_{19}}{\BRF{\tau^-}{h^- 2\pi^0 \nu_\tau\;(\text{ex.~} K^0)}}{\ensuremath{(9.303 \pm 0.097) \cdot 10^{-2}}}{9.303\cdot 10^{-2}}{0.097\cdot 10^{-2}}%
\htquantdef{Gamma19by13}{\frac{\Gamma_{19}}{\Gamma_{13}}}{\frac{\BRF{\tau^-}{h^- 2\pi^0 \nu_\tau\;(\text{ex.~} K^0)}}{\BRF{\tau^-}{h^- \pi^0 \nu_\tau}}}{\ensuremath{0.3587 \pm 0.0044}}{0.3587}{0.0044}%
\htquantdef{Gamma20}{\Gamma_{20}}{\BRF{\tau^-}{\pi^- 2\pi^0 \nu_\tau ~(\text{ex.~}K^0)}}{\ensuremath{(9.240 \pm 0.100) \cdot 10^{-2}}}{9.240\cdot 10^{-2}}{0.100\cdot 10^{-2}}%
\htquantdef{Gamma23}{\Gamma_{23}}{\BRF{\tau^-}{K^- 2\pi^0 \nu_\tau ~(\text{ex.~}K^0)}}{\ensuremath{(6.300 \pm 2.204) \cdot 10^{-4}}}{6.300\cdot 10^{-4}}{2.204\cdot 10^{-4}}%
\htquantdef{Gamma25}{\Gamma_{25}}{\BRF{\tau^-}{h^- \ge{} 3\pi^0 \nu_\tau\;(\text{ex.~} K^0)}}{\ensuremath{(1.233 \pm 0.065) \cdot 10^{-2}}}{1.233\cdot 10^{-2}}{0.065\cdot 10^{-2}}%
\htquantdef{Gamma26}{\Gamma_{26}}{\BRF{\tau^-}{h^- 3\pi^0 \nu_\tau}}{\ensuremath{(1.157 \pm 0.072) \cdot 10^{-2}}}{1.157\cdot 10^{-2}}{0.072\cdot 10^{-2}}%
\htquantdef{Gamma26by13}{\frac{\Gamma_{26}}{\Gamma_{13}}}{\frac{\BRF{\tau^-}{h^- 3\pi^0 \nu_\tau}}{\BRF{\tau^-}{h^- \pi^0 \nu_\tau}}}{\ensuremath{(4.460 \pm 0.277) \cdot 10^{-2}}}{4.460\cdot 10^{-2}}{0.277\cdot 10^{-2}}%
\htquantdef{Gamma27}{\Gamma_{27}}{\BRF{\tau^-}{\pi^- 3\pi^0 \nu_\tau ~(\text{ex.~}K^0)}}{\ensuremath{(1.030 \pm 0.075) \cdot 10^{-2}}}{1.030\cdot 10^{-2}}{0.075\cdot 10^{-2}}%
\htquantdef{Gamma28}{\Gamma_{28}}{\BRF{\tau^-}{K^- 3\pi^0 \nu_\tau ~(\text{ex.~}K^0,\eta)}}{\ensuremath{(4.190 \pm 2.160) \cdot 10^{-4}}}{4.190\cdot 10^{-4}}{2.160\cdot 10^{-4}}%
\htquantdef{Gamma29}{\Gamma_{29}}{\BRF{\tau^-}{h^- 4\pi^0 \nu_\tau\;(\text{ex.~} K^0)}}{\ensuremath{(0.1566 \pm 0.0391) \cdot 10^{-2}}}{0.1566\cdot 10^{-2}}{0.0391\cdot 10^{-2}}%
\htquantdef{Gamma3}{\Gamma_{3}}{\BRF{\tau^-}{\mu^- \bar{\nu}_\mu \nu_\tau}}{\ensuremath{0.17391 \pm 0.00040}}{0.17391}{0.00040}%
\htquantdef{Gamma30}{\Gamma_{30}}{\BRF{\tau^-}{h^- 4\pi^0 \nu_\tau ~(\text{ex.~}K^0,\eta)}}{\ensuremath{(0.1097 \pm 0.0391) \cdot 10^{-2}}}{0.1097\cdot 10^{-2}}{0.0391\cdot 10^{-2}}%
\htquantdef{Gamma31}{\Gamma_{31}}{\BRF{\tau^-}{K^- \ge{}0 \pi^0 \ge{}0 K^0 \ge{}0 \gamma \nu_\tau}}{\ensuremath{(1.548 \pm 0.030) \cdot 10^{-2}}}{1.548\cdot 10^{-2}}{0.030\cdot 10^{-2}}%
\htquantdef{Gamma33}{\Gamma_{33}}{\BRF{\tau^-}{K_S^0 (\text{particles})^- \nu_\tau}}{\ensuremath{(0.9019 \pm 0.0081) \cdot 10^{-2}}}{0.9019\cdot 10^{-2}}{0.0081\cdot 10^{-2}}%
\htquantdef{Gamma34}{\Gamma_{34}}{\BRF{\tau^-}{h^- \bar{K}^0 \nu_\tau}}{\ensuremath{(0.9878 \pm 0.0119) \cdot 10^{-2}}}{0.9878\cdot 10^{-2}}{0.0119\cdot 10^{-2}}%
\htquantdef{Gamma35}{\Gamma_{35}}{\BRF{\tau^-}{\pi^- \bar{K}^0 \nu_\tau}}{\ensuremath{(0.8378 \pm 0.0123) \cdot 10^{-2}}}{0.8378\cdot 10^{-2}}{0.0123\cdot 10^{-2}}%
\htquantdef{Gamma37}{\Gamma_{37}}{\BRF{\tau^-}{K^- K^0 \nu_\tau}}{\ensuremath{(0.1500 \pm 0.0050) \cdot 10^{-2}}}{0.1500\cdot 10^{-2}}{0.0050\cdot 10^{-2}}%
\htquantdef{Gamma38}{\Gamma_{38}}{\BRF{\tau^-}{K^- K^0 \ge{}0 \pi^0 \nu_\tau}}{\ensuremath{(0.3029 \pm 0.0074) \cdot 10^{-2}}}{0.3029\cdot 10^{-2}}{0.0074\cdot 10^{-2}}%
\htquantdef{Gamma39}{\Gamma_{39}}{\BRF{\tau^-}{h^- \bar{K}^0 \pi^0 \nu_\tau}}{\ensuremath{(0.5209 \pm 0.0114) \cdot 10^{-2}}}{0.5209\cdot 10^{-2}}{0.0114\cdot 10^{-2}}%
\htquantdef{Gamma3by5}{\frac{\Gamma_{3}}{\Gamma_{5}}}{\frac{\BRF{\tau^-}{\mu^- \bar{\nu}_\mu \nu_\tau}}{\BRF{\tau^-}{e^- \bar{\nu}_e \nu_\tau}}}{\ensuremath{0.9761 \pm 0.0028}}{0.9761}{0.0028}%
\htquantdef{Gamma40}{\Gamma_{40}}{\BRF{\tau^-}{\pi^- \bar{K}^0 \pi^0 \nu_\tau}}{\ensuremath{(0.3680 \pm 0.0103) \cdot 10^{-2}}}{0.3680\cdot 10^{-2}}{0.0103\cdot 10^{-2}}%
\htquantdef{Gamma42}{\Gamma_{42}}{\BRF{\tau^-}{K^- \pi^0 K^0 \nu_\tau}}{\ensuremath{(0.1528 \pm 0.0070) \cdot 10^{-2}}}{0.1528\cdot 10^{-2}}{0.0070\cdot 10^{-2}}%
\htquantdef{Gamma43}{\Gamma_{43}}{\BRF{\tau^-}{\pi^- \bar{K}^0 \ge{}1 \pi^0 \nu_\tau}}{\ensuremath{(0.3805 \pm 0.0229) \cdot 10^{-2}}}{0.3805\cdot 10^{-2}}{0.0229\cdot 10^{-2}}%
\htquantdef{Gamma44}{\Gamma_{44}}{\BRF{\tau^-}{\pi^- \bar{K}^0 \pi^0 \pi^0 \nu_\tau ~(\text{ex.~}K^0)}}{\ensuremath{(1.245 \pm 2.043) \cdot 10^{-4}}}{1.245\cdot 10^{-4}}{2.043\cdot 10^{-4}}%
\htquantdef{Gamma46}{\Gamma_{46}}{\BRF{\tau^-}{\pi^- K^0 \bar{K}^0 \nu_\tau}}{\ensuremath{(0.1329 \pm 0.0110) \cdot 10^{-2}}}{0.1329\cdot 10^{-2}}{0.0110\cdot 10^{-2}}%
\htquantdef{Gamma47}{\Gamma_{47}}{\BRF{\tau^-}{\pi^- K_S^0 K_S^0 \nu_\tau}}{\ensuremath{(2.359 \pm 0.061) \cdot 10^{-4}}}{2.359\cdot 10^{-4}}{0.061\cdot 10^{-4}}%
\htquantdef{Gamma48}{\Gamma_{48}}{\BRF{\tau^-}{\pi^- K_S^0 K_L^0 \nu_\tau}}{\ensuremath{(8.574 \pm 1.037) \cdot 10^{-4}}}{8.574\cdot 10^{-4}}{1.037\cdot 10^{-4}}%
\htquantdef{Gamma49}{\Gamma_{49}}{\BRF{\tau^-}{\pi^- K^0 \bar{K}^0 \pi^0 \nu_\tau}}{\ensuremath{(2.896 \pm 1.051) \cdot 10^{-4}}}{2.896\cdot 10^{-4}}{1.051\cdot 10^{-4}}%
\htquantdef{Gamma5}{\Gamma_{5}}{\BRF{\tau^-}{e^- \bar{\nu}_e \nu_\tau}}{\ensuremath{0.17817 \pm 0.00041}}{0.17817}{0.00041}%
\htquantdef{Gamma50}{\Gamma_{50}}{\BRF{\tau^-}{\pi^- \pi^0 K_S^0 K_S^0 \nu_\tau}}{\ensuremath{(1.845 \pm 0.206) \cdot 10^{-5}}}{1.845\cdot 10^{-5}}{0.206\cdot 10^{-5}}%
\htquantdef{Gamma51}{\Gamma_{51}}{\BRF{\tau^-}{\pi^- \pi^0 K_S^0 K_L^0 \nu_\tau}}{\ensuremath{(2.527 \pm 1.047) \cdot 10^{-4}}}{2.527\cdot 10^{-4}}{1.047\cdot 10^{-4}}%
\htquantdef{Gamma53}{\Gamma_{53}}{\BRF{\tau^-}{\bar{K}^0 h^- h^- h^+ \nu_\tau}}{\ensuremath{(2.221 \pm 2.024) \cdot 10^{-4}}}{2.221\cdot 10^{-4}}{2.024\cdot 10^{-4}}%
\htquantdef{Gamma54}{\Gamma_{54}}{\BRF{\tau^-}{h^- h^- h^+ \ge{}0 \text{neutrals} \ge{}0 K_L^0 \nu_\tau}}{\ensuremath{0.15201 \pm 0.00059}}{0.15201}{0.00059}%
\htquantdef{Gamma55}{\Gamma_{55}}{\BRF{\tau^-}{h^- h^- h^+ \ge{}0 \text{neutrals} \nu_\tau\;(\text{ex.~} K^0)}}{\ensuremath{0.14573 \pm 0.00056}}{0.14573}{0.00056}%
\htquantdef{Gamma57}{\Gamma_{57}}{\BRF{\tau^-}{h^- h^- h^+ \nu_\tau\;(\text{ex.~} K^0)}}{\ensuremath{(9.448 \pm 0.053) \cdot 10^{-2}}}{9.448\cdot 10^{-2}}{0.053\cdot 10^{-2}}%
\htquantdef{Gamma57by55}{\frac{\Gamma_{57}}{\Gamma_{55}}}{\frac{\BRF{\tau^-}{h^- h^- h^+ \nu_\tau\;(\text{ex.~} K^0)}}{\BRF{\tau^-}{h^- h^- h^+ \ge{}0 \text{neutrals} \nu_\tau\;(\text{ex.~} K^0)}}}{\ensuremath{0.6483 \pm 0.0029}}{0.6483}{0.0029}%
\htquantdef{Gamma58}{\Gamma_{58}}{\BRF{\tau^-}{h^- h^- h^+ \nu_\tau\;(\text{ex.~} K^0, \omega)}}{\ensuremath{(9.418 \pm 0.053) \cdot 10^{-2}}}{9.418\cdot 10^{-2}}{0.053\cdot 10^{-2}}%
\htquantdef{Gamma60}{\Gamma_{60}}{\BRF{\tau^-}{\pi^- \pi^- \pi^+ \nu_\tau\;(\text{ex.~} K^0)}}{\ensuremath{(9.010 \pm 0.051) \cdot 10^{-2}}}{9.010\cdot 10^{-2}}{0.051\cdot 10^{-2}}%
\htquantdef{Gamma62}{\Gamma_{62}}{\BRF{\tau^-}{\pi^- \pi^- \pi^+ \nu_\tau ~(\text{ex.~}K^0,\omega)}}{\ensuremath{(8.980 \pm 0.051) \cdot 10^{-2}}}{8.980\cdot 10^{-2}}{0.051\cdot 10^{-2}}%
\htquantdef{Gamma66}{\Gamma_{66}}{\BRF{\tau^-}{h^- h^- h^+ \pi^0 \nu_\tau\;(\text{ex.~} K^0)}}{\ensuremath{(4.603 \pm 0.051) \cdot 10^{-2}}}{4.603\cdot 10^{-2}}{0.051\cdot 10^{-2}}%
\htquantdef{Gamma69}{\Gamma_{69}}{\BRF{\tau^-}{\pi^- \pi^- \pi^+ \pi^0 \nu_\tau\;(\text{ex.~} K^0)}}{\ensuremath{(4.516 \pm 0.052) \cdot 10^{-2}}}{4.516\cdot 10^{-2}}{0.052\cdot 10^{-2}}%
\htquantdef{Gamma7}{\Gamma_{7}}{\BRF{\tau^-}{h^- \ge{}0 K_L^0 \nu_\tau}}{\ensuremath{0.12026 \pm 0.00054}}{0.12026}{0.00054}%
\htquantdef{Gamma70}{\Gamma_{70}}{\BRF{\tau^-}{\pi^- \pi^- \pi^+ \pi^0 \nu_\tau ~(\text{ex.~}K^0,\omega)}}{\ensuremath{(2.767 \pm 0.071) \cdot 10^{-2}}}{2.767\cdot 10^{-2}}{0.071\cdot 10^{-2}}%
\htquantdef{Gamma74}{\Gamma_{74}}{\BRF{\tau^-}{h^- h^- h^+ \ge{} 2\pi^0 \nu_\tau\;(\text{ex.~} K^0)}}{\ensuremath{(0.5130 \pm 0.0311) \cdot 10^{-2}}}{0.5130\cdot 10^{-2}}{0.0311\cdot 10^{-2}}%
\htquantdef{Gamma76}{\Gamma_{76}}{\BRF{\tau^-}{h^- h^- h^+ 2\pi^0 \nu_\tau\;(\text{ex.~} K^0)}}{\ensuremath{(0.4919 \pm 0.0310) \cdot 10^{-2}}}{0.4919\cdot 10^{-2}}{0.0310\cdot 10^{-2}}%
\htquantdef{Gamma76by54}{\frac{\Gamma_{76}}{\Gamma_{54}}}{\frac{\BRF{\tau^-}{h^- h^- h^+ 2\pi^0 \nu_\tau\;(\text{ex.~} K^0)}}{\BRF{\tau^-}{h^- h^- h^+ \ge{}0 \text{neutrals} \ge{}0 K_L^0 \nu_\tau}}}{\ensuremath{(3.236 \pm 0.202) \cdot 10^{-2}}}{3.236\cdot 10^{-2}}{0.202\cdot 10^{-2}}%
\htquantdef{Gamma77}{\Gamma_{77}}{\BRF{\tau^-}{h^- h^- h^+ 2\pi^0 \nu_\tau ~(\text{ex.~}K^0,\omega,\eta)}}{\ensuremath{(9.734 \pm 3.546) \cdot 10^{-4}}}{9.734\cdot 10^{-4}}{3.546\cdot 10^{-4}}%
\htquantdef{Gamma78}{\Gamma_{78}}{\BRF{\tau^-}{h^- h^- h^+ 3\pi^0 \nu_\tau}}{\ensuremath{(2.109 \pm 0.299) \cdot 10^{-4}}}{2.109\cdot 10^{-4}}{0.299\cdot 10^{-4}}%
\htquantdef{Gamma8}{\Gamma_{8}}{\BRF{\tau^-}{h^- \nu_\tau}}{\ensuremath{0.11509 \pm 0.00054}}{0.11509}{0.00054}%
\htquantdef{Gamma800}{\Gamma_{800}}{\BRF{\tau^-}{\pi^- \omega \nu_\tau}}{\ensuremath{(1.954 \pm 0.065) \cdot 10^{-2}}}{1.954\cdot 10^{-2}}{0.065\cdot 10^{-2}}%
\htquantdef{Gamma801}{\Gamma_{801}}{\BRF{\tau^-}{K^- \phi \nu_\tau (\phi \to KK)}}{\ensuremath{(3.664 \pm 1.360) \cdot 10^{-5}}}{3.664\cdot 10^{-5}}{1.360\cdot 10^{-5}}%
\htquantdef{Gamma802}{\Gamma_{802}}{\BRF{\tau^-}{K^- \pi^- \pi^+ \nu_\tau ~(\text{ex.~}K^0,\omega)}}{\ensuremath{(0.2922 \pm 0.0068) \cdot 10^{-2}}}{0.2922\cdot 10^{-2}}{0.0068\cdot 10^{-2}}%
\htquantdef{Gamma803}{\Gamma_{803}}{\BRF{\tau^-}{K^- \pi^- \pi^+ \pi^0 \nu_\tau ~(\text{ex.~}K^0,\omega,\eta)}}{\ensuremath{(4.101 \pm 1.429) \cdot 10^{-4}}}{4.101\cdot 10^{-4}}{1.429\cdot 10^{-4}}%
\htquantdef{Gamma804}{\Gamma_{804}}{\BRF{\tau^-}{\pi^- K_L^0 K_L^0 \nu_\tau}}{\ensuremath{(2.359 \pm 0.061) \cdot 10^{-4}}}{2.359\cdot 10^{-4}}{0.061\cdot 10^{-4}}%
\htquantdef{Gamma805}{\Gamma_{805}}{\BRF{\tau^-}{a_1^- (\to \pi^- \gamma) \nu_\tau}}{\ensuremath{(4.000 \pm 2.000) \cdot 10^{-4}}}{4.000\cdot 10^{-4}}{2.000\cdot 10^{-4}}%
\htquantdef{Gamma806}{\Gamma_{806}}{\BRF{\tau^-}{\pi^- \pi^0 K_L^0 K_L^0 \nu_\tau}}{\ensuremath{(1.845 \pm 0.206) \cdot 10^{-5}}}{1.845\cdot 10^{-5}}{0.206\cdot 10^{-5}}%
\htquantdef{Gamma80by60}{\frac{\Gamma_{80}}{\Gamma_{60}}}{\frac{\BRF{\tau^-}{K^- \pi^- h^+ \nu_\tau\;(\text{ex.~} K^0)}}{\BRF{\tau^-}{\pi^- \pi^- \pi^+ \nu_\tau\;(\text{ex.~} K^0)}}}{\ensuremath{(4.845 \pm 0.081) \cdot 10^{-2}}}{4.845\cdot 10^{-2}}{0.081\cdot 10^{-2}}%
\htquantdef{Gamma810}{\Gamma_{810}}{\BRF{\tau^-}{2\pi^- \pi^+ 3\pi^0 \nu_\tau ~(\text{ex.~}K^0)}}{\ensuremath{(1.925 \pm 0.298) \cdot 10^{-4}}}{1.925\cdot 10^{-4}}{0.298\cdot 10^{-4}}%
\htquantdef{Gamma811}{\Gamma_{811}}{\BRF{\tau^-}{\pi^- 2\pi^0 \omega \nu_\tau ~(\text{ex.~}K^0)}}{\ensuremath{(7.110 \pm 1.586) \cdot 10^{-5}}}{7.110\cdot 10^{-5}}{1.586\cdot 10^{-5}}%
\htquantdef{Gamma812}{\Gamma_{812}}{\BRF{\tau^-}{2\pi^- \pi^+ 3\pi^0 \nu_\tau ~(\text{ex.~}K^0, \eta, \omega, f_1)}}{\ensuremath{(1.336 \pm 2.682) \cdot 10^{-5}}}{1.336\cdot 10^{-5}}{2.682\cdot 10^{-5}}%
\htquantdef{Gamma81by69}{\frac{\Gamma_{81}}{\Gamma_{69}}}{\frac{\BRF{\tau^-}{K^- \pi^- h^+ \pi^0 \nu_\tau\;(\text{ex.~} K^0)}}{\BRF{\tau^-}{\pi^- \pi^- \pi^+ \pi^0 \nu_\tau\;(\text{ex.~} K^0)}}}{\ensuremath{(1.932 \pm 0.266) \cdot 10^{-2}}}{1.932\cdot 10^{-2}}{0.266\cdot 10^{-2}}%
\htquantdef{Gamma82}{\Gamma_{82}}{\BRF{\tau^-}{K^- \pi^- \pi^+ \ge{}0 \text{neutrals} \nu_\tau}}{\ensuremath{(0.4796 \pm 0.0138) \cdot 10^{-2}}}{0.4796\cdot 10^{-2}}{0.0138\cdot 10^{-2}}%
\htquantdef{Gamma820}{\Gamma_{820}}{\BRF{\tau^-}{3\pi^- 2\pi^+ \nu_\tau ~(\text{ex.~}K^0, \omega)}}{\ensuremath{(8.205 \pm 0.315) \cdot 10^{-4}}}{8.205\cdot 10^{-4}}{0.315\cdot 10^{-4}}%
\htquantdef{Gamma821}{\Gamma_{821}}{\BRF{\tau^-}{3\pi^- 2\pi^+ \nu_\tau ~(\text{ex.~}K^0, \omega, f_1)}}{\ensuremath{(7.685 \pm 0.296) \cdot 10^{-4}}}{7.685\cdot 10^{-4}}{0.296\cdot 10^{-4}}%
\htquantdef{Gamma822}{\Gamma_{822}}{\BRF{\tau^-}{K^- 2\pi^- 2\pi^+ \nu_\tau ~(\text{ex.~}K^0)}}{\ensuremath{(0.596 \pm 1.208) \cdot 10^{-6}}}{0.596\cdot 10^{-6}}{1.208\cdot 10^{-6}}%
\htquantdef{Gamma830}{\Gamma_{830}}{\BRF{\tau^-}{3\pi^- 2\pi^+ \pi^0 \nu_\tau ~(\text{ex.~}K^0)}}{\ensuremath{(1.626 \pm 0.113) \cdot 10^{-4}}}{1.626\cdot 10^{-4}}{0.113\cdot 10^{-4}}%
\htquantdef{Gamma831}{\Gamma_{831}}{\BRF{\tau^-}{2\pi^- \pi^+ \omega \nu_\tau ~(\text{ex.~}K^0)}}{\ensuremath{(8.370 \pm 0.624) \cdot 10^{-5}}}{8.370\cdot 10^{-5}}{0.624\cdot 10^{-5}}%
\htquantdef{Gamma832}{\Gamma_{832}}{\BRF{\tau^-}{3\pi^- 2\pi^+ \pi^0 \nu_\tau ~(\text{ex.~}K^0, \eta, \omega, f_1)}}{\ensuremath{(3.783 \pm 0.873) \cdot 10^{-5}}}{3.783\cdot 10^{-5}}{0.873\cdot 10^{-5}}%
\htquantdef{Gamma833}{\Gamma_{833}}{\BRF{\tau^-}{K^- 2\pi^- 2\pi^+ \pi^0 \nu_\tau ~(\text{ex.~}K^0)}}{\ensuremath{(1.108 \pm 0.566) \cdot 10^{-6}}}{1.108\cdot 10^{-6}}{0.566\cdot 10^{-6}}%
\htquantdef{Gamma85}{\Gamma_{85}}{\BRF{\tau^-}{K^- \pi^- \pi^+ \nu_\tau\;(\text{ex.~} K^0)}}{\ensuremath{(0.2929 \pm 0.0068) \cdot 10^{-2}}}{0.2929\cdot 10^{-2}}{0.0068\cdot 10^{-2}}%
\htquantdef{Gamma88}{\Gamma_{88}}{\BRF{\tau^-}{K^- \pi^- \pi^+ \pi^0 \nu_\tau\;(\text{ex.~} K^0)}}{\ensuremath{(8.113 \pm 1.168) \cdot 10^{-4}}}{8.113\cdot 10^{-4}}{1.168\cdot 10^{-4}}%
\htquantdef{Gamma89}{\Gamma_{89}}{\BRF{\tau^-}{K^- \pi^- \pi^+ \pi^0 \nu_\tau\;(\text{ex.~} K^0, \eta)}}{\ensuremath{(7.758 \pm 1.168) \cdot 10^{-4}}}{7.758\cdot 10^{-4}}{1.168\cdot 10^{-4}}%
\htquantdef{Gamma9}{\Gamma_{9}}{\BRF{\tau^-}{\pi^- \nu_\tau}}{\ensuremath{0.10813 \pm 0.00053}}{0.10813}{0.00053}%
\htquantdef{Gamma910}{\Gamma_{910}}{\BRF{\tau^-}{2\pi^- \pi^+ \eta \nu_\tau ~(\eta \to 3\pi^0) ~(\text{ex.~}K^0)}}{\ensuremath{(7.144 \pm 0.423) \cdot 10^{-5}}}{7.144\cdot 10^{-5}}{0.423\cdot 10^{-5}}%
\htquantdef{Gamma911}{\Gamma_{911}}{\BRF{\tau^-}{\pi^- 2\pi^0 \eta \nu_\tau ~(\eta \to \pi^+ \pi^- \pi^0) ~(\text{ex.~}K^0)}}{\ensuremath{(4.424 \pm 0.867) \cdot 10^{-5}}}{4.424\cdot 10^{-5}}{0.867\cdot 10^{-5}}%
\htquantdef{Gamma92}{\Gamma_{92}}{\BRF{\tau^-}{\pi^- K^- K^+ \ge{}0 \text{neutrals} \nu_\tau}}{\ensuremath{(0.1497 \pm 0.0033) \cdot 10^{-2}}}{0.1497\cdot 10^{-2}}{0.0033\cdot 10^{-2}}%
\htquantdef{Gamma920}{\Gamma_{920}}{\BRF{\tau^-}{\pi^- f_1 \nu_\tau ~(f_1 \to 2\pi^- 2\pi^+)}}{\ensuremath{(5.202 \pm 0.444) \cdot 10^{-5}}}{5.202\cdot 10^{-5}}{0.444\cdot 10^{-5}}%
\htquantdef{Gamma93}{\Gamma_{93}}{\BRF{\tau^-}{\pi^- K^- K^+ \nu_\tau}}{\ensuremath{(0.1436 \pm 0.0027) \cdot 10^{-2}}}{0.1436\cdot 10^{-2}}{0.0027\cdot 10^{-2}}%
\htquantdef{Gamma930}{\Gamma_{930}}{\BRF{\tau^-}{2\pi^- \pi^+ \eta \nu_\tau ~(\eta \to \pi^+\pi^-\pi^0) ~(\text{ex.~}K^0)}}{\ensuremath{(5.010 \pm 0.297) \cdot 10^{-5}}}{5.010\cdot 10^{-5}}{0.297\cdot 10^{-5}}%
\htquantdef{Gamma93by60}{\frac{\Gamma_{93}}{\Gamma_{60}}}{\frac{\BRF{\tau^-}{\pi^- K^- K^+ \nu_\tau}}{\BRF{\tau^-}{\pi^- \pi^- \pi^+ \nu_\tau\;(\text{ex.~} K^0)}}}{\ensuremath{(1.594 \pm 0.030) \cdot 10^{-2}}}{1.594\cdot 10^{-2}}{0.030\cdot 10^{-2}}%
\htquantdef{Gamma94}{\Gamma_{94}}{\BRF{\tau^-}{\pi^- K^- K^+ \pi^0 \nu_\tau}}{\ensuremath{(6.113 \pm 1.829) \cdot 10^{-5}}}{6.113\cdot 10^{-5}}{1.829\cdot 10^{-5}}%
\htquantdef{Gamma944}{\Gamma_{944}}{\BRF{\tau^-}{2\pi^- \pi^+ \eta \nu_\tau ~(\eta \to \gamma\gamma) ~(\text{ex.~}K^0)}}{\ensuremath{(8.615 \pm 0.510) \cdot 10^{-5}}}{8.615\cdot 10^{-5}}{0.510\cdot 10^{-5}}%
\htquantdef{Gamma94by69}{\frac{\Gamma_{94}}{\Gamma_{69}}}{\frac{\BRF{\tau^-}{\pi^- K^- K^+ \pi^0 \nu_\tau}}{\BRF{\tau^-}{\pi^- \pi^- \pi^+ \pi^0 \nu_\tau\;(\text{ex.~} K^0)}}}{\ensuremath{(0.1353 \pm 0.0406) \cdot 10^{-2}}}{0.1353\cdot 10^{-2}}{0.0406\cdot 10^{-2}}%
\htquantdef{Gamma96}{\Gamma_{96}}{\BRF{\tau^-}{K^- K^- K^+ \nu_\tau}}{\ensuremath{(2.156 \pm 0.800) \cdot 10^{-5}}}{2.156\cdot 10^{-5}}{0.800\cdot 10^{-5}}%
\htquantdef{Gamma998}{\Gamma_{998}}{1 - \Gamma_{\text{All}}}{\ensuremath{(9.902 \pm 9.850) \cdot 10^{-4}}}{9.902\cdot 10^{-4}}{9.850\cdot 10^{-4}}%
\htquantdef{Gamma9by5}{\frac{\Gamma_{9}}{\Gamma_{5}}}{\frac{\BRF{\tau^-}{\pi^- \nu_\tau}}{\BRF{\tau^-}{e^- \bar{\nu}_e \nu_\tau}}}{\ensuremath{0.6069 \pm 0.0032}}{0.6069}{0.0032}%
\htquantdef{GammaAll}{\Gamma_{\text{All}}}{\Gamma_{\text{All}}}{\ensuremath{0.99901 \pm 0.00098}}{0.99901}{0.00098}%
\htquantdef{gmubyge_tau}{gmubyge_tau}{}{\ensuremath{1.0018 \pm 0.0014}}{1.0018}{0.0014}%
\htquantdef{gtaubyge_tau}{gtaubyge_tau}{}{\ensuremath{1.0029 \pm 0.0015}}{1.0029}{0.0015}%
\htquantdef{gtaubygmu_fit}{gtaubygmu_fit}{}{\ensuremath{1.0001 \pm 0.0014}}{1.0001}{0.0014}%
\htquantdef{gtaubygmu_K}{gtaubygmu_K}{}{\ensuremath{0.9858 \pm 0.0071}}{0.9858}{0.0071}%
\htquantdef{gtaubygmu_pi}{gtaubygmu_pi}{}{\ensuremath{0.9963 \pm 0.0027}}{0.9963}{0.0027}%
\htquantdef{gtaubygmu_tau}{gtaubygmu_tau}{}{\ensuremath{1.0011 \pm 0.0015}}{1.0011}{0.0015}%
\htquantdef{hcut}{hcut}{}{\ensuremath{(6.58211928 \pm 0.00000015) \cdot 10^{-22}}}{6.58211928\cdot 10^{-22}}{0.00000015\cdot 10^{-22}}%
\htquantdef{KmKzsNu}{KmKzsNu}{}{7.950\cdot 10^{-4}}{7.950\cdot 10^{-4}}{0}%
\htquantdef{KmPizKzsNu}{KmPizKzsNu}{}{7.950\cdot 10^{-4}}{7.950\cdot 10^{-4}}{0}%
\htquantdef{KtoENu}{KtoENu}{}{\ensuremath{(1.5810 \pm 0.0080) \cdot 10^{-5}}}{1.5810\cdot 10^{-5}}{0.0080\cdot 10^{-5}}%
\htquantdef{KtoMuNu}{KtoMuNu}{}{\ensuremath{0.6355 \pm 0.0011}}{0.6355}{0.0011}%
\htquantdef{m_e}{m_e}{}{\ensuremath{0.510998928 \pm 0.000000011}}{0.510998928}{0.000000011}%
\htquantdef{m_K}{m_K}{}{\ensuremath{493.677 \pm 0.016}}{493.677}{0.016}%
\htquantdef{m_mu}{m_mu}{}{\ensuremath{105.6583715 \pm 0.0000035}}{105.6583715}{0.0000035}%
\htquantdef{m_pi}{m_pi}{}{\ensuremath{139.57018 \pm 0.00035}}{139.57018}{0.00035}%
\htquantdef{m_s}{m_s}{}{\ensuremath{93.50 \pm 2.50}}{93.50}{2.50}%
\htquantdef{m_tau}{m_tau}{}{\ensuremath{(1.77682 \pm 0.00016) \cdot 10^{3}}}{1.77682\cdot 10^{3}}{0.00016\cdot 10^{3}}%
\htquantdef{m_W}{m_W}{}{\ensuremath{(8.0399 \pm 0.0015) \cdot 10^{4}}}{8.0399\cdot 10^{4}}{0.0015\cdot 10^{4}}%
\htquantdef{phspf_mebymmu}{phspf_mebymmu}{}{\ensuremath{0.999812949174 \pm 0.000000000015}}{0.999812949174}{0.000000000015}%
\htquantdef{phspf_mebymtau}{phspf_mebymtau}{}{\ensuremath{0.99999933833 \pm 0.00000000012}}{0.99999933833}{0.00000000012}%
\htquantdef{phspf_mmubymtau}{phspf_mmubymtau}{}{\ensuremath{0.9725588 \pm 0.0000048}}{0.9725588}{0.0000048}%
\htquantdef{pi}{pi}{}{3.142}{3.142}{0}%
\htquantdef{PimKmKpNu}{PimKmKpNu}{}{0.1440\cdot 10^{-2}}{0.1440\cdot 10^{-2}}{0}%
\htquantdef{PimKmPipNu}{PimKmPipNu}{}{0.2940\cdot 10^{-2}}{0.2940\cdot 10^{-2}}{0}%
\htquantdef{PimKzsKzlNu}{PimKzsKzlNu}{}{9.774\cdot 10^{-4}}{9.774\cdot 10^{-4}}{0}%
\htquantdef{PimKzsKzsNu}{PimKzsKzsNu}{}{2.310\cdot 10^{-4}}{2.310\cdot 10^{-4}}{0}%
\htquantdef{PimPimPipNu}{PimPimPipNu}{}{9.020\cdot 10^{-2}}{9.020\cdot 10^{-2}}{0}%
\htquantdef{PimPimPipPizNu}{PimPimPipPizNu}{}{4.480\cdot 10^{-2}}{4.480\cdot 10^{-2}}{0}%
\htquantdef{PimPizKzsNu}{PimPizKzsNu}{}{0.1725\cdot 10^{-2}}{0.1725\cdot 10^{-2}}{0}%
\htquantdef{PimPizNu}{PimPizNu}{}{0.2552}{0.2552}{0}%
\htquantdef{pitoENu}{pitoENu}{}{\ensuremath{(1.2300 \pm 0.0040) \cdot 10^{-4}}}{1.2300\cdot 10^{-4}}{0.0040\cdot 10^{-4}}%
\htquantdef{pitoMuNu}{pitoMuNu}{}{\ensuremath{0.99987700 \pm 0.00000040}}{0.99987700}{0.00000040}%
\htquantdef{R_tau}{R_tau}{}{\ensuremath{3.6315 \pm 0.0081}}{3.6315}{0.0081}%
\htquantdef{R_tau_s}{R_tau_s}{}{\ensuremath{0.1618 \pm 0.0026}}{0.1618}{0.0026}%
\htquantdef{R_tau_VA}{R_tau_VA}{}{\ensuremath{3.4697 \pm 0.0080}}{3.4697}{0.0080}%
\htquantdef{rrad_LD_kmu_pimu}{rrad_LD_kmu_pimu}{}{\ensuremath{0.9930 \pm 0.0035}}{0.9930}{0.0035}%
\htquantdef{rrad_LD_tauK_taupi}{rrad_LD_tauK_taupi}{}{\ensuremath{1.0003 \pm 0.0044}}{1.0003}{0.0044}%
\htquantdef{rrad_tau_Knu}{rrad_tau_Knu}{}{\ensuremath{1.02010 \pm 0.00030}}{1.02010}{0.00030}%
\htquantdef{sigmataupmy4s}{sigmataupmy4s}{}{0.9190}{0.9190}{0}%
\htquantdef{tau_K}{tau_K}{}{\ensuremath{(1.2380 \pm 0.0021) \cdot 10^{-8}}}{1.2380\cdot 10^{-8}}{0.0021\cdot 10^{-8}}%
\htquantdef{tau_mu}{tau_mu}{}{\ensuremath{(2.196981 \pm 0.000022) \cdot 10^{-6}}}{2.196981\cdot 10^{-6}}{0.000022\cdot 10^{-6}}%
\htquantdef{tau_pi}{tau_pi}{}{\ensuremath{(2.60330 \pm 0.00050) \cdot 10^{-8}}}{2.60330\cdot 10^{-8}}{0.00050\cdot 10^{-8}}%
\htquantdef{tau_tau}{tau_tau}{}{290.3 \pm 0.5}{290.3}{0.5}%
\htquantdef{Vud}{Vud}{}{0.97425 \pm 0.00022}{0.97425}{0.00022}%
\htquantdef{Vud_moulson_ckm14}{Vud_moulson_ckm14}{}{\ensuremath{0.97417 \pm 0.00021}}{0.97417}{0.00021}%
\htquantdef{Vus}{Vus}{}{\ensuremath{0.2176 \pm 0.0021}}{0.2176}{0.0021}%
\htquantdef{Vus_err_exp}{Vus_err_exp}{}{0.1833\cdot 10^{-2}}{0.1833\cdot 10^{-2}}{0}%
\htquantdef{Vus_err_exp_perc}{Vus_err_exp_perc}{}{0.8425}{0.8425}{0}%
\htquantdef{Vus_err_perc}{Vus_err_perc}{}{0.9527}{0.9527}{0}%
\htquantdef{Vus_err_th}{Vus_err_th}{}{9.680\cdot 10^{-4}}{9.680\cdot 10^{-4}}{0}%
\htquantdef{Vus_err_th_perc}{Vus_err_th_perc}{}{0.44}{0.44}{0}%
\htquantdef{Vus_kl2_moulson_ckm14}{Vus_kl2_moulson_ckm14}{}{\ensuremath{0.22484 \pm 0.00059}}{0.22484}{0.00059}%
\htquantdef{Vus_kl3_moulson_ckm14}{Vus_kl3_moulson_ckm14}{}{\ensuremath{0.22320 \pm 0.00090}}{0.22320}{0.00090}%
\htquantdef{Vus_mism}{Vus_mism}{}{\ensuremath{(-0.7875 \pm 0.2304) \cdot 10^{-2}}}{-0.7875\cdot 10^{-2}}{0.2304\cdot 10^{-2}}%
\htquantdef{Vus_mism_sigma}{Vus_mism_sigma}{}{-3.4}{-3.4}{0}%
\htquantdef{Vus_mism_sigma_abs}{Vus_mism_sigma_abs}{}{3.4}{3.4}{0}%
\htquantdef{Vus_tau}{Vus_tau}{}{\ensuremath{0.2204 \pm 0.0014}}{0.2204}{0.0014}%
\htquantdef{Vus_tau_mism}{Vus_tau_mism}{}{\ensuremath{(-0.5106 \pm 0.1747) \cdot 10^{-2}}}{-0.5106\cdot 10^{-2}}{0.1747\cdot 10^{-2}}%
\htquantdef{Vus_tau_mism_sigma}{Vus_tau_mism_sigma}{}{-2.9}{-2.9}{0}%
\htquantdef{Vus_tau_mism_sigma_abs}{Vus_tau_mism_sigma_abs}{}{2.9}{2.9}{0}%
\htquantdef{Vus_tauKnu}{Vus_tauKnu}{}{\ensuremath{0.2212 \pm 0.0020}}{0.2212}{0.0020}%
\htquantdef{Vus_tauKnu_err_th_perc}{Vus_tauKnu_err_th_perc}{}{0.5760}{0.5760}{0}%
\htquantdef{Vus_tauKnu_mism}{Vus_tauKnu_mism}{}{\ensuremath{(-0.4242 \pm 0.2218) \cdot 10^{-2}}}{-0.4242\cdot 10^{-2}}{0.2218\cdot 10^{-2}}%
\htquantdef{Vus_tauKnu_mism_sigma}{Vus_tauKnu_mism_sigma}{}{-1.9}{-1.9}{0}%
\htquantdef{Vus_tauKnu_mism_sigma_abs}{Vus_tauKnu_mism_sigma_abs}{}{1.9}{1.9}{0}%
\htquantdef{Vus_tauKpi}{Vus_tauKpi}{}{\ensuremath{0.2232 \pm 0.0019}}{0.2232}{0.0019}%
\htquantdef{Vus_tauKpi_err_th_perc}{Vus_tauKpi_err_th_perc}{}{0.4731}{0.4731}{0}%
\htquantdef{Vus_tauKpi_err_th_perc_delta_LD_tauK_Kmu}{Vus_tauKpi_err_th_perc_delta_LD_tauK_Kmu}{}{-0.1090}{-0.1090}{0}%
\htquantdef{Vus_tauKpi_err_th_perc_delta_LD_taupi_pimu}{Vus_tauKpi_err_th_perc_delta_LD_taupi_pimu}{}{6.989\cdot 10^{-2}}{6.989\cdot 10^{-2}}{0}%
\htquantdef{Vus_tauKpi_err_th_perc_f_K_by_f_pi}{Vus_tauKpi_err_th_perc_f_K_by_f_pi}{}{-0.4195}{-0.4195}{0}%
\htquantdef{Vus_tauKpi_err_th_perc_rrad_LD_kmu_pimu}{Vus_tauKpi_err_th_perc_rrad_LD_kmu_pimu}{}{-0.1762}{-0.1762}{0}%
\htquantdef{Vus_tauKpi_mism}{Vus_tauKpi_mism}{}{\ensuremath{(-0.2280 \pm 0.2190) \cdot 10^{-2}}}{-0.2280\cdot 10^{-2}}{0.2190\cdot 10^{-2}}%
\htquantdef{Vus_tauKpi_mism_sigma}{Vus_tauKpi_mism_sigma}{}{-1.0}{-1.0}{0}%
\htquantdef{Vus_tauKpi_mism_sigma_abs}{Vus_tauKpi_mism_sigma_abs}{}{1.0}{1.0}{0}%
\htquantdef{Vus_uni}{Vus_uni}{}{\ensuremath{0.22547 \pm 0.00095}}{0.22547}{0.00095}%
\htquantdef{VusbyVud_moulson_ckm14}{VusbyVud_moulson_ckm14}{}{\ensuremath{0.23080 \pm 0.00060}}{0.23080}{0.00060}%
\htdef{couplingsCorr}{%
$\left( \frac{g_\tau}{g_e} \right)$ &   53\\
$\left( \frac{g_\mu}{g_e} \right)$ &  -49 &   48\\
$\left( \frac{g_\tau}{g_\mu} \right)_\pi$ &   24 &   26 &    2\\
$\left( \frac{g_\tau}{g_\mu} \right)_K$ &   12 &   10 &   -2 &    5\\
 & $\left( \frac{g_\tau}{g_\mu} \right)$ & $\left( \frac{g_\tau}{g_e} \right)$ & $\left( \frac{g_\mu}{g_e} \right)$ & $\left( \frac{g_\tau}{g_\mu} \right)_\pi$}%

\begin{fleqn}
\let\tausection\subsection
\let\tausubsection\subsubsection
\newenvironment{envsmall}%
  {\small}%
  {}
\section{Tau lepton properties}
\label{sec:tau}
We present world averages of a selection of \mtau lepton quantities with
the goal to provide the best up-to-date determinations of the tests of the
universality of the charged weak interaction
(Section~\ref{sec:tau:leptonuniv}) and of the Cabibbo-Kobayashi-Maskawa
(CKM) matrix coefficient \Vus from \mtau decays
(Section~\ref{sec:tau:vus}).  We concentrate our effort in the averages
that most benefit from the adoption of the HFAG
methodology~\cite{Asner:2010qj}, namely a global fit of the \mtau branching
fractions that best exploits the available experimental information. We
also average the \mtau lifetime, in order to use the recent precise Belle
result~\cite{Belous:2013dba}, which has not been used by PDG until the 2014
edition.

All published statistical correlations are used, and a selection of
measurements, particularly the most precise and the most recent, were
studied to take into account the significant systematic dependencies from
external parameters and common systematic sources.

Finally, we report in Section~\ref{sec:tau:lfv} the most up-to-date limits
on the lepton-flavour-violating \mtau branching fractions and we determine
the combined upper limits~\ref{sec:tau:lfv-comb} for the branching
fractions that have multiple experimental results.

\tausection{Branching fractions fit}
\cutname{br-fit.html}
\label{sec:tau:br-fit}

The \mtau branching fractions provide a test for theory predictions based
on the Standard Model (SM) EW and QCD interactions and can be further
elaborated to test the EW charged-current universality for leptons, to
determine the CKM matrix coefficient \Vus and the QCD coupling constant
$\alpha_s$ at the \mtau mass. A global constrained fit of the available
experimental measurements is used to obtain the averages of the \mtau
branching fractions to a complete set of the observed final states,
together with their uncertainties and statistical correlations: these data
summarize the experimental information for further elaborations.
\label{sec:tau:fit}
The fit procedure is functionally equivalent to the one employed in the
former HFAG reports~\cite{Asner:2010qj,Amhis:2012bh} and consists in a minimum \chisq
fit subject to linear and non-linear constraints.

The measurements listed in Table~\ref{tab:tau:br-fit} have been used in a
minimum \chisq fit subject to the constraints that are listed
either in the same table (where some fitted quantities and experimental
measurements are expressed as ratios of fit quantities) or in
Section~\ref{sec:tau:constraints}. The fitted quantities and the measurements
are labelled using the PDG $\Gamma_{n}$ notation, where $n$ is
an integer number, which matches the PDG notation for $n<800$. We use
$n\ge 800$ to denote some additional branching fractions, as documented in the
former HFAG report~\cite{Amhis:2012bh}. The PDG $\Gamma_{n}$ notation does
not maintain the same numbers across editions. We continue using the PDG
2010~\cite{PDG_2010} numbers with the aim to eventually switch to a different stable notation,
probably based on PDG identifiers~\cite{pdg-identifiers-2014}.

The fit output consists in \htuse{QuantNum} quantities, which correspond to
either branching fractions or ratios of linear combinations of branching
fractions. Although the fit treats all quantities in the same way, for the
purpose of describing the results we divide the above quantities in a set
of \htuse{BaseQuantNum} ``base nodes'' that permit the definition of all
the remaining ones as either a sum of ``base nodes'' (\htuse{ConstrNum}
quantities, see Section~\ref{sec:tau:constraints}) or as ratios of linear
combinations of ``base nodes'' (the remaining quantities, see
Table~\ref{tab:tau:br-fit}).

Furthermore we define (see Section~\ref{sec:tau:constraints}) $\Gamma_{110}
= \BR(\tau^- \to X_s^- \nu_\tau)$, the total branching fraction of the \mtau
decays to final states with the strangeness quantum number equal to one, and
$\Gamma_{\text{All}}$, the branching fraction of the \mtau into any
measured final state, which is supposed to be equal to $1$ within the
experimental uncertainty. We define the unitarity residual as $\Gamma_{998}
= 1 -\Gamma_{\text{All}}$.

The fitted \hfagtau averages are reported in
Table~\ref{tab:tau:br-fit}. The fit has $\chi^2/\text{d.o.f.} = \htuse{Chisq}/\htuse{Dof}$,
corresponding to a confidence level $\text{CL} = \htuse{ChisqProb}$. We use a total of
\htuse{MeasNum} measurements to fit the above mentioned \htuse{QuantNum} quantities.
Although the unitarity constraint is not applied, the fit is statistically
consistent with unitarity, and the unitarity residual is
\htuse{Gamma998.gn} =  \htuse{Gamma998.td} = \htuse{Gamma998}.

A scale factor of 5.44 (as in the two previous
reports~\cite{Asner:2010qj,Amhis:2012bh}) has been applied to the published
uncertainties of the two severely inconsistent measurements of
\(\Gamma_{96} = \tau \to KKK\nu\) by \babar and Belle, following the same
procedure as the PDG.

For several old results, for historical reasons, the table reports as
statistical errors the sum in quadrature of the statistical and systematic
errors and as systematic errors zero: this does not affect the fit results
since the systematic errors are treated exactly like the statistical ones.

\tausubsection{Changes with respect to the previous report}

The following additions and changes have been done with respect to the
previous HFAG report~\cite{Amhis:2012bh}.

Published results from two \babar papers and one Belle paper have been
added. The following results have been used from the \htuse{LEES 2012X.collab} high
multiplicity decay \mtau branching fractions paper~\htuse{LEES 2012X.cite}:
{\setlength{\LTleft}{\parindent}%
\begin{longtable}{@{}ll@{}}
\htuse{LEES 2012X.meas}.
\end{longtable}}
\noindent These results supersede the previous \babar results
$\htuse{Gamma136.gn} = \tau^- \to \htuse{Gamma136.td}$~\cite{Aubert:2008nj}
and $\htuse{Gamma103.gn} = \tau^- \to
\htuse{Gamma103.td}$~\cite{Aubert:2005waa}. The following results have been
used from the \htuse{LEES 2012Y.collab} paper on the \mtau branching
fractions with two $K_S$~\htuse{LEES 2012Y.cite}:
{\setlength{\LTleft}{\parindent}%
\begin{longtable}{@{}ll@{}}
\htuse{LEES 2012Y.meas}.
\end{longtable}}
\noindent The following results have been used from the
\htuse{Ryu:2014vpc.collab} paper on the $K_S$ final
states~\htuse{Ryu:2014vpc.cite}:
{\setlength{\LTleft}{\parindent}%
\begin{longtable}{@{}ll@{}}
\htuse{Ryu:2014vpc.meas}.
\end{longtable}}
\noindent These results supersede the preliminary ones~\cite{Ryu:2012pm}
and the former published result on $\tau^- \to \htuse{Gamma35.td}$~\cite{Epifanov:2007rf}.
In order to profit from the measurements of branching fractions with two
$K_S$ in the final state, we discard the inclusive ALEPH measurement on
$\tau^- \to \htuse{Gamma49.td}$~\cite{Barate:1999hj} and we add the exclusive ALEPH
measurement on $\tau^- \to \htuse{Gamma51.td}$~\cite{Barate:1999hj}.

The CLEO result on  $\tau^- \to \htuse{Gamma136.td}$~\cite{Anastassov:2000xu}
has been discarded since it is very correlated with the branching fractions
into six pions measured in the same paper, which are dominated by the $\eta$ and
$\omega$ resonances.

We added the $\tau^- \to \htuse{Gamma128.td}$ measurements by CLEO~\cite{Bartelt:1996iv}
and ALEPH~\cite{Buskulic:1996qs} to complete the list of useful experimental inputs for
this mode.

In order to best integrate the above new results in the global fit, the following
constraints have been added:
{\setlength{\LTleft}{\parindent}%
\begin{tabularx}{\linewidth-\parindent}{@{}lX@{}}
  \htuse{Gamma33.c.constr.eq} \\
  \htuse{Gamma49.c.constr.eq} \\
  \htuse{Gamma78.c.constr.eq} \\
  \htuse{Gamma103.c.constr.eq} \\
  \htuse{Gamma104.c.constr.eq} \\
  \htuse{Gamma806.c.constr.eq} \\
  \htuse{Gamma810.c.constr.eq} \\
  \htuse{Gamma820.c.constr.eq} \\
  \htuse{Gamma830.c.constr.eq} \\
  \htuse{Gamma910.c.constr.eq} \\
  \htuse{Gamma930.c.constr.eq} \\
  \htuse{Gamma944.c.constr.eq}~. \\
\end{tabularx}}
\noindent It was realised that the \htuse{Gamma44.gn} fit quantity, which is
exclusively determined by one single ALEPH
result~\htuse{ALEPH.Gamma44.pub.BARATE.99R,ref}, does actually exclude the
contribution from $K^0\to\pi^0\pi^0$, so we renamed it to
$\htuse{Gamma44.gn} = \htuse{Gamma44.td}$.
The definition of the total branching fraction and of the inclusive
branching fraction of the \mtau lepton into a strange final states have
been updated as follows:
{\setlength{\LTleft}{\parindent}%
\begin{tabularx}{\linewidth-\parindent}{@{}lX@{}}
  \htuse{Gamma110.c.constr.eq} \\
  \htuse{GammaAll.c.constr.eq}~.
\end{tabularx}}
\noindent The total \mtau branching fraction \htuse{GammaAll.gn} definition
includes two modes that have overlapping final states, to a minor extent:
\begin{align*}
  &\htuse{Gamma50.gn} =  \htuse{Gamma50.td} \\
  &\htuse{Gamma132.gn} =  \htuse{Gamma132.td}~.
\end{align*}
\noindent The amount of overlap cannot be disentangled with the presently
available measurements, however we consider it negligible since the
involved branching fractions are small and their overlap is conceivably minor.
An inaccurate constraint used in the two previous reports has been
removed:
{\setlength{\LTleft}{\parindent}%
\begin{tabularx}{\linewidth-\parindent}{@{}lX@{}}
  $\Gamma_{136}$ ={}& $\Gamma_{104}\cdot \Gamma_{\eta \to \pi^+\pi^-\pi^0} +
  \Gamma_{78} \cdot \Gamma_{\eta \to 3\pi^0}$~.
\end{tabularx}}
\noindent The inaccurate constraint had negligible effect on the global
fit in the previous HFAG reports, except for the involved specific
branching ratios.  In particular, the effects on the lepton universality
tests and in the \Vus determination were negligible.

Finally, the constraint parameters (see Section~\ref{sec:tau:constraints})
have been updated to the PDG 2013 results~\cite{PDG_2012}.

\tausubsection{Branching ratio fit results and experimental inputs}
\label{sec:tau:br-fit-results}

Table~\ref{tab:tau:br-fit} reports the \mtau branching ratio fit results
and experimental inputs.

\begin{center}
\begin{envsmall}
\setlength{\LTcapwidth}{0.85\linewidth}
\renewcommand*{\arraystretch}{1.3}%
\ifhevea
\renewcommand{\bar}[1]{\textoverline{#1}}
\else
\begin{citenoleadsp}
\fi
\begin{longtable}{llll}
\caption{HFAG \hfagTauTag branching fractions fit results.\label{tab:tau:br-fit}}%
\\
\hline
\multicolumn{1}{l}{\bfseries \mtau lepton branching fraction} &
\multicolumn{1}{l}{\bfseries Fit value / Exp.} &
\multicolumn{1}{l}{\bfseries HFAG Fit / Ref.} \\
\hline
\endfirsthead
\multicolumn{4}{c}{{\bfseries \tablename\ \thetable{} -- continued from previous page}} \\ \hline
\multicolumn{1}{l}{\bfseries \mtau lepton branching fraction} &
\multicolumn{1}{l}{\bfseries Fit value / Exp.} &
\multicolumn{1}{l}{\bfseries HFAG Fit / Ref.} \\
\hline
\endhead
\endfoot
\endlastfoot
\htuse{BrVal} \\
\hline
\end{longtable}
\ifhevea\else
\end{citenoleadsp}
\fi
\end{envsmall}
\end{center}

\tausubsection{Correlation between base nodes uncertainties}
\label{sec:tau:fitcorr}

The following tables report the correlation coefficients between base nodes,
in percent.

\htuse{BrCorr}

\tausubsection{Equality constraints}
\label{sec:tau:constraints}

We use equality constraints that relate a branching fraction to a sum of
branching fractions. As mentioned above, the \mtau branching fractions are
denoted with $\Gamma_n$ labels. In the constraint relations we use the
values of some non-tau branching fractions, denoted \eg\ with the
self-describing notation $\Gamma_{K_S \to \pi^0\pi^0}$. We also use
probabilities corresponding to modulus square amplitudes describing quantum
mixtures of states such as $K^0$, $\bar{K}^0$, $K_S$, $K_L$, denoted with
\eg\ $\Gamma_{<K^0|K_S>} = |{<}K^0|K_S{>}|^2$.
In the fit, all non-tau quantities are taken from the PDG 2013~\cite{PDG_2012}
fits (when available) or averages, and are used without accounting for their
uncertainties, which are however in general small with respect
to the uncertainties on the \mtau branching fractions.
The \mtau branching fractions are illustrated in Table~\ref{tab:tau:br-fit}.
The equations in the following permit the computation of the values and
uncertainties for branching fractions that are not listed in
Table~\ref{tab:tau:br-fit}, once they are expressed as function of the
quantities that are listed there. The following list does not include the
(non-linear) constraints already introduced in
Section~\ref{sec:tau:br-fit}, and illustrated in
Table~\ref{tab:tau:br-fit}, where some measured branching fractions are
expressed as ratios of ``base'' branching fractions.

\begin{envsmall}
  \setlength\abovedisplayskip{0pt}
  \setlength\belowdisplayshortskip{0pt}
  \ifhevea\renewcommand{\bar}[1]{\textoverline{#1}}\fi
  \htuse{ConstrEqs}
\end{envsmall}

\tausection{\mtau lifetime average}
\cutname{lifetime.html}
\label{sec:tau:lifetime}

When the work for this report started, the PDG did not yet include the 2013
precise Belle \mtau lifetime measurement~\cite{Belous:2013dba}. In order to
obtain the most precise and up-to-date lepton universality tests, we
computed and used an updated \mtau lifetime average. In this case, the HFAG
procedure does not improve the PDG average with additional information or
technical treatment and we obtain the same result as PDG
2014~\cite{PDG_2014} (see Fig.~\ref{fig:tau:tau-lifetime}) by doing a
standard error-weighted average.
With respect to PDG 2013, the uncertainty is now reduced by a factor two.
\begin{figure}[tb]
  \begin{center}
   \ifhevea
    \begin{tabular}{@{}cc@{}}
      \larger\bfseries\ahref{plot-taulife-hfag-summer2014.png}{PNG format} &
      \larger\bfseries\ahref{plot-taulife-hfag-summer2014.pdf}{PDF format} \\
      \multicolumn{2}{c}{\ahref{plot-taulife-hfag-summer2014.png}{%
          \imgsrc[alt="Vus summary plot"]{plot-taulife-hfag-summer2014.png}}}
    \end{tabular}
    \else
    \includegraphics[width=0.75\linewidth,clip]{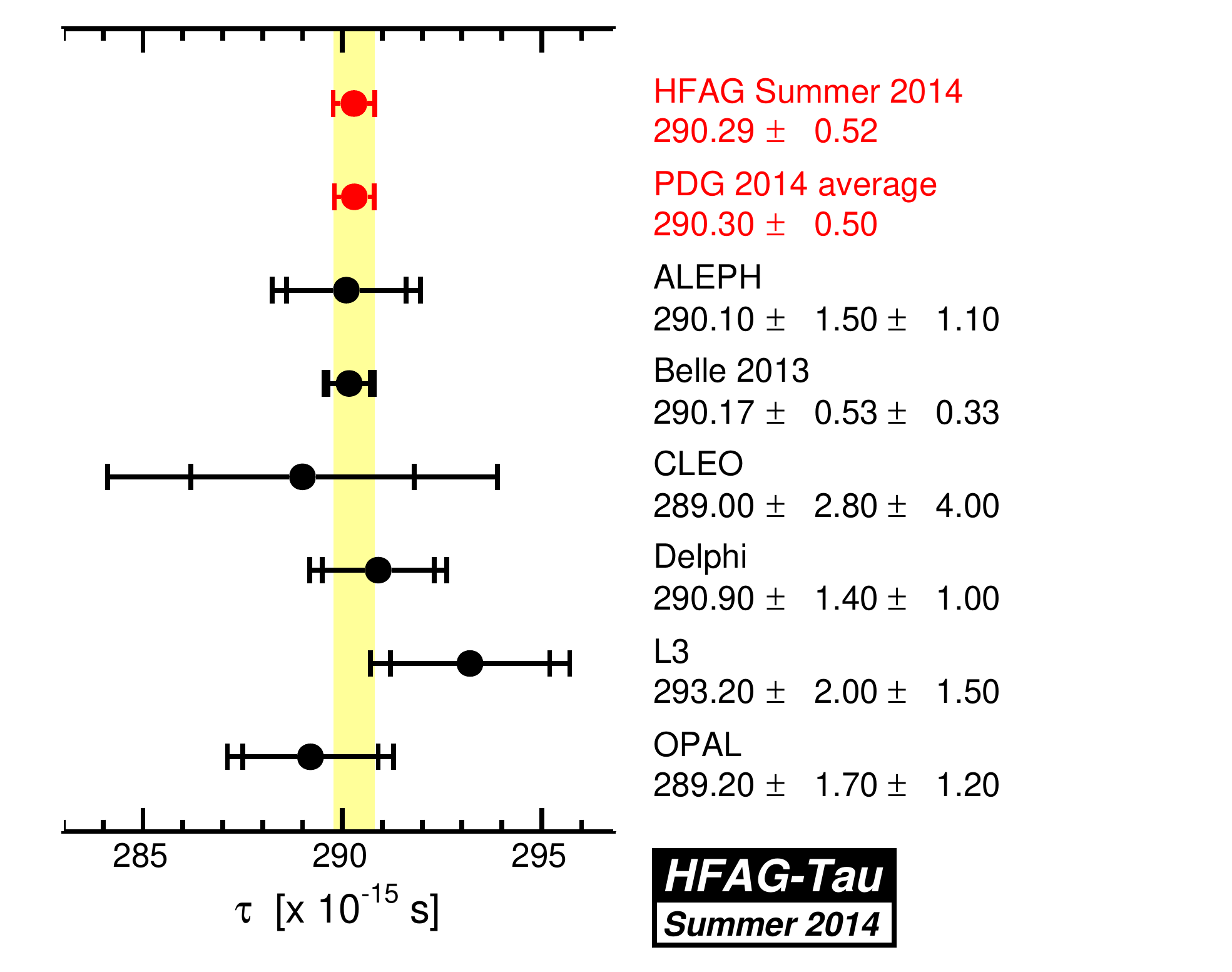}
    \fi
    \caption{\mtau lifetime average.%
      \label{fig:tau:tau-lifetime}%
    }
  \end{center}
\end{figure}

\tausection{Tests of lepton universality}
\cutname{lepton-univ.html}
\label{sec:tau:leptonuniv}

The lepton universality tests probe the Standard Model prediction that the
charged weak current interaction has the same coupling for all leptons.
The precision of such tests has been significantly improved by the
addition of the recent Belle \mtau lifetime
measurement~\cite{Belous:2013dba}, while improvements from the \mtau
branching fraction fit are negligible.
We compute the universality tests like in the previous report by using
proper ratios of the partial widths of a heavier lepton \lepth
decaying to a
lighter lepton \leptl~\cite{Marciano:1988vm},
\begin{align*}
  \Gamma(\lepth \to \nu_{\lepth} \leptl \nub_{\leptl} (\gamma)) =
  \frac{\BR(\lepth \to \nu_{\lepth} \leptl \nub_{\leptl})}{\tau_{\lepth}} =
  \frac {G_{\lepth} G_{\leptl} m^5_{\lepth}}{192 \pi^3}\, f\left(\frac {m^2_{\leptl}}{m^2_{\lepth}}\right)
  \opdelta^{\lepth}_W \opdelta^{\lepth}_\gamma~,
\end{align*}
where
\begin{alignat*}{3}
 G_{\leptl} &= \frac {g^2_{\leptl}}{4 \sqrt{2} M^2_W}~, &\quad&&
 f(x) &= 1 -8x +8x^3 -x^4 -12x^2 \text{ln}x~, \\
 \opdelta^{\lepth}_W &= 1 + \frac {3}{5} \frac {m^2_{\lepth}}{M^2_W}~, &\quad\quad&&
 \opdelta^{\lepth}_\gamma &= 1+\frac {\alpha(m_{\lepth})}{2\pi} \left(\frac {25}{4}-\pi^2\right)~.
\end{alignat*}
We use $\opdelta^\tau_\gamma=1-43.2\cdot 10^{-4}$ and
$\opdelta^\mu_\gamma=1-42.4\cdot 10^{-4}$~\cite{Marciano:1988vm} and $M_W$
from PDG 2013~\cite{PDG_2012}.
We use HFAG 2014 averages and PDG 2013 for the other quantities.
Using pure leptonic processes we obtain
\begin{align*}
  \left( \frac{g_\tau}{g_\mu} \right) = \htuse{gtaubygmu_tau}~,
  && \left( \frac{g_\tau}{g_e} \right) = \htuse{gtaubyge_tau}~,
  && \left( \frac{g_\mu}{g_e} \right) = \htuse{gmubyge_tau}~.
\end{align*}
Using semi-hadronic processes
\begin{align*}
  \left( \frac{g_\tau}{g_\mu} \right)^2 =
  \frac{\BR({\tau \to h \nu_\tau})}{\BR({h \to \mu \bar{\nu}_\mu})}
  \frac{2m_h m^2_{\mu}\tau_h}{(1+\delta_{h})m^3_{\tau}\tau_{\tau}}
  \left( \frac{1-m^2_{\mu}/m^2_h}{1-m^2_h/m^2_{\tau}} \right)^2~,
\end{align*}
where $h$ = $\pi$ or $K$ and the radiative corrections are
$\delta_{\pi} = (\htuse{delta_LD_taupi_pimu})\%$ and
$\delta_{K} = (\htuse{delta_LD_tauK_Kmu})\%$~\cite{Decker:1994dd},
we measure:
\begin{align*}
  \left( \frac{g_\tau}{g_\mu} \right)_\pi &= \htuse{gtaubygmu_pi}~,
  & \left( \frac{g_\tau}{g_\mu} \right)_K = \htuse{gtaubygmu_K}~.
\end{align*}
Similar tests could be performed with decays to electrons, however they are
less precise because the hadron two body decays to electrons are
helicity-suppressed.
Averaging the three \(g_\tau/g_\mu\) ratios we obtain
\begin{align*}
  \left( \frac{g_\tau}{g_\mu} \right)_{\tau{+}\pi{+}K} &= \htuse{gtaubygmu_fit}~,
\end{align*}
accounting for statistical correlations.
Table~\ref{tab:tau:univ-fit-corr} reports the statistical correlation coefficients for the fitted coupling ratios:
\ifhevea\begin{table}\fi
\begin{center}
\ifhevea
\caption{Universality coupling ratios correlation coefficients (\%).\label{tab:tau:univ-fit-corr}}%
\else
\begin{minipage}{\linewidth}
\begin{center}
\captionof{table}{Universality coupling ratios correlation coefficients (\%).}\label{tab:tau:univ-fit-corr}%
\fi
\begin{center}
\renewcommand*{\arraystretch}{1.1}%
\begin{tabular}{lcccc}
\toprule
\htuse{couplingsCorr}
\\\bottomrule
\end{tabular}
\end{center}
\ifhevea\else
\end{center}
\end{minipage}
\fi
\end{center}
\ifhevea\end{table}\fi

\tausection{Universality improved $\BR(\tau \to e \nu \bar{\nu})$ and $\Rhad$}
\cutname{Be_univ_and_Rtau.html}
\label{sec:tau:be-univ-rtau}

We compute two quantities that are used in this report and have been
traditionally used for further elaborations and
tests involving the \mtau branching fractions: the ``universality improved'' experimental
determination of $\BR_e = \BR(\tau \to e \nu \bar{\nu})$, which relies on assuming
that the Standard Model and lepton universality hold, and the ratio \Rhad between the
total branching fraction of the \mtau to hadrons and the universality
improved $\BR_e$, which is the same as the ratio of the two respective
partial widths.

Following Ref.~\cite{Davier:2005xq}, we obtain a more precise experimental
determination of $\BR_e$ using the
\mtau branching fraction to muon and the \mtau lifetime. We average:
\begin{itemize}

\item the $\BR_e$ fit value \htuse{Gamma5.gn},

\item
  the $\BR_e$ determination from the $\BR_\mu = \BR(\tau \to \mu \nu
  \bar{\nu})$ fit value \htuse{Gamma3.gn} assuming that $g_\mu/g_e = 1$
  hence (see also Section~\ref{sec:tau:leptonuniv}) $\BR_e = \BR_\mu \cdot
  f(m^2_e/m^2_\tau)/f(m^2_\mu/m^2_\tau)$,

\item
  the $\BR_e$ determination from the \mtau lifetime assuming that
  $g_\tau/g_\mu =1$ hence $\BR_e = \BR(\mu \to e \bar{\nu}_e
  \nu_\mu)\cdot (\tau_\tau / \tau_\mu) \cdot (m_\tau/m_\mu)^5 \cdot
  f(m^2_e/m^2_\tau)/f(m^2_e/m^2_\mu) \cdot (\delta_\gamma^\tau
  \delta_W^\tau)/(\delta_\gamma^\mu \delta_W^\mu)$ where $\BR(\mu \to e
  \bar{\nu}_e \nu_\mu) = 1$.

\end{itemize}
Accounting for statistical correlations, we obtain
\begin{align*}
  \BR_e^{\text{uni}} = (\htuse{Be_univ})\%.
\end{align*}
The recent Belle \mtau lifetime measurement has brought a significant improvement.
We use $\BR_e^{\text{uni}}$ to obtain the ratio
\begin{align*}
  \Rhad = \frac{\Gamma(\tau \to \text{hadrons})}{\Gamma(\tau\to
    e\nu\bar{\nu})} = \frac{\Gamma_{\text{hadrons}}}{\BR_e^{\text{uni}}} =
  \htuse{R_tau},
\end{align*}
where $\Gamma(\tau \to \text{hadrons})$ and $\Gamma(\tau\to e\nu\bar{\nu})$
indicate the partial widths and $\Gamma_{\text{hadrons}}$ is the total
branching fraction of the \mtau to hadrons, or the total branching fraction
in any measured final state minus the leptonic branching fractions, \ie\
with our notation $\Gamma_{\text{hadrons}} = \htuse{GammaAll.gn} -
\htuse{Gamma3.gn} - \htuse{Gamma5.gn} = (\htuse{B_tau_had_fit})\%$ (see
Section~\ref{sec:tau:br-fit} and Table~\ref{tab:tau:br-fit} for the
definitions of \htuse{GammaAll.gn}, \htuse{Gamma3.gn}, \htuse{Gamma5.gn}).
We underline that this report's definition of $\Gamma_{\text{hadrons}}$
corresponds to summing all \mtau hadronic decay modes, like in the previous
report, rather than -- as done elsewhere -- subtracting the leptonic
branching fractions from unity, \ie\ $\Gamma_{\text{hadrons}} = 1 -
\htuse{Gamma3.gn} - \htuse{Gamma5.gn}$.

\tausection{$\Vus$ measurement}
\cutname{vus.html}
\label{sec:tau:vus}

The CKM matrix element \Vus is most precisely determined from kaon
decays~\cite{Antonelli:2010yf}, and its precision is limited by the
uncertainties of the lattice QCD estimates of $f_+^{K\pi}(0)$ and $f_K/f_\pi$.
Using the \mtau branching fractions, it is possible to determine \Vus in an
alternative way~\cite{Gamiz:2006xx} that does not depend on lattice QCD and
has small theory uncertainties (see Section~\ref{sec:tau:vus:incl}).
Moreover, \Vus can be determined using the \mtau branching fractions
similarly to the kaon case, using the same lattice QCD estimates, in order
to check the overall experimental consistency.

We have updated the CKM coefficient \Vus determinations that we did in the
previous report using the updated data from HFAG 2014 and PDG 2013.

\tausubsection{Inclusive \mtau partial width to strange}
\label{sec:tau:vus:incl}

The \mtau hadronic partial width is the sum of the \mtau partial widths to
strange and to non-strange hadronic final states, $\Gammahad =
\Gammastrange + \Gammanonstrange$.  Dividing any partial width $\Gamma_x$
by the electronic partial width, $\Gamma_e$, we obtain partial width ratios
$R_x$ (which are equal to the respective branching fraction ratios
$\BR_x/\BR_e$) for which $\Rhad = \Rstrange + \Rnonstrange$. In terms of
such ratios, \Vus is measured as~\cite{Gamiz:2006xx}
\begin{align*}
  \VusTauIncl &= \sqrt{\Rstrange/\left[\frac{\Rnonstrange}{\Vud^2} -  \delta R_{\text{theory}}\right]}~,
\end{align*}
where $\delta R_{\text{theory}}$ can be determined in the context of low
energy QCD theory, partly relying on experimental low energy scattering
data. The literature reports several
calculations~\cite{Gamiz:2006xx,Gamiz:2007qs,Maltman:2010hb}. In this
report we use Ref.~\cite{Gamiz:2006xx}, whose estimated uncertainty size is
in between the two other ones. We use the information in that paper and the
PDG 2013 value for the $s$-quark mass $m_s = \htuse{m_s}\,\text{MeV}$~\cite{PDG_2012}
to calculate $\delta R_{\text{theory}} = \htuse{deltaR_su3break}$.

We proceed following the same procedure of the 2012 HFAG
report~\cite{Amhis:2012bh}, using the universality improved
$\BR_e^{\text{uni}} = (\htuse{Be_univ})\%$
(see Section~\ref{sec:tau:be-univ-rtau}) to compute the $R_x$ ratios, and
using the sum of the \mtau branching fractions to strange and
non-strange hadronic final states to compute \Rstrange and \Rnonstrange,
respectively. 

Using the \mtau branching fraction fit results with their uncertainties
and correlations (Section~\ref{sec:tau:br-fit}), we compute $\BRstrange =
(\htuse{B_tau_s_fit})\%$ (see also Table~\ref{tab:tau:vus}) and
$\BRnonstrange = \BR_{\text{hadrons}} - \BRstrange=
(\htuse{B_tau_VA_fit})\%$, where $\BR_{\text{hadrons}} =
\Gamma_{\text{hadrons}}$ defined in section~\ref{sec:tau:be-univ-rtau}. PDG 2013 averages
are used for non-\mtau quantities, including $\Vud = \htuse{Vud}$, which
comes from Ref.~\cite{Hardy:2008gy} like for the previous HFAG report.

We obtain $\VusTauIncl = \htuse{Vus}$, which
is $\htuse{Vus_mism_sigma_abs}\sigma$ lower than the unitarity CKM
prediction $\VusUni = \htuse{Vus_uni}$, from $(\VusUni)^2 = 1 -
\Vud^2$. The \VusTauIncl uncertainty includes a systematic error
contribution of \htuse{Vus_err_th_perc}\% from the theory uncertainty on
$\delta R_{\text{theory}}$. There is no significant change with respect to
the previous HFAG report.

Kim Maltman has computed an alternative theoretical estimate based on Fixed
Order Pertubation Theory and experimental inputs restricted to \mtau
quantities~\cite{Maltman:oct2014}.
The result is  $\delta R_{\text{theory}} = 0.254 \pm 0.038$.
The uncertainty includes an additional contribution to account
for the differences between using Fixed Order Pertubation Theory and Contour
Improved Perturbation Theory. With this alternative value, we would obtain
$\VusTauIncl = 0.2181 \pm 0.0022$, which would be $3.1\sigma$ lower than
the unitarity CKM prediction.

\begin{table}
\begin{center}
\renewcommand*{\arraystretch}{1.3}%
\caption{HFAG \hfagTauTag \mtau branching fractions to strange final states.\label{tab:tau:vus}}%
\ifhevea\renewcommand{\bar}[1]{\textoverline{#1}}\fi
\begin{envsmall}
\begin{center}
\begin{tabular}{llll}
\hline
\multicolumn{1}{c}{\bfseries Branching fraction} &
\multicolumn{1}{c}{\bfseries HFAG \hfagTauTag fit} \\
\hline
\htuse{BrStrangeVal}
\\\hline
\htuse{BrStrangeTotVal}
\\\hline
\end{tabular}
\end{center}
\end{envsmall}
\end{center}
\end{table}

\tausubsection{\Vus from $\BR(\tau \to K\nu) / \BR(\tau \to \pi\nu)$ and from $\BR(\tau \to K\nu)$}

We follow the same procedure of the HFAG 2012 report to compute \Vus from
the ratio of branching fractions $\BFtautoknu/\BFtautopinu =
\htuse{Gamma10by5}$ from the equation
\begin{align*}
\frac{\BFtautoknu}{\BFtautopinu} &=
\frac{f_K^2 \Vus^2}{f_\pi^2 \Vud^2} \frac{\left( 1 - m_K^2/m_\tau^2 \right)^2}{\left( 1 -  m_\pi^2/m_\tau^2 \right)^2}
\frac{\opdelta_{\text{LD}}(\tau^- \to K^-\nut)}{\opdelta_{\text{LD}}(\tau^- \to \pi^-\nut)}~.
\end{align*}
We use $f_K/f_\pi = \htuse{f_K_by_f_pi}$ from the
FLAG 2013 Lattice averages with $N_f=2+1$~\cite{Aoki:2013ldr}.
We compute $\VusTauKpi = \htuse{Vus_tauKpi}$,
$\htuse{Vus_tauKpi_mism_sigma_abs}\sigma$ below the CKM unitarity prediction.

We proceed like in 2012 also to determine \Vus from the branching fraction
$\BFtautoknu$ using
\begin{align*}
  \BR(\tau^- \to K^-\nu_\tau) =
  \frac{G^2_F f^2_K \Vus^2 m^3_{\tau} \tau_{\tau}}{16\pi\hbar} \left (1 - \frac{m_K^2}{m_\tau^2} \right )^2 S_{EW}~.
\end{align*}
We use $f_K = \htuse{f_K}\,\mev$ from FLAG 2013 with
$N_f=2+1$~\cite{Aoki:2013ldr}. We obtain $\VusTauKnu = \htuse{Vus_tauKnu}$,
which is $\htuse{Vus_tauKnu_mism_sigma_abs}\sigma$ below
the CKM unitarity prediction. CODATA 2010 results~\cite{Mohr:2012tt} and
PDG 2013 have been used for the physics constants.

\tausubsection{\Vus from \mtau summary}

\begin{figure}[tb]
  \begin{center}
   \ifhevea
    \begin{tabular}{@{}cc@{}}
      \larger\bfseries\ahref{hfag-tau-vus-plot.png}{PNG format} &
      \larger\bfseries\ahref{hfag-tau-vus-plot.pdf}{PDF format} \\
      \multicolumn{2}{c}{\ahref{hfag-tau-vus-plot.png}{%
          \imgsrc[alt="Vus summary plot"]{hfag-tau-vus-plot.png}}}
    \end{tabular}
    \else
    \includegraphics[width=0.75\linewidth,clip]{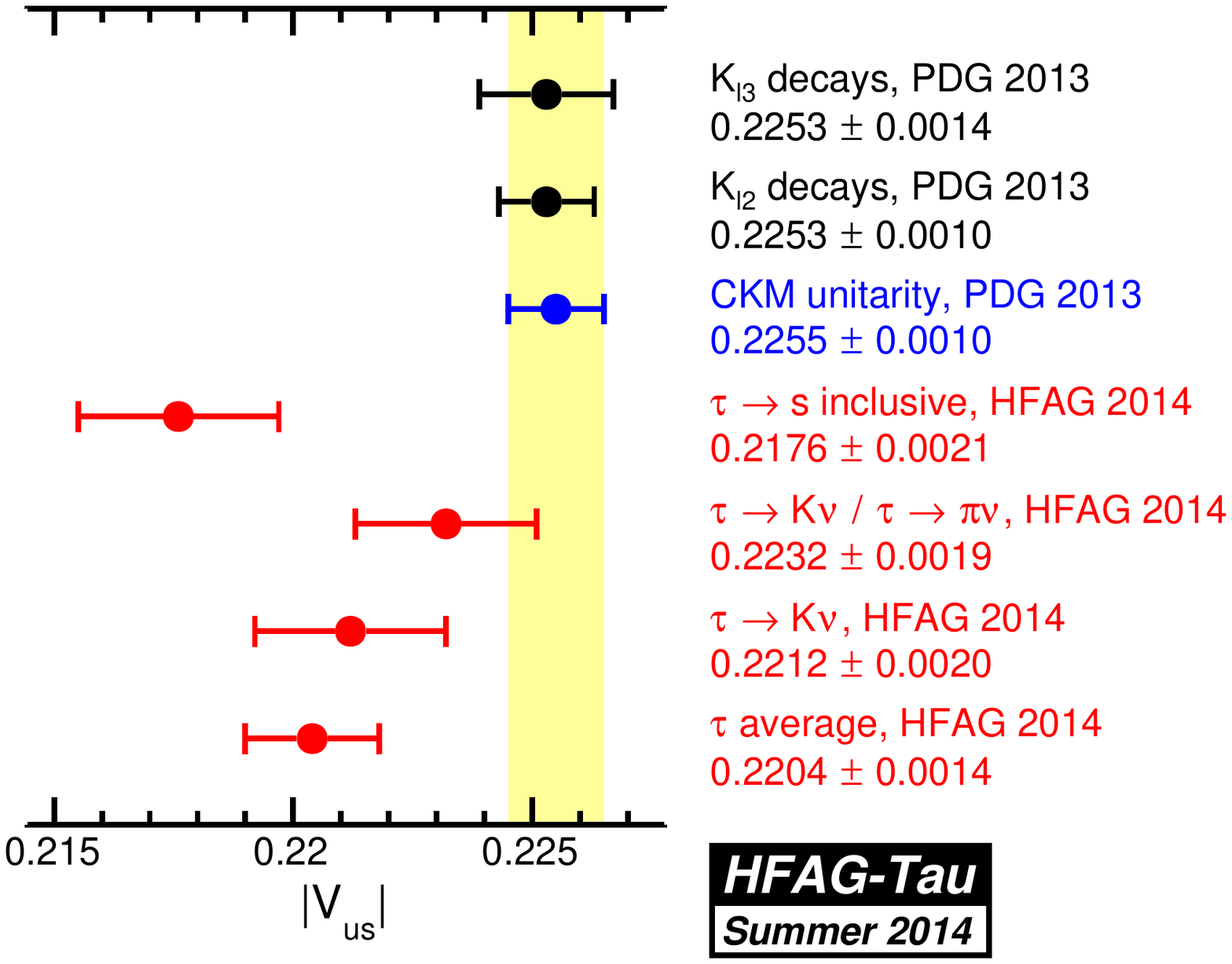}
    \fi
    \caption{\Vus averages of this document compared with the FlaviaNet results~\cite{Antonelli:2010yf}.
      \label{fig:tau:vus-summary}%
    }
  \end{center}
\end{figure}

We summarize the \Vus results reporting the values, the discrepancy with
respect to the \Vus determination from CKM unitarity, and an illustration
of the measurement method:
\begin{alignat*}{6}
  &\VusUni &&= \htuse{Vus_uni.v} &&\pm \htuse{Vus_uni.e} & \quad & & \quad
  & {\smaller\text{from } \sqrt{1 - \Vud^2} \quad\text{(CKM unitarity)}}~, \\
  &\VusTauIncl &&= \htuse{Vus.v} &&\pm \htuse{Vus.e} & \quad & \htuse{Vus_mism_sigma}\sigma &
  & {\smaller\text{from } \Gamma(\tau^- \to X_s^- \nut)}~, \\
  &\VusTauKpi &&= \htuse{Vus_tauKpi.v} &&\pm \htuse{Vus_tauKpi.e} & \quad & \htuse{Vus_tauKpi_mism_sigma}\sigma &
  & {\smaller\text{from } \Gamma(\tauknu)/\Gamma(\taupinu)}~,  \\
  &\VusTauKnu &&= \htuse{Vus_tauKnu.v} &&\pm \htuse{Vus_tauKnu.e} & \quad & \htuse{Vus_tauKnu_mism_sigma}\sigma &
  & {\smaller\text{from } \Gamma(\tauknu)}~.
\end{alignat*}

Averaging the three above \Vus determinations (taking into account all
correlations due to the usage of the fitted \mtau branching fractions and
the other mentioned inputs) we obtain:
\begin{alignat*}{6}
  & \Vus_\tau &&= \htuse{Vus_tau} &\quad &\htuse{Vus_tau_mism_sigma}\sigma \quad
  & \text{average of 3 \Vus \mtau measurements}.
\end{alignat*}
We could not find a published estimate of the correlation of the
uncertainties on $f_K$ and $f_K/f_\pi$, but even if we assume $\pm
100\%$ correlation, the uncertainty on $\Vus_\tau$ does not change
more than about $\pm 5\%$. Figure~\ref{fig:tau:vus-summary} summarizes the
\Vus results.

\tausection{Upper limits on \mtau LFV branching fractions}
\cutname{lfv-limits.html}
\label{sec:tau:lfv}
We report in Table~\ref{tab:tau:lfv-upper-limits} and
Figure~\ref{fig:tau:lfv-limits-plot} the up-to-date upper
limits on the \mtau LFV branching fractions in order to track and
summarize the experimental results in this area, which are sensitive
to physics beyond the Standard Model.
\begin{center}
\begin{longtable}{lcl@{}rll}
\caption{Experimental upper limits on lepton flavor violating \mtau
  decays. The modes are grouped according to the particle content of their final
  states. Modes with lepton number violation are labeled with ``(L)'',
  modes with baryon number violation are labeled with ``(BNV)''.
  \label{tab:tau:lfv-upper-limits}}%
\\
\toprule
\multicolumn{1}{l}{\bfseries Decay mode} &
\multicolumn{1}{l}{\bfseries Category} &
\multicolumn{2}{c}{\bfseries \begin{tabular}{@{}c@{}}90\% CL\\Limit\end{tabular}} &
\multicolumn{1}{l}{\bfseries Exp.} &
\multicolumn{1}{l}{\bfseries Ref.} \\
\midrule
\endfirsthead
\multicolumn{6}{c}{{\bfseries \tablename\ \thetable{} -- continued from previous page}} \\ \midrule
\multicolumn{1}{l}{\bfseries Decay mode} &
\multicolumn{1}{l}{\bfseries Category} &
\multicolumn{2}{c}{\bfseries \begin{tabular}{@{}c@{}}90\% CL\\Limit\end{tabular}} &
\multicolumn{1}{l}{\bfseries Exp.} &
\multicolumn{1}{l}{\bfseries Ref.} \\
\midrule
\endhead
%
%
\begin{ensuredisplaymath}
\Gamma_{156} =  {e^- \gamma} 
\end{ensuredisplaymath}
 &\(l\gamma\) & \( <\; \) & \(12.0 \cdot 10^{-8}\)         & Belle &  \cite{Hayasaka:2007vc} \\
 &            & \( <\; \) & \(3.3 \cdot 10^{-8}\)         & \babar &  \cite{Aubert:2009ag}   \\ 
\begin{ensuredisplaymath}
\Gamma_{157} =  {\mu^- \gamma} 
\end{ensuredisplaymath}
 &            & \( <\; \) & \(4.5 \cdot 10^{-8}\)         & Belle &  \cite{Hayasaka:2007vc} \\
 &            & \( <\; \) & \(4.4 \cdot 10^{-8}\)         & \babar &  \cite{Aubert:2009ag}   \\ 
\midrule
%
%
\begin{ensuredisplaymath}
\Gamma_{158} =  {e^- \pi^0} 
\end{ensuredisplaymath}
 &\(lP^0 \)   & \( <\; \) & \(2.2 \cdot 10^{-8}\)         & Belle & \cite{Hayasaka:2011zz} \\
 &            & \( <\; \) & \(13.0 \cdot 10^{-8}\)         & \babar & \cite{Aubert:2006cz} \\ 
\begin{ensuredisplaymath}
\Gamma_{159} =  {\mu^- \pi^0} 
\end{ensuredisplaymath}
 &            & \( <\; \) & \(2.7 \cdot 10^{-8}\)         & Belle &  \cite{Hayasaka:2011zz}  \\
 &            & \( <\; \) & \(11.0 \cdot 10^{-8}\)         & \babar &  \cite{Aubert:2006cz} \\ 
\begin{ensuredisplaymath}
\Gamma_{162} =  {e^- \eta} 
\end{ensuredisplaymath}
 &            & \( <\; \) & \(4.4 \cdot 10^{-8}\)         & Belle &  \cite{Hayasaka:2011zz}  \\
 &            & \( <\; \) & \(16.0 \cdot 10^{-8}\)         & \babar &  \cite{Aubert:2006cz} \\ 
\begin{ensuredisplaymath}
\Gamma_{163} =  {\mu^- \eta} 
\end{ensuredisplaymath}
 &            & \( <\; \) & \(2.3 \cdot 10^{-8}\)         & Belle &   \cite{Hayasaka:2011zz} \\
 &            & \( <\; \) & \(15.0 \cdot 10^{-8}\)         & \babar &   \cite{Aubert:2006cz} \\ 
\begin{ensuredisplaymath}
\Gamma_{172} =  {e^- \eta'(958)} 
\end{ensuredisplaymath}
 &            & \( <\; \) & \(3.6 \cdot 10^{-8}\)         & Belle &   \cite{Hayasaka:2011zz}  \\
 &            & \( <\; \) & \(24.0 \cdot 10^{-8}\)         & \babar &   \cite{Aubert:2006cz} \\ 
\begin{ensuredisplaymath}
\Gamma_{173} =  {\mu^- \eta'(958)} 
\end{ensuredisplaymath}
 &            & \( <\; \) & \(3.8 \cdot 10^{-8}\)         & Belle &   \cite{Hayasaka:2011zz}  \\
 &            & \( <\; \) & \(14.0 \cdot 10^{-8}\)         & \babar &   \cite{Aubert:2006cz} \\ 
\begin{ensuredisplaymath}
\Gamma_{160} =  {e^- K^0_S} 
\end{ensuredisplaymath}
 &            & \( <\; \) & \(2.6 \cdot 10^{-8}\)         & Belle &  \cite{Miyazaki:2010qb} \\
 &            & \( <\; \) & \(3.3 \cdot 10^{-8}\)         & \babar &  \cite{Aubert:2009ys}   \\ 
\begin{ensuredisplaymath}
\Gamma_{161} =  {\mu^- K^0_S} 
\end{ensuredisplaymath}
 &            & \( <\; \) & \(2.3 \cdot 10^{-8}\)         & Belle &   \cite{Miyazaki:2010qb} \\
 &            & \( <\; \) & \(4.0 \cdot 10^{-8}\)         & \babar &   \cite{Aubert:2009ys}   \\ 
\midrule
%
%
\begin{ensuredisplaymath}
\Gamma_{174} =  {e^- f_0(980)} 
\end{ensuredisplaymath}
 &  \(l S^0\) & \( <\; \) & \(3.2 \cdot 10^{-8}\)         & Belle & \cite{Miyazaki:2008mw}\\
\begin{ensuredisplaymath}
\Gamma_{175} =  {\mu^- f_0(980)} 
\end{ensuredisplaymath}
 &            & \( <\; \) & \(3.4 \cdot 10^{-8}\)         & Belle & \cite{Miyazaki:2008mw}\\  
\midrule
%
%
\begin{ensuredisplaymath}
\Gamma_{164} =  {e^- \rho^0} 
\end{ensuredisplaymath}
 &  \(l V^0\) & \( <\; \) & \(1.8 \cdot 10^{-8}\)         & Belle &  \cite{Miyazaki:2011xe}\\
 &            & \( <\; \) & \(4.6 \cdot 10^{-8}\)         & \babar &  \cite{Aubert:2009ap}  \\ 
\begin{ensuredisplaymath}
\Gamma_{165} =  {\mu^- \rho^0} 
\end{ensuredisplaymath}
 &            & \( <\; \) & \(1.2 \cdot 10^{-8}\)         & Belle &  \cite{Miyazaki:2011xe}\\
 &            & \( <\; \) & \(2.6 \cdot 10^{-8}\)         & \babar &  \cite{Aubert:2009ap}  \\ 
\begin{ensuredisplaymath}
\Gamma_{168} =  {e^- K^*(892)^0} 
\end{ensuredisplaymath}
 &            & \( <\; \) & \(3.2 \cdot 10^{-8}\)         & Belle &  \cite{Miyazaki:2011xe} \\
 &            & \( <\; \) & \(5.9 \cdot 10^{-8}\)         & \babar &  \cite{Aubert:2009ap}   \\ 
\begin{ensuredisplaymath}
\Gamma_{169} =  {\mu^- K^*(892)^0} 
\end{ensuredisplaymath}
 &            & \( <\; \) & \(7.2 \cdot 10^{-8}\)         & Belle &   \cite{Miyazaki:2011xe} \\
 &            & \( <\; \) & \(17.0 \cdot 10^{-8}\)         & \babar &   \cite{Aubert:2009ap}   \\ 
\begin{ensuredisplaymath}
\Gamma_{170} =  {e^- \bar{K}^*(892)^0} 
\end{ensuredisplaymath}
 &            & \( <\; \) & \(3.4 \cdot 10^{-8}\)         & Belle &   \cite{Miyazaki:2011xe} \\
 &            & \( <\; \) & \(4.6 \cdot 10^{-8}\)         & \babar &   \cite{Aubert:2009ap}   \\ 
\begin{ensuredisplaymath}
\Gamma_{171} =  {\mu^- \bar{K}^*(892)^0} 
\end{ensuredisplaymath}
 &            & \( <\; \) & \(7.0 \cdot 10^{-8}\)         & Belle &  \cite{Miyazaki:2011xe} \\
 &            & \( <\; \) & \(7.3 \cdot 10^{-8}\)         & \babar &  \cite{Aubert:2009ap}   \\ 

\begin{ensuredisplaymath}
\Gamma_{176} =  {e^- \phi} 
\end{ensuredisplaymath}
 &            & \( <\; \) & \(3.1 \cdot 10^{-8}\)         & Belle &   \cite{Miyazaki:2011xe} \\
 &            & \( <\; \) & \(3.1 \cdot 10^{-8}\)         & \babar &   \cite{Aubert:2009ap}   \\ 
\begin{ensuredisplaymath}
\Gamma_{177} =  {\mu^- \phi} 
\end{ensuredisplaymath}
 &            & \( <\; \) & \(8.4 \cdot 10^{-8}\)         & Belle &   \cite{Miyazaki:2011xe} \\
 &            & \( <\; \) & \(19.0 \cdot 10^{-8}\)         & \babar &   \cite{Aubert:2009ap}   \\ 
\begin{ensuredisplaymath}
\Gamma_{166} =  {e^- \omega} 
\end{ensuredisplaymath}
 &            & \( <\; \) & \(4.8 \cdot 10^{-8}\)         & Belle &  \cite{Miyazaki:2011xe} \\
 &            & \( <\; \) & \(11.0 \cdot 10^{-8}\)         & \babar &  \cite{Aubert:2007kx}   \\ 
\begin{ensuredisplaymath}
\Gamma_{167} =  {\mu^- \omega} 
\end{ensuredisplaymath}
 &            & \( <\; \) & \(4.7 \cdot 10^{-8}\)         & Belle &  \cite{Miyazaki:2011xe} \\
 &            & \( <\; \) & \(10.0 \cdot 10^{-8}\)         & \babar &  \cite{Aubert:2007kx}   \\ 
\midrule
%
%
\begin{ensuredisplaymath}
\Gamma_{178} =  {e^- e^+ e^-} 
\end{ensuredisplaymath}
 &  \(lll\)   & \( <\; \) & \(2.7 \cdot 10^{-8}\)         & Belle & \cite{Hayasaka:2010np} \\
 &            & \( <\; \) & \(2.9 \cdot 10^{-8}\)         & \babar & \cite{Lees:2010ez}     \\ 
\begin{ensuredisplaymath}
\Gamma_{181} =  {\mu^- e^+ e^-} 
\end{ensuredisplaymath}
 &            & \( <\; \) & \(1.8 \cdot 10^{-8}\)         & Belle & \cite{Hayasaka:2010np} \\
 &            & \( <\; \) & \(2.2 \cdot 10^{-8}\)         & \babar & \cite{Lees:2010ez}     \\ 
\begin{ensuredisplaymath}
\Gamma_{179} =  {e^- \mu^+ \mu^-} 
\end{ensuredisplaymath}
 &            & \( <\; \) & \(2.7 \cdot 10^{-8}\)         & Belle & \cite{Hayasaka:2010np} \\
 &            & \( <\; \) & \(3.2 \cdot 10^{-8}\)         & \babar & \cite{Lees:2010ez}     \\ 
\begin{ensuredisplaymath}
\Gamma_{183} =  {\mu^- \mu^+ \mu^-} 
\end{ensuredisplaymath}
 &            & \( <\; \) & \(2.1 \cdot 10^{-8}\)         & Belle & \cite{Hayasaka:2010np} \\
 &            & \( <\; \) & \(3.3 \cdot 10^{-8}\)         & \babar &
  \cite{Lees:2010ez}     \\ 
 &            & \( <\; \) & \(4.6 \cdot 10^{-8}\)         & \lhcb &
  \cite{Aaij:2014azz}     \\ 

\begin{ensuredisplaymath}
\Gamma_{182} =  {e^- \mu^+ e^-} 
\end{ensuredisplaymath}
 &            & \( <\; \) & \(1.5 \cdot 10^{-8}\)         & Belle & \cite{Hayasaka:2010np} \\
 &            & \( <\; \) & \(1.8 \cdot 10^{-8}\)         & \babar & \cite{Lees:2010ez}     \\ 
\begin{ensuredisplaymath}
\Gamma_{180} =  {\mu^- e^+ \mu^-} 
\end{ensuredisplaymath}
 &            & \( <\; \) & \(1.7 \cdot 10^{-8}\)         & Belle & \cite{Hayasaka:2010np} \\
 &            & \( <\; \) & \(2.6 \cdot 10^{-8}\)         & \babar & \cite{Lees:2010ez}     \\ 
\midrule
%
%
\begin{ensuredisplaymath}
\Gamma_{184} =  {e^- \pi^+ \pi^-} 
\end{ensuredisplaymath}
 &    \(lhh\) & \( <\; \) & \(2.3 \cdot 10^{-8}\)         & Belle &  \cite{Miyazaki:2012mx}\\
 &            & \( <\; \) & \(12.0 \cdot 10^{-8}\)         & \babar &  \cite{Aubert:2005tp}  \\ 
\begin{ensuredisplaymath}
\Gamma_{186} =  {\mu^- \pi^+  \pi^-} 
\end{ensuredisplaymath}
 &            & \( <\; \) & \(2.1 \cdot 10^{-8}\)         & Belle &  \cite{Miyazaki:2012mx} \\
 &            & \( <\; \) & \(29.0 \cdot 10^{-8}\)         & \babar &  \cite{Aubert:2005tp}   \\ 
\begin{ensuredisplaymath}
\Gamma_{188} =  {e^- \pi^+ K^-} 
\end{ensuredisplaymath}
 &            & \( <\; \) & \(3.7 \cdot 10^{-8}\)         & Belle &  \cite{Miyazaki:2012mx} \\
 &            & \( <\; \) & \(32.0 \cdot 10^{-8}\)         & \babar &  \cite{Aubert:2005tp}   \\ 
\begin{ensuredisplaymath}
\Gamma_{194} =  {\mu^- \pi^+  K^-} 
\end{ensuredisplaymath}
 &            & \( <\; \) & \(8.6 \cdot 10^{-8}\)         & Belle &   \cite{Miyazaki:2012mx} \\
 &            & \( <\; \) & \(26.0 \cdot 10^{-8}\)         & \babar &   \cite{Aubert:2005tp}   \\ 
\begin{ensuredisplaymath}
\Gamma_{189} =  {e^- K^+ \pi^-} 
\end{ensuredisplaymath}
 &            & \( <\; \) & \(3.1 \cdot 10^{-8}\)         & Belle &   \cite{Miyazaki:2012mx} \\
 &            & \( <\; \) & \(17.0 \cdot 10^{-8}\)         & \babar &   \cite{Aubert:2005tp}   \\ 
\begin{ensuredisplaymath}
\Gamma_{195} =  {\mu^- K^+  \pi^-} 
\end{ensuredisplaymath}
 &            & \( <\; \) & \(4.5 \cdot 10^{-8}\)         & Belle &   \cite{Miyazaki:2012mx} \\
 &            & \( <\; \) & \(32.0 \cdot 10^{-8}\)         & \babar &   \cite{Aubert:2005tp}   \\ 

\begin{ensuredisplaymath}
\Gamma_{192} =  {e^- K^+ K^-} 
\end{ensuredisplaymath}
 &            & \( <\; \) & \(3.4 \cdot 10^{-8}\)         & Belle &   \cite{Miyazaki:2012mx} \\
 &            & \( <\; \) & \(14.0 \cdot 10^{-8}\)         & \babar &   \cite{Aubert:2005tp}   \\ 
\begin{ensuredisplaymath}
\Gamma_{198} =  {\mu^- K^+  K^-} 
\end{ensuredisplaymath}
 &            & \( <\; \) & \(4.4 \cdot 10^{-8}\)         & Belle &   \cite{Miyazaki:2012mx} \\
 &            & \( <\; \) & \(25.0 \cdot 10^{-8}\)         & \babar &   \cite{Aubert:2005tp}   \\ 
\begin{ensuredisplaymath}
\Gamma_{191} =  {e^- K^0_S K^0_S} 
\end{ensuredisplaymath}
 &            & \( <\; \) & \(7.1 \cdot 10^{-8}\)         & Belle &  \cite{Miyazaki:2010qb}  \\
\begin{ensuredisplaymath}
\Gamma_{197} =  {\mu^- K^0_S  K^0_S} 
\end{ensuredisplaymath}
 &            & \( <\; \) & \(8.0 \cdot 10^{-8}\)         & Belle &   \cite{Miyazaki:2010qb} \\
%
%
\begin{ensuredisplaymath}
\Gamma_{185} =  {e^+ \pi^- \pi^- } 
\end{ensuredisplaymath}
 & (L)           & \( <\; \) & \(2.0 \cdot 10^{-8}\)         & Belle & \cite{Miyazaki:2012mx} \\
 & (L)           & \( <\; \) & \(27.0 \cdot 10^{-8}\)         & \babar & \cite{Aubert:2005tp}   \\ 
\begin{ensuredisplaymath}
\Gamma_{187} =  {\mu^+ \pi^- \pi^-} 
\end{ensuredisplaymath}
 & (L)        & \( <\; \) & \(3.9 \cdot 10^{-8}\)         & Belle &   \cite{Miyazaki:2012mx} \\
 & (L)        & \( <\; \) & \(7.0 \cdot 10^{-8}\)         & \babar &   \cite{Aubert:2005tp}   \\ 
\begin{ensuredisplaymath}
\Gamma_{190} =  {e^+ \pi^- K^- } 
\end{ensuredisplaymath}
 & (L)        & \( <\; \) & \(3.2 \cdot 10^{-8}\)         & Belle &   \cite{Miyazaki:2012mx} \\
 & (L)        & \( <\; \) & \(18.0\cdot 10^{-8}\)         & \babar &   \cite{Aubert:2005tp}   \\ 
\begin{ensuredisplaymath}
\Gamma_{196} =  {\mu^+ \pi^- K^-} 
\end{ensuredisplaymath}
 & (L)        & \( <\; \) & \(4.8 \cdot 10^{-8}\)         & Belle &   \cite{Miyazaki:2012mx} \\
 & (L)        & \( <\; \) & \(22.0 \cdot 10^{-8}\)         & \babar &   \cite{Aubert:2005tp}   \\ 
\begin{ensuredisplaymath}
\Gamma_{193} =  {e^+ K^- K^- } 
\end{ensuredisplaymath}
 &  (L)       & \( <\; \) & \(3.3 \cdot 10^{-8}\)         & Belle &   \cite{Miyazaki:2012mx} \\
 &  (L)       & \( <\; \) & \(15.0 \cdot 10^{-8}\)         & \babar &   \cite{Aubert:2005tp}   \\ 
\begin{ensuredisplaymath}
\Gamma_{199} =  {\mu^+ K^- K^-} 
\end{ensuredisplaymath}
 &  (L)       & \( <\; \) & \(4.7 \cdot 10^{-8}\)         & Belle &   \cite{Miyazaki:2012mx} \\
 &  (L)       & \( <\; \) & \(48.0 \cdot 10^{-8}\)         & \babar &   \cite{Aubert:2005tp}   \\ 
\midrule
%
%
\begin{ensuredisplaymath}
\Gamma_{211} =  { \pi^- \Lambda } 
\end{ensuredisplaymath}
 & BNV & \( <\; \) & \(3.0 \cdot 10^{-8}\)         & Belle & \cite{Hayasaka:2012pj}  \\
 &               & \( <\; \) & \(5.8 \cdot 10^{-8}\)         & \babar &  \cite{Lafferty:2007zz}  \\ 
\begin{ensuredisplaymath}
\Gamma_{212} =  { \pi^- \bar{\Lambda}} 
\end{ensuredisplaymath}
 &            & \( <\; \) & \(2.8 \cdot 10^{-8}\)         & Belle & \cite{Hayasaka:2012pj}  \\
 &            & \( <\; \) & \(5.9 \cdot 10^{-8}\)         & \babar &  \cite{Lafferty:2007zz}  \\ 
\begin{ensuredisplaymath}
\Gamma_{213} =  { K^- \Lambda } 
\end{ensuredisplaymath}
 &            & \( <\; \) & \(4.2 \cdot 10^{-8}\)         & Belle &  \cite{Hayasaka:2012pj} \\
 &            & \( <\; \) & \(15.\cdot 10^{-8}\)         & \babar &  \cite{Lafferty:2007zz} \\ 
\begin{ensuredisplaymath}
\Gamma_{214} =  { K^- \bar{\Lambda}} 
\end{ensuredisplaymath}
 &            & \( <\; \) & \(3.1 \cdot 10^{-8}\)         & Belle & \cite{Hayasaka:2012pj}  \\
 &            & \( <\; \) & \(7.2 \cdot 10^{-8}\)         & \babar & \cite{Lafferty:2007zz}  \\ 
 \begin{ensuredisplaymath}
\Gamma_{215} =  { \proton \mu^- \mu^-} 
\end{ensuredisplaymath}
&            & \( <\; \) & \(44.0 \cdot 10^{-8}\)         & \lhcb & \cite{Aaij:2013fia}  \\
 \begin{ensuredisplaymath}
\Gamma_{216} =  { \bar{\proton} \mu^+ \mu^-} 
\end{ensuredisplaymath}
&            & \( <\; \) & \(33.0 \cdot 10^{-8}\)         & \lhcb & \cite{Aaij:2013fia}  \\
\bottomrule
\end{longtable}
\end{center}

\ifhevea
\tausection{Upper Limits on \mtau LFV Branching Fractions: Summary Plot}
\cutname{lfv-limits-plot.html}
\fi

\begin{figure}[tb]
  \begin{center}
    \ifhevea
    \begin{tabular}{@{}cc@{}}
      \larger\bfseries\ahref{TauLFV_UL_2014001.png}{full size PNG} &
      \larger\bfseries\ahref{TauLFV_UL_2014001.pdf}{PDF format} \\
      \multicolumn{2}{c}{\ahref{TauLFV_UL_2014001.png}{%
          \imgsrc[alt="Tau LFV limits combinations plot" width=720]{TauLFV_UL_2014001.png}}}
    \end{tabular}
    \else
    \includegraphics[angle=270,totalheight=0.9\textheight,clip]{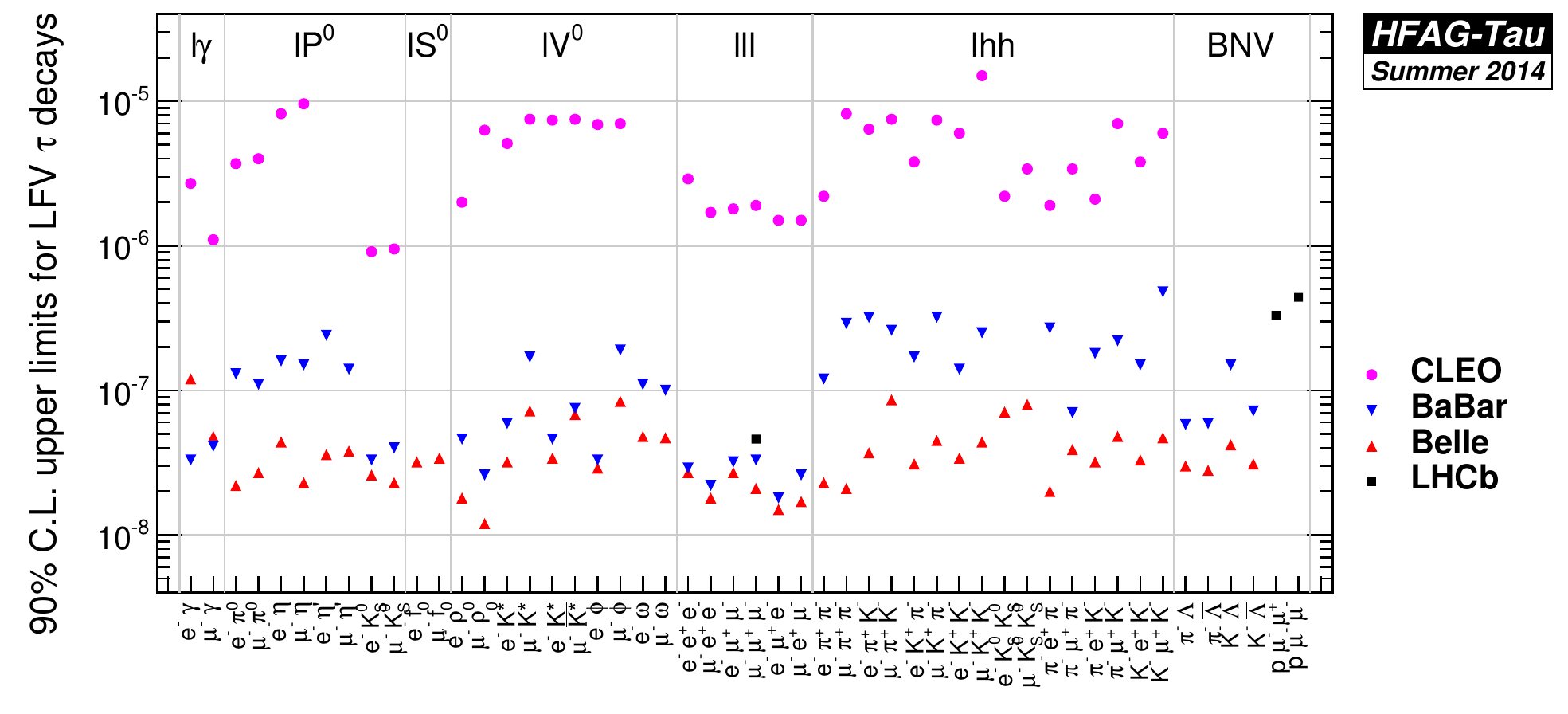}
    \fi
    \caption{Tau lepton-flavor-violating branching fraction upper
      limits summary plot.
      \label{fig:tau:lfv-limits-plot}
    }
  \end{center}
\end{figure}

\tausection{Combination of upper limits on \mtau LFV branching fractions}
\cutname{lfv-combinations.html}
\label{sec:tau:lfv-comb}
\newcommand{\cls}{\ensuremath{\text{CL}_s}\xspace}
\newcommand{\clsb}{\ensuremath{\text{CL}_{s+b}}\xspace}
\newcommand{\clb}{\ensuremath{\text{CL}_b}\xspace}

Combining upper limits is a delicate issue, since there is no standard and
generally agreed procedure. Furthermore, the \mtau LFV searches
published limits are extracted from the data with a variety of
methods, and cannot be directly combined with a uniform procedure. It
is however possible to use a single and effective
upper limits combination procedure for all modes by re-computing the published upper
limits with just one extraction method, using the published
information that documents the upper limit determination:
number of observed candidates, expected background, signal efficiency and
number of analyzed \mtau decays.

We chose to use the \cls method~\cite{Mistlberger:2012rs} to re-compute the
\mtau LFV upper limits, since it is well known and widely used (see the
Statistics review of PDG 2013~\cite{PDG_2012}), and since the
limits computed with the \cls method can be combined in a straightforward
way (see below). The \cls method is based on two hypotheses: signal plus background and
background only. We calculate the observed confidence levels for the two
hypotheses:
\begin{align}
&\clsb = P_{s+b}(Q \leq Q_{obs}) = \int_{- \infty}^{Q_{obs}} \frac{dP_{s+b}}{dQ} dQ,
\label{eq:tau:clspdf1} \\
&\clb = P_{b}(Q \leq Q_{obs}) = \int_{- \infty}^{Q_{obs}} \frac{dP_{b}}{dQ} dQ,
\label{eq:tau:clspdf2}
\end{align}
where \clsb is the confidence level observed for the signal plus background
hypothesis, \clb is the confidence level observed for the background only
hypothesis, $\frac{dP_{s+b}}{dQ}$ and $\frac{dP_{b}}{dQ}$ are the probability
distribution functions (PDFs) for the two corresponding hypothesis and
$Q$ is called the test statistics. The \cls value is defined as the ratio
between the confidence level for the signal plus background hypothesis to
the confidence level for the background hypothesis:
\begin{align}
\cls = \dfrac{\clsb}{\clb}.
\end{align}
When multiple results are combined, the PDFs in
Equations~\ref{eq:tau:clspdf1} and \ref{eq:tau:clspdf2} are the
product of the individual PDFs,
\begin{align}
\cls = \dfrac{\prod_{i=1}^{N_{\text{chan}}}\sum_{n=0}^{n_i} \dfrac{e^{-(s_i+b_i)} (s_i+b_i)^{n}}{n!} }{\prod_{i=1}^{n_{\text{chan}}}  \sum_{n=0}^{n_i} \dfrac{e^{-b_i} b_i^{n}}{n!}}    \dfrac{\prod_{j=1}^{n} s_iS_i(x_{ij})+b_iB_i(x_{ij})}{\prod_{j=1}^{n_i}B_i(x_{ij})}~,
\end{align}
where $N_{\text{chan}}$ is the number of results (or channels), and, for each channel $i$,
$n_i$ is the number of observed candidates, $x_{ij}$ are the values of the
discriminating variables (with index $j$), $s_i$ and $b_i$ are the number
of signal and background events and $S_i$, $B_i$ are the probability
distribution functions of the discriminating variables.
The expected signal $s_i$ is related to the \mtau lepton branching
fraction $\BR(\tau \rightarrow f_i)$ into
the searched final state $f_i$ by $s_i = N_i\epsilon_i\BR(\tau \rightarrow
f_i)$, where $N_i$ is the number of produced \mtau leptons and
$\epsilon_i$ is the detection efficiency for observing the decay $\tau\to
f_i$. For $e^+ e^-$ experiments,
$N_i = 2\mathcal{L}_i\sigma_{\tau\tau}$, where $\mathcal{L}_i$ is the
integrated luminosity and $\sigma_{\tau\tau}$ is the
\mtau pair production cross section $\sigma(e^+ e^- \rightarrow \tau^+
\tau^-)$~\cite{Banerjee:2007is}.
In experiments where \mtau leptons are produced in more complex multiple
reactions, the effective $N_i$ is typically estimated with Monte Carlo simulations
calibrated with related data yields.

The extraction of the upper limits is performed using the code provided by
Tom Junk~\cite{junk:2007:cdfnote}. The systematic uncertainties are modeled
in the Monte Carlo toy experiments by convolving the $S_i$ and $B_i$
PDFs with with Gaussian distributions corresponding to the nuisance
parameters. 

Table~\ref{tab:tau:lfv-upper-limits-comb} reports the re-computed limits
for the $B$ factories results as well as the corresponding HFAG
combination. Since there is negligible gain in combining limits of very
different strength, the combinations do not include the CLEO searches and we do not
combine results for modes where the best limit is more than an order of
magnitude better than the other limits.
Figure~\ref{fig:tau:lfv-limits-plot-average} reports the re-computed \mtau
LFV searches upper limits and their combination.

\htdef{g156.babar.lumi}{524}%
\htdef{g156.babar.xsec}{0.919}%
\htdef{g156.babar.eff}{$3.90 \pm 0.30$}%
\htdef{g156.babar.bkg}{$1.60 \pm 0.40$}%
\htdef{g156.babar.evs}{0}%
\htdef{g156.babar.row}{\htuse{g156.babar.lumi} & \htuse{g156.babar.xsec} & \htuse{g156.babar.eff} & \htuse{g156.babar.bkg} & \htuse{g156.babar.evs}}%
\htdef{g171.babar.lumi}{451}%
\htdef{g171.babar.xsec}{0.919}%
\htdef{g171.babar.eff}{$4.10 \pm 0.30$}%
\htdef{g171.babar.bkg}{$1.72 \pm 0.17$}%
\htdef{g171.babar.evs}{1}%
\htdef{g171.babar.row}{\htuse{g171.babar.lumi} & \htuse{g171.babar.xsec} & \htuse{g171.babar.eff} & \htuse{g171.babar.bkg} & \htuse{g171.babar.evs}}%
\htdef{g176.babar.lumi}{451}%
\htdef{g176.babar.xsec}{0.919}%
\htdef{g176.babar.eff}{$6.40 \pm 0.20$}%
\htdef{g176.babar.bkg}{$0.68 \pm 0.12$}%
\htdef{g176.babar.evs}{0}%
\htdef{g176.babar.row}{\htuse{g176.babar.lumi} & \htuse{g176.babar.xsec} & \htuse{g176.babar.eff} & \htuse{g176.babar.bkg} & \htuse{g176.babar.evs}}%
\htdef{g177.babar.lumi}{451}%
\htdef{g177.babar.xsec}{0.919}%
\htdef{g177.babar.eff}{$5.20 \pm 0.30$}%
\htdef{g177.babar.bkg}{$2.76 \pm 0.16$}%
\htdef{g177.babar.evs}{6}%
\htdef{g177.babar.row}{\htuse{g177.babar.lumi} & \htuse{g177.babar.xsec} & \htuse{g177.babar.eff} & \htuse{g177.babar.bkg} & \htuse{g177.babar.evs}}%
\htdef{g166.babar.lumi}{451}%
\htdef{g166.babar.xsec}{0.919}%
\htdef{g166.babar.eff}{$2.96 \pm 0.13$}%
\htdef{g166.babar.bkg}{$0.35 \pm 0.06$}%
\htdef{g166.babar.evs}{0}%
\htdef{g166.babar.row}{\htuse{g166.babar.lumi} & \htuse{g166.babar.xsec} & \htuse{g166.babar.eff} & \htuse{g166.babar.bkg} & \htuse{g166.babar.evs}}%
\htdef{g167.babar.lumi}{451}%
\htdef{g167.babar.xsec}{0.919}%
\htdef{g167.babar.eff}{$2.56 \pm 0.16$}%
\htdef{g167.babar.bkg}{$0.73 \pm 0.03$}%
\htdef{g167.babar.evs}{0}%
\htdef{g167.babar.row}{\htuse{g167.babar.lumi} & \htuse{g167.babar.xsec} & \htuse{g167.babar.eff} & \htuse{g167.babar.bkg} & \htuse{g167.babar.evs}}%
\htdef{g164.belle.lumi}{854}%
\htdef{g164.belle.xsec}{0.919}%
\htdef{g164.belle.eff}{$7.58 \pm 0.41$}%
\htdef{g164.belle.bkg}{$0.29 \pm 0.15$}%
\htdef{g164.belle.evs}{0}%
\htdef{g164.belle.row}{\htuse{g164.belle.lumi} & \htuse{g164.belle.xsec} & \htuse{g164.belle.eff} & \htuse{g164.belle.bkg} & \htuse{g164.belle.evs}}%
\htdef{g165.belle.lumi}{854}%
\htdef{g165.belle.xsec}{0.919}%
\htdef{g165.belle.eff}{$7.09 \pm 0.37$}%
\htdef{g165.belle.bkg}{$1.48 \pm 0.35$}%
\htdef{g165.belle.evs}{0}%
\htdef{g165.belle.row}{\htuse{g165.belle.lumi} & \htuse{g165.belle.xsec} & \htuse{g165.belle.eff} & \htuse{g165.belle.bkg} & \htuse{g165.belle.evs}}%
\htdef{g168.belle.lumi}{854}%
\htdef{g168.belle.xsec}{0.919}%
\htdef{g168.belle.eff}{$4.37 \pm 0.24$}%
\htdef{g168.belle.bkg}{$0.29 \pm 0.14$}%
\htdef{g168.belle.evs}{0}%
\htdef{g168.belle.row}{\htuse{g168.belle.lumi} & \htuse{g168.belle.xsec} & \htuse{g168.belle.eff} & \htuse{g168.belle.bkg} & \htuse{g168.belle.evs}}%
\htdef{g169.belle.lumi}{854}%
\htdef{g169.belle.xsec}{0.919}%
\htdef{g169.belle.eff}{$3.39 \pm 0.19$}%
\htdef{g169.belle.bkg}{$0.53 \pm 0.20$}%
\htdef{g169.belle.evs}{1}%
\htdef{g169.belle.row}{\htuse{g169.belle.lumi} & \htuse{g169.belle.xsec} & \htuse{g169.belle.eff} & \htuse{g169.belle.bkg} & \htuse{g169.belle.evs}}%
\htdef{g170.belle.lumi}{854}%
\htdef{g170.belle.xsec}{0.919}%
\htdef{g170.belle.eff}{$4.41 \pm 0.25$}%
\htdef{g170.belle.bkg}{$0.08 \pm 0.08$}%
\htdef{g170.belle.evs}{0}%
\htdef{g170.belle.row}{\htuse{g170.belle.lumi} & \htuse{g170.belle.xsec} & \htuse{g170.belle.eff} & \htuse{g170.belle.bkg} & \htuse{g170.belle.evs}}%
\htdef{g157.babar.lumi}{524}%
\htdef{g157.babar.xsec}{0.919}%
\htdef{g157.babar.eff}{$6.10 \pm 0.50$}%
\htdef{g157.babar.bkg}{$3.60 \pm 0.70$}%
\htdef{g157.babar.evs}{2}%
\htdef{g157.babar.row}{\htuse{g157.babar.lumi} & \htuse{g157.babar.xsec} & \htuse{g157.babar.eff} & \htuse{g157.babar.bkg} & \htuse{g157.babar.evs}}%
\htdef{g171.belle.lumi}{854}%
\htdef{g171.belle.xsec}{0.919}%
\htdef{g171.belle.eff}{$3.60 \pm 0.20$}%
\htdef{g171.belle.bkg}{$0.45 \pm 0.17$}%
\htdef{g171.belle.evs}{1}%
\htdef{g171.belle.row}{\htuse{g171.belle.lumi} & \htuse{g171.belle.xsec} & \htuse{g171.belle.eff} & \htuse{g171.belle.bkg} & \htuse{g171.belle.evs}}%
\htdef{g176.belle.lumi}{854}%
\htdef{g176.belle.xsec}{0.919}%
\htdef{g176.belle.eff}{$4.18 \pm 0.25$}%
\htdef{g176.belle.bkg}{$0.47 \pm 0.19$}%
\htdef{g176.belle.evs}{0}%
\htdef{g176.belle.row}{\htuse{g176.belle.lumi} & \htuse{g176.belle.xsec} & \htuse{g176.belle.eff} & \htuse{g176.belle.bkg} & \htuse{g176.belle.evs}}%
\htdef{g177.belle.lumi}{854}%
\htdef{g177.belle.xsec}{0.919}%
\htdef{g177.belle.eff}{$3.21 \pm 0.19$}%
\htdef{g177.belle.bkg}{$0.06 \pm 0.06$}%
\htdef{g177.belle.evs}{1}%
\htdef{g177.belle.row}{\htuse{g177.belle.lumi} & \htuse{g177.belle.xsec} & \htuse{g177.belle.eff} & \htuse{g177.belle.bkg} & \htuse{g177.belle.evs}}%
\htdef{g166.belle.lumi}{854}%
\htdef{g166.belle.xsec}{0.919}%
\htdef{g166.belle.eff}{$2.92 \pm 0.18$}%
\htdef{g166.belle.bkg}{$0.30 \pm 0.14$}%
\htdef{g166.belle.evs}{0}%
\htdef{g166.belle.row}{\htuse{g166.belle.lumi} & \htuse{g166.belle.xsec} & \htuse{g166.belle.eff} & \htuse{g166.belle.bkg} & \htuse{g166.belle.evs}}%
\htdef{g167.belle.lumi}{854}%
\htdef{g167.belle.xsec}{0.919}%
\htdef{g167.belle.eff}{$2.38 \pm 0.14$}%
\htdef{g167.belle.bkg}{$0.72 \pm 0.18$}%
\htdef{g167.belle.evs}{0}%
\htdef{g167.belle.row}{\htuse{g167.belle.lumi} & \htuse{g167.belle.xsec} & \htuse{g167.belle.eff} & \htuse{g167.belle.bkg} & \htuse{g167.belle.evs}}%
\htdef{g178.babar.lumi}{472}%
\htdef{g178.babar.xsec}{0.919}%
\htdef{g178.babar.eff}{$8.60 \pm 0.20$}%
\htdef{g178.babar.bkg}{$0.12 \pm 0.02$}%
\htdef{g178.babar.evs}{0}%
\htdef{g178.babar.row}{\htuse{g178.babar.lumi} & \htuse{g178.babar.xsec} & \htuse{g178.babar.eff} & \htuse{g178.babar.bkg} & \htuse{g178.babar.evs}}%
\htdef{g179.babar.lumi}{472}%
\htdef{g179.babar.xsec}{0.919}%
\htdef{g179.babar.eff}{$6.40 \pm 0.40$}%
\htdef{g179.babar.bkg}{$0.54 \pm 0.14$}%
\htdef{g179.babar.evs}{0}%
\htdef{g179.babar.row}{\htuse{g179.babar.lumi} & \htuse{g179.babar.xsec} & \htuse{g179.babar.eff} & \htuse{g179.babar.bkg} & \htuse{g179.babar.evs}}%
\htdef{g180.babar.lumi}{472}%
\htdef{g180.babar.xsec}{0.919}%
\htdef{g180.babar.eff}{$10.20 \pm 0.60$}%
\htdef{g180.babar.bkg}{$0.03 \pm 0.02$}%
\htdef{g180.babar.evs}{0}%
\htdef{g180.babar.row}{\htuse{g180.babar.lumi} & \htuse{g180.babar.xsec} & \htuse{g180.babar.eff} & \htuse{g180.babar.bkg} & \htuse{g180.babar.evs}}%
\htdef{g181.babar.lumi}{472}%
\htdef{g181.babar.xsec}{0.919}%
\htdef{g181.babar.eff}{$8.80 \pm 0.50$}%
\htdef{g181.babar.bkg}{$0.64 \pm 0.19$}%
\htdef{g181.babar.evs}{0}%
\htdef{g181.babar.row}{\htuse{g181.babar.lumi} & \htuse{g181.babar.xsec} & \htuse{g181.babar.eff} & \htuse{g181.babar.bkg} & \htuse{g181.babar.evs}}%
\htdef{g182.babar.lumi}{472}%
\htdef{g182.babar.xsec}{0.919}%
\htdef{g182.babar.eff}{$12.70 \pm 0.70$}%
\htdef{g182.babar.bkg}{$0.34 \pm 0.12$}%
\htdef{g182.babar.evs}{0}%
\htdef{g182.babar.row}{\htuse{g182.babar.lumi} & \htuse{g182.babar.xsec} & \htuse{g182.babar.eff} & \htuse{g182.babar.bkg} & \htuse{g182.babar.evs}}%
\htdef{g156.belle.lumi}{535}%
\htdef{g156.belle.xsec}{0.919}%
\htdef{g156.belle.eff}{$3.00 \pm 0.10$}%
\htdef{g156.belle.bkg}{$5.14 \pm 3.30$}%
\htdef{g156.belle.evs}{5}%
\htdef{g156.belle.row}{\htuse{g156.belle.lumi} & \htuse{g156.belle.xsec} & \htuse{g156.belle.eff} & \htuse{g156.belle.bkg} & \htuse{g156.belle.evs}}%
\htdef{g183.babar.lumi}{472}%
\htdef{g183.babar.xsec}{0.919}%
\htdef{g183.babar.eff}{$6.60 \pm 0.60$}%
\htdef{g183.babar.bkg}{$0.44 \pm 0.17$}%
\htdef{g183.babar.evs}{0}%
\htdef{g183.babar.row}{\htuse{g183.babar.lumi} & \htuse{g183.babar.xsec} & \htuse{g183.babar.eff} & \htuse{g183.babar.bkg} & \htuse{g183.babar.evs}}%
\htdef{g178.belle.lumi}{782}%
\htdef{g178.belle.xsec}{0.919}%
\htdef{g178.belle.eff}{$6.00 \pm 0.59$}%
\htdef{g178.belle.bkg}{$0.21 \pm 0.15$}%
\htdef{g178.belle.evs}{0}%
\htdef{g178.belle.row}{\htuse{g178.belle.lumi} & \htuse{g178.belle.xsec} & \htuse{g178.belle.eff} & \htuse{g178.belle.bkg} & \htuse{g178.belle.evs}}%
\htdef{g179.belle.lumi}{782}%
\htdef{g179.belle.xsec}{0.919}%
\htdef{g179.belle.eff}{$6.10 \pm 0.58$}%
\htdef{g179.belle.bkg}{$0.10 \pm 0.04$}%
\htdef{g179.belle.evs}{0}%
\htdef{g179.belle.row}{\htuse{g179.belle.lumi} & \htuse{g179.belle.xsec} & \htuse{g179.belle.eff} & \htuse{g179.belle.bkg} & \htuse{g179.belle.evs}}%
\htdef{g180.belle.lumi}{782}%
\htdef{g180.belle.xsec}{0.919}%
\htdef{g180.belle.eff}{$10.10 \pm 0.77$}%
\htdef{g180.belle.bkg}{$0.02 \pm 0.02$}%
\htdef{g180.belle.evs}{0}%
\htdef{g180.belle.row}{\htuse{g180.belle.lumi} & \htuse{g180.belle.xsec} & \htuse{g180.belle.eff} & \htuse{g180.belle.bkg} & \htuse{g180.belle.evs}}%
\htdef{g181.belle.lumi}{782}%
\htdef{g181.belle.xsec}{0.919}%
\htdef{g181.belle.eff}{$9.30 \pm 0.73$}%
\htdef{g181.belle.bkg}{$0.04 \pm 0.04$}%
\htdef{g181.belle.evs}{0}%
\htdef{g181.belle.row}{\htuse{g181.belle.lumi} & \htuse{g181.belle.xsec} & \htuse{g181.belle.eff} & \htuse{g181.belle.bkg} & \htuse{g181.belle.evs}}%
\htdef{g182.belle.lumi}{782}%
\htdef{g182.belle.xsec}{0.919}%
\htdef{g182.belle.eff}{$11.50 \pm 0.89$}%
\htdef{g182.belle.bkg}{$0.01 \pm 0.01$}%
\htdef{g182.belle.evs}{0}%
\htdef{g182.belle.row}{\htuse{g182.belle.lumi} & \htuse{g182.belle.xsec} & \htuse{g182.belle.eff} & \htuse{g182.belle.bkg} & \htuse{g182.belle.evs}}%
\htdef{g183.belle.lumi}{782}%
\htdef{g183.belle.xsec}{0.919}%
\htdef{g183.belle.eff}{$7.60 \pm 0.56$}%
\htdef{g183.belle.bkg}{$0.13 \pm 0.20$}%
\htdef{g183.belle.evs}{0}%
\htdef{g183.belle.row}{\htuse{g183.belle.lumi} & \htuse{g183.belle.xsec} & \htuse{g183.belle.eff} & \htuse{g183.belle.bkg} & \htuse{g183.belle.evs}}%
\htdef{g211.babar.lumi}{237}%
\htdef{g211.babar.xsec}{0.919}%
\htdef{g211.babar.eff}{$12.20 \pm 8.50$}%
\htdef{g211.babar.bkg}{$0.56 \pm 0.56$}%
\htdef{g211.babar.evs}{0}%
\htdef{g211.babar.row}{\htuse{g211.babar.lumi} & \htuse{g211.babar.xsec} & \htuse{g211.babar.eff} & \htuse{g211.babar.bkg} & \htuse{g211.babar.evs}}%
\htdef{g212.babar.lumi}{237}%
\htdef{g212.babar.xsec}{0.919}%
\htdef{g212.babar.eff}{$12.28 \pm 8.50$}%
\htdef{g212.babar.bkg}{$0.42 \pm 0.42$}%
\htdef{g212.babar.evs}{0}%
\htdef{g212.babar.row}{\htuse{g212.babar.lumi} & \htuse{g212.babar.xsec} & \htuse{g212.babar.eff} & \htuse{g212.babar.bkg} & \htuse{g212.babar.evs}}%
\htdef{g213.babar.lumi}{237}%
\htdef{g213.babar.xsec}{0.919}%
\htdef{g213.babar.eff}{$9.47 \pm 0.66$}%
\htdef{g213.babar.bkg}{$0.12 \pm 0.12$}%
\htdef{g213.babar.evs}{1}%
\htdef{g213.babar.row}{\htuse{g213.babar.lumi} & \htuse{g213.babar.xsec} & \htuse{g213.babar.eff} & \htuse{g213.babar.bkg} & \htuse{g213.babar.evs}}%
\htdef{g157.belle.lumi}{535}%
\htdef{g157.belle.xsec}{0.919}%
\htdef{g157.belle.eff}{$5.07 \pm 0.20$}%
\htdef{g157.belle.bkg}{$13.90 \pm 5.00$}%
\htdef{g157.belle.evs}{10}%
\htdef{g157.belle.row}{\htuse{g157.belle.lumi} & \htuse{g157.belle.xsec} & \htuse{g157.belle.eff} & \htuse{g157.belle.bkg} & \htuse{g157.belle.evs}}%
\htdef{g214.babar.lumi}{237}%
\htdef{g214.babar.xsec}{0.919}%
\htdef{g214.babar.eff}{$10.63 \pm 0.74$}%
\htdef{g214.babar.bkg}{$0.26 \pm 0.26$}%
\htdef{g214.babar.evs}{0}%
\htdef{g214.babar.row}{\htuse{g214.babar.lumi} & \htuse{g214.babar.xsec} & \htuse{g214.babar.eff} & \htuse{g214.babar.bkg} & \htuse{g214.babar.evs}}%
\htdef{g211.belle.lumi}{906}%
\htdef{g211.belle.xsec}{0.919}%
\htdef{g211.belle.eff}{$4.39 \pm 0.36$}%
\htdef{g211.belle.bkg}{$0.31 \pm 0.18$}%
\htdef{g211.belle.evs}{0}%
\htdef{g211.belle.row}{\htuse{g211.belle.lumi} & \htuse{g211.belle.xsec} & \htuse{g211.belle.eff} & \htuse{g211.belle.bkg} & \htuse{g211.belle.evs}}%
\htdef{g212.belle.lumi}{906}%
\htdef{g212.belle.xsec}{0.919}%
\htdef{g212.belle.eff}{$4.80 \pm 0.39$}%
\htdef{g212.belle.bkg}{$0.21 \pm 0.15$}%
\htdef{g212.belle.evs}{0}%
\htdef{g212.belle.row}{\htuse{g212.belle.lumi} & \htuse{g212.belle.xsec} & \htuse{g212.belle.eff} & \htuse{g212.belle.bkg} & \htuse{g212.belle.evs}}%
\htdef{g213.belle.lumi}{906}%
\htdef{g213.belle.xsec}{0.919}%
\htdef{g213.belle.eff}{$3.16 \pm 0.27$}%
\htdef{g213.belle.bkg}{$0.42 \pm 0.19$}%
\htdef{g213.belle.evs}{0}%
\htdef{g213.belle.row}{\htuse{g213.belle.lumi} & \htuse{g213.belle.xsec} & \htuse{g213.belle.eff} & \htuse{g213.belle.bkg} & \htuse{g213.belle.evs}}%
\htdef{g214.belle.lumi}{906}%
\htdef{g214.belle.xsec}{0.919}%
\htdef{g214.belle.eff}{$4.11 \pm 0.35$}%
\htdef{g214.belle.bkg}{$0.31 \pm 0.14$}%
\htdef{g214.belle.evs}{0}%
\htdef{g214.belle.row}{\htuse{g214.belle.lumi} & \htuse{g214.belle.xsec} & \htuse{g214.belle.eff} & \htuse{g214.belle.bkg} & \htuse{g214.belle.evs}}%
\htdef{g160.babar.lumi}{469}%
\htdef{g160.babar.xsec}{0.919}%
\htdef{g160.babar.eff}{$9.10 \pm 1.73$}%
\htdef{g160.babar.bkg}{$0.59 \pm 0.25$}%
\htdef{g160.babar.evs}{1}%
\htdef{g160.babar.row}{\htuse{g160.babar.lumi} & \htuse{g160.babar.xsec} & \htuse{g160.babar.eff} & \htuse{g160.babar.bkg} & \htuse{g160.babar.evs}}%
\htdef{g161.babar.lumi}{469}%
\htdef{g161.babar.xsec}{0.919}%
\htdef{g161.babar.eff}{$6.14 \pm 0.20$}%
\htdef{g161.babar.bkg}{$0.30 \pm 0.18$}%
\htdef{g161.babar.evs}{1}%
\htdef{g161.babar.row}{\htuse{g161.babar.lumi} & \htuse{g161.babar.xsec} & \htuse{g161.babar.eff} & \htuse{g161.babar.bkg} & \htuse{g161.babar.evs}}%
\htdef{g160.belle.lumi}{671}%
\htdef{g160.belle.xsec}{0.919}%
\htdef{g160.belle.eff}{$10.20 \pm 0.67$}%
\htdef{g160.belle.bkg}{$0.18 \pm 0.18$}%
\htdef{g160.belle.evs}{0}%
\htdef{g160.belle.row}{\htuse{g160.belle.lumi} & \htuse{g160.belle.xsec} & \htuse{g160.belle.eff} & \htuse{g160.belle.bkg} & \htuse{g160.belle.evs}}%
\htdef{g161.belle.lumi}{671}%
\htdef{g161.belle.xsec}{0.919}%
\htdef{g161.belle.eff}{$10.70 \pm 0.73$}%
\htdef{g161.belle.bkg}{$0.35 \pm 0.21$}%
\htdef{g161.belle.evs}{0}%
\htdef{g161.belle.row}{\htuse{g161.belle.lumi} & \htuse{g161.belle.xsec} & \htuse{g161.belle.eff} & \htuse{g161.belle.bkg} & \htuse{g161.belle.evs}}%
\htdef{g164.babar.lumi}{451}%
\htdef{g164.babar.xsec}{0.919}%
\htdef{g164.babar.eff}{$7.31 \pm 0.20$}%
\htdef{g164.babar.bkg}{$1.32 \pm 0.17$}%
\htdef{g164.babar.evs}{1}%
\htdef{g164.babar.row}{\htuse{g164.babar.lumi} & \htuse{g164.babar.xsec} & \htuse{g164.babar.eff} & \htuse{g164.babar.bkg} & \htuse{g164.babar.evs}}%
\htdef{g165.babar.lumi}{451}%
\htdef{g165.babar.xsec}{0.919}%
\htdef{g165.babar.eff}{$4.52 \pm 0.40$}%
\htdef{g165.babar.bkg}{$2.04 \pm 0.19$}%
\htdef{g165.babar.evs}{0}%
\htdef{g165.babar.row}{\htuse{g165.babar.lumi} & \htuse{g165.babar.xsec} & \htuse{g165.babar.eff} & \htuse{g165.babar.bkg} & \htuse{g165.babar.evs}}%
\htdef{g168.babar.lumi}{451}%
\htdef{g168.babar.xsec}{0.919}%
\htdef{g168.babar.eff}{$8.00 \pm 0.20$}%
\htdef{g168.babar.bkg}{$1.65 \pm 0.23$}%
\htdef{g168.babar.evs}{2}%
\htdef{g168.babar.row}{\htuse{g168.babar.lumi} & \htuse{g168.babar.xsec} & \htuse{g168.babar.eff} & \htuse{g168.babar.bkg} & \htuse{g168.babar.evs}}%
\htdef{g169.babar.lumi}{451}%
\htdef{g169.babar.xsec}{0.919}%
\htdef{g169.babar.eff}{$4.60 \pm 0.40$}%
\htdef{g169.babar.bkg}{$1.79 \pm 0.21$}%
\htdef{g169.babar.evs}{4}%
\htdef{g169.babar.row}{\htuse{g169.babar.lumi} & \htuse{g169.babar.xsec} & \htuse{g169.babar.eff} & \htuse{g169.babar.bkg} & \htuse{g169.babar.evs}}%
\htdef{g170.babar.lumi}{451}%
\htdef{g170.babar.xsec}{0.919}%
\htdef{g170.babar.eff}{$7.80 \pm 0.20$}%
\htdef{g170.babar.bkg}{$2.76 \pm 0.28$}%
\htdef{g170.babar.evs}{2}%
\htdef{g170.babar.row}{\htuse{g170.babar.lumi} & \htuse{g170.babar.xsec} & \htuse{g170.babar.eff} & \htuse{g170.babar.bkg} & \htuse{g170.babar.evs}}%

\begin{center}\smaller
\begin{longtable}{lcl@{}rlrrrrrr}
\caption{Combinations of upper limits on lepton flavor violating \mtau decay
  modes. The table includes, for each experimental result, the published information that
  has been used to re-compute the limit with the \cls method, \ie\
  the integrated luminosity, the cross-section for \mtau lepton pairs production,
  the detection efficiency, the number of expected background events and the number
  of observed events. Since the \lhcb collaboration published a limit determined
  with the \cls method, in this case only the published limit is reported.
  The table finally reports the combined limit that is obtained by combining the
  re-computed limits. The modes are grouped according to the particle content of their final
  states. Modes with lepton number violation are labeled with ``(L)'',
  modes with baryon number violation are labeled with ``(BNV)''.
\label{tab:tau:lfv-upper-limits-comb}}%
\\
\toprule
\multicolumn{1}{l}{\bfseries Decay mode} &
\multicolumn{1}{c}{\bfseries Cat.} &
\multicolumn{2}{c}{\bfseries \begin{tabular}{@{}c@{}}90\% CL\\Limit\end{tabular}} &
\multicolumn{1}{c}{\bfseries Exp.} &
\multicolumn{1}{c}{\bfseries \begin{tabular}{@{}c@{}}{\cal L}\\($\text{fb}^{-1}$)\end{tabular}} &
\multicolumn{1}{c}{\bfseries \begin{tabular}{@{}c@{}}$\sigma_{\tau\tau}$\\(nb)\end{tabular}} &
\multicolumn{1}{c}{\bfseries \begin{tabular}{@{}c@{}}efficiency\\(\%)\end{tabular}} &
\multicolumn{1}{c}{\bfseries $N_{\text{bkg}}$} &
\multicolumn{1}{c}{\bfseries $N_{\text{obs}}$} \\
\midrule
\endfirsthead
\multicolumn{5}{c}{{\bfseries \tablename\ \thetable{} -- continued from previous page}} \\
\midrule
\multicolumn{1}{l}{\bfseries Decay mode} &
\multicolumn{1}{c}{\bfseries Cat.} &
\multicolumn{2}{c}{\bfseries \begin{tabular}{@{}c@{}}90\% CL\\Limit\end{tabular}} &
\multicolumn{1}{c}{\bfseries Exp.} &
\multicolumn{1}{c}{\bfseries \begin{tabular}{@{}c@{}}{\cal L}\\($\text{fb}^{-1}$)\end{tabular}} &
\multicolumn{1}{c}{\bfseries \begin{tabular}{@{}c@{}}$\sigma_{\tau\tau}$\\(nb)\end{tabular}} &
\multicolumn{1}{c}{\bfseries \begin{tabular}{@{}c@{}}efficiency\\(\%)\end{tabular}} &
\multicolumn{1}{c}{\bfseries $N_{\text{bkg}}$} &
\multicolumn{1}{c}{\bfseries $N_{\text{obs}}$} \\
\midrule
\endhead
%
%
\begin{ensuredisplaymath}
\Gamma_{156} =  {e^- \gamma} 
\end{ensuredisplaymath}
 &\(l\gamma\) & \( <\; \) &  \(5.4 \cdot 10^{-8}\)        & HFAG \\
 &            &&& Belle & \htuse{g156.belle.row} \\
 &            &&& \babar  \\ 
\begin{ensuredisplaymath}
\Gamma_{157} =  {\mu^- \gamma} 
\end{ensuredisplaymath}
 &            & \( <\; \) &  \(5.0 \cdot 10^{-8}\)        & HFAG \\
 &            &&& Belle & \htuse{g157.belle.row} \\
 &            &&& \babar & \htuse{g157.babar.row} \\ 
\midrule
%
\begin{ensuredisplaymath}
\Gamma_{160} =  {e^- K^0_S} 
\end{ensuredisplaymath}
 & \(lP^0 \)  & \( <\; \) & \(1.4 \cdot 10^{-8}\)         & HFAG  \\
 &            &&& Belle  & \htuse{g160.belle.row} \\
 &            &&& \babar   & \htuse{g160.babar.row} \\ 
\begin{ensuredisplaymath}
\Gamma_{161} =  {\mu^- K^0_S} 
\end{ensuredisplaymath}
 &            & \( <\; \) & \(1.5 \cdot 10^{-8}\)         & HFAG  \\
&            &&& Belle  & \htuse{g161.belle.row} \\
 &            &&& \babar   & \htuse{g161.babar.row} \\ 
\midrule
%
%
\begin{ensuredisplaymath}
\Gamma_{164} =  {e^- \rho^0} 
\end{ensuredisplaymath}
 &  \(l V^0\) & \( <\; \) & \(1.5 \cdot 10^{-8}\)         & HFAG  \\
 &            &&& Belle& \htuse{g164.belle.row} \\
 &            &&& \babar   & \htuse{g164.babar.row} \\ 
\begin{ensuredisplaymath}
\Gamma_{165} =  {\mu^- \rho^0} 
\end{ensuredisplaymath}
 &            & \( <\; \) & \(1.5 \cdot 10^{-8}\)         & HFAG  \\
 &            &&& Belle& \htuse{g165.belle.row} \\
 &            &&& \babar & \htuse{g165.babar.row} \\ 
\begin{ensuredisplaymath}
\Gamma_{168} =  {e^- K^*(892)^0} 
\end{ensuredisplaymath}
 &            & \( <\; \) & \(2.3 \cdot 10^{-8}\)         & HFAG \\
 &            &&& Belle & \htuse{g168.belle.row} \\
 &            &&& \babar   & \htuse{g168.babar.row} \\ 
\begin{ensuredisplaymath}
\Gamma_{169} =  {\mu^- K^*(892)^0} 
\end{ensuredisplaymath}
 &            & \( <\; \) & \(6.0 \cdot 10^{-8}\)         & HFAG \\
 &            &&& Belle & \htuse{g169.belle.row} \\
 &            &&& \babar   & \htuse{g169.babar.row} \\ 
\begin{ensuredisplaymath}
\Gamma_{170} =  {e^- \bar{K}^*(892)^0} 
\end{ensuredisplaymath}
 &            & \( <\; \) & \(2.2 \cdot 10^{-8}\)         & HFAG \\
 &            &&& Belle & \htuse{g170.belle.row} \\
 &            &&& \babar   & \htuse{g170.babar.row} \\ 
\begin{ensuredisplaymath}
\Gamma_{171} =  {\mu^- \bar{K}^*(892)^0} 
\end{ensuredisplaymath}
 &            & \( <\; \) & \(4.2 \cdot 10^{-8}\)         & HFAG  \\
 &            &&& Belle  & \htuse{g171.belle.row} \\
 &            &&& \babar & \htuse{g171.babar.row} \\ 

\begin{ensuredisplaymath}
\Gamma_{176} =  {e^- \phi} 
\end{ensuredisplaymath}
 &            & \( <\; \) & \(2.0 \cdot 10^{-8}\)         & HFAG \\
 &            &&& Belle & \htuse{g176.belle.row} \\
 &            &&& \babar   & \htuse{g176.babar.row} \\ 
\begin{ensuredisplaymath}
\Gamma_{177} =  {\mu^- \phi} 
\end{ensuredisplaymath}
 &            & \( <\; \) &\(6.8 \cdot 10^{-8}\)          & HFAG \\
 &            &&& Belle  & \htuse{g177.belle.row} \\
 &            &&& \babar   & \htuse{g177.babar.row} \\ 
\begin{ensuredisplaymath}
\Gamma_{166} =  {e^- \omega} 
\end{ensuredisplaymath}
 &            & \( <\; \) & \(3.3 \cdot 10^{-8}\)         & HFAG \\
 &            &&& Belle  & \htuse{g166.belle.row} \\
 &            &&& \babar    & \htuse{g166.babar.row} \\ 
\begin{ensuredisplaymath}
\Gamma_{167} =  {\mu^- \omega} 
\end{ensuredisplaymath}
 &            & \( <\; \) & \(4.0 \cdot 10^{-8}\)         & HFAG \\
 &            &&& Belle  & \htuse{g167.belle.row} \\
 &            &&& \babar   & \htuse{g167.babar.row} \\ 
\midrule
%
%
\begin{ensuredisplaymath}
\Gamma_{178} =  {e^- e^+ e^-} 
\end{ensuredisplaymath}
 &  \(lll\)   & \( <\; \) & \(1.4 \cdot 10^{-8}\)         & HFAG \\
 &            &&& Belle  & \htuse{g178.belle.row} \\
 &            &&& \babar    & \htuse{g178.babar.row} \\ 
\begin{ensuredisplaymath}
\Gamma_{181} =  {\mu^- e^+ e^-} 
\end{ensuredisplaymath}
 &            & \( <\; \) & \(1.1 \cdot 10^{-8}\)         & HFAG \\
 &            &&& Belle & \htuse{g181.belle.row} \\
 &            &&& \babar    & \htuse{g181.babar.row} \\ 
\begin{ensuredisplaymath}
\Gamma_{179} =  {e^- \mu^+ \mu^-} 
\end{ensuredisplaymath}
 &            & \( <\; \) & \(1.6 \cdot 10^{-8}\)         & HFAG \\
 &            &&& Belle  & \htuse{g179.belle.row} \\
 &            &&& \babar     & \htuse{g179.babar.row} \\ 
\begin{ensuredisplaymath}
\Gamma_{183} =  {\mu^- \mu^+ \mu^-} 
\end{ensuredisplaymath}
 &            & \( <\; \) & \(1.2 \cdot 10^{-8}\)         & HFAG \\
 &            &&& Belle& \htuse{g183.belle.row} \\
 &            &&& \babar  & \htuse{g183.babar.row} \\ 
 &            & \( <\; \) & \(4.6 \cdot 10^{-8}\)         & \lhcb   \\ 

\begin{ensuredisplaymath}
\Gamma_{182} =  {e^- \mu^+ e^-} 
\end{ensuredisplaymath}
 &            & \( <\; \) & \(8.4 \cdot 10^{-9}\)         & HFAG \\
 &            &&& Belle  & \htuse{g182.belle.row} \\
 &            &&& \babar    & \htuse{g182.babar.row} \\ 
\begin{ensuredisplaymath}
\Gamma_{180} =  {\mu^- e^+ \mu^-} 
\end{ensuredisplaymath}
 &            & \( <\; \) & \(9.8 \cdot 10^{-9}\)         & HFAG \\
 &            &&& Belle & \htuse{g180.belle.row} \\
 &            &&& \babar     & \htuse{g180.babar.row} \\ 
 %
%
\midrule
\begin{ensuredisplaymath}
\Gamma_{211} =  { \pi^- \Lambda } 
\end{ensuredisplaymath}
& BNV & \( <\; \) & \(1.9 \cdot 10^{-8}\)                 & HFAG  \\
&                &&& Belle  & \htuse{g211.belle.row} \\
 &               &&& \babar     & \htuse{g211.babar.row} \\  
\begin{ensuredisplaymath}
\Gamma_{212} =  { \pi^- \bar{\Lambda}} 
\end{ensuredisplaymath}
 &            & \( <\; \) & \(1.8 \cdot 10^{-9}\)         & HFAG \\
 &            &&& Belle  & \htuse{g212.belle.row} \\
 &            &&& \babar     & \htuse{g212.babar.row} \\  
\begin{ensuredisplaymath}
\Gamma_{213} =  { K^- \Lambda } 
\end{ensuredisplaymath}
 &            & \( <\; \) & \(3.7 \cdot 10^{-9}\)         & HFAG \\
 &            &&& Belle  & \htuse{g213.belle.row} \\
 &            &&& \babar     & \htuse{g213.babar.row} \\  
\begin{ensuredisplaymath}
\Gamma_{214} =  { K^- \bar{\Lambda}} 
\end{ensuredisplaymath}
 &            & \( <\; \) & \(2.0 \cdot 10^{-9}\)         & HFAG \\
 &            &&& Belle & \htuse{g214.belle.row} \\
 &            &&& \babar     & \htuse{g214.babar.row} \\  
\bottomrule
\end{longtable}
\end{center}

\ifhevea
\tausection{Combination of Upper Limits on \mtau LFV Branching Fractions: Summary Plot}
\cutname{lfv-combinations-plot.html}
\fi

\begin{figure}[tb]
  \begin{center}
    \ifhevea
    \begin{tabular}{@{}cc@{}}
      \larger\bfseries\ahref{TauLFV_UL_2014001_averaged.png}{full size PNG} &
      \larger\bfseries\ahref{TauLFV_UL_2014001_averaged.pdf}{PDF format} \\
      \multicolumn{2}{c}{\ahref{TauLFV_UL_2014001_averaged.png}{%
          \imgsrc[alt="Tau LFV limits combinations plot" width=720]{TauLFV_UL_2014001_averaged.png}}}
    \end{tabular}
    \else
    \includegraphics[angle=270,totalheight=0.86\textheight,clip]{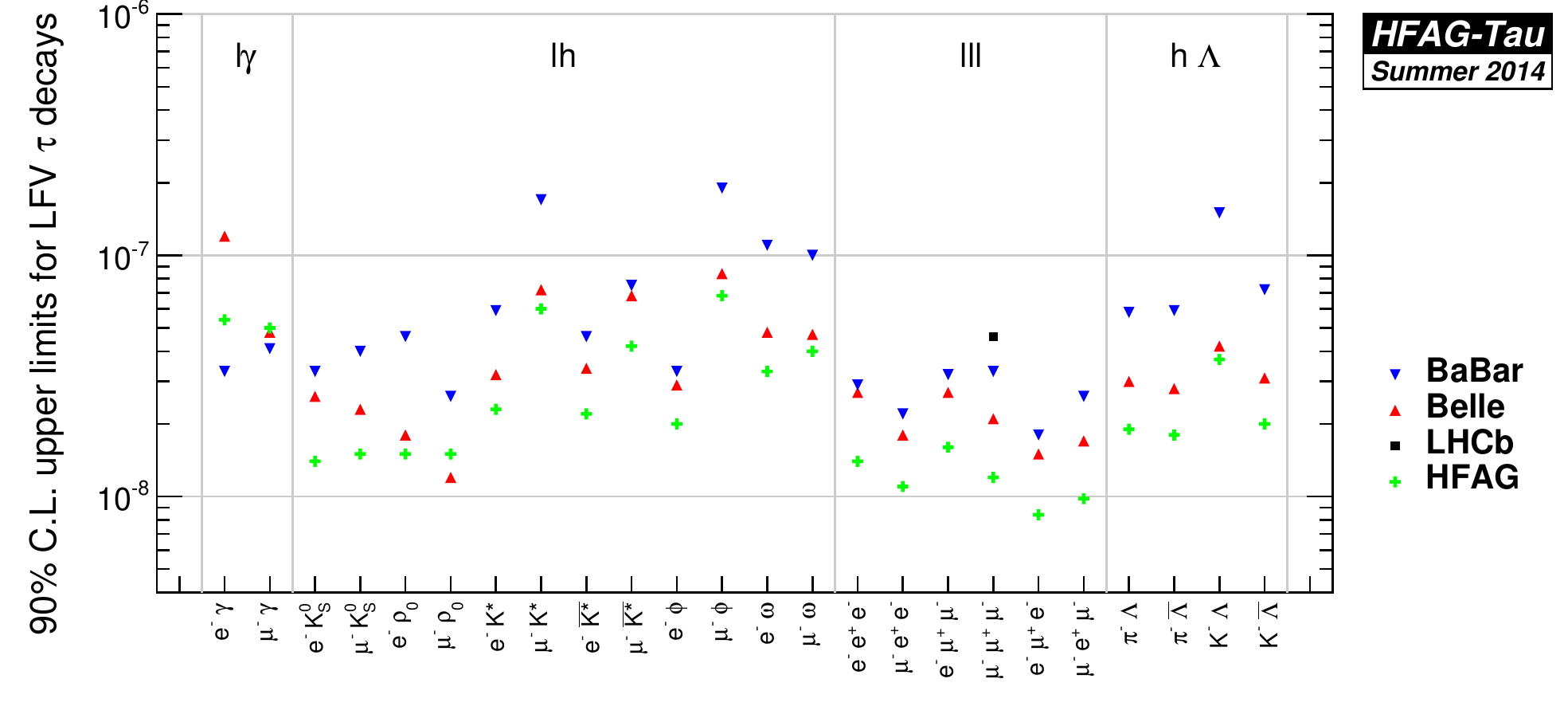}
    \fi
    \caption{Tau lepton-flavour-violating branching fraction upper limits
      combinations summary plot. For each channel we report the HFAG
      combined limit, and the experimental published limits. In some cases,
      the combined limit is weaker than the limit published by a single
      experiment. This arises since the \cls method used in the
      combination can be more conservative compared to other legitimate
      methods, especially when the number of observed events fluctuates below the
      expected background. 
      \label{fig:tau:lfv-limits-plot-average}
    }
  \end{center}
\end{figure}

\let\tausection\subsection
\let\cite\citeOld
\end{fleqn}



\clearpage
\clearpage

\section{Summary}
\label{sec:summary}

This article provides updated world averages for 
$b$-hadron properties using results available through summer 2014. 
A small selection of highlights of the results described in Sections
\ref{sec:life_mix}-\ref{sec:tau} is given in 
Table~\ref{tab_summary}.

\begin{longtable}{|l|c|}
\caption{
  Selected world averages.} 
\label{tab_summary}
\endfirsthead
\multicolumn{2}{c}{Selected world averages -- continued from previous page.}
\endhead
\endfoot
\endlastfoot
\hline
 {\bf\boldmath \b-hadron fractions} &   \\
 ~~$f^{+-}/f^{00}$ in \Ups decays  & \hfagFF \\ 
 ~~\fBs in \Upsfive decays & \hfagFSFIVE \\
 ~~\fBs, \fbb in $Z$ decays & \hfagZFBS, \hfagZFBB \\
 ~~\fBs, \fbb at Tevatron & \hfagTFBS, \hfagTFBB \\
\hline
 {\bf\boldmath \b-hadron lifetimes} &   \\
 ~~$\tau(\Bd)$         & \hfagTAUBD \\
 ~~$\tau(\Bu)$         & \hfagTAUBU \\
 ~~$\bar{\tau}(\Bs) = 1/\Gs$  & \hfagTAUBSMEANC \\
 ~~$\tau(\B_{s\rm L})$ & \hfagTAUBSLCON \\
 ~~$\tau(B_{s\rm H})$  & \hfagTAUBSHCON \\
 ~~$\tau(\Bc)$         & \hfagTAUBC \\
 ~~$\tau(\Lb)$         & \hfagTAULB \\
 ~~$\tau(\Xibd)$       & \hfagTAUXBD  \\
 ~~$\tau(\Xibu)$       & \hfagTAUXBU  \\
 ~~$\tau(\Omegab)$     & \hfagTAUOB   \\
\hline
 {\bf\boldmath \Bd\ and \Bs\ mixing / \CP violation parameters} &   \\
 ~~\dmd &  \hfagDMDWU \\
 ~~\DGGd  & \hfagSDGDGD \\
 ~~$|q/p|_{\particle{d}}$ & \hfagQPDB  \\
 ~~\dms  &  \hfagDMS \\
 ~~\DGs & \hfagDGSCON \\
 ~~$|q/p|_{\particle{s}}$ & \hfagQPS   \\
 ~~\phiccbars  & \hfagPHISCOMB \\
\hline
{\bf Parameters related to Unitarity Triangle angles} & \\
 ~~ $\stwob \equiv \sin\! 2\phi_1$ & $0.682 \pm 0.019$ \\
 ~~ $\beta \equiv \phi_1$          & $\left( 21.5 \,^{+0.8}_{-0.7} \right)^\circ$ \\
 ~~ $-\etacp S_{\phi \KS}$       & $0.74\,^{+0.11}_{-0.13}$ \\
 ~~ $-\etacp S_{\etapr \Kz}$       & $\phantom{-}0.63 \pm 0.06$ \\
 ~~ $-\etacp S_{\KS \KS \KS}$       & $\phantom{-}0.72 \pm 0.19$ \\
 ~~ $-\etacp S_{\Kp \Km \KS}$       & $0.68\,^{+0.09}_{-0.10}$ \\
~~  $\phi_s(\phi\phi)$              & $-0.17 \pm 0.15 \pm 0.03 \, {\rm rad}$ \\
 ~~ $-\etacp S_{\jpsi \piz}$       & $\phantom{-}0.93 \pm 0.15$ \\
 ~~ $S_{K^* \gamma}$       & $-0.16 \pm 0.22$ \\
 ~~ $S_{\pi^+\pi^-}$               & $-0.66 \pm 0.06$ \\  
 ~~ $C_{\pi^+\pi^-}$               & $-0.31 \pm 0.05$ \\  
 ~~ $S_{\rho^+\rho^-}$       & $-0.05 \pm 0.17$ \\
 ~~ $a(D^{*\pm}\pi^{\mp})$       & $-0.039 \pm 0.010$ \\
 ~~ $A^{}_{CP}(B\ra D^{}_{CP+}K)$       & $\phantom{-}0.19 \pm 0.03$ \\
 ~~ $A_{\rm ADS}(B\ra D^{}_{K\pi}K)$       & $-0.54 \pm 0.12$ \\
 ~~ $R_{\rm ADS}(B\ra D^{}_{K\pi}K)$       & $0.0153 \pm 0.0017$ \\
\hline
{\bf\boldmath Semileptonic \B decay parameters} & \\
 ~~${\cal B}(\Bzb\to D^{*+}\ell^-\nub)$ & $(4.93\pm 0.11)\%$\\
 ~~${\cal B}(\B^-\to D^{*0}\ell^-\nub)$ & $(5.69\pm 0.19)\%$\\
 ~~$\eta_{\rm EW}{\cal F}(1)\vcb$ & $(35.81\pm 0.45)\times 10^{-3}$\\
 ~~$\vcb$ from $\bar B\to D^*\ell^-\bar\nu_\ell$ & $(38.94\pm
 0.49_{\rm exp}\pm 0.58_{\rm th})\times 10^{-3}$\\
\hline
 ~~${\cal B}(\Bzb\to D^+\ell^-\nub)$ & $(2.19\pm 0.12)\%$\\
 ~~${\cal B}(\B^-\to D^0\ell^-\nub)$ & $(2.27\pm 0.11)\%$\\
 ~~$\eta_{\rm EW}{\cal G}(1)\vcb$ & $(42.65 \pm 1.53)\times 10^{-3}$\\
 ~~$\vcb$ from $\bar B\to D\ell^-\bar\nu_\ell$ & $(39.45\pm 1.42_{\rm
 exp}\pm 0.88_{\rm th})\times 10^{-3}$\\
\hline
 ~~${\cal B}(\bar B\to X_c\ell^-\bar\nu_\ell)$ & $(10.65\pm 0.16)\%$\\
 ~~${\cal B}(\bar B\to X\ell^-\bar\nu_\ell)$ & $(10.86\pm 0.16)\%$\\
 ~~$\vcb$ from $\bar B\to X\ell^-\bar\nu_\ell$ & $(42.46\pm
 0.88)\times 10^{-3}$\\
\hline
 ~~${\cal B}(\Bb\to\pi\ell^-\nub)$ & $(1.45\pm 0.05)\times 10^{-4}$\\
 ~~$\vub$ from $\Bb\to\pi\ell^-\nub$ & $(3.28\pm 0.29)\times
 10^{-3}$\\
 ~~$\vub$ from $\Bb\to X_u\ell^-\nub$ & $(4.45\pm 0.16_{\rm exp}\pm
 0.22_{\rm th})\times 10^{-3}$\\
\hline
{\bf\boldmath Rare \B decays} &   \\
 ~~ ${\cal B}(\Bs \to \mu^+\mu^-)$ & $2.8^{+0.07}_{-0.06} \times 10^{-9}$ \\
 ~~ ${\cal B}(\Bz \to \mu^+\mu^-)$ & $0.39^{+0.16}_{-0.14} \times 10^{-9}$ \\
 ~~ Different observables in $\Bz \to K^{*0}\mu^+\mu^-$ decays & \multirow{2}{*}{See Sec.~\ref{sec:rare-radll}} \\
 ~~ ~~~~ in bins of $q^2 = m^2(\mu^+\mu^-)$ & \\
 ~~ Up-down asymmetry in $\Bp\to\Kp\pim\pim\gamma$ decays & \multirow{2}{*}{See Sec.~\ref{sec:rare-radll}} \\
 ~~ ~~~~ in bins of $m(\Kp\pim\pim)$ & \\
 ~~ ${\cal B}(B \to X_s \gamma)$ & $(3.43 \pm 0.21 \pm 0.07) \times 10^{-4}$ \\
 ~~ ${\cal B}(\Bp \to \tau^+ \nu)$ & $(1.14 \pm 0.22) \times 10^{-4}$ \\
 ~~ $A_{\CP}(\particle{\Bd\to K^+\pi^-})$ & $(-0.082 \pm 0.006)$\\
 ~~ $A_{\CP}(\particle{B^+\to K^+\pi^0})$ & $(0.040 \pm 0.021)$ \\
 ~~ $A_{\CP}(\particle{\Bs\to K^-\pi^+})$ & $(0.26 \pm 0.04)$ \\
\hline
 {\bf\boldmath $D^0$ mixing and \CP violation parameters} &   \\
 ~~$x$ &  $(0.41\,^{+0.14}_{-0.15})\%$  \\
 ~~$y$ &  $(0.63\,^{+0.07}_{-0.08})\%$  \\
 ~~$A^{}_D$ &  $(-0.71\,^{+0.92}_{-0.95})\%$  \\
 ~~$|q/p|$ & $0.93\,^{+0.09}_{-0.08}$  \\
 ~~$\phi$ &  $(-8.7\,^{+8.7}_{-9.1})^\circ$  \\
\hline
 ~~$x^{}_{12}$ (no direct \CP violation) &  $(0.43\,^{+0.14}_{-0.15})\%$  \\
 ~~$y^{}_{12}$ (no direct \CP violation) &  $(0.60\,\pm 0.07)\%$  \\
 ~~$\phi^{}_{12}$ (no direct \CP violation) &  $(0.9\,^{+1.9}_{-1.7})^\circ$  \\
\hline
~~$a^{\rm ind}_{CP}$ & $(0.01 \pm 0.05)\%$ \\
~~$\Delta a^{\rm dir}_{CP}$ & $(-0.25 \pm 0.10)\%$ \\
\hline
 {\bf\boldmath Semileptonic/Leptonic $D$ decays} &   \\
 ~~$f^{}_D$     & $(203.7\,\pm 4.9)$~MeV  \\
 ~~$f^{}_{D_s}$  & $(257.4\,\pm 4.6)$~MeV  \\
 ~~$|V^{}_{cd}|$ & $0.219\,\pm 0.006$  \\
 ~~$|V^{}_{cs}|$ & $0.998\,\pm 0.020$  \\
\hline
{\bf\boldmath $\tau$ parameters, lepton universality, and $|V_{us}|$} &   \\
~~ $g^{}_\mu/g^{}_e$           & \htuse{gmubyge_tau} \\
~~ $g^{}_{\tau}/g^{}_{\mu}$    & \htuse{gtaubyge_tau} \\
~~ $g^{}_{\tau}/g^{}_{e}$      & \htuse{gtaubyge_tau} \\
~~ ${\cal B}_e^{\text{uni}}$   & \htuse{Be_univ}\% \\
~~ $R_{\text{had}}$            & \htuse{R_tau} \\
~~ $|V_{us}|$ from inclusive sum of strange branching fractions                                   & \htuse{Vus} \\
~~ $|V_{us}|$ from ${\cal{B}}(\tau^- \to K^-\nu^{}_\tau)/ {\cal{B}}(\tau^- \to \pi^-\nu^{}_\tau)$ & \htuse{Vus_tauKpi} \\ 
~~ $|V_{us}|$ from ${\cal{B}}(\tau^-\to K^-\nu^{}_\tau)$                                          & \htuse{Vus_tauKnu} \\
~~ $|V_{us}|$ \mtau average                                                                         & \htuse{Vus_tau} \\
\hline
\end{longtable}

The \b-hadron lifetime and mixing averages have made substantial 
progress in precision in the past two years, since the previous 
version of this writeup~\cite{Amhis:2012bh}. In total 60 new results 
(of which almost two thirds are from LHCb, with the rest from CDF,
\dzero, Belle, ATLAS, CMS and \babar)
have been incorporated in these averages.
Our knowledge of the lifetime of each individual weakly decaying \b-hadron species 
has improved by a factor 2--3, with the exception of the
\Xibd (factor 7 improvement), the \Xibu (first measurement), and 
the \b baryons with more than one heavy quark (no measurement yet). 
Impressive precisions of 0.3\% are achieved for the most common species, 
and the hierarchy
$\tau(\Bu)> \tau(B_{s\rm H})> \tau(\Bd) \sim \bar{\tau}(\Bs) > \tau(\Lb) > \tau(B_{s\rm L}) > \tau(\Bc)$
is established. 
Similarly the precision on the decay width differences in the \Bd and \Bs systems has improved by about a factor of two, confirming a sizable value of \DGs in agreement with the Standard Model (SM) prediction.
The results are not yet precise enough to distinguish the small (as expected) value of \DGd from zero.
As to the mass differences, only the uncertainty on \dms has substantially decreased (by a factor 2) 
reaching now the per mil level. Several new results on \CP violation in mixing have been 
published, including the final like-sign dimuon and inclusive muon analyses from \dzero (with a claimed 
$3.6\,\sigma$ overall deviation from the SM), but the averages of the individual \Bd and \Bs asymmetries 
remain well consistent both with zero and with their small SM predictions. 
Very impressive progress has been achieved in the measurement of
\CP violation induced by \Bs mixing in the $b\to c\bar{c}s$ transition, through the 
time-dependent angular analysis of tagged $\Bs\to \jpsi K^+K^-$ and $\Bs\to \jpsi \pi^+\pi^-$ decays,
as well as the time-dependent analysis of tagged  $\Bs\to D_s^+ D_s^-$ decays; 
despite the large improvement in sensitivity, the measured weak phase
$\phiccbars = \hfagPHISCOMB$ remains compatible with zero and with the SM prediction (\hfagPHISSM), 
but its experimental uncertainty (which has reduced by a factor 2.5) is now reaching the level of the SM central value.
Many measurements are still dominated by statistical uncertainties and will improve once new data from the LHC becomes available.

The measurement of $\sin 2\beta \equiv \sin 2\phi_1$ from $b \to
c\bar{c}s$ transitions such as $\Bz \to \jpsi\KS$ has reached $<3\,\%$
precision: $\sin 2\beta \equiv \sin 2\phi_1 = 0.682 \pm 0.019$.
Measurements of the same parameter using different quark-level processes
provide a consistency test of the Standard Model and allow insight into
possible new physics.  
All results among hadronic $b \to s$ penguin dominated decays of \Bz mesons are currently consistent with the Standard Model expectations.  
Recently, first measurements of \CP violation parameters in $\Bs \to \phi\phi$ have become available, allowing a similar comparison to the value of $\phiccbars$; again, results are consistent with the SM expectation (which in this case is very close to zero).
Among measurements related to the Unitarity Triangle angle $\alpha \equiv \phi_2$, results from the $\rho\rho$ system allow constraints at the level of $\approx
6^\circ$.  
These remain the strongest constraints, although recent results from all of \babar, \belle and LHCb have led to significantly improved precision of the \CP violation parameters in $\Bz \to \pip\pim$ decays.
Knowledge of the third angle $\gamma \equiv \phi_3$ also continues to improve.  
The world average values of the parameters in $B \to DK$ decays now show significant direct \CP\ violation effects, and determinations of $\gamma$ from the individual experiments now approach the level of $10^\circ$ precision.
The precision is expected to improve further as more data and more decays modes, for example time-dependent analysis of $\Bs \to D_s^{\mp}\Kpm$ decays and Dalitz plot analysis of $\Bz \to D\Kp\pim$ decays, are added.

Regarding semileptonic $B$-meson decays, only a few new experimental
results have appeared since the last update.
The changes in the numerical values are mainly due to improvements in the theoretical calculations: the form factor normalization in $B\to D^*\ell\nu$ is now known with 1.5\% precision, 
and calculations of inclusive decays $B\to X_c\ell\nu$ have reached NNLO precision. 
For both $|V_{cb}|$ and $|V_{ub}|$, the discrepancy between determinations from exclusive and inclusive decays is now at the level of about 3$\sigma$ -- a puzzle that must be addressed by the next generation of flavour experiments. 
Finally, the anomalous results obtained in $B\to D^{(*)}\tau\nu$ decays 
await experimental confirmation.

The most important new measurements of rare $b$ hadron decays are coming from the LHC. 
Precision measurements of $\Bs$ decays are particularly noteworthy. 
CMS and LHCb have both been constantly improving their restrictive
limits for the decays $B^0_{(s)}\to\mumu$. 
They recently published a combined analysis that allowed the first observation of the $\Bs\to\mumu$ decay to be obtained, and provided three standard deviations evidence of the $\Bz\to\mumu$ decay. 
These results are compatible with the SM predictions, and yield constrains on the parameter space of new physics models.
Atlas, CMS and LHCb have also performed angular analyses of the $\Bz\to\Kstarz\mumu$ decay, complementing and extending earlier results from \babar and Belle. 
One of the observables measured by LHCb, $P_5^{\prime}$, differs from the SM prediction by 3.7 standard deviations in one of the $m^2_{\mumu}$ intervals.
Updated measurements from LHCb, and results from other experiments on this observable, are keenly anticipated.
Among the $CP$ violating observables in rare decays, the ``$K\pi$ puzzle'' persists, and important new results have appeared in three-body decays.
LHCb has produced many other results on a wide variety of decays, including $b$ baryon decays. 
Belle and \babar\ continue to produce new results though their output rates are dwindling.  
It will still be some years before we see new results from the upgraded SuperKEKB $B$ factory and the Belle~II experiment.

Many $b$ to charm results from LHCb are included in our report for the  
first time this year, combining with results from \babar, Belle and CDF to yield a total of 632 measurements reported in 216 papers.
The huge combined sample of $b$ hadrons allows measurements of decays to states with open or hidden charm content with unprecedented precision.
Since our previous report, there is a dramatically increased number of results on decays of the $\Bc$ meson and of $b$ baryons.

In the charm sector, $\Dz$--$\Dzb$ mixing is now well-established.
Measurements of 45 separate observables from five experiments are input 
into a global fit for 10 underlying parameters, and the no-mixing 
hypothesis is excluded at a confidence level exceeding $11.5\sigma$. 
The mixing parameters $x$ and $y$ differ from zero by 
$2.4\sigma$ and $9.4\sigma$, respectively. The central values are 
consistent with mixing arising from long-distance processes, as
predicted by theory; thus it will be difficult to identify 
new physics from mixing alone. The WA value for the observable $y_{\CP}$ 
is positive, which indicates that the \CP-even state is shorter-lived 
as in the $\Kz$--$\Kzb$ system. However, $x$ also appears to be 
positive, which implies that the \CP-even state is heavier, 
unlike in the $\Kz$--$\Kzb$ system. 
%
In the $\Dz$--$\Dzb$ system, 
there is no evidence for \CP violation arising from mixing ($|q/p|\neq 1$) or 
from a phase difference between the mixing amplitude and 
a direct decay amplitude ($\phi\neq 0$). An initial evidence for direct
\CP violation in $D^0\ra K^+K^-$ and $D^0\ra \pi^+\pi^-$ decays is currently unconfirmed. 
Inputting the relevant measurements into a global 
fit gives $\Delta a^{\rm dir}_{CP}\neq 0$ with a significance 
of $2.4\sigma$.

The \mtau branching fraction fit has been updated using results from two
\babar papers and one from Belle, and updating the constraints on
the experimental branching fraction measurements.
The precise Belle measurement of the \mtau lifetime reduced the uncertainty on the PDG 2013 average by a factor two. 
We have updated the lepton universality tests, which give significantly improved precision thanks to the smaller uncertainty on the \mtau lifetime.
We have also updated the three determinations of \Vus using the \mtau HFAG averages; there is no significant change with respect to the HFAG 2012 report.
We updated the list of upper limits on \mtau Lepton-Flavour-Violating
decays and, for the first time, we combine the available experimental
information to compute world-average limits.

\section{Acknowledgments}

We are grateful for the strong support of the 
ATLAS, \babar, \belle, BES, CLEO, CDF, CMS, \dzero\ and LHCb collaborations,
without whom this compilation of results and world averages would not have  
been possible. The success of these experiments in turn would 
not have been possible without the excellent operations of the 
CESR, PEP-II, KEKB, Tevatron, BEPC and LHC accelerators, and fruitful 
collaborations between the accelerator groups and the experiments.
We also recognise the interplay between theoretical and experimental
communities that has provided a stimulus for many of the measurements in this
document.

Our averages and this compilation have benefitted greatly from 
contributions to the Heavy Flavor Averaging Group from numerous
individuals. We especially thank David Kirkby, Yoshihide Sakai, 
Simon Eidelman, Soeren Prell, and Gianluca Cavoto for their
past leadership of HFAG. 
We are grateful to Paolo Gambino and Vera L\"{u}th for assistance with
averages that appear in Chapter~\ref{sec:slbdecays};
to David Asner, David Cassel, Milind Purohit, and 
Patrick Roudeau for significant contributions to 
Chapter~\ref{sec:charm_physics};
and to Michel Davier for providing valuable input to Chapter~\ref{sec:tau}.
We also acknowledge the computing resources and support provided to HFAG by
SLAC.

\clearpage


\bibliographystyle{HFAGutphys}
\raggedright
\setlength{\parskip}{0pt}
\setlength{\itemsep}{0pt plus 0.3ex}
\begin{small}
\bibliography{Summer14,life_mix/life_mix,cp_uta,slbdecays/slb_ref,rare/rare,b2charm/b2charm,charm/charm_refs,tau/tau-refs,tau/tau-refs-pdg}

\begin{mcitethebibliography}{1000}
\bibitem{Aaltonen:2011qsa_cont}
 we consider that these results supersede those from the old CDF note
  7386~\cite{CDFnote7386:2005}, although the lifetime analysis of one of the
  modes ($B_s \to D_s\pi\pi\pi$) has not been updated\relax
\mciteBstWouldAddEndPuncttrue
\mciteSetBstMidEndSepPunct{\mcitedefaultmidpunct}
{\mcitedefaultendpunct}{\mcitedefaultseppunct}\relax
\EndOfBibitem
\end{mcitethebibliography}
\end{small}

\end{document}